\documentclass[useAMS,usenatbib]{mn2e}

\usepackage{latexsym}
\usepackage{afterpage}
\usepackage{longtable}
\usepackage{graphicx}
\usepackage{natbib}
\usepackage{psfrag}
\usepackage{rotating}
\usepackage{amssymb}

\def\degsq{$\deg^2$ }
\def\Hzero{$H_0$}
\def\OM{$\Omega_M$}
\def\OL{$\Omega_{\Lambda}$}

\def\SoneG{$S_{\rm 1.4GHz}$}

\def\SoneM{$S_{\rm 178MHz}$}

\def\AJ{{\it Astron. J.}}
\def\ApJ{{\it Astroph. J.}}
\def\ApJS{{\it Astroph. J. Sup.}}
\def\MNRAS{{\it Mon. Not. R. Astr. Soc.}}
\def\AA{{\it Astron. Astrophys.}}

\title[The CoNFIG Catalogue II.]{The CoNFIG Catalogue - II. Comparison of Space Densities in the FR Dichotomy.}

\author[Melanie A. Gendre, P. N. Best and
  J. V. Wall]{M. A. Gendre$^{1}$\thanks{E-mail: mgendre@phas.ubc.ca},
  P. N. Best$^{2}$ and J. V. Wall$^{1}$\\ 
$^{1}$Department of Physics and Astronomy, The University of British
  Colombia, 6224 Agricultural Rd, Vancouver, BC, V6T 1Z1, Canada\\
$^{2}$Institute for Astronomy, Royal Observatory, Blackford Hill,
Edinburgh EH9 3HJ, United Kingdom}

\begin{document}

\date{Accepted . Received ; in original form }

\pagerange{\pageref{firstpage}--\pageref{lastpage}} \pubyear{}

\maketitle

\label{firstpage}

%Abstract
\begin{abstract}
This paper focuses on a comparison of the space densities of FRI and
FRII sources at different epochs, with a particular focus on FRI
sources.\\
First, we present the concluding steps in constructing the Combined
NVSS-FIRST Galaxy catalogue (CoNFIG), including new VLA observations,
optical identifications and redshift estimates. The final catalogue
consists of 859 sources over 4 samples (CoNFIG-1, 2, 3 and 4 with flux
density limits of \SoneG~=1.3, 0.8, 0.2 and 0.05~Jy respectively). It is
95.7\% complete in radio morphology classification and 74.3\% of the
sources have redshift data.\\
Combining CoNFIG with complementary samples, the distribution and
evolution of FRI and FRII sources are investigated. We find that FRI
sources undergo mild evolution and that, at the same radio luminosity,
FRI and FRII sources show similar space density enhancements in
various redshift ranges, possibly implying a common evolution.
\end{abstract}

%Keywords
\begin{keywords}
Surveys - Radio Continuum: Galaxies - Galaxies: Active - Galaxies:
Statistics - Galaxies: Luminosity Function
\end{keywords}

%%%%%%%%%%%%%%%%%%%%%%%%%%%%%%%%%%%%%%%%%%%%%%%%%%%%%%%%%%%%%%%%%%%%%%%%%%%
\section{Introduction}

%Definition of the FR classes 
\indent
Radio AGN are classified in various ways such as luminosity, spectral
type or morphology. The Fanaroff-Riley (FR) classification
\citep{FR74} provides a classification of extended radio sources. The
FRI objects have the highest surface brightness along the jets and
core, reside in moderately rich cluster environments \citep{Hill91}
and include sources with irregular structure \citep{Parma92}. In
contrast, FRII sources show the highest surface brightness at the lobe
extremities, as well as more collimated jets, are found in more
isolated environments and generally display stronger emission lines
\citep{Rawlings89,Baum89}.\\
\indent
The FRI/FRII dichotomy is based purely on the appearance of the radio
objects and, although some hypotheses exist \citep[e.g.][]{Bick95},
the mechanisms differentiating the two populations are still
unknown. If sources with different FR classes undergo different
evolution, this might imply that their fundamental characteristics,
such as the black hole spin or jet composition, are different
too. However, the cut between FRI and FRII is somewhat ambiguous:
hybrid sources showing jets FRI-like on one side and FRII-like on the
other have been observed \citep{Capetti95}.\\

\indent
In an initial modeling of the space density of radio AGN,
\cite{Wall97} and \cite{Jack99} assumed that the cosmic evolution of
radio loud AGN was based on a division of the radio sources into a
high-luminosity component ($P_{\rm 178-MHz}>10^{25}\, {\rm W
  Hz^{-1}sr^{-1}}$) corresponding to FRIIs and a low-luminosity
component showing no cosmic evolution, corresponding to FRIs. With the
advent of large-scale redshift surveys for nearby galaxies, many
authors, including \cite{Snellen01}, \cite{Willott01}, \cite{Sadler02}
and \cite{Rigby08}, found significant evolution for low power sources
-- but mild evolution in comparison with that of the high-luminosity
sources. \cite{Rigby08} argued that if FRIs and FRIIs have similar
evolution, the dual-population scheme could be reduced to a
single-population model. Their sample was however confined to a small
number of low flux density sources.\\
\indent
A dedicated study and comparison of FRI and FRII sources and their
evolution, using large samples of sources of each type, is the key to
understanding these populations and determining if the FR
classification is valid or if a different classification, such as
whether they display high or low excitation emission lines, is
physically more relevant.

\medskip
%Feedback
A further motivation for studying the cosmic behaviour of radio AGNs
is to assess their contribution to feedback processes. The current
paradigm for galaxy formation, hierarchical build-up in a Cold Dark
Matter (CDM) universe, implies that the most massive galaxies in the
local Universe ought to be the largest and bluest and have the highest
star forming rate of all galaxies. Yet, observations show that they
are red, old galaxies, and the bulk of star formation is observed at
earlier epochs. This is known as downsizing, first described by
\cite{Cowie96}.\\
\indent
AGN negative feedback, in which the ignition of the nucleus in a
star-forming galaxy ejects the gas into the inter-galactic medium, is
a possible way to understand this phenomenon. AGNs jets could indeed
be responsible for reducing or even stopping star formation, breaking
the hierarchical buildup \citep{Silk98,Granato01,Quilis01}. AGN can also
have a positive feedback effect, whereby pressure from the jets
compresses the inter-stellar medium and induces star formation
\citep{vanB04,Klamer04}. However, modeling of jet power and its
relation to star formation have shown that the overall effect is a
decrease in star formation rate \citep{Anto08}.\\
\indent
If AGN feedback from the heating and ejection of gas in the ISM
possibly suppresses star formation, it is reasonable to think that it
should be related to the energy output from the jets. \cite{Best06}
studied the output energy from AGNs and concluded that heating
dissipated in the host galaxy is dominated by low-luminosity radio
sources, which tend to be confined predominantly to the size of the
galaxy and its halo. Such sources also stay `on' for a longer period
of time than high luminosity sources, allowing heat to be supplied
pseudo-continuously. \cite{Scha09} investigated the relation between
the amount of molecular gas and AGN activity in galaxies and concluded
that a low luminosity AGN episode was sufficient to suppress residual
star formation in early type galaxies.\\
\indent
Establishing the potential space-density behaviour of radio AGN is
thus important in studying the precise role of the feedback
mechanisms. Could feedback be linked to source morphology as it is to
luminosity? Do FRI sources have a higher impact on star formation rate
than typically more powerful FRII sources?

%Intro to CoNFIG
\medskip
\indent
The lack of a large comprehensive catalogue of
morphologically-classified radio sources has been a limiting factor in
all these studies. This is the goal of the CoNFIG catalogue, which we
propose to use in modeling the radio luminosity function of AGN.\\
\indent
This is the second paper in a series studying extended radio galaxies
and their role in AGN feedback. Paper I \citep{PaperI} outlined the
initial construction of the Combined NVSS-FIRST Galaxy (CoNFIG)
sample. This paper describes the complete catalogue, including optical
identifications and redshift estimates, as well as a preliminary study
of FRI and FRII space densities.\\
\indent
The structure of this paper is as follows. The construction of the
catalogue is explained in \S\ref{catdef} while \S\ref{morpho}
describes how the morphologies were determined. Optical
identifications and redshift information are discussed in
\S\ref{IDandz} and an overall summary of the catalogue is given in
\S\ref{CatSum}, along with the introduction of complementary datasets
that will be used in the modelling. \S\ref{Stat} describes the
morphology-dependent luminosity distributions and the FRI/FRII source
counts, as well as cosmic evolution of the radio luminosity
function. Finally, \S\ref{Sum} summarizes the results.\\
\indent
Throughout this paper, we assume a standard $\Lambda$CDM
cosmology with \Hzero=70 km s$^{-1}$ Mpc$^{-1}$, \OM=0.3 and \OL=0.7.

%%%%%%%%%%%%%%%%%%%%%%%%%%%%%%%%%%%%%%%%%%%%%%%%%%%%%%%%%%%%%%%%%%%%%%%%%%%
\section{The CoNFIG catalogue}\label{catdef}

\subsection{Catalogue definition}

\begin{figure}
    \centerline{
      \includegraphics[angle=270,scale=0.3]{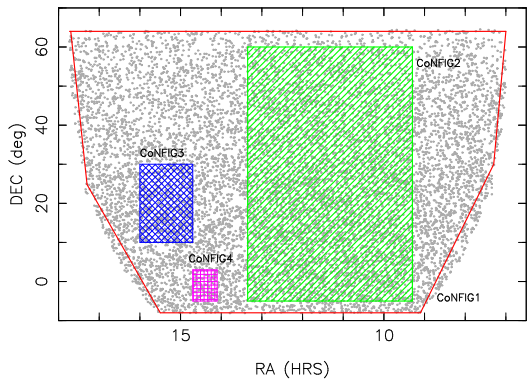}}
    \caption[Samples Regions]{\label{SampReg}Map of the sample
      regions. Each sample is located in the North field of FIRST
      (grey area). CoNFIG-1 (red contour), CoNFIG-2 (green hatched),
      CoNFIG-3 (blue diagonally cross-hatched) and CoNFIG-4 (pink
      vertically cross-hatched) have flux density limits of 1.3, 0.8,
      0.2 and 0.05~Jy respectively. Definition of the regions and
      details of the samples can be found in Tables~\ref{Reg} and
      \ref{Chartab}.}
\end{figure}

The CoNFIG Catalogue consists of 4 samples, CoNFIG-1, 2, 3 and 4,
which include all sources selected from the NVSS catalogue with
\SoneG~$\ge$1.3, 0.8, 0.2 and 0.05~Jy respectively in defined areas
(see Fig. \ref{SampReg} and Table~\ref{Reg}).\\
%NVSS
\indent
The NRAO-VLA Sky Survey \citep[NVSS;][]{Condon98} is a 1.4-GHz
continuum survey covering the entire sky north of $\delta=-40^{\circ}$
(corresponding to an area of 10.3~sr). The completeness
limit is $\sim$2.5~mJy/beam with an rms of $\sim$0.45~mJy/beam. The
catalogue from the survey contains over 1.8 million sources, implying
a surface density of $\sim$50 sources per square degree. It was
carried out with the Very Large Array (VLA) in D and DnC configuration
(the D configuration being the most compact VLA configuration with a
maximum antenna separation of $\sim$1 km), providing an angular
resolution of about 45 arcsec FWHM.\\
\indent
Since the median angular size of faint extra-galactic sources at the
CoNFIG flux density levels is $\lesssim$10 arcsec \citep{Condon98},
most sources in NVSS are unresolved, and the flux density measurements
are quite accurate. However, the large beam size does not reveal
precise structure of sources or determine positions accurate
enough to establish unambiguous optical counterparts.\\
\indent
Very large sources resolved in NVSS within the initial samples, such
as a few 3CRR sources \citep{Laing83}, need to be considered. In some
of these cases, two or more NVSS `sources' with \SoneG $> S_{lim}$ are
actually components of a much larger resolved source. Multi-component
sources in which each component has \SoneG $< S_{lim}$ but with a
total flux density \SoneG $\ge S_{lim}$, also need to be
considered. For this purpose, NVSS sources with \SoneG $< S_{lim}$ were
selected and, if any other source in the catalogue was located within
4 arcmin of the listed source, the combination was set aside as a
candidate extended source. The final decision on whether or not the
sources were actually components of a resolved source was made by
visual inspection of the NVSS contour plots.\\
\indent
A summary of each sample is given in Table~\ref{Chartab}. Because the area
of the CoNFIG-2, 3 and 4 samples overlap with CoNFIG-1, all statistics
estimated from CoNFIG 2, 3 and 4 use only sources with $S_{lim}
<$ \SoneG $<$ 1.3~Jy.

%------------------------------
%------------------------------

\subsection{Spectral indices}\label{specind}
In order to compute the radio luminosity, the spectral
index $\alpha$ (defined as $S_{\nu} \propto \nu^{\alpha}$) of
each source needs to be determined. To achieve this, flux densities at
different frequencies for each source were compiled and the spectral
index computed following the relation: 
\begin{eqnarray}
  \alpha=\frac{\Delta \log(S)}{\Delta \log(\nu)}
\end{eqnarray}
A summary of the different frequencies and corresponding surveys used
to retrieve the flux density information can be found in
\cite{PaperI}. We were able to compute the low-frequency spectral index
(with $178~{\rm MHz} \le \nu \le 1.4~{\rm GHz}$) for 99.6\%, 97.7\%,
89.3\% and 52.7\% of the sources in CoNFIG-1, 2, 3 and 4
respectively.

%------------------------------
\noindent
\begin{table}
  \caption[]{\label{Reg}Region corners for the CoNFIG samples (\{RA;
    DEC\} in \{HRS; deg\}) as shown in Fig.\ref{SampReg}.}
%  \centerline{
    \begin{tabular}{|cccc|}
      \hline
      \bf C-1 & \bf C-2 & \bf C-3 & {\bf C-4}\\
      \hline
      \{17.7; 64.0\} & \{\phantom{1}9.30; 60.0\} & \{14.7; 30.0\} & \{14.1; \phantom{$-$}3.0\} \\
      \{\phantom{1}7.0; 64.0\} & \{13.35; 60.0\} & \{16.0; 30.0\} & \{14.7; \phantom{$-$}3.0\} \\
      \{\phantom{1}7.3; 30.0\} & \{13.35; $-$5.0\} & \{16.0; 10.0\} & \{14.7; $-$3.5\} \\
      \{17.3; 24.8\} & \{\phantom{1}9.30; $-$5.0\} & \{14.7; 10.0\} & \{14.1; $-$3.5\} \\
      \{15.5; $-$8.0\} & & & \\
      \{\phantom{1}9.1; $-$8.0\} & & & \\
      \hline
  \end{tabular}%}
\end{table}
%------------------------------
%------------------------------
\begin{table}
  \caption[]{\label{Chartab}Characteristics of the CoNFIG samples, as
    described in section \ref{catdef}.}
  \centerline{
    \begin{tabular}{|ccccc|}
      \hline
      & $S_{lim}$ & Area & \# of & \# not \\
      & (Jy) & (\degsq) &  sources & in C1 \\
      \hline
      \bf C-1 & 1.30 & 4924 & 273 & - \\
      \bf C-2 & 0.80 & 2915 & 243 & 132 ({\it 54.3\%})\\
      \bf C-3 & 0.20 & \phantom{0}370 & 286 & 270 ({\it 94.4\%})\\
      \bf C-4 & 0.05 & \phantom{00}52 & 185 & 184 ({\it 99.4\%})\\
      \hline
  \end{tabular}}
\end{table}
%------------------------------
%%%%%%%%%%%%%%%%%%%%%%%%%%%%%%%%%%%%%%%%%%%%%%%%%%%%%%%%%%%%%%%%%%%%%%%%%%%
\section{Morphology}\label{morpho}

\subsection{Initial classification}\label{IniMorph}
\begin{figure}
  \begin{minipage}{8.0cm}
    \centerline{
     \includegraphics[angle=270,scale=0.2]{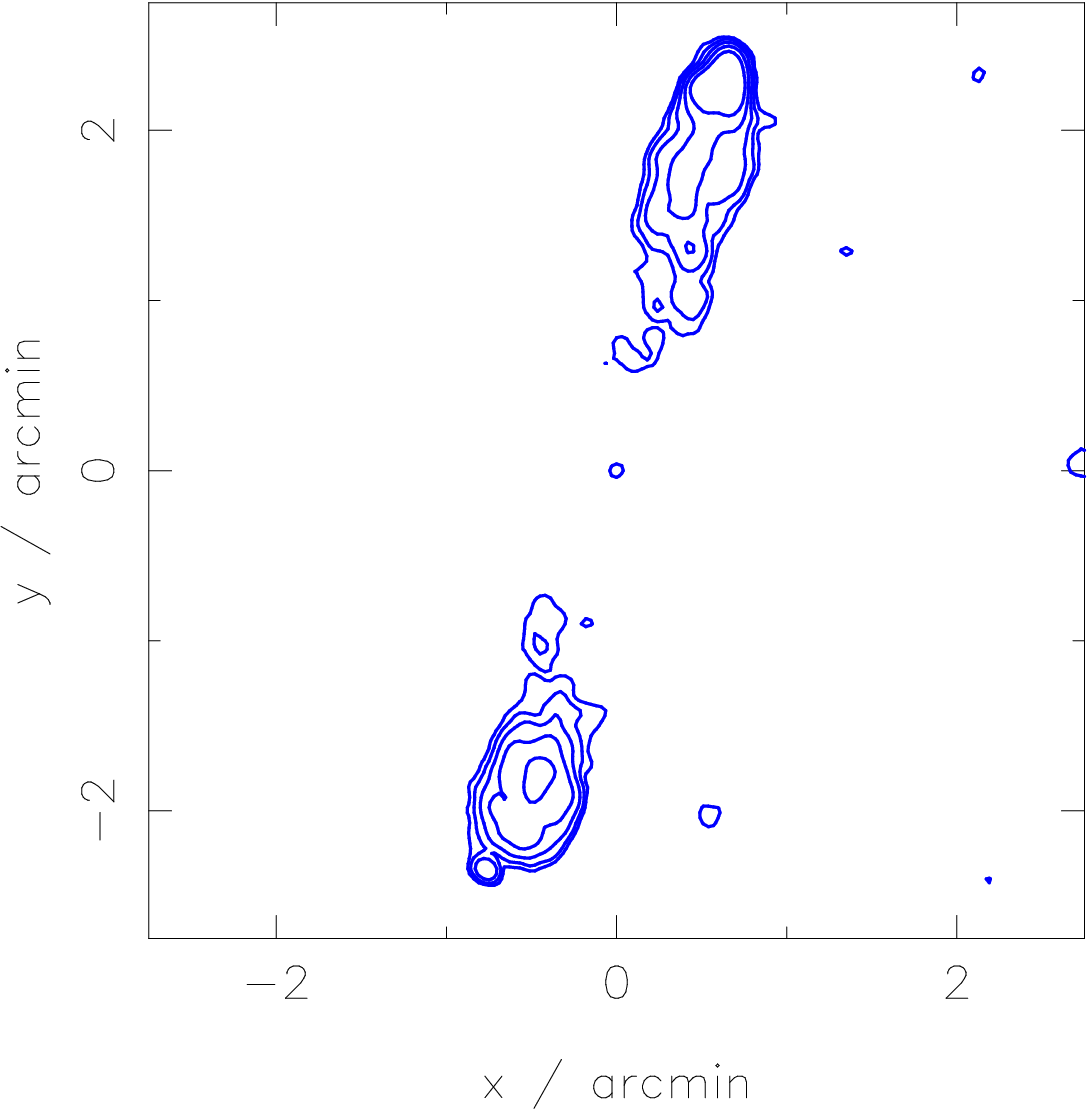}}
    \caption[]{\label{contour2}FIRST contour plot of a characteristic
      example of an FRII source, 3C~223. The hot spots are located at
      the ends of the aligned lobes.\\} 
  \end{minipage}
  \vfill
  \vfill
  \begin{minipage}{8.0cm}
    \centerline{
      \includegraphics[angle=270,scale=0.2]{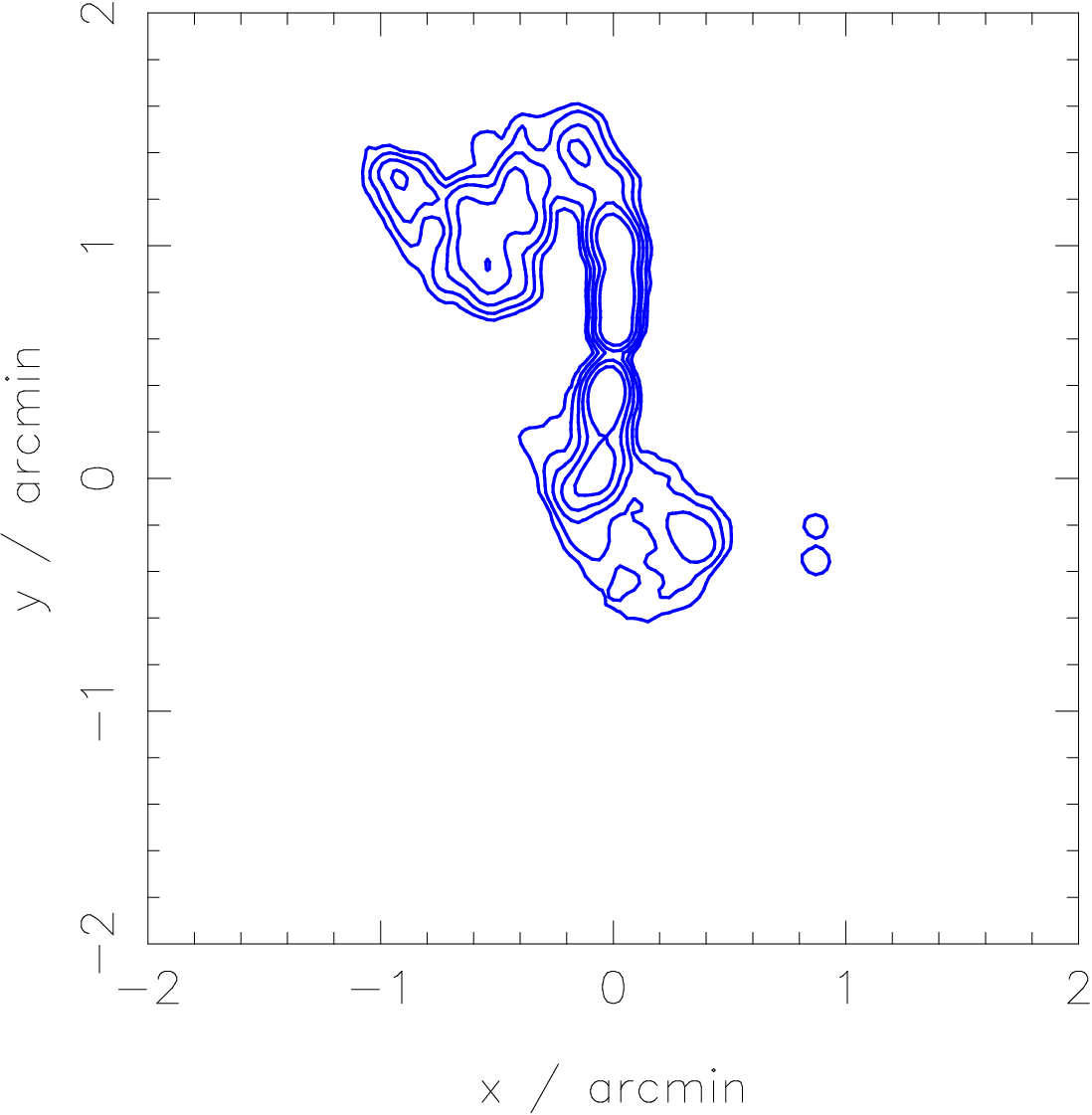}}
    \caption[]{\label{contour1}FIRST contour plot of a characteristic
      example of an FRI source, 3C~272.1 (M84). The regions of highest
      surface brightness are located along the jets.\\}
  \end{minipage}
\end{figure}

The initial morphologies were determined either from previously
referenced work or following the procedure described by \cite{PaperI},
primarily by looking at FIRST and NVSS contour plots.\\
%FIRST
\indent
The Faint Images of the Radio Sky at Twenty-cm survey
\citep[FIRST;][]{White97} is another 1.4-GHz continuum survey with the
VLA, covering an area of $\sim$9030~\degsq including the North Galactic
Pole. The completeness limit is $\sim$1~mJy/beam with a typical rms of
0.15~mJy/beam. The survey yielded $\sim$811,000 sources, implying a surface
density of $\sim$90 sources per square degree. It was carried out in B
configuration (the B configuration having a maximum antenna separation
of $\sim$10 km), which provides an angular resolution of about 5
arcsec FWHM. This survey complements the NVSS survey well, providing a
beam size small enough to resolve the structure of most nearby
extended radio sources and source positions to better than 1 arcsec to
enable cross-waveband identification.\\
\indent
If the FIRST/NVSS contour plot displays distinct hot spots at the edge of
the lobes (as in Figure~\ref{contour2}), and the lobes are aligned,
the source was classified as FRII. Sources with collimated jets
showing hot spots near the core and jets were classified as FRI (see
Figure~\ref{contour1}). Wide angle tail sources
\citep[WAT;][]{Leahy93} as well as most irregular-looking sources
\citep{Parma92} were also classified as FRI. Sources of size smaller
than 1 arcsec or previously classified as QSOs were classified as
`compact' while extended sources for which the FRI/FRII classification
was impossible to determine were classified as `uncertain'.

%------------------------------ 
%------------------------------ 

\subsection{VLA observations}\label{VLAObs}
%AG800
In addition to the observations described by \cite{PaperI}, radio
observations of 213 extended CoNFIG sources with previously uncertain
morphological classification were made at 1.4-GHz using the VLA in A
configuration. These observations included polarization measurements
for 31 sources as preliminary work for a possible study of the
morphology-dependent polarized source count.\\
\indent
The A-configuration, the most extended VLA configuration with a
maximum antenna separation of $\sim$36 km, provides a
synthesized beam of 1.4 arcsec FWHM at 1.4-GHz. Three frequency bands were used:
(1) two IFs of 1464.9 and 1385.1-MHz, with a bandwidth of 50-MHz (2)
two IFs of 1372.5 and 1422.5-MHz, with a bandwidth of 25-MHz and (3)
two IFs of 1425.5 and 1397.5-MHz, with a bandwidth of
25-MHz. Frequency bands 2 and 3 were used in the polarization
measurements.\\
\indent
The exposure time was computed for each source such as to provide a
signal-to-noise ratio of at least 5, and the exposures were split into
two or three separate integrations to improve \textit{uv} coverage.
The primary calibrator 3C286 (1331+305) was observed several times
during the run. Nearby secondary calibrators were observed
approximately every 30 min to provide phase calibration. All data
were reduced using standard procedures incorporated within the AIPS
software provided by NRAO. 

\subsection{Final classification}

62.5\% of sources in the CoNFIG sample were classified either as FRI
(I) or FRII (II). Following the unified model of AGN \citep{Jack99}
core-jet sources were classified as FRII. Hybrid sources, showing jets
FRI-like on one side and FRII-like on the other \citep{Capetti95},
were classified according to the characteristics of the most prominent
jet. Extended sources for which FRI/FRII identification was ambiguous
were classified as uncertain (U). Sources with size smaller than 3
arcsec were classified as compact (C) or (C*), depending on whether or
not the source was confirmed compact from the VLBA calibrator list
\citep[see ][]{Beasley02,Fomalont03,Petrov06,Kovalev07} or the
Pearson-Readhead survey \citep{Pearson88}. Finally, sources of type
(S*) correspond to confirmed compact sources which show a steep
($\alpha \le -0.6$) spectral index. These are probably compact
steep-spectrum (CSS) sources.\\
\indent
The final classification for each source is shown in
Appendix~\ref{datatables} and the distribution of morphological types
is presented in Table~\ref{Morphtab}. Contour plots of extended
sources, including the VLA observation presented in \S\ref{VLAObs},
are presented in Appendix~\ref{contour}. In order to study the
evolution of the space density of FRI and FRII sources accurately,
each extended source was assigned a sub-classification - confirmed (c)
or possible (p) - depending on how clearly the source showed either
FRI or FRII characteristics.\\
\indent
The complete catalogue consists of 859 sources, with 71 (8.3\%) FRIs
(50 confirmed, 21 possible), 466 (54.2\%) FRIIs (390 confirmed, 76
possible), 285 (33.2\%) compact sources and 37 (4.3\%) uncertain
sources.\\

%------------------------------
\begin{table}
  \caption[]{\label{Morphtab}Morphology of the sources in the CoNFIG
    samples. The morphology of each source was determined by looking at
    FIRST and NVSS contour plots or from VLA observations as described
    in \S\ref{IniMorph} and \S\ref{VLAObs}. Sources of size smaller
    than 3 arcsec or previously classified as QSOs were classified as
    `compact' (C) while extended sources for which the FRI/FRII
    classification was impossible to determine were classified as
    `uncertain' (U). In each cases, the corresponding percentage of
    sample is given in {\it italic}.}
  \medskip
  \medskip
  \centerline{
    \begin{tabular}{|cccccc|}
      \hline
      & \bf C-1 & \bf C-2 & \bf C-3 & \bf C-4 & {\bf Tot.}\\
      & \multicolumn{5}{c}{{\it \% of sample}}\\
      \hline
      FRI   & \phantom{0.}25  & \phantom{00.}7 & \phantom{0.}22  & \phantom{0.}17  &
      \phantom{0.}71 \\
            & {\it \phantom{0}9.2} & {\it \phantom{0}5.3} & {\it
        \phantom{0}8.1} & {\it \phantom{0}9.2} & {\it \phantom{0}8.3}\\
      FRII  & \phantom{.}149 & \phantom{.0}75 & \phantom{.}152 &
      \phantom{.0}90 &  \phantom{.}466\\
            & {\it 54.6} & {\it 56.8} & {\it 56.3} & {\it 48.9} & {\it 54.2}\\
      C     & \phantom{0.}86 & \phantom{.0}47 & \phantom{0.}88 &
      \phantom{.0}64 & \phantom{.}285\\
            & {\it 31.5} & {\it 35.6} & {\it 32.6} & {\it 34.8} & {\it 33.2}\\
      U     & \phantom{0.}13 & \phantom{00.}3  & \phantom{00.}8 &
      \phantom{.0}13 & \phantom{0.}37\\
            & {\it \phantom{0}4.8} & {\it \phantom{0}2.3} & {\it
        \phantom{0}3.0} & {\it \phantom{0}7.1} & {\it \phantom{0}4.3}\\
      \hline
  \end{tabular}}
\end{table}
%------------------------------

%%%%%%%%%%%%%%%%%%%%%%%%%%%%%%%%%%%%%%%%%%%%%%%%%%%%%%%%%%%%%%

\section{Optical identifications and redshifts}\label{IDandz}

A preliminary search for counterparts was performed using the unified
catalogue of radio objects of
\cite{Kimball08}\footnote{http://www.astro.washington.edu/akimball/radiocat/},
and optical identifications were obtained, principally from the Sloan
Digital Sky Survey \citep[SDSS; ][]{SDSS}.\\
\indent
%SDSS
The SDSS, with the 2.5 meter telescope at Apache Point Observatory,
New Mexico, has imaged one quarter of the entire sky in {\it ugriz}
magnitudes\footnote{The limiting magnitudes at the detection limit
  given in \cite{SDSSDR7} correspond to a 95\% detection repeatability
  for point sources. However, for galaxies, these are typically
  between half a magnitude and a magnitude brighter at the same signal
  to noise ratio (from SDSS project book:
  http://www.astro.princeton.edu/PBOOK/camera/camera.htm)}, as well as
performing a spectroscopic redshift survey. The seventh data release
\citep[DR7; ][]{SDSSDR7} imaging survey contains a total of 357
million objects over 11,663~\degsq while the spectroscopic survey
contains 1.6 million objects over 9380~\degsq.\\
\indent
$K_s$ band photometric information was  obtained from 2MASS. The Two
Micron All Sky Survey \citep[2MASS; ][]{2MASSref} is a near-infrared
survey using 1.3~m telescopes at Mount Hopkins in Arizona and CTIO in
Chile. It aimed at imaging the entire sky in {\it J}, {\it H} and {\it
  $K_S$} magnitudes. The now-complete catalogue, divided into a point
source and an extended source (semi-major axis $>$10'' in size)
catalogue, contains 472 million sources over 99.998\% of the sky.\\
\indent
A summary of the number of identified optical counterparts is given in
Table~\ref{SDSStab}.

%------------------------------
\begin{table}
  \caption[]{\label{SDSStab}Numbers of SDSS and 2MASS optical
    identifications for the CoNFIG samples. In each cases, the
    corresponding percentage of sample is given in {\it italic}.}
  \medskip
  \medskip
  \centerline{
    \begin{tabular}{|cccccc|c|}
      \hline
      & \multicolumn{5}{c}{\bf SDSS} & \bf 2MASS\\
      & All & FRI & FRII & C & U & All\\
      & \multicolumn{6}{c}{{\it \% of sample}}\\
      \hline
      \bf C-1 & \phantom{.}233 & \phantom{9.}25 & \phantom{.}125 &
      \phantom{0.}73 & \phantom{0.}10 & \phantom{.}117\\ 
       & {\it 85.3} &  {\it 100} &  {\it 83.9} &  {\it 84.9} & {\it
        76.9} &  {\it 42.9} \\ 
      \bf C-2 & \phantom{.}108 & \phantom{00.}6 & \phantom{0.}62 &
      \phantom{0.}37 & \phantom{00.}3 & \phantom{0.}44 \\  
              & {\it 81.8} &  {\it 85.7} &  {\it 82.7} &  {\it 78.7} &
              {\it 100} &  {\it 33.3} \\ 
      \bf C-3 & \phantom{.}190 & \phantom{.0}20 & \phantom{.}111 &
      \phantom{0.}53 & \phantom{00.}6 & \phantom{0.}47 \\  
              & {\it 70.4} &  {\it 90.9} &  {\it 73.0} &  {\it 60.2} &
              {\it 75.0} &  {\it 17.4} \\ 
      \bf C-4 & \phantom{.}110  & \phantom{.0}17 & \phantom{0.}52 &
      \phantom{0.}37 & \phantom{00.}4 & \phantom{0.}22 \\  
              & {\it 59.8} &  {\it \phantom{.}100} &  {\it 57.8} &
              {\it 57.8} &  {\it 30.8} &  {\it 12.0} \\ 
      &&&&&&\\
      \bf Tot. & \phantom{.}641 & \phantom{.0}68 & \phantom{.}350 &
      \phantom{.}200 & \phantom{.0}23 & \phantom{.}230\\ 
              & {\it 74.6} &  {\it 95.8} &  {\it 75.1} &  {\it 70.2} &
              {\it 62.2} &  {\it 26.8} \\ 
      \hline
  \end{tabular}}
\end{table}
%------------------------------

\subsection{Spectroscopic and photometric redshifts}

%------------------------------ 
\begin{figure*}
    \centerline{
      \includegraphics[angle=270,scale=0.7]{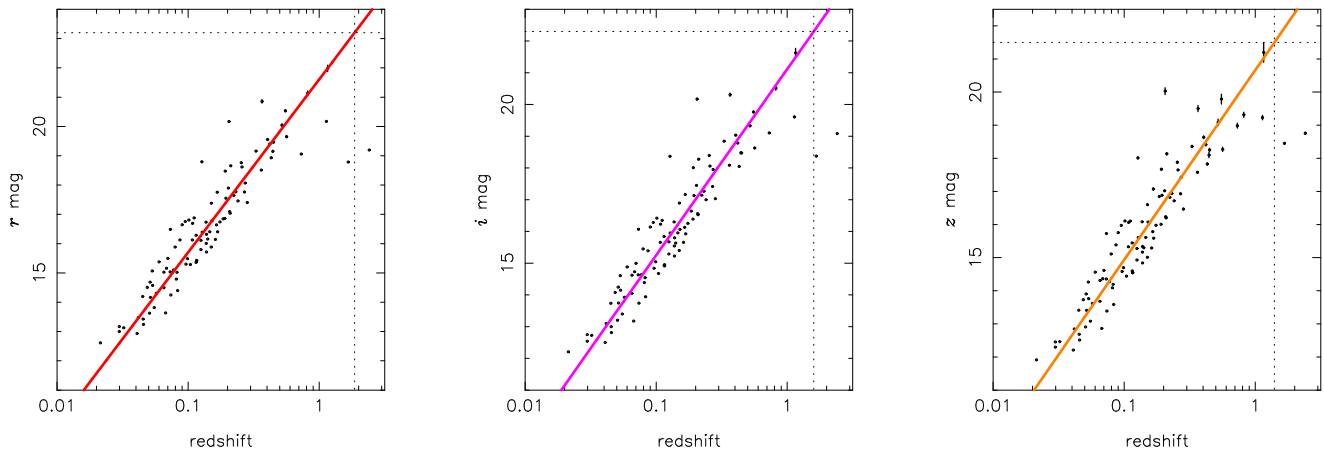}}
    \caption[]{\label{IZ}The SDSS magnitude-redshift relations were
      computed by finding the best fit (solid lines) to data
      from CoNFIG non-QSO sources having both spectroscopic redshift and
      SDSS magnitude information (dots). The relations
      (Equ.\ref{Izequ}-\ref{Rzequ}) were used to estimate photometric
      redshifts for sources not in the {\it photoz2} catalogue, but
      with an SDSS counterpart.\\}
\end{figure*}
%------------------------------ 

\begin{figure}
    \centerline{
      \includegraphics[angle=270,scale=0.35]{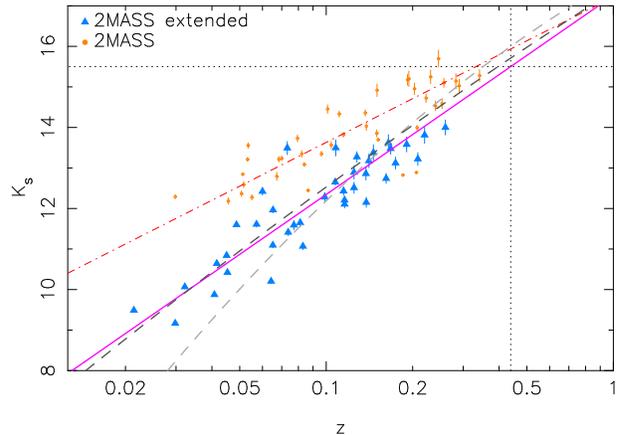}}
    \caption[]{\label{KsZ}The $K_S$-z relation was computed by finding
      the best fit (solid pink and dot-dashed red lines respectively) to data
      from CoNFIG non-QSO sources having both spectroscopic redshift
      and $K_S$-band information from the 2MASS extended (blue triangles)
      and point source (orange dots) catalogues. The relations
      (Equ.\ref{Kzequ}-\ref{Kzequ2}) were used to estimate photometric
      redshifts for sources with a magnitude $K_S \le 15.5$, which
      corresponds to an upper estimated redshift limit of z=0.43
      (dotted lines) from the extended source relation. For
      comparison, the K-z relations from CENSORS \citep{Brookes06} and
      \citep{Willott03} are shown in light and dark grey dashed lines
      respectively.}
\end{figure}

Spectroscopic redshifts were obtained for 45.5\% of the catalogue (see
Table~\ref{redtab}) using either the
SIMBAD\footnote{http://simbad.u-strasbg.fr/simbad} database or the
SDSS DR7 catalogue.\\ 
\indent
Because redshift information is essential to computing space
densities and examining their evolution, we estimated redshifts for
sources with no spectroscopic data available.
\indent
For a number of sources with an SDSS counterpart identified but with
no spectroscopic information available, photometric redshifts
were retrieved from the SDSS {\it photoz2} catalogue \citep{Oya08},
which covers SDSS galaxies with {\it r}$\le$22.0. For other galaxies
(excluding the 285 sources identified as `compact', which are most
likely QSOs), redshifts were estimated using a magnitude-redshift
relationship computed from SDSS-identified CoNFIG non-compact
(i.e. non-QSO) sources with spectroscopic redshifts:
\begin{eqnarray}\label{Izequ}
log({\rm z})= -3.599 + 0.170 {\it i}\\
log({\rm z})= -3.609 + 0.175 {\it z}\label{Zzequ}\\
log({\rm z})= -3.660 + 0.169 {\it r}\label{Rzequ}
\end{eqnarray}
The relations are shown in Fig.~\ref{IZ} and were used to estimate
photometric redshifts for 73 sources.\\
\indent
$K_S$-z relations were also obtained using data from CoNFIG
non-compact sources having both spectroscopic redshifts and $K_S$-band
information from the 2MASS extended and point source catalogues.
\begin{eqnarray}\label{Kzequ}
log({\rm z})= -3.515 + 0.204 K_S \ \ \ {\rm 2MASS \ extended \ sources}\\
log({\rm z})= -4.800 + 0.279 K_S \ \ \ \ \ \ {\rm \ \ 2MASS \ point \ sources}\label{Kzequ2}
\end{eqnarray}
The relations, shown in Fig.~\ref{KsZ}, provide good estimates of
redshifts up to $K_S=15.5$. They were used to estimate photometric
redshifts for 6 sources which had no SDSS spectroscopic or photometric
redshifts available but had 2MASS counterparts ($K_S \le 15.5 $).\\

\begin{figure*}
    \centerline{
      \includegraphics[scale=0.8]{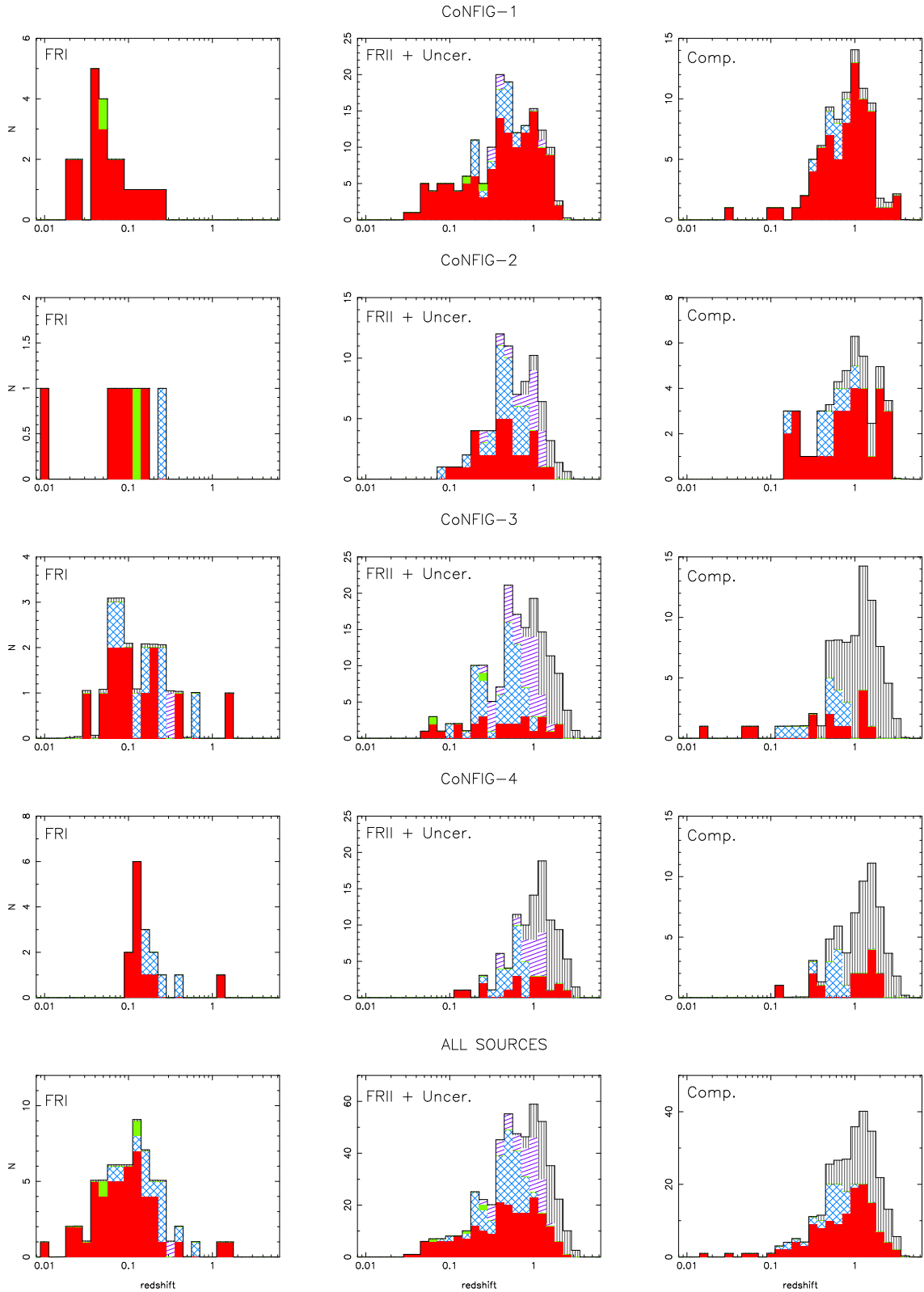}}
    \caption[]{\label{Zdist}Redshift distribution of the sources in
      the CoNFIG catalogue for each morphological type. Sources with
      spectroscopic, photometric {\it photoz2}, $K_S$-z estimated and
      SDSS {\it mag}-z estimated redshifts are represented by the red
      solid, blue cross-hatched, green solid and purple diagonally
      hatched columns. The estimated contribution from sources with no
      redshift information available (\S\ref{noZest}) is shown in
      black vertically hatched columns.}
\end{figure*}

\indent
Overall, 74.3\% of the sources in the CoNFIG catalogue have
spectroscopic or photometric redshift information available, with mean
and median redshifts of ${\rm z}_{mean}$~=~0.714 and ${\rm
  z}_{med}$~=~0.588. The redshift distributions, by samples and
morphological types, is shown in Fig.~\ref{Zdist}.

%------------------------------
\begin{table}
  \caption[]{\label{redtab}Distribution and ranges of
    redshifts for the CoNFIG samples. Spectroscopic redshifts are
    retrieved either from the SIMBAD database or from the SDSS
    catalogue. Photometric redshifts are either obtained from the SDSS
    {\it photoz2} catalogue or estimated using either the SDSS {\it mag}-z relation
    defined by Equ.~\ref{Izequ}-\ref{Rzequ} or the $K_S$-z relation defined by
    Equ.~\ref{Kzequ}-\ref{Kzequ2}. In each cases, the corresponding
    percentage of sample is given in {\it italic}.}
  \medskip
  \medskip
  \centerline{
    \begin{tabular}{|llcccc|}
      \hline
      & & \bf C-1 & \bf C-2 & \bf C-3 & {\bf C-4}\\
      \hline
      \hline
      \multicolumn{6}{l}{\bf Total Number of Sources in Sample} \\
       & & 273 &132 & 270 & 184\\
      \hline
      \hline
      \multicolumn{6}{l}{\bf Redshift types} \\
      \multicolumn{6}{l}{\it \% of sample} \\
      \hline
      \hline
      Spectro. & & \phantom{.}226 & \phantom{0.}67 & \phantom{0.}54 & \phantom{0.}44\\
      &          & {\it 82.8} & {\it 58.8 } & {\it 20.0} & {\it 23.9}\\
      \hline
      Photo.   & {\it photoz2} & \phantom{0.}29 & \phantom{0.}33 & \phantom{0.}71 & \phantom{0.}35\\ 
      && {\it 10.6} & {\it 25.0} & {\it 26.3} & {\it 19.0}\\
               & sdss {\it mag}-z& \phantom{00.}5 & \phantom{0.}13 & \phantom{0.}38 & \phantom{0.}17\\
      && {\it \phantom{0}1.8} & {\it \phantom{0}5.3} & {\it 13.3} & {\it \phantom{0}9.2}\\
               & $K_S$-z & \phantom{00.}3 & \phantom{00.}1 & \phantom{00.}2 & \phantom{00.}0 \\
      && {\it \phantom{0}1.1} & {\it \phantom{0}0.8} & {\it \phantom{0}0.7} & {\it \phantom{0}0.0}\\
      \hline
      Total    &       & \phantom{.}263 & \phantom{.}114 & \phantom{.}165 & \phantom{0.}96 \\
      && {\it 96.3} & {\it 86.4} & {\it 61.1} & {\it 52.2}\\
               & FRI   & \phantom{0}25 & \phantom{00}7 & \phantom{0.}21 & \phantom{0.}17 \\
     && {\it 100} & {\it 100} & {\it 95.4} & {\it 100}\\
               & FRII  & \phantom{.}145 & \phantom{0.}65 & \phantom{.}112 & \phantom{0.}52 \\
     && {\it 97.3} & {\it 86.7} & {\it 73.7} & {\it 57.8}\\
               & C & \phantom{0.}80 & \phantom{0.}39 & \phantom{0.}26 & \phantom{0.}23 \\
     && {\it 93.0} & {\it 83.0} & {\it 29.5} & {\it 35.9}\\
               & U & \phantom{0.}13 & \phantom{00.}3 & \phantom{00.}6 & \phantom{00.}4 \\
     && {\it 100} & {\it 100} & {\it 75.0} & {\it 30.8}\\
      \hline
      \hline
      \multicolumn{2}{l}{\bf Redshift ranges} &\bf C-1 & \bf C-2 & \bf C-3 & {\bf C-4}\\
      \hline
      \hline
             & min. & 0.003 & 0.011 & 0.018 & 0.006 \\
      All    & max. & 3.530 & 2.707 & 2.408 & 2.677 \\
             & mean & 0.711 & 0.760 & 0.623 & 0.828 \\
             & med. & 0.555 & 0.599 & 0.564 & 0.695 \\
      \hline                
             & min. & 0.003 & 0.011 & 0.032 & 0.006 \\
      FRI    & max. & 0.269 & 0.309 & 1.847 & 1.531 \\
             & mean & 0.071 & 0.128 & 0.264 & 0.261 \\
             & med. & 0.049 & 0.099 & 0.116 & 0.150 \\
      \hline                
             & min. & 0.036 & 0.098 & 0.062 & 0.138 \\
      FRII   & max. & 2.183 & 1.711 & 2.408 & 2.677 \\
             & mean & 0.637 & 0.660 & 0.674 & 0.938 \\
             & med. & 0.523 & 0.566 & 0.604 & 0.800 \\
      \hline                
             & min. & 0.034 & 0.160 & 0.018 & 0.133 \\
      C      & max. & 3.530 & 2.707 & 1.764 & 2.235 \\
             & mean & 1.024 & 1.050 & 0.665 & 1.026 \\
             & med. & 0.880 & 0.795 & 0.580 & 0.725 \\
      \hline
  \end{tabular}}
\end{table}
%------------------------------
%------------------------------ 
\subsection{Sources with no redshift information}\label{noZest}

A total of 221 sources in the CoNFIG catalogue, mostly in CoNFIG-3 and
4, have no redshift information available. 104 of these sources are of
morphological type I, II or U (we will ignore sources of type C for
the time being, being only interested in the study of extended radio
sources). One way to include these sources in the space density
modelling is to assign an estimated redshift to each of them (by the
procedures described below), compute the RLF, repeat the procedure and
average the results.\\

\indent
Based on SDSS non-detection we can determine a lower redshift limit
for these sources. The {\it i}-band being effectively the deepest SDSS
band for objects with the typical colours of high-redshift radio
galaxies, Equ.\ref{Izequ} was used to determine the lower limit,
yielding a value of ${\rm z}_{lim}$~$\simeq$~1.0. To account for the
spread in the {\it i}-z relation, the estimate of the limit was drawn
randomly from a Gaussian of variance 0.1, centered on ${\rm
  z}_{lim}$=1.0.\\
\indent
Our next step was to use the (admittedly naive) assumption that the
redshift of the radio source could be estimated from the
distribution of measured or estimated redshifts for sources of similar
flux density. For each of the 113 sources, we derived the sample of
sources with redshift information available and flux densities within
the range of a tenth to ten times the flux density of the source with
no redshift. The redshift distribution of this sample was computed and
fit with a polynomial; the region of this polynomial above the
calculated redshift limit was then normalized to determine the
redshift probability distribution for the source.\\ 
\indent
To complete the catalogue redshift distributions, we determined that
each source with no redshift will contribute a fraction to each
redshift bin, following its assigned probability distribution. For
space densities computation (see \S\ref{Stat}), approximated redshifts
were assigned to each source by making random realizations following
the probability distribution, repeating the process in a Monte-Carlo
manner.\\

Because most of the approximate redshifts are greater than z=0.3, the
redshift upper-limit used to define the local universe, the results of
the local radio luminosity function (LRLF) are completely unaffected
by redshift uncertainties. As the redshift lower-limits used in the
computation of the approximate redshifts are mostly z$\ge$1.0, results
out to z$\sim$1.0 are also not significantly affected. Over the range
1.0$\le$z$\le$2.0, the results are likely to be
impacted. Nevertheless, the fact that the redshift distribution is
well determined over that range implies that the impact is perhaps not
severe. Beyond z=2.0, results would be unreliable as the redshift
distribution is not well determined and the use of approximate
redshifts may have introduced significant biases.\\

\section{Catalogue Summary and Complementary Samples}\label{CatSum}

The CoNFIG catalogue (Appendix~\ref{ConfSamp}) consists of 859 sources
over 4 samples, CoNFIG-1, 2, 3 and 4 with flux density limits
\SoneG=1.3, 0.8, 0.2 and 0.05~Jy respectively. Spectral indices were
computed for 86.0\% of the sources using flux densities at different
frequencies for each source. The catalogue is 95.7\% complete for
radio morphologies and 74.3\% complete for redshift information.\\
\indent
Sources were morphologically classified into 6 categories, using NVSS,
FIRST and VLA 1.4GHz A-configuration observation contour plots as well
as previously referenced information. Sources of type I and II
correspond to \cite{FR74} morphologies; extended sources for which
FRI/FRII identification was uncertain were classified as type U;
sources with size smaller than 1 arcsec were classified as C type or
C*-type, depending on whether or not the source was confirmed compact;
sources of S* type correspond to confirmed compact sources which show
a steep spectral index.\\
\indent
Optical counterparts were obtained from the SDSS and 2MASS catalogues
for 74.6\% and 26.8\% of the sources respectively. Spectroscopic
redshift information was retrieved from SDSS and the SIMBAD database,
while photometric redshifts (or redshift estimates) were compiled from
the SDSS {\it photoz2} catalogue, or using the $K_S$-z or SDSS {\it
  mag}-z relations (Equ.\ref{Izequ}-\ref{Kzequ2}).\\

To improve the flux density coverage of the catalogue, three
complementary samples were appended (Appendix \ref{3CR},
\ref{CENSORSsamp} and \ref{EmmaSamp}): 
\begin{itemize} 
\item The 3CRR (Third Cambridge Revised) catalogue \citep{Laing83} is
  complete to \SoneM=10~Jy and contains 173 sources over an area of
  4.2~sr. The conversion from \SoneM \ to \SoneG \  with $\alpha=0.8$
  yield a flux density limit of \SoneG$\approx$1.92~Jy. In order to
  maximize the completeness of the sample at 1.4-GHz, we increased the
  flux density limit to \SoneG=3.5~Jy. The compiled spectral indices
  were used in the conversion for each 3CRR source. After excluding
  sources already present in the CoNFIG samples, 38 sources were
  selected to complement the CoNFIG catalogue. All sources were
  morphologically classified, either using the classification of
  \cite{Laing83} or following the method described in \S\ref{morpho},
  and the sample includes 8 FRI, 24 FRII and 6 compact sources.
\item The CENSORS (Combined EIS-NVSS Survey Of Radio Sources) sample
  \citep{Best03} is complete to \SoneG=7.2~mJy and contains 136 sources 
  selected from NVSS over the 6 \degsq of the ESO Imaging Survey (EIS)
  Patch D. The sample has spectroscopic redshifts for 68\% of the
  sources, and optical or near-IR identifications (giving redshift
  estimates) for almost all of the remainder.\\
  \indent
  Little radio morphological classification of the CENSORS sources has
  been done as the image resolution is often not high enough to
  identify the source morphology. For this reason, the VLA observation
  program described in \S\ref{VLAObs} also included 40 CENSORS
  sources, allowing us to morphologically classify 84.5\% of the
  CENSORS sources. The sample includes 13 FRI, 64 FRII, 38 compact and
  21 uncertain sources. 
\item The Lynx \& Hercules sample \citep{Rigby07} is complete to
  a catalogue flux limit of \SoneG=0.5~mJy ,from radio images with
  inital flux density limits of 0.07-0.09 mJy/bm  It contains 81
  sources within an area of 0.6~\degsq. It is complete in redshift
  estimation (49\% spectroscopic and 51\% photometric) and 95.6\% of
  the sample members have morphological classification, including 57
  FRI, 18 FRII and 6 uncertain sources.
\end{itemize}

The final list, including the complementary samples, contains 1114
sources and is 75.9\% complete for redshift information and 94.2\%
complete for radio morphologies. It includes a total of 136 FRI (78
confirmed, 58 possible) and 571 FRII (477 confirmed, 94 possible)
sources, making it one of the largest, most comprehensive databases of
morphologically-classified radio sources and an important tool in the
study of AGN space-densities.

%------------------------------ 
%------------------------------ 

%------------------------------ 
\begin{sidewaystable*}
  \centering
  \small
  \begin{tabular}{lllllllllllllllllll}
    \multicolumn{19}{c}{{\bf Table 6.} CoNFIG-1 Data Table (example)}\\
    (1) & \multicolumn{2}{c}{(2)} & (3) & (4) &(5)& \multicolumn{2}{c}{(6)} & (7) & (8) & (9) & \multicolumn{2}{c}{(10)} & (11) & (12) & (13) & (14) & (15) & (16) \\
    \hline
    1 & 07 13 38.15 & +43 49 17.20 & B0710+439    & \phantom{00}2011.4 & \phantom{$-$}0.82 & C* &   & 0.5180 & 0.0010 & S & 07 13 38.10 & +43 49 17.00$^o$ &      &      &      &      &      &     \\
    2 & 07 14 24.80 & +35 34 39.90 & B0711+35     & \phantom{00}1467.1 & \phantom{$-$}0.41 & C* &   & 1.6260 &        & S & 07 14 24.82 & +35 34 39.80$^o$ &      &      &      &      &      &     \\
    3 & 07 16 41.09 & +53 23 10.30 & 4C 53.16     & \phantom{00}1501.4 & $-$0.63 & II & p & 0.0643 & 0.0001 & S & 07 16 41.21 & +53 23 09.60 &      &      &      &      &      & 10.2$^e$\\
    4 & 07 35 55.54 & +33 07 09.60 & 4C 33.21     & \phantom{00}2473.1 & $-$0.56 & C  &   & 0.7010 & 0.0939 & P & 07 35 55.57 & +33 07 09.59 & 21.9 & 21.1 & 20.5 & 19.6 & 19.7 &     \\
    5 & 07 41 10.70 & +31 12 00.40 & J0741+3111   & \phantom{00}2284.3 & \phantom{$-$}0.38 & C* &   & 0.6300 & 0.0014 & S & 07 41 10.71 & +31 12 00.22 & 17.0 & 16.5 & 16.6 & 16.6 & 16.7 & 14.0\\
    6 & 07 45 42.13 & +31 42 52.60 & 4C 31.30     & \phantom{00}1357.8 & $-$0.55 & II & c & 0.4608 & 0.0004 & S & 07 45 41.67 & +31 42 56.70 & 15.7 & 15.5 & 15.6 & 15.4 & 15.3 & 12.9\\
    7 & 07 49 48.10 & +55 54 21.00 & DA 240       & \phantom{00}1660.4 & $-$0.44 & II & c & 0.0360 & 0.0001 & S & 07 48 34.70 & +55 48 59.00$^o$ &      &      &      &      &      &     \\
    8 & 07 58 28.60 & +37 47 13.80 & NGC 2484     & \phantom{00}2717.9 & $-$0.68 & I  & c & 0.0408 & 0.0002 & S & 07 58 28.11 & +37 47 11.87 & 15.7 & 13.8 & 12.9 & 12.5 & 12.2 & \phantom{0}9.9$^e$\\
    9 & 07 59 47.26 & +37 38 50.20 & 4C 37.21     & \phantom{00}1691.2 & $-$0.84 & II & c &        &        &   &             &              &      &      &      &      &      &     \\
    10 & 08 01 35.32 & +50 09 43.00 & TXS 0757+503 & \phantom{00}1471.7 & $-$1.02 & II & p & 0.4855 & 0.0419 & P & 08 01 35.35 & +50 09 43.99 & 22.2 & 21.6 & 20.6 & 20.0 & 19.3 &     \\
    11 & 08 05 31.31 & +24 10 21.30 & 3C 192       & \phantom{00}5330.6 & $-$0.67 & II & c & 0.0600 &        & S & 08 05 35.00 & +24 09 50.36 & 18.1 & 16.2 & 15.4 & 14.9 & 14.6 & 12.4$^e$\\
    12 & 08 10 03.67 & +42 28 04.00 & 3C 194       & \phantom{00}2056.6 & $-$0.78 & II & c & 1.1840 &        & S & 08 10 03.60 & +42 28 04.00$^o$ &      &      &      &      &      &     \\
    13 & 08 12 59.48 & +32 43 05.60 & 4C 32.24     & \phantom{00}1522.5 & $-$0.68 & II & c & 0.4306 & 0.0047 & I & 08 13  0.27 & +32 42 43.71 & 22.0 & 20.7 & 19.8 & 19.0 & 18.4 &     \\
    14 & 08 13 36.07 & +48 13 01.90 & 3C 196       & \phantom{0}15010.0 & $-$0.75 & II & c & 0.8710 &        & S & 08 13 36.07 & +48 13 02.64 & 18.6 & 17.9 & 17.7 & 17.5 & 17.3 & 14.8\\
    15 & 08 19 47.55 & +52 32 29.50 & 4C 52.18     & \phantom{00}2104.2 & $-$0.62 & II & c & 0.1890 &        & S & 08 19 47.51 & +52 32 27.13 & 20.7 & 19.0 & 17.9 & 17.4 & 17.0 &     \\
    16 & 08 21 33.77 & +47 02 35.70 & 3C 197.1     & \phantom{00}1787.1 & $-$0.75 & II & c & 0.1280 & 0.0012 & S & 08 21 33.61 & +47 02 37.36 & 19.1 & 17.7 & 16.8 & 16.3 & 16.0 & 13.5$^e$\\
    17 & 08 21 44.02 & +17 48 20.50 & 4C 17.44     & \phantom{00}1875.1 & $-$0.57 & C  &   & 0.2960 & 0.0002 & S & 08 21 44.02 & +17 48 20.30 & 21.0 & 19.6 & 18.1 & 17.6 & 17.2 & 14.9\\
    18 & 08 23 24.72 & +22 23 03.70 & 4C 22.21     & \phantom{00}2272.4 & $-$0.34 & C* &   & 2.2103 & 0.0013 & S & 08 23 24.76 & +22 23 03.30 & 20.3 & 19.8 & 19.3 & 18.9 & 18.5 & 15.4\\
    19 & 08 24 47.27 & +55 52 42.60 & 4C 56.16A    & \phantom{00}1449.4 & $-$0.25 & C* &   & 1.4181 & 0.0016 & S & 08 24 47.24 & +55 52 42.71 & 18.2 & 18.1 & 17.9 & 17.8 & 17.8 &     \\
    20 & 08 24 55.43 & +39 16 41.80 & 4C 39.23     & \phantom{00}1480.8 & $-$0.56 & C* &   & 1.2160 & 0.0010 & S & 08 24 55.48 & +39 16 41.92 & 18.3 & 18.1 & 17.8 & 17.6 & 17.3 & 14.2\\
    21 & 08 27 25.40 & +29 18 44.80 & 3C 200       & \phantom{00}2043.1 & $-$0.92 & II & c & 0.4580 &        & S & 08 27 25.38 & +29 18 45.04 & 21.7 & 20.4 & 19.1 & 18.5 & 18.0 & 15.3\\
    22 & 08 31 10.00 & +37 42 09.90 & 4C 37.24     & \phantom{00}2259.6 & $-$0.65 & C  &   & 0.9188 & 0.0014 & S & 08 31 10.01 & +37 42 09.58 & 19.2 & 18.7 & 18.6 & 18.6 & 18.4 &     \\
    23 & 08 33 18.80 & +51 03 07.80 & 4C 51.25     & \phantom{00}1313.5 & $-$0.81 & II & c & 0.5621 & 0.0419 & P & 08 33 18.72 & +51 03 06.88 & 26.9 & 22.3 & 20.5 & 19.6 & 19.2 &     \\
    24 & 08 34 48.37 & +17 00 46.10 & 3C 202       & \phantom{00}1882.8 & $-$0.72 & II & c & 0.6237 & 0.1740 & P & 08 34 48.22 & +17 00 42.44 & 23.0 & 22.4 & 21.7 & 21.1 & 21.9 &     \\
    25 & 08 34 54.91 & +55 34 21.00 & 4C 55.16     & \phantom{00}8283.1 & $-$0.01 & C  &   & 0.2412 & 0.0014 & S & 08 34 54.90 & +55 34 21.11 & 19.6 & 17.9 & 16.7 & 16.1 & 15.8 &     \\
    26 & 08 37 53.51 & +44 50 54.60 & 4C 45.17     & \phantom{00}1528.9 & $-$0.60 & II & c & 0.2072 & 0.0009 & S & 08 37 52.76 & +44 50 25.95 & 20.3 & 18.4 & 17.1 & 16.6 & 16.2 & 14.0\\
    27 & 08 39 06.50 & +57 54 13.40 & 3C 205       & \phantom{00}2257.7 & $-$0.86 & II & c & 1.5360 &        & S & 08 39 06.54 & +57 54 17.06 & 17.9 & 17.4 & 17.0 & 16.6 & 16.5 & 14.5\\
    28 & 08 40 47.70 & +13 12 23.90 & 3C 207       & \phantom{00}2613.0 & $-$0.81 & II & c & 0.6804 & 0.0010 & S & 08 40 47.59 & +13 12 23.62 & 18.7 & 18.1 & 18.0 & 17.9 & 17.7 & 15.0\\
    29 & 08 43 31.63 & +42 15 29.70 & B3 0840+424A & \phantom{00}1409.7 & $-$0.41 & C* &   & 0.8393 & 0.1758 & P & 08 43 31.64 & +42 15 29.38 & 25.6 & 22.6 & 21.3 & 21.3 & 20.2 &     \\
    30 & 08 47 53.83 & +53 52 36.80 & NGC 2656     & \phantom{00}1542.3 & $-$0.58 & I  & p & 0.0453 & 0.0002 & S & 08 47 53.07 & +53 52 34.25 & 16.3 & 14.3 & 13.4 & 13.0 & 12.7 & 10.4$^e$\\
    31 & 08 47 57.00 & +31 48 40.50 & 4C 31.32     & \phantom{00}1482.0 & $-$0.52 & II & c & 0.0673 & 0.0003 & S & 08 47 59.05 & +31 47 08.34 & 16.5 & 14.5 & 13.6 & 13.2 & 12.9 & 12.8\\
    32 & 08 53 08.83 & +13 52 55.30 & 3C 208       & \phantom{00}2364.3 & $-$0.97 & II & c & 1.1115 & 0.0014 & S & 08 53 08.61 & +13 52 54.84 & 17.9 & 17.9 & 17.6 & 17.6 & 17.8 &     \\
    33 & 08 54 39.35 & +14 05 52.10 & 3C 208.1     & \phantom{00}2163.8 & $-$0.71 & II$^c$ & p & 1.0200 &        & S & 08 54 39.32 & +14 05 51.86 & 19.8 & 19.6 & 19.3 & 19.3 & 19.2 &     \\
    34 & 08 54 48.87 & +20 06 30.70 & PKS 0851+202 & \phantom{00}1511.8 & \phantom{$-$}0.21 & C* &   & 0.4190 & 0.0016 & S & 08 54 48.87 & +20 06 30.71 & 16.4 & 15.8 & 15.4 & 15.0 & 14.7 & 11.8$^e$\\
    35 & 08 57 40.64 & +34 04 06.40 & 3C 211       & \phantom{00}1798.4 & $-$0.77 & II & c & 0.4789 & 0.0618 & P & 08 57 40.28 & +34 04 04.91 & 22.8 & 21.9 & 20.6 & 19.8 & 19.4 &     \\
    36 & 08 58 10.07 & +27 50 50.80 & 3C 210       & \phantom{00}1807.8 & $-$0.83 & II & c & 1.1690 &        & S & 08 58 10.04 & +27 50 54.17 & 23.3 & 22.6 & 21.5 & 21.0 & 20.2 &     \\
    37 & 08 58 41.51 & +14 09 43.80 & 3C 212       & \phantom{00}2370.8 & $-$0.87 & II & c & 1.0430 &        & S & 08 58 41.45 & +14 09 44.78 & 21.0 & 20.0 & 19.1 & 18.8 & 18.6 & 15.3\\
    38 & 09 01 05.40 & +29 01 45.70 & 3C 213.1     & \phantom{00}2003.4 & $-$0.58 & II & p & 0.1940 & 0.0002 & S & 09 01 05.26 & +29 01 46.92 & 20.1 & 18.5 & 17.6 & 17.1 & 16.9 & 15.2\\
    39 & 09 03 04.04 & +46 51 04.70 & 4C 47.29     & \phantom{00}1754.9 & $-$0.39 & C* &   & 1.4710 & 0.0024 & S & 09 03 04.01 & +46 51 04.21 & 19.3 & 19.3 & 18.9 & 18.7 & 18.7 &     \\
    40 & 09 06 31.88 & +16 46 13.00 & 3C 215       & \phantom{00}1586.2 & $-$0.95 & II & c & 0.4115 & 0.0003 & S & 09 06 31.80 & +16 46 12.00 & 25.1 & 22.3 & 22.6 & 20.8 & 21.0 & 15.5\\
    41 & 09 07 34.92 & +41 34 53.80 & 4C 41.19     & \phantom{00}1394.5 & $-$0.70 & II & c & 0.4783 & 0.0379 & P & 09 07 33.18 & +41 34 44.10 & 21.9 & 20.5 & 19.1 & 18.3 & 18.0 &     \\
    42 & 09 08 50.56 & +37 48 20.20 & 3C 217       & \phantom{00}2086.4 & $-$0.95 & II & c & 0.8980 &        & S & 09 08 50.67 & +37 48 19.69 & 22.3 & 22.2 & 21.2 & 20.3 & 19.8 &     \\
    43 & 09 09 33.53 & +42 53 47.40 & 3C 216       & \phantom{00}4233.8 & $-$0.77 & S* &   & 0.6700 & 0.0014 & S & 09 09 33.50 & +42 53 46.50 & 19.9 & 19.3 & 18.7 & 18.3 & 18.0 & 14.6\\
    44 & 09 12 04.00 & +16 18 29.70 & 4C 16.27     & \phantom{00}1374.6 & $-$0.77 & II & c & 0.9182 & 0.4087 & P & 09 12 03.99 & +16 18 29.99 & 21.6 & 21.7 & 21.8 & 22.0 & 22.5 &     \\
    45 & 09 14 04.83 & +17 15 52.40 & 4C 17.48     & \phantom{00}1527.3 & $-$0.74 & II & c & 0.5395 & 0.0291 & P & 09 14  5.21 & +17 15 54.35 & 24.3 & 21.0 & 19.7 & 18.6 & 18.4 &     \\
    46 & 09 21 07.54 & +45 38 45.70 & 3C 219       & \phantom{00}8101.6 & $-$0.79 & II & c & 0.1744 & 0.0012 & S & 09 21 08.62 & +45 38 57.39 & 19.2 & 17.9 & 16.7 & 16.3 & 16.0 & 13.1$^e$\\
    47 & 09 22 49.93 & +53 02 21.20 & 4C 53.18     & \phantom{00}1597.8 & $-$0.77 & II & c & 0.5974 & 0.1361 & P & 09 22 49.90 & +53 02 21.00 & 23.5 & 25.3 & 21.9 & 20.9 & 20.2 &     \\
    48 & 09 27 03.04 & +39 02 20.70 & 4C 39.25     & \phantom{00}2884.6 & $-$0.29 & C* &   & 0.6967 & 0.0019 & S & 09 27 03.01 & +39 02 20.87 & 17.0 & 16.7 & 16.6 & 16.7 & 16.6 & 14.0\\
    49 & 09 30 33.45 & +36 01 23.60 & 3C 220.2     & \phantom{00}1875.1 & $-$0.68 & II & c & 1.1570 & 0.0013 & S & 09 30 33.54 & +36 01 25.18 & 18.6 & 18.3 & 17.8 & 17.6 & 17.6 & 15.8\\
    \hline
  \end{tabular}
\end{sidewaystable*}

%%%%%%%%%%%%%%%%%%%%%%%%%%%%%%%%%%%%%%%%%%%%%%%%%%%%%%%%%%%%%%%%%%%%%%%%%%%
\section{Source statistics and evolution}\label{Stat}

%------------------------------ 
\begin{figure*}
    \centerline{
      \includegraphics[angle=270,scale=0.5]{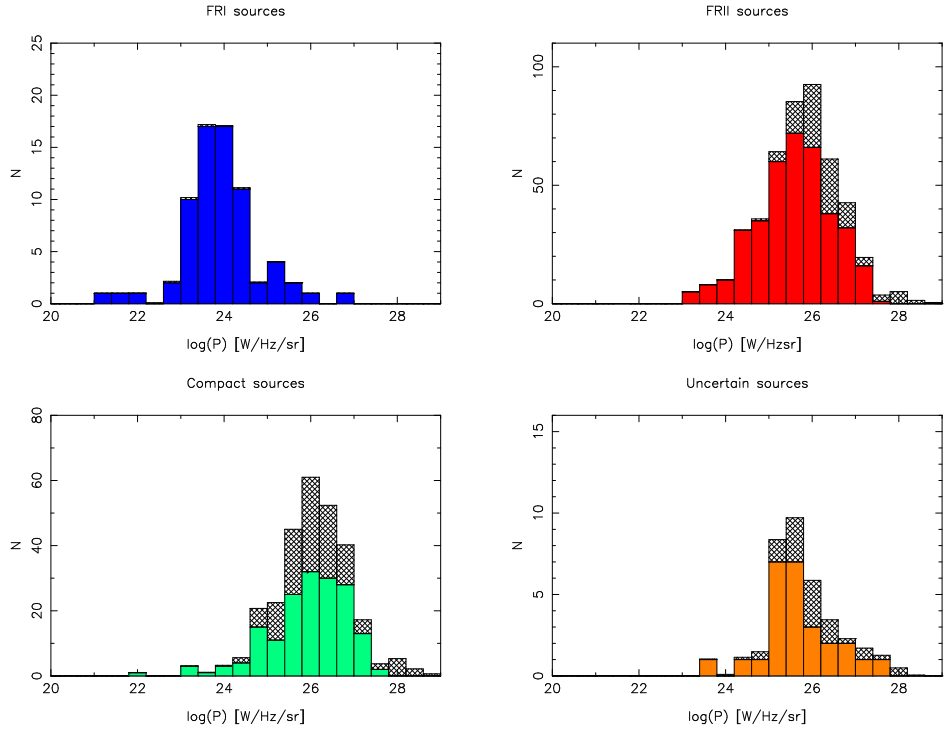}}
    \caption[]{\label{Ldist}Luminosity distributions for compact and
      extended (FRI, FRII and uncertain) sources. The cross-hatched
      columns represent the estimated contribution to each luminosity
      bin of sources with no redshift information available, following
      the method presented in \S\ref{noZest}}
\end{figure*}
%-----------------------------------------------------------
\begin{figure*}
    \centerline{
      \includegraphics[angle=270,scale=0.5]{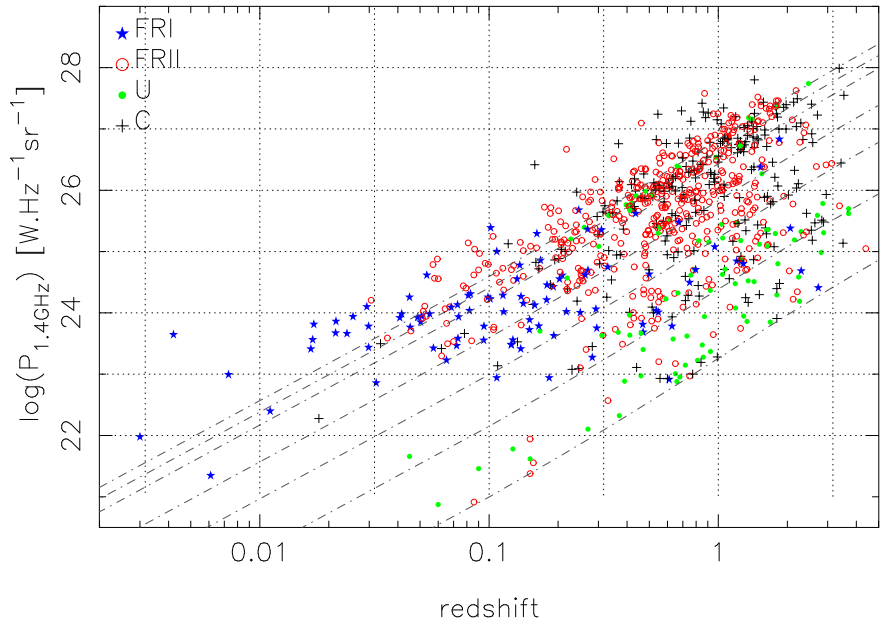}}
    \caption[]{\label{PzPlot}P-z plane coverage for the four
      CoNFIG samples, as well as the 3CRR, CENSORS and  Lynx \&
      Hercules samples, by radio-morphological type (limited only to
      sources with estimated redshifts). The dot-dashed lines show the
      survey limits for each sample. Sources are identified by their
      radio morphological classification: FRIs, FRIIs, uncertain and
      compact sources are represented by blue stars, red circles,
      green dots and black crosses respectively.}
\end{figure*}
\begin{figure*}
    \centerline{
      \includegraphics[angle=270,scale=0.5]{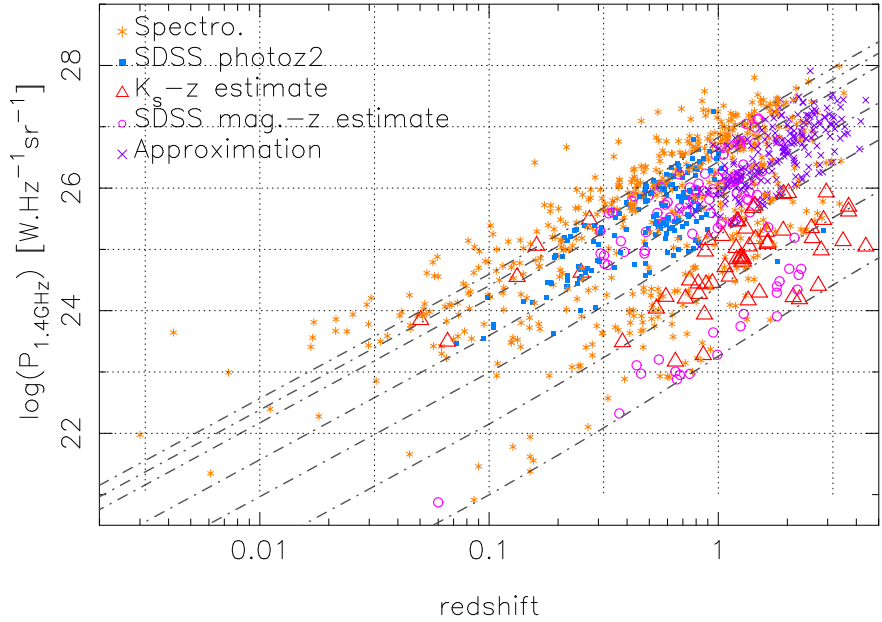}}
    \caption[]{\label{PzPlotZtype}P-z plane coverage for
      the four CoNFIG samples, as well as the 3CRR, CENSORS and  Lynx
      \& Hercules samples, by redshift type. The dot-dashed lines show
      the survey limits for each sample. Sources are identified by
      their redshift type: spectroscopic, SDSS {\it photoz2}
      photometric, $K_S$-z estimated and SDSS {\it mag}-z estimated redshift
      are represented by orange asterisks, blue squares, red triangles
      and pink circles respectively. Sources with approximated
      redshifts, as described in \S\ref{noZest}, are represented by
      purple crosses.}
\end{figure*}
%-----------------------------------------------------------

The main goal of CoNFIG is to produce a comprehensive catalogue of
morphologically-classified radio sources to be used in the modeling of
the radio luminosity function of AGN, in order to investigate their
evolution and the role of the different types in feedback
processes. For this purpose, we computed the luminosity distributions
and source counts based on morphological classification, to be
used in the RLF modeling.

\subsection{Luminosity distribution and the P-z plane} 
The luminosity distribution is computed for each morphological type
(FRI, FRII, C and U) for sources with available redshift information,
using the 1.4-GHz flux density and spectral index values of each
source. When the latter was unavailable, a value of $\alpha=-0.8$ was
used. This introduced a minimal bias in the results, since extended
sources in the CoNFIG samples have a median spectral index of $-$0.75
and less than 6\% of them have $\alpha \ge -0.5$. Finally, sources
with no redshift information were included, with redshifts as
estimated in \S\ref{noZest}, and the resulting distributions are shown
in Fig.~\ref{Ldist}.\\

A wide coverage of the P-z plane is essential to any modeling of the
radio luminosity function \citep{Rawlings02}. The combination of
CoNFIG, 3CRR, CENSORS and the Lynx \& Hercules samples covers a large
range of luminosity and redshift (Fig.~\ref{PzPlot} and
\ref{PzPlotZtype}), providing a powerful basis from which to study FRI
and FRII sources.

\subsection{Source counts}\label{FRSC}

\begin{figure}
    \centerline{
      \includegraphics[angle=270,scale=0.35]{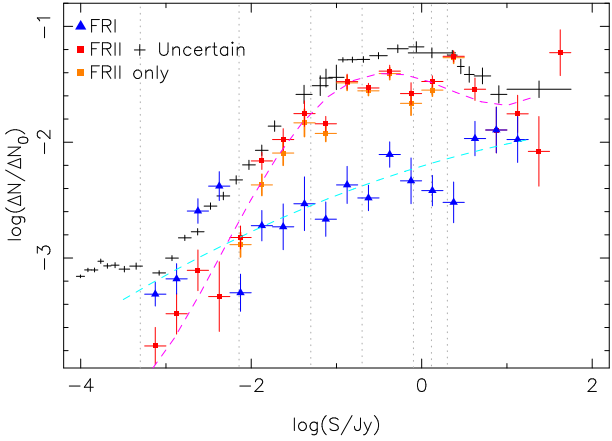}}
    \caption[Source Count]{\label{SC}Relative differential source
      counts $\Delta N/\Delta N_0$ for FRI (blue triangles) and
      FRII+uncertain (red squares) sources. The count for FRII
      sources only is shown in orange squares, to illustrate
      the possible bias due to the inclusion of uncertain sources in
      the FRII group. Here, the normalization is given by $\Delta N_0
      = 3618 \Delta (S^{-1.5})$ \citep{Jack99} and the error bars
      correspond to $\sqrt{N}$ where N is the number of objects in
      each bin. The counts are fitted by a polynomial (dashed lines)
      to indicate the shapes of the counts. A 1.4-GHz source count,
      compiled from the data of \cite{BDFL}, \cite{Machalski78},
      \cite{Hopkins03} and \cite{Prandoni01}, is represented with
      crosses for comparison.}
\end{figure}

The morphologically-dependent source counts (Fig.~\ref{SC}) were
compiled as described by \cite{PaperI}, from the CoNFIG and
complementary samples.\\
\indent
As seen in Fig.~\ref{Ldist}, uncertain sources (which are
extended but uncertain to whether they are FRI or FRII) have a
luminosity distribution closer to that of FRII sources than FRI
sources. Thus, we make the assumption to include uncertain sources
into the FRII morphology group for the source count. This inclusion
does not make any significant change from the source count of FRII
sources only.\\

The FRII sources dominate the total count, except at low
flux densities (log\SoneG$\lesssim -1.6$), where the FRI sources
suddenly take over, constituting a significant portion of the mJy and
sub-mJy sources in contrast to FRII sources. Since most of the FRI
count at low flux densities is composed of low-luminosity sources at
low redshift, our results show that FRI objects must undergo some mild
evolution. This is consistent with the results of \cite{Sadler07}, who
studied low power sources in the 2SLAQ survey \citep{Richards05} and
found evidence that FRIs undergo significant evolution over
$z<0.7$. Our results also show that FRIs undergo less evolution than
FRIIs, and they do not participate much in the source-count
``evolution bump'' around \SoneG$\sim$1~Jy. This is in agreement with
previous investigations stretching back to \cite{Long66}.

%-----------------------------------------------------------
%-----------------------------------------------------------

%%%%%%%%%%%%%%%%%%%%%%%%%%%%%%%%%%%%%%%%%%%%%%%%%%%%%%%%%%%%%%%%%%%%%%%%%%%
\subsection{The local FRI/FRII RLFs}\label{localI-II}

The radio luminosity functions (RLF) were computed using the
$1/V_{max}$ technique, in which, for each P-z bin, the space density
of sources is given by:
\begin{eqnarray}\label{EqVmax}
  \rho = \sum_{i=1}^N\frac{1}{V_i}\ \ \ \ \ \ \ \ \ \ \ 
  \sigma^2 = \sum_{i=1}^N\frac{1}{V_i^2}
\end{eqnarray}
where $V_i$ is the largest volume in which the source could be observe
in bin {\it i}.\\

\begin{figure}
    \centerline{
      \includegraphics[angle=270,scale=0.35]{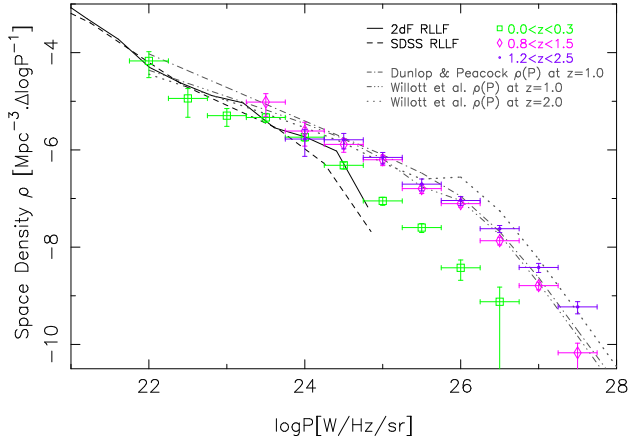}}
    \caption[]{\label{RLF}Luminosity function $\rho(P)$ computed from
      the four CoNFIG samples, as well as the 3CRR, CENSORS and Lynx
      \& Hercules samples. The radio local luminosity function (LRLF)
      for z$<$0.3 for all extended sources is represented by green
      open squares. The LRLF is consistent with both the LRLF of the
      2dF survey \citep{Sadler02} and the SDSS \citep{Best05}, shown
      by solid and dotted lines respectively. In addition, the
      luminosity functions at z=1.0 (in the interval z=[0.8;1.5]) and
      at z=2.0 (in the interval z=[1.2;2.5]) are displayed by pink
      diamonds and purple dots respectively. For comparison, the
      modelled RLFs from \cite{DP90} and \cite{Willott01} at z=1.0 are
      displayed in dot-dashed and triple-dot-dashed lines
      respectively.\\\\}
\end{figure}

The general local radio luminosity function, defined here as the RLF
for z$\le$0.3, is displayed in Fig.~\ref{RLF}. It is consistent with
both the LRLF of the 2dF survey \citep{Sadler02} and the SDSS
\citep{Best05}, and extends to significantly larger luminosities,
because of the larger area covered by our bright samples.\\
\indent
In addition, the luminosity functions at z=1.0 (in the interval
z=[0.8;1.5]) and at z=2.0 (in the interval z=[1.2;2.5]) were computed
and compared with modelled RLFs from \cite{DP90} and \cite{Willott01}
at z=1.0. The CoNFIG RLF agrees well with both models.\\

%-----------------------------------------------------------
\begin{figure}
    \centerline{
      \includegraphics[angle=270,scale=0.35]{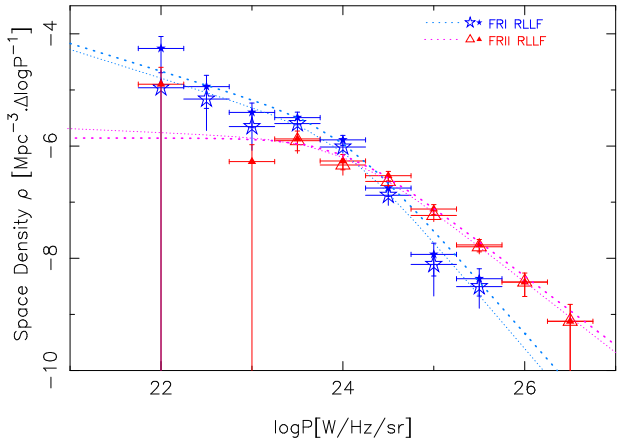}}
    \caption[]{\label{RLFI-II}Local luminosity function $\rho(P)$ for
      FRIs and FRIIs, represented by blue stars and red triangles
      respectively. The values for both confirmed and possible
      FRI/FRII are shown by filled symbols and thick error bars,
      whereas the values for confirmed FRI/FRII only are displayed
      with open symbols and thin error bars. Data are fitted by a
      broken power law, described by Equ.\ref{rhofit}.}
\end{figure}
%-----------------------------------------------------------
\begin{figure}
    \centerline{
      \includegraphics[angle=270,scale=0.35]{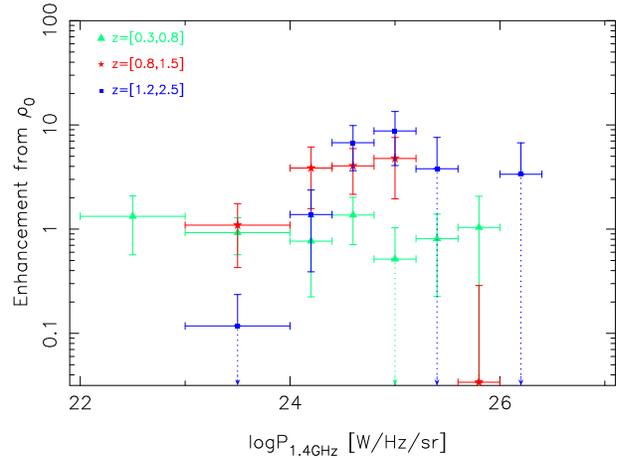}}
    \caption[]{\label{bootFRI}Space density enhancement for
      confirmed+possible FRI sources for different redshift bins:
      z=[0.3:0.8] in green triangles, z=[0.8:1.5] in red stars and
      z=[1.2:2.5] in blue squares.\\
      An enhancement of a factor of 7 to 10 is seen at z=1.0 for high
      luminosity sources ($logP_{1.4~GHz} \ge$24.5~W/Hz/sr), in
      agreement with \cite{Rigby08}. This enhancement appears to
      continue to higher redshifts.}
\end{figure}
%-----------------------------------------------------------
\begin{figure}
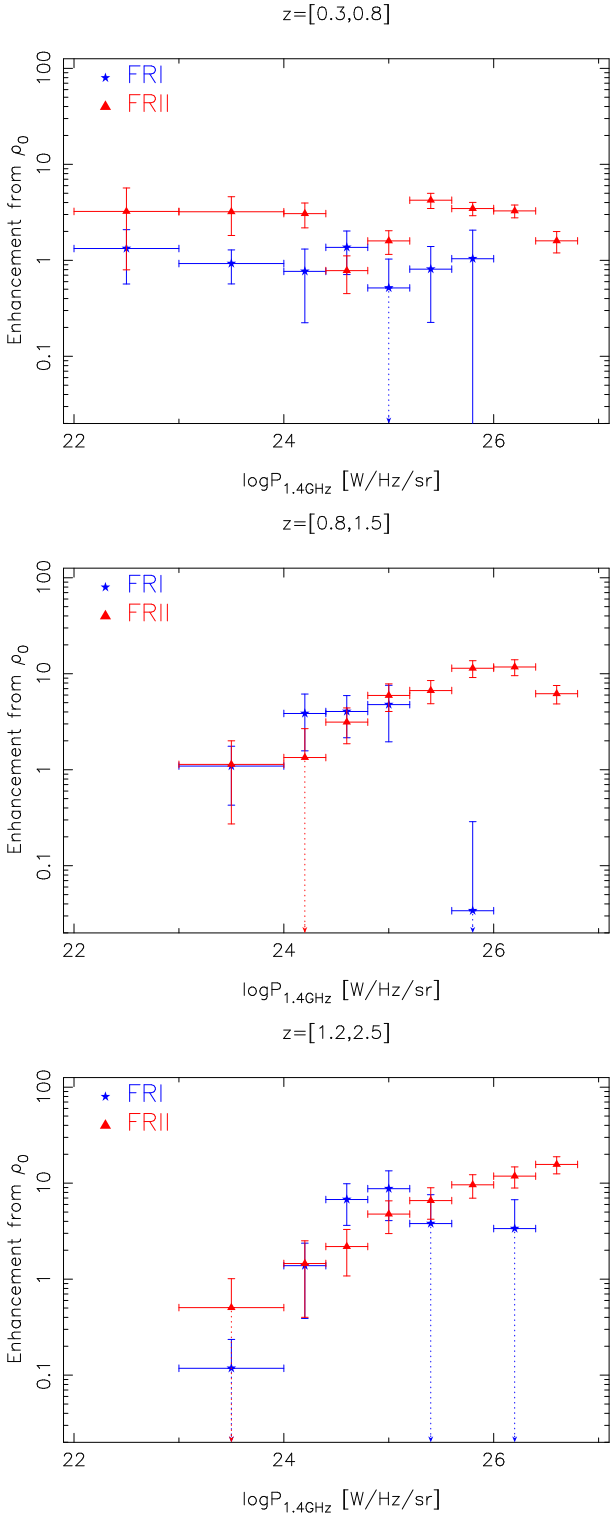

  \begin{minipage}{8.0cm}
    \centerline{
      \includegraphics[angle=270,scale=0.35]{Figures/FRIvsFRII1.ps}}
  \end{minipage}
  \vfill
  \medskip
  \begin{minipage}{8.0cm}
    \centerline{
      \includegraphics[angle=270,scale=0.35]{Figures/FRIvsFRII2.ps}}
  \end{minipage}
  \vfill
  \medskip
  \begin{minipage}{8.0cm}
    \centerline{
      \includegraphics[angle=270,scale=0.35]{Figures/FRIvsFRII3.ps}}
    \caption[]{\label{bootFRIvsII}Comparison of the space density
      enhancement between confirmed+possible FRI (blue stars) and FRII
      (red triangles) sources, for different redshift bins (z=[0.3:0.8],
      z=[0.8:1.5] and z=[1.2:2.5]). For FRIs with $logP_{1.4~GHz}
      \ge$26.0 and FRIIs with $logP_{1.4~GHz} \le$23.0 and
      $logP_{1.4~GHz} \ge$27.0, the value of the LRLF was extrapolated
      from the power law fit described in \S\ref{I-II}. Both
      populations show similar enhancement history, hinting at a
      common mechanism governing the luminosity-dependent evolution.}
  \end{minipage}
\end{figure}
%-----------------------------------------------------------

The LRLFs for each population were computed and fitted, using the
method of least-squares, by a broken power law, to provide parametric
representations:
\begin{eqnarray}\label{rhofit}
\rho(P)=\rho_0 \left [\left (\frac{P}{P^*} \right )^{\alpha}+\left
  (\frac{P}{P^*}\right )^{\beta}
  \right ]^{-1}
\end{eqnarray}
where log($P^*$)=24.0 is the break luminosity, determined by visual
inspection of the LRLFs. These LRLF models are plotted in
Fig.~\ref{RLFI-II}.\\
\indent
The FRI and FRII LRLFs show apparent differences, such as the
flattening of the FRII LRLF at lower powers and the steeper slope of
the FRI LRLF at higher power.Overall, these LRLFs suggest that, locally, FRI
and FRII sources constitute two distinct populations. However, these
local space densities do not indicate any sharp luminosity divide
between FRIs and FRIIs: at higher power ($logP_{1.4GHz} \gtrsim$ 25.0)
the FRII LRLF is only a factor of $\sim$3-4 higher than for FRIs and
the two population show a large degree of overlap at intermediate
powers.\\
\indent
Because most of the approximate redshifts (\S~\ref{noZest}) are
greater than z=0.3, the results of the LRLF are completely unaffected
by redshift uncertainties.

\subsection{FRI/FRII evolution}\label{I-II}

The RLF for combined confirmed and possible sources for each
population was then computed for different redshift bins (z=[0.3;0.8],
z=[0.8;1.5] and z=[1.2;2.5]). In order to account for data with no
redshift information, the random redshift assignment technique
described in \S\ref{noZest} was used. This process was repeated 1000
times and the final RLF was computed by averaging the results.\\

For each population, the space-density enhancement above the local
value was computed. FRI sources (Fig.\ref{bootFRI}) show an
enhancement of a factor of 7 to 10 in the interval z=[0.8;1.5]
for high luminosity sources ($logP_{1.4GHz}\ge$24.0~W/Hz/sr), in
agreement with the results of \cite{Rigby08}. This enhancement remains
present at redshifts up to 2.5.\\
\indent
A comparison of the space-density enhancement for FRI and FRII sources
in the same redshift bins is shown in Fig.\ref{bootFRIvsII}. The
overall behaviour of the enhancement with luminosity of FRI and FRII
sources is very similar, with little or no enhancement in the
interval z=[0.3;0.8] and up to a factor of 10 enhancement for higher
luminosity sources in higher redshift bins. Both populations show
similar enhancement history, hinting at a common mechanism governing
the luminosity-dependent evolution. The RLFs were also computed for
confirmed source only, and show the same overall trends.\\

\begin{figure}
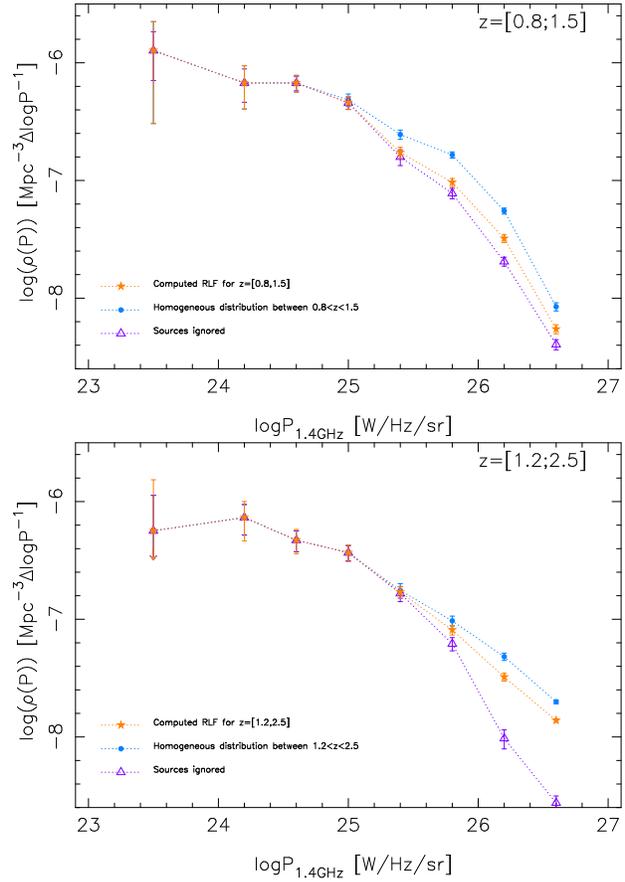

  \begin{minipage}{8.0cm}
    \centerline{
      \includegraphics[angle=270,scale=0.35]{Figures/ExtremeRLF_FRII_z1.ps}}
  \end{minipage}
  \begin{minipage}{8.0cm}
    \centerline{
      \includegraphics[angle=270,scale=0.35]{Figures/ExtremeRLF_FRII_z2.ps}}
    \caption[]{\label{Extrem} FRII RLFs in the ranges z=[0.8;1.5] and
      z=[1.2;2.5]where the approximate redshifts were drawn using
      various distributions, in which all sources with no redshift
      were either approximated as described in \S~\ref{noZest}
      (orange stars), distributed homogeneously in the given range
      (blue dots) or ignored (purple triangles).}
    \end{minipage}
\end{figure}

In Fig.~\ref{Extrem}, we investigated the impact of the approximate
redshift selection method. We compared RLFs in the ranges z=[0.8;1.5]
and z=[1.2;2.5] from the CoNFIG FRII sub-sample, where the approximate
redshifts were drawn using the distributions in which all sources with
no redshift were either distributed homogeneously within the given
range (to estimate the maximum space densities) or ignored(equivalent
to setting all of them outside this range, hence giving minimum space
densities)\\
\indent
In the range z=[0.8;1.5], the RLFs computed using approximate
redshifts distributed homogeneously and ignored differ by a factor of
~2.5, which is comparable to the size of the error estimates in the
LRLF and RLF computed in this paper. The data and method therefore
give a reasonably reliable estimate of the RLF in this redshift range,
across all radio powers.\\ 
\noindent
In the range z=[1.2;2.5] the approximate redshifts distribution method
used in this paper gives results close to the maximal density
calculated, whilst the minimal density lies significantly below this
at high radio powers. This is because most of the approximated
redshifts lie in this redshift range (as expected since the sources
have $z_{lim} ~ 1$) so the minimal density method provides a significant
underestimate. The data allow an acceptable estimate of the RLF in
this redshift range, but at higher powers ($logP_{1.4GHz} \ge 26.0$)
significant uncertainties remain.

%-----------------------------------------------------------
%-----------------------------------------------------------

%%%%%%%%%%%%%%%%%%%%%%%%%%%%%%%%%%%%%%%%%%%%%%%%%%%%%%%%%%%%%%%%%%%%%%%%%%%
\section{Summary}\label{Sum}
In this paper, we first described the latest steps in the construction
of the CoNFIG catalogue, including new VLA observations. The catalogue
now consists of 859 sources in 4 samples (CoNFIG-1, 2, 3 and 4 with flux
density limits \SoneG=1.3, 0.8, 0.2 and 0.05~Jy respectively)
and is 95.7\% complete for radio morphologies. 74.3\% of the sources
have redshift information. Optical counterpart identifications were
obtained from the SDSS and 2MASS catalogues for 74.6\% and 26.8\% of
the sources respectively. Spectroscopic redshift information was
retrieved from SDSS and the SIMBAD database, while photometric
redshifts (or redshift estimates) were compiled from the SDSS {\it
  photoz2} catalogue, or using a $K_S$-z or SDSS {\it mag}-z relation.\\
\indent
Combining CoNFIG with 3CRR, CENSORS and the Lynx \& Hercules samples,
the comparative distribution and evolution of FRI and FRII sources were
investigated. The conclusions of this preliminary study are as follows:
\begin{itemize}
\item The FRII sources dominate the total count, except at low
  flux densities (log\SoneG$\lesssim -1.6$), where the FRI sources
  suddenly take over, constituting a significant portion of the mJy and
  sub-mJy sources in contrast to FRII sources.
\item The FRI and FRII LRLFs show apparent differences, suggesting
  that, locally, FRI and FRII sources constitute two distinct
  populations. However, they do not indicate any sharp luminosity
  divide between FRIs and FRIIs.
\item The FRI RLF shows an enhancement of a factor 7-10 over the local
  value, which continues to higher redshifts. This result is in
  agreement with the findings of \cite{Rigby08}.
\item The comparison of space density enhancement between FRI and FRII
  sources at various redshifts does not show any significant differences,
  suggesting a common mechanism governing the luminosity-dependent
  evolution.
\end{itemize}

%%%%%%%%%%%%%%%%%%%%%%%%%%%%%%%%%%%%%%%%%%%%%%%%%%%%%%%%%%%%%%%%%%%%%%%%%%%

%Acknowledgments
\section*{Acknowledgements}
This work was supported by the National Sciences and Engineering
Research Council of Canada (MAG and JVW). PNB is grateful for support
from the Leverhulme Trust.\\ 
\indent
The National Radio Astronomy Observatory is a facility of the National
Science Foundation operated under cooperative agreement by Associated
Universities, Inc.\\
\indent
This research has made use of the SIMBAD database,operated at CDS,
Strasbourg, France.\\
\indent
This publication makes use of SDSS data products. The SDSS Web Site is
http://www.sdss.org/. Funding for the SDSS and SDSS-II has been
provided by the Alfred P. Sloan Foundation, the Participating
Institutions, the National Science Foundation, the U.S. Department of
Energy, the National Aeronautics and Space Administration, the
Japanese Monbukagakusho, the Max Planck Society, and the Higher
Education Funding Council for England. The SDSS Web Site is
http://www.sdss.org/.\\
\indent
This publication makes use of data products from the Two Micron All
Sky Survey, which is a joint project of the University of
Massachusetts and the Infrared Processing and Analysis
Center/California Institute of Technology, funded by the National
Aeronautics and Space Administration and the National Science
Foundation.\\

%%%%%%%%%%%%%%%%% Bibliography %%%%%%%%%%%%%%%%%%%%%%%%%%%%%%%%%%%%%%%%%%%%

%%%%%%%%%%%%%%%%%%%%%%%%%%%%%%%%%%%%%%%%%%%%%%%%%%%%%%%%%%%%%%%%%%%%%%%%%%%%%
\onecolumn
\appendix

\section{The CoNFIG Catalogue}\label{datatables}
\subsection{CoNFIG Samples}\label{ConfSamp}

Data for the four CoNFIG samples.The RA and DEC gives the NVSS
position of the source.\\

\noindent
{\bf Columns:}\\
(1)\phantom{0} CoNFIG number.\\ 
(2)\phantom{0} NVSS radio position RA and DEC. For sources with
several NVSS component, the coordinates correspond to one of the
component.\\ 
(3)\phantom{0} Name.\\ 
(4)\phantom{0} Flux density \SoneG in mJy.\\ 
(5)\phantom{0} Spectral index $\alpha$.\\ The spectral index $\alpha$
(where $S^{\alpha}_{\nu} \propto \nu^{\alpha}$) corresponds to
$\alpha_{low}$ as defined in \S2.2.\\  
(6)\phantom{0} Morphological type.\\
Designations I and II are \cite{FR74} types. The sources of C* type
are confirmed compact sources from the VLBA calibrator list \citep[see
][]{Beasley02,Fomalont03,Petrov06,Kovalev07} or the Pearson-Readhead
survey \citep{Pearson88}. Sources of S* type are confirmed compact
sources which show a steep ($\alpha \le -0.6$) spectral index. These
are probably CSS sources. Sources of U type have uncertain
morphology. They look compact with a steep spectral index, but are
most likely extended.\\ 
In addition to the main morphological classification, extended sources
of type I and II are assigned a sub-classification (confirmed - c - or
possible - p) depending on how clearly the source showed either FRI or
FRII characteristics.\\
Superscripts w denotes Wide Angle Tail sources and c possible core-jet sources.\\
(7)\phantom{0} Redshift.\\ 
(8)\phantom{0} Error in redshift.\\ 
(9)\phantom{0} Redshift type.\\
S - spectroscopic redshift; P - SDSS {\it photoz2} photometric
redshift; K - 2MASS $K_S$-z estimate; I - SDSS {\it i}-z estimate; Z -
SDSS {\it z}-z estimate; R - SDSS {\it  r}-z estimate; G - SDSS {\it
  g}-z estimate.\\
(10) Optical counterpart RA and DEC. Superscript o denotes sources
for which the identification was found in another catalogue than SDSS
or 2MASS, such as \citep{Veron83}.\\ 
(11) SDSS {\it u} magnitude.\\ 
(12) SDSS {\it g} magnitude.\\ 
(13) SDSS {\it r} magnitude.\\ 
(14) SDSS {\it i} magnitude.\\ 
(15) SDSS {\it z} magnitude.\\ 
(16) 2MASS {\it $K_S$} magnitude. Superscript e denotes sources from
the 2MASS extended catalogue.

%%%%%%%%%%%%%%%%%%%%%%%%%%%%%%%%%%%%%%%%%%%%%%%%%%%%%%%%%%%%%%%%%%%%%%%%%%%%%
\begin{sidewaystable}
 \centering
 \small
\begin{tabular}{lllllllllllllllllll}
 \multicolumn{19}{c}{CoNFIG-1 Data Table}\\
 (1) & \multicolumn{2}{c}{(2)} & (3) & (4) &(5)& \multicolumn{2}{c}{(6)} & (7) & (8) & (9) & \multicolumn{2}{c}{(10)} & (11) & (12) & (13) & (14) & (15) & (16) \\
 \hline
 1 & 07 13 38.15 & +43 49 17.20 & B0710+439 & \phantom{00}2011.4 & \phantom{$-$}0.82 & C* & & 0.5180 & 0.0010 & S & 07 13 38.10 & +43 49 17.00$^o$ & & & & & & \\
 2 & 07 14 24.80 & +35 34 39.90 & B0711+35 & \phantom{00}1467.1 & \phantom{$-$}0.41 & C* & & 1.6260 & & S & 07 14 24.82 & +35 34 39.80$^o$ & & & & & & \\
 3 & 07 16 41.09 & +53 23 10.30 & 4C 53.16 & \phantom{00}1501.4 & $-$0.63 & II & p & 0.0643 & 0.0001 & S & 07 16 41.21 & +53 23 09.60 & & & & & & 10.2$^e$\\
 4 & 07 35 55.54 & +33 07 09.60 & 4C 33.21 & \phantom{00}2473.1 & $-$0.56 & C & & 0.7010 & 0.0939 & P & 07 35 55.57 & +33 07 09.59 & 21.9 & 21.1 & 20.5 & 19.6 & 19.7 & \\
 5 & 07 41 10.70 & +31 12 00.40 & J0741+3111 & \phantom{00}2284.3 & \phantom{$-$}0.38 & C* & & 0.6300 & 0.0014 & S & 07 41 10.71 & +31 12 00.22 & 17.0 & 16.5 & 16.6 & 16.6 & 16.7 & 14.0\\
 6 & 07 45 42.13 & +31 42 52.60 & 4C 31.30 & \phantom{00}1357.8 & $-$0.55 & II & c & 0.4608 & 0.0004 & S & 07 45 41.67 & +31 42 56.70 & 15.7 & 15.5 & 15.6 & 15.4 & 15.3 & 12.9\\
\hline
 \end{tabular}
\end{sidewaystable}

\begin{sidewaystable}
 \centering
 \small
 \begin{tabular}{lllllllllllllllllll}
 \multicolumn{19}{c}{CoNFIG-1 Data Table}\\
 (1) & \multicolumn{2}{c}{(2)} & (3) & (4) &(5)& \multicolumn{2}{c}{(6)} & (7) & (8) & (9) & \multicolumn{2}{c}{(10)} & (11) & (12) & (13) & (14) & (15) & (16) \\
 \hline
 50 & 09 39 50.20 & +35 55 53.10 & 3C 223 & \phantom{00}3719.0 & $-$0.65 & II & c & 0.1368 & 0.0008 & S & 09 39 52.76 & +35 53 58.92 & 19.4 & 17.8 & 16.7 & 16.3 & 16.1 & 14.4\\
 51 & 09 41 23.62 & +39 44 14.10 & 3C 223.1 & \phantom{00}1976.8 & $-$0.54 & II & c & 0.1075 & & S & 09 41 24.02 & +39 44 41.86 & 18.7 & 17.1 & 16.1 & 15.7 & 15.3 & 12.7$^e$\\
 52 & 09 42 08.40 & +13 51 52.20 & 3C 225A & \phantom{00}1338.5 & & II & c & 1.5650 & & S & 09 42 08.06 & +13 51 54.97 & 20.4 & 19.3 & 18.8 & 18.5 & 18.5 & \\
 53 & 09 42 15.35 & +13 45 49.60 & 3C 225 & \phantom{00}3336.4 & $-$1.01 & II & p & 0.5800 & & S & 09 42 15.39 & +13 45 50.51 & 21.9 & 21.3 & 20.0 & 19.3 & 19.3 & \\
 54 & 09 43 12.74 & +02 43 27.50 & 4C 02.29 & \phantom{00}1331.5 & $-$0.43 & II & c & 0.5920 & 0.0002 & S & 09 43 12.81 & +02 43 25.86 & 23.3 & 22.4 & 20.9 & 19.8 & 19.7 & \\
 55 & 09 44 16.40 & +09 46 19.20 & 3C 226 & \phantom{00}2393.7 & $-$0.93 & II & c & 0.8178 & & S & 09 44 16.52 & +09 46 16.90 & 21.6 & 22.0 & 21.1 & 20.5 & 19.3 & \\
\hline
 \end{tabular}
\end{sidewaystable}

\begin{sidewaystable}
 \centering
 \small
 \begin{tabular}{lllllllllllllllllll}
 \multicolumn{19}{c}{CoNFIG-1 Data Table}\\
 (1) & \multicolumn{2}{c}{(2)} & (3) & (4) &(5)& \multicolumn{2}{c}{(6)} & (7) & (8) & (9) & \multicolumn{2}{c}{(10)} & (11) & (12) & (13) & (14) & (15) & (16) \\
 \hline
 99 & 11 34 38.46 & +43 28 00.50 & 4C 43.22 & \phantom{00}1567.1 & $-$0.70 & II & c & 0.5724 & & S & 11 34 38.47 & +43 28 00.30 & 22.8 & 22.3 & 21.0 & 19.9 & 19.8 & \\
 100 & 11 35 55.93 & +42 58 44.80 & TXS 1133+432 & \phantom{00}1448.8 & \phantom{$-$}0.48 & C & & & & & & & & & & & & \\
 101 & 11 37 16.95 & +61 20 38.40 & 4C 61.23 & \phantom{00}1314.0 & $-$0.55 & II & c & 0.1110 & 0.0008 & S & 11 37 21.36 & +61 20 01.16 & 19.2 & 17.8 & 16.9 & 16.3 & 16.1 & 14.3\\
 102 & 11 40 27.69 & +12 03 07.60 & 4C 12.42 & \phantom{00}1527.0 & $-$0.62 & I & c & 0.0812 & 0.0002 & S & 11 40 27.74 & +12 03 08.32 & 17.7 & 15.7 & 14.8 & 14.4 & 14.1 & 11.6$^e$\\
 103 & 11 40 49.54 & +59 12 26.00 & 4C 59.16 & \phantom{00}2179.4 & $-$0.56 & C & & 0.8162 & 0.1229 & P & 11 40 49.57 & +59 12 25.60 & 23.7 & 22.3 & 21.9 & 21.0 & 20.4 & \\
 104 & 11 41 08.23 & +01 14 17.70 & 4C 01.32 & \phantom{00}2690.8 & $-$0.60 & II & c & 0.4430 & 0.0020 & S & 11 41 08.31 & +01 14 18.17 & 21.0 & 20.5 & 19.2 & 18.5 & 18.1 & \\
\hline
 \end{tabular}
\end{sidewaystable}

\begin{sidewaystable}
 \centering
 \small
 \begin{tabular}{lllllllllllllllllll}
 \multicolumn{19}{c}{CoNFIG-1 Data Table}\\
 (1) & \multicolumn{2}{c}{(2)} & (3) & (4) &(5)& \multicolumn{2}{c}{(6)} & (7) & (8) & (9) & \multicolumn{2}{c}{(10)} & (11) & (12) & (13) & (14) & (15) & (16) \\
 \hline
 148 & 12 54 11.68 & +27 37 32.70 & 3C 277.3 & \phantom{00}2923.9 & $-$0.68 & II & c & 0.0858 & 0.0001 & S & 12 54 12.01 & +27 37 33.90 & 18.4 & 16.5 & 15.6 & 15.1 & 14.8 & \\
 149 & 12 56 11.15 & $-$05 47 20.10 & 3C 279 & \phantom{00}9711.2 & $-$0.37 & C* & & 0.5362 & 0.0004 & S & 12 56 11.17 & $-$05 47 21.50 & & & & & & 10.9\\
 150 & 12 56 57.38 & +47 20 19.80 & 3C 280 & \phantom{00}5099.6 & $-$0.75 & II & c & 0.9960 & & S & 12 56 57.77 & +47 20 19.90 & 22.1 & 21.7 & 21.4 & 20.7 & 19.8 & \\
 151 & 13 00 32.87 & +40 09 09.20 & 3C 280.1 & \phantom{00}1368.9 & $-$1.05 & II & c & 1.6670 & 0.0020 & S & 13 00 33.30 & +40 09 07.70 & 19.0 & 18.8 & 18.8 & 18.4 & 18.4 & \\
 152 & 13 05 36.05 & +08 55 15.90 & 4C 09.45 & \phantom{00}1461.8 & $-$0.94 & II & c & 1.4090 & 0.0020 & S & 13 05 35.92 & +08 55 14.95 & 21.6 & 21.8 & 21.4 & 21.3 & 22.5 & \\
 153 & 13 09 49.66 & $-$00 12 36.60 & 4C 00.46 & \phantom{00}1636.7 & $-$0.85 & II & p & 0.4190 & 0.0010 & S & 13 09 49.75 & $-$00 12 35.60 & 21.0 & 20.1 & 19.4 & 18.8 & 18.4 & \\
\hline
 \end{tabular}
\end{sidewaystable}

\begin{sidewaystable}
 \centering
 \small
 \begin{tabular}{lllllllllllllllllll}
 \multicolumn{19}{c}{CoNFIG-1 Data Table}\\
 (1) & \multicolumn{2}{c}{(2)} & (3) & (4) &(5)& \multicolumn{2}{c}{(6)} & (7) & (8) & (9) & \multicolumn{2}{c}{(10)} & (11) & (12) & (13) & (14) & (15) & (16) \\
 \hline
 197 & 14 43 01.45 & +52 01 38.20 & 3C 303 & \phantom{00}2543.0 & $-$0.73 & II & c & 0.1410 & 0.0010 & S & 14 43 00.99 & +52 01 39.40 & 26.0 & 21.4 & 20.8 & 20.7 & 20.5 & \\
 198 & 14 45 16.48 & +09 58 36.00 & TXS 1442+101 & \phantom{00}2417.6 & \phantom{$-$}0.21 & C & & 3.5295 & 0.0007 & S & 14 45 16.47 & +09 58 36.12 & 19.4 & 18.5 & 17.8 & 17.8 & 17.7 & 15.8\\
 199 & 14 48 39.98 & +00 18 17.90 & 4C 00.52 & \phantom{00}1651.5 & $-$0.65 & U & & 0.4381 & 0.0003 & S & 14 48 39.94 & +00 18 16.70 & 23.0 & 20.7 & 19.0 & 18.3 & 17.9 & \\
 200 & 14 49 21.74 & +63 16 13.90 & 3C 305 & \phantom{00}3006.0 & $-$0.81 & I & c & 0.0416 & 0.0001 & S & 14 49 21.57 & +63 16 13.98 & 16.0 & 14.2 & 13.5 & 13.1 & 12.8 & 10.6$^e$\\
 201 & 14 55 01.43 & $-$04 20 22.50 & 4C -04.53 & \phantom{00}2104.3 & $-$0.81 & II & c & 0.4403 & & S & 14 55 01.43 & $-$04 21 04.50 & & & & & & \\
 202 & 15 04 09.27 & +60 00 55.50 & 3C 311 & \phantom{00}1553.1 & $-$0.78 & C & & 1.0220 & & S & 15 04 09.22 & +60 00 55.37 & 20.3 & 20.2 & 19.9 & 19.5 & 19.9 & \\
\hline
 \end{tabular}
\end{sidewaystable}

\begin{sidewaystable}
 \centering
 \small
 \begin{tabular}{lllllllllllllllllll}
 \multicolumn{19}{c}{CoNFIG-1 Data Table}\\
 (1) & \multicolumn{2}{c}{(2)} & (3) & (4) &(5)& \multicolumn{2}{c}{(6)} & (7) & (8) & (9) & \multicolumn{2}{c}{(10)} & (11) & (12) & (13) & (14) & (15) & (16) \\
 \hline
 246 & 16 25 57.66 & +41 34 41.20 & 4C 41.32 & \phantom{00}1677.4 & $-$0.20 & C* & & 2.5500 & & S & 16 25 57.68 & +41 34 40.48 & 24.5 & 22.7 & 22.0 & 21.6 & 21.4 & \\
 247 & 16 28 03.57 & +27 41 36.10 & 3C 341 & \phantom{00}1998.6 & $-$0.82 & II & c & 0.4480 & & S & 16 28 03.98 & +27 41 39.34 & 21.4 & 20.6 & 19.5 & 18.5 & 18.3 & \\
 248 & 16 28 38.34 & +39 33 04.70 & 3C 338 & \phantom{00}3678.7 & $-$1.23 & I$^w$ & c & 0.0298 & & S & 16 28 38.23 & +39 33 04.38 & 15.8 & 13.9 & 13.0 & 12.5 & 12.3 & \phantom{0}9.2$^e$\\
 249 & 16 28 53.30 & +44 19 05.20 & 3C 337 & \phantom{00}3155.8 & $-$0.67 & II & c & 0.6300 & & S & 16 28 52.58 & +44 19 06.80 & 24.0 & 23.0 & 21.5 & 20.2 & 19.5 & \\
 250 & 16 29 37.52 & +23 20 13.40 & 3C 340 & \phantom{00}2599.0 & $-$0.68 & II & c & 0.7750 & & S & 16 29 36.59 & +23 20 12.77 & 23.3 & 22.3 & 21.4 & 20.7 & 19.6 & \\
 251 & 16 31 45.29 & +11 56 03.30 & 4C 12.59 & \phantom{00}1733.7 & $-$0.47 & C & & 1.7920 & & S & 16 31 45.17 & +11 56 03.34 & 18.6 & 18.4 & 18.3 & 18.1 & 17.9 & \\
\hline
 \end{tabular}
\end{sidewaystable}

\begin{sidewaystable}
 \centering
 \small
 \begin{tabular}{lllllllllllllllllll}
 \multicolumn{19}{c}{CoNFIG-2 Data Table}\\
 (1) & \multicolumn{2}{c}{(2)} & (3) & (4) &(5)& \multicolumn{2}{c}{(6)} & (7) & (8) & (9) & \multicolumn{2}{c}{(10)} & (11) & (12) & (13) & (14) & (15) & (16) \\
 \hline
 1 & 09 20 11.16 & +17 53 25.00 & 4C 18.29 & \phantom{0}1070.6 & $-$0.88 & C & & 0.4155 & 0.3312 & P & 09 20 11.12 & +17 53 24.76 & 22.9 & 23.3 & 22.3 & 21.4 & 21.0 & \\
 2 & 09 20 58.48 & +44 41 53.70 & B0917+449 & \phantom{0}1017.2 & $-$0.15 & C* & & 2.1890 & 0.0017 & S & 09 20 58.46 & +44 41 54.02 & 18.6 & 18.0 & 17.9 & 17.7 & 17.4 & 15.4\\
 4 & 09 21 46.55 & +37 54 10.10 & 4C 38.29 & \phantom{00}826.4 & $-$0.88 & II$^c$ & c & 1.1080 & 0.0010 & S & 09 21 46.51 & +37 54 08.71 & 20.7 & 19.7 & 18.8 & 18.4 & 18.3 & 15.4\\
 8 & 09 30 54.27 & +58 55 16.60 & 4C 59.10 & \phantom{0}1082.9 & $-$0.91 & U & & 0.5128 & 0.1626 & P & 09 30 54.49 & +58 55 17.83 & 22.5 & 21.8 & 21.0 & 20.6 & 20.5 & \\
 9 & 09 34 15.80 & +49 08 21.00 & 0930+493 & \phantom{00}800.5 & \phantom{$-$}0.37 & C* & & 2.5820 & 0.0090 & S & 09 34 15.77 & +49 08 21.84 & 19.9 & 19.3 & 19.1 & 19.1 & 19.0 & \\
 10 & 09 35 04.06 & +08 41 37.30 & 4C 08.31 & \phantom{0}1037.6 & $-$0.74 & II & p & 0.3578 & 0.0312 & P & 09 35 04.06 & +08 41 37.86 & 24.5 & 21.4 & 19.5 & 19.0 & 18.4 & \\
\hline
 \end{tabular}
\end{sidewaystable}

\begin{sidewaystable}
 \centering
 \small
 \begin{tabular}{lllllllllllllllllll}
 \multicolumn{19}{c}{CoNFIG-2 Data Table}\\
 (1) & \multicolumn{2}{c}{(2)} & (3) & (4) &(5)& \multicolumn{2}{c}{(6)} & (7) & (8) & (9) & \multicolumn{2}{c}{(10)} & (11) & (12) & (13) & (14) & (15) & (16) \\
 \hline
 83 & 10 51 48.80 & +21 19 52.80 & 4C 21.28 & \phantom{0}1253.1 & $-$0.29 & C* & & 1.3000 & 0.0050 & S & 10 51 48.79 & +21 19 52.32 & 18.6 & 18.8 & 18.3 & 18.4 & 18.5 & 15.7\\
 85 & 10 57 15.77 & +00 12 03.80 & TXS 1054+004 & \phantom{00}898.1 & $-$0.43 & C* & & & & & 10 57 15.77 & +00 12 03.70 & 24.9 & 24.3 & 23.2 & 21.8 & 21.1 & \\
 89 & 11 00 02.02 & +30 27 42.00 & 3C 248 & \phantom{00}991.4 & $-$1.05 & II & c & 0.8313 & 0.1300 & P & 11 00 04.62 & +30 27 07.75 & 21.7 & 21.5 & 20.9 & 20.5 & 19.7 & \\
 91 & 11 02 24.97 & $-$02 35 34.10 & 4C-02.44 & \phantom{00}838.4 & $-$0.72 & C & & & & & & & & & & & & \\
 92 & 11 02 26.19 & +55 50 03.30 & 4C 56.18 & \phantom{0}1206.4 & $-$0.93 & U & & 0.4059 & 0.0453 & P & 11 02 26.16 & +55 50 05.96 & 24.8 & 21.6 & 20.0 & 19.3 & 19.0 & \\
 93 & 11 05 26.17 & +20 52 17.40 & TXS 1102+211 & \phantom{00}985.6 & $-$0.66 & C & & & & & & & & & & & & \\
\hline
 \end{tabular}
\end{sidewaystable}

\begin{sidewaystable}
 \centering
 \small
 \begin{tabular}{lllllllllllllllllll}
 \multicolumn{19}{c}{CoNFIG-2 Data Table}\\
 (1) & \multicolumn{2}{c}{(2)} & (3) & (4) &(5)& \multicolumn{2}{c}{(6)} & (7) & (8) & (9) & \multicolumn{2}{c}{(10)} & (11) & (12) & (13) & (14) & (15) & (16) \\
 \hline
171 & 12 03 21.95 & +04 14 17.70 & AO 1200+04 & \phantom{0}1146.1 & $-$0.21 & C* & & 1.2118 & 0.0014 & S & 12 03 21.94 & +04 14 19.07 & 18.9 & 18.8 & 18.5 & 18.4 & 18.5 & \\
173 & 12 04 18.46 & +52 02 18.80 & WSTB 14W50 & \phantom{00}962.3 & $-$0.28 & C & & 0.6510 & 0.2832 & P & 12 04 18.37 & +52 02 18.92 & 26.4 & 22.6 & 21.4 & 21.9 & 21.6 & \\
174 & 12 04 52.31 & +29 29 12.40 & 4C 29.46 & \phantom{0}1230.6 & $-$0.77 & II & c & 0.8608 & 0.0622 & P & 12 04 52.78 & +29 30 04.28 & 20.7 & 19.9 & 19.2 & 18.4 & 17.6 & 14.2\\
180 & 12 14 47.73 & $-$01 00 12.10 & 4C-00.48 & \phantom{0}1092.5 & $-$0.84 & II & c & 1.4449 & 0.1148 & R & 12 14 48.71 & $-$01 00 16.30 & 24.2 & 22.8 & 22.5 & 22.6 & 21.6 & \\
181 & 12 15 14.69 & +17 30 02.20 & 4C 17.54 & \phantom{0}1010.2 & $-$0.68 & S* & & 0.5831 & 0.2324 & P & 12 15 14.72 & +17 30 02.34 & 22.1 & 22.2 & 21.4 & 21.2 & 20.3 & \\
185 & 12 17 56.90 & +25 29 27.20 & 3C 269 & \phantom{00}866.8 & $-$0.83 & II & c & 0.3056 & 0.0011 & I & 12 17 56.88 & +25 29 27.56 & 18.5 & 18.1 & 18.1 & 18.1 & 18.1 & 15.9\\
\hline
 \end{tabular}
\end{sidewaystable}

\begin{sidewaystable}
 \centering
 \small
 \begin{tabular}{lllllllllllllllllll}
 \multicolumn{19}{c}{CoNFIG-3 Data Table}\\
 (1) & \multicolumn{2}{c}{(2)} & (3) & (4) &(5)& \multicolumn{2}{c}{(6)} & (7) & (8) & (9) & \multicolumn{2}{c}{(10)} & (11) & (12) & (13) & (14) & (15) & (16) \\
 \hline
 1 & 14 41 28.26 & +10 55 24.70 & J1441+1055 & \phantom{0}422.7 & $-$0.78 & C & & & & & 14 41 28.47 & +10 55 25.18 & 23.2 & 24.5 & 23.9 & 21.6 & 20.6 & \\
 2 & 14 42 03.43 & +25 05 04.30 & TXS 1439+252 & \phantom{0}204.7 & & II & c & 0.824 & 0.1123 & P & 14 42 04.01 & +25 04 13.20 & 22.7 & 22.7 & 21.5 & 20.7 & 19.7 & \\
 3 & 14 42 03.93 & +13 29 17.60 & 4C 13.53 & \phantom{0}706.0 & $-$0.41 & C & & & & & 14 42 04.04 & +13 29 16.04 & 18.7 & 18.5 & 18.2 & 18.0 & 17.9 & \\
 4 & 14 42 22.99 & +14 54 58.20 & TXS 1440+151 & \phantom{0}202.6 & $-$0.97 & II & c & & & & & & & & & & & \\
 5 & 14 42 27.11 & +14 31 39.90 & TXS 1440+147 & \phantom{0}224.1 & $-$0.95 & II & c & 1.0425 & 0.3905 & P & 14 42 27.23 & +14 31 38.82 & 22.3 & 22.4 & 21.9 & 22.8 & 20.9 & \\
 6 & 14 42 48.37 & +11 44 15.00 & TXS 1440+119 & \phantom{0}200.7 & & II & c & 0.6102 & 0.0882 & P & 14 42 48.36 & +11 44 15.72 & 23.5 & 22.6 & 21.5 & 20.4 & 20.3 & \\
\hline
 \end{tabular}
\end{sidewaystable}

\begin{sidewaystable}
 \centering
 \small
 \begin{tabular}{lllllllllllllllllll}
 \multicolumn{19}{c}{CoNFIG-3 Data Table}\\
 (1) & \multicolumn{2}{c}{(2)} & (3) & (4) &(5)& \multicolumn{2}{c}{(6)} & (7) & (8) & (9) & \multicolumn{2}{c}{(10)} & (11) & (12) & (13) & (14) & (15) & (16) \\
 \hline
 51 & 14 54 20.86 & +16 24 25.10 & B1452+166 & \phantom{0}228.4 & & C* & & & & & 14 54 20.85 & +16 24 24.41 & 19.3 & 18.9 & 18.4 & 18.2 & 18.0 & \\
 52 & 14 54 22.75 & +25 39 55.50 & TXS 1452+258 & \phantom{0}232.4 & $-$0.71 & II & p & 0.2186 & 0.0865 & P & 14 54 22.84 & +25 39 55.73 & 22.4 & 20.9 & 19.7 & 19.3 & 19.0 & \\
 53 & 14 54 29.16 & +20 15 26.10 & TXS 1452+204 & \phantom{0}237.5 & $-$1.03 & II & c & & & & & & & & & & & \\
 54 & 14 54 32.30 & +29 55 58.50 & TXS 1452+301 & \phantom{0}730.7 & $-$0.54 & C* & & 0.5800 & & S & 14 54 32.30 & +29 55 58.01 & 20.5 & 20.3 & 20.3 & 19.6 & 19.5 & \\
 55 & 14 54 42.58 & +27 32 11.30 & TXS 1452+277 & \phantom{0}211.1 & $-$0.80 & II & c & 0.5321 & 0.0363 & P & 14 54 42.23 & +27 32 11.84 & 26.3 & 22.3 & 20.2 & 19.3 & 19.1 & \\
 56 & 14 55 07.32 & +14 12 22.20 & TXS 1452+144 & \phantom{0}246.9 & $-$0.76 & II & c & 0.2538 & 0.0729 & P & 14 55 07.36 & +14 12 21.96 & 24.5 & 22.1 & 21.0 & 20.2 & 19.7 & \\
\hline
 \end{tabular}
\end{sidewaystable}

\begin{sidewaystable}
 \centering
 \small
 \begin{tabular}{lllllllllllllllllll}
 \multicolumn{19}{c}{CoNFIG-3 Data Table}\\
 (1) & \multicolumn{2}{c}{(2)} & (3) & (4) &(5)& \multicolumn{2}{c}{(6)} & (7) & (8) & (9) & \multicolumn{2}{c}{(10)} & (11) & (12) & (13) & (14) & (15) & (16) \\
 \hline
101 & 15 05 00.08 & +11 58 43.90 & TXS 1502+121 & \phantom{0}205.9 & $-$0.91 & C & & & & & 15 05 00.11 & +11 58 44.08 & 20.8 & 20.5 & 20.3 & 20.0 & 20.0 & \\
102 & 15 06 19.33 & +20 27 41.50 & TXS 1504+206 & \phantom{0}294.9 & $-$0.96 & II & c & 0.4960 & 0.0402 & P & 15 06 19.40 & +20 27 42.73 & 26.5 & 21.9 & 20.3 & 19.3 & 18.9 & \\
103 & 15 07 09.23 & +16 07 16.70 & WB 1504+1618 & \phantom{0}513.1 & $-$0.93 & II & c & 0.6007 & 0.0074 & I & 15 07 09.43 & +16 07 16.68 & 24.8 & 21.3 & 20.2 & 19.8 & 19.5 & \\
104 & 15 07 21.88 & +10 18 46.30 & B1507+105 & \phantom{0}403.2 & $-$0.43 & C & & 0.0781 & 0.0002 & S & 15 07 21.88 & +10 18 44.96 & 18.5 & 16.2 & 15.2 & 14.8 & 14.5 & 13.0\\
105 & 15 07 39.40 & +11 04 01.00 & GB6 B1505+113& \phantom{0}226.4 & $-$0.65 & II & c & 0.4755 & 0.0001 & S & 15 07 39.50 & +11 04 03.72 & 18.3 & 17.9 & 17.9 & 17.6 & 17.4 & 15.0\\
106 & 15 07 41.89 & +12 08 04.30 & 4C 12.54 & \phantom{0}421.0 & $-$1.04 & II & c & & & & & & & & & & & \\
\hline
 \end{tabular}
\end{sidewaystable}

\begin{sidewaystable}
 \centering
 \small
 \begin{tabular}{lllllllllllllllllll}
 \multicolumn{19}{c}{CoNFIG-3 Data Table}\\
 (1) & \multicolumn{2}{c}{(2)} & (3) & (4) &(5)& \multicolumn{2}{c}{(6)} & (7) & (8) & (9) & \multicolumn{2}{c}{(10)} & (11) & (12) & (13) & (14) & (15) & (16) \\
 \hline
151 & 15 17 24.70 & +17 29 28.30 & TXS 1515+176 & \phantom{0}226.0 & $-$0.42 & II & c & 0.2419 & 0.0388 & P & 15 17 25.01 & +17 29 21.37 & 19.9 & 19.2 & 18.2 & 17.8 & 17.5 & 15.3\\
152 & 15 17 49.52 & +14 27 59.30 & TXS 1515+146 & \phantom{0}201.8 & $-$0.71 & II & c & & & & & & & & & & & \\
153 & 15 17 51.37 & +10 00 26.90 & 4C 10.41 & \phantom{0}229.5 & $-$1.48 & II & c & 0.8129 & 0.0725 & P & 15 17 51.03 & +09 59 36.72 & 22.7 & 21.9 & 21.1 & 20.3 & 19.3 & \\
154 & 15 17 53.32 & +26 48 39.50 & TXS 1515+269 & \phantom{0}284.1 & $-$0.67 & II & c & 0.9560 & 0.0389 & I & 15 17 53.24 & +26 48 41.69 & 25.6 & 25.3 & 22.1 & 21.0 & 19.8 & \\
155 & 15 18 05.11 & +19 42 43.90 & TXS 1515+198 & \phantom{0}423.2 & $-$1.00 & II & p & 0.7581 & 0.1275 & P & 15 18 05.15 & +19 42 44.42 & 22.0 & 21.5 & 21.0 & 20.4 & 19.8 & \\
156 & 15 18 14.06 & +15 49 32.60 & TXS 1515+160 & \phantom{0}219.3 & $-$0.93 & II & c & 1.1964 & 0.0567 & I & 15 18 13.99 & +15 49 32.04 & 23.0 & 23.8 & 22.8 & 21.6 & 20.5 & \\
\hline
 \end{tabular}
\end{sidewaystable}

\begin{sidewaystable}
 \centering
 \small
 \begin{tabular}{lllllllllllllllllll}
 \multicolumn{19}{c}{CoNFIG-3 Data Table}\\
 (1) & \multicolumn{2}{c}{(2)} & (3) & (4) &(5)& \multicolumn{2}{c}{(6)} & (7) & (8) & (9) & \multicolumn{2}{c}{(10)} & (11) & (12) & (13) & (14) & (15) & (16) \\
 \hline
201 & 15 32 28.50 & +24 15 29.90 & J153233.19 & \phantom{0}328.5 & & II & c & 0.5640 & & S & 15 32 33.18 & +24 15 25.81 & 20.6 & 20.2 & 19.7 & 18.6 & 18.3 & \\
202 & 15 32 34.03 & +20 06 38.80 & 4C 20.36 & \phantom{0}577.7 & $-$0.88 & II & c & & & & & & & & & & & \\
203 & 15 32 44.30 & +28 03 46.40 & B2 1530+28 & \phantom{0}381.0 & $-$0.43 & I & c & 0.0732 & & S & 15 32 44.74 & +28 03 47.30 & 19.3 & 17.4 & 16.5 & 16.1 & 15.7 & 13.5$^e$\\
204 & 15 32 50.67 & +22 41 35.20 & TXS 1530+228 & \phantom{0}222.9 & $-$0.61 & C & & & & & & & & & & & & \\
205 & 15 32 58.12 & +15 56 05.80 & TXS 1530+161 & \phantom{0}284.5 & $-$0.88 & II & c & & & & & & & & & & & \\
206 & 15 33 14.49 & +15 16 42.20 & 4C 15.49 & \phantom{0}947.5 & $-$0.69 & C & & & & & & & & & & & & \\
\hline
 \end{tabular}
\end{sidewaystable}

\begin{sidewaystable}
 \centering
 \small
 \begin{tabular}{lllllllllllllllllll}
 \multicolumn{19}{c}{CoNFIG-3 Data Table}\\
 (1) & \multicolumn{2}{c}{(2)} & (3) & (4) &(5)& \multicolumn{2}{c}{(6)} & (7) & (8) & (9) & \multicolumn{2}{c}{(10)} & (11) & (12) & (13) & (14) & (15) & (16) \\
 \hline
251 & 15 50 38.88 & +18 39 01.20 & TXS 1548+188 & \phantom{0}214.5 & $-$0.94 & II & p & & & & & & & & & & & \\
252 & 15 50 43.51 & +11 20 58.50 & 4C 11.50 & \phantom{0}828.3 & $-$0.74 & I & p & 0.4360 & & S & 15 50 43.59 & +11 20 47.47 & 18.6 & 18.5 & 18.6 & 18.0 & 17.9 & 15.1\\
253 & 15 51 11.66 & +26 06 14.30 & TXS 1549+262 & \phantom{0}455.3 & $-$0.88 & II & c & 0.6965 & 0.0941 & P & 15 51 11.61 & +26 06 12.67 & 25.1 & 22.8 & 21.6 & 20.4 & 20.0 & \\
255 & 15 51 36.81 & +10 36 22.70 & TXS 1549+107 & \phantom{0}530.7 & $-$0.66 & II & c & 0.3984 & 0.0204 & P & 15 51 40.30 & +10 35 48.66 & 22.5 & 19.6 & 18.0 & 17.3 & 16.8 & \\
256 & 15 51 45.36 & +12 43 25.10 & TXS 1549+128 & \phantom{0}250.1 & $-$0.93 & C & & & & & & & & & & & & \\
257 & 15 52 10.54 & +18 42 07.80 & TXS 1549+188 & \phantom{0}274.7 & $-$0.81 & U & & 0.5555 & 0.0510 & P & 15 52 10.47 & +18 42 05.26 & 25.0 & 23.2 & 21.1 & 20.1 & 19.6 & \\
\hline
 \end{tabular}
\end{sidewaystable}

\begin{sidewaystable}
 \centering
 \small
 \begin{tabular}{lllllllllllllllllll}
 \multicolumn{19}{c}{CoNFIG-4 Data Table}\\
 (1) & \multicolumn{2}{c}{(2)} & (3) & (4) &(5)& \multicolumn{2}{c}{(6)} & (7) & (8) & (9) & \multicolumn{2}{c}{(10)} & (11) & (12) & (13) & (14) & (15) & (16) \\
 \hline
 1 & 14 08 18.51 & +00 13 09.50 & J140818+00 & 113.8 & $-$0.93 & C & & 1.2377 & 0.0050 & S & 14 08 18.46 & +00 13 09.26 & 21.8 & 21.5 & 20.9 & 21.0 & 20.9 & \\
 2 & 14 08 28.14 & +02 25 48.70 & TXS 1405+026 & 243.7 & $-$0.64 & I & p & 0.1782 & 0.0438 & P & 14 08 28.42 & +02 25 50.34 & 20.8 & 19.4 & 18.4 & 17.9 & 17.5 & \\
 3 & 14 08 32.29 & +00 31 33.90 & 1408+0050 & \phantom{0}61.2 & & II & c & 1.6734 & 0.0025 & S & 14 08 32.65 & +00 31 38.53 & 19.8 & 19.7 & 19.6 & 19.4 & 19.4 & \\
 4 & 14 08 32.70 & $-$01 31 19.50 & J140832+00 & \phantom{0}73.3 & & C & & & & & & & & & & & & \\
 5 & 14 08 33.31 & +01 16 22.10 & TXS 1406+015 & 601.3 & $-$0.60 & U & & & & & & & & & & & & \\
 6 & 14 08 37.71 & +02 48 29.40 & 1408+0281 & \phantom{0}77.0 & & II & c & & & & & & & & & & & \\
\hline
 \end{tabular}
\end{sidewaystable}

\begin{sidewaystable}
 \centering
 \small
 \begin{tabular}{lllllllllllllllllll}
 \multicolumn{19}{c}{CoNFIG-4 Data Table}\\
 (1) & \multicolumn{2}{c}{(2)} & (3) & (4) &(5)& \multicolumn{2}{c}{(6)} & (7) & (8) & (9) & \multicolumn{2}{c}{(10)} & (11) & (12) & (13) & (14) & (15) & (16) \\
 \hline
 51 & 14 15 30.47 & +02 23 01.30 & TXS 1412+026 & 153.9 & $-$0.79 & II & p & & & & & & & & & & & \\
 52 & 14 15 53.96 & +02 10 31.90 & 1415+0217 & \phantom{0}86.5 & & C & & 0.5991 & 0.0753 & P & 14 15 53.89 & +02 10 34.14 & 22.2 & 22.7 & 20.4 & 19.7 & 19.6 & \\
 53 & 14 16 04.11 & +00 29 15.40 & TXS 1413+007 & 107.2 & $-$1.11 & II & c & 0.9250 & 0.0702 & Z & 14 16 04.22 & +00 29 13.99 & 25.0 & 24.1 & 24.3 & 24.0 & 20.5 & \\
 54 & 14 16 09.42 & $-$01 25 13.60 & TXS 1413-011 & 114.2 & $-$0.67 & II & p & 0.6789 & 0.0838 & P & 14 16 09.66 & $-$01 25 14.74 & 23.5 & 23.7 & 21.7 & 20.6 & 19.7 & \\
 55 & 14 16 13.74 & +02 19 22.50 & 1416+0219 & 256.2 & & I$^w$ & c & 0.1583 & 0.0011 & S & 14 16 13.37 & +02 19 07.82 & 18.9 & 17.8 & 16.8 & 16.3 & 16.1 & 13.1$^e$\\
 56 & 14 16 43.04 & $-$02 56 11.30 & J141643-02 & \phantom{0}93.6 & & C & & 0.1330 & 0.0002 & S & 14 16 43.03 & $-$02 56 11.98 & 20.4 & 18.3 & 17.3 & 16.8 & 16.4 & 13.9$^e$\\
\hline
 \end{tabular}
\end{sidewaystable}

\begin{sidewaystable}
 \centering
 \small
 \begin{tabular}{lllllllllllllllllll}
 \multicolumn{19}{c}{CoNFIG-4 Data Table}\\
 (1) & \multicolumn{2}{c}{(2)} & (3) & (4) &(5)& \multicolumn{2}{c}{(6)} & (7) & (8) & (9) & \multicolumn{2}{c}{(10)} & (11) & (12) & (13) & (14) & (15) & (16) \\
 \hline
101 & 14 26 49.60 & $-$00 47 18.30 & J142649-00 & \phantom{0}87.7 & & C & & 1.8200 & 0.0050 & S & 14 26 49.59 & $-$00 47 17.84 & 19.3 & 19.2 & 19.3 & 19.1 & 19.2 & \\
102 & 14 26 49.84 & +00 55 59.90 & 1426+0093 & \phantom{0}57.9 & & II & p & & & & & & & & & & & \\
103 & 14 26 55.09 & $-$02 49 20.90 & J142655-02 & \phantom{0}60.3 & & C & & 2.2350 & 0.0050 & S & 14 26 55.11 & $-$02 49 20.46 & 20.2 & 19.6 & 19.4 & 19.2 & 18.9 & \\
104 & 14 26 55.47 & $-$02 15 45.10 & 1426-0226 & 167.1 & & C & & & & & 14 26 55.48 & $-$02 15 44.21 & 22.3 & 22.1 & 21.3 & 20.9 & 20.1 & \\
105 & 14 27 11.20 & $-$01 52 12.70 & 1427-0187 & \phantom{0}52.0 & $-$1.51 & U & & & & & & & & & & & & \\
106 & 14 27 39.70 & +00 21 44.60 & TXS 1425+005 & 110.3 & $-$1.28 & II & c & & & & & & & & & & & \\
\hline
 \end{tabular}
\end{sidewaystable}

\begin{sidewaystable}
 \centering
 \small
 \begin{tabular}{lllllllllllllllllll}
 \multicolumn{19}{c}{CoNFIG-4 Data Table}\\
 (1) & \multicolumn{2}{c}{(2)} & (3) & (4) &(5)& \multicolumn{2}{c}{(6)} & (7) & (8) & (9) & \multicolumn{2}{c}{(10)} & (11) & (12) & (13) & (14) & (15) & (16) \\
 \hline
151 & 14 34 52.84 & +02 36 03.00 & TXS 1432+028 & 322.1 & $-$0.81 & II & c & & & & & & & & & & & \\
152 & 14 35 21.30 & $-$02 40 51.70 & 1435-0268 & \phantom{0}62.9 & & I$^w$ & c & 0.4999 & 0.0242 & P & 14 35 21.02 & $-$02 40 56.70 & 26.3 & 21.9 & 19.9 & 18.9 & 18.4 & \\
153 & 14 35 23.20 & +02 25 42.80 & 1435+0243 & \phantom{0}71.5 & & II & c & 1.0272 & 0.0655 & I & 14 35 23.69 & +02 25 45.05 & 22.7 & 23.1 & 22.8 & 21.2 & 20.5 & \\
154 & 14 35 41.65 & $-$01 47 25.90 & TXS 1433-015 & 132.2 & $-$0.95 & II & c & 1.4496 & 0.1145 & I & 14 35 40.75 & $-$01 47 24.20 & 22.1 & 22.0 & 22.1 & 22.1 & 21.0 & \\
155 & 14 36 09.04 & +01 48 49.20 & 1436+0181 & \phantom{0}52.7 & & C & & & & & 14 36 09.20 & +01 48 47.74 & 21.7 & 21.4 & 21.0 & 20.6 & 20.4 & \\
156 & 14 36 30.26 & +00 35 19.80 & TXS 1433+008 & 123.0 & $-$0.43 & C & & & & & & & & & & & & \\
\hline
 \end{tabular}
\end{sidewaystable}

\begin{table}
  \subsection{Complementary Samples: 3CRR Sample}\label{3CR}
  \begin{center}
    \small
    \begin{tabular}{cclcclc}
      \multicolumn{7}{c}{3CRR Data Table}\\
      \hline
      \multicolumn{2}{c}{Radio position (J2000)}& Name & \SoneM &
      $\alpha$ & Type & redsh. \\
      RA & DEC & & (Jy) & & & \\
      \hline
      00 38 13.76 &  +32 53 39.9 & 3C19    & \phantom{00}3.6 & $-$0.63 & II & 0.4820\\
      00 40 20.08 &  +51 47 10.2 & 3C20    & \phantom{0}11.0 & $-$0.66 & II & 0.3500\\
      01 04 39.10 &  +32 08 43.3 & 3C31    & \phantom{00}5.2 & $-$0.57 & I  & 0.0167\\
      01 06 14.54 &  +13 04 14.8 & 3C33    & \phantom{0}11.3 & $-$0.76 & II & 0.0595\\
      01 06 06.48 &  +72 55 59.2 & 3C33.1  & \phantom{00}3.6 & $-$0.62 & II & 0.1810\\
\hline
    \end{tabular}
  \end{center}

  \subsection{Complementary Samples: CENSORS}\label{CENSORSsamp}
  \begin{center}
    \small
    \begin{tabular}{cclclc}
      \multicolumn{6}{c}{CENSORS Data Table}\\
      \hline
      Name & \multicolumn{2}{c}{Radio position (J2000)}& \SoneG &
      Type & redsh. \\
      & RA & DEC & & (mJy) & \\    
      \hline
       CENSORS-001 & 09 51 29.07 & -20 50 30.10 & 659.5 & II & 1.1550\\
       CENSORS-002 & 09 46 50.21 & -20 20 44.40 & 452.3 & C  & 0.9130\\
       CENSORS-003 & 09 50 31.39 & -21 02 44.80 & 355.3 & C  & 0.7900\\
       CENSORS-004 & 09 49 53.60 & -21 56 18.40 & 283.0 & II & 1.0130\\
       CENSORS-005 & 09 53 44.42 & -21 36 02.50 & 244.7 & II & 1.5880\\
\hline
    \end{tabular}
  \end{center}
\end{table}

\begin{table}
  \begin{center}
    \small
    \begin{tabular}{cclclc}
      \multicolumn{6}{c}{CENSORS Data Table}\\
      \hline
      Name & \multicolumn{2}{c}{Radio position (J2000)}& \SoneG &
      Type & redsh. \\
      & RA & DEC & & (mJy) & \\    
      \hline
       CENSORS-033 & 09 53 04.71 & -20 44 09.80 &\phantom{0}34.3 & II & 1.2030\\
       CENSORS-034 & 09 47 53.55 & -21 47 19.60 &\phantom{0}34.2 & C  & 1.3190\\
       CENSORS-035 & 09 54 52.43 & -21 19 29.00 &\phantom{0}34.1 & II & 0.4730\\
       CENSORS-036 & 09 49 33.23 & -21 27 08.30 &\phantom{0}32.3 & C  & 1.4850\\
       CENSORS-037 & 09 49 19.44 & -21 51 35.40 &\phantom{0}31.8 & II & 0.5110\\
\hline
    \end{tabular}
  \end{center}
\end{table}

\begin{table}
  \begin{center}
    \small
    \begin{tabular}{cclclc}
      \multicolumn{6}{c}{CENSORS Data Table}\\
      \hline
      Name & \multicolumn{2}{c}{Radio position (J2000)}& \SoneG &
      Type & redsh. \\
      & RA & DEC & & (mJy) & \\    
      \hline
       CENSORS-097 & 09 54 36.32 & -21 44 26.60 &\phantom{0}12.0 & II & 1.6350\\
       CENSORS-098 & 09 49 35.13 & -21 58 10.50 &\phantom{0}11.8 & U  & 1.6350\\
       CENSORS-099 & 09 57  2.25 & -21 56 51.80 &\phantom{0}11.6 & C  & 0.7380\\
       CENSORS-100 & 09 50 48.57 & -21 54 57.10 &\phantom{0}11.5 & II & 1.2880\\
       CENSORS-101 & 09 52 50.38 & -21 31 48.00 &\phantom{0}11.4 & U  & 1.0430\\
\hline
    \end{tabular}
  \end{center}
\end{table}

%---------------------------------------------------------
\begin{table}
  \subsection{Complementary Samples: Lynx \& Hercules Sample}\label{EmmaSamp}
  \begin{center}
    \small
    \begin{tabular}{cclclc}
      \multicolumn{6}{c}{Lynx \& Hercules Data Table}\\
      \hline
      \multicolumn{2}{c}{Radio position (J2000)}& Name & \SoneG &
      Type & redsh. \\
      RA & DEC & & (mJy) & & \\          
      \hline
      08 43 40.72 &  +44 39 24.7 & 60w067   & \phantom{0}1.4 & I  & 1.800\\
      08 43 46.86 &  +44 35 49.7 & 60w071   & \phantom{0}0.9 & II & 1.250\\
      08 43 52.89 &  +44 24 29.0 & 60w084   & \phantom{0}1.8 & I  & 0.127\\
      08 44 03.58 &  +44 38 10.2 & 55w165a  & \phantom{0}0.6 & I  & 0.680\\
      08 44 04.06 &  +44 31 19.4 & 55w116   & \phantom{0}1.2 & I  & 0.851\\
\hline
    \end{tabular}
  \end{center}
\end{table}

\begin{table}
  \begin{center}
  \small
  \begin{tabular}{cclclc}
    \multicolumn{6}{c}{Lynx \& Hercules Data Table}\\          
      \hline
      \multicolumn{2}{c}{Radio position (J2000)}& Name & \SoneG &
      Type & redsh. \\
      RA & DEC & & (mJy) & & \\            
      \hline
      17 20 55.82 &  +49 41 02.2 & 53w076   & \phantom{0}1.9 & I  & 0.390\\
      17 21 01.32 &  +49 48 34.0 & 53w077   & \phantom{0}6.5 & II & 0.800\\
      17 21 05.43 &  +49 56 56.0 & 66w047   & \phantom{0}0.6 & I  & 0.370\\
      17 21 11.25 &  +49 58 32.4 & 66w049   & \phantom{0}1.4 & II & 0.950\\
      17 21 18.17 &  +50 03 35.2 & 53w078   & \phantom{0}0.7 & I  & 0.270\\
      17 21 22.75 &  +50 10 31.0 & 53w079   &   11.7 & U  & 0.548\\
      17 21 37.48 &  +49 55 36.8 & 53w080   &   25.9 & II & 0.546\\
      17 21 37.86 &  +49 57 57.6 & 53w081   &   12.1 & U  & 2.060\\
      17 21 37.64 &  +50 08 27.4 & 53w082   & \phantom{0}2.5 & II & 2.040\\
      17 21 48.23 &  +49 47 07.3 & 66w058   & \phantom{0}1.9 & U  & 2.300\\
      17 21 48.95 &  +50 02 39.7 & 53w083   & \phantom{0}5.0 & U  & 0.628\\
      17 21 50.43 &  +49 48 30.5 & 53w084   & \phantom{0}0.7 & U  & 2.730\\
      17 21 52.48 &  +49 54 34.1 & 53w085   & \phantom{0}4.5 & I  & 1.350\\
      17 21 56.42 &  +49 53 39.8 & 53w086a  & \phantom{0}1.6 & I  & 0.460\\
      17 21 57.65 &  +49 53 33.8 & 53w086b  & \phantom{0}2.4 & I  & 0.730\\
      17 21 59.10 &  +50 08 42.9 & 53w087   & \phantom{0}5.6 & I  & 3.700\\
      17 21 58.90 &  +50 11 52.7 & 53w088   &   14.1 & I  & 1.773\\
      17 22 01.05 &  +50 06 54.7 & 53w089   & \phantom{0}3.0 & I  & 0.635\\       
    \hline
  \end{tabular}
    \end{center}
\end{table}

\begin{figure}
  \section{Contour Plots}\label{contour}
  CoNFIG-1: NVSS (red) and FIRST (blue) or VLA observation
  \cite{PaperI} (green) contours of extended sources, against
  Supercosmos Sky Survey background. The pink square, purple stars and
  orange triangle point to the NVSS, FIRST and optical identification
  coordinates.\\
  CoNFIG-2, 3 and 4: NVSS (red) and FIRST (blue) or VLA 1.4GHz
  A-configuration observation (purple) contours of extended sources in
  the CoNFIG catalogue, against Supercosmos Sky Survey background. The
  pink square and green star point to the NVSS and optical
  identification coordinates.\\
  CENSORS:VLA 1.4GHz A-configuration observation (purple) contours of
  sources in the CENSORS sample, against Supercosmos Sky Survey
  background. The pink square points to the radio centroid (or the
  catalogued radio source coordinates).\\
  \begin{center}
    {\bf CoNFIG-1}\\
    \begin{minipage}{3cm}      
      \mbox{}
      \centerline{\includegraphics[scale=0.26,angle=270]{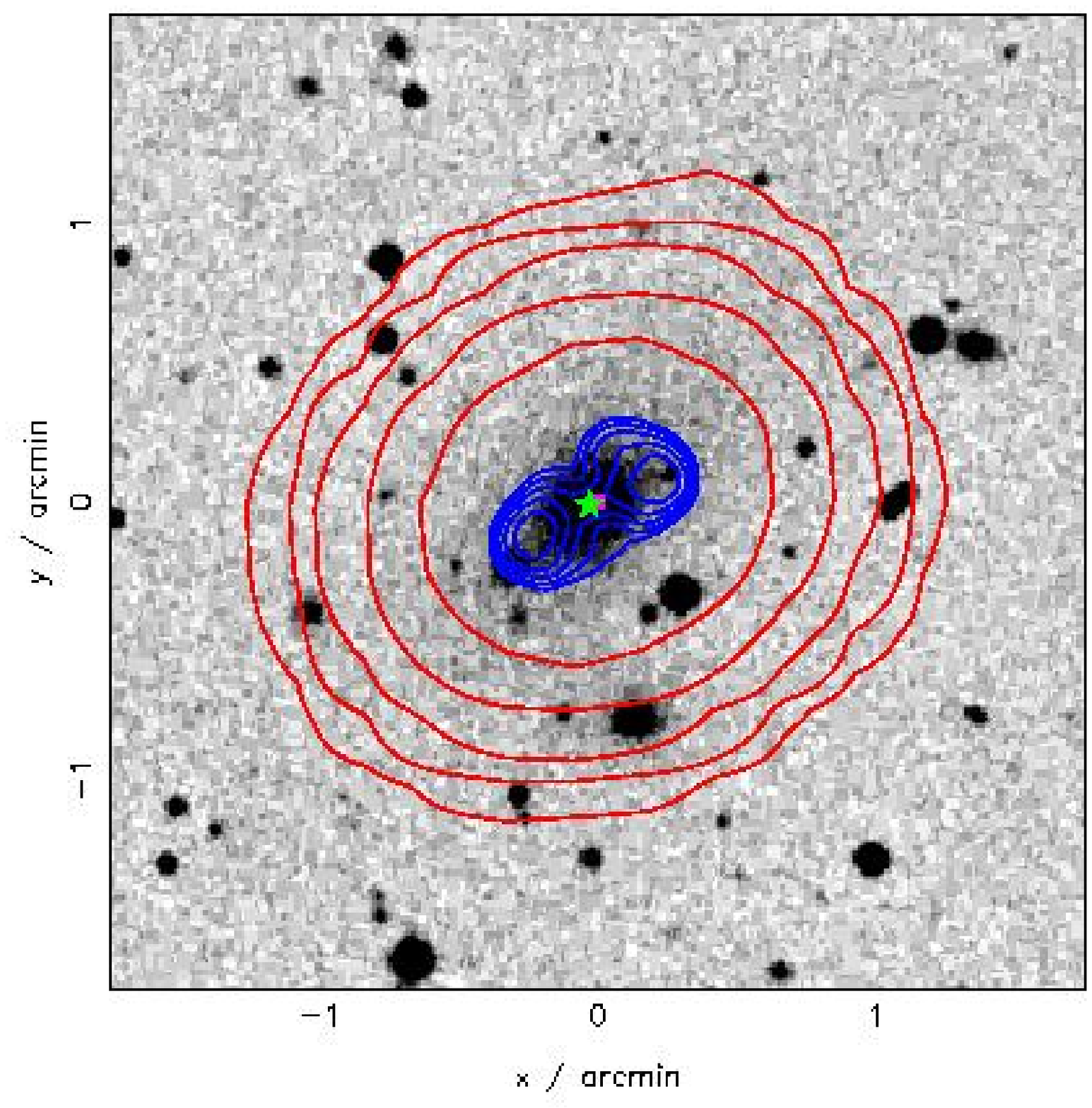}}
      \centerline{C1-003: 4C 53.16}
    \end{minipage}
    \hspace{3cm}
    \begin{minipage}{3cm}
      \mbox{}
      \centerline{\includegraphics[scale=0.26,angle=270]{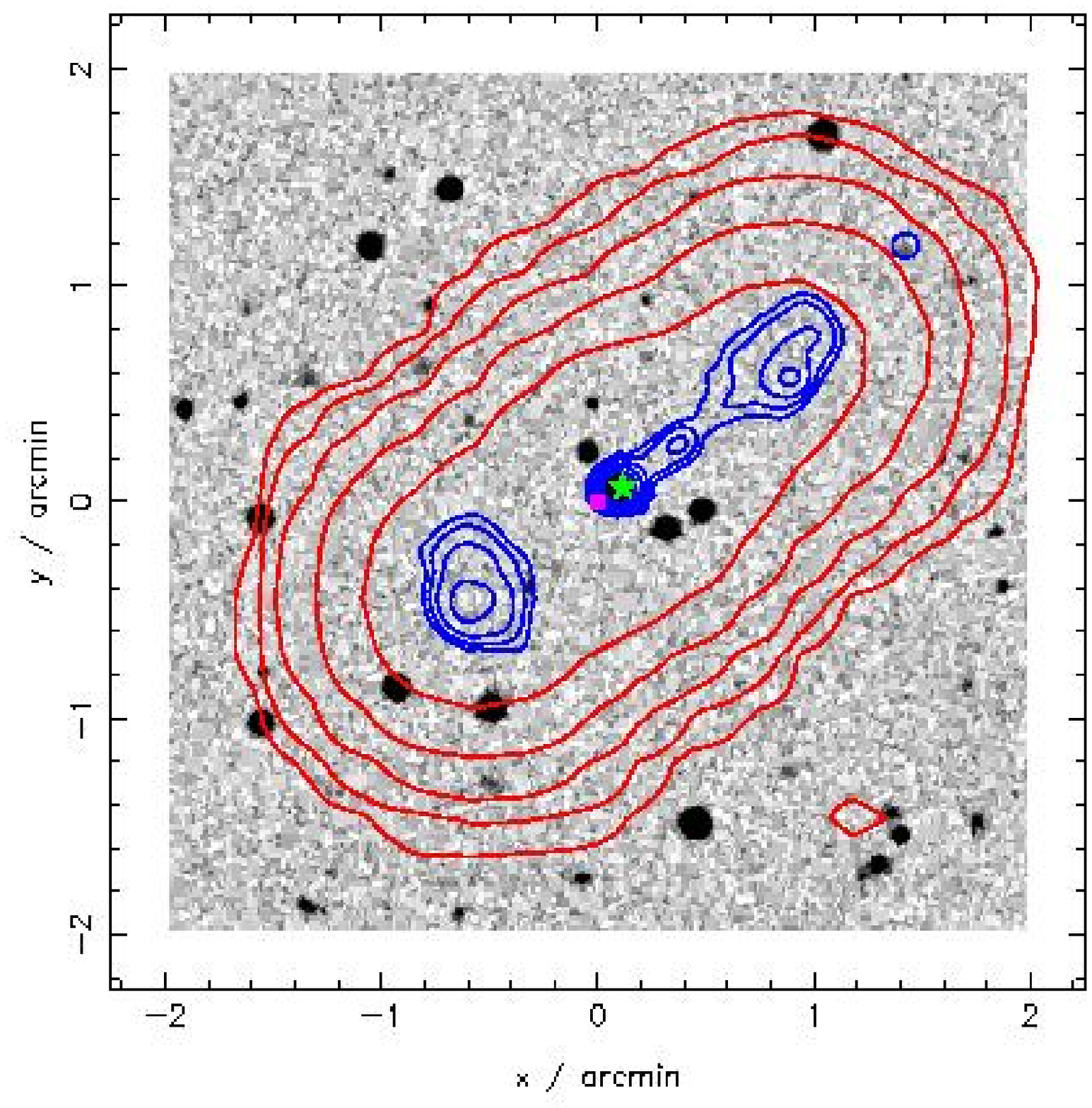}}
      \centerline{C1-006: 4C 31.30}
    \end{minipage}
    \hspace{3cm}
    \begin{minipage}{3cm}
      \mbox{}
      \centerline{\includegraphics[scale=0.26,angle=270]{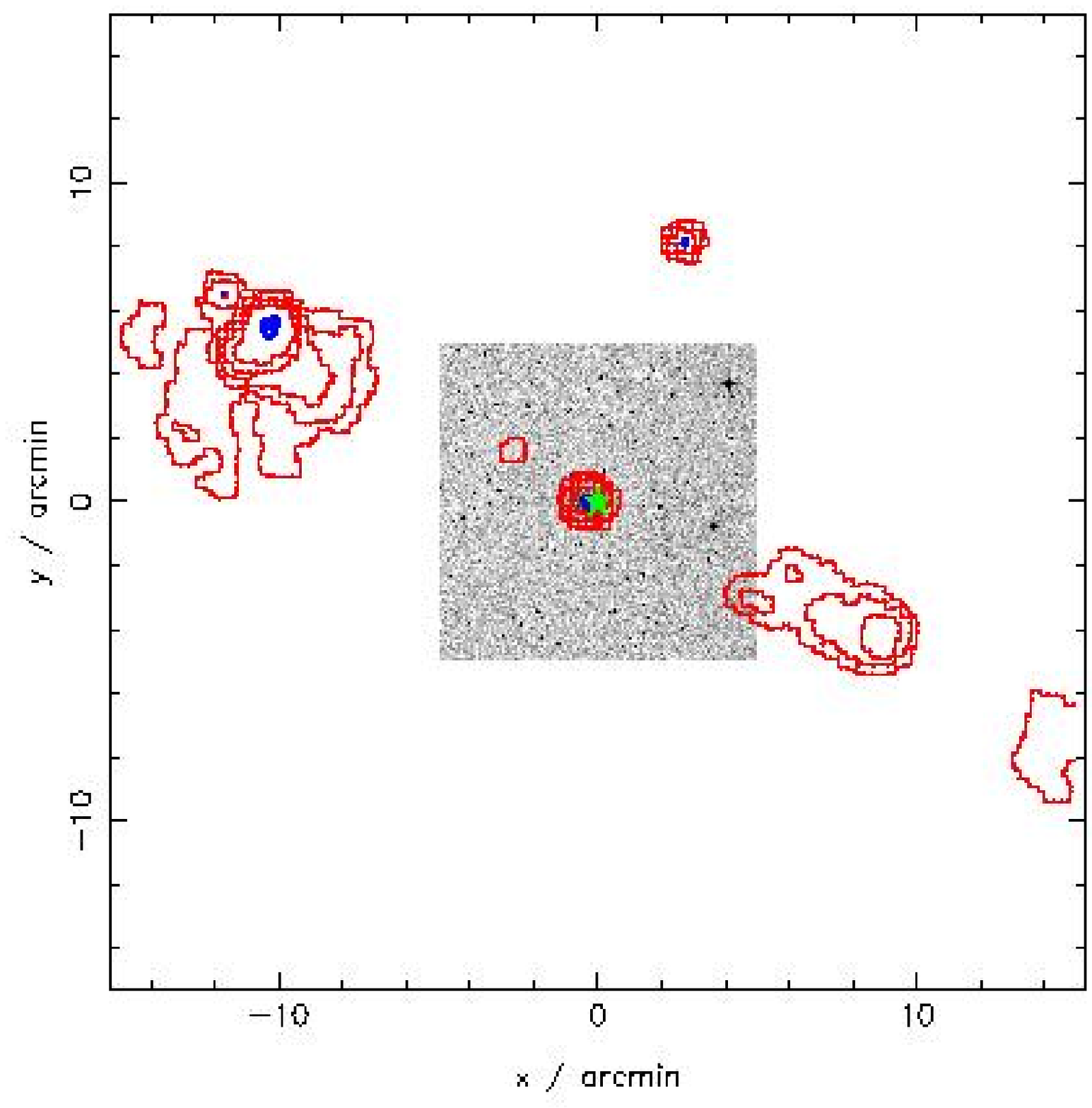}}
      \centerline{C1-007: DA 240}
    \end{minipage}
    \vfill
    \begin{minipage}{3cm}      
      \mbox{}
      \centerline{\includegraphics[scale=0.26,angle=270]{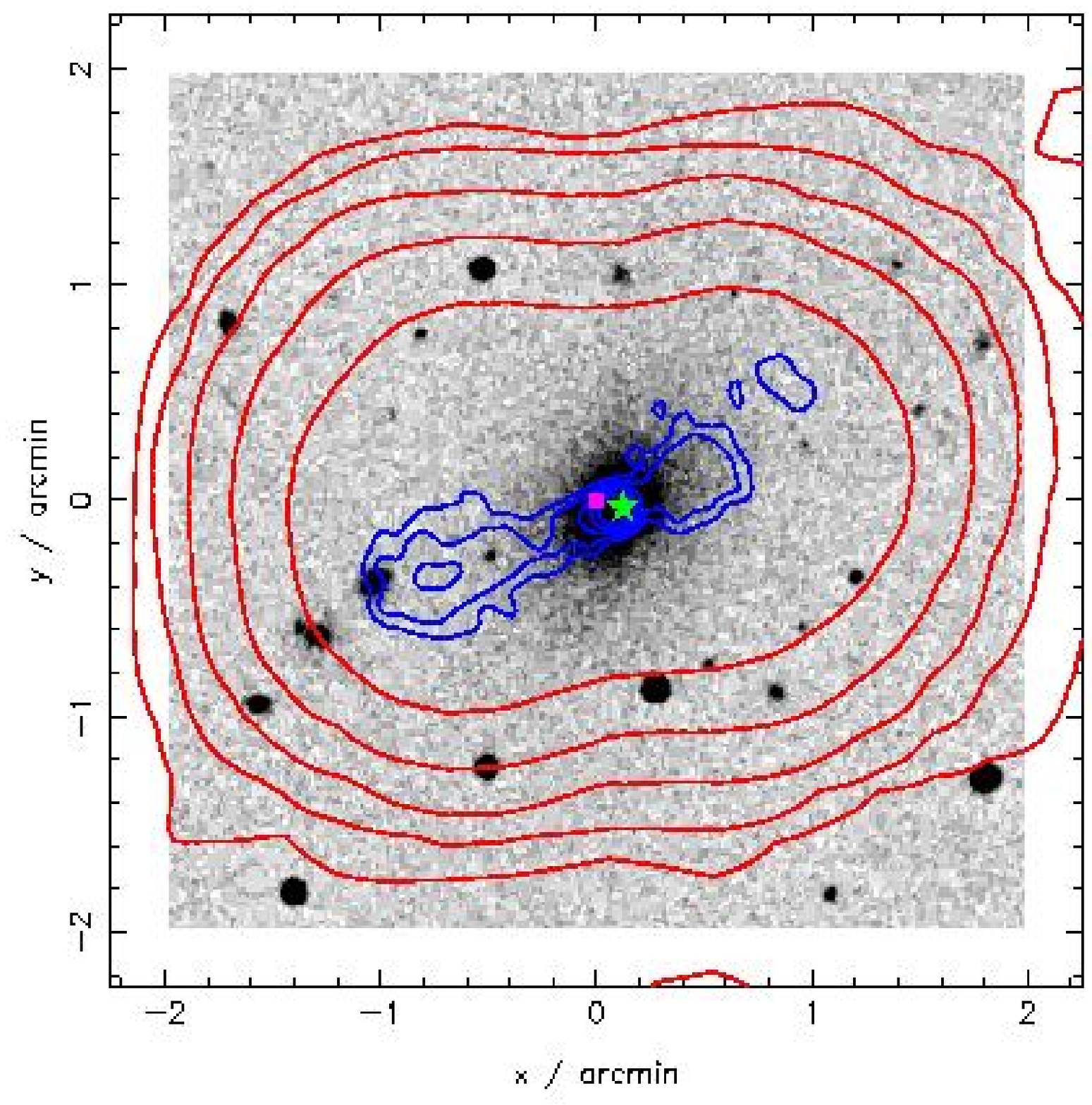}}
      \centerline{C1-008: NGC 2484}
    \end{minipage}
    \hspace{3cm}
    \begin{minipage}{3cm}
      \mbox{}
      \centerline{\includegraphics[scale=0.26,angle=270]{Contours/C1/009.ps}}
      \centerline{C1-009: 4C 37.21}
    \end{minipage}
    \hspace{3cm}
    \begin{minipage}{3cm}
      \mbox{}
      \centerline{\includegraphics[scale=0.26,angle=270]{Contours/C1/010.ps}}
      \centerline{C1-010: TXS 0757+503}
    \end{minipage}
    \vfill
    \begin{minipage}{3cm}      
      \mbox{}
      \centerline{\includegraphics[scale=0.26,angle=270]{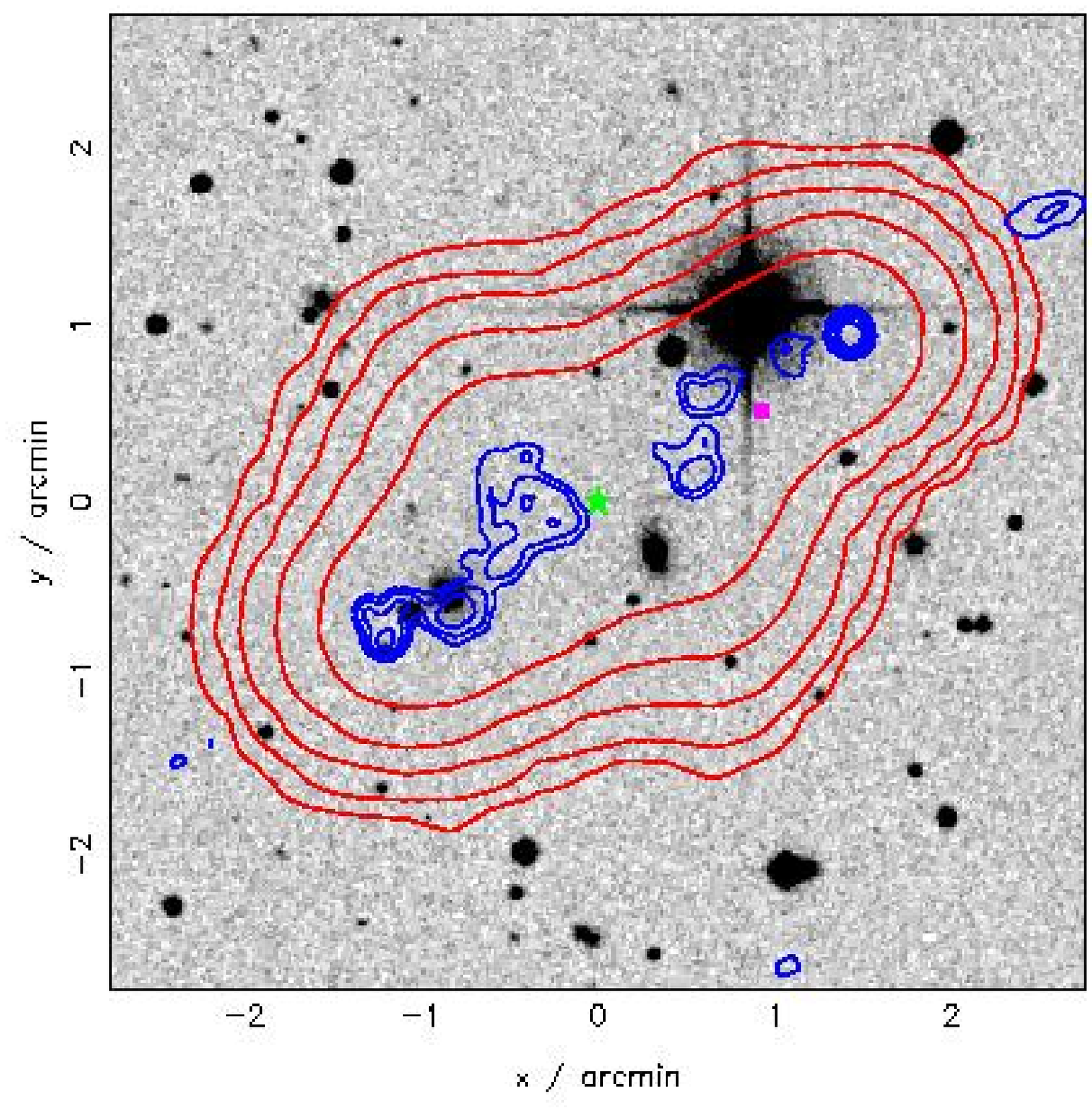}}
      \centerline{C1-011: 3C 192}
    \end{minipage}
    \hspace{3cm}
    \begin{minipage}{3cm}
      \mbox{}
      \centerline{\includegraphics[scale=0.26,angle=270]{Contours/C1/012.ps}}
      \centerline{C1-012: 3C 194}
    \end{minipage}
    \hspace{3cm}
    \begin{minipage}{3cm}
      \mbox{}
      \centerline{\includegraphics[scale=0.26,angle=270]{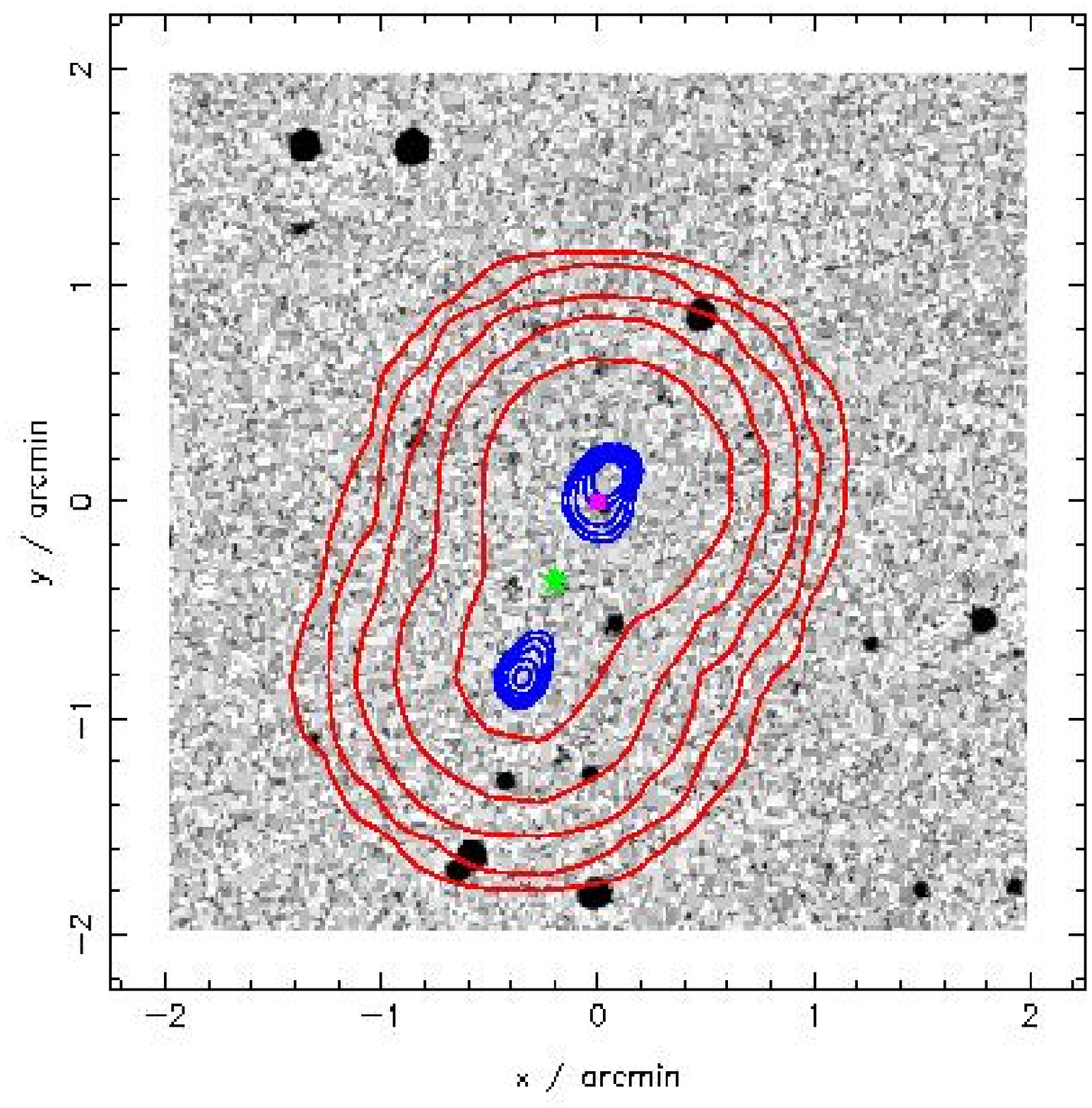}}
      \centerline{C1-013: 4C 32.24}
    \end{minipage}
  \end{center}
\end{figure}

\begin{figure}
  \begin{center}
    {\bf CoNFIG-1}\\  
  \begin{minipage}{3cm}      
      \mbox{}
      \centerline{\includegraphics[scale=0.26,angle=270]{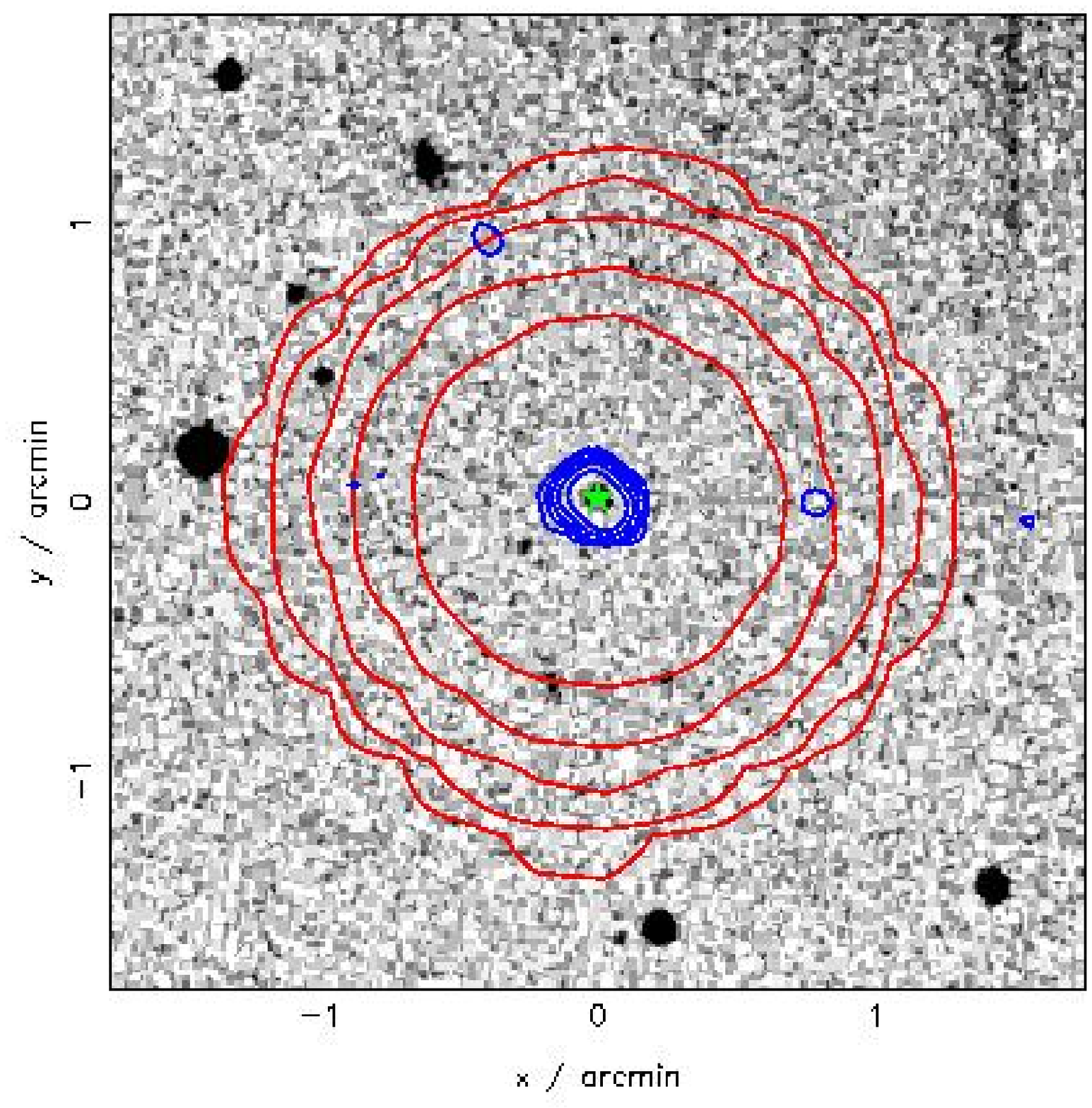}}
      \centerline{C1-014: 3C 196}
    \end{minipage}
    \hspace{3cm}
    \begin{minipage}{3cm}
      \mbox{}
      \centerline{\includegraphics[scale=0.26,angle=270]{Contours/C1/015.ps}}
      \centerline{C1-015: 4C 52.18}
    \end{minipage}
    \hspace{3cm}
    \begin{minipage}{3cm}
      \mbox{}
      \centerline{\includegraphics[scale=0.26,angle=270]{Contours/C1/016.ps}}
      \centerline{C1-016: 3C 197.1}
    \end{minipage}
    \vfill
    \begin{minipage}{3cm}      
      \mbox{}
      \centerline{\includegraphics[scale=0.26,angle=270]{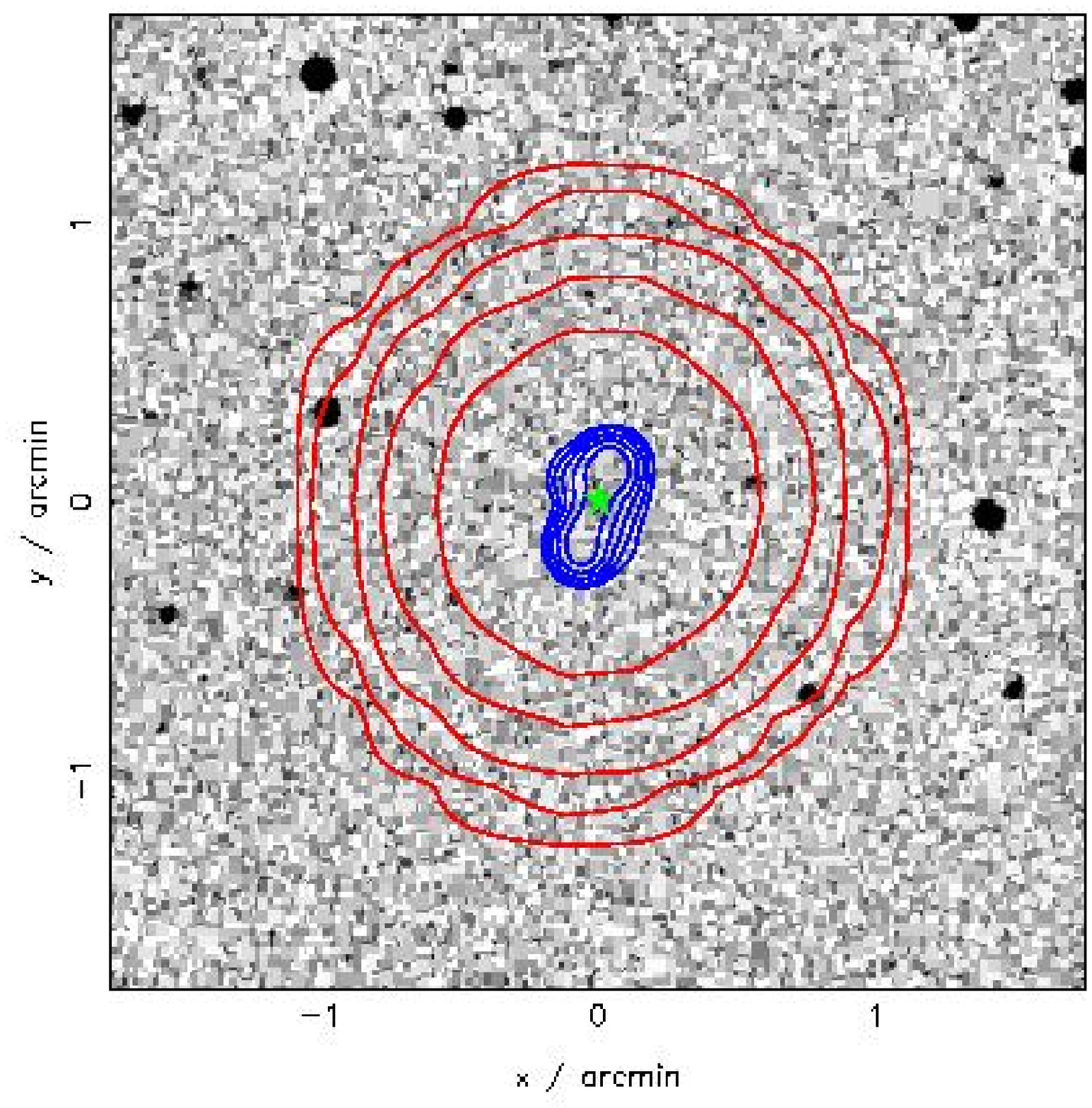}}
      \centerline{C1-021: 3C 200}
    \end{minipage}
    \hspace{3cm}
    \begin{minipage}{3cm}
      \mbox{}
      \centerline{\includegraphics[scale=0.26,angle=270]{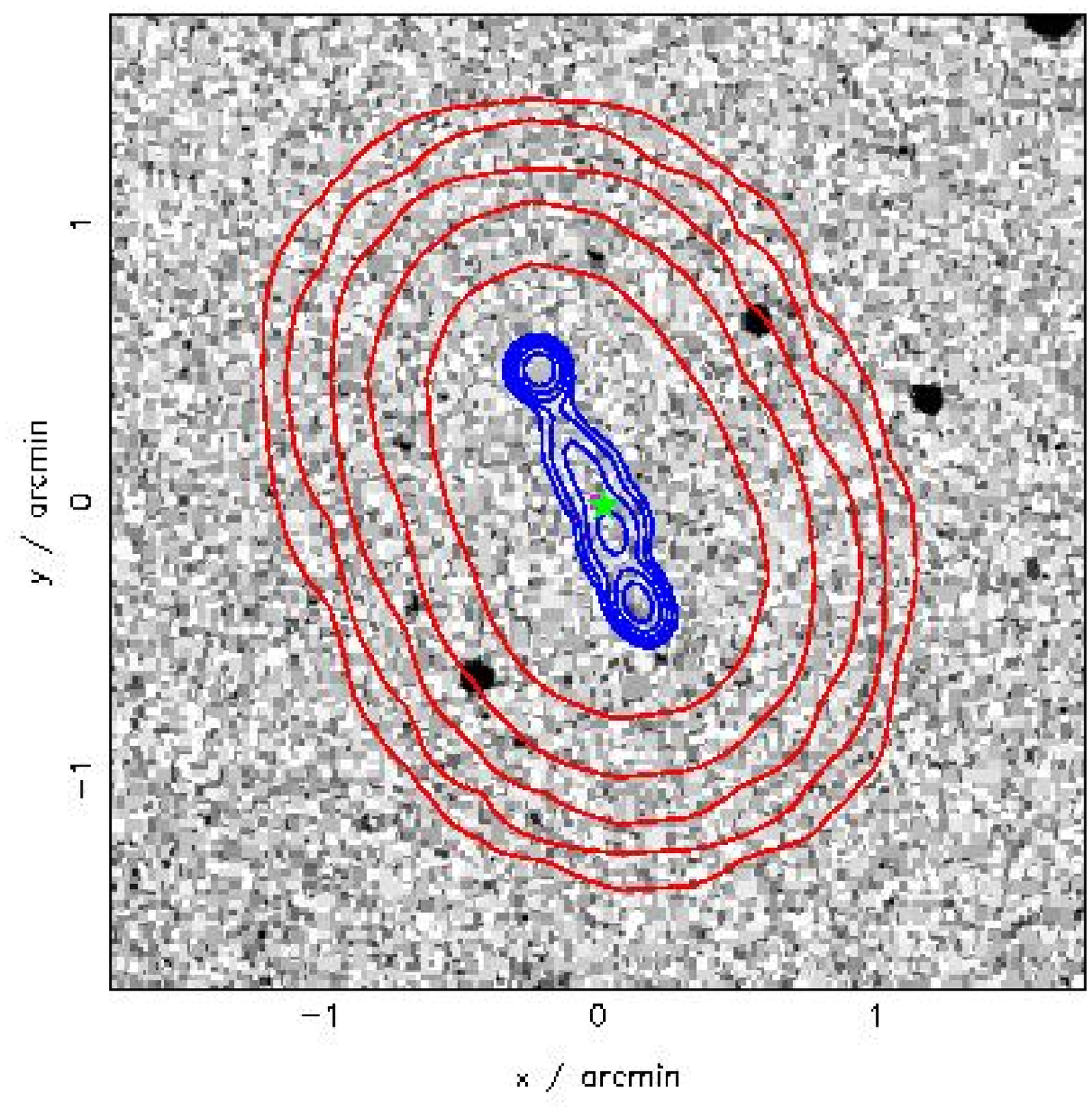}}
      \centerline{C1-023: 4C 51.25}
    \end{minipage}
    \hspace{3cm}
    \begin{minipage}{3cm}
      \mbox{}
      \centerline{\includegraphics[scale=0.26,angle=270]{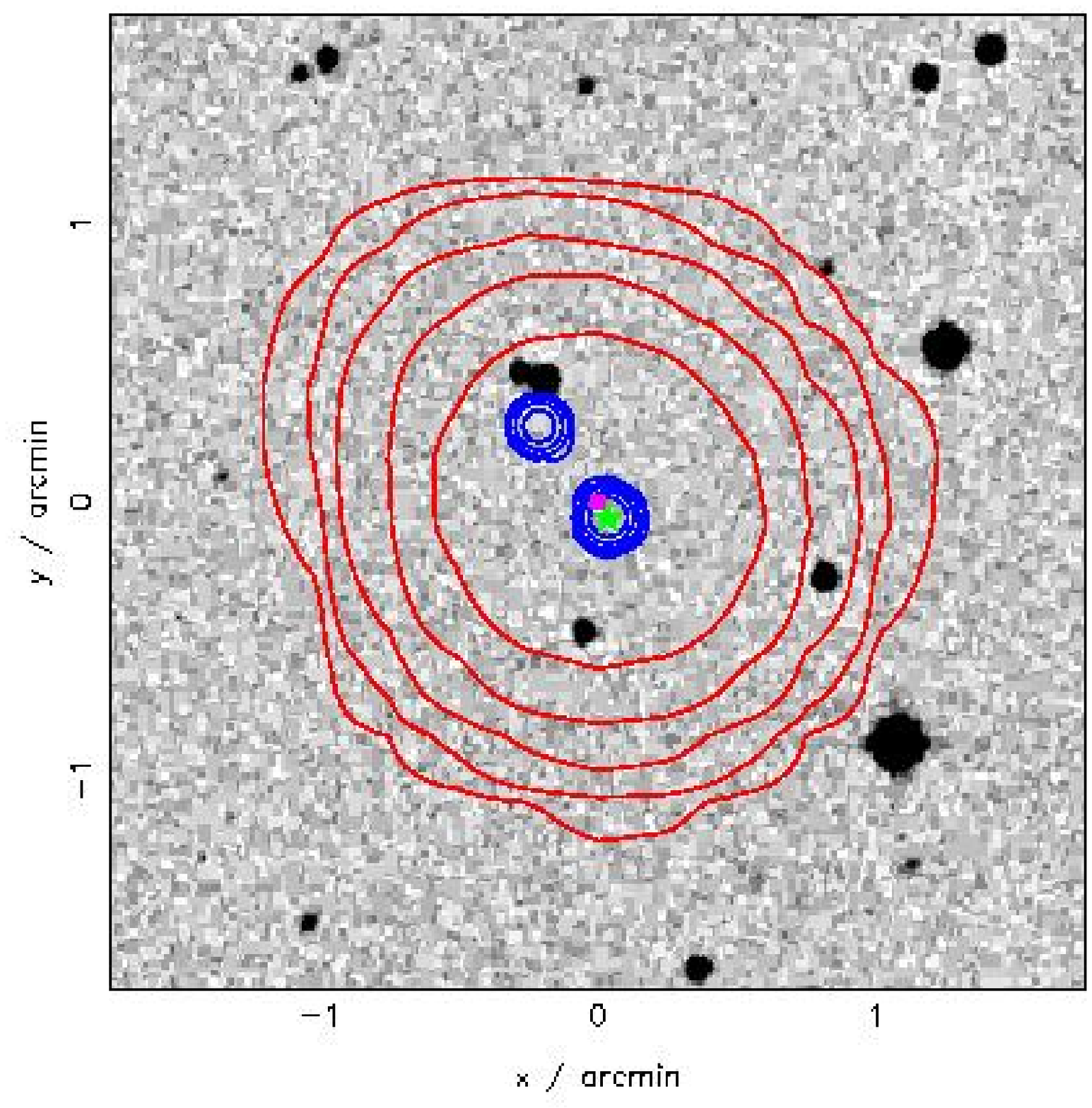}}
      \centerline{C1-024: 3C 202}
    \end{minipage}
    \vfill
    \begin{minipage}{3cm}      
      \mbox{}
      \centerline{\includegraphics[scale=0.26,angle=270]{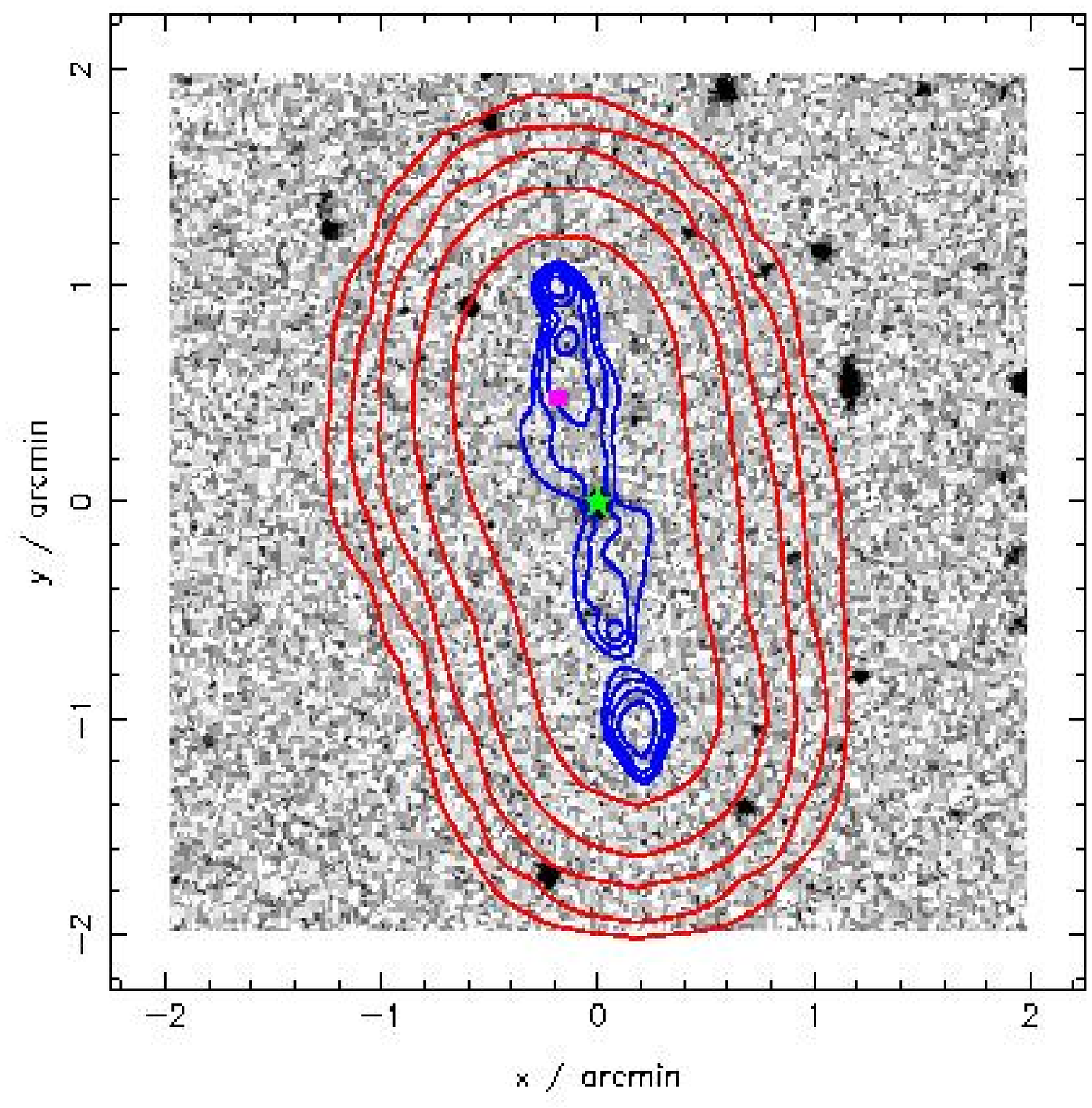}}
      \centerline{C1-026: 4C 45.17}
    \end{minipage}
    \hspace{3cm}
    \begin{minipage}{3cm}
      \mbox{}
      \centerline{\includegraphics[scale=0.26,angle=270]{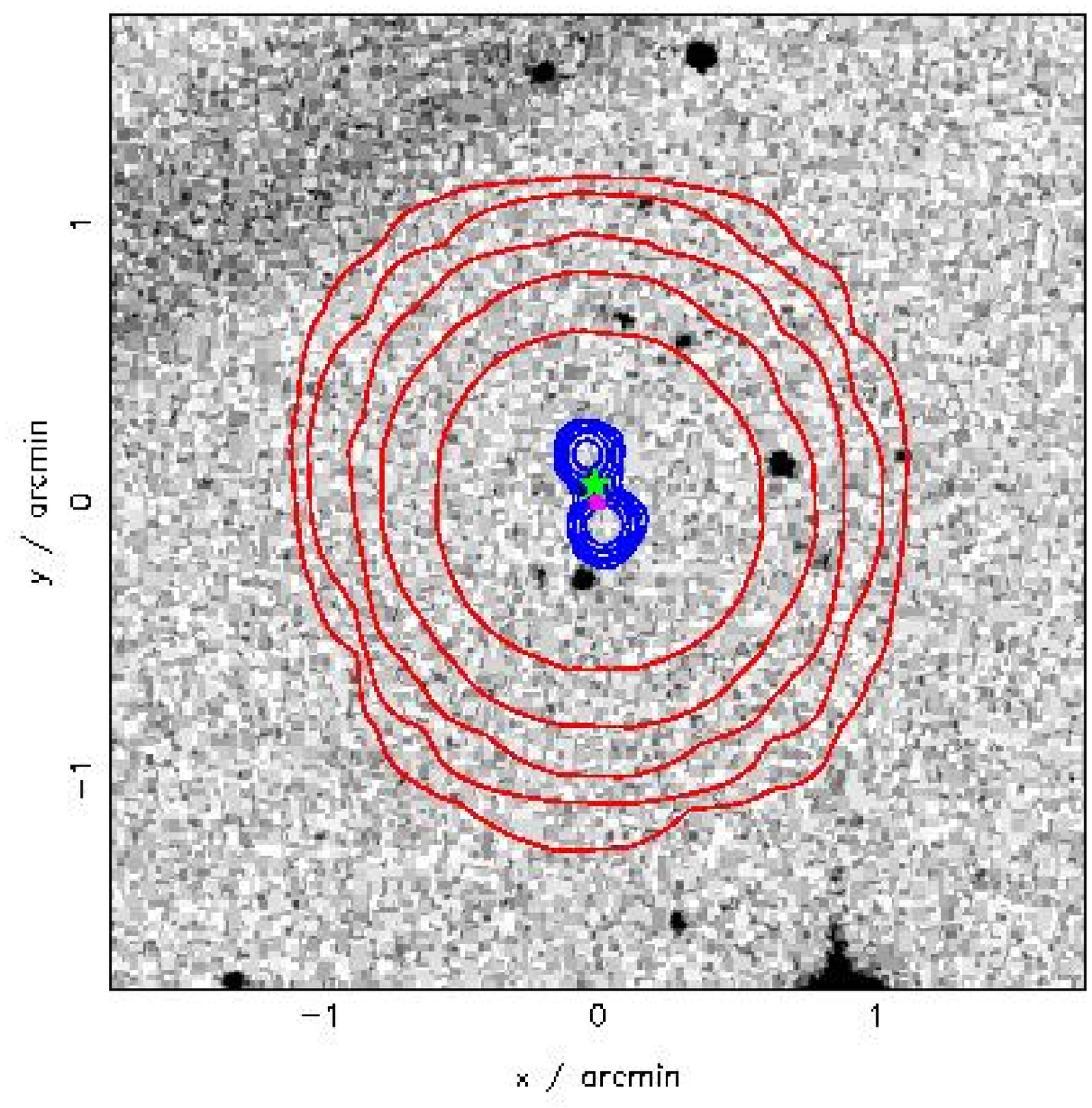}}
      \centerline{C1-027: 3C 205}
    \end{minipage}
    \hspace{3cm}
    \begin{minipage}{3cm}
      \mbox{}
      \centerline{\includegraphics[scale=0.26,angle=270]{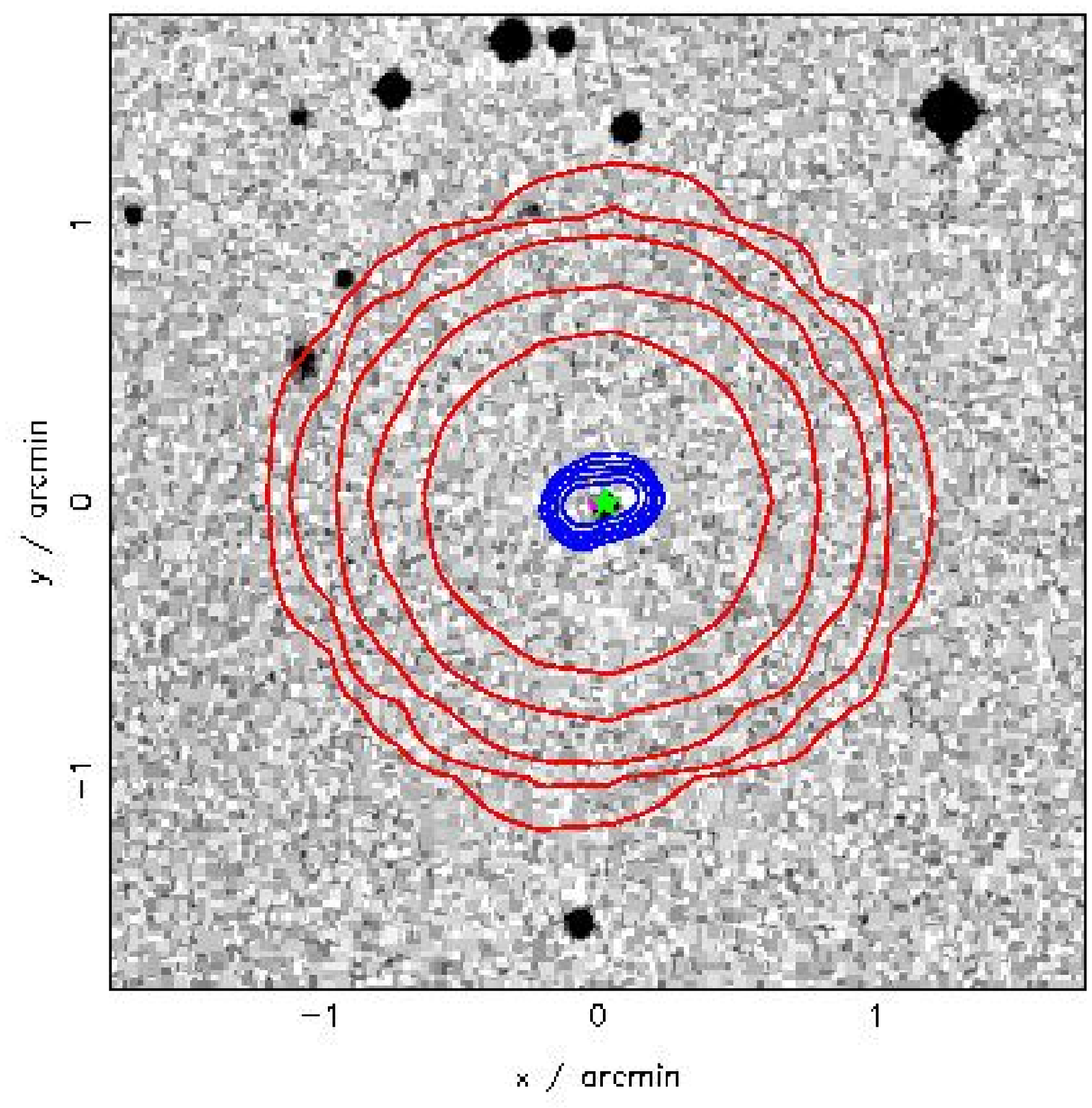}}
      \centerline{C1-028: 3C 207}
    \end{minipage}
    \vfill
    \begin{minipage}{3cm}     
      \mbox{}
      \centerline{\includegraphics[scale=0.26,angle=270]{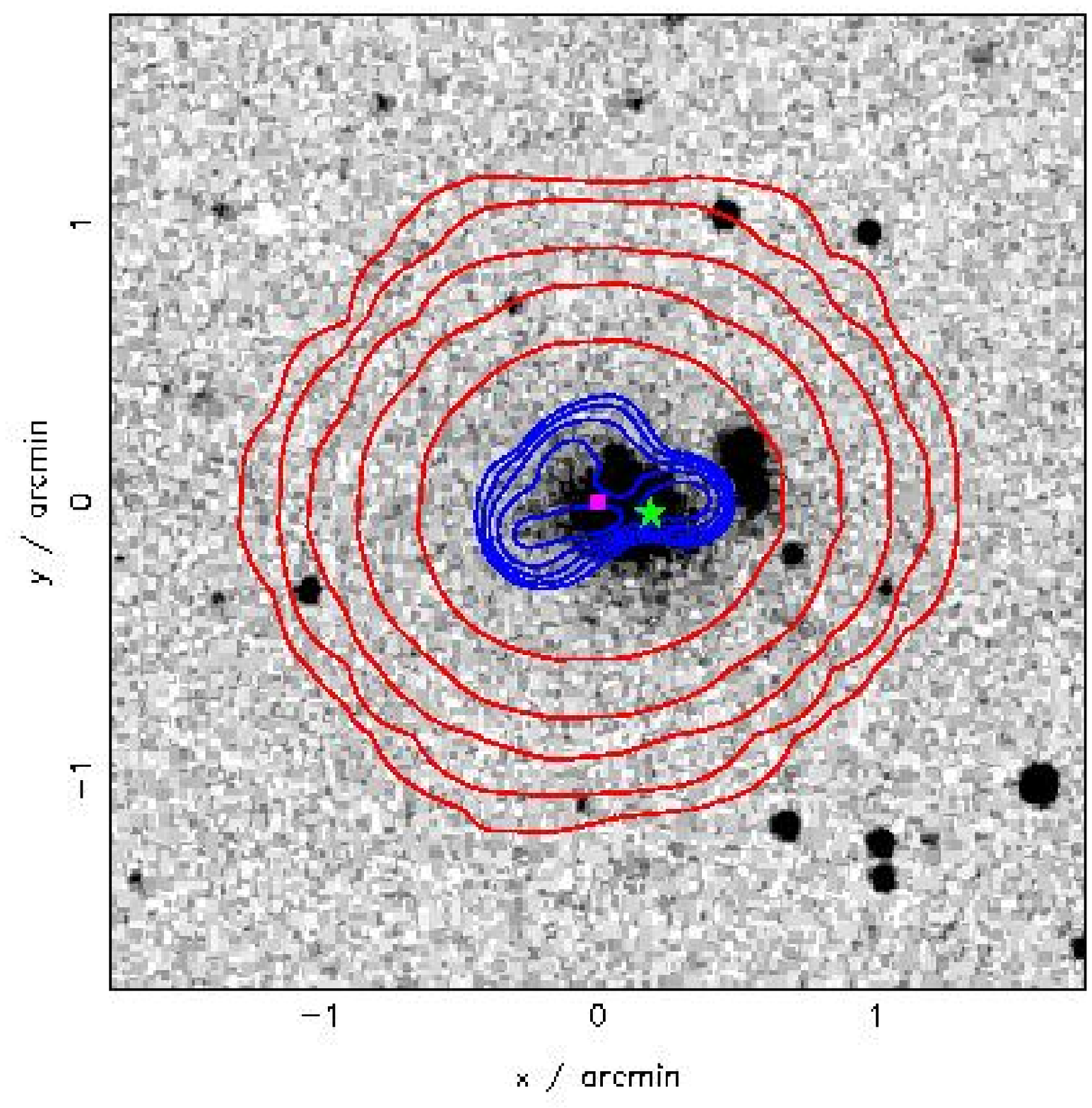}}
      \centerline{C1-030: NGC 2656}
    \end{minipage}
    \hspace{3cm}
    \begin{minipage}{3cm}
      \mbox{}
      \centerline{\includegraphics[scale=0.26,angle=270]{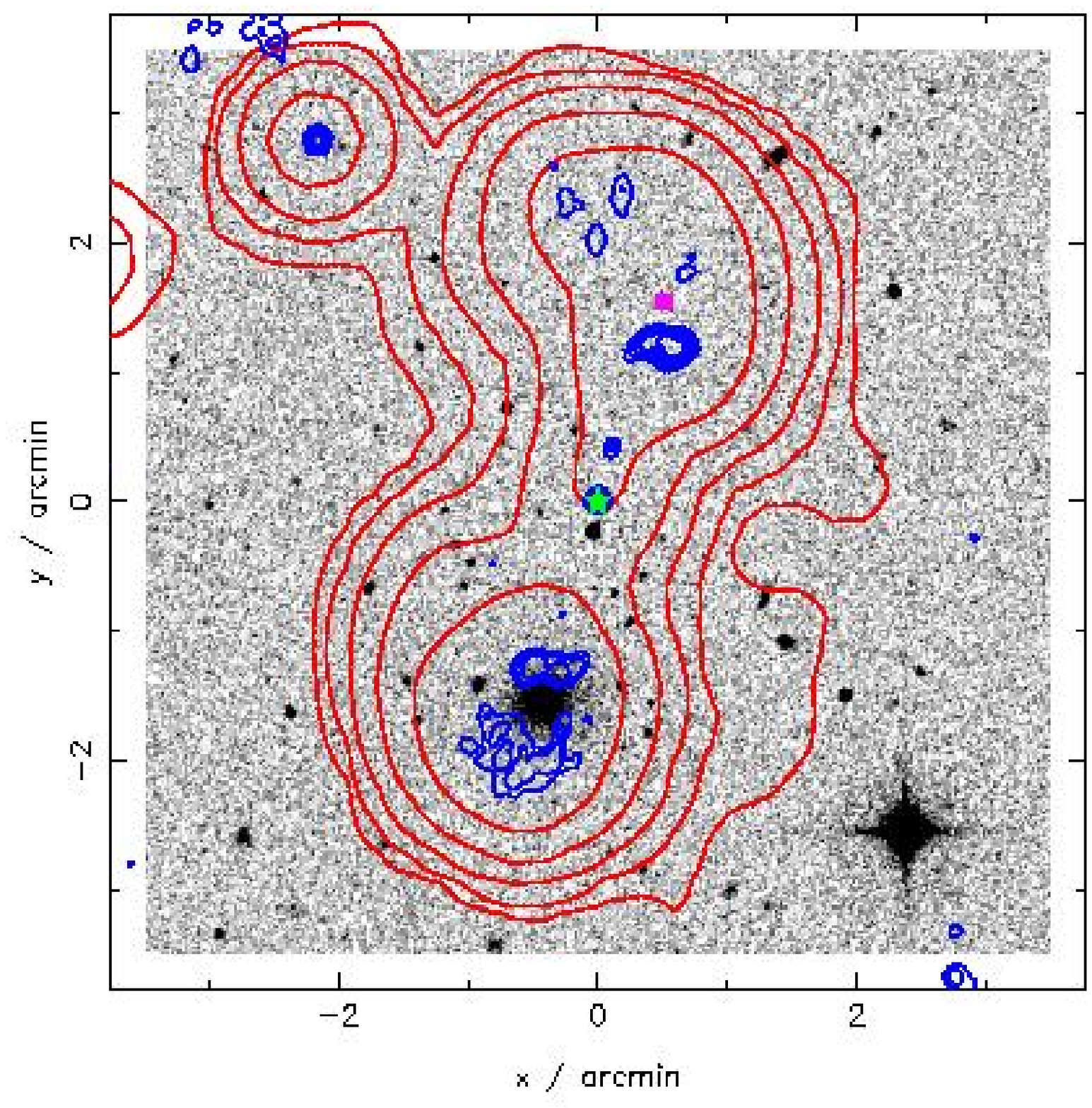}}
      \centerline{C1-031: 4C 31.32}
    \end{minipage}
    \hspace{3cm}
    \begin{minipage}{3cm}
      \mbox{}
      \centerline{\includegraphics[scale=0.26,angle=270]{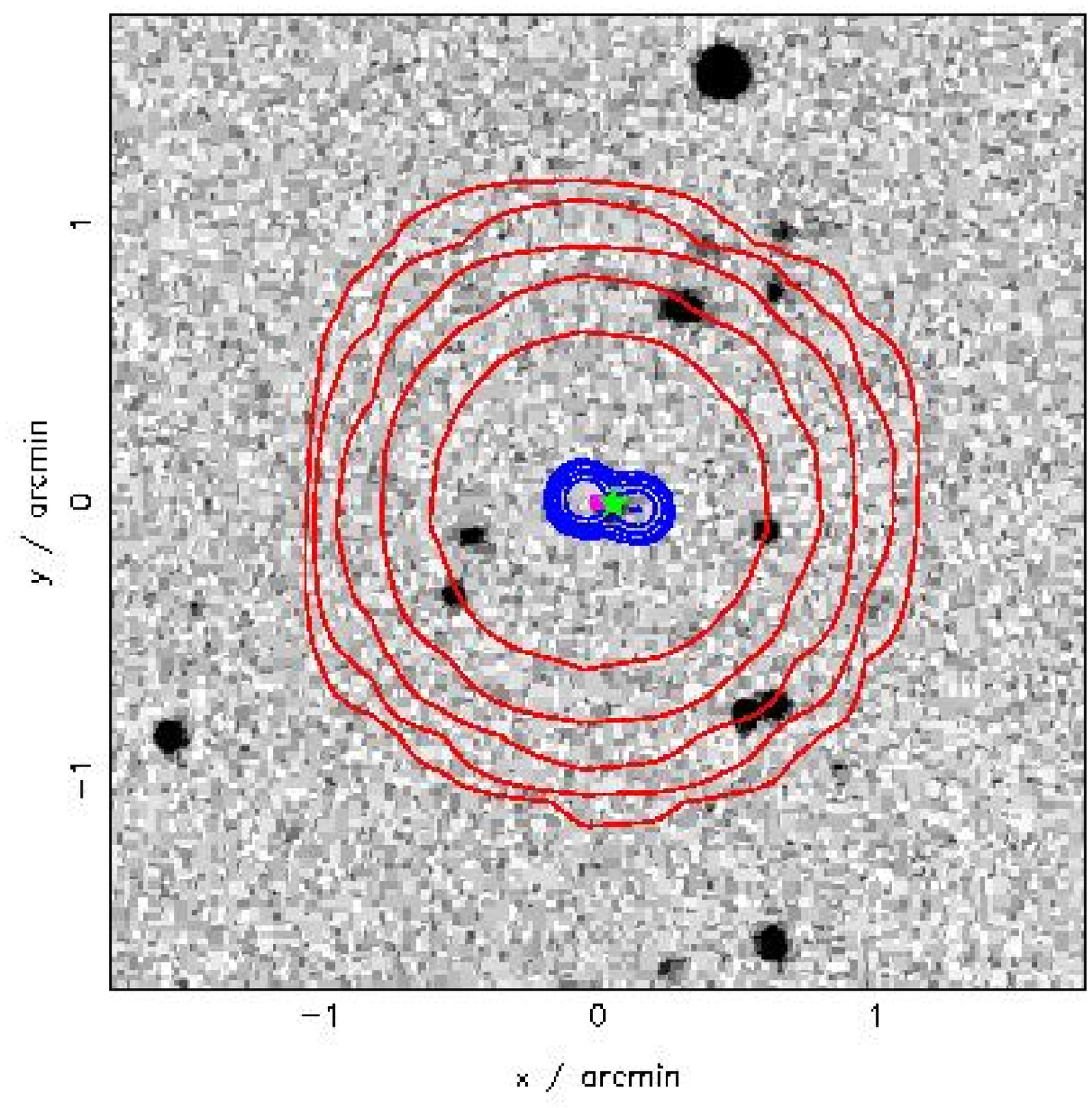}}
      \centerline{C1-032: 3C 208}
    \end{minipage}
  \end{center}
\end{figure}

\begin{figure}
  \begin{center}
    {\bf CoNFIG-1}\\  
  \begin{minipage}{3cm}      
      \mbox{}
      \centerline{\includegraphics[scale=0.26,angle=270]{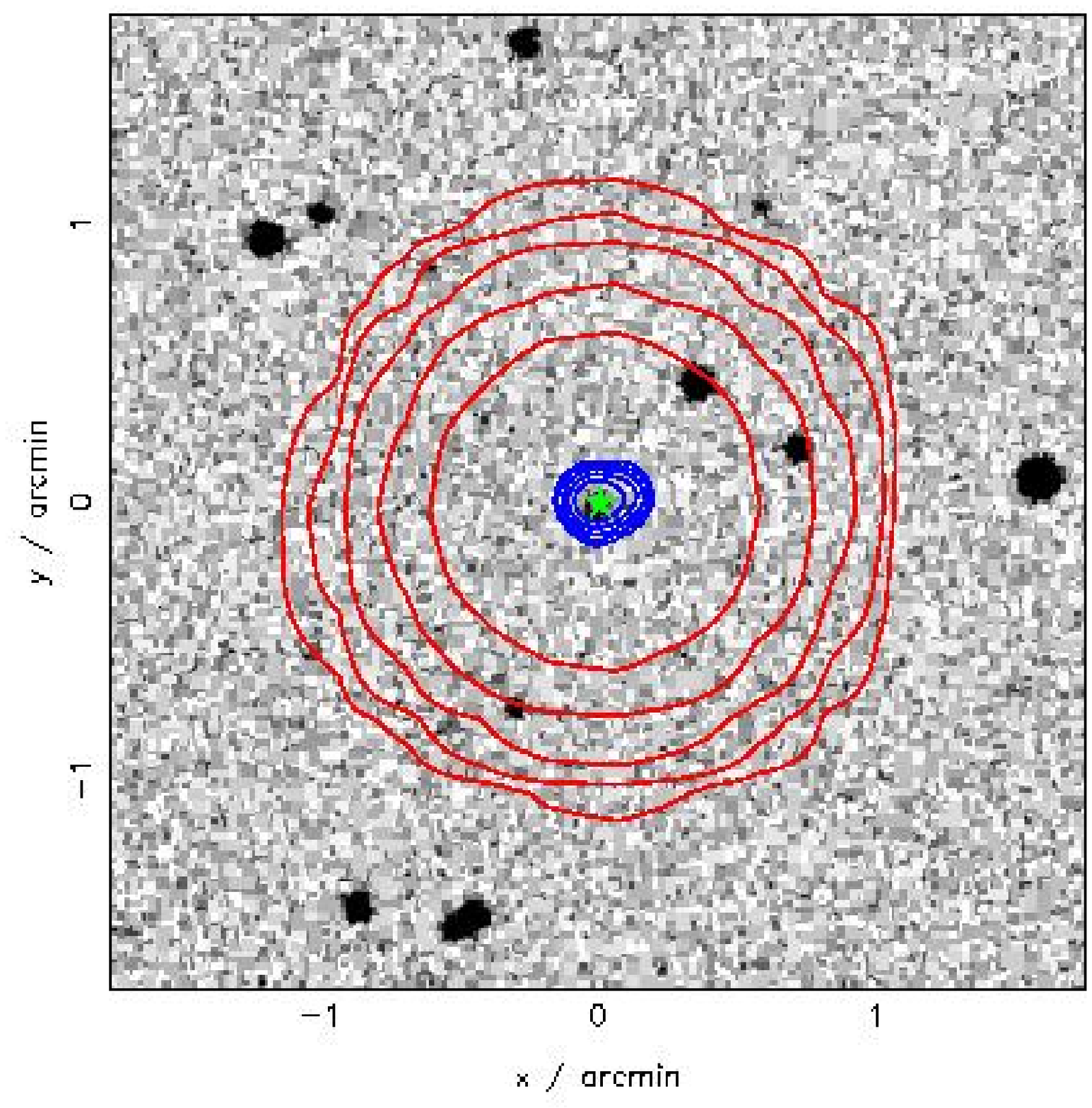}}
      \centerline{C1-033: 3C 208.1}
    \end{minipage}
    \hspace{3cm}
    \begin{minipage}{3cm}
      \mbox{}
      \centerline{\includegraphics[scale=0.26,angle=270]{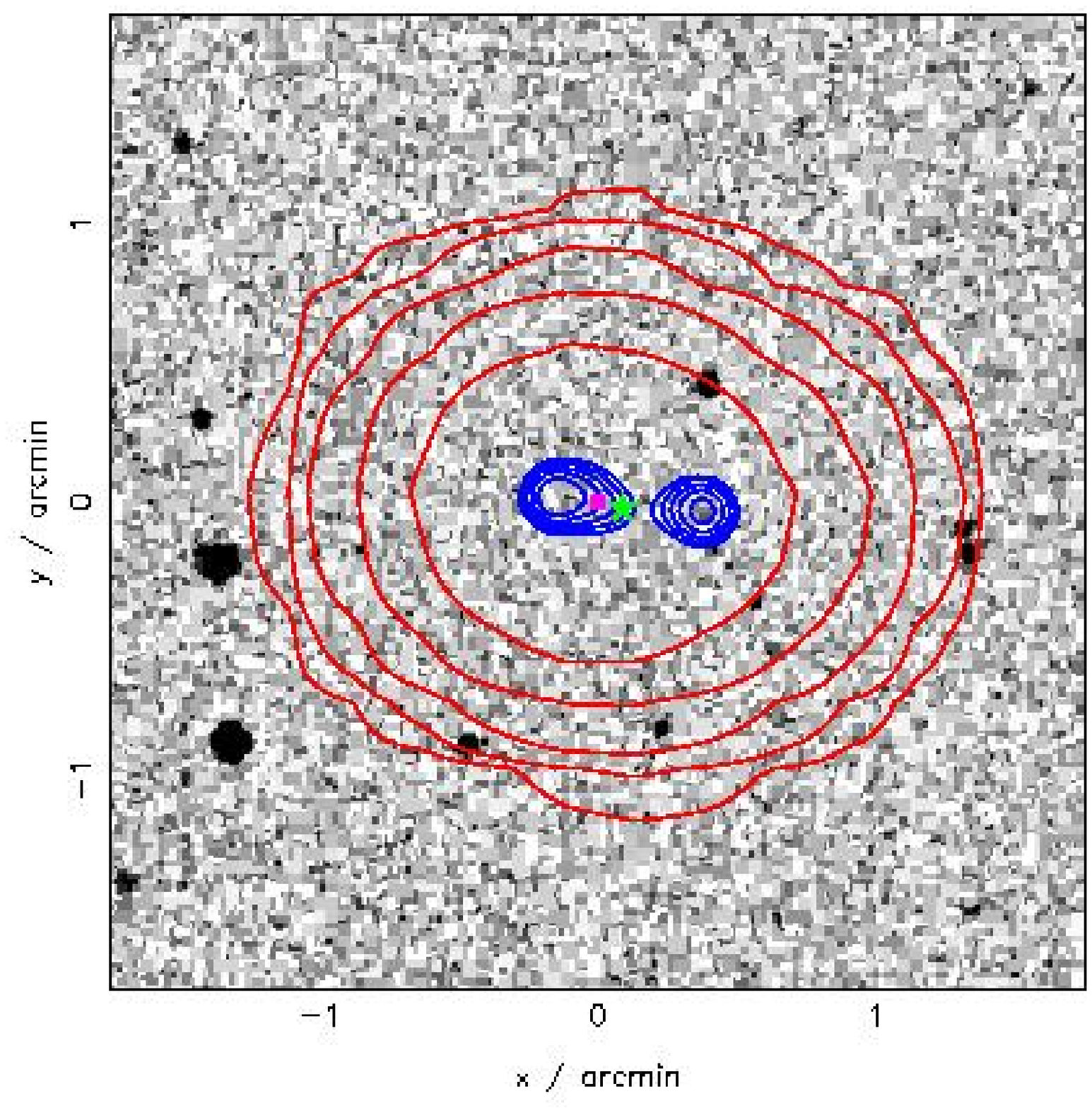}}
      \centerline{C1-035: 3C 211}
    \end{minipage}
    \hspace{3cm}
    \begin{minipage}{3cm}
      \mbox{}
      \centerline{\includegraphics[scale=0.26,angle=270]{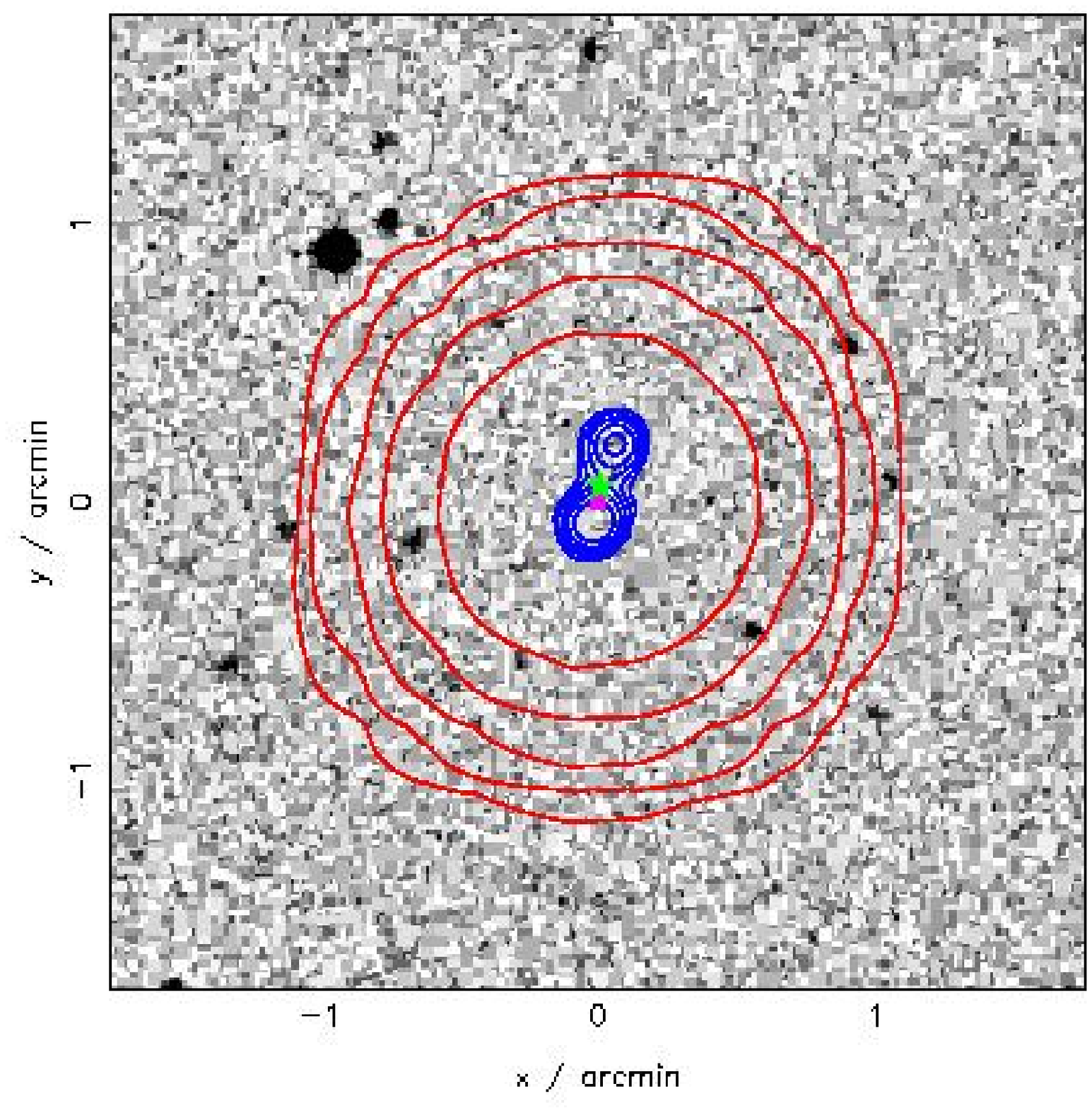}}
      \centerline{C1-036: 3C 210}
    \end{minipage}
    \vfill
    \begin{minipage}{3cm}      
      \mbox{}
      \centerline{\includegraphics[scale=0.26,angle=270]{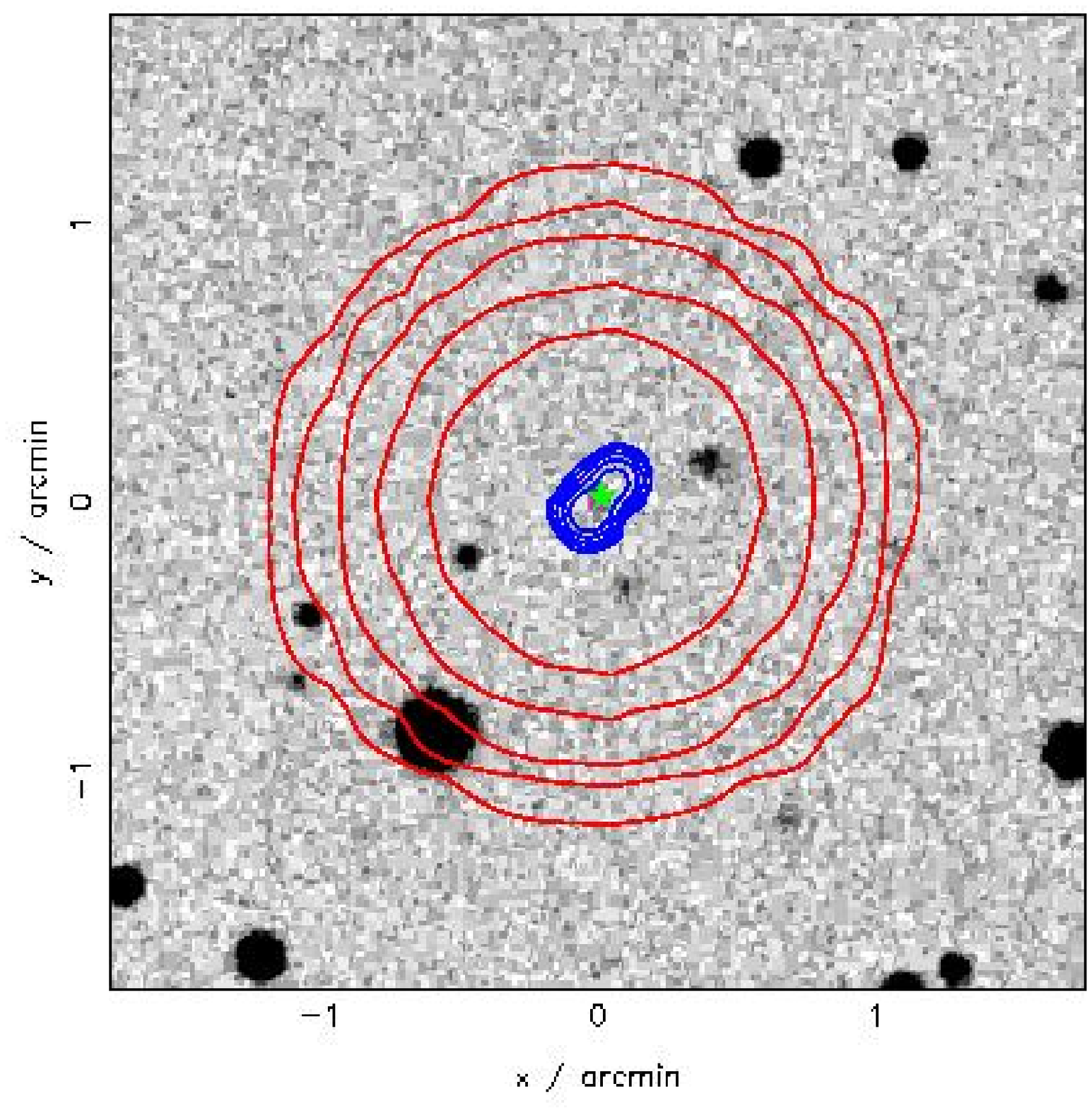}}
      \centerline{C1-037: 3C 212}
    \end{minipage}
    \hspace{3cm}
    \begin{minipage}{3cm}
      \mbox{}
      \centerline{\includegraphics[scale=0.26,angle=270]{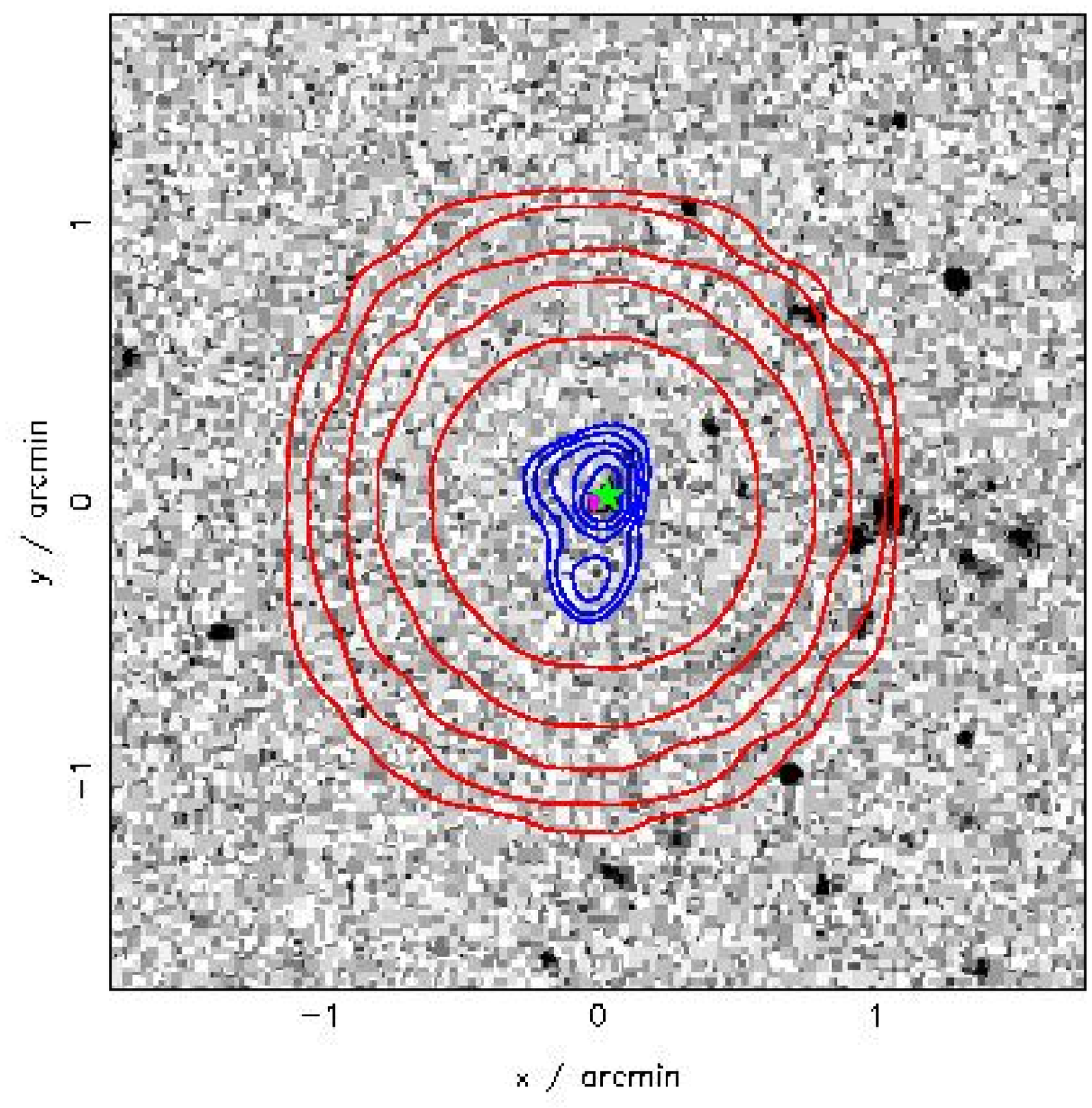}}
      \centerline{C1-038: 3C 213.1}
    \end{minipage}
    \hspace{3cm}
    \begin{minipage}{3cm}
      \mbox{}
      \centerline{\includegraphics[scale=0.26,angle=270]{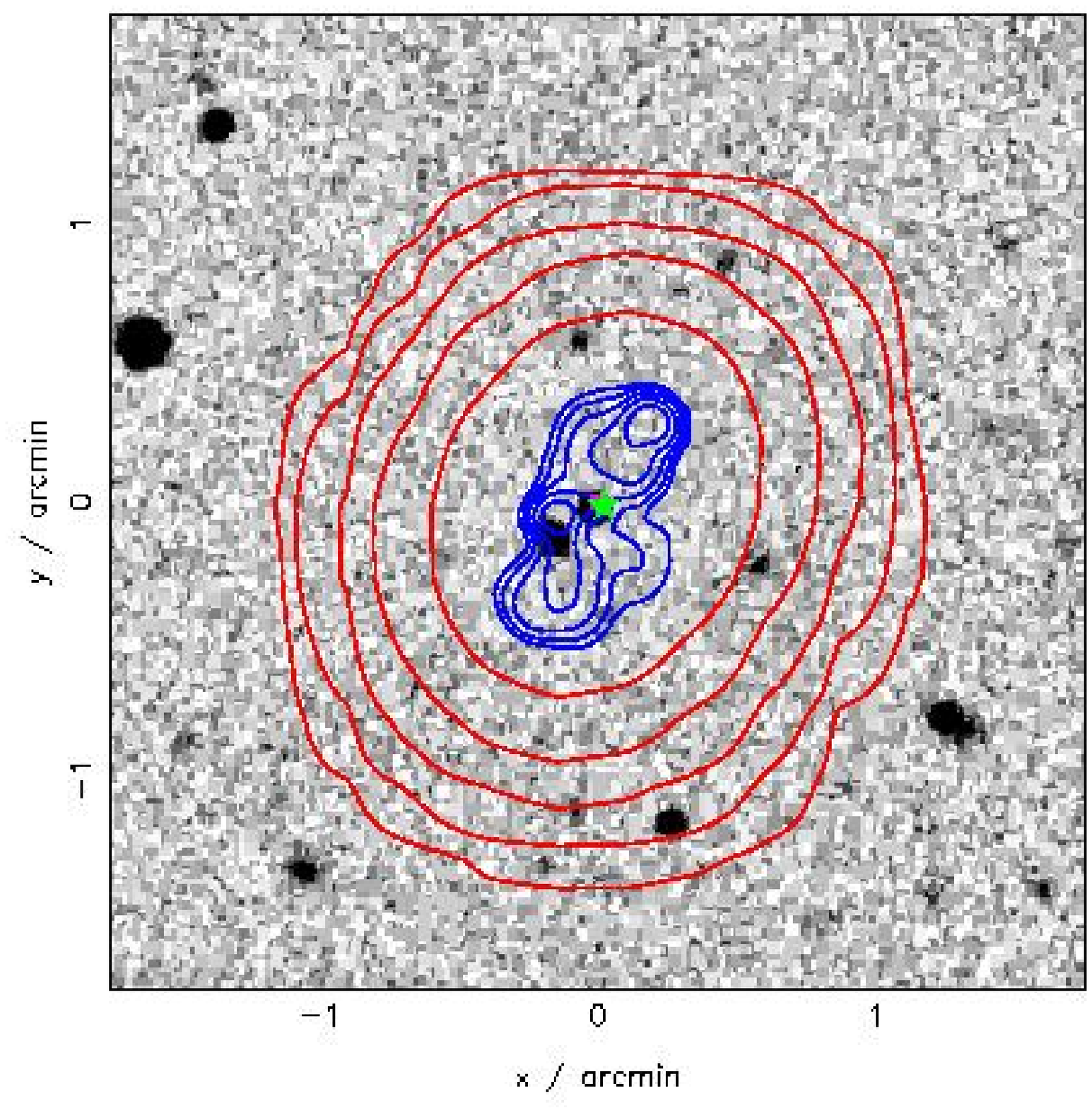}}
      \centerline{C1-040: 3C 215}
    \end{minipage}
    \vfill
    \begin{minipage}{3cm}     
      \mbox{}
      \centerline{\includegraphics[scale=0.26,angle=270]{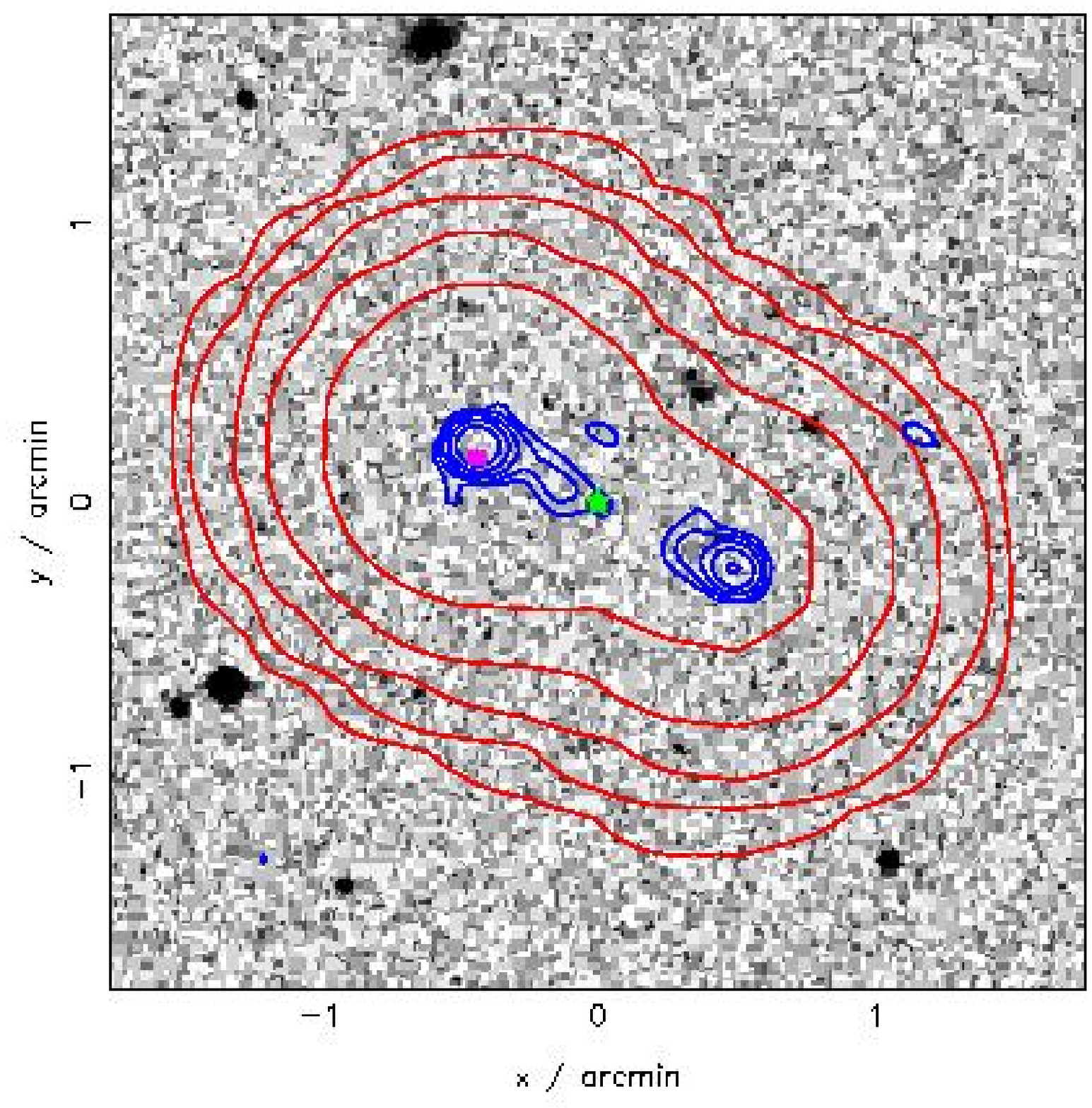}}
      \centerline{C1-041: 4C 41.19}
    \end{minipage}
    \hspace{3cm}
    \begin{minipage}{3cm}
      \mbox{}
      \centerline{\includegraphics[scale=0.26,angle=270]{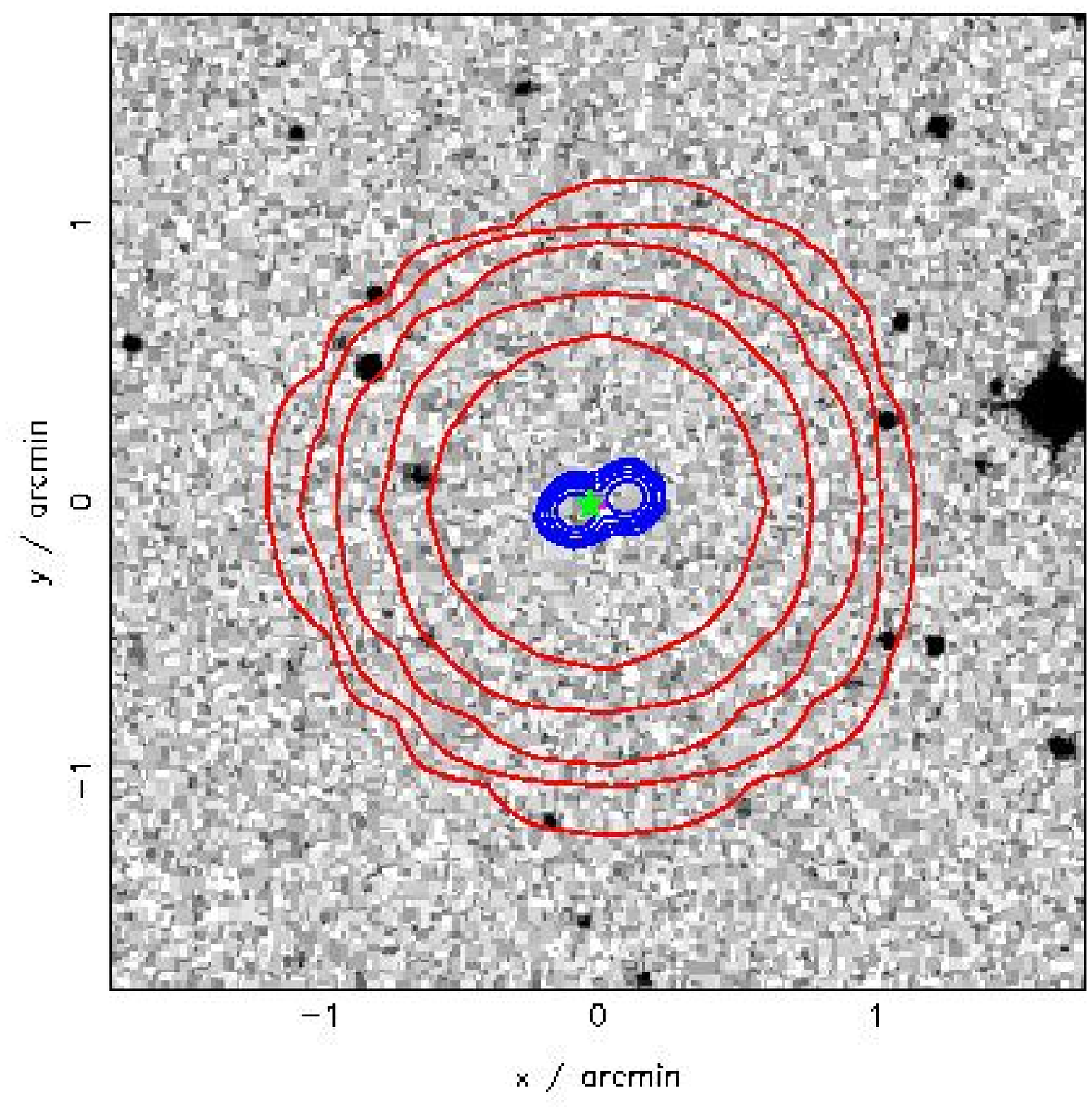}}
      \centerline{C1-042: 3C 217}
    \end{minipage}
    \hspace{3cm}
    \begin{minipage}{3cm}
      \mbox{}
      \centerline{\includegraphics[scale=0.26,angle=270]{Contours/C1/044.ps}}
      \centerline{C1-044: 4C 16.27}
    \end{minipage}
    \vfill
    \begin{minipage}{3cm}     
      \mbox{}
      \centerline{\includegraphics[scale=0.26,angle=270]{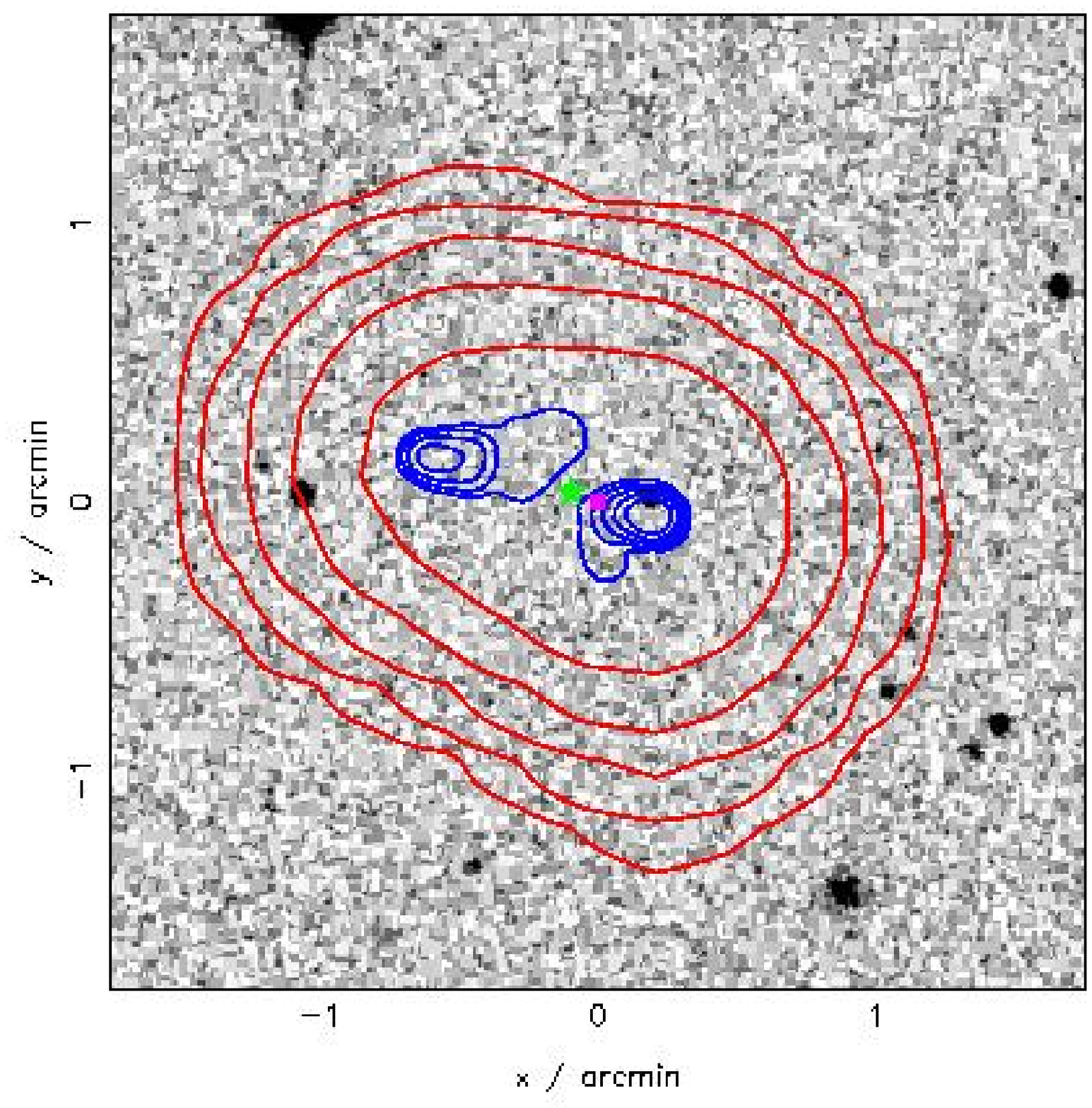}}
      \centerline{C1-045: 4C 17.48}
    \end{minipage}
    \hspace{3cm}
    \begin{minipage}{3cm}
      \mbox{}
      \centerline{\includegraphics[scale=0.26,angle=270]{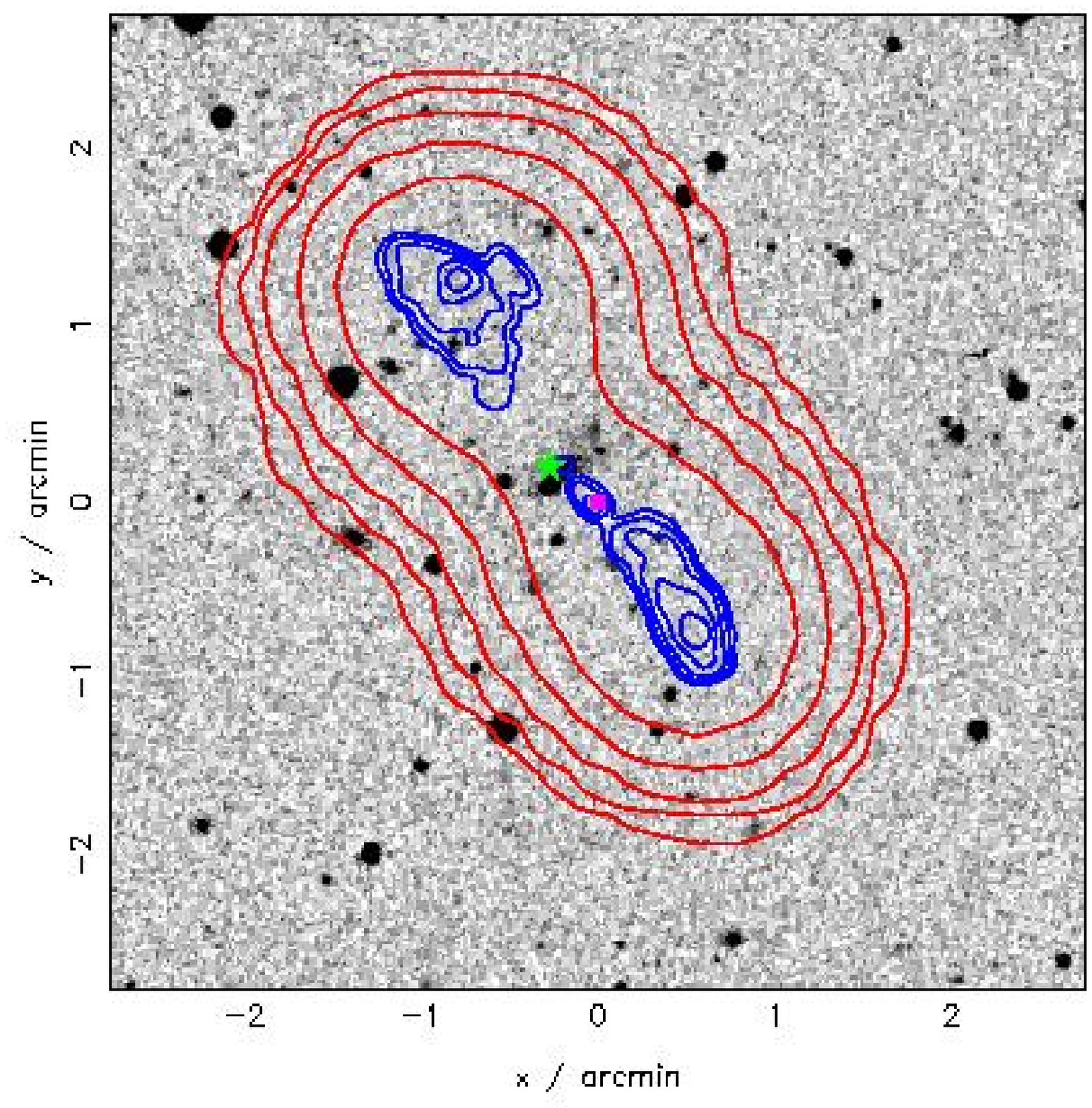}}
      \centerline{C1-046: 3C 219}
    \end{minipage}
    \hspace{3cm}
    \begin{minipage}{3cm}
      \mbox{}
      \centerline{\includegraphics[scale=0.26,angle=270]{Contours/C1/047.ps}}
      \centerline{C1-047: 4C 53.18}
    \end{minipage}
  \end{center}
\end{figure}

\begin{figure}
  \begin{center}
    {\bf CoNFIG-1}\\  
  \begin{minipage}{3cm}      
      \mbox{}
      \centerline{\includegraphics[scale=0.26,angle=270]{Contours/C1/049.ps}}
      \centerline{C1-049: 3C 220.2}
    \end{minipage}
    \hspace{3cm}
    \begin{minipage}{3cm}
      \mbox{}
      \centerline{\includegraphics[scale=0.26,angle=270]{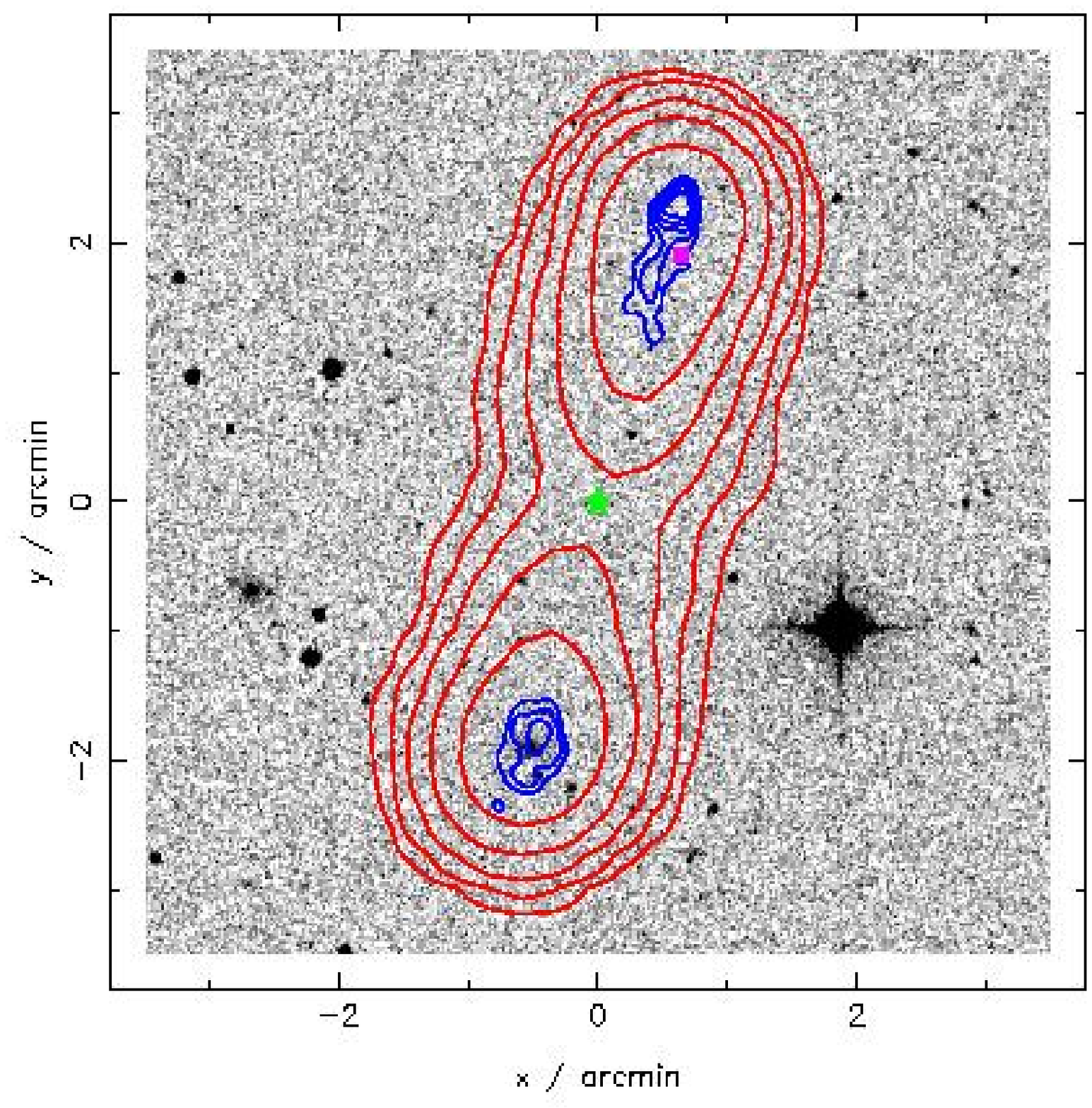}}
      \centerline{C1-050: 3C 223}
    \end{minipage}
    \hspace{3cm}
    \begin{minipage}{3cm}
      \mbox{}
      \centerline{\includegraphics[scale=0.26,angle=270]{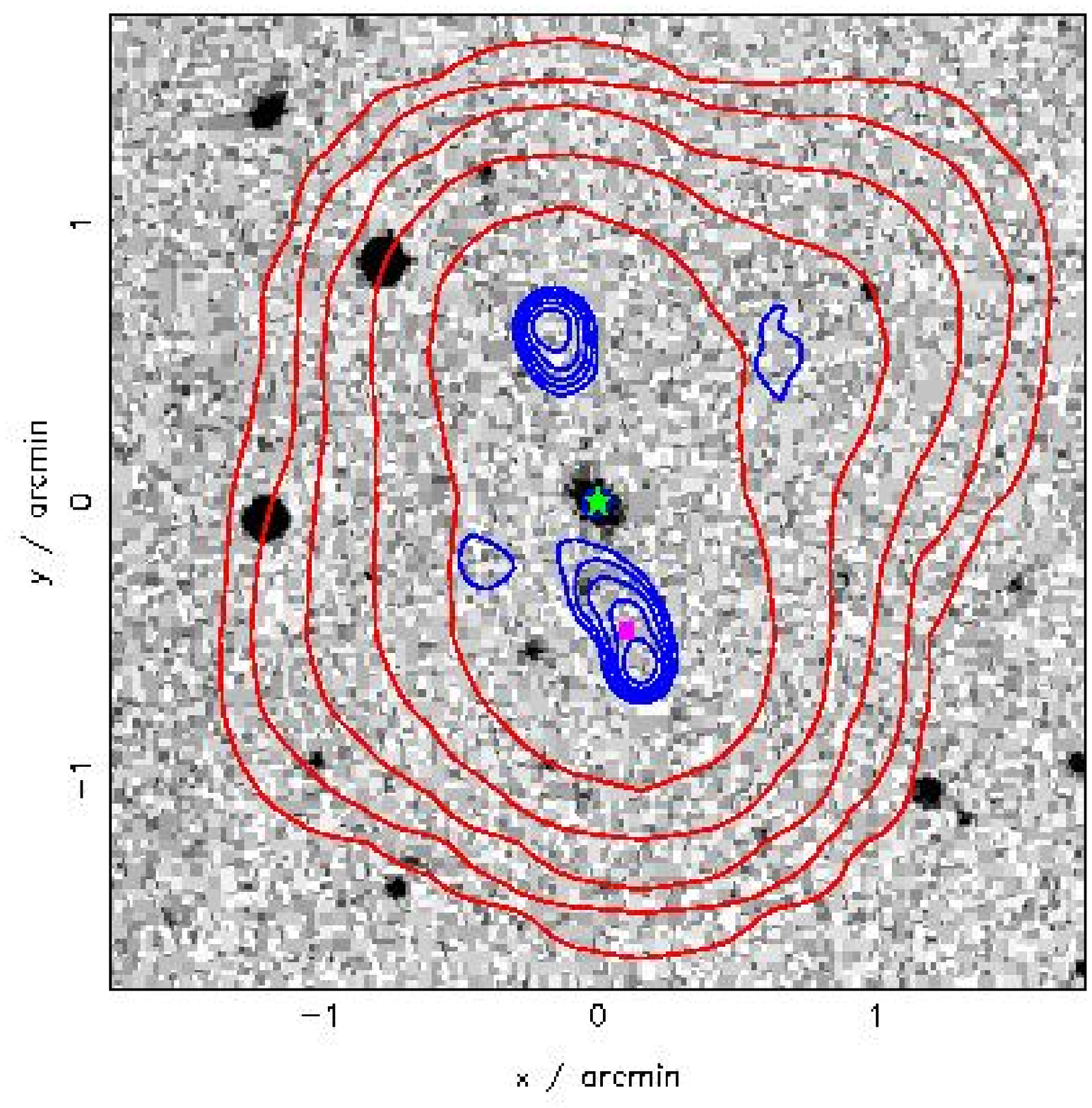}}
      \centerline{C1-051: 3C 223.1}
    \end{minipage}
    \vfill
    \begin{minipage}{3cm}      
      \mbox{}
      \centerline{\includegraphics[scale=0.26,angle=270]{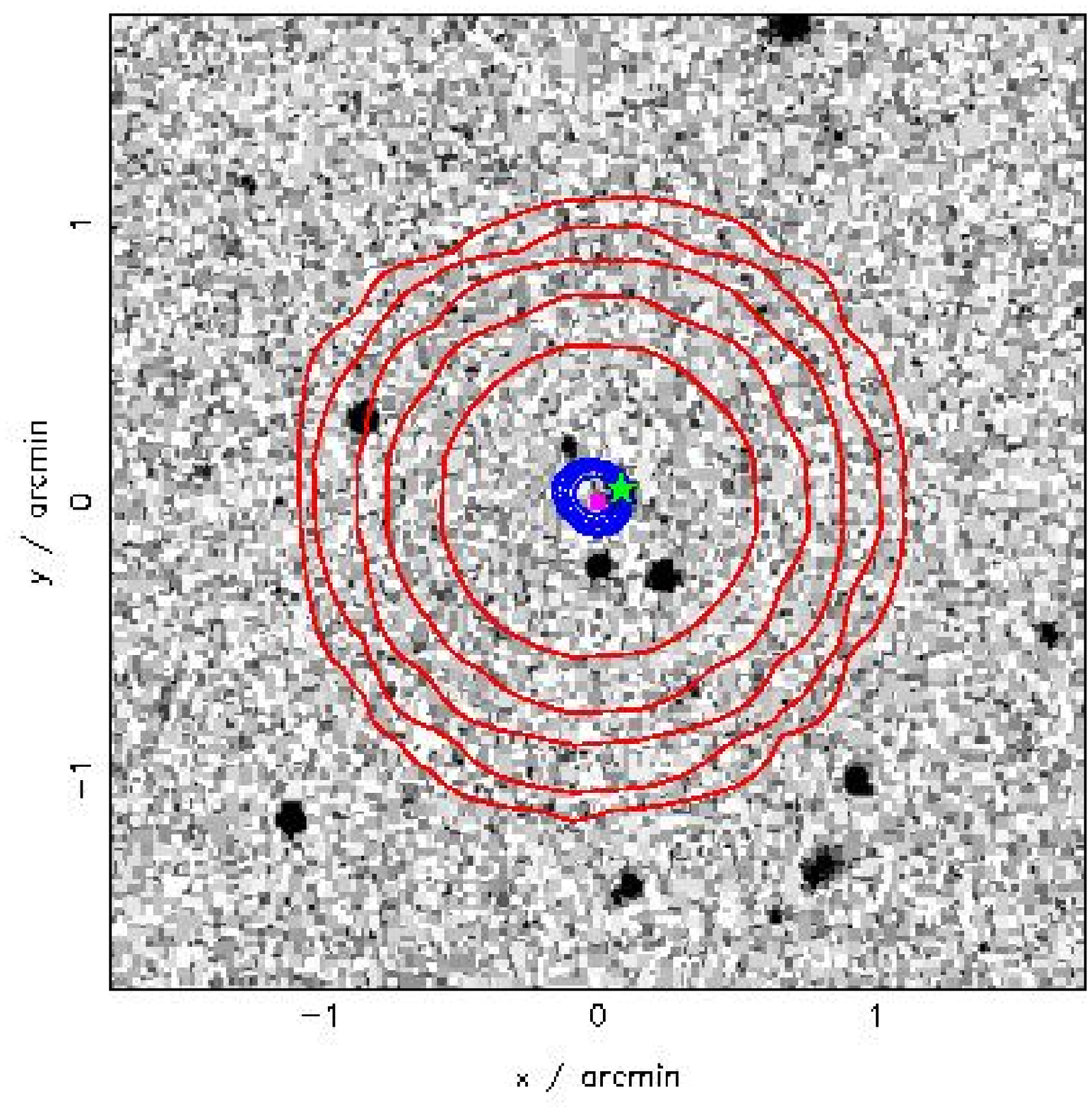}}
      \centerline{C1-052: 3C 225A}
    \end{minipage}
    \hspace{3cm}
    \begin{minipage}{3cm}
      \mbox{}
      \centerline{\includegraphics[scale=0.26,angle=270]{Contours/C1/053.ps}}
      \centerline{C1-053: 3C 225}
    \end{minipage}
    \hspace{3cm}
    \begin{minipage}{3cm}
      \mbox{}
      \centerline{\includegraphics[scale=0.26,angle=270]{Contours/C1/054.ps}}
      \centerline{C1-054: 4C 02.29}
    \end{minipage}
    \vfill
    \begin{minipage}{3cm}     
      \mbox{}
      \centerline{\includegraphics[scale=0.26,angle=270]{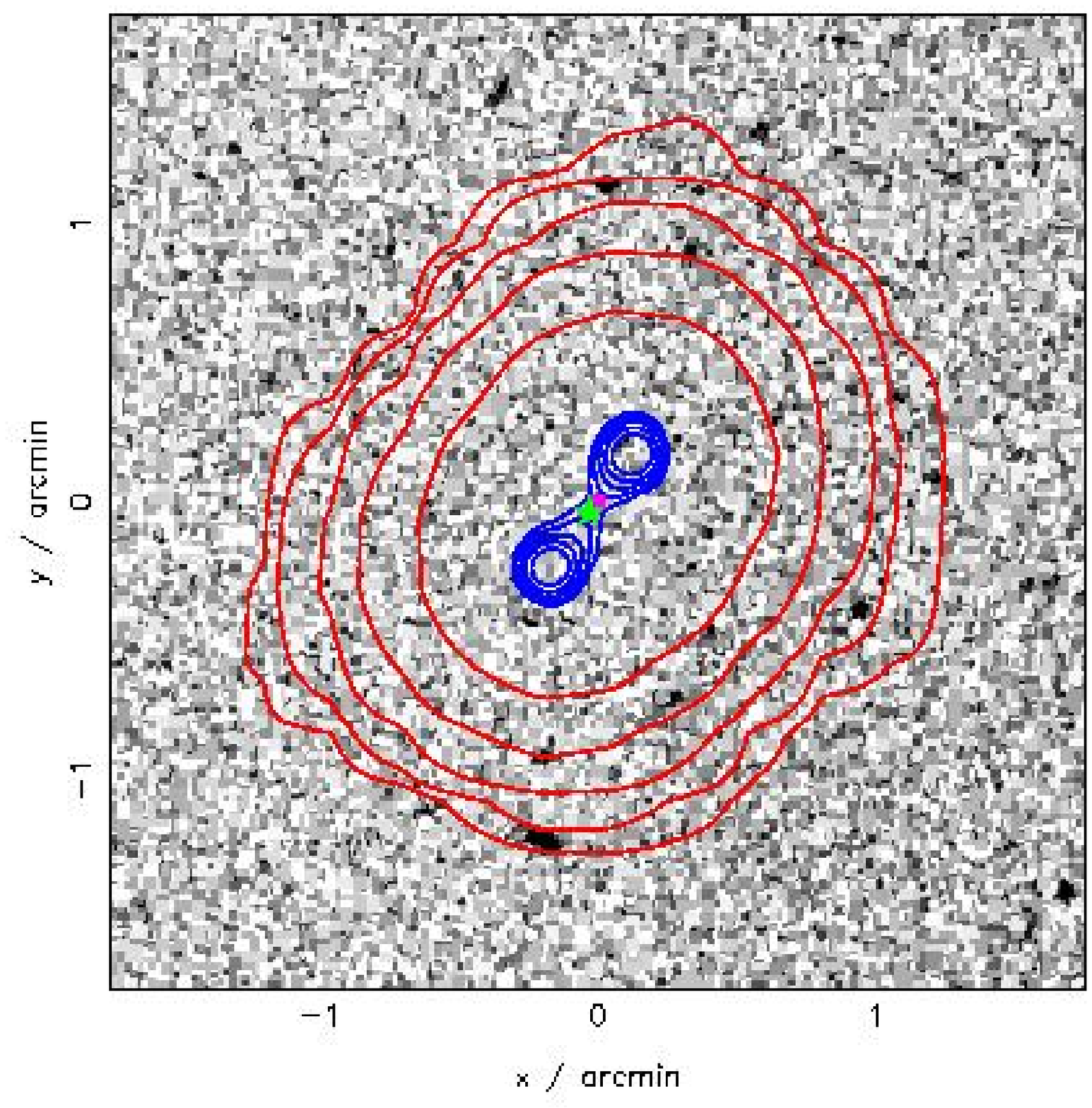}}
      \centerline{C1-055: 3C 226}
    \end{minipage}
    \hspace{3cm}
    \begin{minipage}{3cm}
      \mbox{}
      \centerline{\includegraphics[scale=0.26,angle=270]{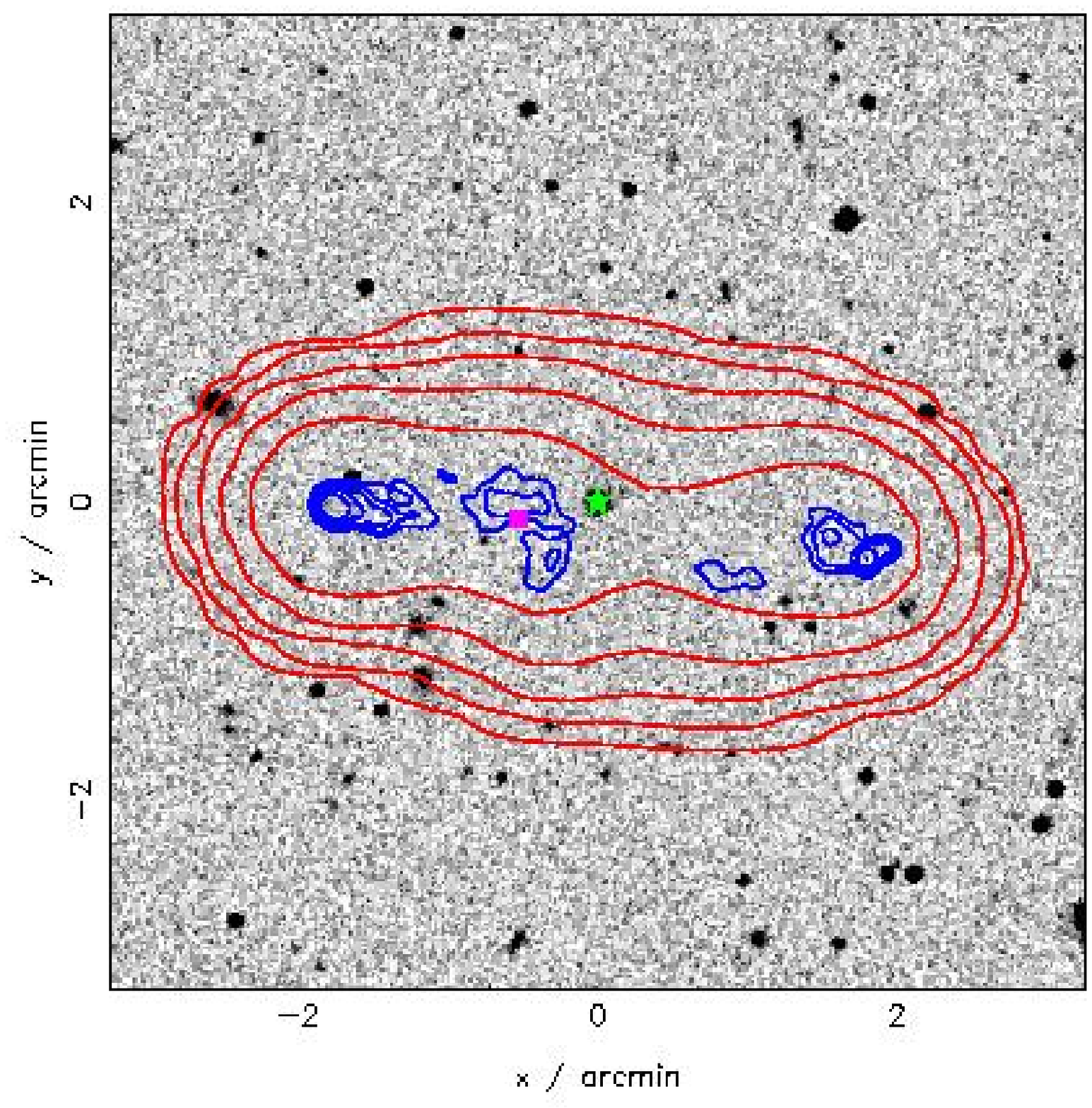}}
      \centerline{C1-056: 3C 227}
    \end{minipage}
    \hspace{3cm}
    \begin{minipage}{3cm}
      \mbox{}
      \centerline{\includegraphics[scale=0.26,angle=270]{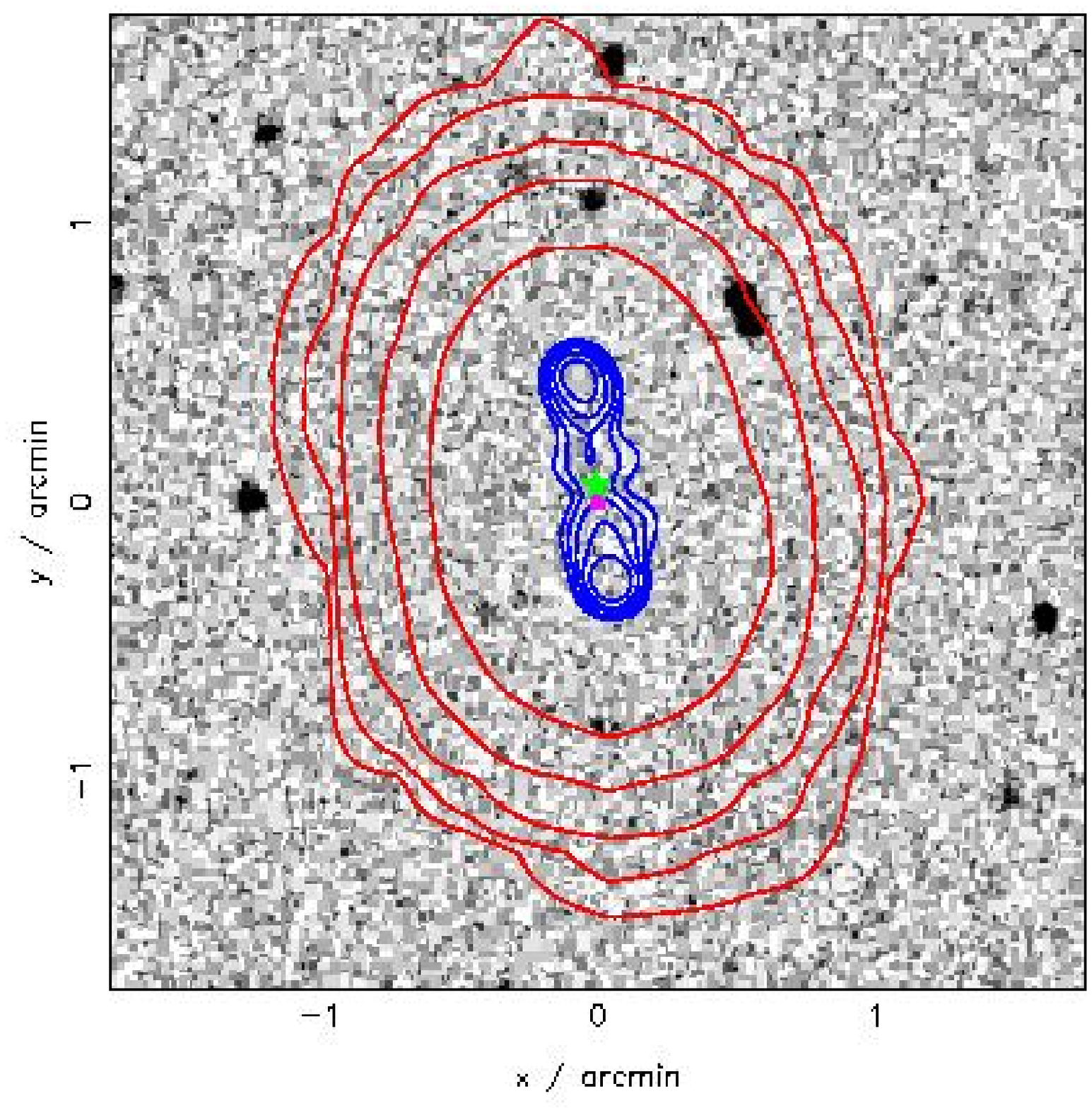}}
      \centerline{C1-058: 3C 228}
    \end{minipage}
    \vfill
    \begin{minipage}{3cm}     
      \mbox{}
      \centerline{\includegraphics[scale=0.26,angle=270]{Contours/C1/059.ps}}
      \centerline{C1-059: 3C 230}
    \end{minipage}
    \hspace{3cm}
    \begin{minipage}{3cm}
      \mbox{}
      \centerline{\includegraphics[scale=0.26,angle=270]{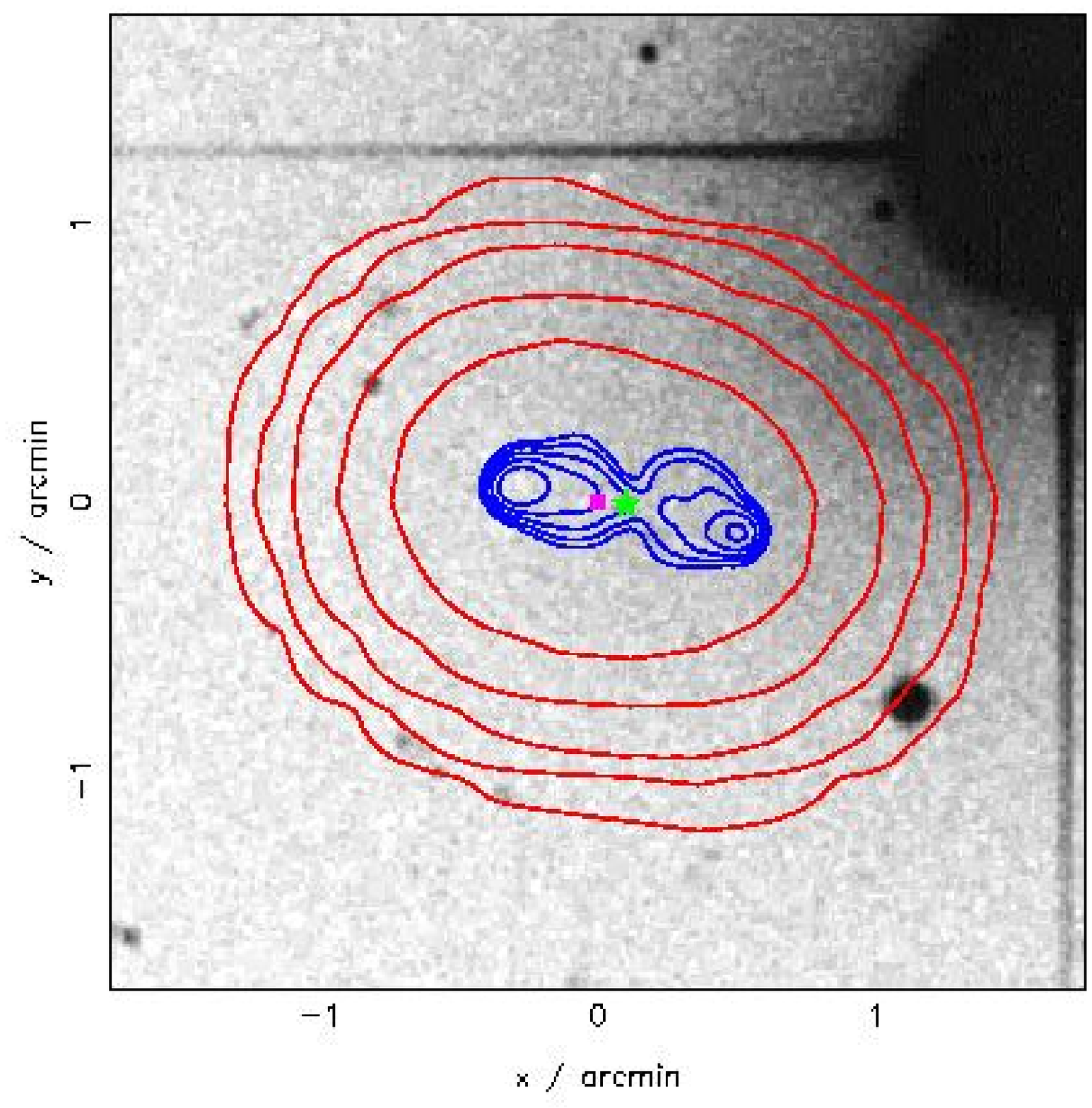}}
      \centerline{C1-060: 3C 229}
    \end{minipage}
    \hspace{3cm}
    \begin{minipage}{3cm}
      \mbox{}
      \centerline{\includegraphics[scale=0.26,angle=270]{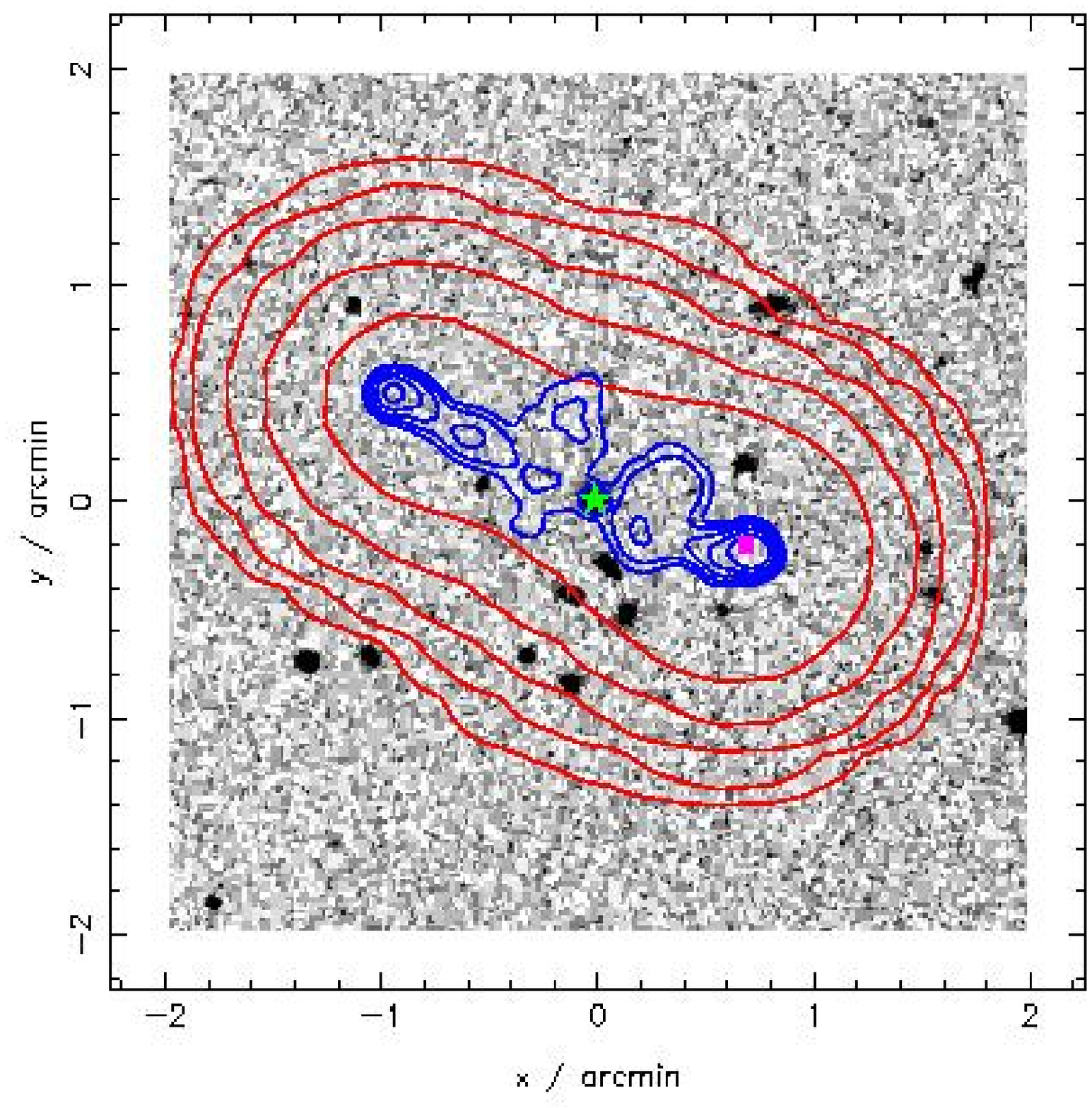}}
      \centerline{C1-063: 3C 234}
    \end{minipage}
  \end{center}
\end{figure}

\begin{figure}
  \begin{center}
    {\bf CoNFIG-1}\\  
  \begin{minipage}{3cm}      
      \mbox{}
      \centerline{\includegraphics[scale=0.26,angle=270]{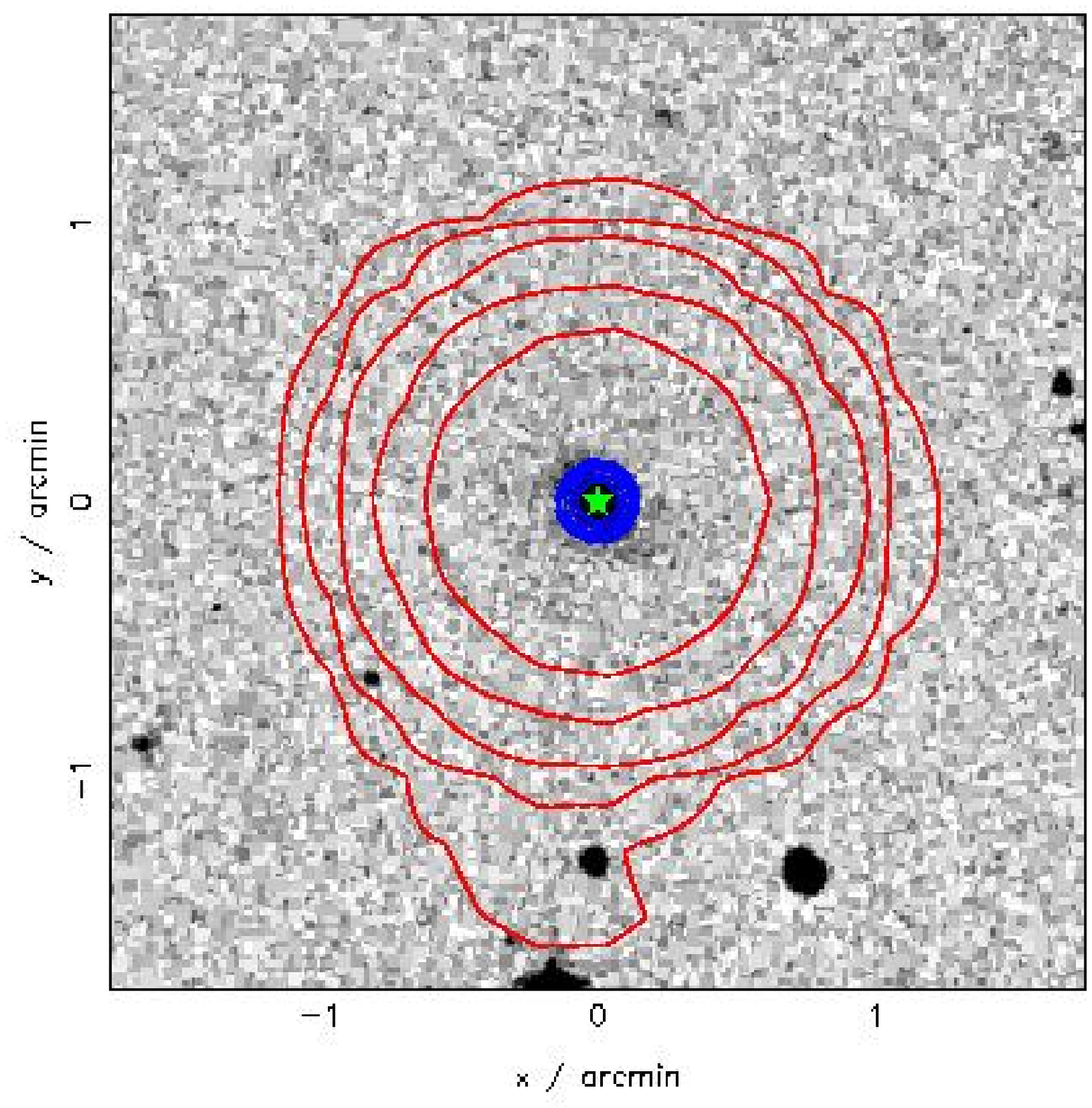}}
      \centerline{C1-064: 3C 236}
    \end{minipage}
    \hspace{3cm}
    \begin{minipage}{3cm}
      \mbox{}
      \centerline{\includegraphics[scale=0.26,angle=270]{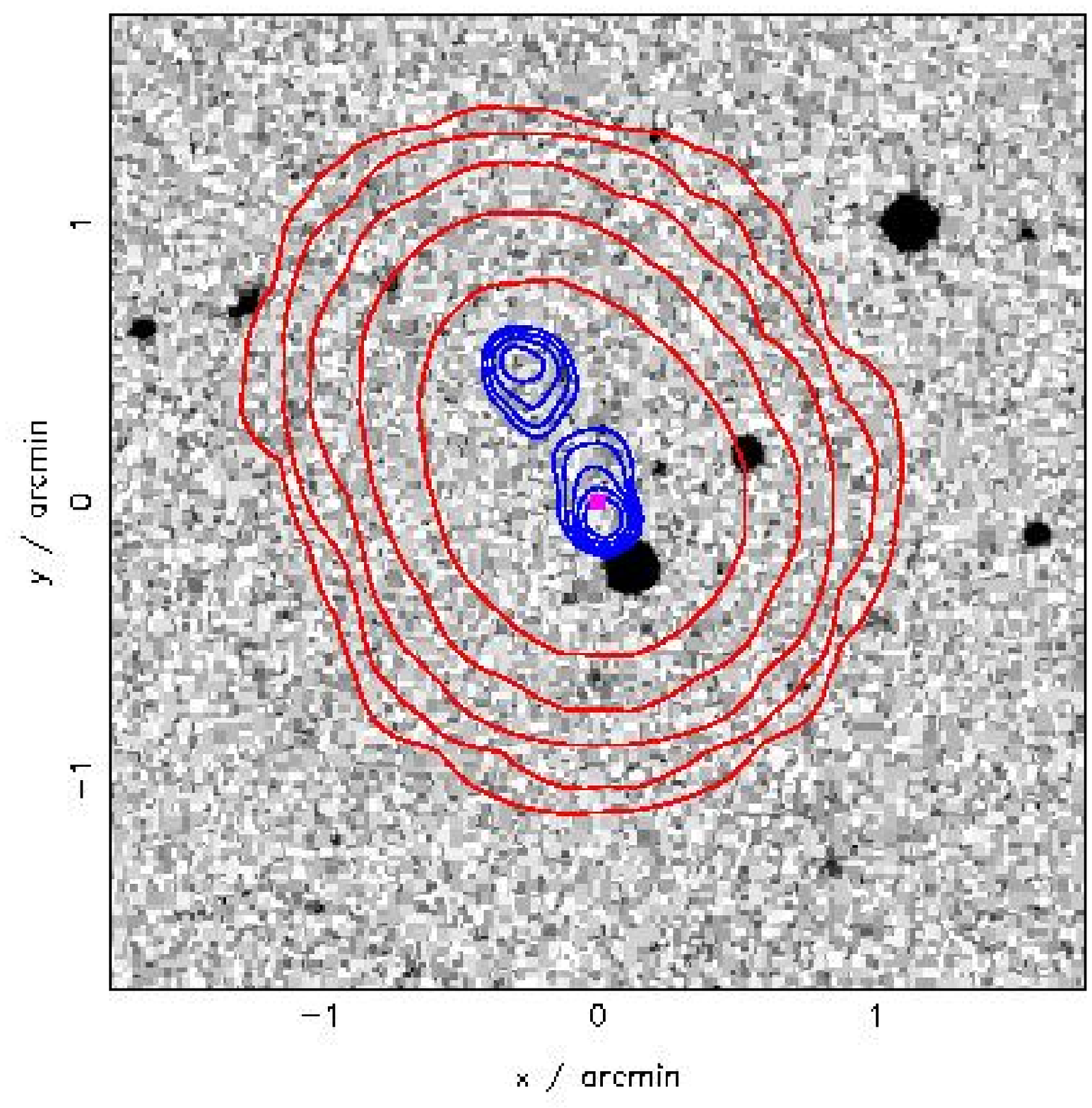}}
      \centerline{C1-065: 4C 44.19}
    \end{minipage}
    \hspace{3cm}
    \begin{minipage}{3cm}
      \mbox{}
      \centerline{\includegraphics[scale=0.26,angle=270]{Contours/C1/067.ps}}
      \centerline{C1-067: 3C 238}
    \end{minipage}
    \vfill
    \begin{minipage}{3cm}      
      \mbox{}
      \centerline{\includegraphics[scale=0.26,angle=270]{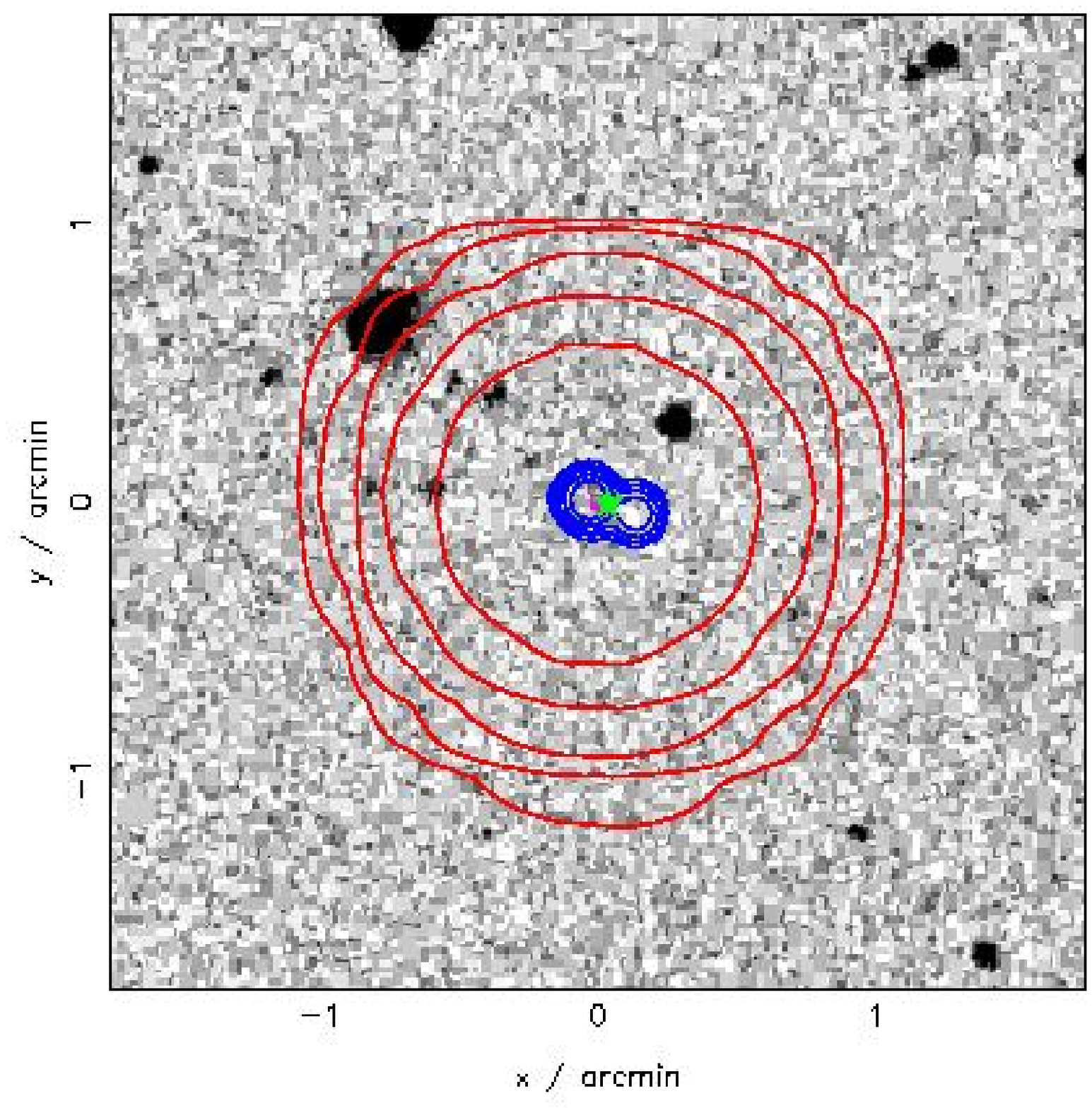}}
      \centerline{C1-068: 3C 239}
    \end{minipage}
    \hspace{3cm}
    \begin{minipage}{3cm}
      \mbox{}
      \centerline{\includegraphics[scale=0.26,angle=270]{Contours/C1/069.ps}}
      \centerline{C1-069: 4C 39.29}
    \end{minipage}
    \hspace{3cm}
    \begin{minipage}{3cm}
      \mbox{}
      \centerline{\includegraphics[scale=0.26,angle=270]{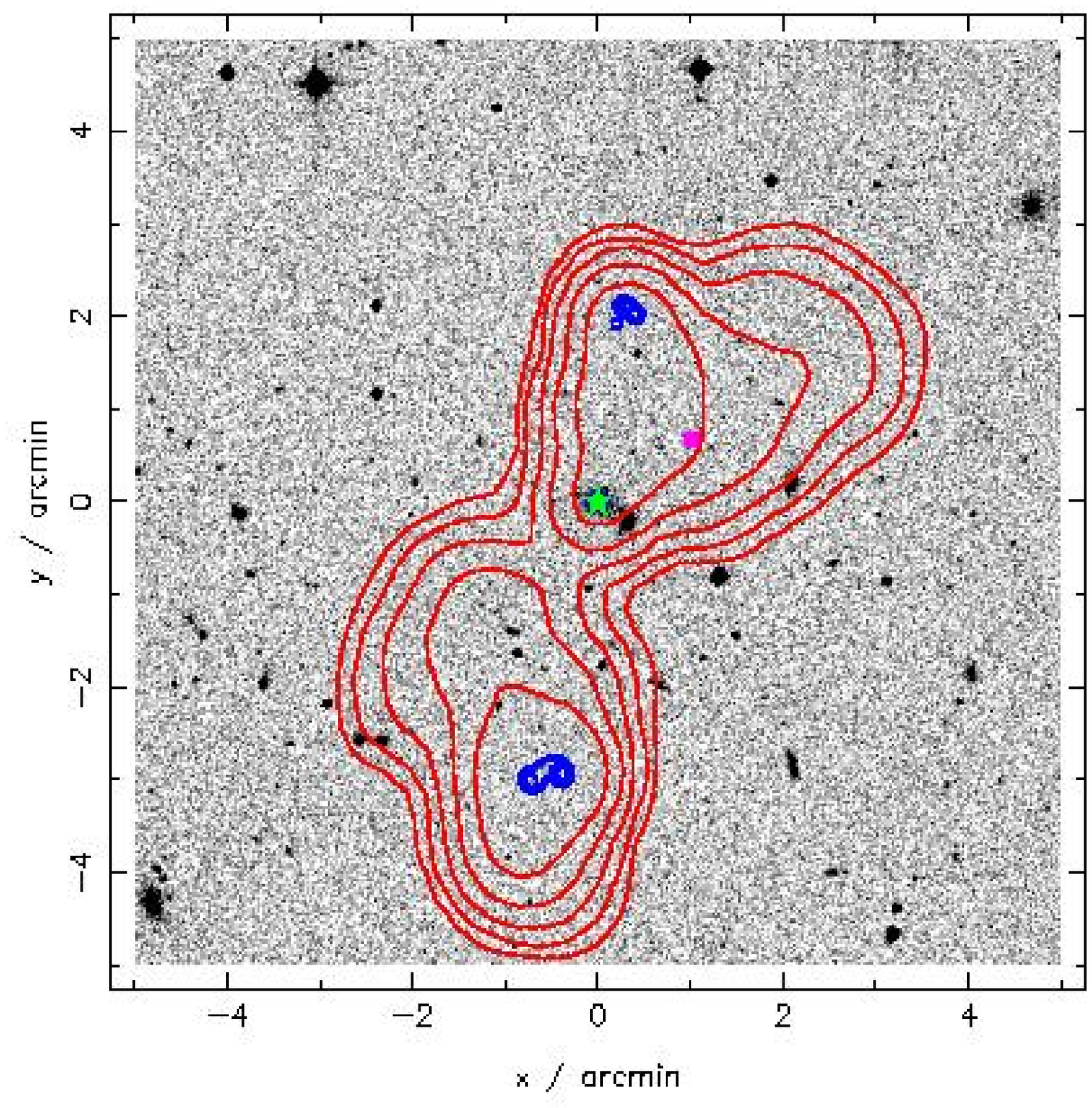}}
      \centerline{C1-070: 4C 48.29A}
    \end{minipage}
    \vfill
    \begin{minipage}{3cm}     
      \mbox{}
      \centerline{\includegraphics[scale=0.26,angle=270]{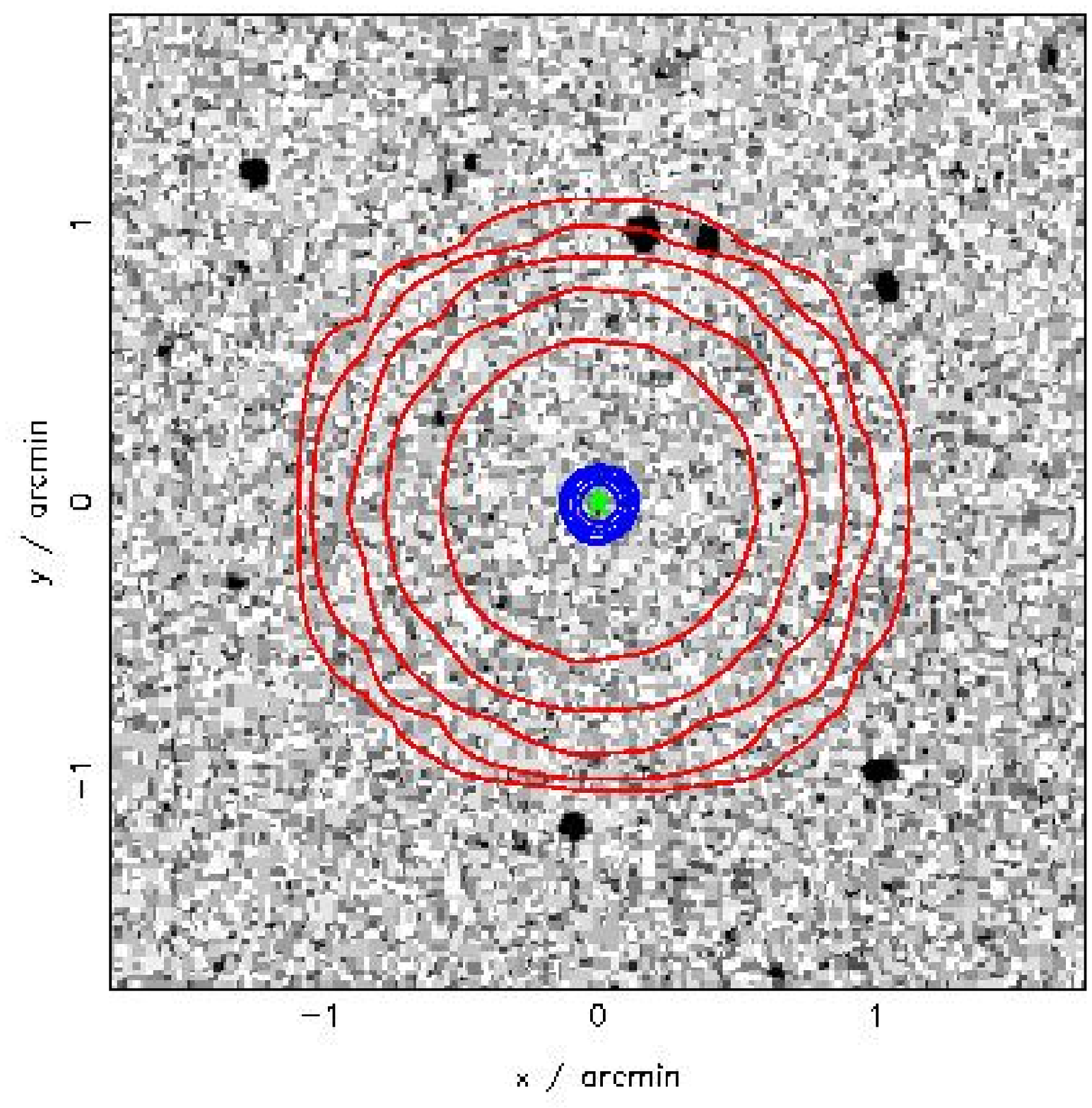}}
      \centerline{C1-071: 3C 241}
    \end{minipage}
    \hspace{3cm}
    \begin{minipage}{3cm}
      \mbox{}
      \centerline{\includegraphics[scale=0.26,angle=270]{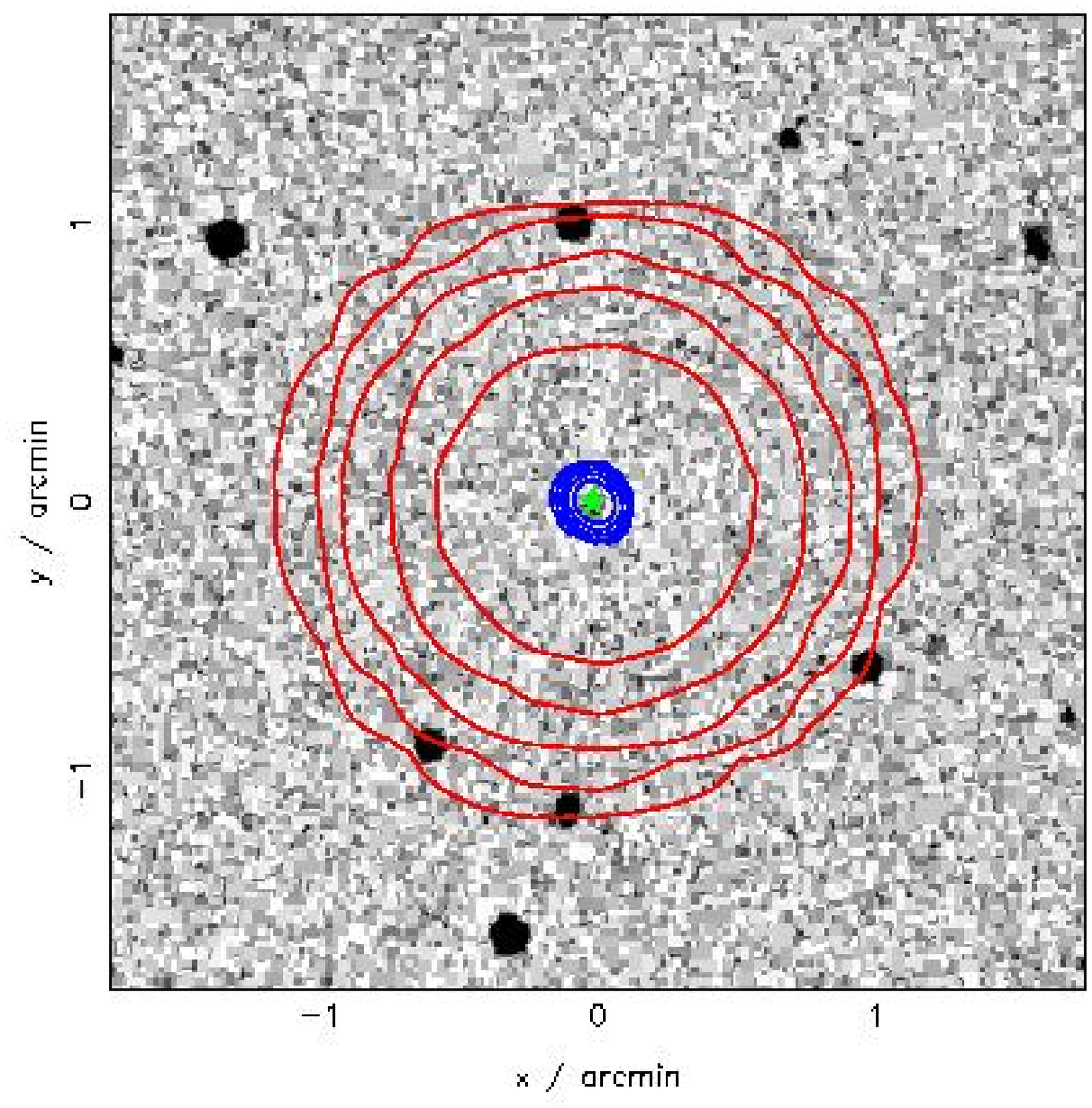}}
      \centerline{C1-072: 4C 59.13}
    \end{minipage}
    \hspace{3cm}
    \begin{minipage}{3cm}
      \mbox{}
      \centerline{\includegraphics[scale=0.26,angle=270]{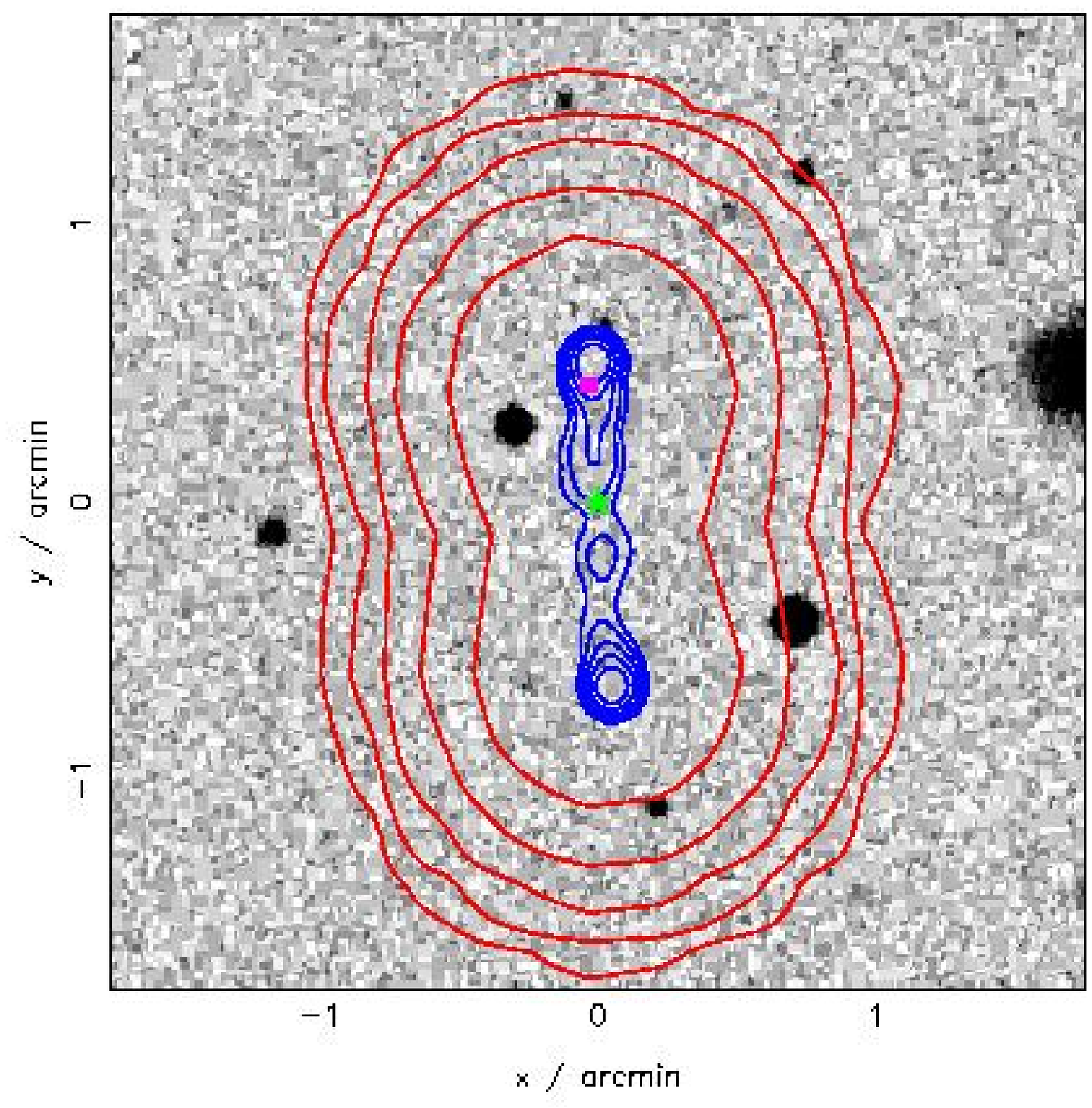}}
      \centerline{C1-073: 4C 46.21}
    \end{minipage}
    \vfill
    \begin{minipage}{3cm}     
      \mbox{}
      \centerline{\includegraphics[scale=0.26,angle=270]{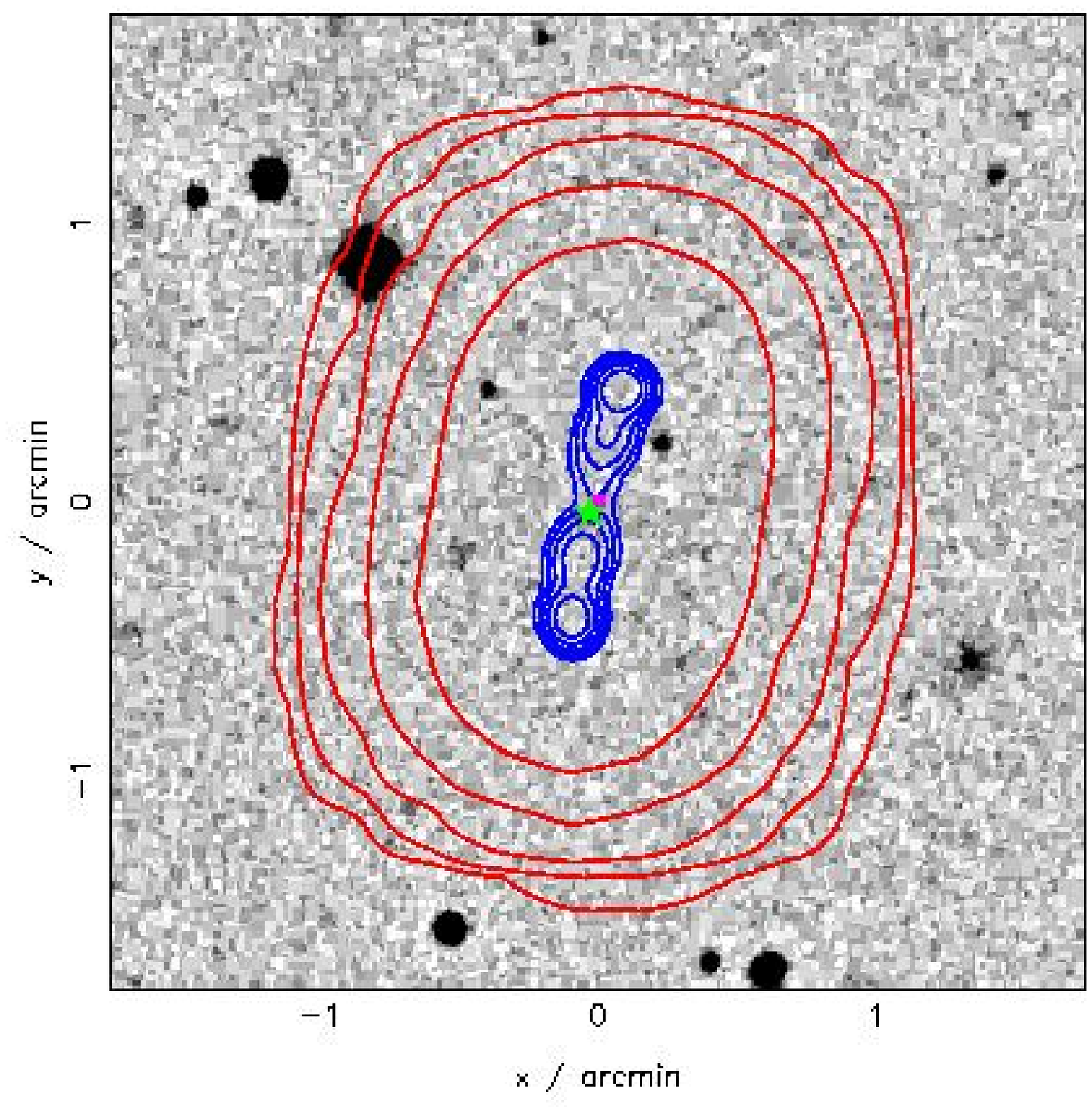}}
      \centerline{C1-074: 3C 244.1}
    \end{minipage}
    \hspace{3cm}
    \begin{minipage}{3cm}
      \mbox{}
      \centerline{\includegraphics[scale=0.26,angle=270]{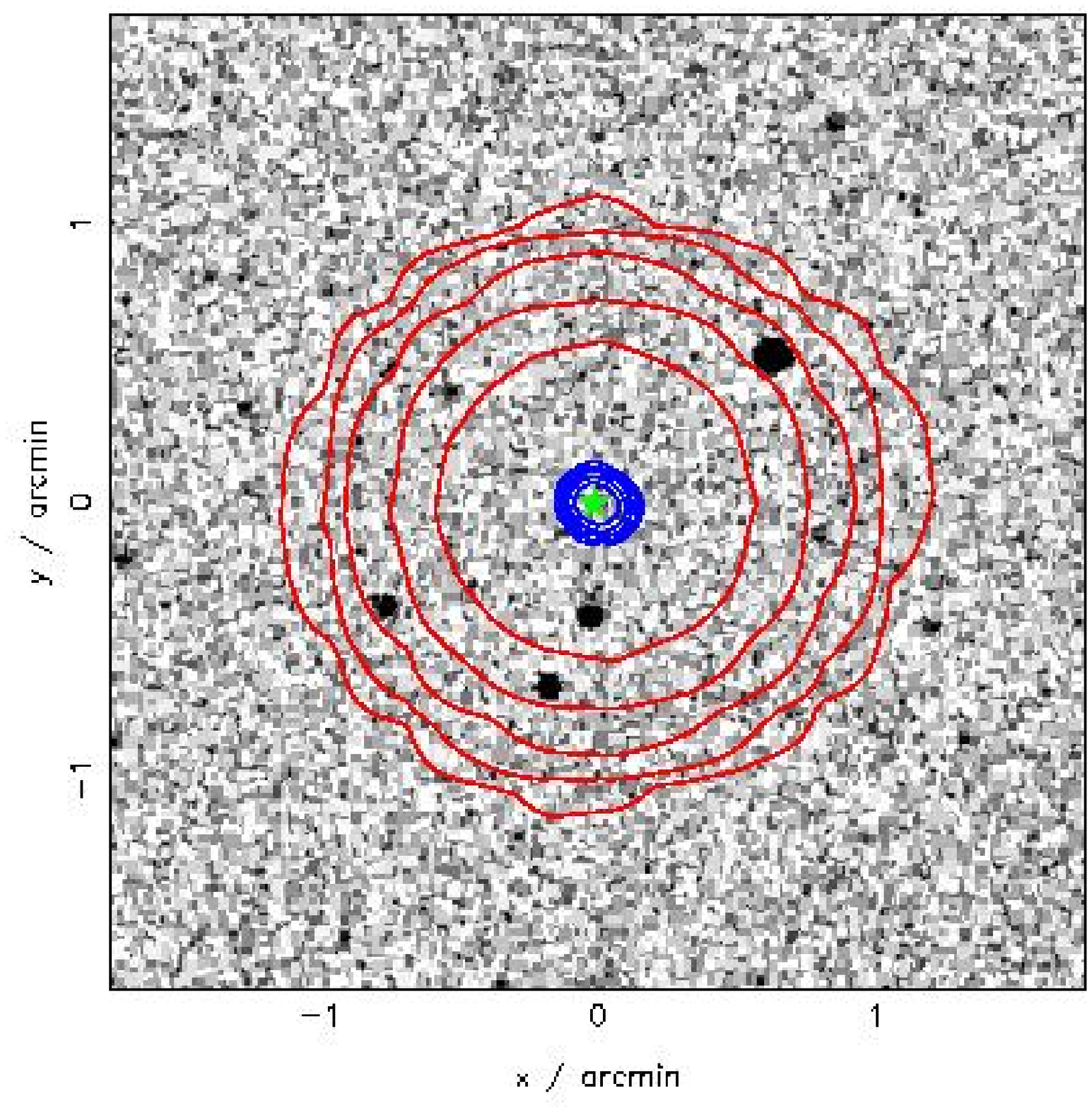}}
      \centerline{C1-075: 4C 50.30}
    \end{minipage}
    \hspace{3cm}
    \begin{minipage}{3cm}
      \mbox{}
      \centerline{\includegraphics[scale=0.26,angle=270]{Contours/C1/078.ps}}
      \centerline{C1-078: 4C 03.18}
    \end{minipage}
  \end{center}
\end{figure}

\begin{figure}
  \begin{center}
    {\bf CoNFIG-1}\\  
  \begin{minipage}{3cm}      
      \mbox{}
      \centerline{\includegraphics[scale=0.26,angle=270]{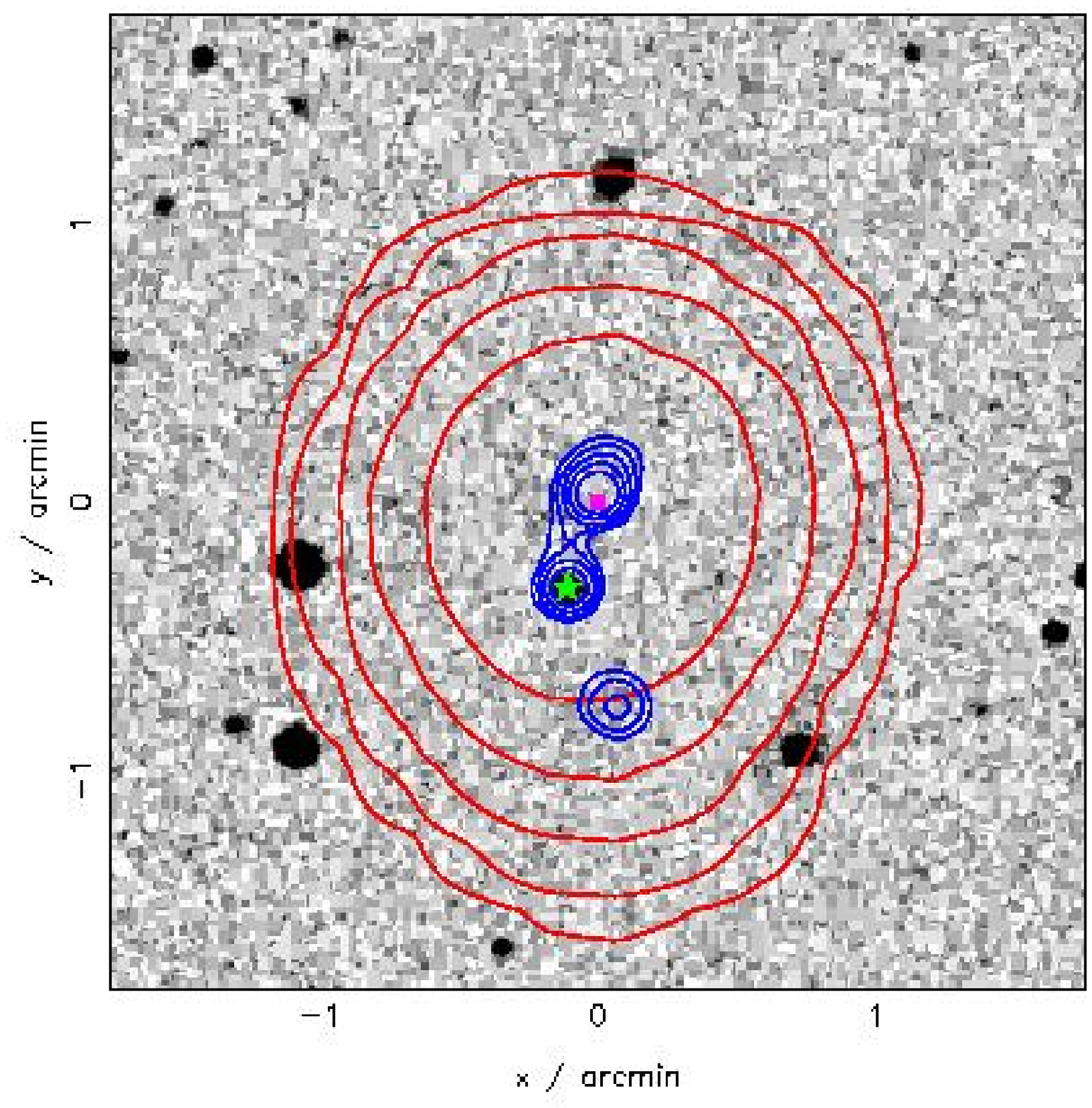}}
      \centerline{C1-081: 4C 20.24}
    \end{minipage}
    \hspace{3cm}
    \begin{minipage}{3cm}
      \mbox{}
      \centerline{\includegraphics[scale=0.26,angle=270]{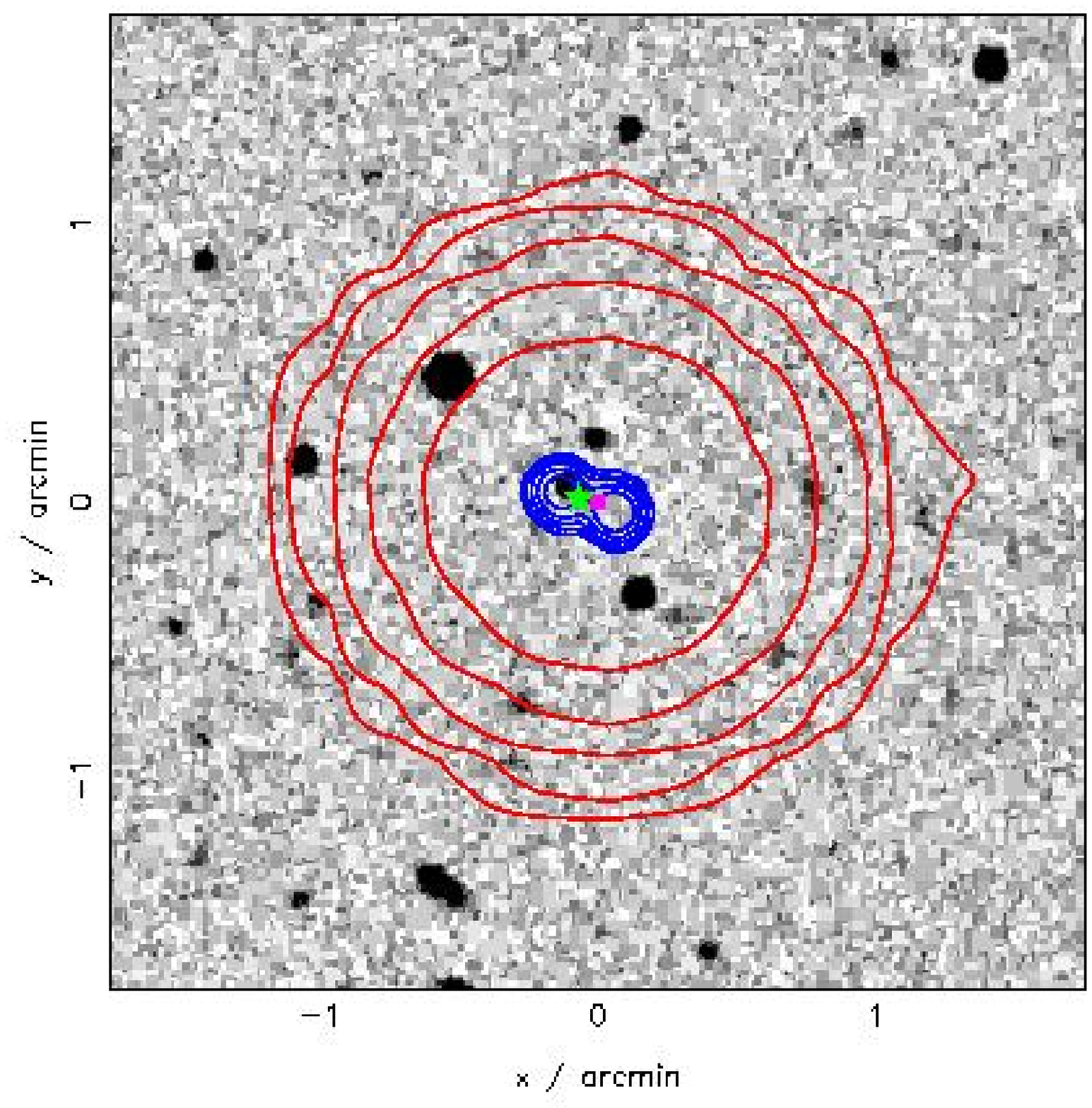}}
      \centerline{C1-083: 3C 247}
    \end{minipage}
    \hspace{3cm}
    \begin{minipage}{3cm}
      \mbox{}
      \centerline{\includegraphics[scale=0.26,angle=270]{Contours/C1/084.ps}}
      \centerline{C1-084: 3C 249}
    \end{minipage}
    \vfill
    \begin{minipage}{3cm}      
      \mbox{}
      \centerline{\includegraphics[scale=0.26,angle=270]{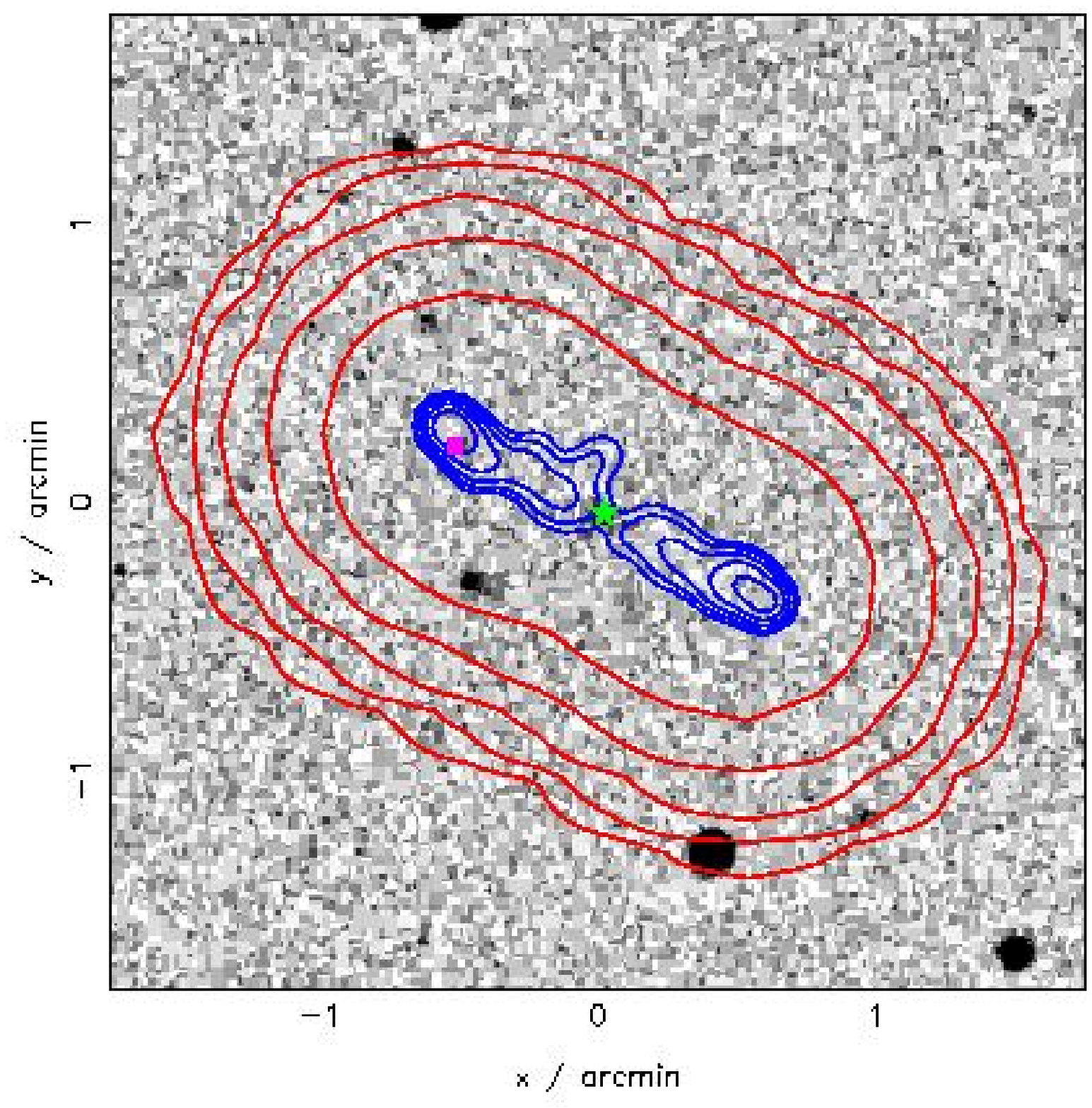}}
      \centerline{C1-087: 4C 37.29}
    \end{minipage}
    \hspace{3cm}
    \begin{minipage}{3cm}
      \mbox{}
      \centerline{\includegraphics[scale=0.26,angle=270]{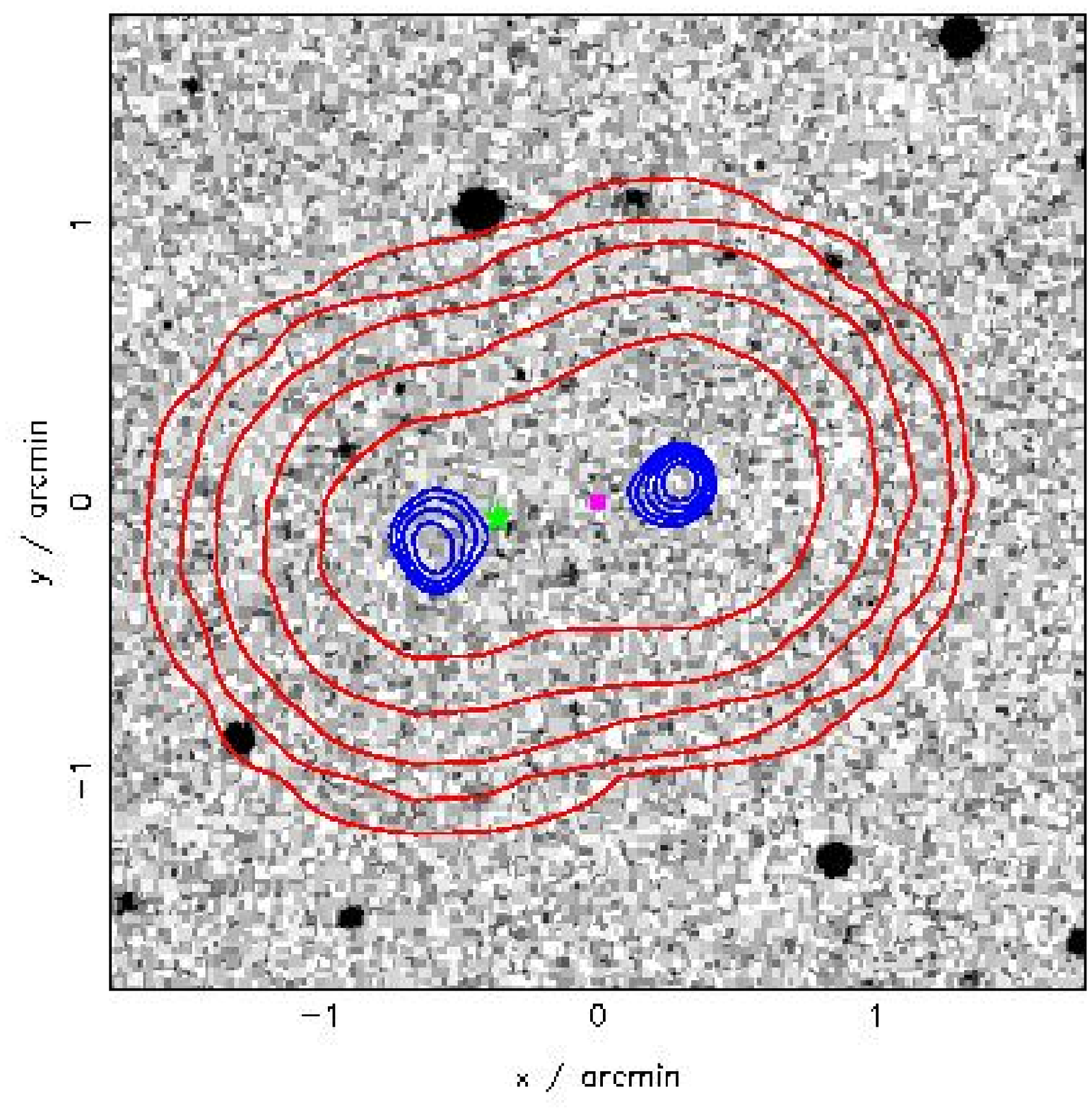}}
      \centerline{C1-088: 3C 252}
    \end{minipage}
    \hspace{3cm}
    \begin{minipage}{3cm}
      \mbox{}
      \centerline{\includegraphics[scale=0.26,angle=270]{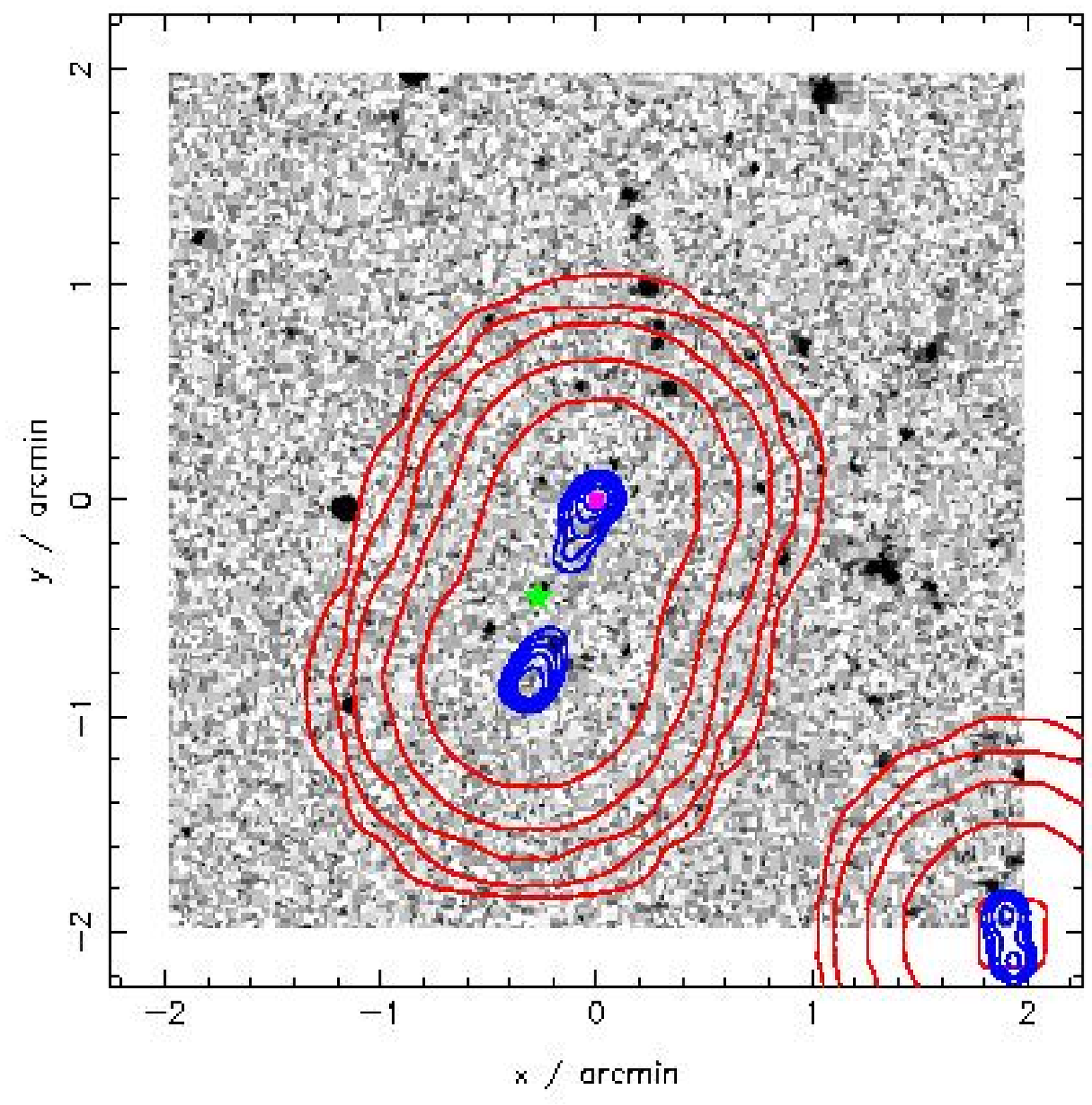}}
      \centerline{C1-089: 4C 43.21}
    \end{minipage}
    \vfill
    \begin{minipage}{3cm}     
      \mbox{}
      \centerline{\includegraphics[scale=0.26,angle=270]{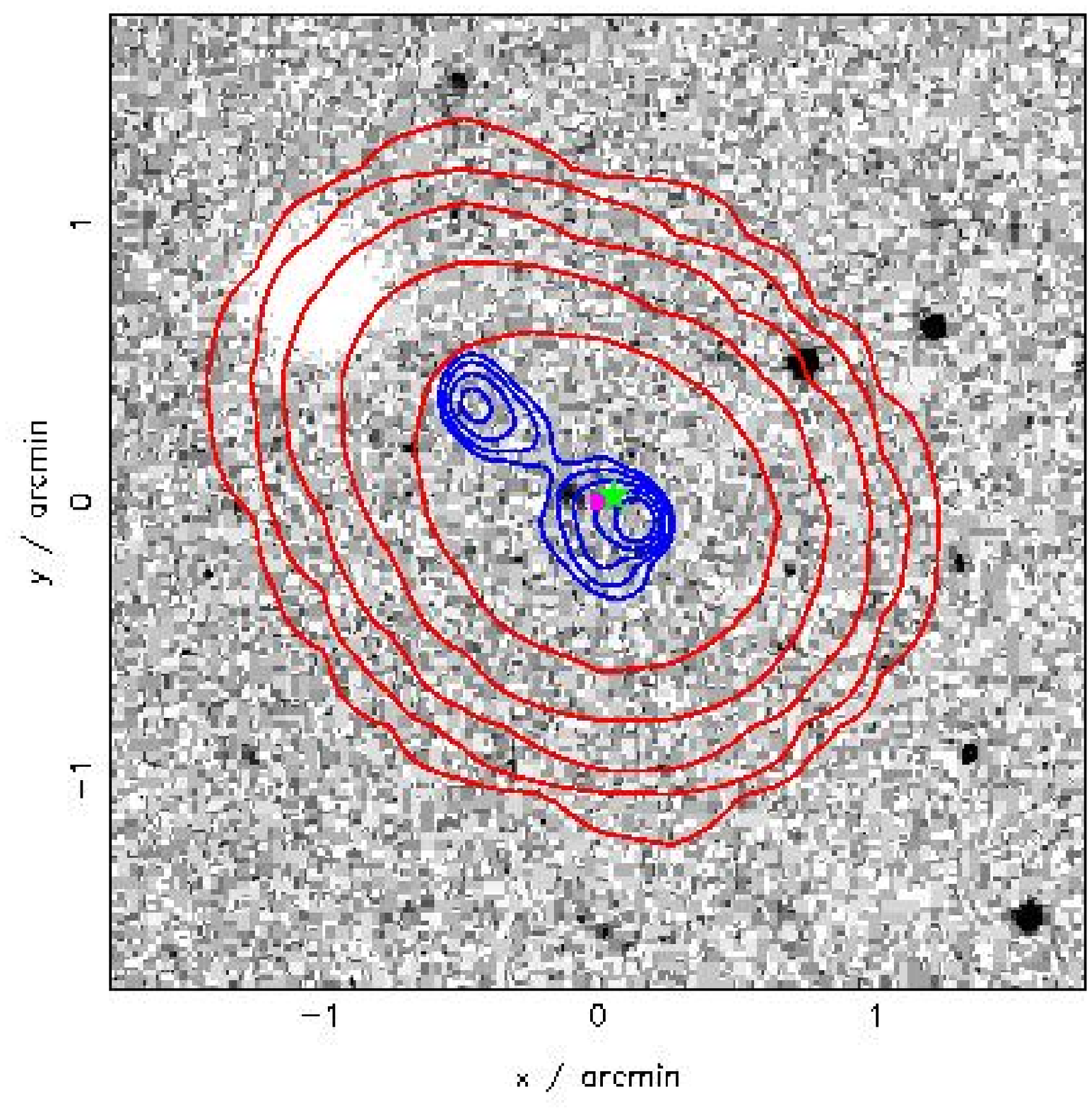}}
      \centerline{C1-090: 3C 253}
    \end{minipage}
    \hspace{3cm}
    \begin{minipage}{3cm}
      \mbox{}
      \centerline{\includegraphics[scale=0.26,angle=270]{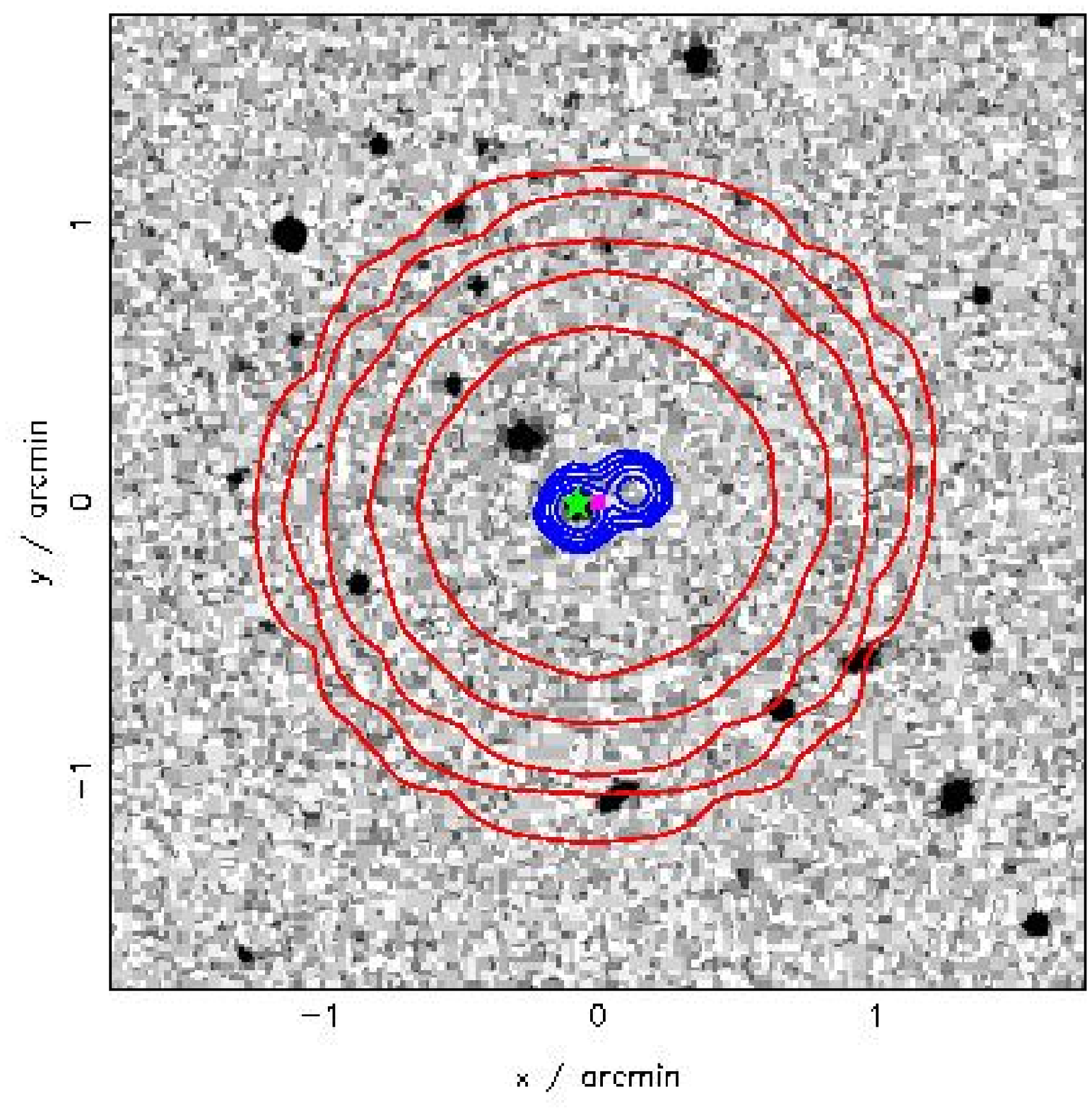}}
      \centerline{C1-091: 3C 254}
    \end{minipage}
    \hspace{3cm}
    \begin{minipage}{3cm}
      \mbox{}
      \centerline{\includegraphics[scale=0.26,angle=270]{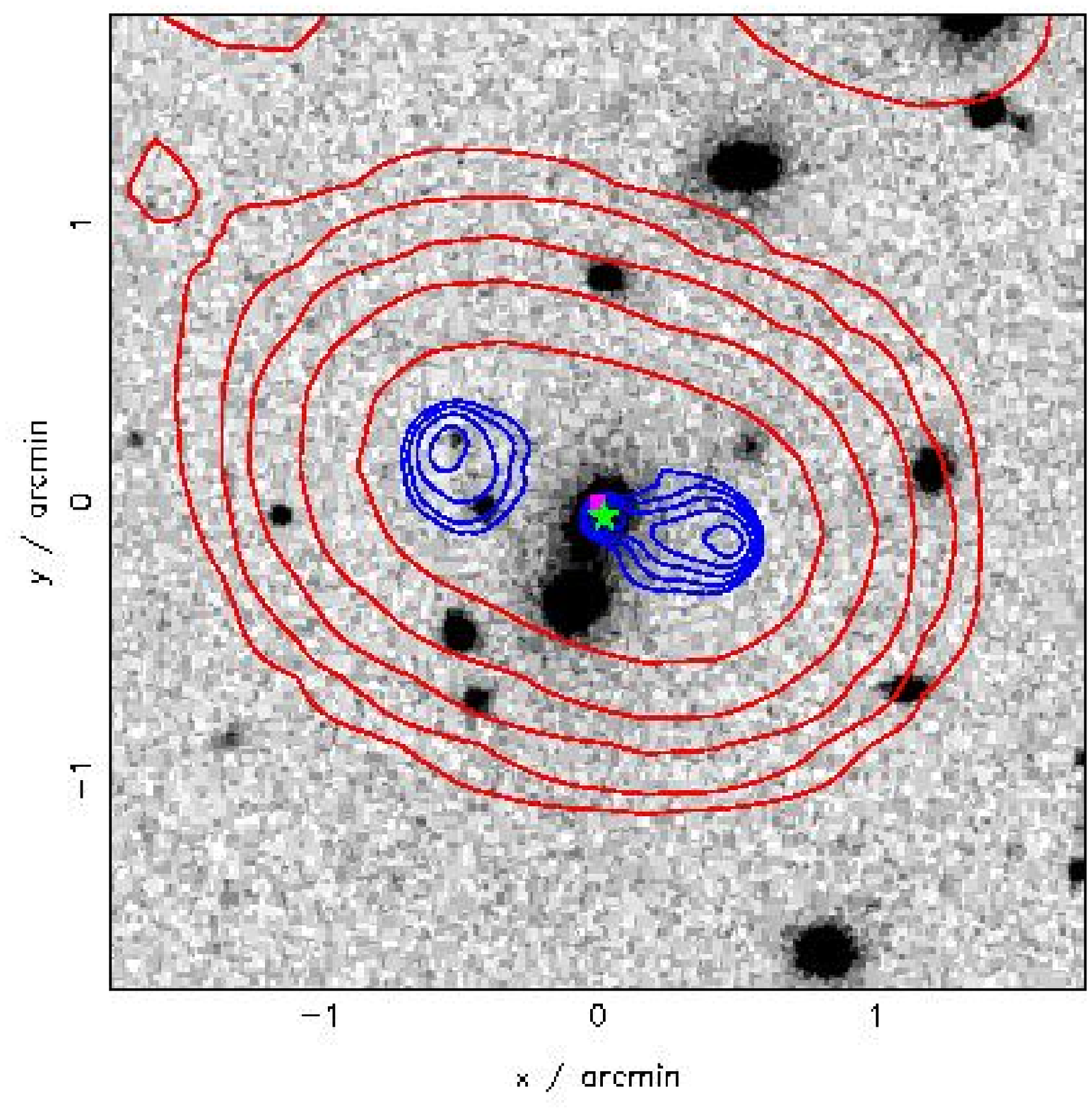}}
      \centerline{C1-092: 4C 29.41}
    \end{minipage}
    \vfill
    \begin{minipage}{3cm}     
      \mbox{}
      \centerline{\includegraphics[scale=0.26,angle=270]{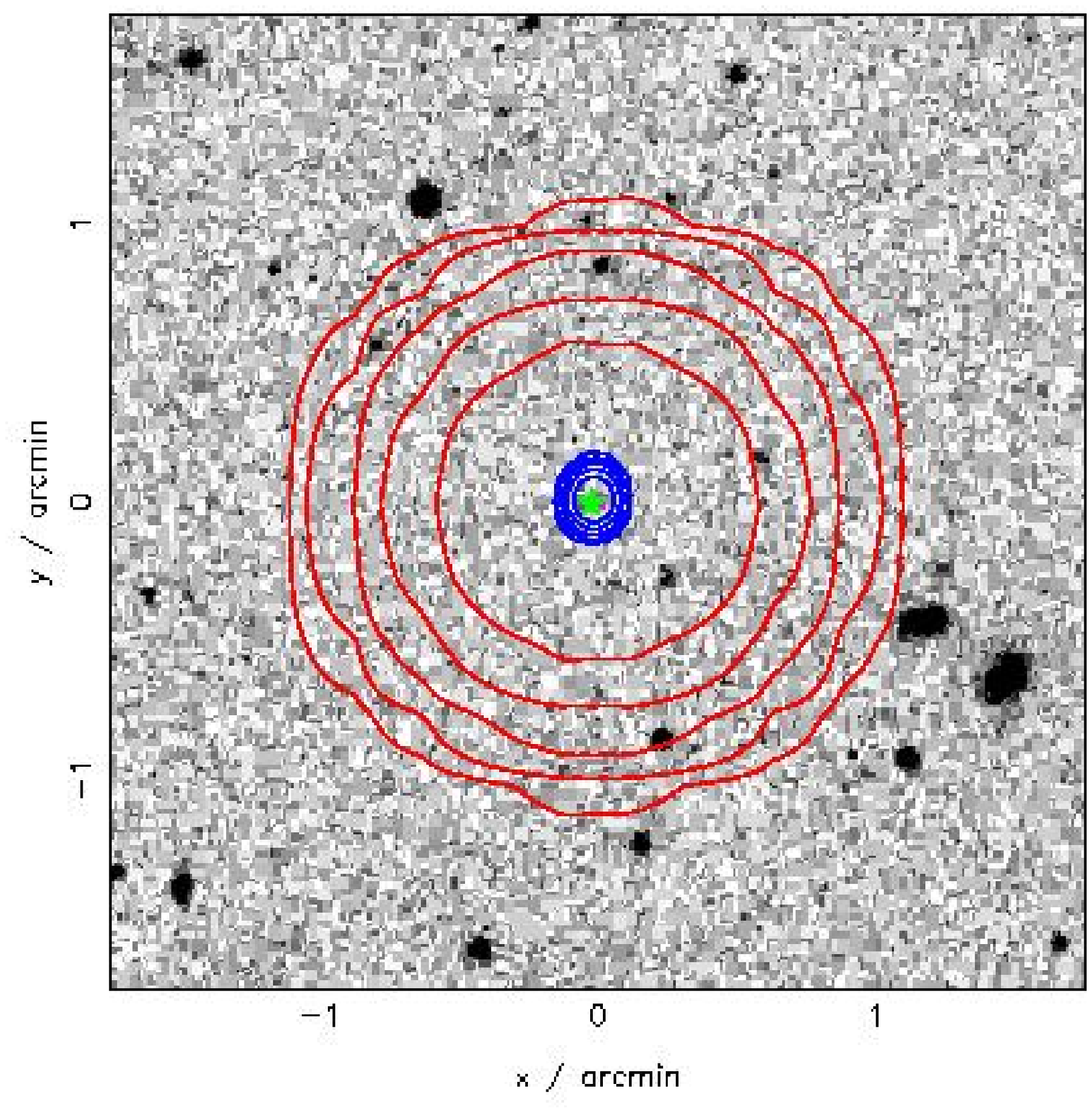}}
      \centerline{C1-093: 3C 255}
    \end{minipage}
    \hspace{3cm}
    \begin{minipage}{3cm}
      \mbox{}
      \centerline{\includegraphics[scale=0.26,angle=270]{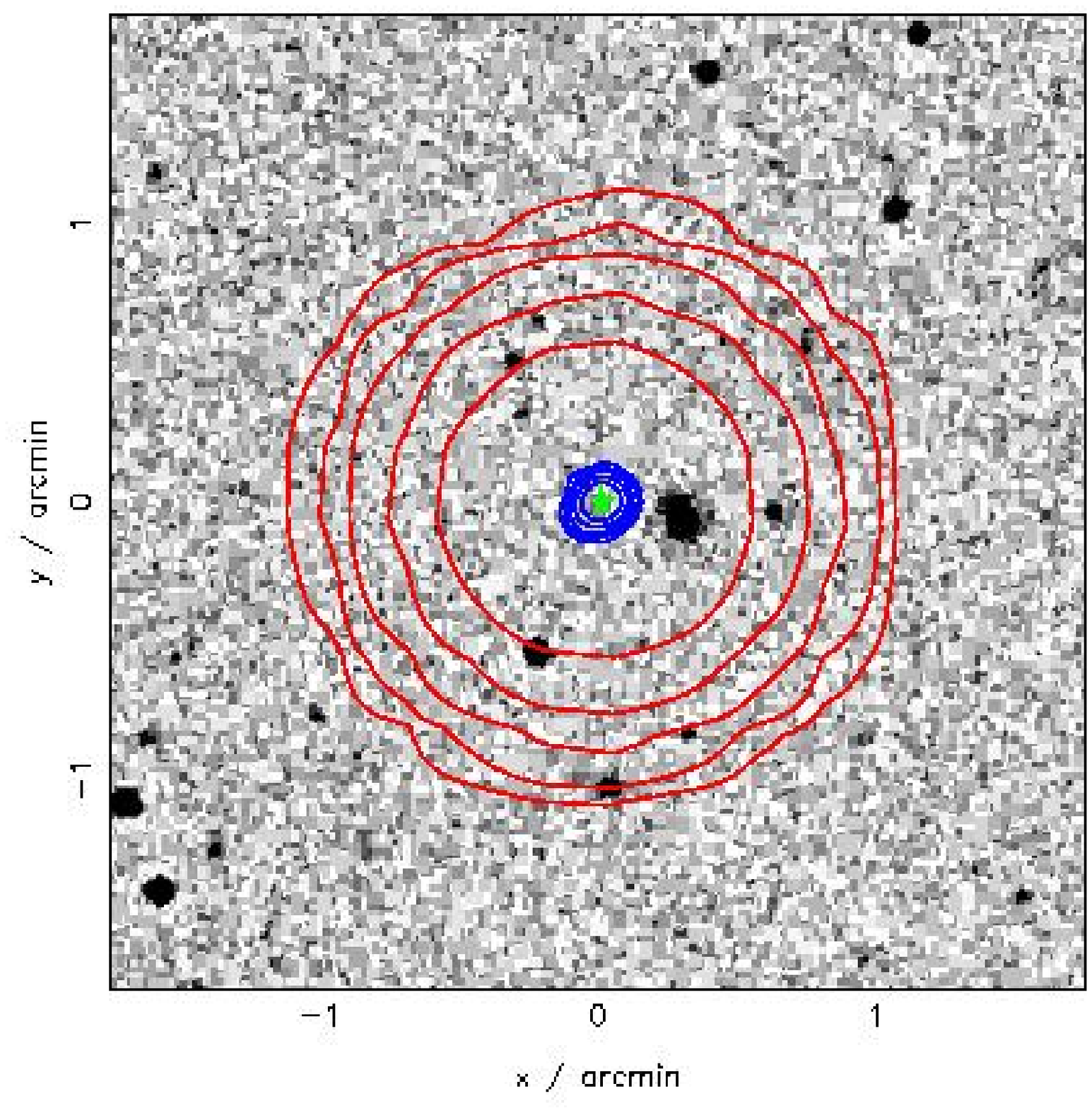}}
      \centerline{C1-095: 3C 256}
    \end{minipage}
    \hspace{3cm}
    \begin{minipage}{3cm}
      \mbox{}
      \centerline{\includegraphics[scale=0.26,angle=270]{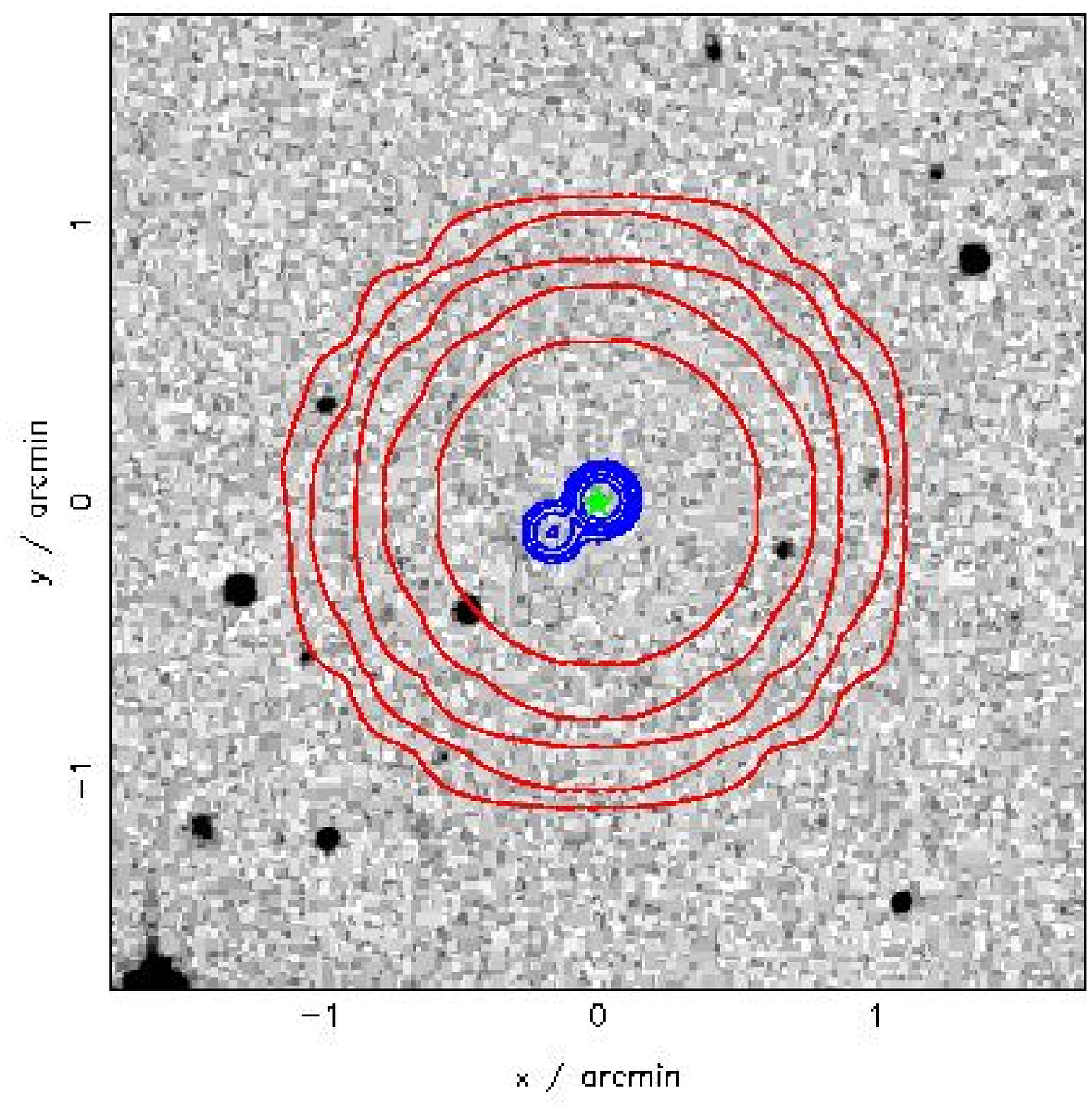}}
      \centerline{C1-096: 3C 257}
    \end{minipage}
  \end{center}
\end{figure}

\begin{figure}
  \begin{center}
    {\bf CoNFIG-1}\\  
  \begin{minipage}{3cm}      
      \mbox{}
      \centerline{\includegraphics[scale=0.26,angle=270]{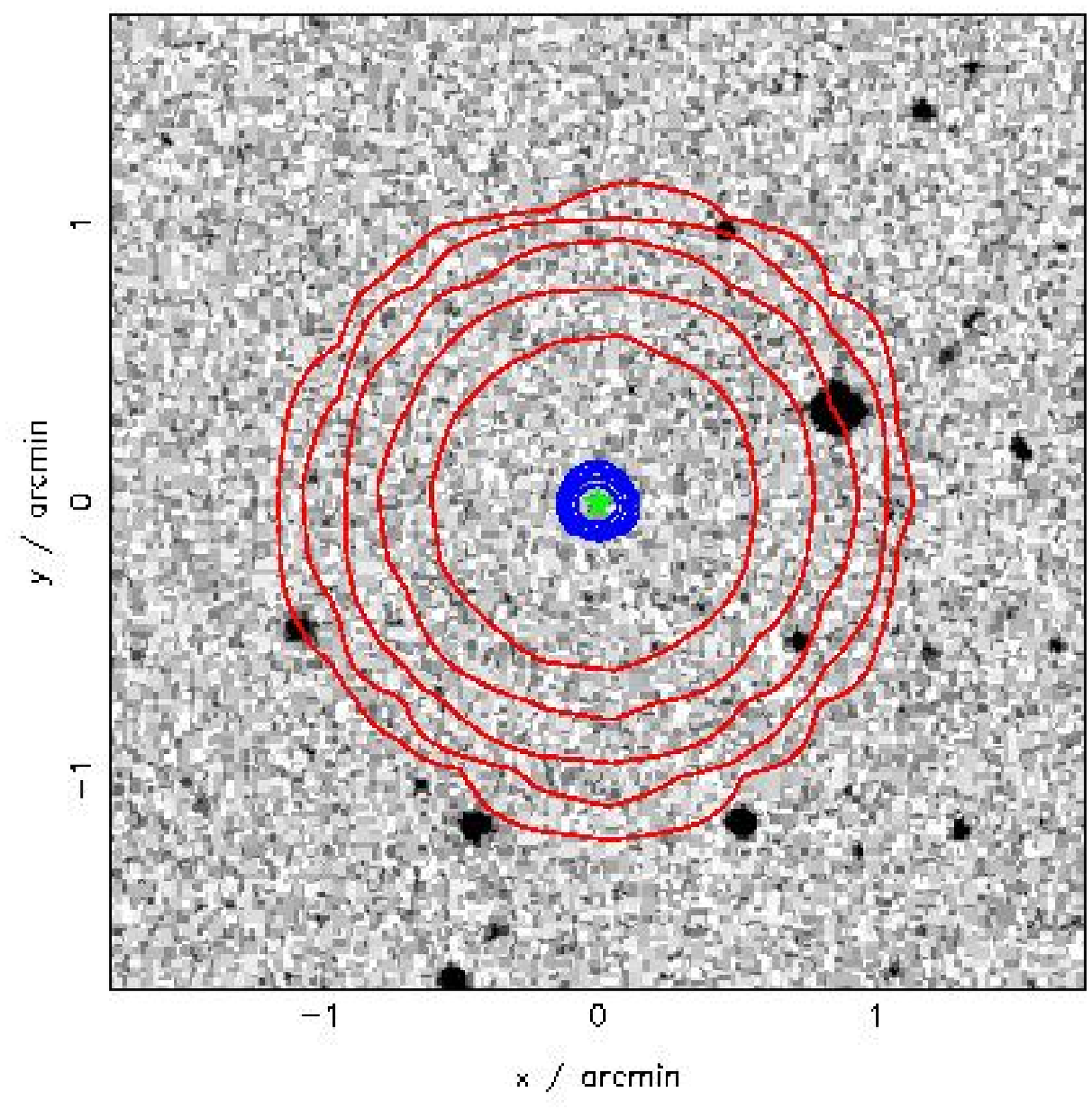}}
      \centerline{C1-098: TXS 1128+455}
    \end{minipage}
    \hspace{3cm}
    \begin{minipage}{3cm}
      \mbox{}
      \centerline{\includegraphics[scale=0.26,angle=270]{Contours/C1/099.ps}}
      \centerline{C1-099: 4C 43.22}
    \end{minipage}
    \hspace{3cm}
    \begin{minipage}{3cm}
      \mbox{}
      \centerline{\includegraphics[scale=0.26,angle=270]{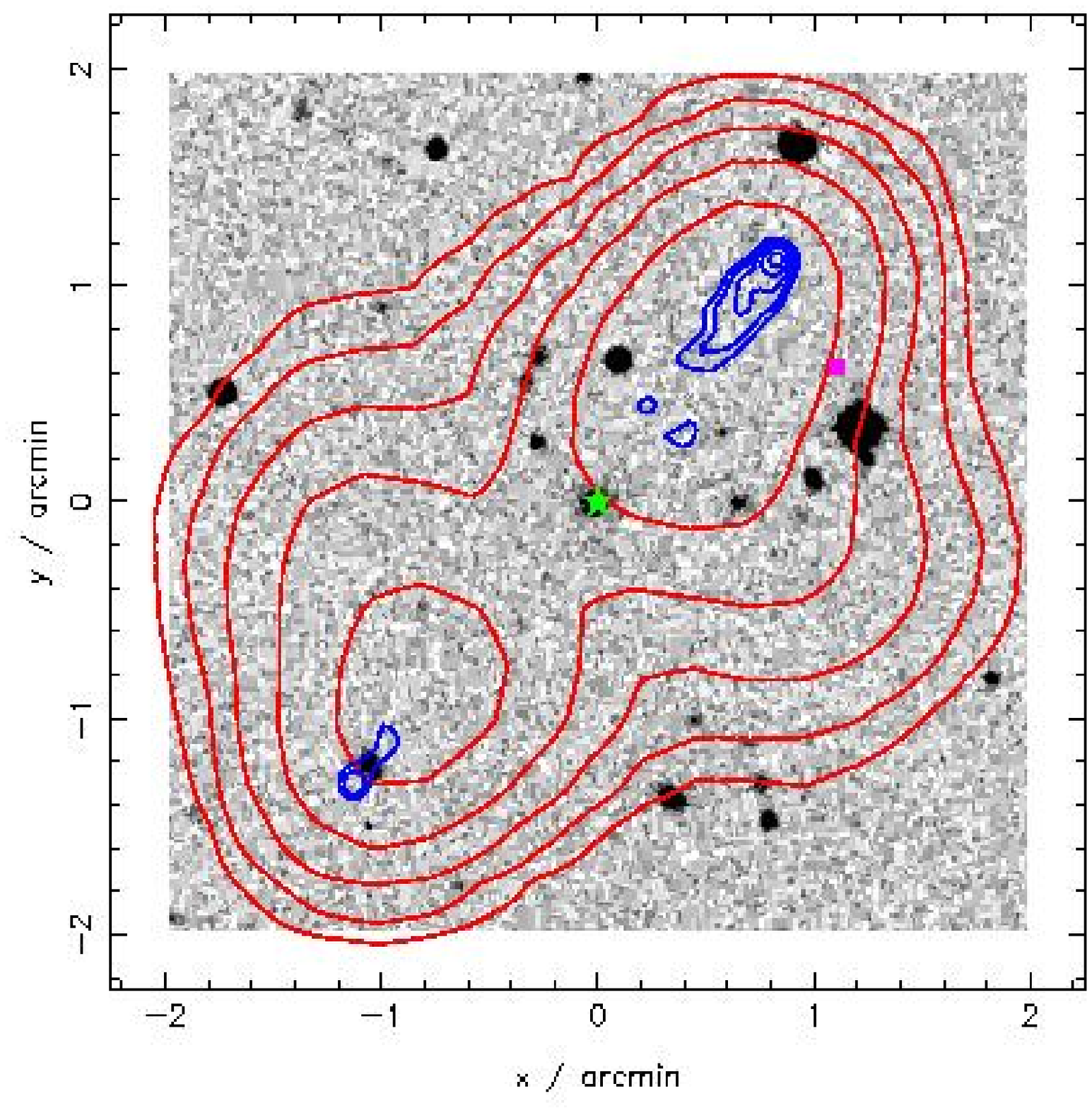}}
      \centerline{C1-101: 4C 61.23}
    \end{minipage}
    \vfill
    \begin{minipage}{3cm}      
      \mbox{}
      \centerline{\includegraphics[scale=0.26,angle=270]{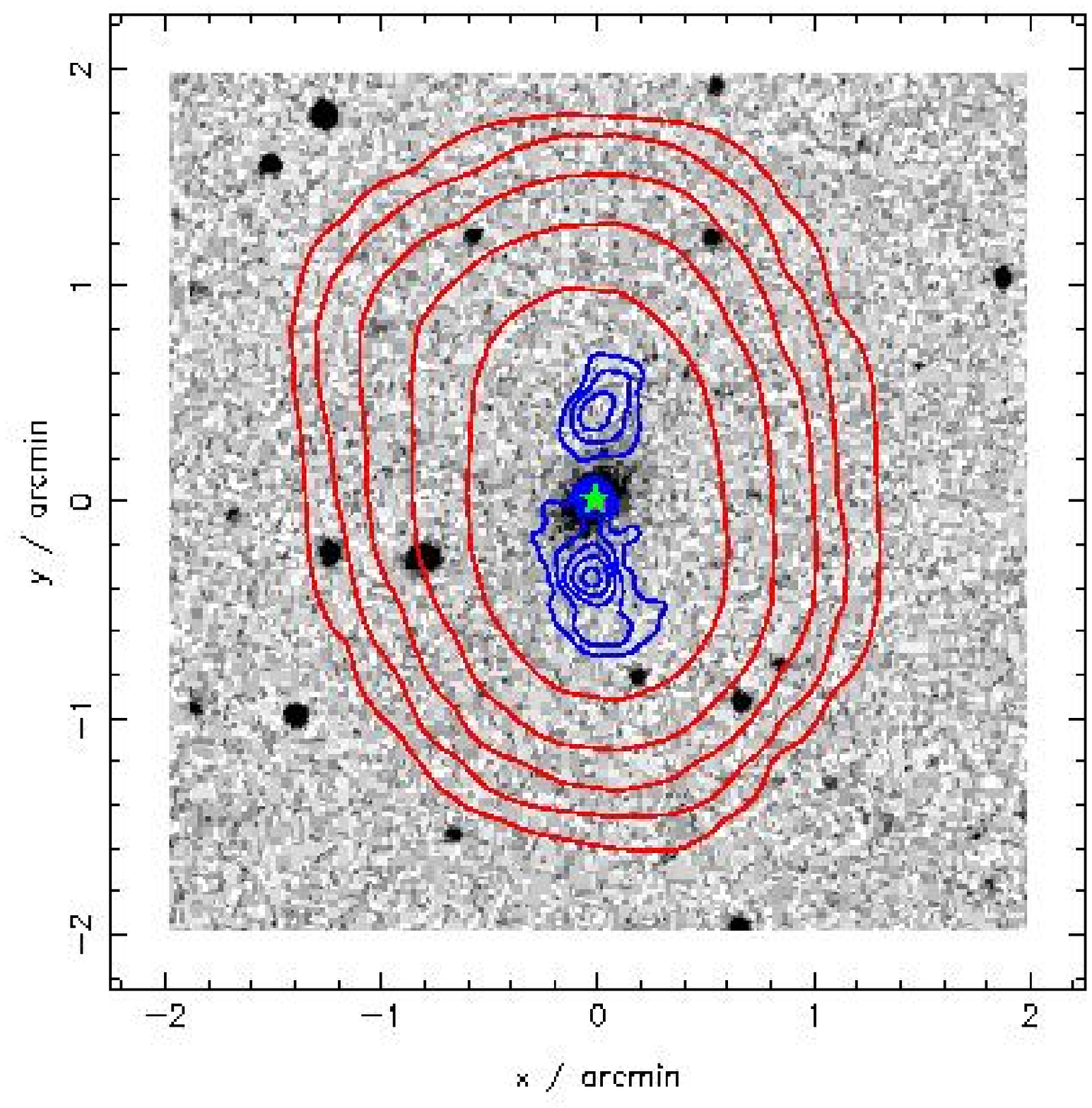}}
      \centerline{C1-102: 4C 12.42}
    \end{minipage}
    \hspace{3cm}
    \begin{minipage}{3cm}
      \mbox{}
      \centerline{\includegraphics[scale=0.26,angle=270]{Contours/C1/104.ps}}
      \centerline{C1-104: 4C 01.32}
    \end{minipage}
    \hspace{3cm}
    \begin{minipage}{3cm}
      \mbox{}
      \centerline{\includegraphics[scale=0.26,angle=270]{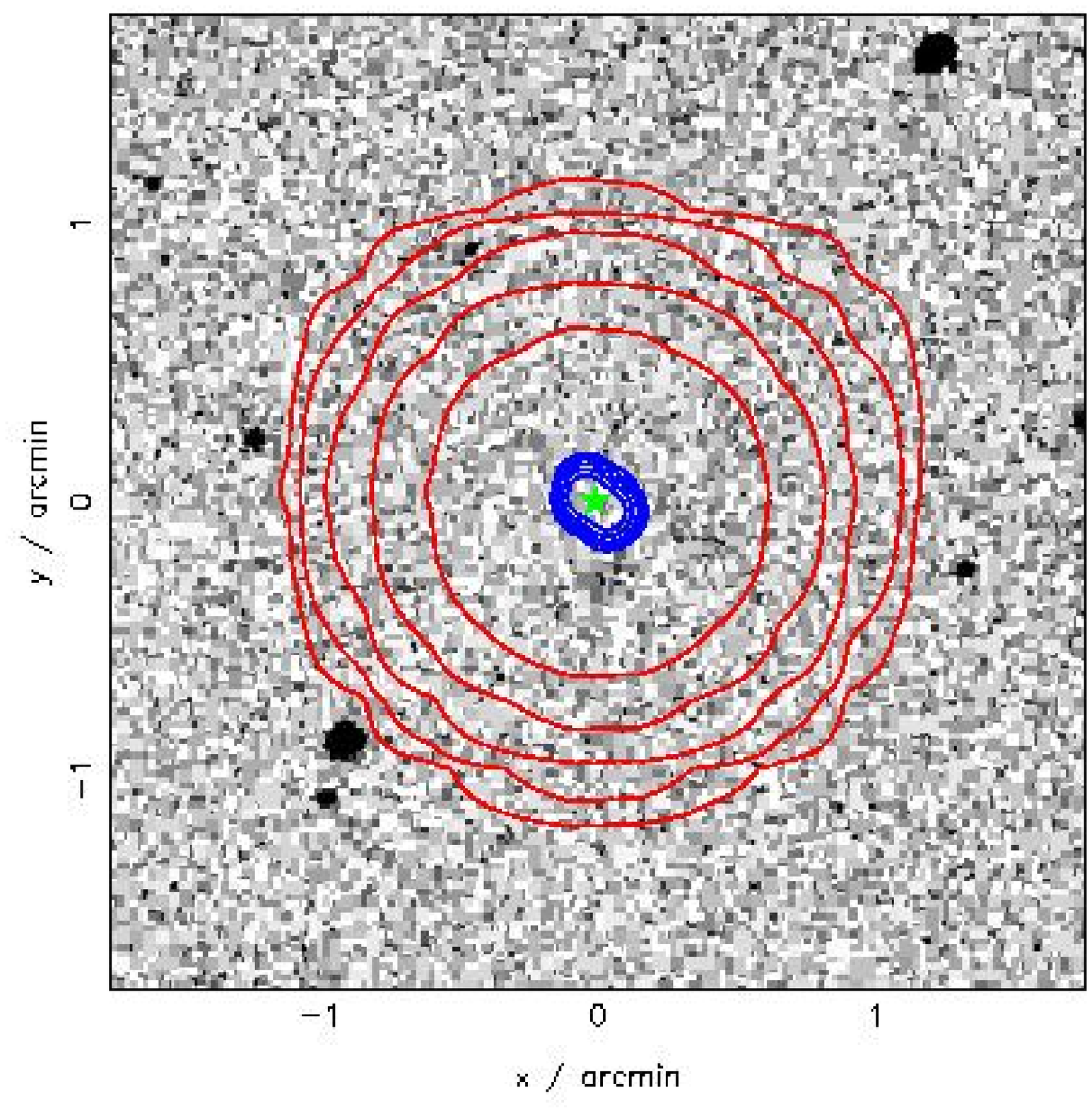}}
      \centerline{C1-105: 3C 263.1}
    \end{minipage}
    \vfill
    \begin{minipage}{3cm}     
      \mbox{}
      \centerline{\includegraphics[scale=0.26,angle=270]{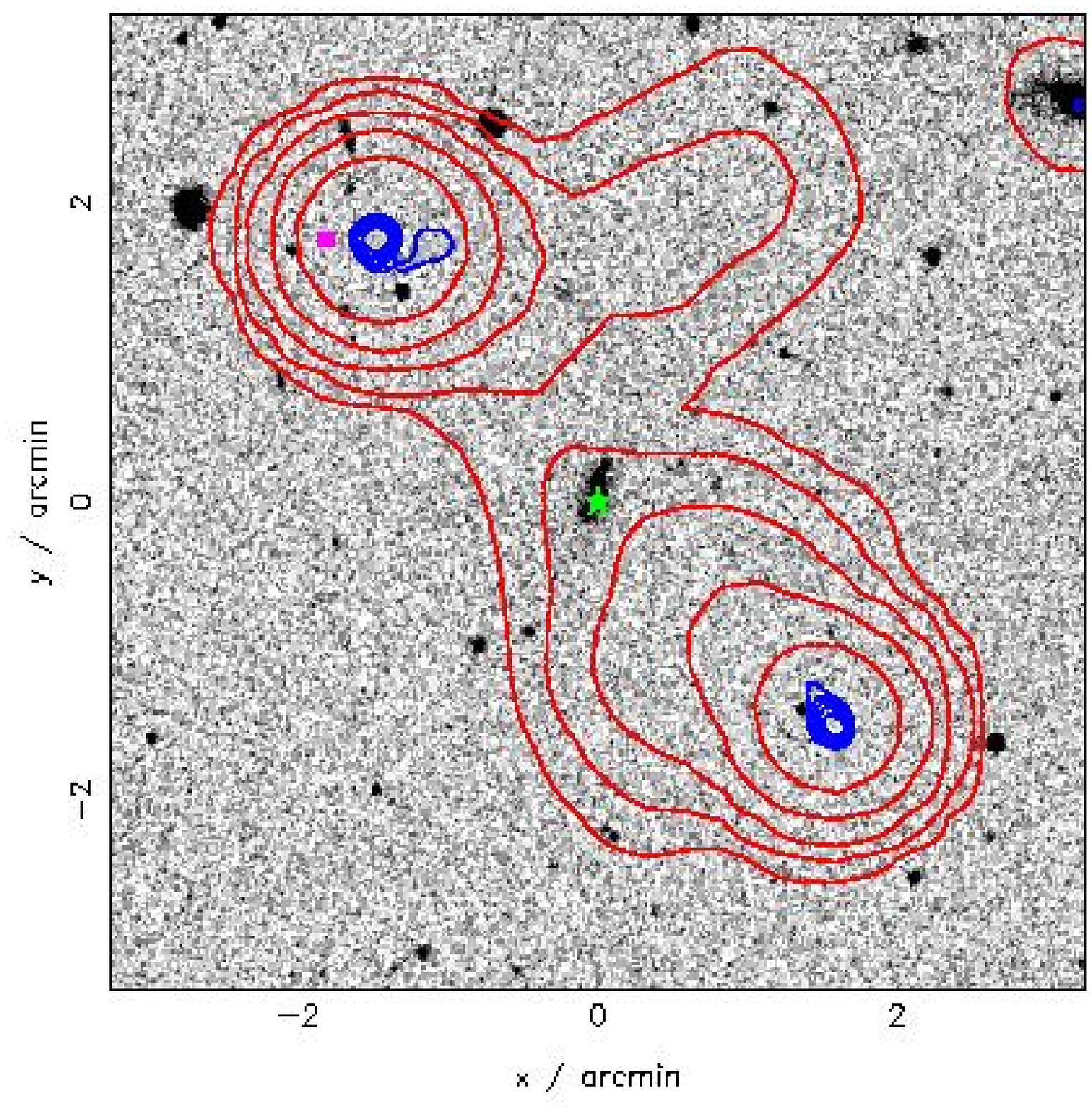}}
      \centerline{C1-106: 4C 37.32}
    \end{minipage}
    \hspace{3cm}
    \begin{minipage}{3cm}
      \mbox{}
      \centerline{\includegraphics[scale=0.26,angle=270]{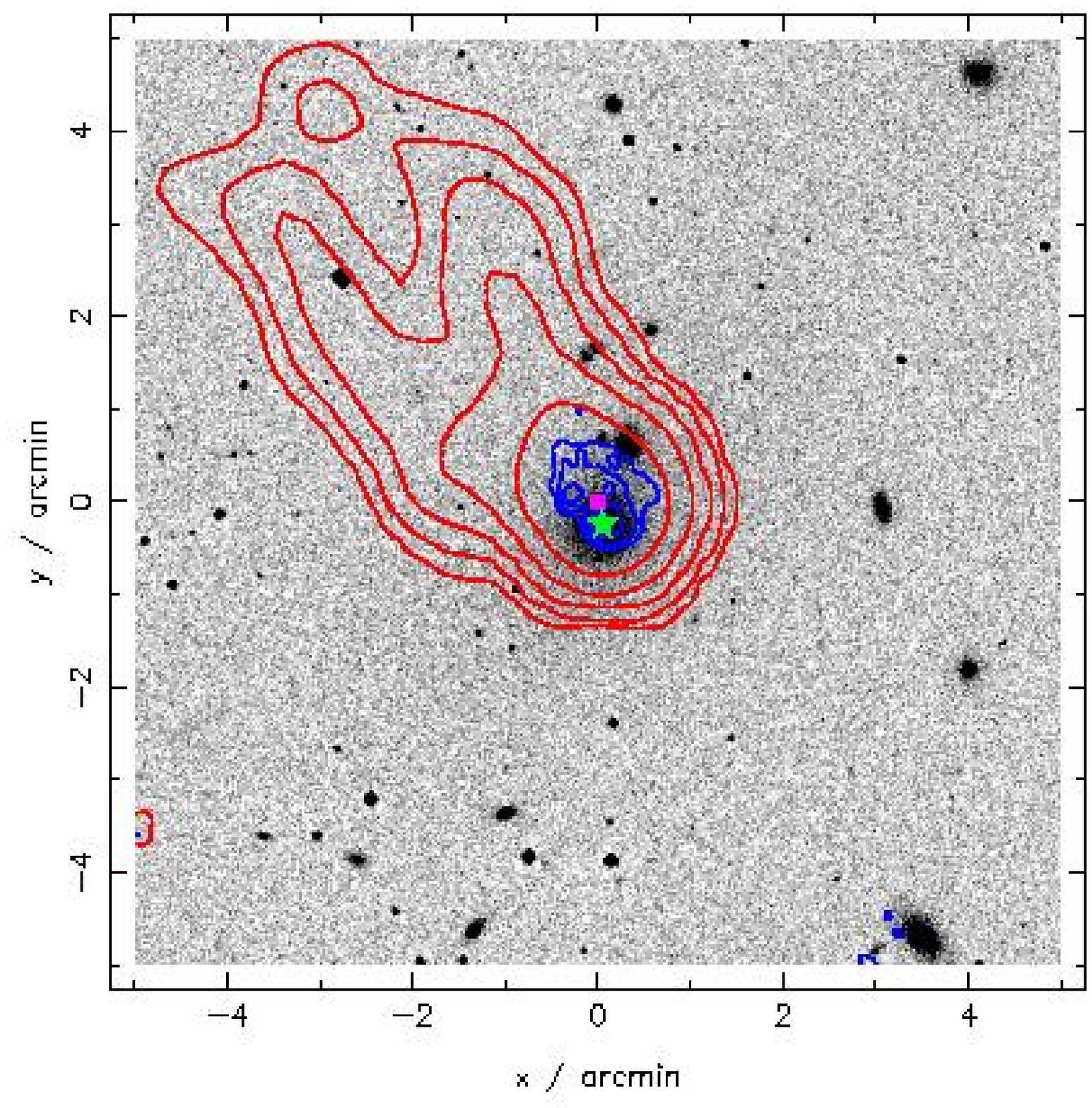}}
      \centerline{C1-107: 3C 264}
    \end{minipage}
    \hspace{3cm}
    \begin{minipage}{3cm}
      \mbox{}
      \centerline{\includegraphics[scale=0.26,angle=270]{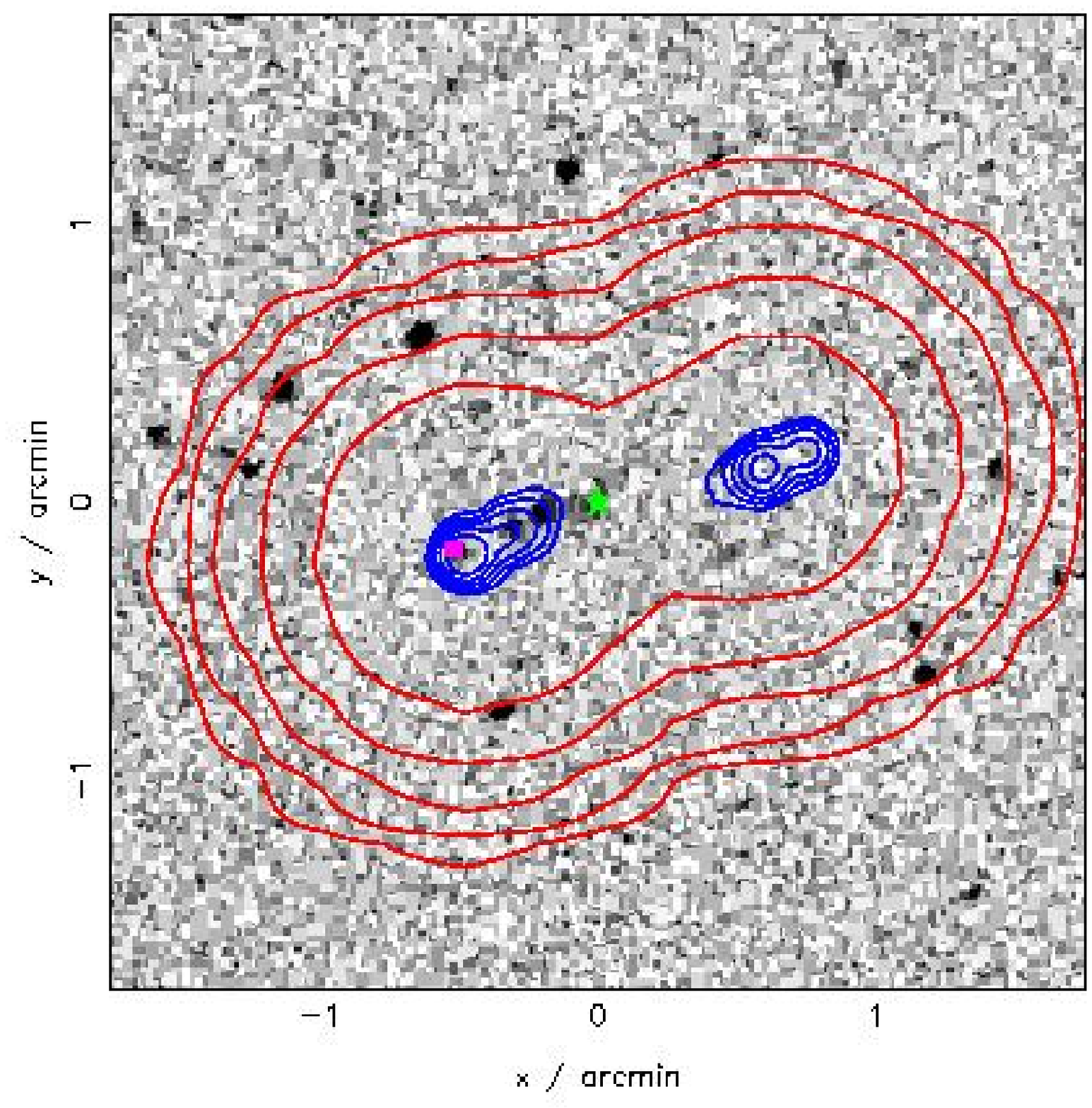}}
      \centerline{C1-108: 3C 265}
    \end{minipage}
    \vfill
    \begin{minipage}{3cm}     
      \mbox{}
      \centerline{\includegraphics[scale=0.26,angle=270]{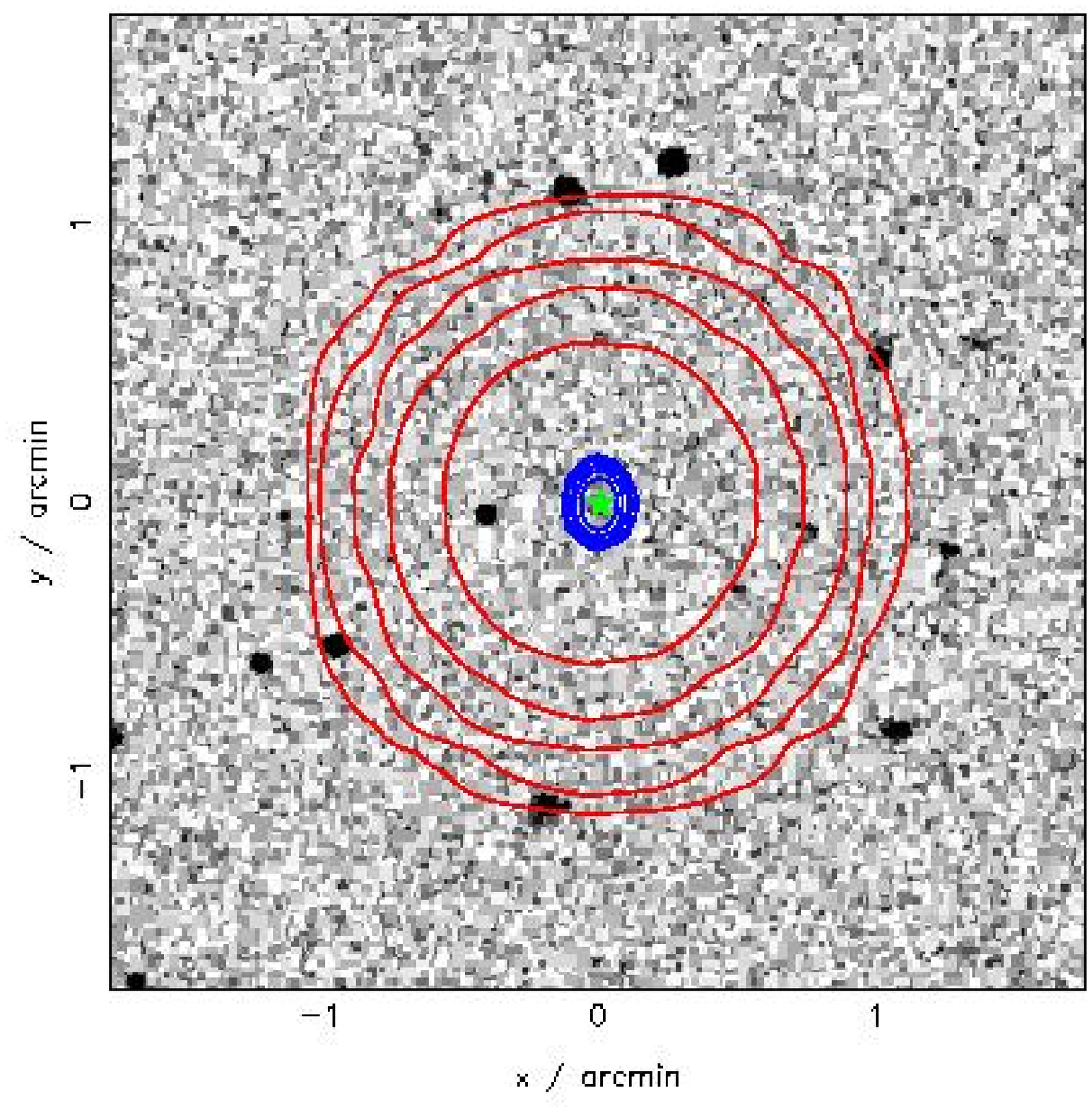}}
      \centerline{C1-109: 3C 266}
    \end{minipage}
    \hspace{3cm}
    \begin{minipage}{3cm}
      \mbox{}
      \centerline{\includegraphics[scale=0.26,angle=270]{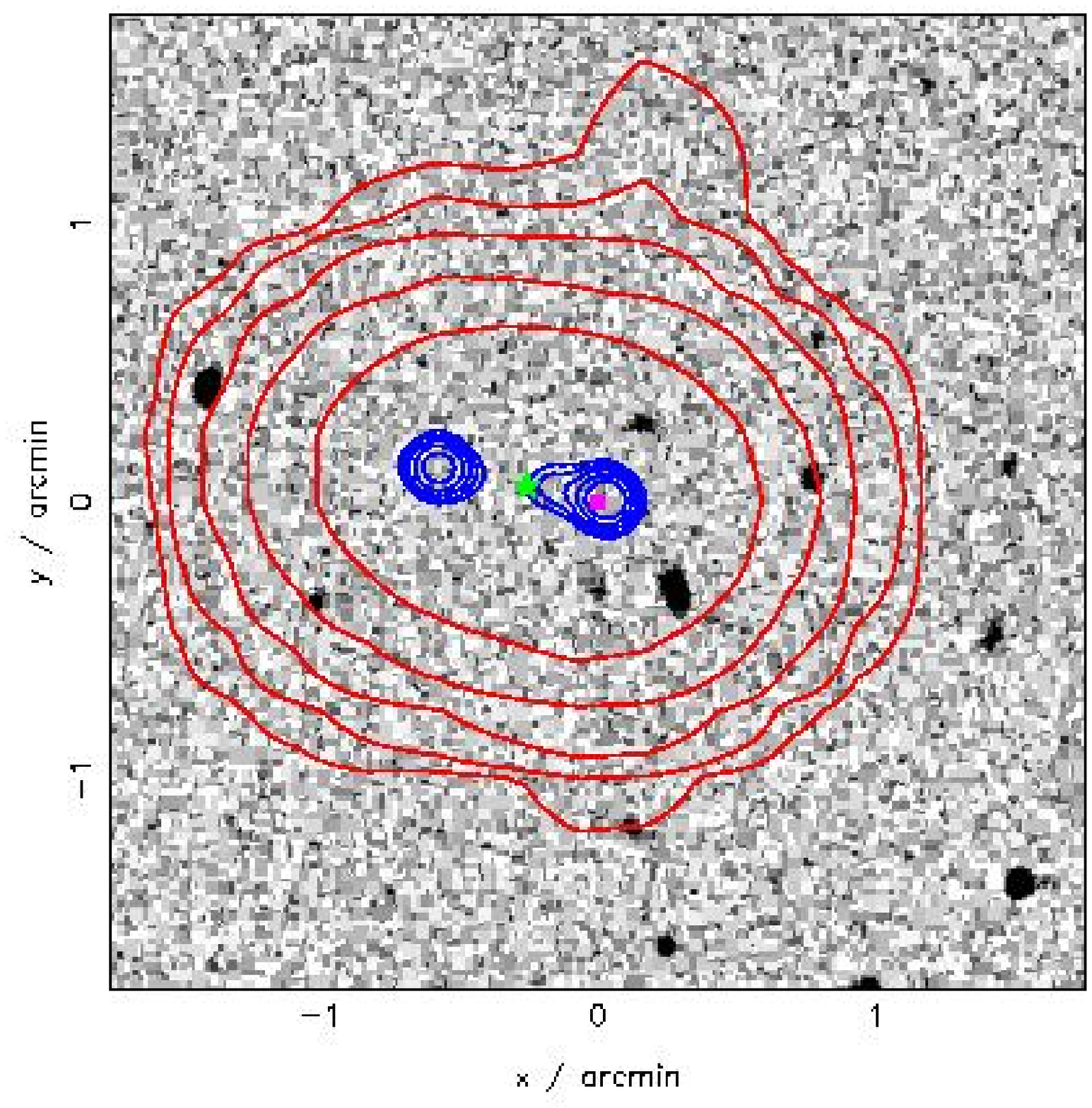}}
      \centerline{C1-110: 3C 267}
    \end{minipage}
    \hspace{3cm}
    \begin{minipage}{3cm}
      \mbox{}
      \centerline{\includegraphics[scale=0.26,angle=270]{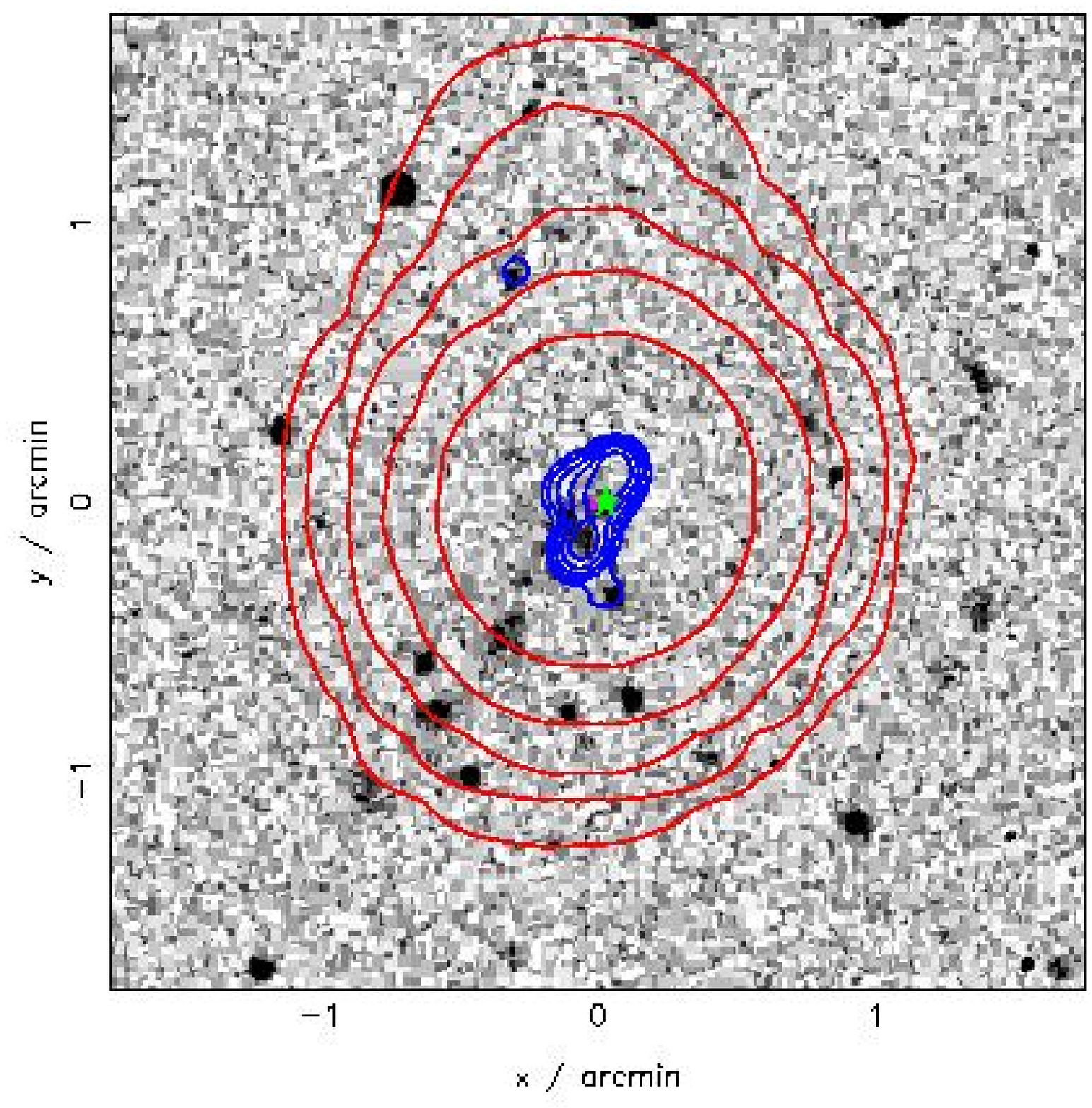}}
      \centerline{C1-113: 4C 29.44}
    \end{minipage}
  \end{center}
\end{figure}

\begin{figure}
  \begin{center}
    {\bf CoNFIG-1}\\  
  \begin{minipage}{3cm}      
      \mbox{}
      \centerline{\includegraphics[scale=0.26,angle=270]{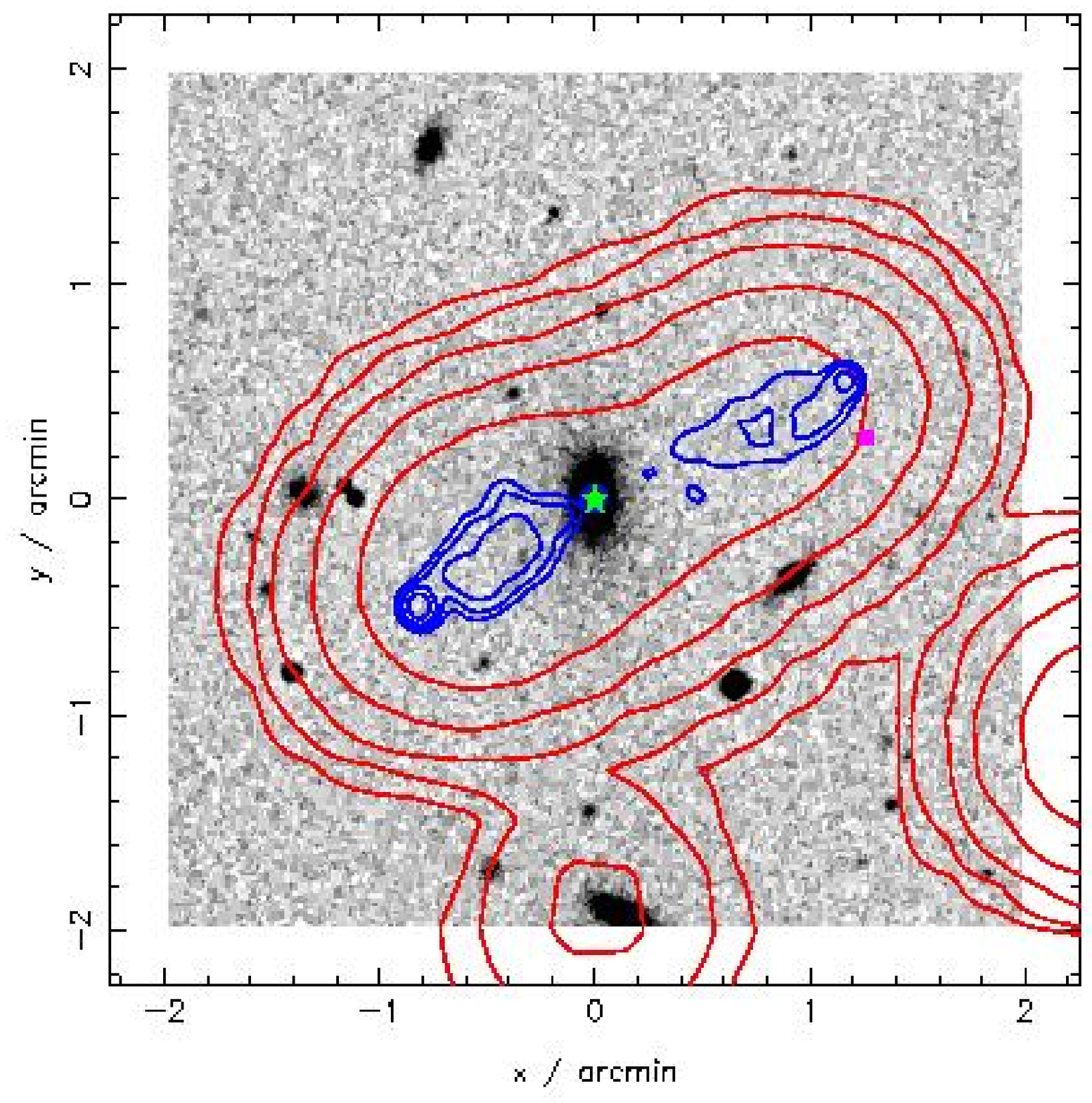}}
      \centerline{C1-114: 4C 55.22}
    \end{minipage}
    \hspace{3cm}
    \begin{minipage}{3cm}
      \mbox{}
      \centerline{\includegraphics[scale=0.26,angle=270]{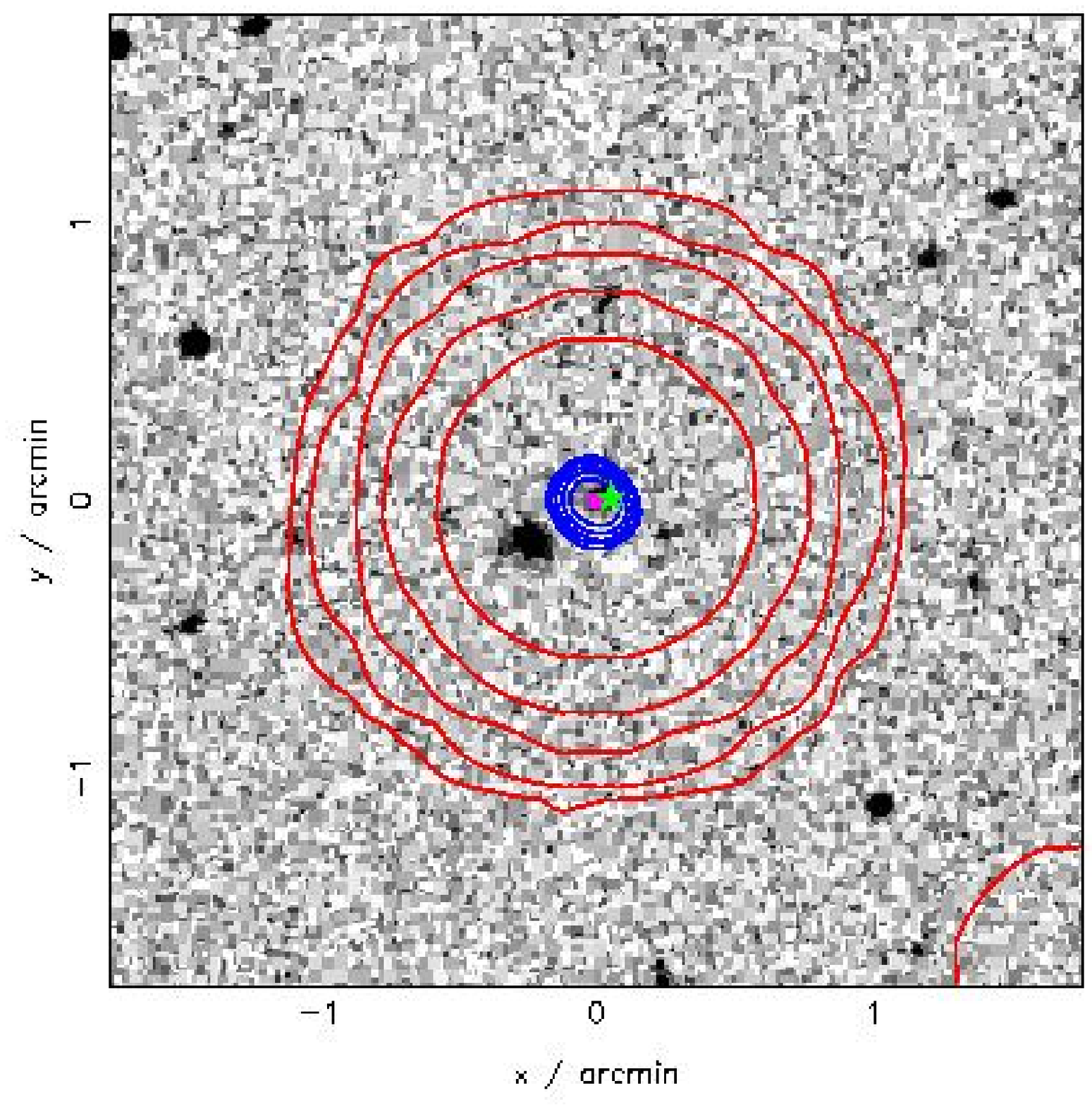}}
      \centerline{C1-115: 4C 59.17}
    \end{minipage}
    \hspace{3cm}
    \begin{minipage}{3cm}
      \mbox{}
      \centerline{\includegraphics[scale=0.26,angle=270]{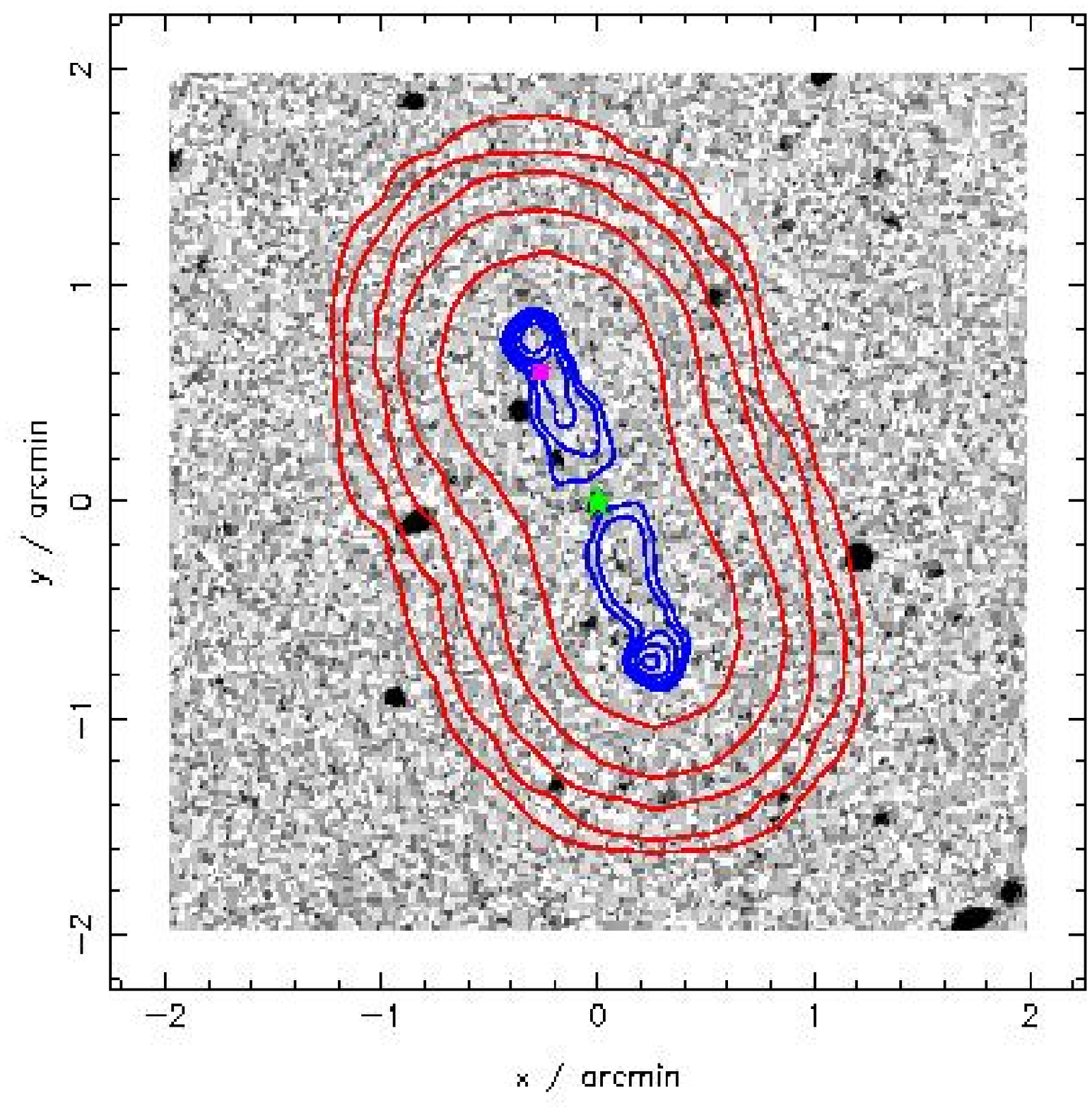}}
      \centerline{C1-119: 3C 268.2}
    \end{minipage}
    \vfill
    \begin{minipage}{3cm}      
      \mbox{}
      \centerline{\includegraphics[scale=0.26,angle=270]{Contours/C1/120.ps}}
      \centerline{C1-120: 4C -04.40}
    \end{minipage}
    \hspace{3cm}
    \begin{minipage}{3cm}
      \mbox{}
      \centerline{\includegraphics[scale=0.26,angle=270]{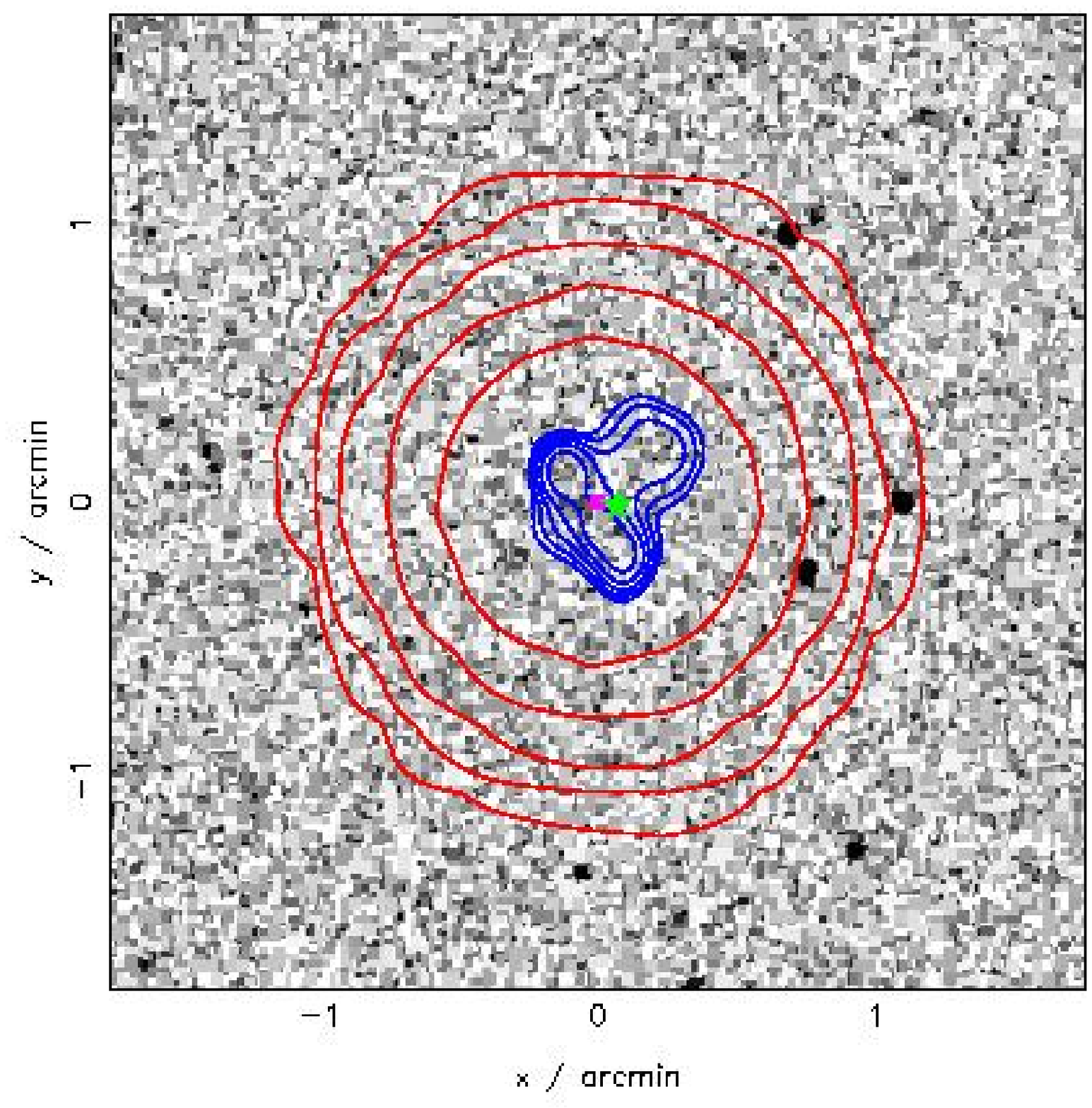}}
      \centerline{C1-121: 4C 04.40}
    \end{minipage}
    \hspace{3cm}
    \begin{minipage}{3cm}
      \mbox{}
      \centerline{\includegraphics[scale=0.26,angle=270]{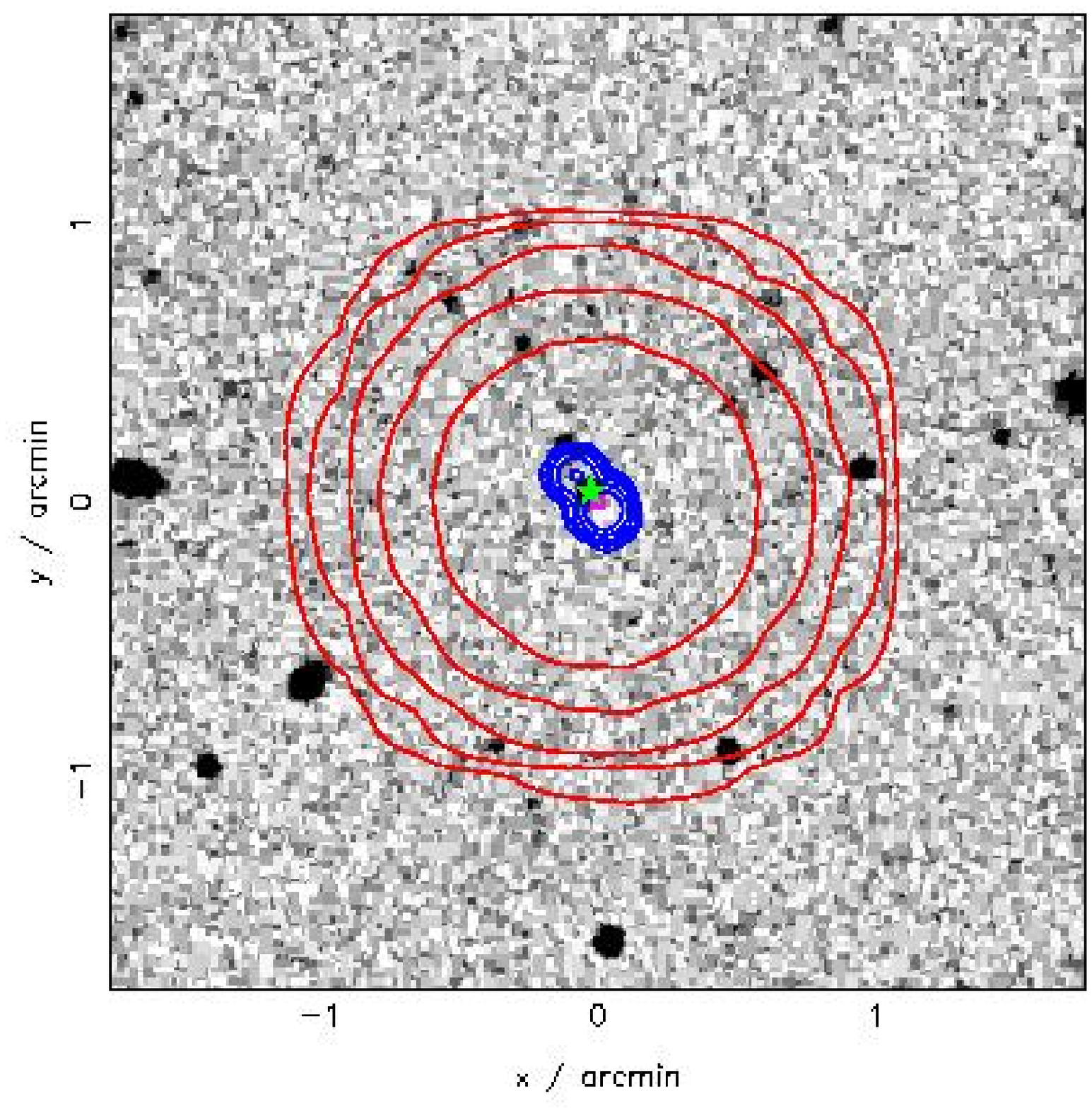}}
      \centerline{C1-122: 3C 268.4}
    \end{minipage}
    \vfill
    \begin{minipage}{3cm}     
      \mbox{}
      \centerline{\includegraphics[scale=0.26,angle=270]{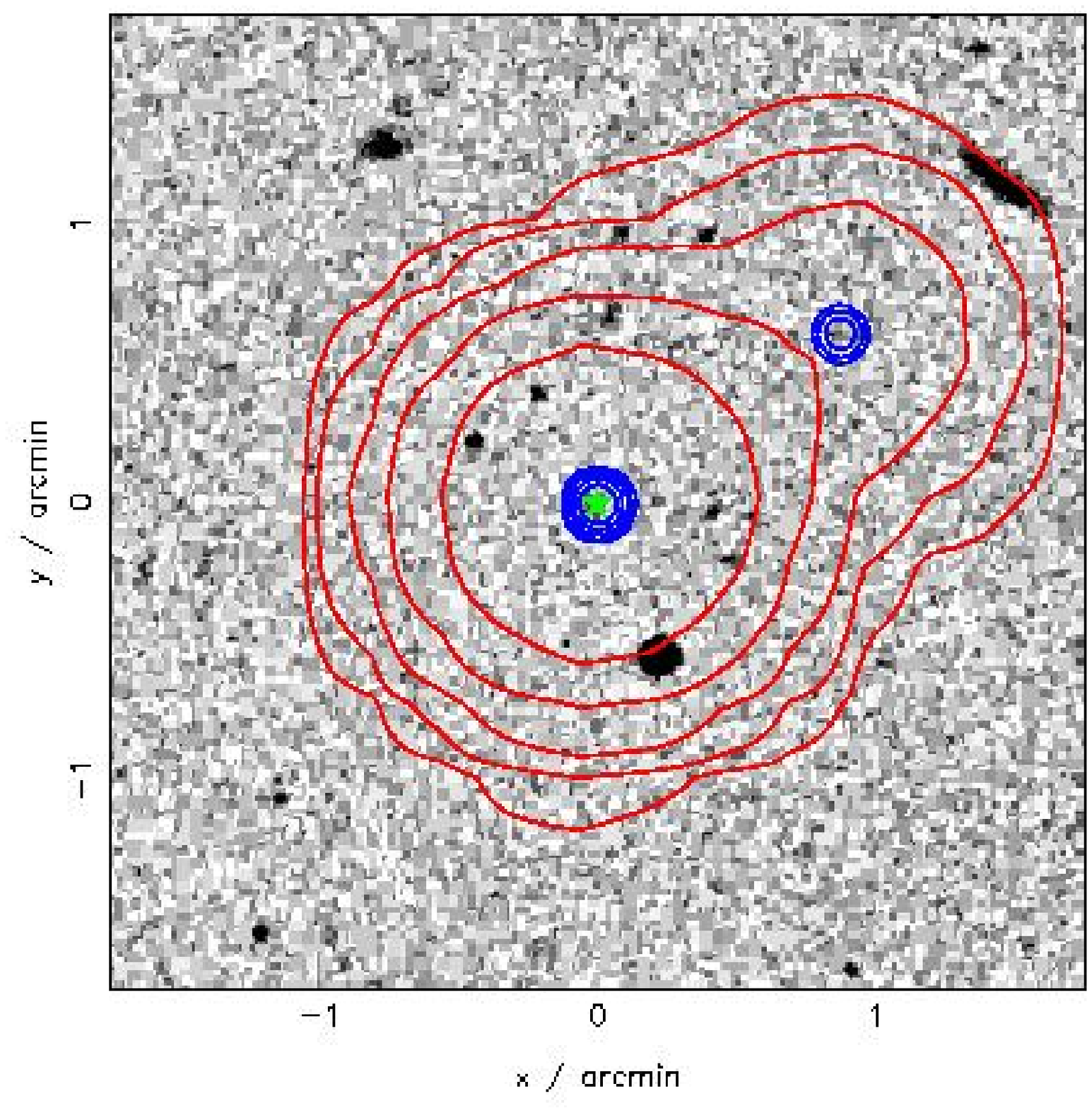}}
      \centerline{C1-123: 4C 20.27}
    \end{minipage}
    \hspace{3cm}
    \begin{minipage}{3cm}
      \mbox{}
      \centerline{\includegraphics[scale=0.26,angle=270]{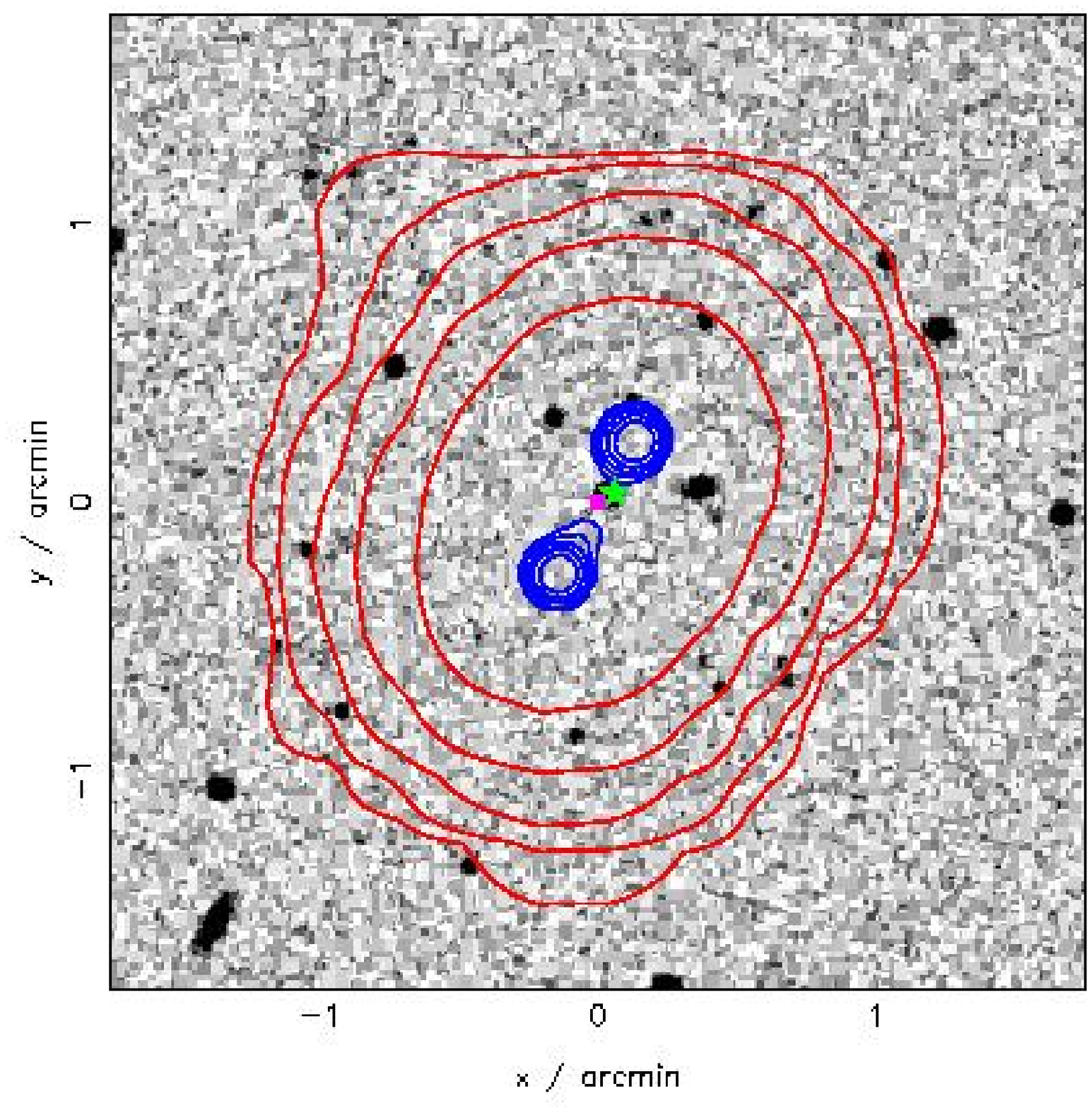}}
      \centerline{C1-126: 4C 53.24}
    \end{minipage}
    \hspace{3cm}
    \begin{minipage}{3cm}
      \mbox{}
      \centerline{\includegraphics[scale=0.26,angle=270]{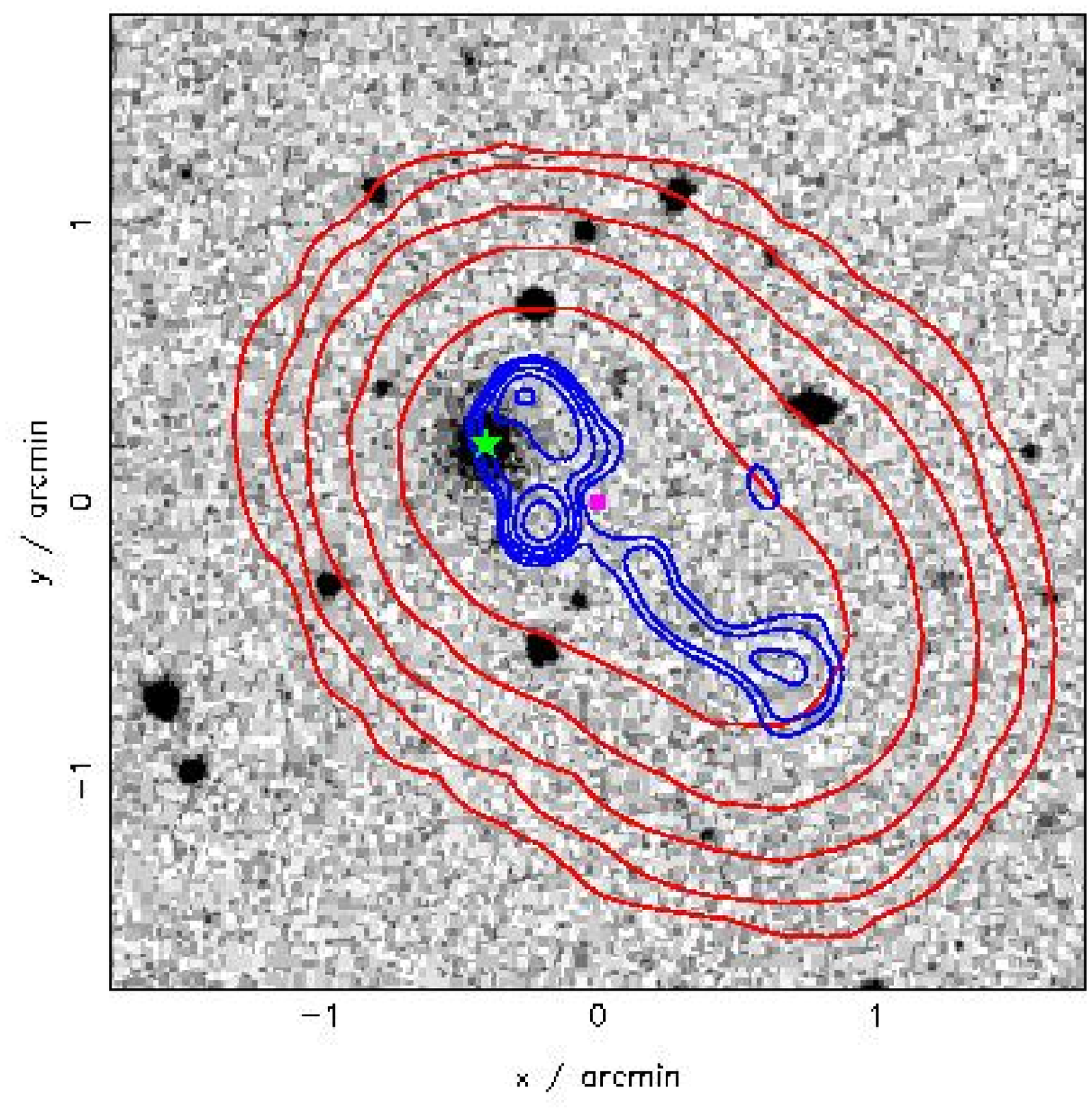}}
      \centerline{C1-128: 4C 04.41}
    \end{minipage}
    \vfill
    \begin{minipage}{3cm}     
      \mbox{}
      \centerline{\includegraphics[scale=0.26,angle=270]{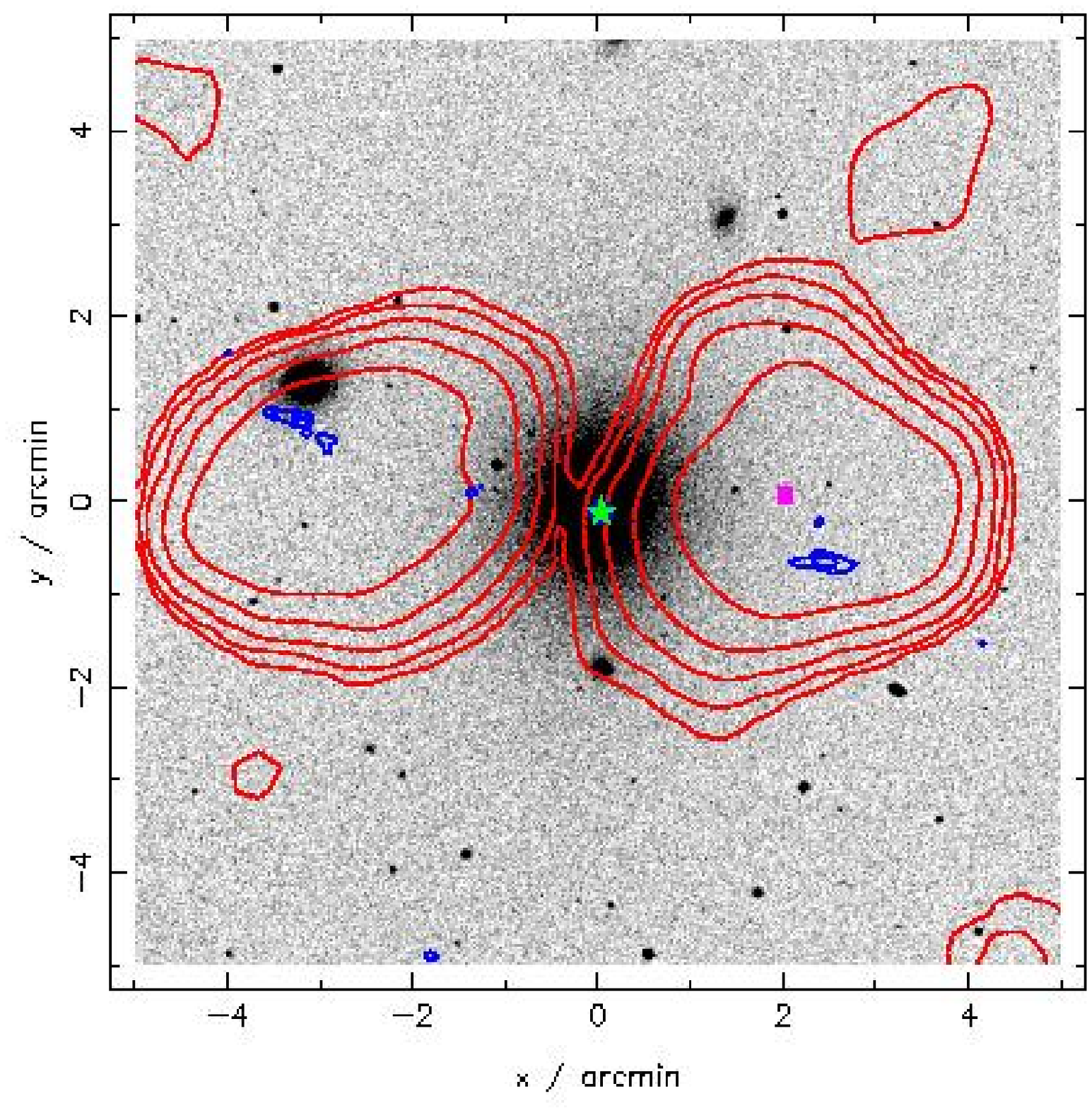}}
      \centerline{C1-129: 3C 270}
    \end{minipage}
    \hspace{3cm}
    \begin{minipage}{3cm}
      \mbox{}
      \centerline{\includegraphics[scale=0.26,angle=270]{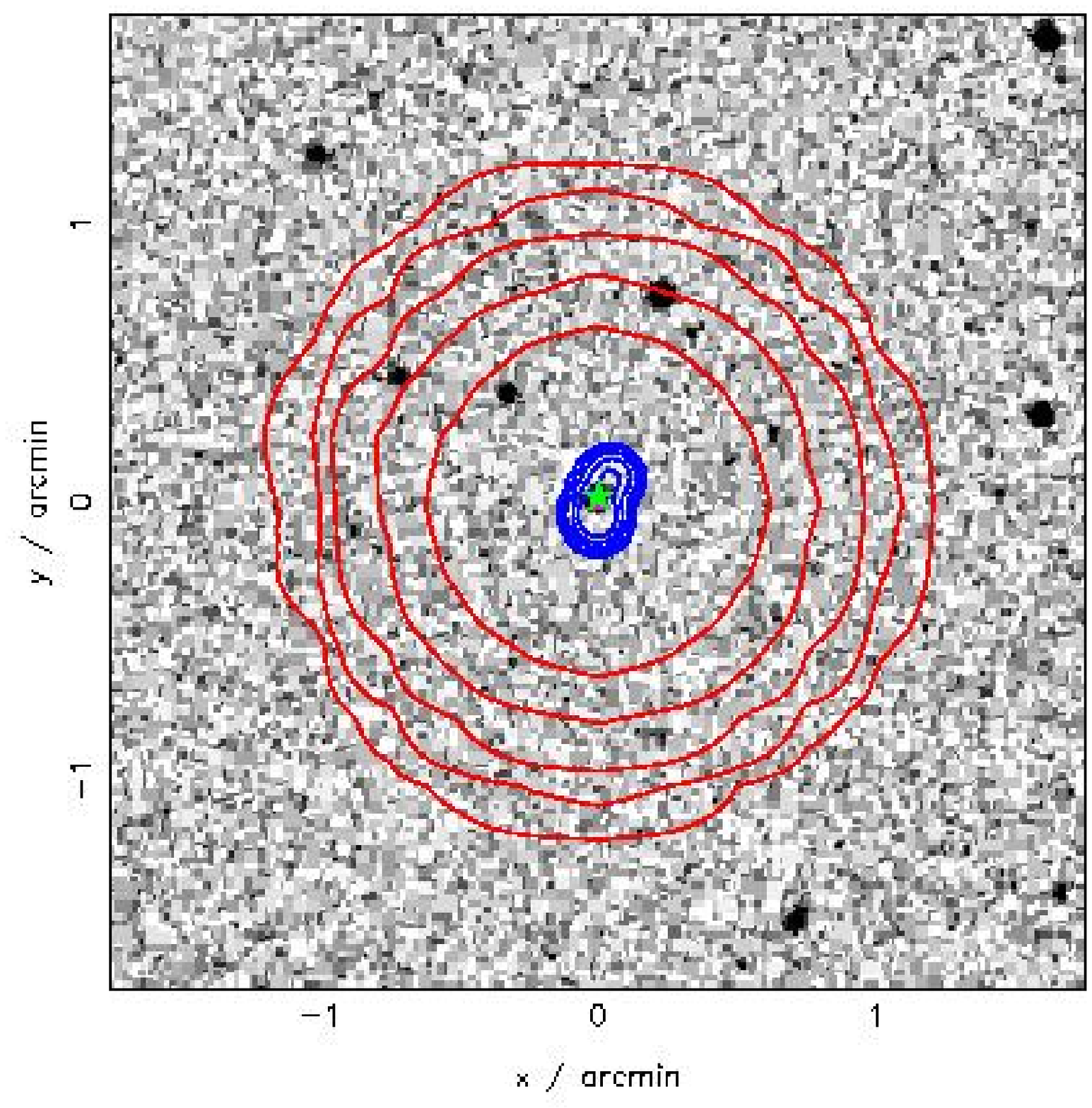}}
      \centerline{C1-130: 3C 270.1}
    \end{minipage}
    \hspace{3cm}
    \begin{minipage}{3cm}
      \mbox{}
      \centerline{\includegraphics[scale=0.26,angle=270]{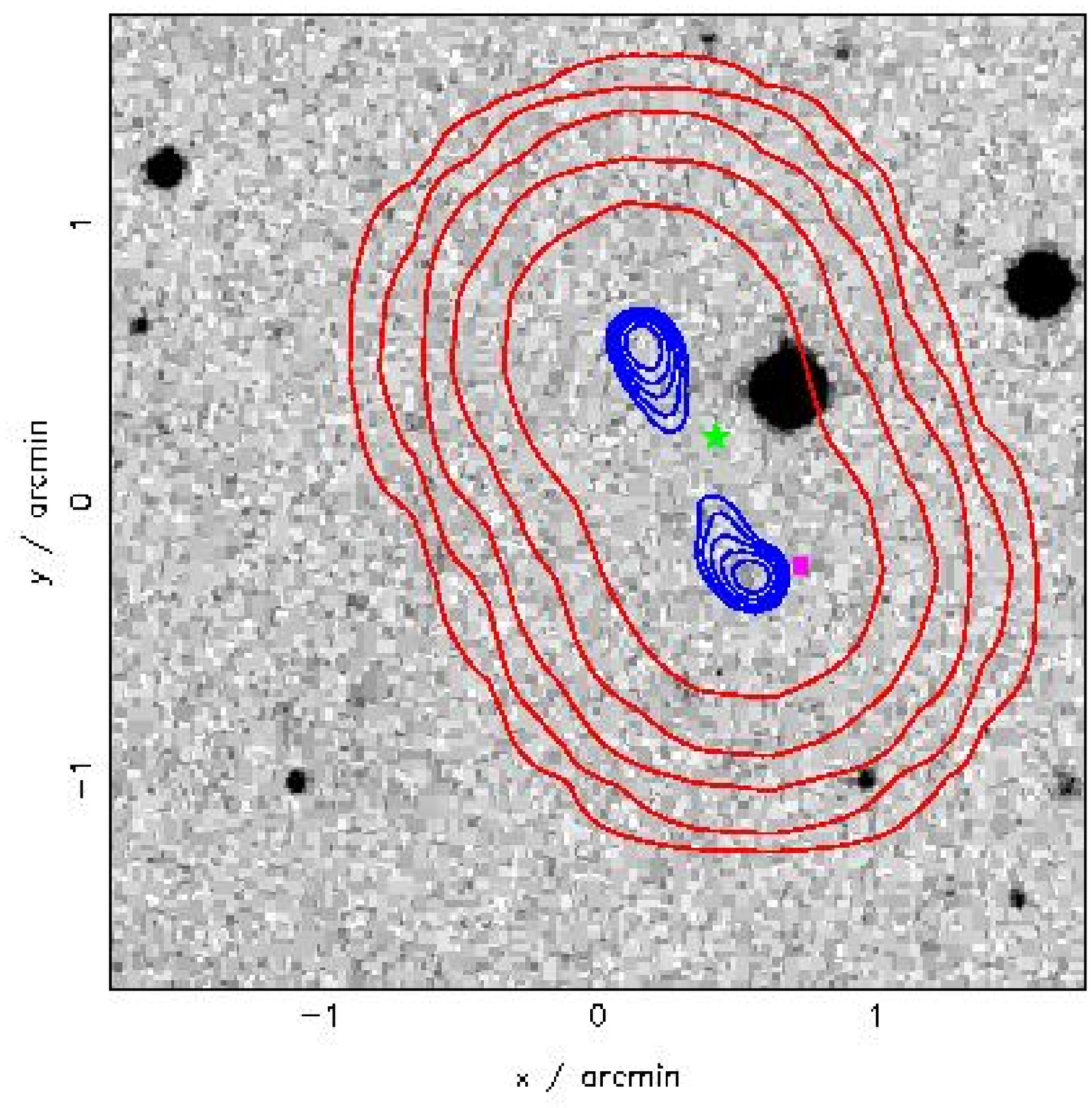}}
      \centerline{C1-131: 3C 272}
    \end{minipage}
  \end{center}
\end{figure}

\begin{figure}
  \begin{center}
    {\bf CoNFIG-1}\\  
  \begin{minipage}{3cm}      
      \mbox{}
      \centerline{\includegraphics[scale=0.26,angle=270]{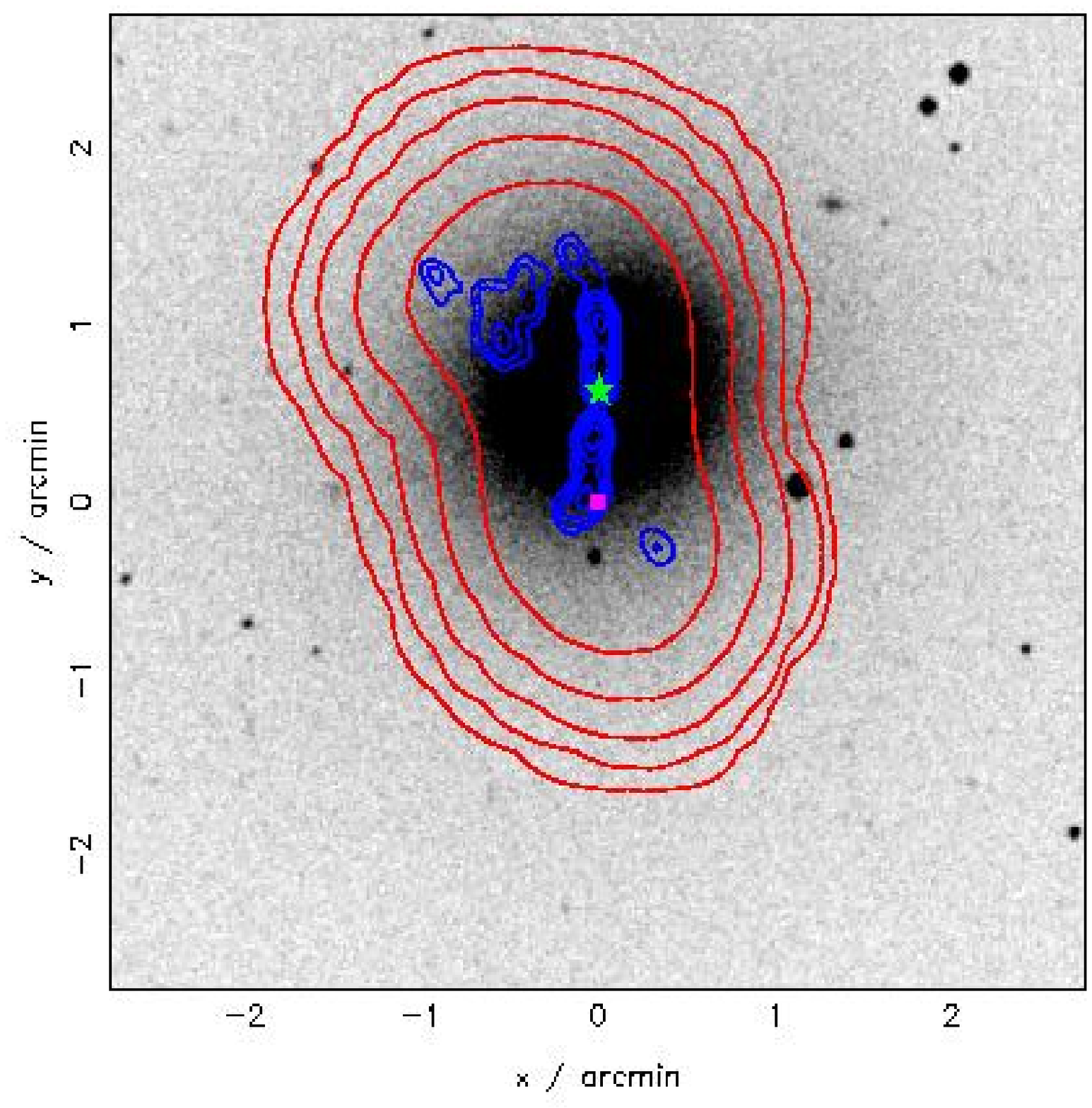}}
      \centerline{C1-133: M84}
    \end{minipage}
    \hspace{3cm}
    \begin{minipage}{3cm}
      \mbox{}
      \centerline{\includegraphics[scale=0.26,angle=270]{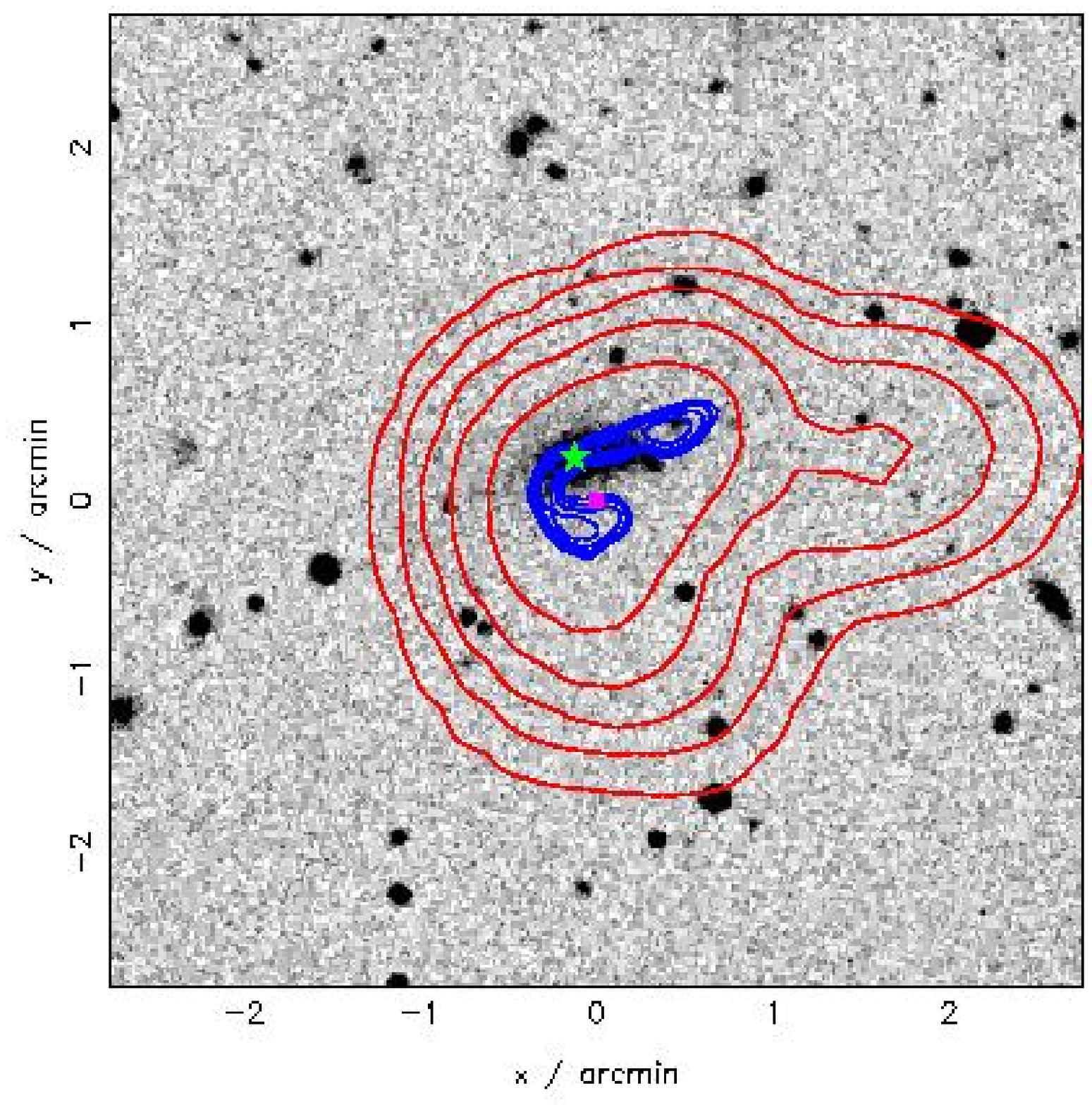}}
      \centerline{C1-136: PKS 1227+119}
    \end{minipage}
    \hspace{3cm}
    \begin{minipage}{3cm}
      \mbox{}
      \centerline{\includegraphics[scale=0.26,angle=270]{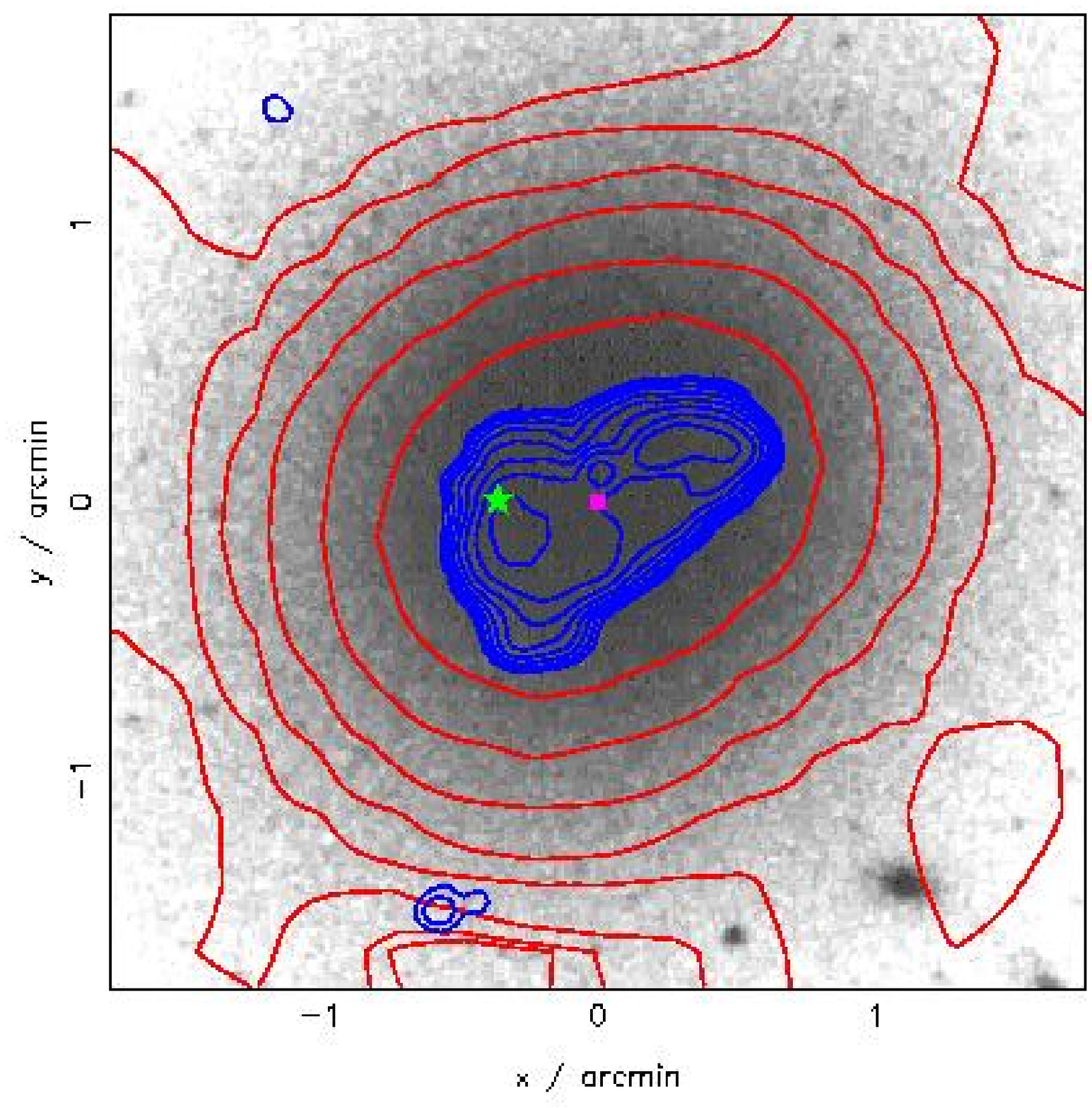}}
      \centerline{C1-137: M87}
    \end{minipage}
    \vfill
    \begin{minipage}{3cm}      
      \mbox{}
      \centerline{\includegraphics[scale=0.26,angle=270]{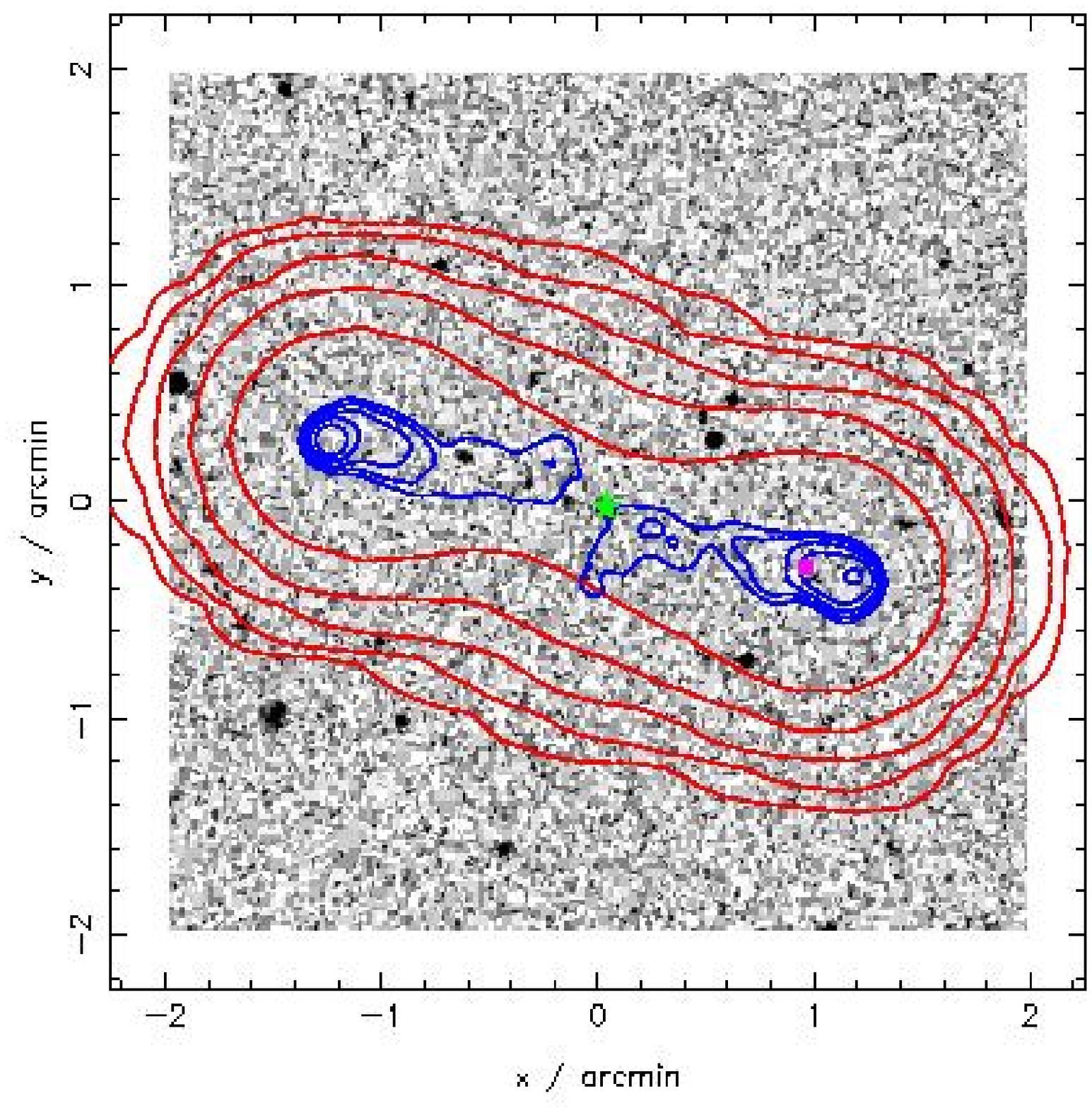}}
      \centerline{C1-139: 3C 274.1}
    \end{minipage}
    \hspace{3cm}
    \begin{minipage}{3cm}
      \mbox{}
      \centerline{\includegraphics[scale=0.26,angle=270]{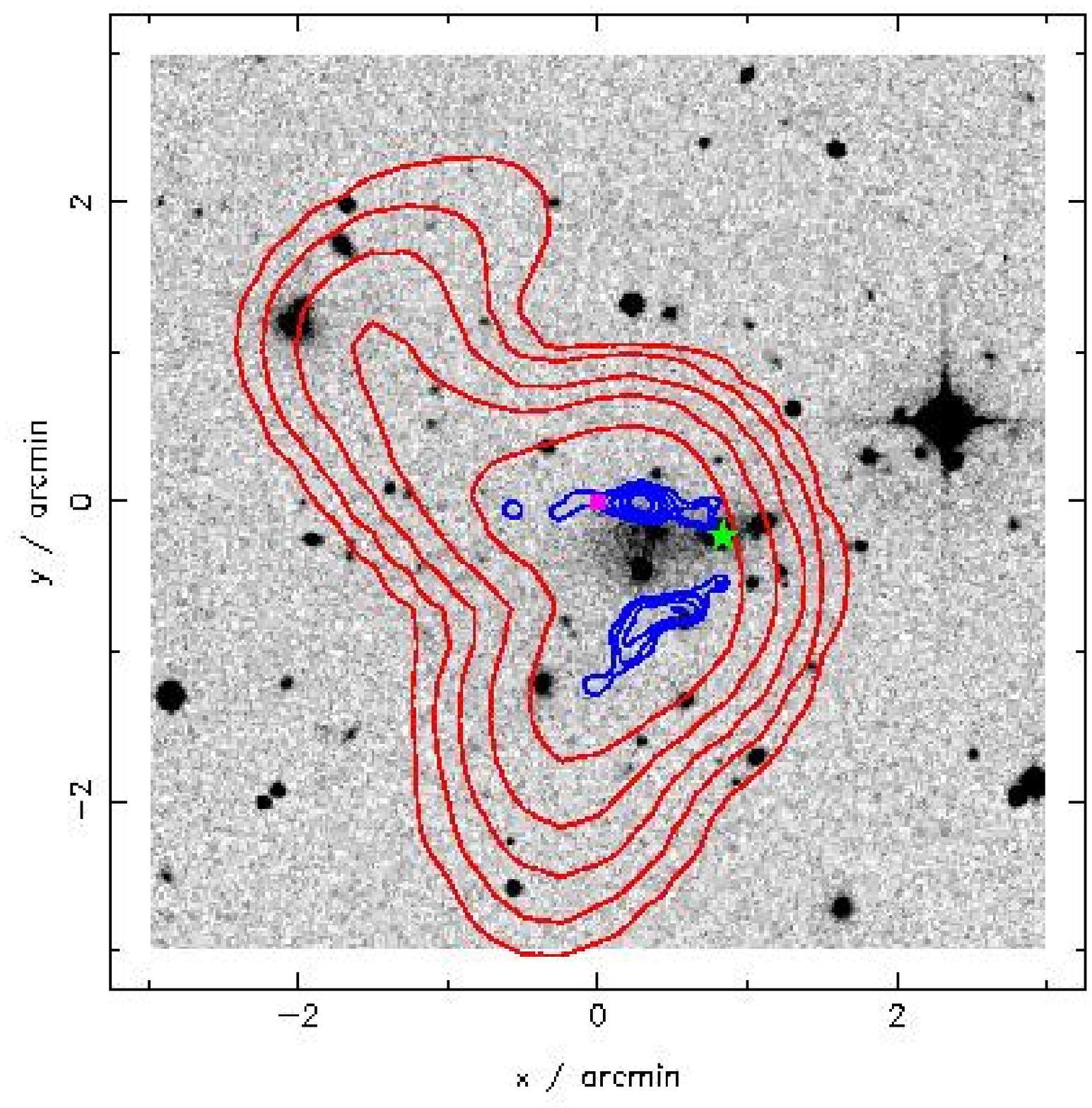}}
      \centerline{C1-140: 4C 16.33}
    \end{minipage}
    \hspace{3cm}
    \begin{minipage}{3cm}
      \mbox{}
      \centerline{\includegraphics[scale=0.26,angle=270]{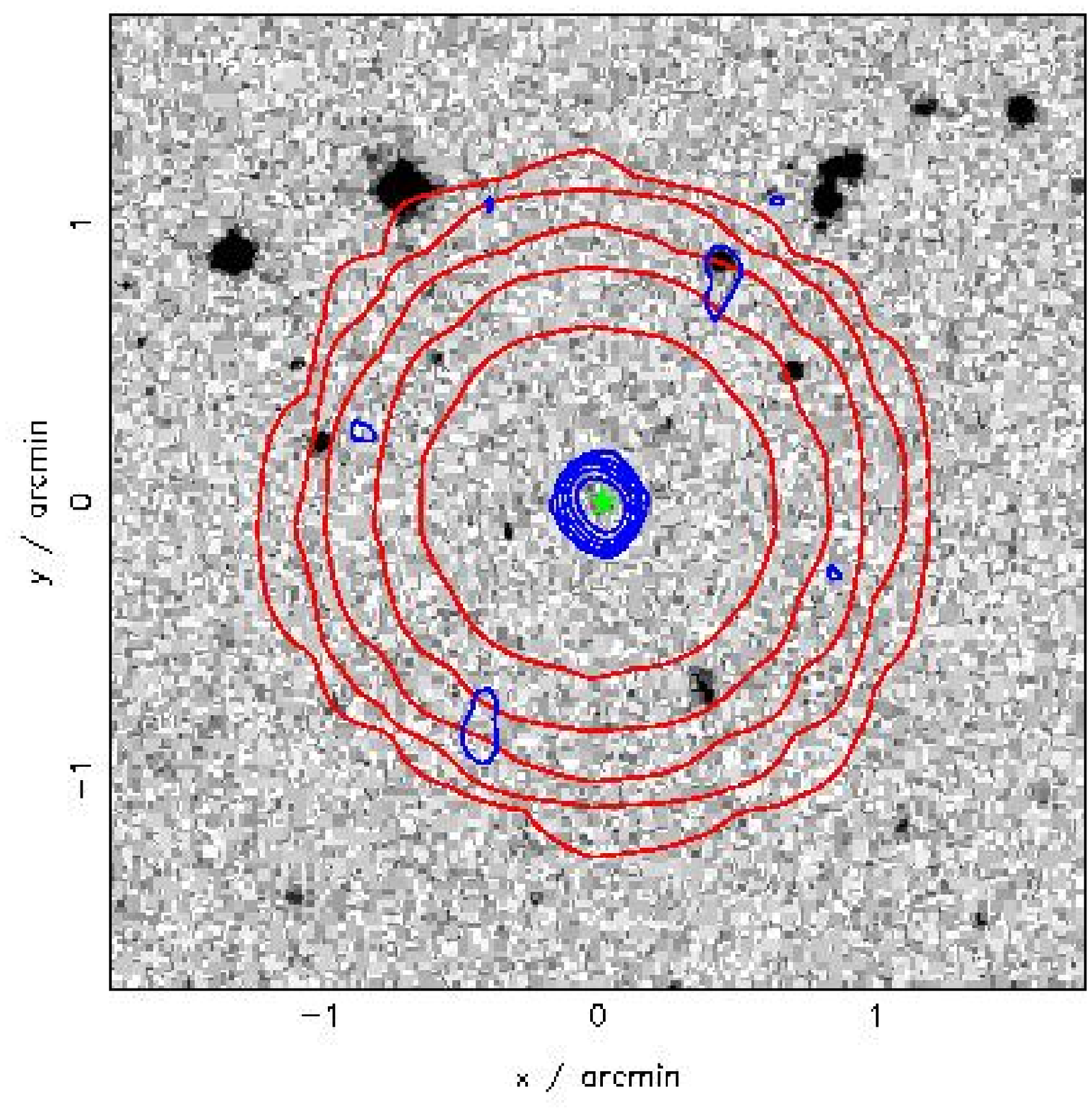}}
      \centerline{C1-141: 3C 275}
    \end{minipage}
    \vfill
    \begin{minipage}{3cm}     
      \mbox{}
      \centerline{\includegraphics[scale=0.26,angle=270]{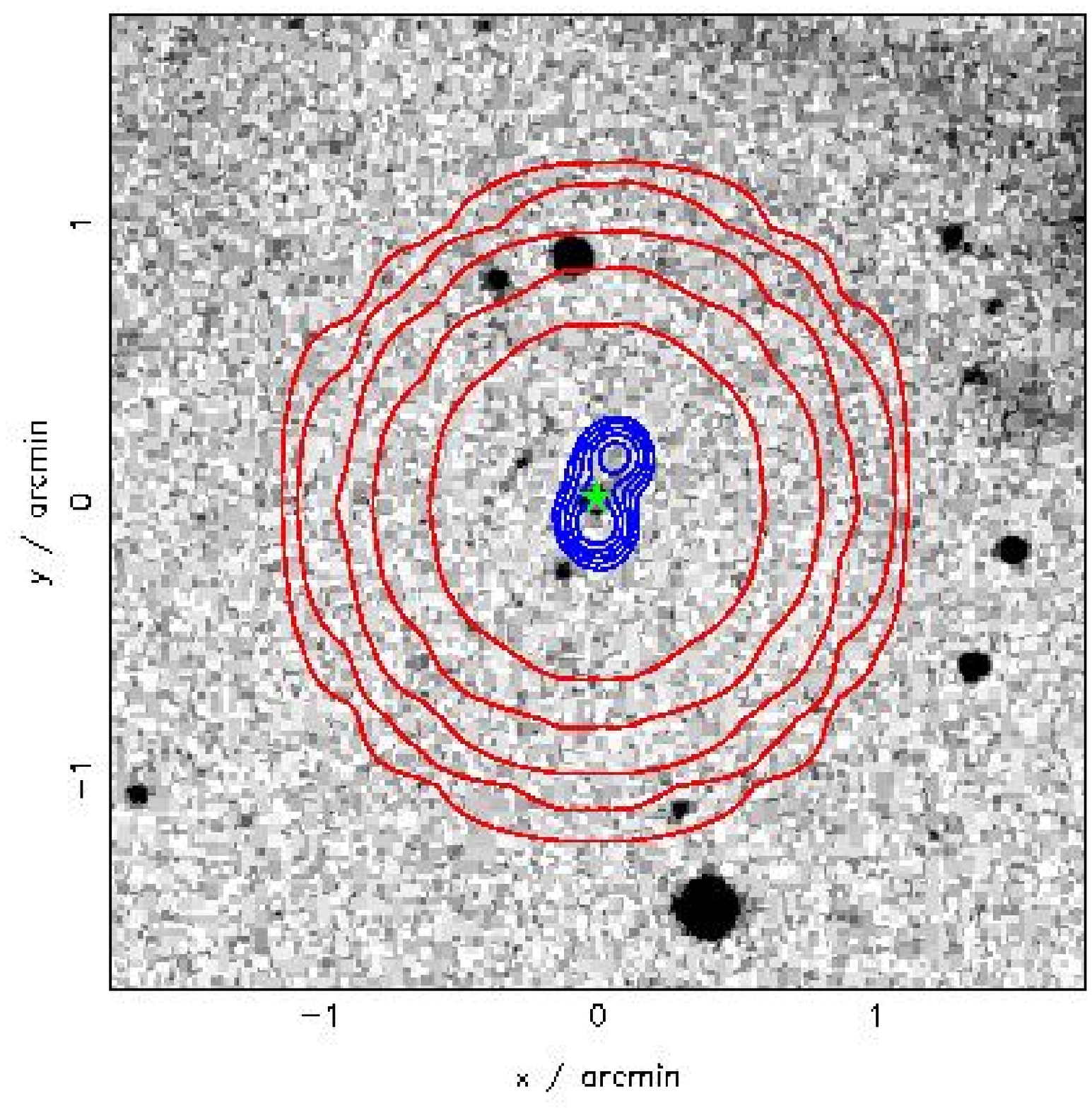}}
      \centerline{C1-142: 3C 275.1}
    \end{minipage}
    \hspace{3cm}
    \begin{minipage}{3cm}
      \mbox{}
      \centerline{\includegraphics[scale=0.26,angle=270]{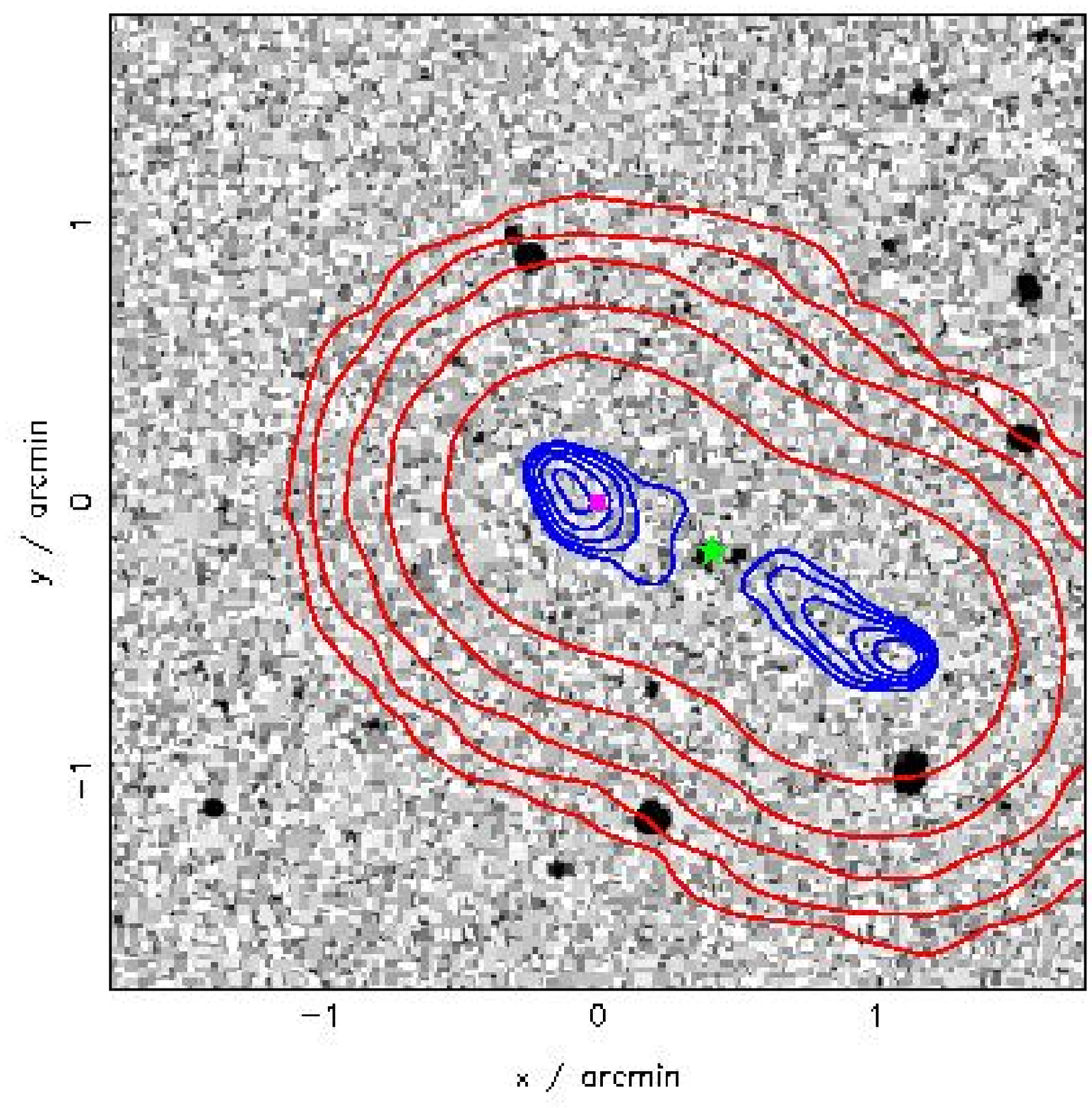}}
      \centerline{C1-144: 4C 09.44}
    \end{minipage}
    \hspace{3cm}
    \begin{minipage}{3cm}
      \mbox{}
      \centerline{\includegraphics[scale=0.26,angle=270]{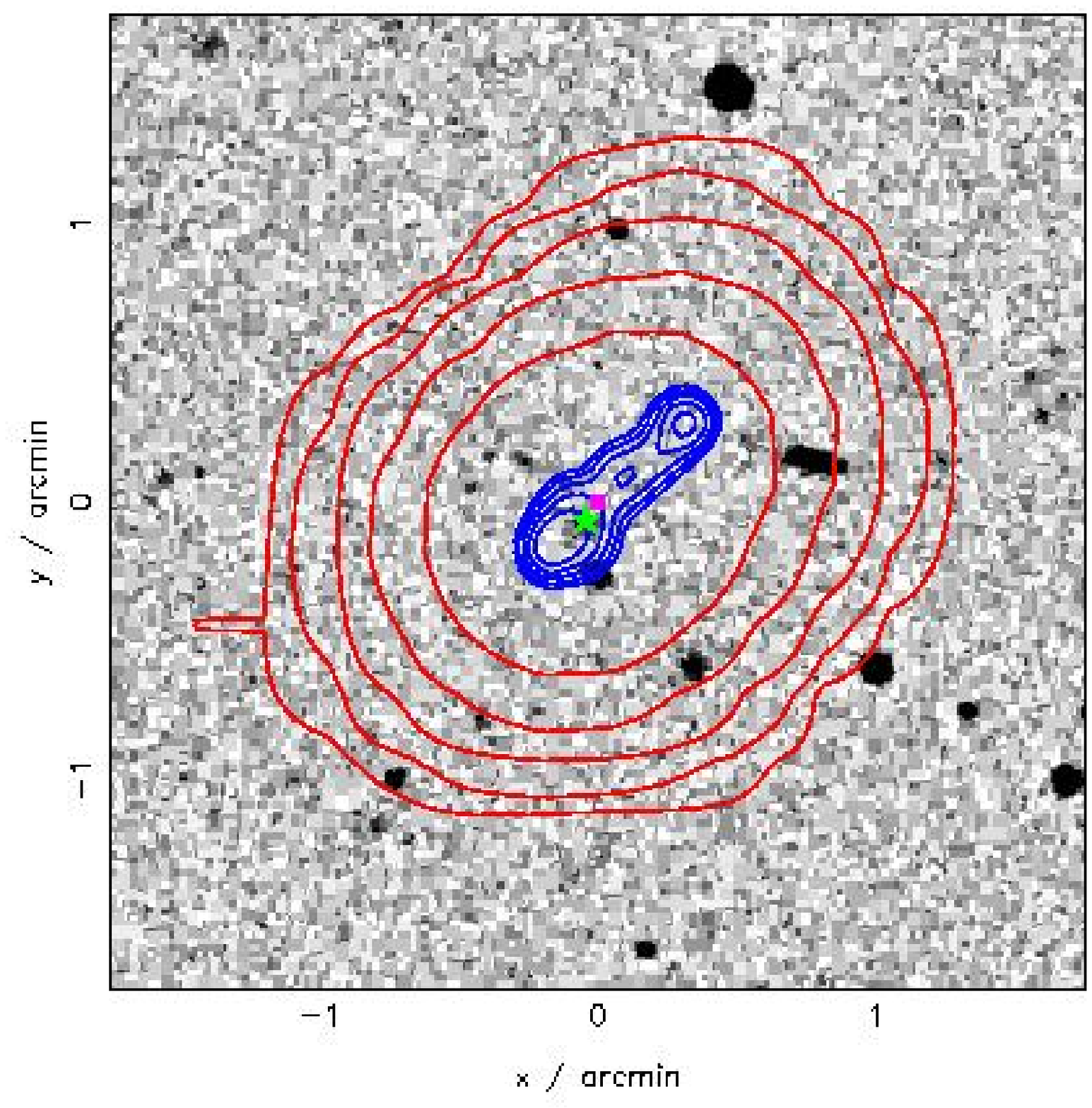}}
      \centerline{C1-146: 4C 02.34}
    \end{minipage}
    \vfill
    \begin{minipage}{3cm}     
      \mbox{}
      \centerline{\includegraphics[scale=0.26,angle=270]{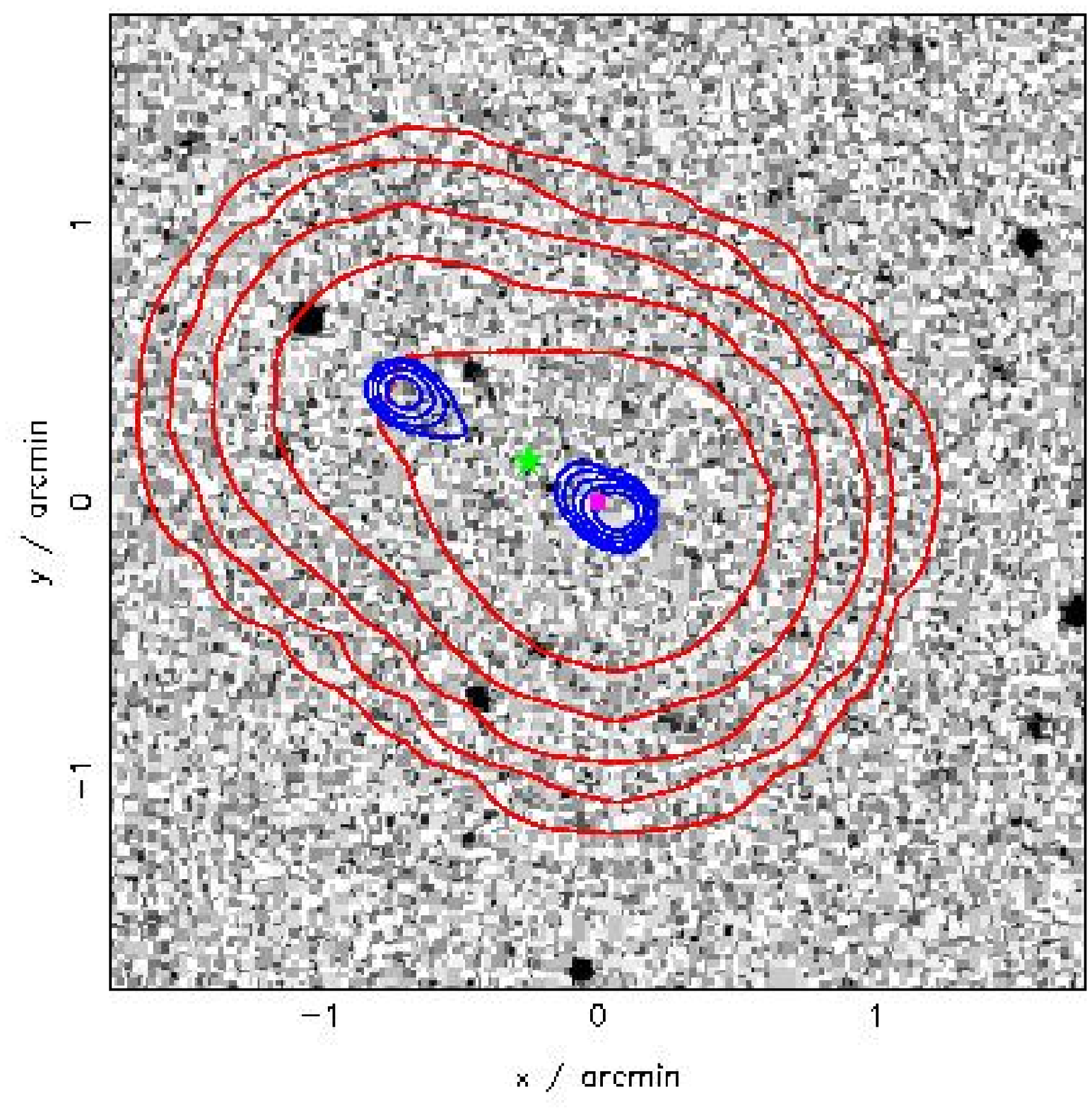}}
      \centerline{C1-147: 3C 277.2}
    \end{minipage}
    \hspace{3cm}
    \begin{minipage}{3cm}
      \mbox{}
      \centerline{\includegraphics[scale=0.26,angle=270]{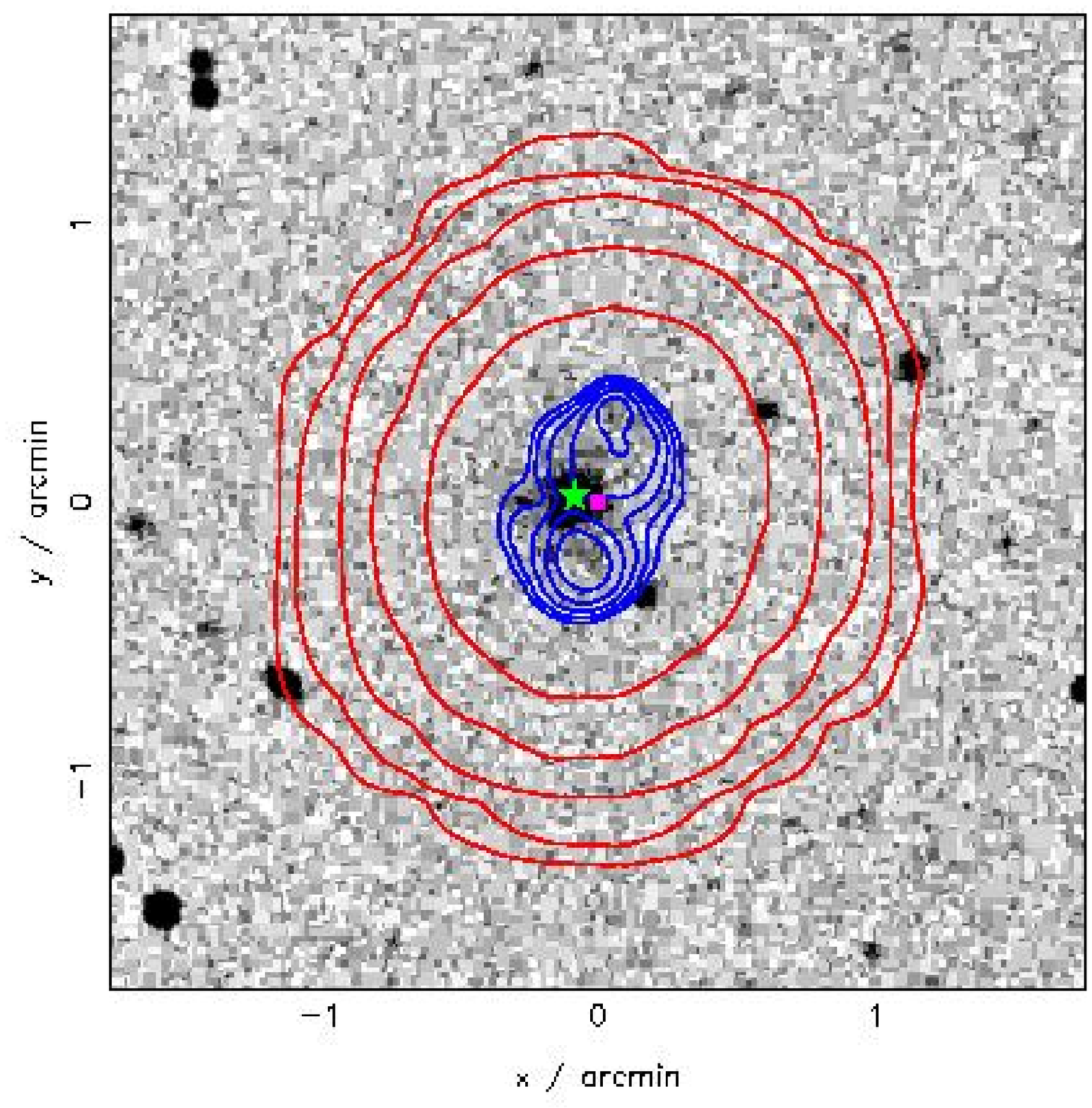}}
      \centerline{C1-148: 3C 277.3}
    \end{minipage}
    \hspace{3cm}
    \begin{minipage}{3cm}
      \mbox{}
      \centerline{\includegraphics[scale=0.26,angle=270]{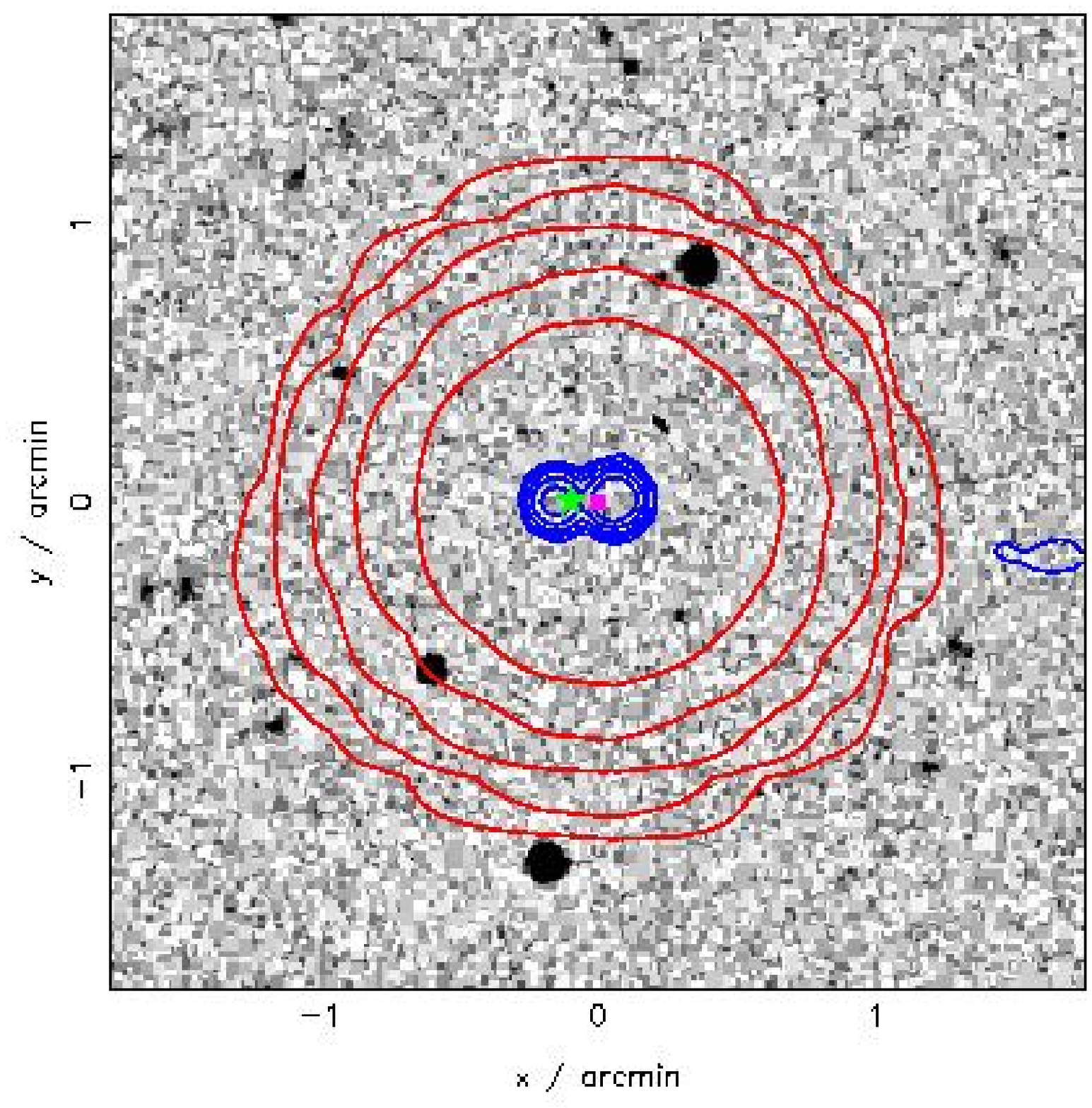}}
      \centerline{C1-150: 3C 280}
    \end{minipage}
  \end{center}
\end{figure}

\begin{figure}
  \begin{center}
    {\bf CoNFIG-1}\\  
  \begin{minipage}{3cm}      
      \mbox{}
      \centerline{\includegraphics[scale=0.26,angle=270]{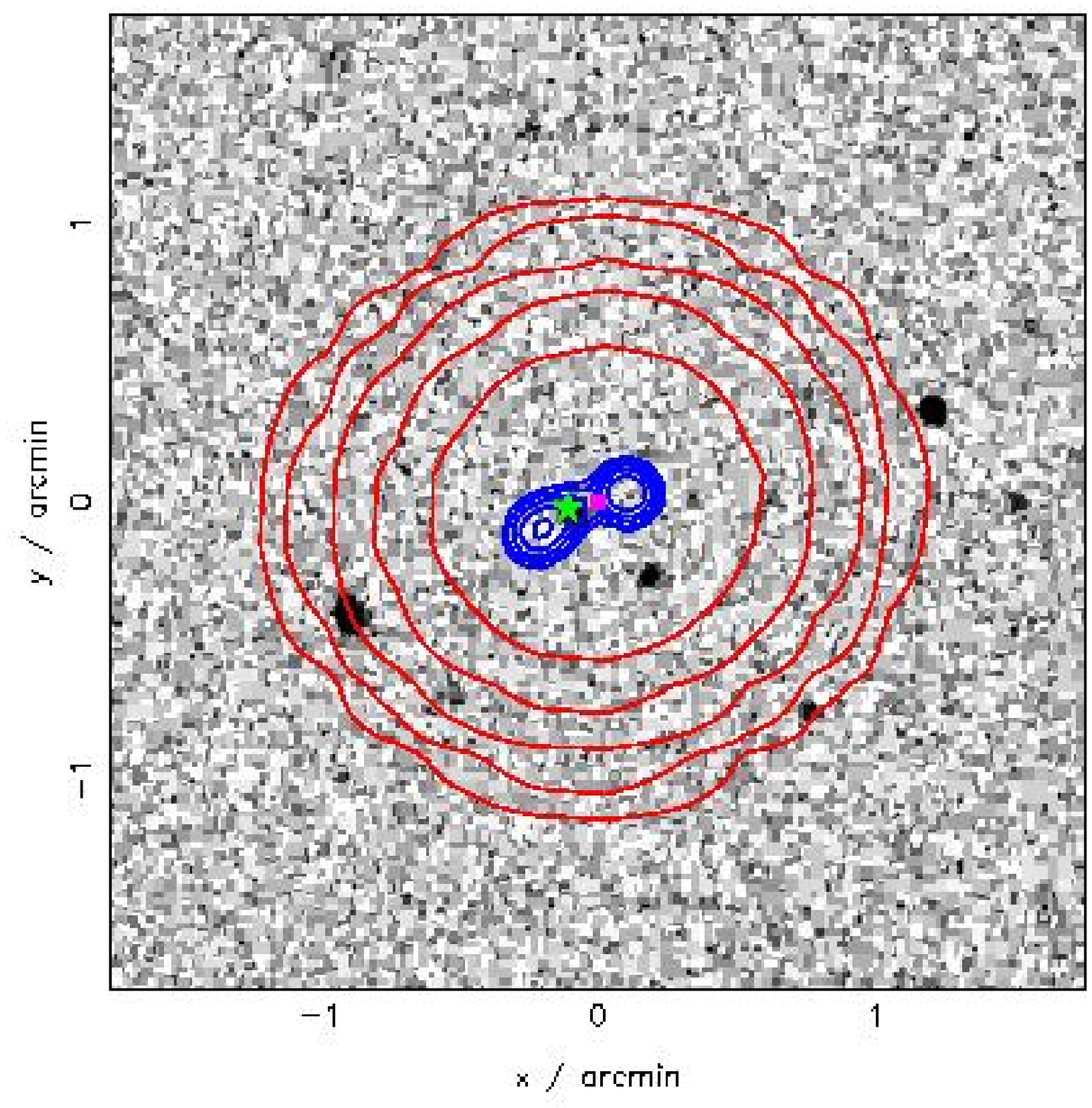}}
      \centerline{C1-151: 3C 280.1}
    \end{minipage}
    \hspace{3cm}
    \begin{minipage}{3cm}
      \mbox{}
      \centerline{\includegraphics[scale=0.26,angle=270]{Contours/C1/152.ps}}
      \centerline{C1-152: 4C 09.45}
    \end{minipage}
    \hspace{3cm}
    \begin{minipage}{3cm}
      \mbox{}
      \centerline{\includegraphics[scale=0.26,angle=270]{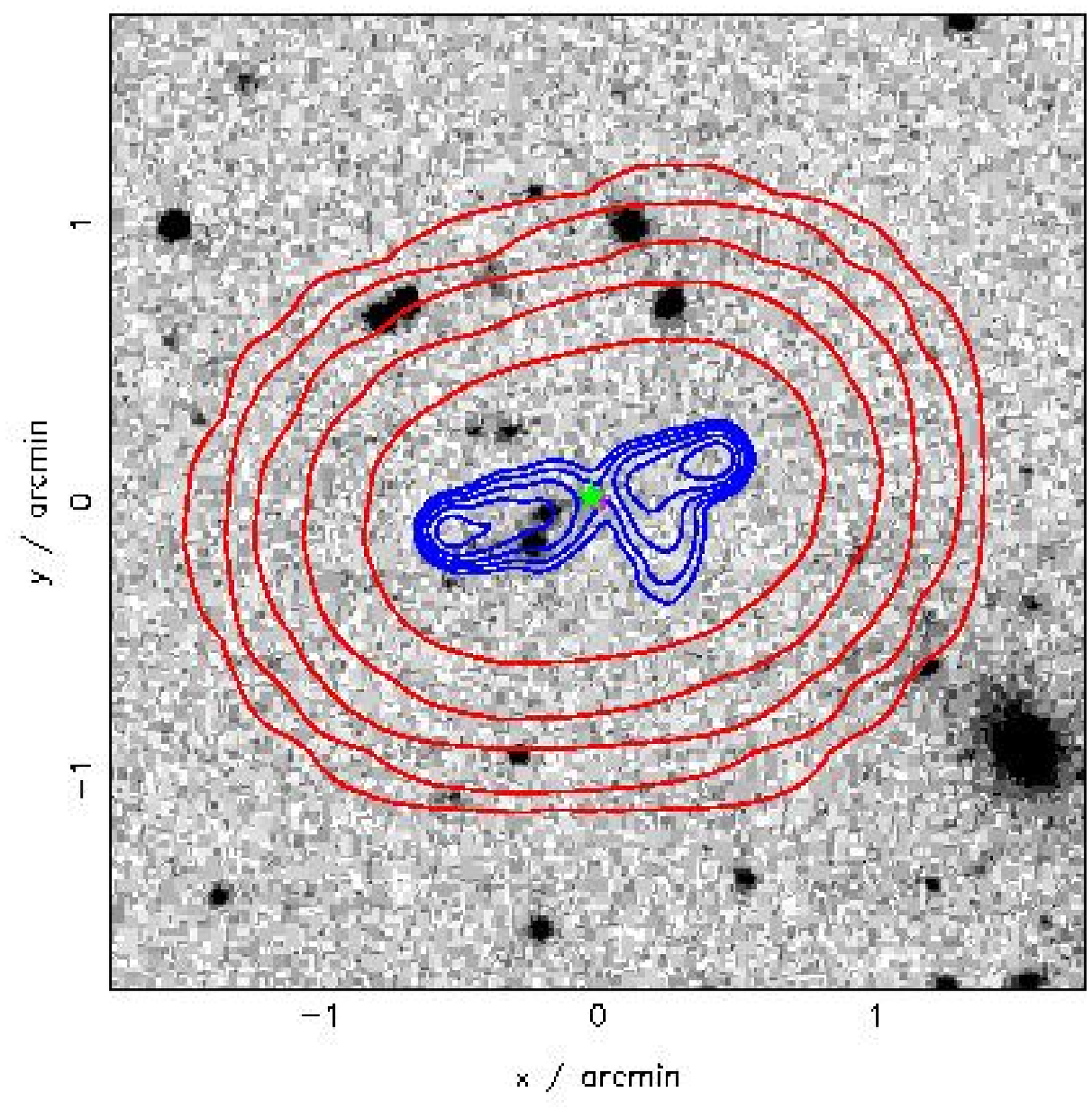}}
      \centerline{C1-153: 4C 00.46}
    \end{minipage}
    \vfill
    \begin{minipage}{3cm}      
      \mbox{}
      \centerline{\includegraphics[scale=0.26,angle=270]{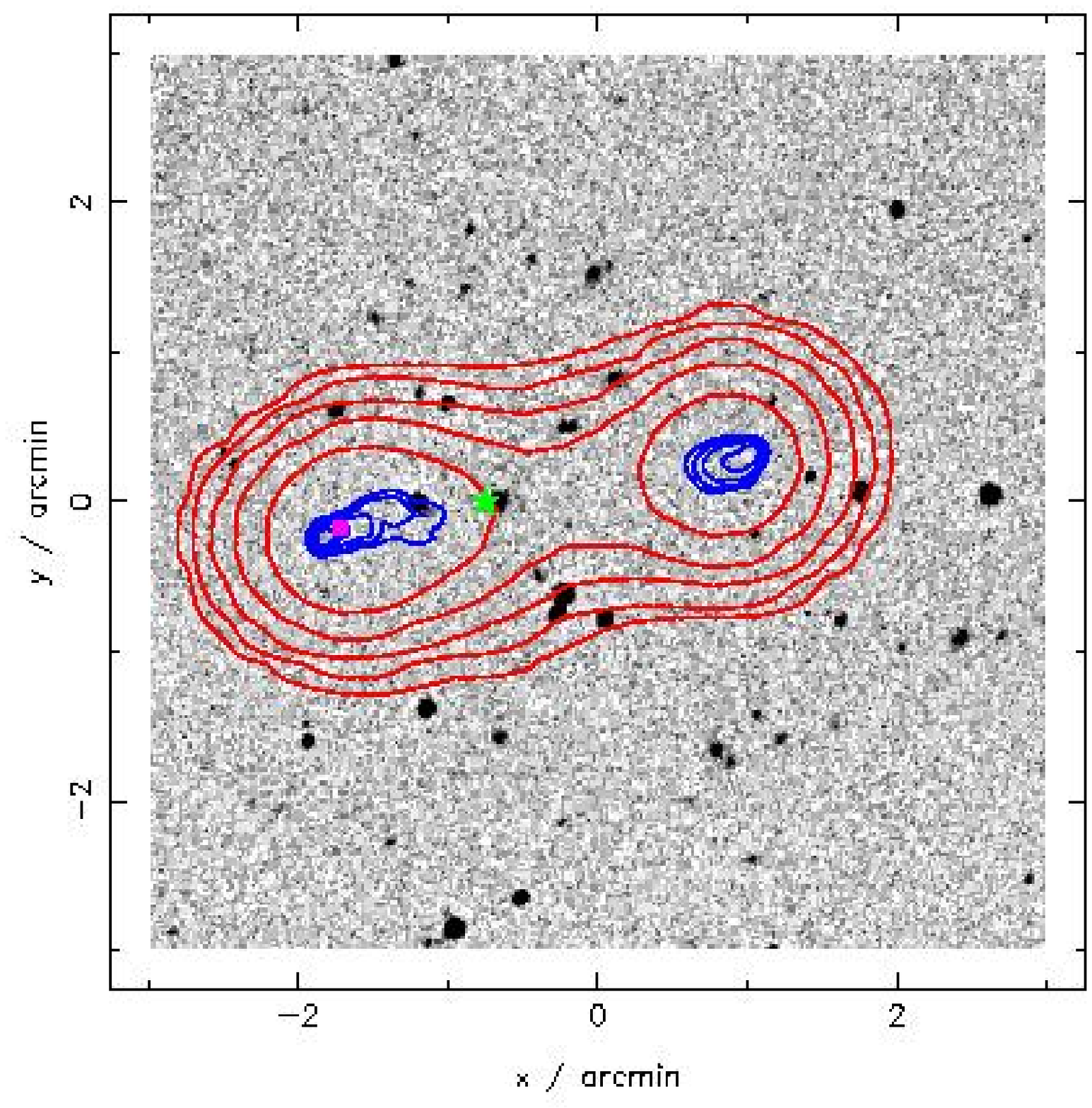}}
      \centerline{C1-155: 3C 284}
    \end{minipage}
    \hspace{3cm}
    \begin{minipage}{3cm}
      \mbox{}
      \centerline{\includegraphics[scale=0.26,angle=270]{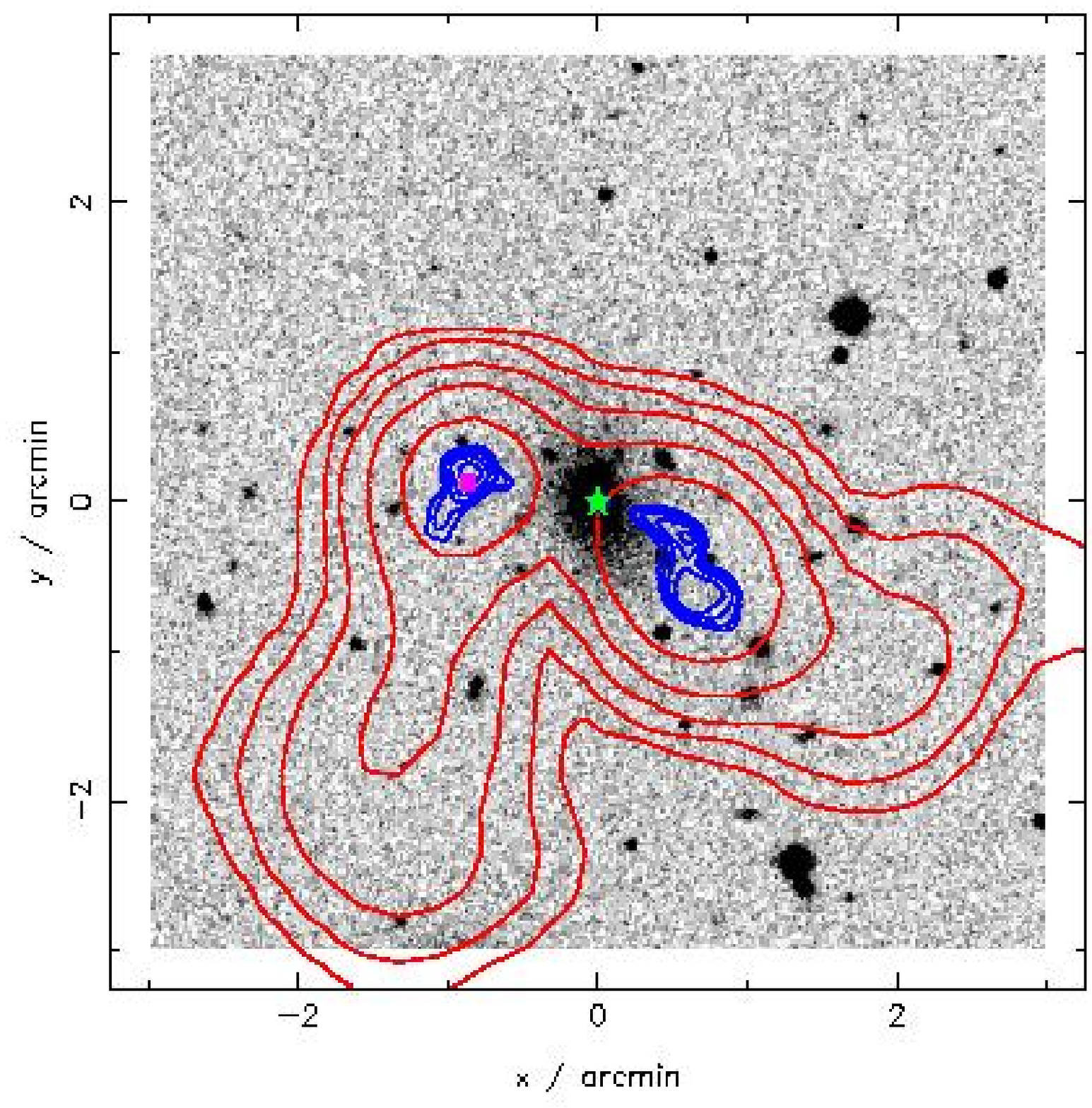}}
      \centerline{C1-157: 4C 07.32}
    \end{minipage}
    \hspace{3cm}
    \begin{minipage}{3cm}
      \mbox{}
      \centerline{\includegraphics[scale=0.26,angle=270]{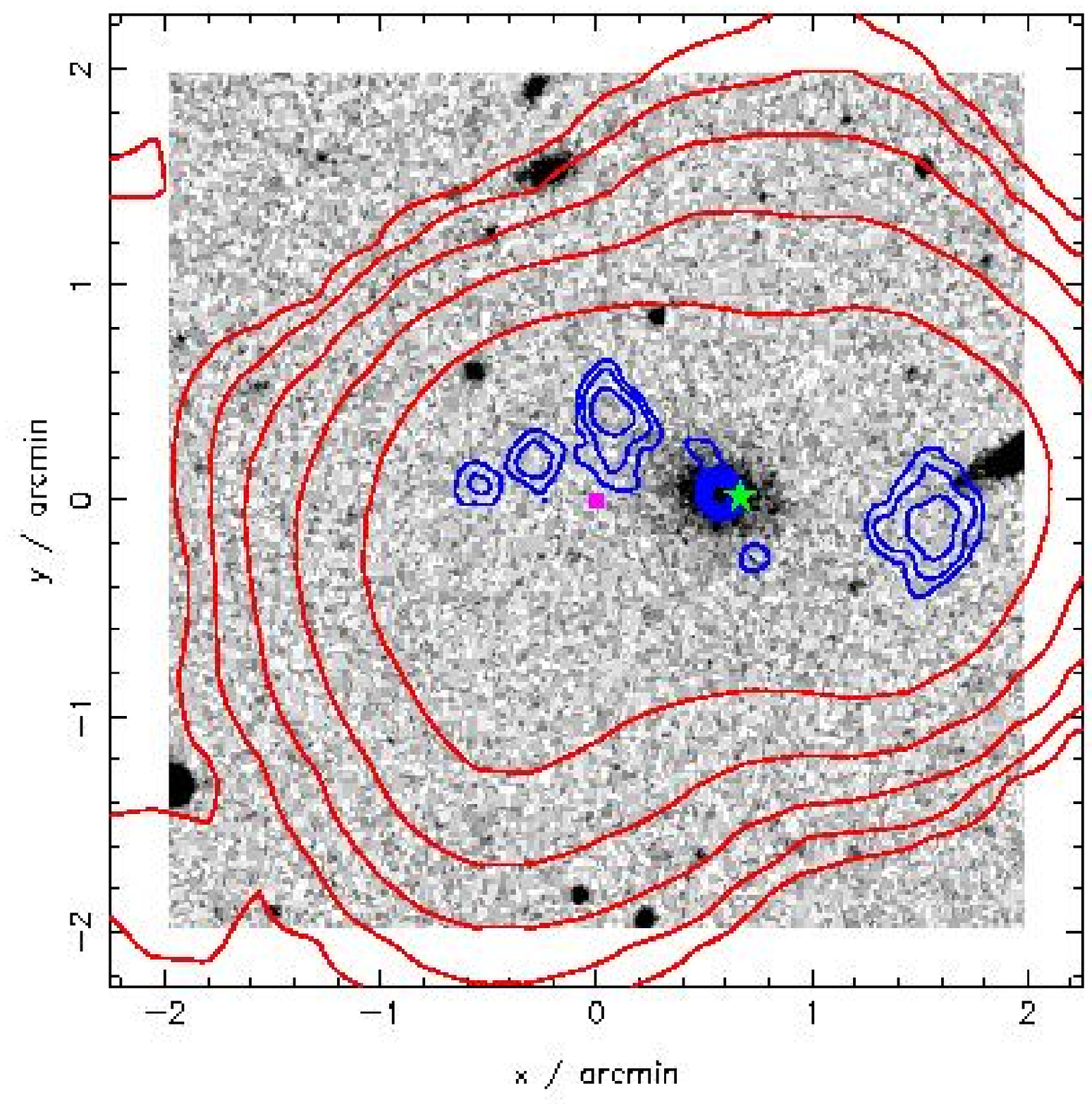}}
      \centerline{C1-158: 4C 29.47}
    \end{minipage}
    \vfill
    \begin{minipage}{3cm}     
      \mbox{}
      \centerline{\includegraphics[scale=0.26,angle=270]{Contours/C1/160.ps}}
      \centerline{C1-160: 4C 17.56}
    \end{minipage}
    \hspace{3cm}
    \begin{minipage}{3cm}
      \mbox{}
      \centerline{\includegraphics[scale=0.26,angle=270]{Contours/C1/161.ps}}
      \centerline{C1-161: 4C 11.45}
    \end{minipage}
    \hspace{3cm}
    \begin{minipage}{3cm}
      \mbox{}
      \centerline{\includegraphics[scale=0.26,angle=270]{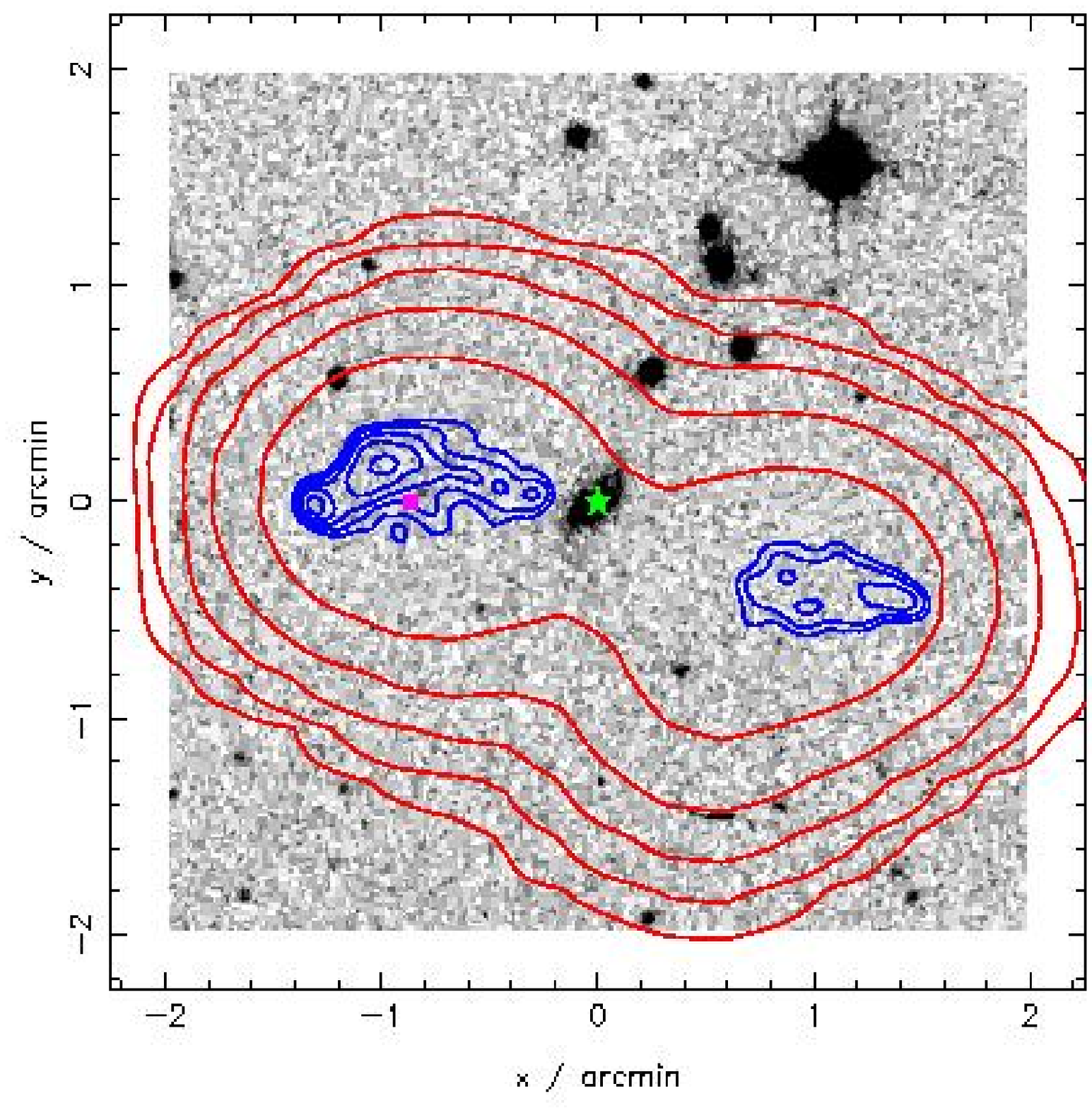}}
      \centerline{C1-162: 3C 285}
    \end{minipage}
    \vfill
    \begin{minipage}{3cm}     
      \mbox{}
      \centerline{\includegraphics[scale=0.26,angle=270]{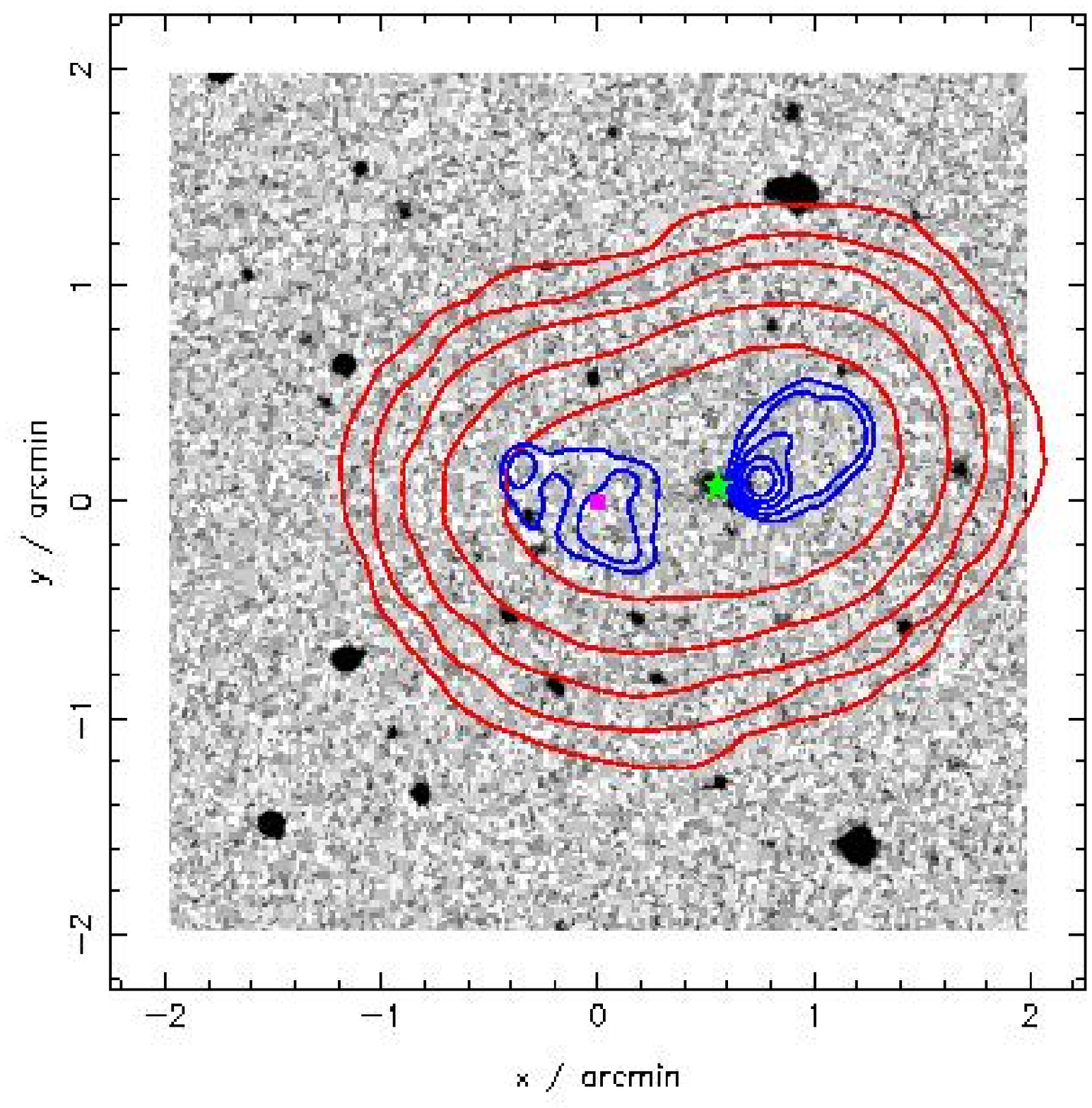}}
      \centerline{C1-163: 4C 03.27}
    \end{minipage}
    \hspace{3cm}
    \begin{minipage}{3cm}
      \mbox{}
      \centerline{\includegraphics[scale=0.26,angle=270]{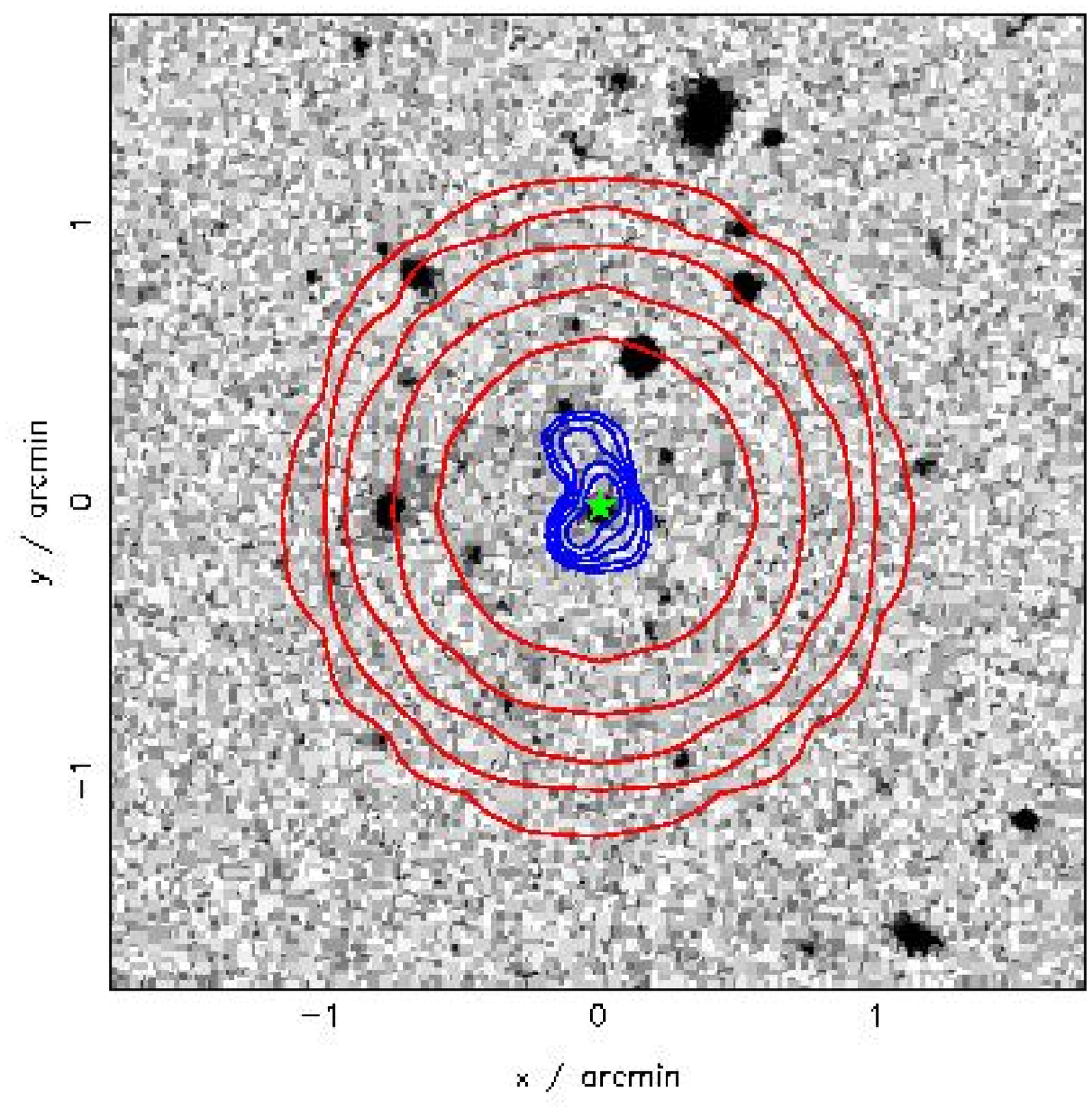}}
      \centerline{C1-165: 4C 32.44B}
    \end{minipage}
    \hspace{3cm}
    \begin{minipage}{3cm}
      \mbox{}
      \centerline{\includegraphics[scale=0.26,angle=270]{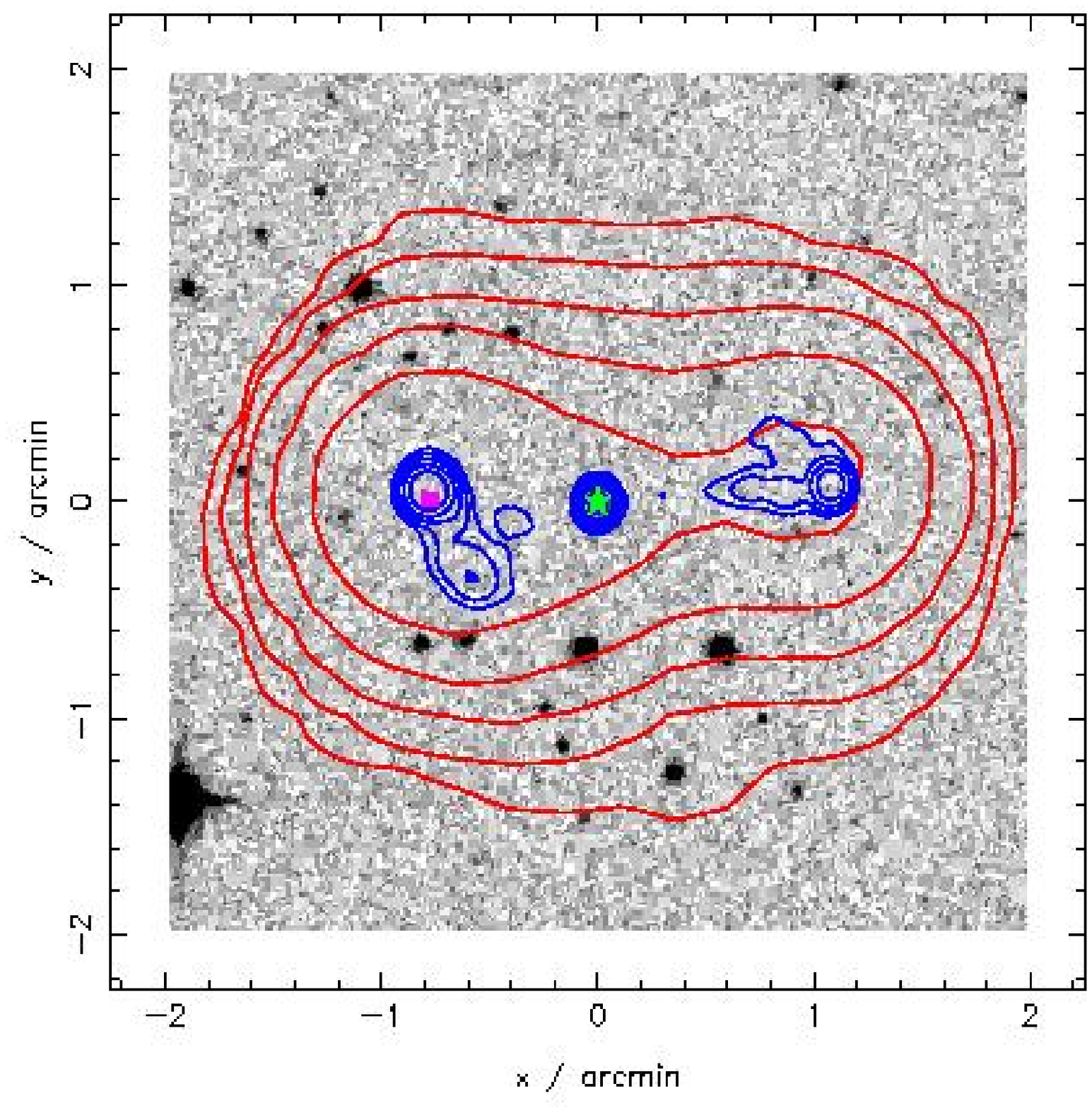}}
      \centerline{C1-168: 3C 287.1}
    \end{minipage}
  \end{center}
\end{figure}

\begin{figure}
  \begin{center}
    {\bf CoNFIG-1}\\  
  \begin{minipage}{3cm}      
      \mbox{}
      \centerline{\includegraphics[scale=0.26,angle=270]{Contours/C1/169.ps}}
      \centerline{C1-169: 4C -06.35}
    \end{minipage}
    \hspace{3cm}
    \begin{minipage}{3cm}
      \mbox{}
      \centerline{\includegraphics[scale=0.26,angle=270]{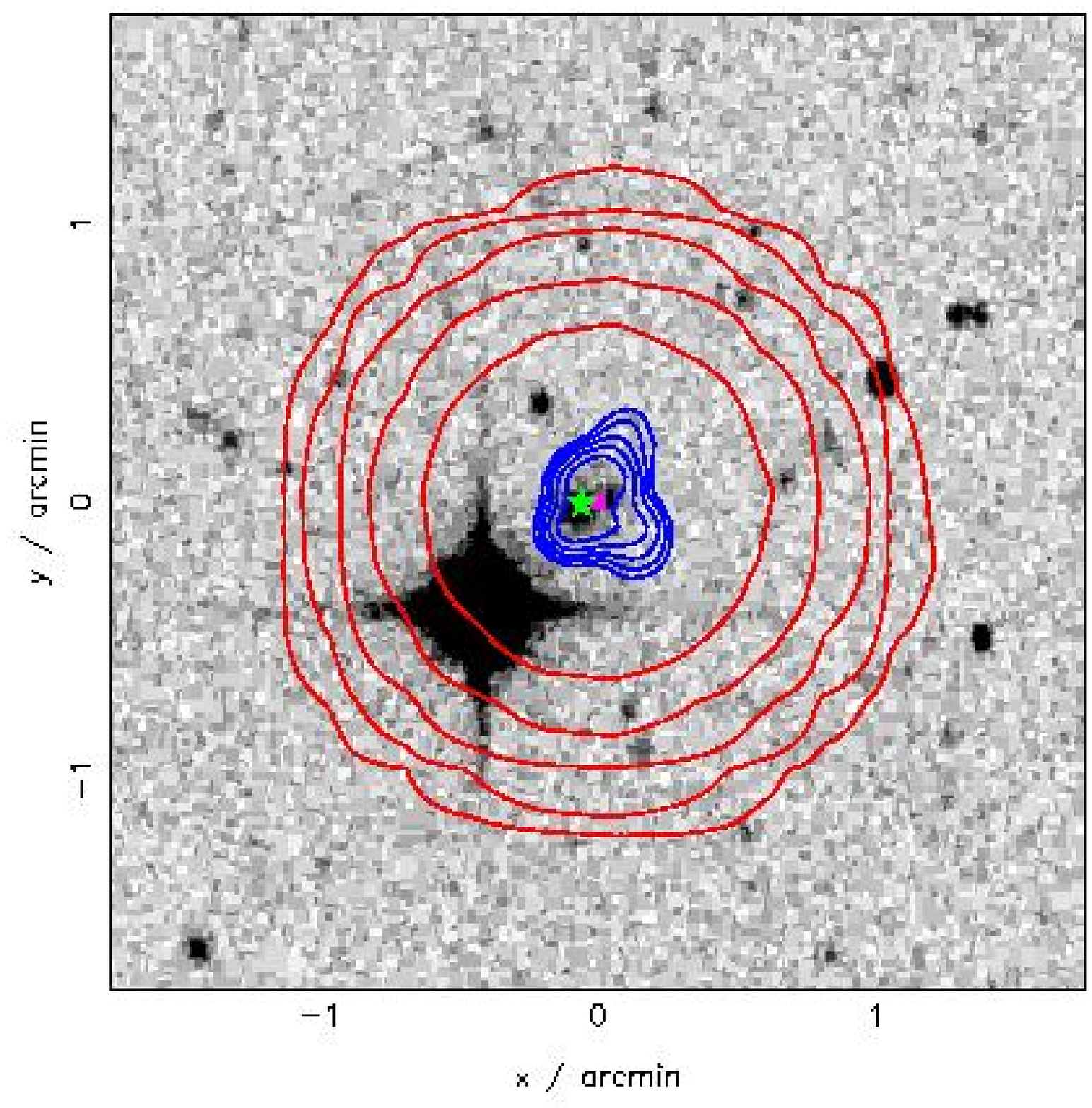}}
      \centerline{C1-170: 3C 288}
    \end{minipage}
    \hspace{3cm}
    \begin{minipage}{3cm}
      \mbox{}
      \centerline{\includegraphics[scale=0.26,angle=270]{Contours/C1/171.ps}}
      \centerline{C1-171: 3C 288.1}
    \end{minipage}
    \vfill
    \begin{minipage}{3cm}      
      \mbox{}
      \centerline{\includegraphics[scale=0.26,angle=270]{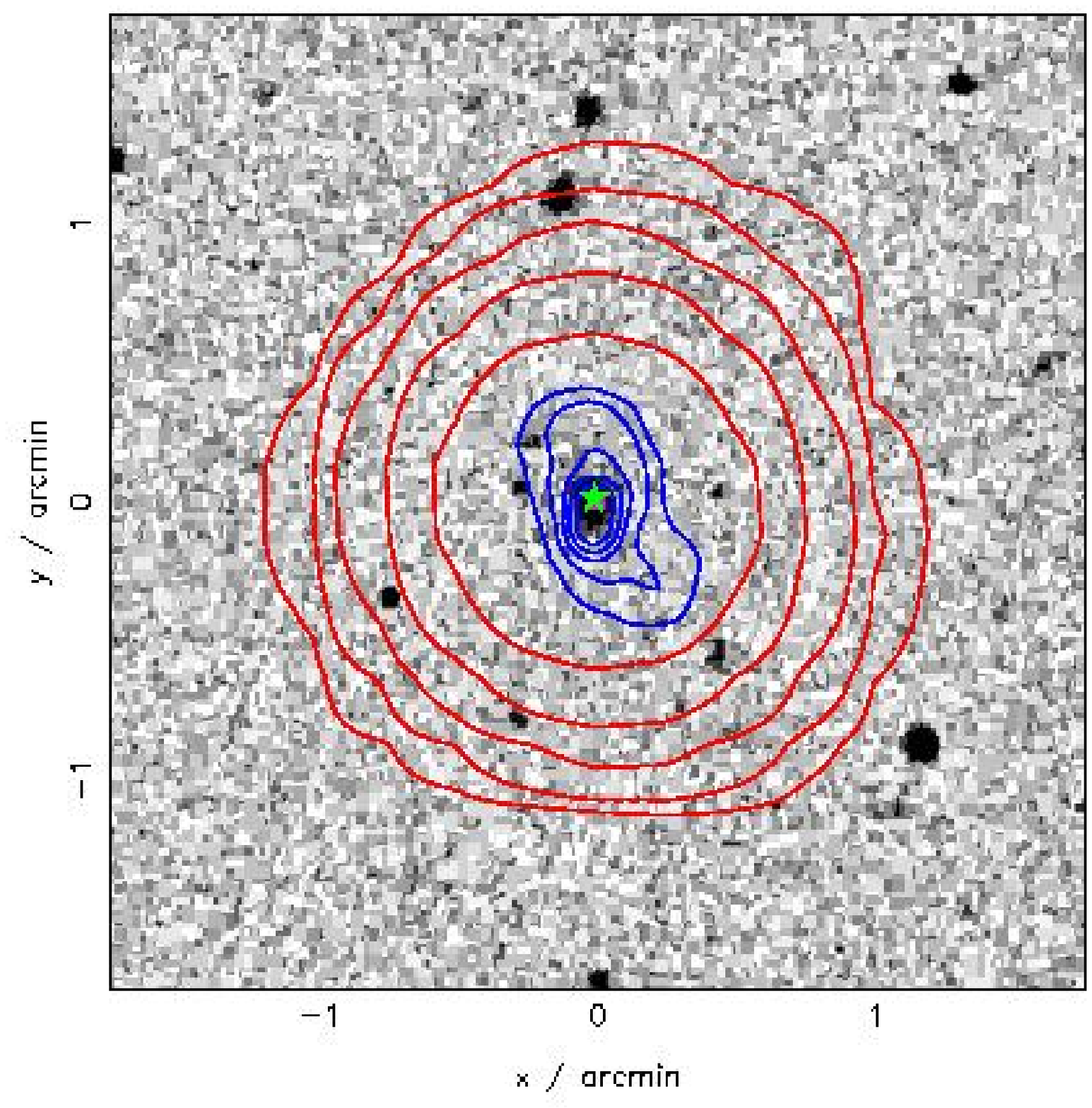}}
      \centerline{C1-172: 4C 05.57}
    \end{minipage}
    \hspace{3cm}
    \begin{minipage}{3cm}
      \mbox{}
      \centerline{\includegraphics[scale=0.26,angle=270]{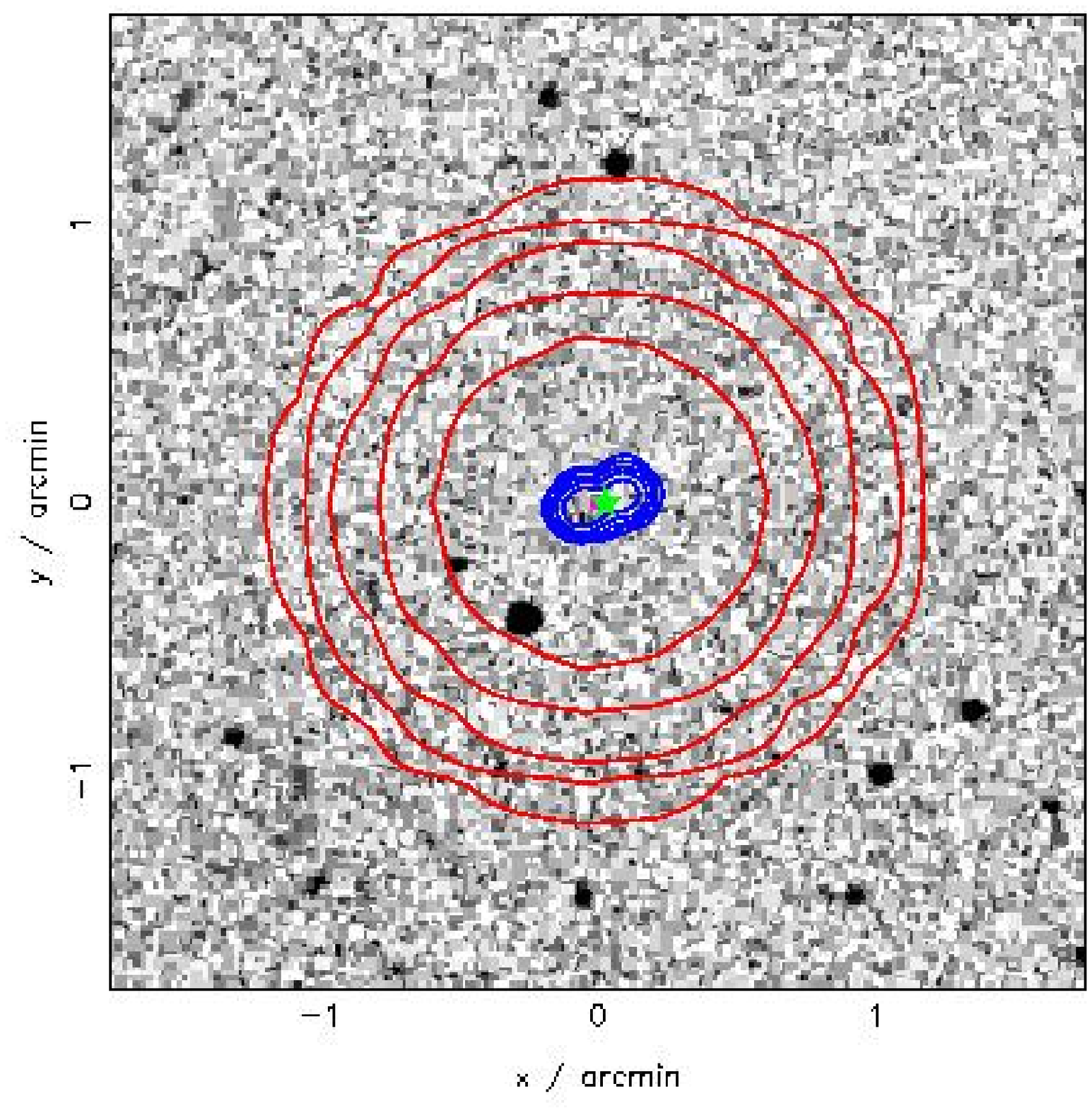}}
      \centerline{C1-174: 3C 289}
    \end{minipage}
    \hspace{3cm}
    \begin{minipage}{3cm}
      \mbox{}
      \centerline{\includegraphics[scale=0.26,angle=270]{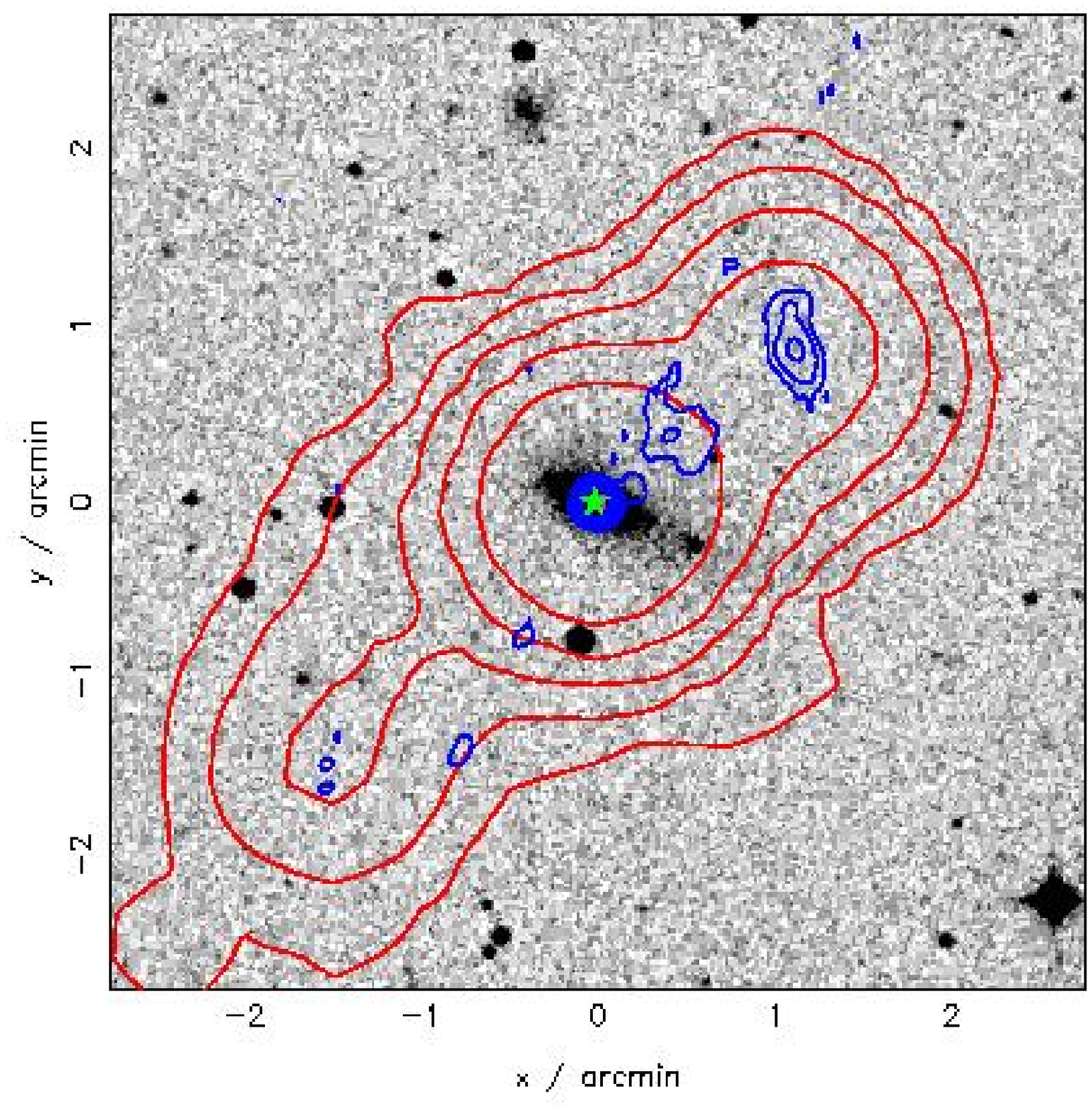}}
      \centerline{C1-176: 3C 293}
    \end{minipage}
    \vfill
    \begin{minipage}{3cm}     
      \mbox{}
      \centerline{\includegraphics[scale=0.26,angle=270]{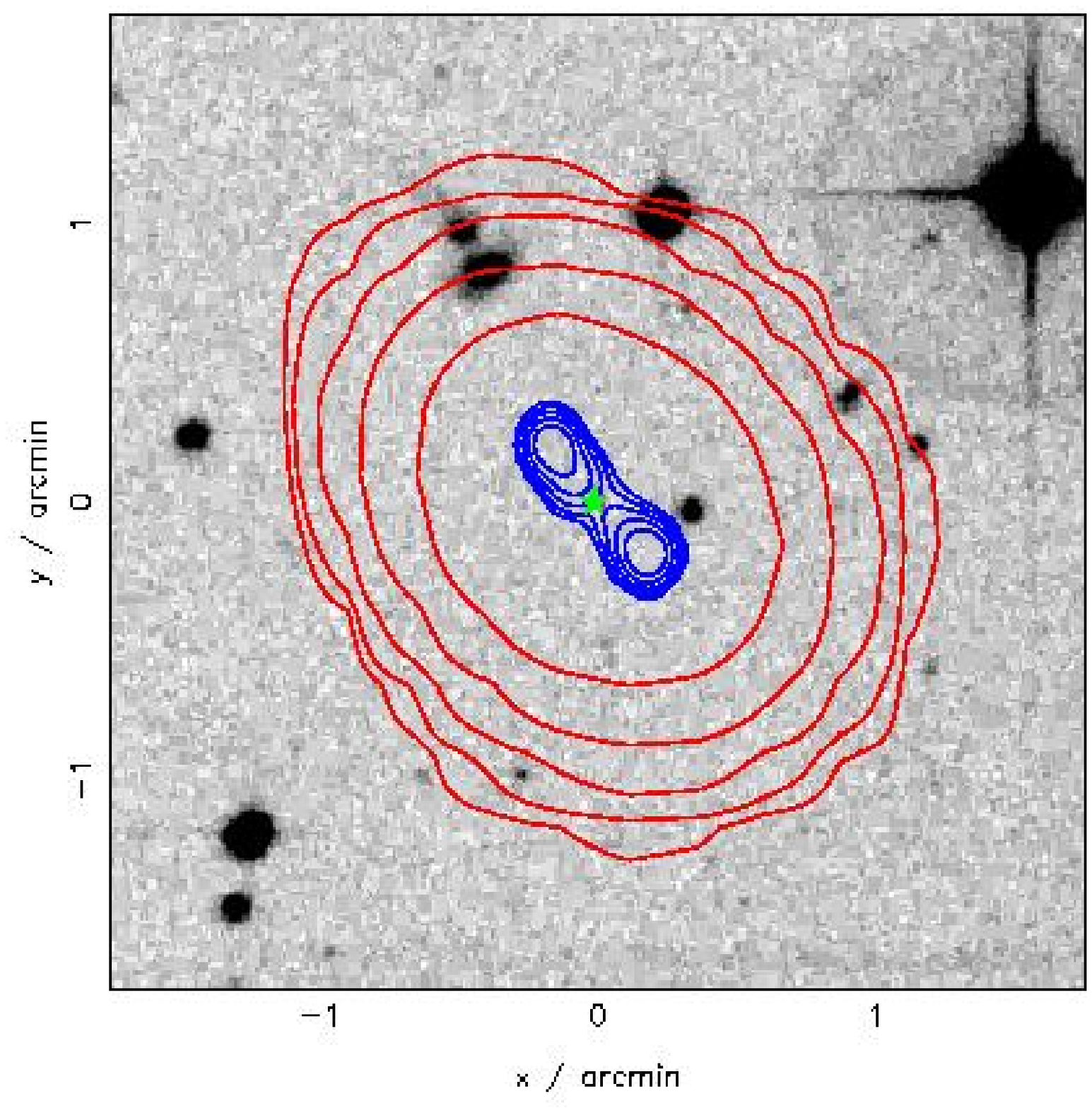}}
      \centerline{C1-178: 4C 01.39}
    \end{minipage}
    \hspace{3cm}
    \begin{minipage}{3cm}
      \mbox{}
      \centerline{\includegraphics[scale=0.26,angle=270]{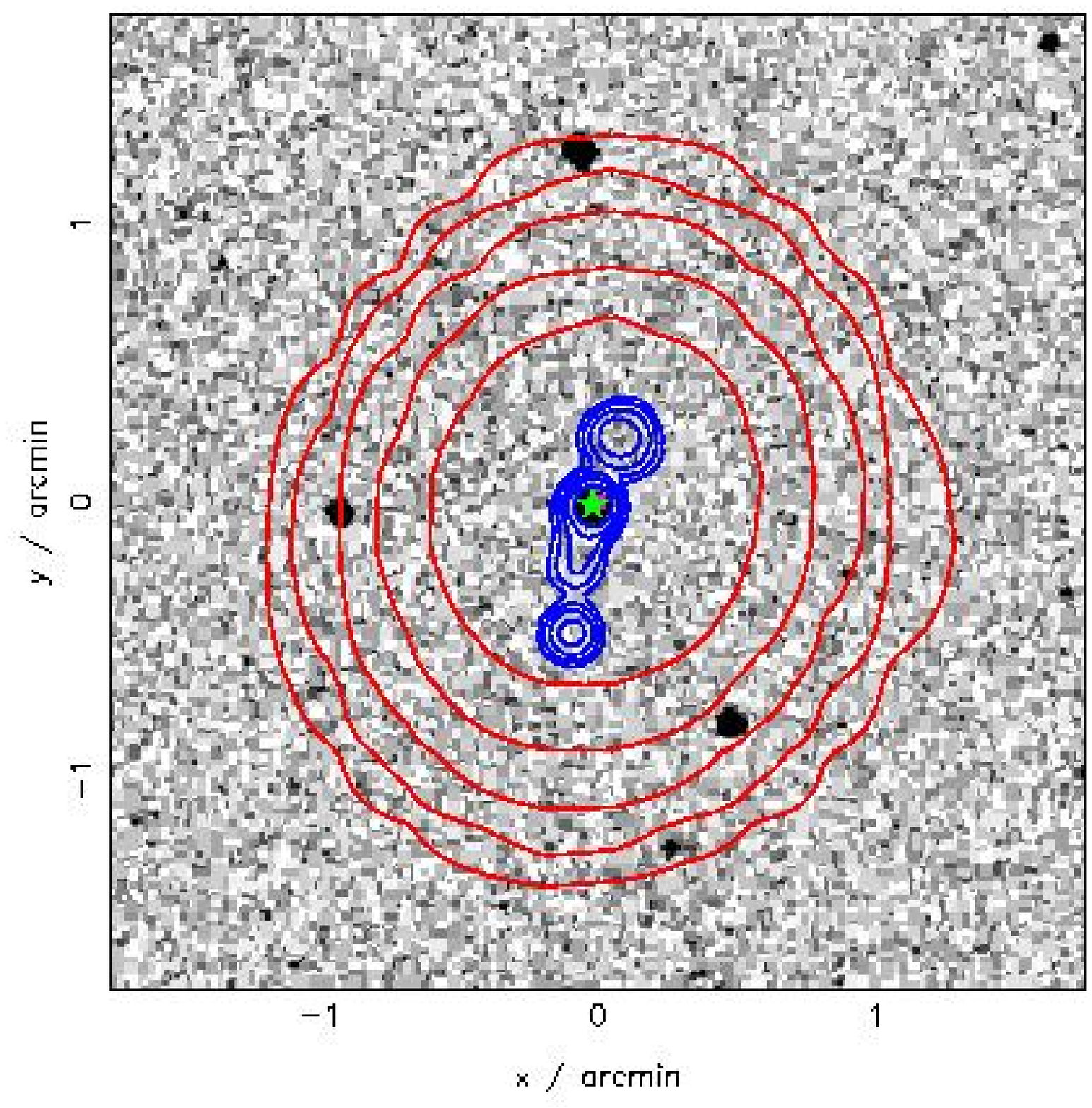}}
      \centerline{C1-179: 4C 19.44}
    \end{minipage}
    \hspace{3cm}
    \begin{minipage}{3cm}
      \mbox{}
      \centerline{\includegraphics[scale=0.26,angle=270]{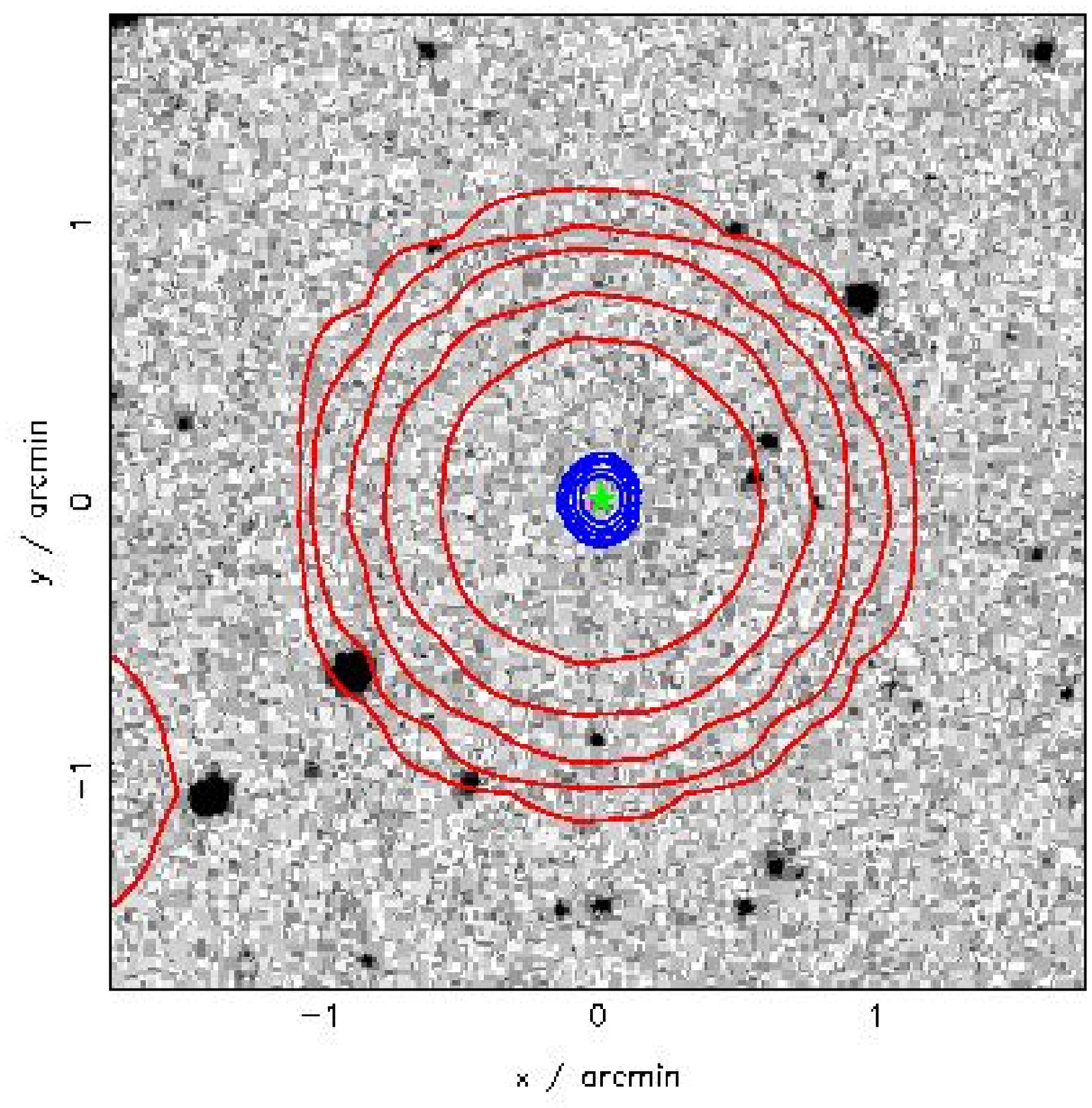}}
      \centerline{C1-180: PKS 1355+01}
    \end{minipage}
    \vfill
    \begin{minipage}{3cm}     
      \mbox{}
      \centerline{\includegraphics[scale=0.26,angle=270]{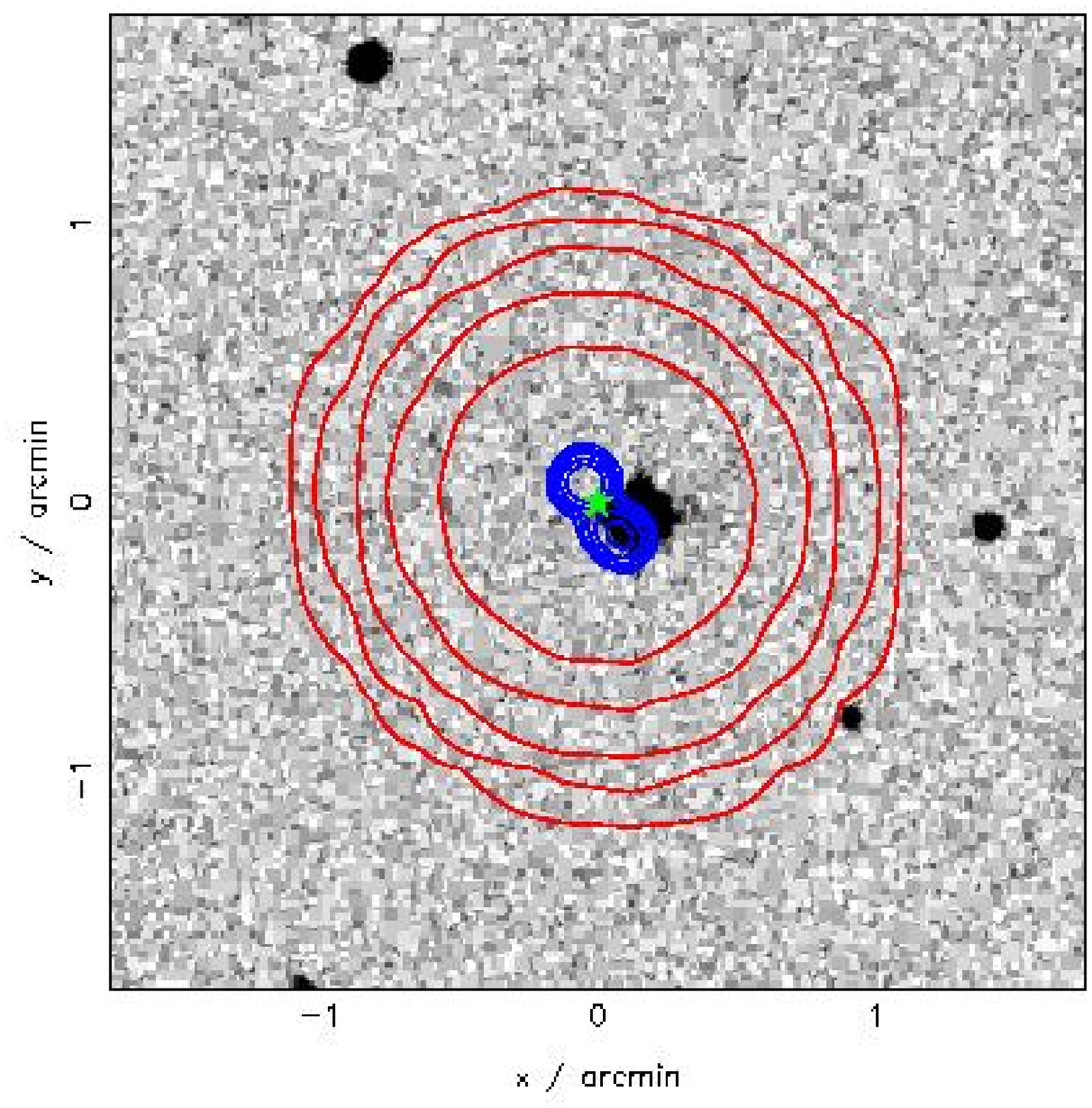}}
      \centerline{C1-182: 3C 294}
    \end{minipage}
    \hspace{3cm}
    \begin{minipage}{3cm}
      \mbox{}
      \centerline{\includegraphics[scale=0.26,angle=270]{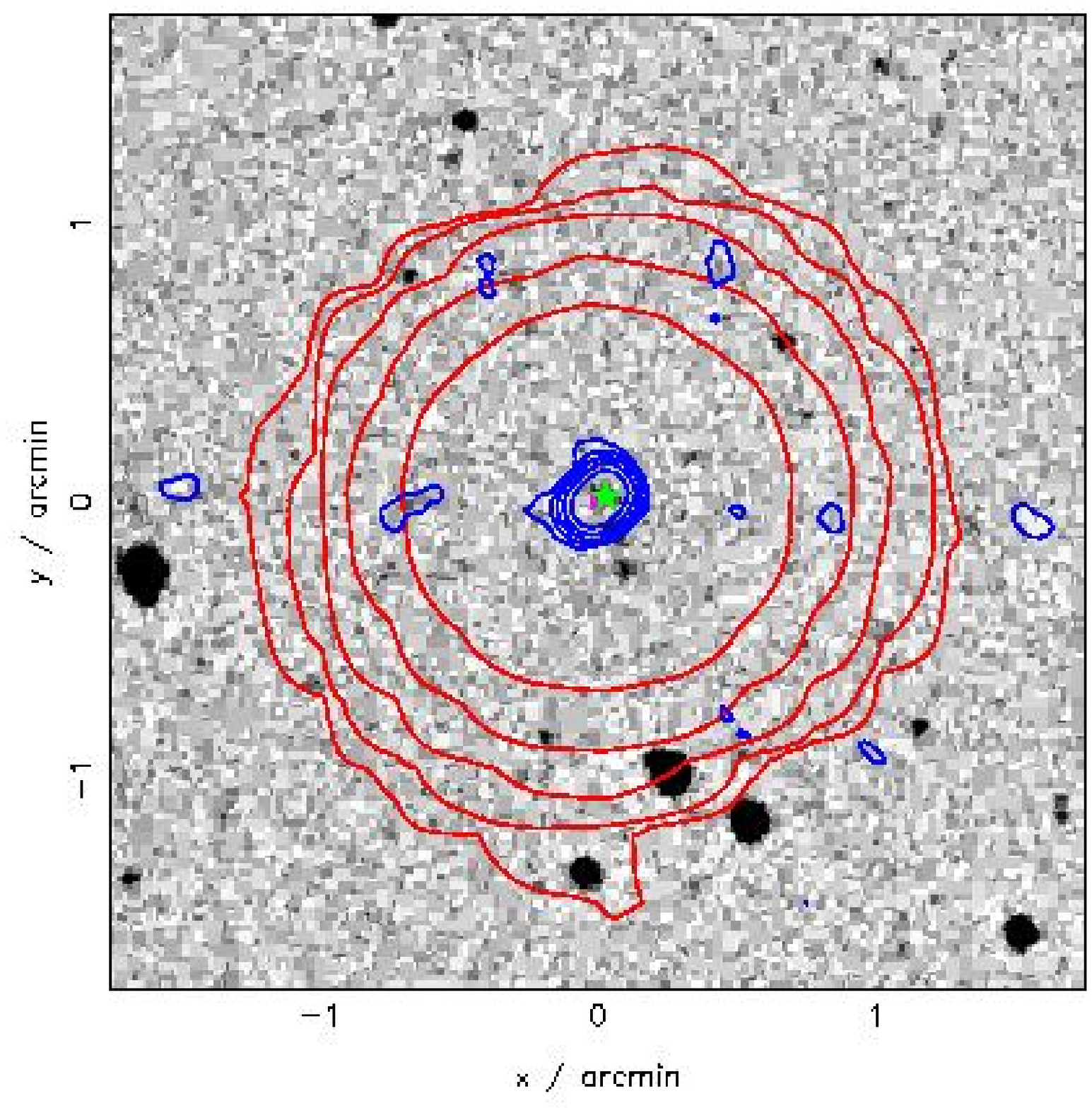}}
      \centerline{C1-183: 3C 295}
    \end{minipage}
    \hspace{3cm}
    \begin{minipage}{3cm}
      \mbox{}
      \centerline{\includegraphics[scale=0.26,angle=270]{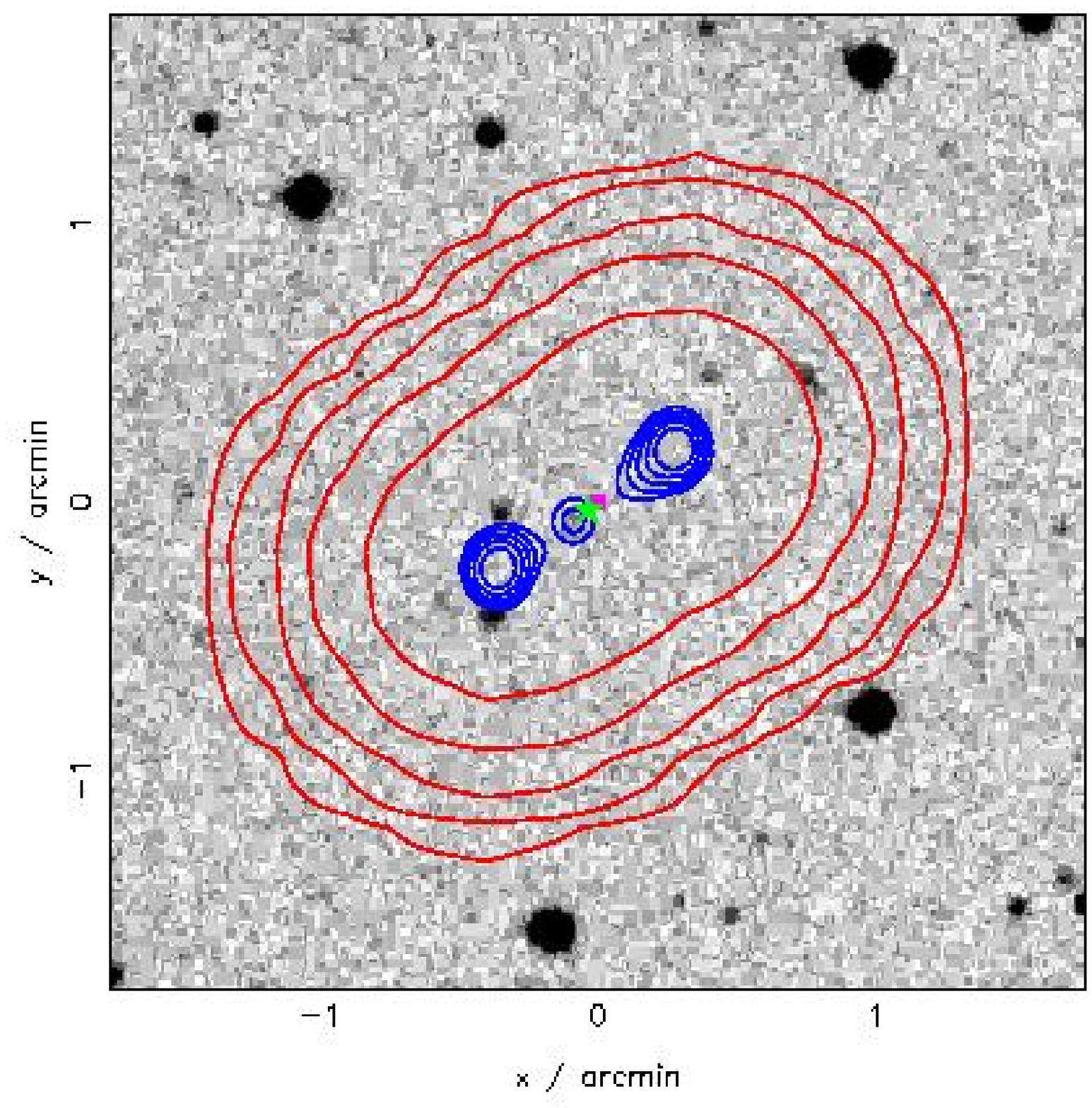}}
      \centerline{C1-184: 4C -05.60}
    \end{minipage}
  \end{center}
\end{figure}

\begin{figure}
  \begin{center}
    {\bf CoNFIG-1}\\  
  \begin{minipage}{3cm}      
      \mbox{}
      \centerline{\includegraphics[scale=0.26,angle=270]{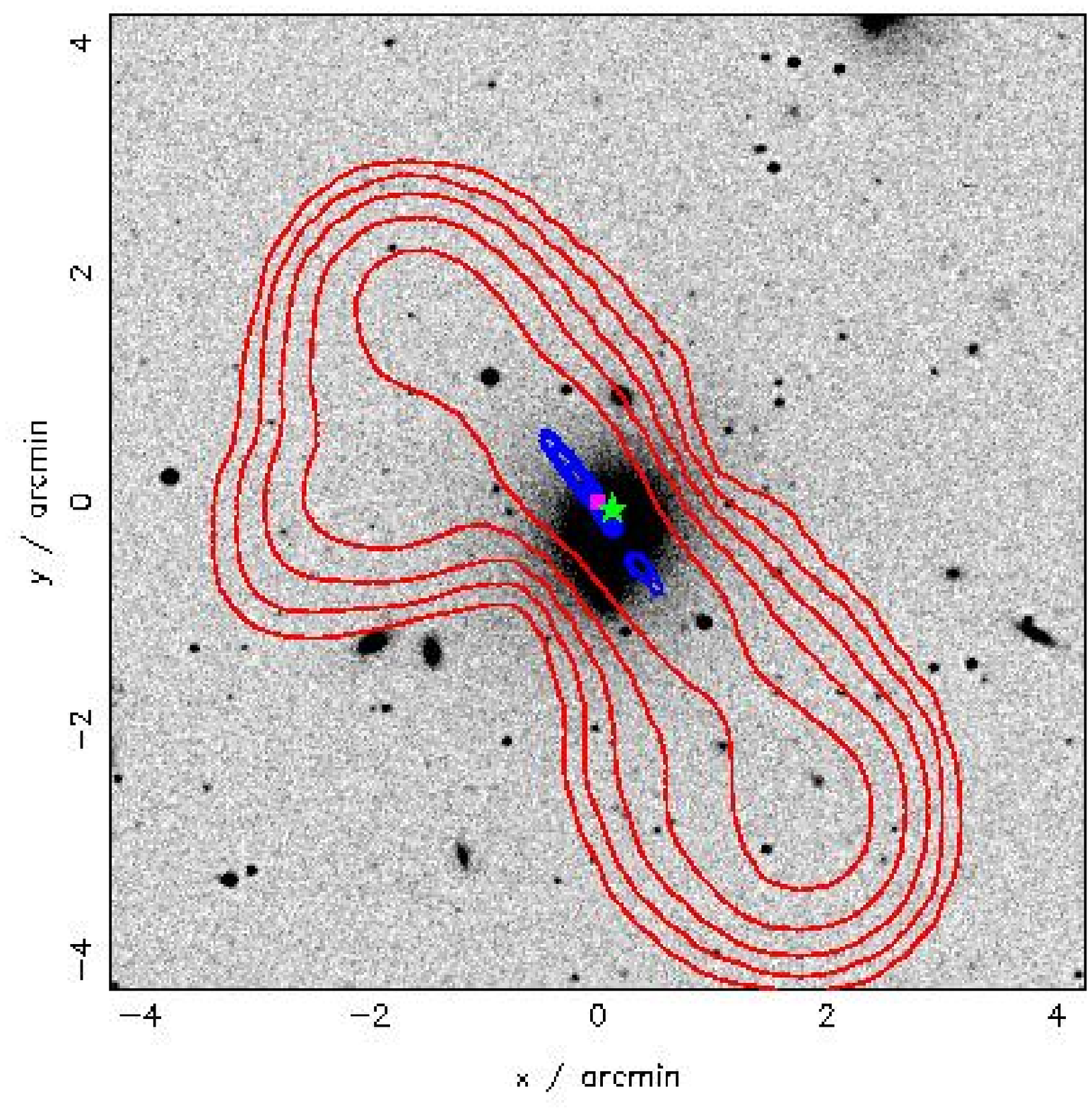}}
      \centerline{C1-186: NGC 5532}
    \end{minipage}
    \hspace{3cm}
    \begin{minipage}{3cm}
      \mbox{}
      \centerline{\includegraphics[scale=0.26,angle=270]{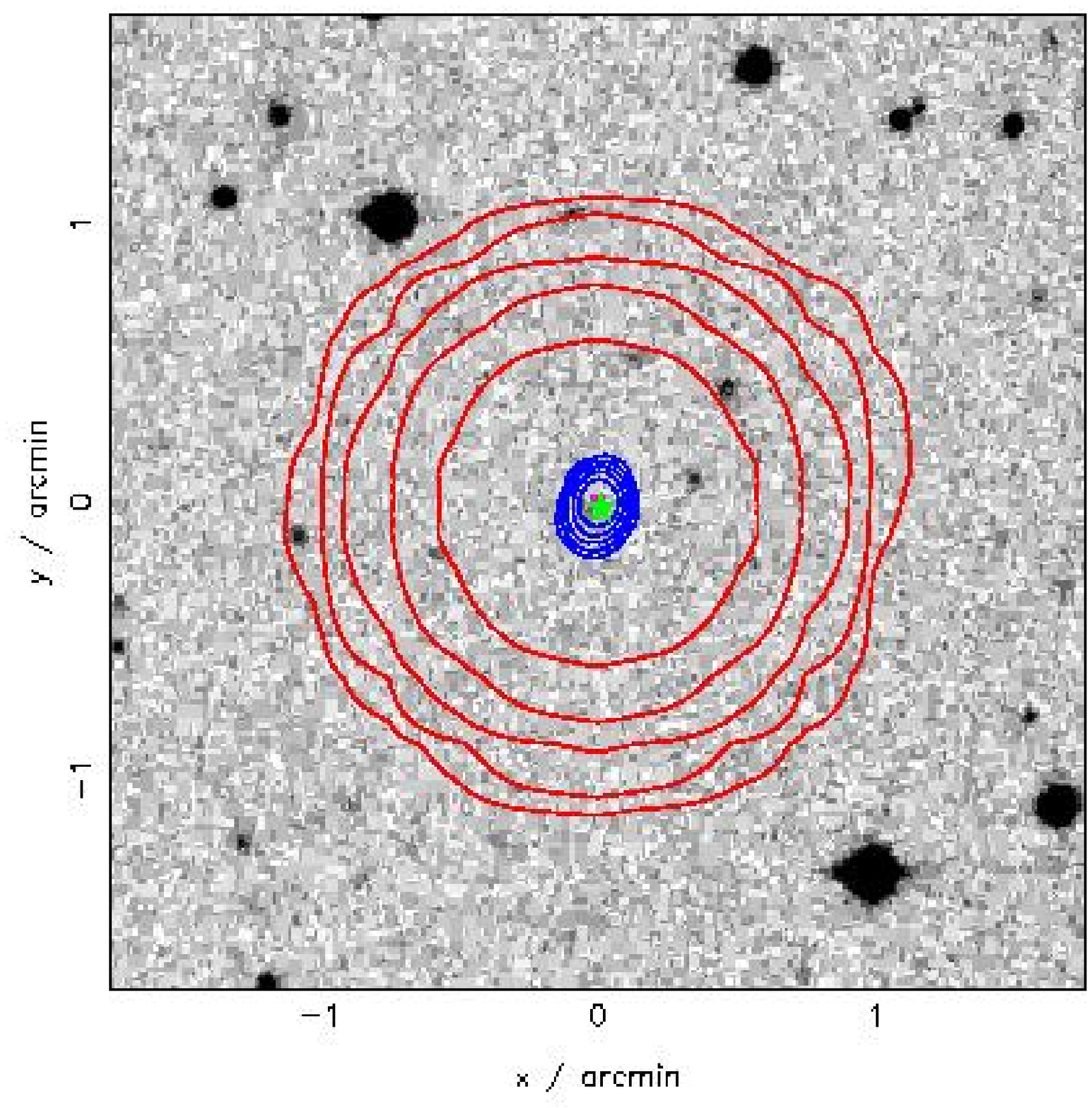}}
      \centerline{C1-187: 3C 297}
    \end{minipage}
    \hspace{3cm}
    \begin{minipage}{3cm}
      \mbox{}
      \centerline{\includegraphics[scale=0.26,angle=270]{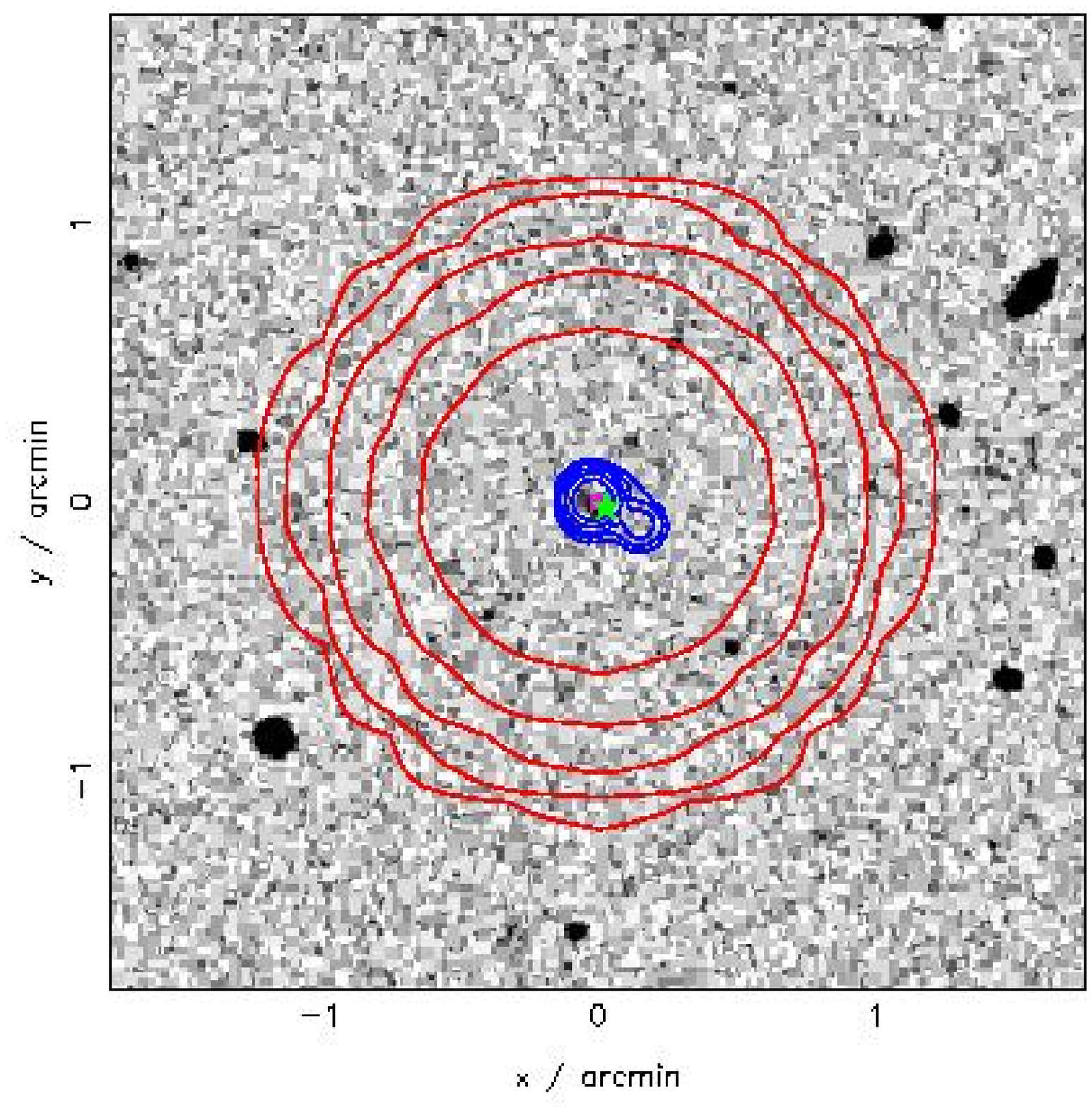}}
      \centerline{C1-189: 3C 299}
    \end{minipage}
    \vfill
    \begin{minipage}{3cm}      
      \mbox{}
      \centerline{\includegraphics[scale=0.26,angle=270]{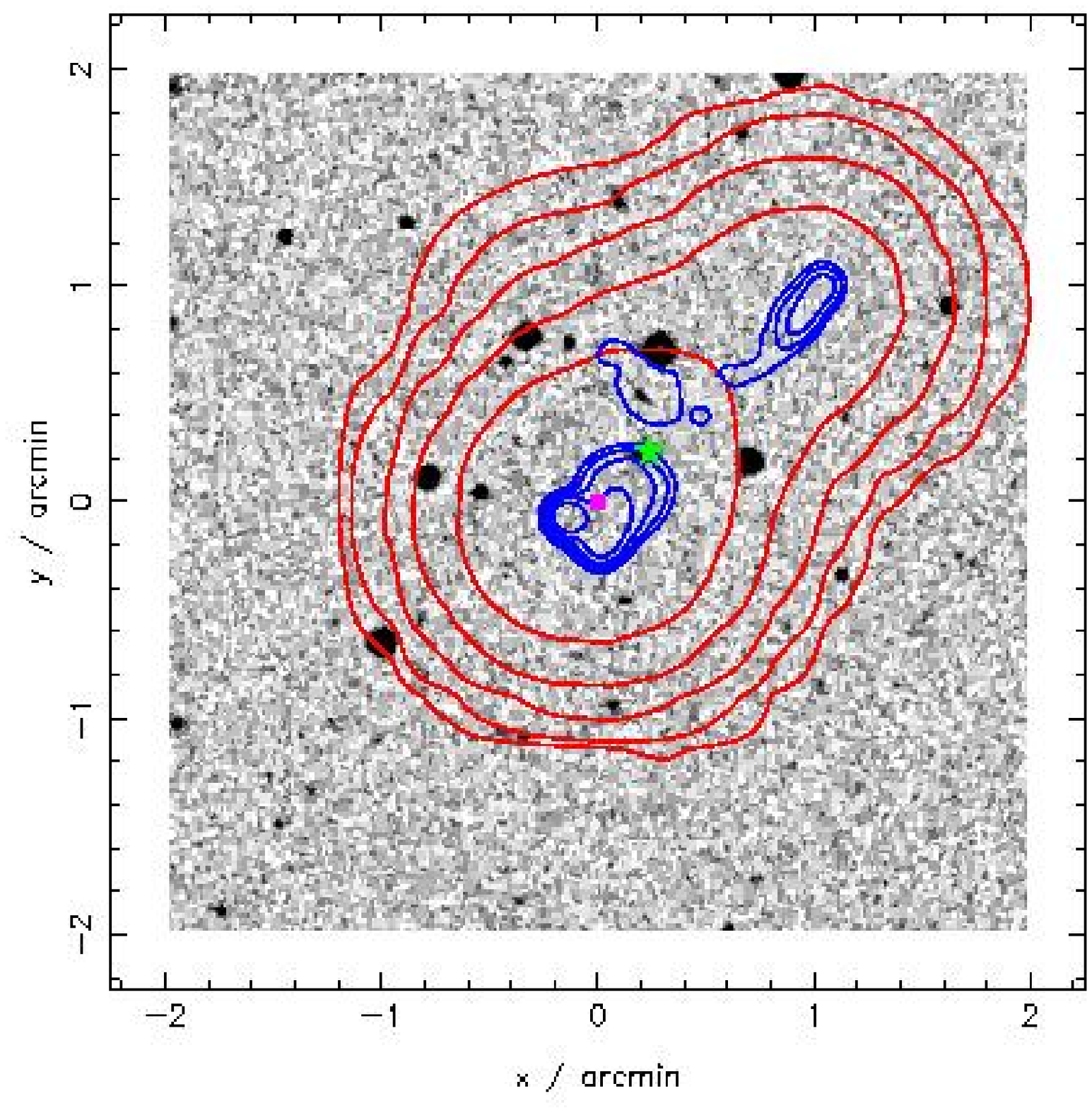}}
      \centerline{C1-190: 3C 300}
    \end{minipage}
    \hspace{3cm}
    \begin{minipage}{3cm}
      \mbox{}
      \centerline{\includegraphics[scale=0.26,angle=270]{Contours/C1/191.ps}}
      \centerline{C1-191: 4C 20.33}
    \end{minipage}
    \hspace{3cm}
    \begin{minipage}{3cm}
      \mbox{}
      \centerline{\includegraphics[scale=0.26,angle=270]{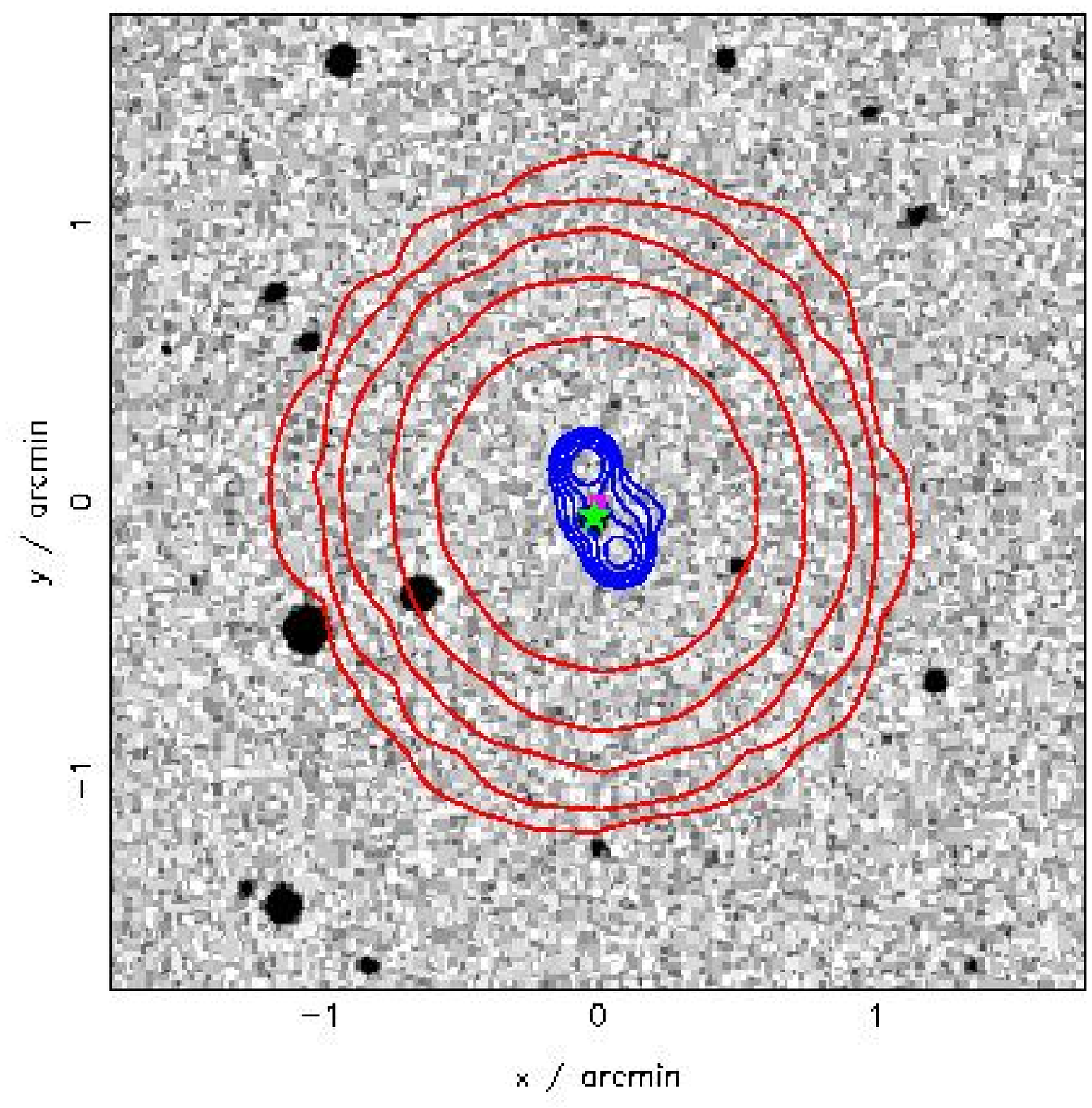}}
      \centerline{C1-192: 4C 24.31}
    \end{minipage}
    \vfill
    \begin{minipage}{3cm}     
      \mbox{}
      \centerline{\includegraphics[scale=0.26,angle=270]{Contours/C1/193.ps}}
      \centerline{C1-193: 3C 300.1}
    \end{minipage}
    \hspace{3cm}
    \begin{minipage}{3cm}
      \mbox{}
      \centerline{\includegraphics[scale=0.26,angle=270]{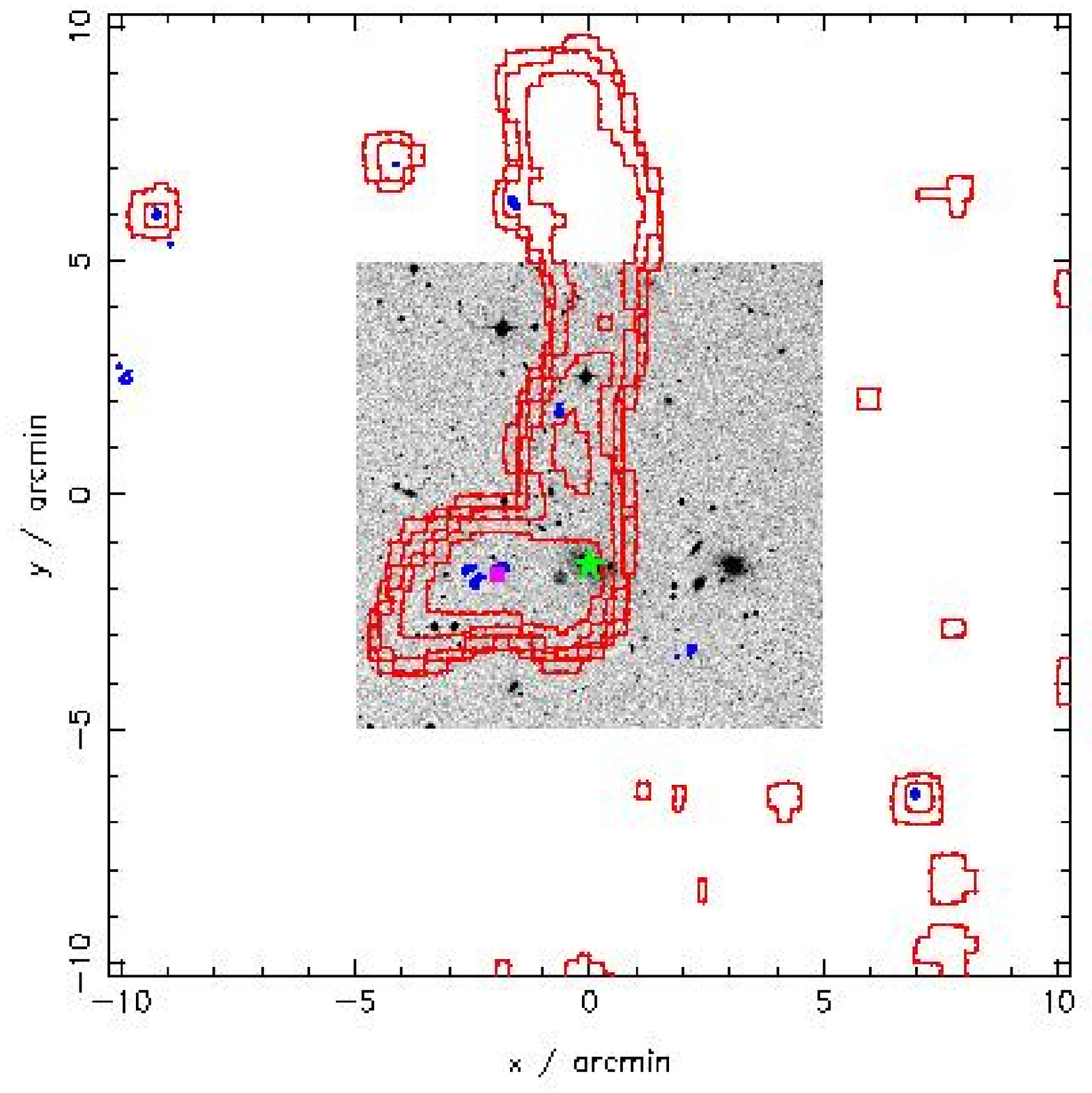}}
      \centerline{C1-194: 4C 07.36}
    \end{minipage}
    \hspace{3cm}
    \begin{minipage}{3cm}
      \mbox{}
      \centerline{\includegraphics[scale=0.26,angle=270]{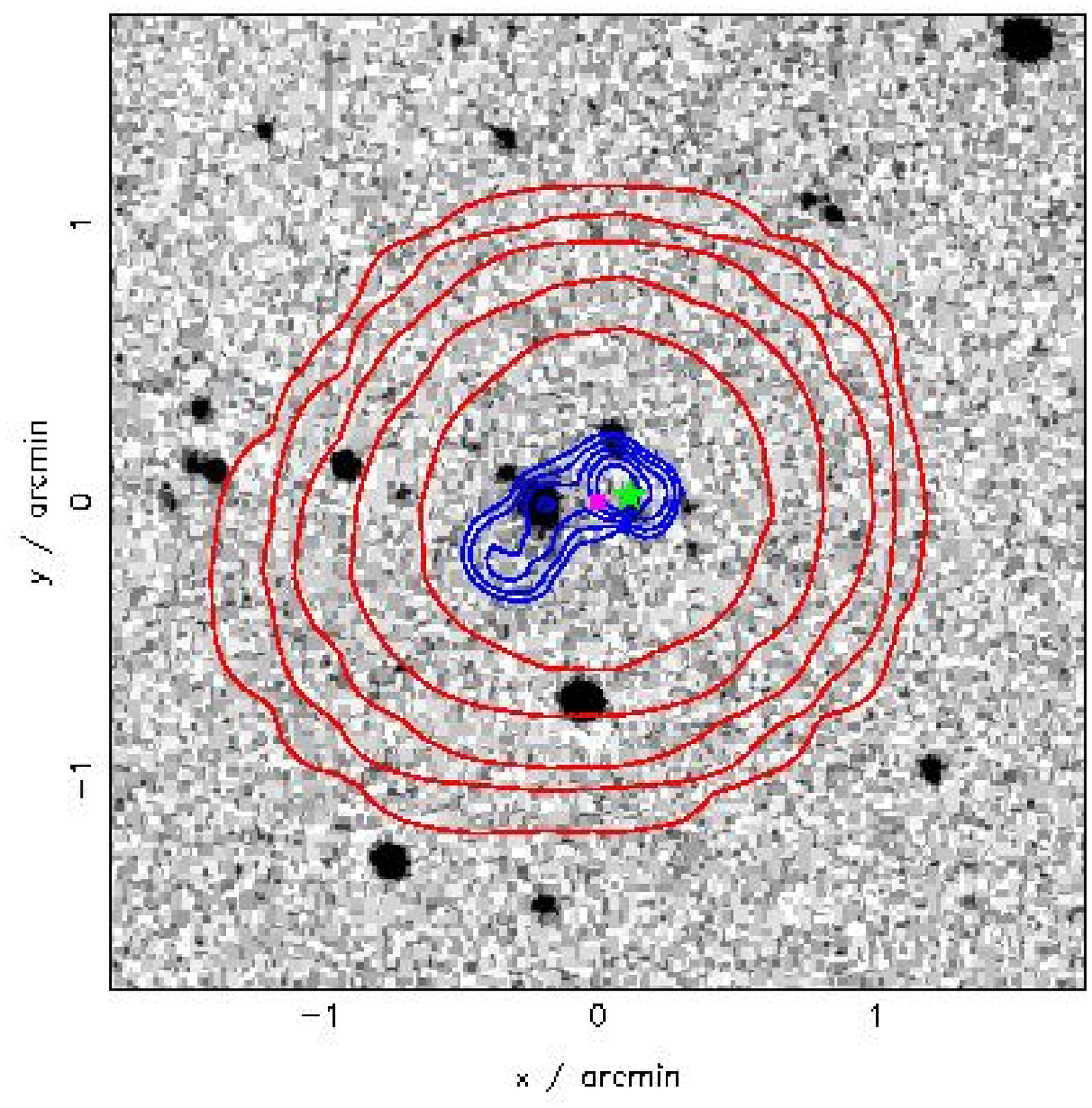}}
      \centerline{C1-197: 3C 303}
    \end{minipage}
    \vfill
    \begin{minipage}{3cm}     
      \mbox{}
      \centerline{\includegraphics[scale=0.26,angle=270]{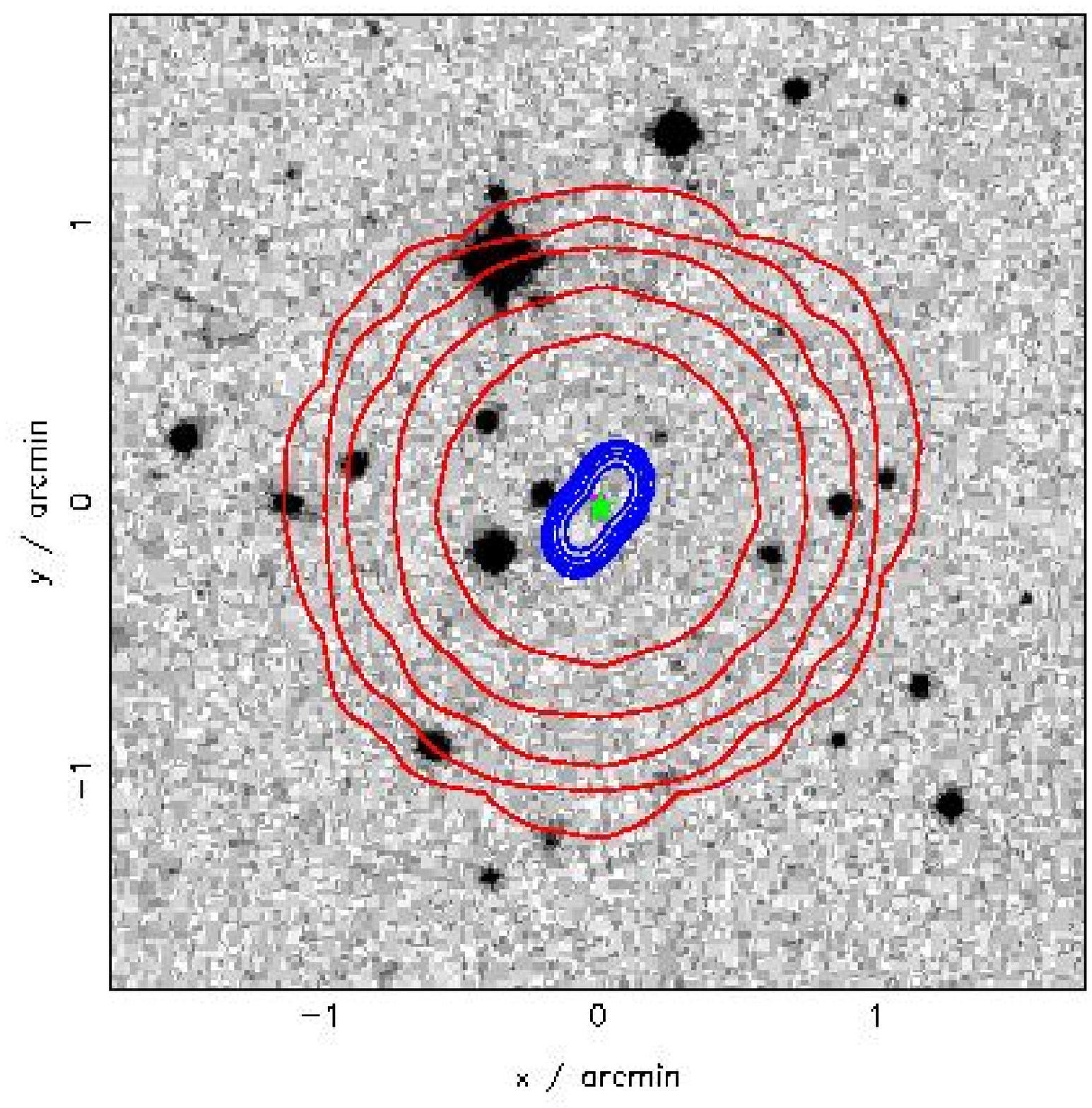}}
      \centerline{C1-199: 4C 00.52}
    \end{minipage}
    \hspace{3cm}
    \begin{minipage}{3cm}
      \mbox{}
      \centerline{\includegraphics[scale=0.26,angle=270]{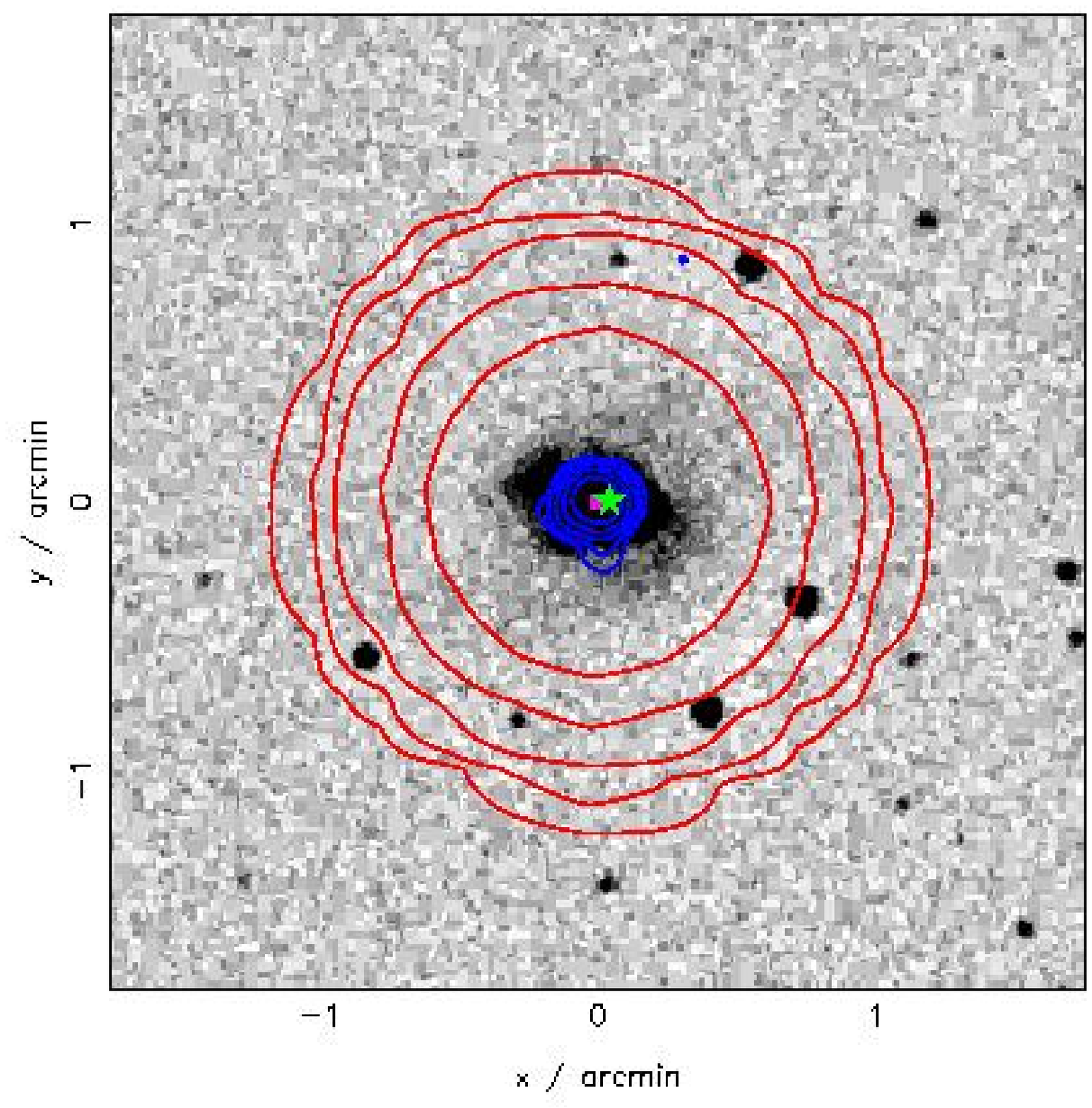}}
      \centerline{C1-200: 3C 305}
    \end{minipage}
    \hspace{3cm}
    \begin{minipage}{3cm}
      \mbox{}
      \centerline{\includegraphics[scale=0.26,angle=270]{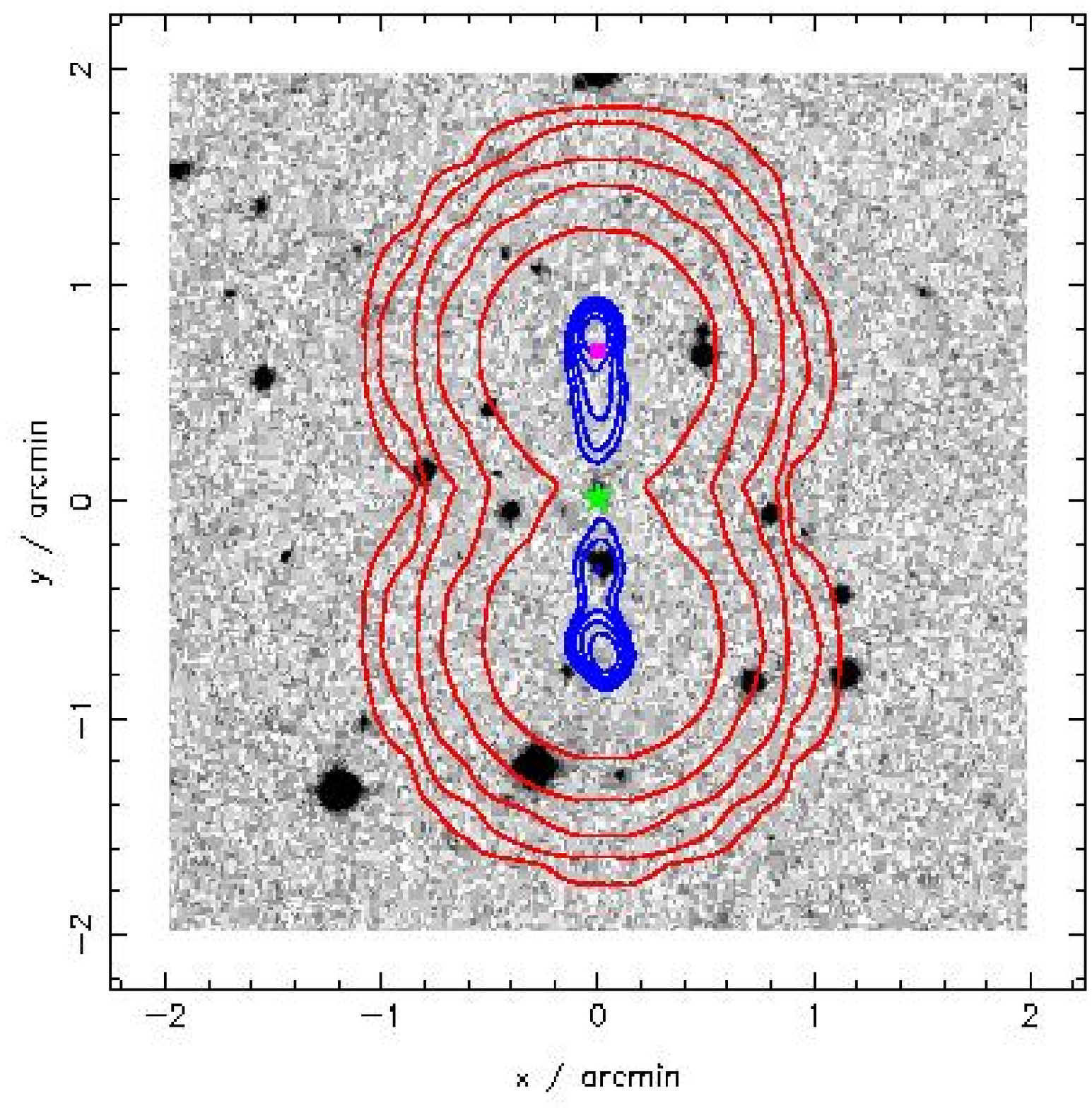}}
      \centerline{C1-201: 4C -04.53}
    \end{minipage}
  \end{center}
\end{figure}

\begin{figure}
  \begin{center}
    {\bf CoNFIG-1}\\  
  \begin{minipage}{3cm}      
      \mbox{}
      \centerline{\includegraphics[scale=0.26,angle=270]{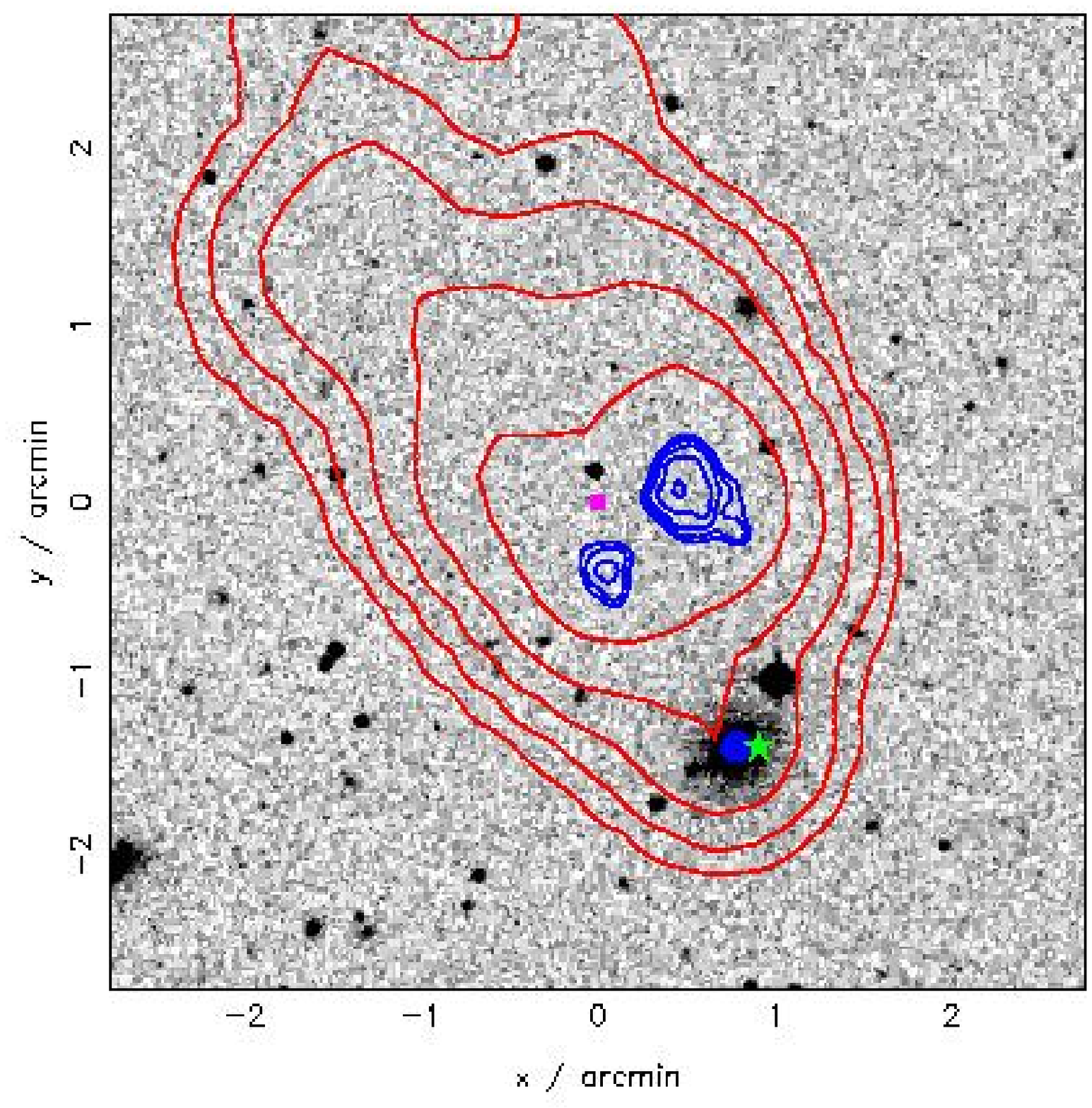}}
      \centerline{C1-203: B2 1502+28}
    \end{minipage}
    \hspace{3cm}
    \begin{minipage}{3cm}
      \mbox{}
      \centerline{\includegraphics[scale=0.26,angle=270]{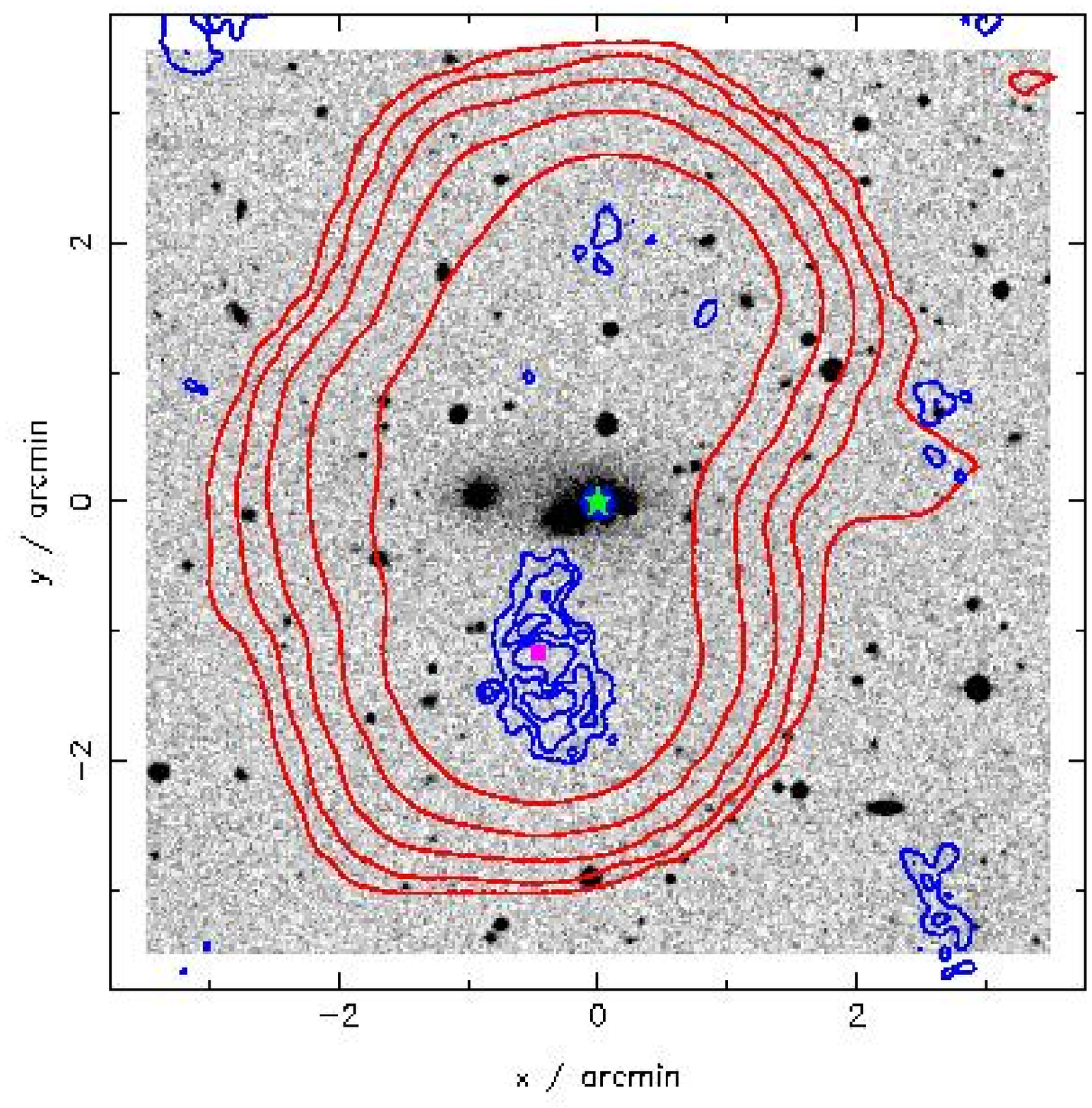}}
      \centerline{C1-205: 3C 310}
    \end{minipage}
    \hspace{3cm}
    \begin{minipage}{3cm}
      \mbox{}
      \centerline{\includegraphics[scale=0.26,angle=270]{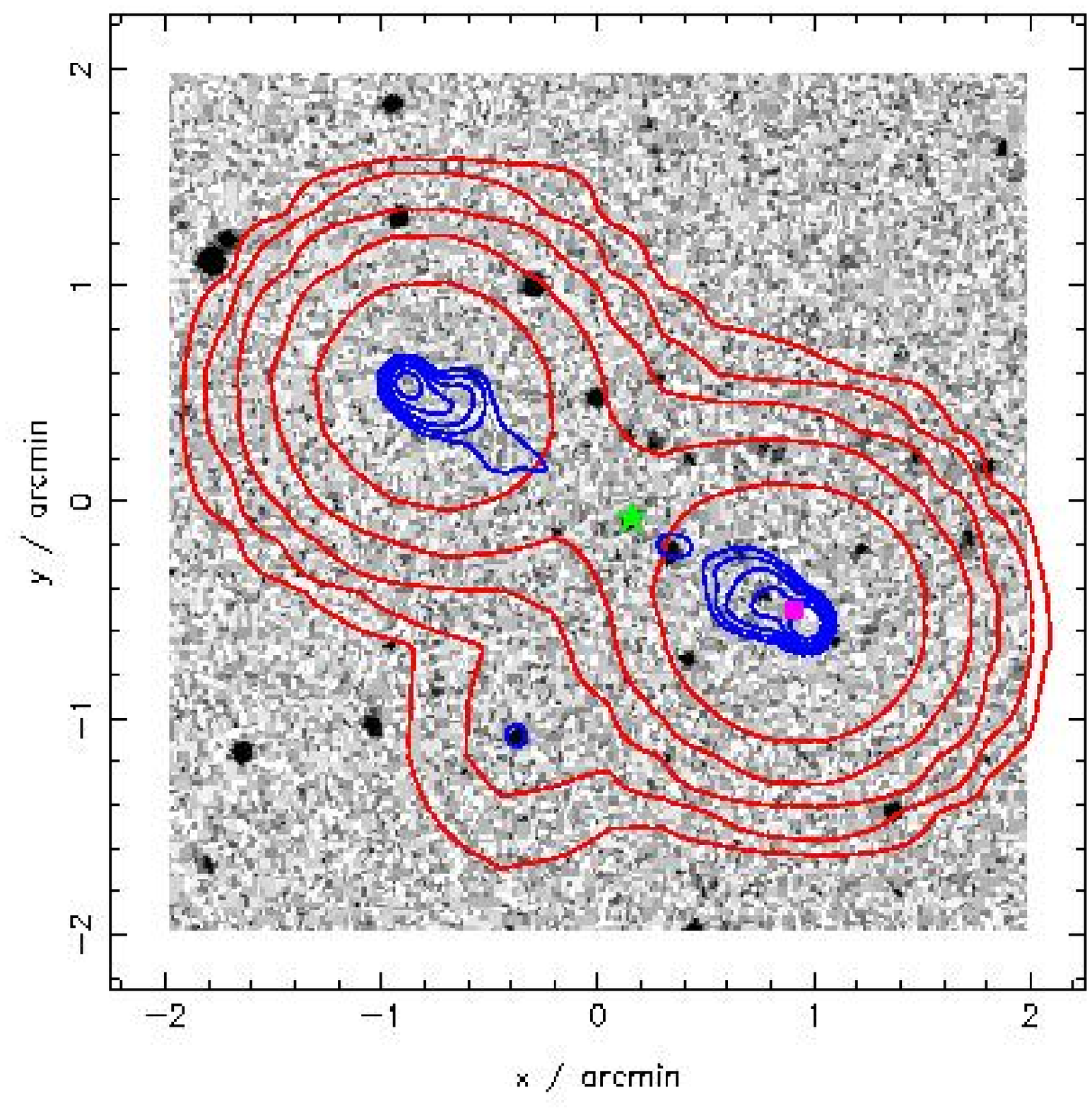}}
      \centerline{C1-207: 3C 313}
    \end{minipage}
    \vfill
    \begin{minipage}{3cm}      
      \mbox{}
      \centerline{\includegraphics[scale=0.26,angle=270]{Contours/C1/208.ps}}
      \centerline{C1-208: 4C 01.42}
    \end{minipage}
    \hspace{3cm}
    \begin{minipage}{3cm}
      \mbox{}
      \centerline{\includegraphics[scale=0.26,angle=270]{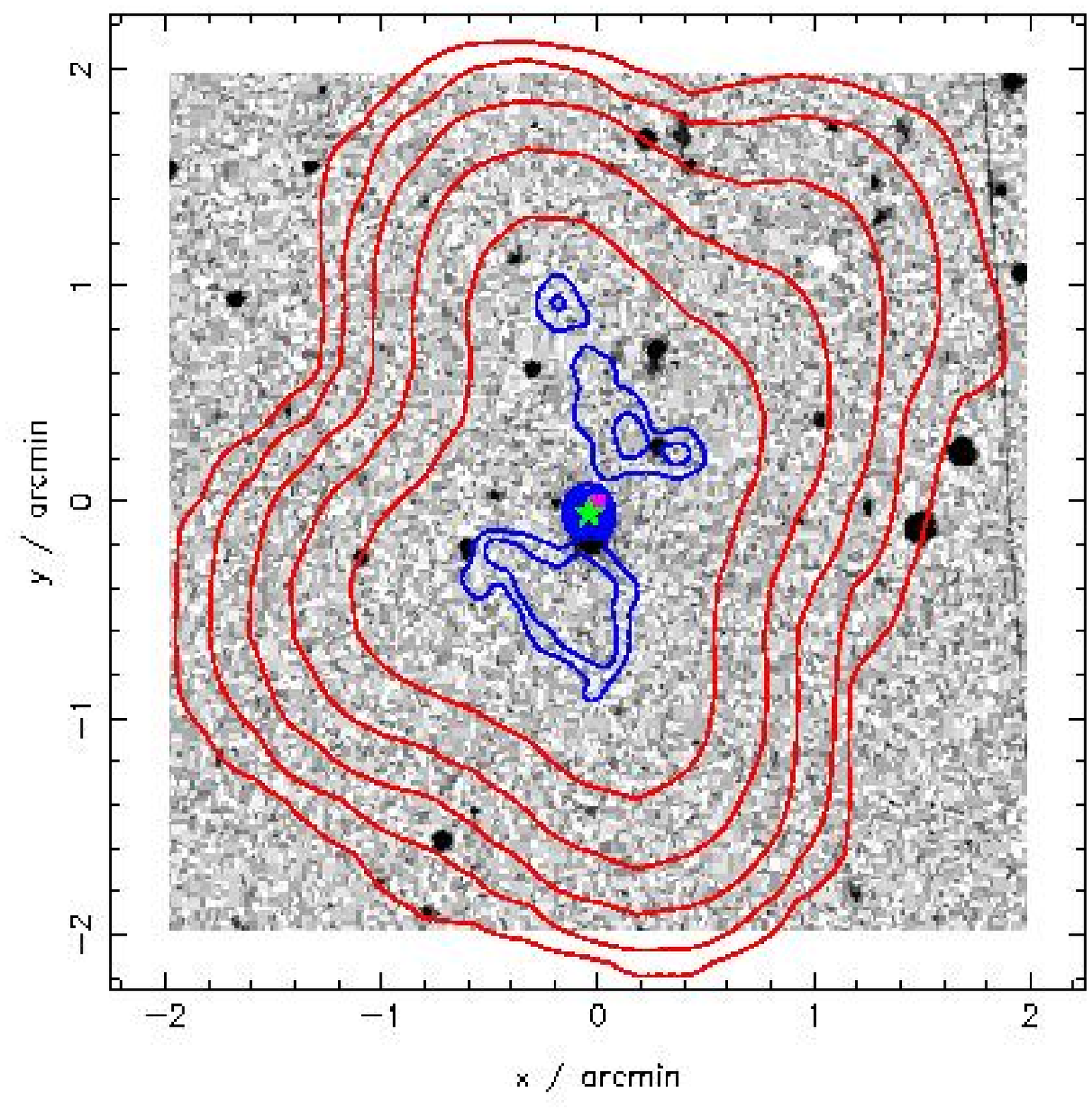}}
      \centerline{C1-209: 3C 315}
    \end{minipage}
    \hspace{3cm}
    \begin{minipage}{3cm}
      \mbox{}
      \centerline{\includegraphics[scale=0.26,angle=270]{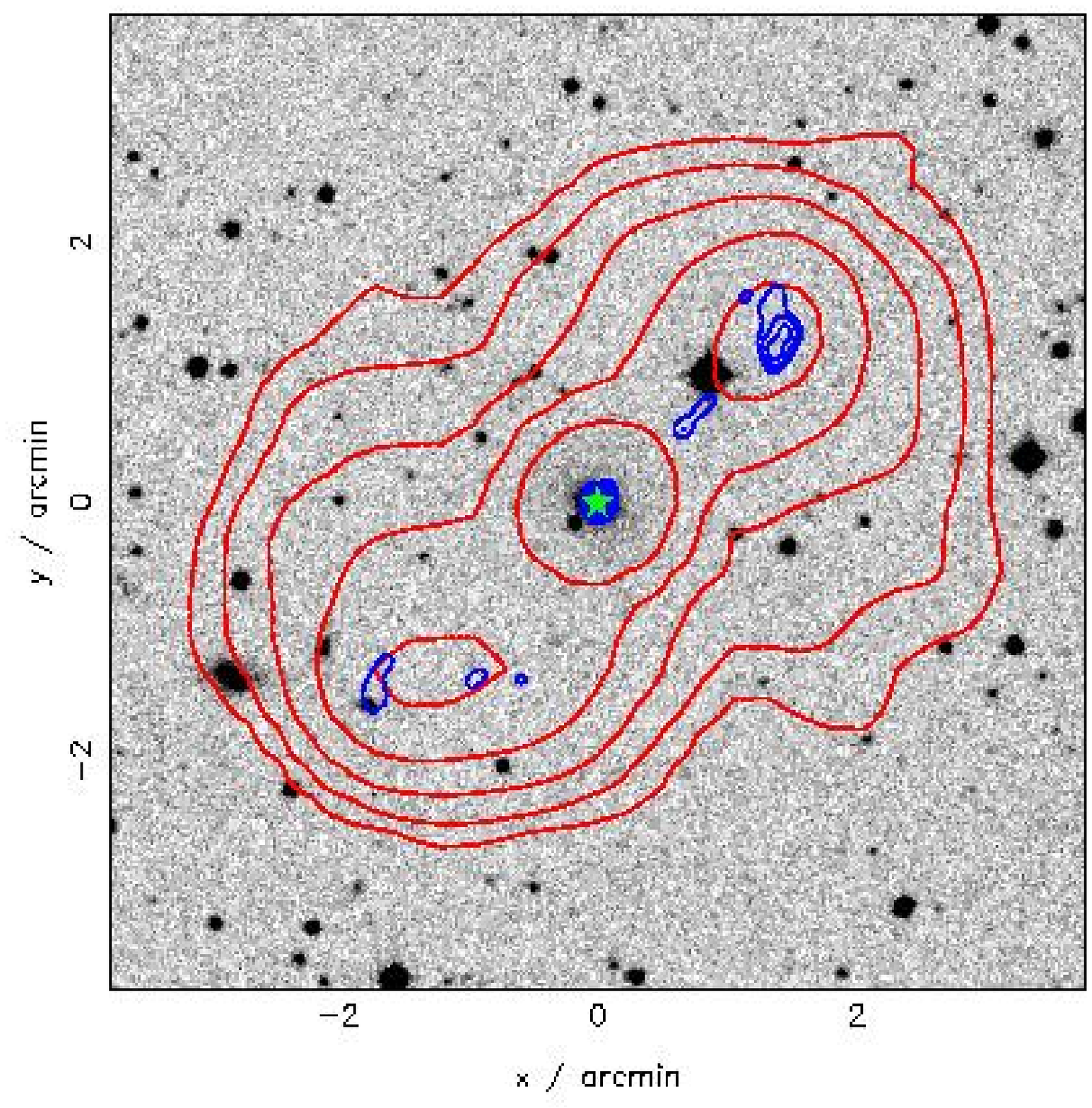}}
      \centerline{C1-211: 4C 00.56}
    \end{minipage}
    \vfill
    \begin{minipage}{3cm}     
      \mbox{}
      \centerline{\includegraphics[scale=0.26,angle=270]{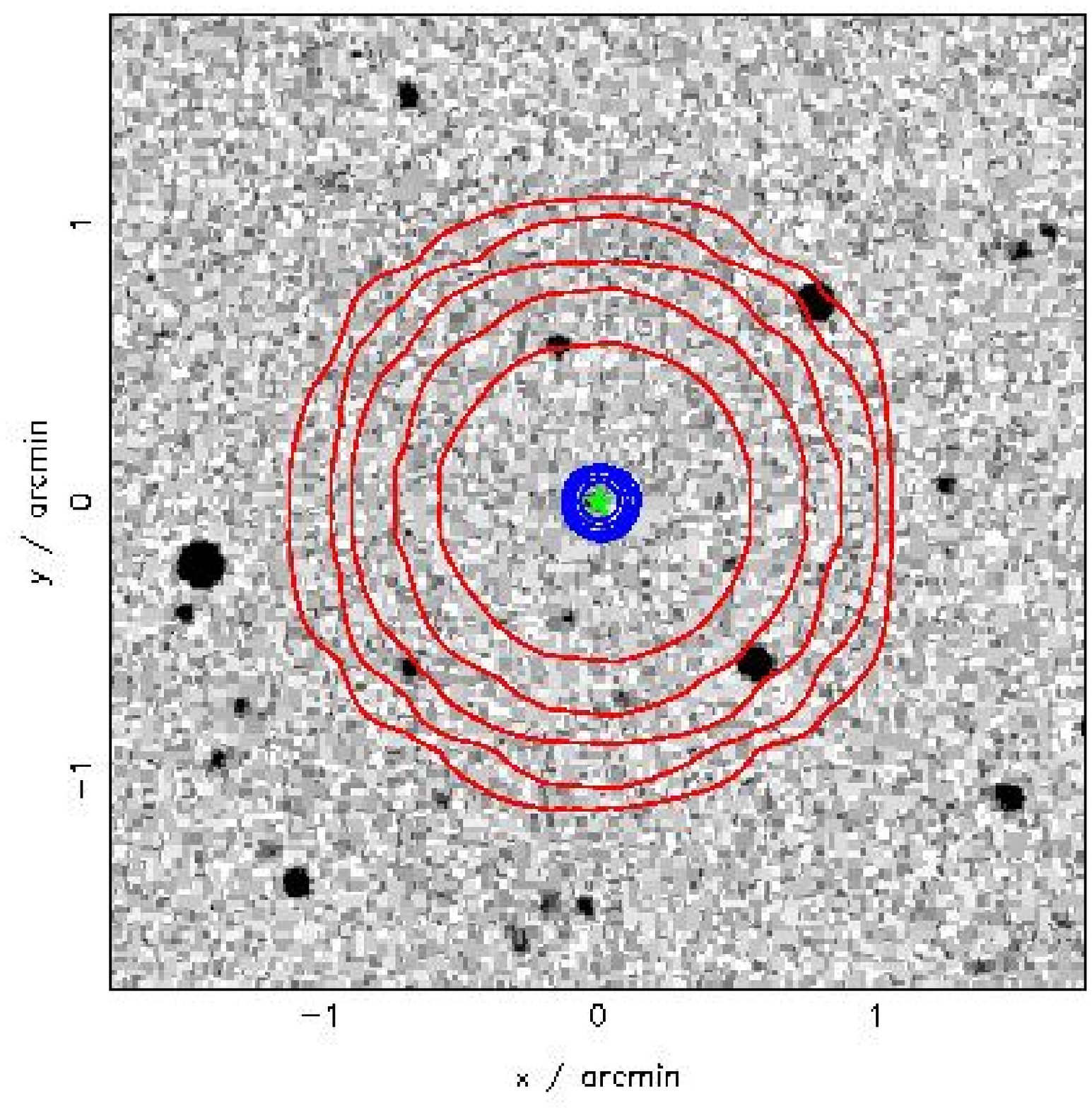}}
      \centerline{C1-213: 3C 316}
    \end{minipage}
    \hspace{3cm}
    \begin{minipage}{3cm}
      \mbox{}
      \centerline{\includegraphics[scale=0.26,angle=270]{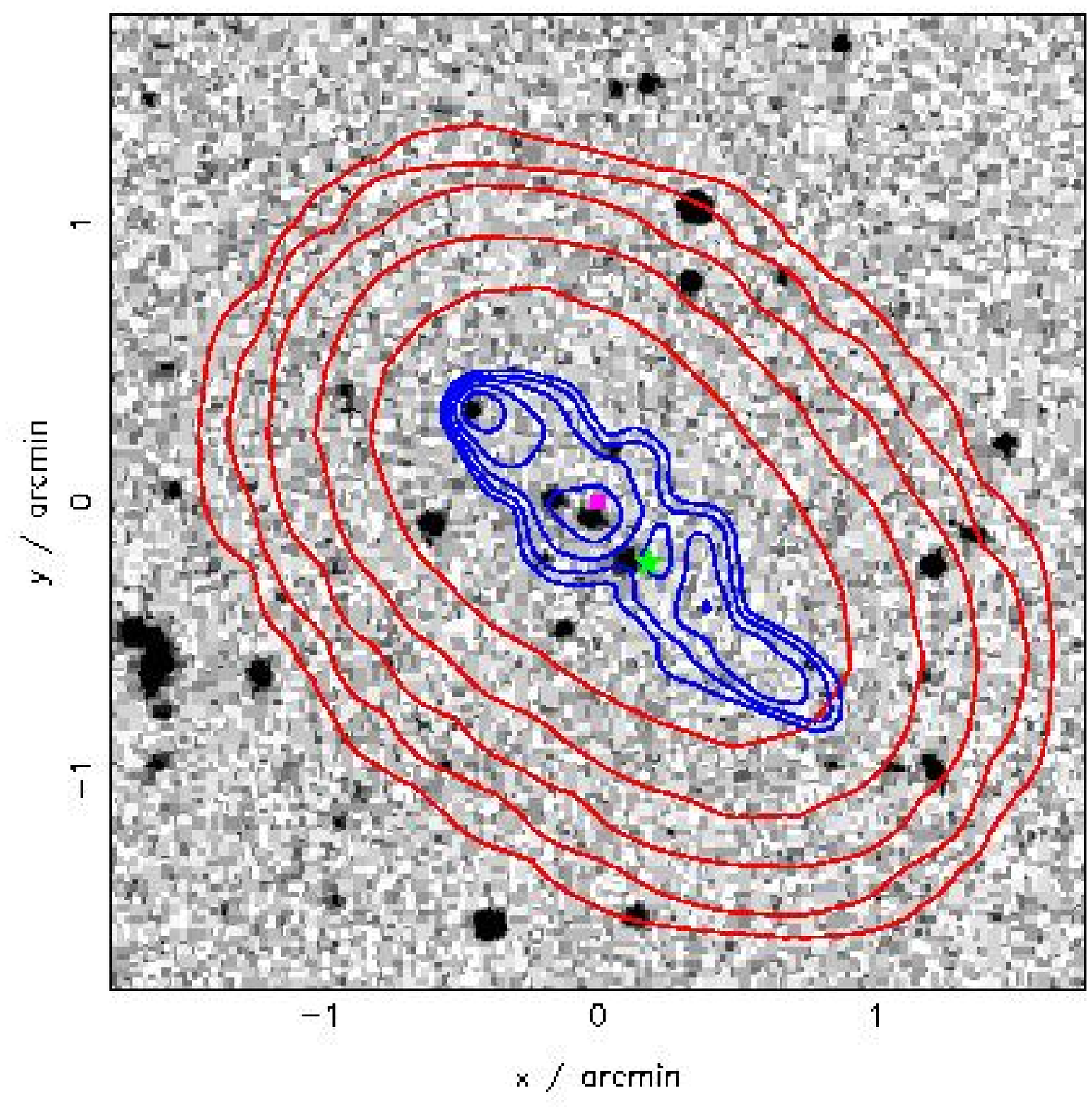}}
      \centerline{C1-216: 3C 319}
    \end{minipage}
    \hspace{3cm}
    \begin{minipage}{3cm}
      \mbox{}
      \centerline{\includegraphics[scale=0.26,angle=270]{Contours/C1/218.ps}}
      \centerline{C1-218: 3C 320}
    \end{minipage}
    \vfill
    \begin{minipage}{3cm}     
      \mbox{}
      \centerline{\includegraphics[scale=0.26,angle=270]{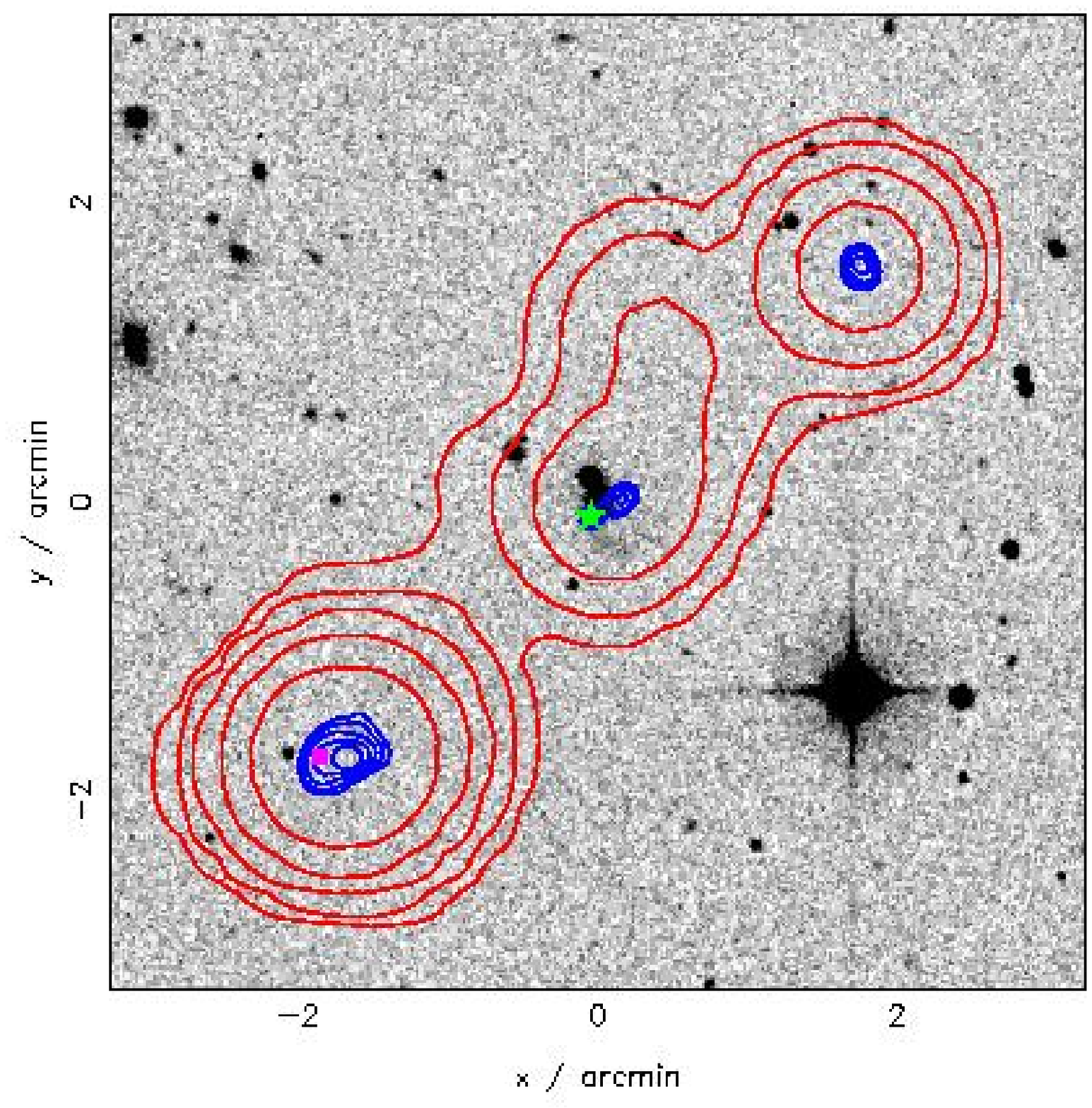}}
      \centerline{C1-219: 3C 321}
    \end{minipage}
    \hspace{3cm}
    \begin{minipage}{3cm}
      \mbox{}
      \centerline{\includegraphics[scale=0.26,angle=270]{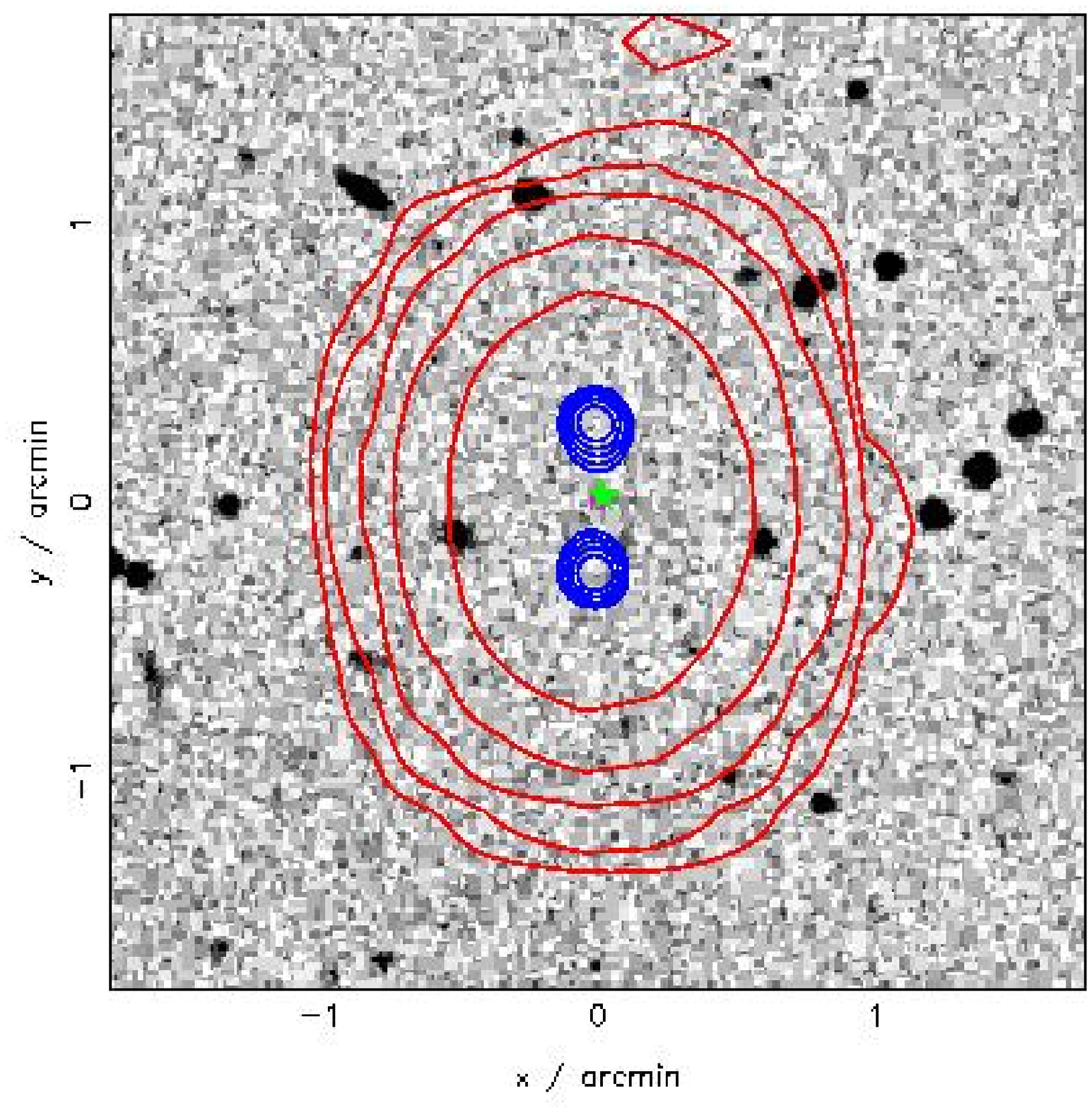}}
      \centerline{C1-221: 3C 322}
    \end{minipage}
    \hspace{3cm}
    \begin{minipage}{3cm}
      \mbox{}
      \centerline{\includegraphics[scale=0.26,angle=270]{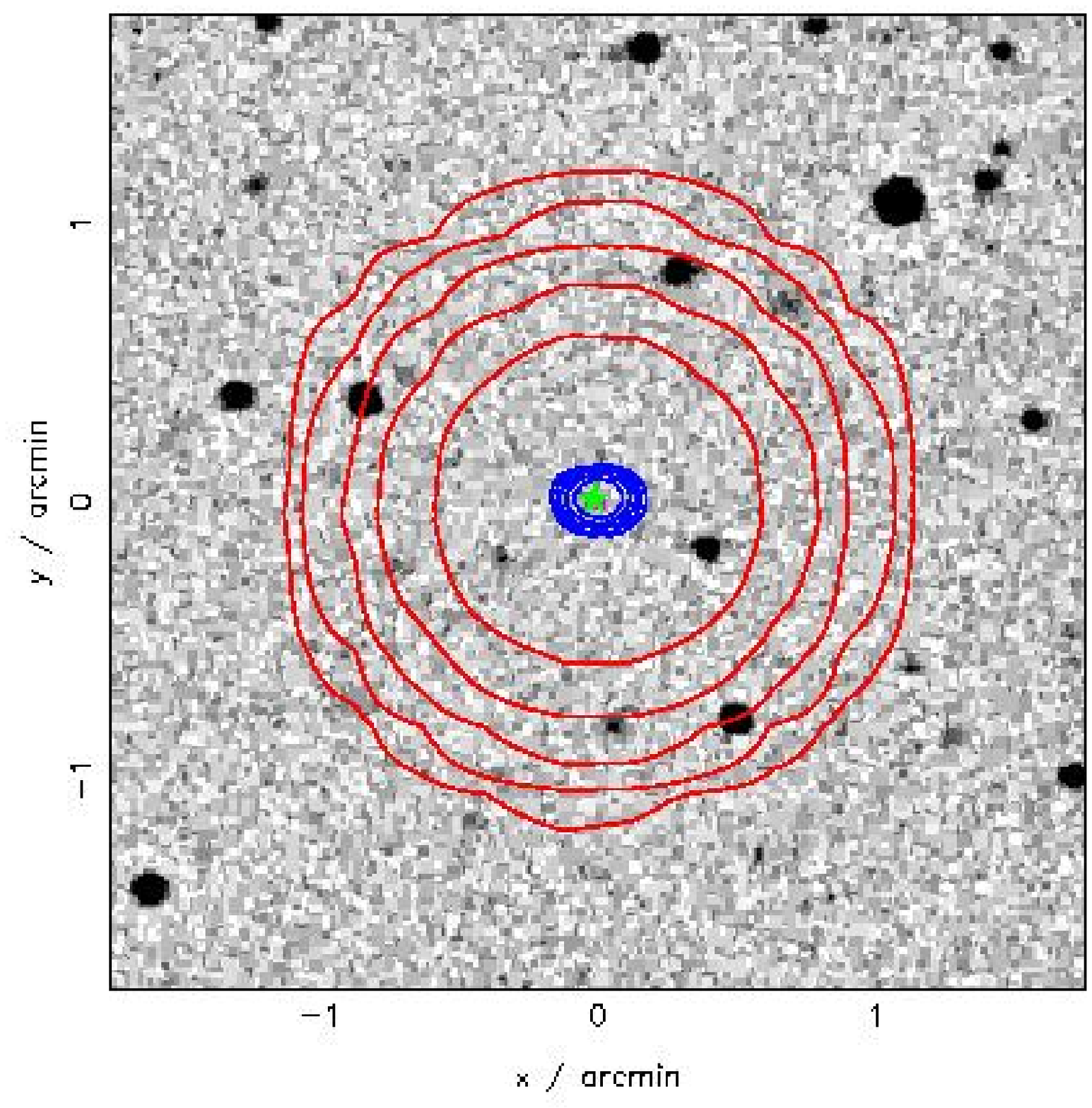}}
      \centerline{C1-222: 4C 13.56} 
    \end{minipage}
  \end{center}
\end{figure}

\begin{figure}
  \begin{center}
    {\bf CoNFIG-1}\\  
  \begin{minipage}{3cm}      
      \mbox{}
      \centerline{\includegraphics[scale=0.26,angle=270]{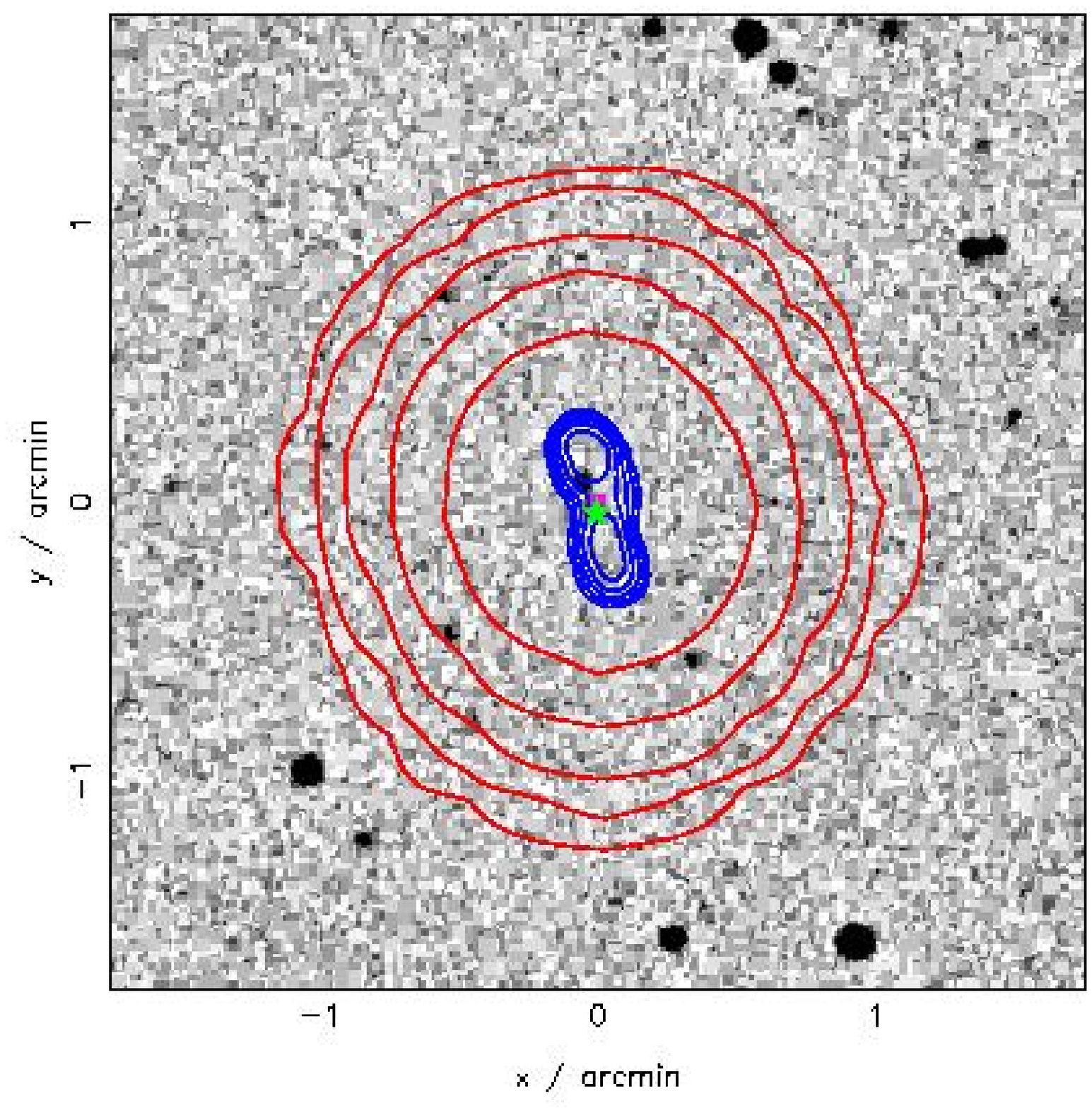}}
      \centerline{C1-224: 3C 323}
    \end{minipage}
    \hspace{3cm}
    \begin{minipage}{3cm}
      \mbox{}
      \centerline{\includegraphics[scale=0.26,angle=270]{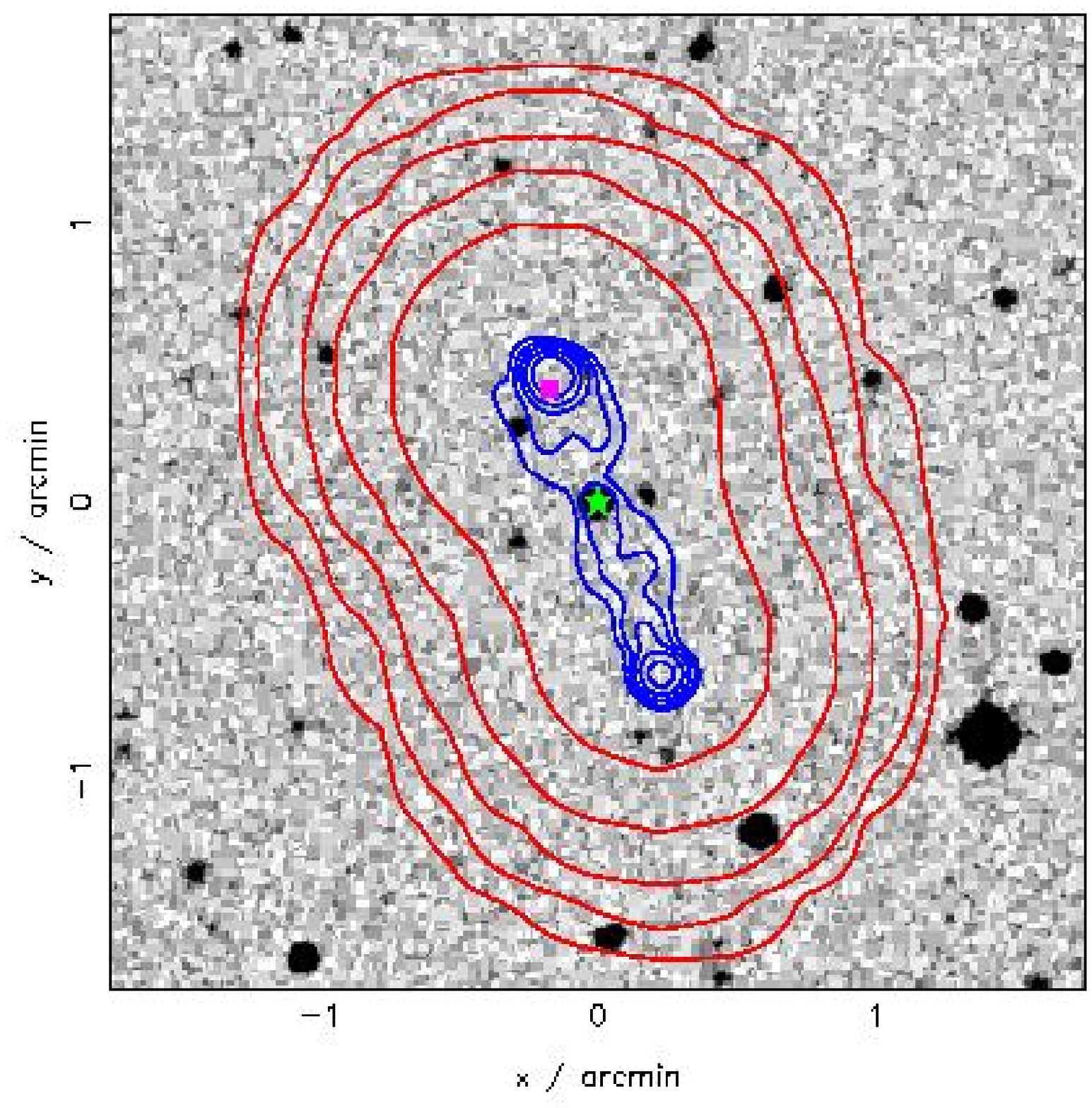}}
      \centerline{C1-226: 3C 323.1}
    \end{minipage}
    \hspace{3cm}
    \begin{minipage}{3cm}
      \mbox{}
      \centerline{\includegraphics[scale=0.26,angle=270]{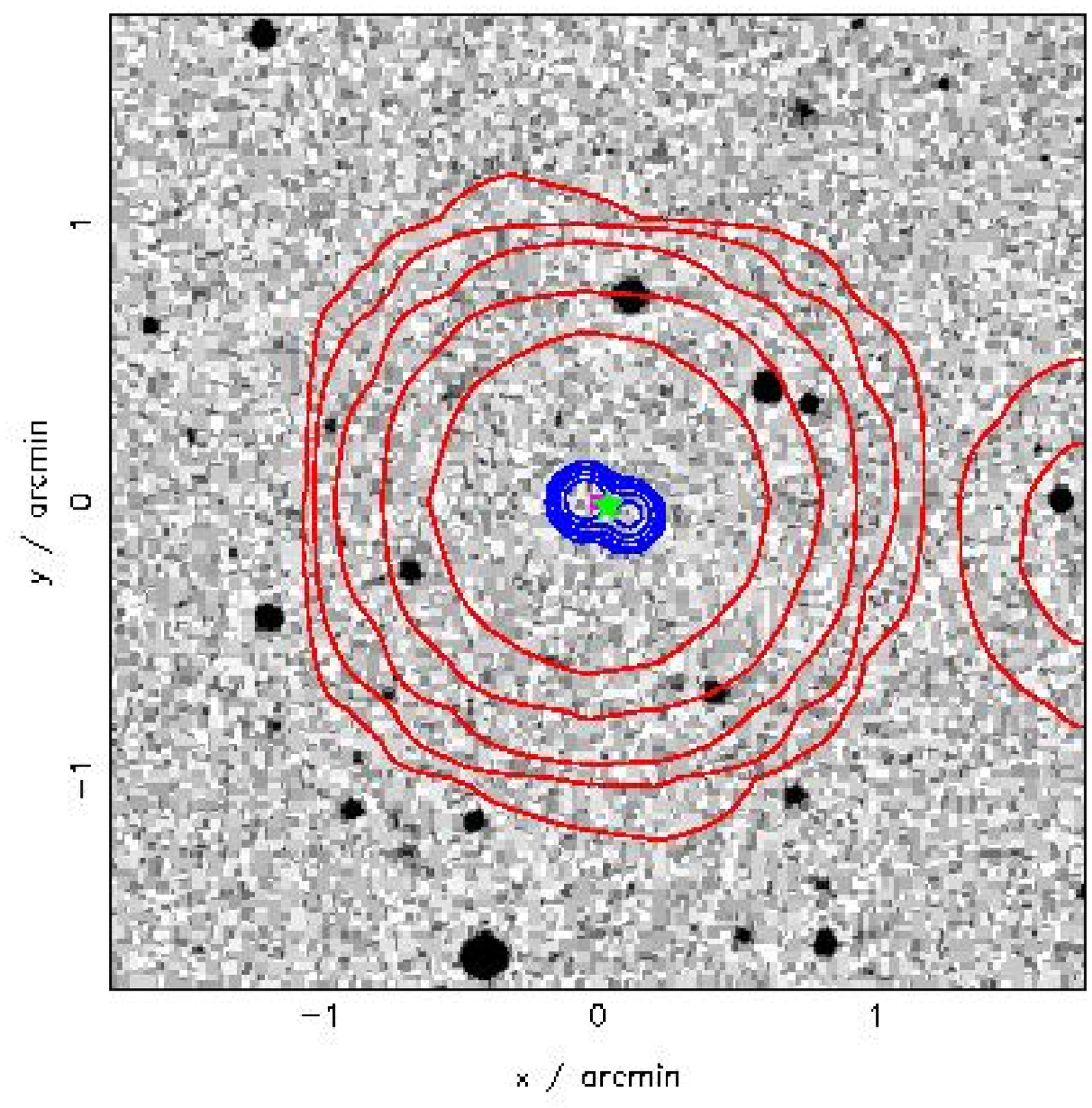}}
      \centerline{C1-227: 3C 324}
    \end{minipage}
    \vfill
    \begin{minipage}{3cm}      
      \mbox{}
      \centerline{\includegraphics[scale=0.26,angle=270]{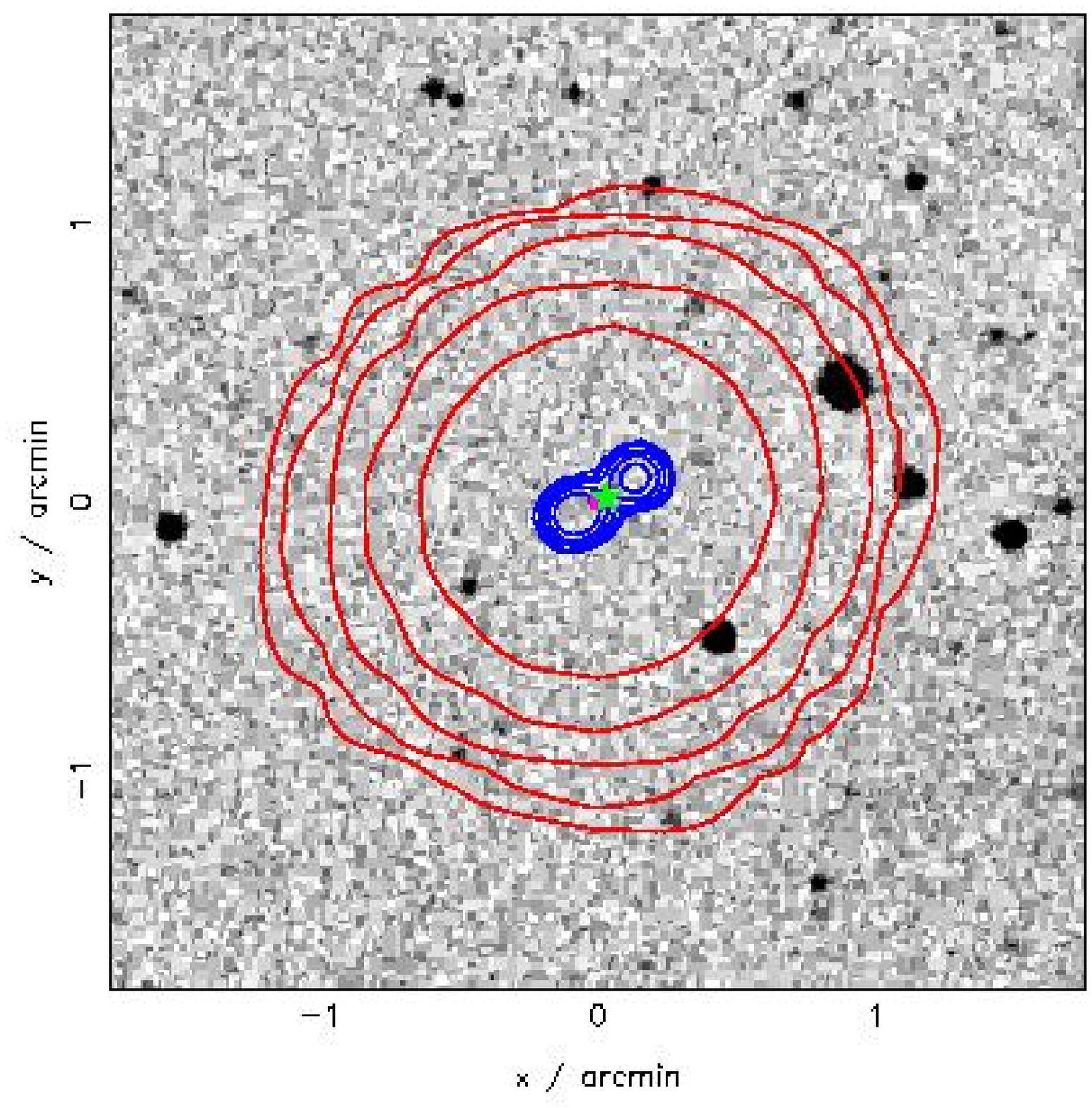}}
      \centerline{C1-228: 3C 325}
    \end{minipage}
    \hspace{3cm}
    \begin{minipage}{3cm}
      \mbox{}
      \centerline{\includegraphics[scale=0.26,angle=270]{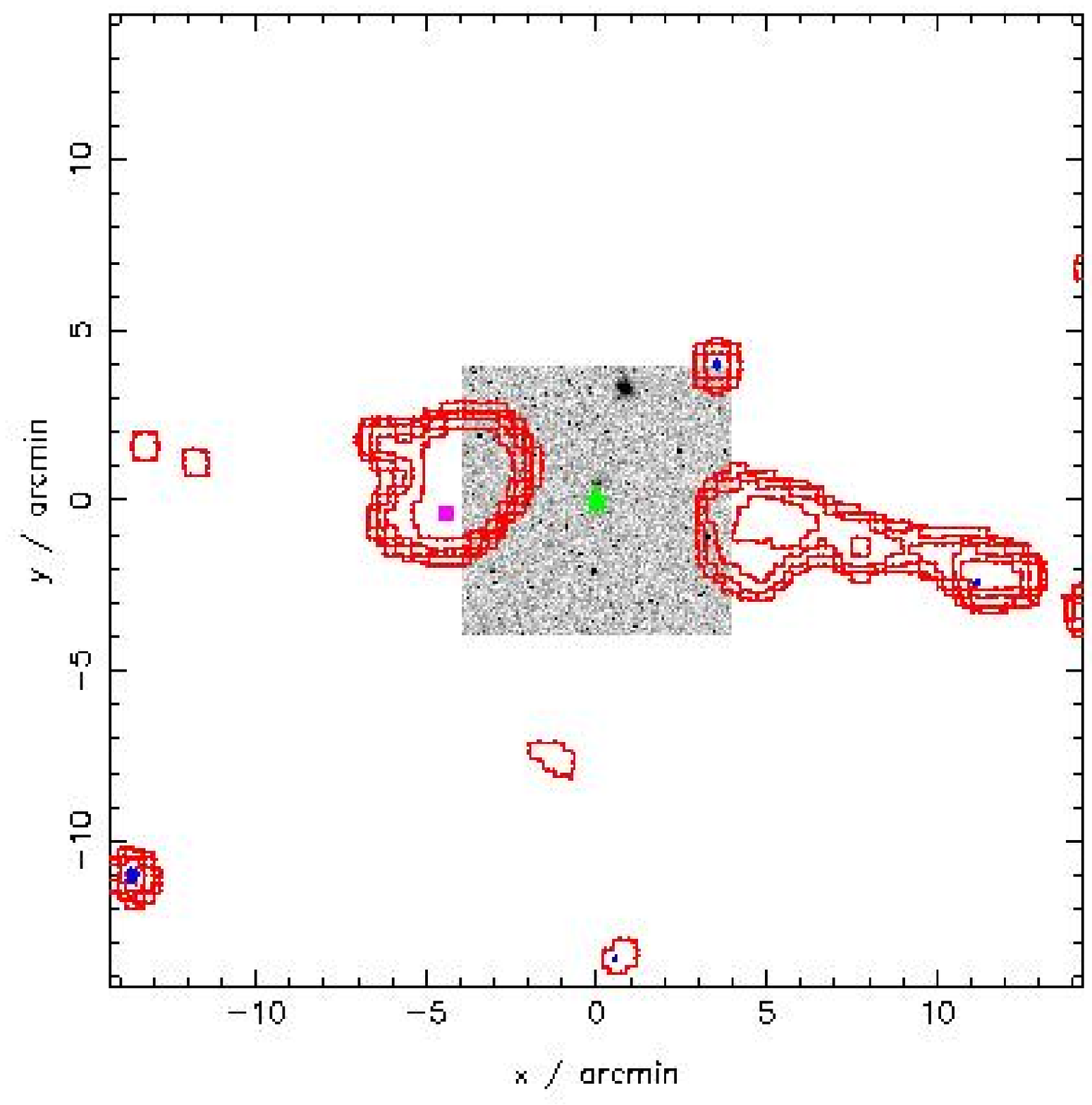}}
      \centerline{C1-230: 3C 326}
    \end{minipage}
    \hspace{3cm}
    \begin{minipage}{3cm}
      \mbox{}
      \centerline{\includegraphics[scale=0.26,angle=270]{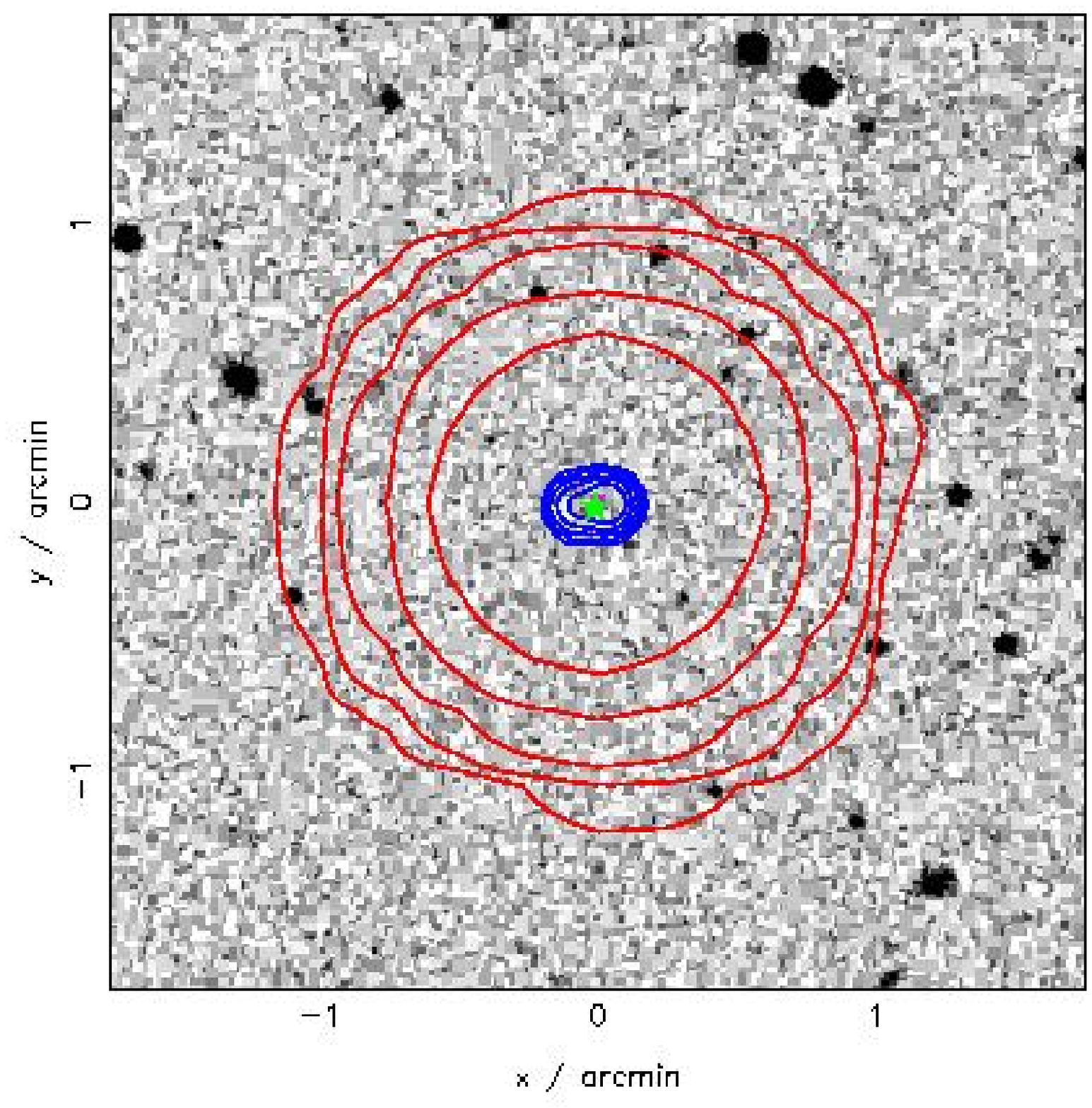}}
      \centerline{C1-231: 3C 326.1}
    \end{minipage}
    \vfill
    \begin{minipage}{3cm}     
      \mbox{}
      \centerline{\includegraphics[scale=0.26,angle=270]{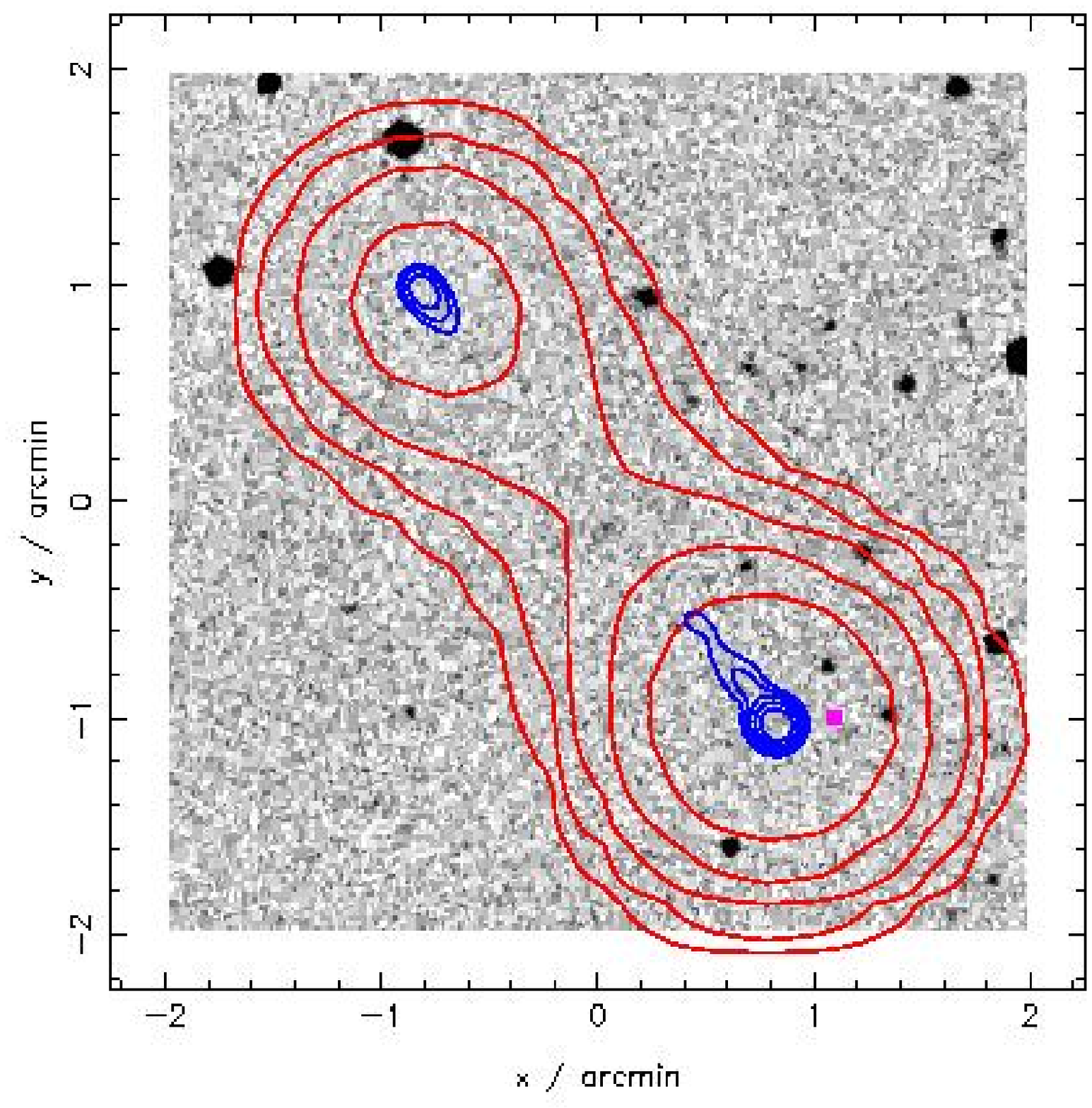}}
      \centerline{C1-232: 4C 43.35}
    \end{minipage}
    \hspace{3cm}
    \begin{minipage}{3cm}
      \mbox{}
      \centerline{\includegraphics[scale=0.26,angle=270]{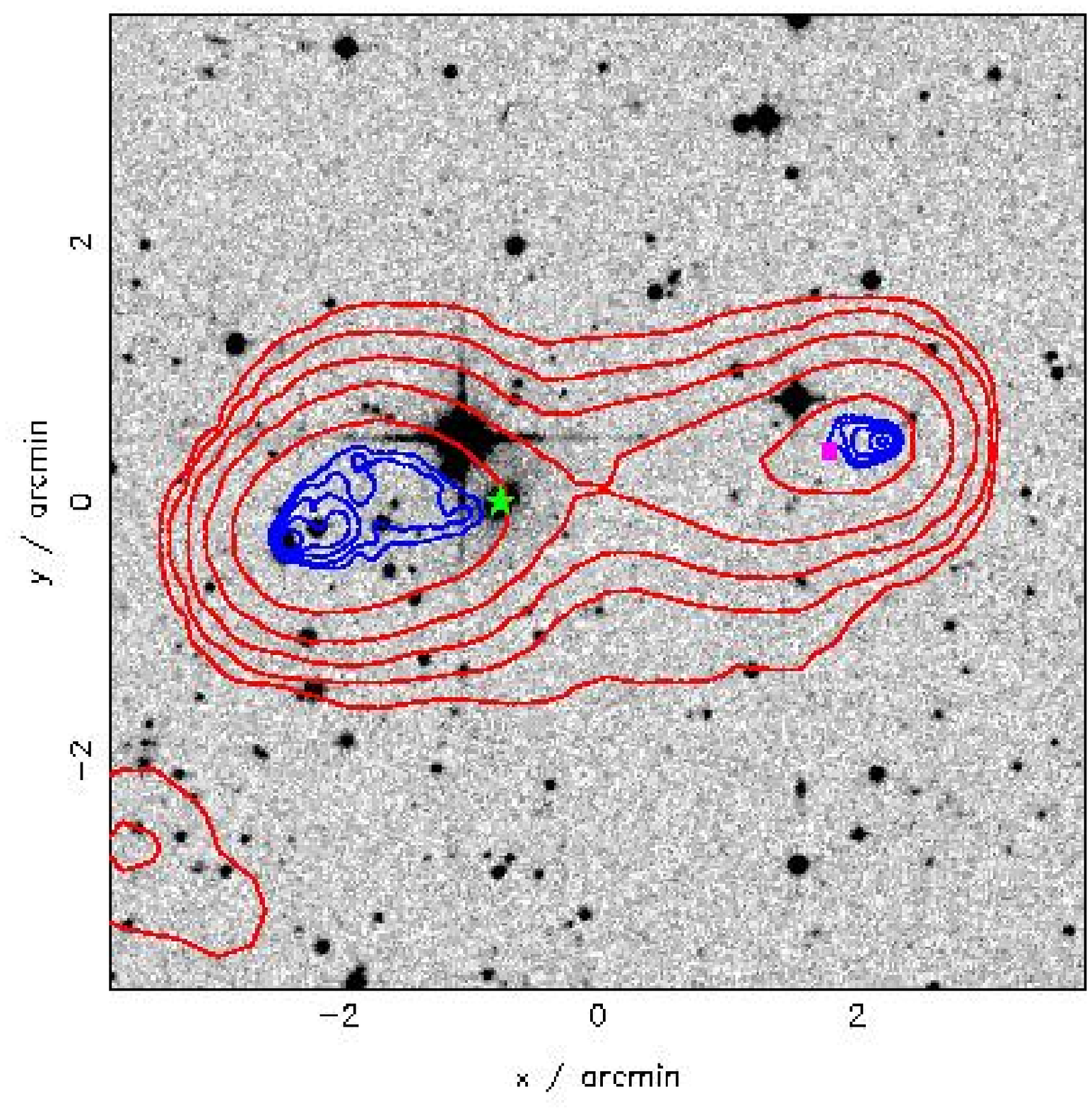}}
      \centerline{C1-234: 3C 327}
    \end{minipage}
    \hspace{3cm}
    \begin{minipage}{3cm}
      \mbox{}
      \centerline{\includegraphics[scale=0.26,angle=270]{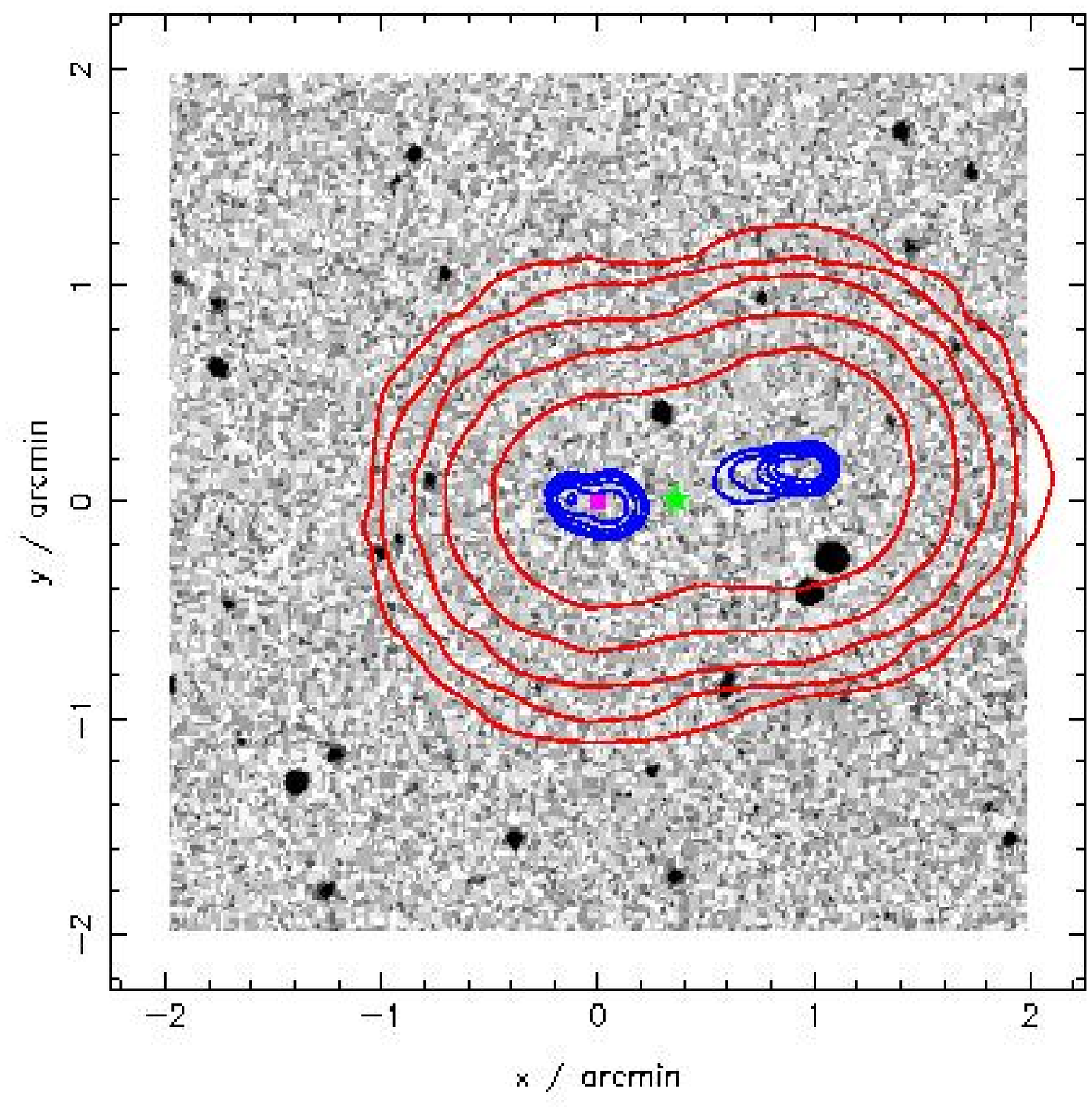}}
      \centerline{C1-237: 3C 329}
    \end{minipage}
    \vfill
    \begin{minipage}{3cm}     
      \mbox{}
      \centerline{\includegraphics[scale=0.26,angle=270]{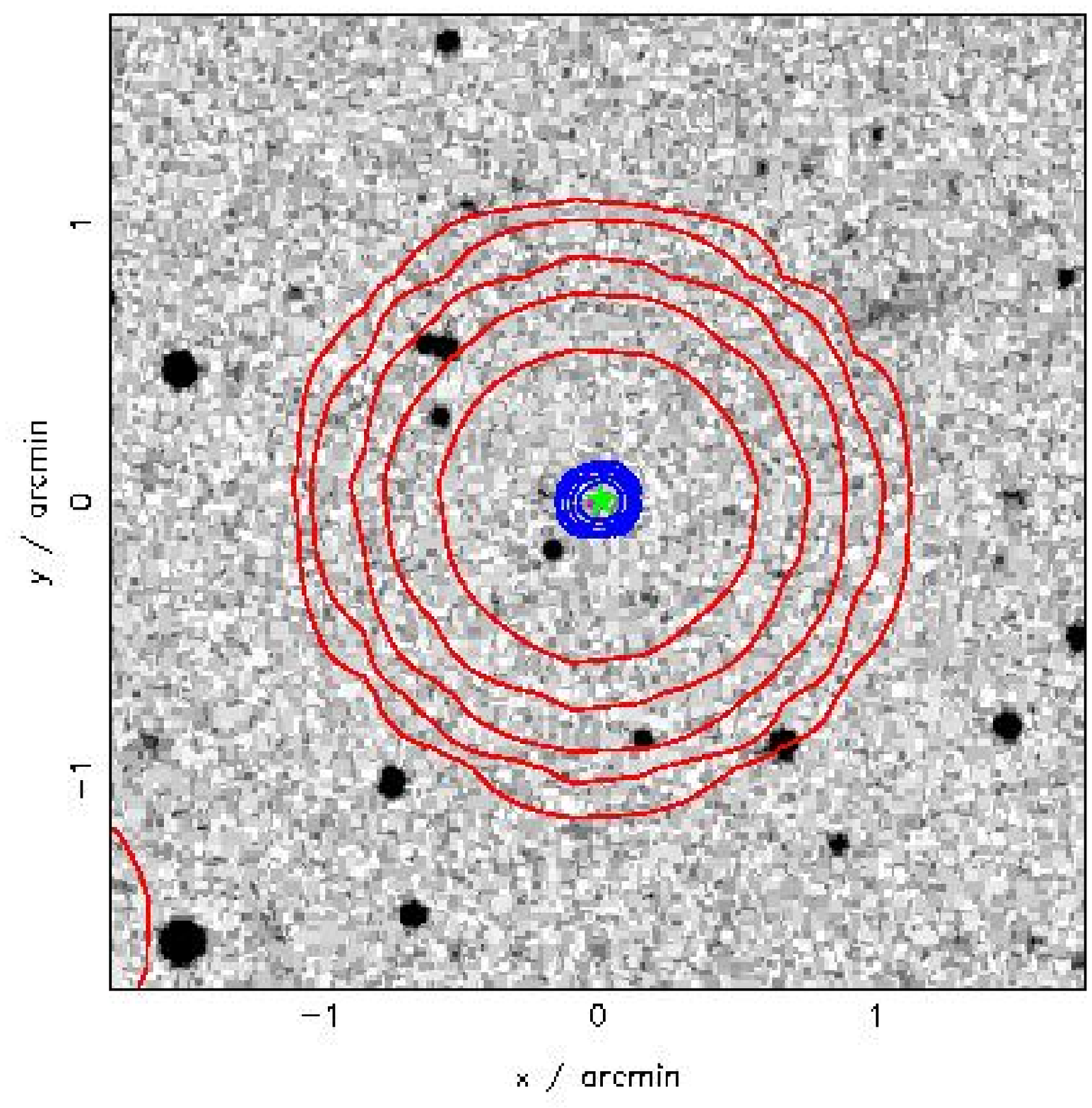}}
      \centerline{C1-238: 3C 331}
    \end{minipage}
    \hspace{3cm}
    \begin{minipage}{3cm}
      \mbox{}
      \centerline{\includegraphics[scale=0.26,angle=270]{Contours/C1/241.ps}}
      \centerline{C1-241: 3C 333}
    \end{minipage}
    \hspace{3cm}
    \begin{minipage}{3cm}
      \mbox{}
      \centerline{\includegraphics[scale=0.26,angle=270]{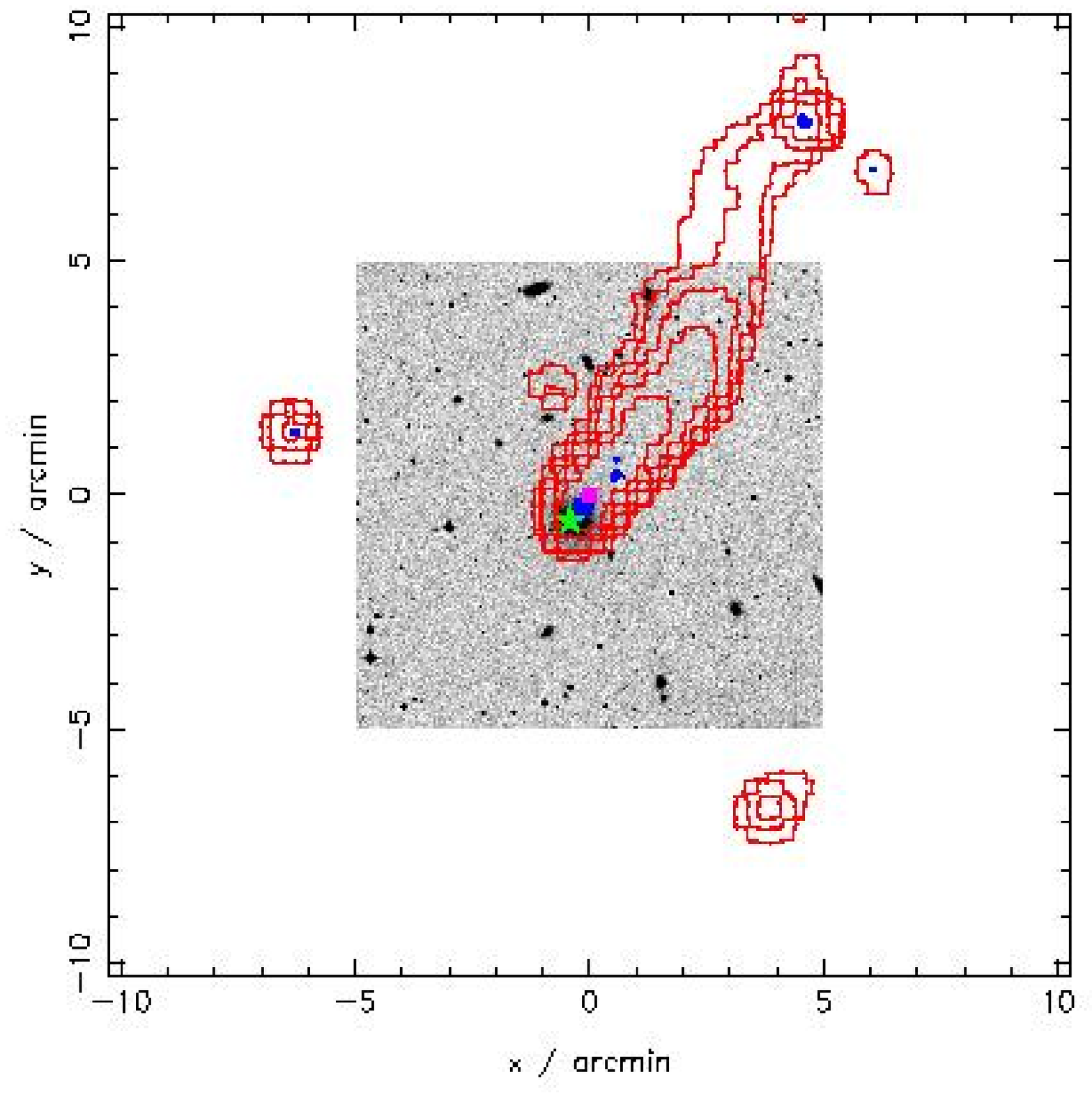}}
      \centerline{C1-242: NGC 6109}
    \end{minipage}
  \end{center}
\end{figure}

\begin{figure}
  \begin{center}
    {\bf CoNFIG-1}\\  
  \begin{minipage}{3cm}      
      \mbox{}
      \centerline{\includegraphics[scale=0.26,angle=270]{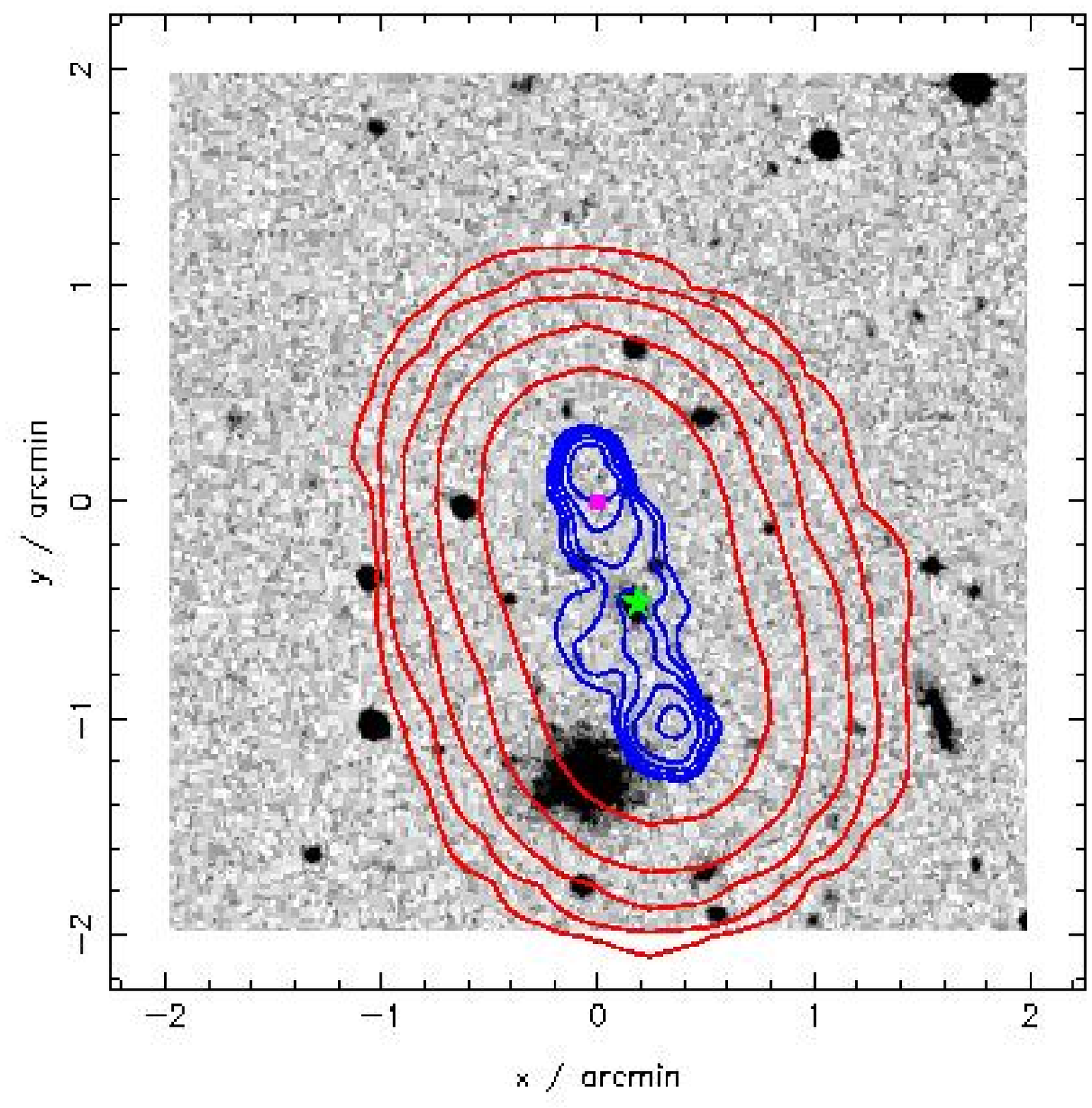}}
      \centerline{C1-243: 3C 332}
    \end{minipage}
    \hspace{3cm}
    \begin{minipage}{3cm}
      \mbox{}
      \centerline{\includegraphics[scale=0.26,angle=270]{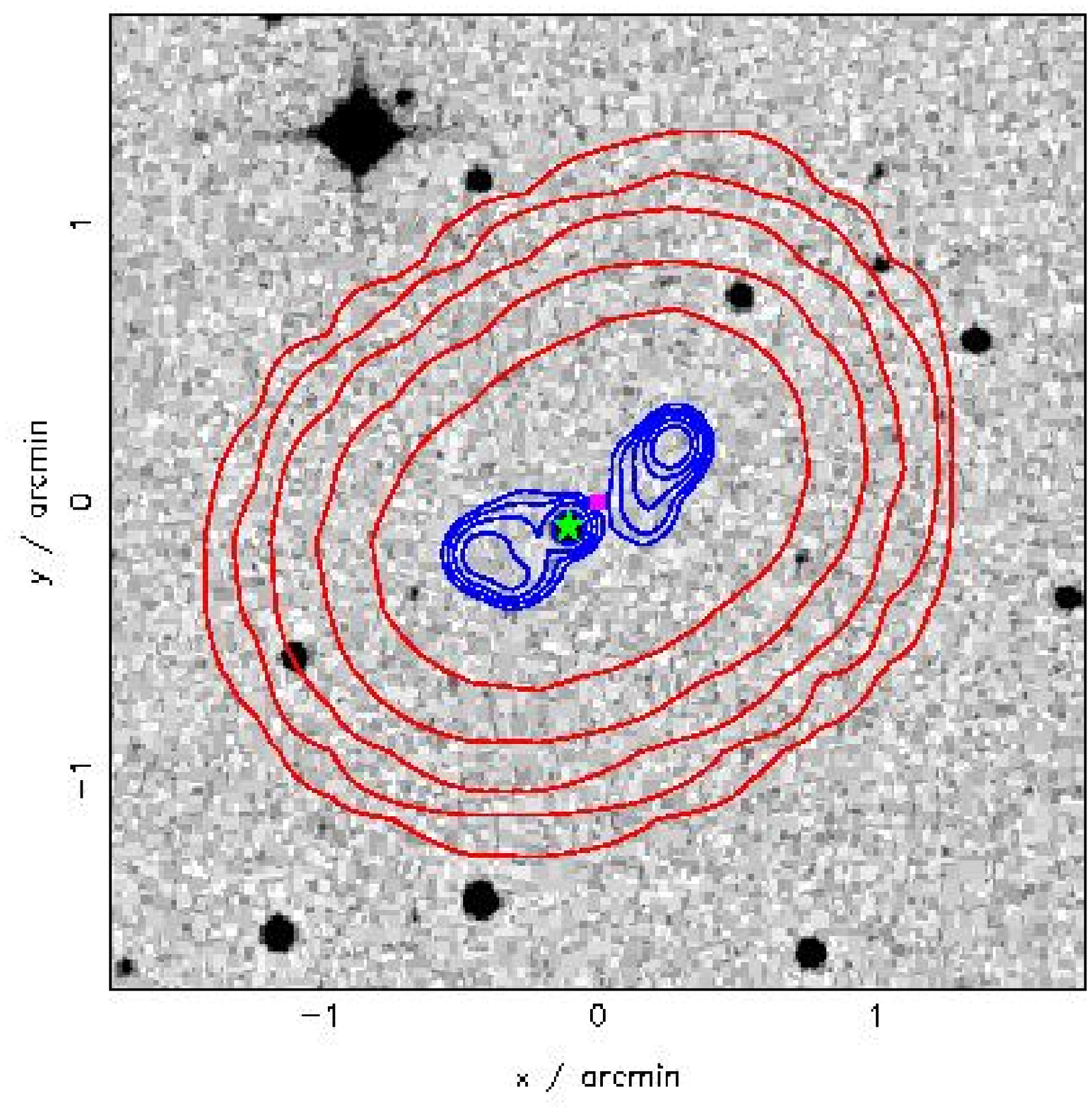}}
      \centerline{C1-244: 3C 334}
    \end{minipage}
    \hspace{3cm}
    \begin{minipage}{3cm}
      \mbox{}
      \centerline{\includegraphics[scale=0.26,angle=270]{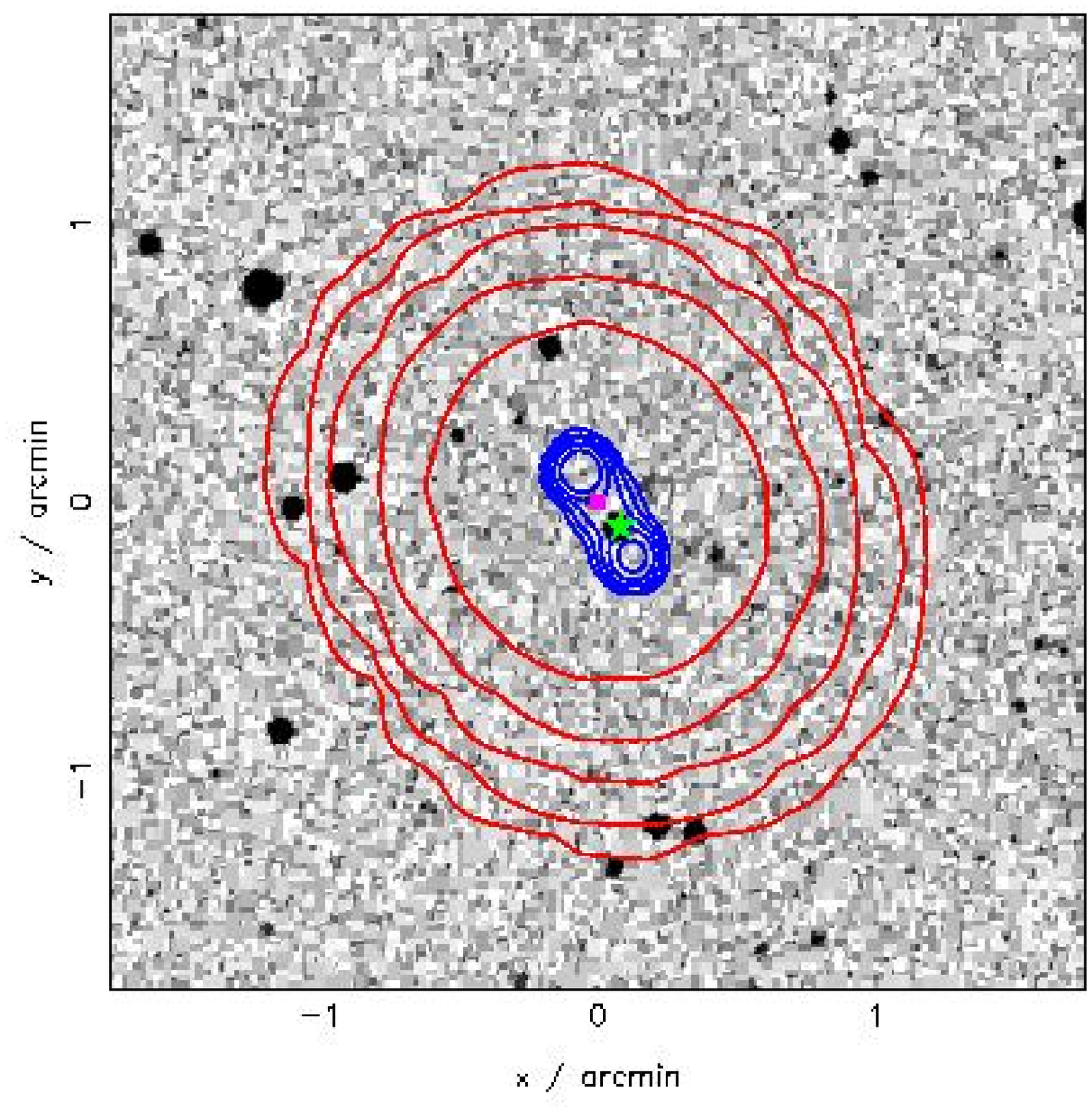}}
      \centerline{C1-245:3C 336}
    \end{minipage}
    \vfill
    \begin{minipage}{3cm}      
      \mbox{}
      \centerline{\includegraphics[scale=0.26,angle=270]{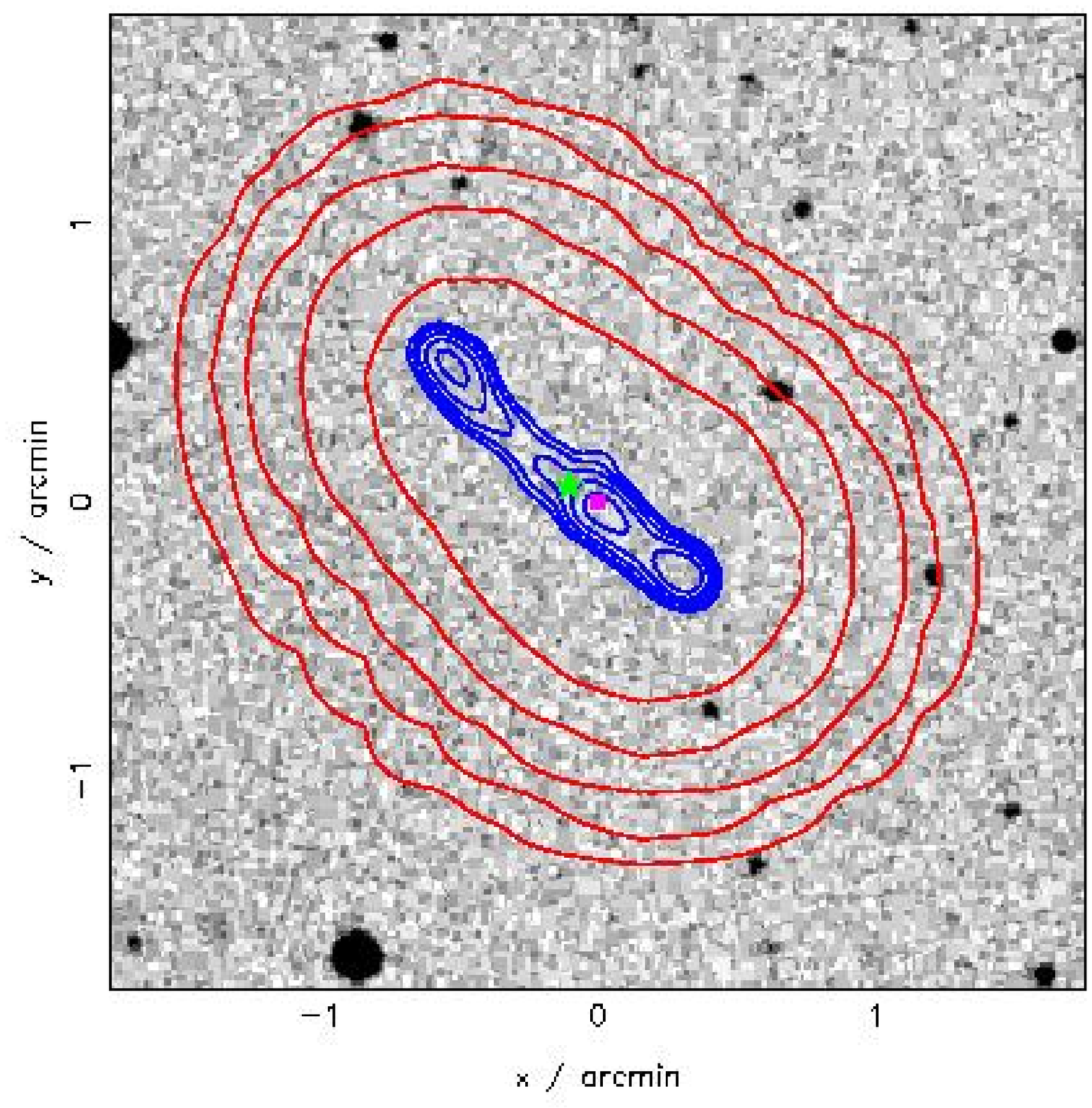}}
      \centerline{C1-247: 3C 341}
    \end{minipage}
    \hspace{3cm}
    \begin{minipage}{3cm}
      \mbox{}
      \centerline{\includegraphics[scale=0.26,angle=270]{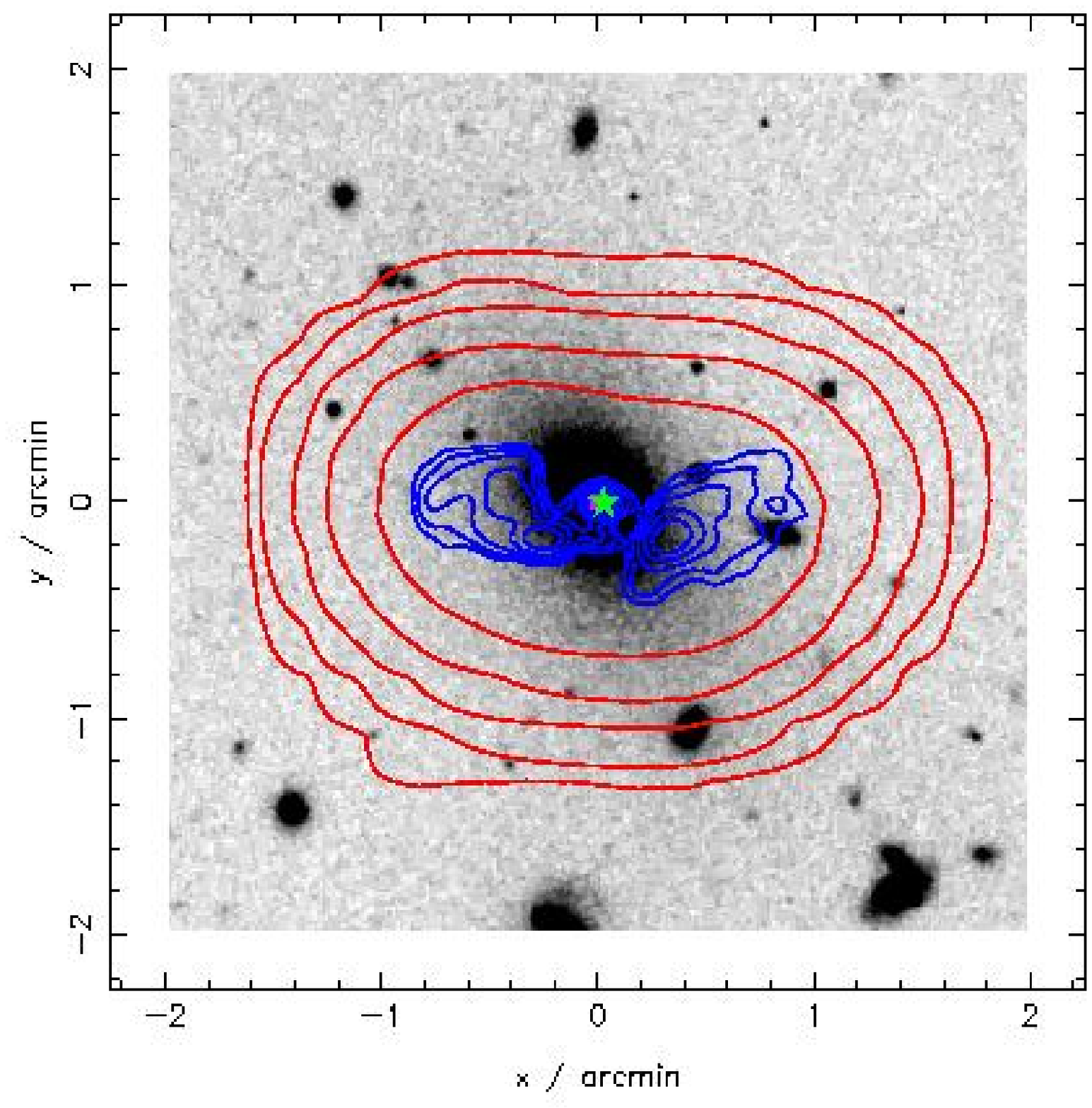}}
      \centerline{C1-248: 3C 338}
    \end{minipage}
    \hspace{3cm}
    \begin{minipage}{3cm}
      \mbox{}
      \centerline{\includegraphics[scale=0.26,angle=270]{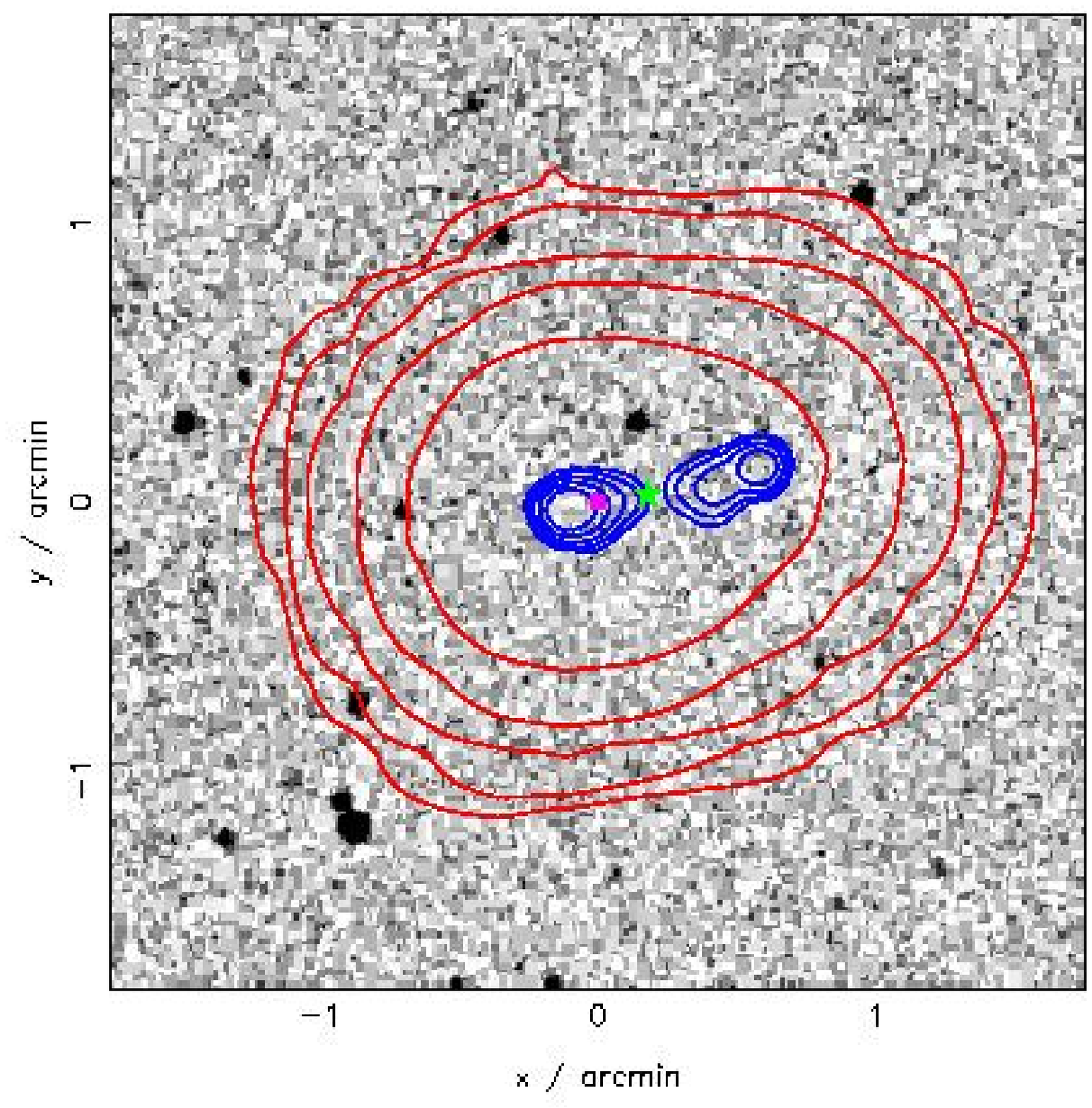}}
      \centerline{C1-249: 3C 337}
    \end{minipage}
    \vfill
    \begin{minipage}{3cm}     
      \mbox{}
      \centerline{\includegraphics[scale=0.26,angle=270]{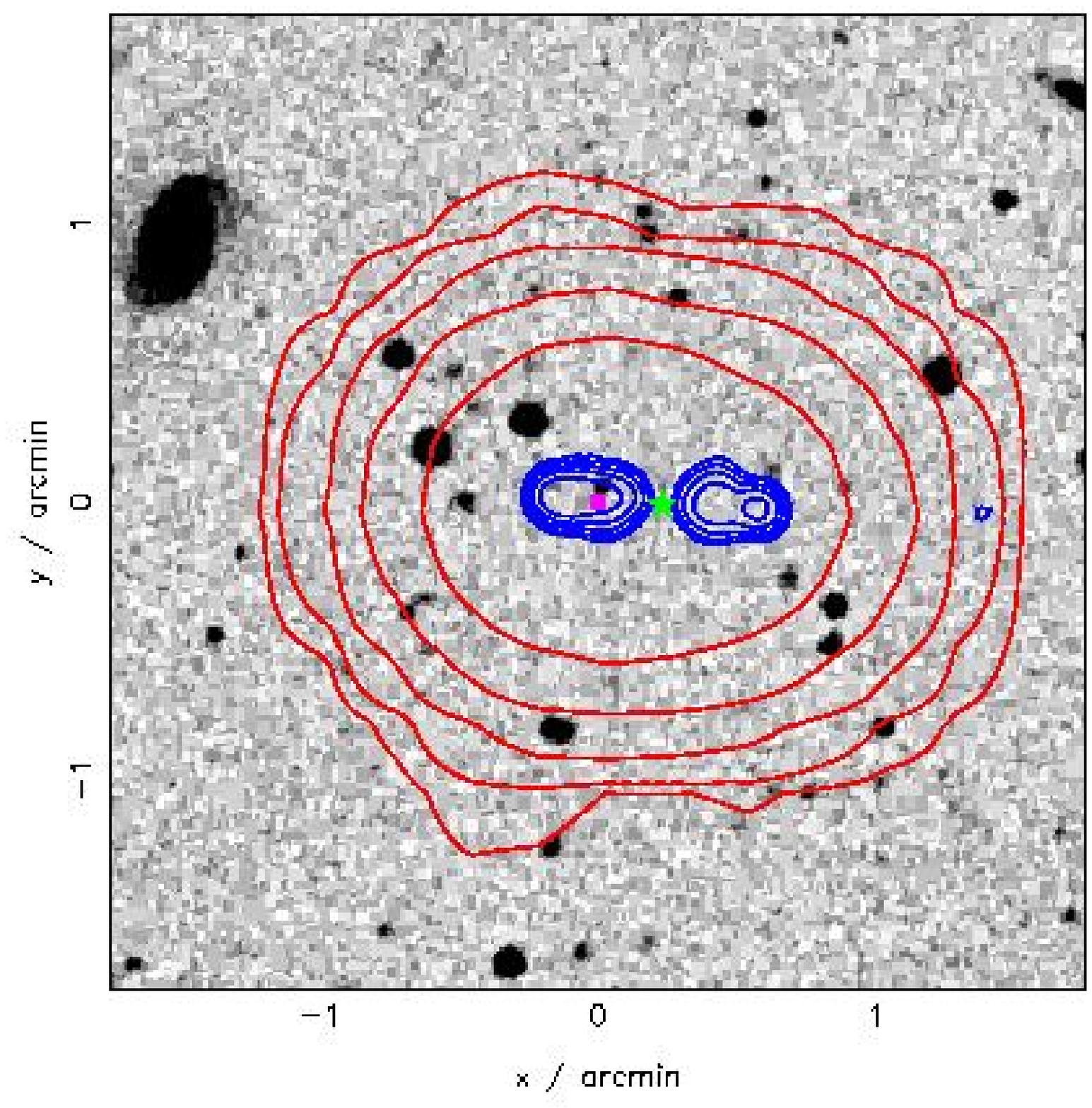}}
      \centerline{C1-250: 3C 340}
    \end{minipage}
    \hspace{3cm}
    \begin{minipage}{3cm}
      \mbox{}
      \centerline{\includegraphics[scale=0.26,angle=270]{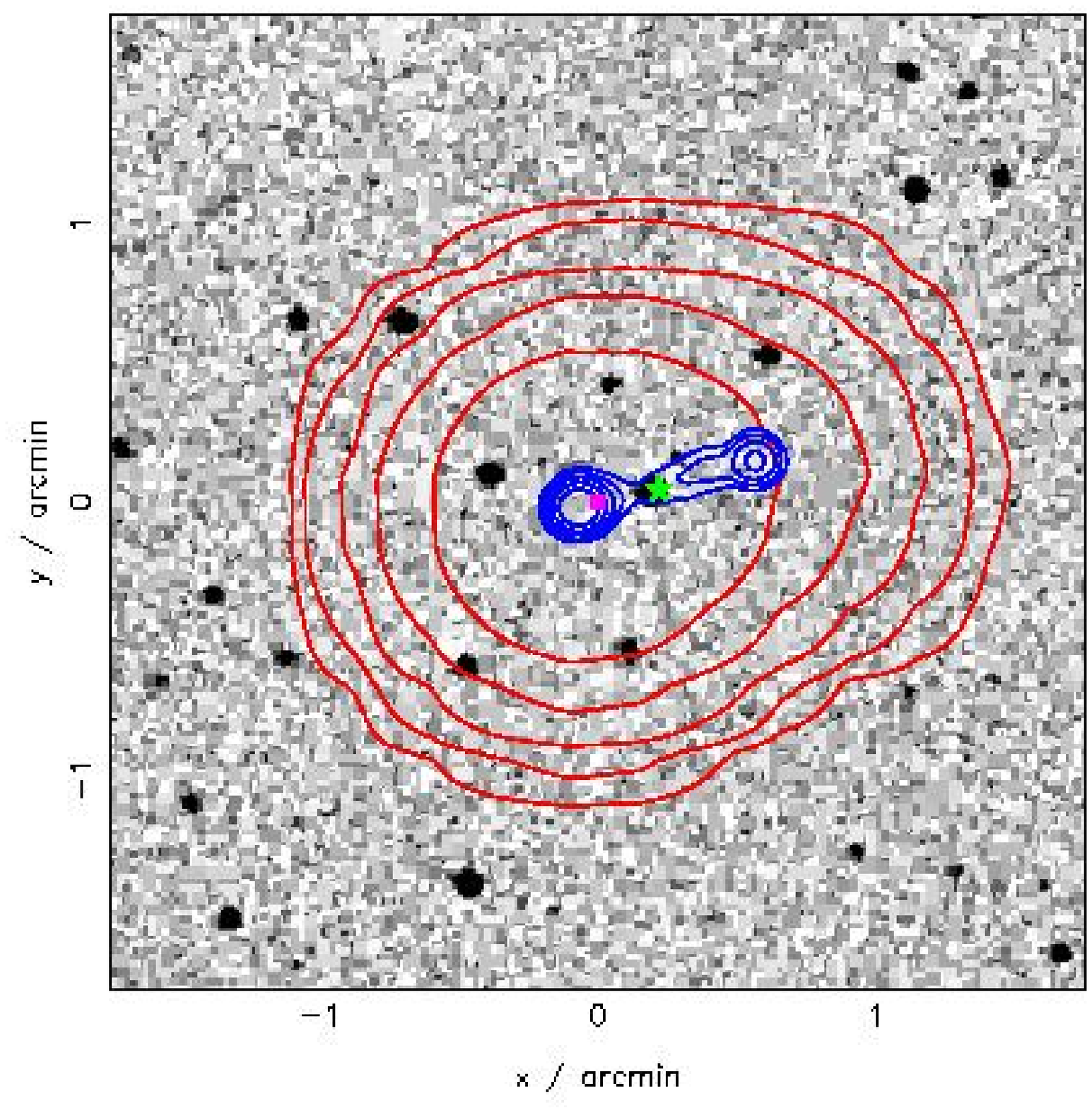}}
      \centerline{C1-254: 3C 342}
    \end{minipage}
    \hspace{3cm}
    \begin{minipage}{3cm}
      \mbox{}
      \centerline{\includegraphics[scale=0.26,angle=270]{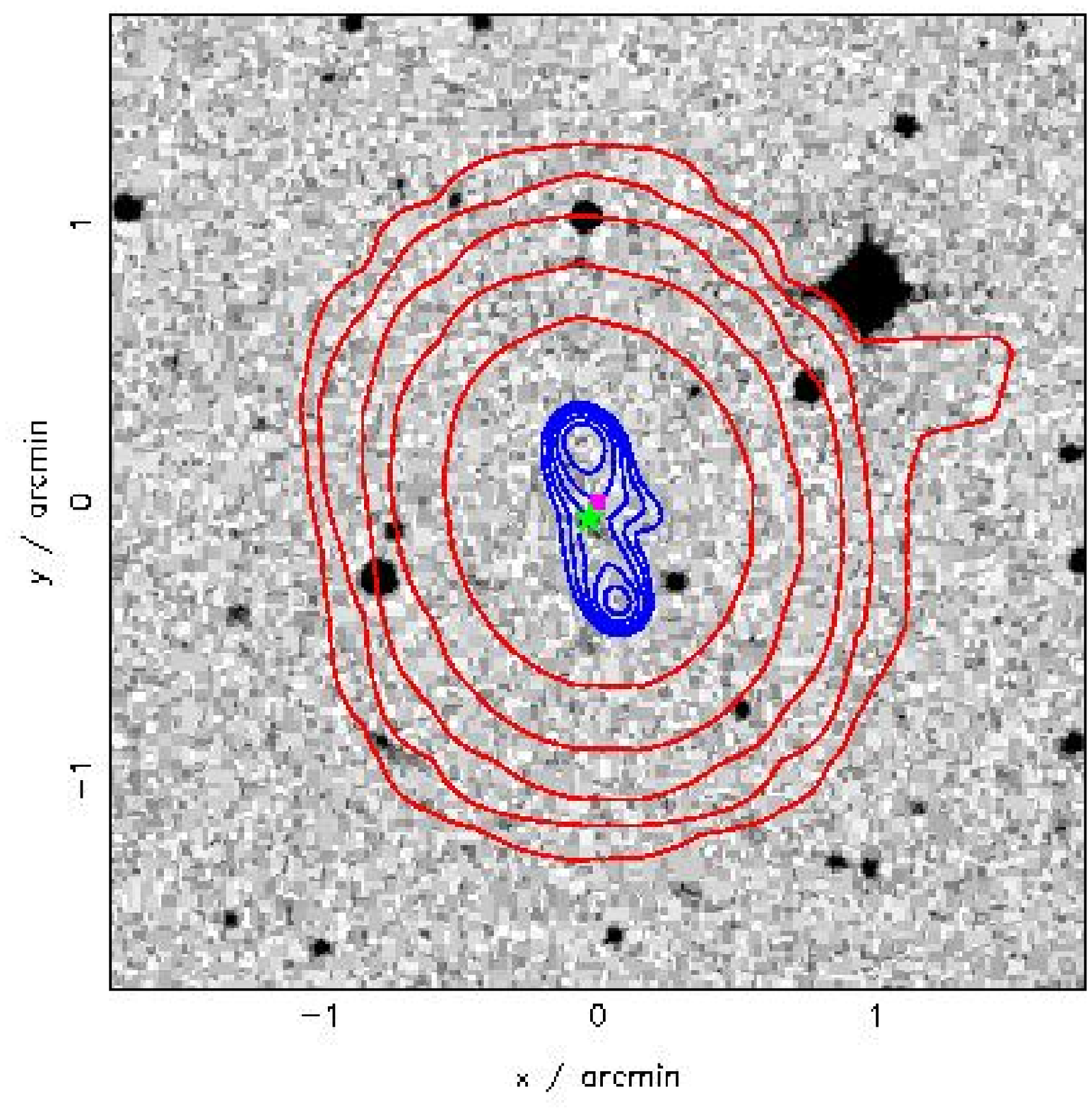}}
      \centerline{C1-257: 3C 344}
    \end{minipage}
    \vfill
    \begin{minipage}{3cm}     
      \mbox{}
      \centerline{\includegraphics[scale=0.26,angle=270]{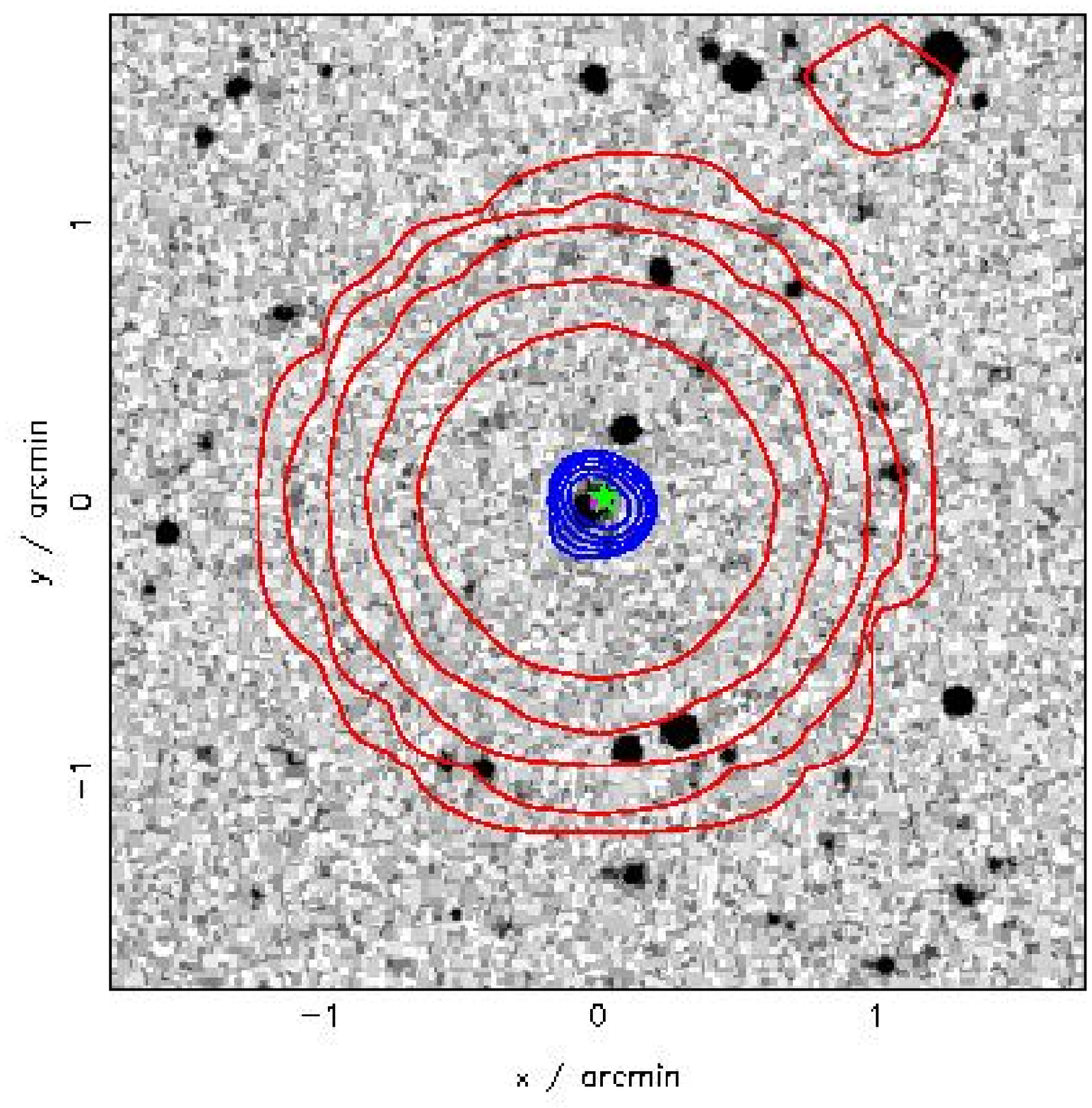}}
      \centerline{C1-258: 3C 346}
    \end{minipage}
    \hspace{3cm}
    \begin{minipage}{3cm}
      \mbox{}
      \centerline{\includegraphics[scale=0.26,angle=270]{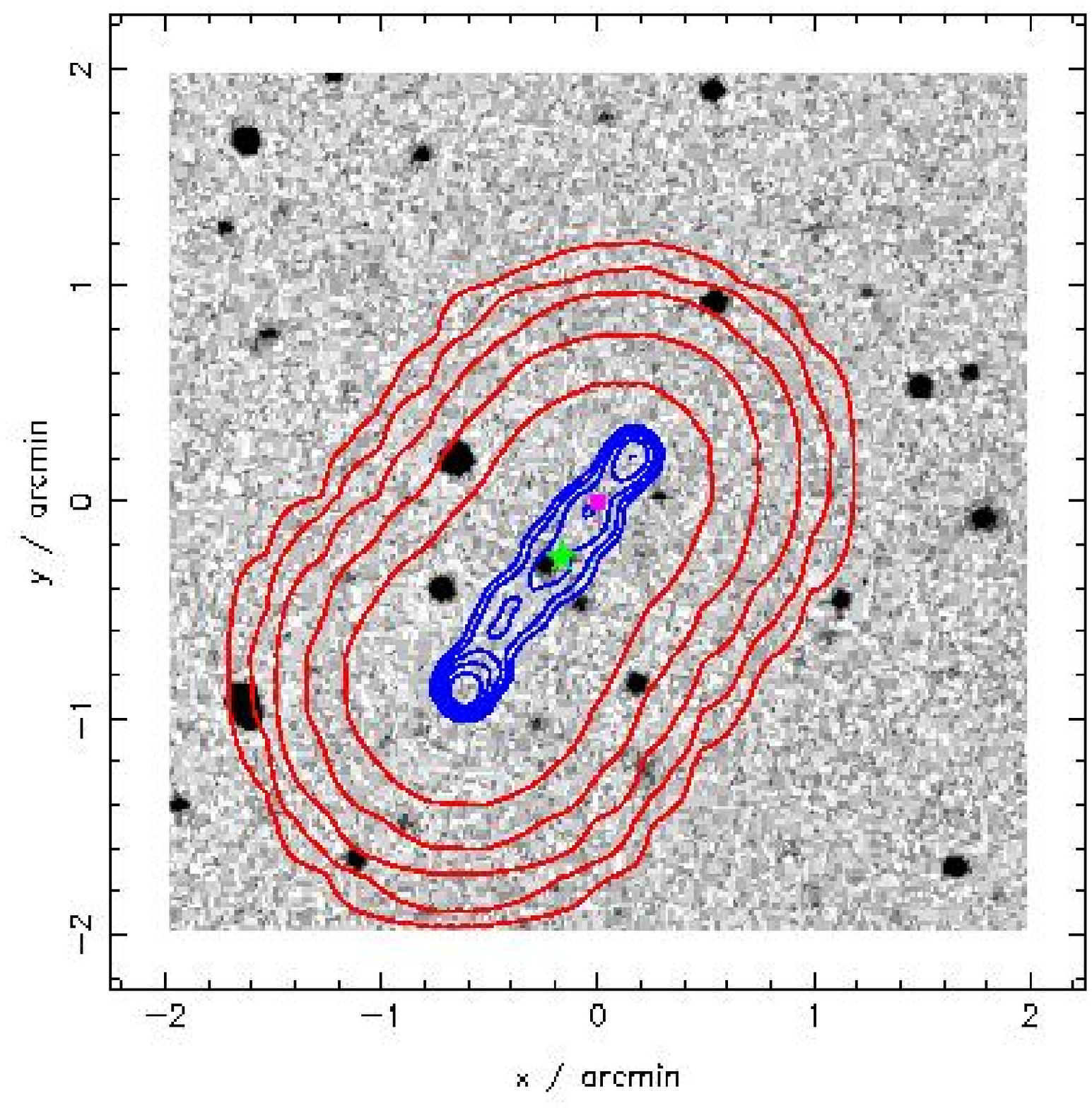}}
      \centerline{C1-261: 3C 349}
    \end{minipage}
    \hspace{3cm}
    \begin{minipage}{3cm}
      \mbox{}
      \centerline{\includegraphics[scale=0.26,angle=270]{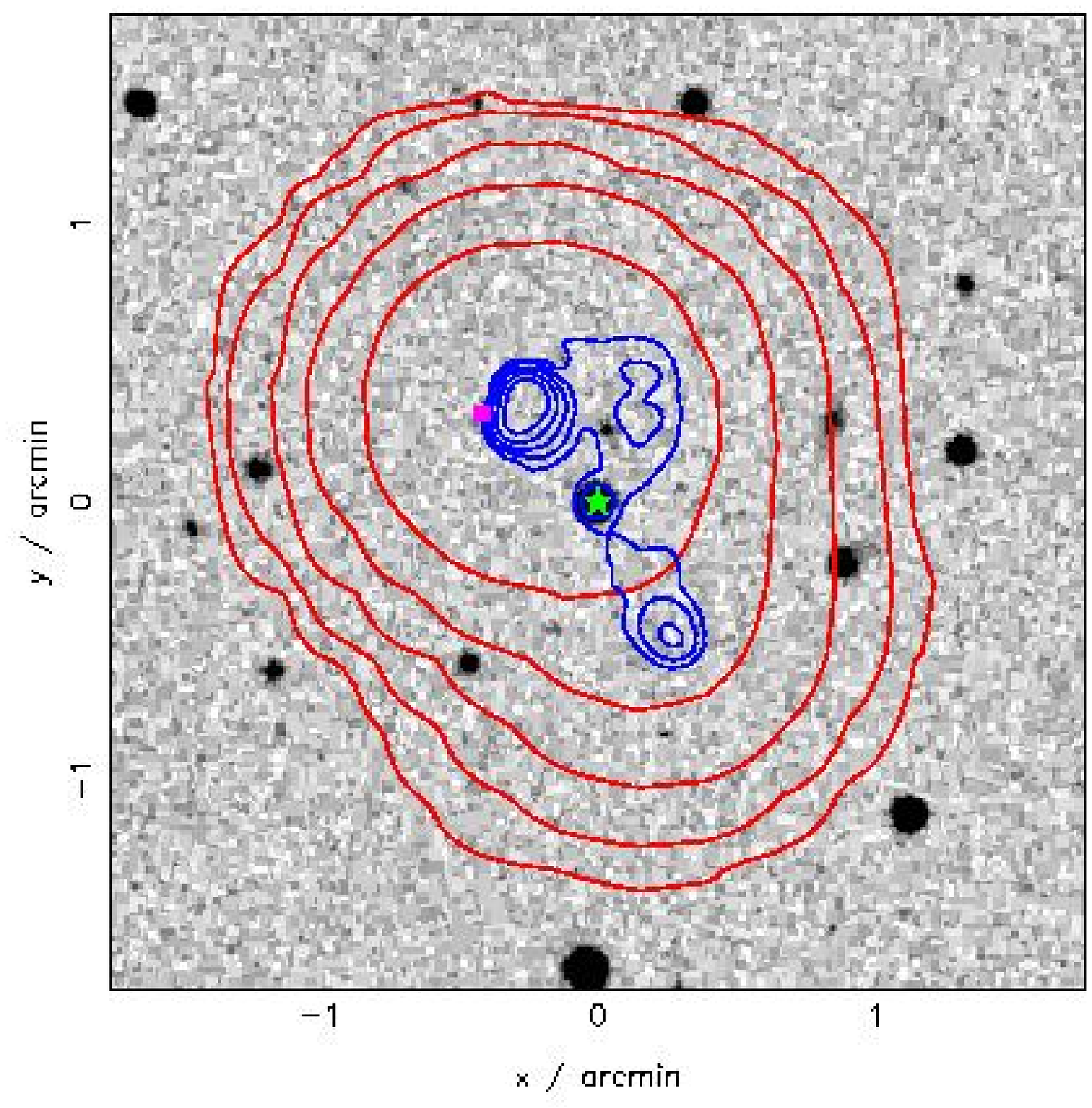}}
      \centerline{C1-263: 3C 351}
    \end{minipage}
  \end{center}
\end{figure}

\begin{figure}
  \begin{center}
    {\bf CoNFIG-1}\\  
  \begin{minipage}{3cm}      
      \mbox{}
      \centerline{\includegraphics[scale=0.26,angle=270]{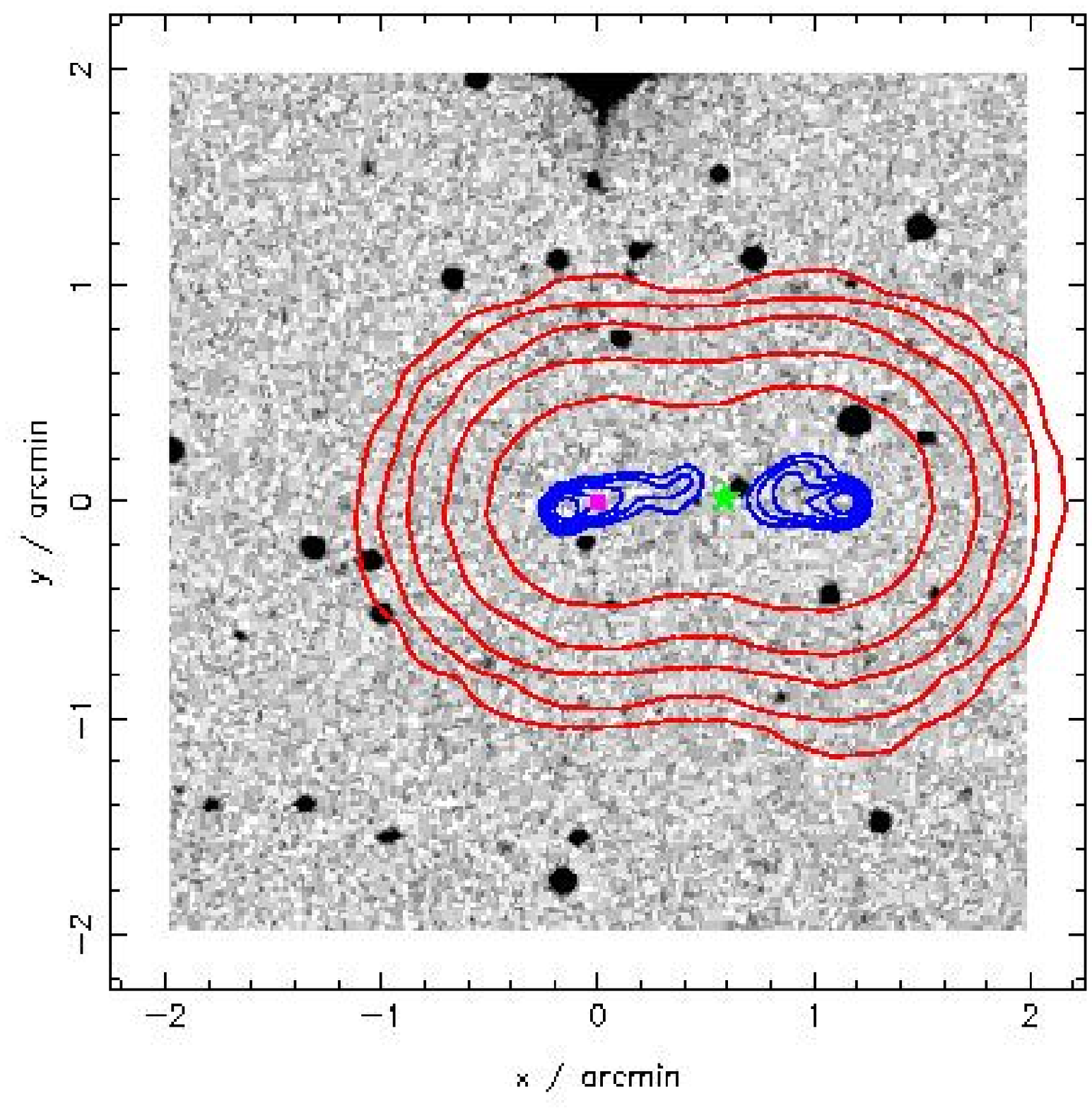}}
      \centerline{C1-264: 3C 350}
    \end{minipage}
    \hspace{3cm}
    \begin{minipage}{3cm}
      \mbox{}
      \centerline{\includegraphics[scale=0.26,angle=270]{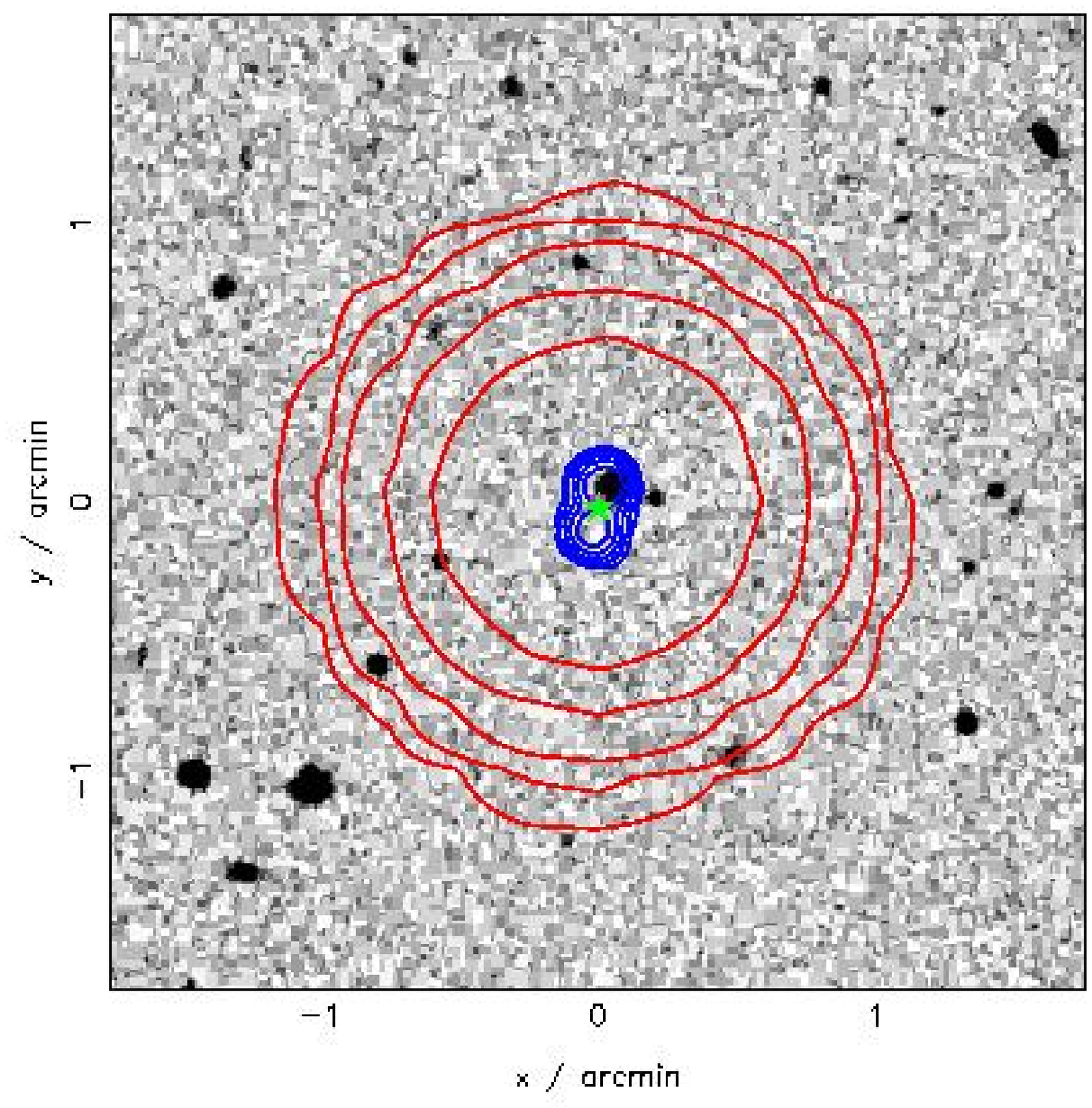}}
      \centerline{C1-265: 3C 352}
    \end{minipage}
    \hspace{3cm}
    \begin{minipage}{3cm}
      \mbox{}
      \centerline{\includegraphics[scale=0.26,angle=270]{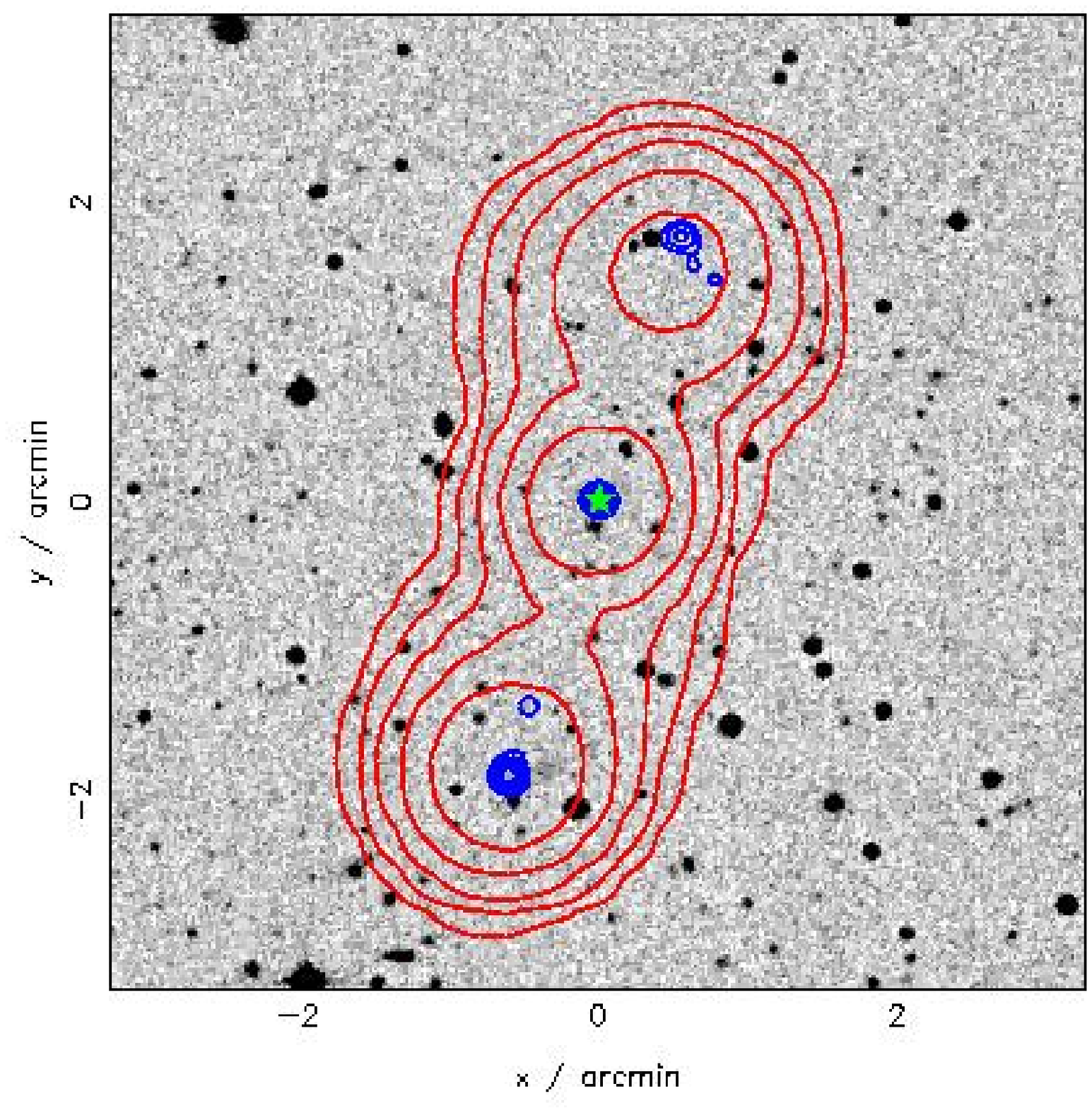}}
      \centerline{C1-266: 4C 34.47}
    \end{minipage}
    \vfill
    \begin{minipage}{3cm}      
      \mbox{}
      \centerline{\includegraphics[scale=0.26,angle=270]{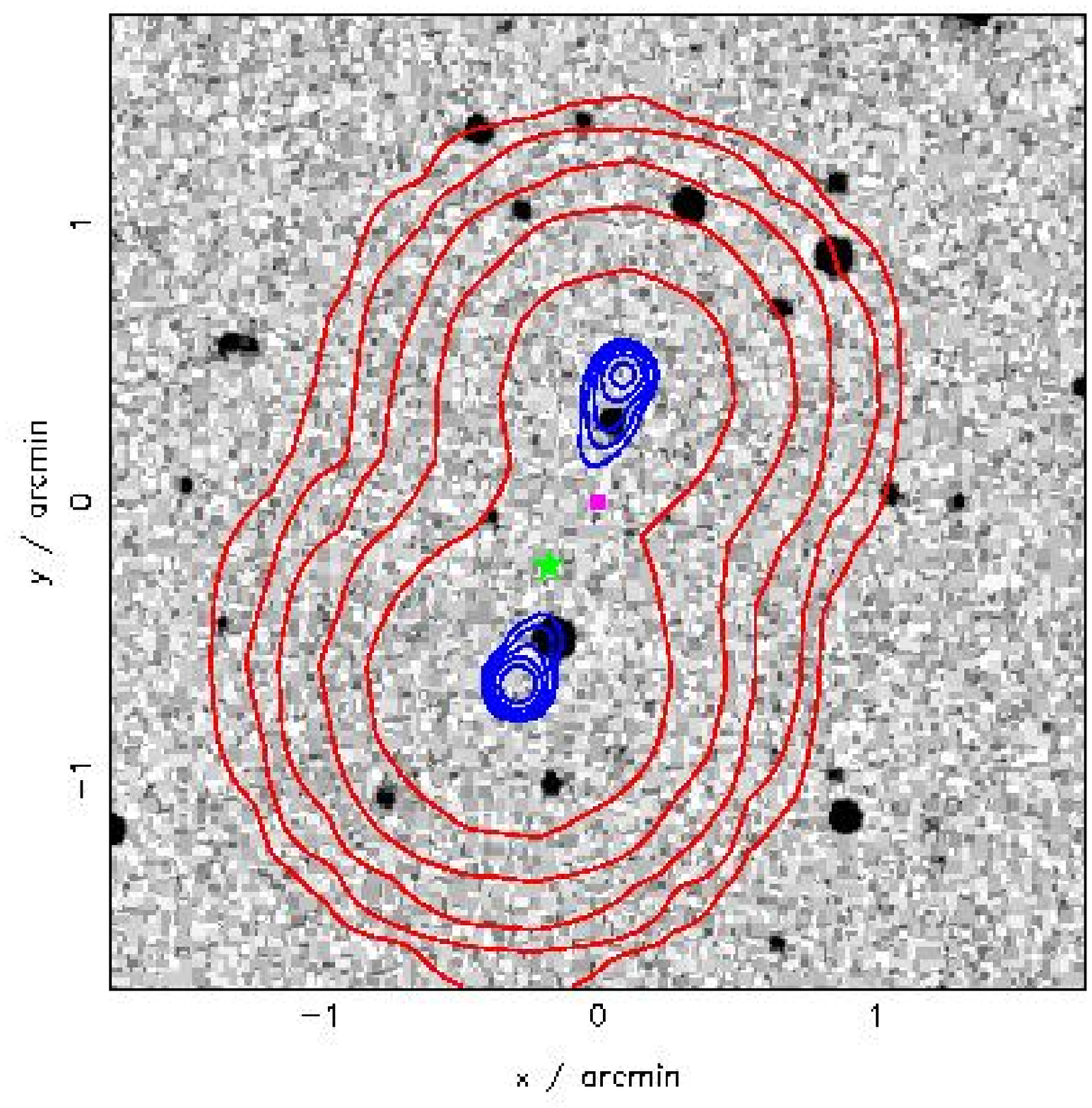}}
      \centerline{C1-267: 3C 356}
    \end{minipage}
    \hspace{3cm}
    \begin{minipage}{3cm}
      \mbox{}
      \centerline{\includegraphics[scale=0.26,angle=270]{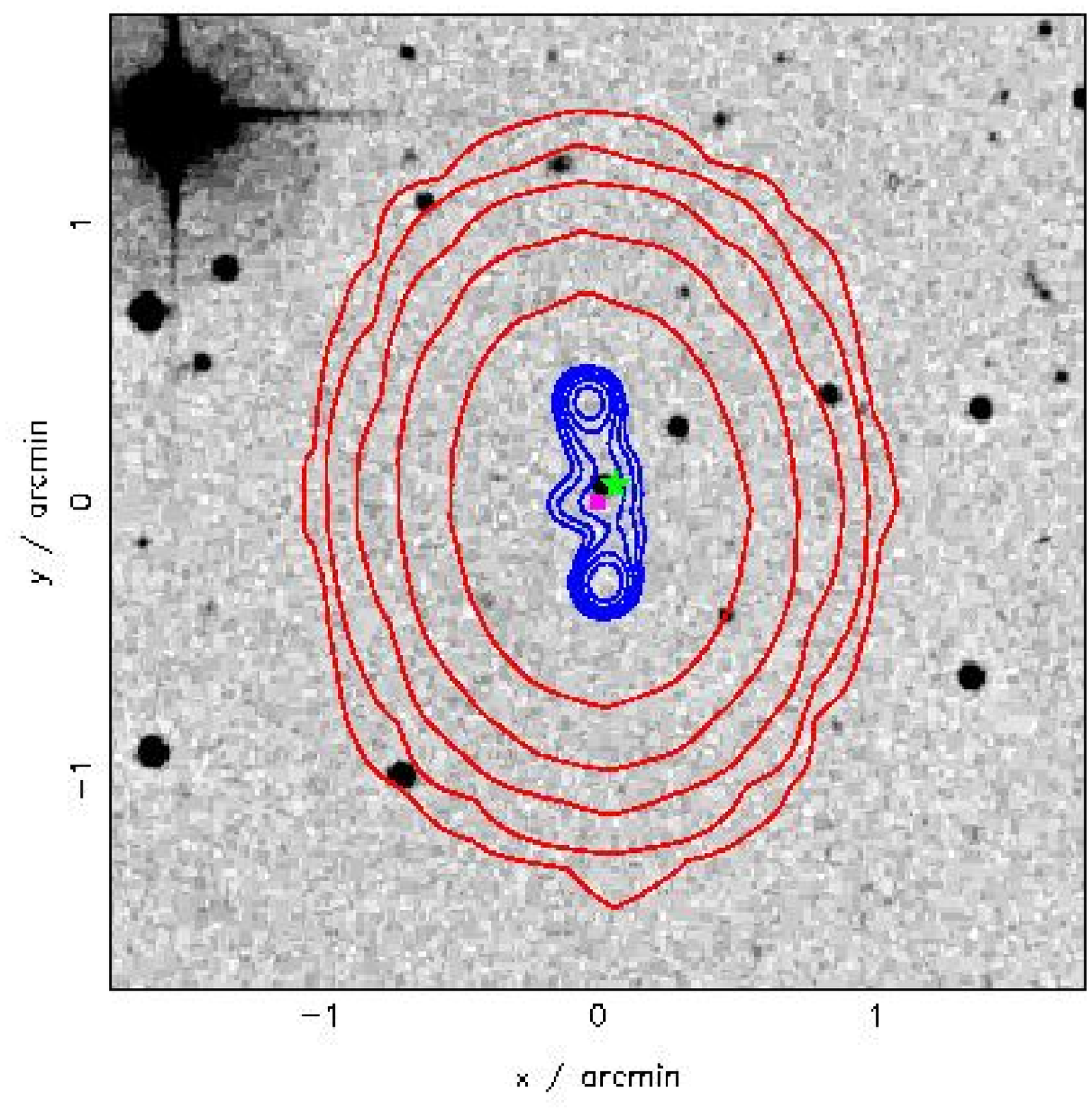}}
      \centerline{C1-268: 4C 61.34}
    \end{minipage}
    \hspace{3cm}
    \begin{minipage}{3cm}
      \mbox{}
      \centerline{\includegraphics[scale=0.26,angle=270]{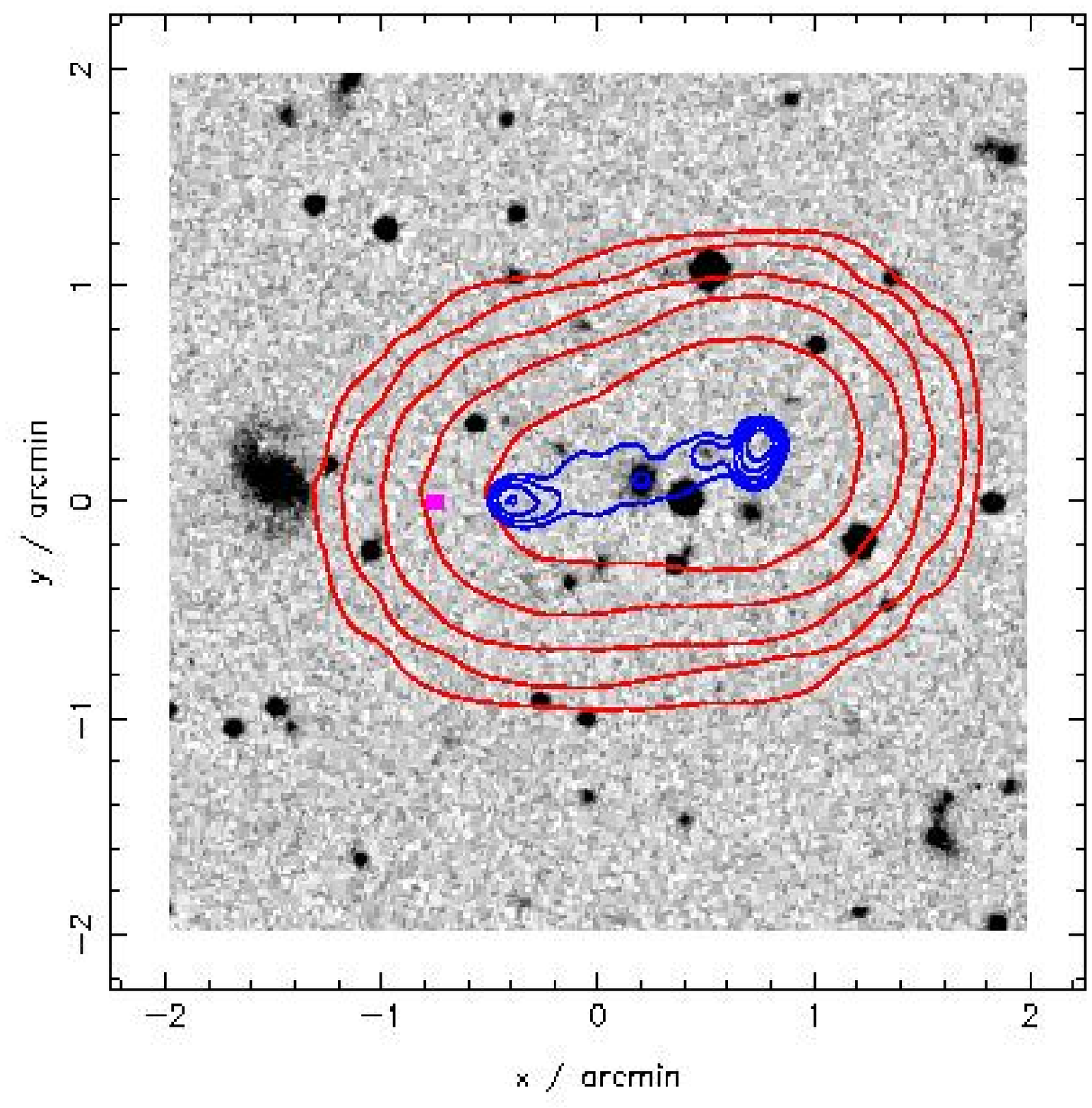}}
      \centerline{C1-269: 4C 45.13}
    \end{minipage}
    \vfill
    \begin{minipage}{3cm}     
      \mbox{}
      \centerline{\includegraphics[scale=0.26,angle=270]{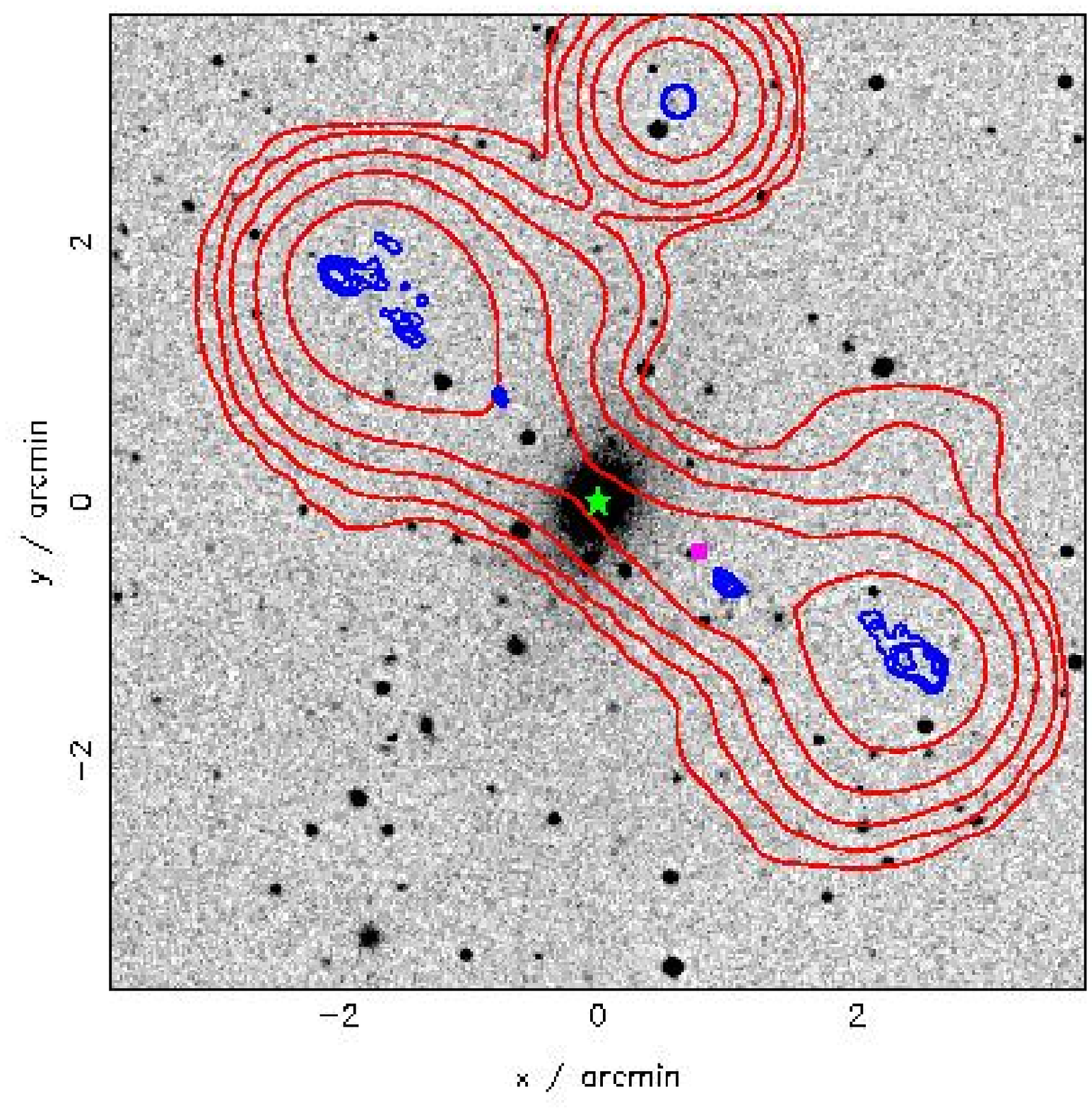}}
      \centerline{C1-270: 3C 306}
    \end{minipage}
    \hspace{3cm}
    \begin{minipage}{3cm}
      \mbox{}
      \centerline{\includegraphics[scale=0.26,angle=270]{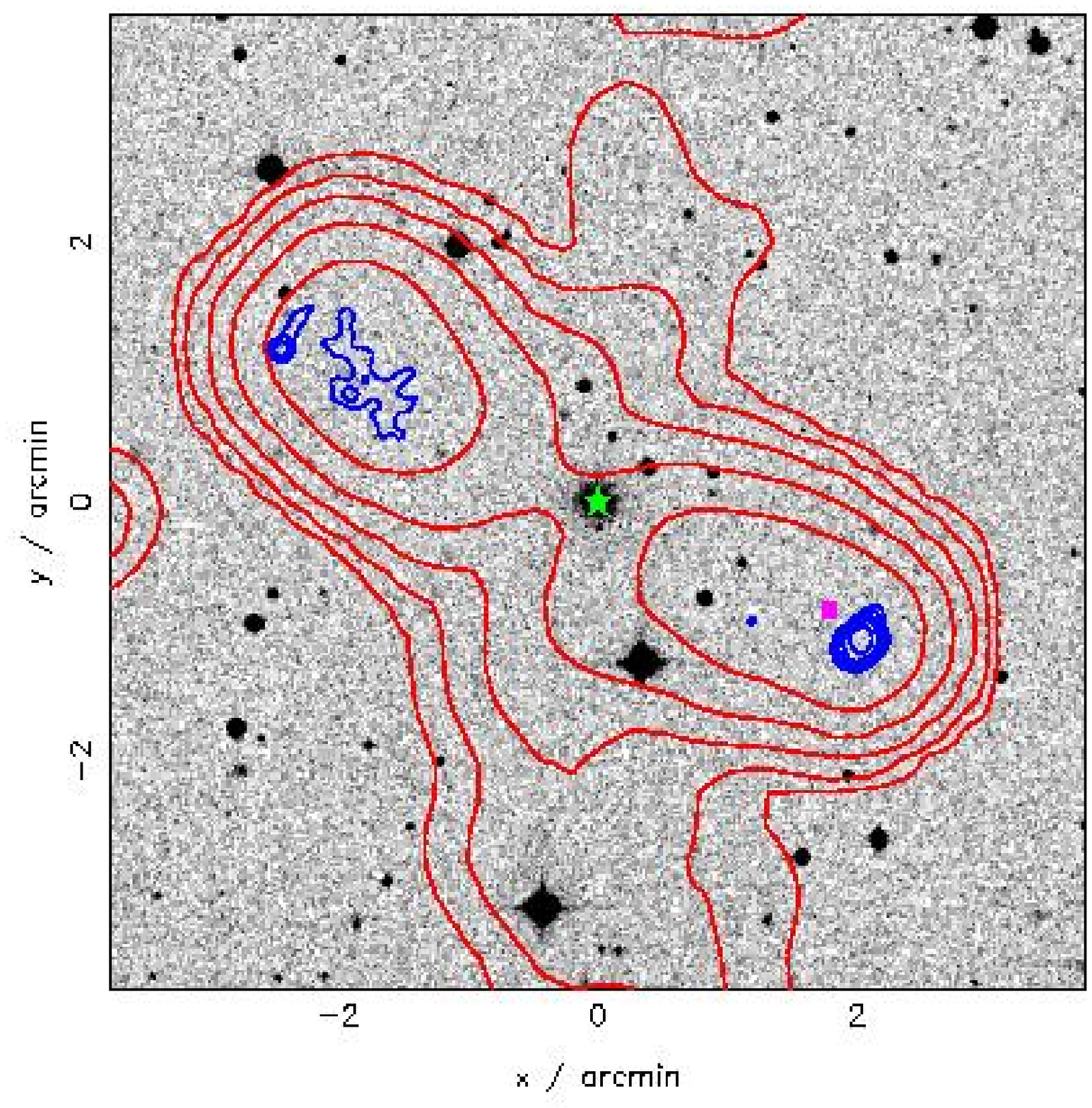}}
      \centerline{C1-271: 4C 32.25A}
    \end{minipage}
    \hspace{3cm}
    \begin{minipage}{3cm}
      \mbox{}
      \centerline{\includegraphics[scale=0.26,angle=270]{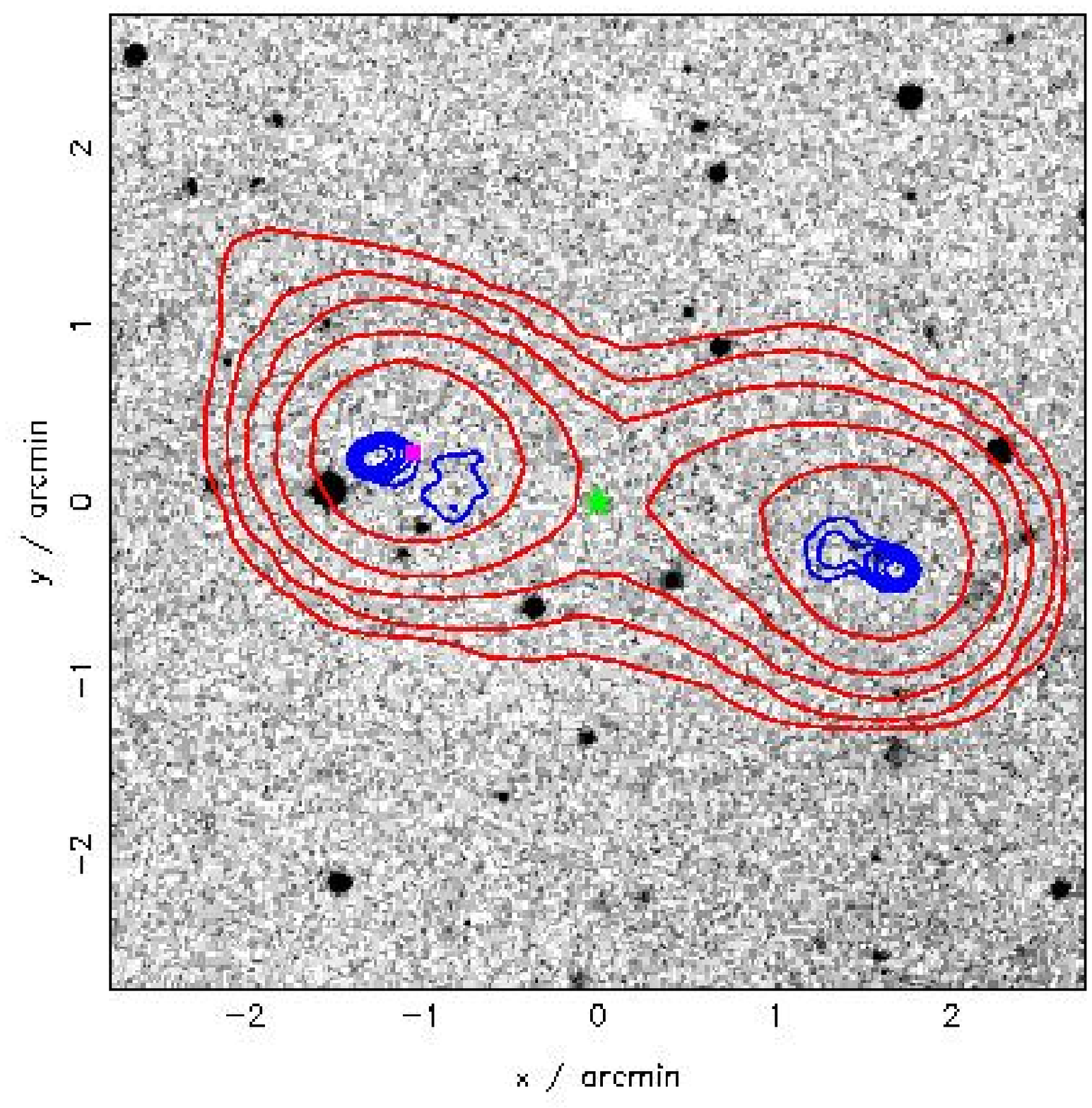}}
      \centerline{C1-272: 4C 06.32}
    \end{minipage}
    \vfill
    \begin{minipage}{3cm}     
      \mbox{}
      \centerline{\includegraphics[scale=0.26,angle=270]{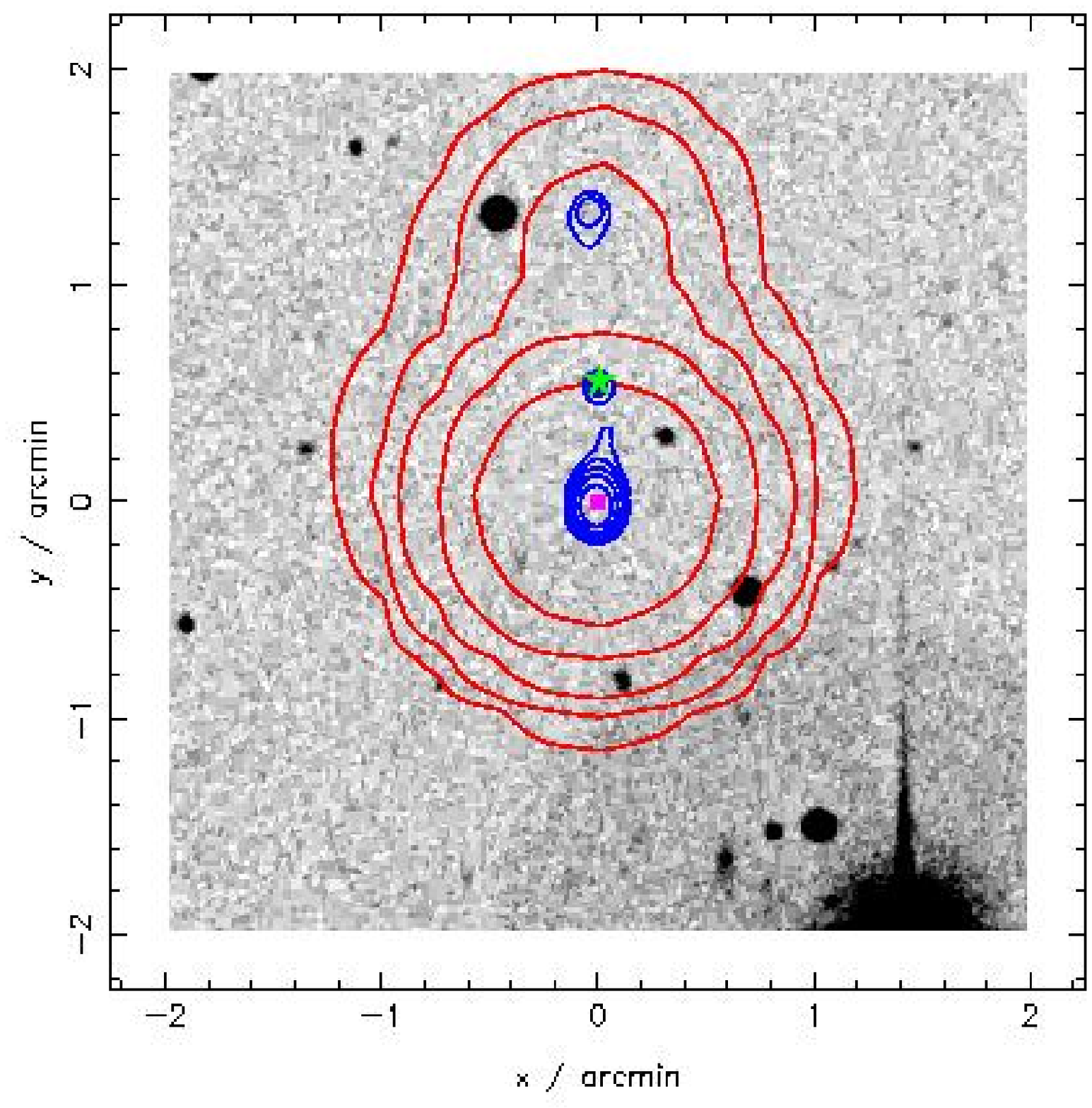}}
      \centerline{C1-273: 4C 20.29}
    \end{minipage}
  \end{center}
\end{figure}

\begin{figure}
  \begin{center}
  {\bf CoNFIG-2}\\
    \begin{minipage}{3cm}      
      \mbox{}
      \centerline{\includegraphics[scale=0.26,angle=270]{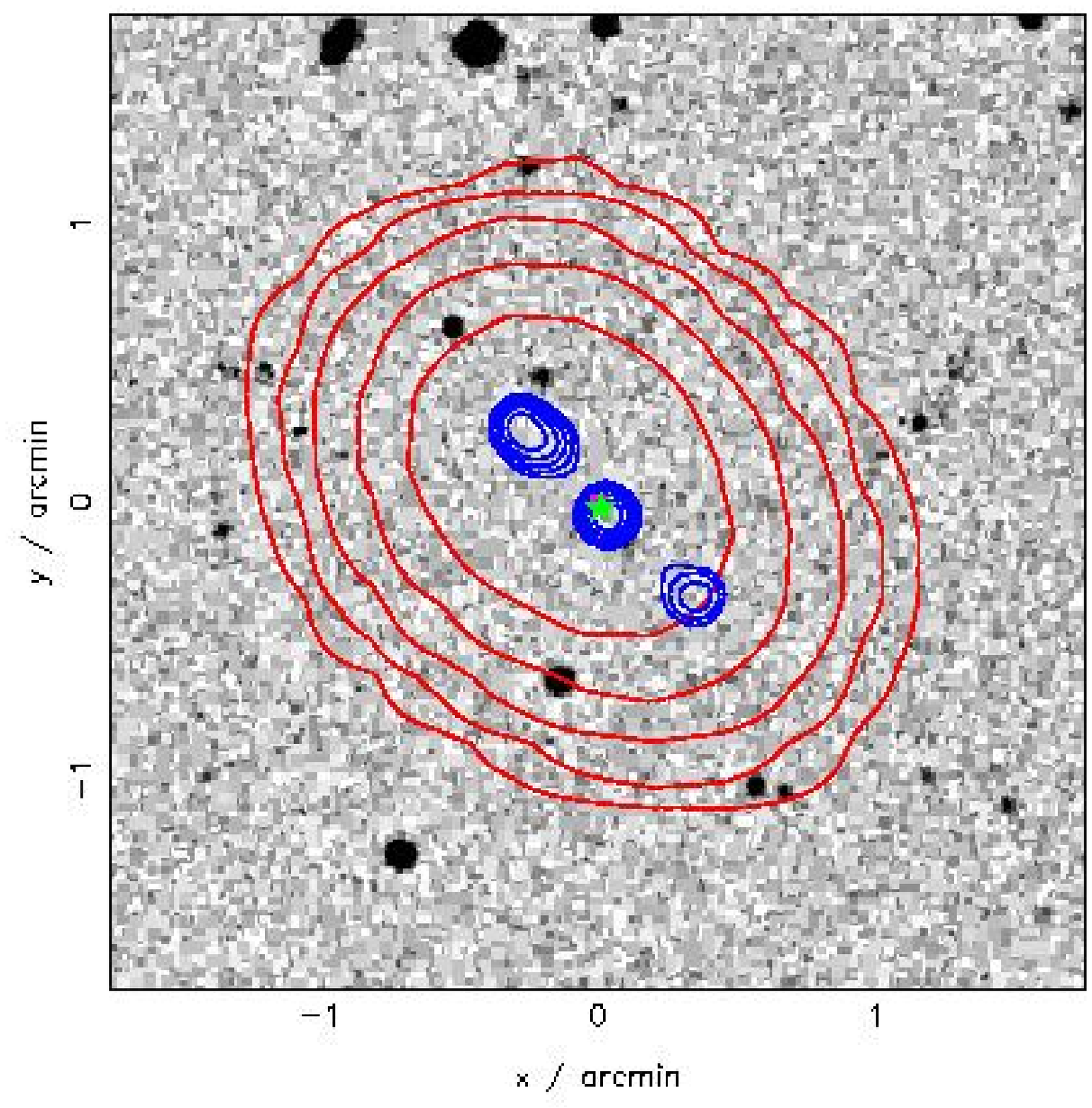}}
      \centerline{C2-004: 4C 38.29}
    \end{minipage}
    \hspace{3cm}
    \begin{minipage}{3cm}
      \mbox{}
      \centerline{\includegraphics[scale=0.26,angle=270]{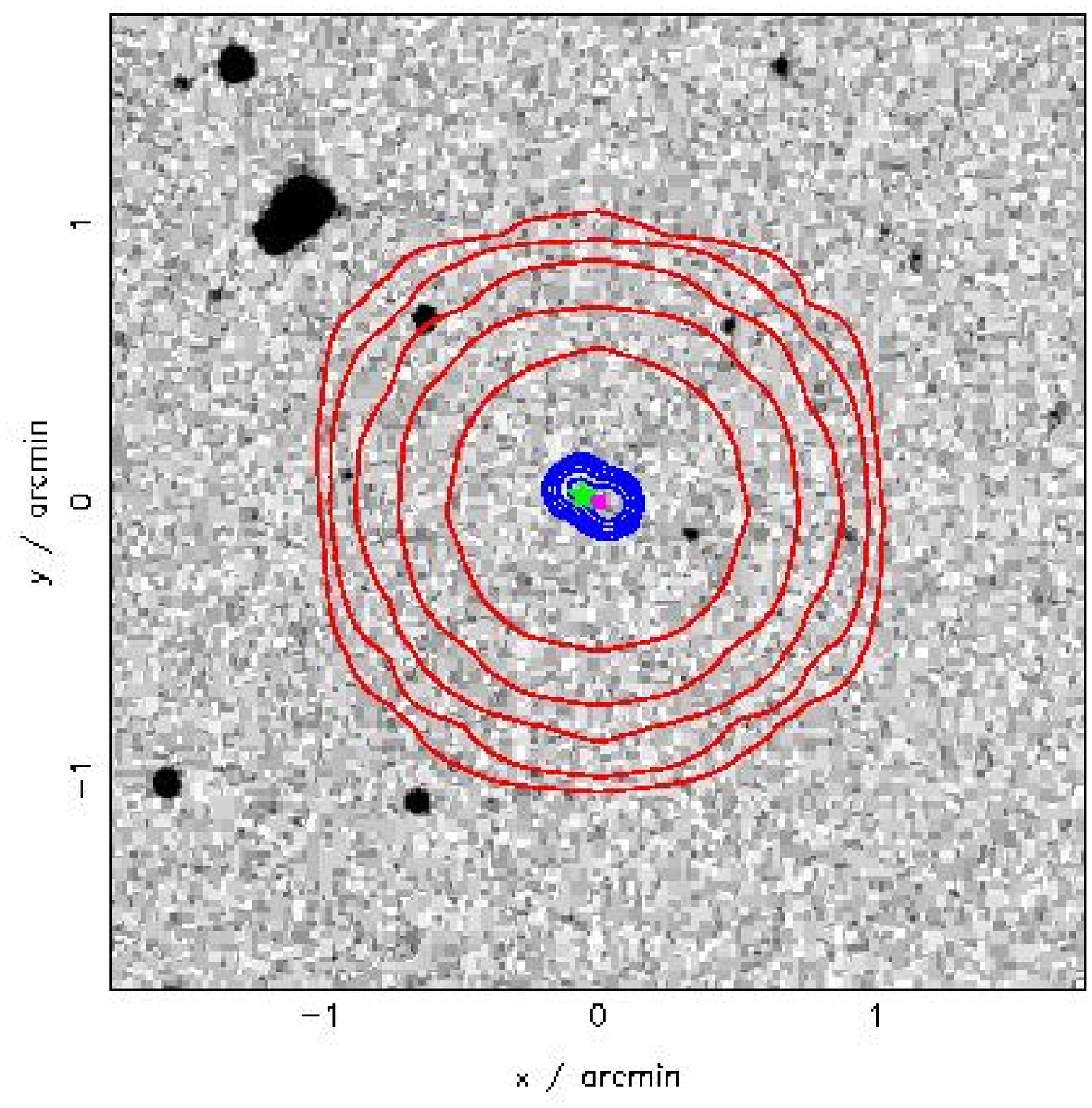}}
      \centerline{C2-008: 4C 59.10}
    \end{minipage}
    \hspace{3cm}
    \begin{minipage}{3cm}
      \mbox{}
      \centerline{\includegraphics[scale=0.26,angle=270]{Contours/C2/010.ps}}
      \centerline{C2-010: 4C 08.31}
    \end{minipage}
    \vfill
    \begin{minipage}{3cm}     
      \mbox{}
      \centerline{\includegraphics[scale=0.26,angle=270]{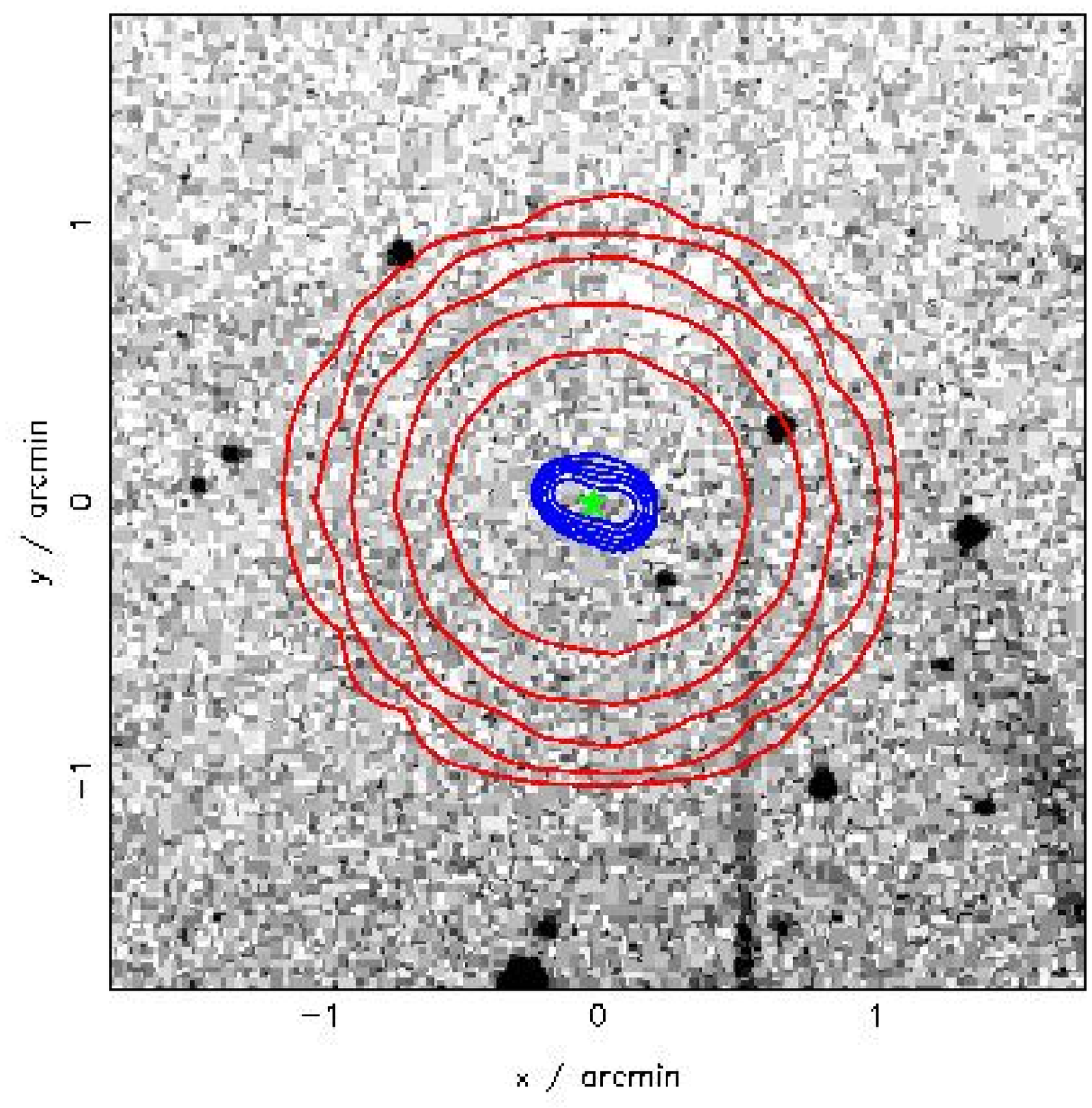}}
      \centerline{C2-011: 3C 221}
    \end{minipage}
    \hspace{3cm}
    \begin{minipage}{3cm}
      \mbox{}
      \centerline{\includegraphics[scale=0.26,angle=270]{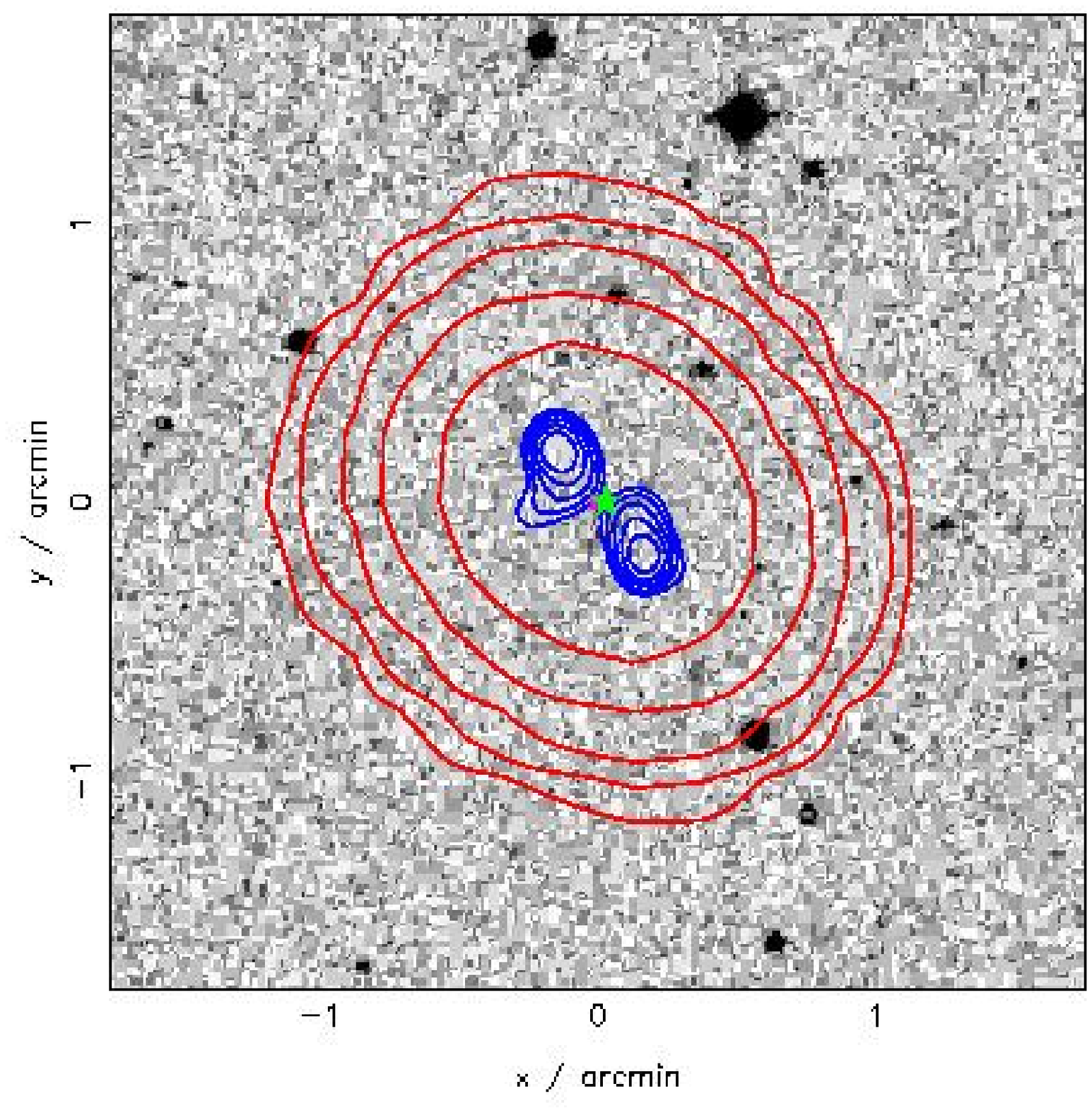}}
      \centerline{C2-014: 4C -01.19}
    \end{minipage}
    \hspace{3cm}
    \begin{minipage}{3cm}
      \mbox{}
      \centerline{\includegraphics[scale=0.26,angle=270]{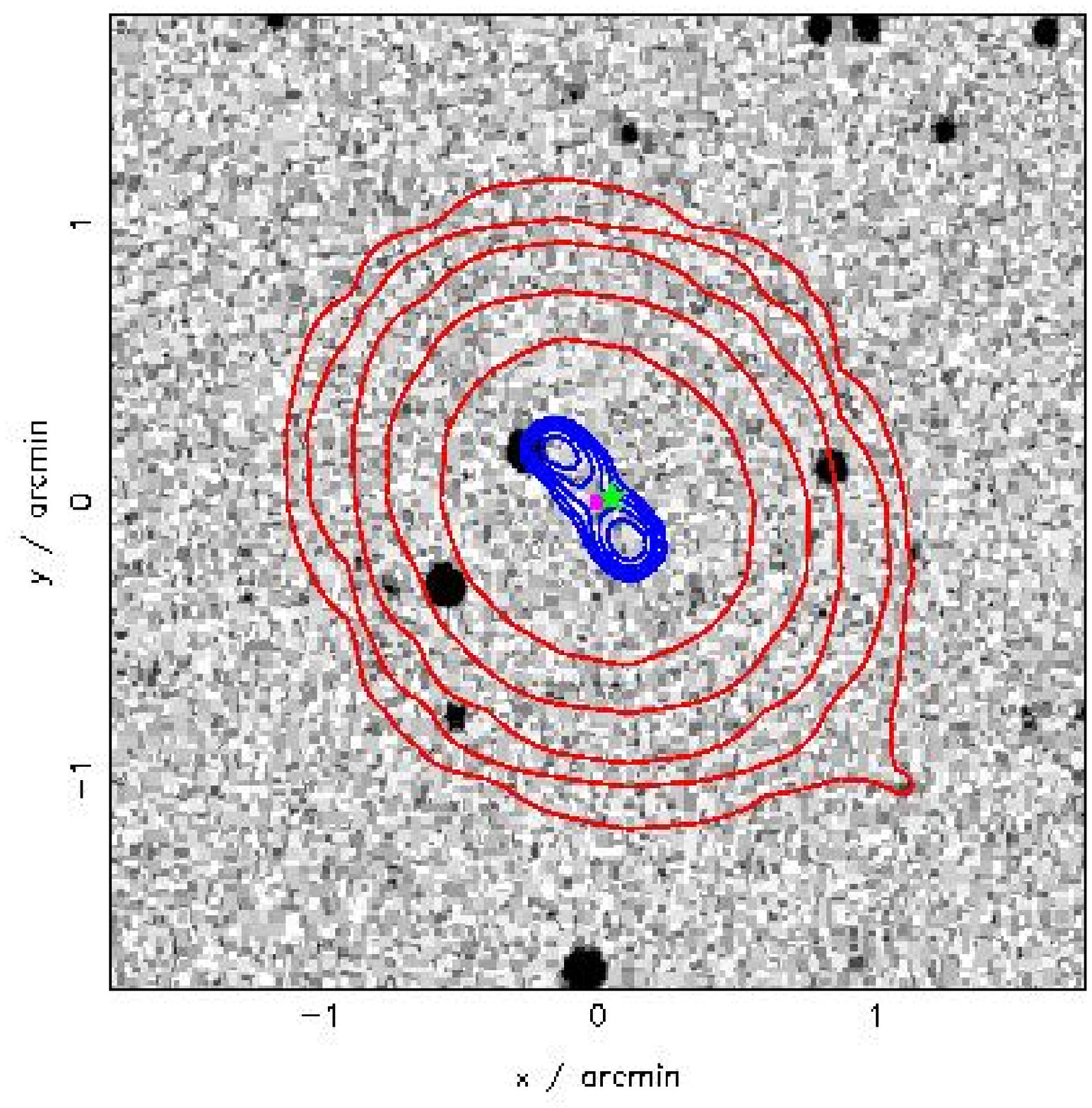}}
      \centerline{C2-021: 4C 17.49}
    \end{minipage}
    \vfill
    \begin{minipage}{3cm}     
      \mbox{}
      \centerline{\includegraphics[scale=0.26,angle=270]{Contours/C2/023.ps}}
      \centerline{C2-023: 4C 00.31}
    \end{minipage}
    \hspace{3cm}
    \begin{minipage}{3cm}
      \mbox{}
      \centerline{\includegraphics[scale=0.26,angle=270]{Contours/C2/029.ps}}
      \centerline{C2-029: 4C 25.29}
    \end{minipage}
    \hspace{3cm}
    \begin{minipage}{3cm}
      \mbox{}
      \centerline{\includegraphics[scale=0.26,angle=270]{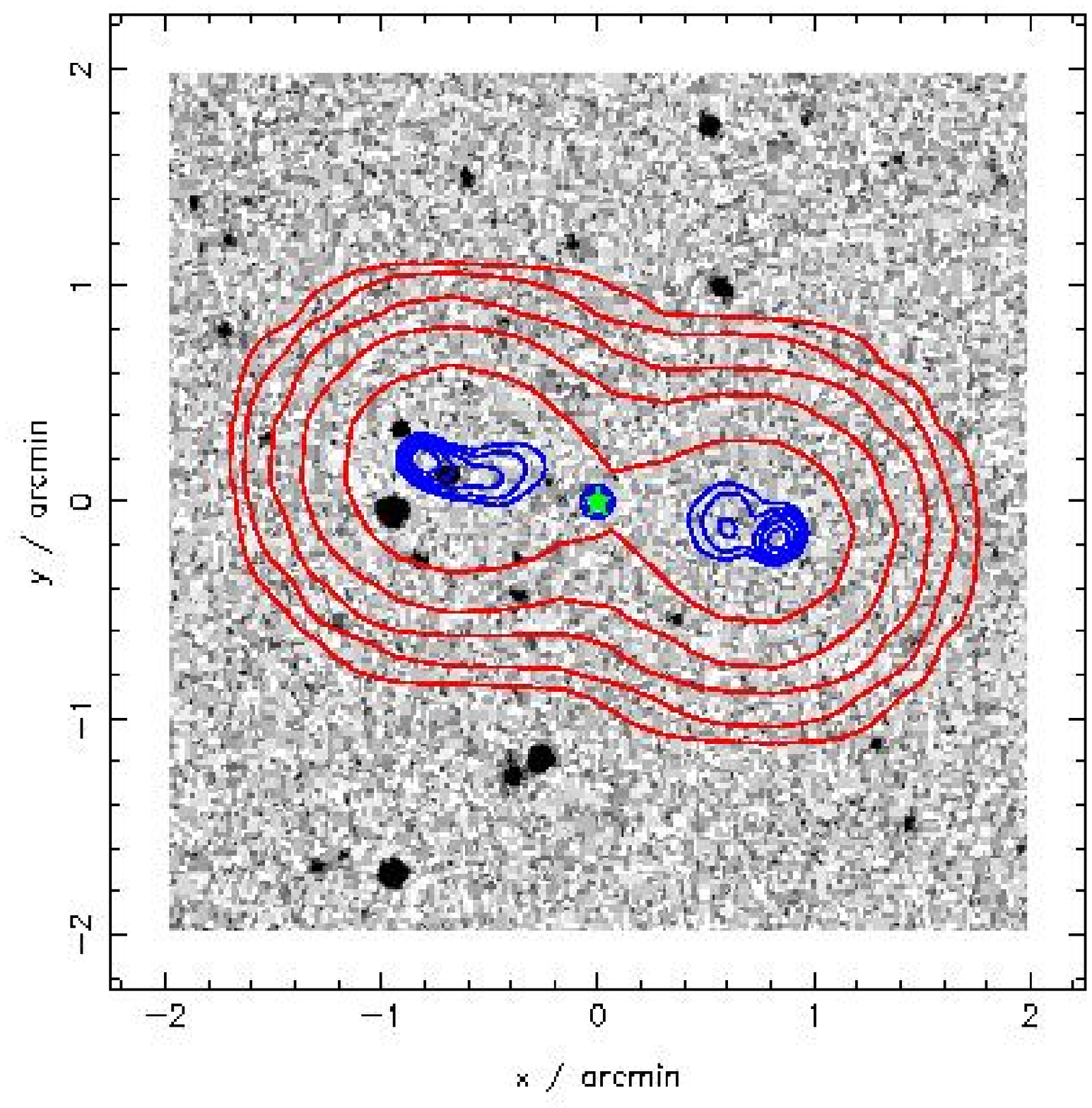}}
      \centerline{C2-031: 4C 21.26}
    \end{minipage}
    \vfill
    \begin{minipage}{3cm}      
      \mbox{}
      \centerline{\includegraphics[scale=0.26,angle=270]{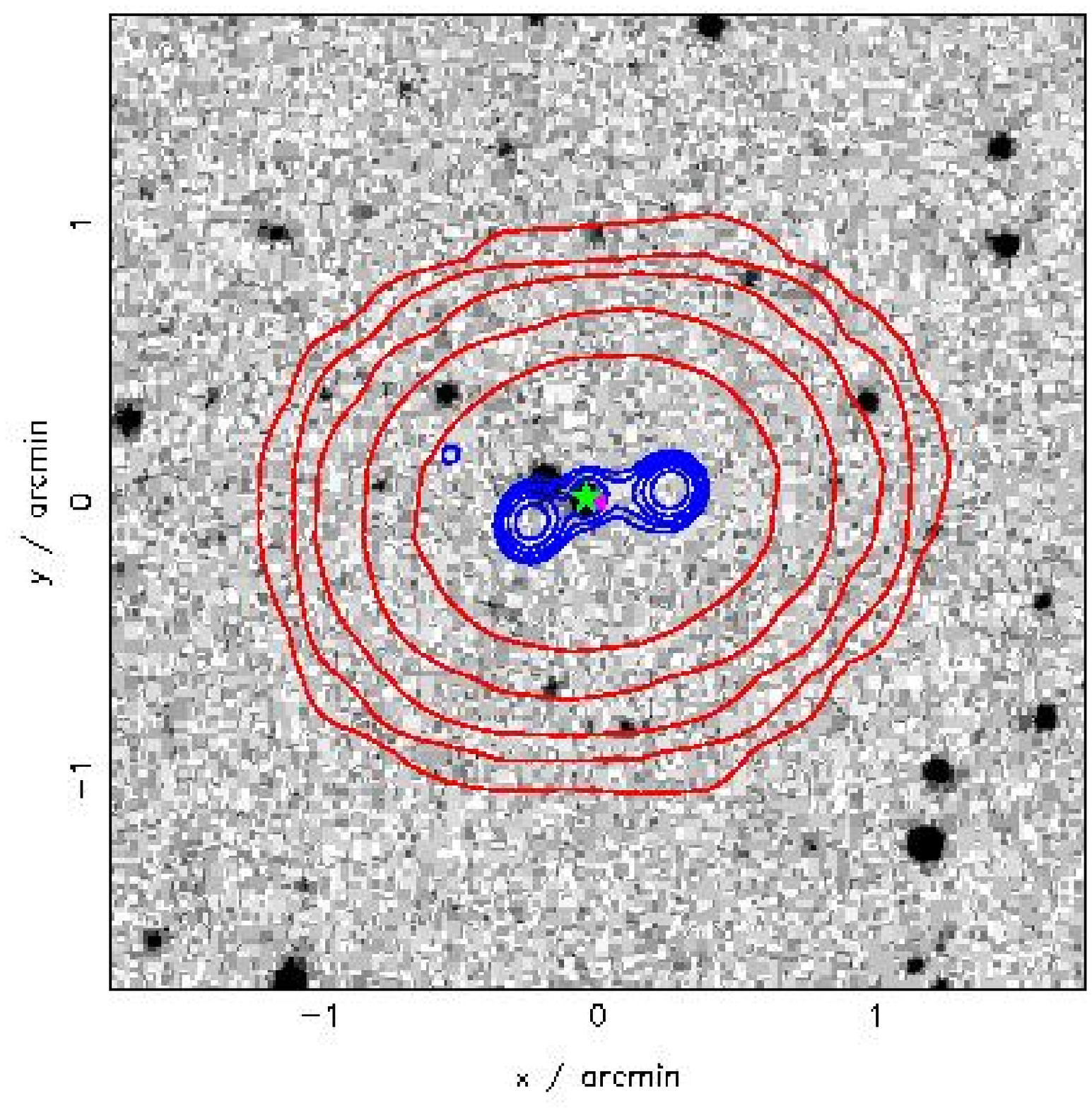}}
      \centerline{C2-036: 4C 00.34}
    \end{minipage}
    \hspace{3cm}
    \begin{minipage}{3cm}
      \mbox{}
      \centerline{\includegraphics[scale=0.26,angle=270]{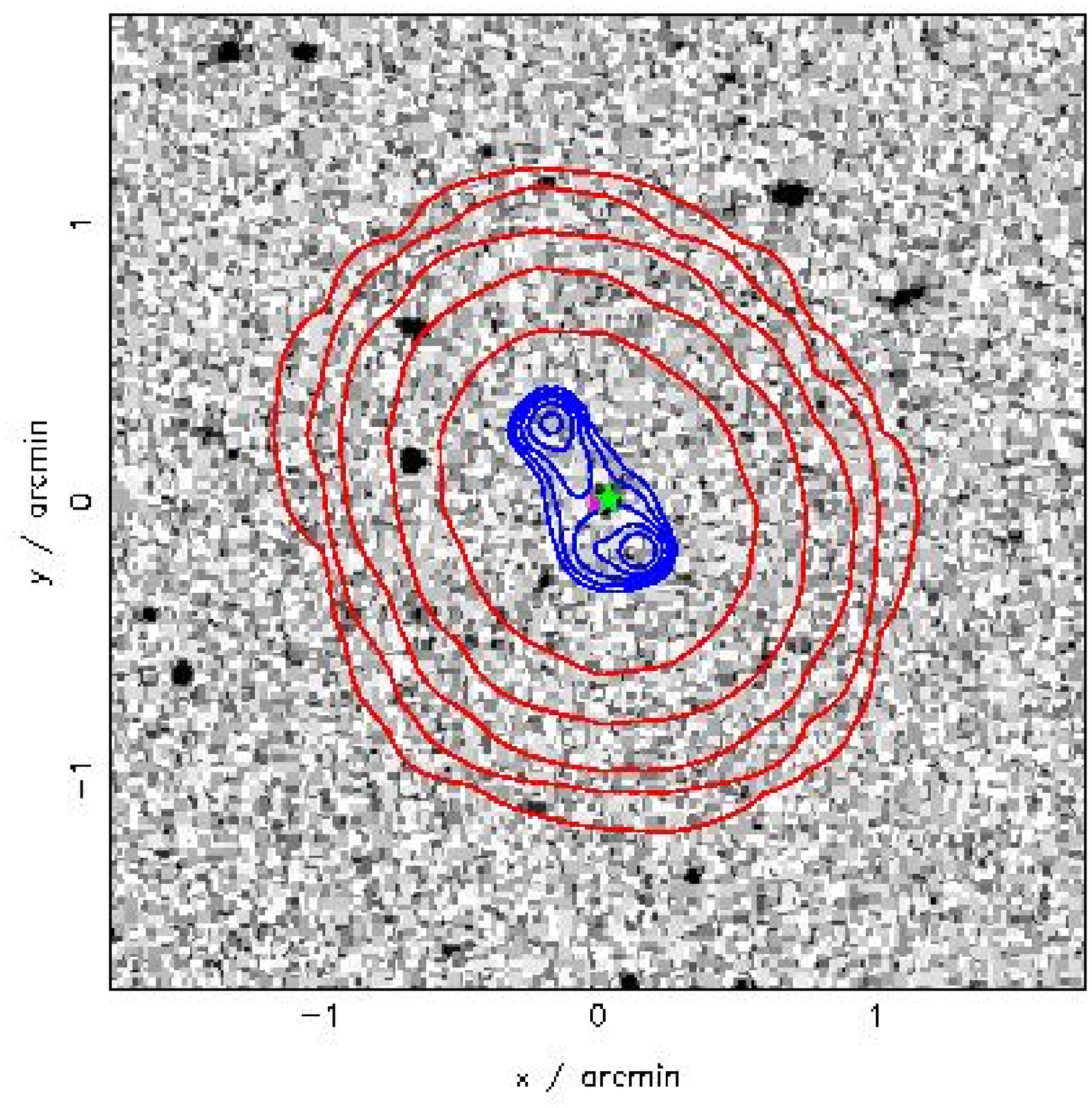}}
      \centerline{C2-037: 4C 22.25}
    \end{minipage}
    \hspace{3cm}
    \begin{minipage}{3cm}
      \mbox{}
      \centerline{\includegraphics[scale=0.26,angle=270]{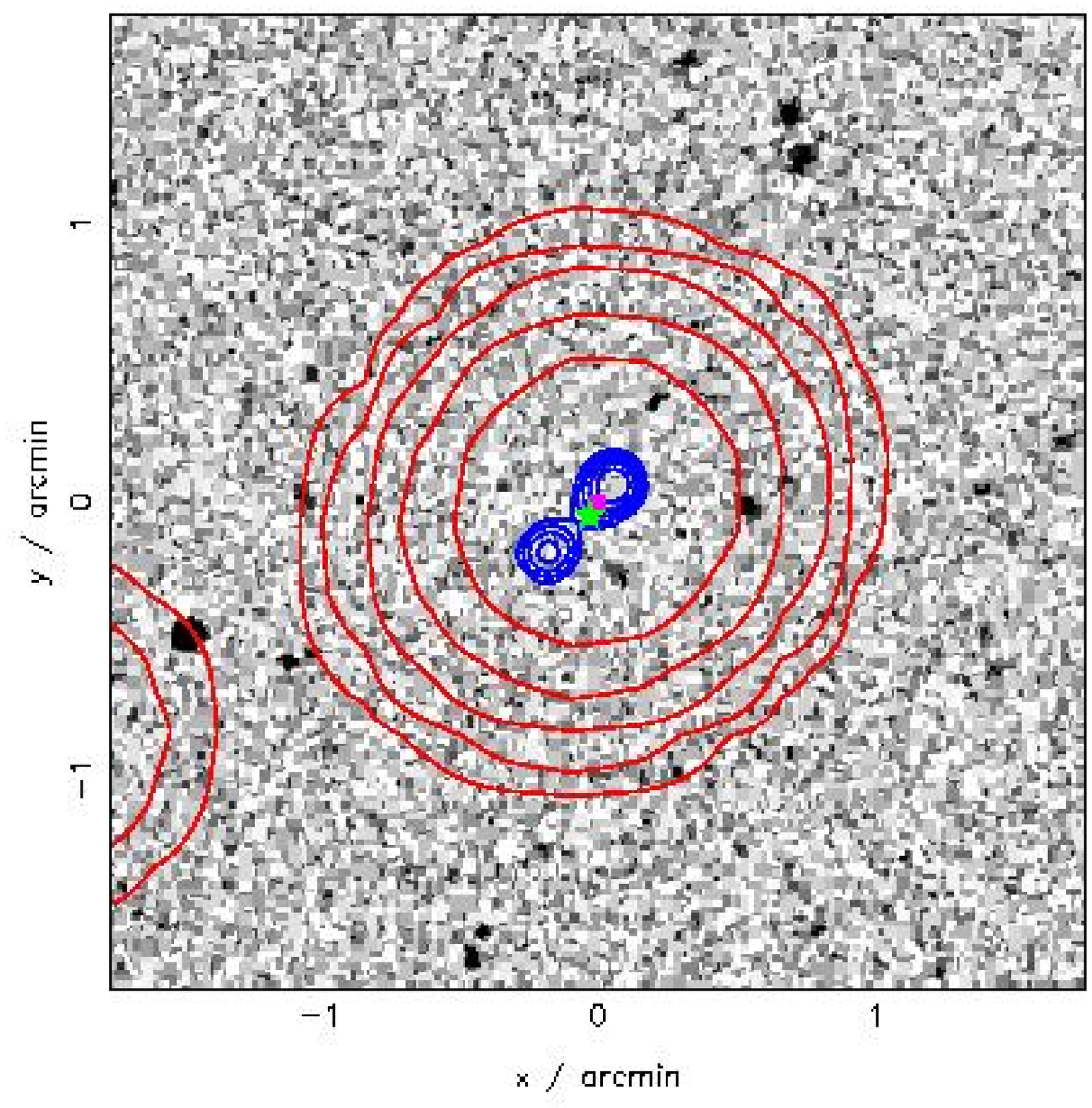}}
      \centerline{C2-038: 4C 14.35}
    \end{minipage}
  \end{center}
\end{figure}

\begin{figure}
  \begin{center}
    {\bf CoNFIG-2}\\  
  \begin{minipage}{3cm}      
      \mbox{}
      \centerline{\includegraphics[scale=0.26,angle=270]{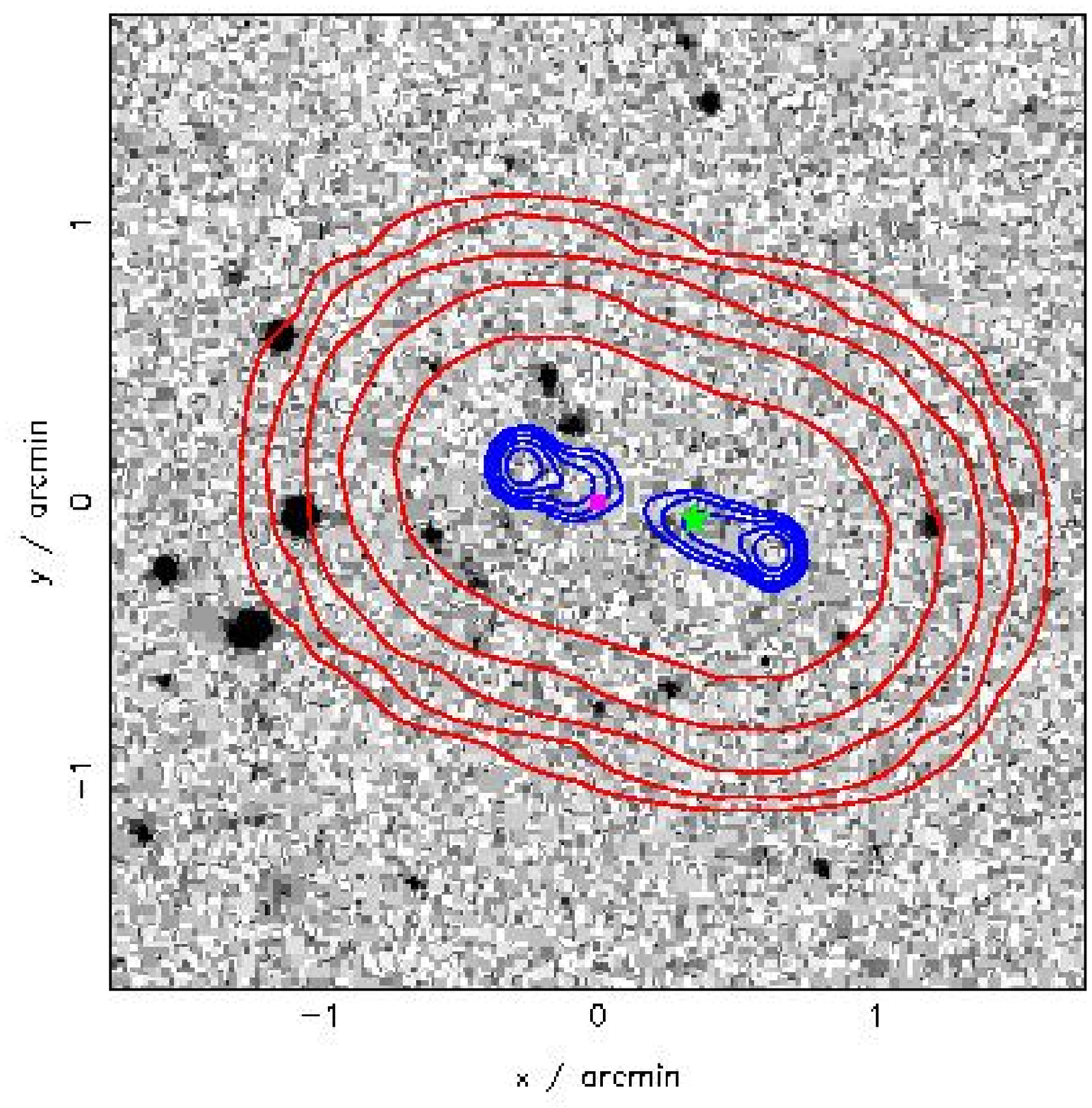}}
      \centerline{C2-039: 4C -00.37}
    \end{minipage}
    \hspace{3cm}
    \begin{minipage}{3cm}
      \mbox{}
      \centerline{\includegraphics[scale=0.26,angle=270]{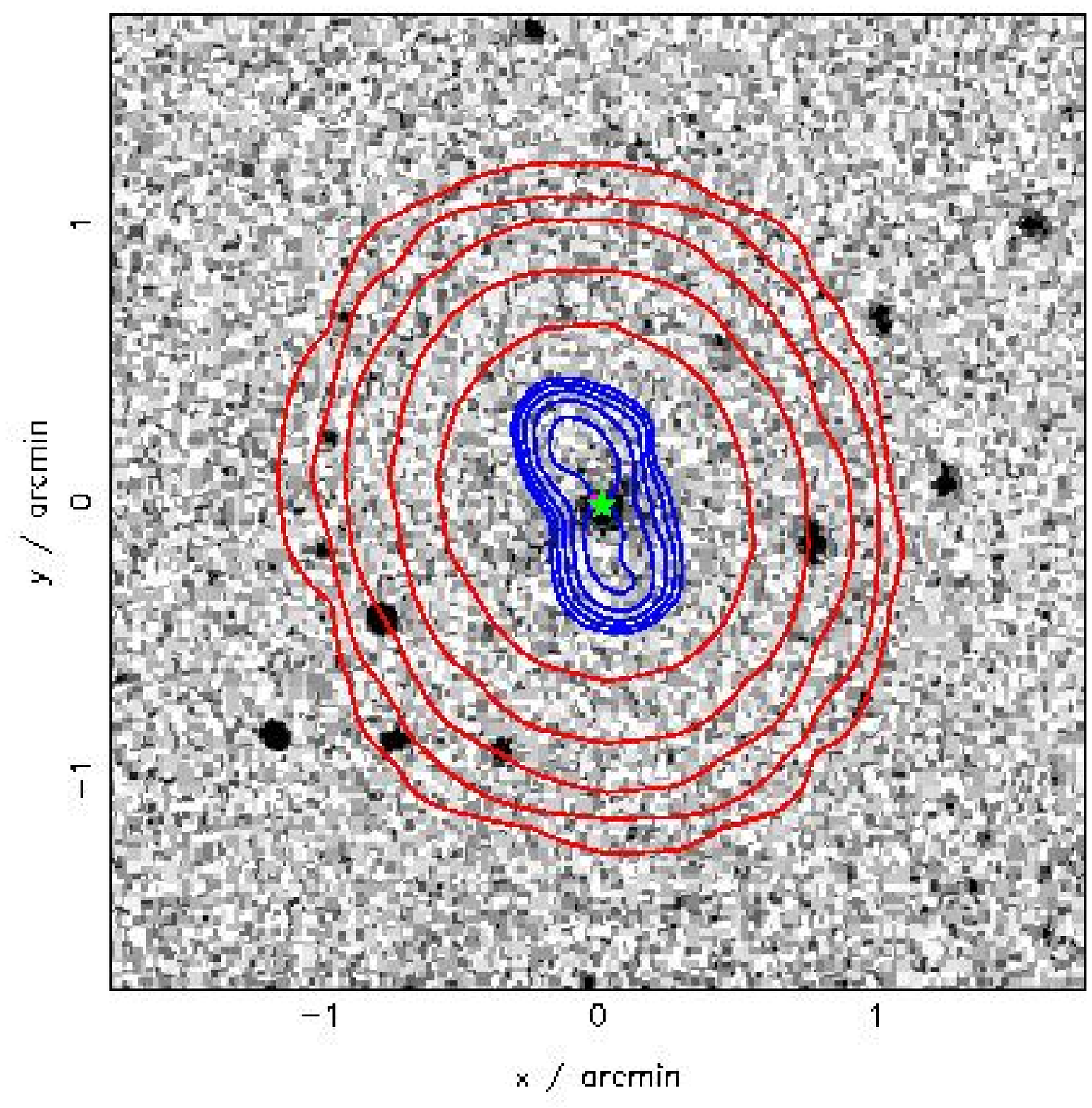}}
      \centerline{C2-041: 4C 20.20}
    \end{minipage}
    \hspace{3cm}
    \begin{minipage}{3cm}
      \mbox{}
      \centerline{\includegraphics[scale=0.26,angle=270]{Contours/C2/042.ps}}
      \centerline{C2-042: 4C 32.34}
    \end{minipage}
    \vfill
    \begin{minipage}{3cm}     
      \mbox{}
      \centerline{\includegraphics[scale=0.26,angle=270]{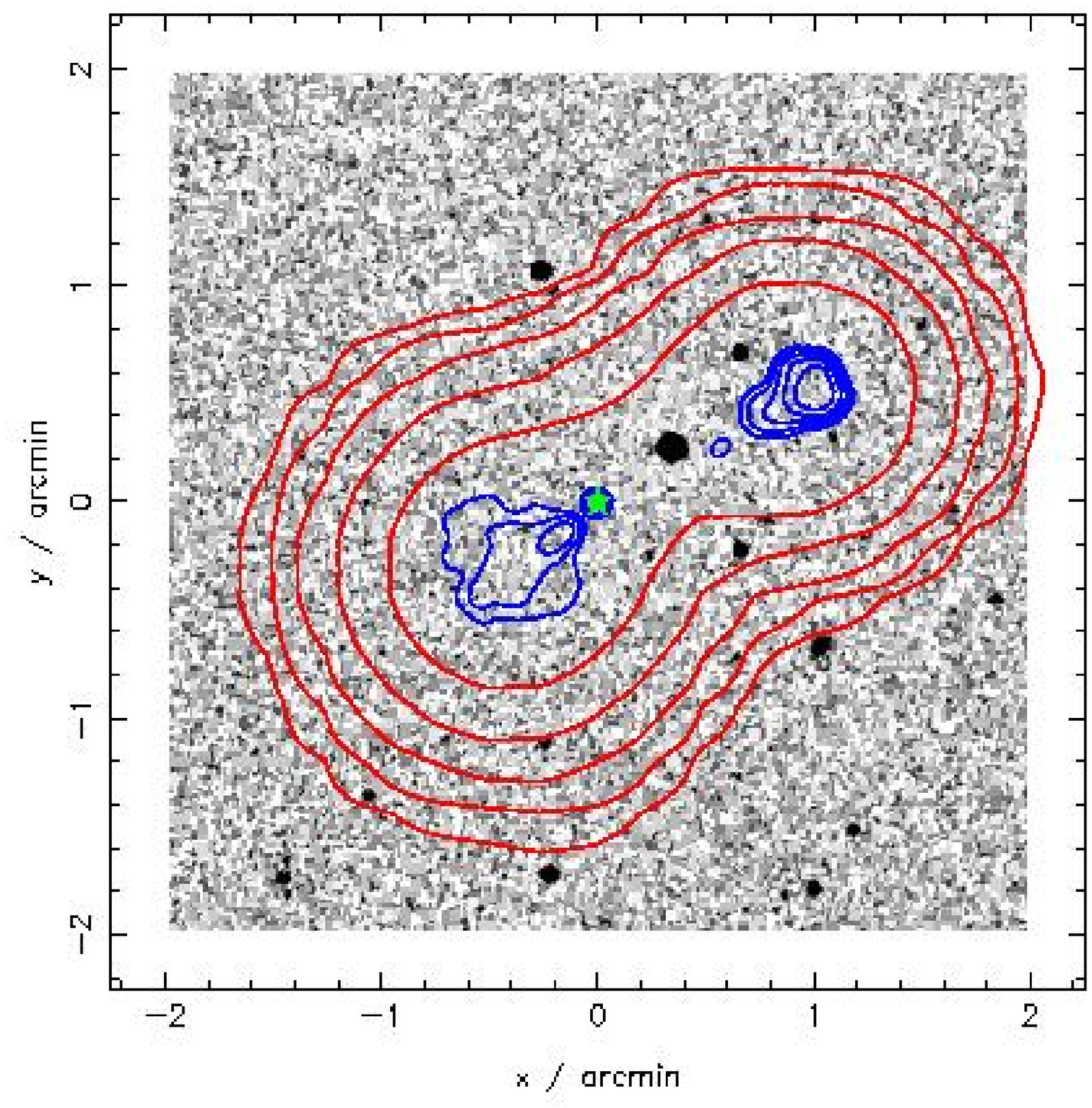}}
      \centerline{C2-045: 4C 13.41}
    \end{minipage}
    \hspace{3cm}
    \begin{minipage}{3cm}
      \mbox{}
      \centerline{\includegraphics[scale=0.26,angle=270]{Contours/C2/047.ps}}
      \centerline{C2-047: 4C 59.11}
    \end{minipage}
    \hspace{3cm}
    \begin{minipage}{3cm}
      \mbox{}
      \centerline{\includegraphics[scale=0.26,angle=270]{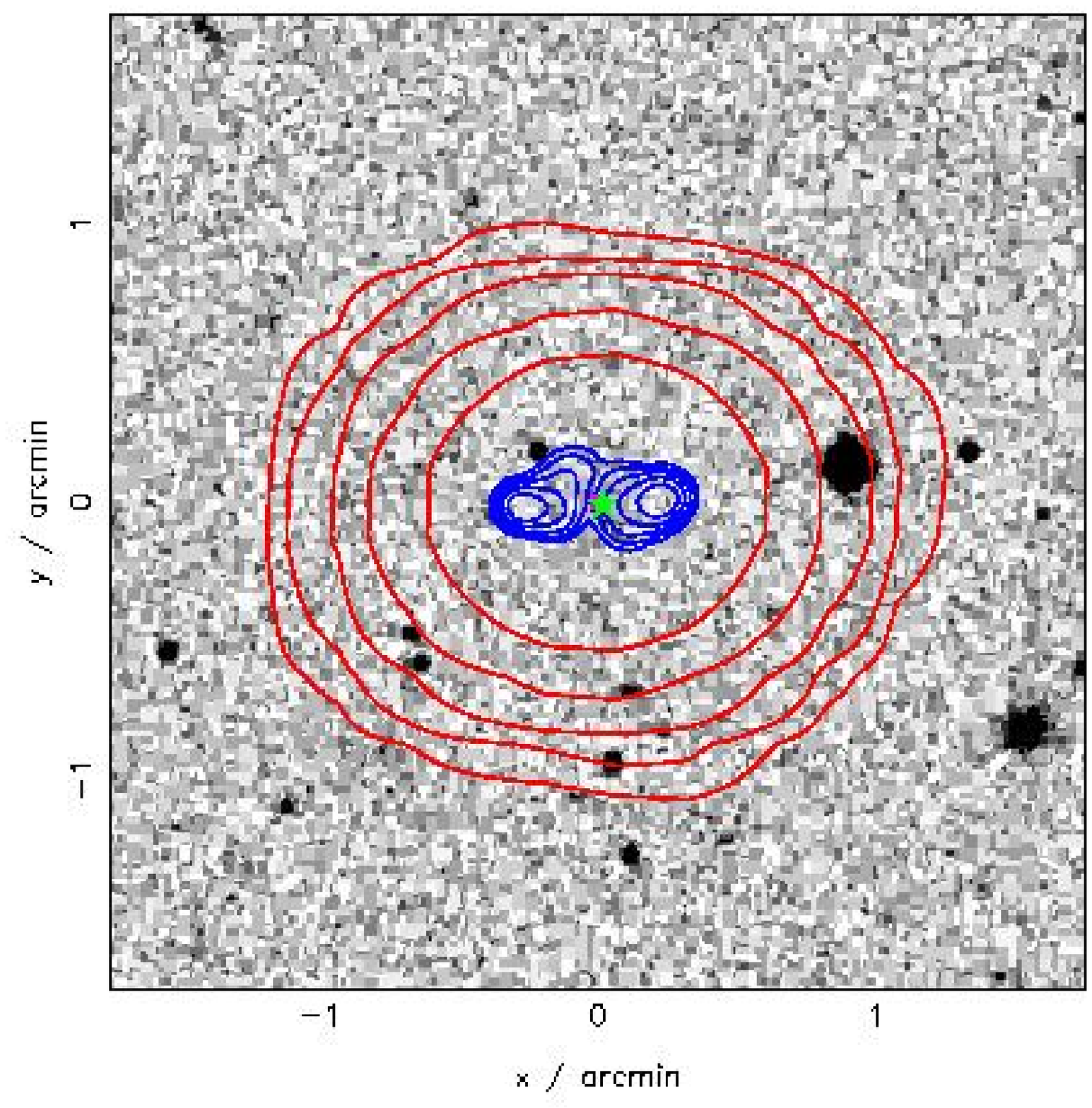}}
      \centerline{C2-052: 4C 11.34}
    \end{minipage}
    \vfill
    \begin{minipage}{3cm}     
      \mbox{}
      \centerline{\includegraphics[scale=0.26,angle=270]{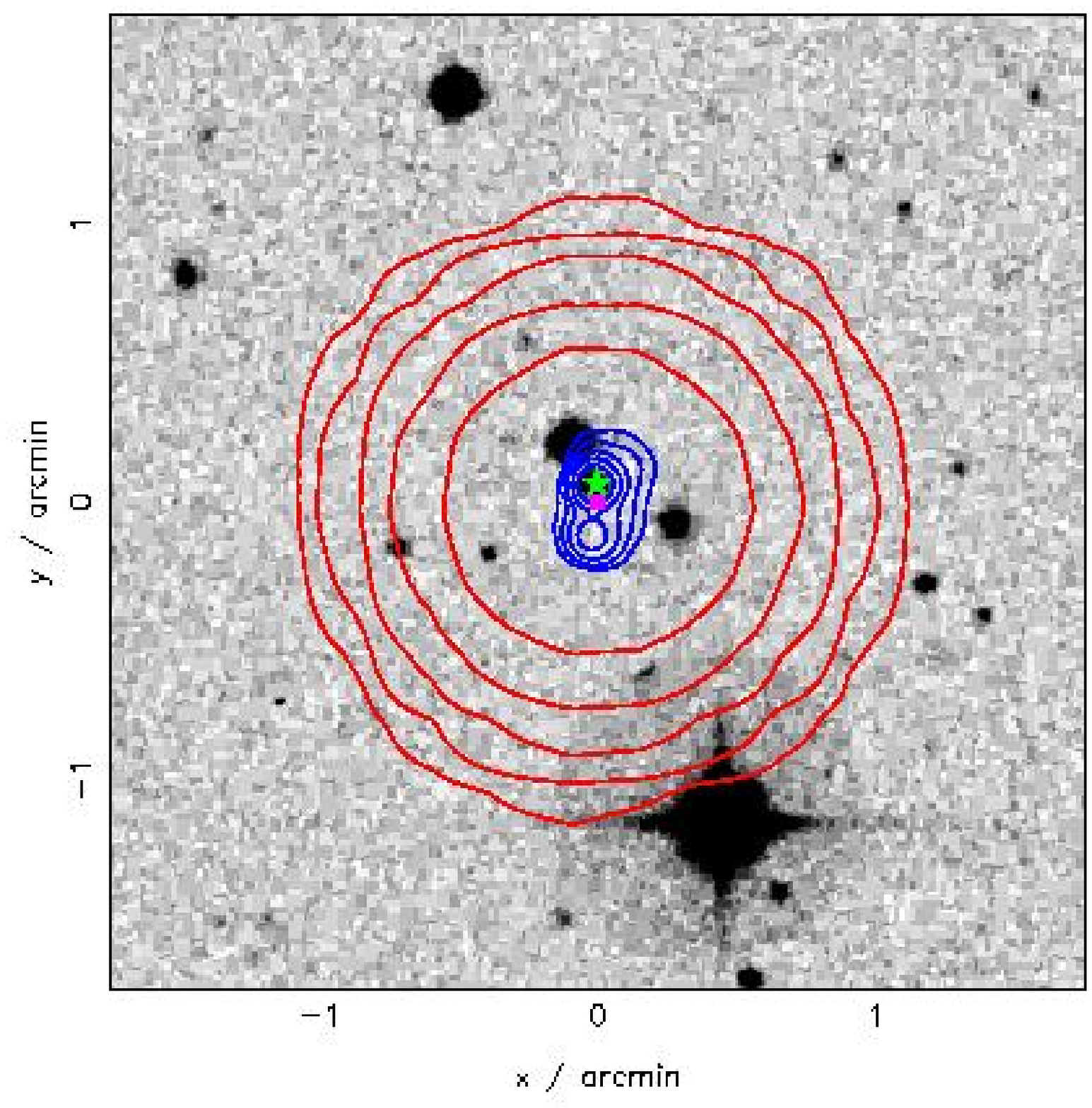}}
      \centerline{C2-053: 4C 23.24}
    \end{minipage}
    \hspace{3cm}
    \begin{minipage}{3cm}
      \mbox{}
      \centerline{\includegraphics[scale=0.26,angle=270]{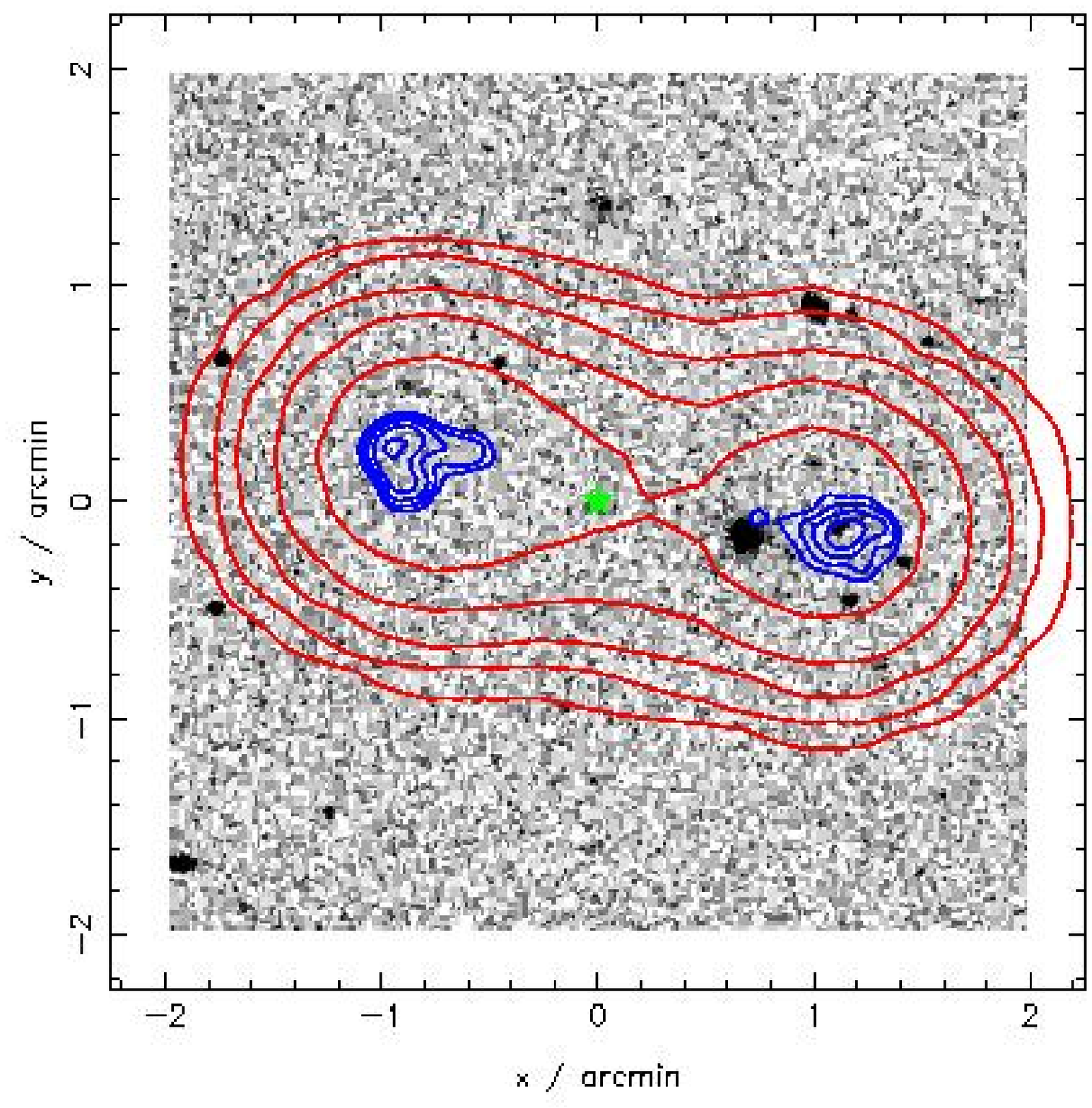}}
      \centerline{C2-055: 4C 41.22}
    \end{minipage}
    \hspace{3cm}
    \begin{minipage}{3cm}
      \mbox{}
      \centerline{\includegraphics[scale=0.26,angle=270]{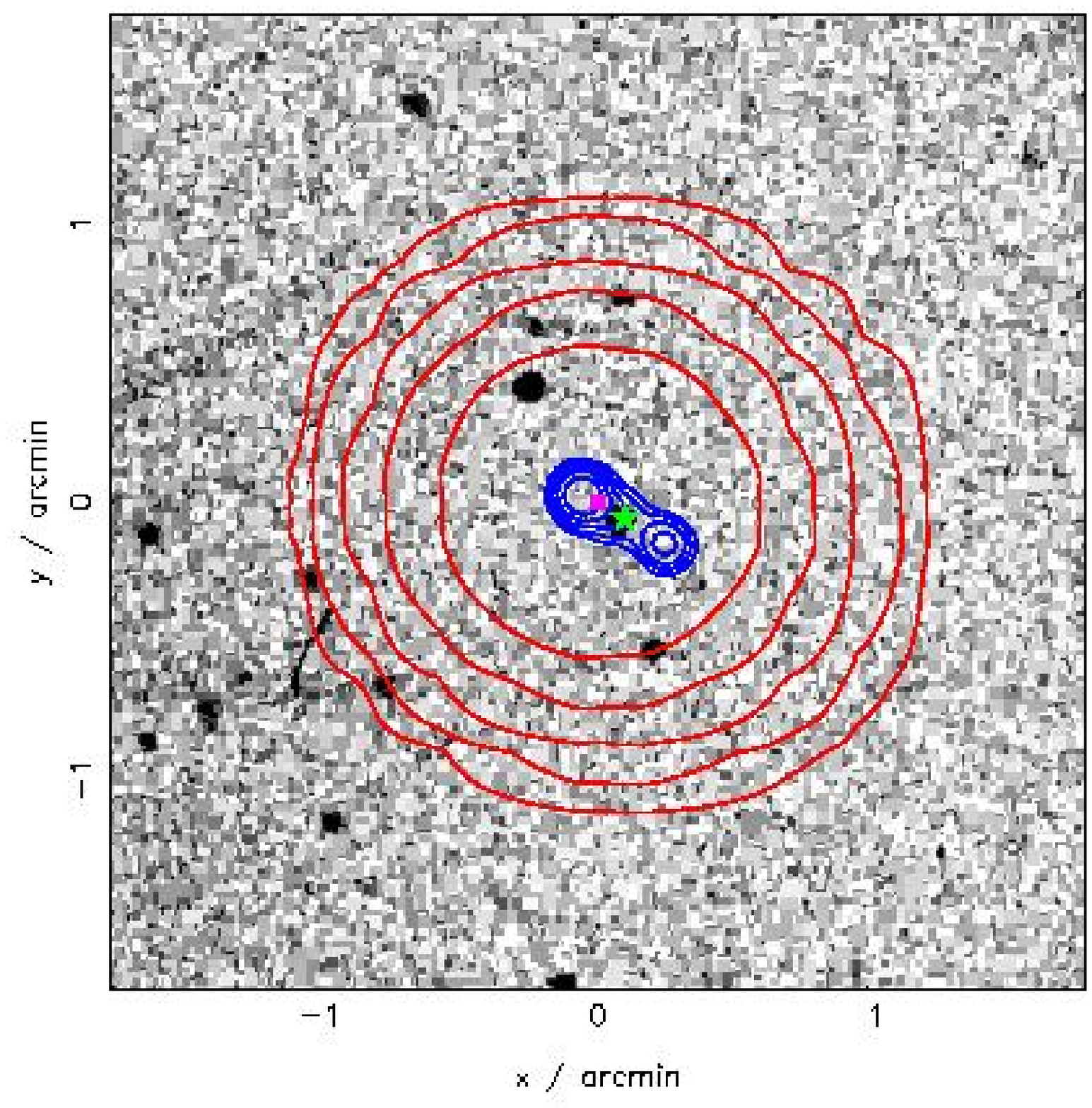}}
      \centerline{C2-057: 3C 240}
    \end{minipage}
    \vfill
    \begin{minipage}{3cm}      
      \mbox{}
      \centerline{\includegraphics[scale=0.26,angle=270]{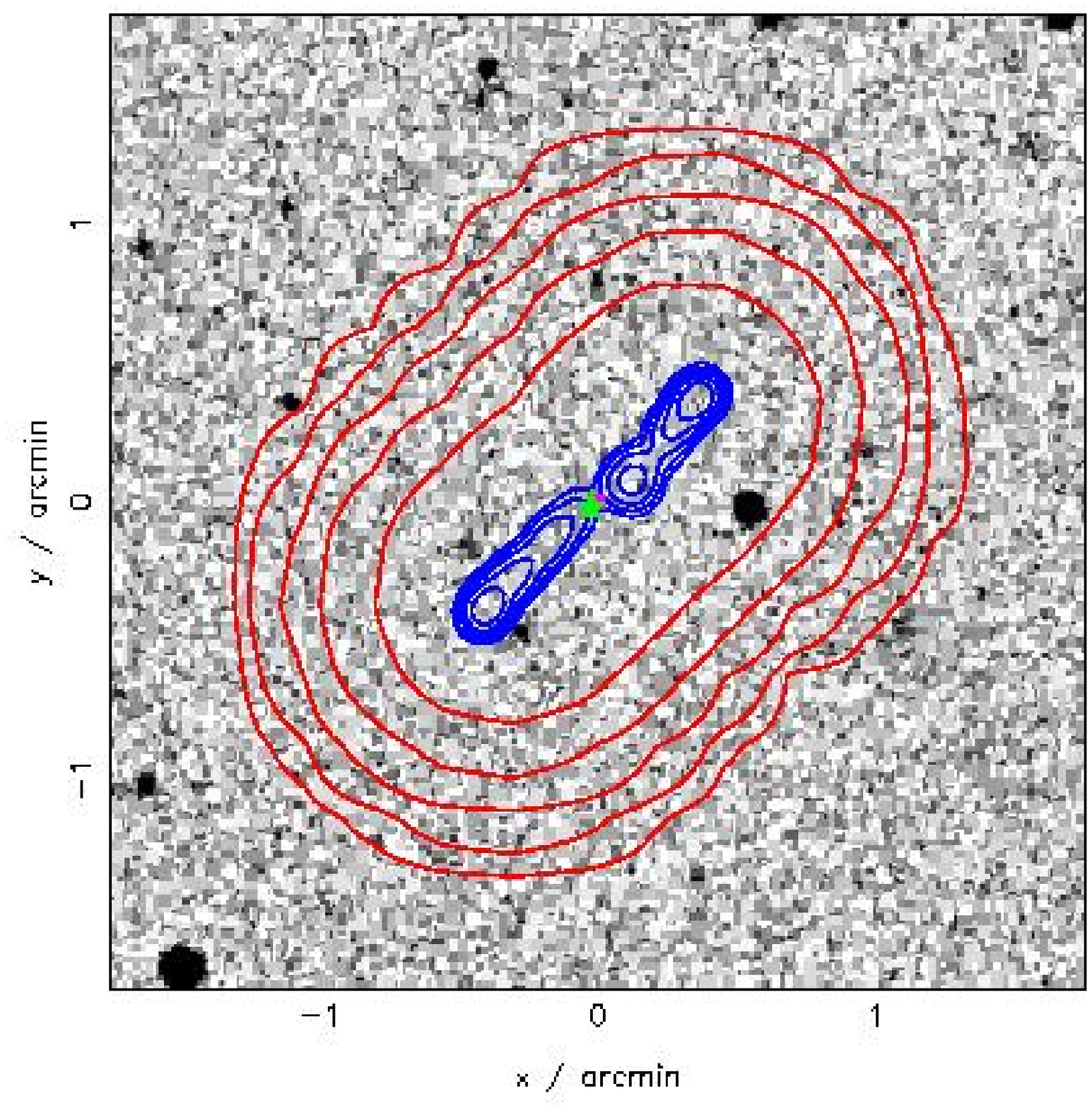}}
      \centerline{C2-063: 3C 242}
    \end{minipage}
    \hspace{3cm}
    \begin{minipage}{3cm}
      \mbox{}
      \centerline{\includegraphics[scale=0.26,angle=270]{Contours/C2/064.ps}}
      \centerline{C2-064: 4C 43.19}
    \end{minipage}
    \hspace{3cm}
    \begin{minipage}{3cm}
      \mbox{}
      \centerline{\includegraphics[scale=0.26,angle=270]{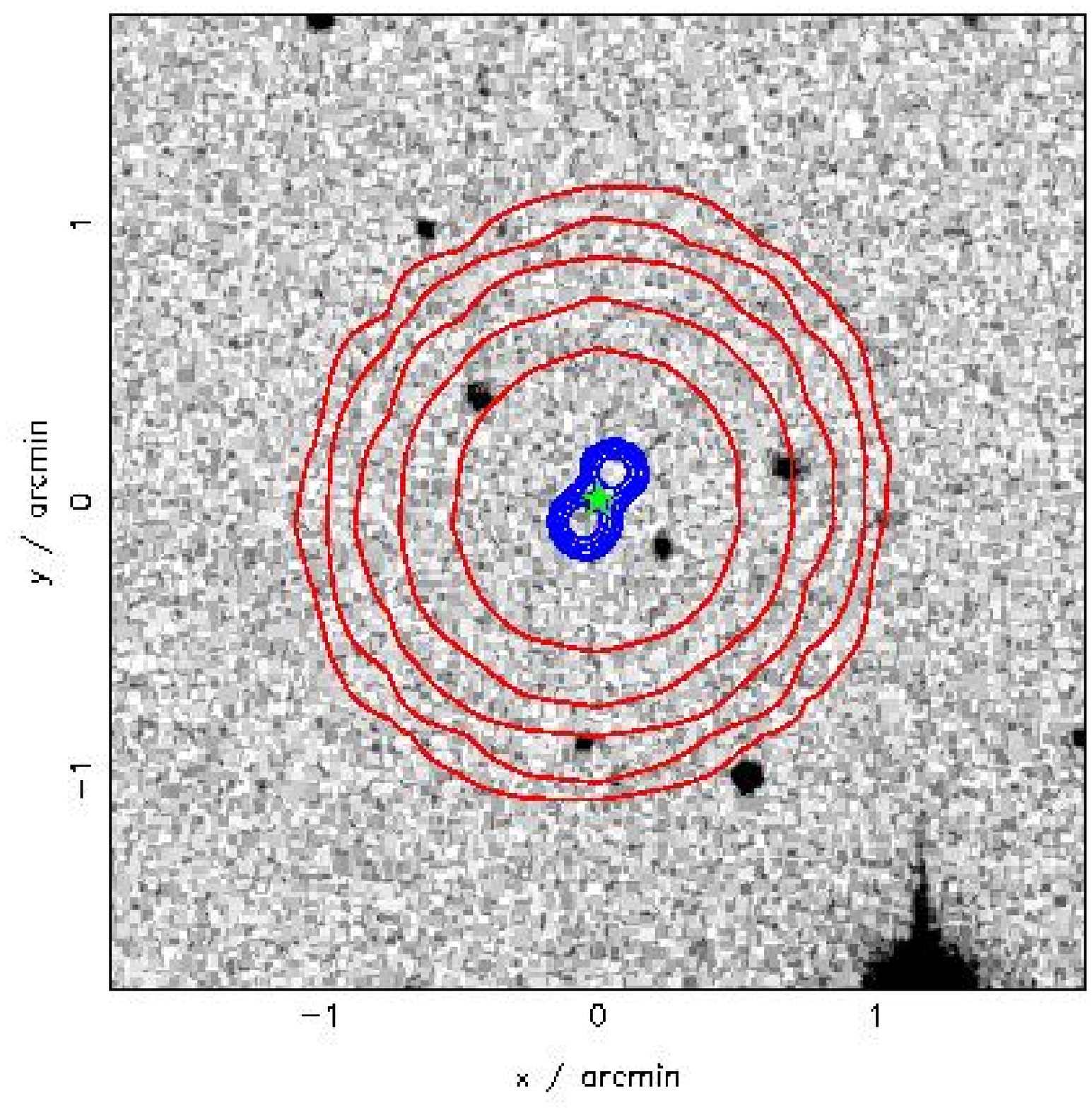}}
      \centerline{C2-065: 3C 243}
    \end{minipage}
  \end{center}
\end{figure}

\begin{figure}
  \begin{center}
    {\bf CoNFIG-2}\\  
  \begin{minipage}{3cm}      
      \mbox{}
      \centerline{\includegraphics[scale=0.26,angle=270]{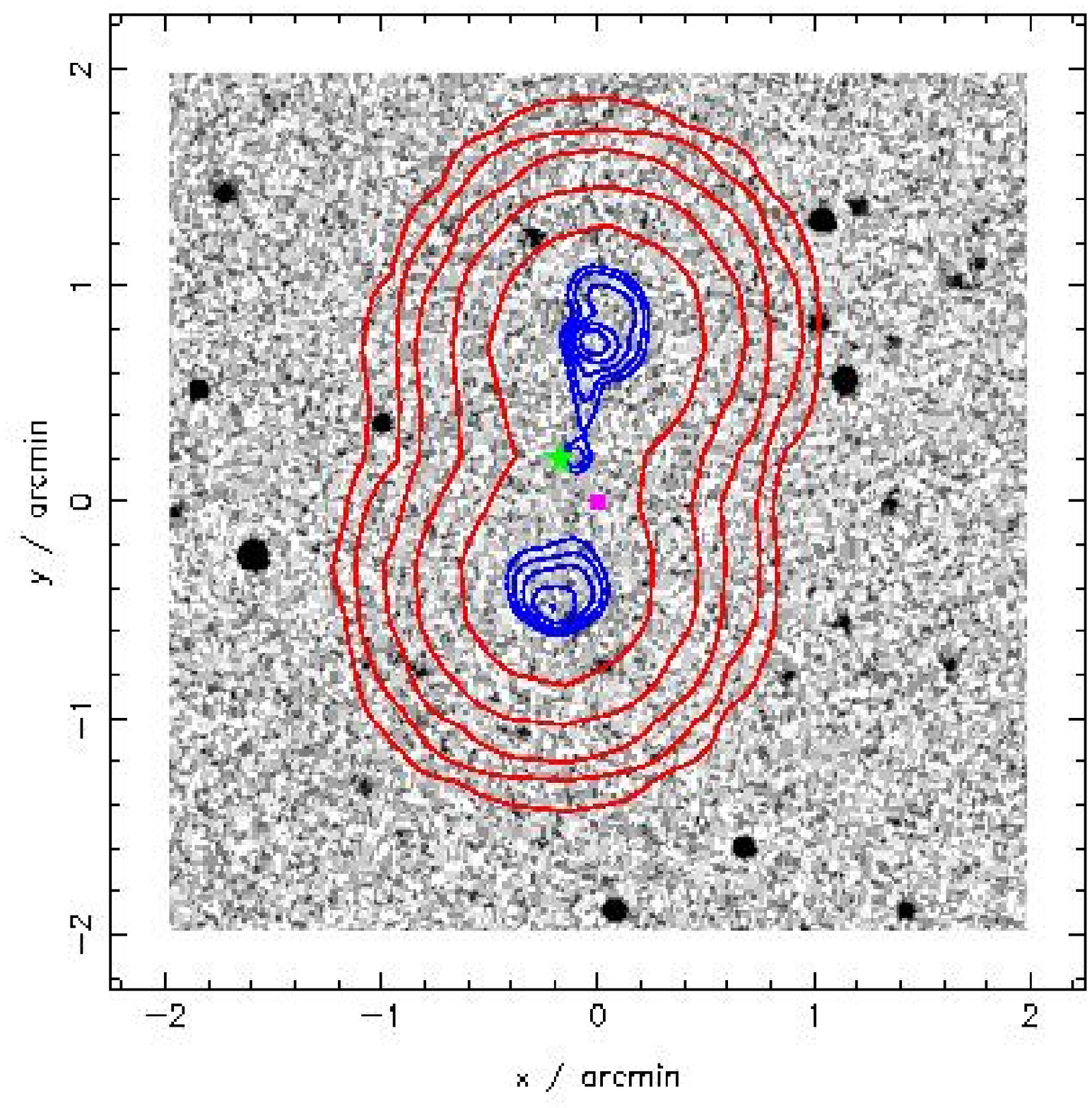}}
      \centerline{C2-067: 3C 244}
    \end{minipage}
    \hspace{3cm}
    \begin{minipage}{3cm}
      \mbox{}
      \centerline{\includegraphics[scale=0.26,angle=270]{Contours/C2/069.ps}}
      \centerline{C2-069: 4C 00.35}
    \end{minipage}
    \hspace{3cm}
    \begin{minipage}{3cm}
      \mbox{}
      \centerline{\includegraphics[scale=0.26,angle=270]{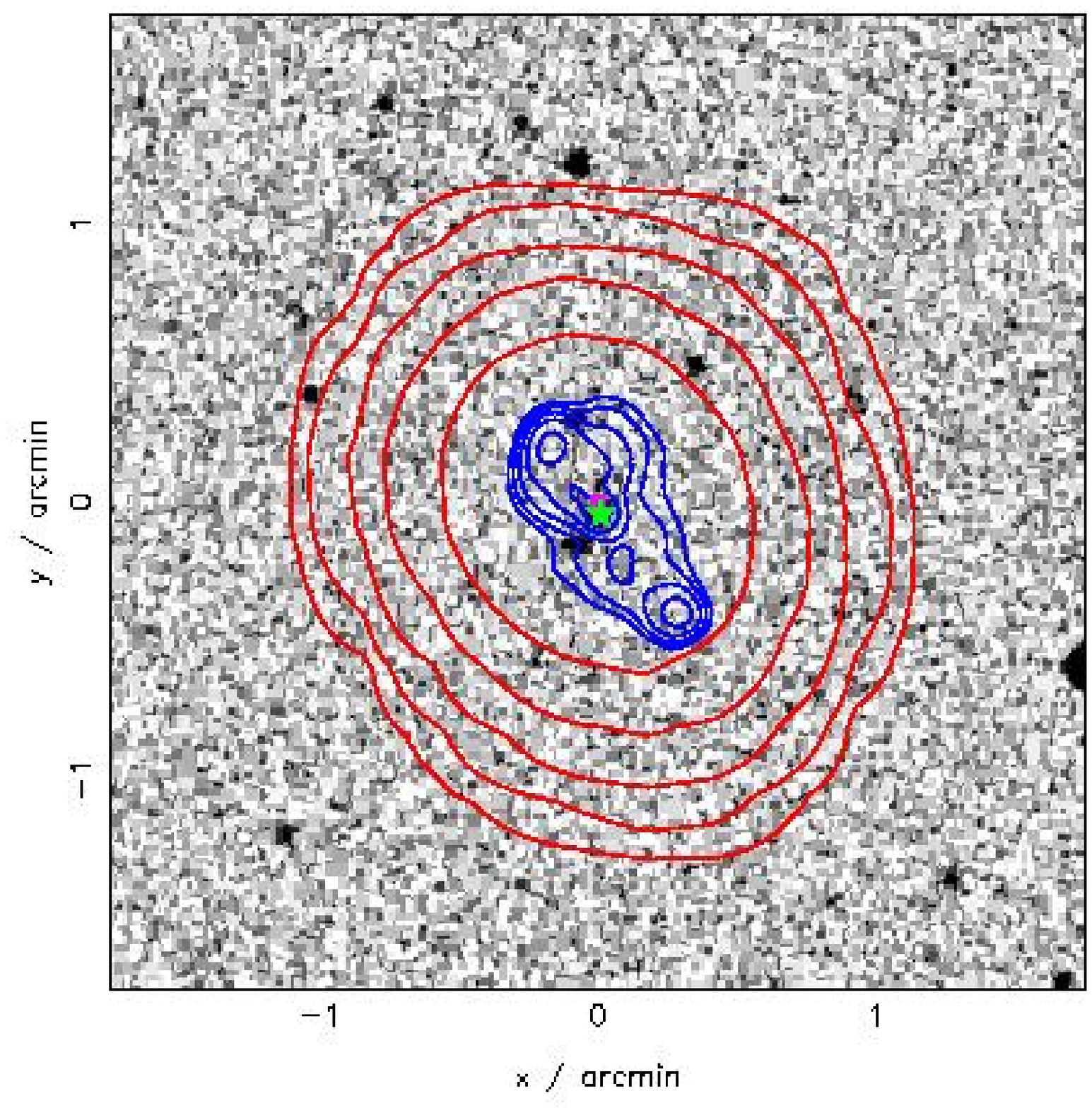}}
      \centerline{C2-070: 4C 52.22}
    \end{minipage}
    \vfill
    \begin{minipage}{3cm}     
      \mbox{}
      \centerline{\includegraphics[scale=0.26,angle=270]{Contours/C2/071.ps}}
      \centerline{C2-071: 4C 17.50}
    \end{minipage}
    \hspace{3cm}
    \begin{minipage}{3cm}
      \mbox{}
      \centerline{\includegraphics[scale=0.26,angle=270]{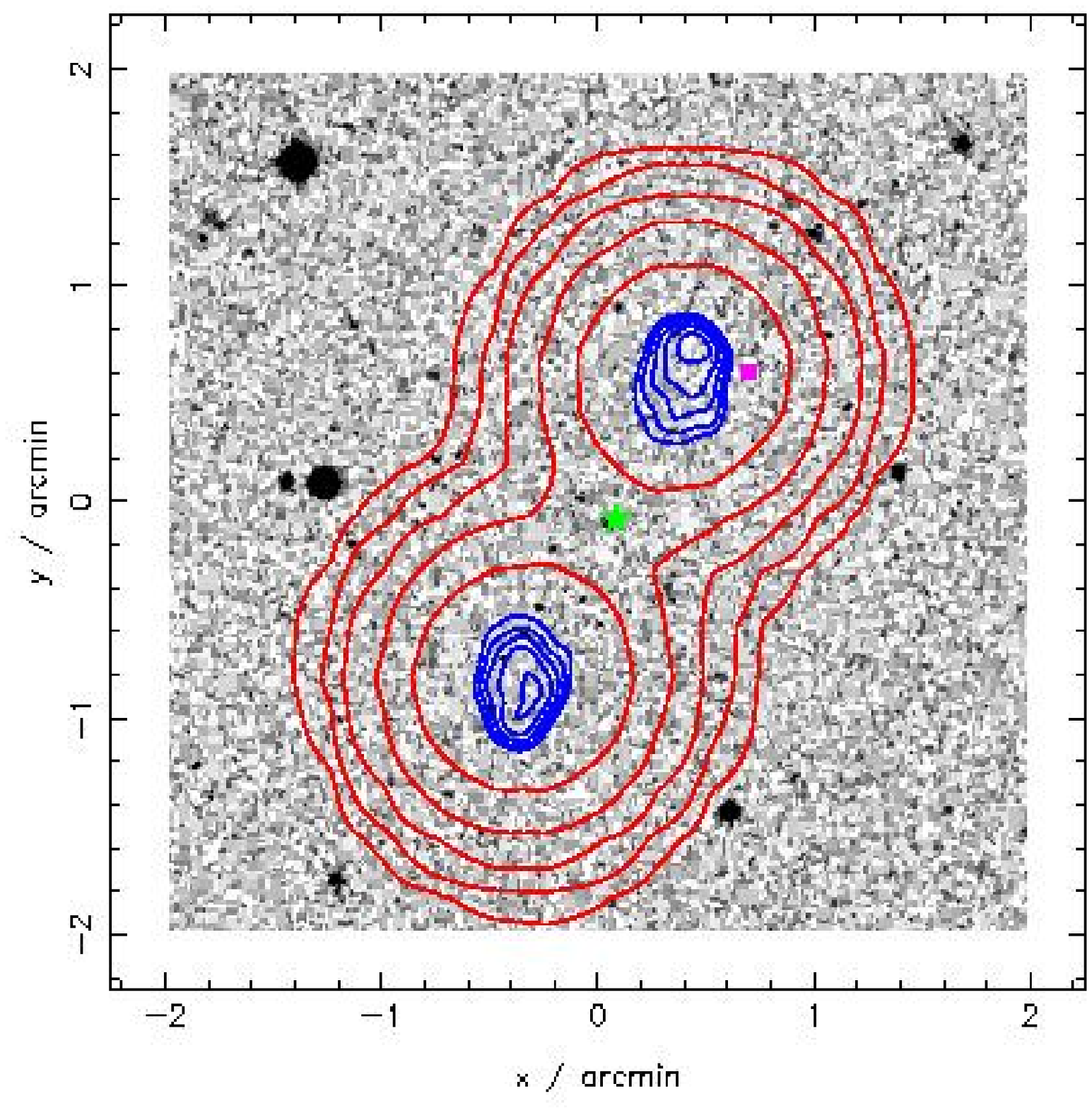}}
      \centerline{C2-079: 4C 55.21}
    \end{minipage}
    \hspace{3cm}
    \begin{minipage}{3cm}
      \mbox{}
      \centerline{\includegraphics[scale=0.26,angle=270]{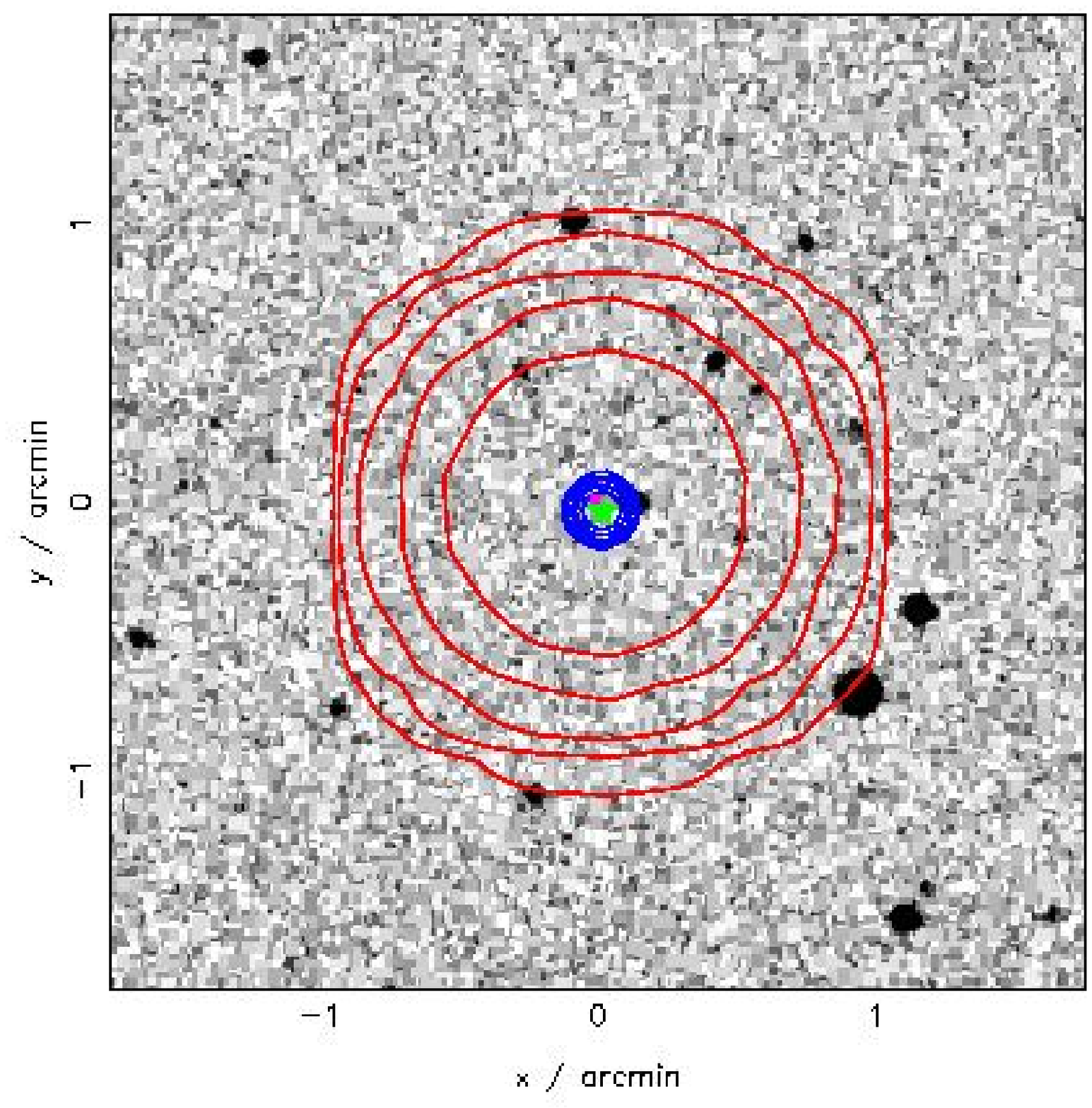}}
      \centerline{C2-080: 4C 15.34}
    \end{minipage}
    \vfill
    \begin{minipage}{3cm}     
      \mbox{}
      \centerline{\includegraphics[scale=0.26,angle=270]{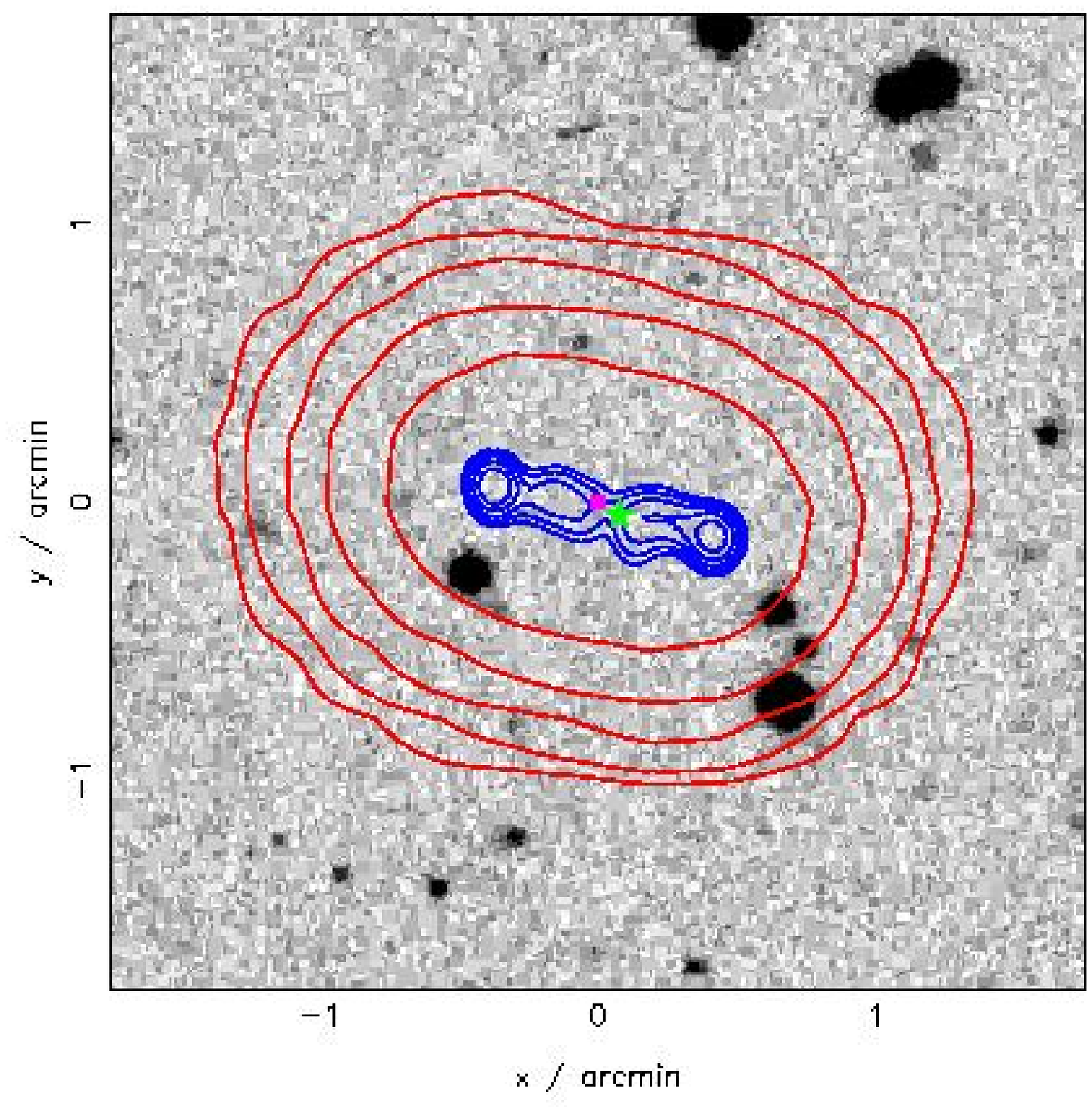}}
      \centerline{C2-082: 4C -02.43}
    \end{minipage}
    \hspace{3cm}
    \begin{minipage}{3cm}
      \mbox{}
      \centerline{\includegraphics[scale=0.26,angle=270]{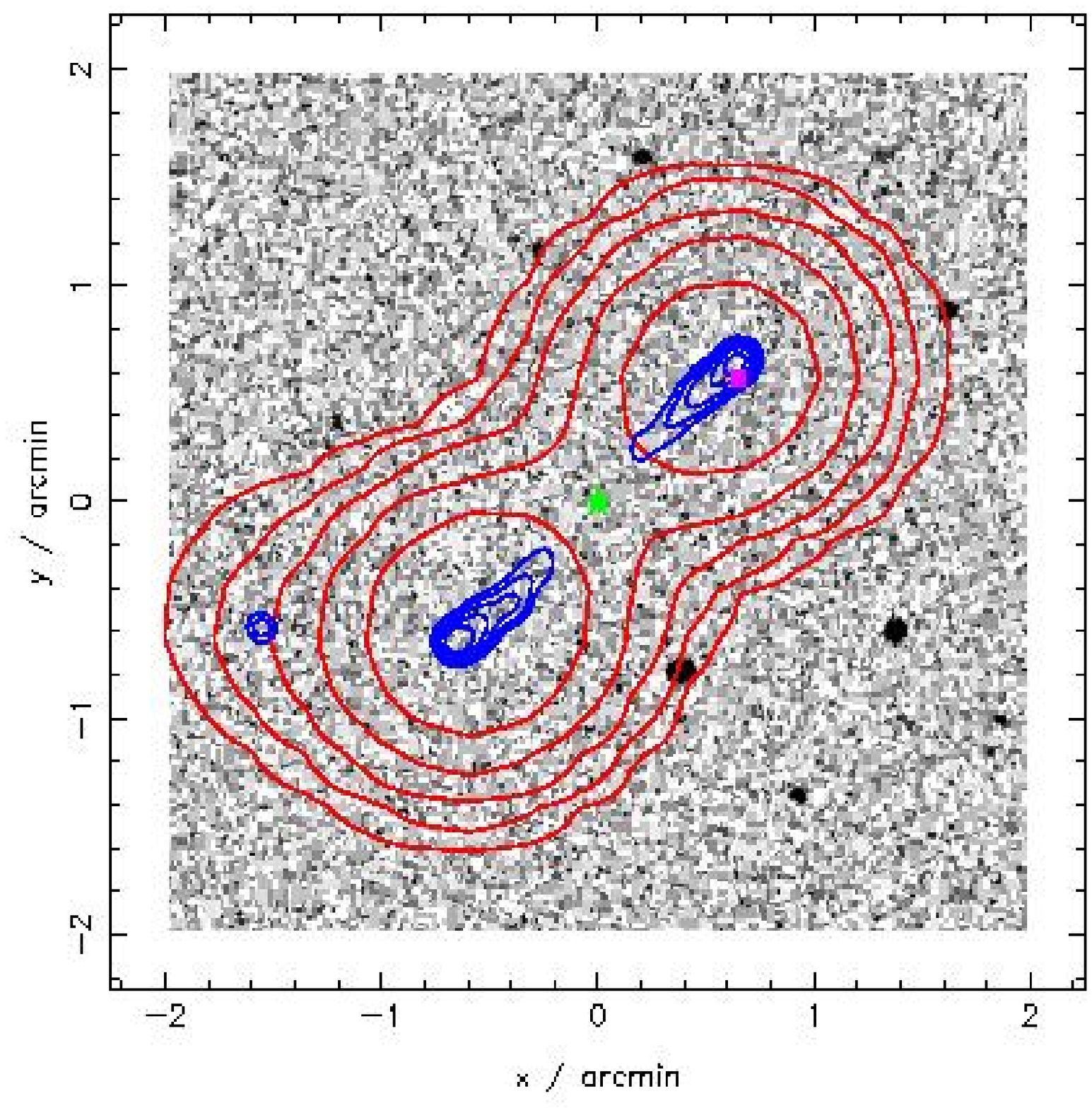}}
      \centerline{C2-089: 3C 248}
    \end{minipage}
    \hspace{3cm}
    \begin{minipage}{3cm}
      \mbox{}
      \centerline{\includegraphics[scale=0.26,angle=270]{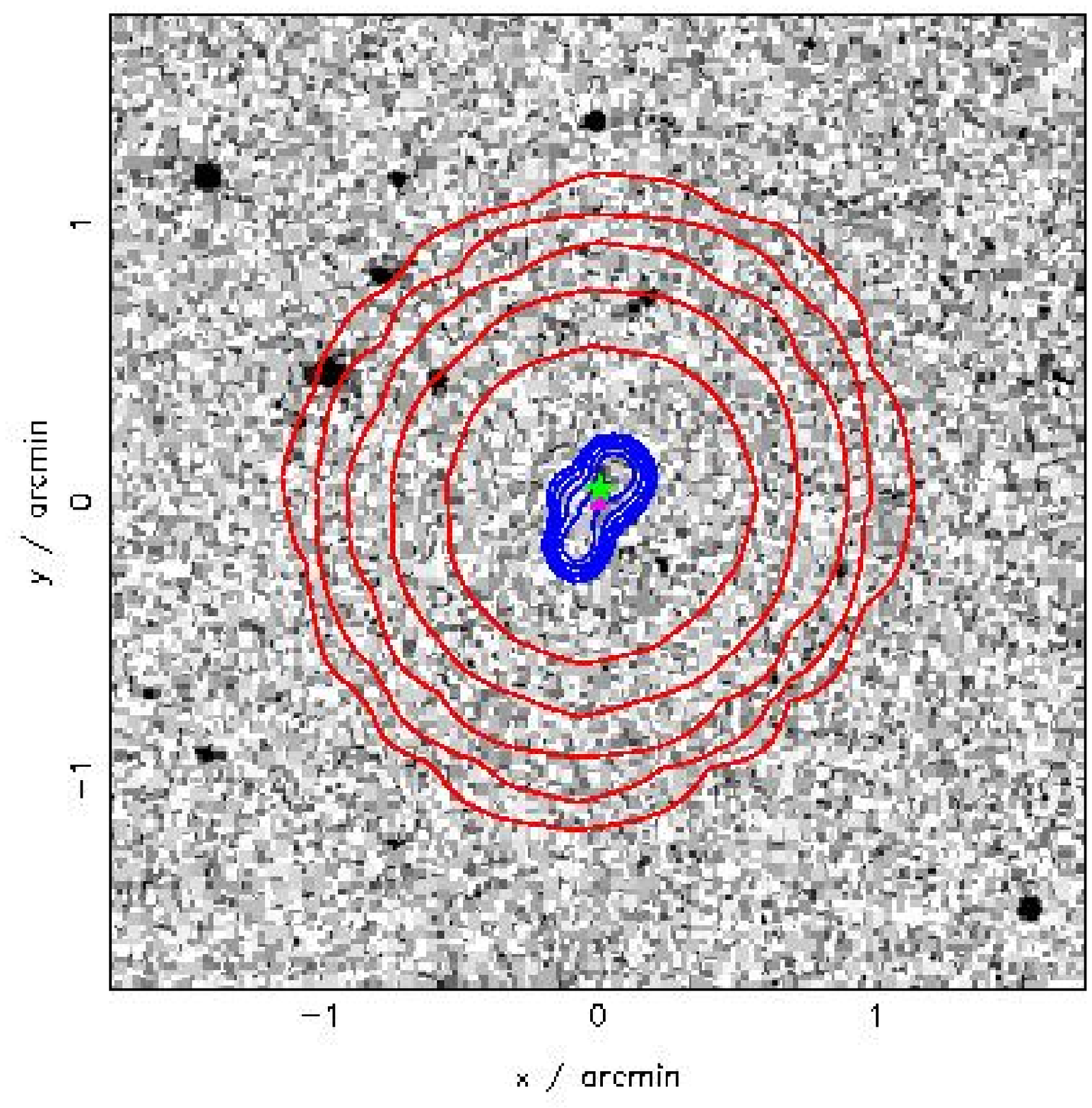}}
      \centerline{C2-092: 4C 56.18}
    \end{minipage}
    \vfill
    \begin{minipage}{3cm}      
      \mbox{}
      \centerline{\includegraphics[scale=0.26,angle=270]{Contours/C2/094.ps}}
      \centerline{C2-094: 4C -00.43}
    \end{minipage}
    \hspace{3cm}
    \begin{minipage}{3cm}
      \mbox{}
      \centerline{\includegraphics[scale=0.26,angle=270]{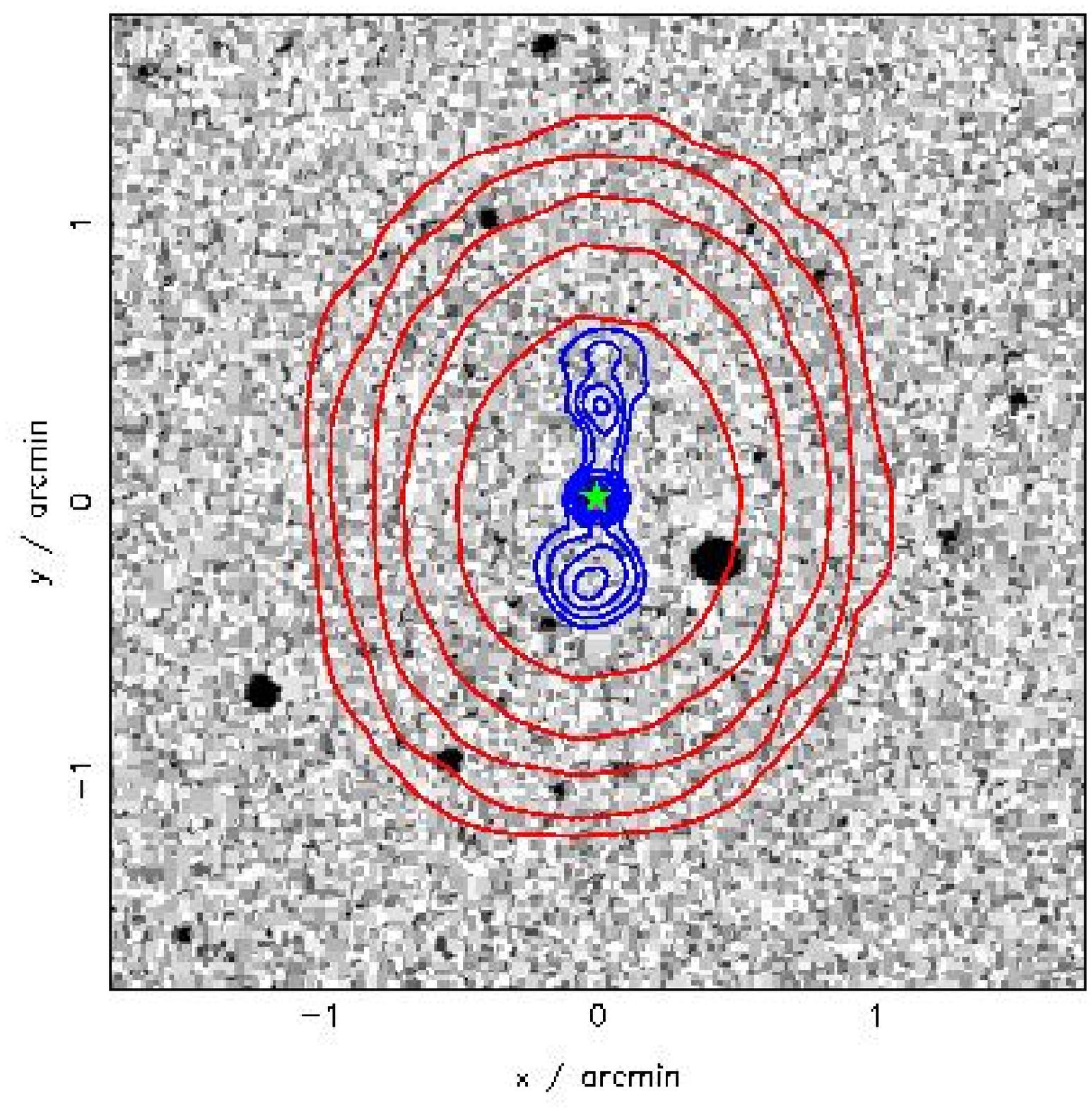}}
      \centerline{C2-095: 4C 16.30}
    \end{minipage}
    \hspace{3cm}
    \begin{minipage}{3cm}
      \mbox{}
      \centerline{\includegraphics[scale=0.26,angle=270]{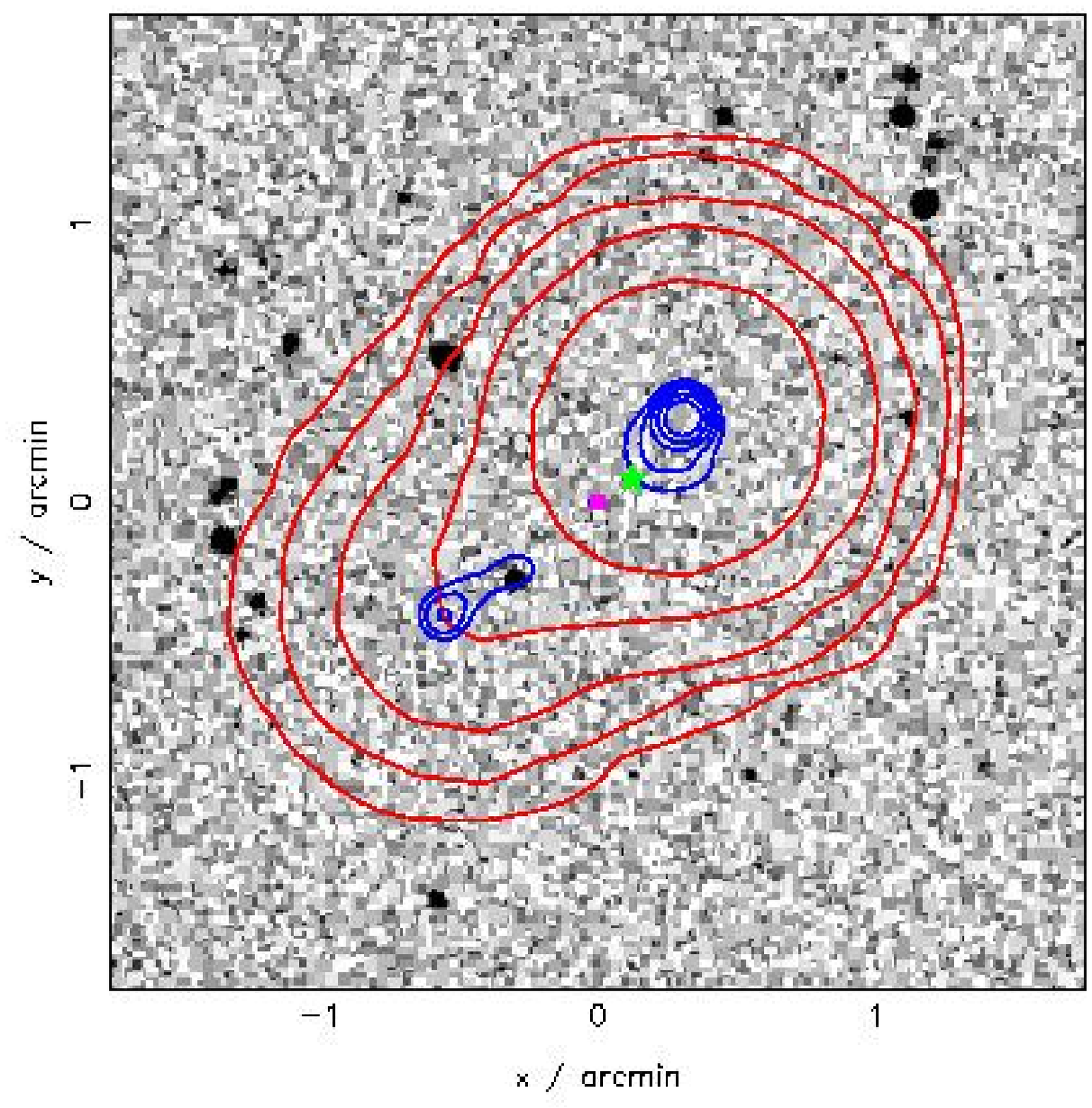}}
      \centerline{C2-097: 3C 251}
    \end{minipage}
  \end{center}
\end{figure}

\begin{figure}
  \begin{center}
    {\bf CoNFIG-2}\\  
  \begin{minipage}{3cm}      
      \mbox{}
      \centerline{\includegraphics[scale=0.26,angle=270]{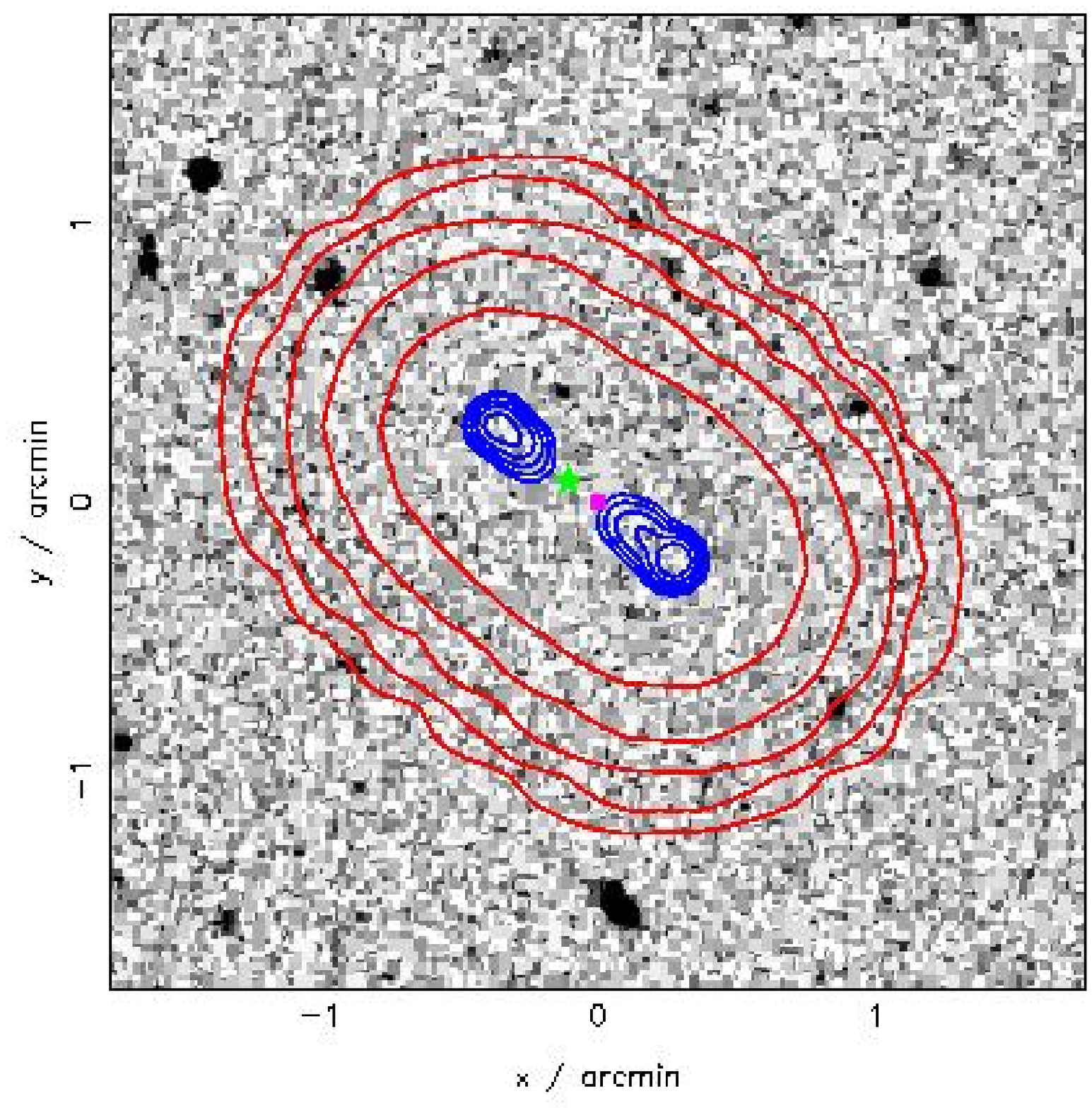}}
      \centerline{C2-098: 3C 250}
    \end{minipage}
    \hspace{3cm}
    \begin{minipage}{3cm}
      \mbox{}
      \centerline{\includegraphics[scale=0.26,angle=270]{Contours/C2/103.ps}}
      \centerline{C2-103: 4C 03.21}
    \end{minipage}
    \hspace{3cm}
    \begin{minipage}{3cm}
      \mbox{}
      \centerline{\includegraphics[scale=0.26,angle=270]{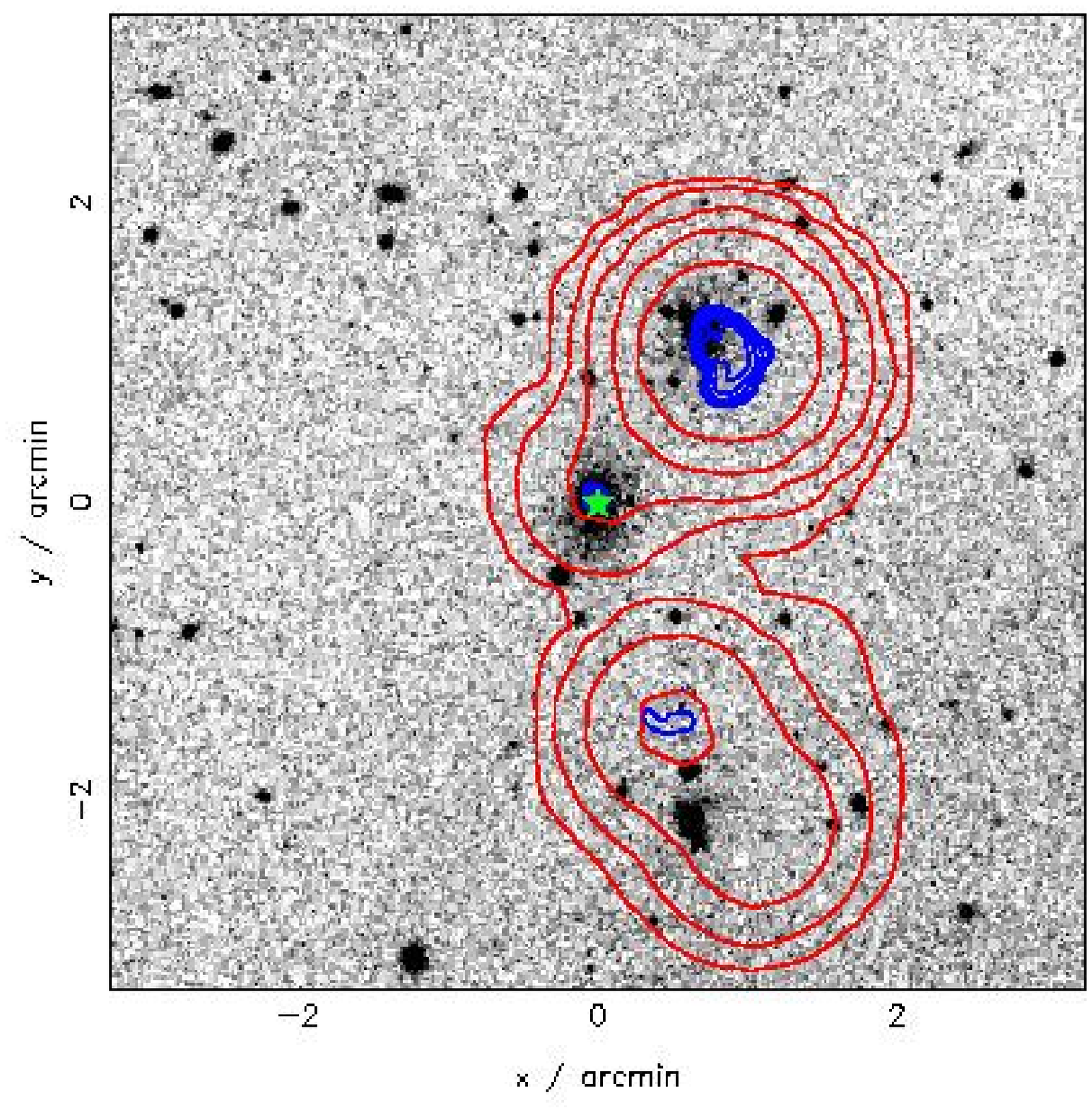}}
      \centerline{C2-105: 4C 41.23}
    \end{minipage}
    \vfill
    \begin{minipage}{3cm}     
      \mbox{}
      \centerline{\includegraphics[scale=0.26,angle=270]{Contours/C2/110.ps}}
      \centerline{C2-110: 4C -02.46}
    \end{minipage}
    \hspace{3cm}
    \begin{minipage}{3cm}
      \mbox{}
      \centerline{\includegraphics[scale=0.26,angle=270]{Contours/C2/111.ps}}
      \centerline{C2-111: TXS 1115+536}
    \end{minipage}
    \hspace{3cm}
    \begin{minipage}{3cm}
      \mbox{}
      \centerline{\includegraphics[scale=0.26,angle=270]{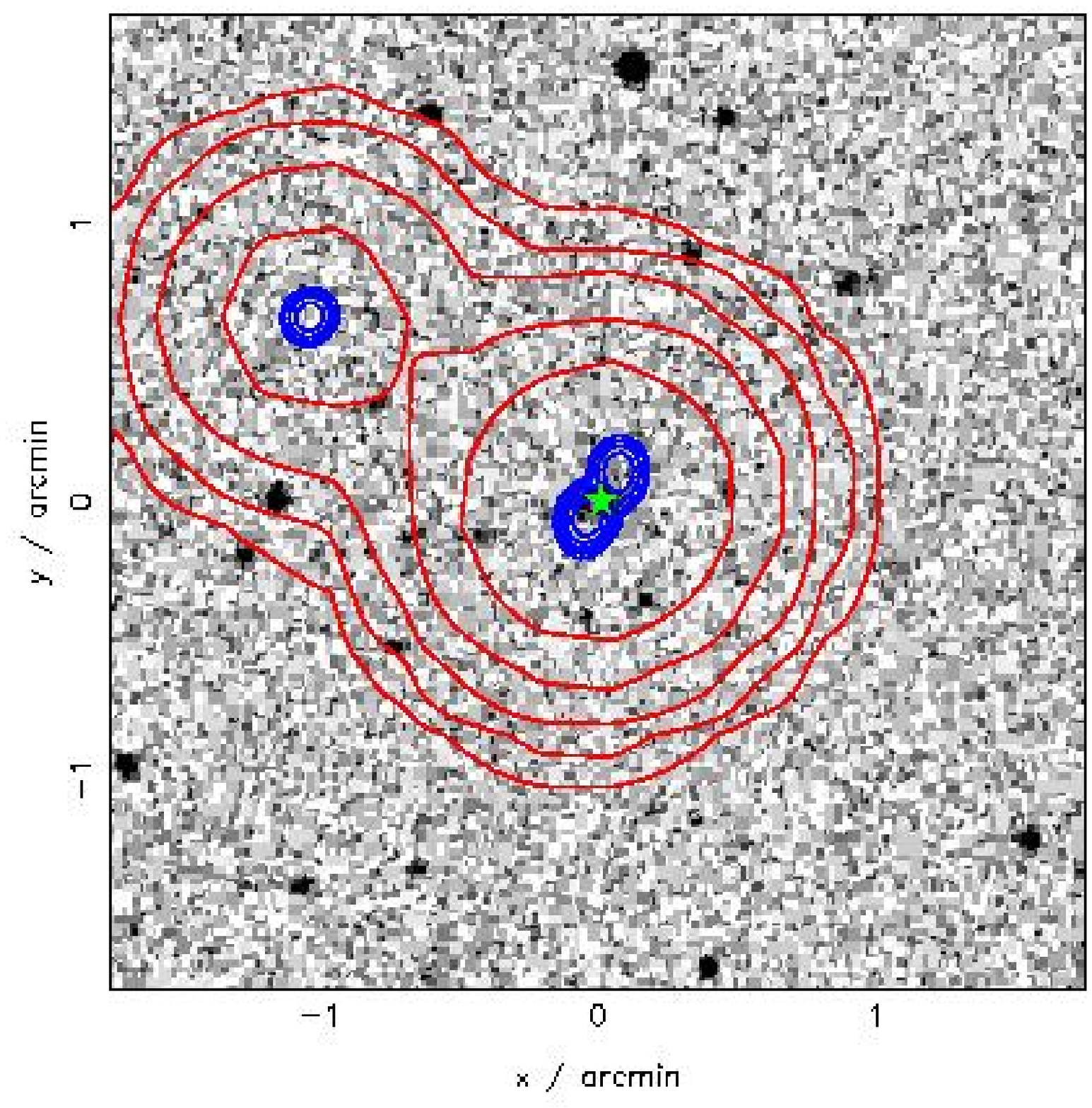}}
      \centerline{C2-117: 4C 05.50}
    \end{minipage}
    \vfill
    \begin{minipage}{3cm}     
      \mbox{}
      \centerline{\includegraphics[scale=0.26,angle=270]{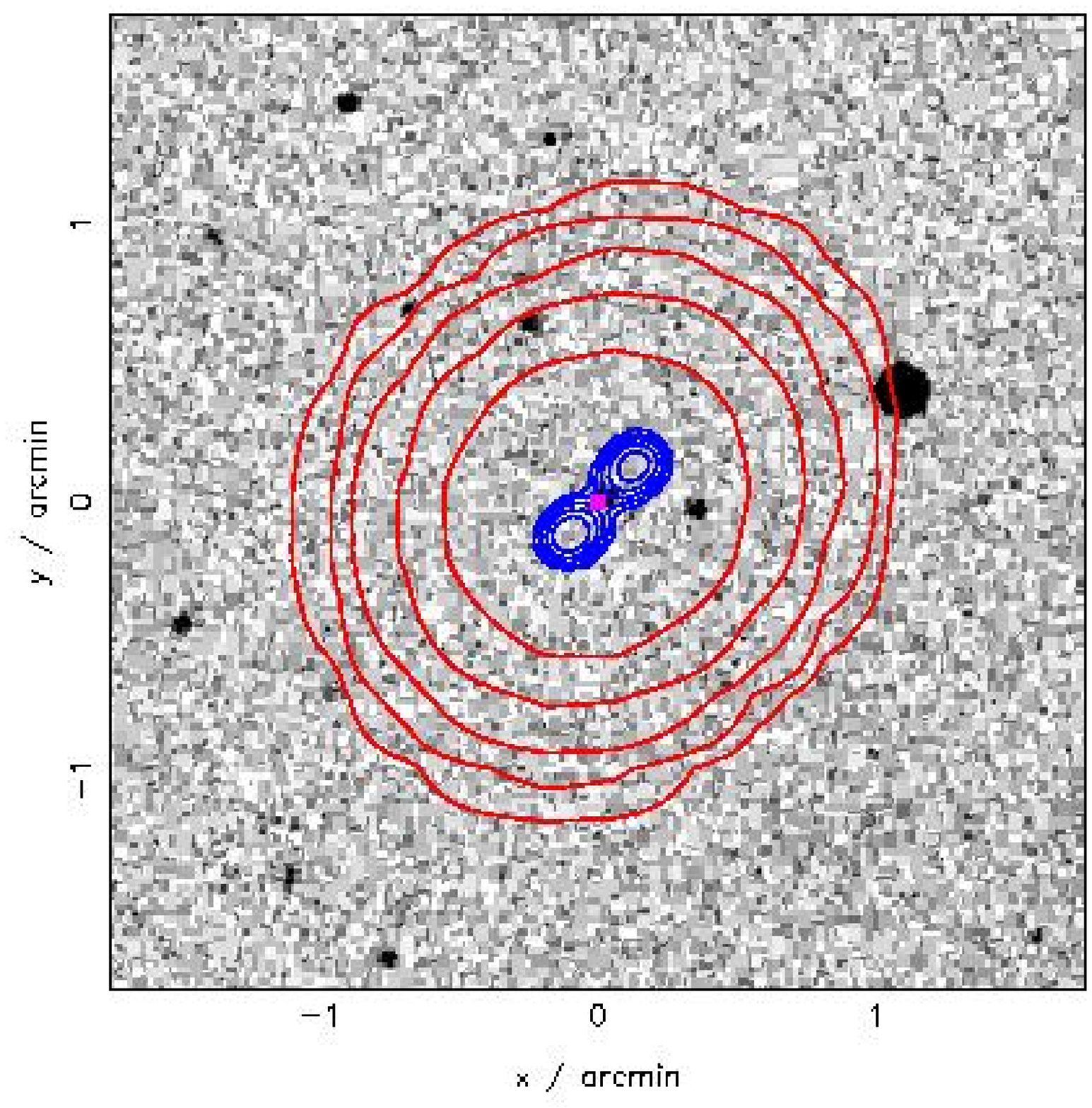}}
      \centerline{C2-120: 4C 30.21}
    \end{minipage}
    \hspace{3cm}
    \begin{minipage}{3cm}
      \mbox{}
      \centerline{\includegraphics[scale=0.26,angle=270]{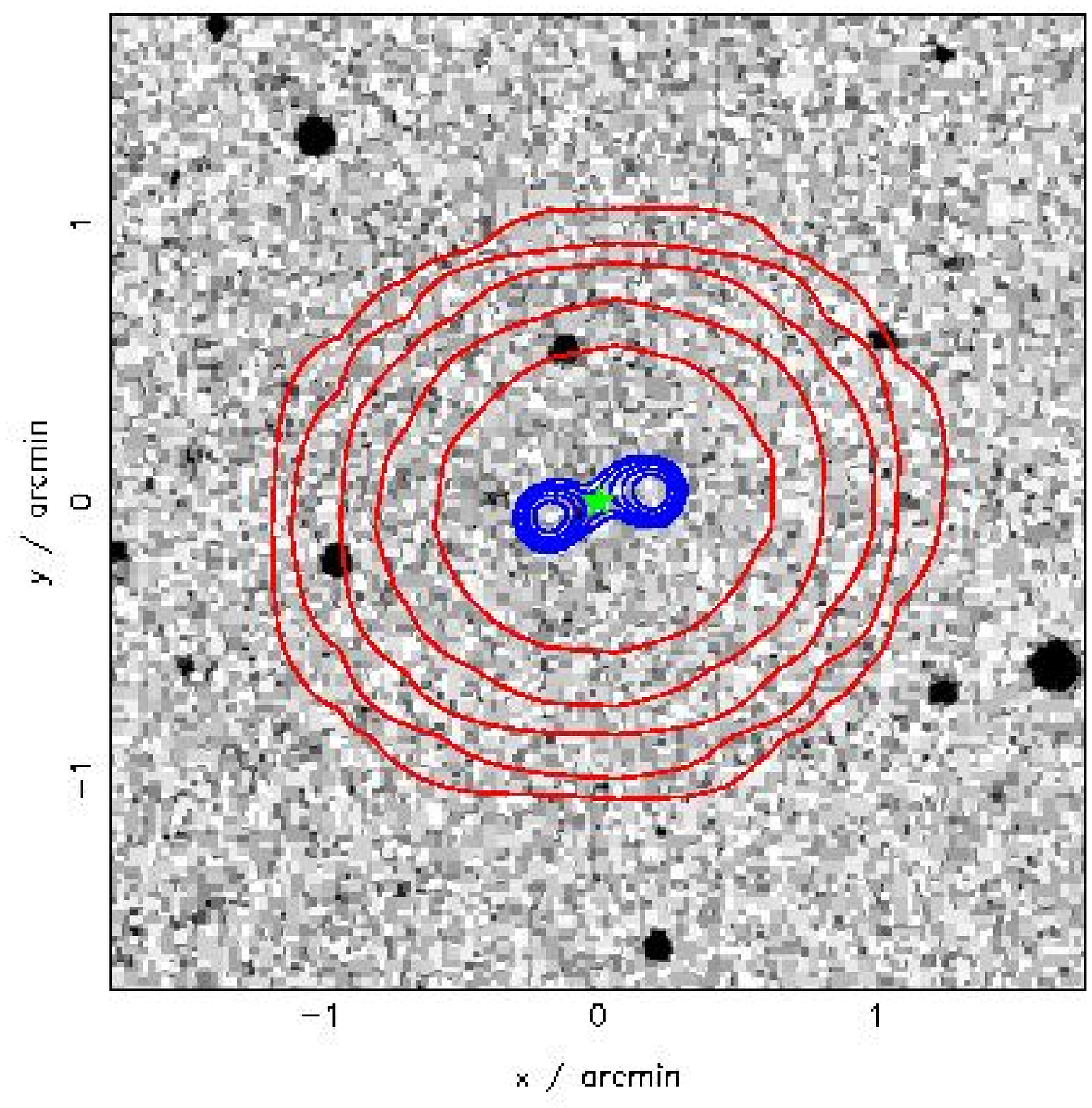}}
      \centerline{C2-122: 4C 12.41}
    \end{minipage}
    \hspace{3cm}
    \begin{minipage}{3cm}
      \mbox{}
      \centerline{\includegraphics[scale=0.26,angle=270]{Contours/C2/123.ps}}
      \centerline{C2-123: 4C 00.40}
    \end{minipage}
    \vfill
    \begin{minipage}{3cm}      
      \mbox{}
      \centerline{\includegraphics[scale=0.26,angle=270]{Contours/C2/126.ps}}
      \centerline{C2-126: 4C 10.33}
    \end{minipage}
    \hspace{3cm}
    \begin{minipage}{3cm}
      \mbox{}
      \centerline{\includegraphics[scale=0.26,angle=270]{Contours/C2/127.ps}}
      \centerline{C2-127: 4C 33.27}
    \end{minipage}
    \hspace{3cm}
    \begin{minipage}{3cm}
      \mbox{}
      \centerline{\includegraphics[scale=0.26,angle=270]{Contours/C2/128.ps}}
      \centerline{C2-128: TXS 1130+504}
    \end{minipage}
  \end{center}
\end{figure}

\begin{figure}
  \begin{center}
    {\bf CoNFIG-2}\\  
  \begin{minipage}{3cm}      
      \mbox{}
      \centerline{\includegraphics[scale=0.26,angle=270]{Contours/C2/130.ps}}
      \centerline{C2-130: 3C 261}
    \end{minipage}
    \hspace{3cm}
    \begin{minipage}{3cm}
      \mbox{}
      \centerline{\includegraphics[scale=0.26,angle=270]{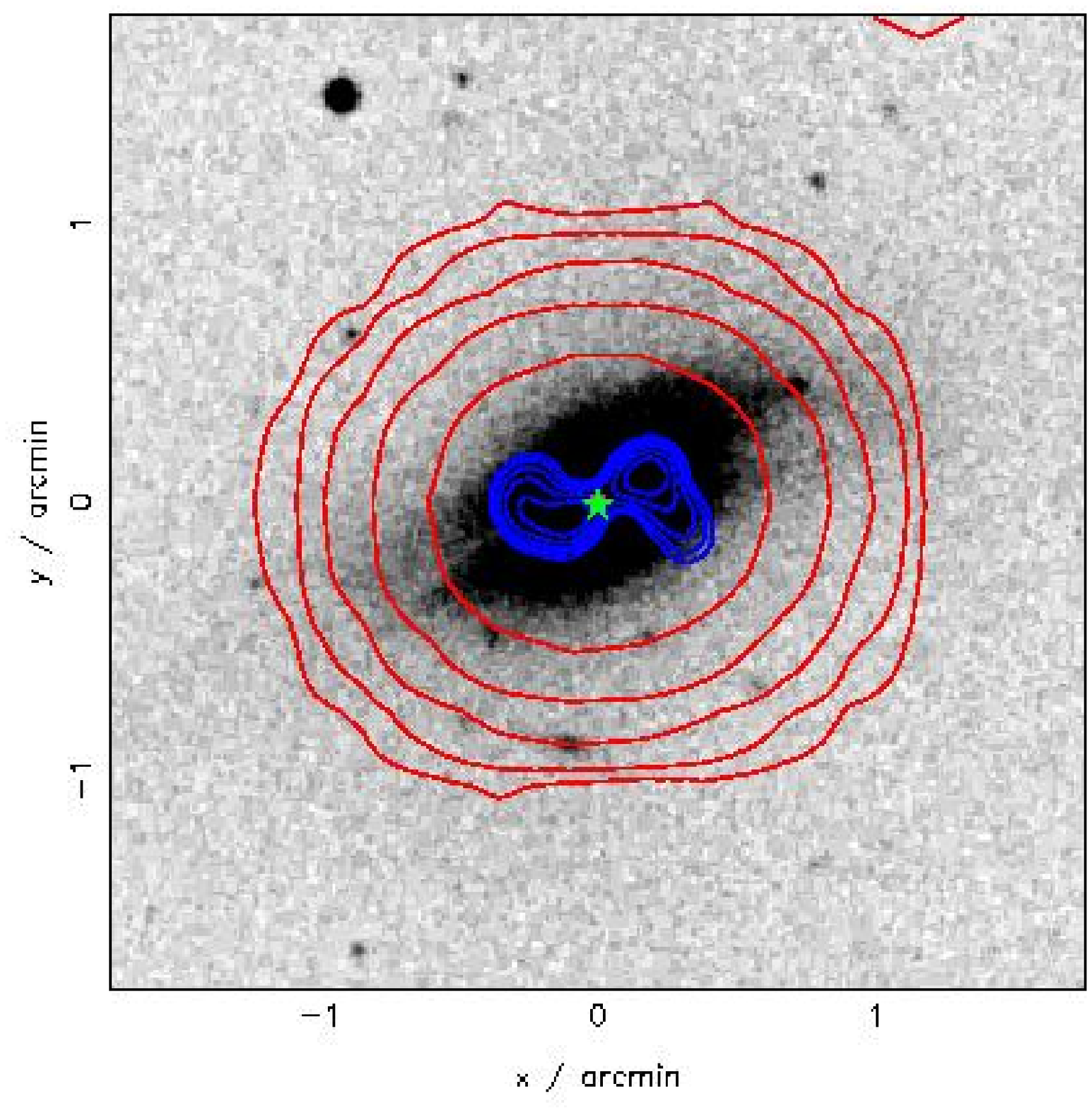}}
      \centerline{C2-134: 4C 17.52}
    \end{minipage}
    \hspace{3cm}
    \begin{minipage}{3cm}
      \mbox{}
      \centerline{\includegraphics[scale=0.26,angle=270]{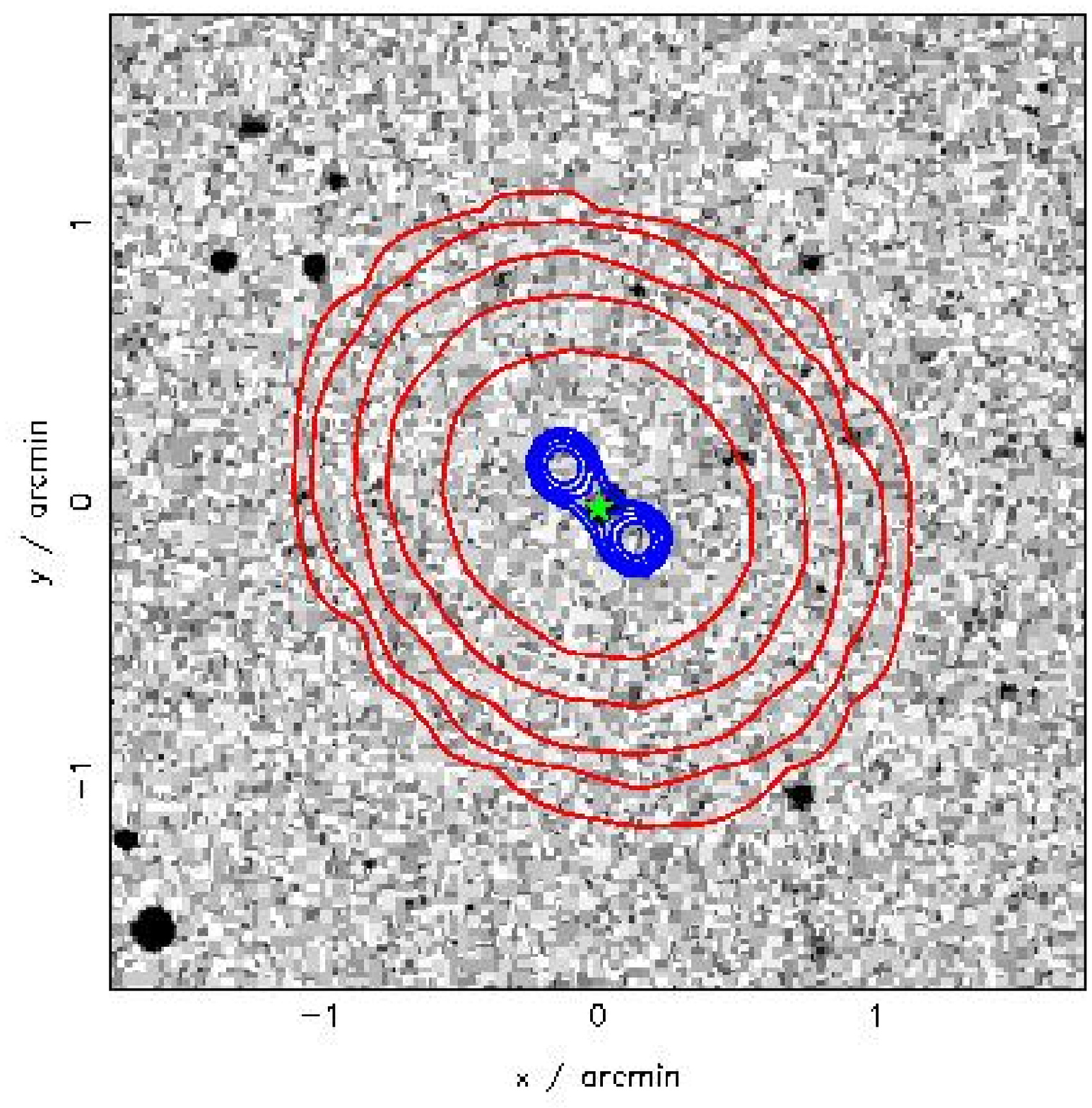}}
      \centerline{C2-138: TXS 1140+217}
    \end{minipage}
    \vfill
    \begin{minipage}{3cm}     
      \mbox{}
      \centerline{\includegraphics[scale=0.26,angle=270]{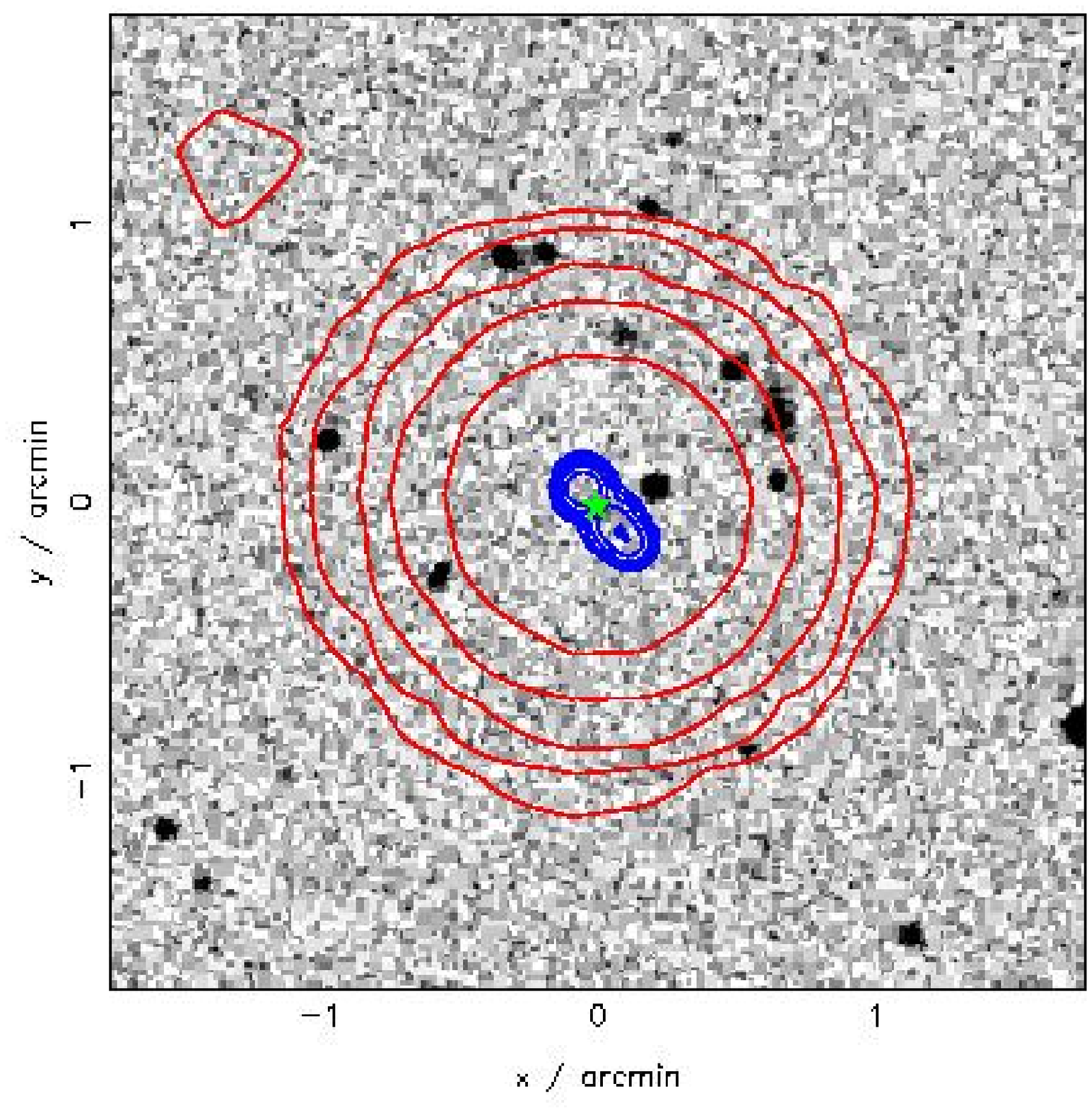}}
      \centerline{C2-139: 4C 49.21}
    \end{minipage}
    \hspace{3cm}
    \begin{minipage}{3cm}
      \mbox{}
      \centerline{\includegraphics[scale=0.26,angle=270]{Contours/C2/141.ps}}
      \centerline{C2-141: 4C 46.23}
    \end{minipage}
    \hspace{3cm}
    \begin{minipage}{3cm}
      \mbox{}
      \centerline{\includegraphics[scale=0.26,angle=270]{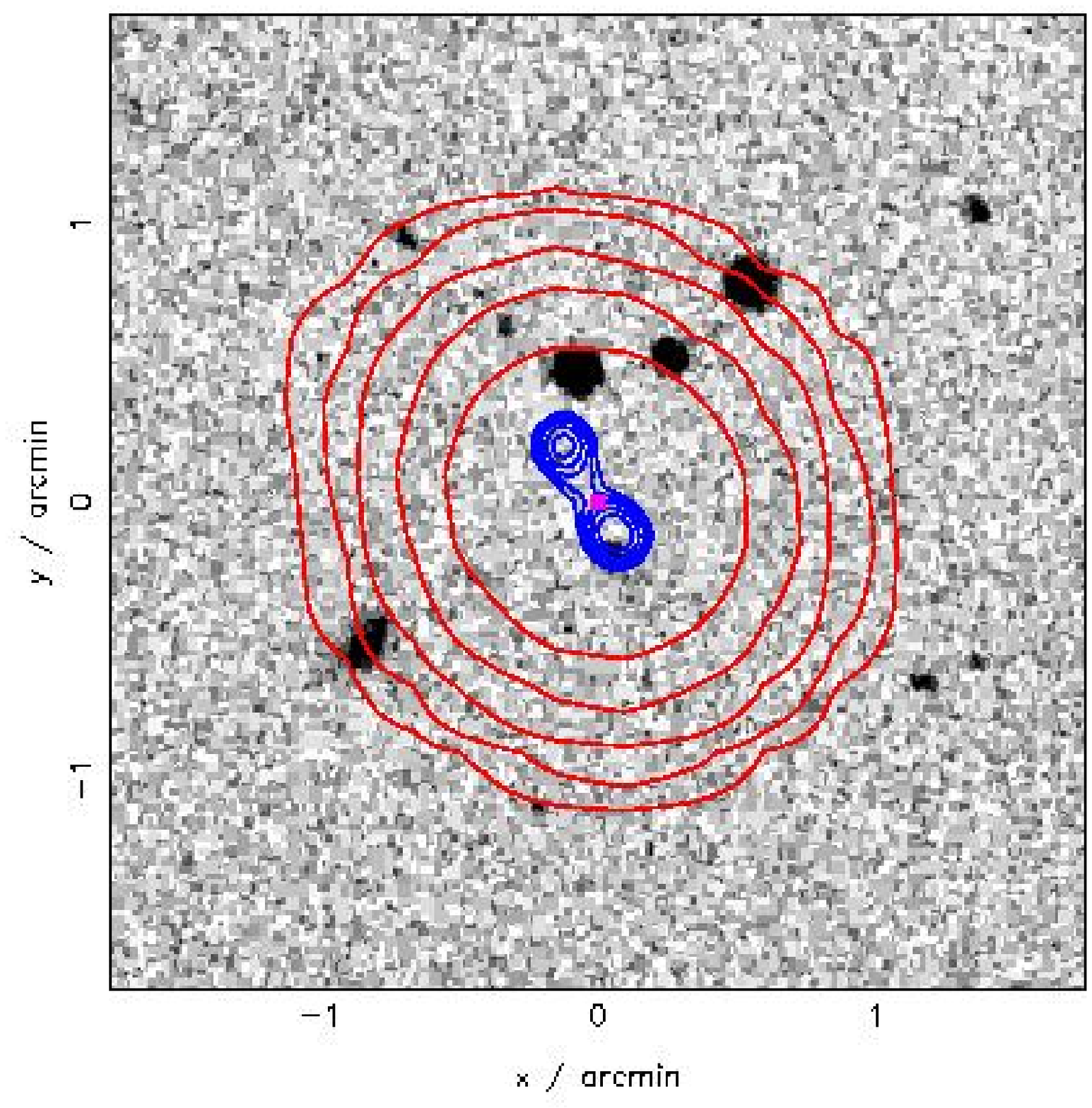}}
      \centerline{C2-142: 4C 30.23}
    \end{minipage}
    \vfill
    \begin{minipage}{3cm}     
      \mbox{}
      \centerline{\includegraphics[scale=0.26,angle=270]{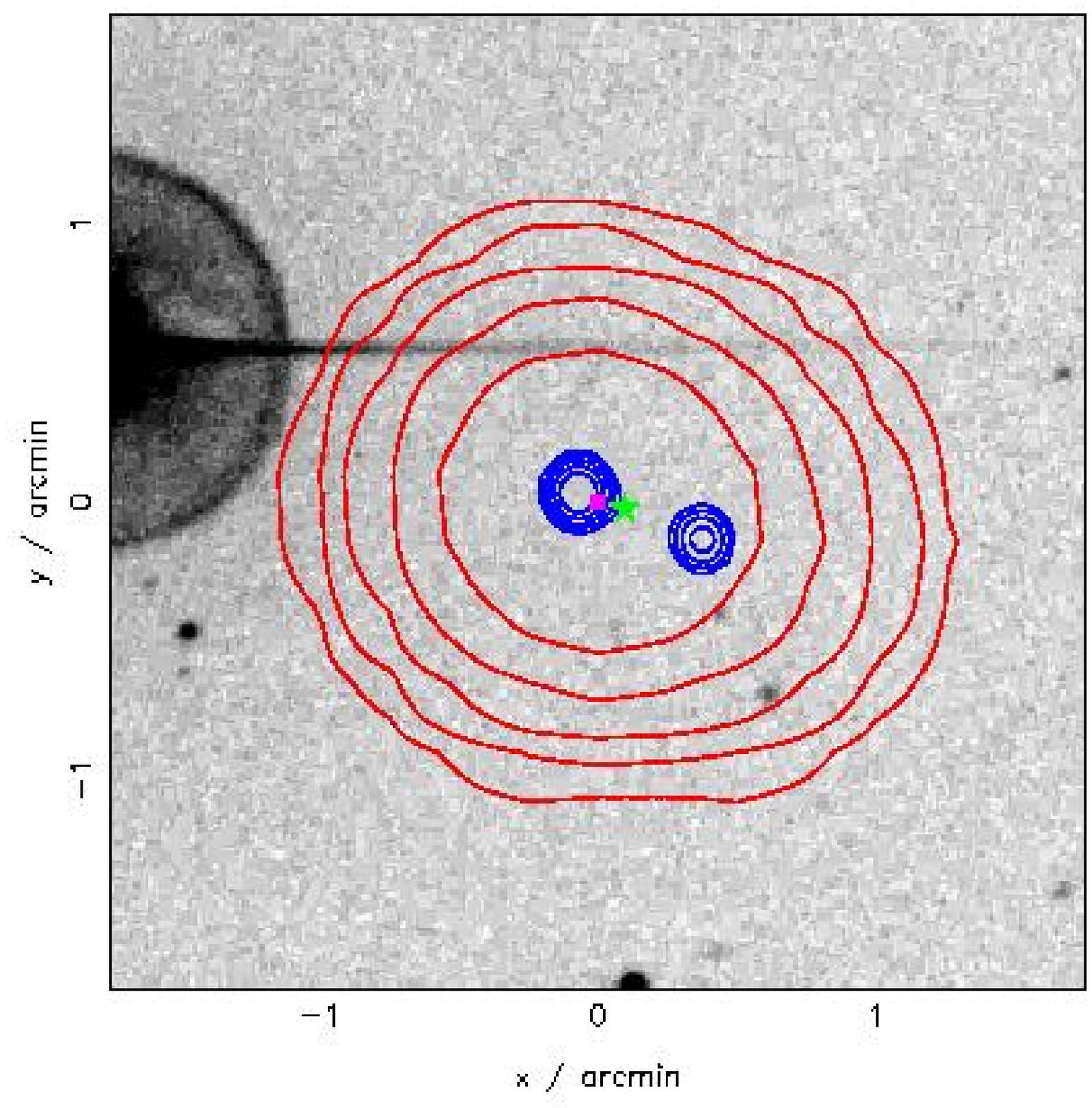}}
      \centerline{C2-144: 4C -00.46}
    \end{minipage}
    \hspace{3cm}
    \begin{minipage}{3cm}
      \mbox{}
      \centerline{\includegraphics[scale=0.26,angle=270]{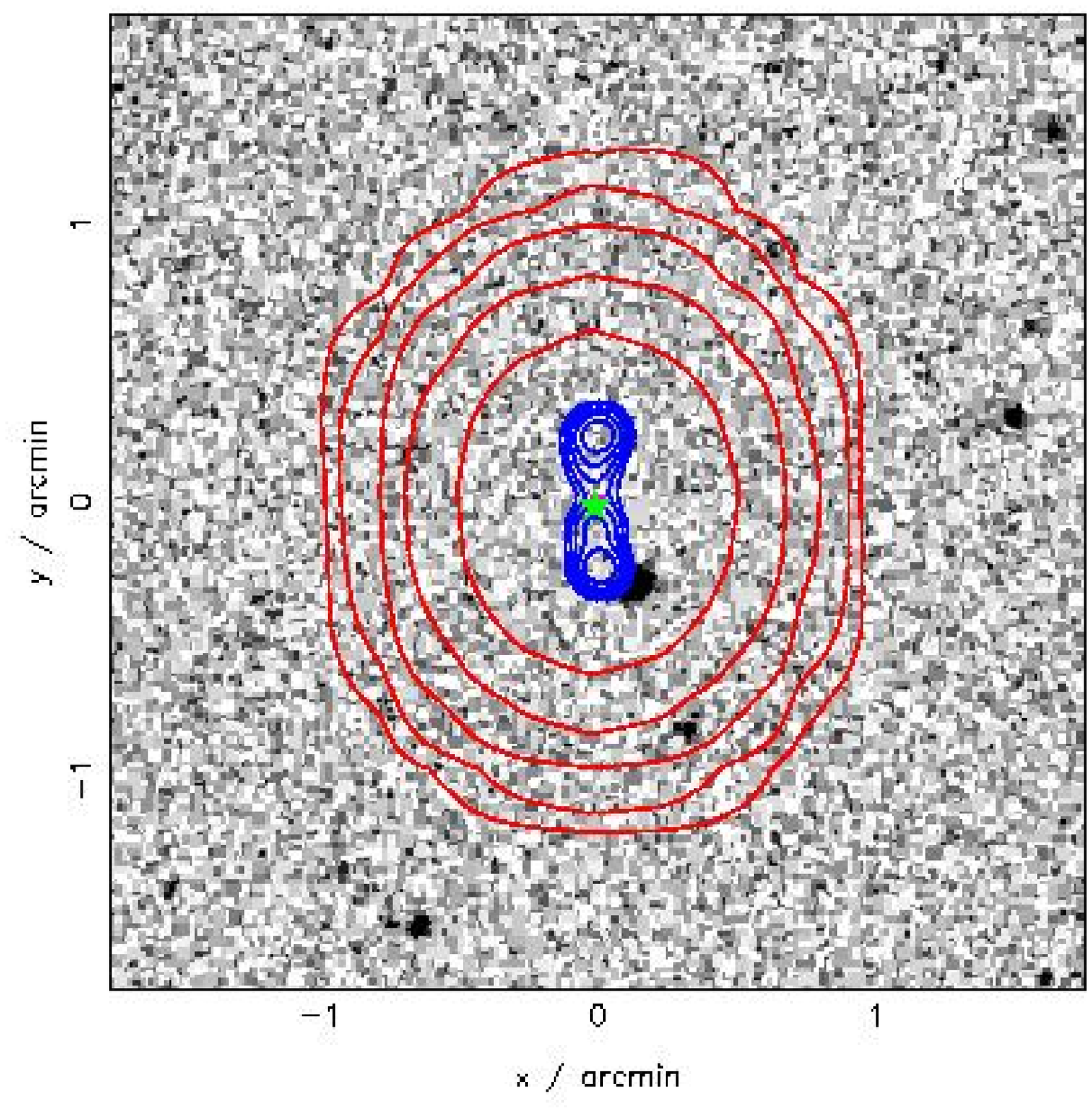}}
      \centerline{C2-148: 4C 25.36}
    \end{minipage}
    \hspace{3cm}
    \begin{minipage}{3cm}
      \mbox{}
      \centerline{\includegraphics[scale=0.26,angle=270]{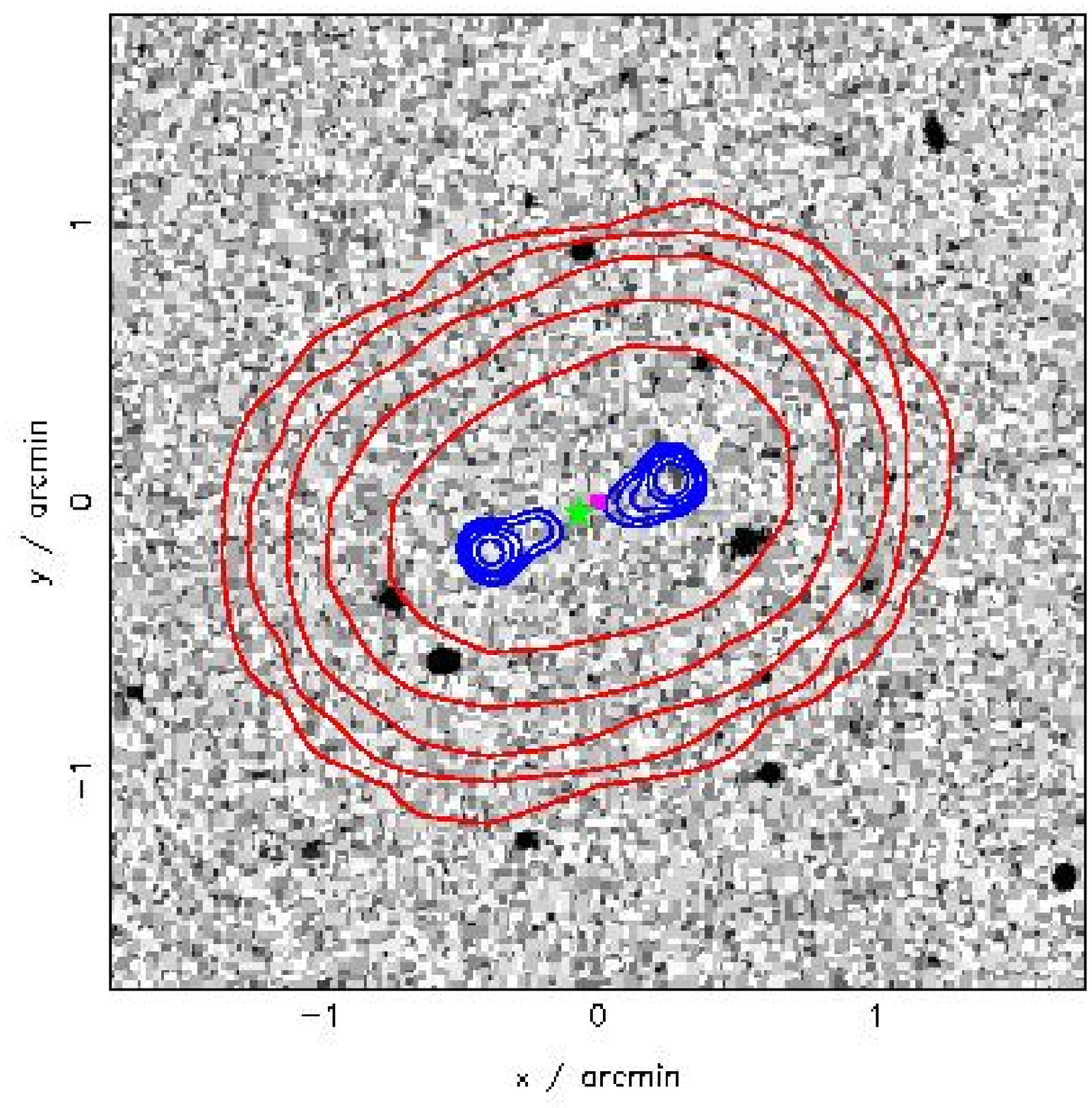}}
      \centerline{C2-149: 4C 05.53}
    \end{minipage}
    \vfill
    \begin{minipage}{3cm}      
      \mbox{}
      \centerline{\includegraphics[scale=0.26,angle=270]{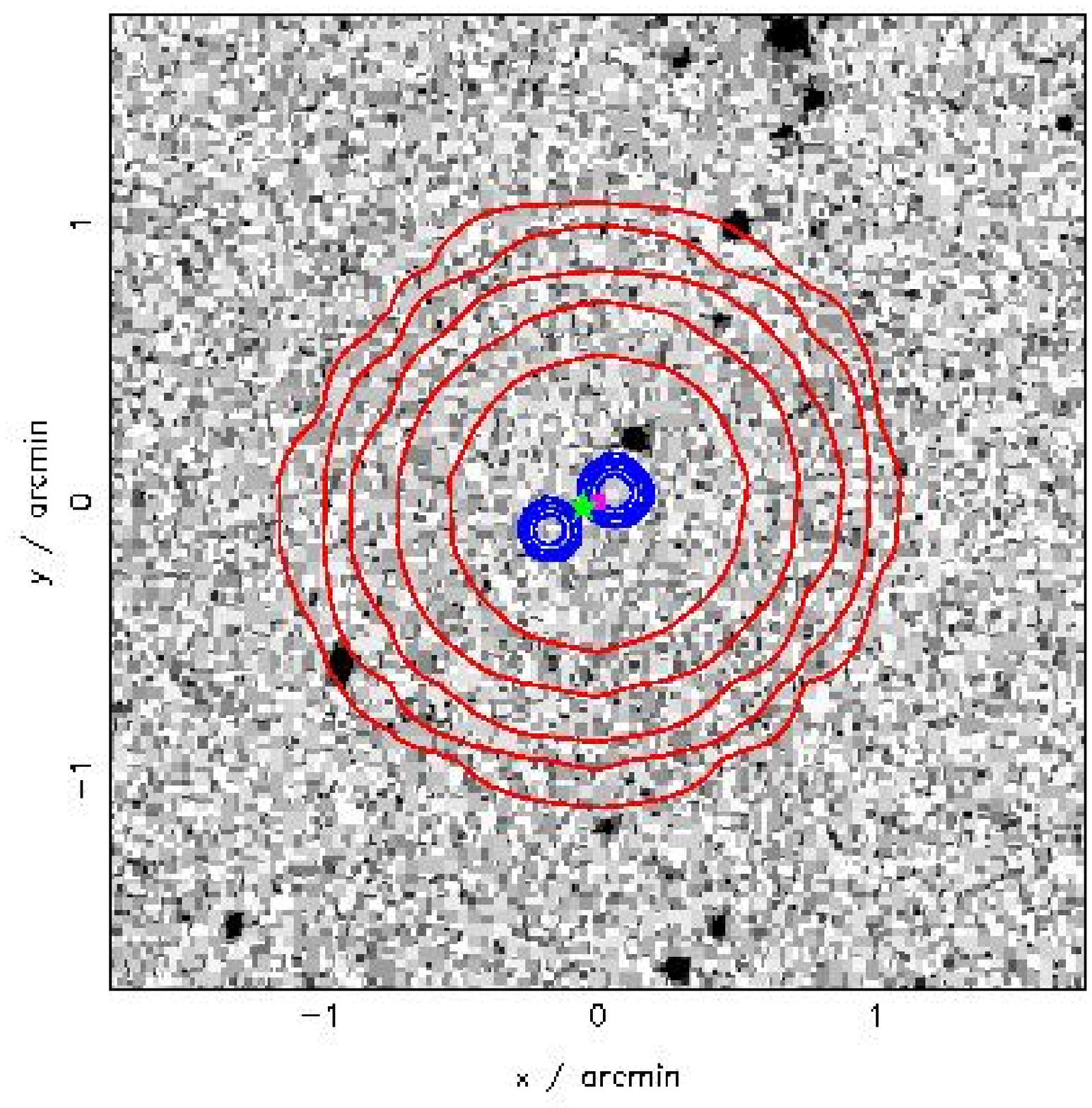}}
      \centerline{C2-152: 4C 11.40}
    \end{minipage}
    \hspace{3cm}
    \begin{minipage}{3cm}
      \mbox{}
      \centerline{\includegraphics[scale=0.26,angle=270]{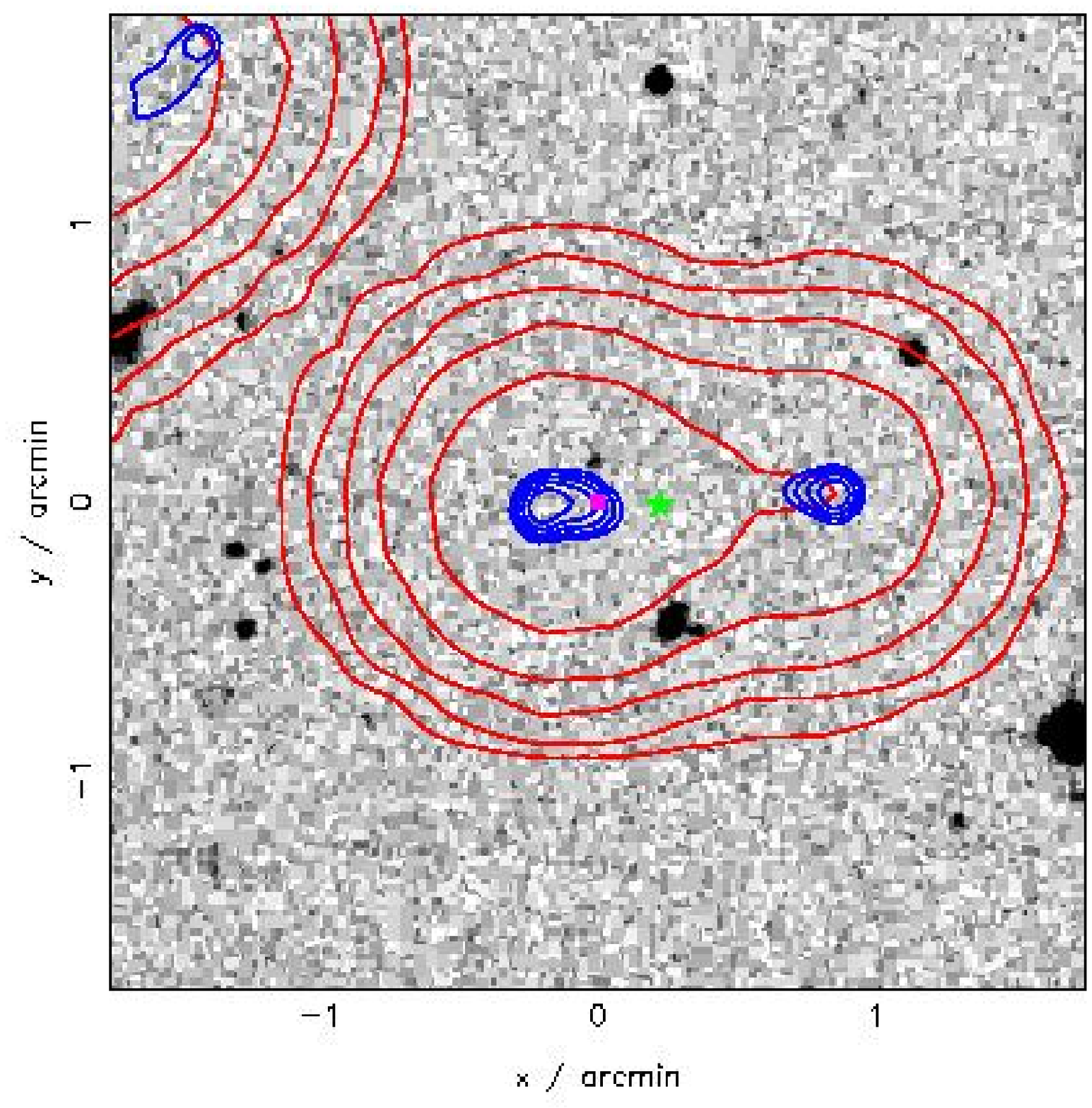}}
      \centerline{C2-157: TXS 1152+551}
    \end{minipage}
    \hspace{3cm}
    \begin{minipage}{3cm}
      \mbox{}
      \centerline{\includegraphics[scale=0.26,angle=270]{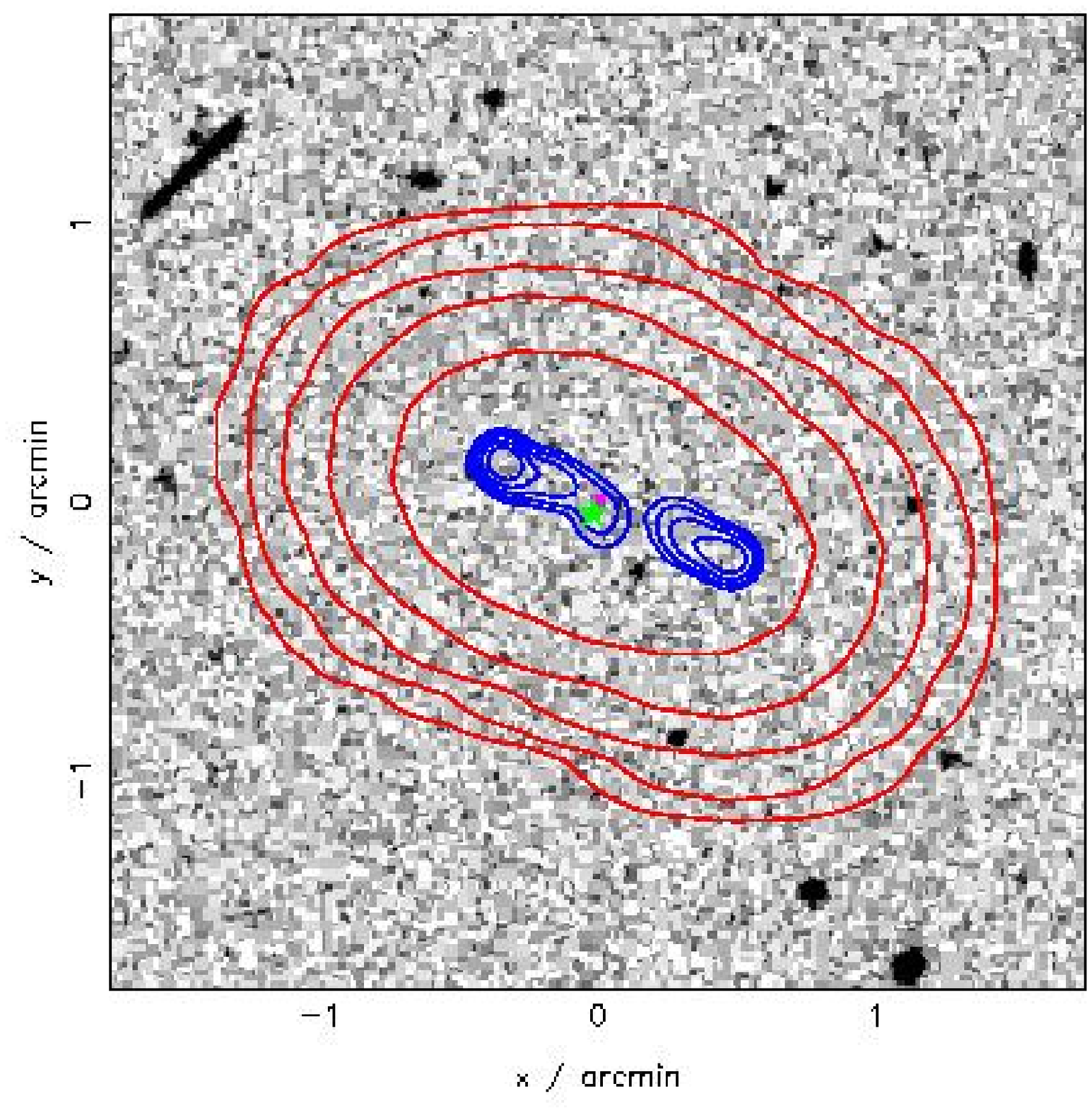}}
      \centerline{C2-168: 4C 25.38}
    \end{minipage}
  \end{center}
\end{figure}

\begin{figure}
  \begin{center}
    {\bf CoNFIG-2}\\  
  \begin{minipage}{3cm}      
      \mbox{}
      \centerline{\includegraphics[scale=0.26,angle=270]{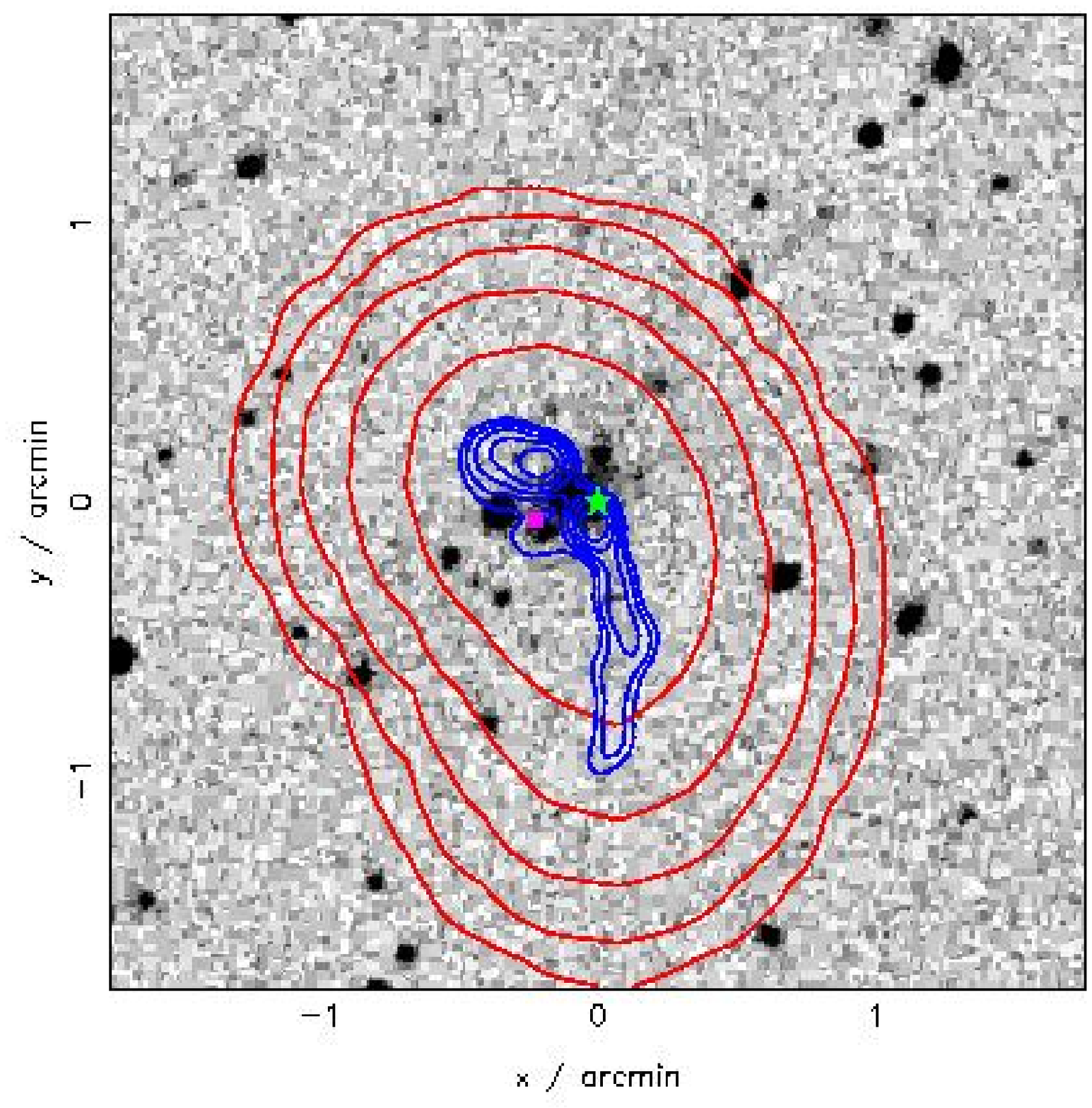}}
      \centerline{C2-169: 4C 58.23}
    \end{minipage}
    \hspace{3cm}
    \begin{minipage}{3cm}
      \mbox{}
      \centerline{\includegraphics[scale=0.26,angle=270]{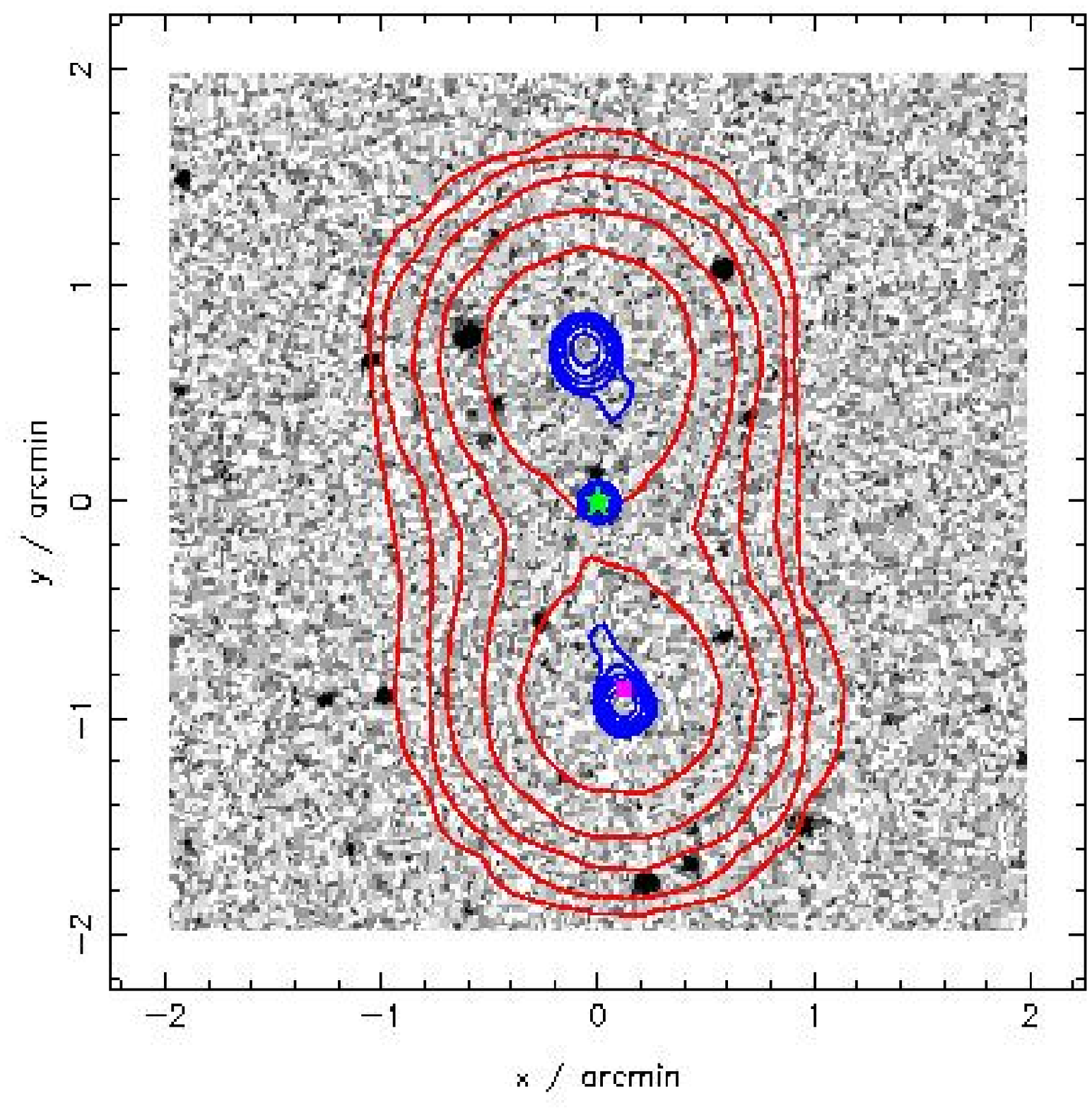}}
      \centerline{C2-174: 4C 29.46}
    \end{minipage}
    \hspace{3cm}
    \begin{minipage}{3cm}
      \mbox{}
      \centerline{\includegraphics[scale=0.26,angle=270]{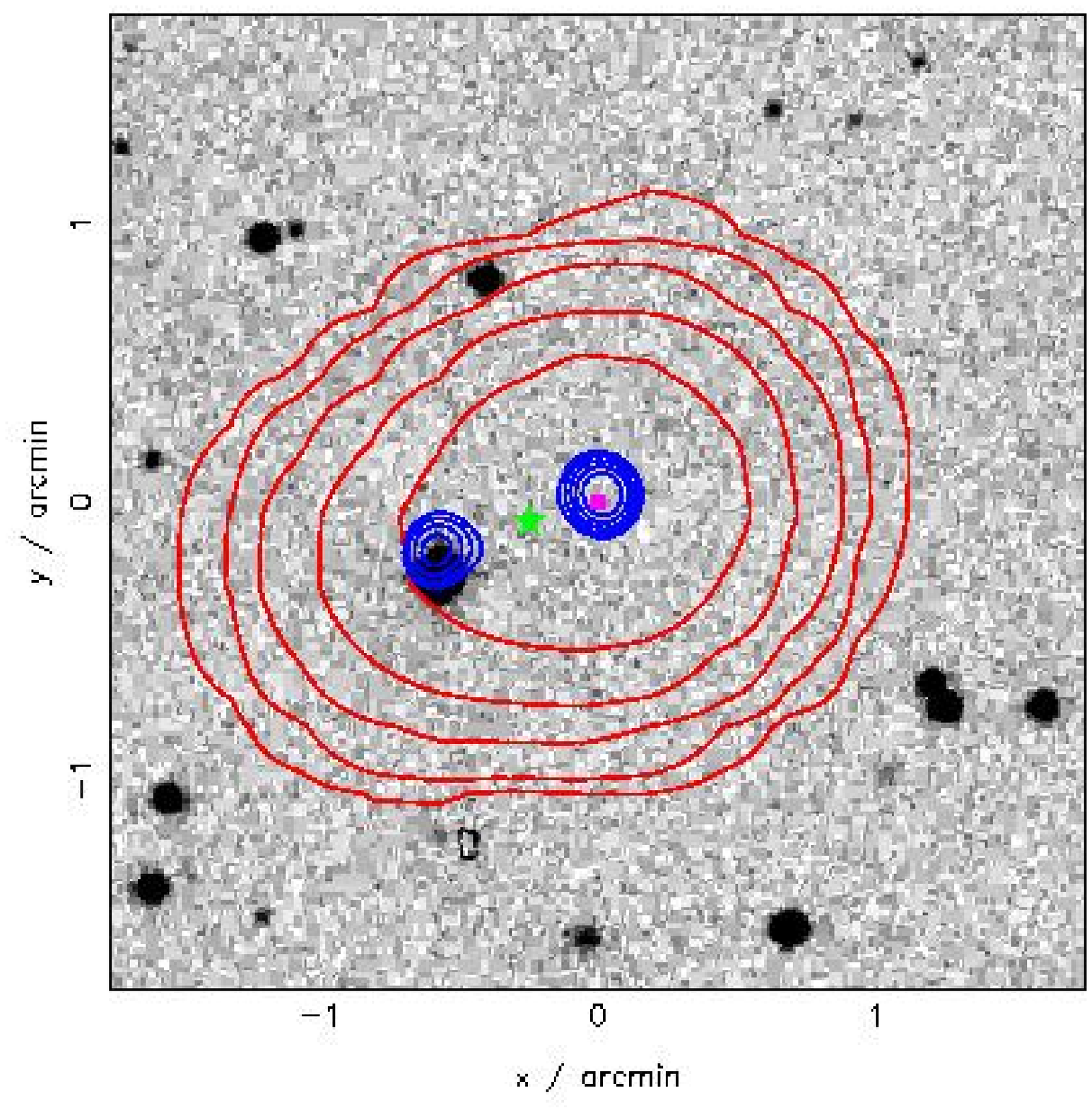}}
      \centerline{C2-180: 4C -00.48}
    \end{minipage}
    \vfill
    \begin{minipage}{3cm}     
      \mbox{}
      \centerline{\includegraphics[scale=0.26,angle=270]{Contours/C2/185.ps}}
      \centerline{C2-185: 3C 269}
    \end{minipage}
    \hspace{3cm}
    \begin{minipage}{3cm}
      \mbox{}
      \centerline{\includegraphics[scale=0.26,angle=270]{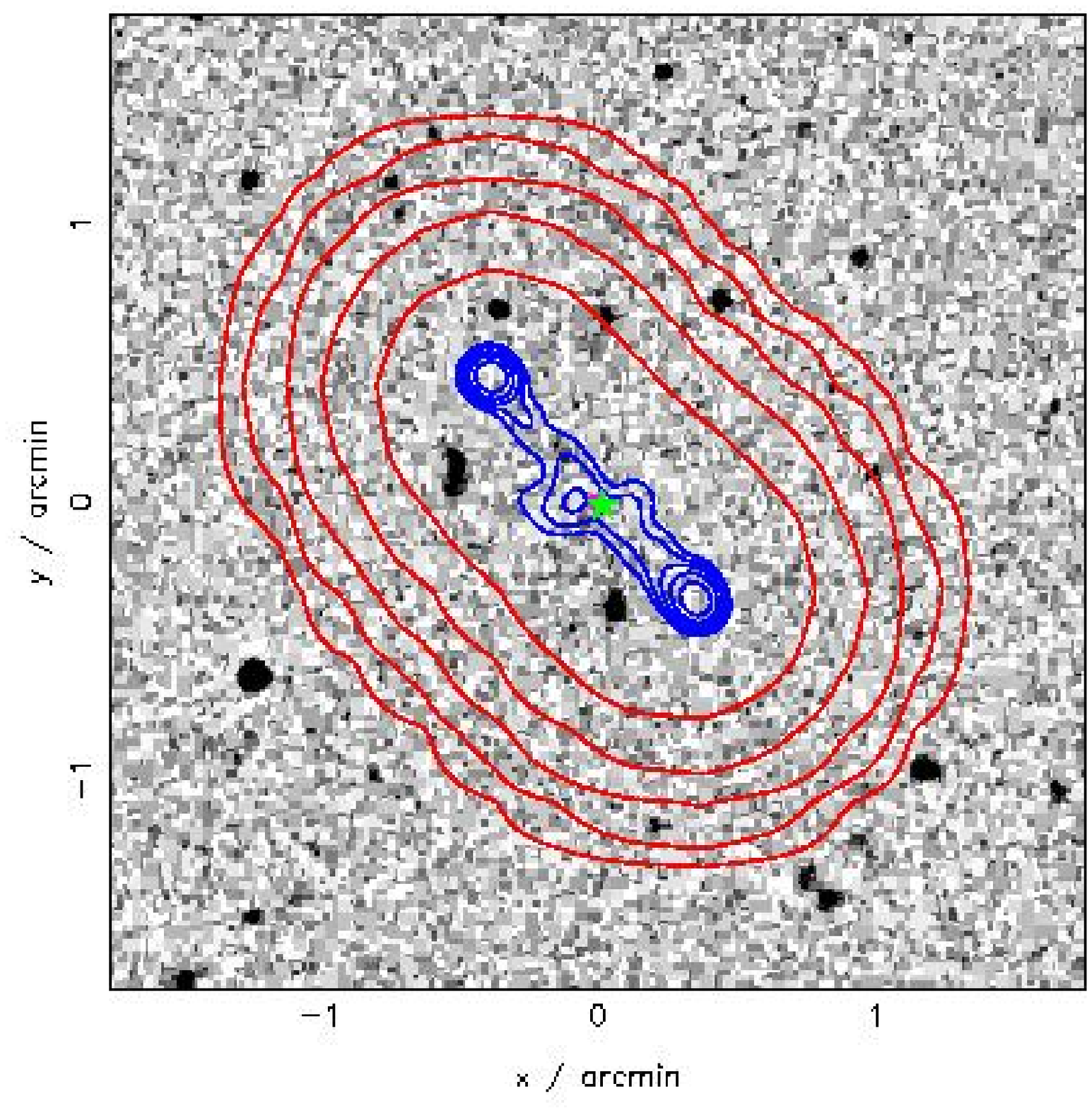}}
      \centerline{C2-186: 4C 20.28}
    \end{minipage}
    \hspace{3cm}
    \begin{minipage}{3cm}
      \mbox{}
      \centerline{\includegraphics[scale=0.26,angle=270]{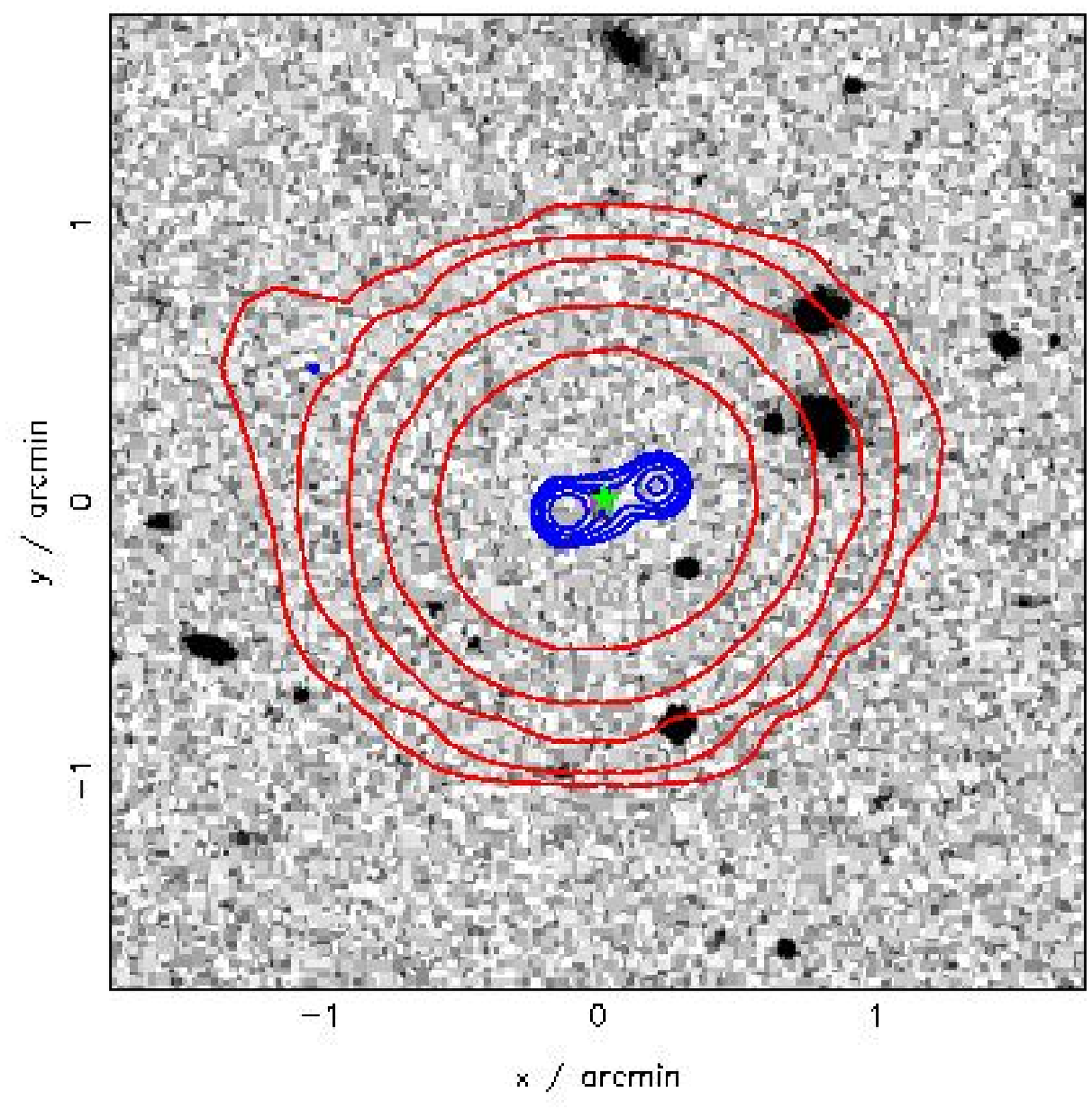}}
      \centerline{C2-188: 4C 09.41}
    \end{minipage}
    \vfill
    \begin{minipage}{3cm}     
      \mbox{}
      \centerline{\includegraphics[scale=0.26,angle=270]{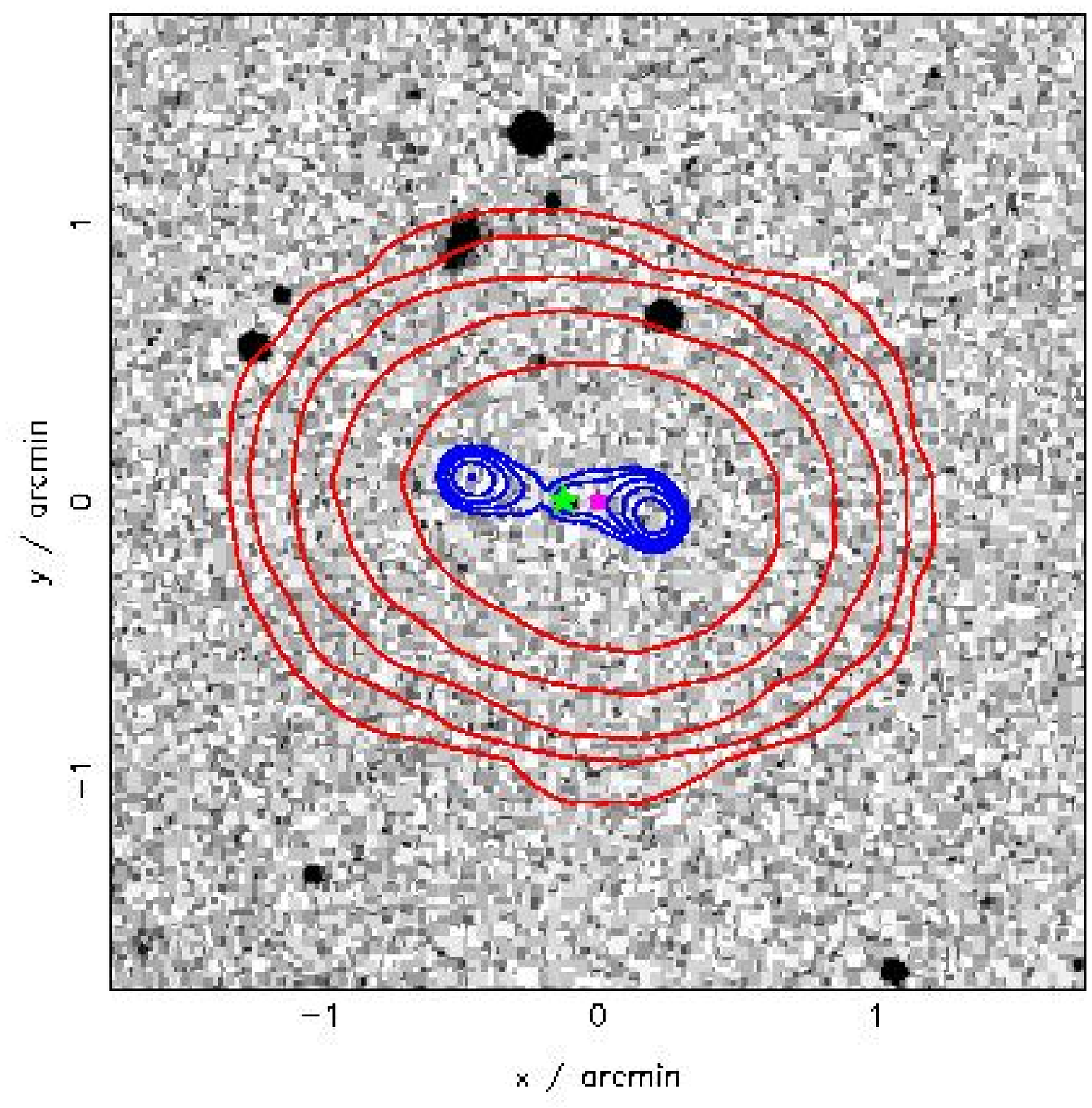}}
      \centerline{C2-190: 4C 31.40}
    \end{minipage}
    \hspace{3cm}
    \begin{minipage}{3cm}
      \mbox{}
      \centerline{\includegraphics[scale=0.26,angle=270]{Contours/C2/196.ps}}
      \centerline{C2-196: TXS 1223+099}
    \end{minipage}
    \hspace{3cm}
    \begin{minipage}{3cm}
      \mbox{}
      \centerline{\includegraphics[scale=0.26,angle=270]{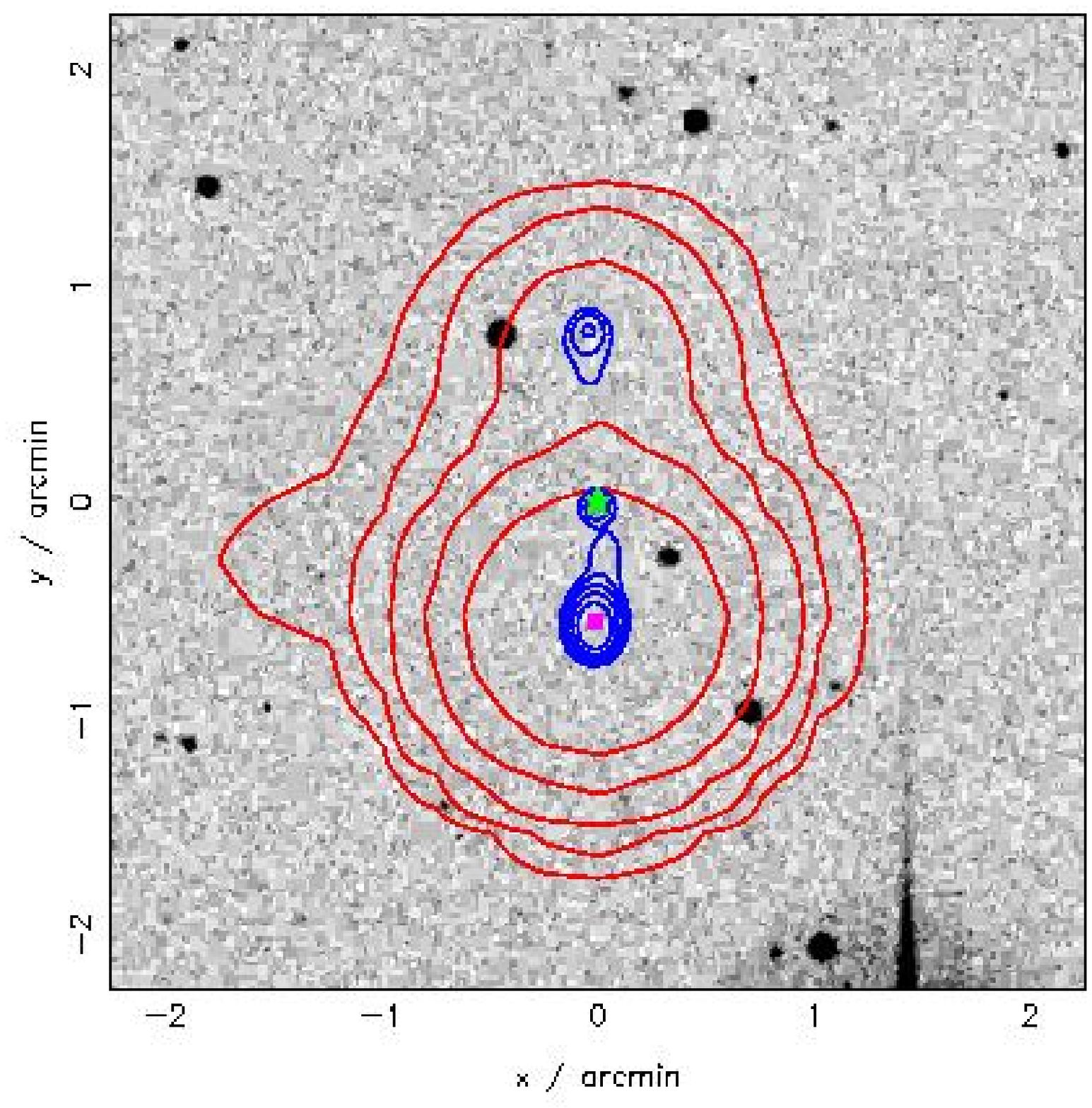}}
      \centerline{C2-198: 4C 20.29}
    \end{minipage}
    \vfill
    \begin{minipage}{3cm}      
      \mbox{}
      \centerline{\includegraphics[scale=0.26,angle=270]{Contours/C2/204.ps}}
      \centerline{C2-204: TXS 1229-013}
    \end{minipage}
    \hspace{3cm}
    \begin{minipage}{3cm}
      \mbox{}
      \centerline{\includegraphics[scale=0.26,angle=270]{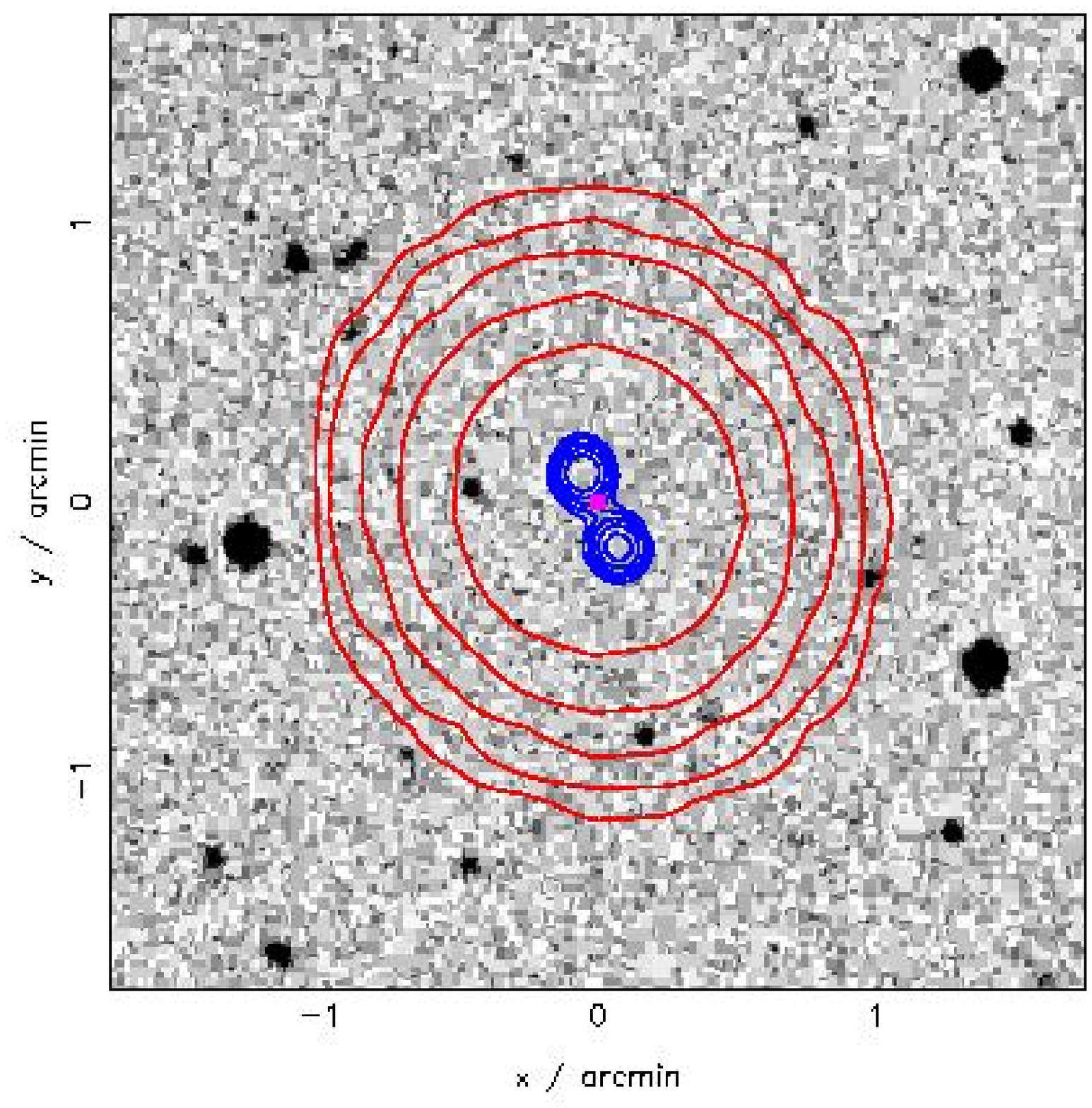}}
      \centerline{C2-207: 4C 37.34}
    \end{minipage}
    \hspace{3cm}
    \begin{minipage}{3cm}
      \mbox{}
      \centerline{\includegraphics[scale=0.26,angle=270]{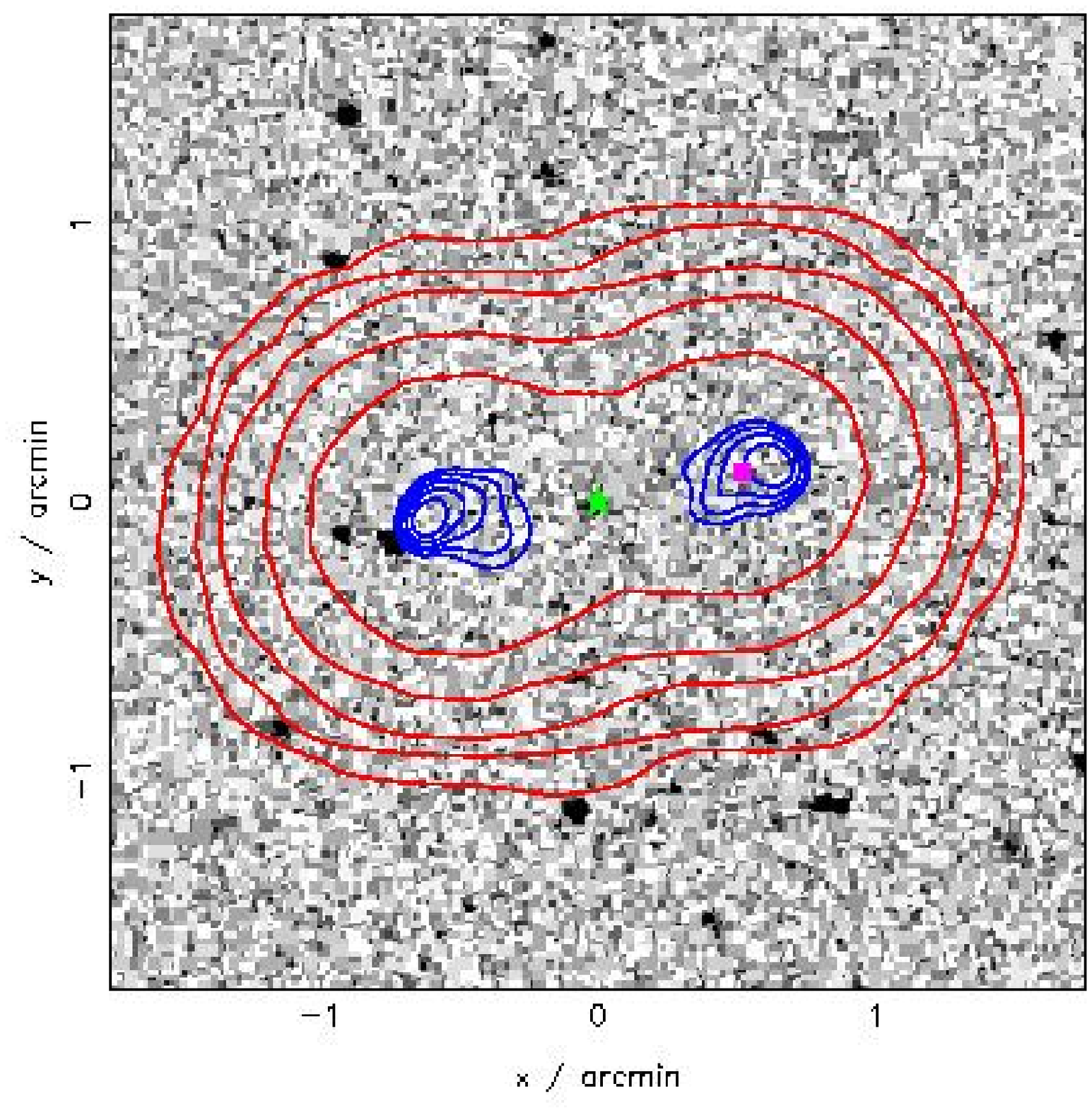}}
      \centerline{C2-208: 4C 05.54}
    \end{minipage}
  \end{center}
\end{figure}

\begin{figure}
  \begin{center}
    {\bf CoNFIG-2}\\  
  \begin{minipage}{3cm}      
      \mbox{}
      \centerline{\includegraphics[scale=0.26,angle=270]{Contours/C2/209.ps}}
      \centerline{C2-209: 4C 32.40}
    \end{minipage}
    \hspace{3cm}
    \begin{minipage}{3cm}
      \mbox{}
      \centerline{\includegraphics[scale=0.26,angle=270]{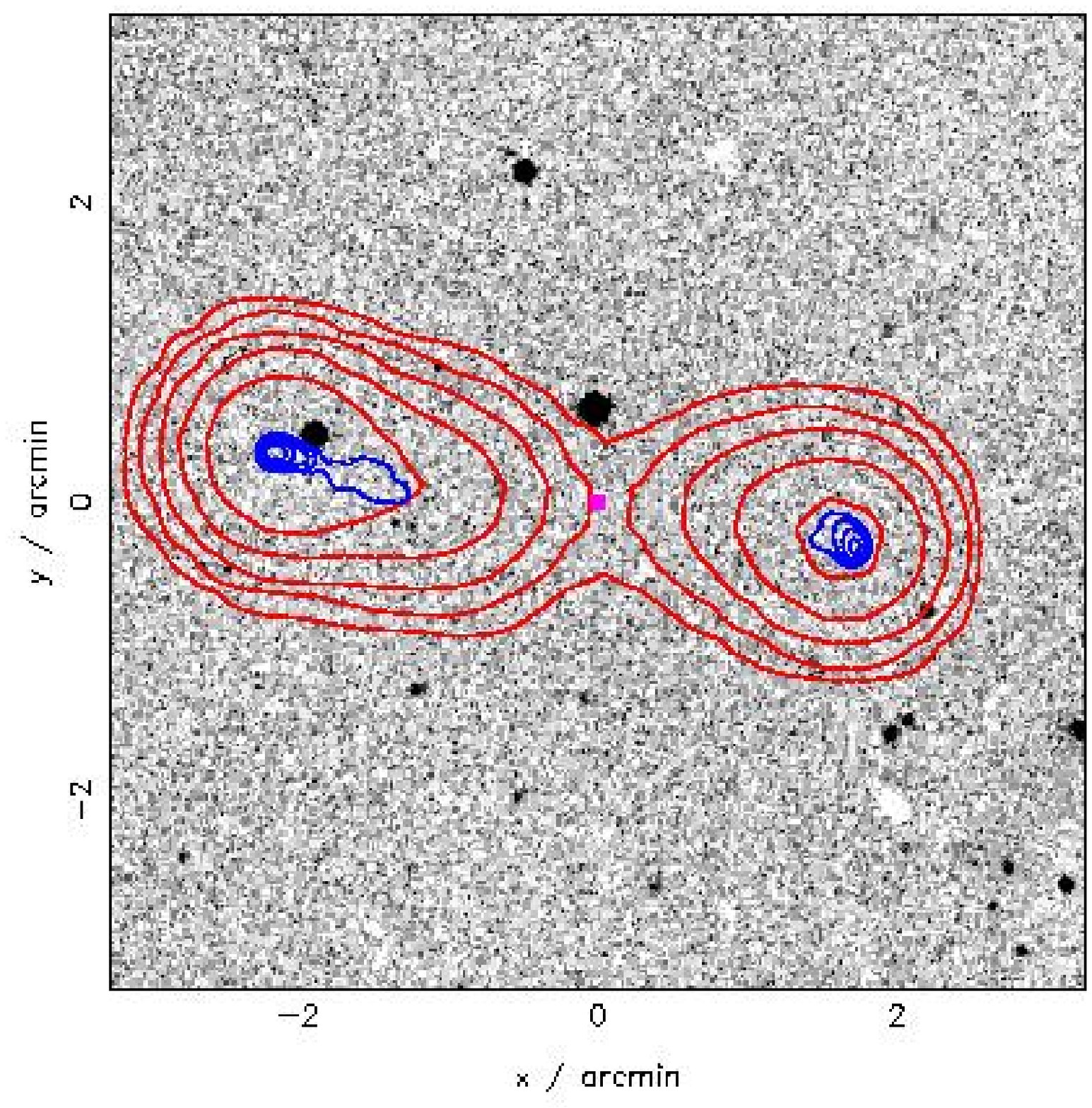}}
      \centerline{C2-210: TXS 1239+577}
    \end{minipage}
    \hspace{3cm}
    \begin{minipage}{3cm}
      \mbox{}
      \centerline{\includegraphics[scale=0.26,angle=270]{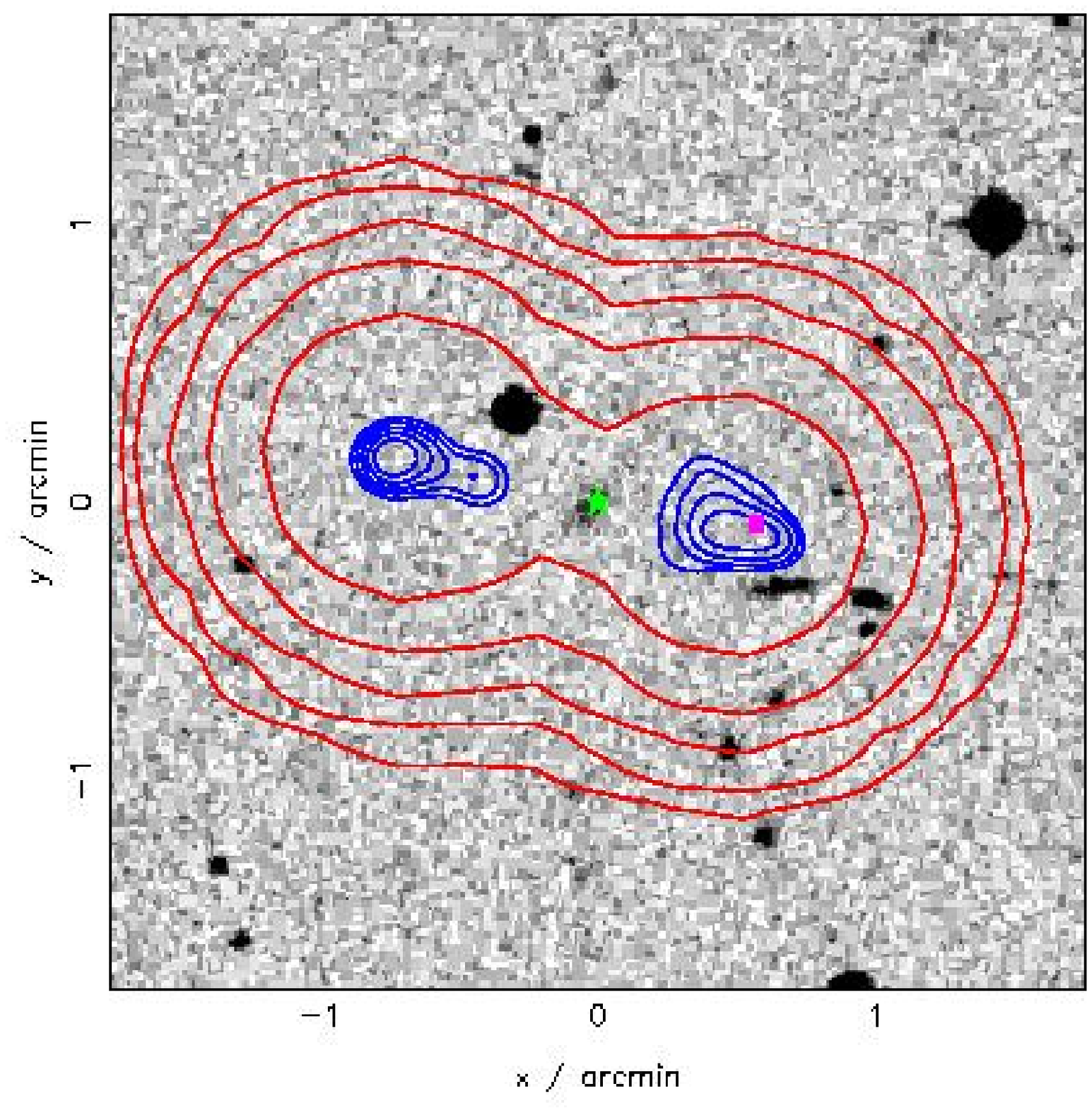}}
      \centerline{C2-215: 4C 33.30}
    \end{minipage}
    \vfill
    \begin{minipage}{3cm}     
      \mbox{}
      \centerline{\includegraphics[scale=0.26,angle=270]{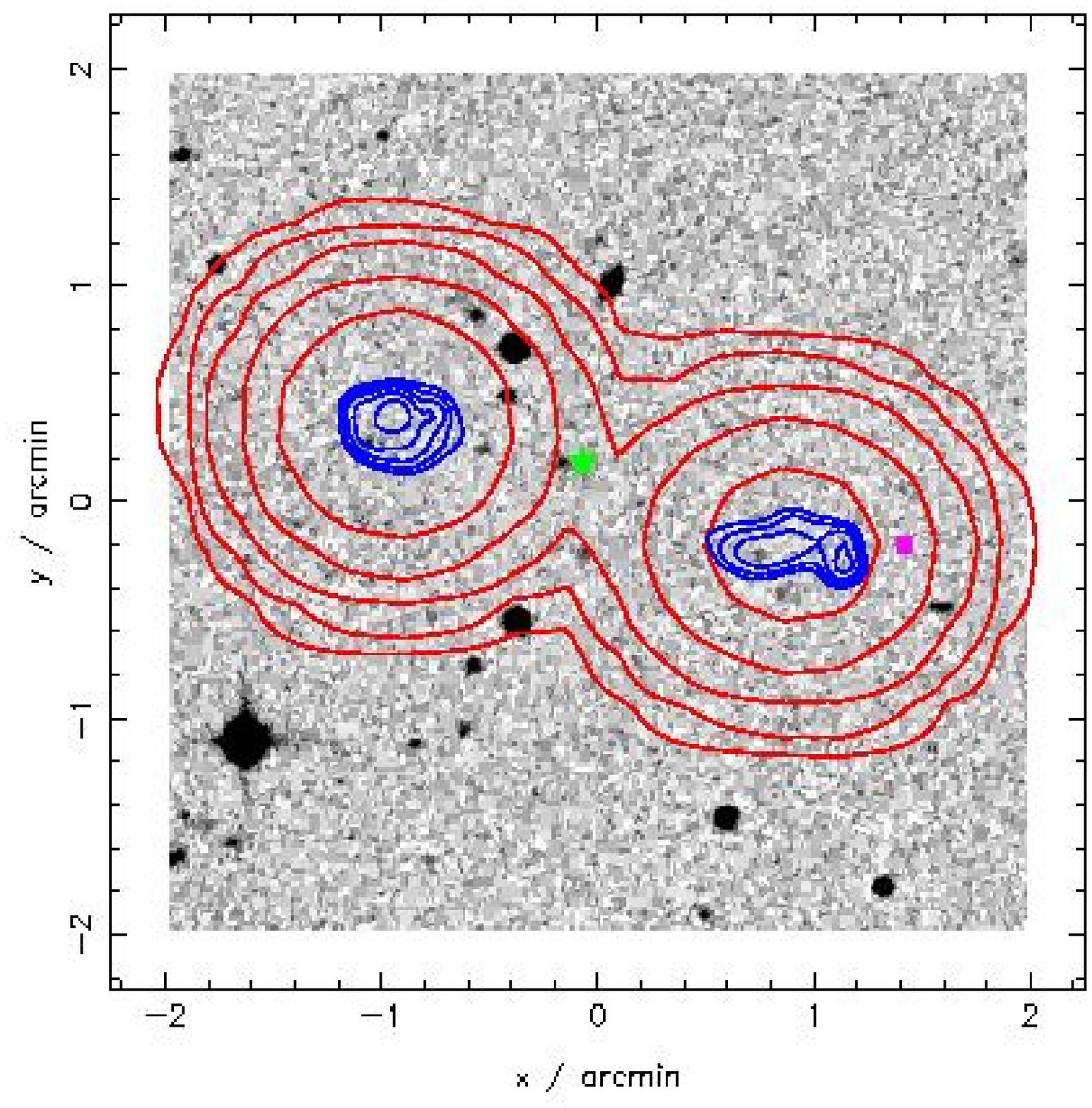}}
      \centerline{C2-216: 3C 277}
    \end{minipage}
    \hspace{3cm}
    \begin{minipage}{3cm}
      \mbox{}
      \centerline{\includegraphics[scale=0.26,angle=270]{Contours/C2/218.ps}}
      \centerline{C2-218: TXS 1249+530}
    \end{minipage}
    \hspace{3cm}
    \begin{minipage}{3cm}
      \mbox{}
      \centerline{\includegraphics[scale=0.26,angle=270]{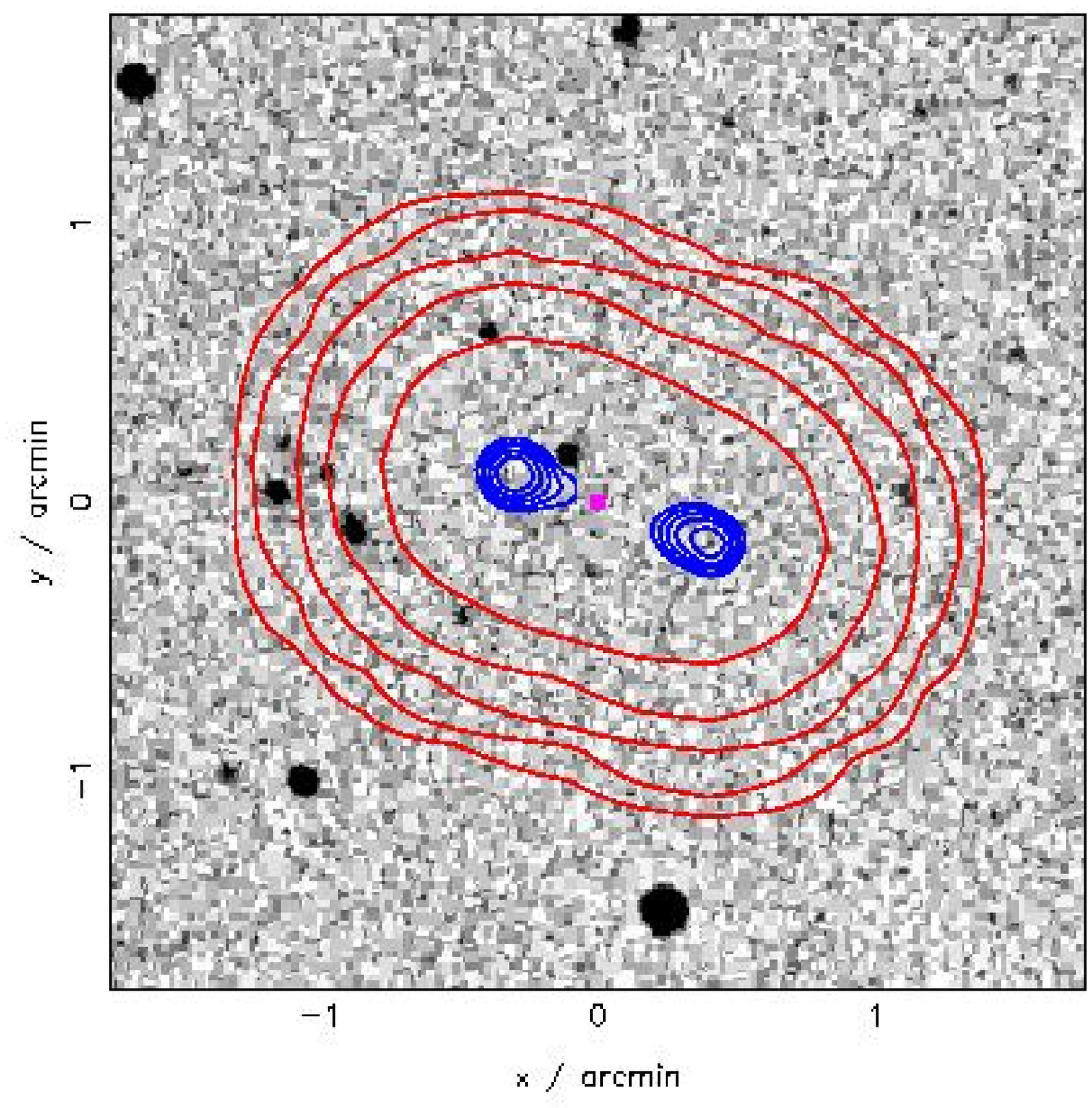}}
      \centerline{C2-219: 3C 276}
    \end{minipage}
    \vfill
    \begin{minipage}{3cm}     
      \mbox{}
      \centerline{\includegraphics[scale=0.26,angle=270]{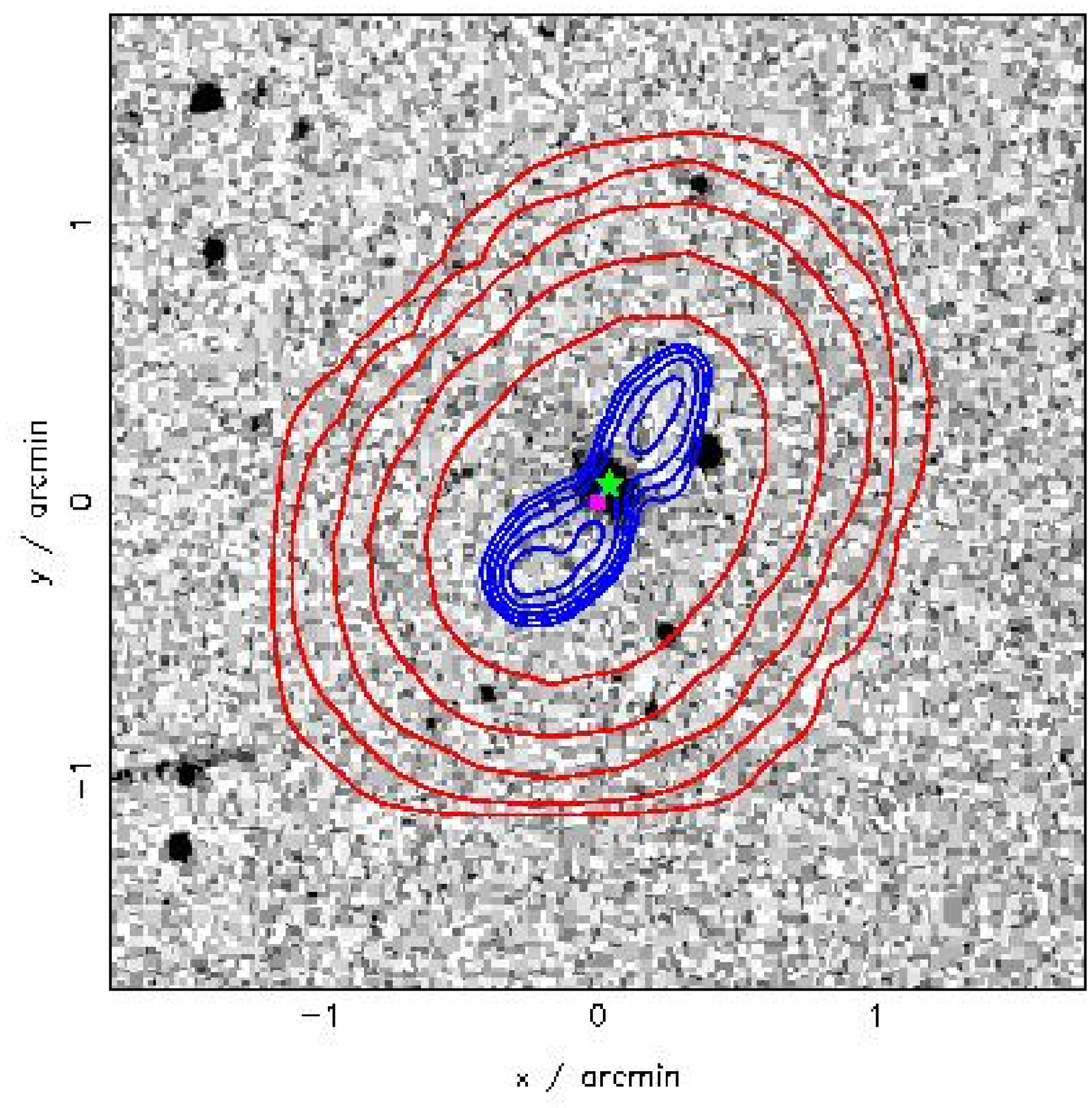}}
      \centerline{C2-220: TXS 1249+035}
    \end{minipage}
    \hspace{3cm}
    \begin{minipage}{3cm}
      \mbox{}
      \centerline{\includegraphics[scale=0.26,angle=270]{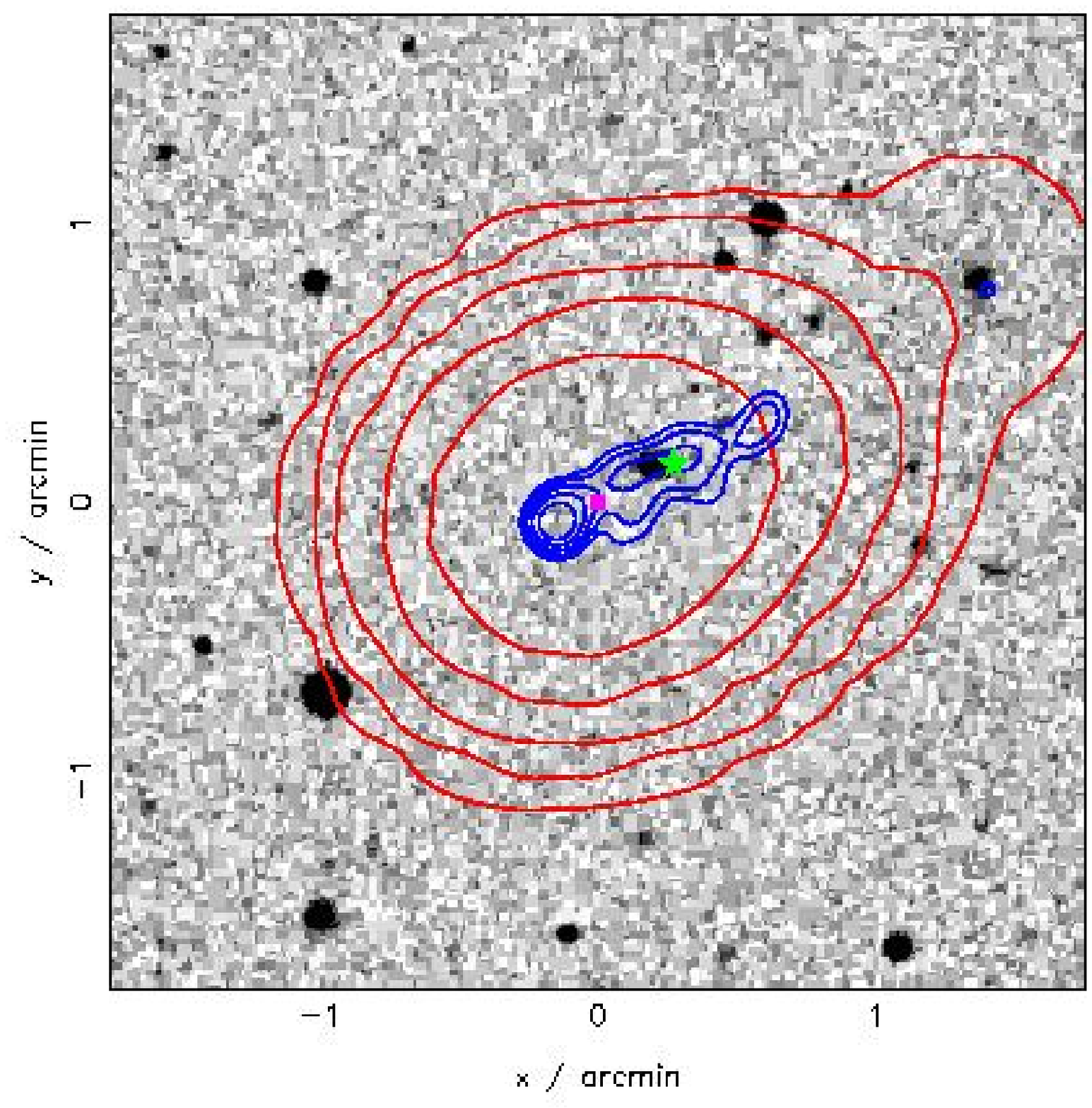}}
      \centerline{C2-226: 4C 44.22}
    \end{minipage}
    \hspace{3cm}
    \begin{minipage}{3cm}
      \mbox{}
      \centerline{\includegraphics[scale=0.26,angle=270]{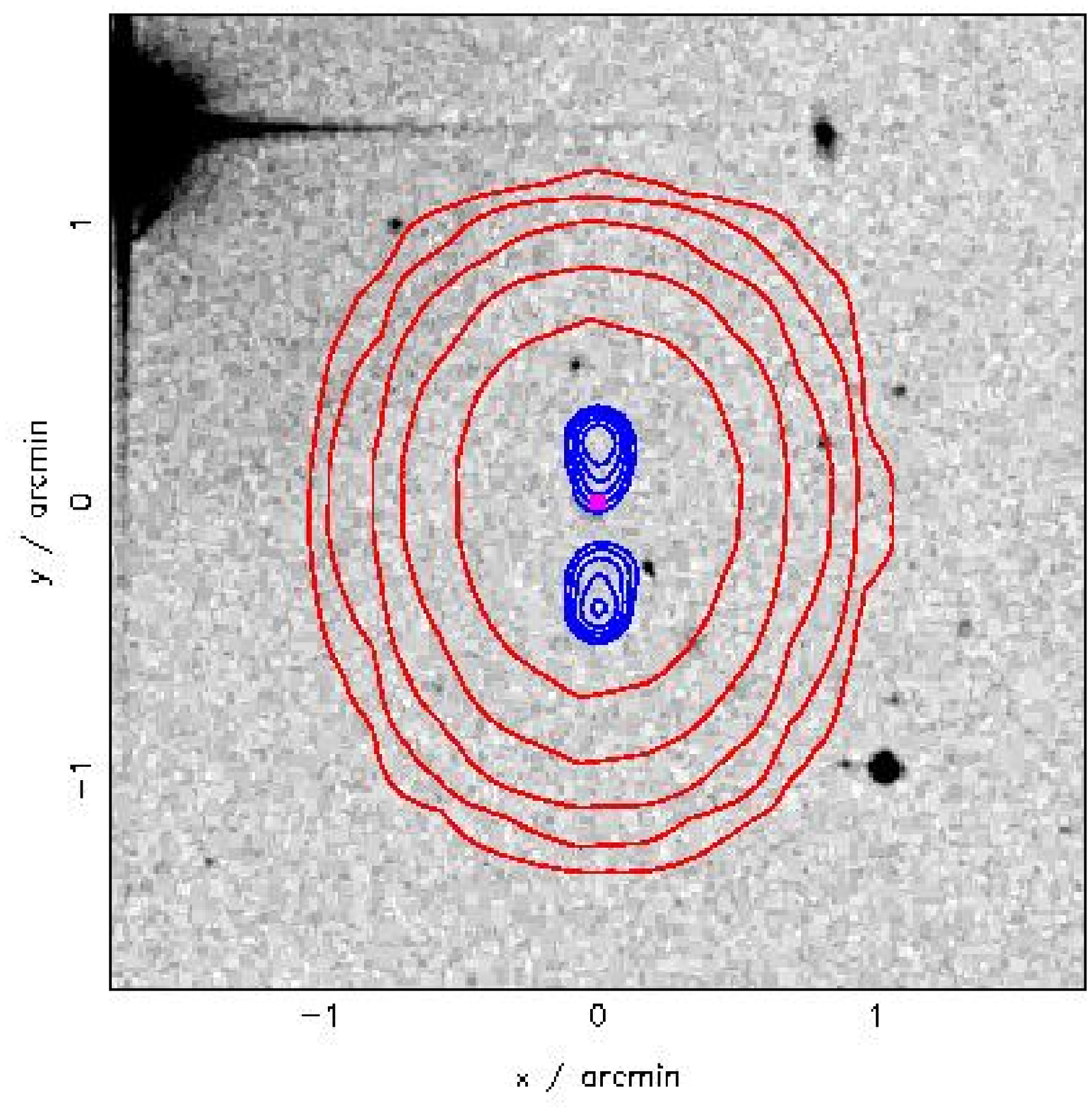}}
      \centerline{C2-229: 4C 54.30}
    \end{minipage}
    \vfill
    \begin{minipage}{3cm}      
      \mbox{}
      \centerline{\includegraphics[scale=0.26,angle=270]{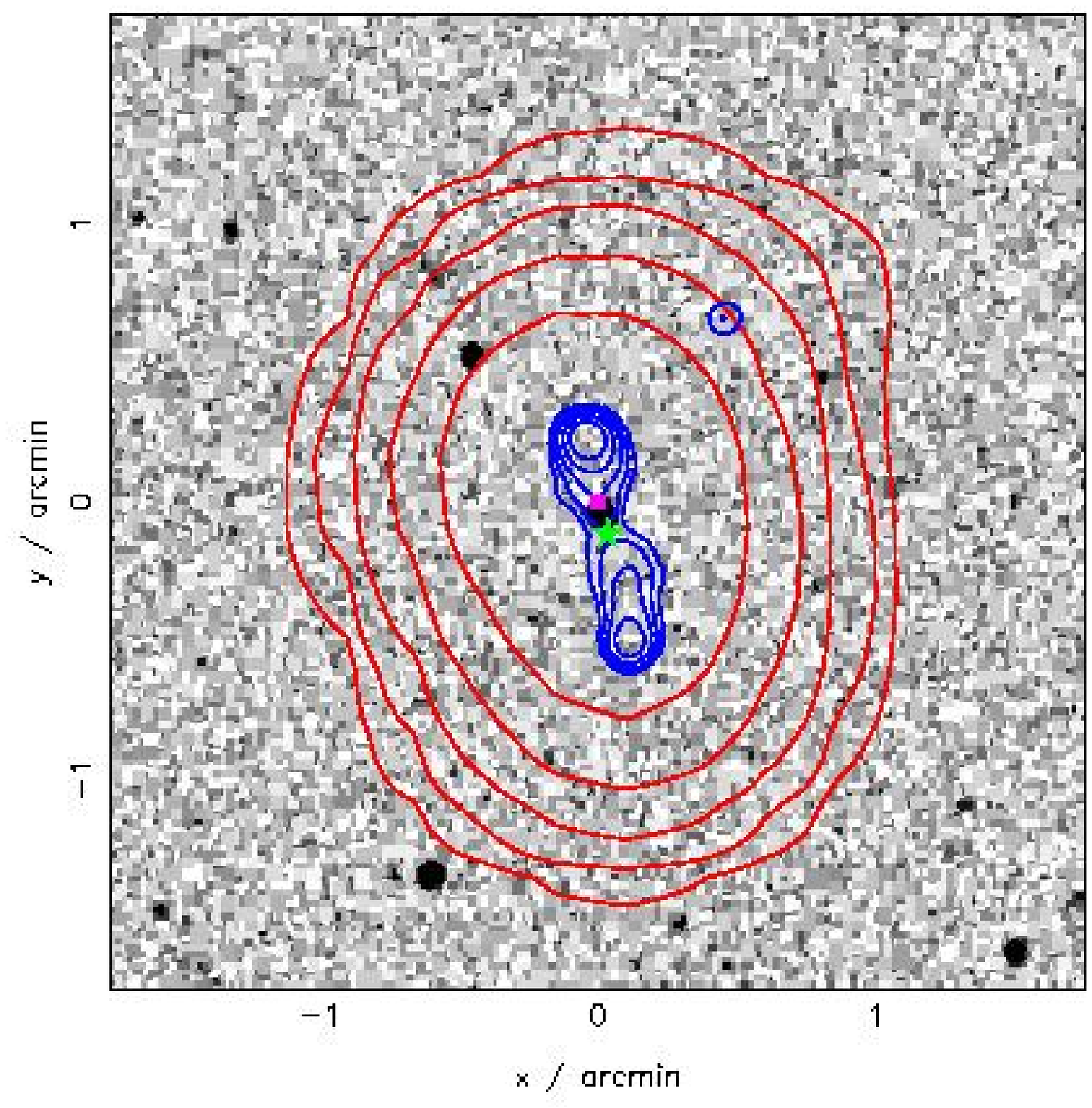}}
      \centerline{C2-232: 3C 281}
    \end{minipage}
    \hspace{3cm}
    \begin{minipage}{3cm}
      \mbox{}
      \centerline{\includegraphics[scale=0.26,angle=270]{Contours/C2/238.ps}}
      \centerline{C2-238: 4C 20.31}
    \end{minipage}
    \hspace{3cm}
    \begin{minipage}{3cm}
      \mbox{}
      \centerline{\includegraphics[scale=0.26,angle=270]{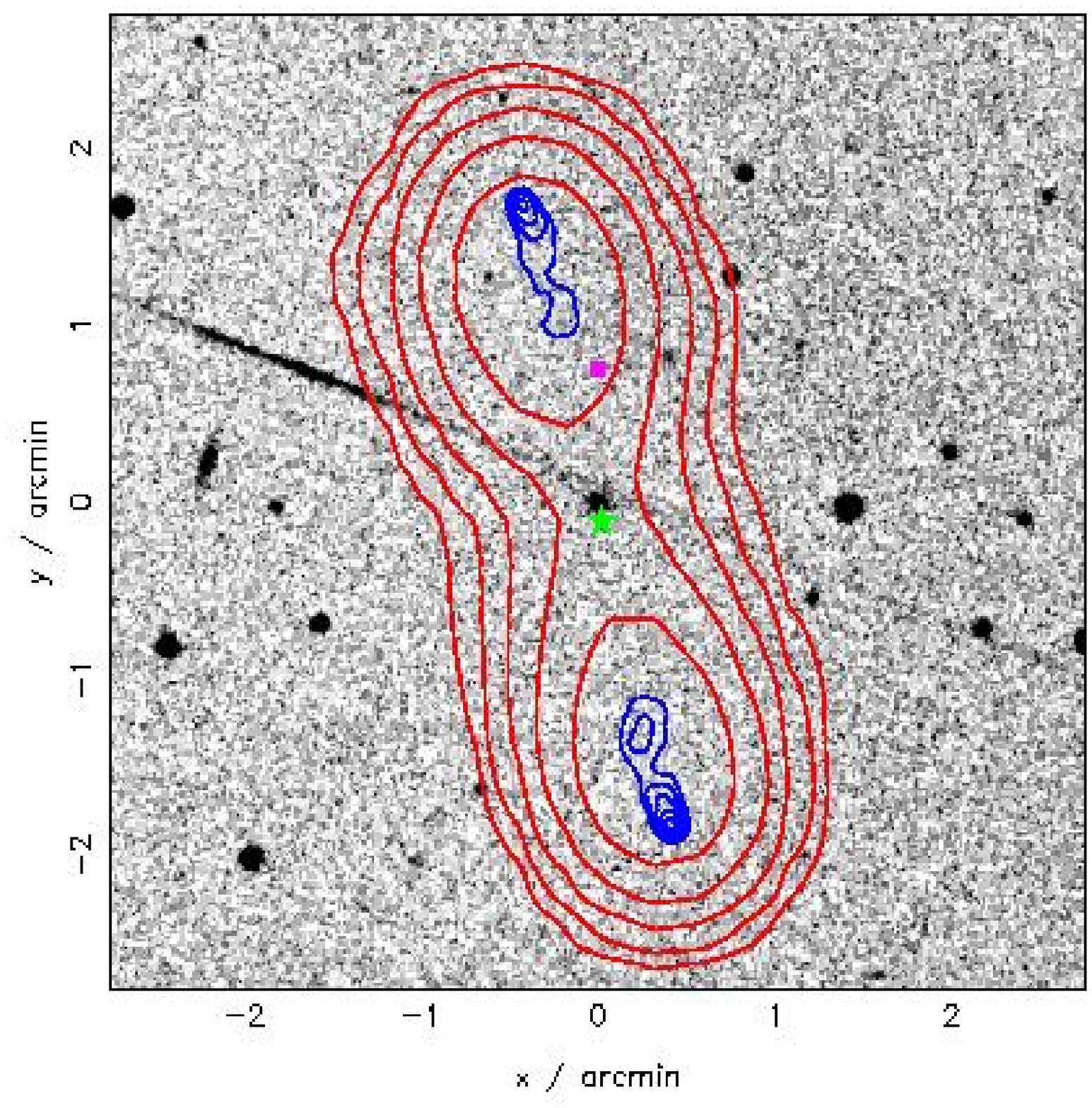}}
      \centerline{C2-239: 4C 08.38}
    \end{minipage}
  \end{center}
\end{figure}

\begin{figure}
  \begin{center}
    {\bf CoNFIG-2}\\  
  \begin{minipage}{3cm}      
      \mbox{}
      \centerline{\includegraphics[scale=0.26,angle=270]{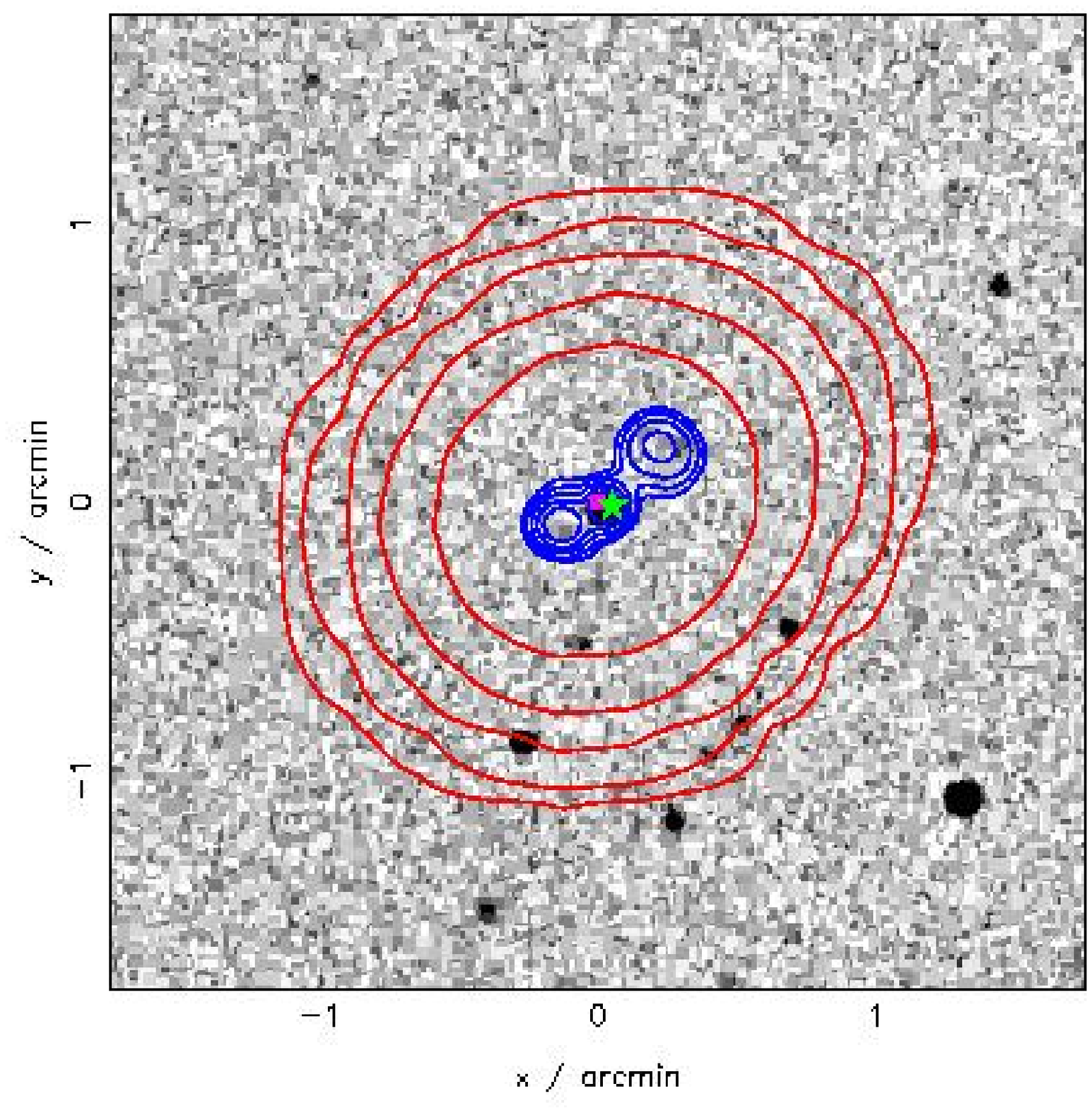}}
      \centerline{C2-243: 4C 52.27}
    \end{minipage}
  \end{center}
\end{figure}

\begin{figure}
  \begin{center}
  {\bf CoNFIG-3}\\
    \begin{minipage}{3cm}      
      \mbox{}
      \centerline{\includegraphics[scale=0.26,angle=270]{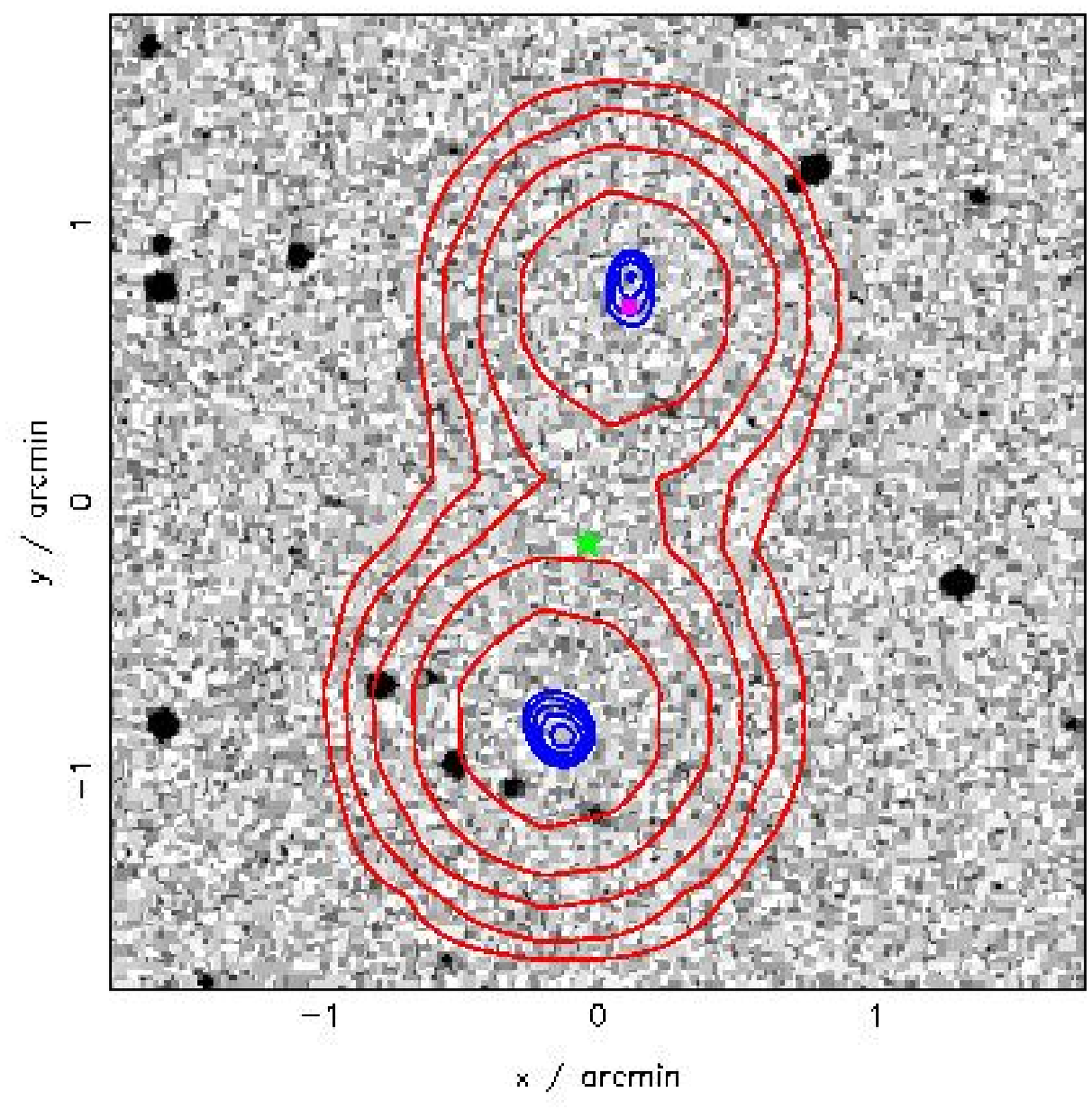}}
      \centerline{C3-002: TXS 1439+252}
    \end{minipage}
    \hspace{3cm}
    \begin{minipage}{3cm}
      \mbox{}
      \centerline{\includegraphics[scale=0.26,angle=270]{Contours/C3/004.ps}}
      \centerline{C3-004: TXS 1440+151}
    \end{minipage}
    \hspace{3cm}
    \begin{minipage}{3cm}
      \mbox{}
      \centerline{\includegraphics[scale=0.26,angle=270]{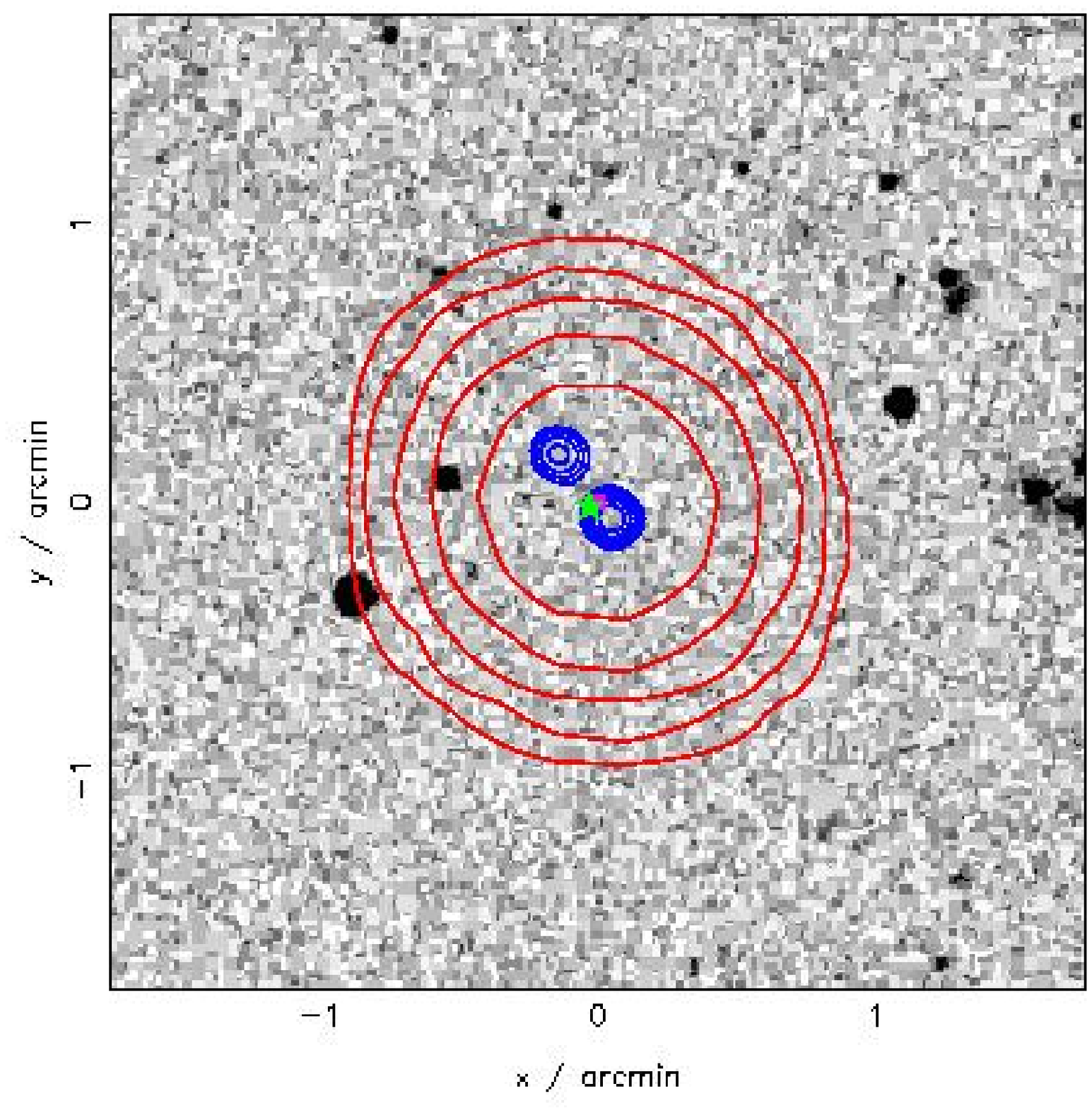}}
      \centerline{C3-005: TXS 1440+147}
    \end{minipage}
    \vfill
    \begin{minipage}{3cm}     
      \mbox{}
      \centerline{\includegraphics[scale=0.26,angle=270]{Contours/C3/006.ps}}
      \centerline{C3-006: TXS 1440+119}
    \end{minipage}
    \hspace{3cm}
    \begin{minipage}{3cm}
      \mbox{}
      \centerline{\includegraphics[scale=0.26,angle=270]{Contours/C3/007.ps}}
      \centerline{C3-007: TXS 1440+163}
    \end{minipage}
    \hspace{3cm}
    \begin{minipage}{3cm}
      \mbox{}
      \centerline{\includegraphics[scale=0.26,angle=270]{Contours/C3/008.ps}}
      \centerline{C3-008: TXS 1440+189}
    \end{minipage}
    \vfill
    \begin{minipage}{3cm}     
      \mbox{}
      \centerline{\includegraphics[scale=0.26,angle=270]{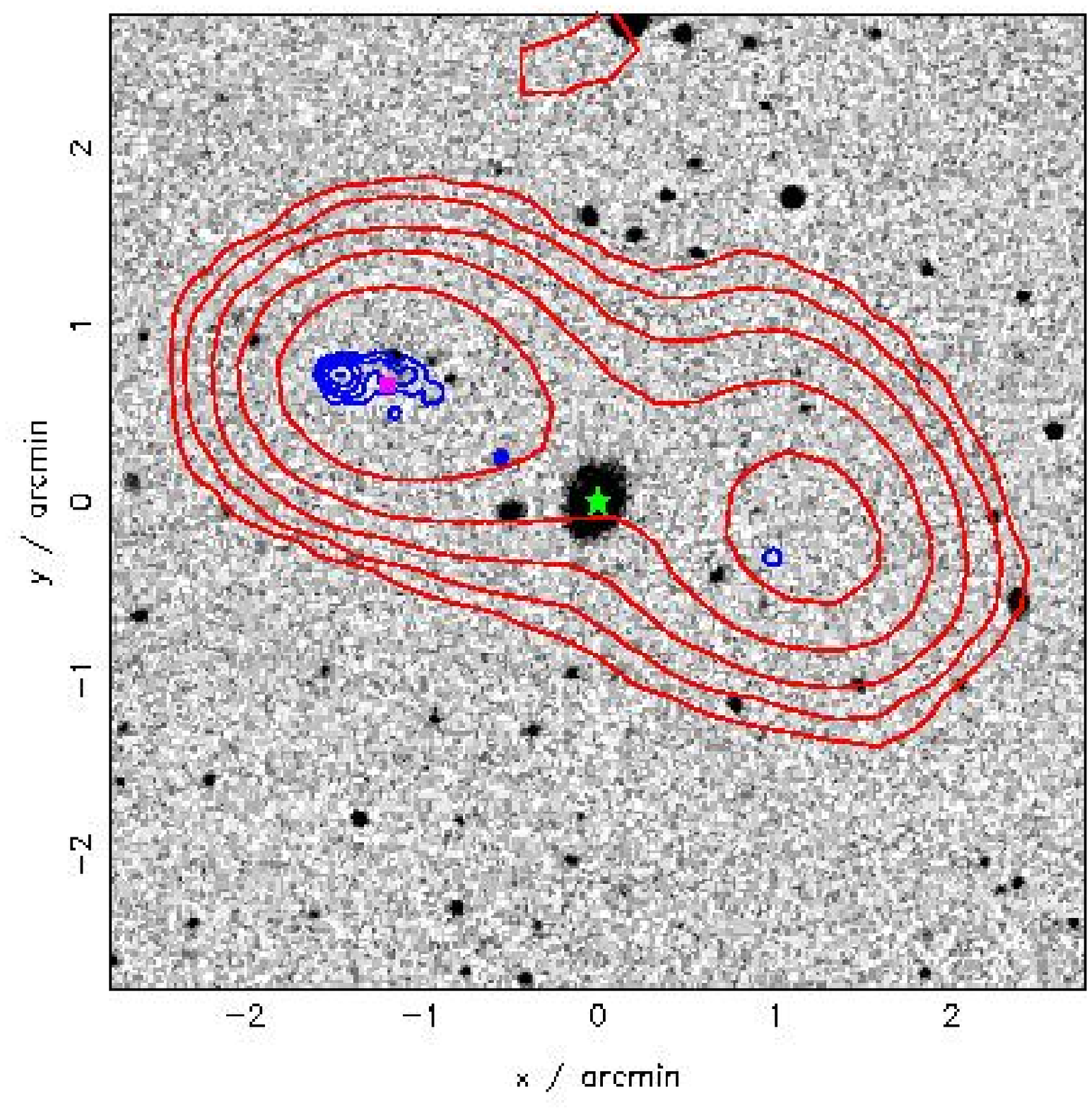}}
      \centerline{C3-012: GB6 1441+2614}
    \end{minipage}
    \hspace{3cm}
    \begin{minipage}{3cm}
      \mbox{}
      \centerline{\includegraphics[scale=0.26,angle=270]{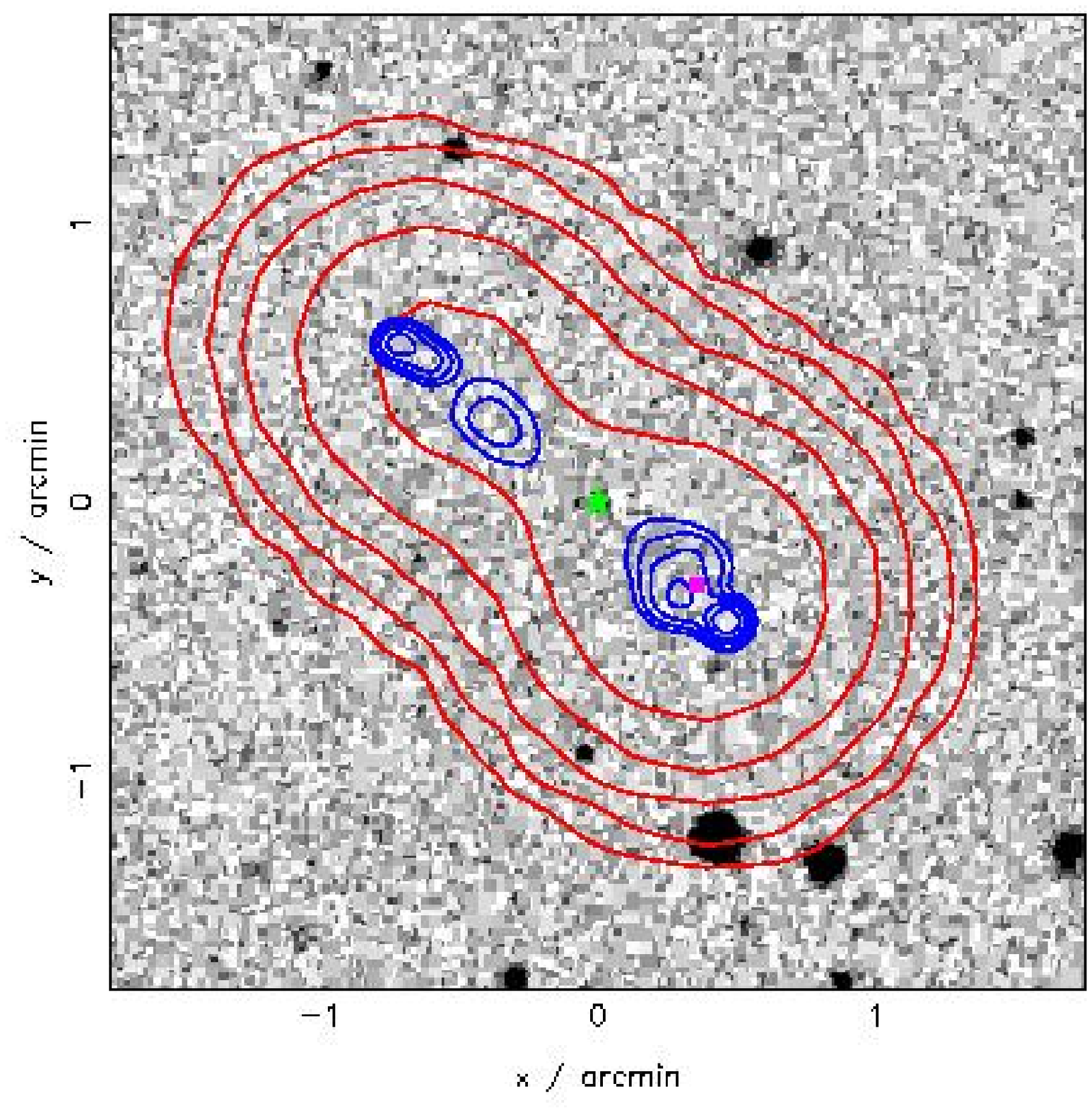}}
      \centerline{C3-014: 4C 14.54}
    \end{minipage}
    \hspace{3cm}
    \begin{minipage}{3cm}
      \mbox{}
      \centerline{\includegraphics[scale=0.26,angle=270]{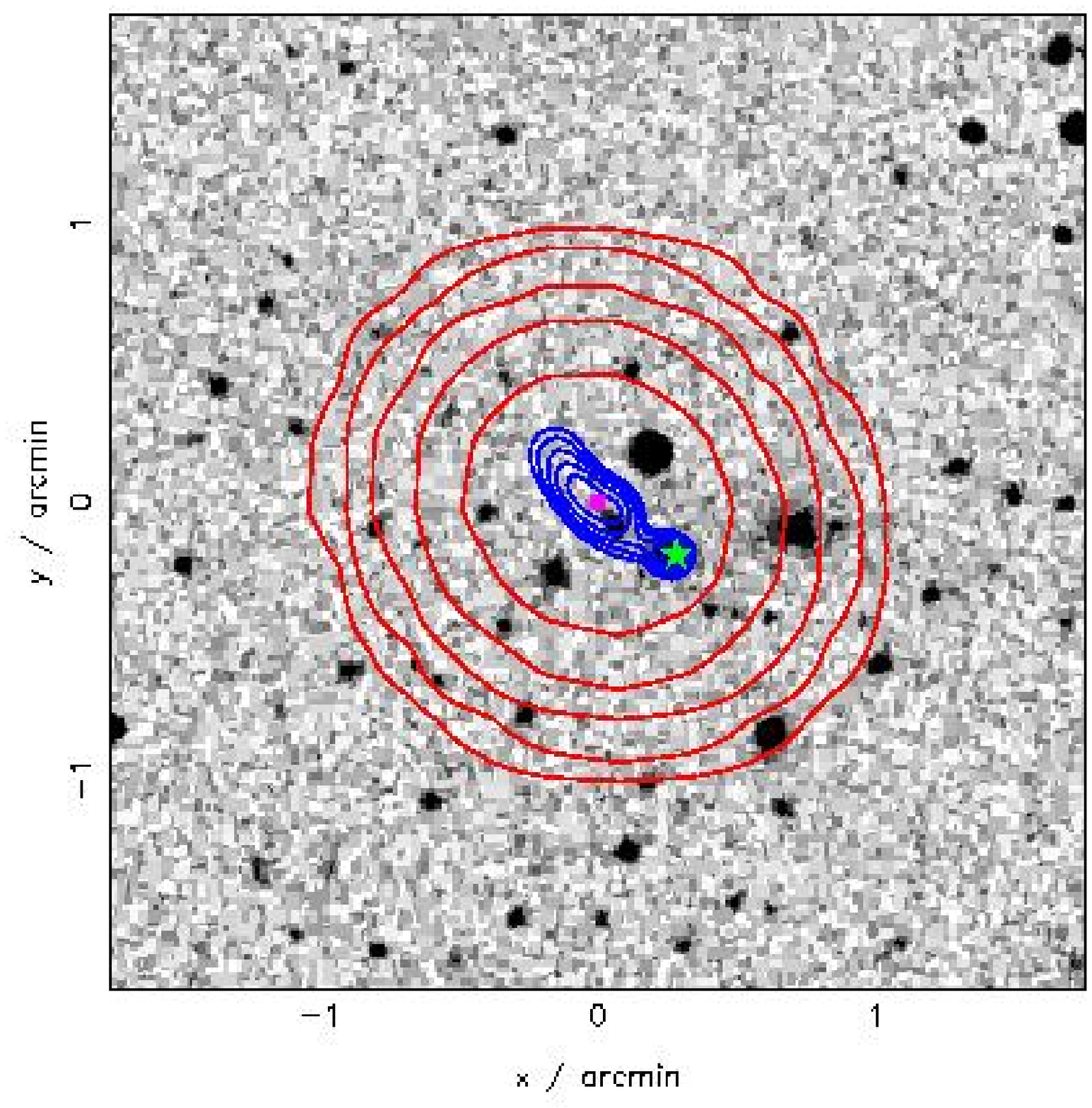}}
      \centerline{C3-015: GB6 1442+195}
    \end{minipage}
    \vfill
    \begin{minipage}{3cm}      
      \mbox{}
      \centerline{\includegraphics[scale=0.26,angle=270]{Contours/C3/016.ps}}
      \centerline{C3-016: GB6 1442+117}
    \end{minipage}
    \hspace{3cm}
    \begin{minipage}{3cm}
      \mbox{}
      \centerline{\includegraphics[scale=0.26,angle=270]{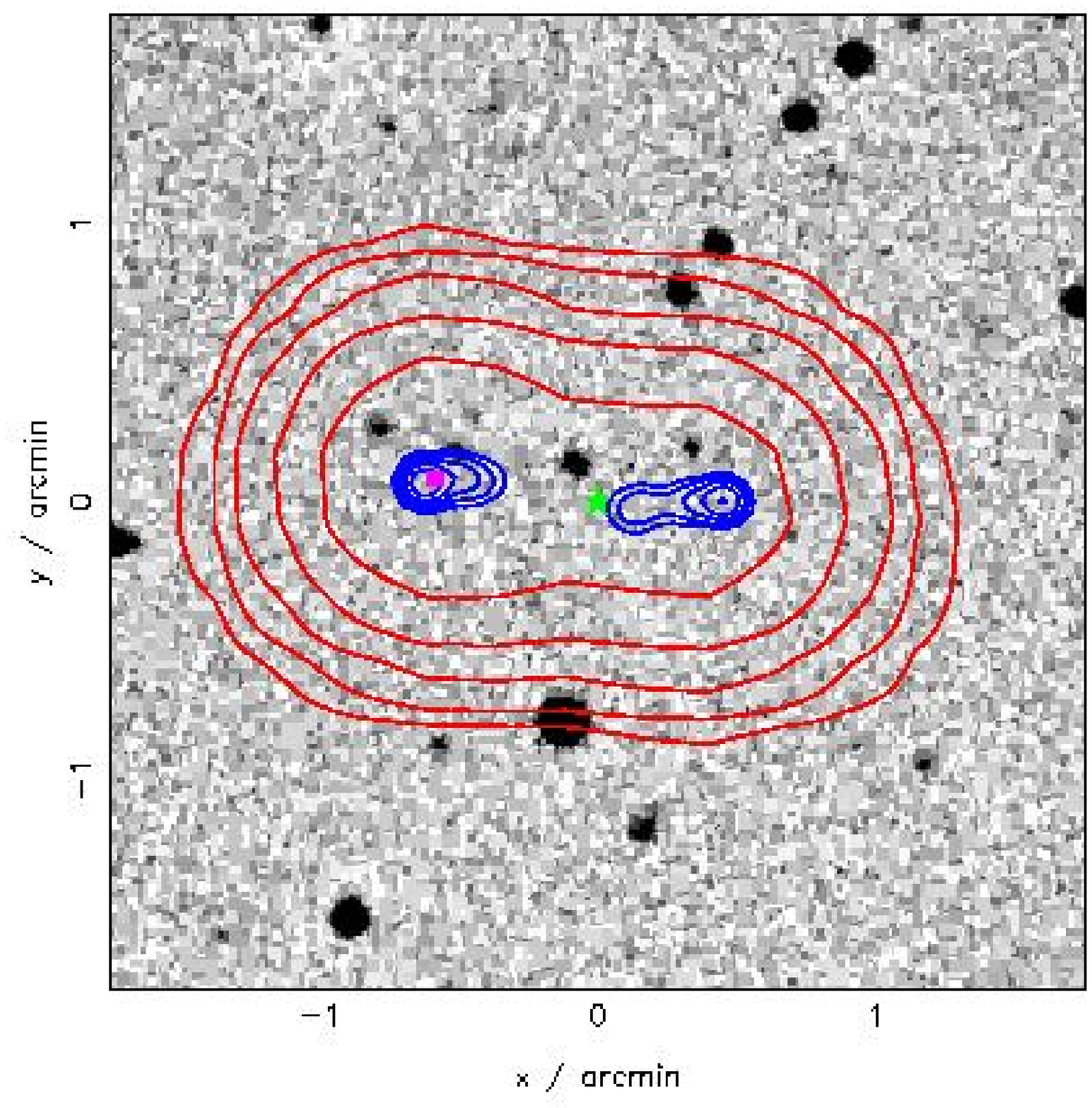}}
      \centerline{C3-017: 4C 16.41}
    \end{minipage}
    \hspace{3cm}
    \begin{minipage}{3cm}
      \mbox{}
      \centerline{\includegraphics[scale=0.26,angle=270]{Contours/C3/019.ps}}
      \centerline{C3-019: 4C 26.44}
    \end{minipage}
  \end{center}
\end{figure}

\begin{figure}
  \begin{center}
    {\bf CoNFIG-3}\\  
  \begin{minipage}{3cm}      
      \mbox{}
      \centerline{\includegraphics[scale=0.26,angle=270]{Contours/C3/020.ps}}
      \centerline{C3-020: TXS 1443+232}
    \end{minipage}
    \hspace{3cm}
    \begin{minipage}{3cm}
      \mbox{}
      \centerline{\includegraphics[scale=0.26,angle=270]{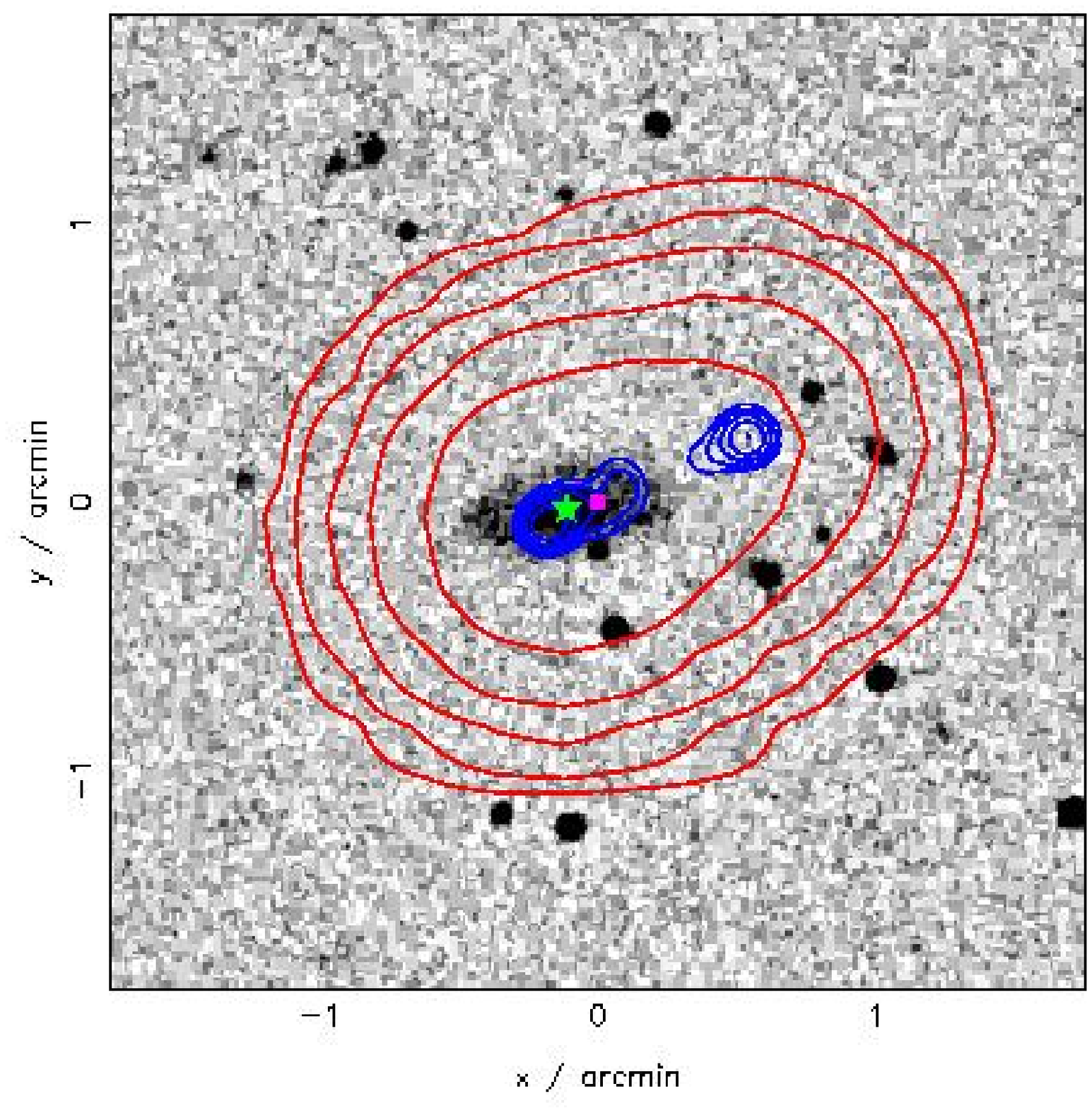}}
      \centerline{C3-021: 4C 17.60}
    \end{minipage}
    \hspace{3cm}
    \begin{minipage}{3cm}
      \mbox{}
      \centerline{\includegraphics[scale=0.26,angle=270]{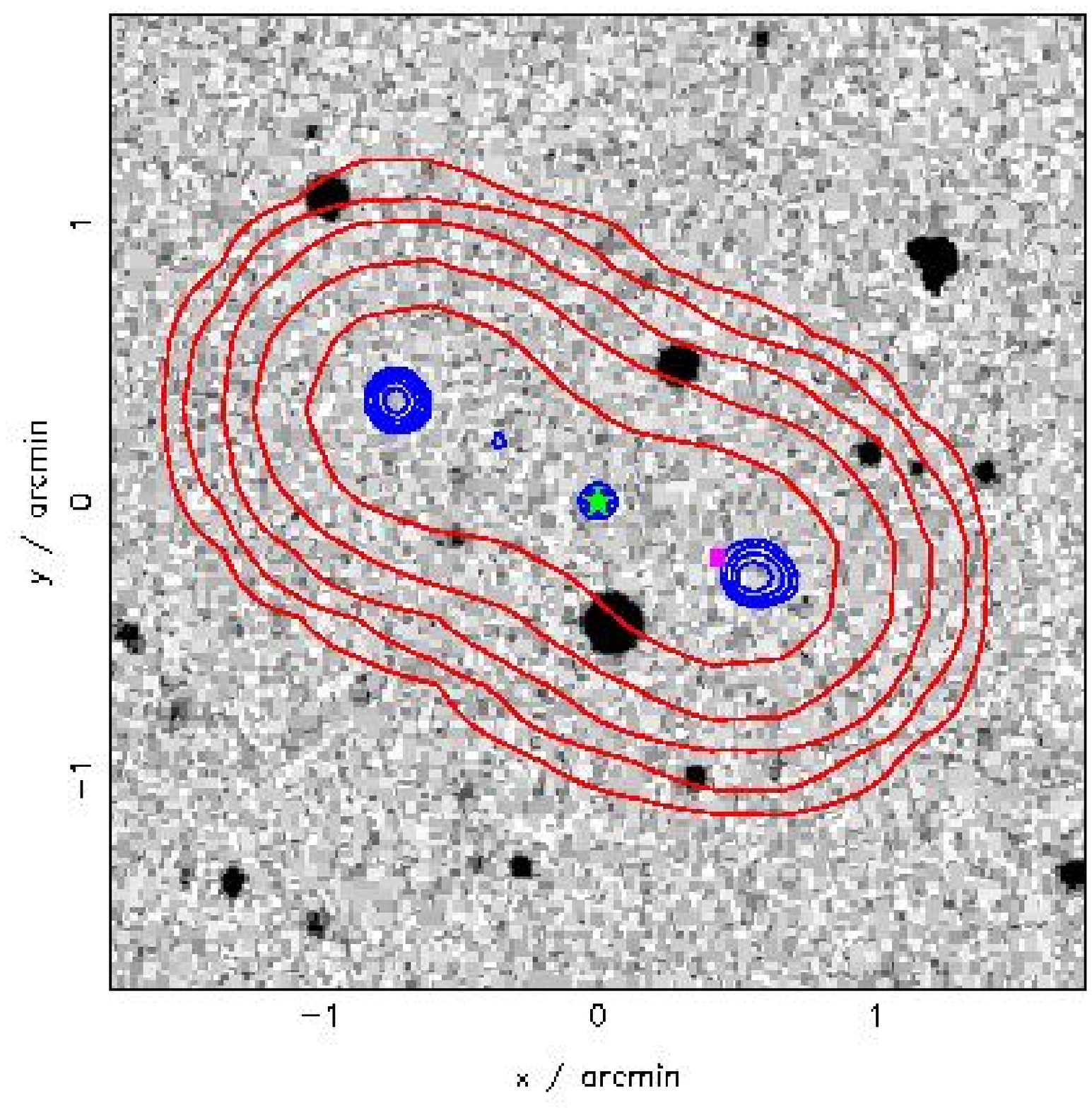}}
      \centerline{C3-022: TXS 1443+125}
    \end{minipage}
    \vfill
    \begin{minipage}{3cm}     
      \mbox{}
      \centerline{\includegraphics[scale=0.26,angle=270]{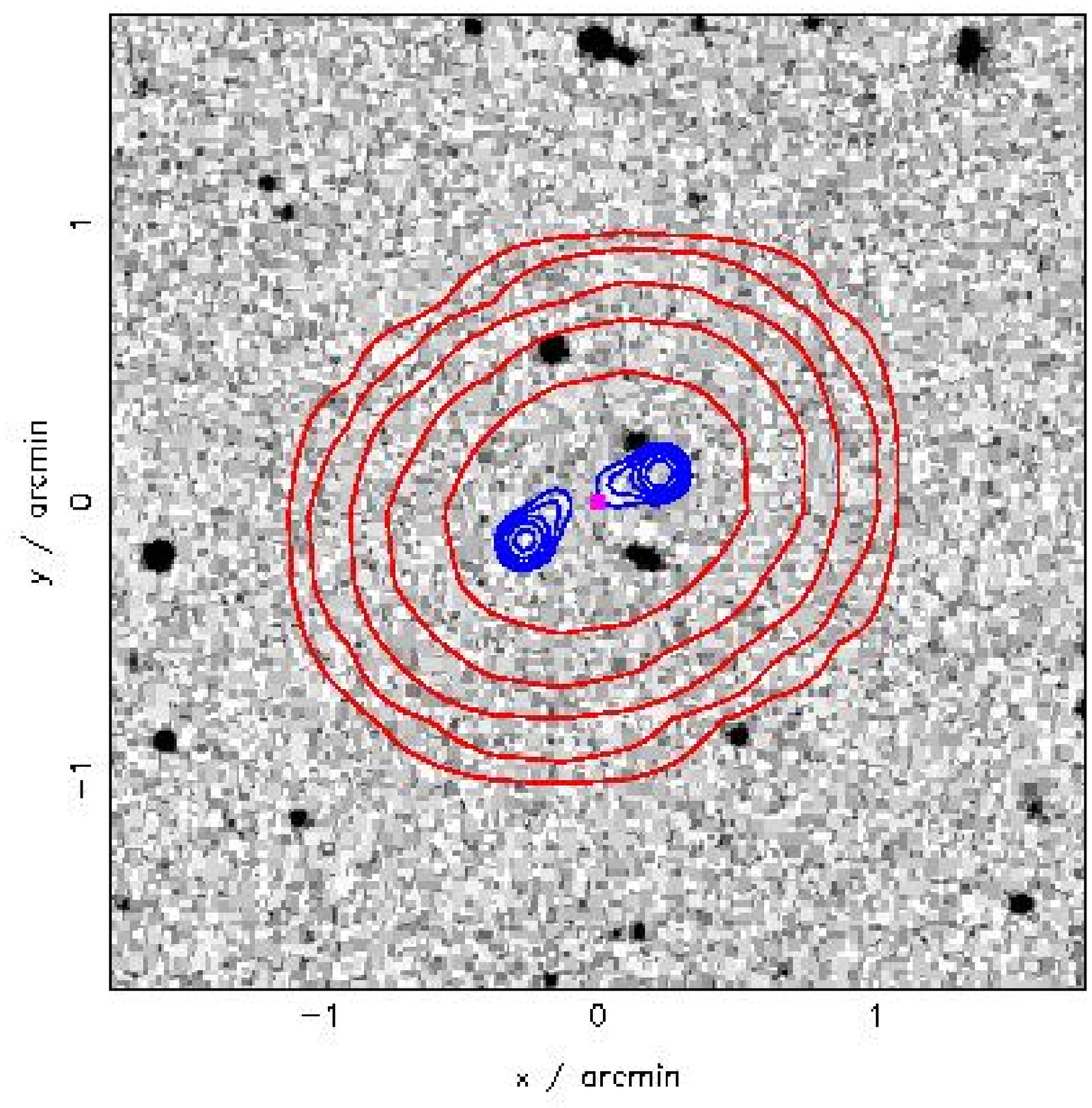}}
      \centerline{C3-023: TXS 1444+254}
    \end{minipage}
    \hspace{3cm}
    \begin{minipage}{3cm}
      \mbox{}
      \centerline{\includegraphics[scale=0.26,angle=270]{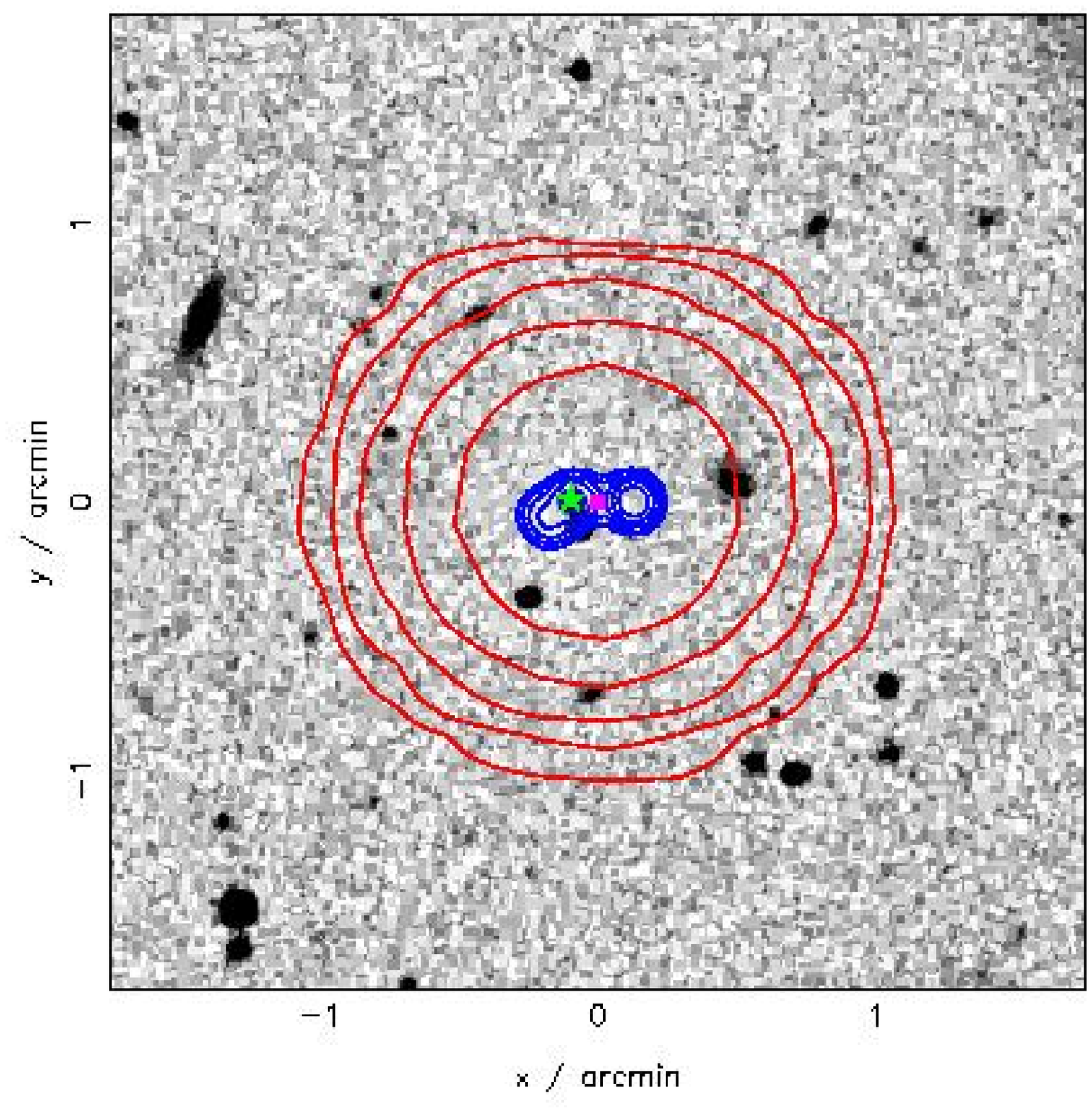}}
      \centerline{C3-026: 4C 21.42}
    \end{minipage}
    \hspace{3cm}
    \begin{minipage}{3cm}
      \mbox{}
      \centerline{\includegraphics[scale=0.26,angle=270]{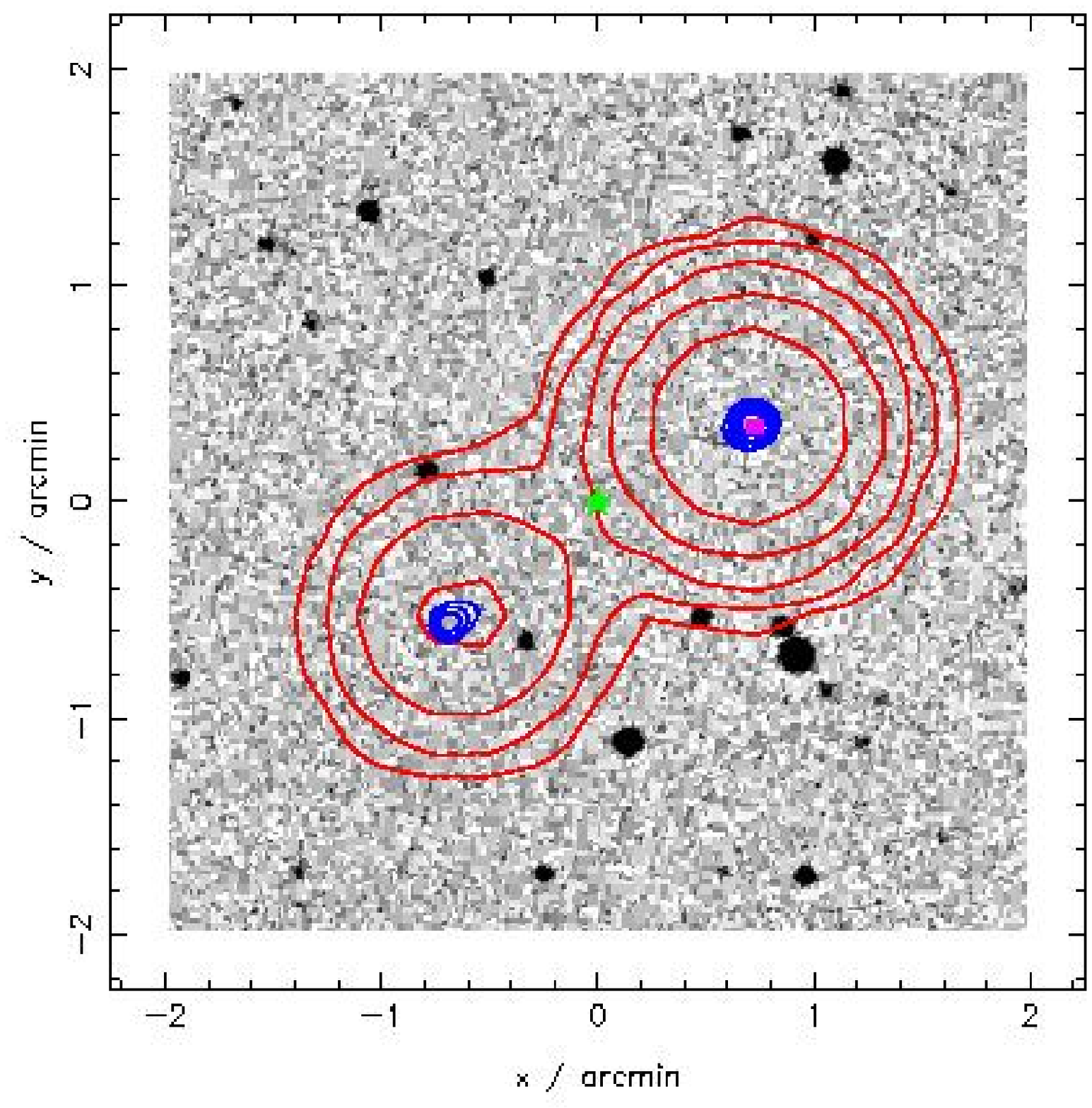}}
      \centerline{C3-029: 4C 16.42}
    \end{minipage}
    \vfill
    \begin{minipage}{3cm}     
      \mbox{}
      \centerline{\includegraphics[scale=0.26,angle=270]{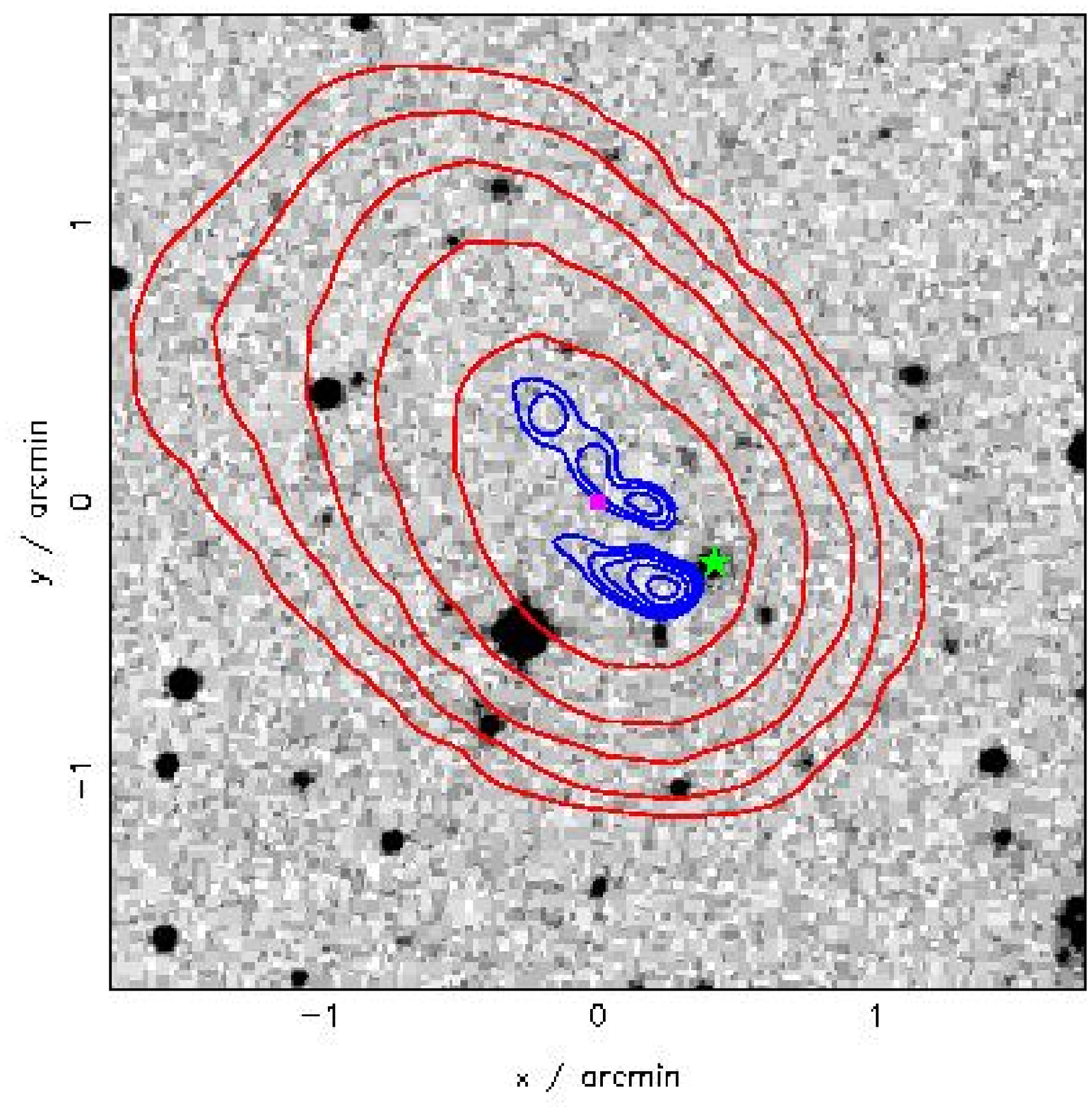}}
      \centerline{C3-030: WB 1445+1459}
    \end{minipage}
    \hspace{3cm}
    \begin{minipage}{3cm}
      \mbox{}
      \centerline{\includegraphics[scale=0.26,angle=270]{Contours/C3/031.ps}}
      \centerline{C3-031: TXS 1445+167}
    \end{minipage}
    \hspace{3cm}
    \begin{minipage}{3cm}
      \mbox{}
      \centerline{\includegraphics[scale=0.26,angle=270]{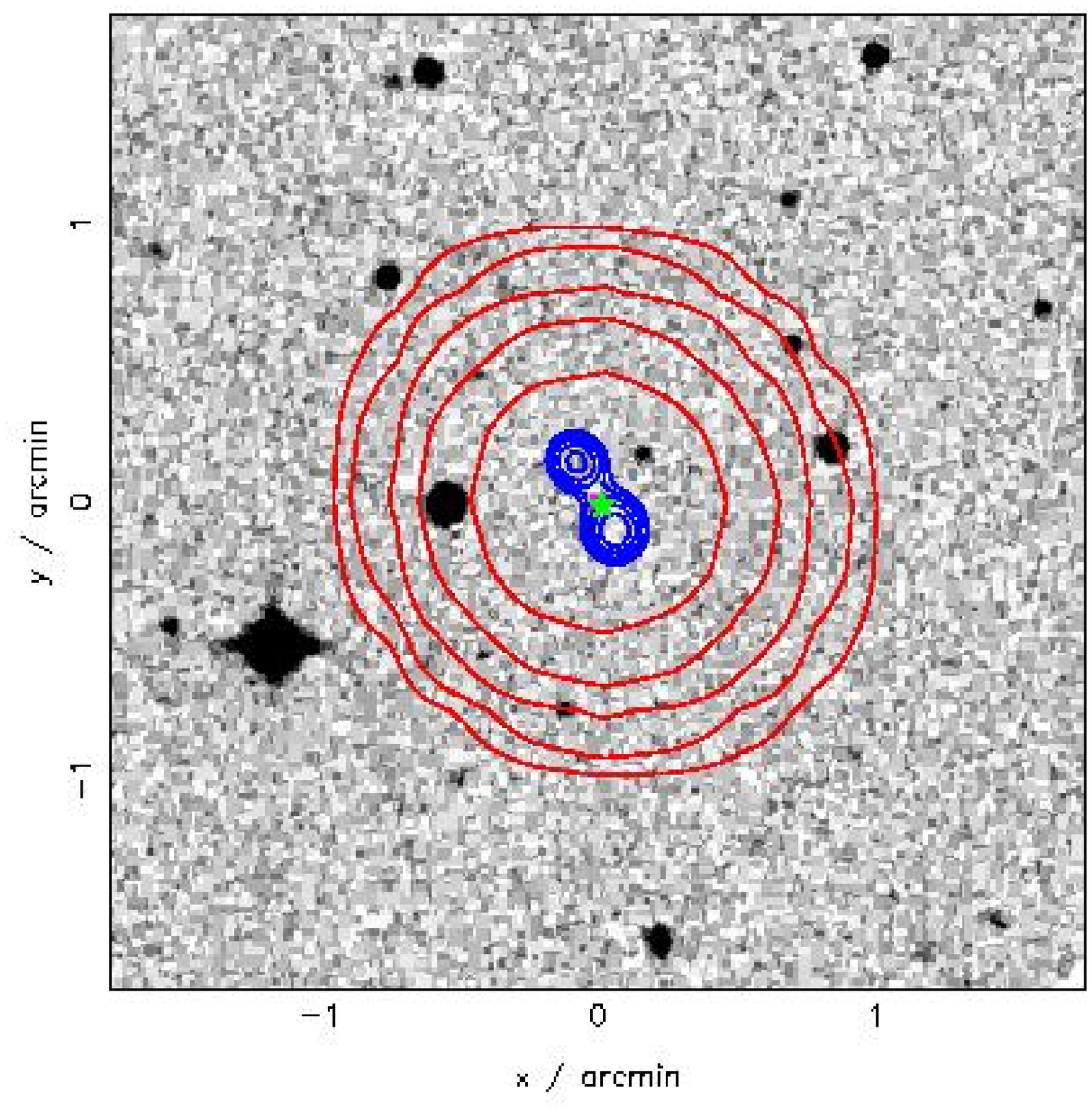}}
      \centerline{C3-033: TXS 1446+177}
    \end{minipage}
    \vfill
    \begin{minipage}{3cm}      
      \mbox{}
      \centerline{\includegraphics[scale=0.26,angle=270]{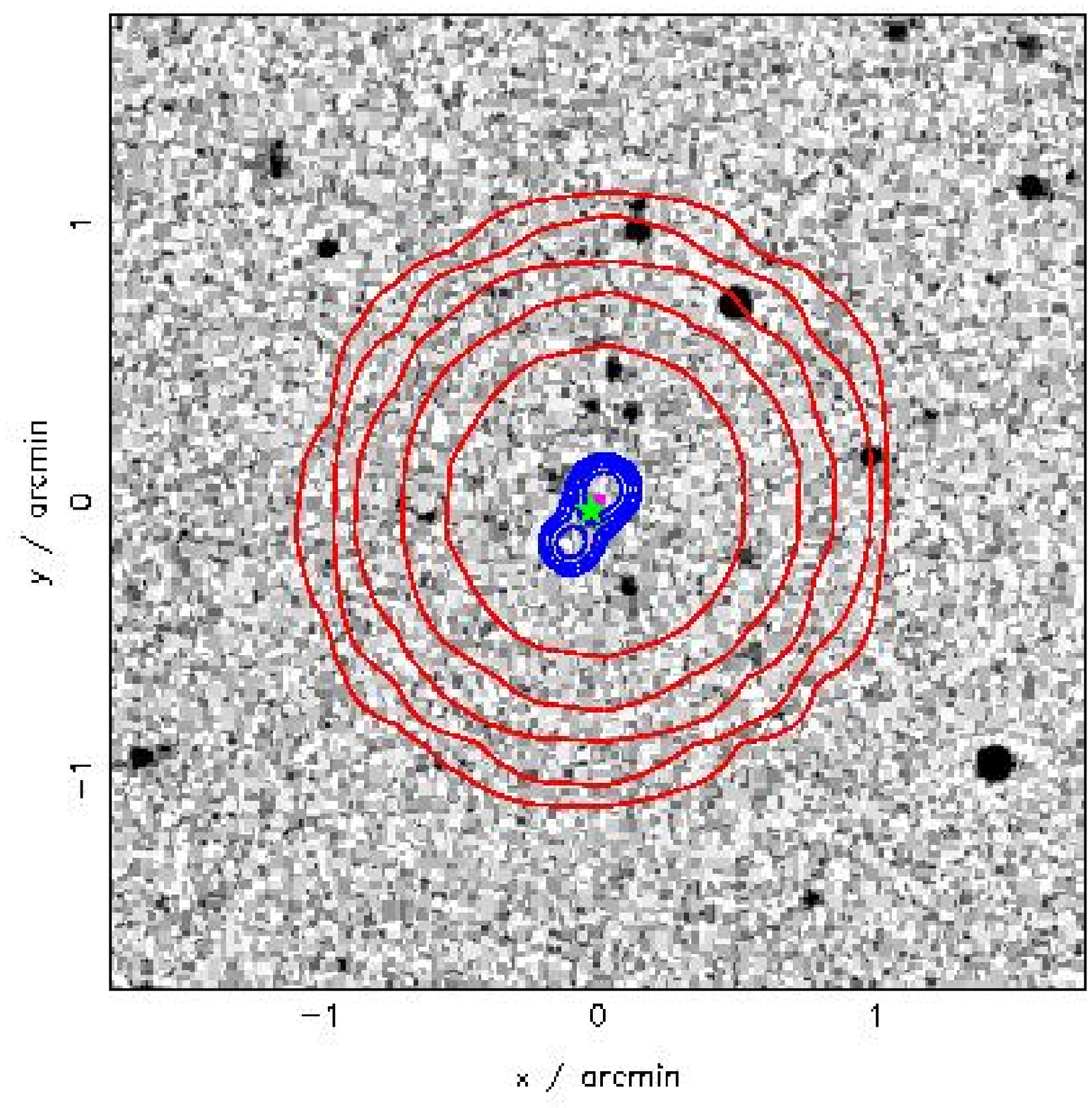}}
      \centerline{C3-034: 3C 304}
    \end{minipage}
    \hspace{3cm}
    \begin{minipage}{3cm}
      \mbox{}
      \centerline{\includegraphics[scale=0.26,angle=270]{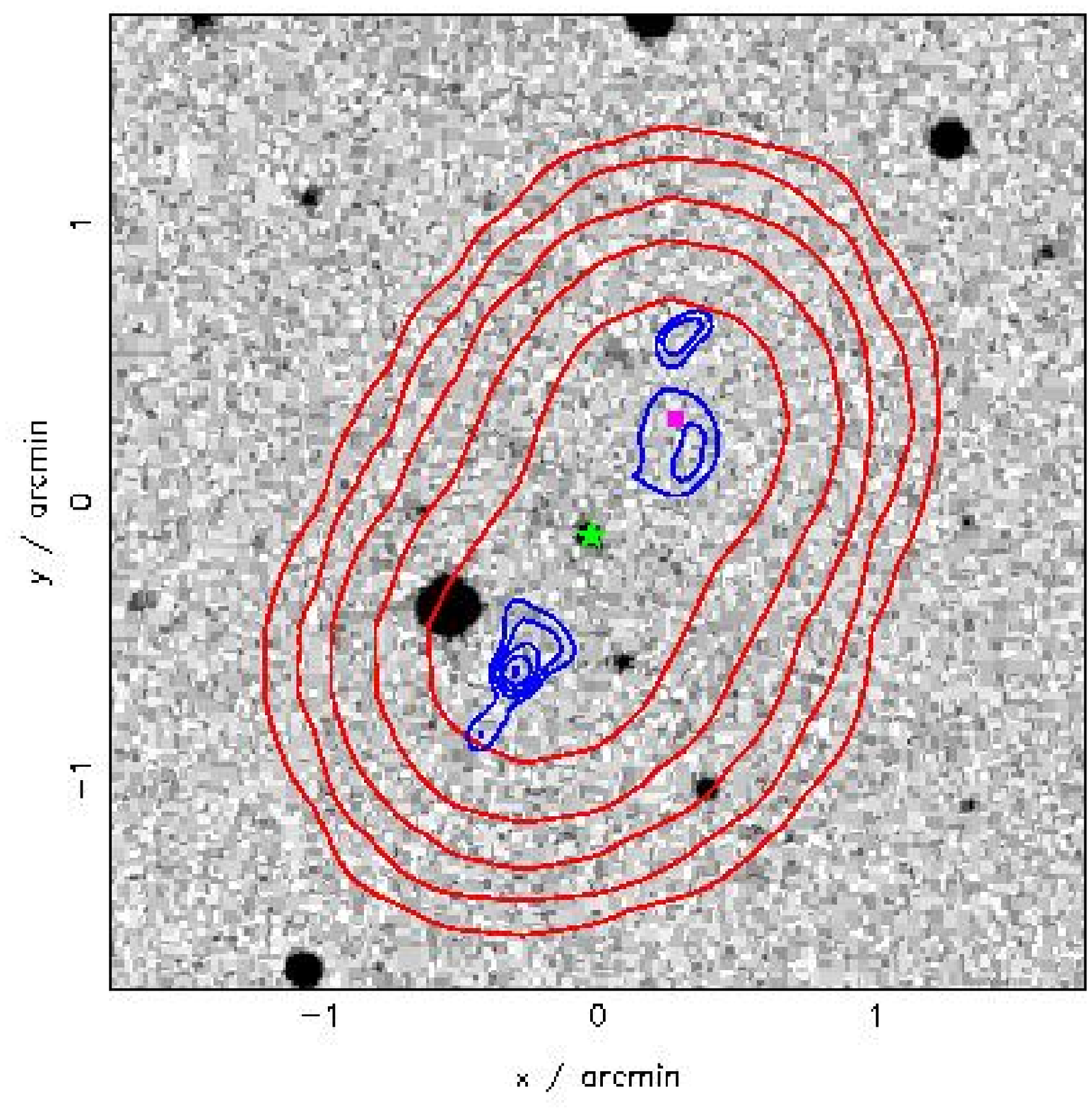}}
      \centerline{C3-035: TXS 1447+213}
    \end{minipage}
    \hspace{3cm}
    \begin{minipage}{3cm}
      \mbox{}
      \centerline{\includegraphics[scale=0.26,angle=270]{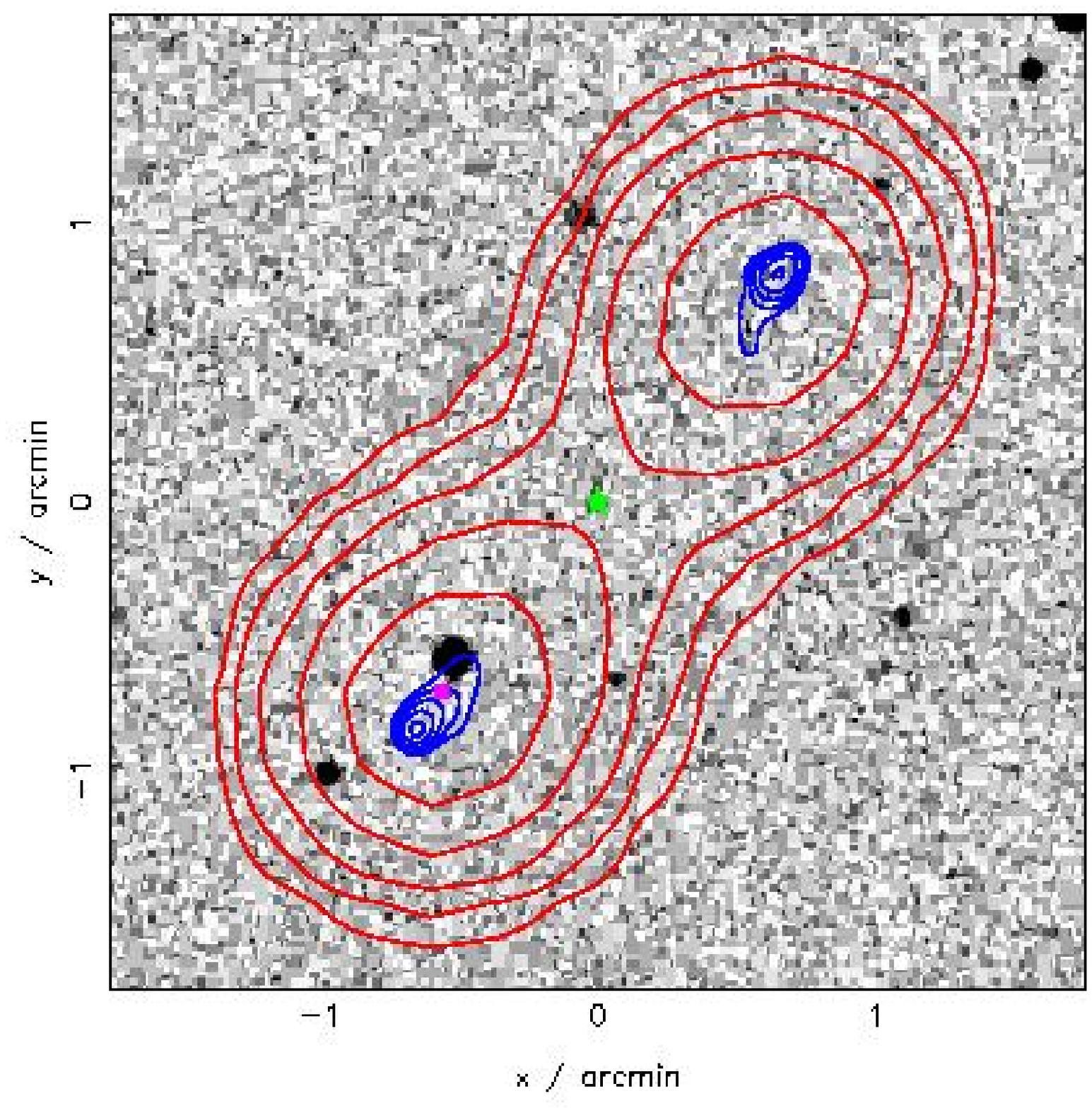}}
      \centerline{C3-036: TXS 1447+224}
    \end{minipage}
  \end{center}
\end{figure}

\begin{figure}
  \begin{center}
    {\bf CoNFIG-3}\\  
  \begin{minipage}{3cm}      
      \mbox{}
      \centerline{\includegraphics[scale=0.26,angle=270]{Contours/C3/038.ps}}
      \centerline{C3-038: TXS 1448+164}
    \end{minipage}
    \hspace{3cm}
    \begin{minipage}{3cm}
      \mbox{}
      \centerline{\includegraphics[scale=0.26,angle=270]{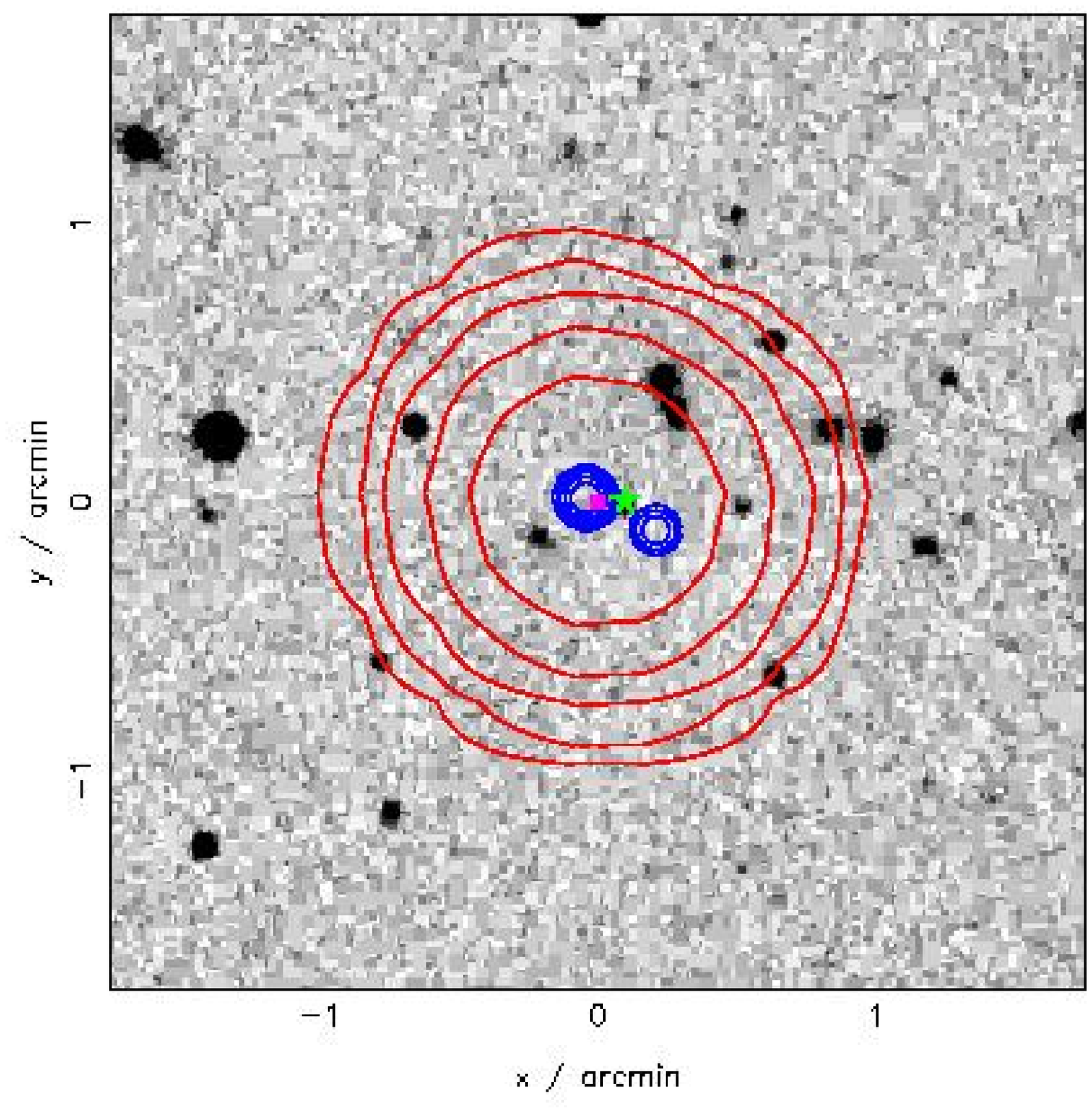}}
      \centerline{C3-040: 4C 14.55}
    \end{minipage}
    \hspace{3cm}
    \begin{minipage}{3cm}
      \mbox{}
      \centerline{\includegraphics[scale=0.26,angle=270]{Contours/C3/045.ps}}
      \centerline{C3-045: TXS 1451+191}
    \end{minipage}
    \vfill
    \begin{minipage}{3cm}     
      \mbox{}
      \centerline{\includegraphics[scale=0.26,angle=270]{Contours/C3/046.ps}}
      \centerline{C3-046: TXS 1451+292}
    \end{minipage}
    \hspace{3cm}
    \begin{minipage}{3cm}
      \mbox{}
      \centerline{\includegraphics[scale=0.26,angle=270]{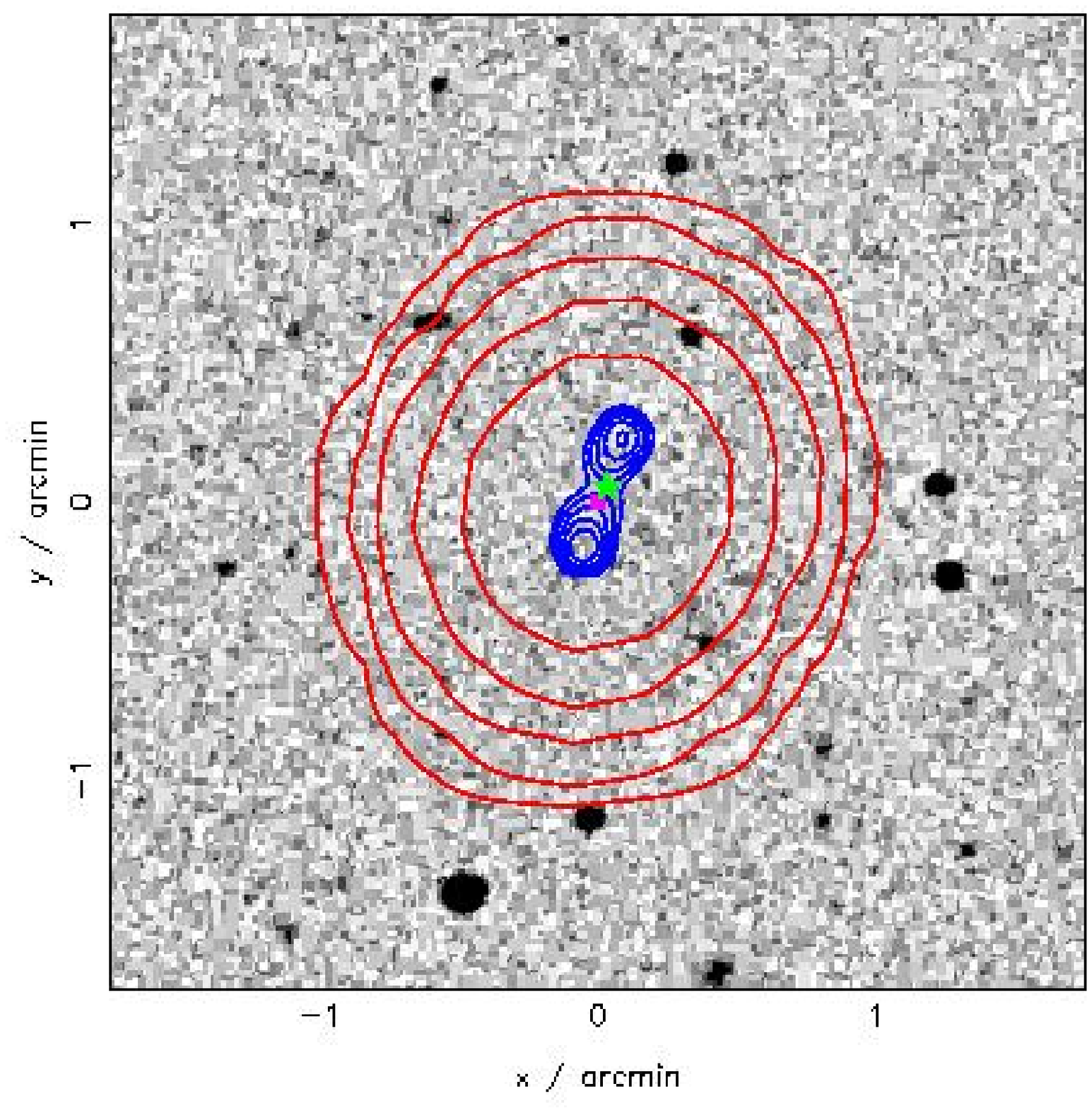}}
      \centerline{C3-049: TXS 1451+118}
    \end{minipage}
    \hspace{3cm}
    \begin{minipage}{3cm}
      \mbox{}
      \centerline{\includegraphics[scale=0.26,angle=270]{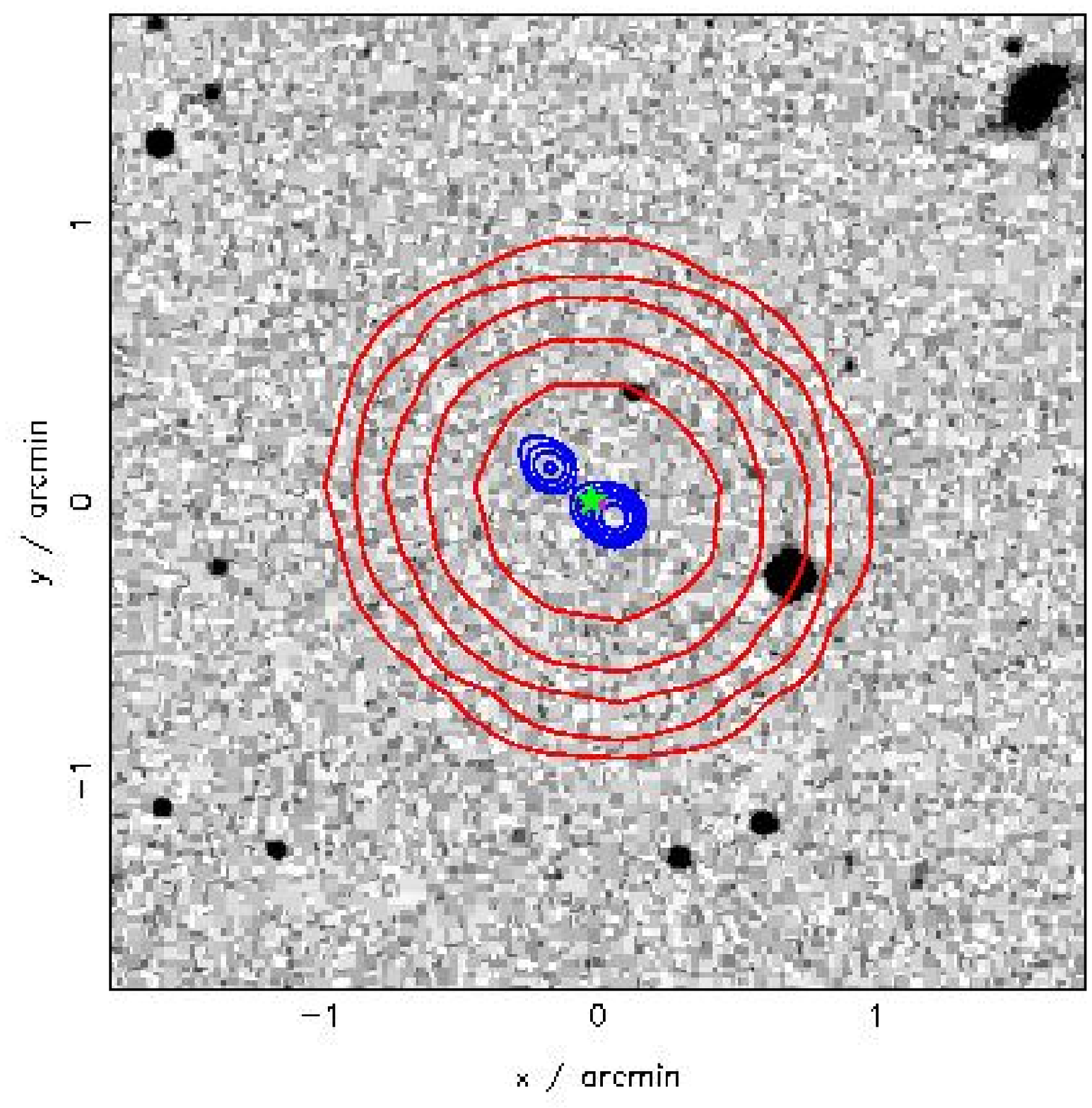}}
      \centerline{C3-052: TXS 1452+258}
    \end{minipage}
    \vfill
    \begin{minipage}{3cm}     
      \mbox{}
      \centerline{\includegraphics[scale=0.26,angle=270]{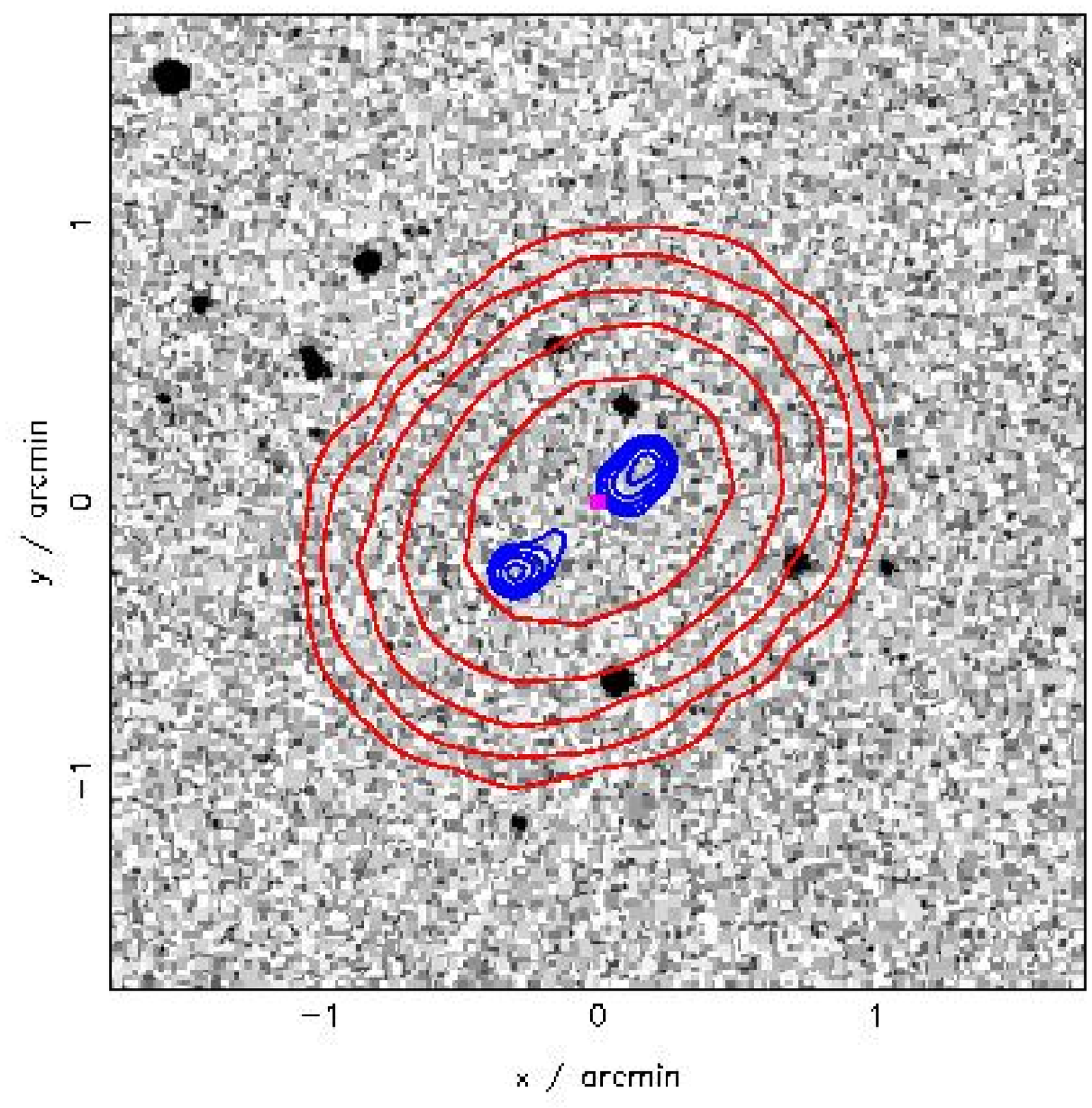}}
      \centerline{C3-053: TXS 1452+204}
    \end{minipage}
    \hspace{3cm}
    \begin{minipage}{3cm}
      \mbox{}
      \centerline{\includegraphics[scale=0.26,angle=270]{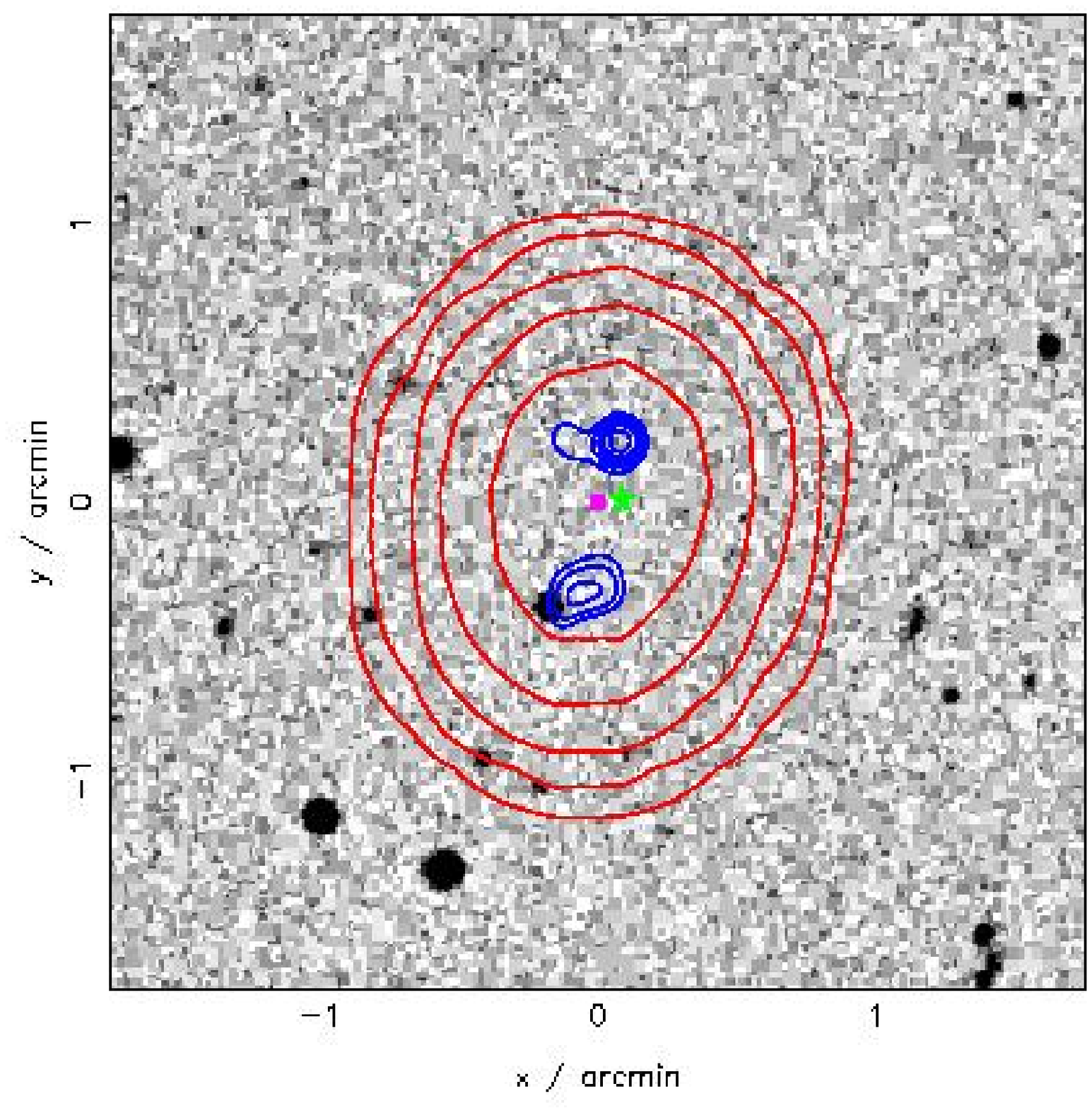}}
      \centerline{C3-055: TXS 1452+277}
    \end{minipage}
    \hspace{3cm}
    \begin{minipage}{3cm}
      \mbox{}
      \centerline{\includegraphics[scale=0.26,angle=270]{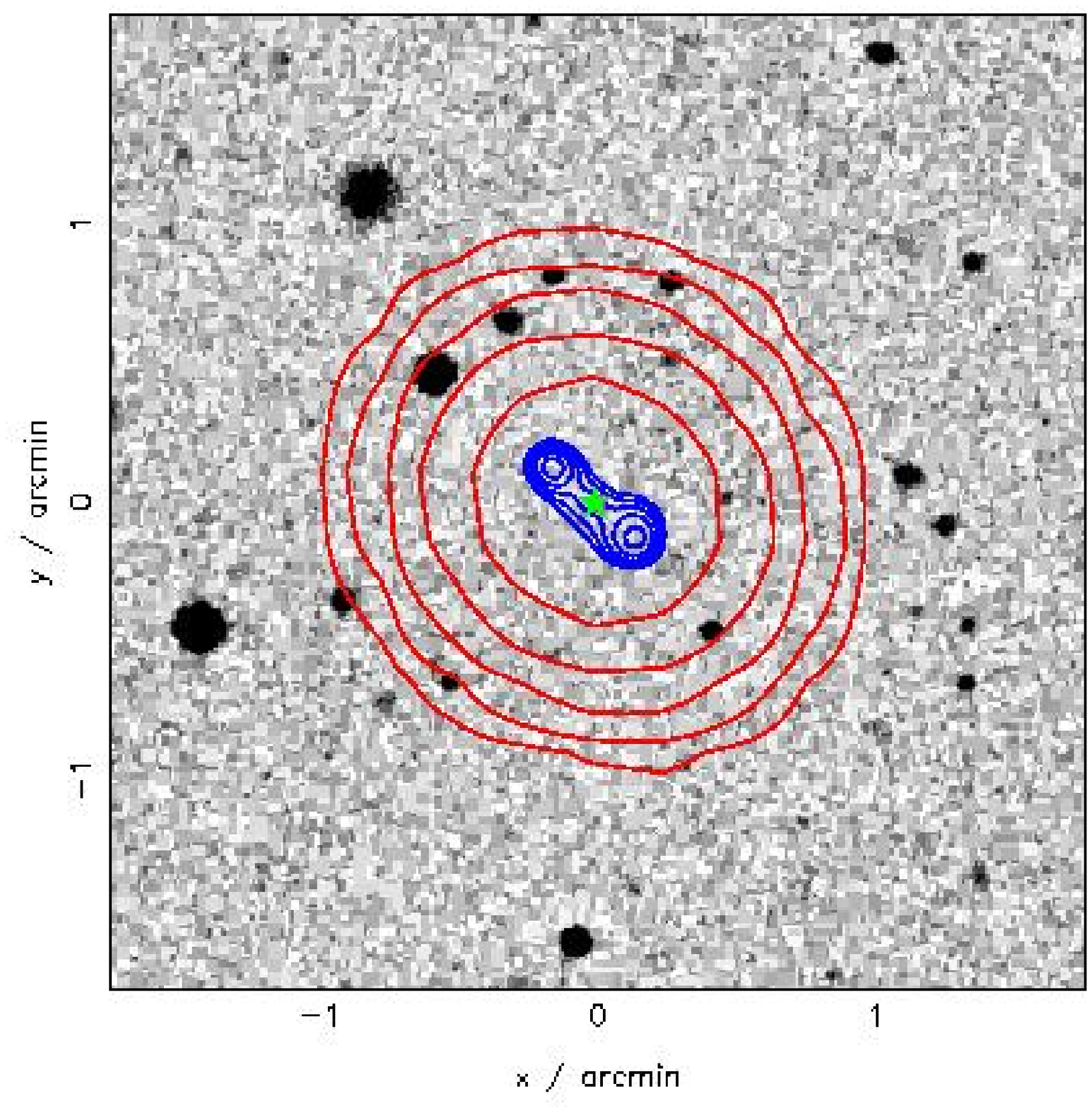}}
      \centerline{C3-056: TXS 1452+144}
    \end{minipage}
    \vfill
    \begin{minipage}{3cm}      
      \mbox{}
      \centerline{\includegraphics[scale=0.26,angle=270]{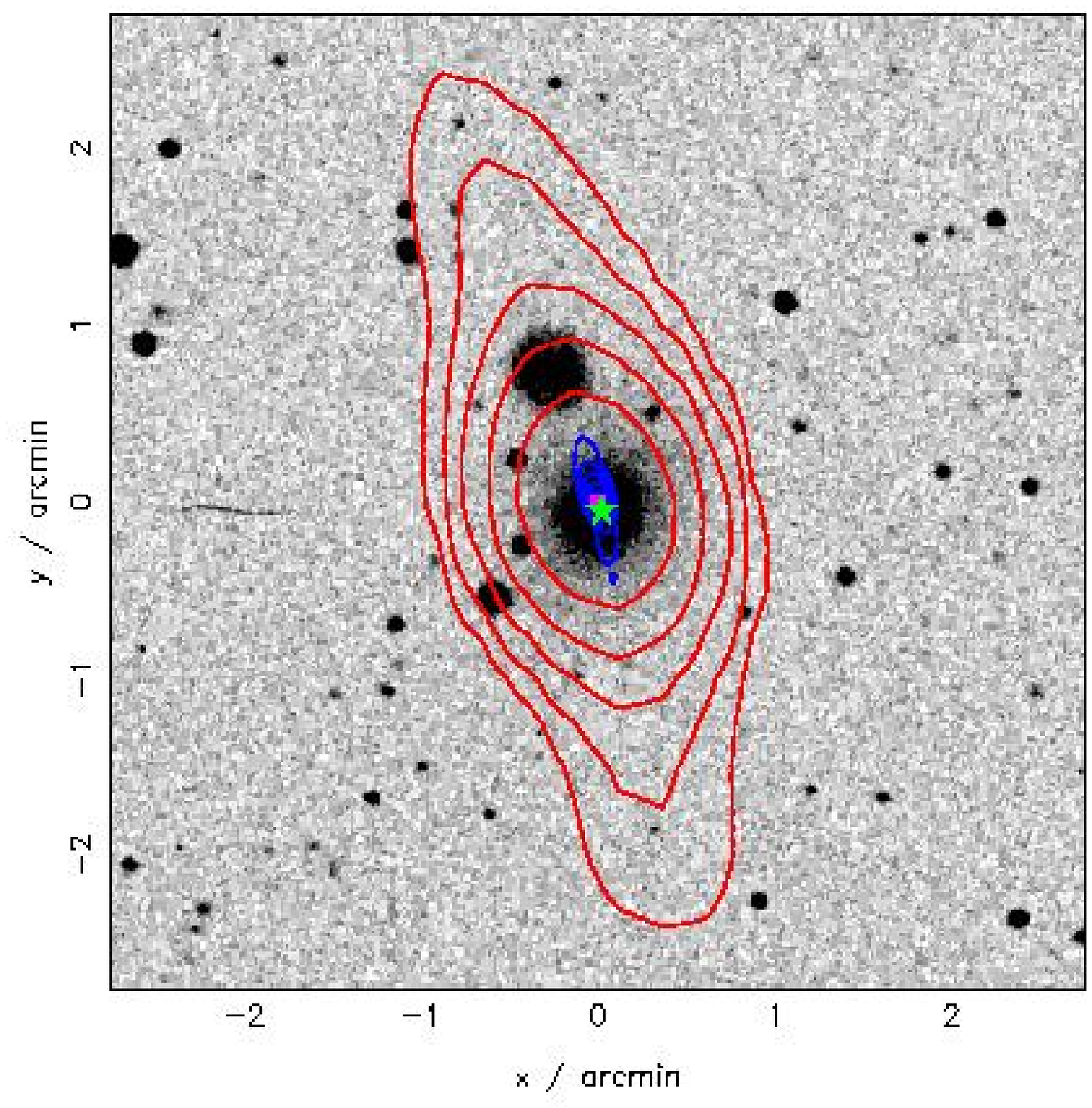}}
      \centerline{C3-057: NGC 5782}
    \end{minipage}
    \hspace{3cm}
    \begin{minipage}{3cm}
      \mbox{}
      \centerline{\includegraphics[scale=0.26,angle=270]{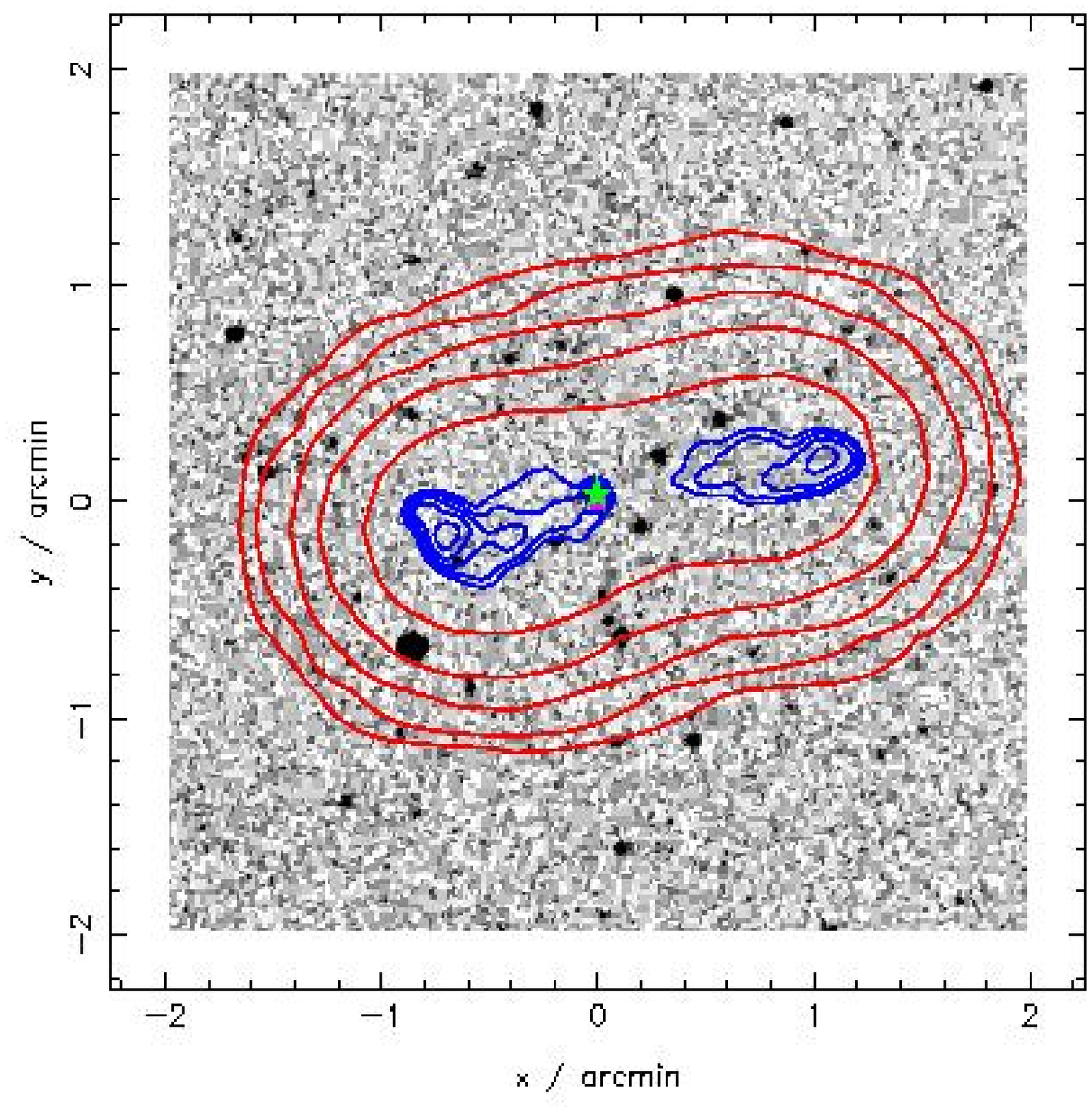}}
      \centerline{C3-058: 4C 16.43}
    \end{minipage}
    \hspace{3cm}
    \begin{minipage}{3cm}
      \mbox{}
      \centerline{\includegraphics[scale=0.26,angle=270]{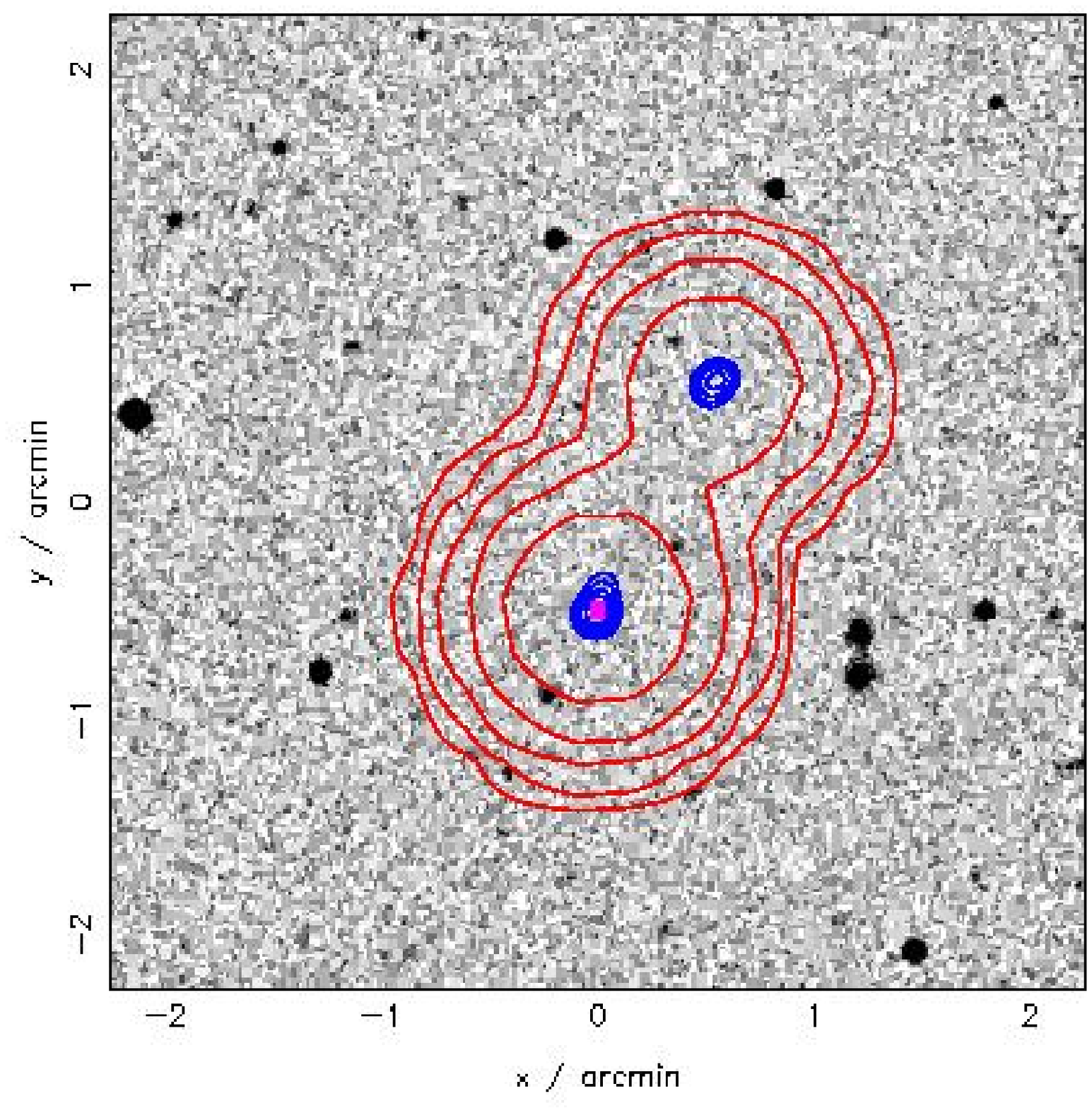}}
      \centerline{C3-059: TXS 1454+268}
    \end{minipage}
  \end{center}
\end{figure}

\begin{figure}
  \begin{center}
    {\bf CoNFIG-3}\\  
  \begin{minipage}{3cm}      
      \mbox{}
      \centerline{\includegraphics[scale=0.26,angle=270]{Contours/C3/060.ps}}
      \centerline{C3-060: TXS 1454+132}
    \end{minipage}
    \hspace{3cm}
    \begin{minipage}{3cm}
      \mbox{}
      \centerline{\includegraphics[scale=0.26,angle=270]{Contours/C3/061.ps}}
      \centerline{C3-061: 4C 18.39}
    \end{minipage}
    \hspace{3cm}
    \begin{minipage}{3cm}
      \mbox{}
      \centerline{\includegraphics[scale=0.26,angle=270]{Contours/C3/062.ps}}
      \centerline{C3-062: TXS 1454+244}
    \end{minipage}
    \vfill
    \begin{minipage}{3cm}     
      \mbox{}
      \centerline{\includegraphics[scale=0.26,angle=270]{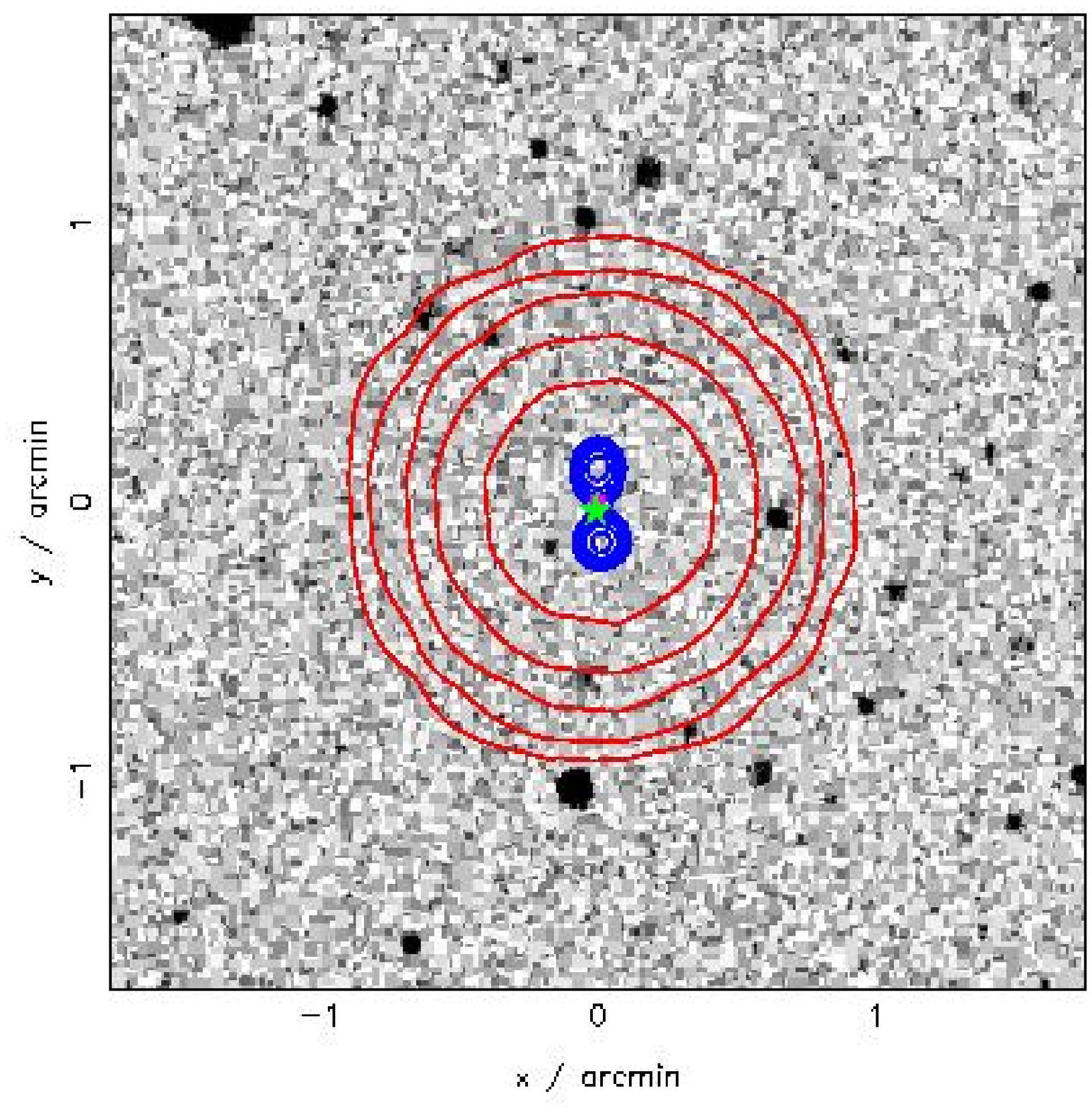}}
      \centerline{C3-063: 7C 1454+2753}
    \end{minipage}
    \hspace{3cm}
    \begin{minipage}{3cm}
      \mbox{}
      \centerline{\includegraphics[scale=0.26,angle=270]{Contours/C3/064.ps}}
      \centerline{C3-064: TXS 1454+139}
    \end{minipage}
    \hspace{3cm}
    \begin{minipage}{3cm}
      \mbox{}
      \centerline{\includegraphics[scale=0.26,angle=270]{Contours/C3/065.ps}}
      \centerline{C3-065: TXS 1454+271}
    \end{minipage}
    \vfill
    \begin{minipage}{3cm}     
      \mbox{}
      \centerline{\includegraphics[scale=0.26,angle=270]{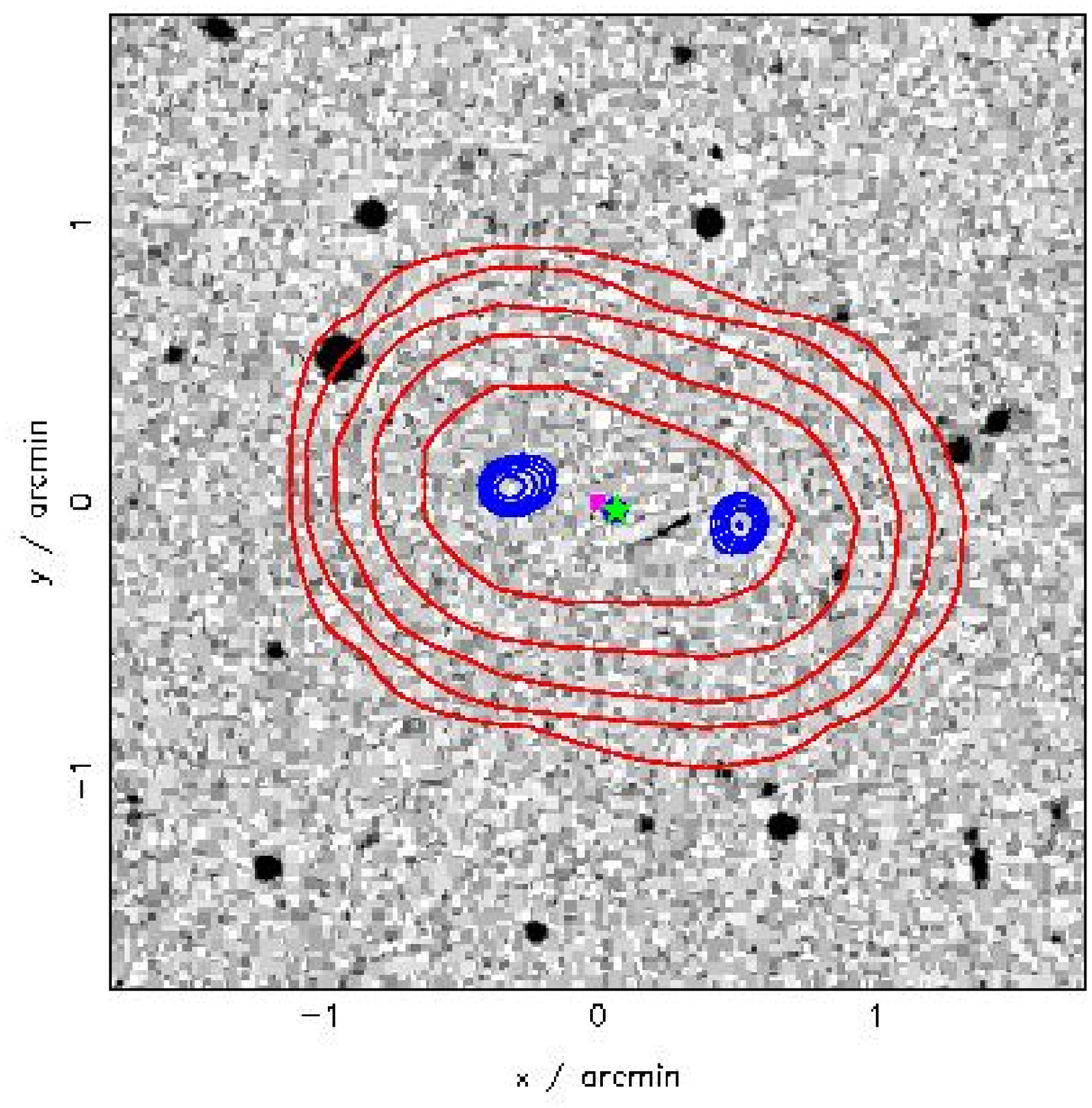}}
      \centerline{C3-066: TXS 1455+251}
    \end{minipage}
    \hspace{3cm}
    \begin{minipage}{3cm}
      \mbox{}
      \centerline{\includegraphics[scale=0.26,angle=270]{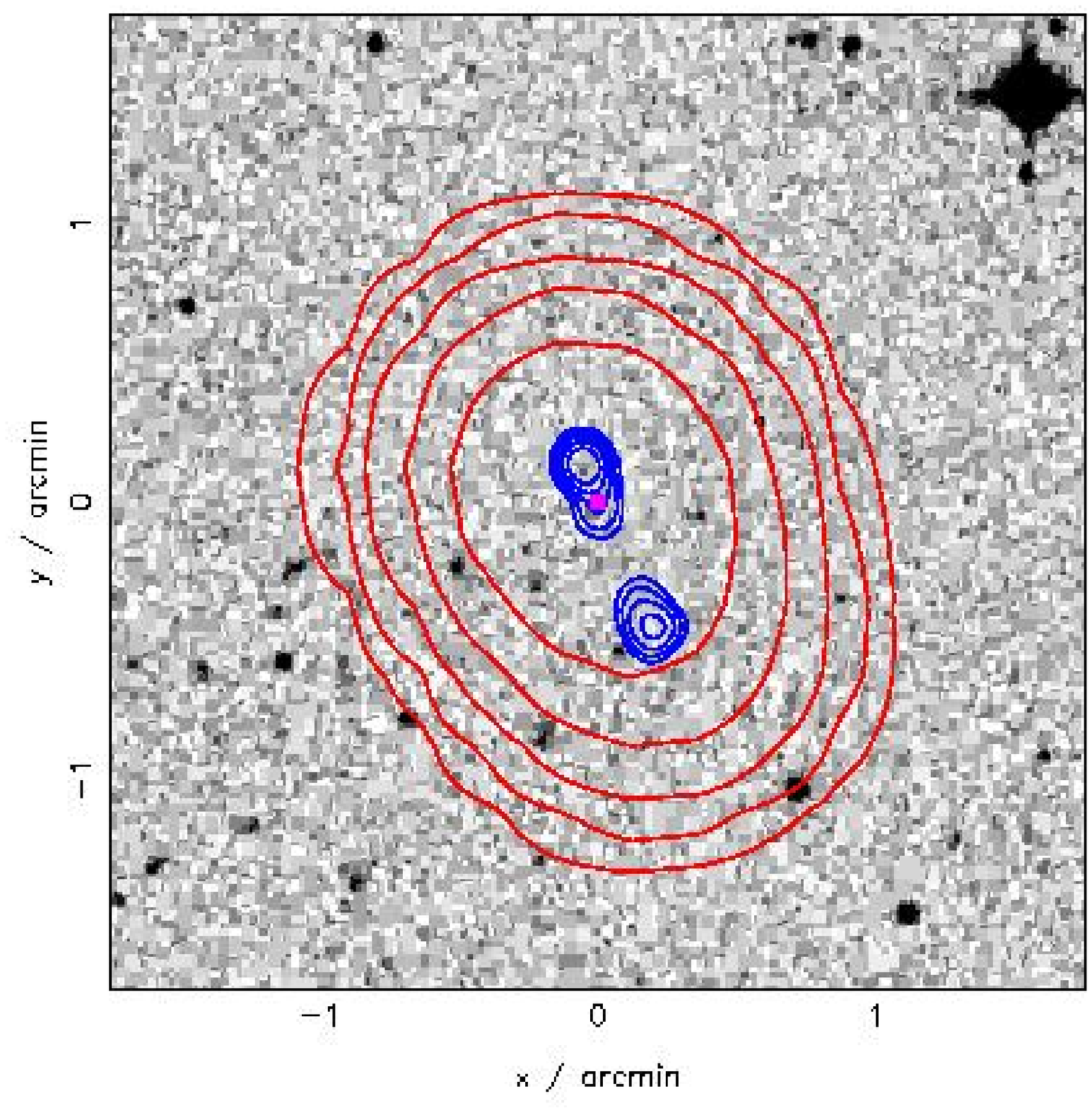}}
      \centerline{C3-068: TXS 1455+253}
    \end{minipage}
    \hspace{3cm}
    \begin{minipage}{3cm}
      \mbox{}
      \centerline{\includegraphics[scale=0.26,angle=270]{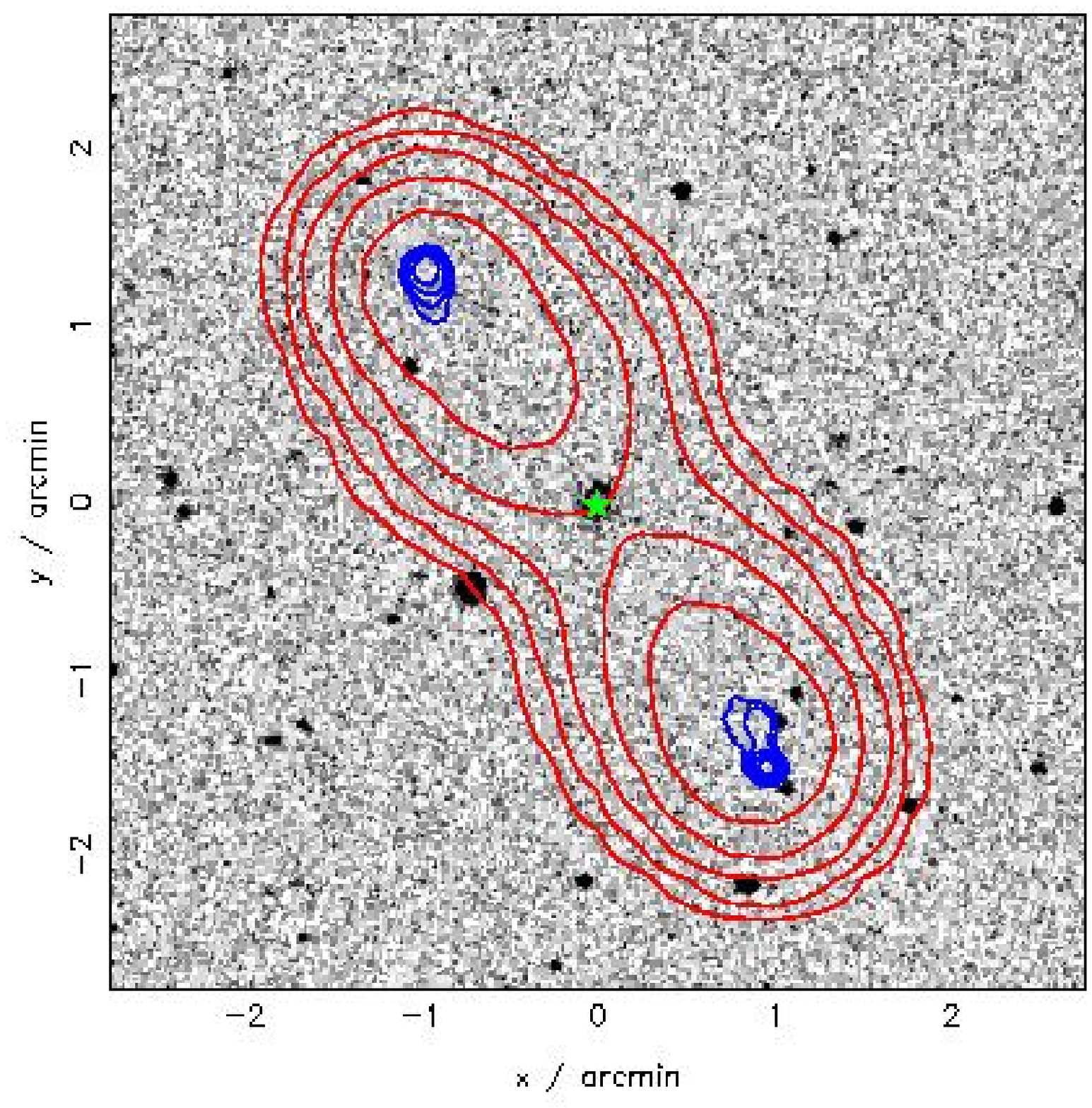}}
      \centerline{C3-069: 4C 28.38}
    \end{minipage}
    \vfill
    \begin{minipage}{3cm}      
      \mbox{}
      \centerline{\includegraphics[scale=0.26,angle=270]{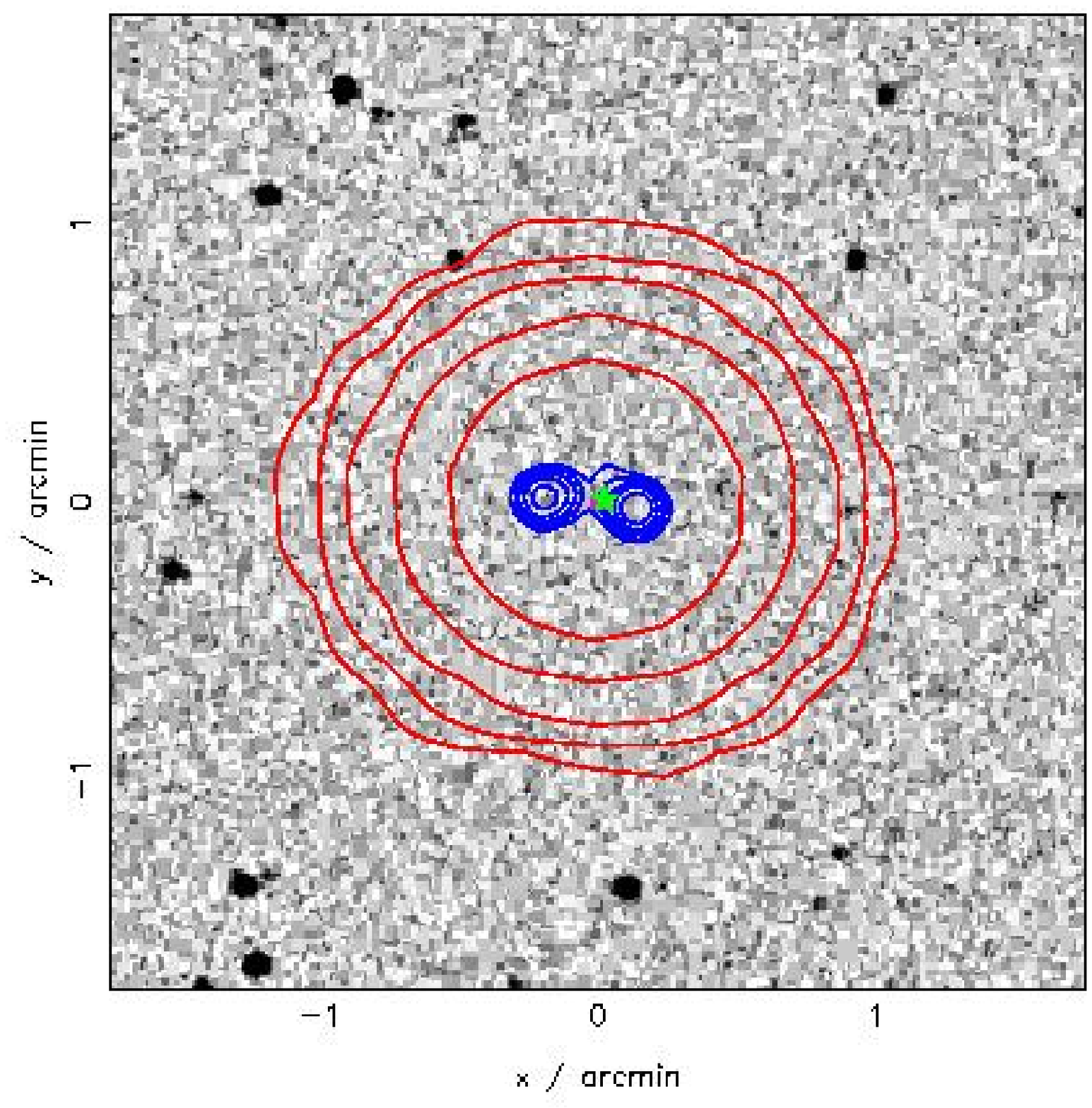}}
      \centerline{C3-070: 4C 11.47}
    \end{minipage}
    \hspace{3cm}
    \begin{minipage}{3cm}
      \mbox{}
      \centerline{\includegraphics[scale=0.26,angle=270]{Contours/C3/073.ps}}
      \centerline{C3-073: 4C 14.56}
    \end{minipage}
    \hspace{3cm}
    \begin{minipage}{3cm}
      \mbox{}
      \centerline{\includegraphics[scale=0.26,angle=270]{Contours/C3/074.ps}}
      \centerline{C3-074: 4C 18.40}
    \end{minipage}
  \end{center}
\end{figure}

\begin{figure}
  \begin{center}
    {\bf CoNFIG-3}\\  
  \begin{minipage}{3cm}      
      \mbox{}
      \centerline{\includegraphics[scale=0.26,angle=270]{Contours/C3/075.ps}}
      \centerline{C3-075: TXS 1456+143}
    \end{minipage}
    \hspace{3cm}
    \begin{minipage}{3cm}
      \mbox{}
      \centerline{\includegraphics[scale=0.26,angle=270]{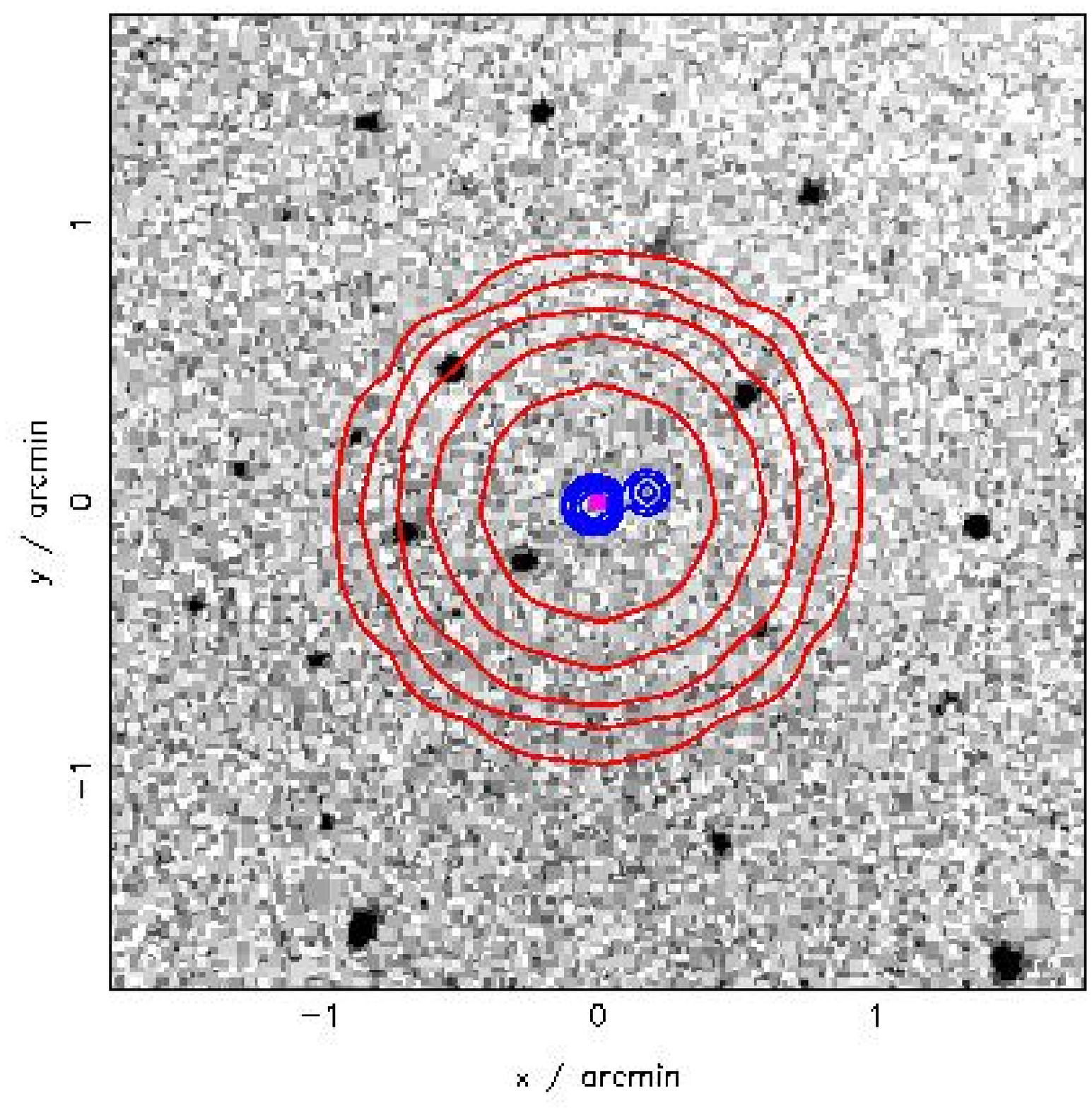}}
      \centerline{C3-076: TXS 1456+251}
    \end{minipage}
    \hspace{3cm}
    \begin{minipage}{3cm}
      \mbox{}
      \centerline{\includegraphics[scale=0.26,angle=270]{Contours/C3/077.ps}}
      \centerline{C3-077: TXS 1457+241}
    \end{minipage}
    \vfill
    \begin{minipage}{3cm}     
      \mbox{}
      \centerline{\includegraphics[scale=0.26,angle=270]{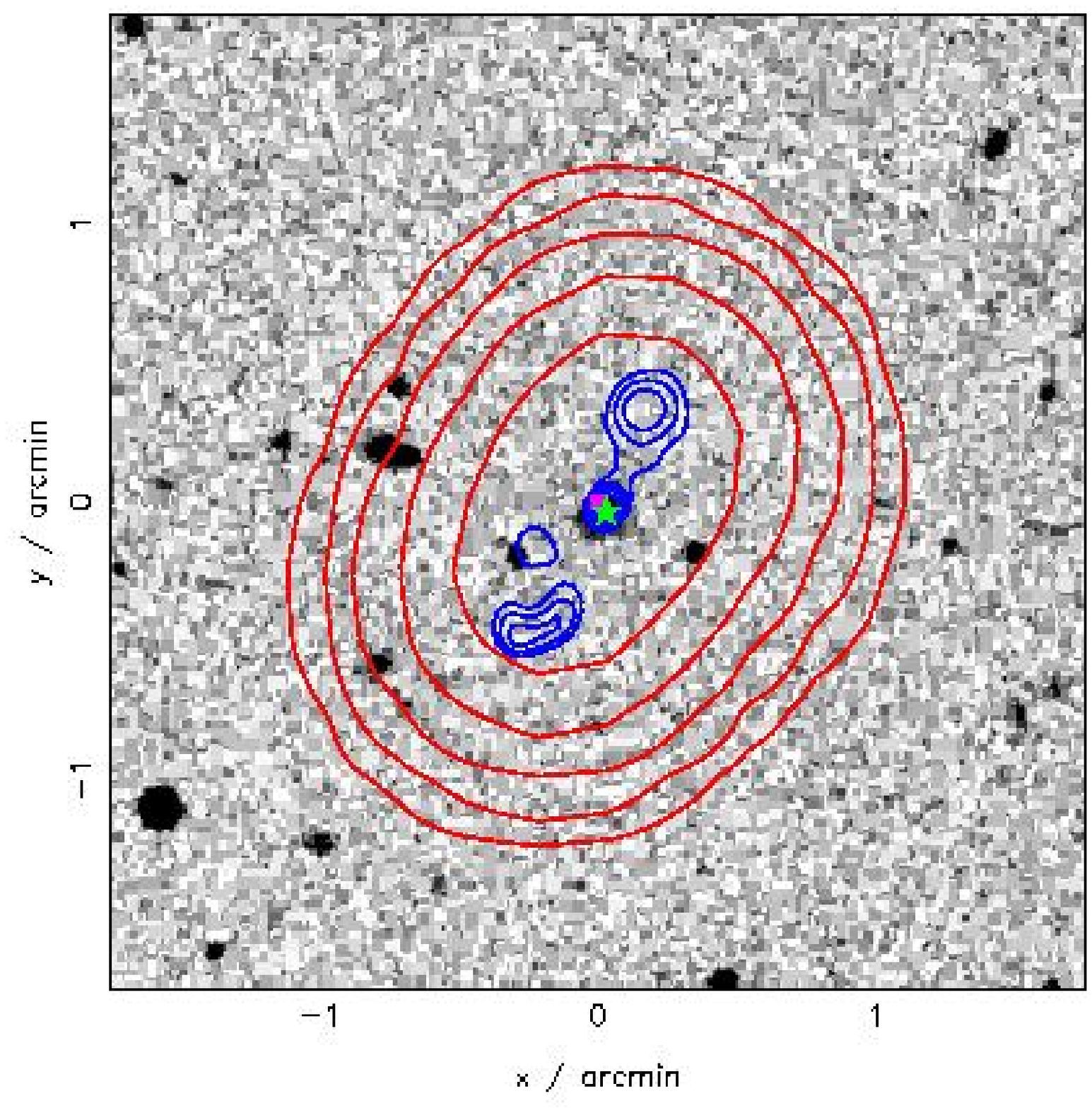}}
      \centerline{C3-078: B2 1457+29}
    \end{minipage}
    \hspace{3cm}
    \begin{minipage}{3cm}
      \mbox{}
      \centerline{\includegraphics[scale=0.26,angle=270]{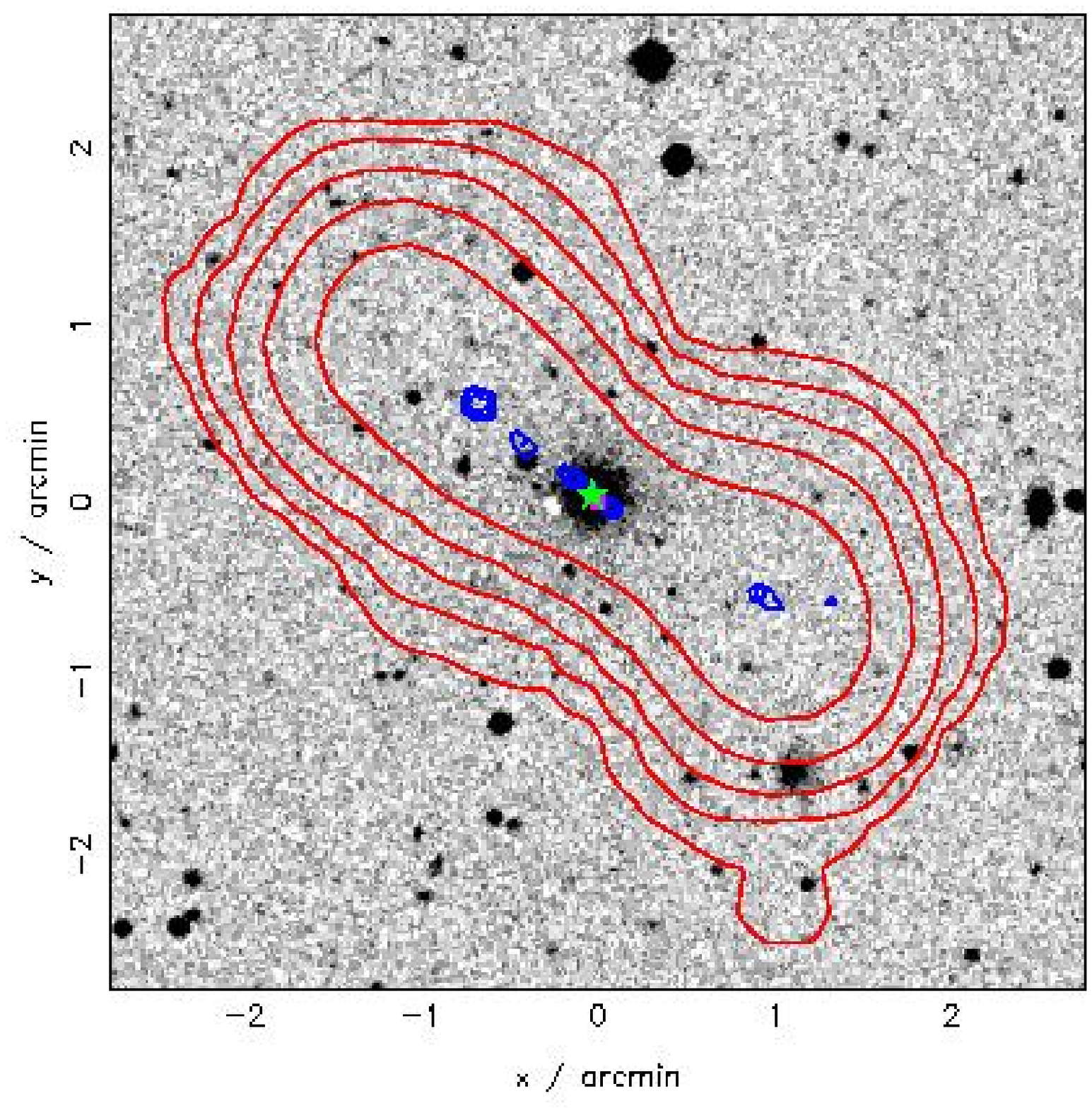}}
      \centerline{C3-079: TXS 1458+204}
    \end{minipage}
    \hspace{3cm}
    \begin{minipage}{3cm}
      \mbox{}
      \centerline{\includegraphics[scale=0.26,angle=270]{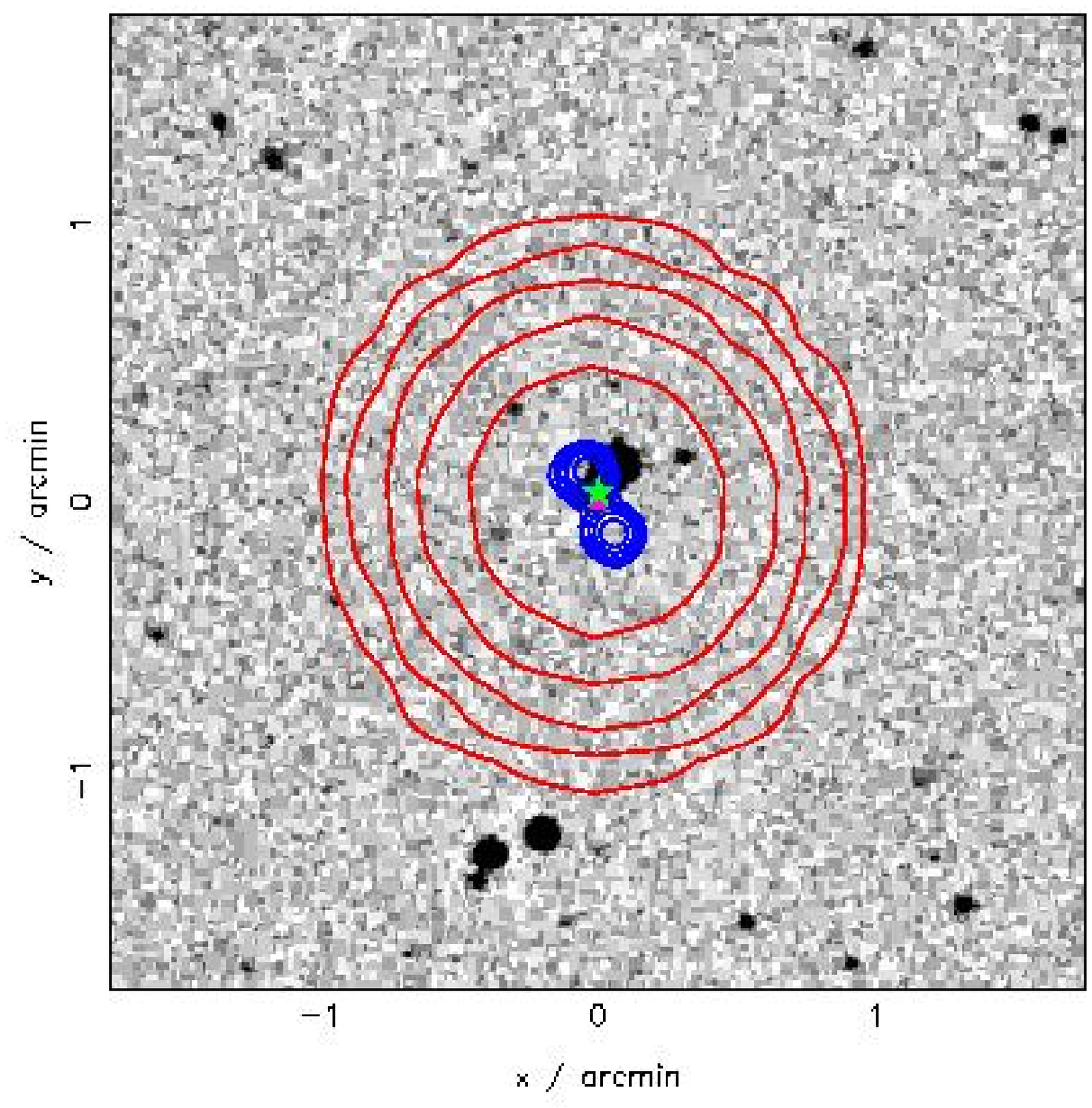}}
      \centerline{C3-080: 4C 14.57}
    \end{minipage}
    \vfill
    \begin{minipage}{3cm}     
      \mbox{}
      \centerline{\includegraphics[scale=0.26,angle=270]{Contours/C3/081.ps}}
      \centerline{C3-081: TXS 1458+178}
    \end{minipage}
    \hspace{3cm}
    \begin{minipage}{3cm}
      \mbox{}
      \centerline{\includegraphics[scale=0.26,angle=270]{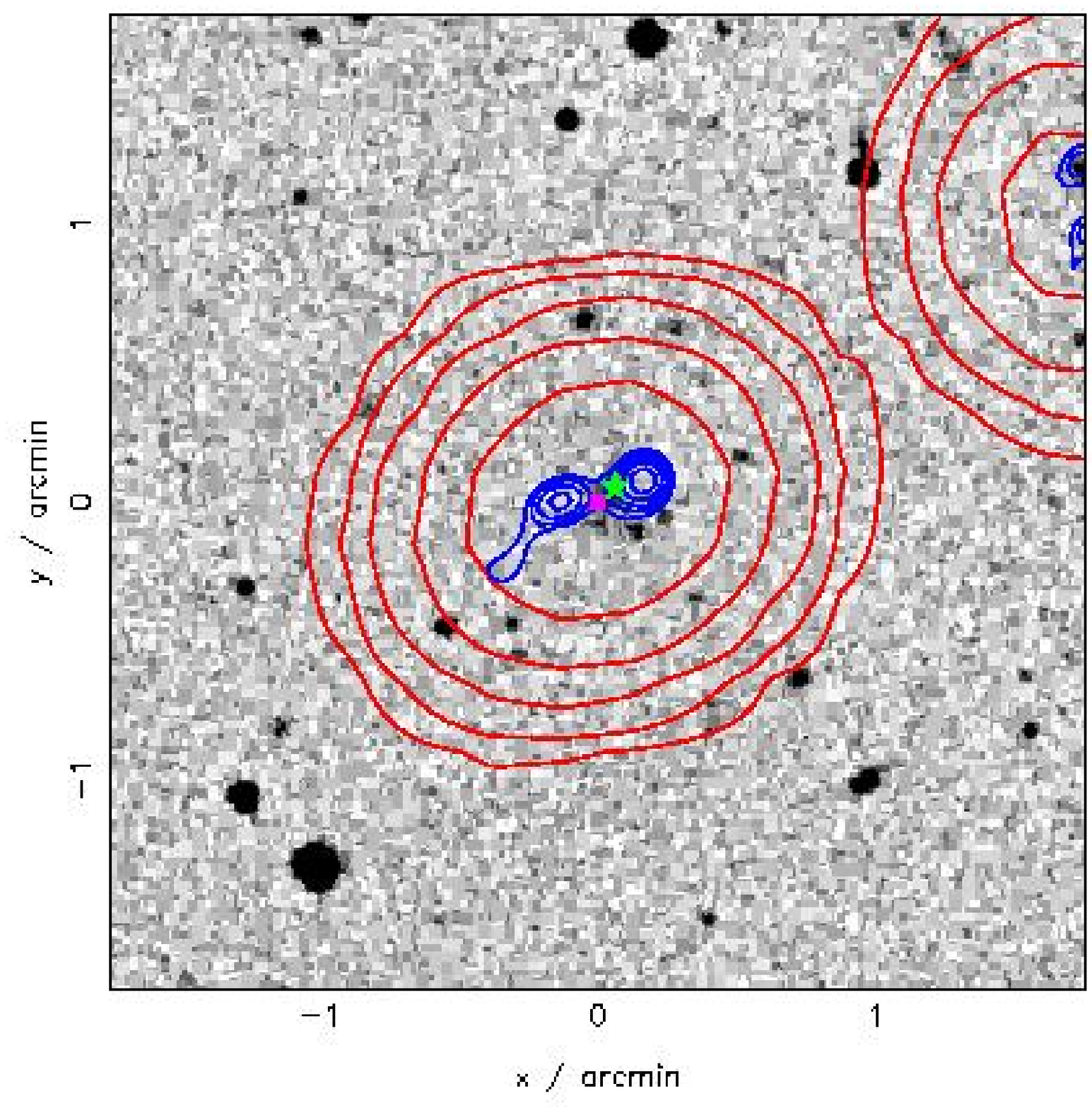}}
      \centerline{C3-082: 4C 21.44}
    \end{minipage}
    \hspace{3cm}
    \begin{minipage}{3cm}
      \mbox{}
      \centerline{\includegraphics[scale=0.26,angle=270]{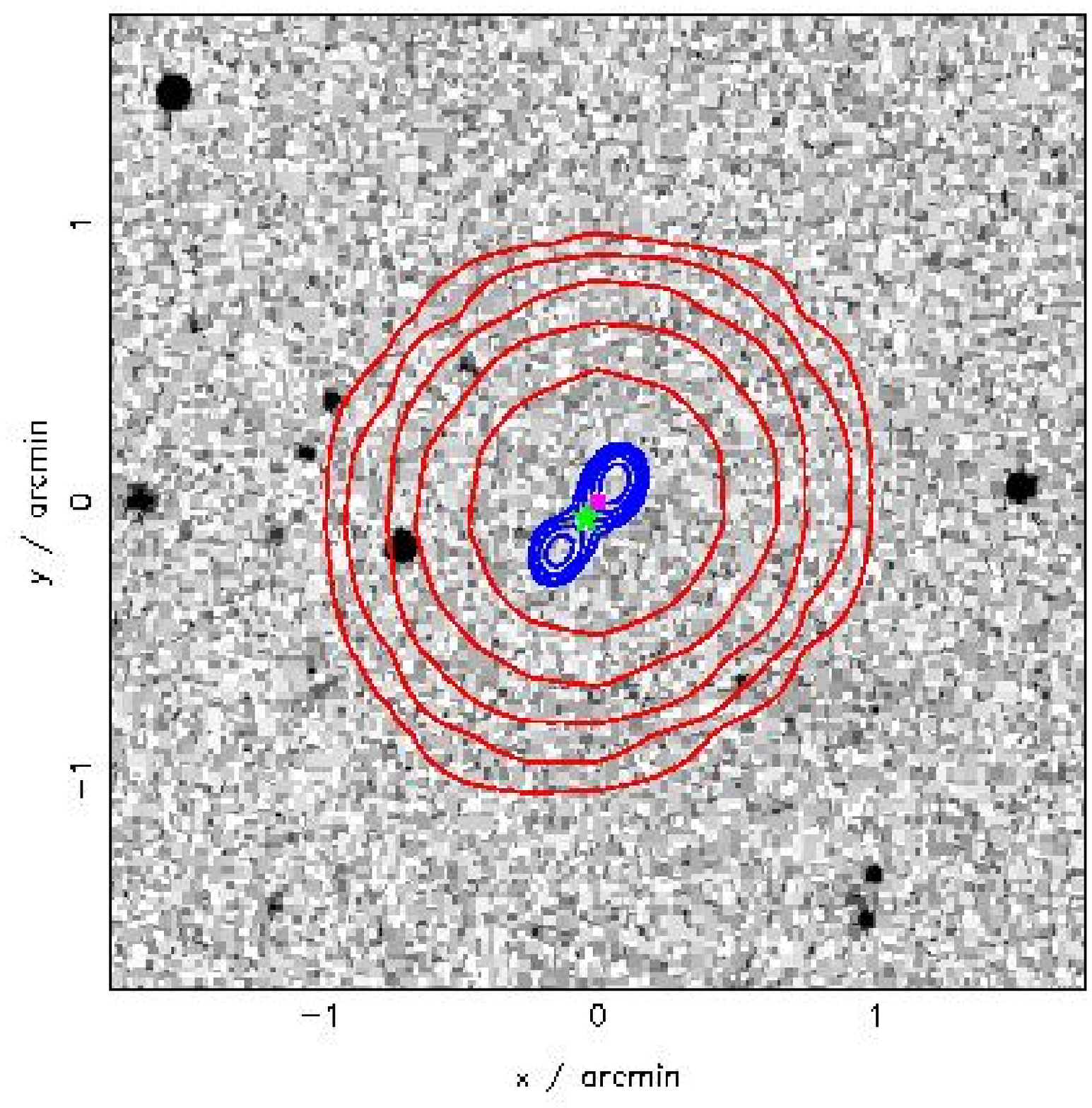}}
      \centerline{C3-083: TXS 1459+279}
    \end{minipage}
    \vfill
    \begin{minipage}{3cm}      
      \mbox{}
      \centerline{\includegraphics[scale=0.26,angle=270]{Contours/C3/085.ps}}
      \centerline{C3-085: TXS 1459+194}
    \end{minipage}
    \hspace{3cm}
    \begin{minipage}{3cm}
      \mbox{}
      \centerline{\includegraphics[scale=0.26,angle=270]{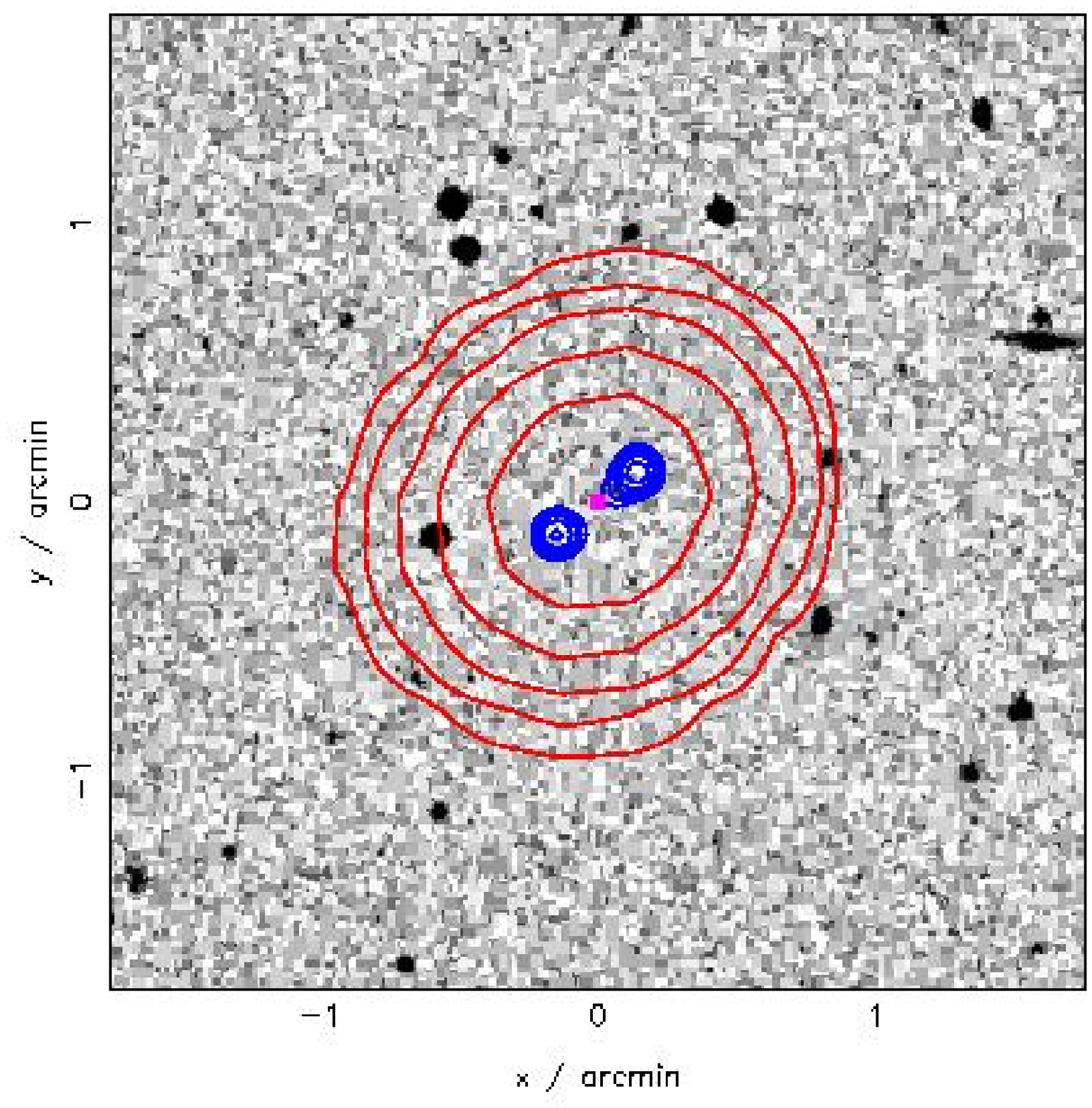}}
      \centerline{C3-086: BWE 1459+2451}
    \end{minipage}
    \hspace{3cm}
    \begin{minipage}{3cm}
      \mbox{}
      \centerline{\includegraphics[scale=0.26,angle=270]{Contours/C3/087.ps}}
      \centerline{C3-087: TXS 1459+133}
    \end{minipage}
  \end{center}
\end{figure}

\begin{figure}
  \begin{center}
    {\bf CoNFIG-3}\\  
  \begin{minipage}{3cm}      
      \mbox{}
      \centerline{\includegraphics[scale=0.26,angle=270]{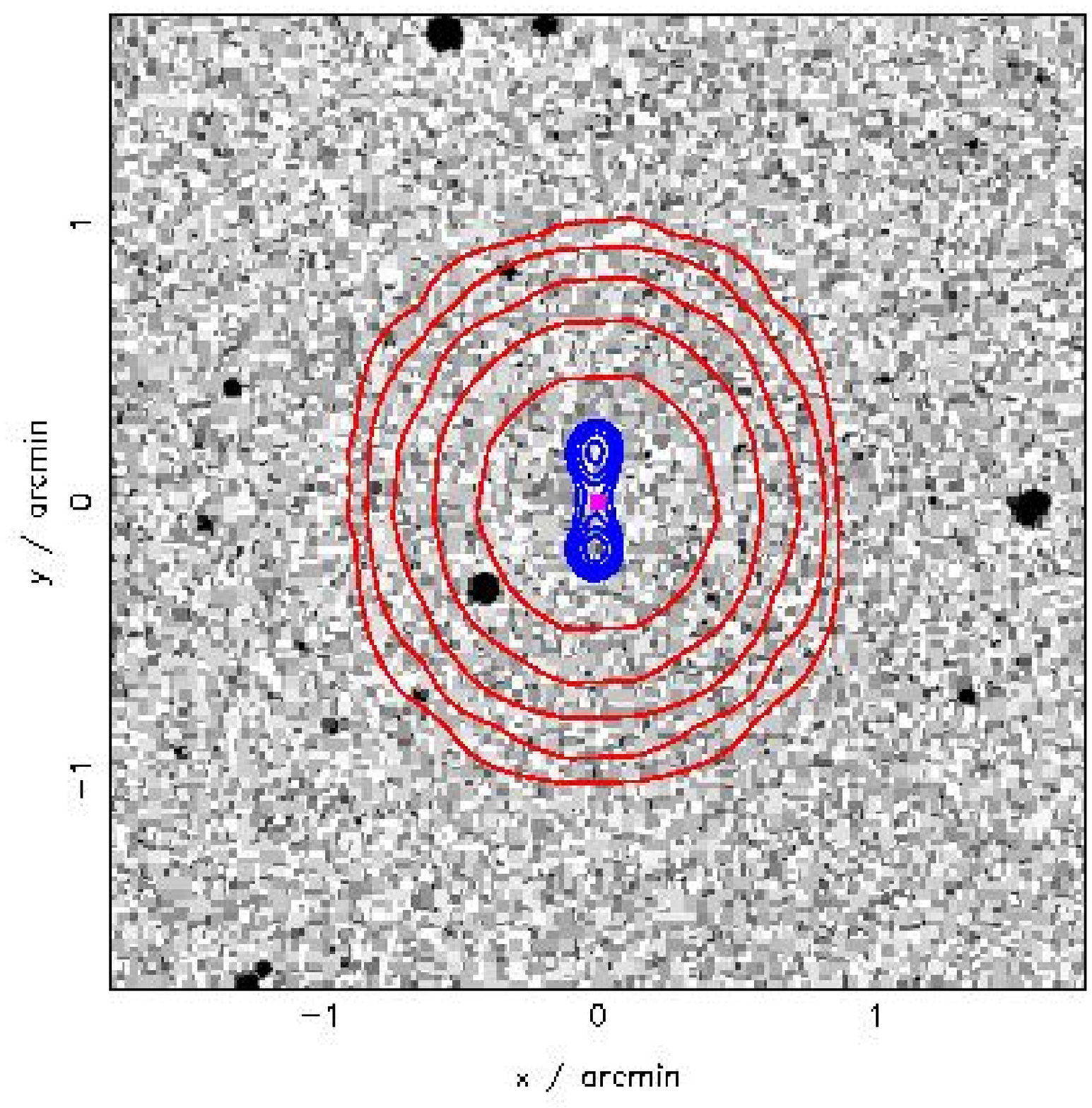}}
      \centerline{C3-088: TXS 1500+259}
    \end{minipage}
    \hspace{3cm}
    \begin{minipage}{3cm}
      \mbox{}
      \centerline{\includegraphics[scale=0.26,angle=270]{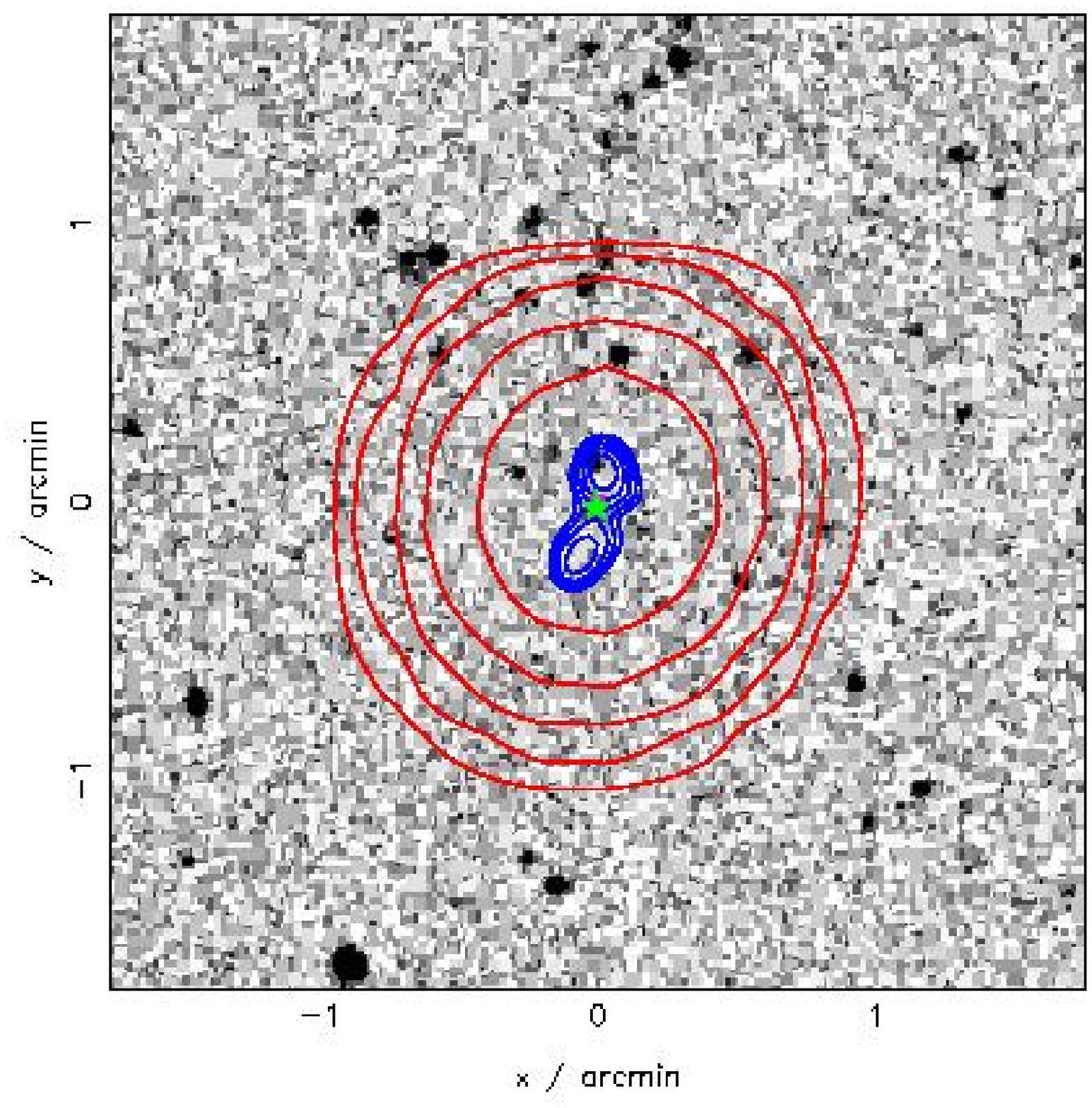}}
      \centerline{C3-089: TXS 1500+185}
    \end{minipage}
    \hspace{3cm}
    \begin{minipage}{3cm}
      \mbox{}
      \centerline{\includegraphics[scale=0.26,angle=270]{Contours/C3/090.ps}}
      \centerline{C3-090: TXS 1500+128}
    \end{minipage}
    \vfill
    \begin{minipage}{3cm}     
      \mbox{}
      \centerline{\includegraphics[scale=0.26,angle=270]{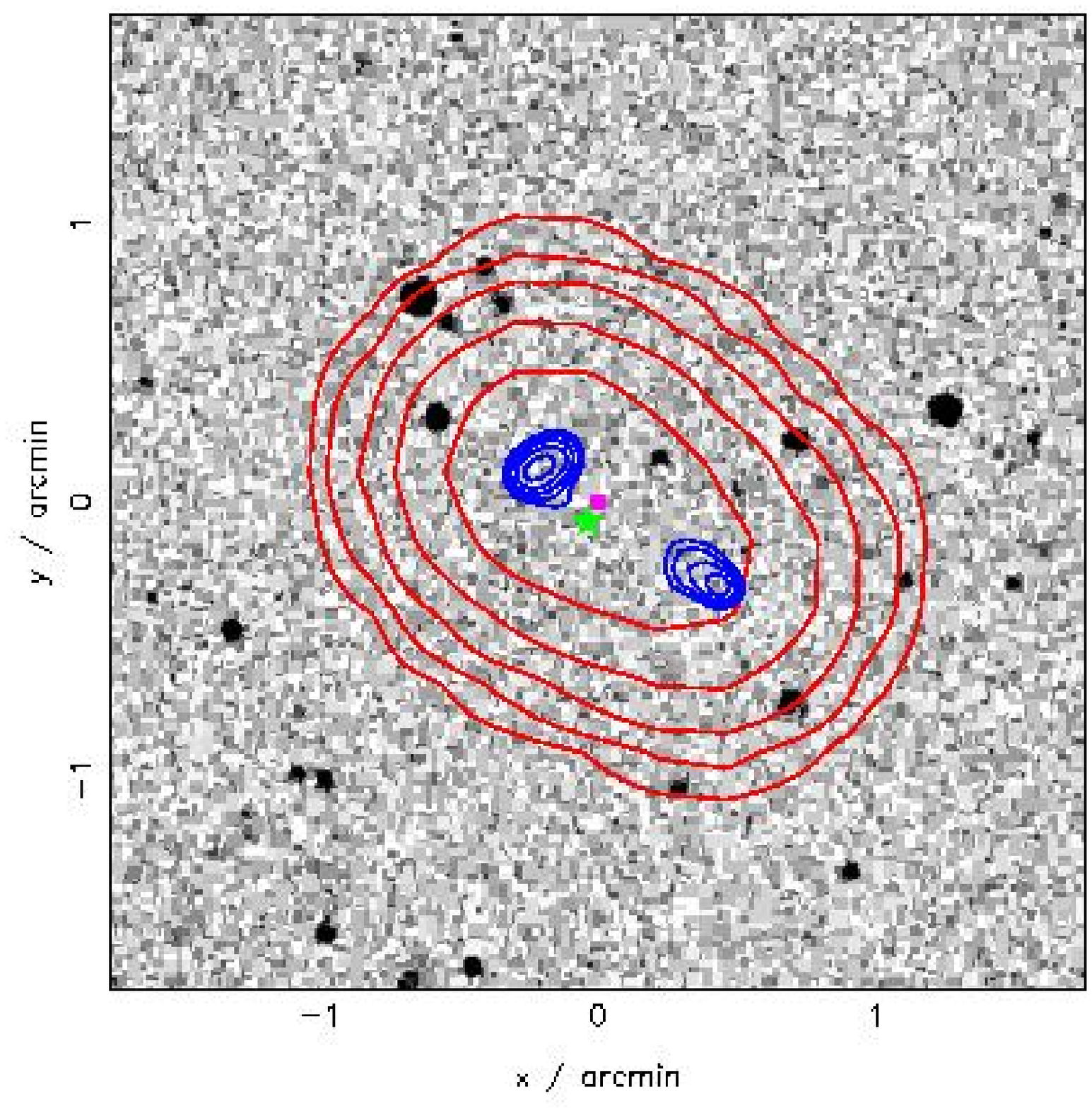}}
      \centerline{C3-091: TXS 1501+197}
    \end{minipage}
    \hspace{3cm}
    \begin{minipage}{3cm}
      \mbox{}
      \centerline{\includegraphics[scale=0.26,angle=270]{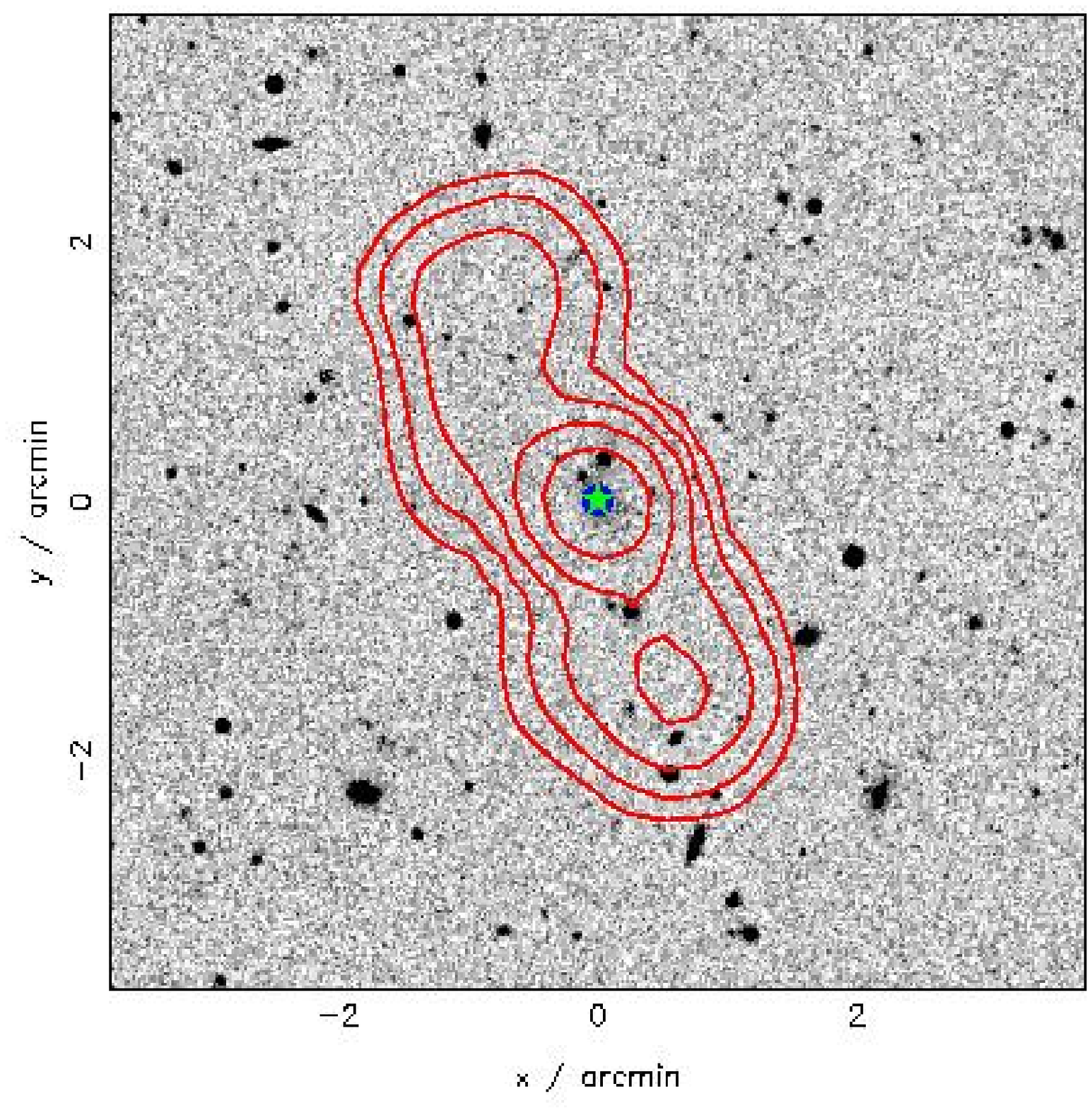}}
      \centerline{C3-093: MRC 1501+104}
    \end{minipage}
    \hspace{3cm}
    \begin{minipage}{3cm}
      \mbox{}
      \centerline{\includegraphics[scale=0.26,angle=270]{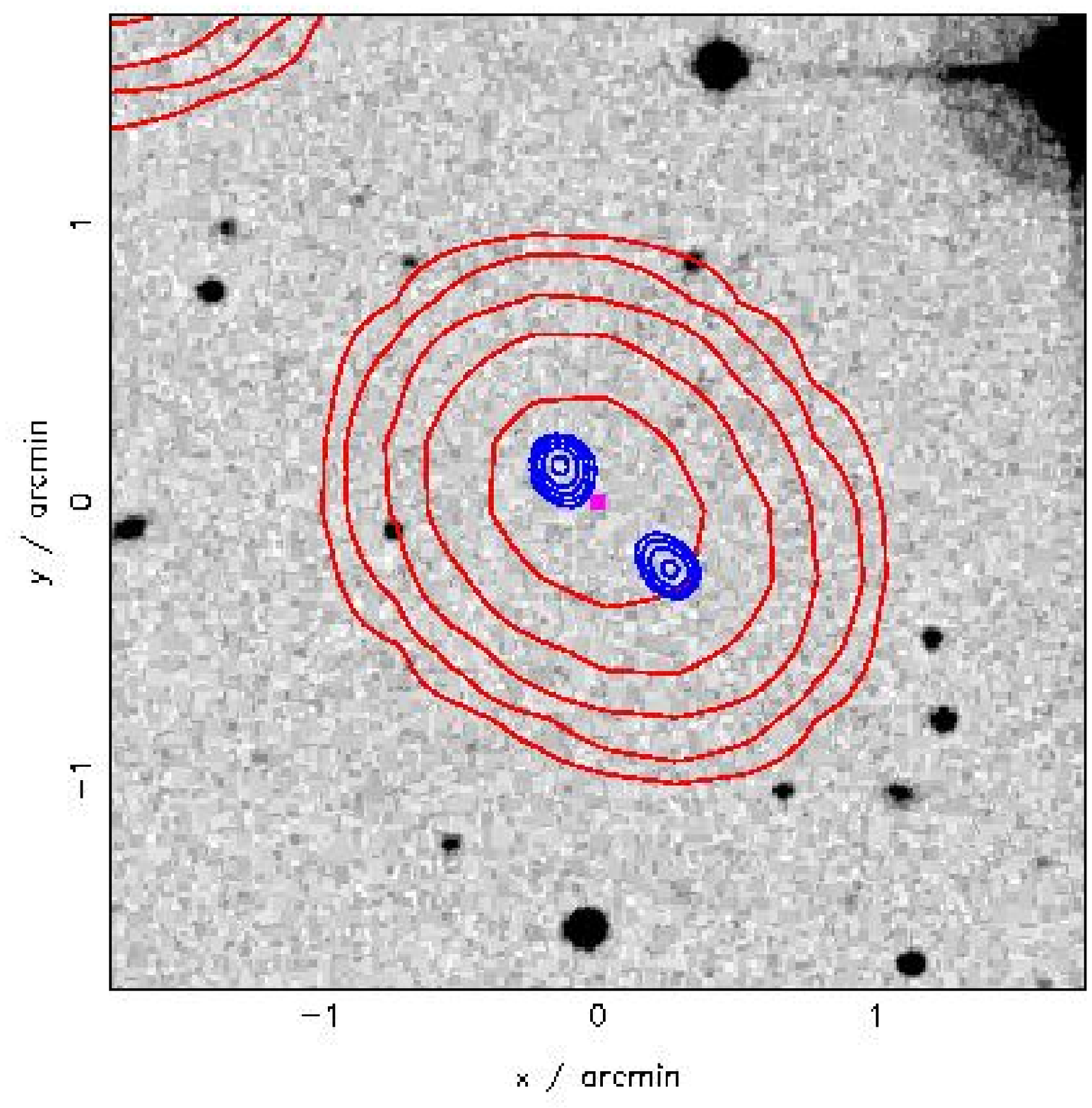}}
      \centerline{C3-094: TXS 1501+126}
    \end{minipage}
    \vfill
    \begin{minipage}{3cm}     
      \mbox{}
      \centerline{\includegraphics[scale=0.26,angle=270]{Contours/C3/095.ps}}
      \centerline{C3-095: 1503+1251}
    \end{minipage}
    \hspace{3cm}
    \begin{minipage}{3cm}
      \mbox{}
      \centerline{\includegraphics[scale=0.26,angle=270]{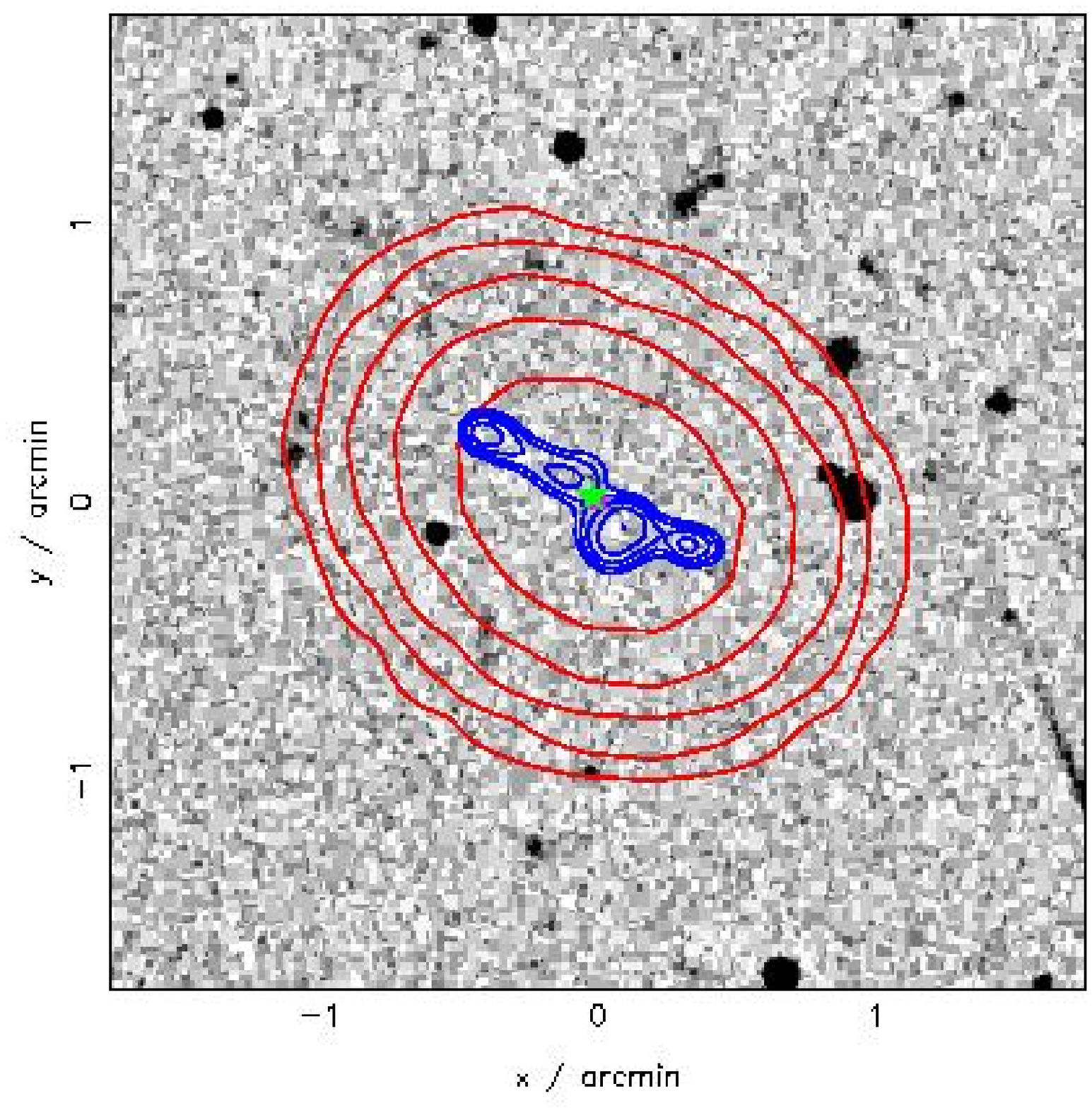}}
      \centerline{C3-102: TXS 1504+206}
    \end{minipage}
    \hspace{3cm}
    \begin{minipage}{3cm}
      \mbox{}
      \centerline{\includegraphics[scale=0.26,angle=270]{Contours/C3/103.ps}}
      \centerline{C3-103: WB 1504+1618}
    \end{minipage}
    \vfill
    \begin{minipage}{3cm}      
      \mbox{}
      \centerline{\includegraphics[scale=0.26,angle=270]{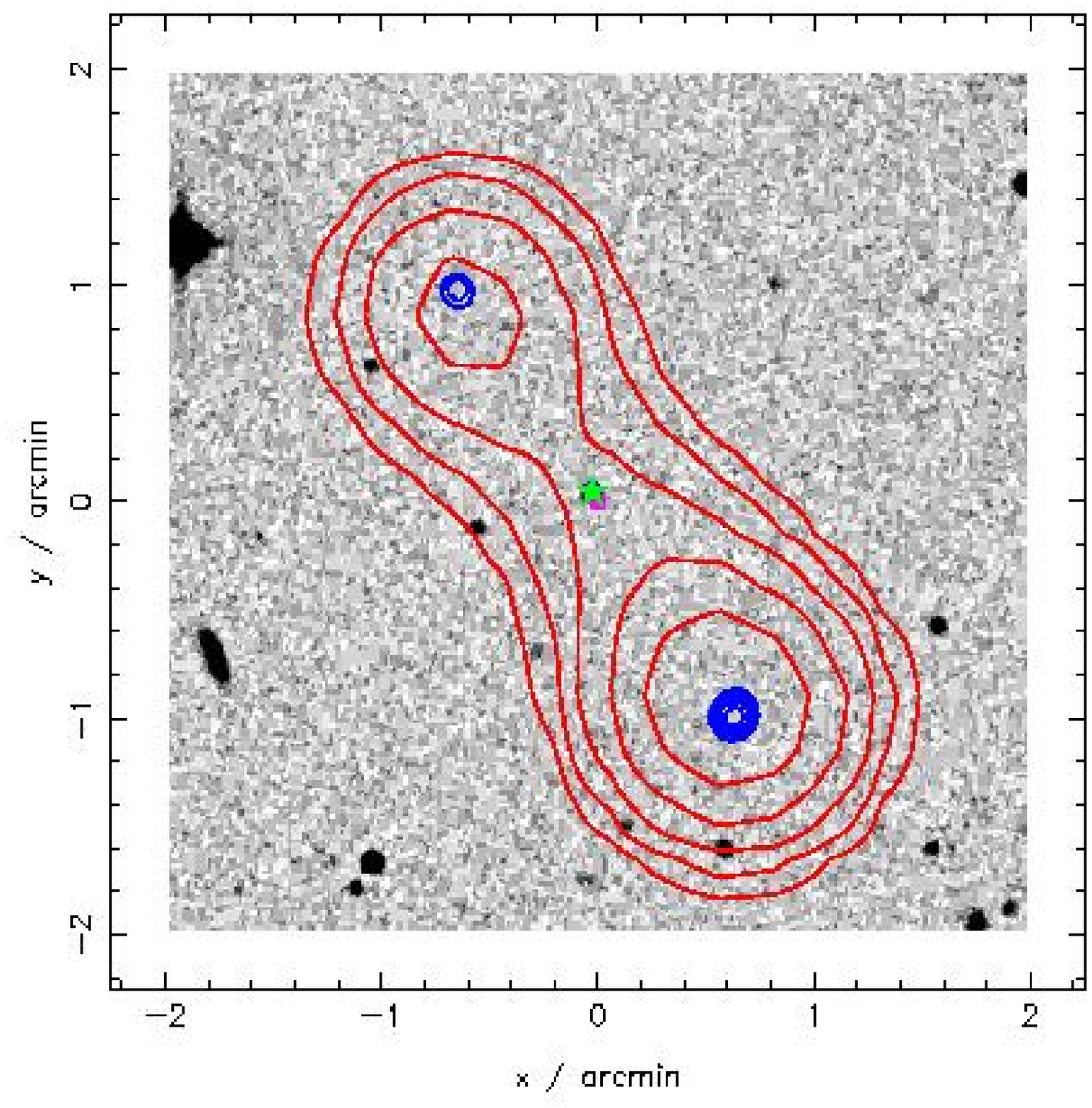}}
      \centerline{C3-105: GB6 B1505+113}
    \end{minipage}
    \hspace{3cm}
    \begin{minipage}{3cm}
      \mbox{}
      \centerline{\includegraphics[scale=0.26,angle=270]{Contours/C3/106.ps}}
      \centerline{C3-106: 4C 12.54}
    \end{minipage}
    \hspace{3cm}
    \begin{minipage}{3cm}
      \mbox{}
      \centerline{\includegraphics[scale=0.26,angle=270]{Contours/C3/107.ps}}
      \centerline{C3-107: TXS 1505+247}
    \end{minipage}
  \end{center}
\end{figure}

\begin{figure}
  \begin{center}
    {\bf CoNFIG-3}\\  
  \begin{minipage}{3cm}      
      \mbox{}
      \centerline{\includegraphics[scale=0.26,angle=270]{Contours/C3/108.ps}}
      \centerline{C3-108: TXS 1505+190}
    \end{minipage}
    \hspace{3cm}
    \begin{minipage}{3cm}
      \mbox{}
      \centerline{\includegraphics[scale=0.26,angle=270]{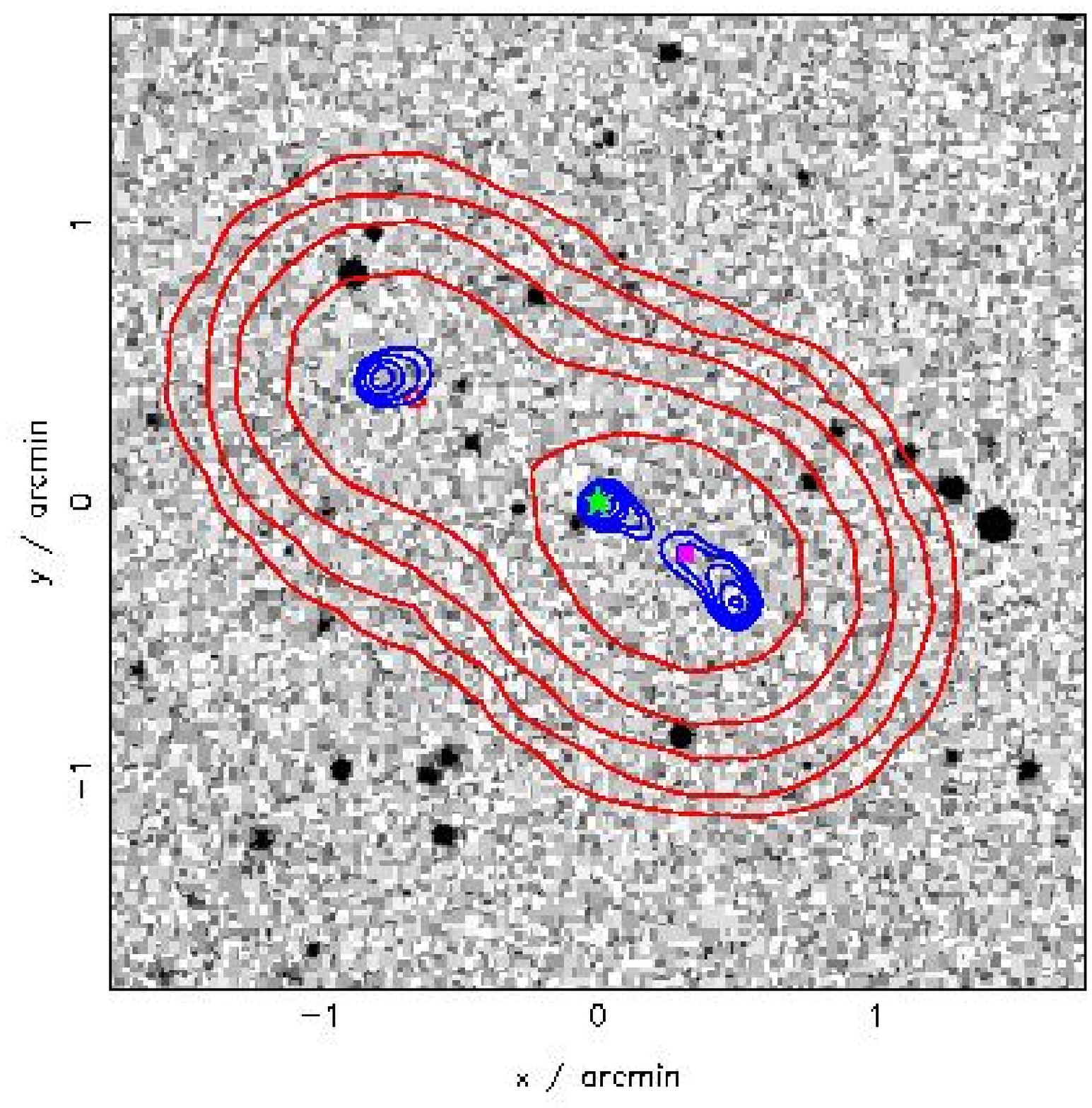}}
      \centerline{C3-111: TXS 1506+245}
    \end{minipage}
    \hspace{3cm}
    \begin{minipage}{3cm}
      \mbox{}
      \centerline{\includegraphics[scale=0.26,angle=270]{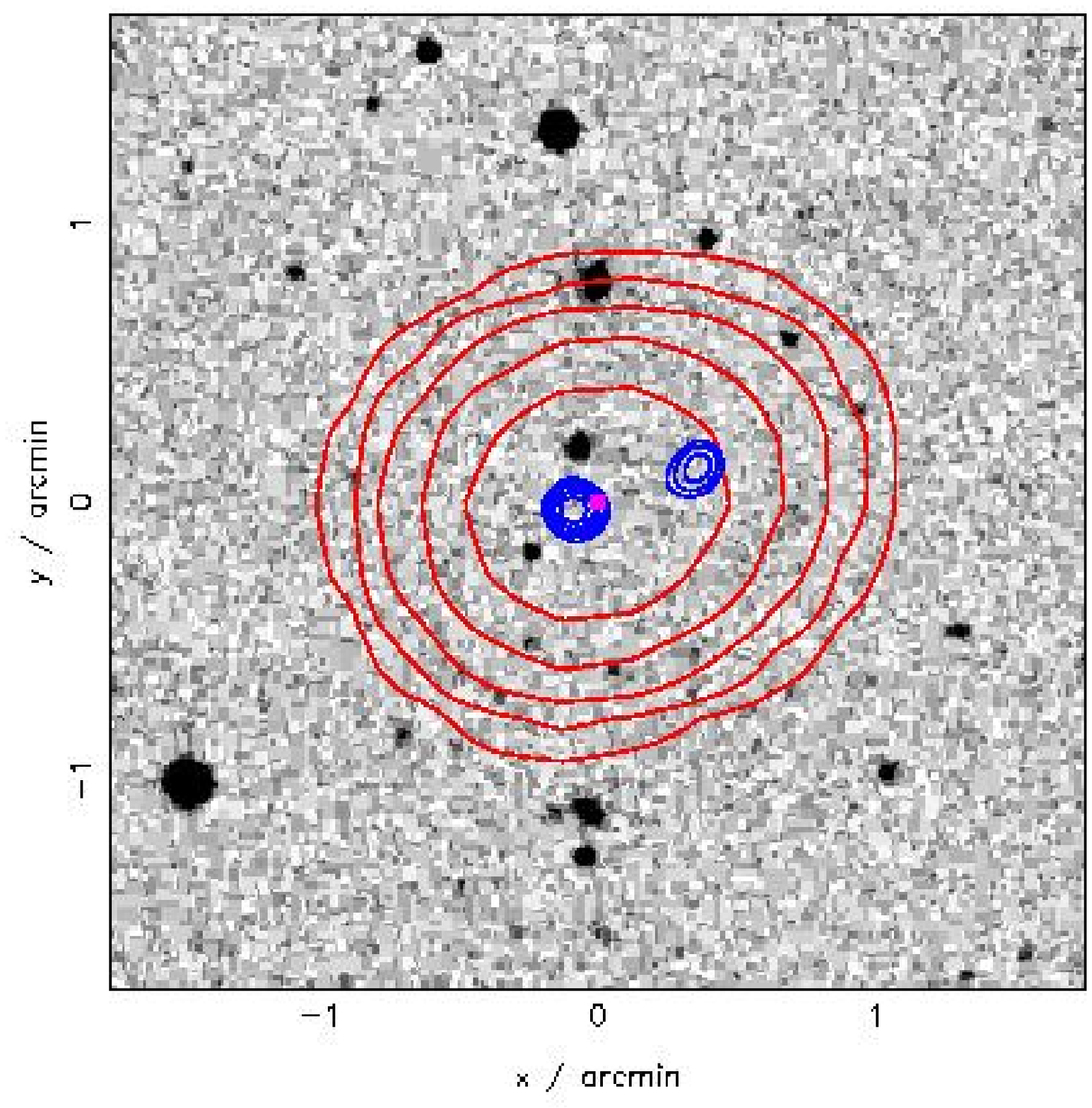}}
      \centerline{C3-114: TXS 1506+171}
    \end{minipage}
    \vfill
    \begin{minipage}{3cm}     
      \mbox{}
      \centerline{\includegraphics[scale=0.26,angle=270]{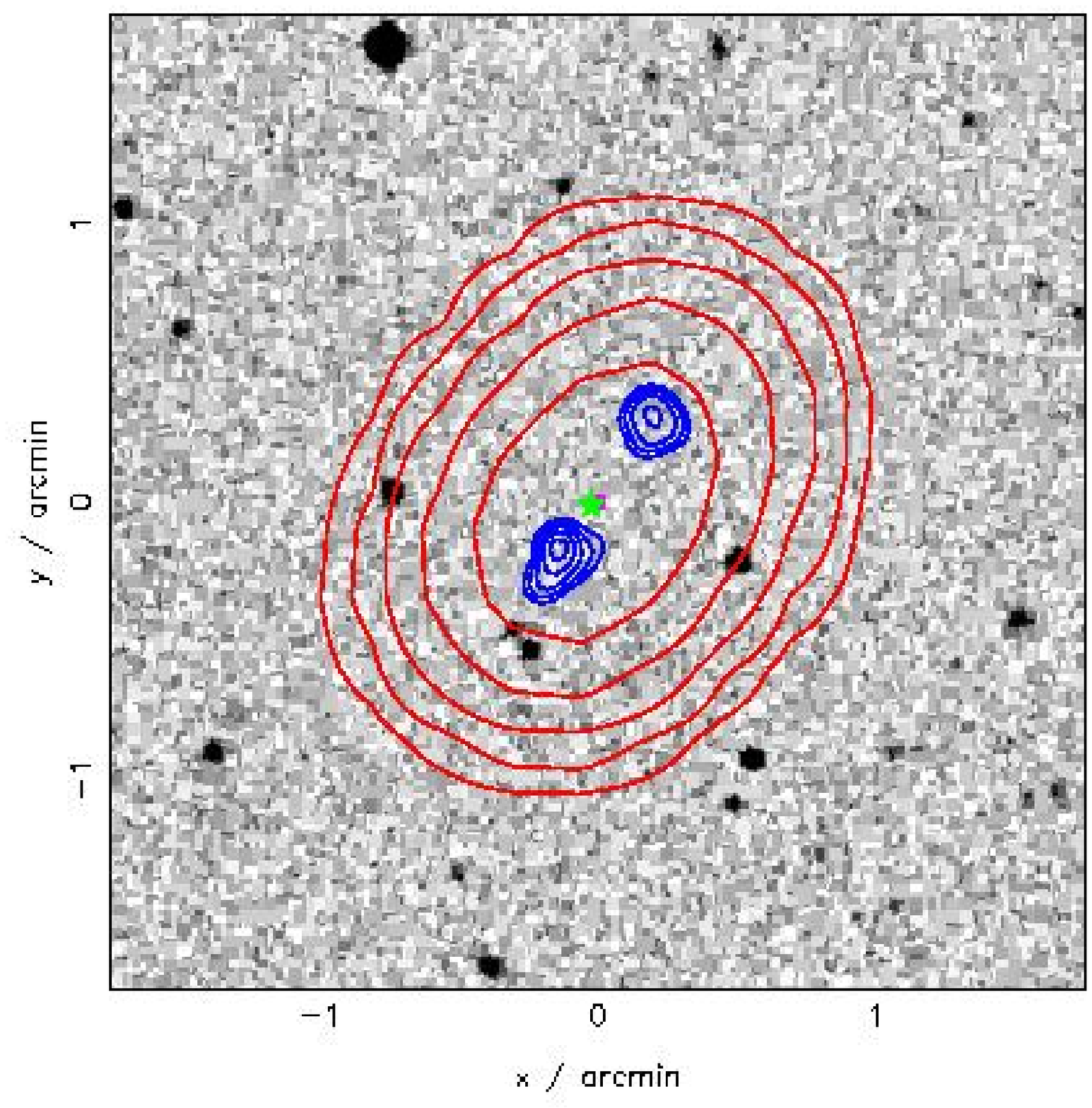}}
      \centerline{C3-115: TXS 1507+298}
    \end{minipage}
    \hspace{3cm}
    \begin{minipage}{3cm}
      \mbox{}
      \centerline{\includegraphics[scale=0.26,angle=270]{Contours/C3/118.ps}}
      \centerline{C3-118:TXS 1507+235}
    \end{minipage}
    \hspace{3cm}
    \begin{minipage}{3cm}
      \mbox{}
      \centerline{\includegraphics[scale=0.26,angle=270]{Contours/C3/119.ps}}
      \centerline{C3-119: TXS 1508+205}
    \end{minipage}
    \vfill
    \begin{minipage}{3cm}     
      \mbox{}
      \centerline{\includegraphics[scale=0.26,angle=270]{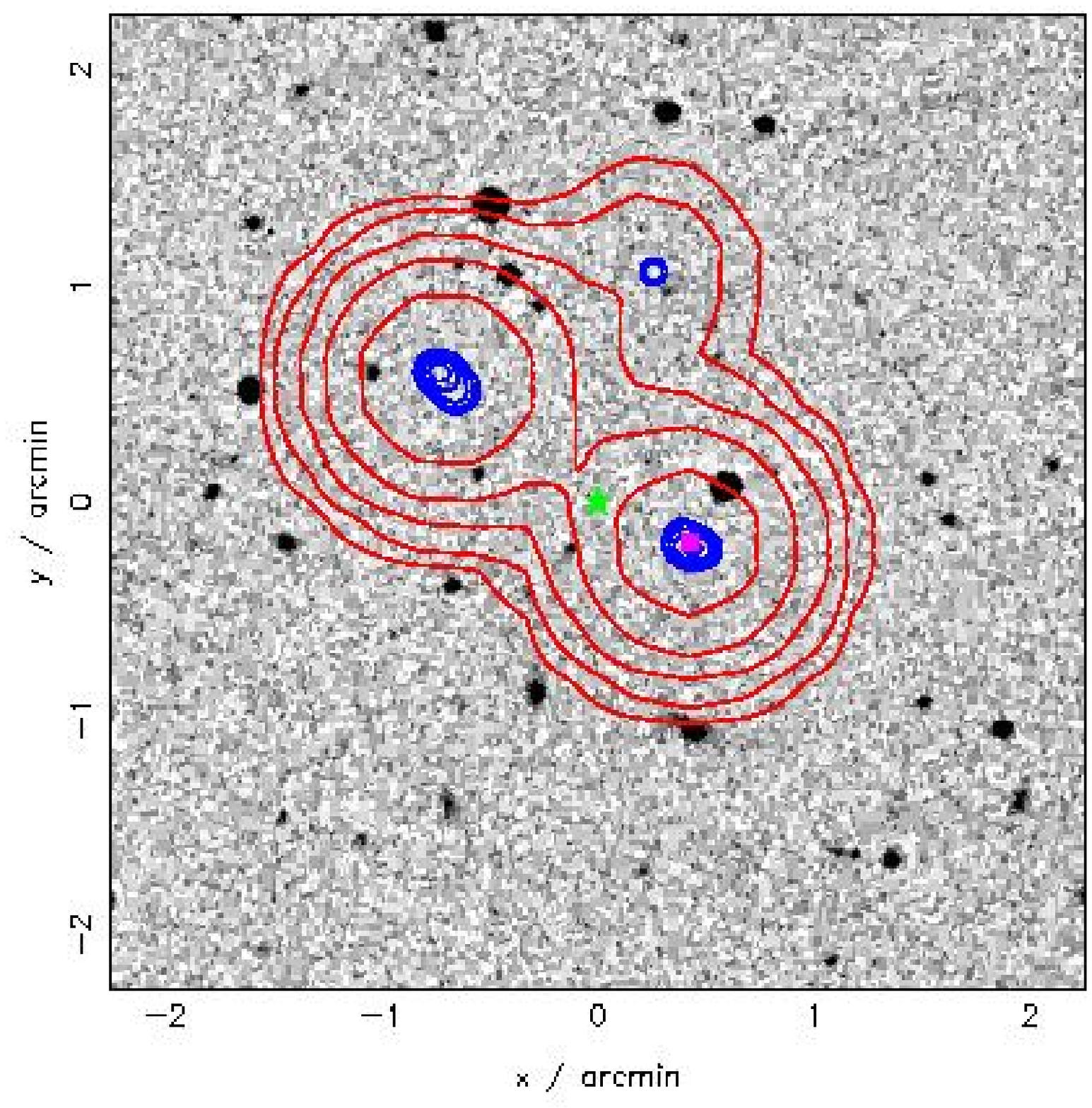}}
      \centerline{C3-120: TXS 1508+128}
    \end{minipage}
    \hspace{3cm}
    \begin{minipage}{3cm}
      \mbox{}
      \centerline{\includegraphics[scale=0.26,angle=270]{Contours/C3/121.ps}}
      \centerline{C3-121:TXS 1508+148}
    \end{minipage}
    \hspace{3cm}
    \begin{minipage}{3cm}
      \mbox{}
      \centerline{\includegraphics[scale=0.26,angle=270]{Contours/C3/122.ps}}
      \centerline{C3-122: TXS 1508+108}
    \end{minipage}
    \vfill
    \begin{minipage}{3cm}      
      \mbox{}
      \centerline{\includegraphics[scale=0.26,angle=270]{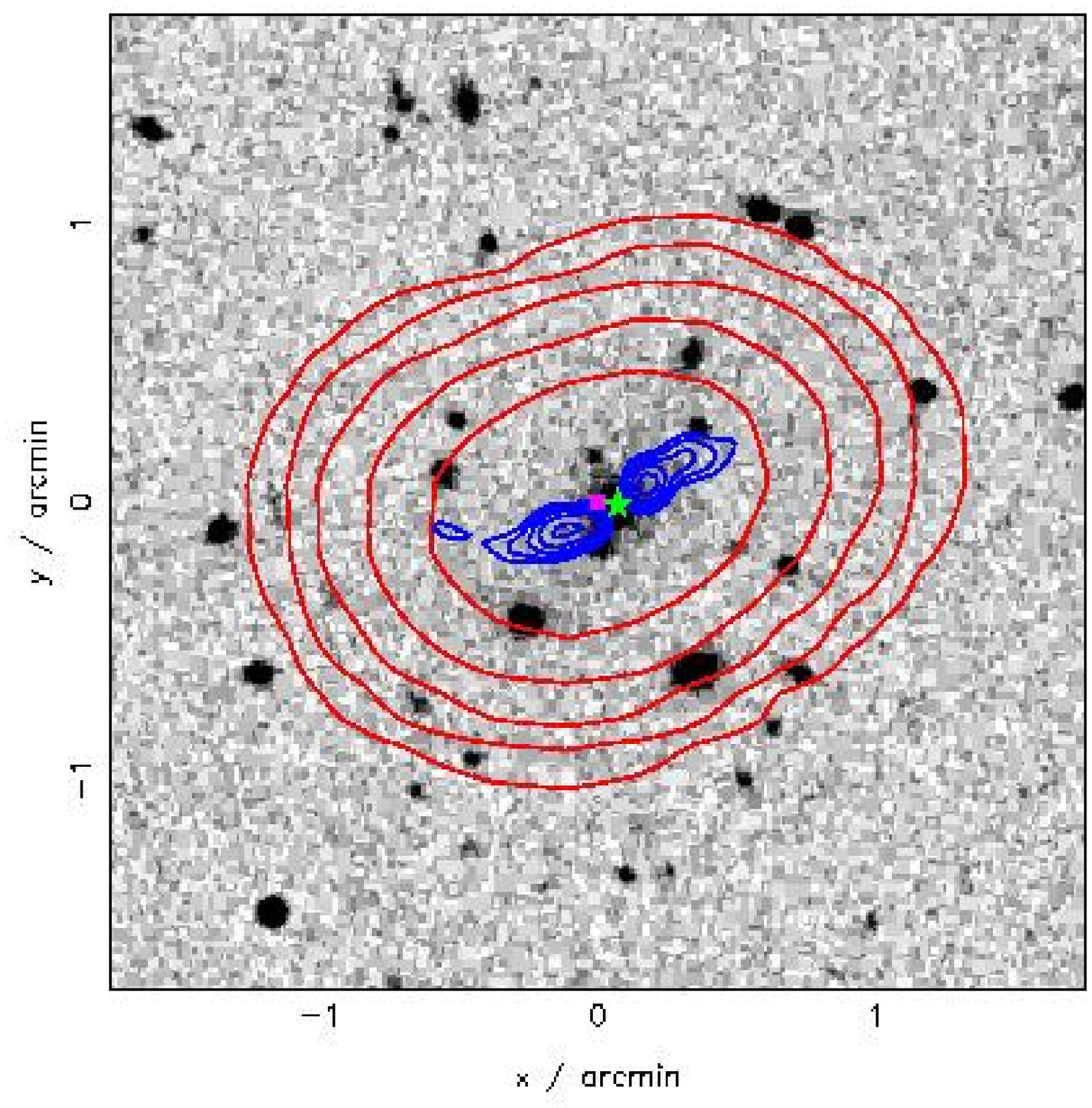}}
      \centerline{C3-125: Cul 1508+182}
    \end{minipage}
    \hspace{3cm}
    \begin{minipage}{3cm}
      \mbox{}
      \centerline{\includegraphics[scale=0.26,angle=270]{Contours/C3/127.ps}}
      \centerline{C3-127: 4C 10.40}
    \end{minipage}
    \hspace{3cm}
    \begin{minipage}{3cm}
      \mbox{}
      \centerline{\includegraphics[scale=0.26,angle=270]{Contours/C3/128.ps}}
      \centerline{C3-128: TXS 1509+28}
    \end{minipage}
  \end{center}
\end{figure}

\begin{figure}
  \begin{center}
    {\bf CoNFIG-3}\\  
  \begin{minipage}{3cm}      
      \mbox{}
      \centerline{\includegraphics[scale=0.26,angle=270]{Contours/C3/129.ps}}
      \centerline{C3-129: TXS 1509+213}
    \end{minipage}
    \hspace{3cm}
    \begin{minipage}{3cm}
      \mbox{}
      \centerline{\includegraphics[scale=0.26,angle=270]{Contours/C3/130.ps}}
      \centerline{C3-130: TXS 1509+229}
    \end{minipage}
    \hspace{3cm}
    \begin{minipage}{3cm}
      \mbox{}
      \centerline{\includegraphics[scale=0.26,angle=270]{Contours/C3/131.ps}}
      \centerline{C3-131: 4C 15.45}
    \end{minipage}
    \vfill
    \begin{minipage}{3cm}     
      \mbox{}
      \centerline{\includegraphics[scale=0.26,angle=270]{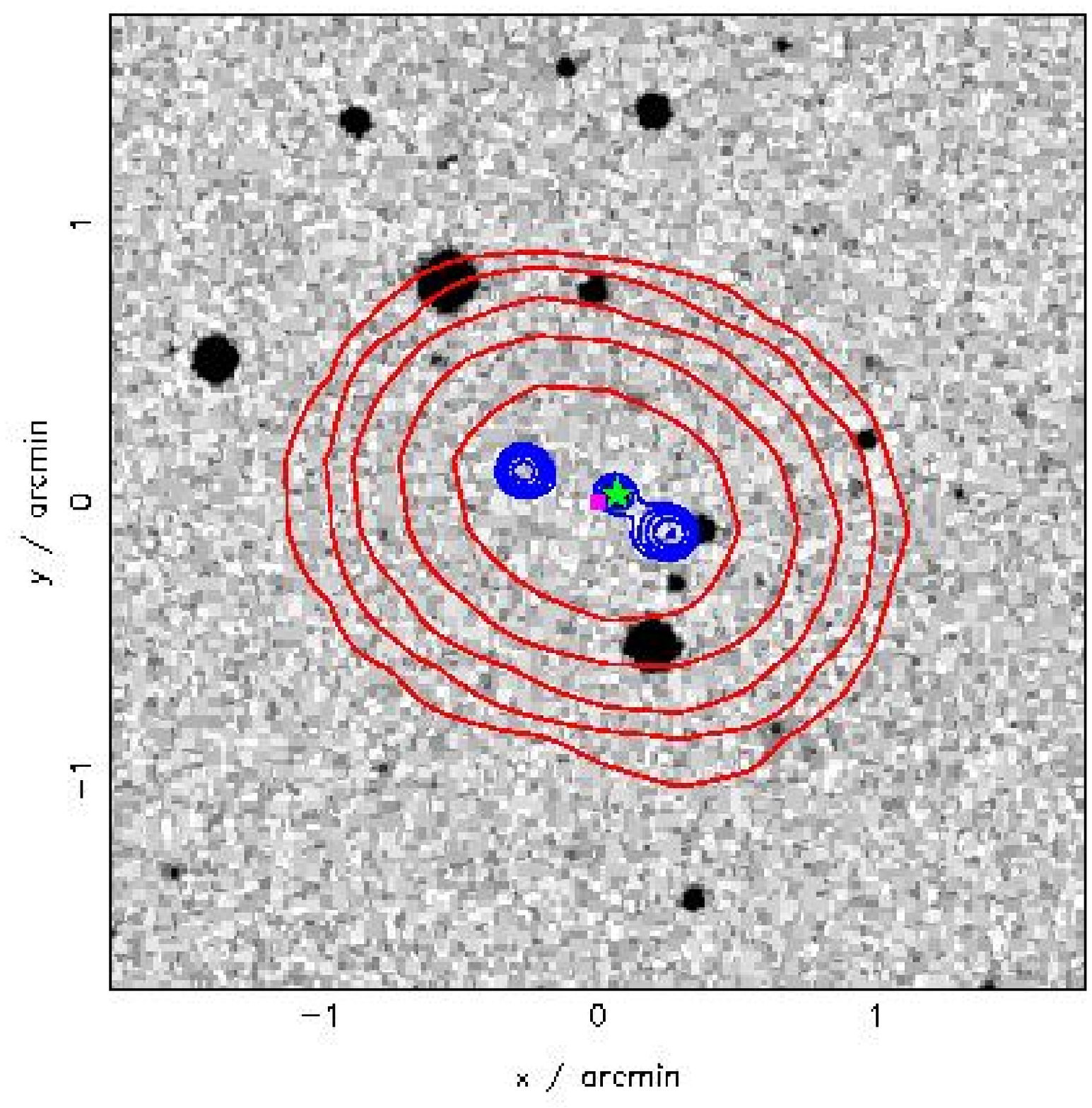}}
      \centerline{C3-134: TXS 1511+103}
    \end{minipage}
    \hspace{3cm}
    \begin{minipage}{3cm}
      \mbox{}
      \centerline{\includegraphics[scale=0.26,angle=270]{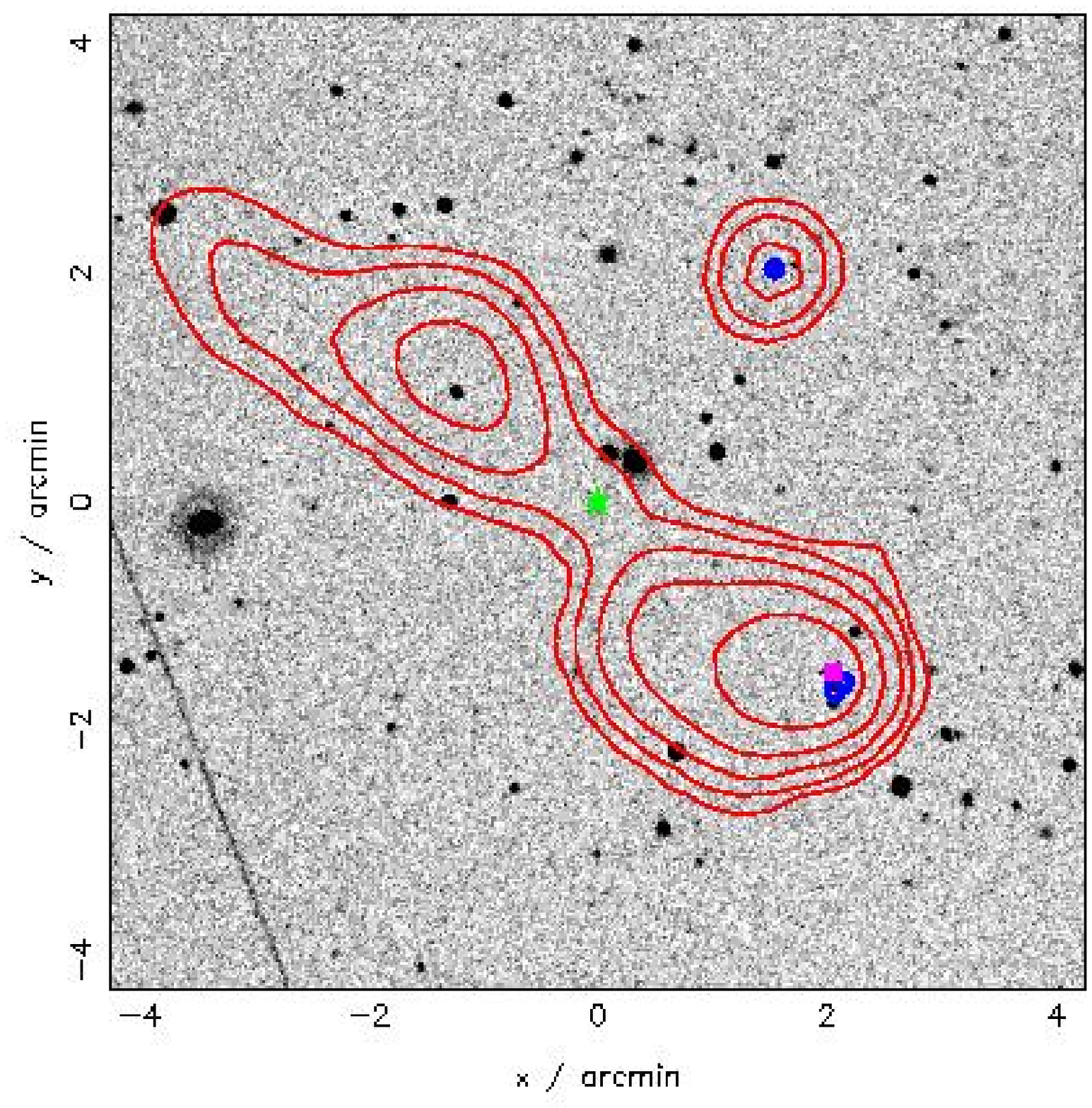}}
      \centerline{C3-137: 7C 1511+2422}
    \end{minipage}
    \hspace{3cm}
    \begin{minipage}{3cm}
      \mbox{}
      \centerline{\includegraphics[scale=0.26,angle=270]{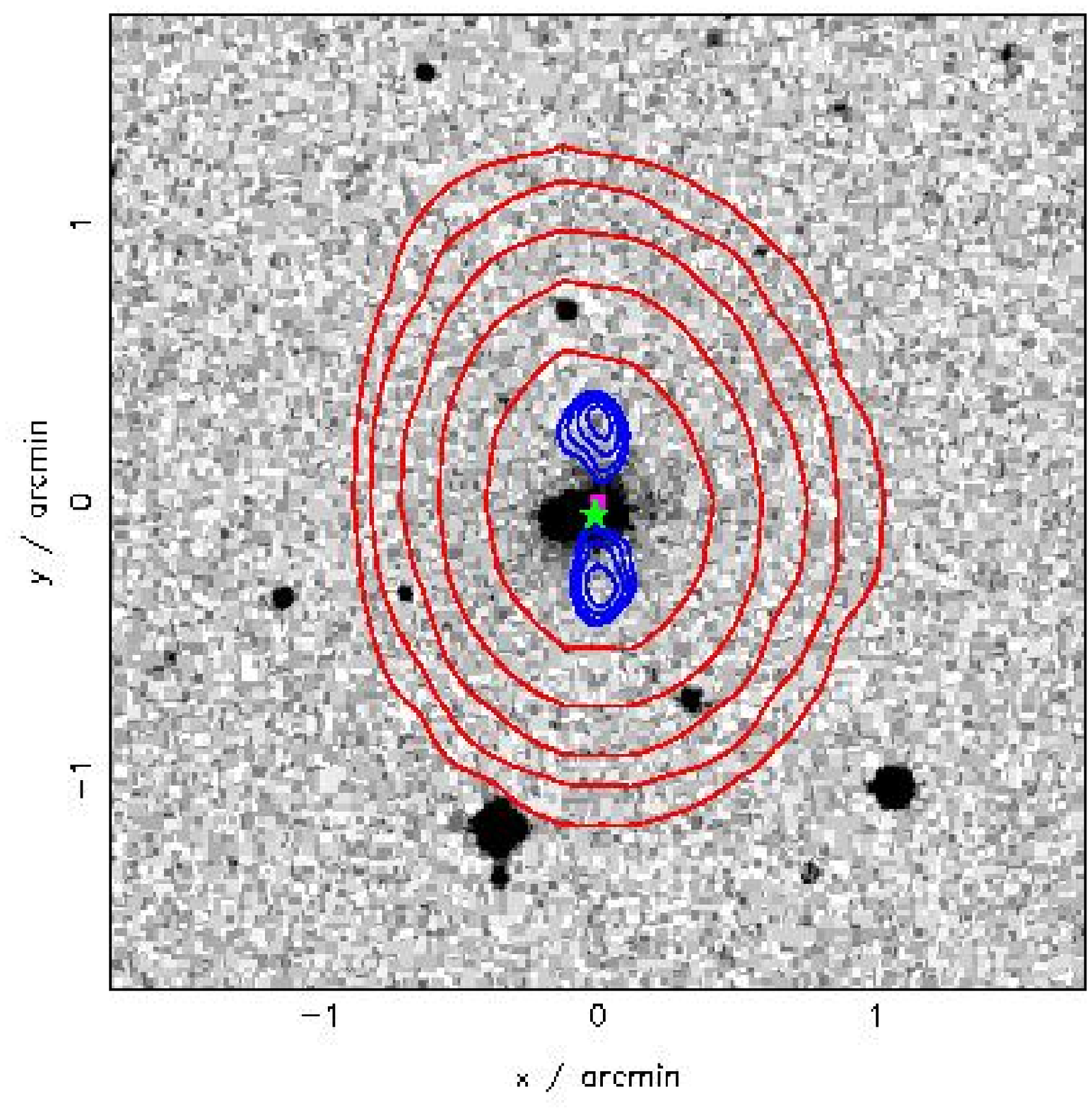}}
      \centerline{C3-139: 7C 1512+2337}
    \end{minipage}
    \vfill
    \begin{minipage}{3cm}     
      \mbox{}
      \centerline{\includegraphics[scale=0.26,angle=270]{Contours/C3/140.ps}}
      \centerline{C3-140: TXS 1511+158}
    \end{minipage}
    \hspace{3cm}
    \begin{minipage}{3cm}
      \mbox{}
      \centerline{\includegraphics[scale=0.26,angle=270]{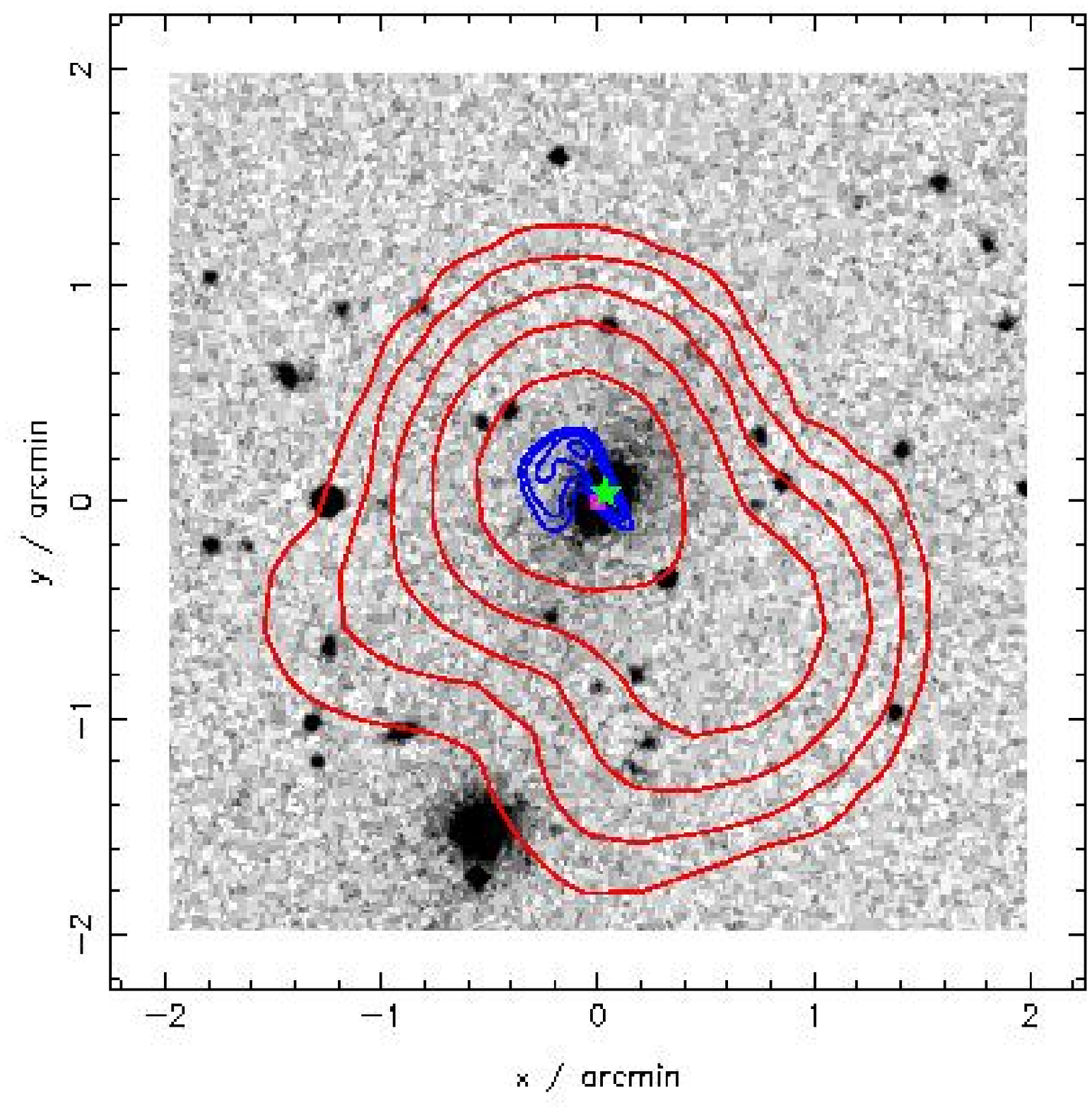}}
      \centerline{C3-142: TXS 1512+104}
    \end{minipage}
    \hspace{3cm}
    \begin{minipage}{3cm}
      \mbox{}
      \centerline{\includegraphics[scale=0.26,angle=270]{Contours/C3/143.ps}}
      \centerline{C3-143: TXS 1512+227}
    \end{minipage}
    \vfill
    \begin{minipage}{3cm}      
      \mbox{}
      \centerline{\includegraphics[scale=0.26,angle=270]{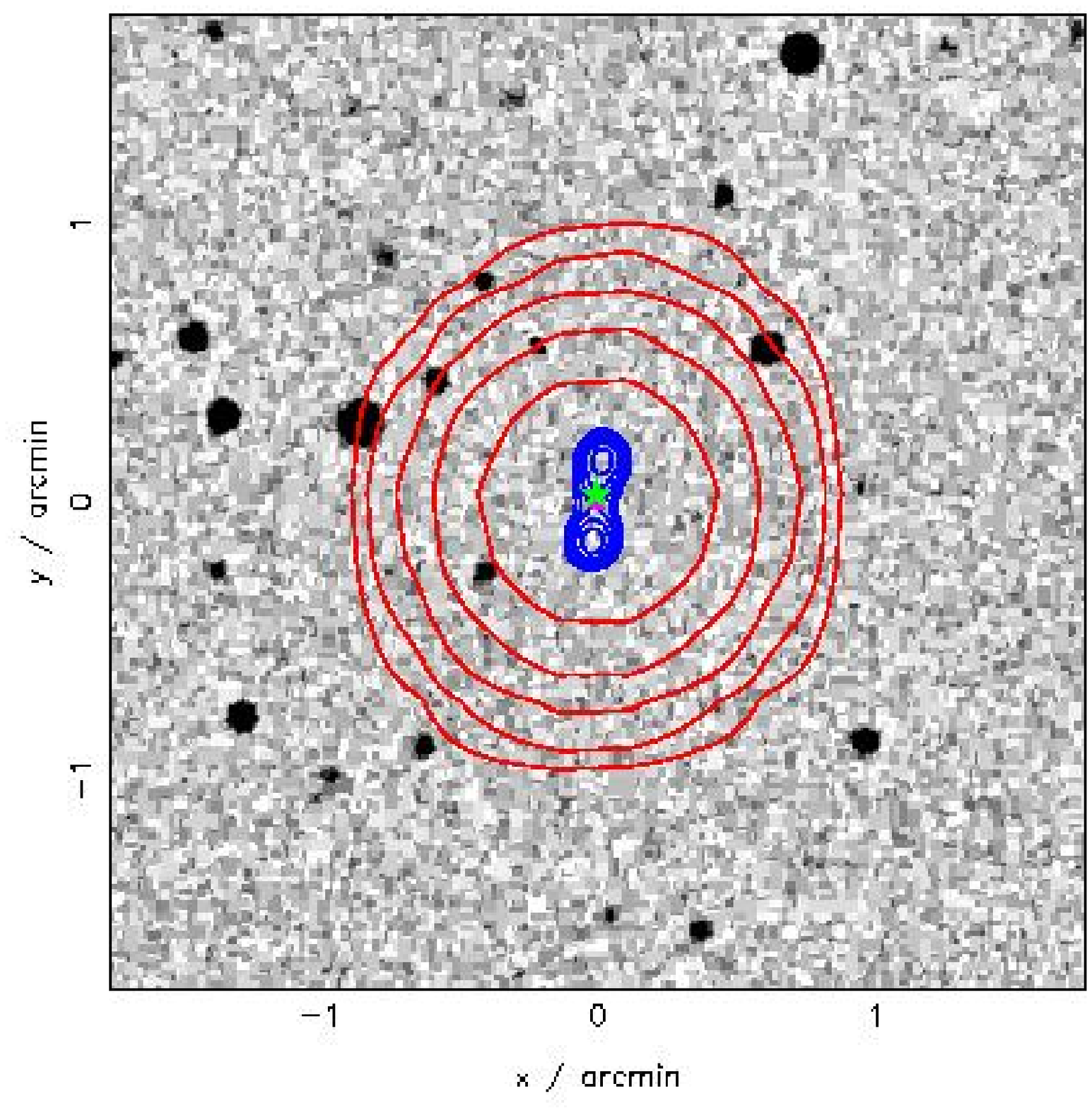}}
      \centerline{C3-144: TXS 1512+104B}
    \end{minipage}
    \hspace{3cm}
    \begin{minipage}{3cm}
      \mbox{}
      \centerline{\includegraphics[scale=0.26,angle=270]{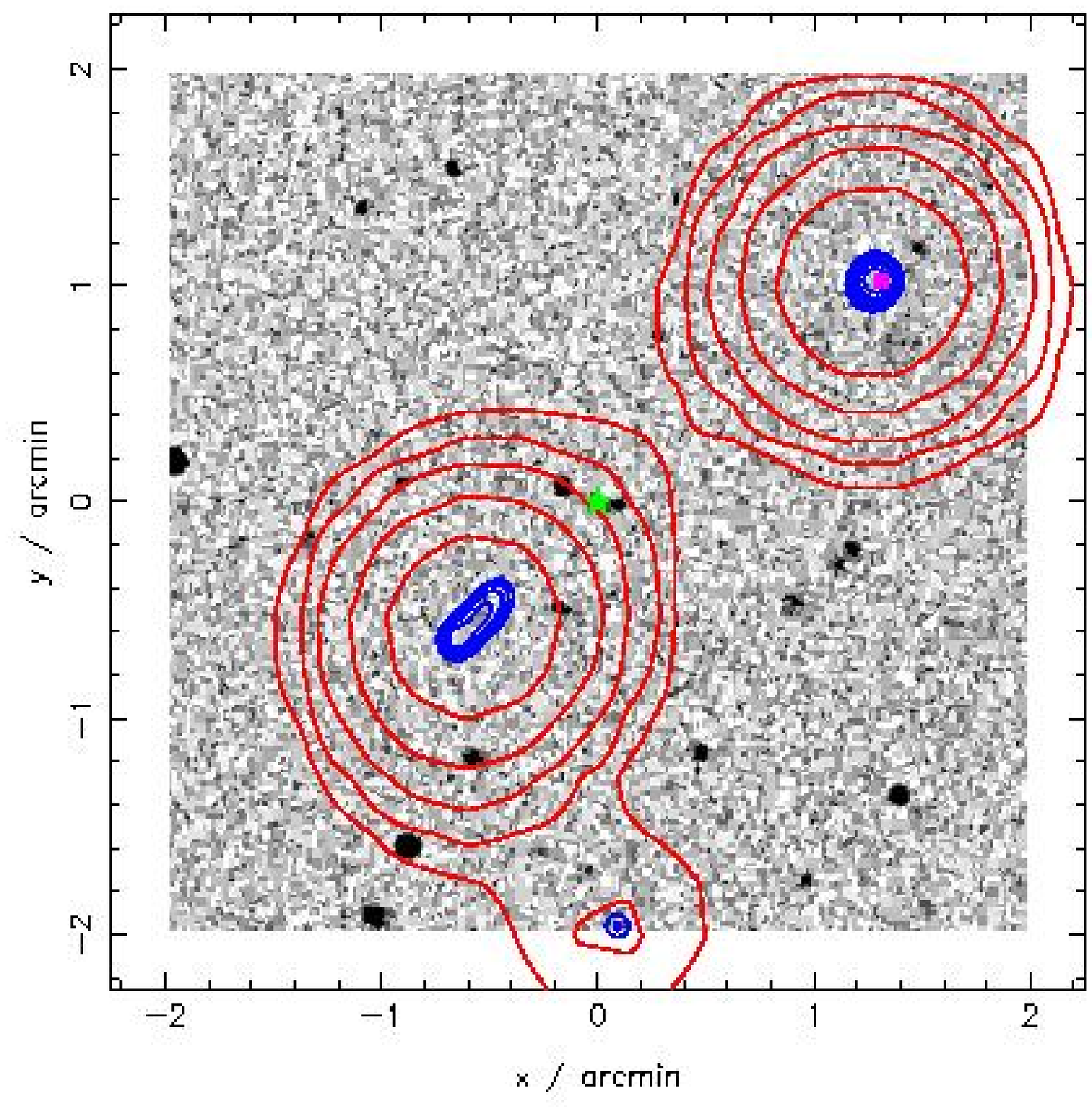}}
      \centerline{C3-146: TXS 1513+144}
    \end{minipage}
    \hspace{3cm}
    \begin{minipage}{3cm}
      \mbox{}
      \centerline{\includegraphics[scale=0.26,angle=270]{Contours/C3/149.ps}}
      \centerline{C3-149: TXS 1514+215}
    \end{minipage}
  \end{center}
\end{figure}

\begin{figure}
  \begin{center}
    {\bf CoNFIG-3}\\  
  \begin{minipage}{3cm}      
      \mbox{}
      \centerline{\includegraphics[scale=0.26,angle=270]{Contours/C3/150.ps}}
      \centerline{C3-150: TXS 1515+301}
    \end{minipage}
    \hspace{3cm}
    \begin{minipage}{3cm}
      \mbox{}
      \centerline{\includegraphics[scale=0.26,angle=270]{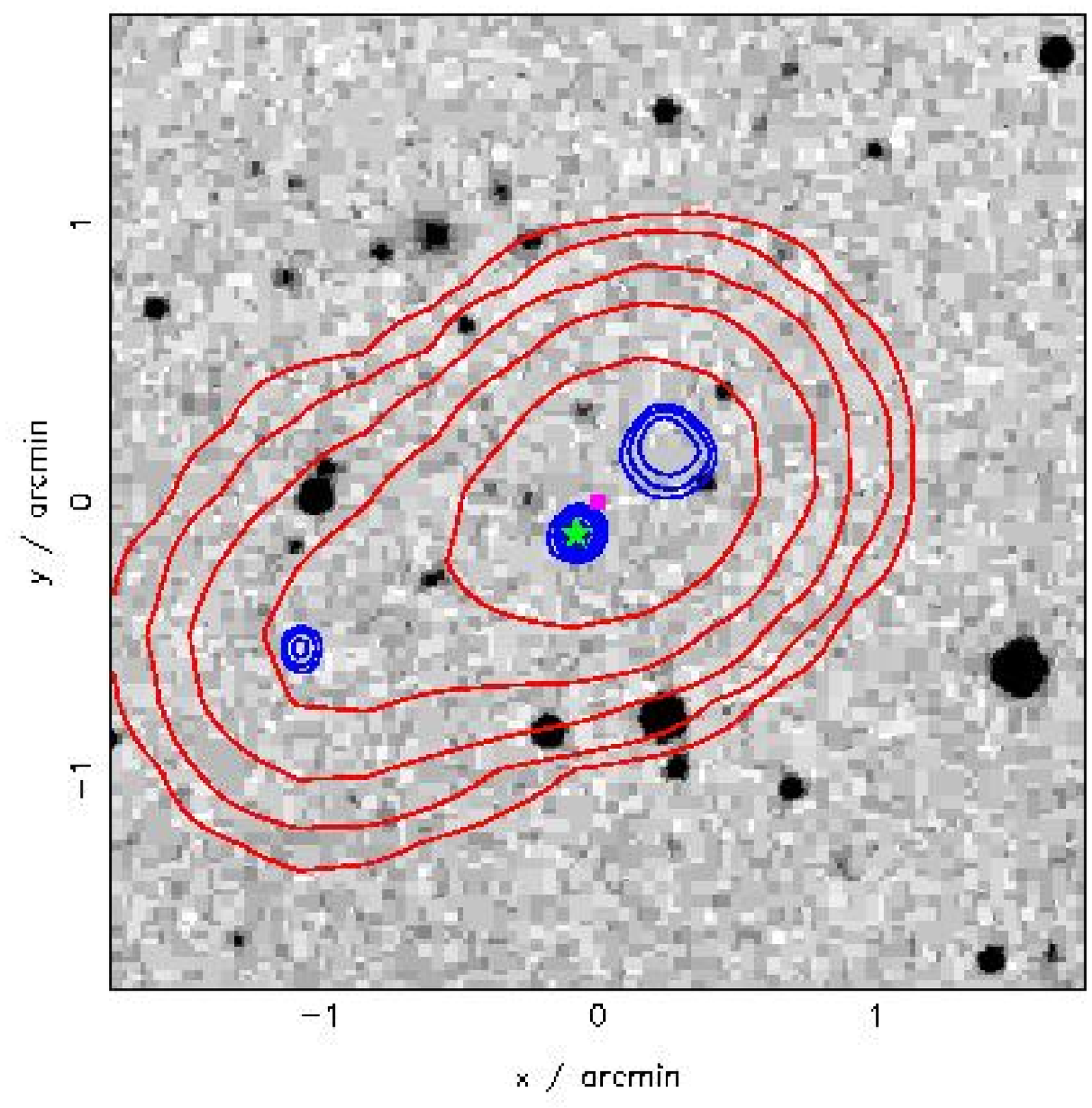}}
      \centerline{C3-151: TXS 1515+176}
    \end{minipage}
    \hspace{3cm}
    \begin{minipage}{3cm}
      \mbox{}
      \centerline{\includegraphics[scale=0.26,angle=270]{Contours/C3/152.ps}}
      \centerline{C3-152:TXS 1515+146}
    \end{minipage}
    \vfill
    \begin{minipage}{3cm}     
      \mbox{}
      \centerline{\includegraphics[scale=0.26,angle=270]{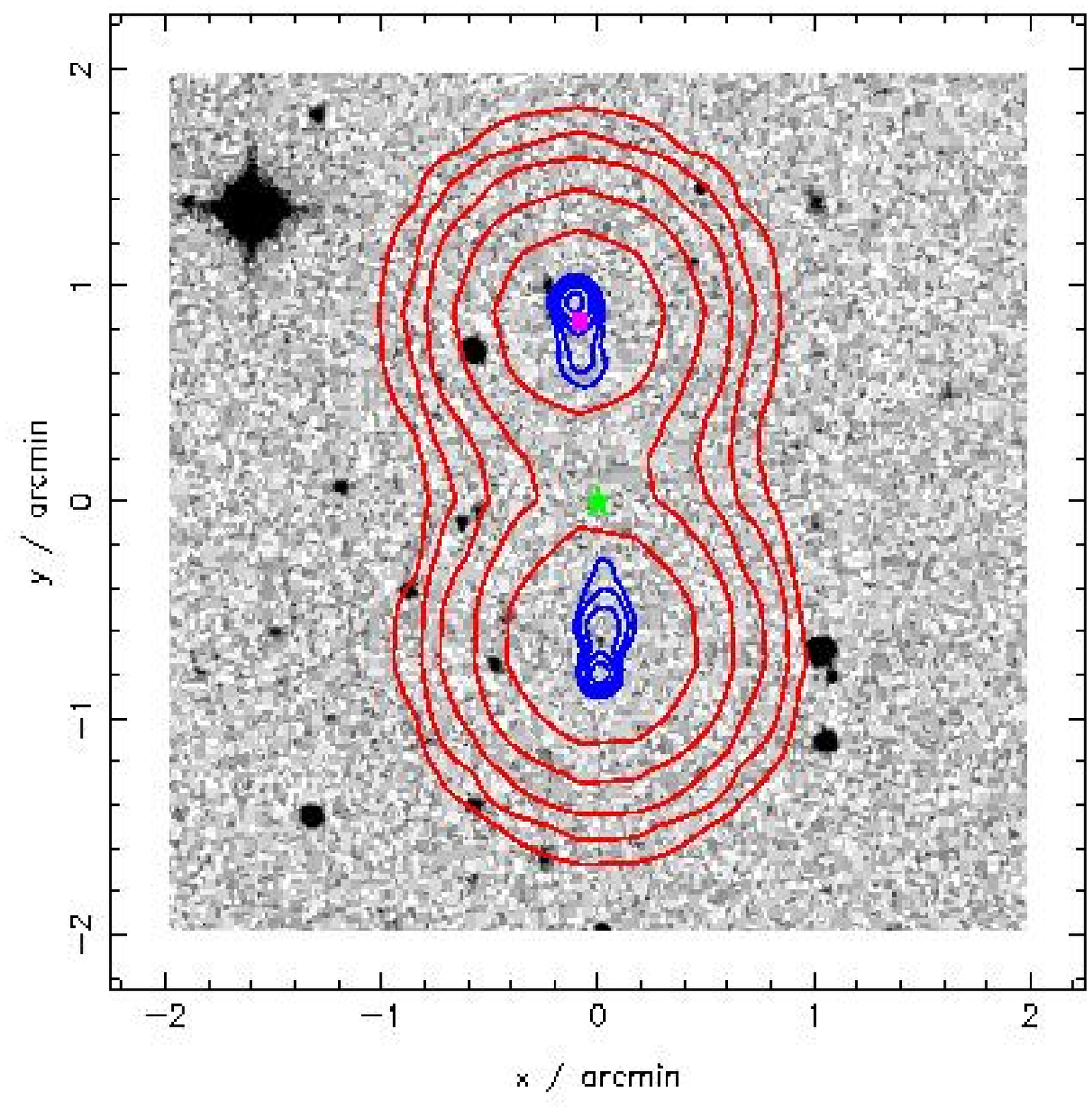}}
      \centerline{C3-153: 4C 10.41}
    \end{minipage}
    \hspace{3cm}
    \begin{minipage}{3cm}
      \mbox{}
      \centerline{\includegraphics[scale=0.26,angle=270]{Contours/C3/154.ps}}
      \centerline{C3-154: TXS 1515+269}
    \end{minipage}
    \hspace{3cm}
    \begin{minipage}{3cm}
      \mbox{}
      \centerline{\includegraphics[scale=0.26,angle=270]{Contours/C3/155.ps}}
      \centerline{C3-155: TXS 1515+198}
    \end{minipage}
    \vfill
    \begin{minipage}{3cm}     
      \mbox{}
      \centerline{\includegraphics[scale=0.26,angle=270]{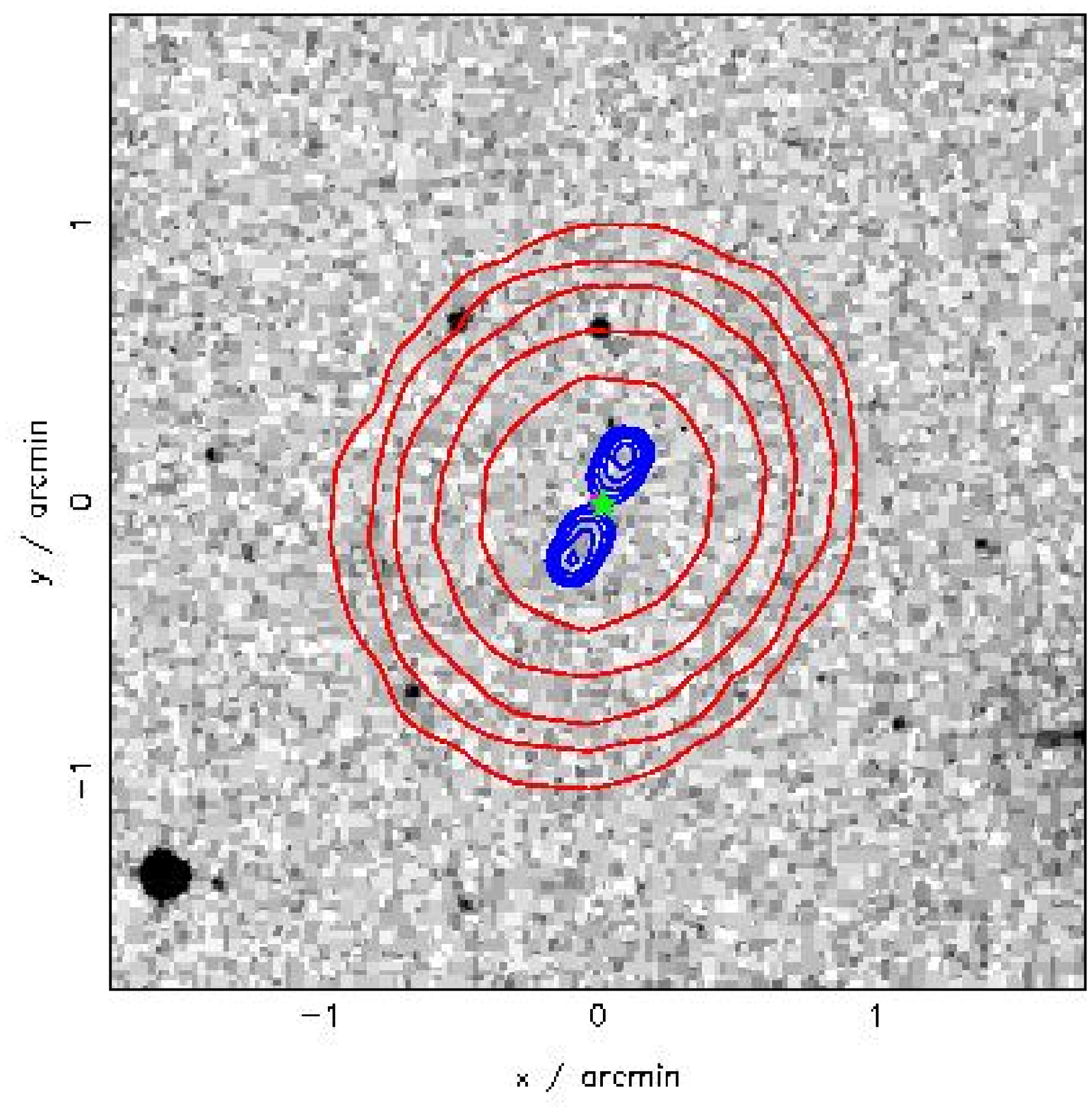}}
      \centerline{C3-156: TXS 1515+160}
    \end{minipage}
    \hspace{3cm}
    \begin{minipage}{3cm}
      \mbox{}
      \centerline{\includegraphics[scale=0.26,angle=270]{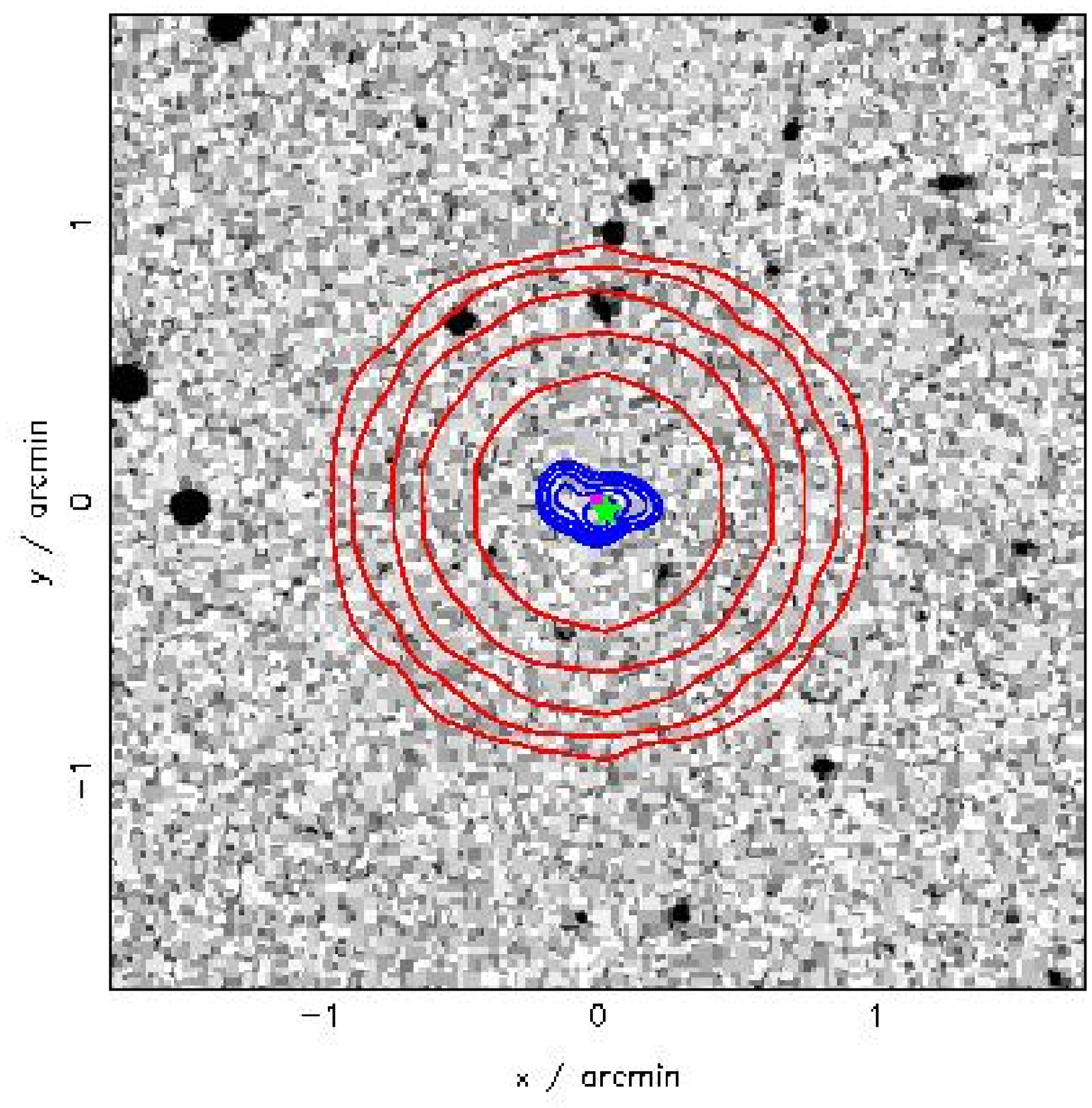}}
      \centerline{C3-160: 4C 24.33}
    \end{minipage}
    \hspace{3cm}
    \begin{minipage}{3cm}
      \mbox{}
      \centerline{\includegraphics[scale=0.26,angle=270]{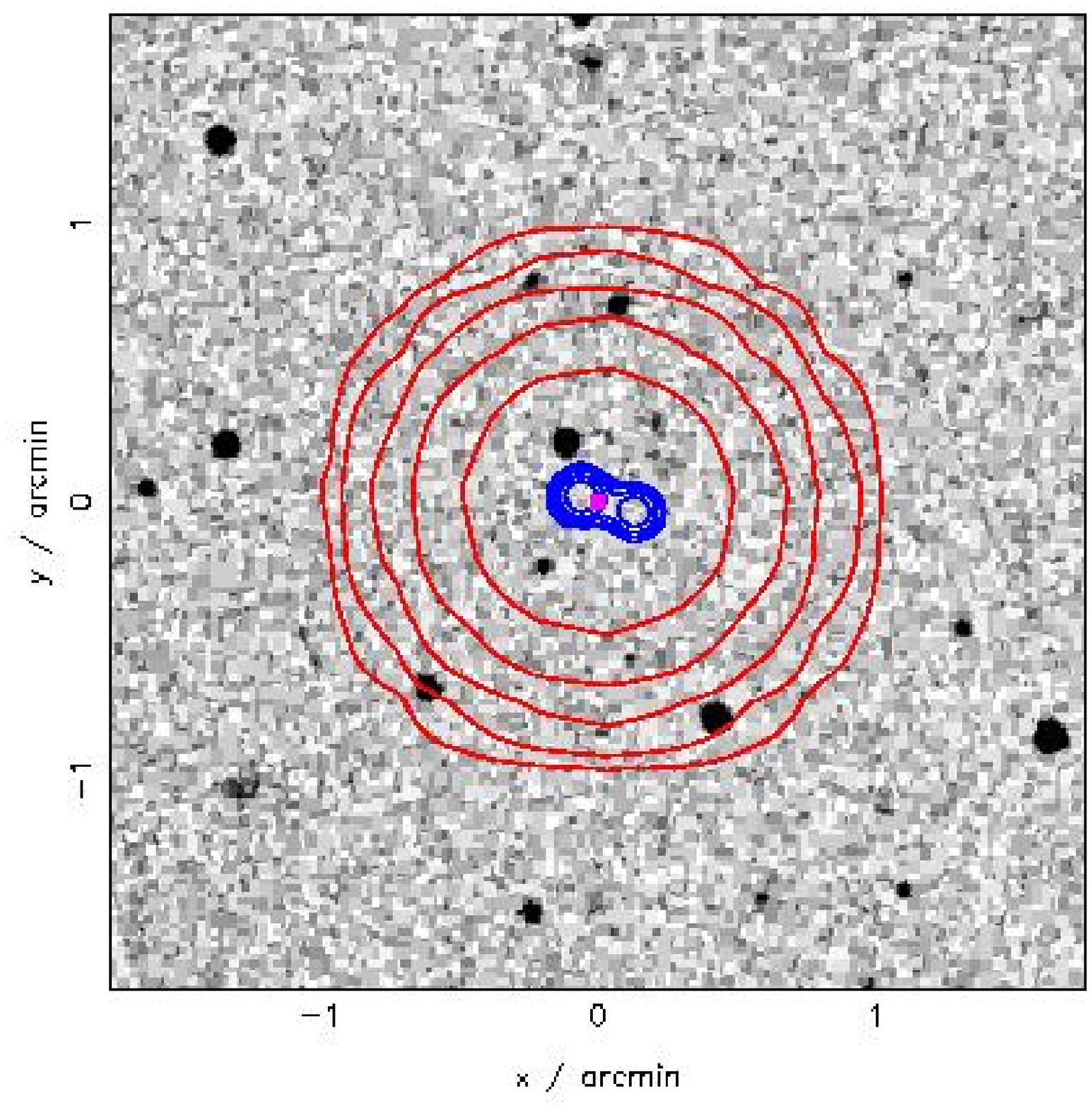}}
      \centerline{C3-163: 4C 15.47} 
    \end{minipage}
    \vfill
    \begin{minipage}{3cm}      
      \mbox{}
      \centerline{\includegraphics[scale=0.26,angle=270]{Contours/C3/164.ps}}
      \centerline{C3-164: TXS 1519+228}
    \end{minipage}
    \hspace{3cm}
    \begin{minipage}{3cm}
      \mbox{}
      \centerline{\includegraphics[scale=0.26,angle=270]{Contours/C3/165.ps}}
      \centerline{C3-165: TXS 1519+153} 
    \end{minipage}
    \hspace{3cm}
    \begin{minipage}{3cm}
      \mbox{}
      \centerline{\includegraphics[scale=0.26,angle=270]{Contours/C3/166.ps}}
      \centerline{C3-166: TXS 1519+108}
    \end{minipage}
  \end{center}
\end{figure}

\begin{figure}
  \begin{center}
    {\bf CoNFIG-3}\\  
  \begin{minipage}{3cm}      
      \mbox{}
      \centerline{\includegraphics[scale=0.26,angle=270]{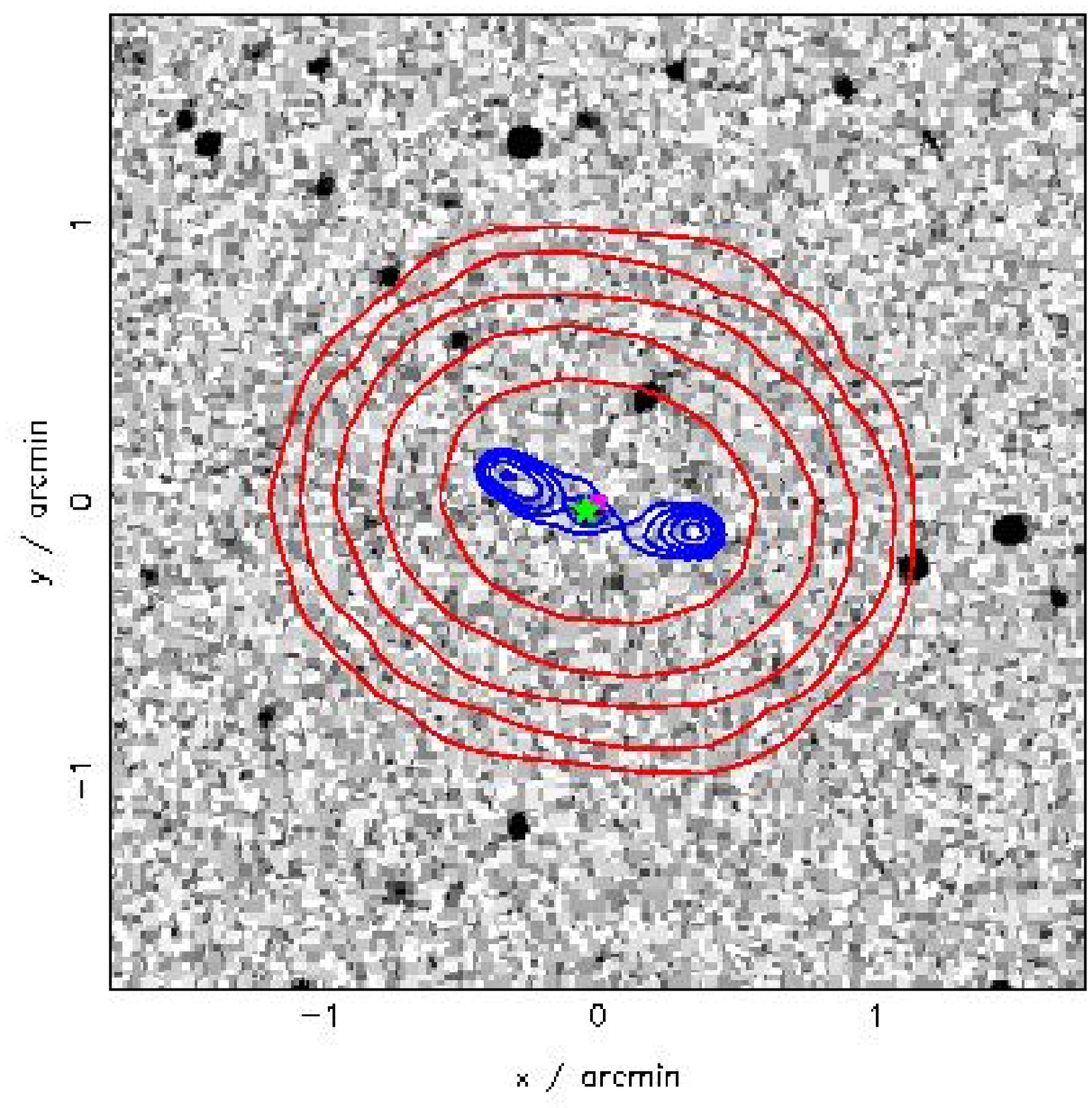}}
      \centerline{C3-167: TXS 1519+103}
    \end{minipage}
    \hspace{3cm}
    \begin{minipage}{3cm}
      \mbox{}
      \centerline{\includegraphics[scale=0.26,angle=270]{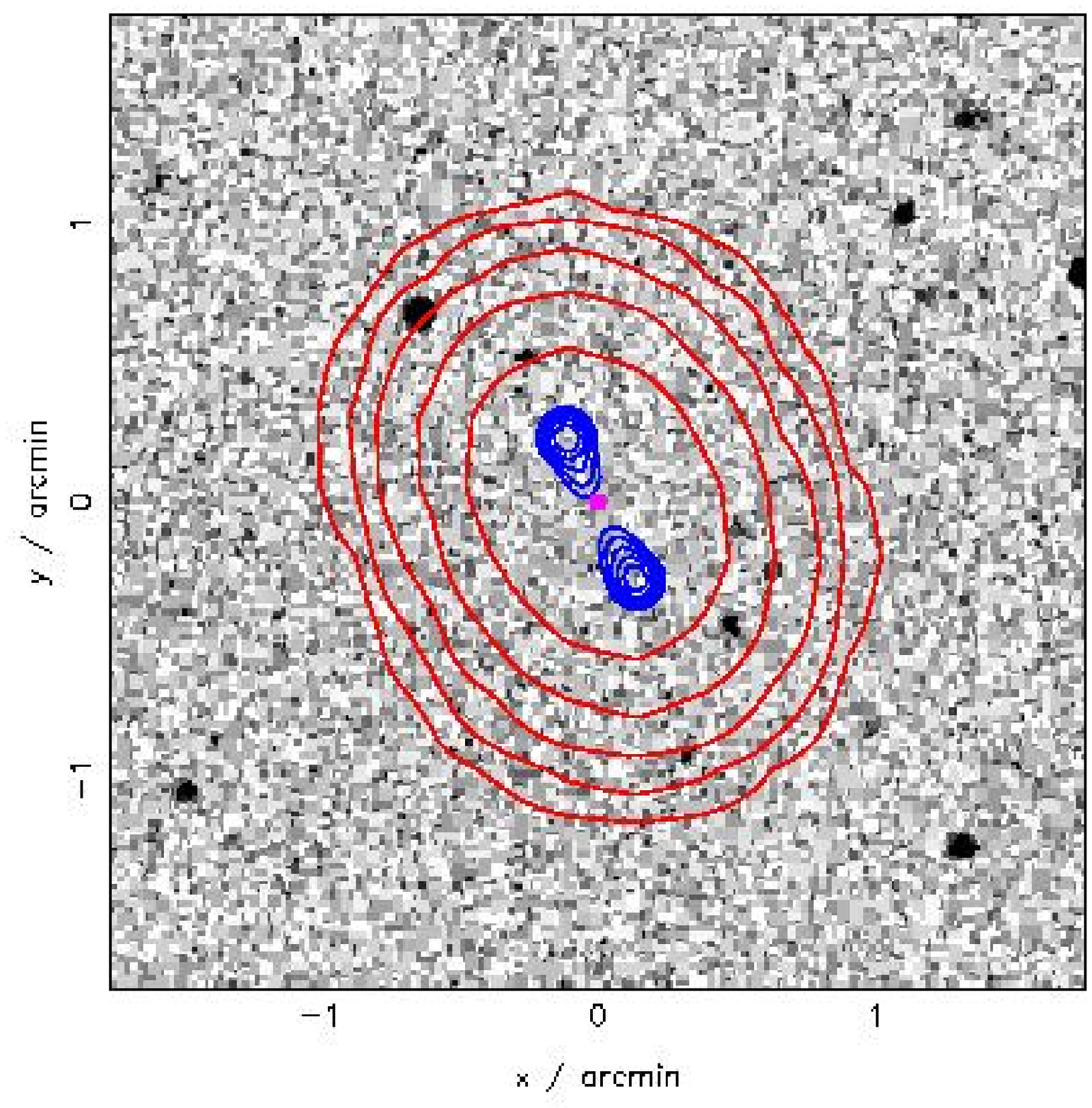}}
      \centerline{C3-169: TXS 1520+221}
    \end{minipage}
    \hspace{3cm}
    \begin{minipage}{3cm}
      \mbox{}
      \centerline{\includegraphics[scale=0.26,angle=270]{Contours/C3/171.ps}}
      \centerline{C3-171: 4C 27.31}
    \end{minipage}
    \vfill
    \begin{minipage}{3cm}     
      \mbox{}
      \centerline{\includegraphics[scale=0.26,angle=270]{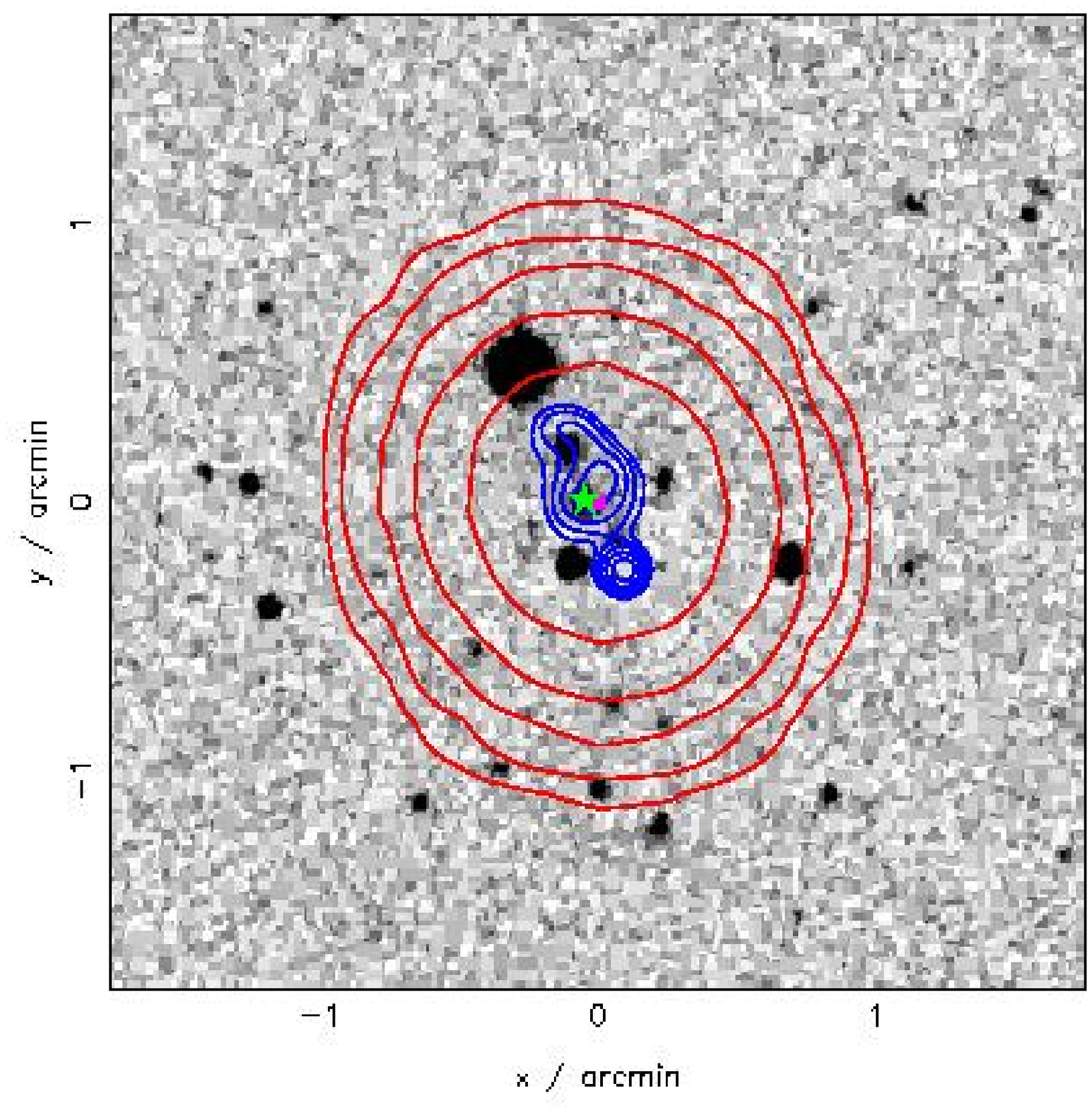}}
      \centerline{C3-172: TXS 1521+116}
    \end{minipage}
    \hspace{3cm}
    \begin{minipage}{3cm}
      \mbox{}
      \centerline{\includegraphics[scale=0.26,angle=270]{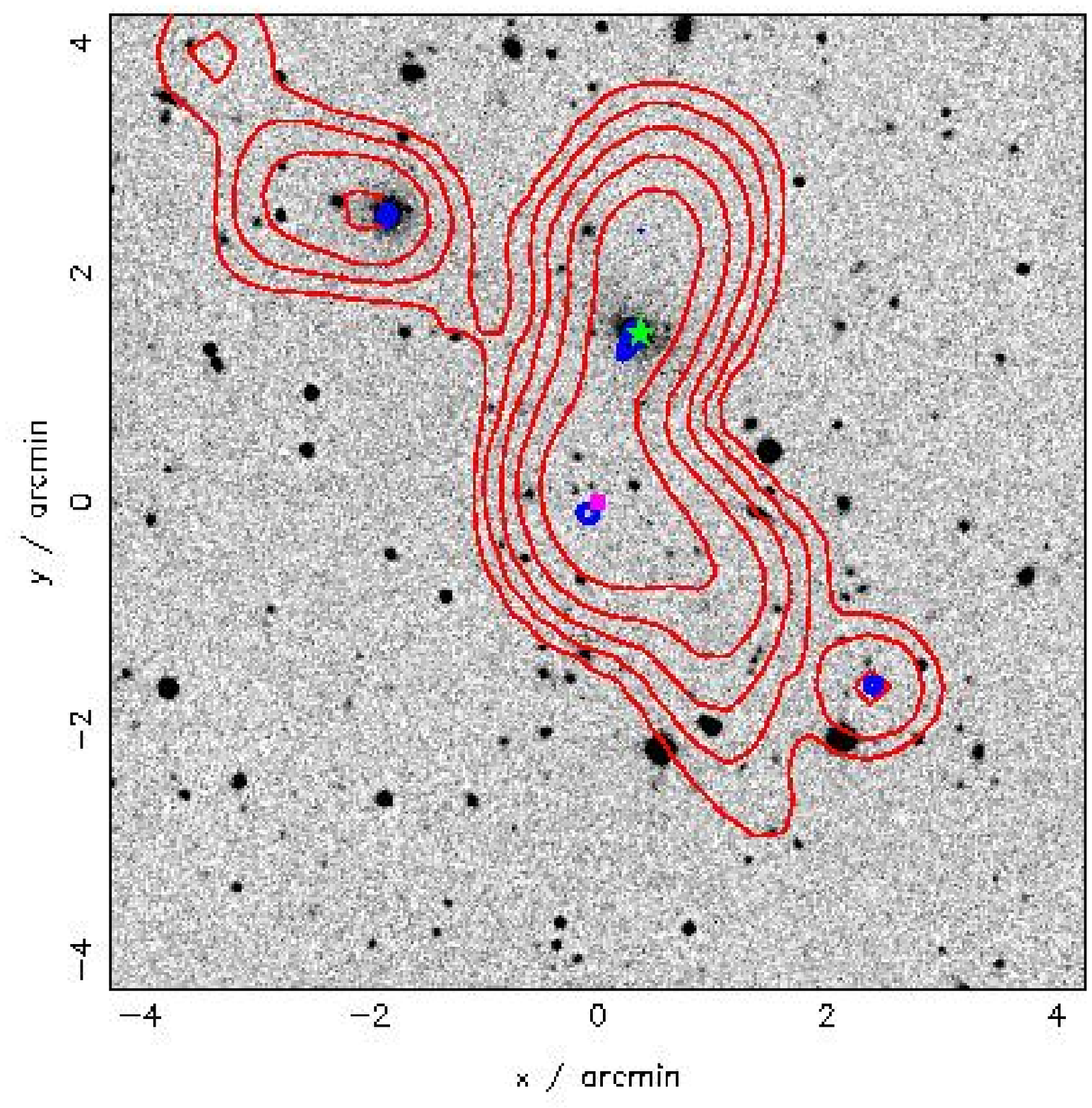}}
      \centerline{C3-173: 4C 28.39}
    \end{minipage}
    \hspace{3cm}
    \begin{minipage}{3cm}
      \mbox{}
      \centerline{\includegraphics[scale=0.26,angle=270]{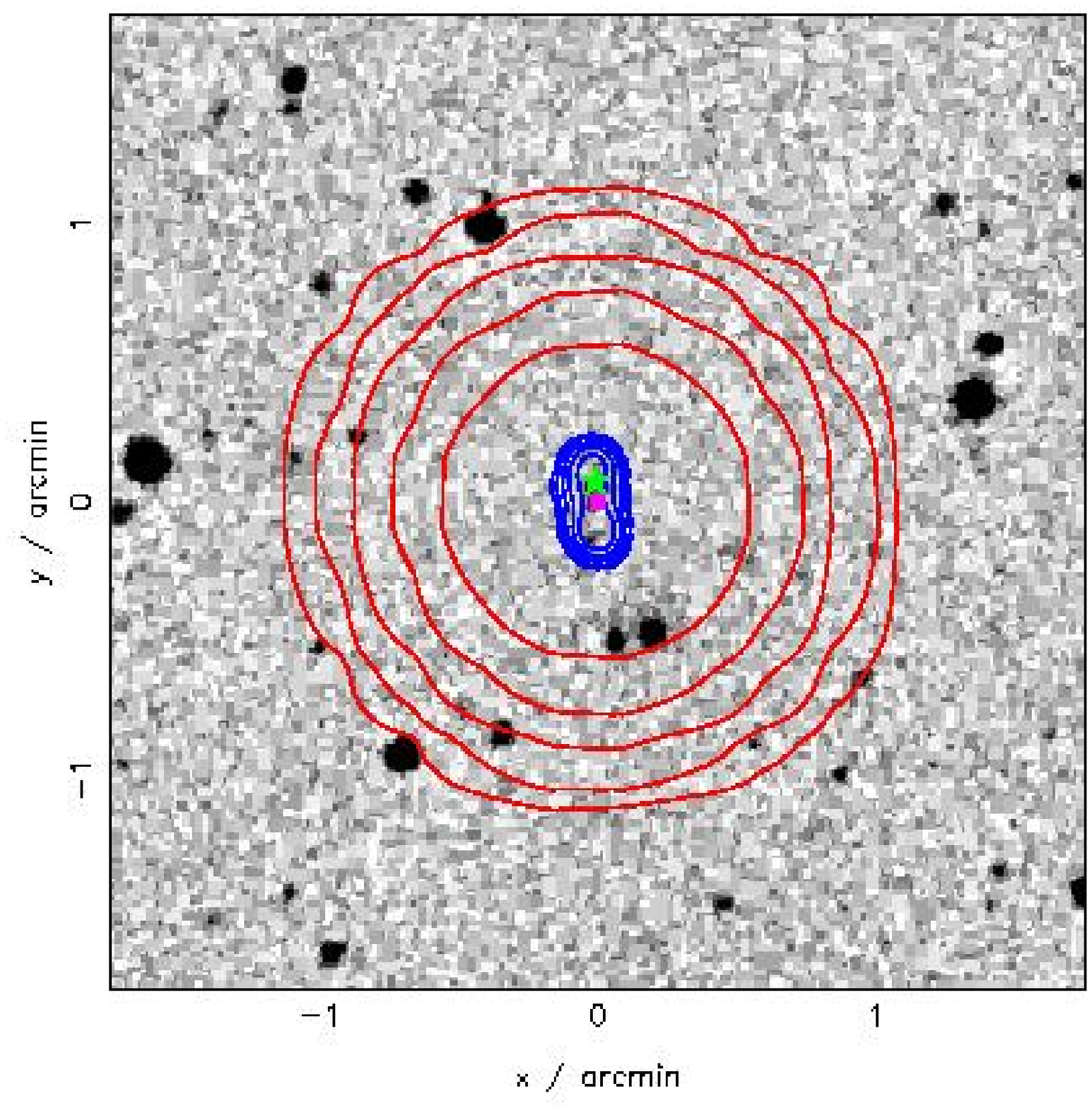}}
      \centerline{C3-174: 4C 11.49}
    \end{minipage}
    \vfill
    \begin{minipage}{3cm}     
      \mbox{}
      \centerline{\includegraphics[scale=0.26,angle=270]{Contours/C3/177.ps}}
      \centerline{C3-177: TXS 1521+195}
    \end{minipage}
    \hspace{3cm}
    \begin{minipage}{3cm}
      \mbox{}
      \centerline{\includegraphics[scale=0.26,angle=270]{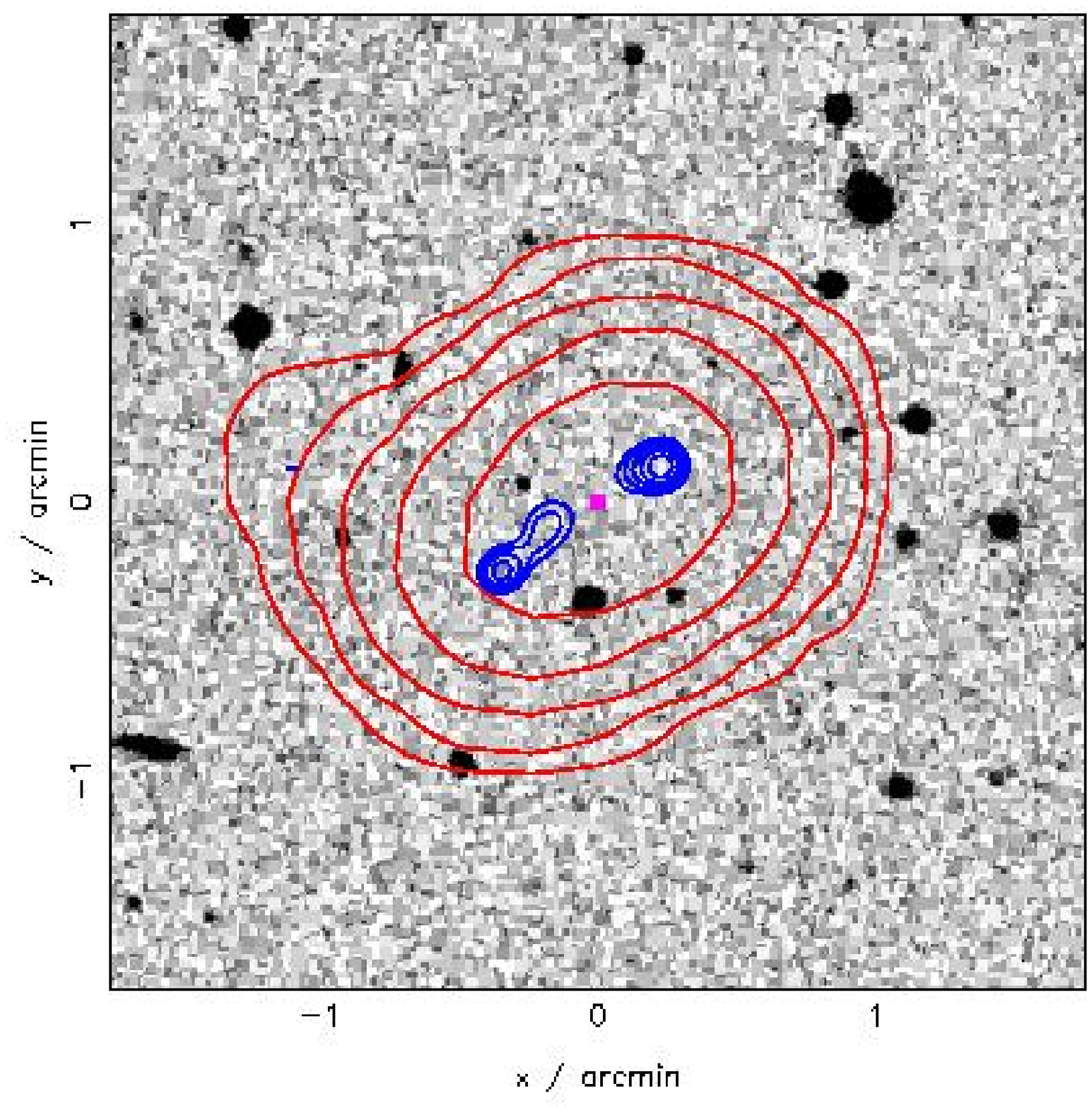}}
      \centerline{C3-179: TXS 1522+281}
    \end{minipage}
    \hspace{3cm}
    \begin{minipage}{3cm}
      \mbox{}
      \centerline{\includegraphics[scale=0.26,angle=270]{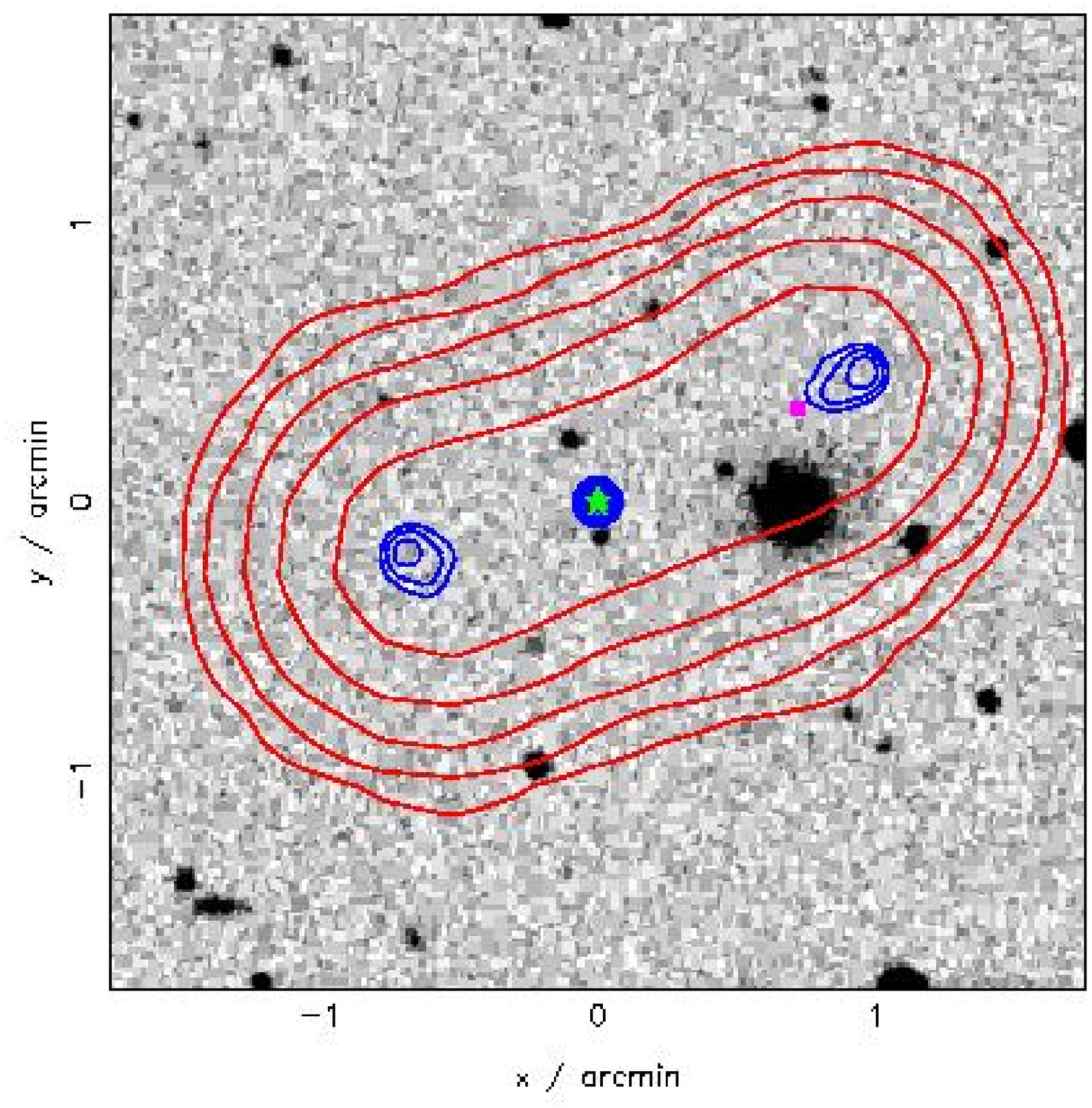}}
      \centerline{C3-181: BWE 1522+1303}
    \end{minipage}
    \vfill
    \begin{minipage}{3cm}      
      \mbox{}
      \centerline{\includegraphics[scale=0.26,angle=270]{Contours/C3/184.ps}}
      \centerline{C3-184: TXS 1524+149}
    \end{minipage}
    \hspace{3cm}
    \begin{minipage}{3cm}
      \mbox{}
      \centerline{\includegraphics[scale=0.26,angle=270]{Contours/C3/185.ps}}
      \centerline{C3-185: BWE 1524+1302}
    \end{minipage}
    \hspace{3cm}
    \begin{minipage}{3cm}
      \mbox{}
      \centerline{\includegraphics[scale=0.26,angle=270]{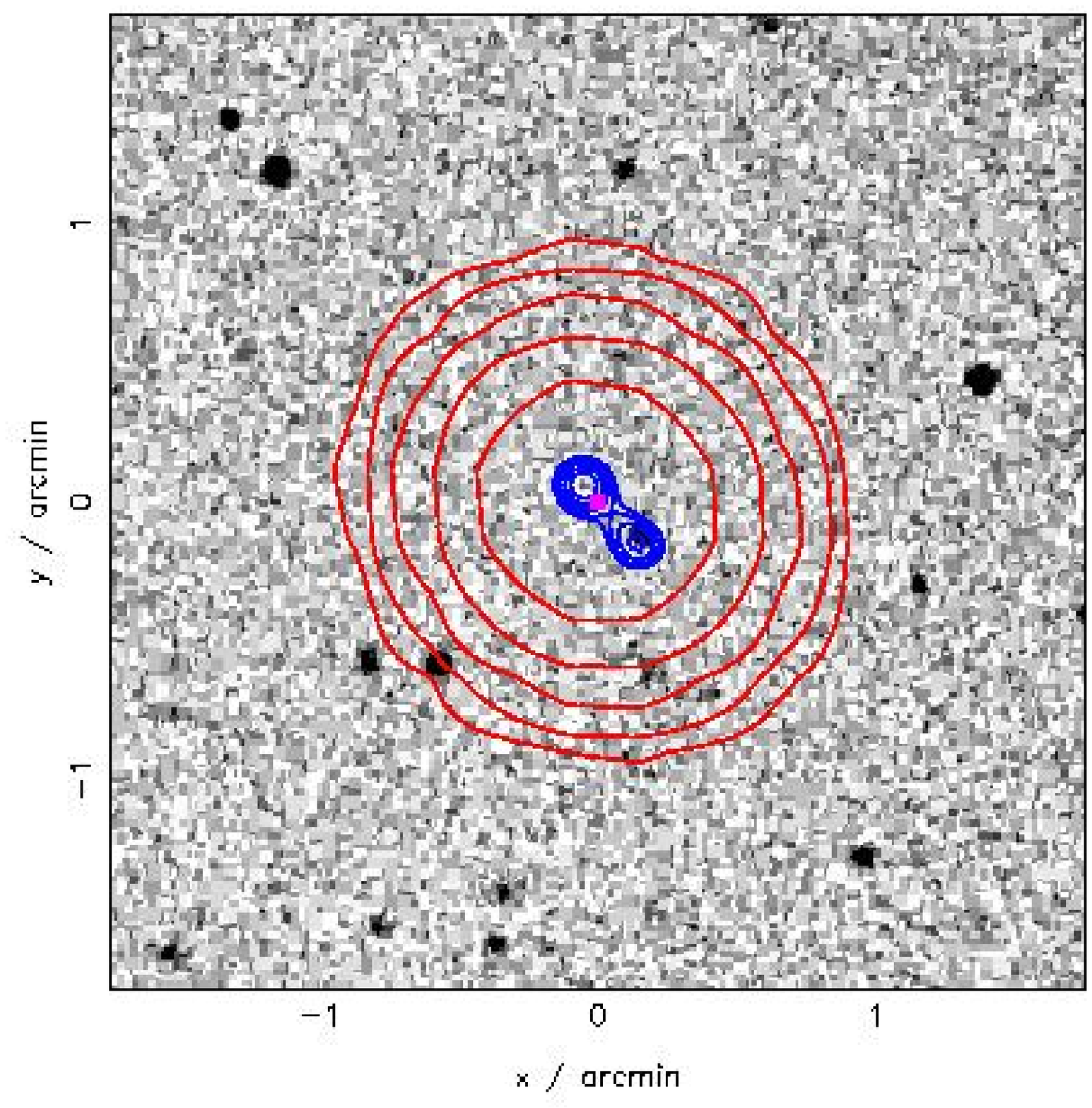}}
      \centerline{C3-186: TXS 1525+210}
    \end{minipage}
  \end{center}
\end{figure}

\begin{figure}
  \begin{center}
    {\bf CoNFIG-3}\\  
  \begin{minipage}{3cm}      
      \mbox{}
      \centerline{\includegraphics[scale=0.26,angle=270]{Contours/C3/187.ps}}
      \centerline{C3-187: 4C 12.55}
    \end{minipage}
    \hspace{3cm}
    \begin{minipage}{3cm}
      \mbox{}
      \centerline{\includegraphics[scale=0.26,angle=270]{Contours/C3/189.ps}}
      \centerline{C3-189: TXS 1525+290}
    \end{minipage}
    \hspace{3cm}
    \begin{minipage}{3cm}
      \mbox{}
      \centerline{\includegraphics[scale=0.26,angle=270]{Contours/C3/190.ps}}
      \centerline{C3-190: TXS 1525+227}
    \end{minipage}
    \vfill
    \begin{minipage}{3cm}     
      \mbox{}
      \centerline{\includegraphics[scale=0.26,angle=270]{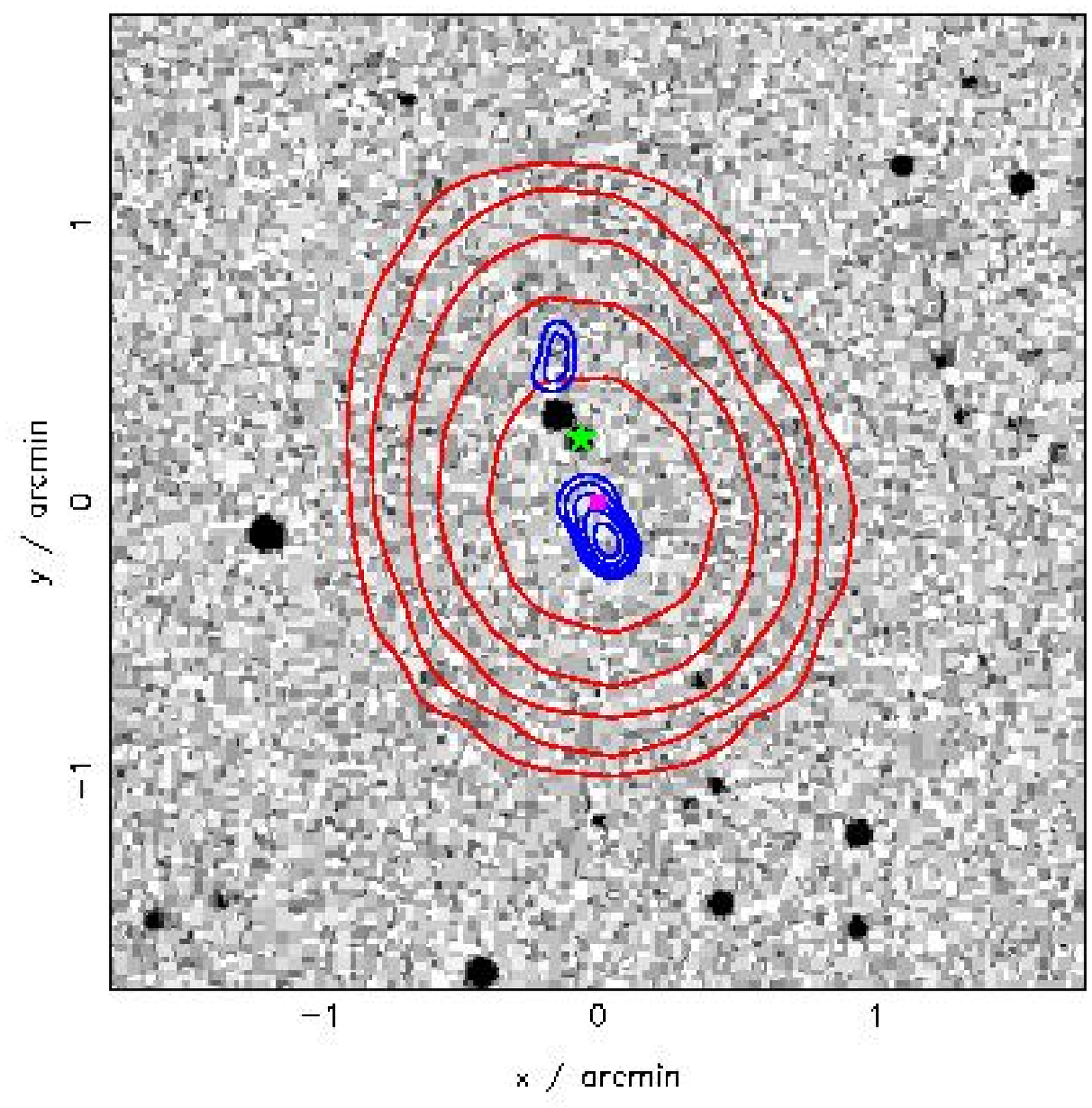}}
      \centerline{C3-191: TXS 1525+135}
    \end{minipage}
    \hspace{3cm}
    \begin{minipage}{3cm}
      \mbox{}
      \centerline{\includegraphics[scale=0.26,angle=270]{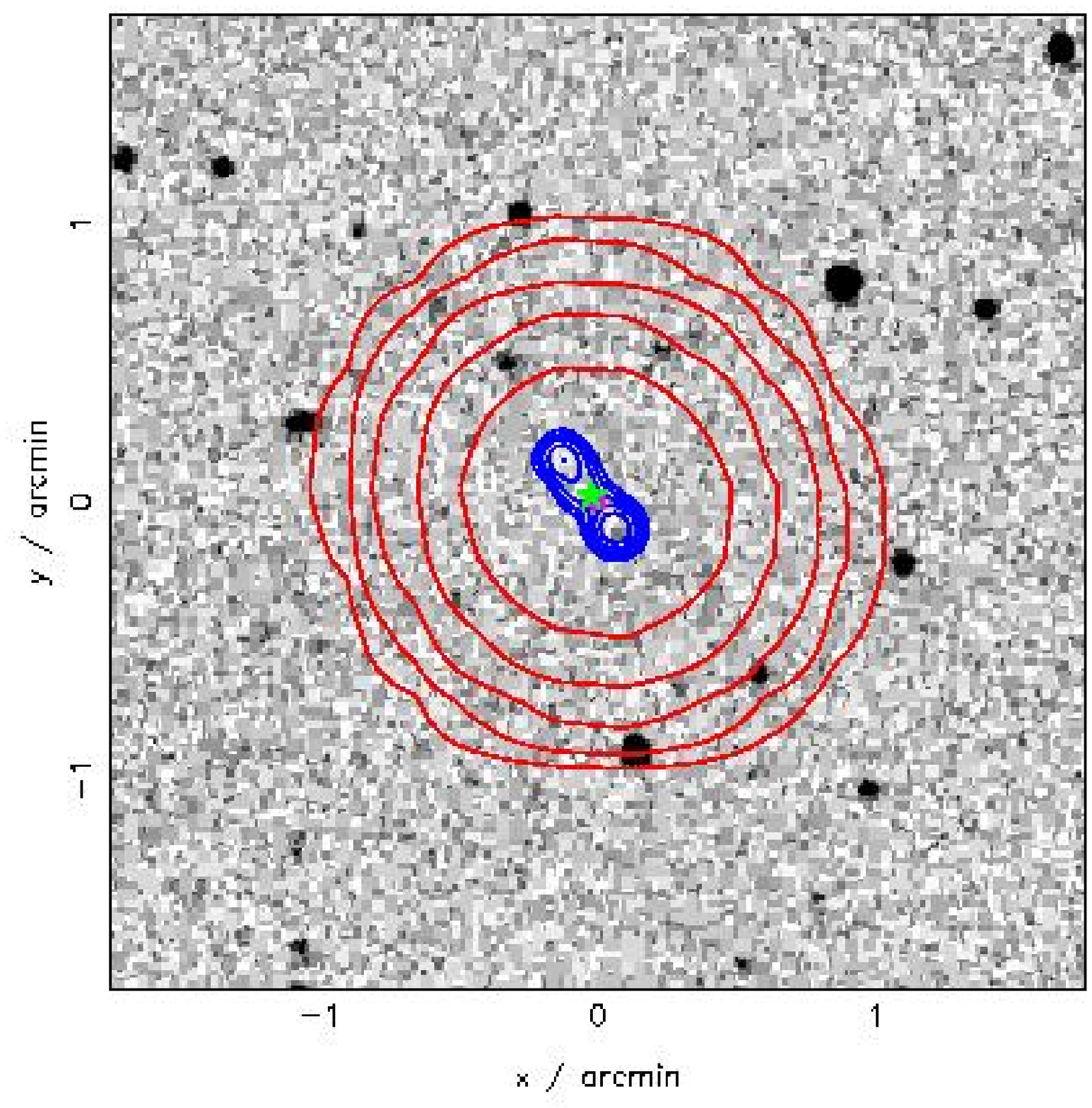}}
      \centerline{C3-192: TXS 1526+173}
    \end{minipage}
    \hspace{3cm}
    \begin{minipage}{3cm}
      \mbox{}
      \centerline{\includegraphics[scale=0.26,angle=270]{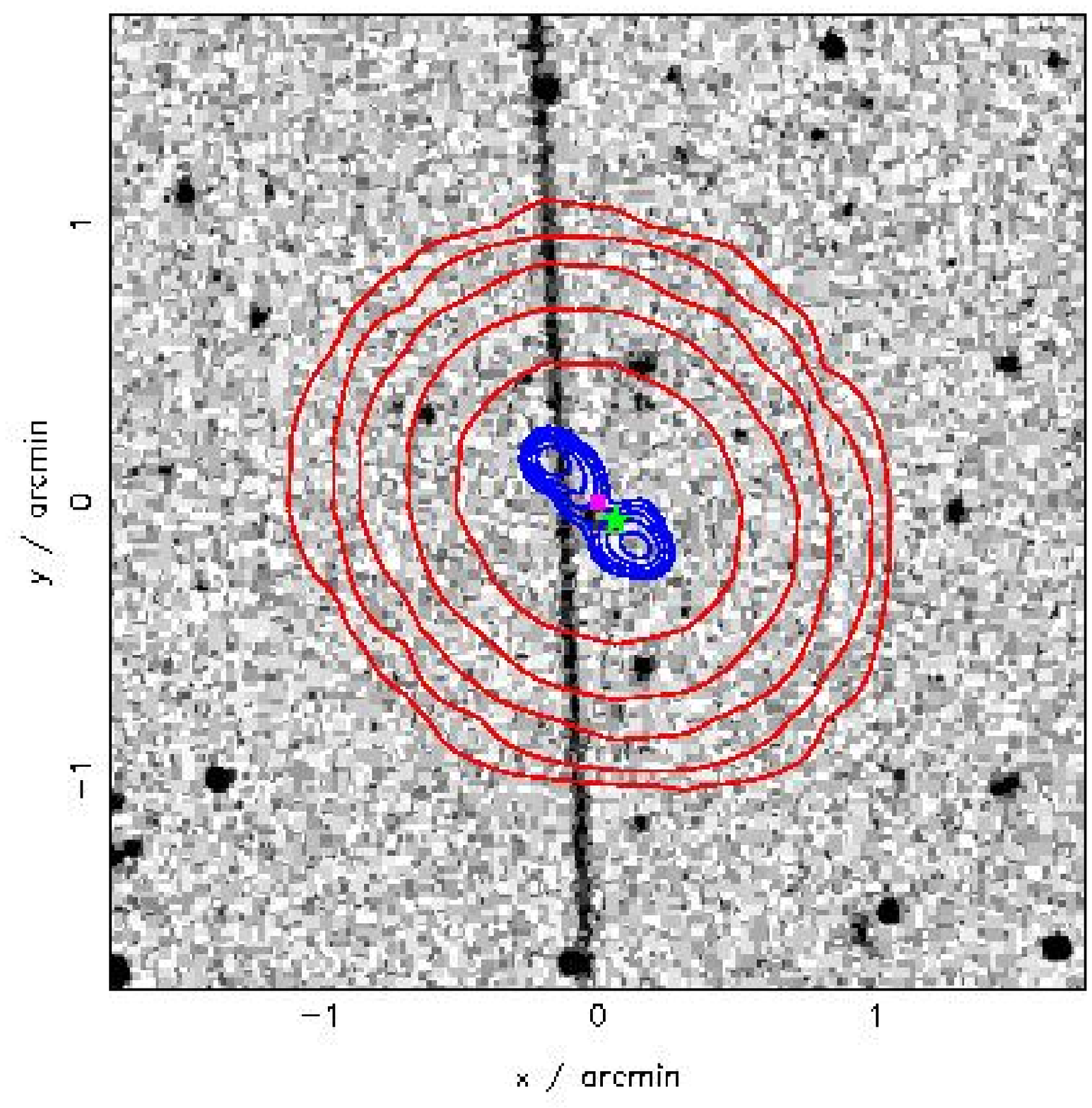}}
      \centerline{C3-193: 4C 15.48}
    \end{minipage}
    \vfill
    \begin{minipage}{3cm}     
      \mbox{}
      \centerline{\includegraphics[scale=0.26,angle=270]{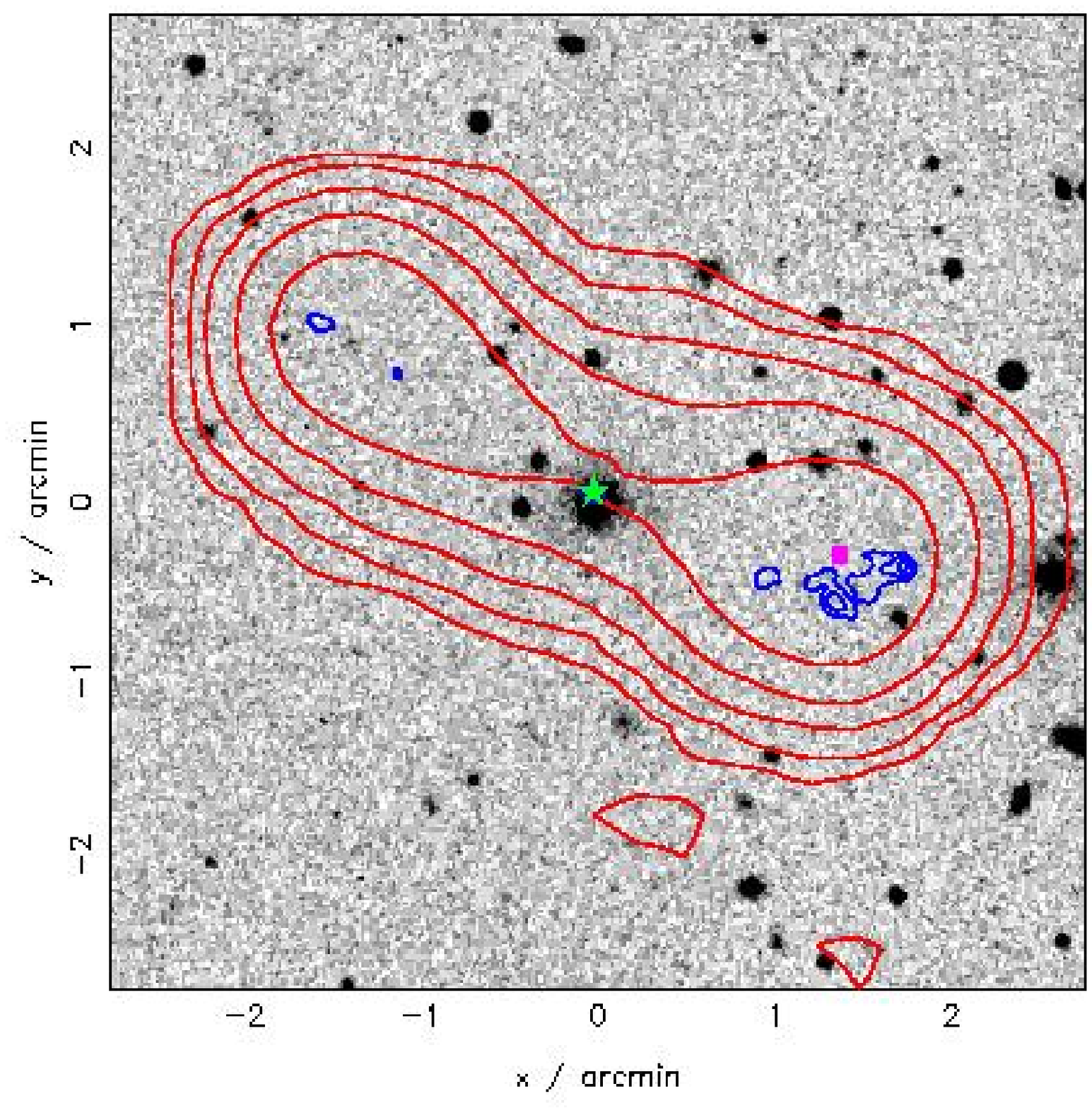}}
      \centerline{C3-195: 7C 1528+2910}
    \end{minipage}
    \hspace{3cm}
    \begin{minipage}{3cm}
      \mbox{}
      \centerline{\includegraphics[scale=0.26,angle=270]{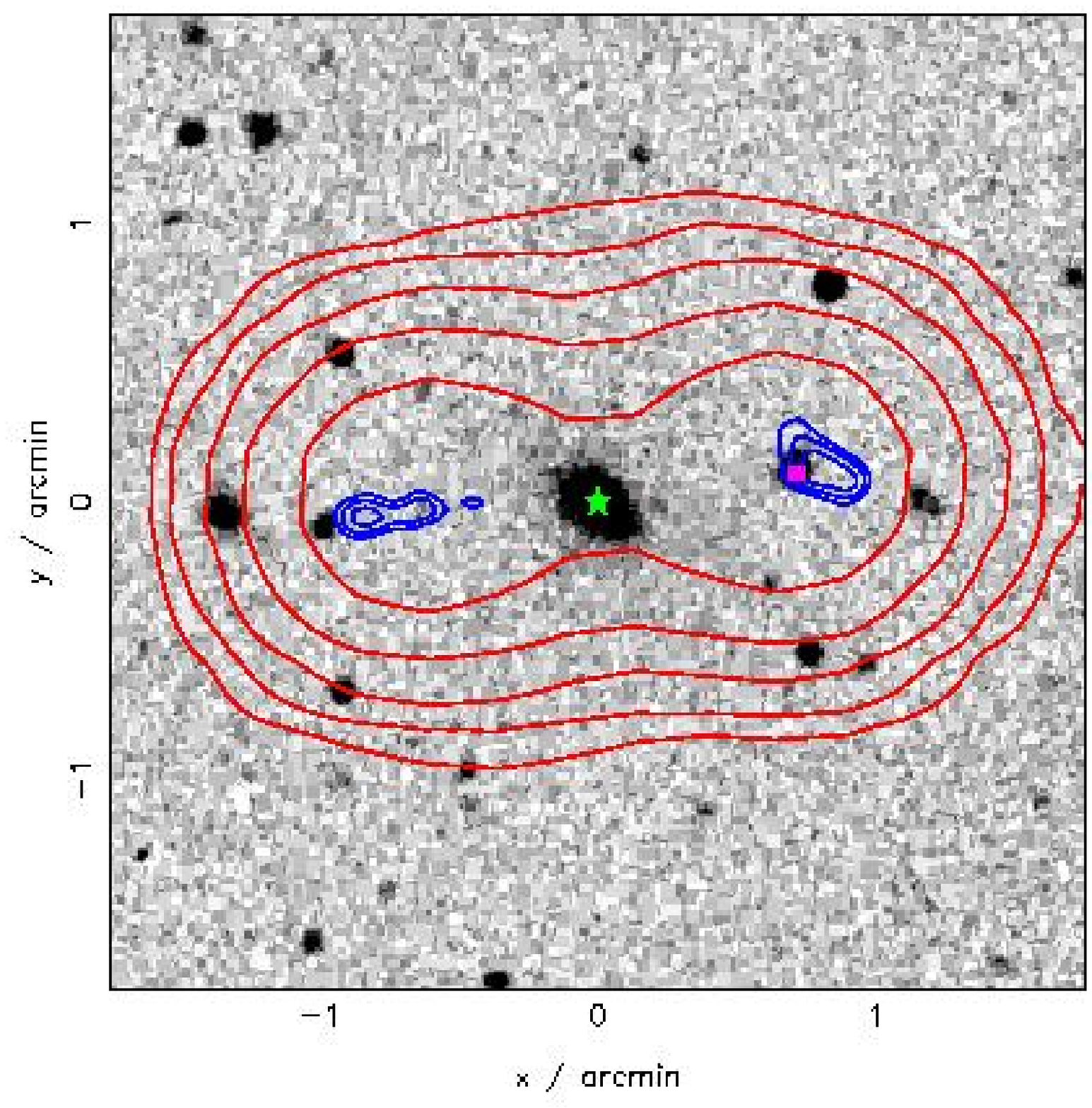}}
      \centerline{C3-196: TXS 1527+234}
    \end{minipage}
    \hspace{3cm}
    \begin{minipage}{3cm}
      \mbox{}
      \centerline{\includegraphics[scale=0.26,angle=270]{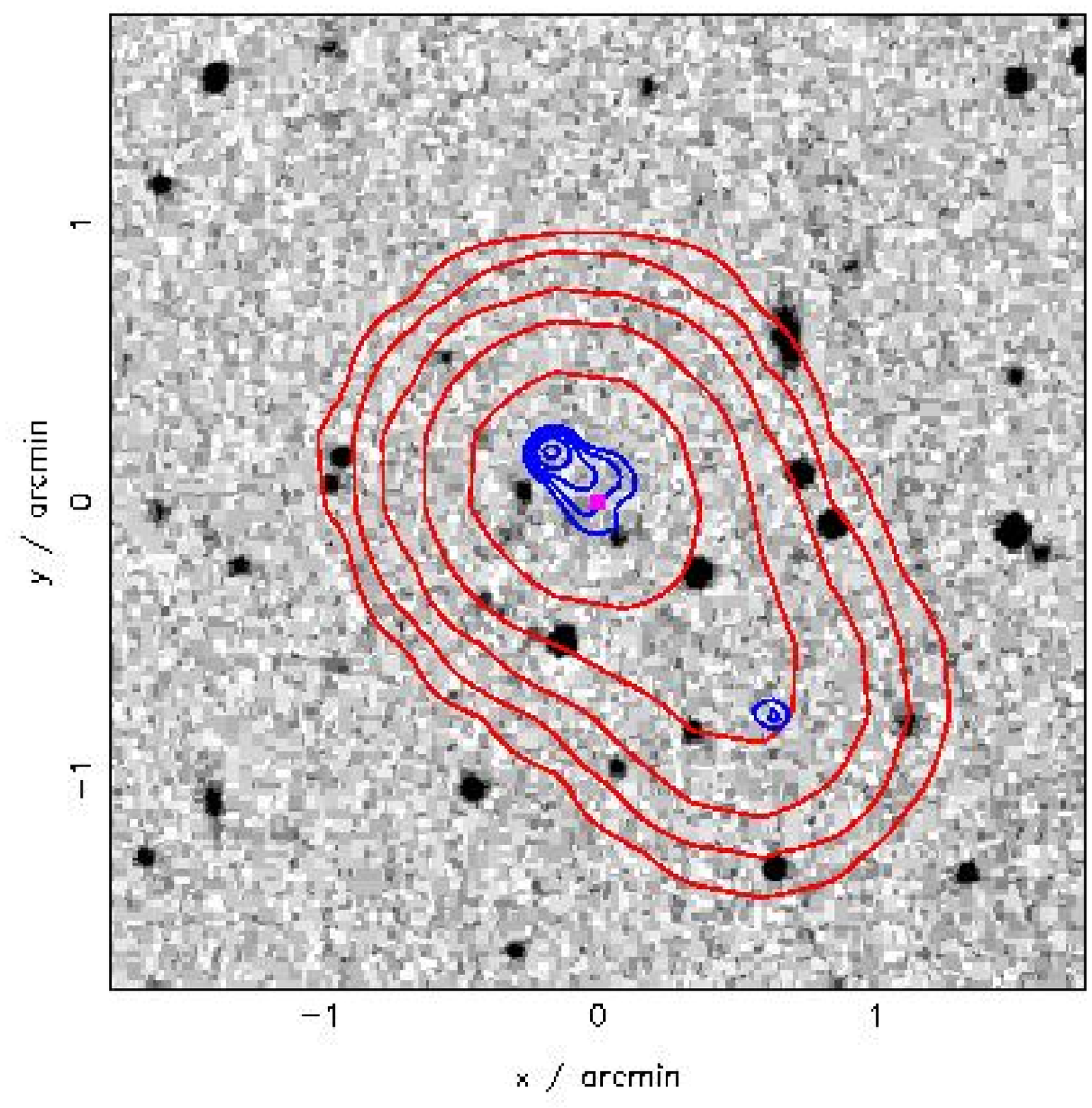}}
      \centerline{C3-199: TXS 1529+110}
    \end{minipage}
    \vfill
    \begin{minipage}{3cm}      
      \mbox{}
      \centerline{\includegraphics[scale=0.26,angle=270]{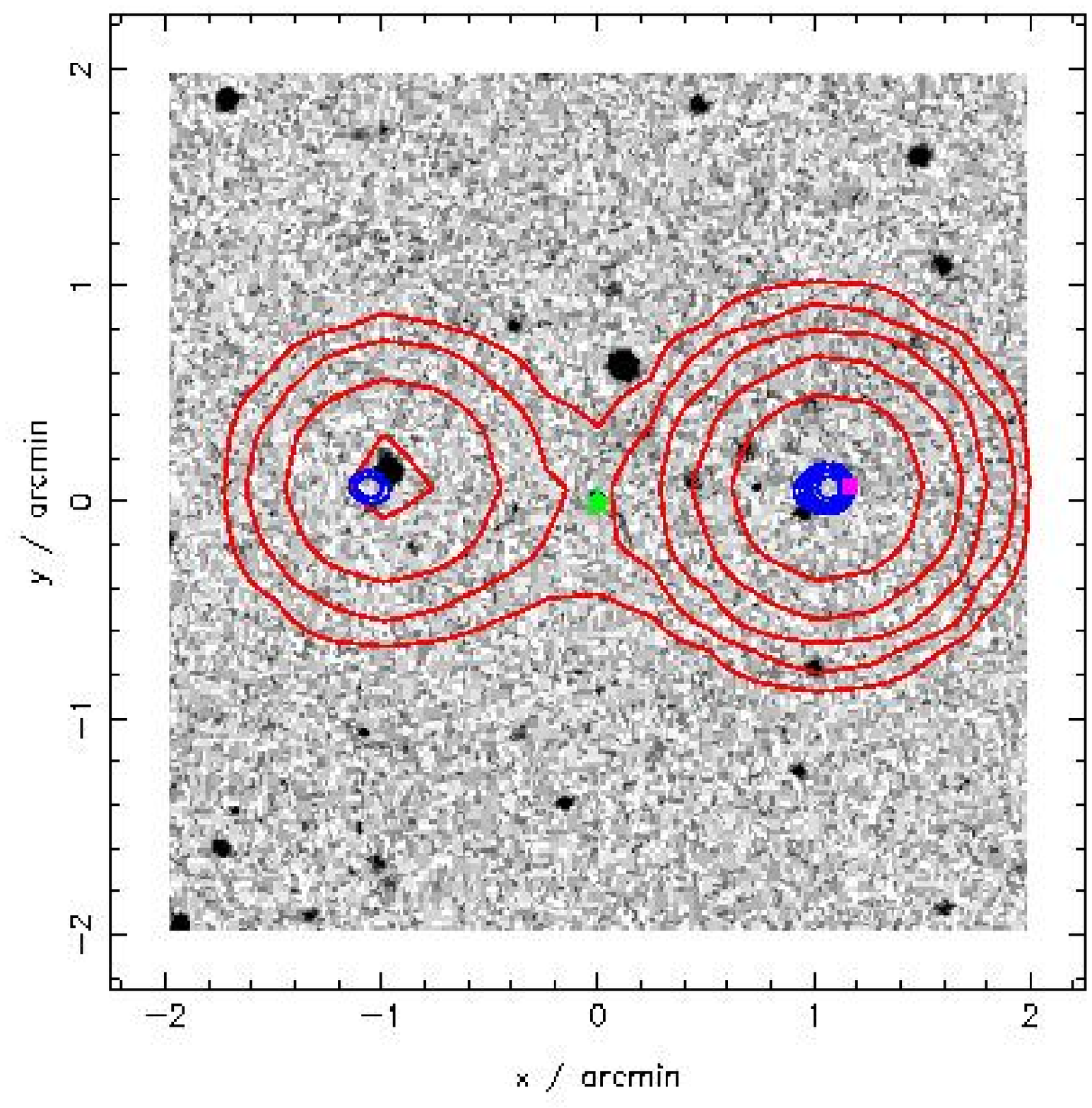}}
      \centerline{C3-201: J153233.19}
    \end{minipage}
    \hspace{3cm}
    \begin{minipage}{3cm}
      \mbox{}
      \centerline{\includegraphics[scale=0.26,angle=270]{Contours/C3/202.ps}}
      \centerline{C3-202: 4C 20.36}
    \end{minipage}
    \hspace{3cm}
    \begin{minipage}{3cm}
      \mbox{}
      \centerline{\includegraphics[scale=0.26,angle=270]{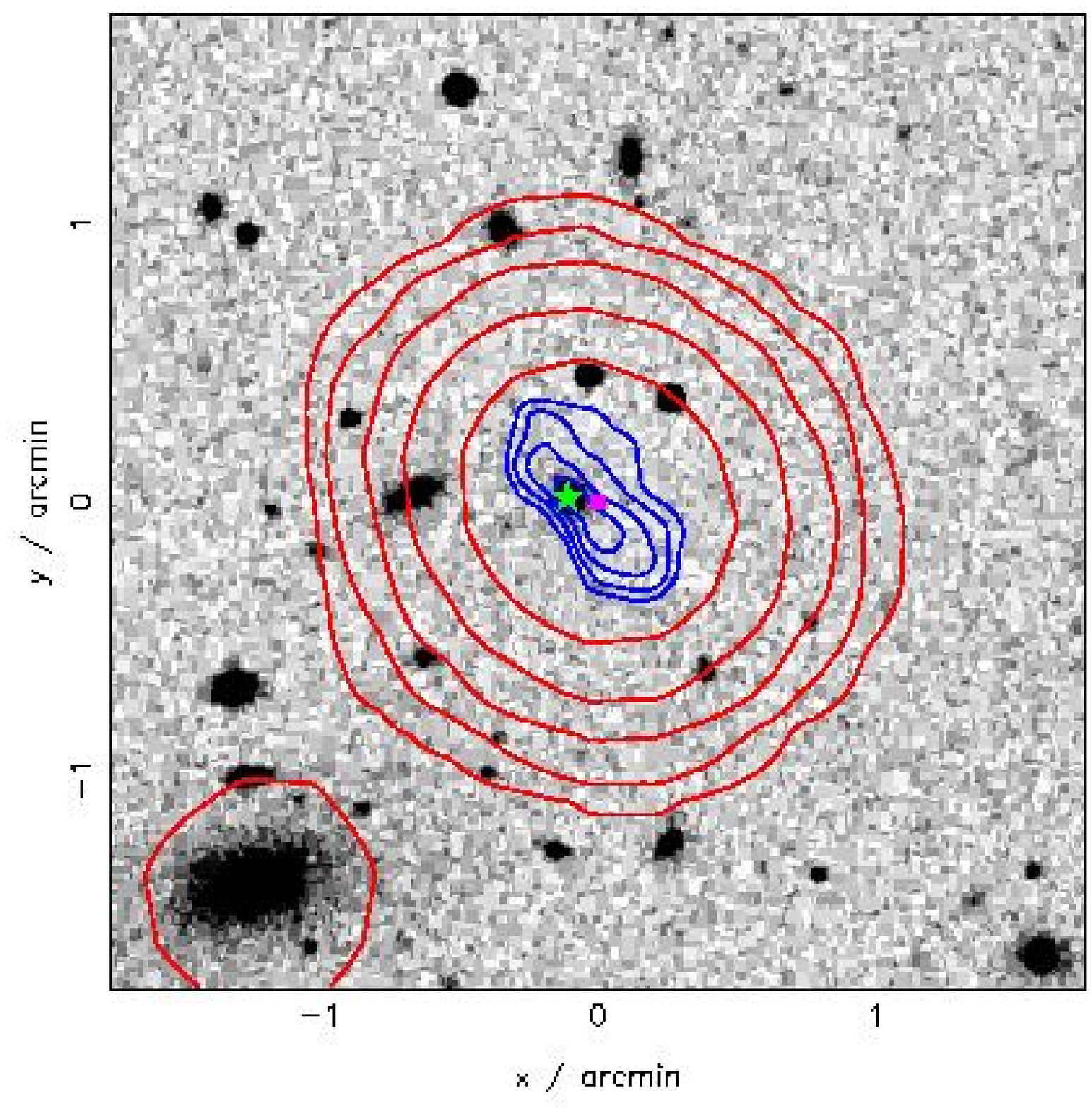}}
      \centerline{C3-203: B2 1530+28}
    \end{minipage}
  \end{center}
\end{figure}

\begin{figure}
  \begin{center}
    {\bf CoNFIG-3}\\  
  \begin{minipage}{3cm}      
      \mbox{}
      \centerline{\includegraphics[scale=0.26,angle=270]{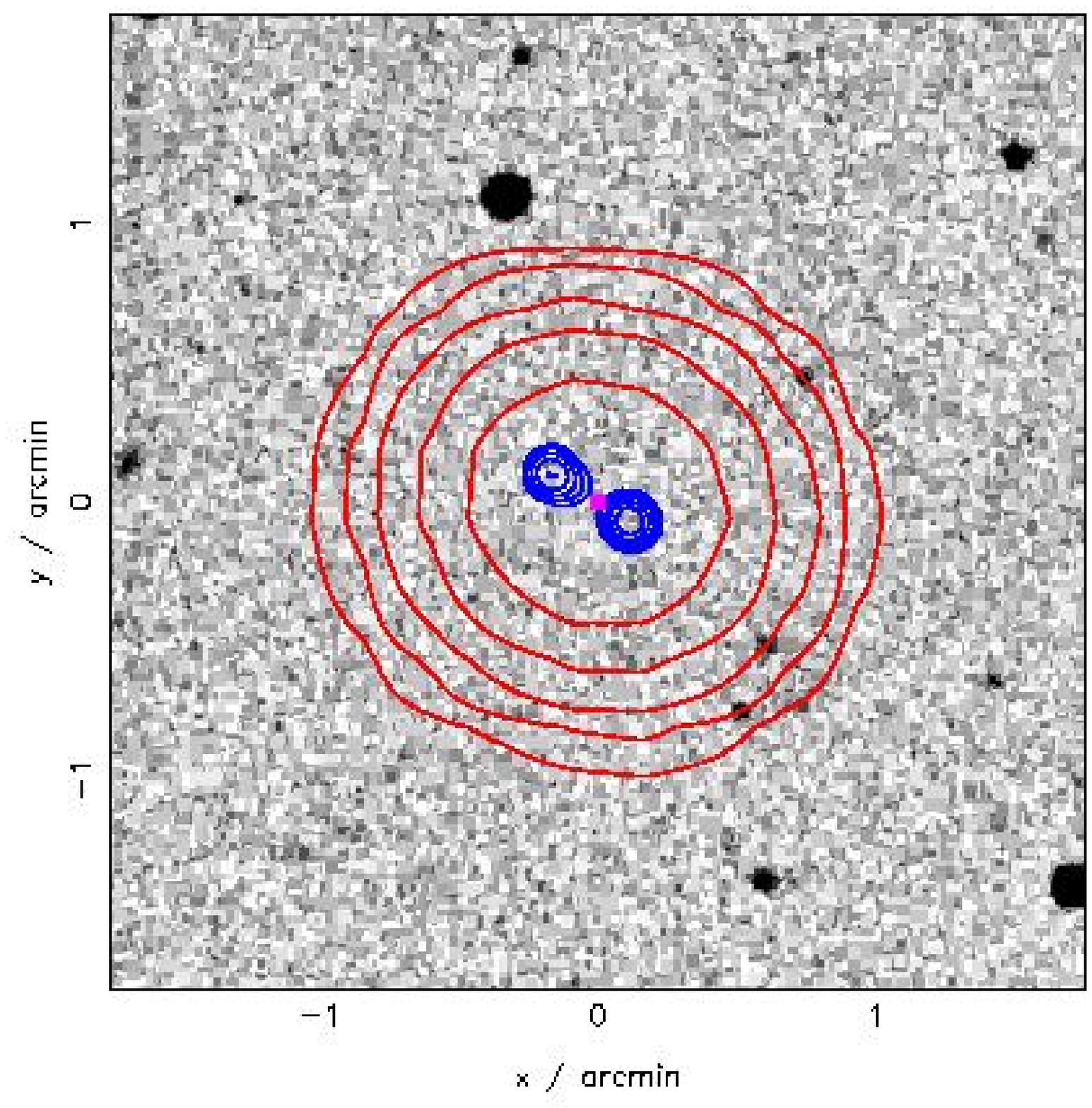}}
      \centerline{C3-205: TXS 1530+161}
    \end{minipage}
    \hspace{3cm}
    \begin{minipage}{3cm}
      \mbox{}
      \centerline{\includegraphics[scale=0.26,angle=270]{Contours/C3/207.ps}}
      \centerline{C3-207: 4C 13.55}
    \end{minipage}
    \hspace{3cm}
    \begin{minipage}{3cm}
      \mbox{}
      \centerline{\includegraphics[scale=0.26,angle=270]{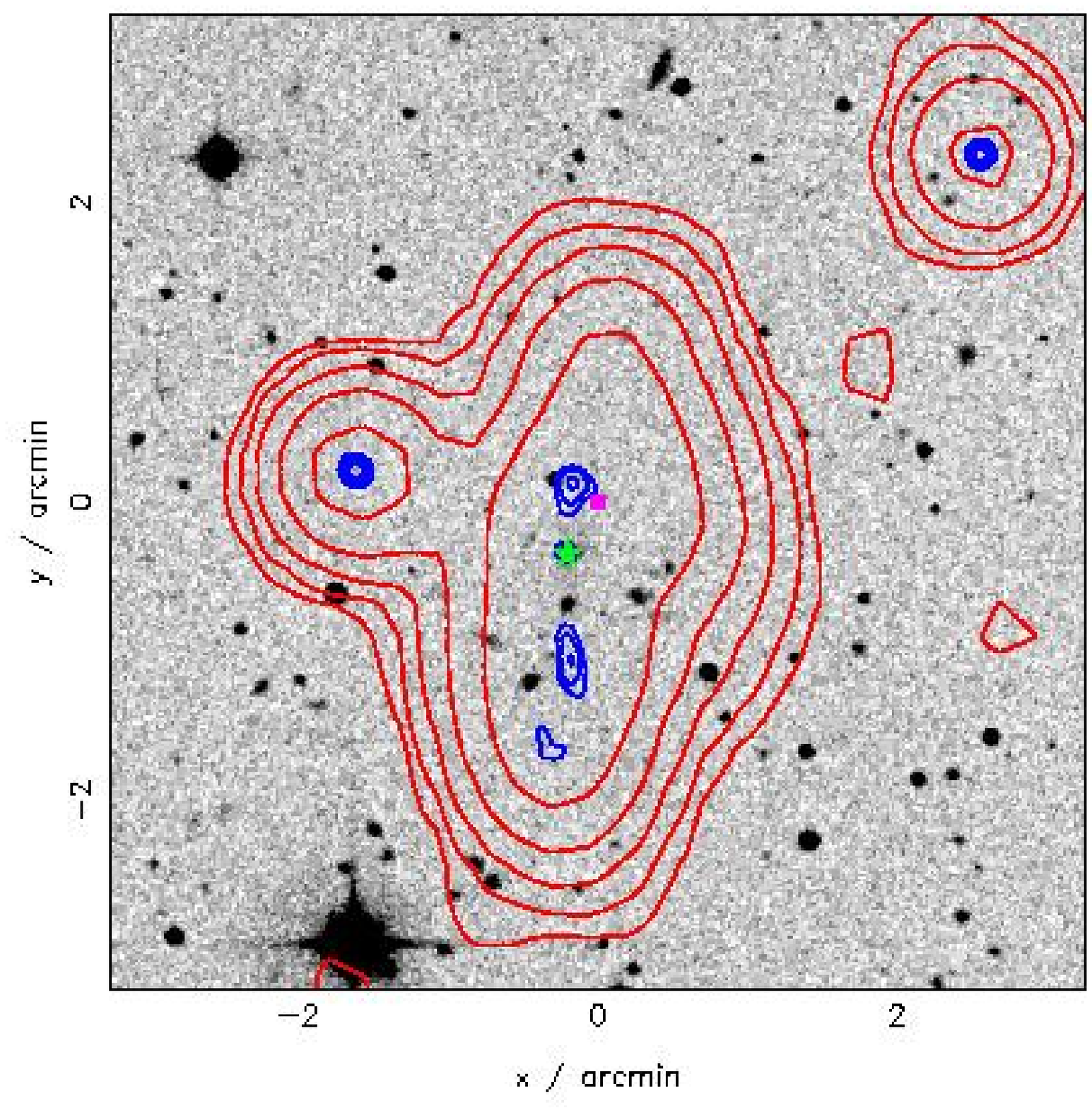}}
      \centerline{C3-208: Cul 1531+104}
    \end{minipage}
    \vfill
    \begin{minipage}{3cm}     
      \mbox{}
      \centerline{\includegraphics[scale=0.26,angle=270]{Contours/C3/209.ps}}
      \centerline{C3-209: TXS 1532+139}
    \end{minipage}
    \hspace{3cm}
    \begin{minipage}{3cm}
      \mbox{}
      \centerline{\includegraphics[scale=0.26,angle=270]{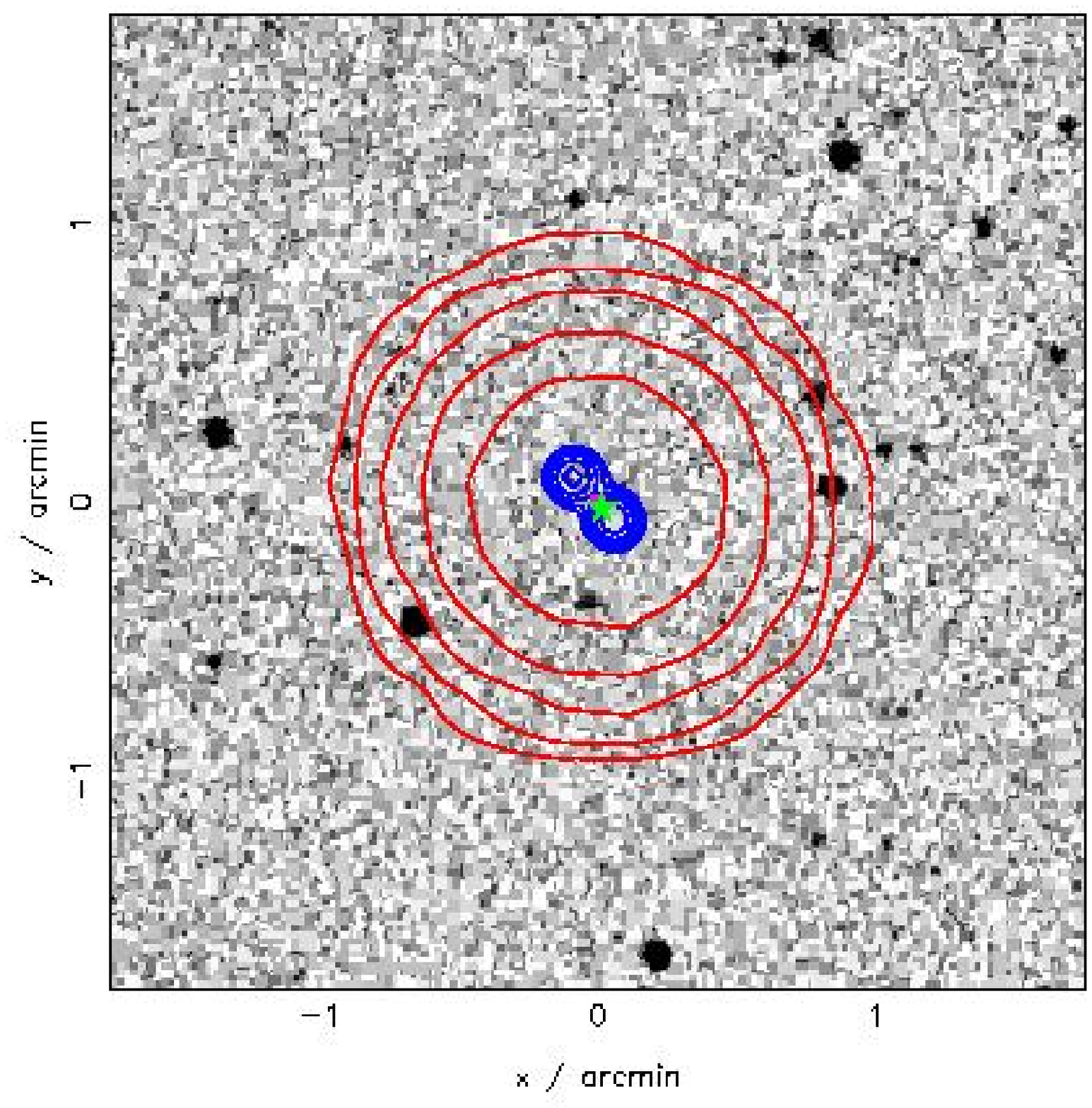}}
      \centerline{C3-212: TXS 1533+280}
    \end{minipage}
    \hspace{3cm}
    \begin{minipage}{3cm}
      \mbox{}
      \centerline{\includegraphics[scale=0.26,angle=270]{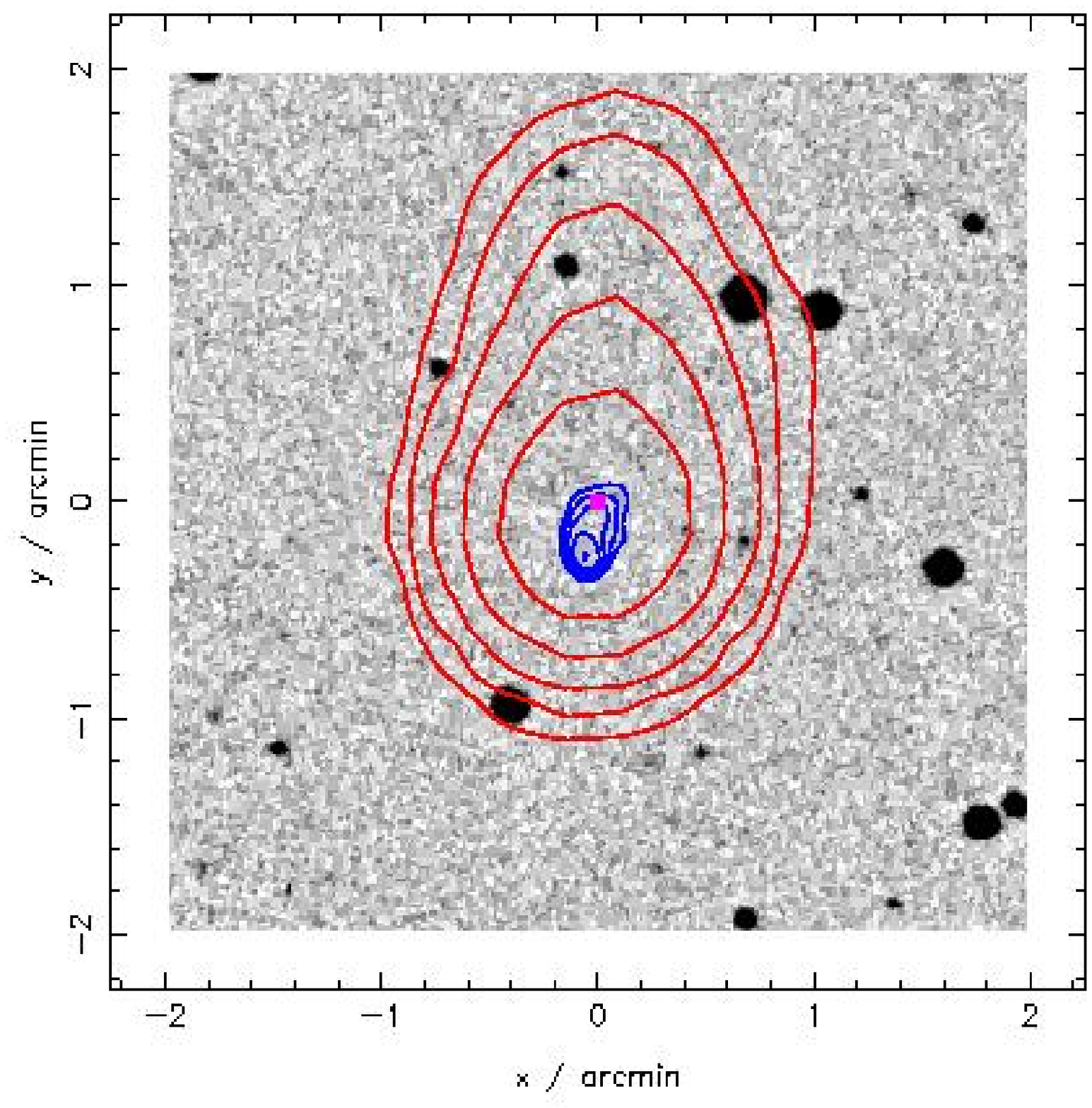}}
      \centerline{C3-213: TXS 1533+142}
    \end{minipage}
    \vfill
    \begin{minipage}{3cm}     
      \mbox{}
      \centerline{\includegraphics[scale=0.26,angle=270]{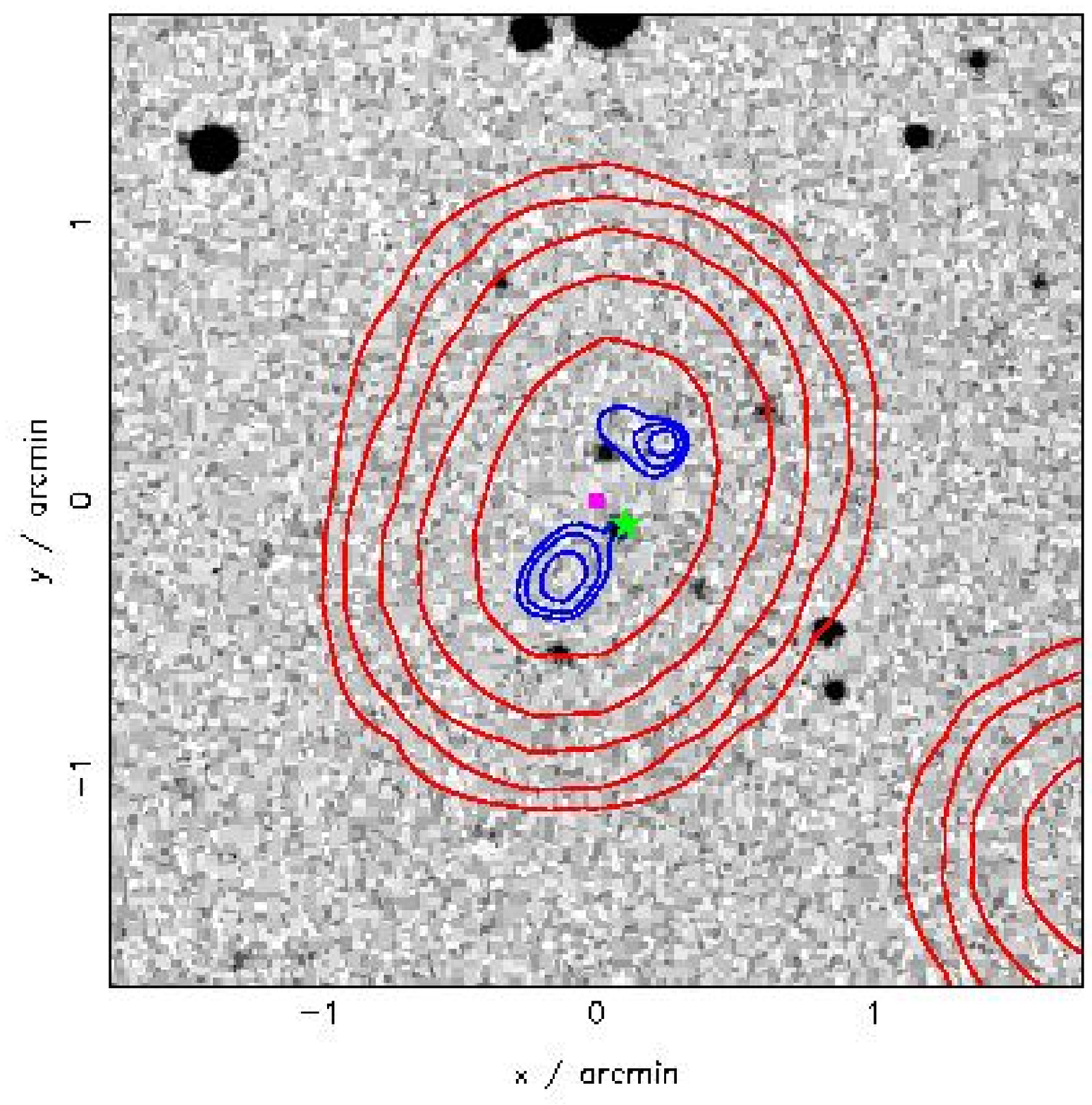}}
      \centerline{C3-216: TXS 1534+269}
    \end{minipage}
    \hspace{3cm}
    \begin{minipage}{3cm}
      \mbox{}
      \centerline{\includegraphics[scale=0.26,angle=270]{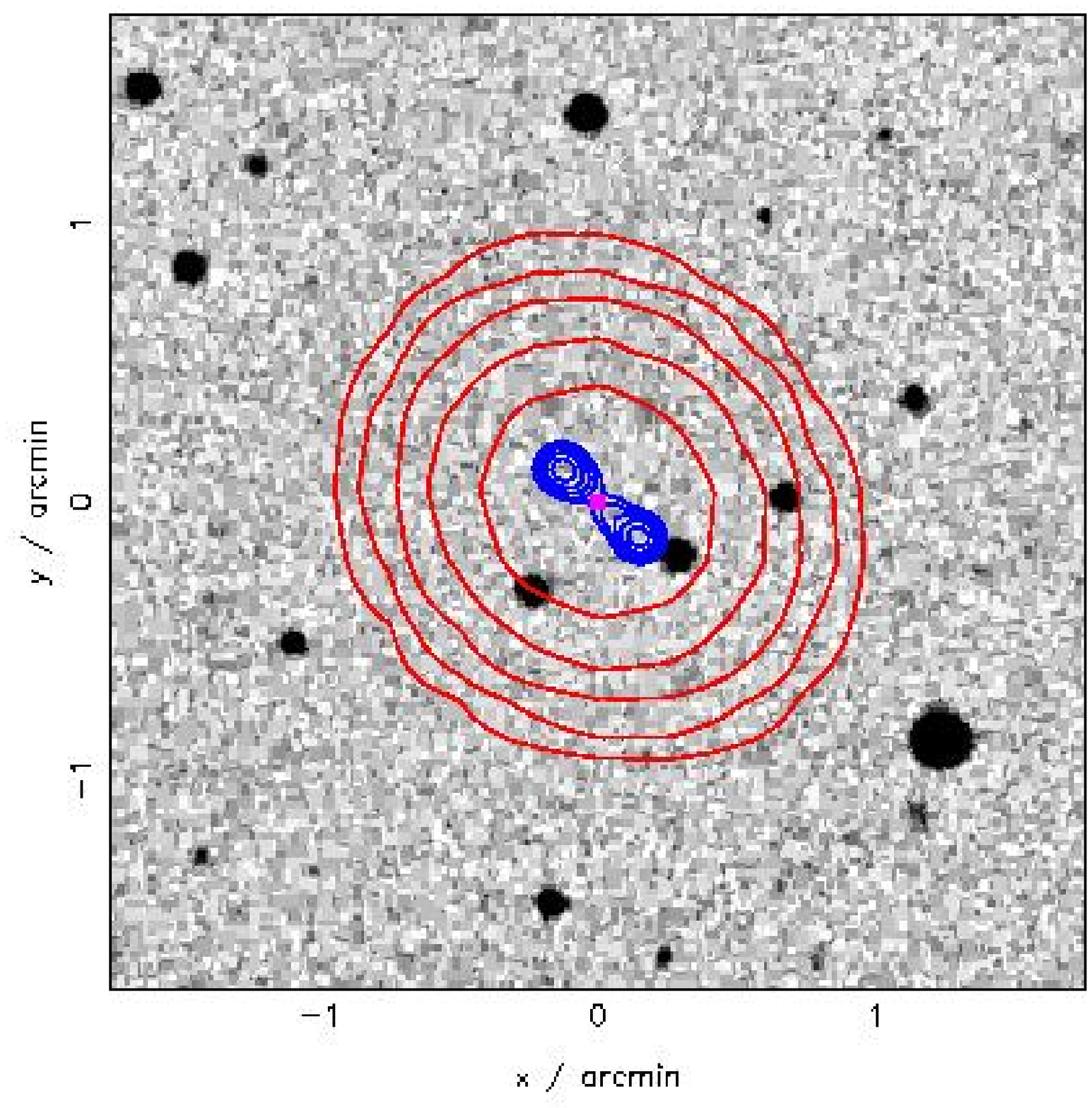}}
      \centerline{C3-217: TXS 1534+145}
    \end{minipage}
    \hspace{3cm}
    \begin{minipage}{3cm}
      \mbox{}
      \centerline{\includegraphics[scale=0.26,angle=270]{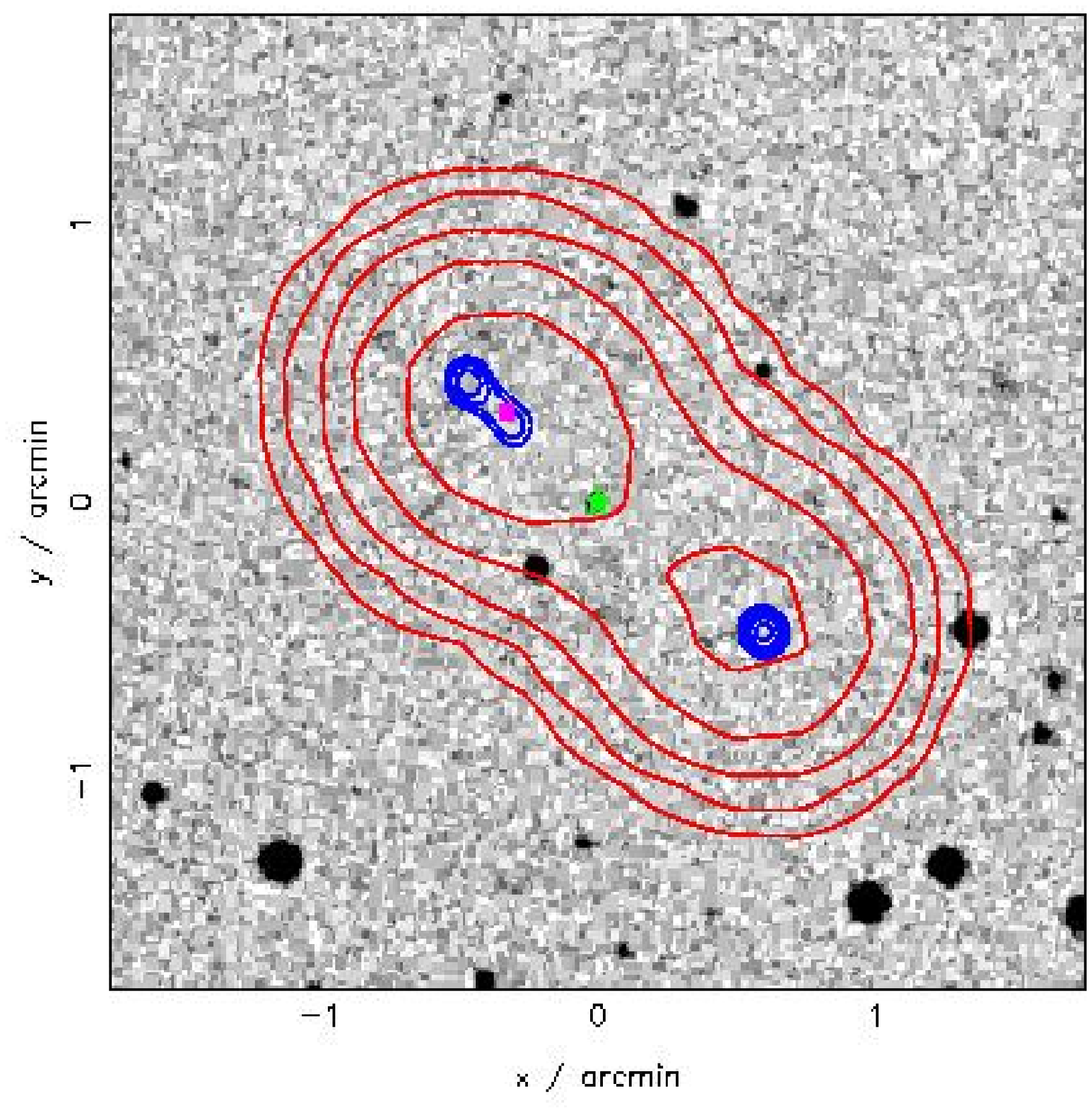}}
      \centerline{C3-221: TXS 1536+144}
    \end{minipage}
    \vfill
    \begin{minipage}{3cm}      
      \mbox{}
      \centerline{\includegraphics[scale=0.26,angle=270]{Contours/C3/223.ps}}
      \centerline{C3-223: TXS 1537+145}
    \end{minipage}
    \hspace{3cm}
    \begin{minipage}{3cm}
      \mbox{}
      \centerline{\includegraphics[scale=0.26,angle=270]{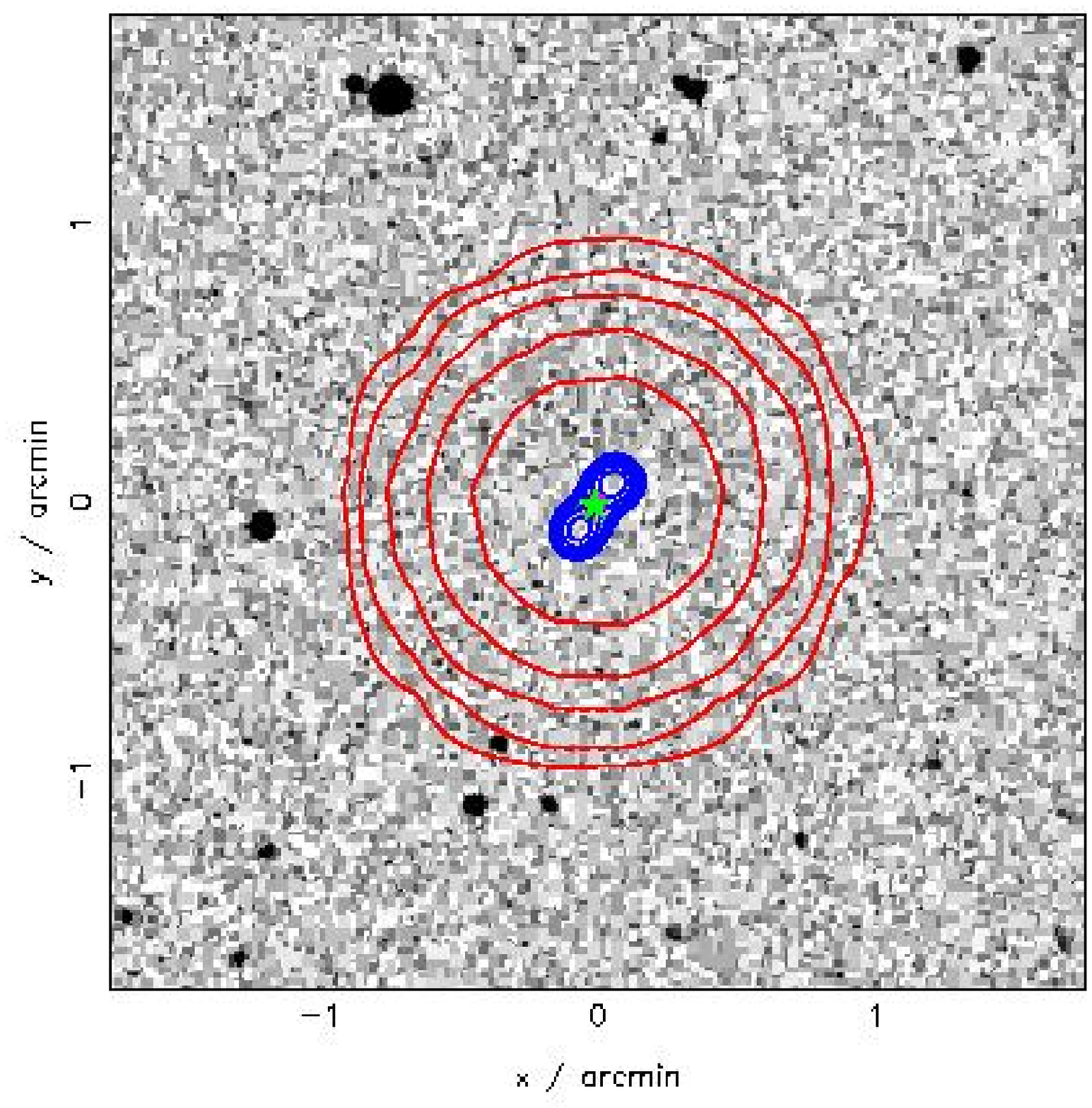}}
      \centerline{C3-225: TXS 1538+182}
    \end{minipage}
    \hspace{3cm}
    \begin{minipage}{3cm}
      \mbox{}
      \centerline{\includegraphics[scale=0.26,angle=270]{Contours/C3/227.ps}}
      \centerline{C3-227: 4C 18.43}
    \end{minipage}
  \end{center}
\end{figure}

\begin{figure}
  \begin{center}
    {\bf CoNFIG-3}\\  
  \begin{minipage}{3cm}      
      \mbox{}
      \centerline{\includegraphics[scale=0.26,angle=270]{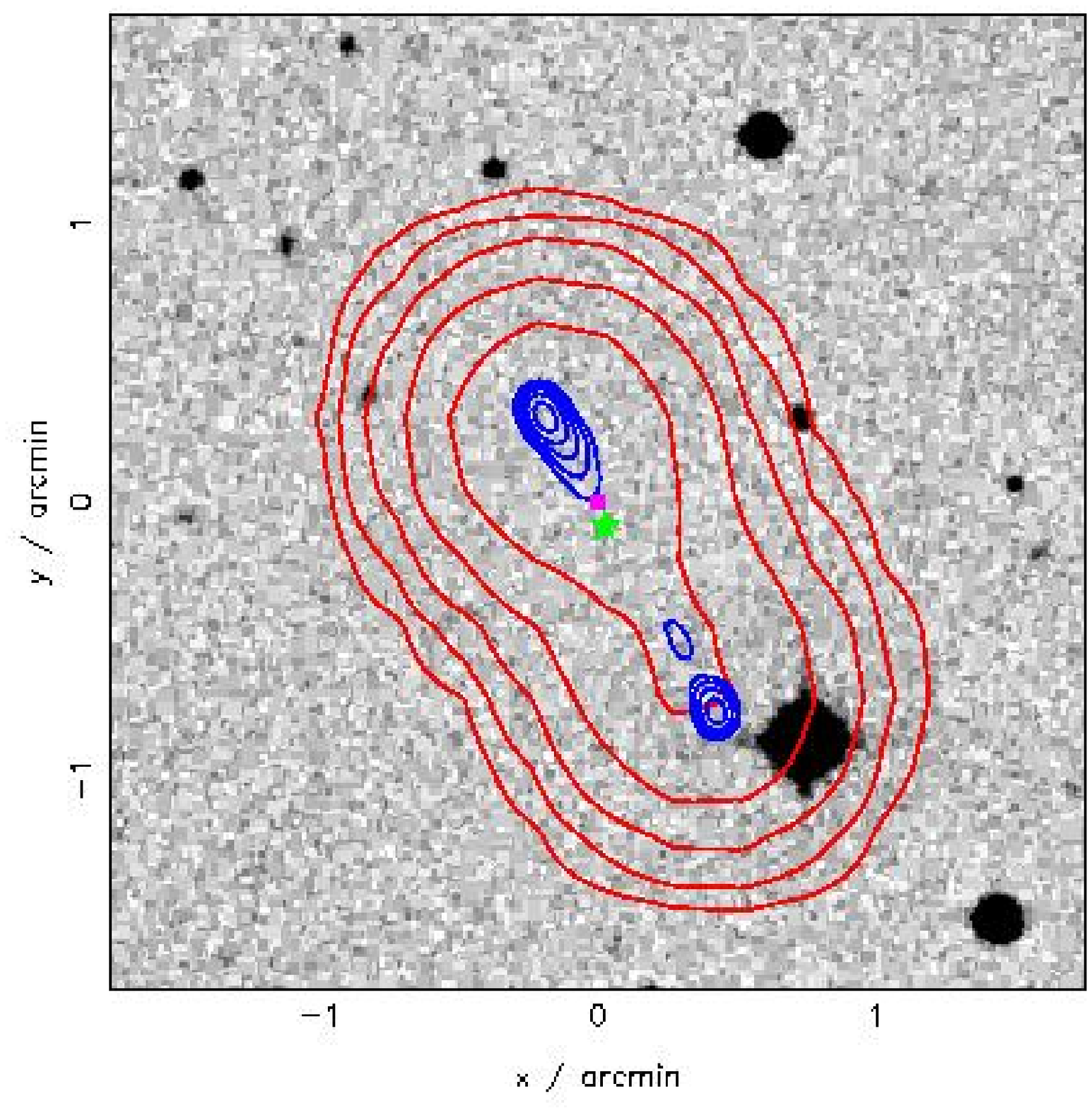}}
      \centerline{C3-228: TXS 1540+241}
    \end{minipage}
    \hspace{3cm}
    \begin{minipage}{3cm}
      \mbox{}
      \centerline{\includegraphics[scale=0.26,angle=270]{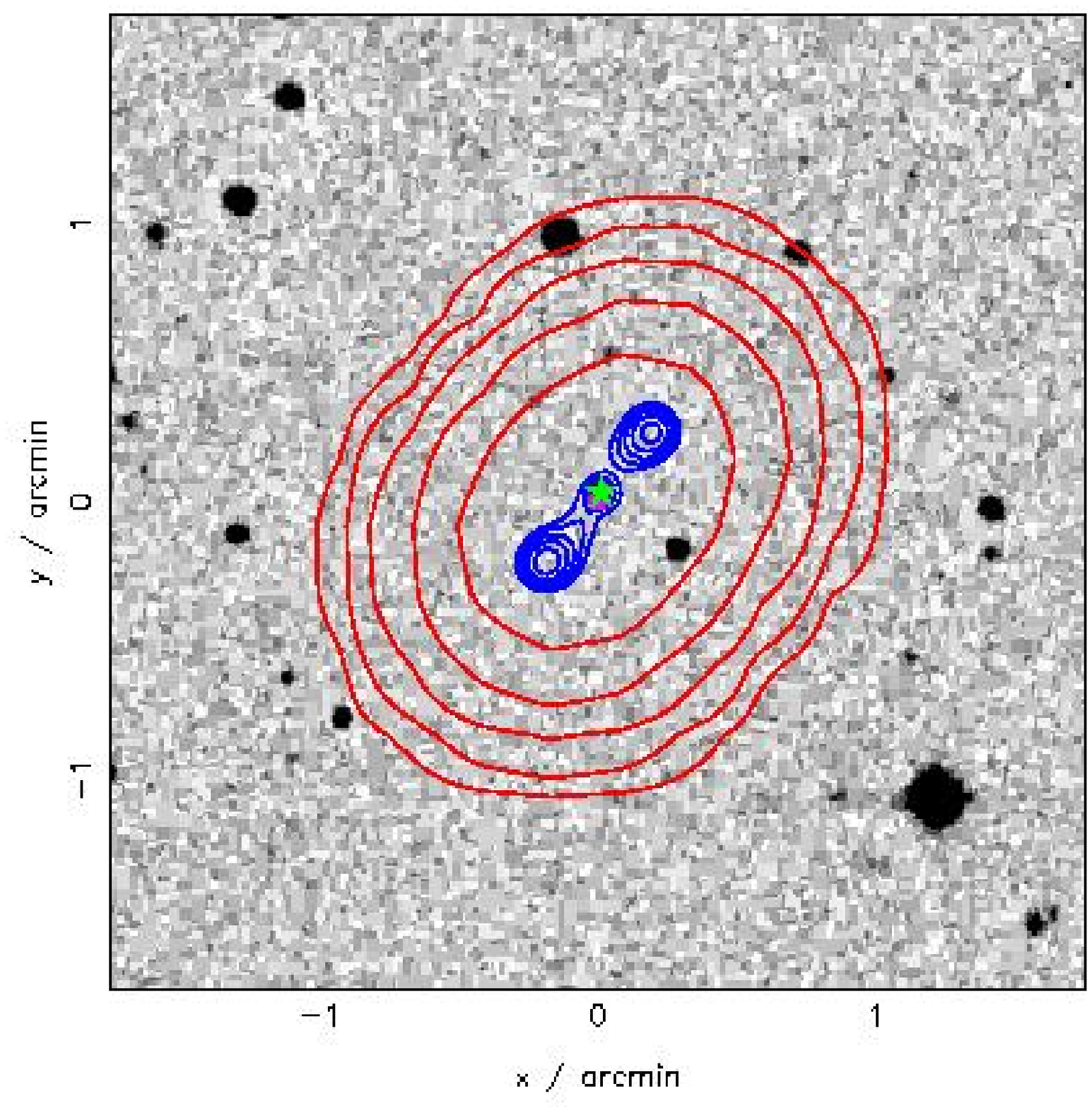}}
      \centerline{C3-229: GB6 B1540+11}
    \end{minipage}
    \hspace{3cm}
    \begin{minipage}{3cm}
      \mbox{}
      \centerline{\includegraphics[scale=0.26,angle=270]{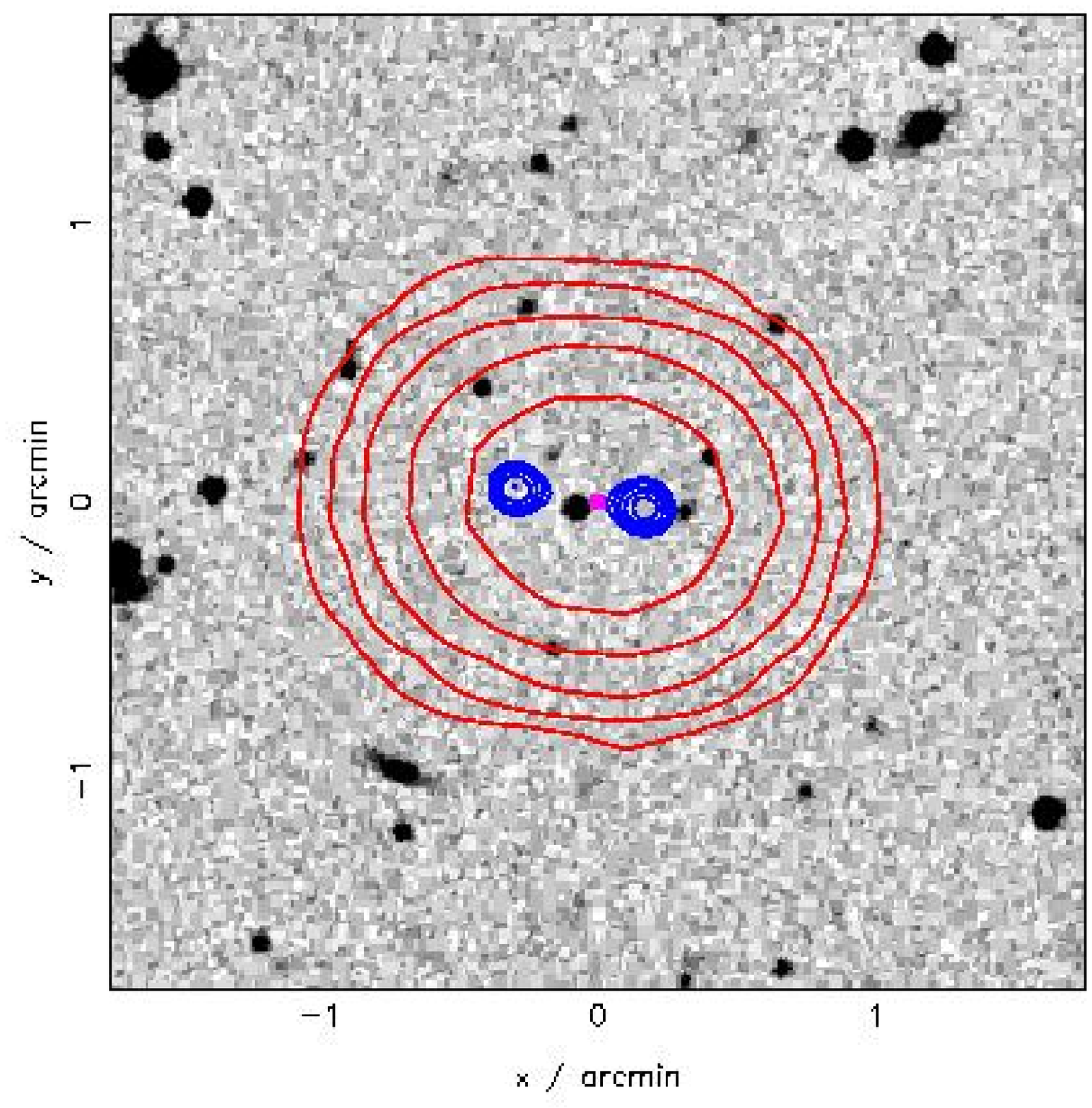}}
      \centerline{C3-230: TXS 1541+219}
    \end{minipage}
    \vfill
    \begin{minipage}{3cm}     
      \mbox{}
      \centerline{\includegraphics[scale=0.26,angle=270]{Contours/C3/231.ps}}
      \centerline{C3-231: TXS 1541+230}
    \end{minipage}
    \hspace{3cm}
    \begin{minipage}{3cm}
      \mbox{}
      \centerline{\includegraphics[scale=0.26,angle=270]{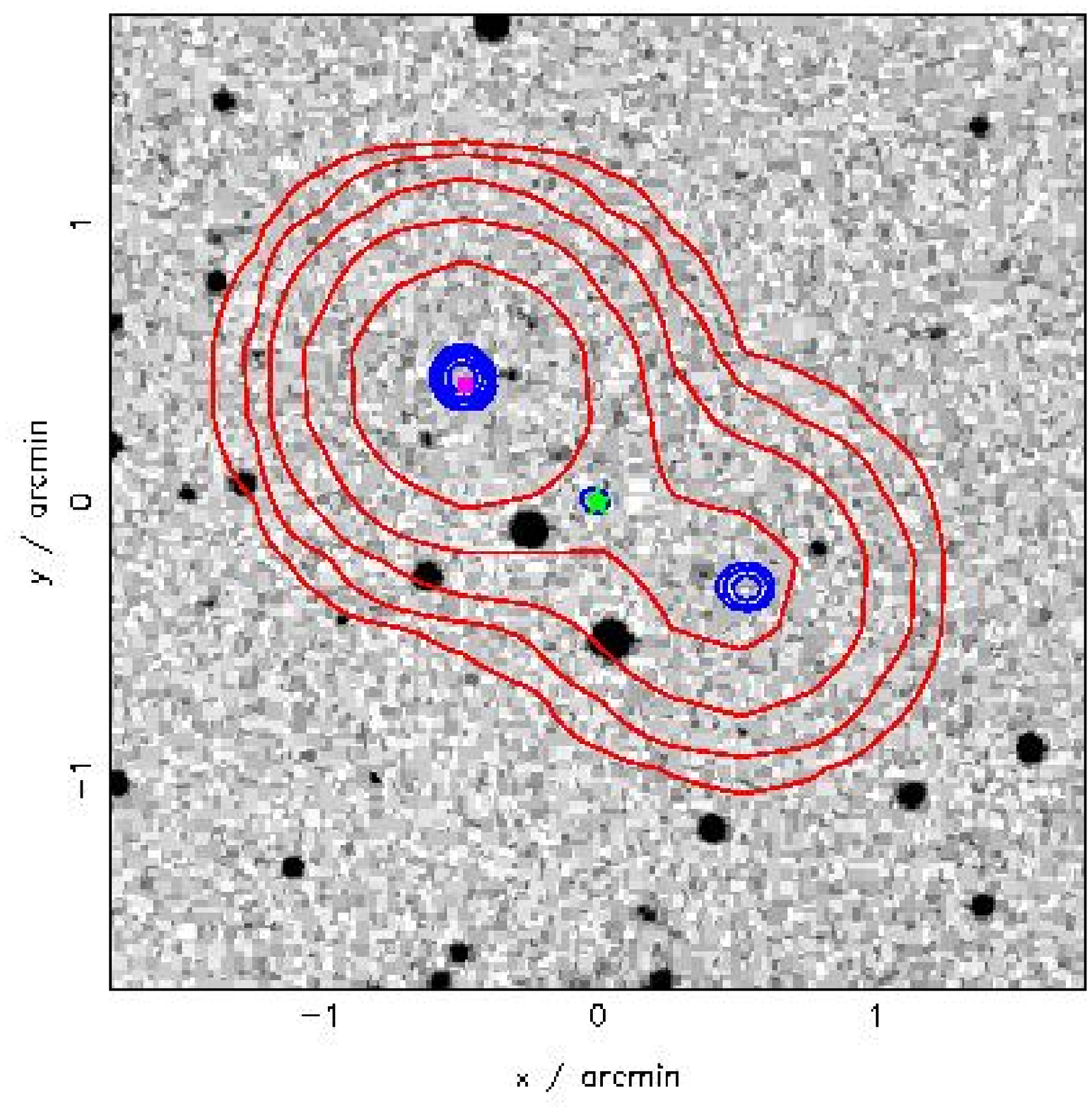}}
      \centerline{C3-232: TXS 1541+136}
    \end{minipage}
    \hspace{3cm}
    \begin{minipage}{3cm}
      \mbox{}
      \centerline{\includegraphics[scale=0.26,angle=270]{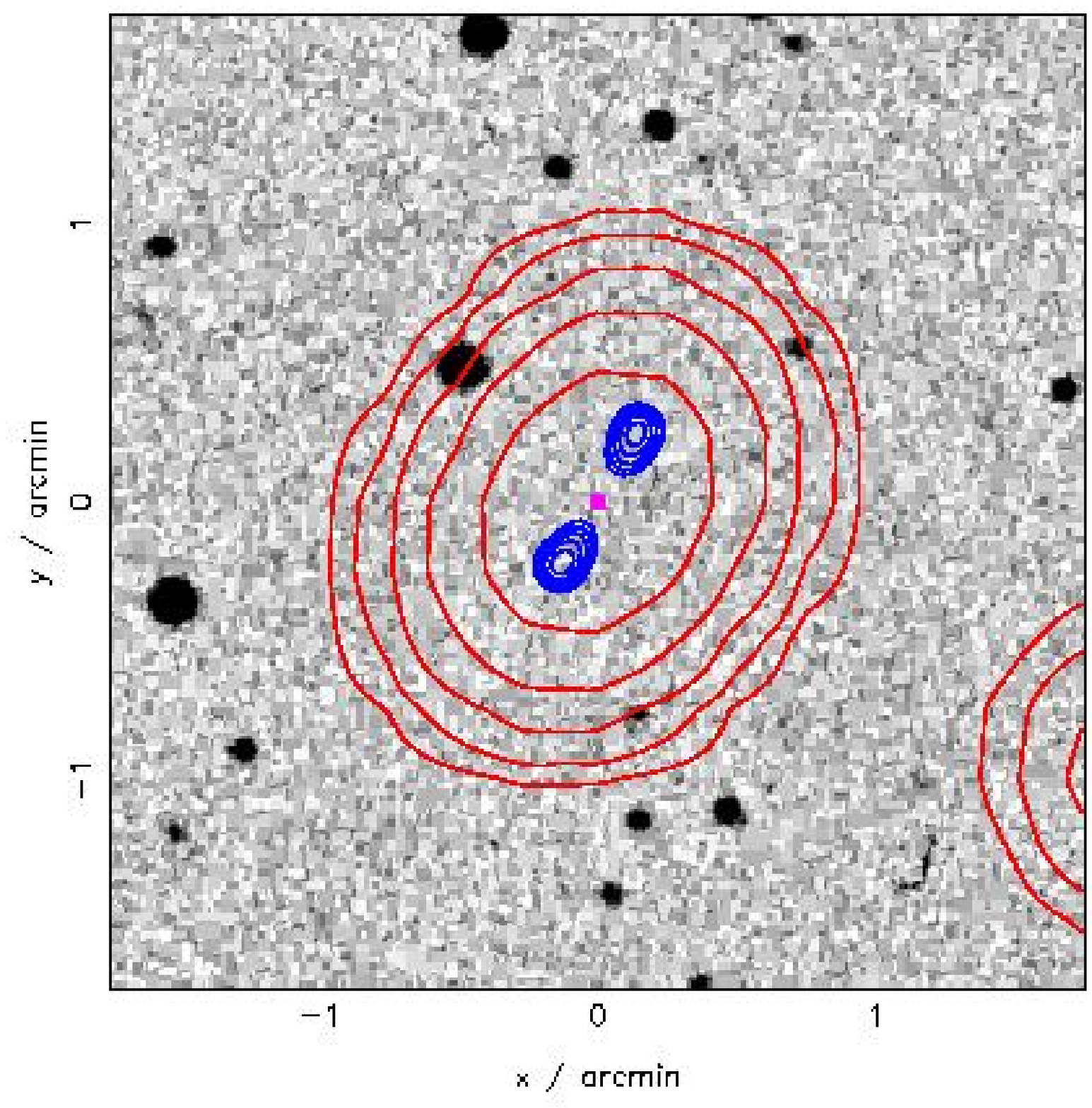}}
      \centerline{C3-234: TXS 1541+143}
    \end{minipage}
    \vfill
    \begin{minipage}{3cm}     
      \mbox{}
      \centerline{\includegraphics[scale=0.26,angle=270]{Contours/C3/236.ps}}
      \centerline{C3-236: 4C 19.51}
    \end{minipage}
    \hspace{3cm}
    \begin{minipage}{3cm}
      \mbox{}
      \centerline{\includegraphics[scale=0.26,angle=270]{Contours/C3/238.ps}}
      \centerline{C3-238: TXS 1543+180}
    \end{minipage}
    \hspace{3cm}
    \begin{minipage}{3cm}
      \mbox{}
      \centerline{\includegraphics[scale=0.26,angle=270]{Contours/C3/239.ps}}
      \centerline{C3-239: TXS 1544+279}
    \end{minipage}
    \vfill
    \begin{minipage}{3cm}      
      \mbox{}
      \centerline{\includegraphics[scale=0.26,angle=270]{Contours/C3/240.ps}}
      \centerline{C3-240: TXS 1544+221}
    \end{minipage}
    \hspace{3cm}
    \begin{minipage}{3cm}
      \mbox{}
      \centerline{\includegraphics[scale=0.26,angle=270]{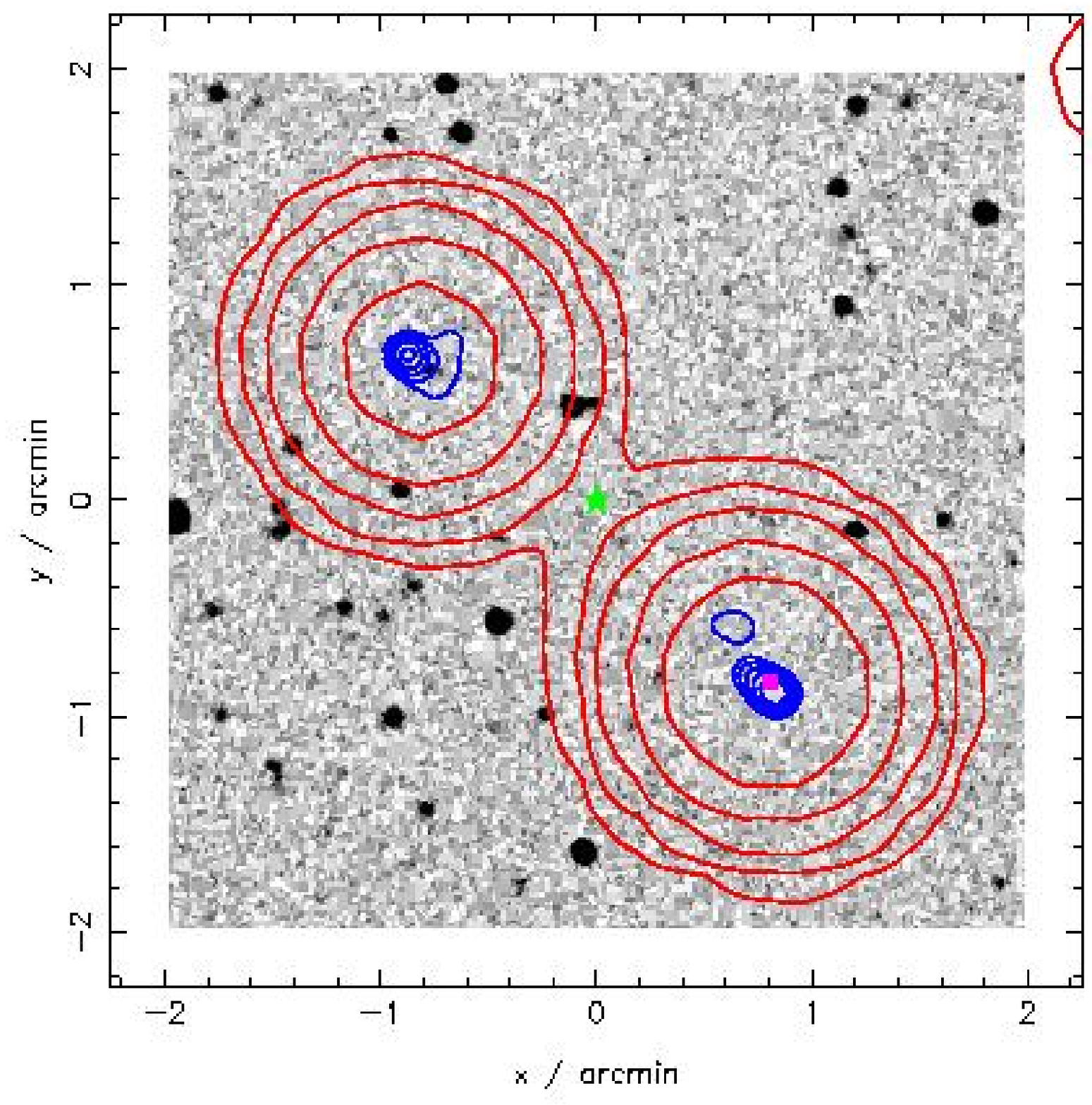}}
      \centerline{C3-241: GB6 1544+1152}
    \end{minipage}
    \hspace{3cm}
    \begin{minipage}{3cm}
      \mbox{}
      \centerline{\includegraphics[scale=0.26,angle=270]{Contours/C3/242.ps}}
      \centerline{C3-242: 4C 18.44}
    \end{minipage}
  \end{center}
\end{figure}

\begin{figure}
  \begin{center}
    {\bf CoNFIG-3}\\  
  \begin{minipage}{3cm}      
      \mbox{}
      \centerline{\includegraphics[scale=0.26,angle=270]{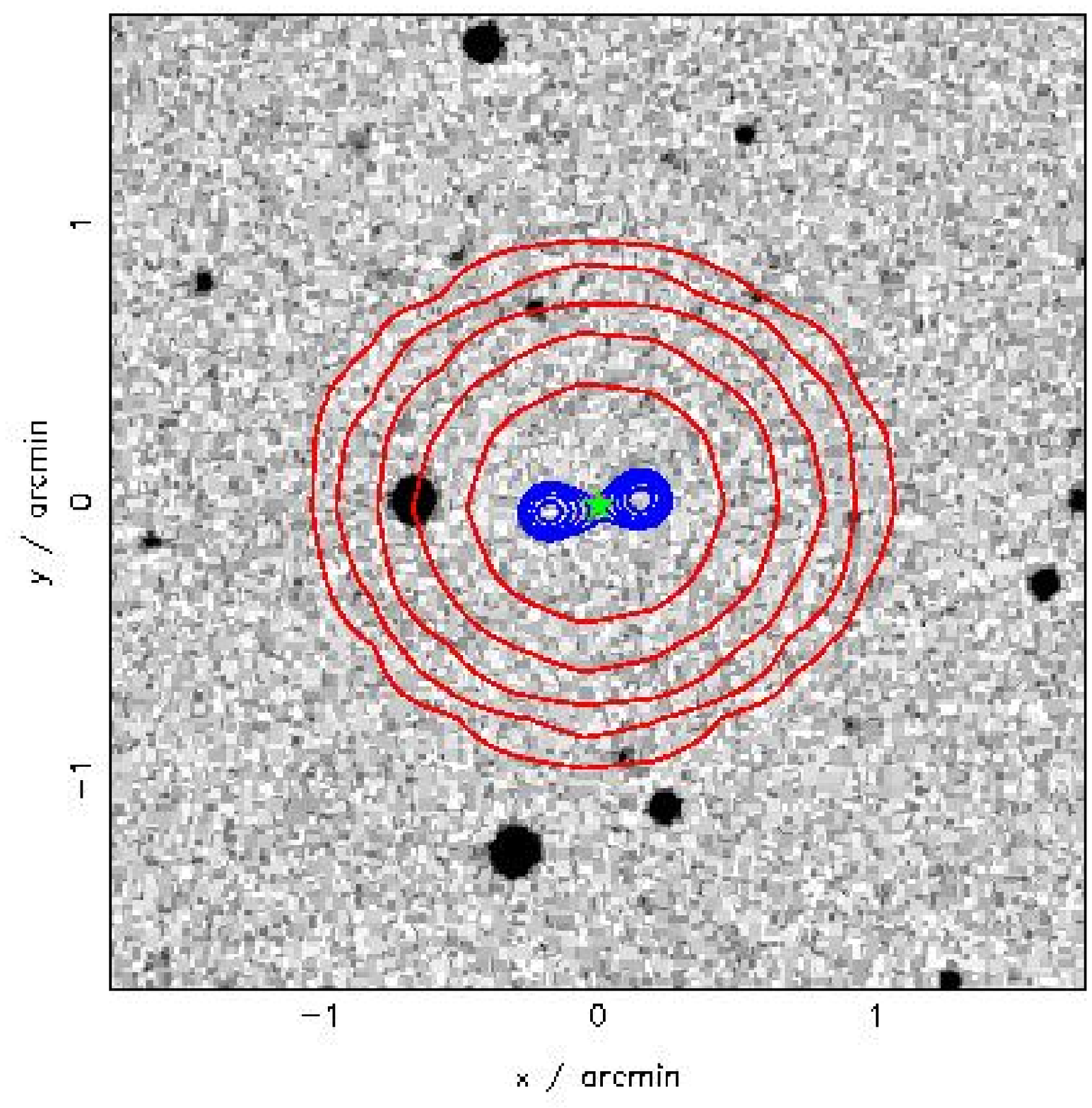}}
      \centerline{C3-243: TXS 1545+279}
    \end{minipage}
    \hspace{3cm}
    \begin{minipage}{3cm}
      \mbox{}
      \centerline{\includegraphics[scale=0.26,angle=270]{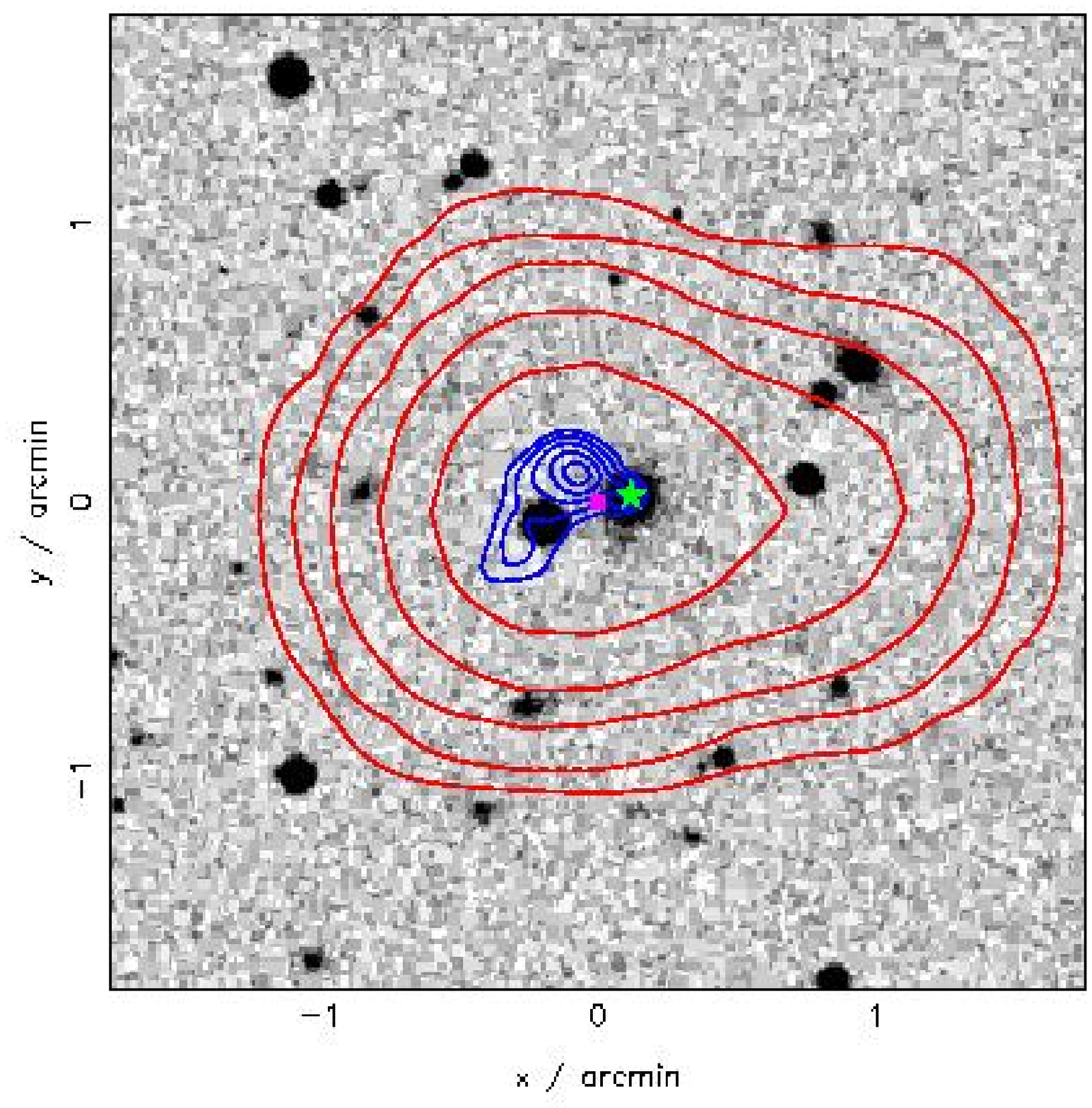}}
      \centerline{C3-244: BWE 1545+1505}
    \end{minipage}
    \hspace{3cm}
    \begin{minipage}{3cm}
      \mbox{}
      \centerline{\includegraphics[scale=0.26,angle=270]{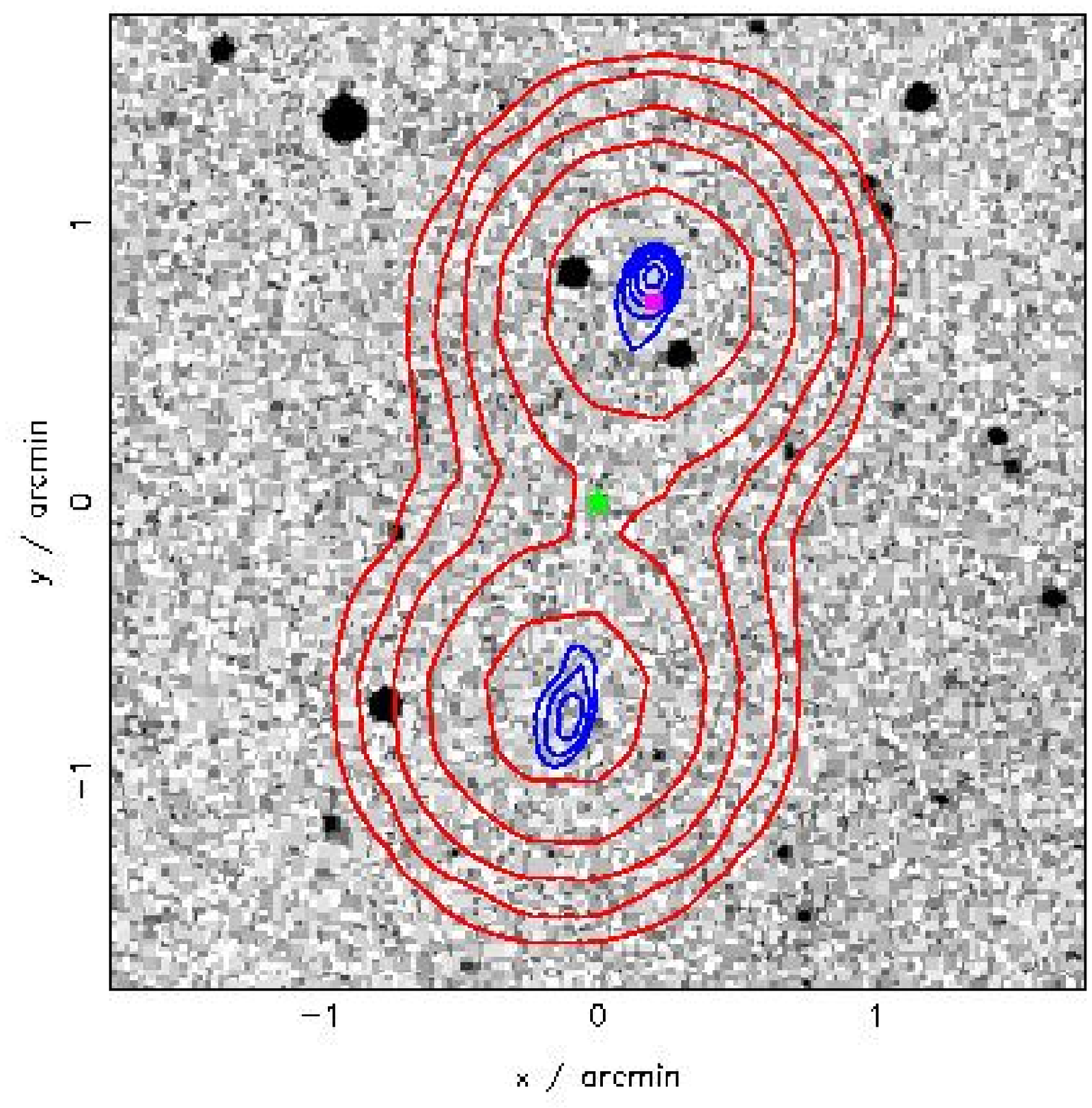}}
      \centerline{C3-246: TXS 1546+268}
    \end{minipage}
    \vfill
    \begin{minipage}{3cm}     
      \mbox{}
      \centerline{\includegraphics[scale=0.26,angle=270]{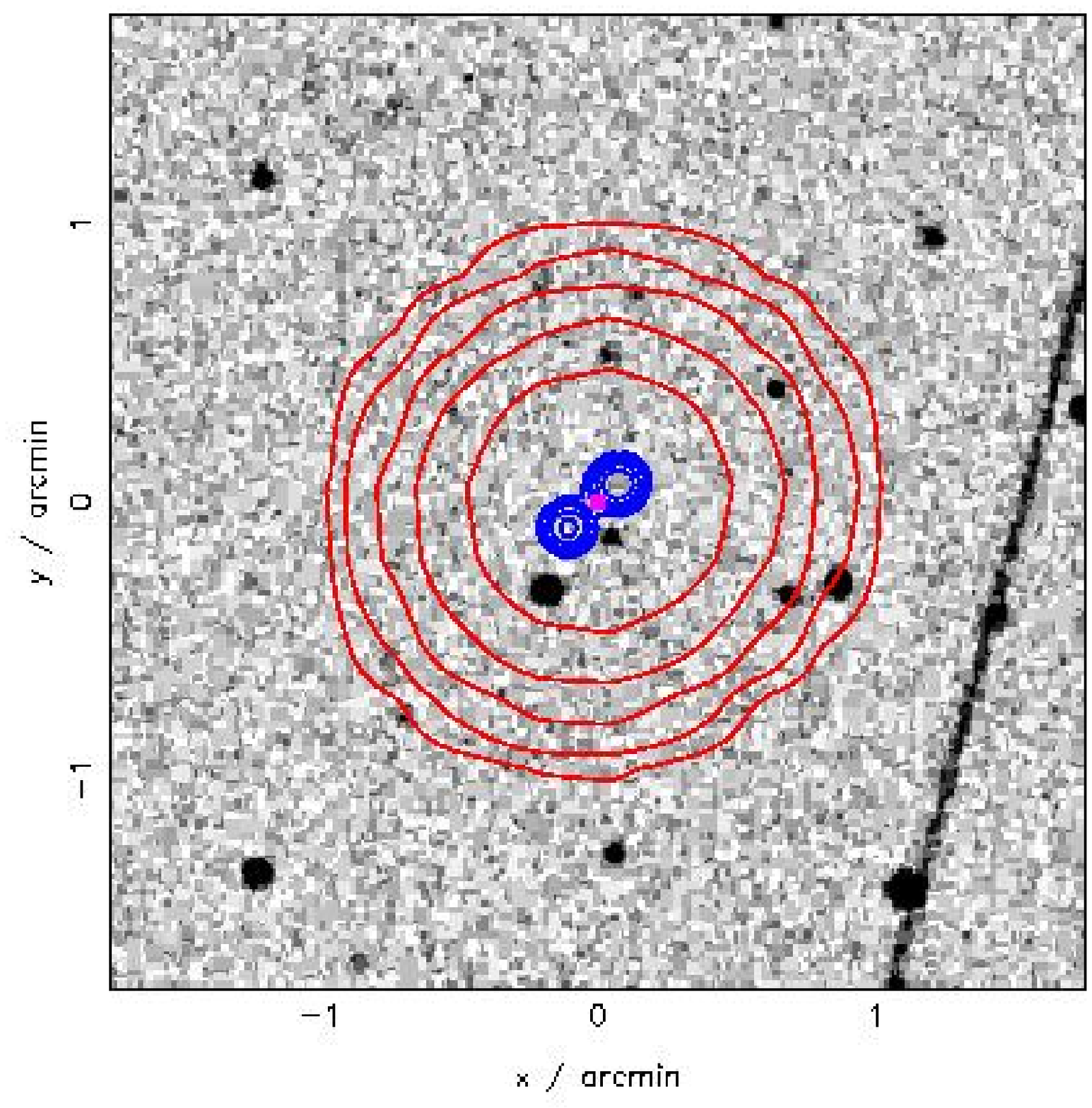}}
      \centerline{C3-247: 4C 15.51}
    \end{minipage}
    \hspace{3cm}
    \begin{minipage}{3cm}
      \mbox{}
      \centerline{\includegraphics[scale=0.26,angle=270]{Contours/C3/248.ps}}
      \centerline{C3-248: 4C 18.45}
    \end{minipage}
    \hspace{3cm}
    \begin{minipage}{3cm}
      \mbox{}
      \centerline{\includegraphics[scale=0.26,angle=270]{Contours/C3/250.ps}}
      \centerline{C3-250: TXS 1548+274}
    \end{minipage}
    \vfill
    \begin{minipage}{3cm}     
      \mbox{}
      \centerline{\includegraphics[scale=0.26,angle=270]{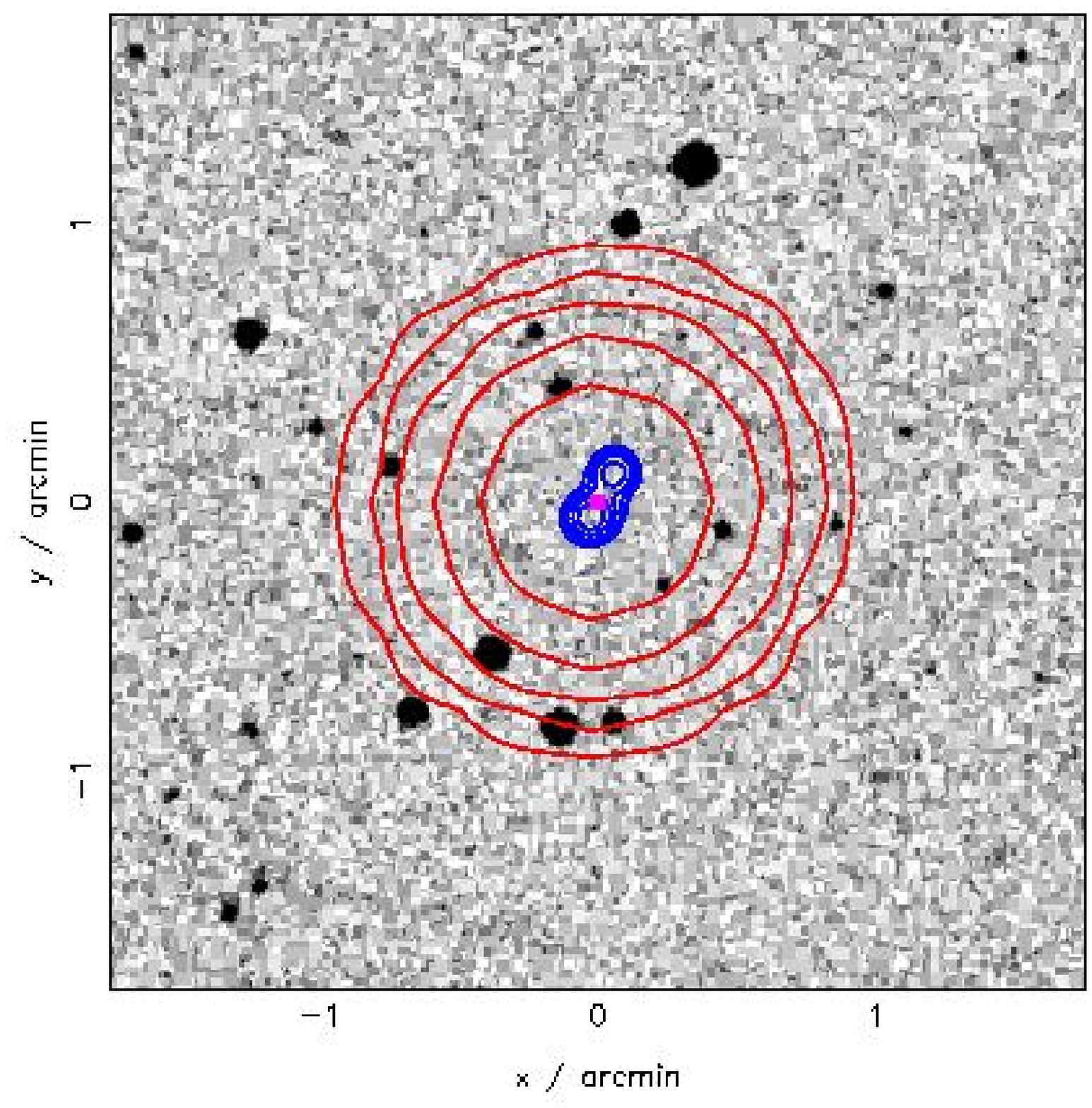}}
      \centerline{C3-251: TXS 1548+188}
    \end{minipage}
    \hspace{3cm}
    \begin{minipage}{3cm}
      \mbox{}
      \centerline{\includegraphics[scale=0.26,angle=270]{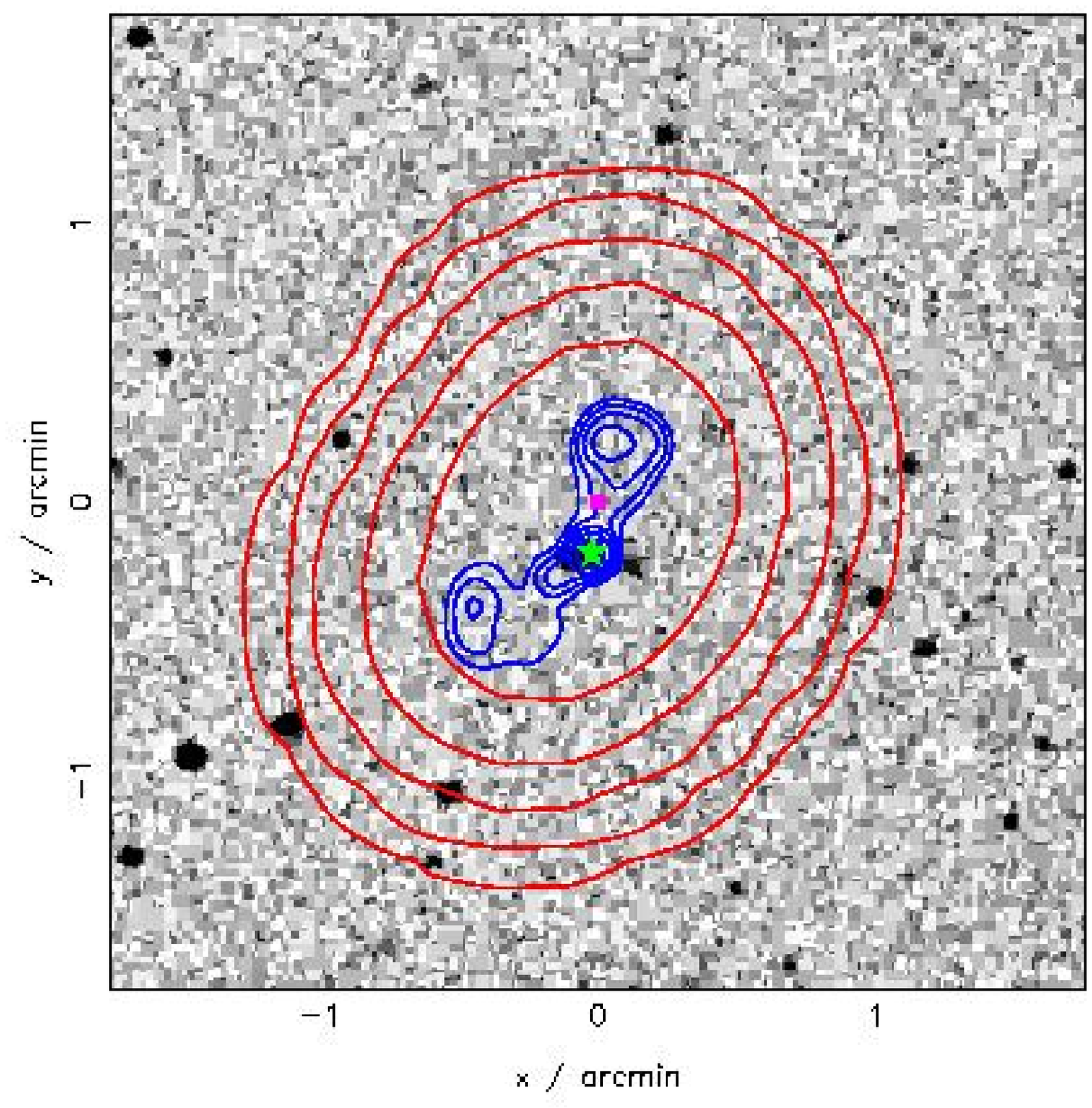}}
      \centerline{C3-252: 4C 11.50}
    \end{minipage}
    \hspace{3cm}
    \begin{minipage}{3cm}
      \mbox{}
      \centerline{\includegraphics[scale=0.26,angle=270]{Contours/C3/253.ps}}
      \centerline{C3-253: TXS 1549+262}
    \end{minipage}
    \vfill
    \begin{minipage}{3cm}      
      \mbox{}
      \centerline{\includegraphics[scale=0.26,angle=270]{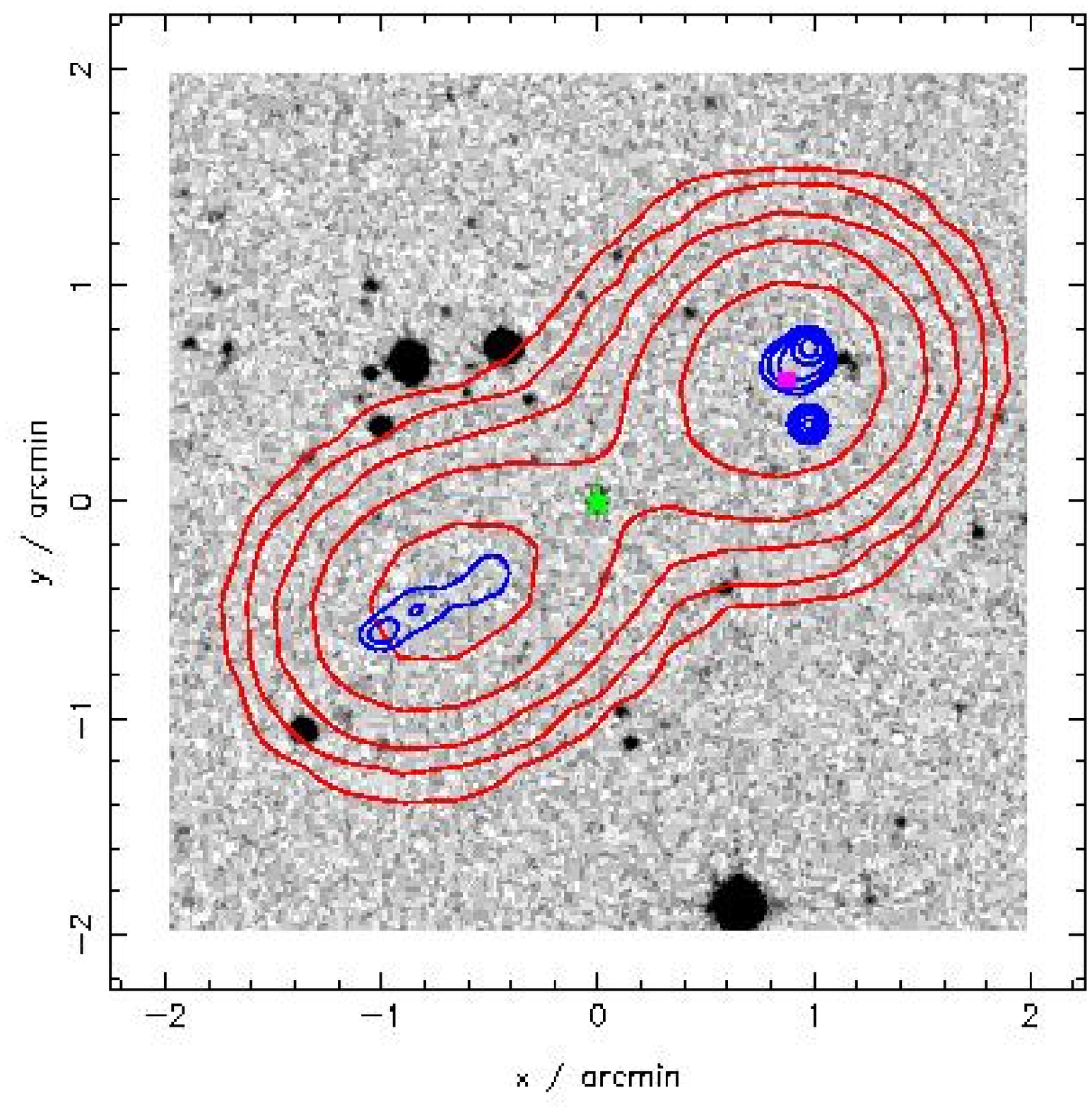}}
      \centerline{C3-255: TXS 1549+107}
    \end{minipage}
    \hspace{3cm}
    \begin{minipage}{3cm}
      \mbox{}
      \centerline{\includegraphics[scale=0.26,angle=270]{Contours/C3/257.ps}}
      \centerline{C3-257: TXS 1549+188}
    \end{minipage}
    \hspace{3cm}
    \begin{minipage}{3cm}
      \mbox{}
      \centerline{\includegraphics[scale=0.26,angle=270]{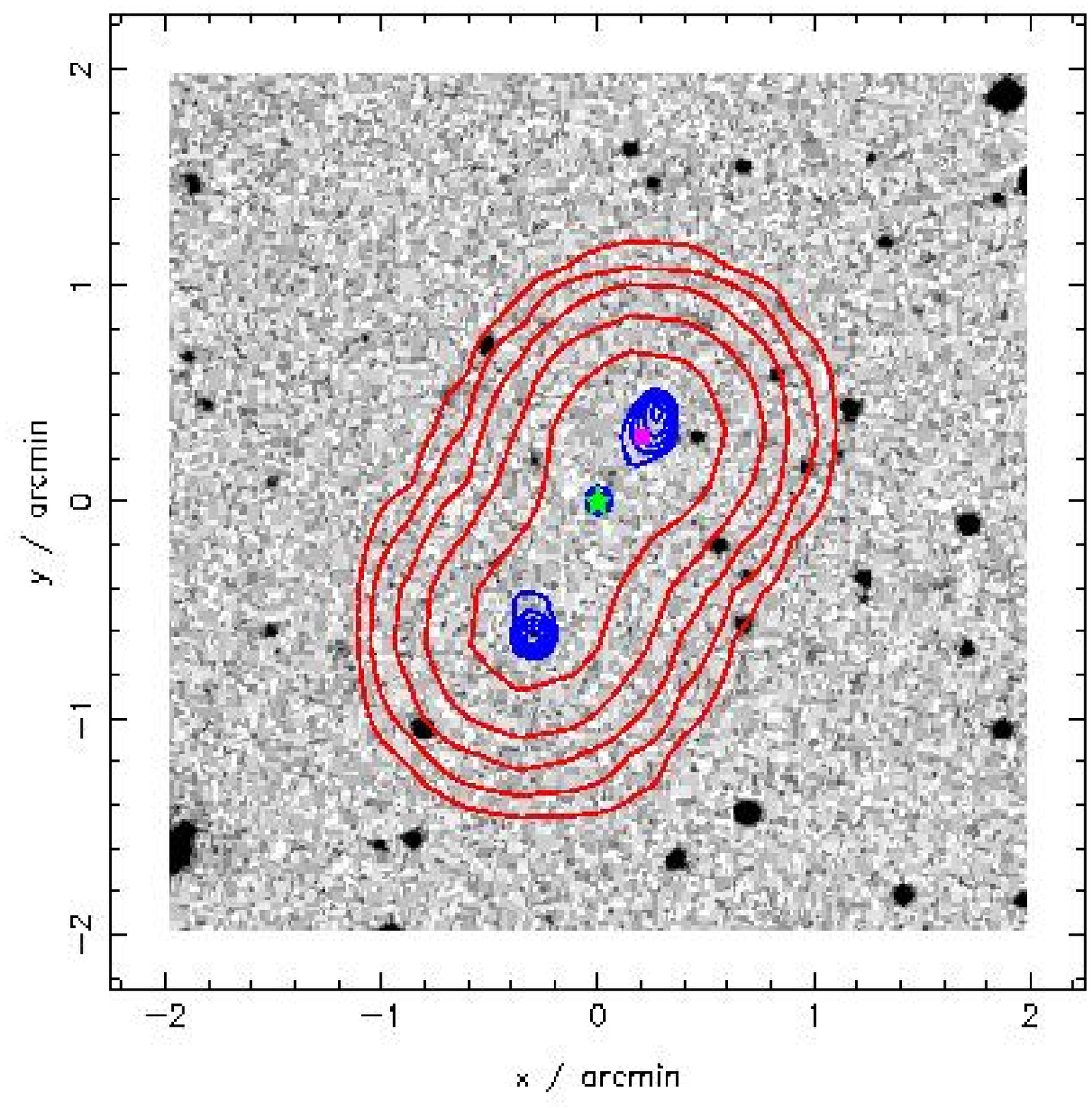}}
      \centerline{C3-261: J1553+1401}
    \end{minipage}
  \end{center}
\end{figure}

\begin{figure}
  \begin{center}
    {\bf CoNFIG-3}\\  
  \begin{minipage}{3cm}      
      \mbox{}
      \centerline{\includegraphics[scale=0.26,angle=270]{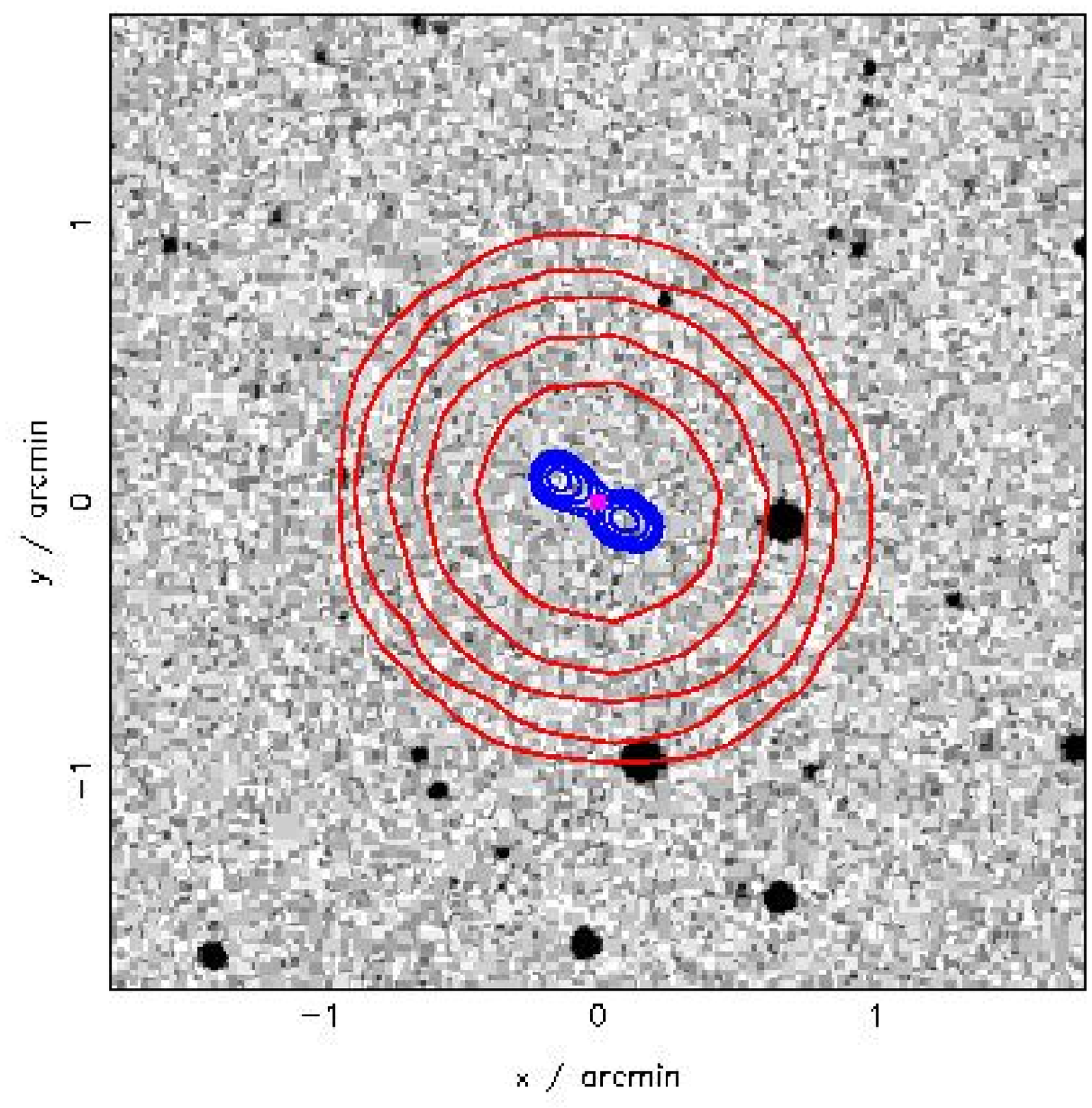}}
      \centerline{C3-262: TXS 1550+211}
    \end{minipage}
    \hspace{3cm}
    \begin{minipage}{3cm}
      \mbox{}
      \centerline{\includegraphics[scale=0.26,angle=270]{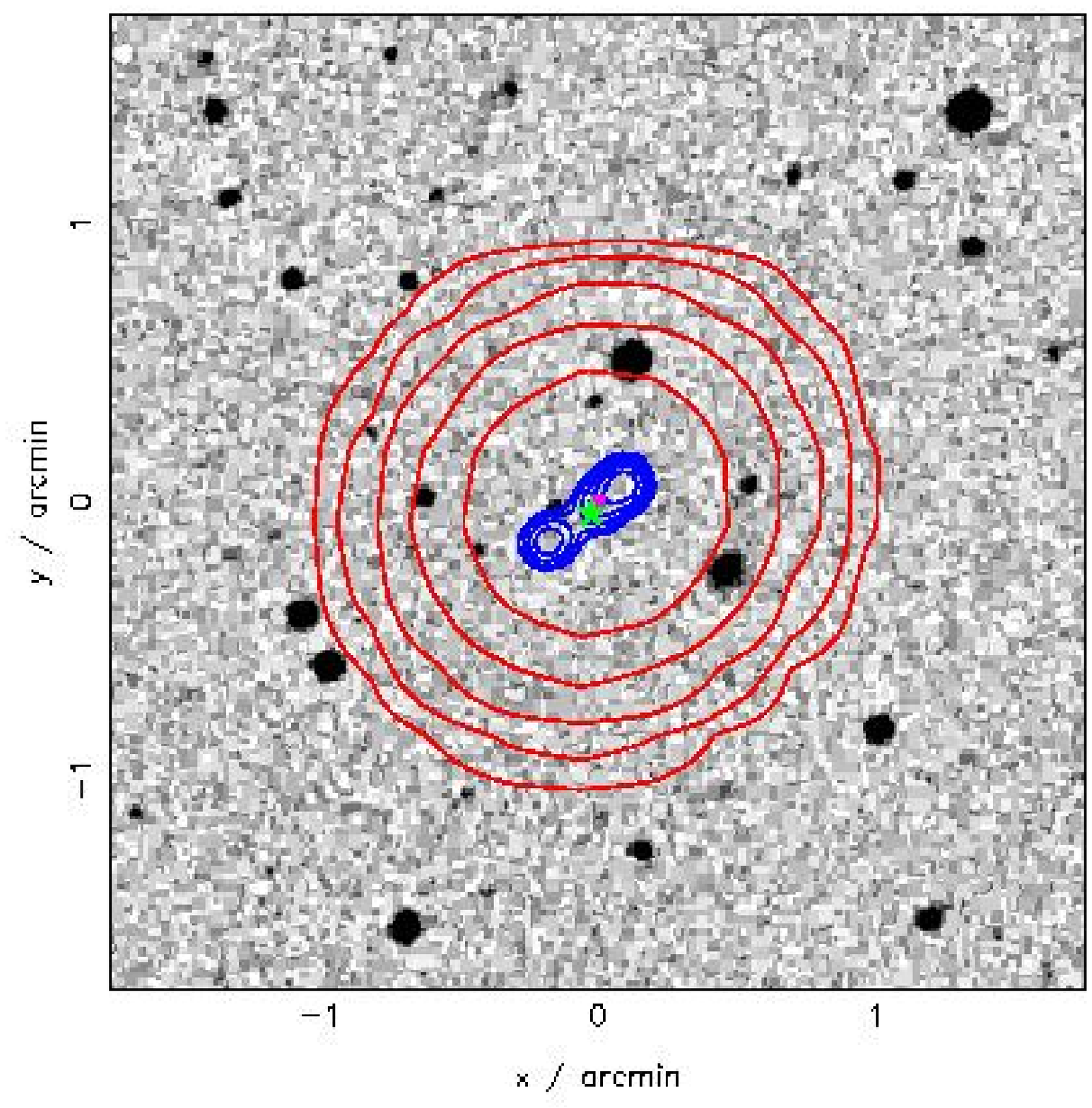}}
      \centerline{C3-263: TXS 1551+179}
    \end{minipage}
    \hspace{3cm}
    \begin{minipage}{3cm}
      \mbox{}
      \centerline{\includegraphics[scale=0.26,angle=270]{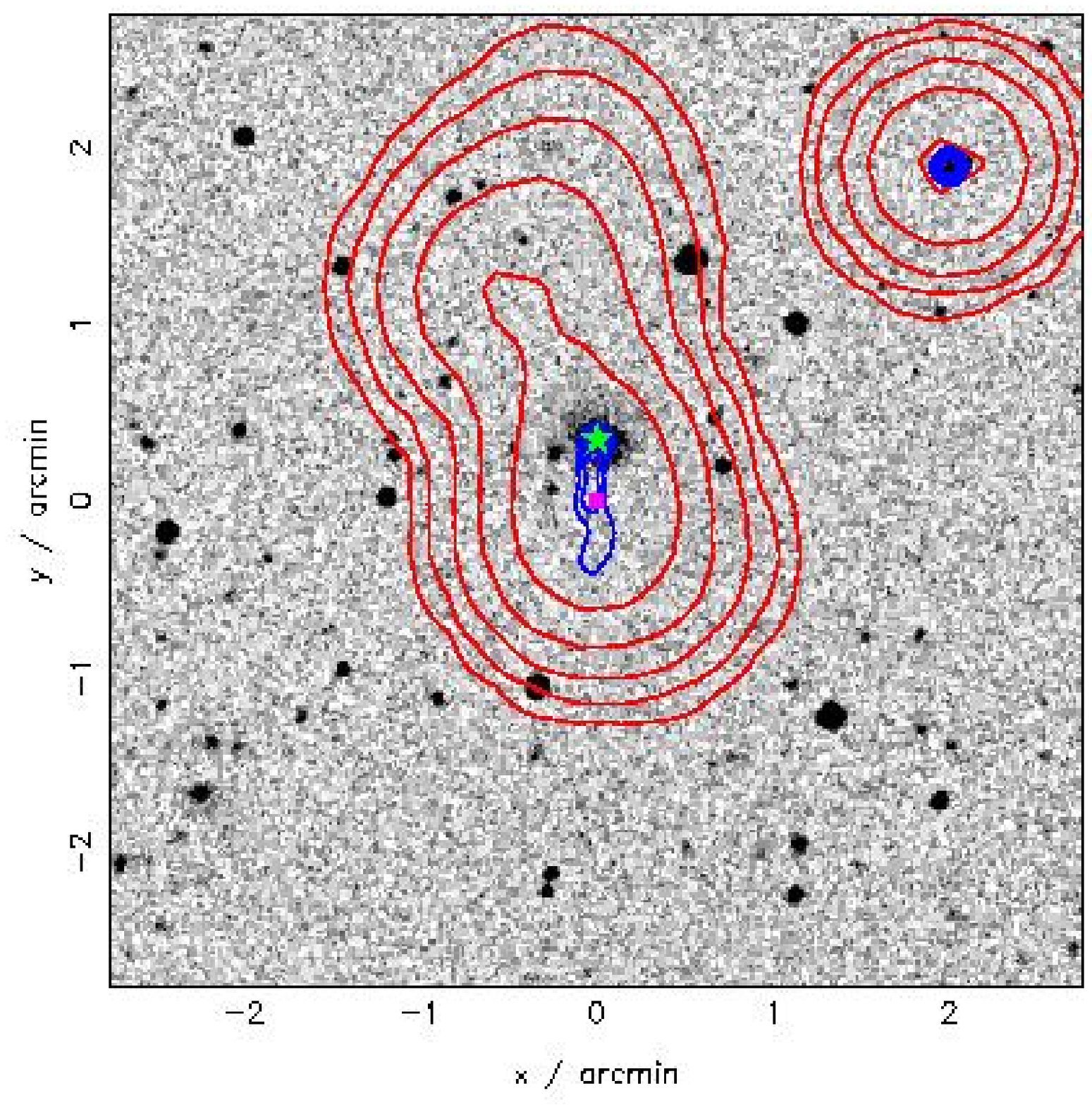}}
      \centerline{C3-266: 4C 23.42}
    \end{minipage}
    \vfill
    \begin{minipage}{3cm}     
      \mbox{}
      \centerline{\includegraphics[scale=0.26,angle=270]{Contours/C3/267.ps}}
      \centerline{C3-267: TXS 1551+251}
    \end{minipage}
    \hspace{3cm}
    \begin{minipage}{3cm}
      \mbox{}
      \centerline{\includegraphics[scale=0.26,angle=270]{Contours/C3/268.ps}}
      \centerline{C3-268: TXS 1551+221}
    \end{minipage}
    \hspace{3cm}
    \begin{minipage}{3cm}
      \mbox{}
      \centerline{\includegraphics[scale=0.26,angle=270]{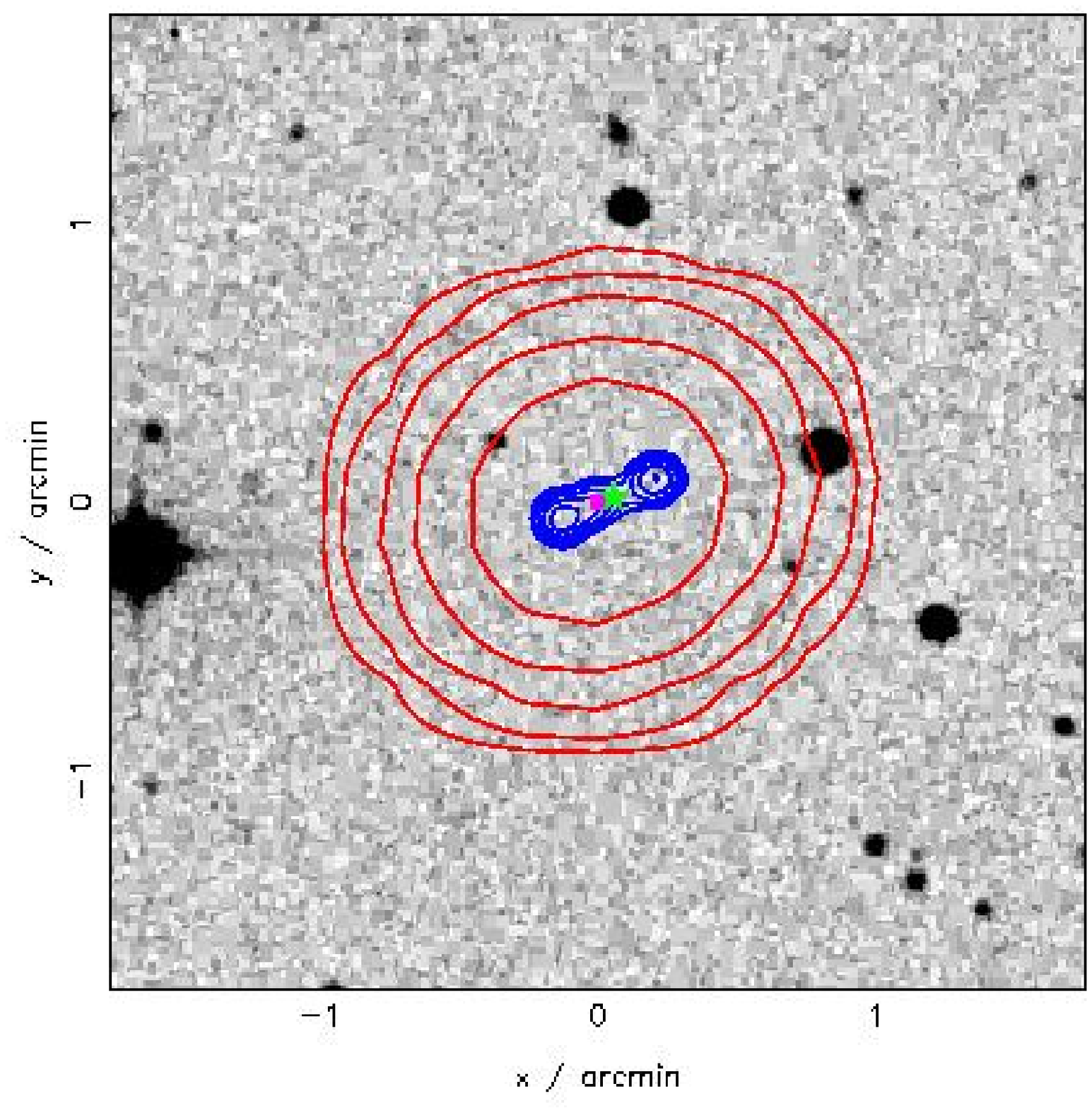}}
      \centerline{C3-271: TXS 1552+151}
    \end{minipage}
    \vfill
    \begin{minipage}{3cm}     
      \mbox{}
      \centerline{\includegraphics[scale=0.26,angle=270]{Contours/C3/273.ps}}
      \centerline{C3-273: TXS 1553+279}
    \end{minipage}
    \hspace{3cm}
    \begin{minipage}{3cm}
      \mbox{}
      \centerline{\includegraphics[scale=0.26,angle=270]{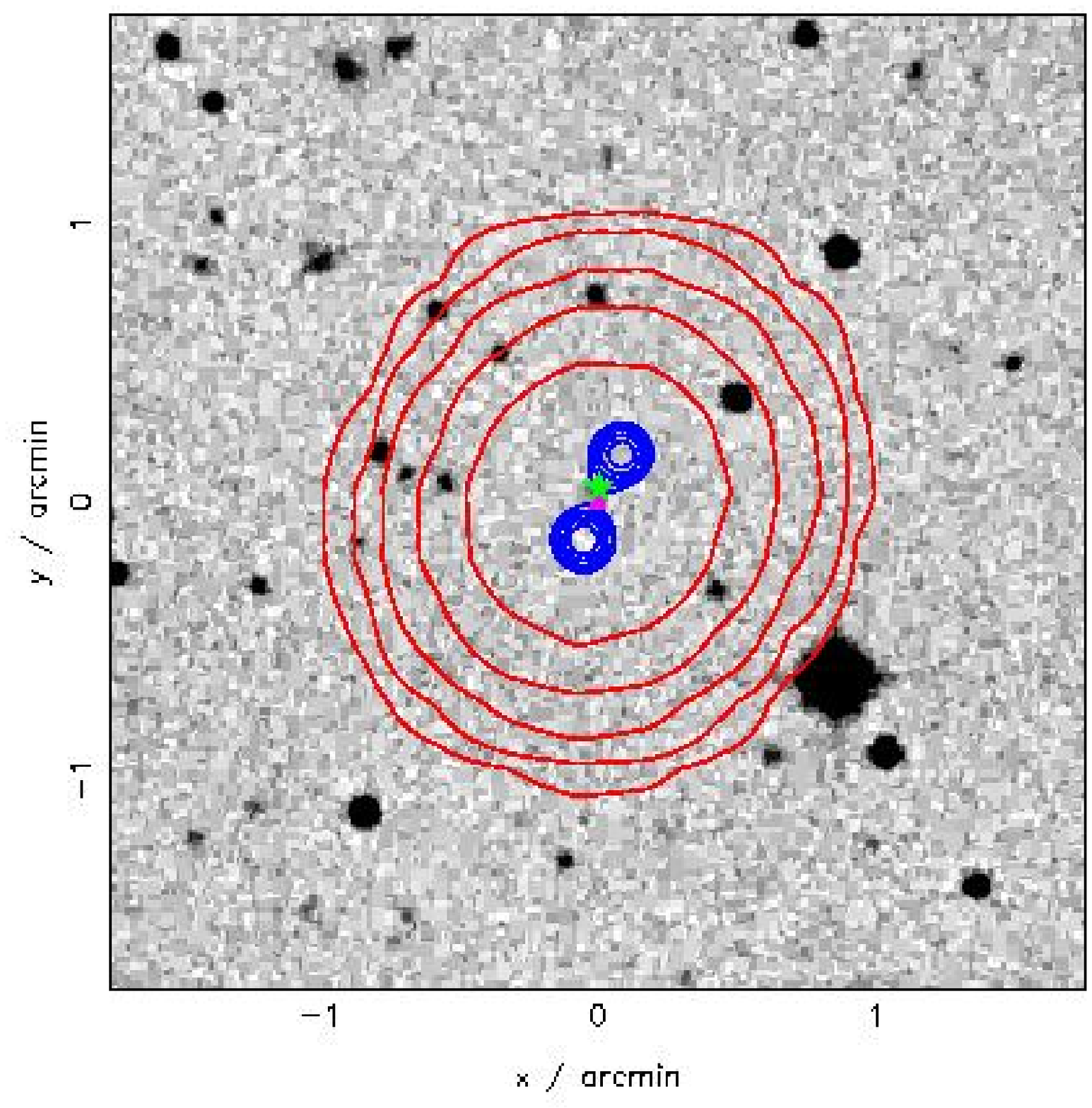}}
      \centerline{C3-276: 4C 24.35}
    \end{minipage}
    \hspace{3cm}
    \begin{minipage}{3cm}
      \mbox{}
      \centerline{\includegraphics[scale=0.26,angle=270]{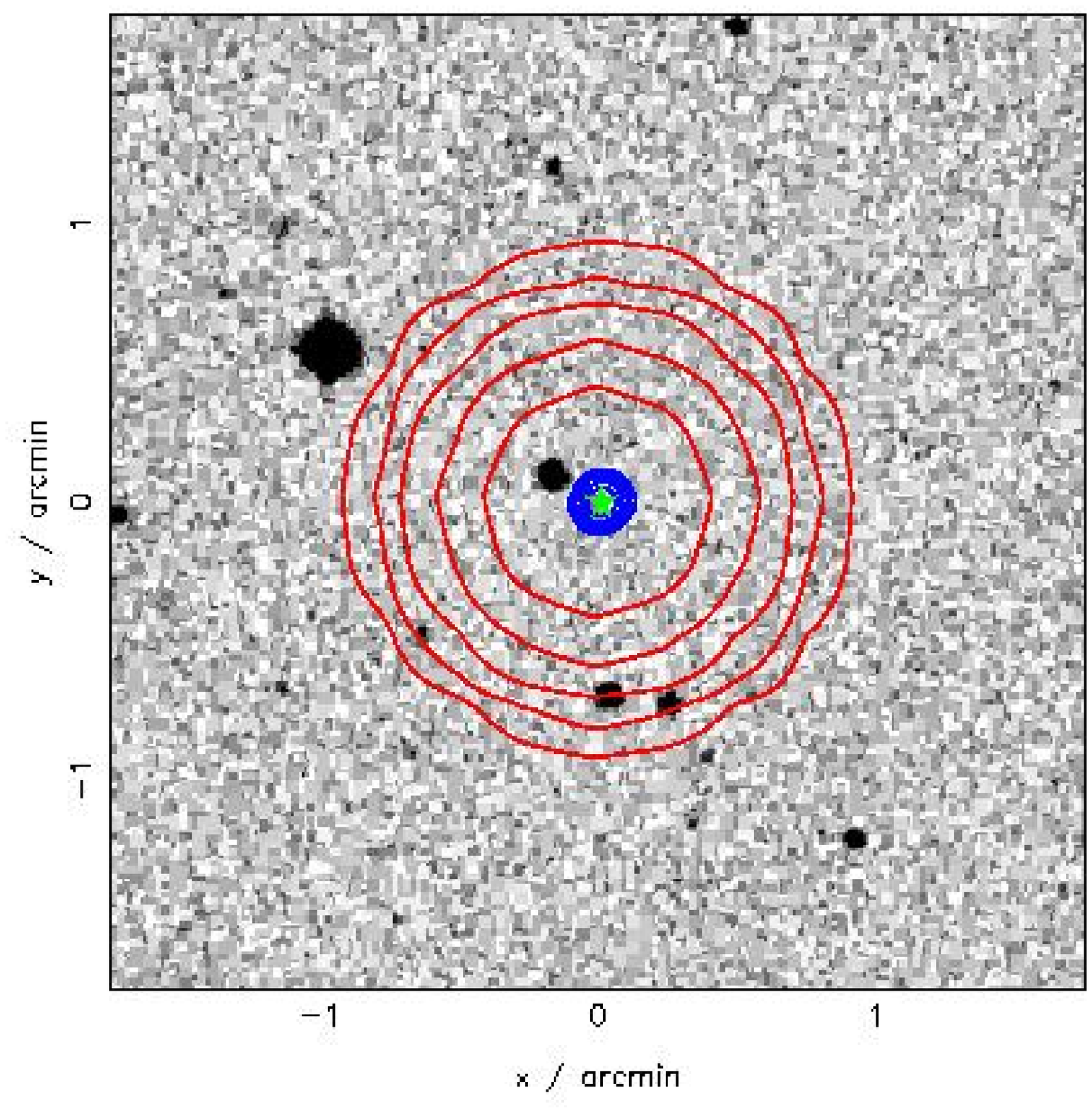}}
      \centerline{C3-277: TXS 1553+134}
    \end{minipage}
    \vfill
    \begin{minipage}{3cm}      
      \mbox{}
      \centerline{\includegraphics[scale=0.26,angle=270]{Contours/C3/279.ps}}
      \centerline{C3-279: TXS 1554+222}
    \end{minipage}
    \hspace{3cm}
    \begin{minipage}{3cm}
      \mbox{}
      \centerline{\includegraphics[scale=0.26,angle=270]{Contours/C3/281.ps}}
      \centerline{C3-281: TXS 1554+144}
    \end{minipage}
    \hspace{3cm}
    \begin{minipage}{3cm}
      \mbox{}
      \centerline{\includegraphics[scale=0.26,angle=270]{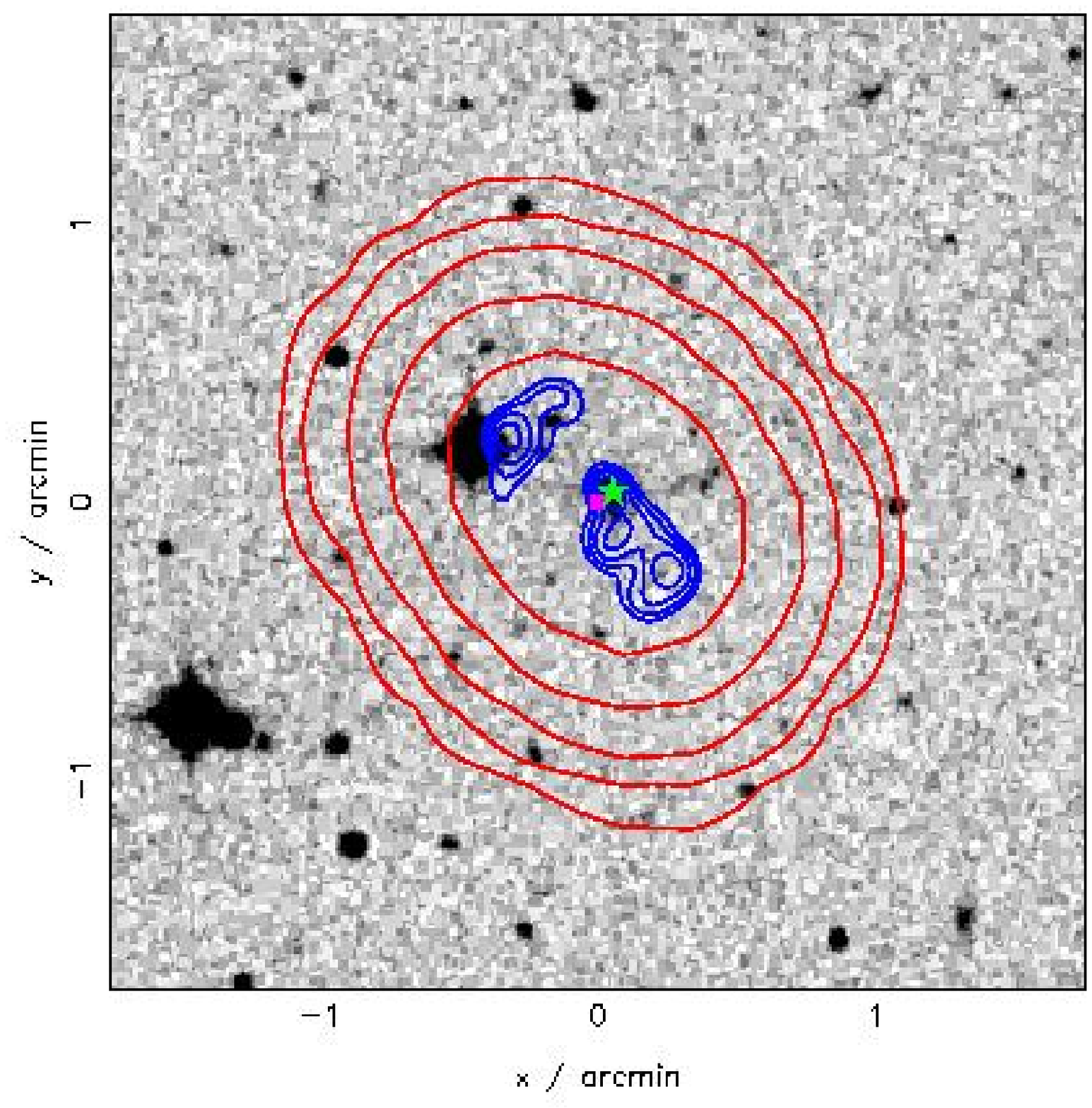}}
      \centerline{C3-282: 4C 10.44}
    \end{minipage}
  \end{center}
\end{figure}

\begin{figure}
  \begin{center}
    {\bf CoNFIG-3}\\  
  \begin{minipage}{3cm}      
      \mbox{}
      \centerline{\includegraphics[scale=0.26,angle=270]{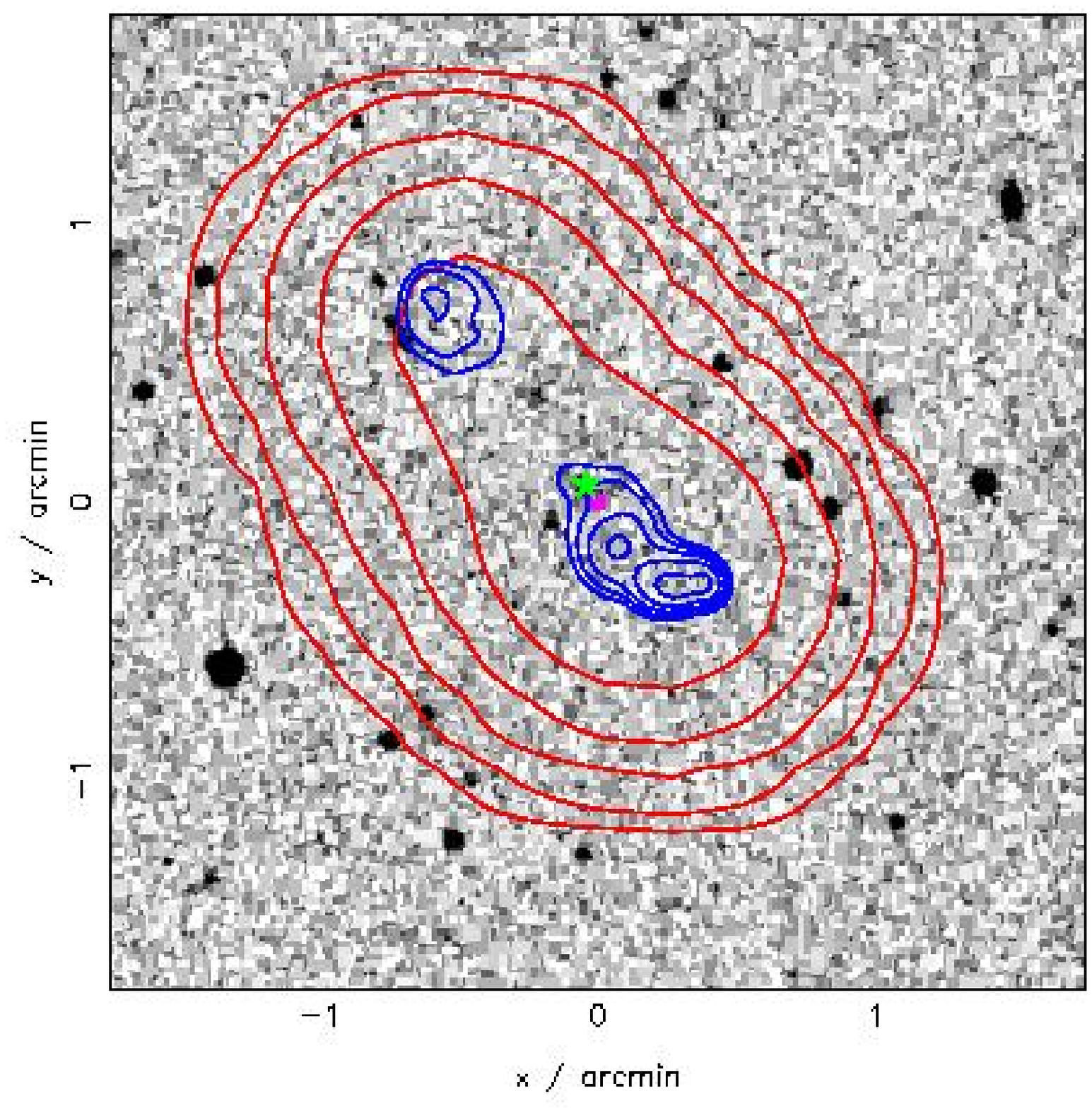}}
      \centerline{C3-284: 4C 12.56}
    \end{minipage}
    \hspace{3cm}
    \begin{minipage}{3cm}
      \mbox{}
      \centerline{\includegraphics[scale=0.26,angle=270]{Contours/C3/285.ps}}
      \centerline{C3-285: 4C 11.51}
    \end{minipage}
  \end{center}
\end{figure}

\begin{figure}
  \begin{center}
  {\bf CoNFIG-4}\\
    \begin{minipage}{3cm}      
      \mbox{}
      \centerline{\includegraphics[scale=0.26,angle=270]{Contours/C4/002.ps}}
      \centerline{C4-002: TXS 1405+026}
    \end{minipage}
    \hspace{3cm}
    \begin{minipage}{3cm}
      \mbox{}
      \centerline{\includegraphics[scale=0.26,angle=270]{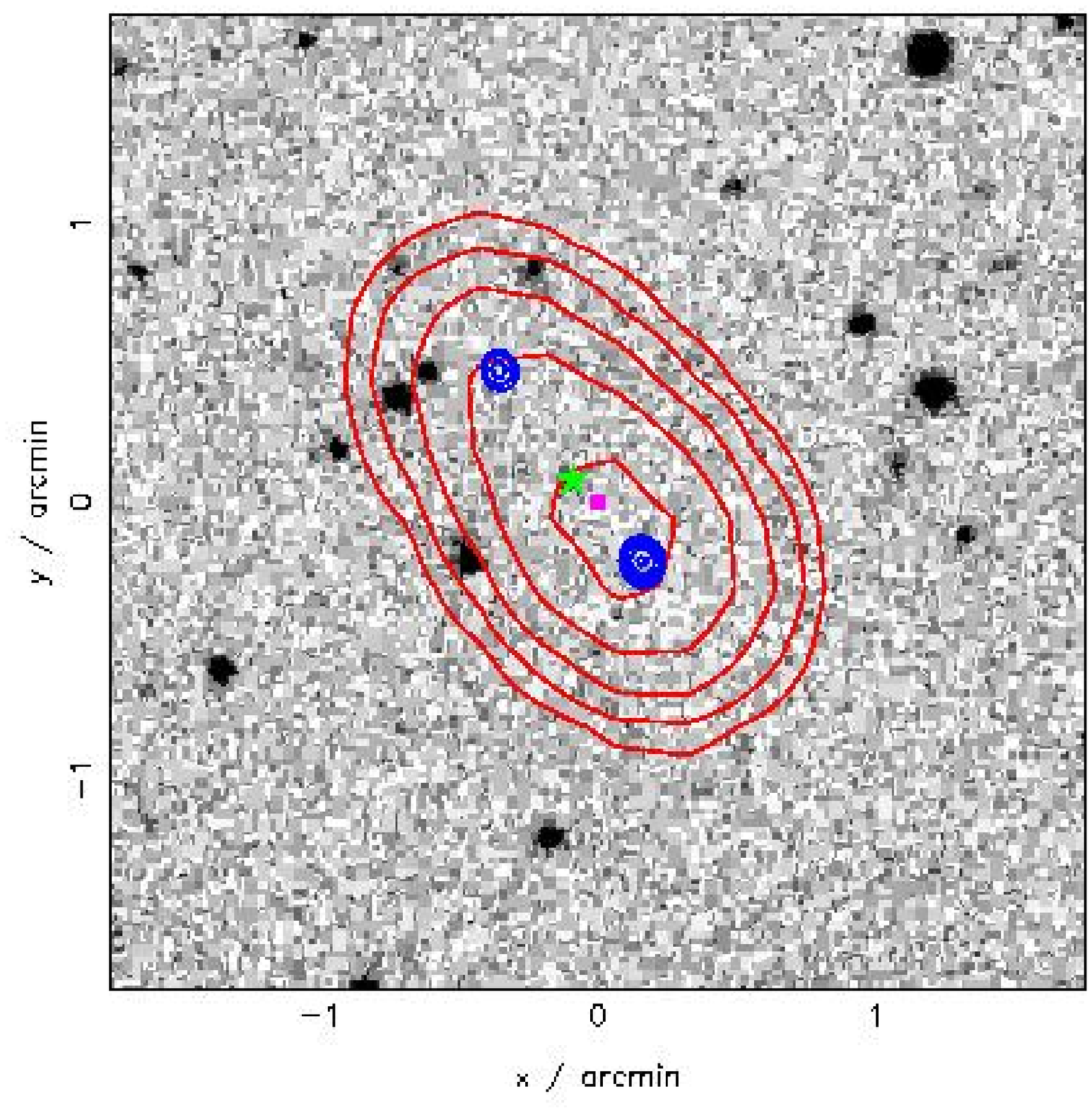}}
      \centerline{C4-003: 1408+0050}
    \end{minipage}
    \hspace{3cm}
    \begin{minipage}{3cm}
      \mbox{}
      \centerline{\includegraphics[scale=0.26,angle=270]{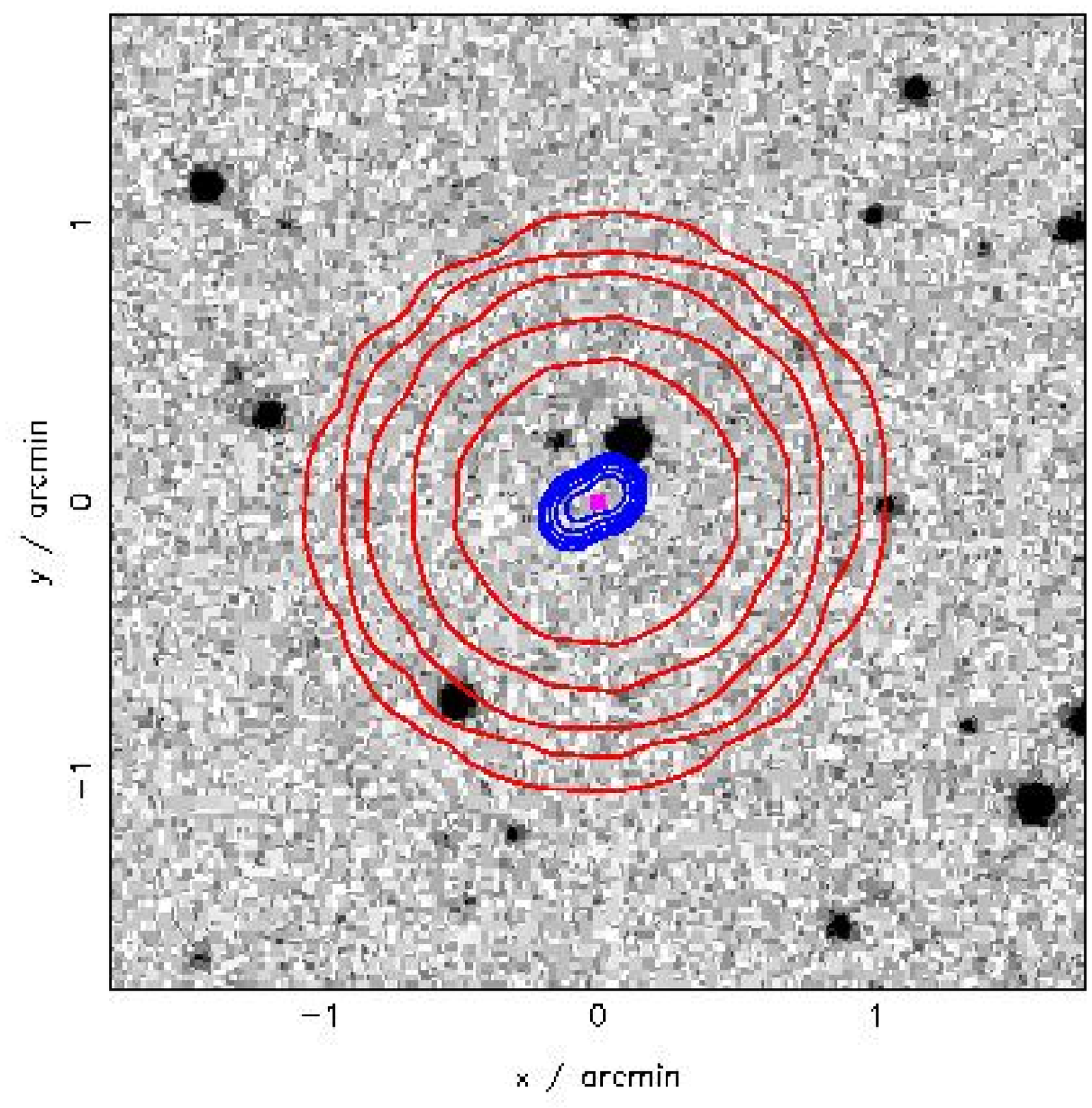}}
      \centerline{C4-005: TXS 1406+015}
    \end{minipage}
    \vfill
    \begin{minipage}{3cm}     
      \mbox{}
      \centerline{\includegraphics[scale=0.26,angle=270]{Contours/C4/006.ps}}
      \centerline{C4-006: 1408+0281}
    \end{minipage}
    \hspace{3cm}
    \begin{minipage}{3cm}
      \mbox{}
      \centerline{\includegraphics[scale=0.26,angle=270]{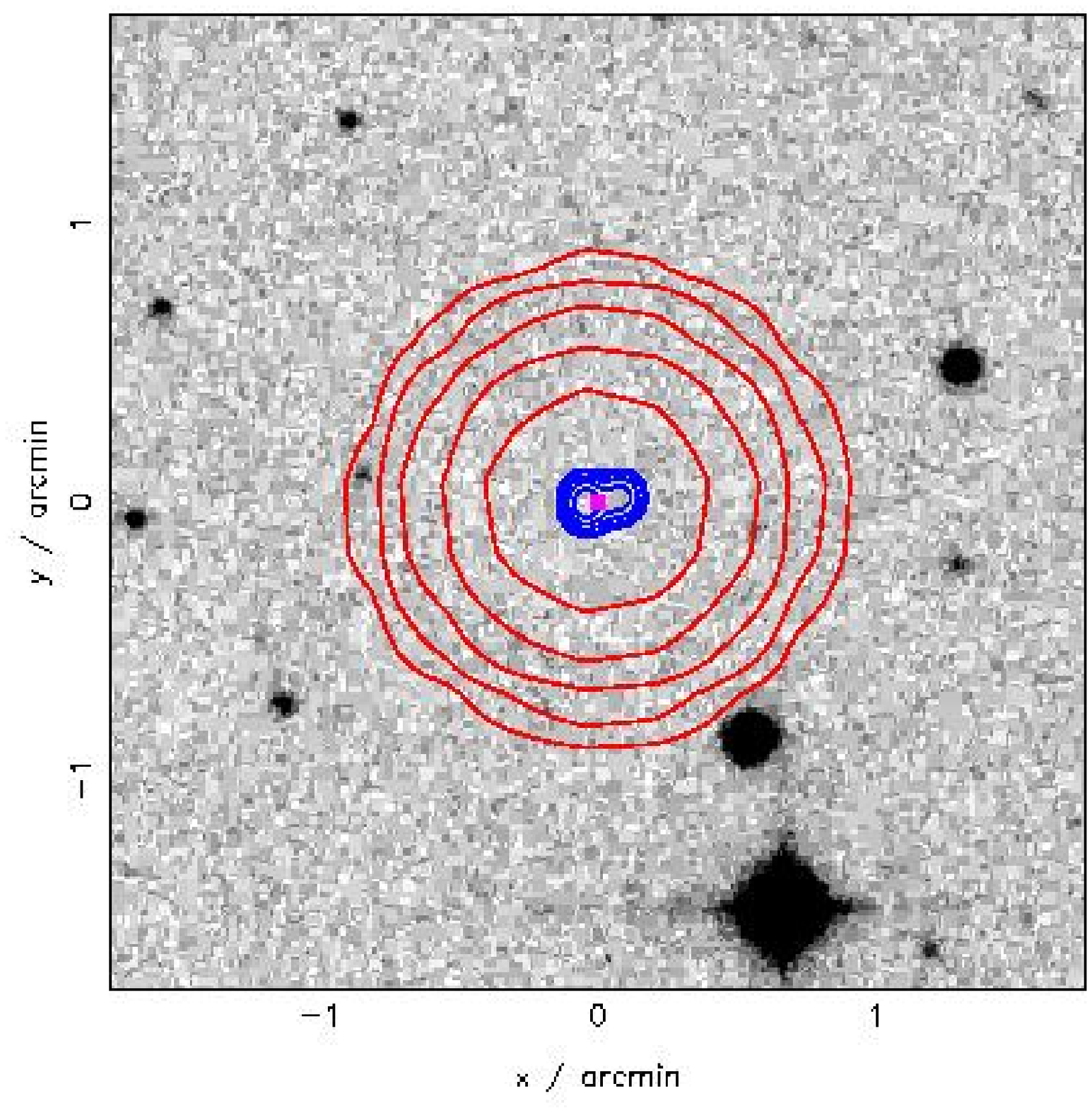}}
      \centerline{C4-007: TXS 1406+018}
    \end{minipage}
    \hspace{3cm}
    \begin{minipage}{3cm}
      \mbox{}
      \centerline{\includegraphics[scale=0.26,angle=270]{Contours/C4/008.ps}}
      \centerline{C4-008: 1408+0271}
    \end{minipage}
    \vfill
    \begin{minipage}{3cm}     
      \mbox{}
      \centerline{\includegraphics[scale=0.26,angle=270]{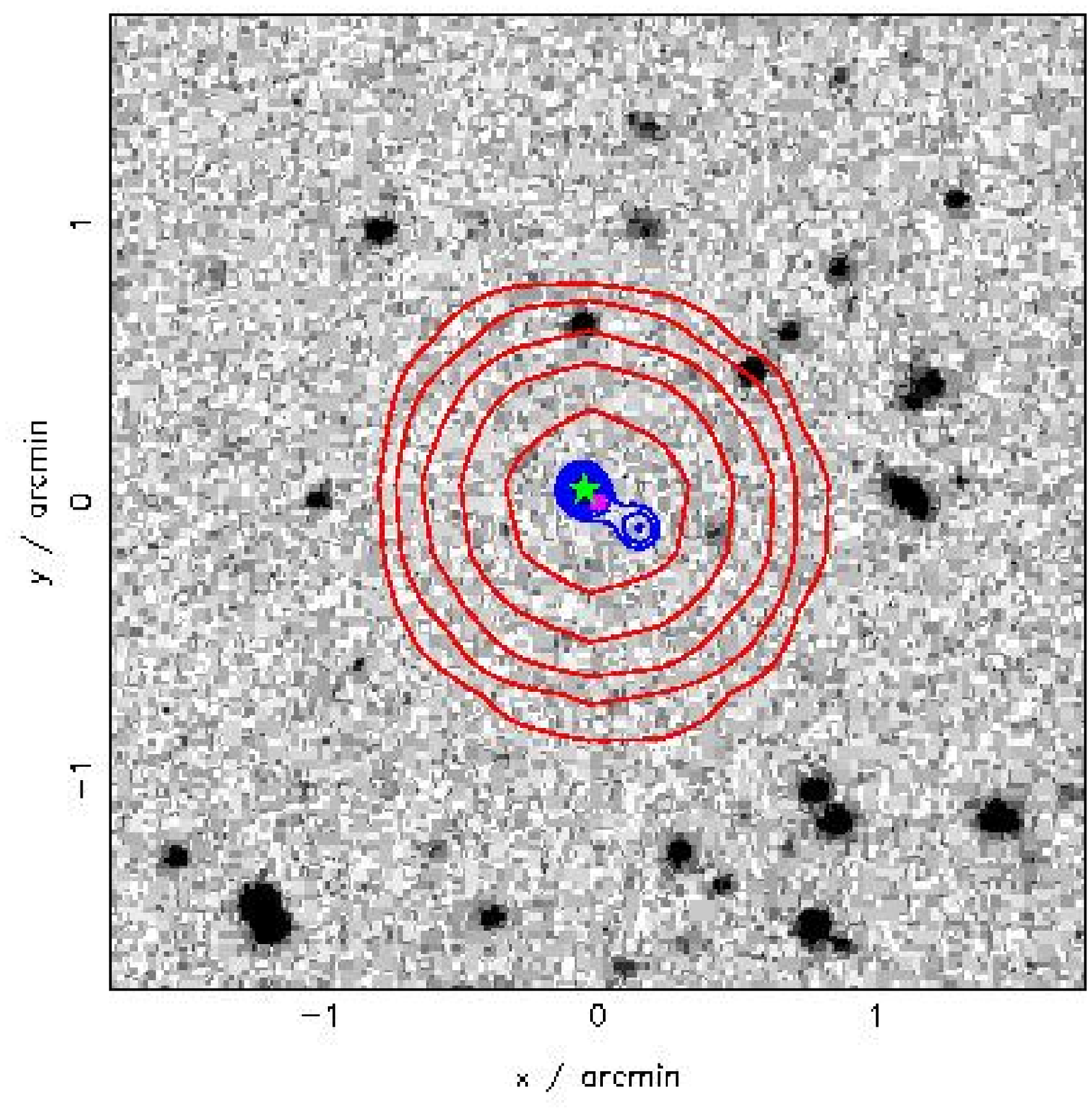}}
      \centerline{C4-011: J140929-01}
    \end{minipage}
    \hspace{3cm}
    \begin{minipage}{3cm}
      \mbox{}
      \centerline{\includegraphics[scale=0.26,angle=270]{Contours/C4/012.ps}}
      \centerline{C4-012: TXS 1406-007}
    \end{minipage}
    \hspace{3cm}
    \begin{minipage}{3cm}
      \mbox{}
      \centerline{\includegraphics[scale=0.26,angle=270]{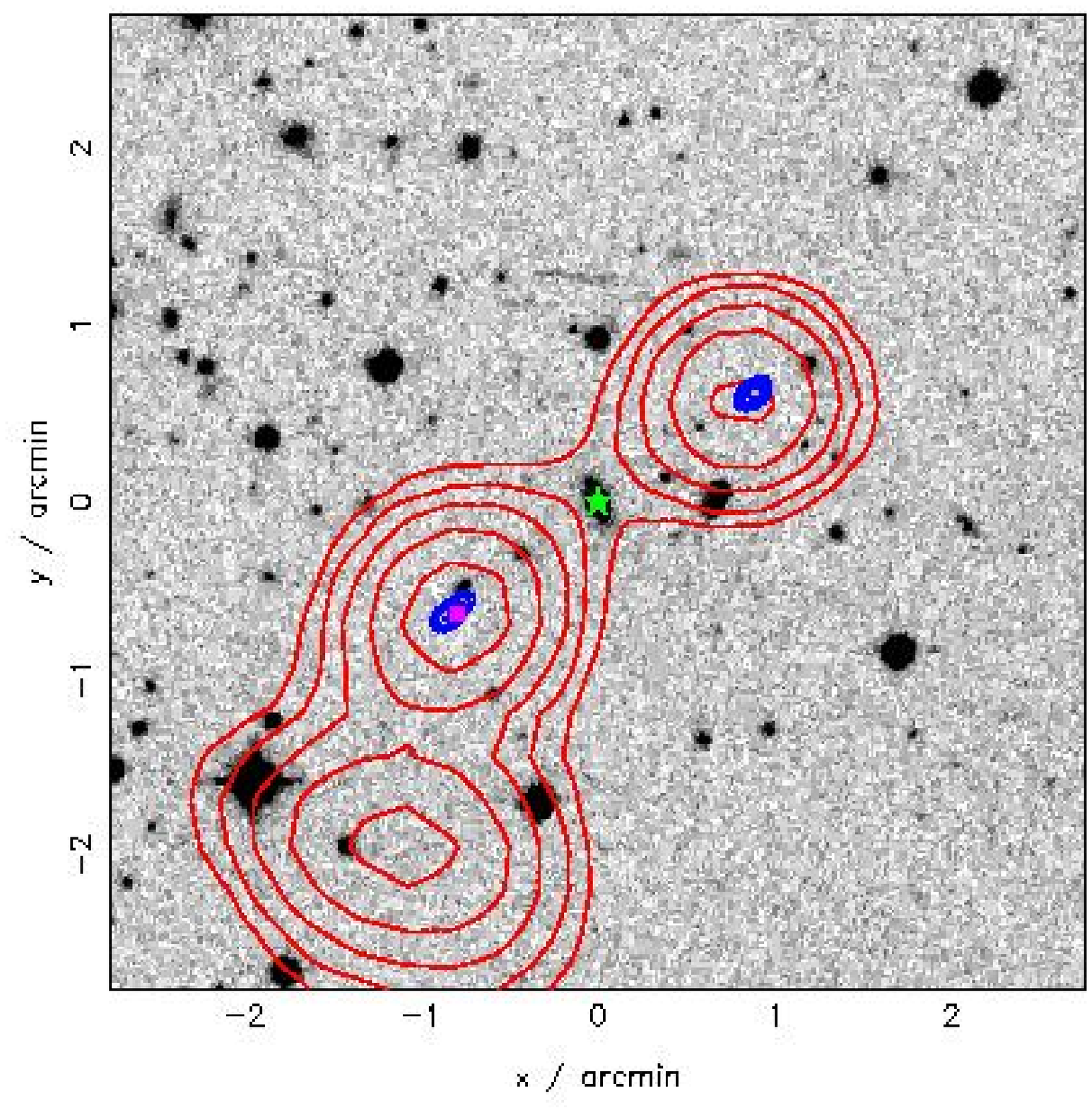}}
      \centerline{C4-014: 1409-0307}
    \end{minipage}
    \vfill
    \begin{minipage}{3cm}      
      \mbox{}
      \centerline{\includegraphics[scale=0.26,angle=270]{Contours/C4/015.ps}}
      \centerline{C4-015: B1407-0231}
    \end{minipage}
    \hspace{3cm}
    \begin{minipage}{3cm}
      \mbox{}
      \centerline{\includegraphics[scale=0.26,angle=270]{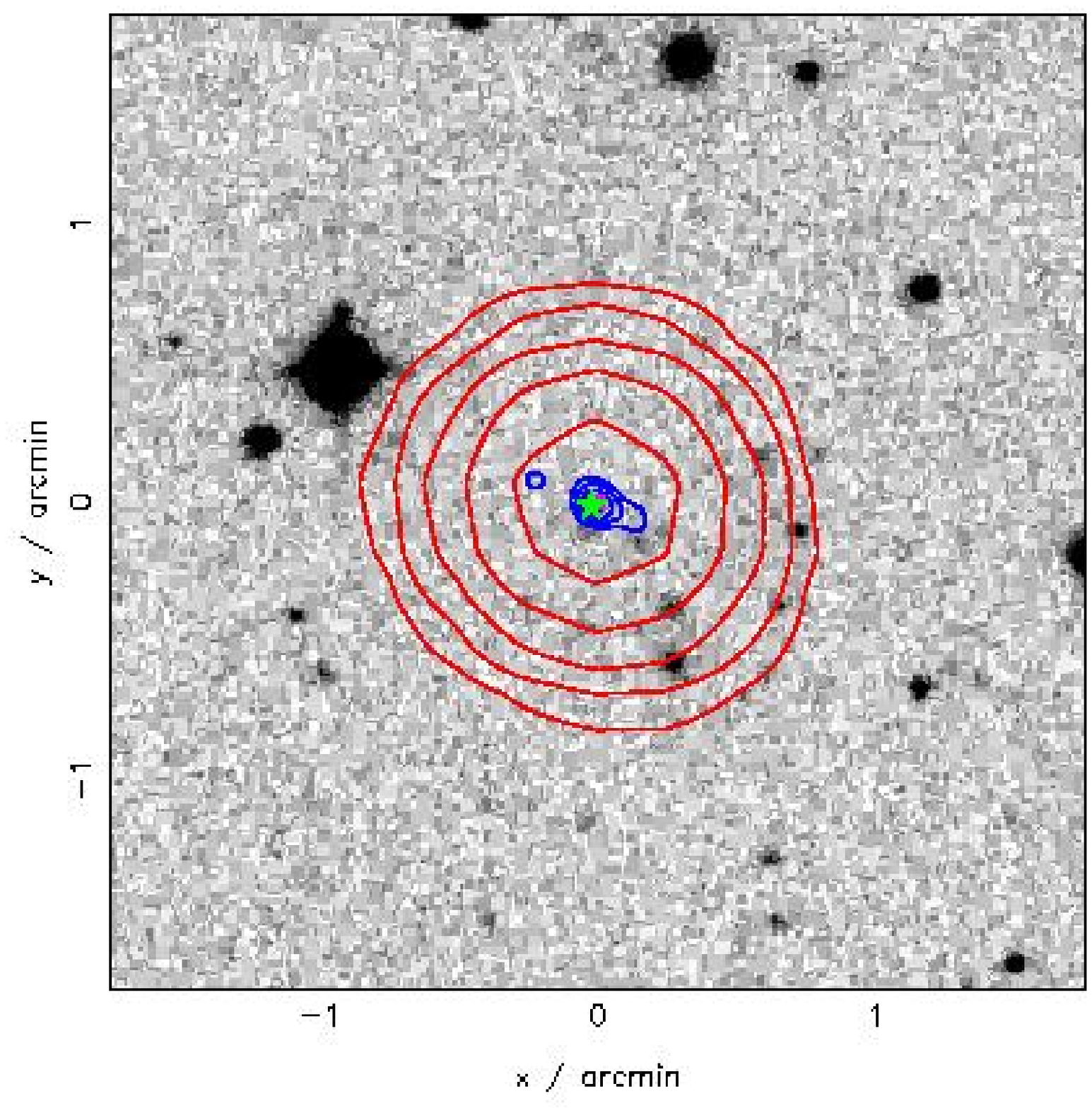}}
      \centerline{C4-016: 1409-0135}
    \end{minipage}
    \hspace{3cm}
    \begin{minipage}{3cm}
      \mbox{}
      \centerline{\includegraphics[scale=0.26,angle=270]{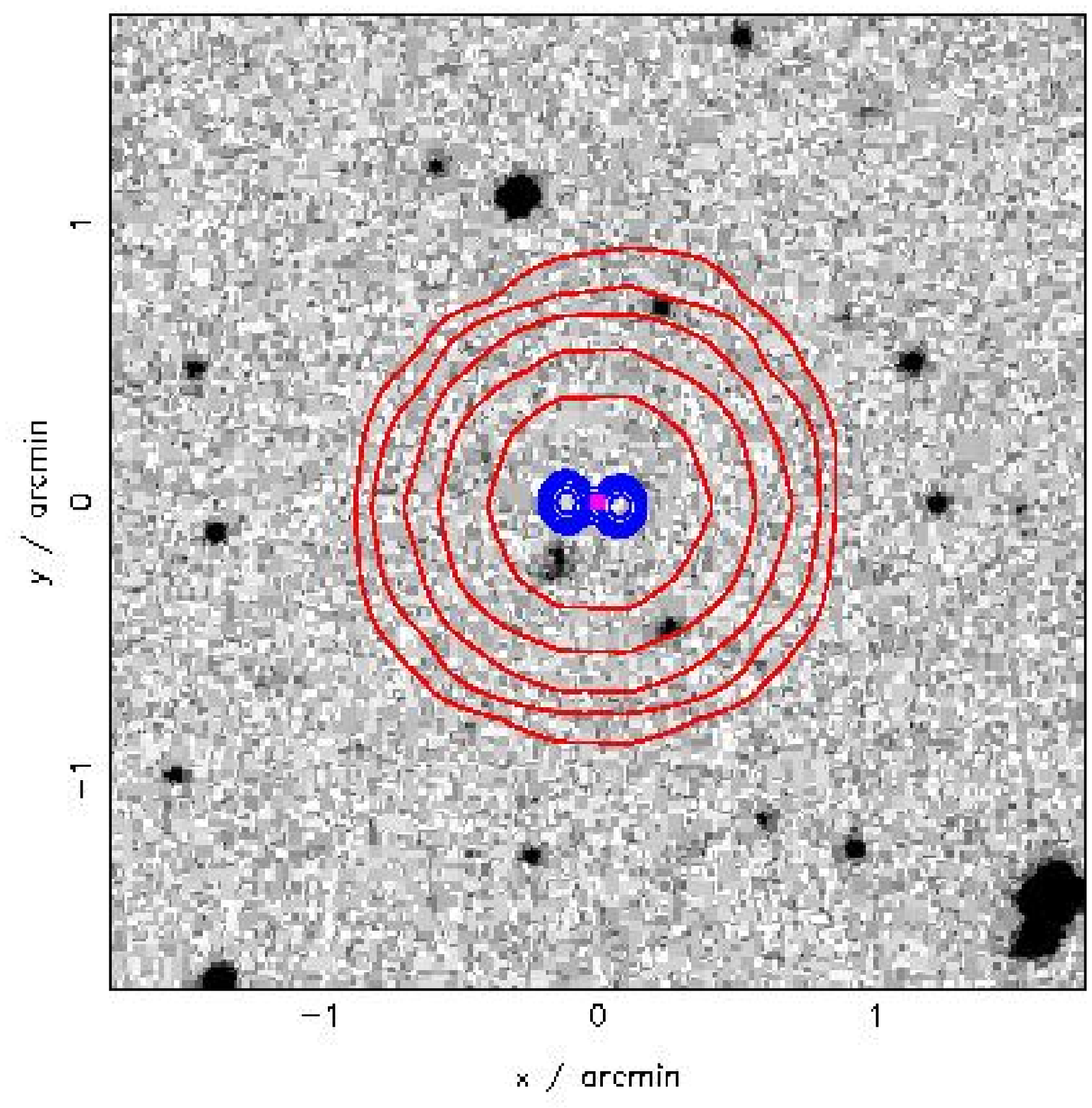}}
      \centerline{C4-017: TXS 1407-009}
    \end{minipage}
  \end{center}
\end{figure}

\begin{figure}
  \begin{center}
    {\bf CoNFIG-4}\\  
  \begin{minipage}{3cm}      
      \mbox{}
      \centerline{\includegraphics[scale=0.26,angle=270]{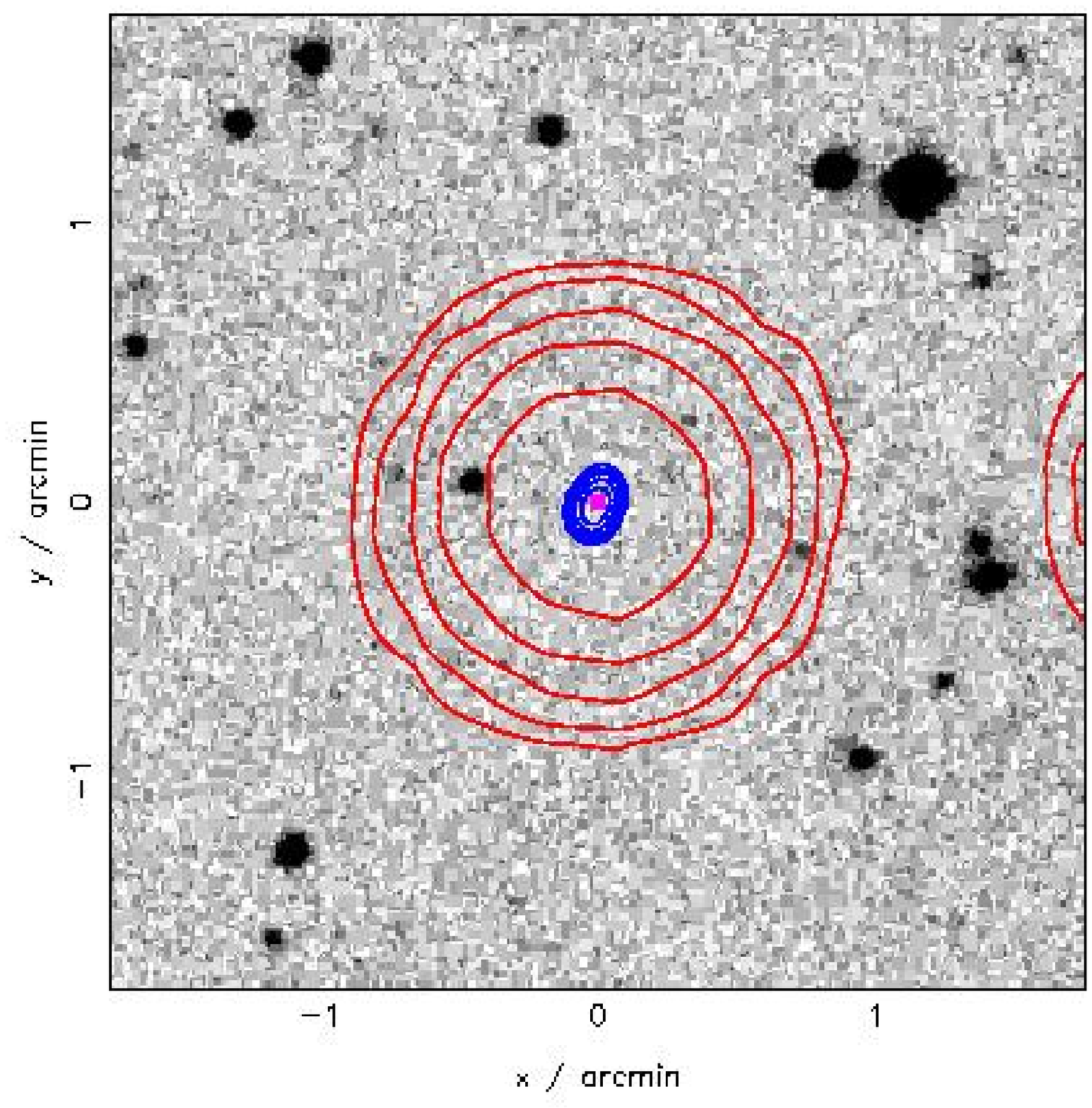}}
      \centerline{C4-023: TXS 1408-004}
    \end{minipage}
    \hspace{3cm}
    \begin{minipage}{3cm}
      \mbox{}
      \centerline{\includegraphics[scale=0.26,angle=270]{Contours/C4/024.ps}}
      \centerline{C4-024: B1408-0246}
    \end{minipage}
    \hspace{3cm}
    \begin{minipage}{3cm}
      \mbox{}
      \centerline{\includegraphics[scale=0.26,angle=270]{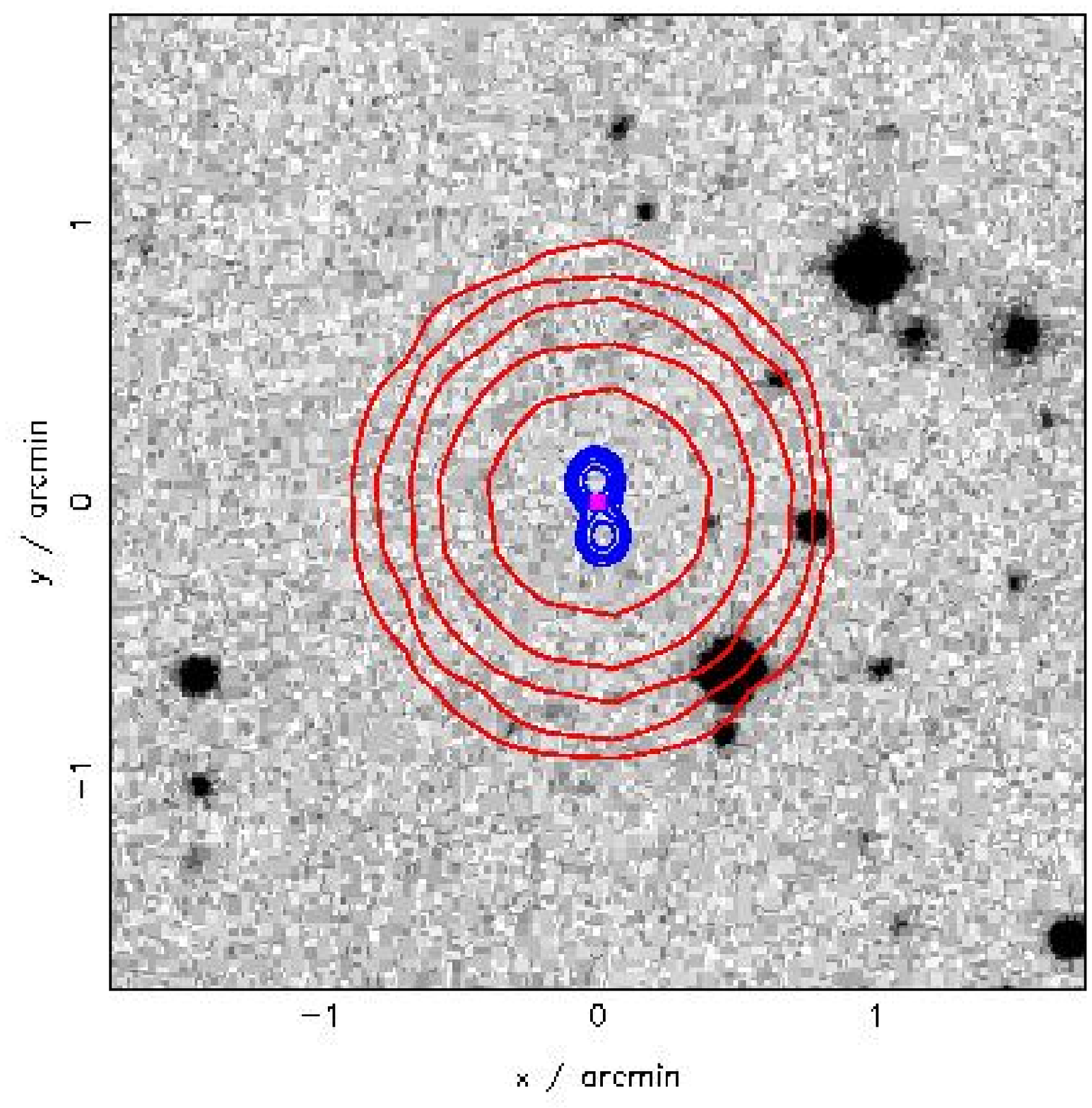}}
      \centerline{C4-026: TXS 1408+016}
    \end{minipage}
    \vfill
    \begin{minipage}{3cm}     
      \mbox{}
      \centerline{\includegraphics[scale=0.26,angle=270]{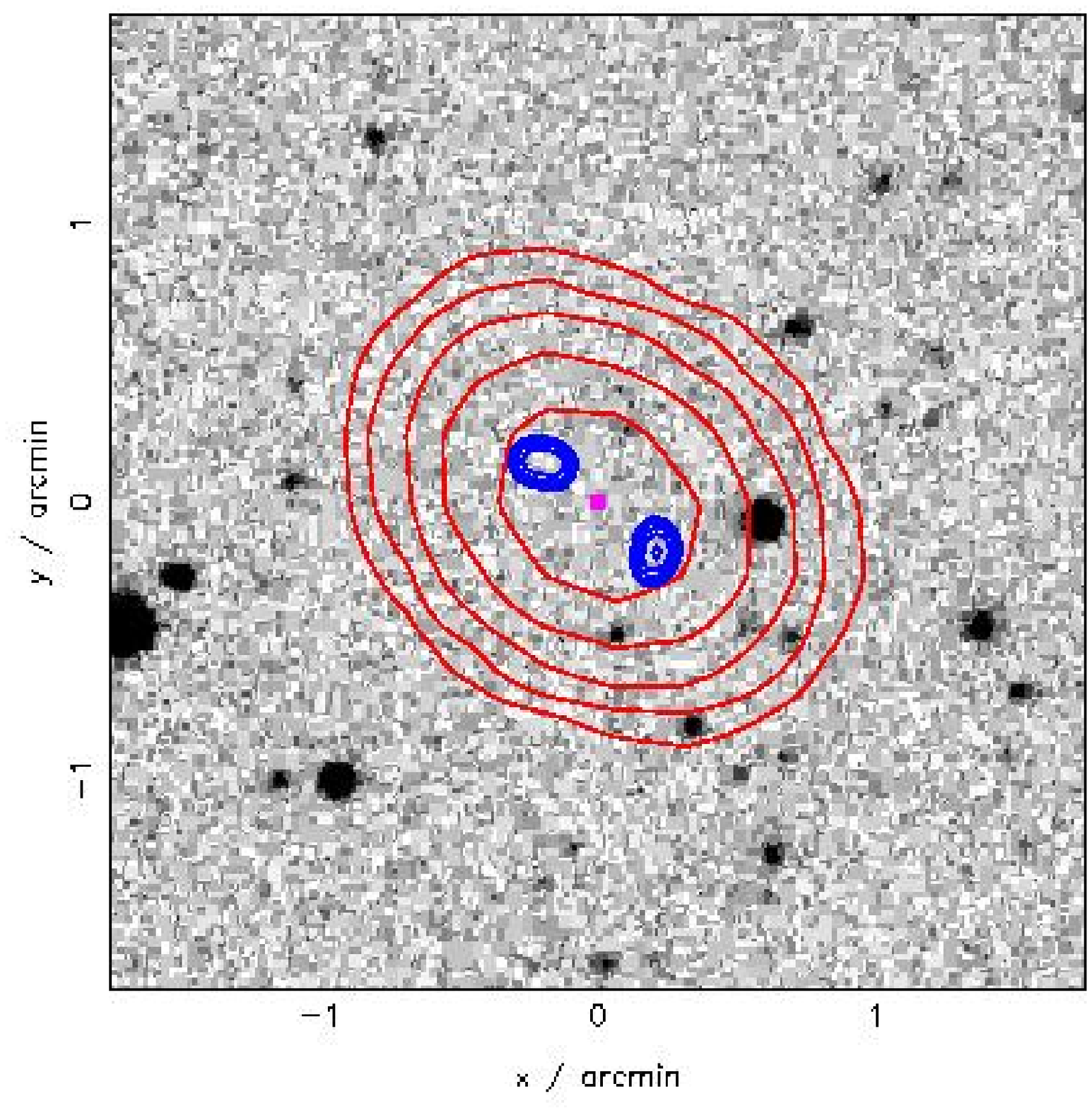}}
      \centerline{C4-027: TXS 1408-003} 
    \end{minipage}
    \hspace{3cm}
    \begin{minipage}{3cm}
      \mbox{}
      \centerline{\includegraphics[scale=0.26,angle=270]{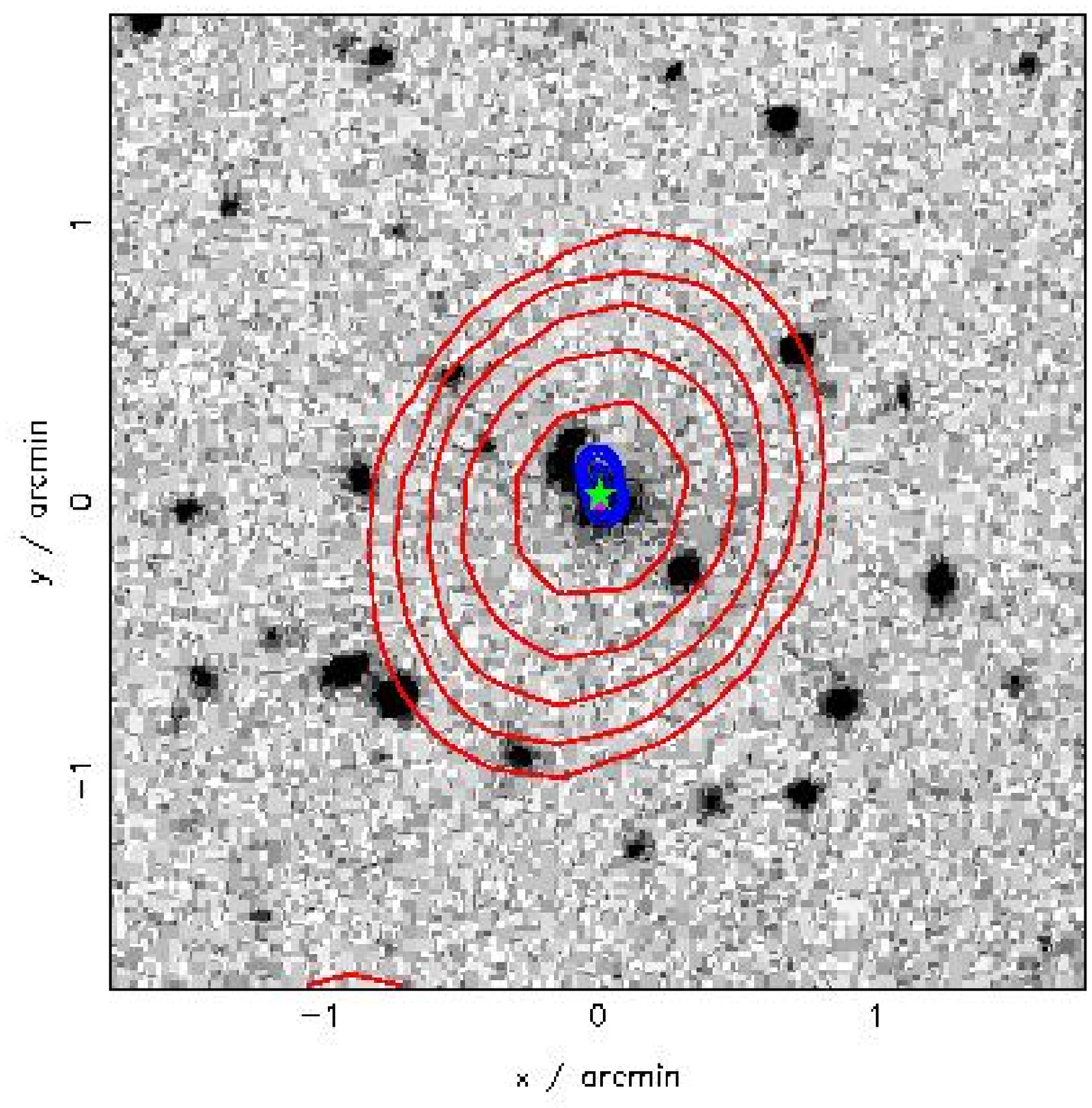}}
      \centerline{C4-028: 1411+0229}
    \end{minipage}
    \hspace{3cm}
    \begin{minipage}{3cm}
      \mbox{}
      \centerline{\includegraphics[scale=0.26,angle=270]{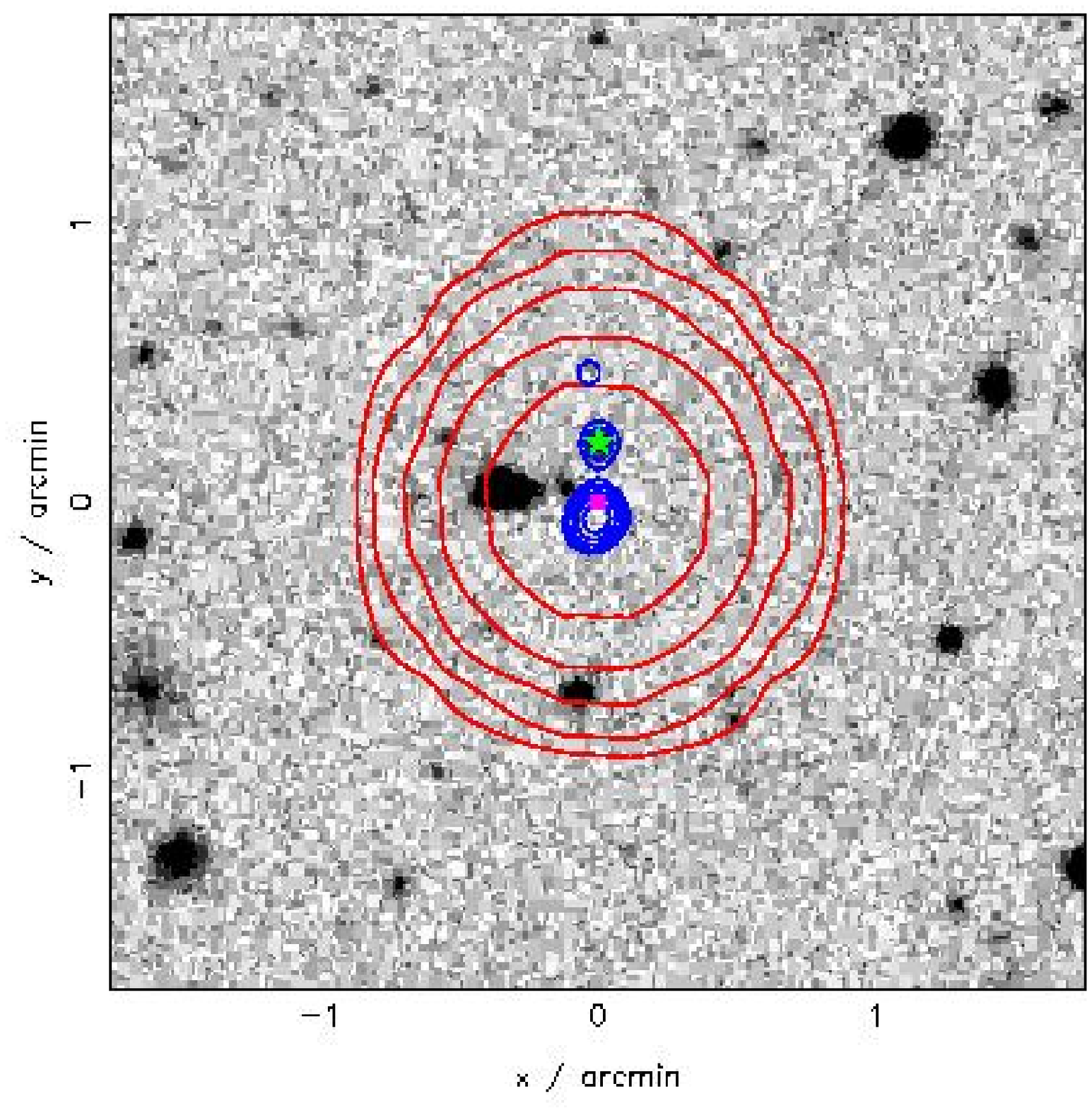}}
      \centerline{C4-029: TXS 1408+009}
    \end{minipage}
    \vfill
    \begin{minipage}{3cm}     
      \mbox{}
      \centerline{\includegraphics[scale=0.26,angle=270]{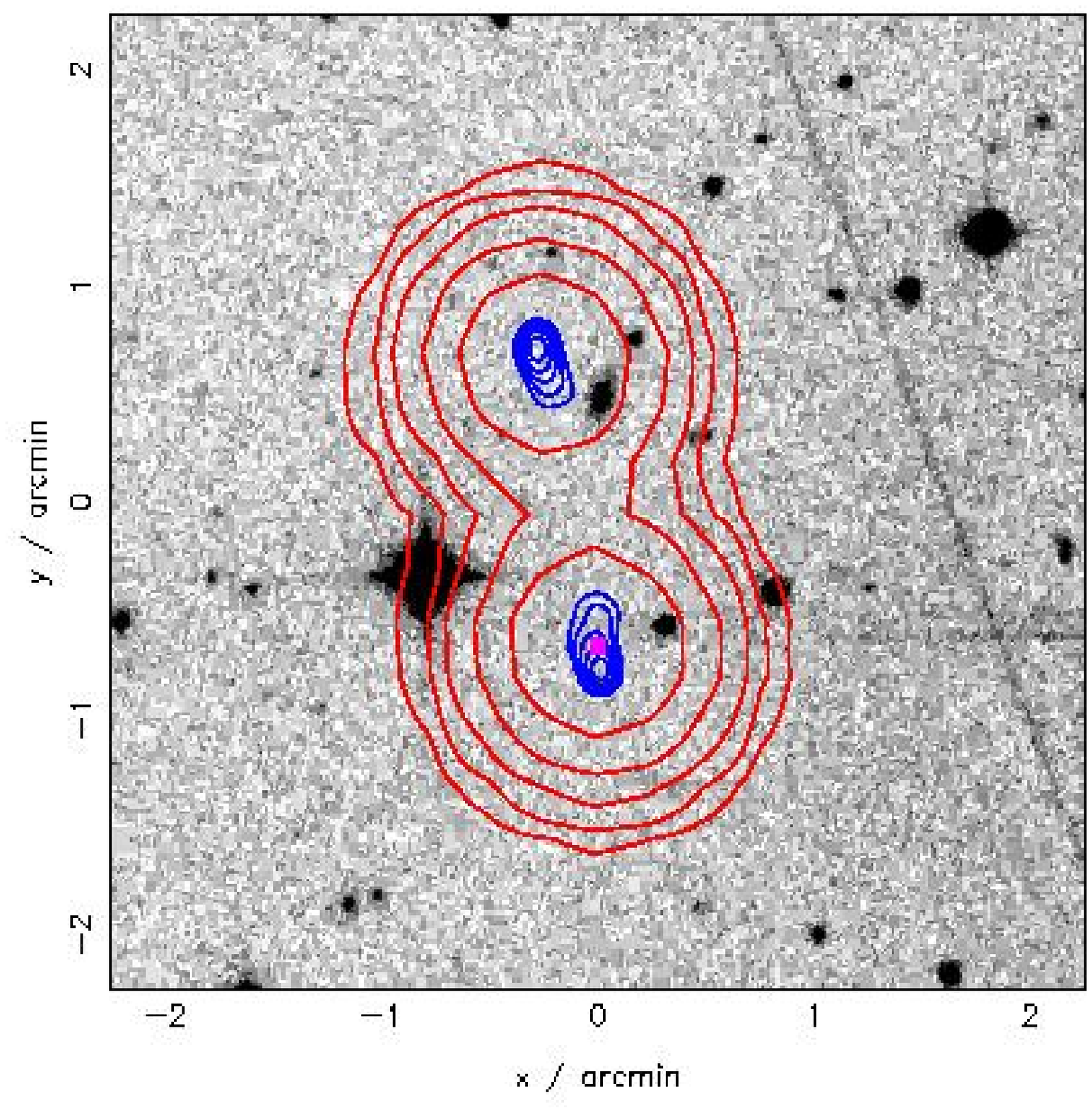}}
      \centerline{C4-030: MRC 1408-030}
    \end{minipage}
    \hspace{3cm}
    \begin{minipage}{3cm}
      \mbox{}
      \centerline{\includegraphics[scale=0.26,angle=270]{Contours/C4/032.ps}}
      \centerline{C4-032: TXS 1409-030}
    \end{minipage}
    \hspace{3cm}
    \begin{minipage}{3cm}
      \mbox{}
      \centerline{\includegraphics[scale=0.26,angle=270]{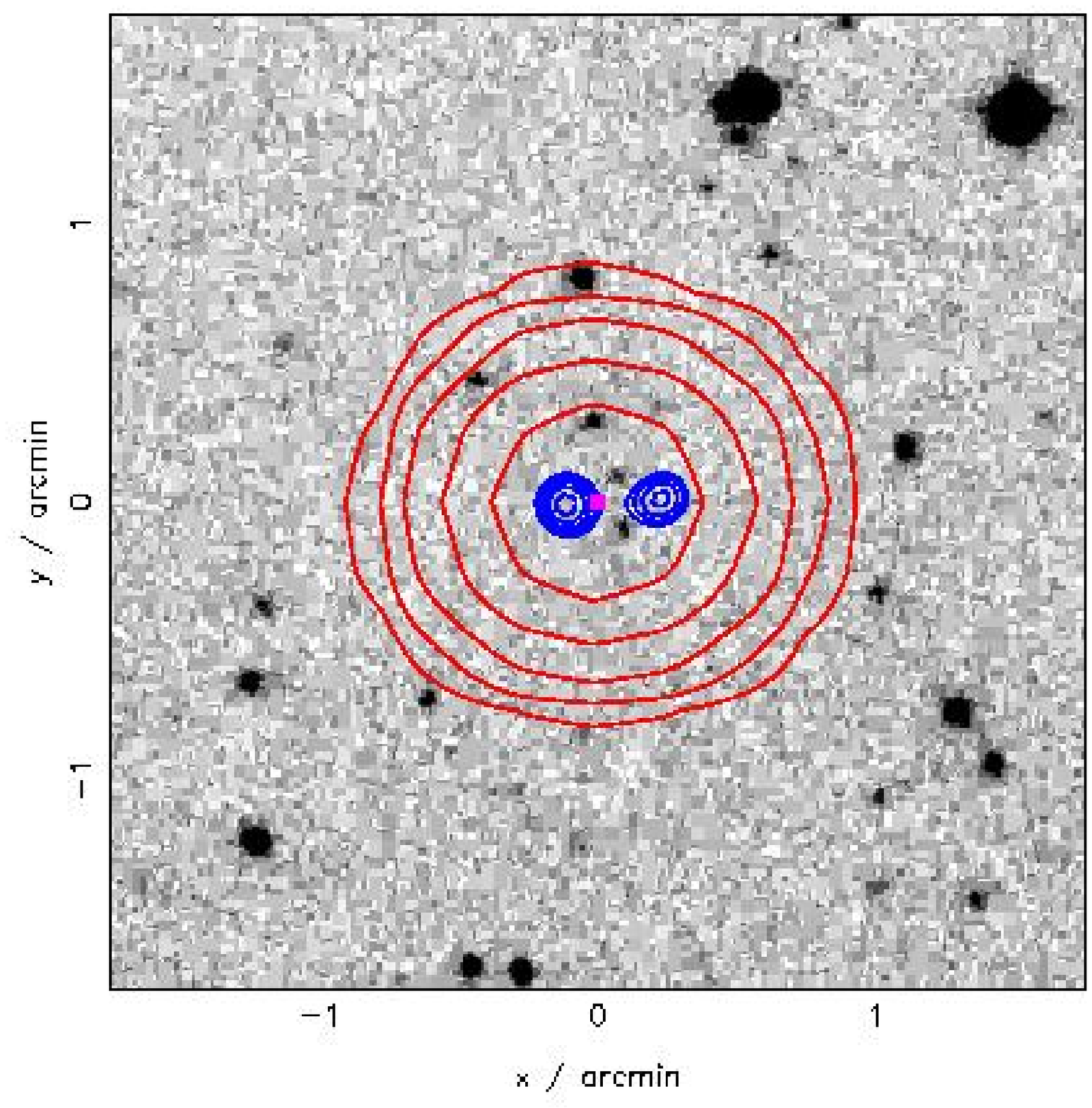}}
      \centerline{C4-033: TXS 1410+027}
    \end{minipage}
    \vfill
    \begin{minipage}{3cm}      
      \mbox{}
      \centerline{\includegraphics[scale=0.26,angle=270]{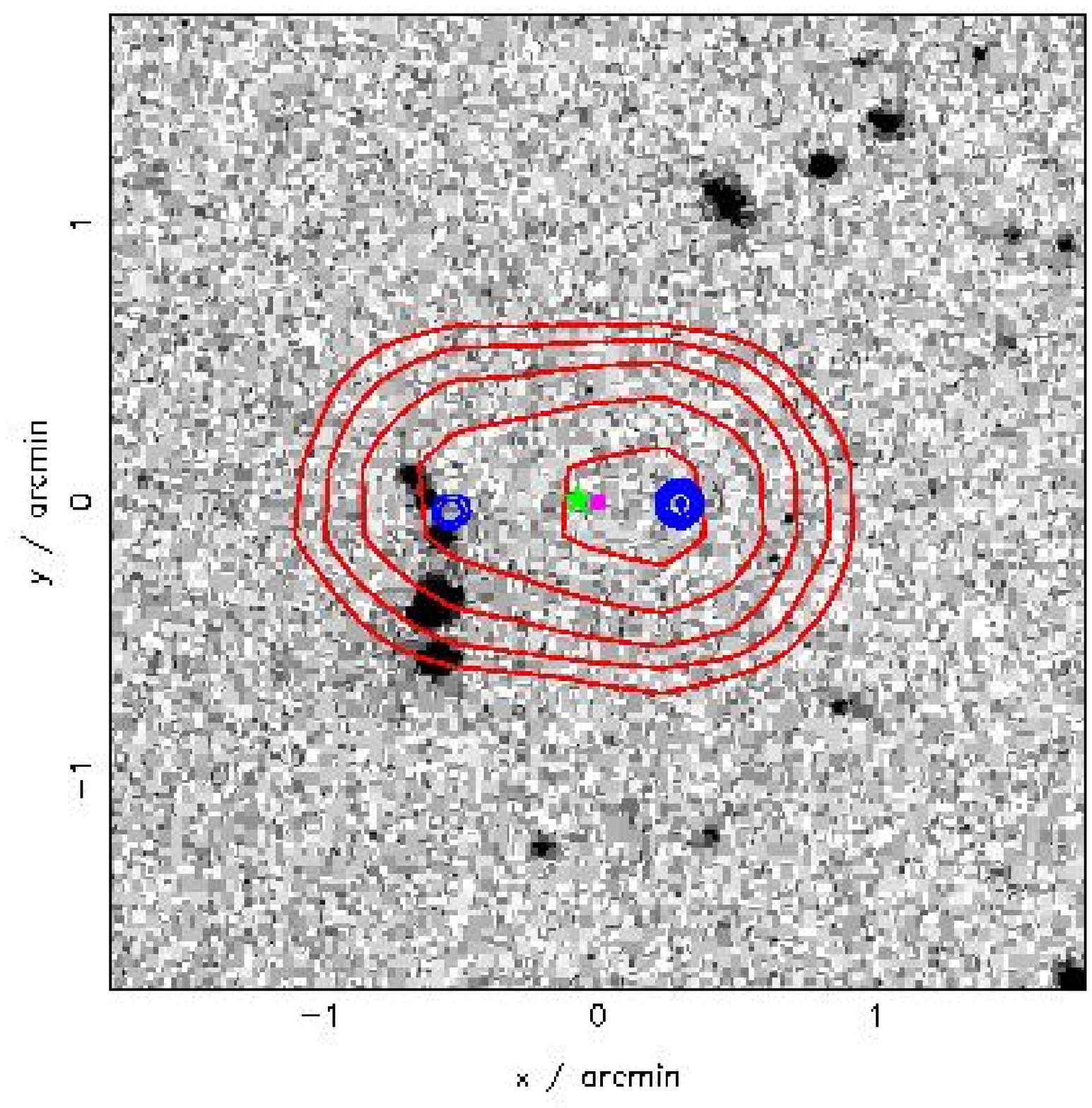}}
      \centerline{C4-035: 1412-0075}
    \end{minipage}
    \hspace{3cm}
    \begin{minipage}{3cm}
      \mbox{}
      \centerline{\includegraphics[scale=0.26,angle=270]{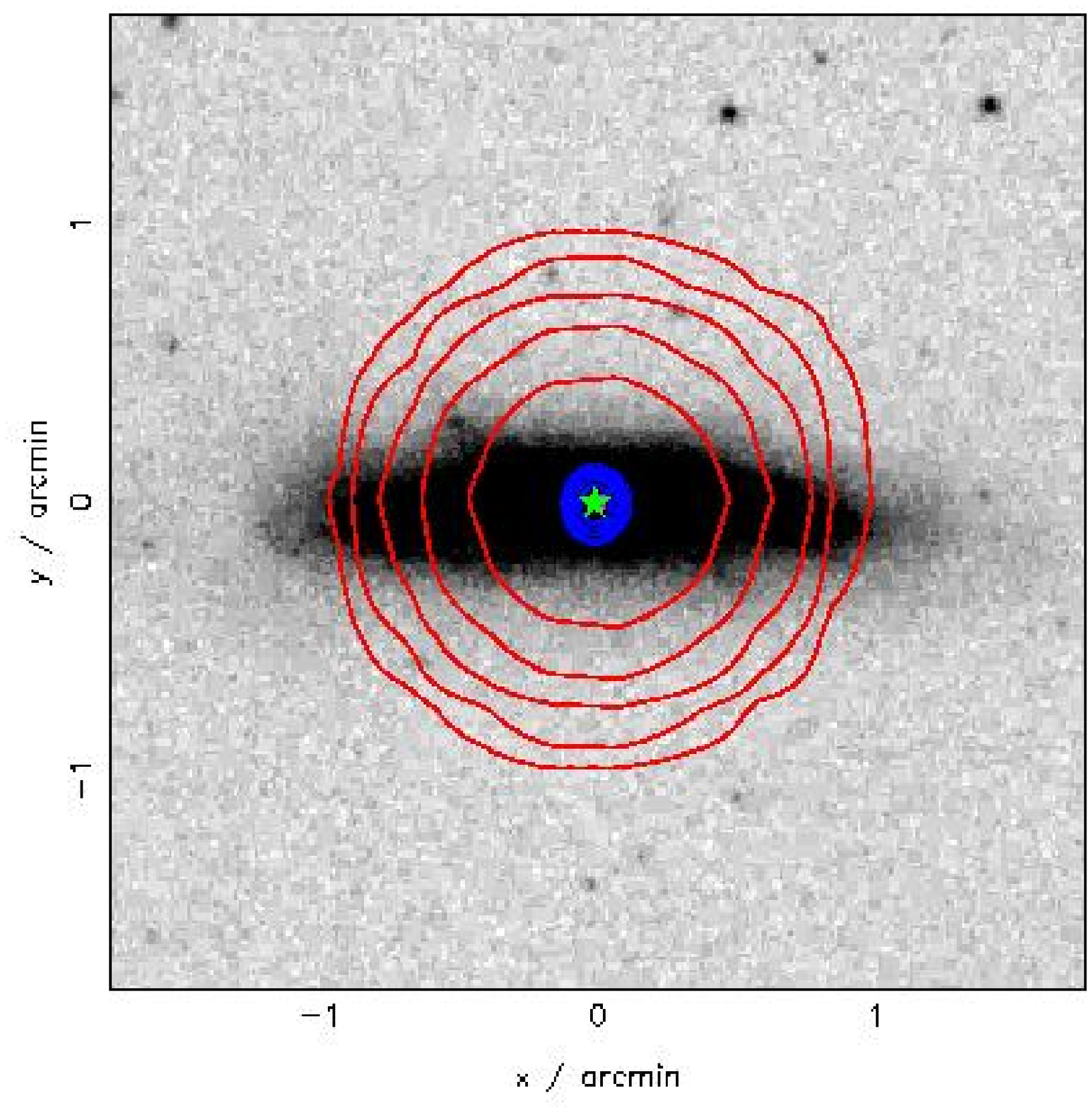}}
      \centerline{C4-036: NGC 5506}
    \end{minipage}
    \hspace{3cm}
    \begin{minipage}{3cm}
      \mbox{}
      \centerline{\includegraphics[scale=0.26,angle=270]{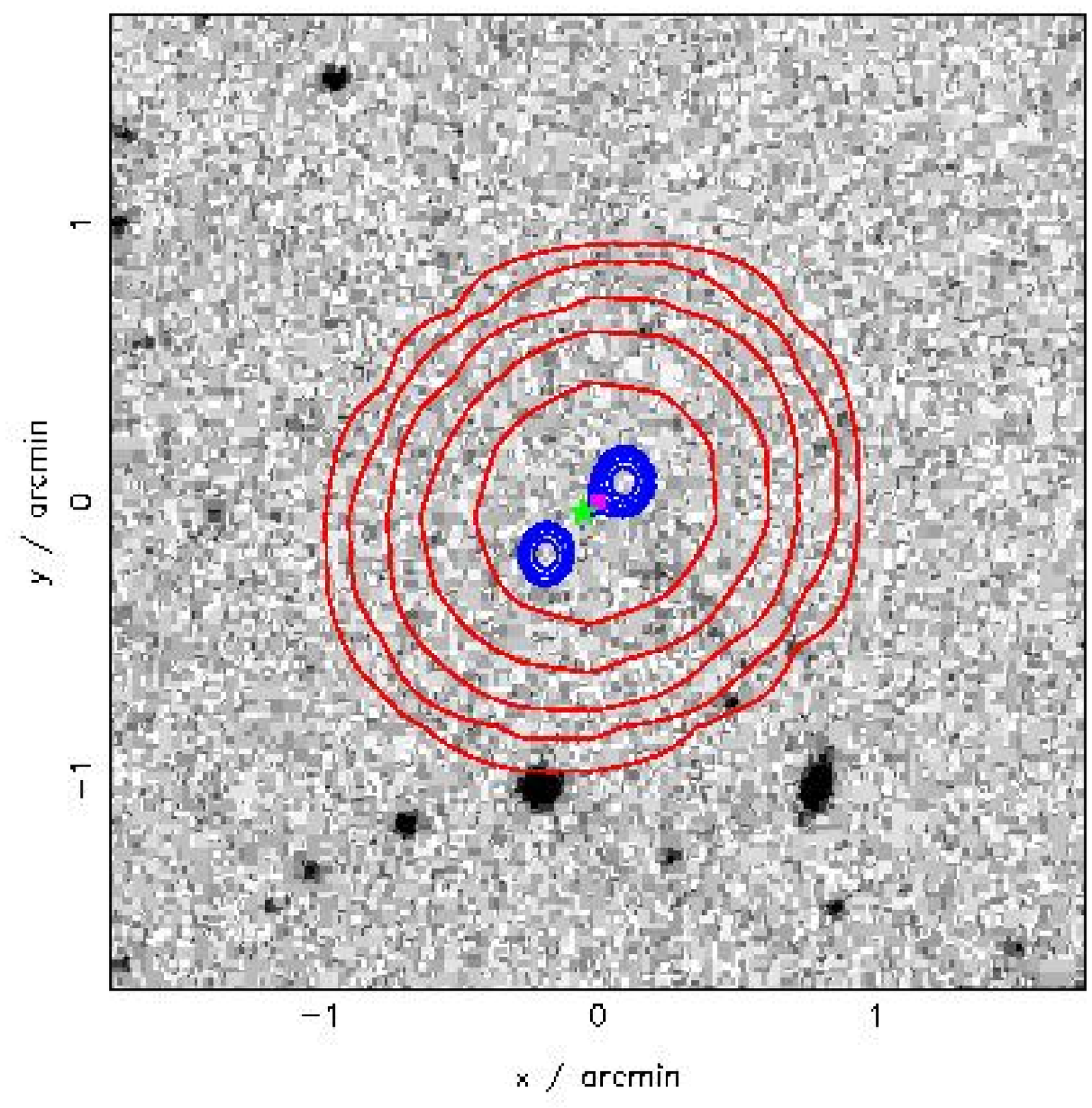}}
      \centerline{C4-037: 4C -00.54}
    \end{minipage}
  \end{center}
\end{figure}

\begin{figure}
  \begin{center}
    {\bf CoNFIG-4}\\  
  \begin{minipage}{3cm}      
      \mbox{}
      \centerline{\includegraphics[scale=0.26,angle=270]{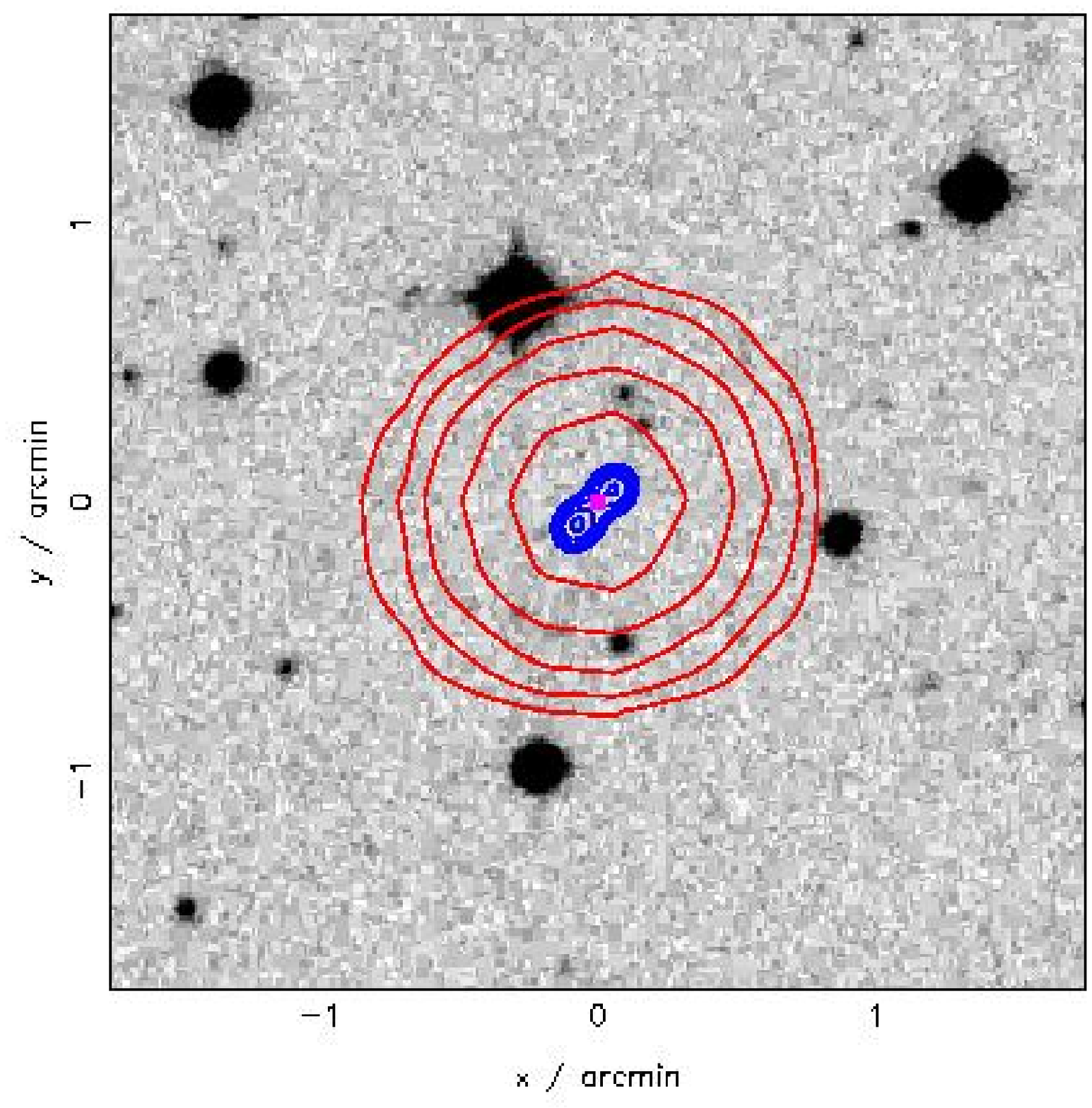}}
      \centerline{C4-038: TXS 1410-015}
    \end{minipage}
    \hspace{3cm}
    \begin{minipage}{3cm}
      \mbox{}
      \centerline{\includegraphics[scale=0.26,angle=270]{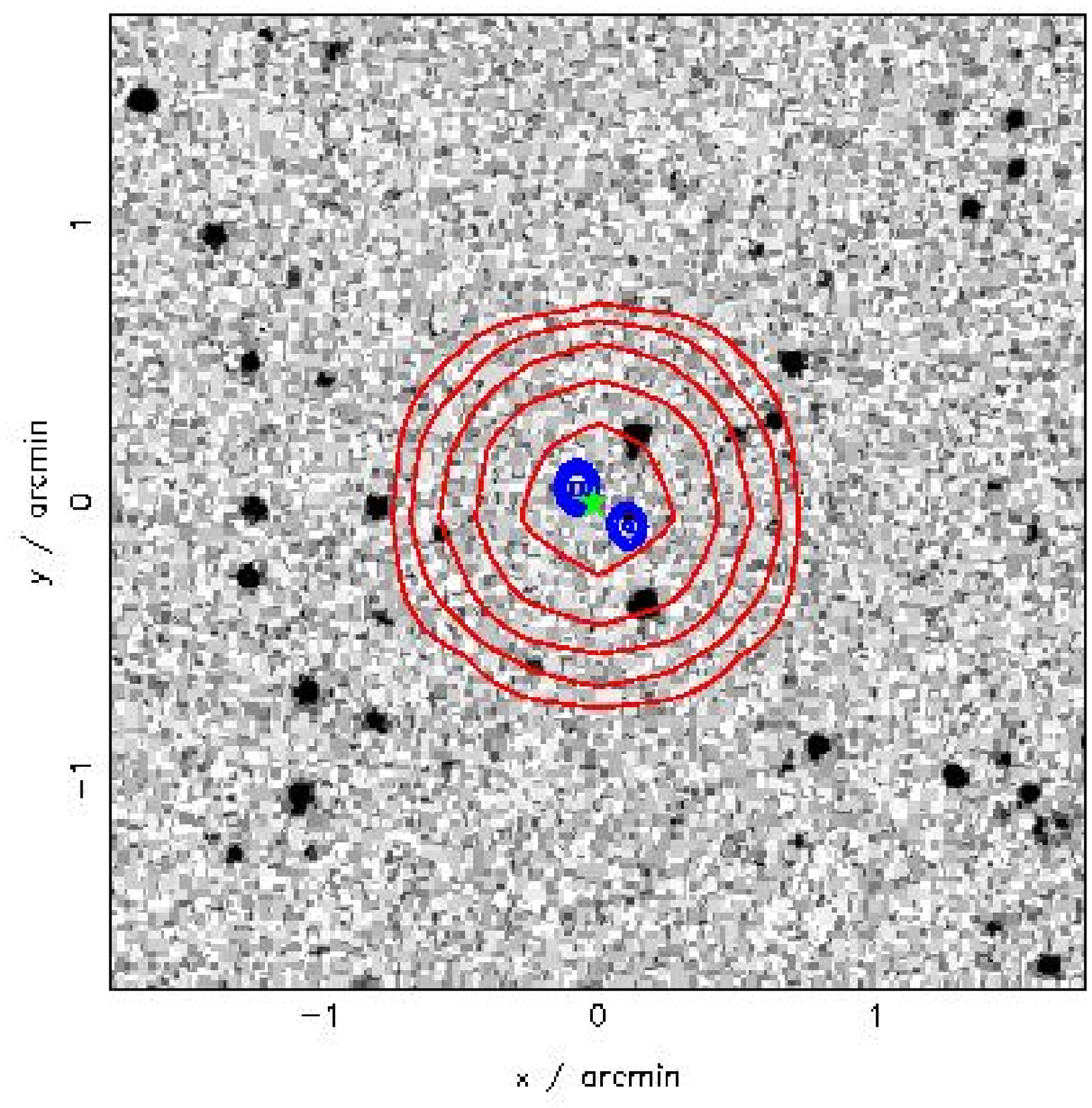}}
      \centerline{C4-039: TXS 1410+028}
    \end{minipage}
    \hspace{3cm}
    \begin{minipage}{3cm}
      \mbox{}
      \centerline{\includegraphics[scale=0.26,angle=270]{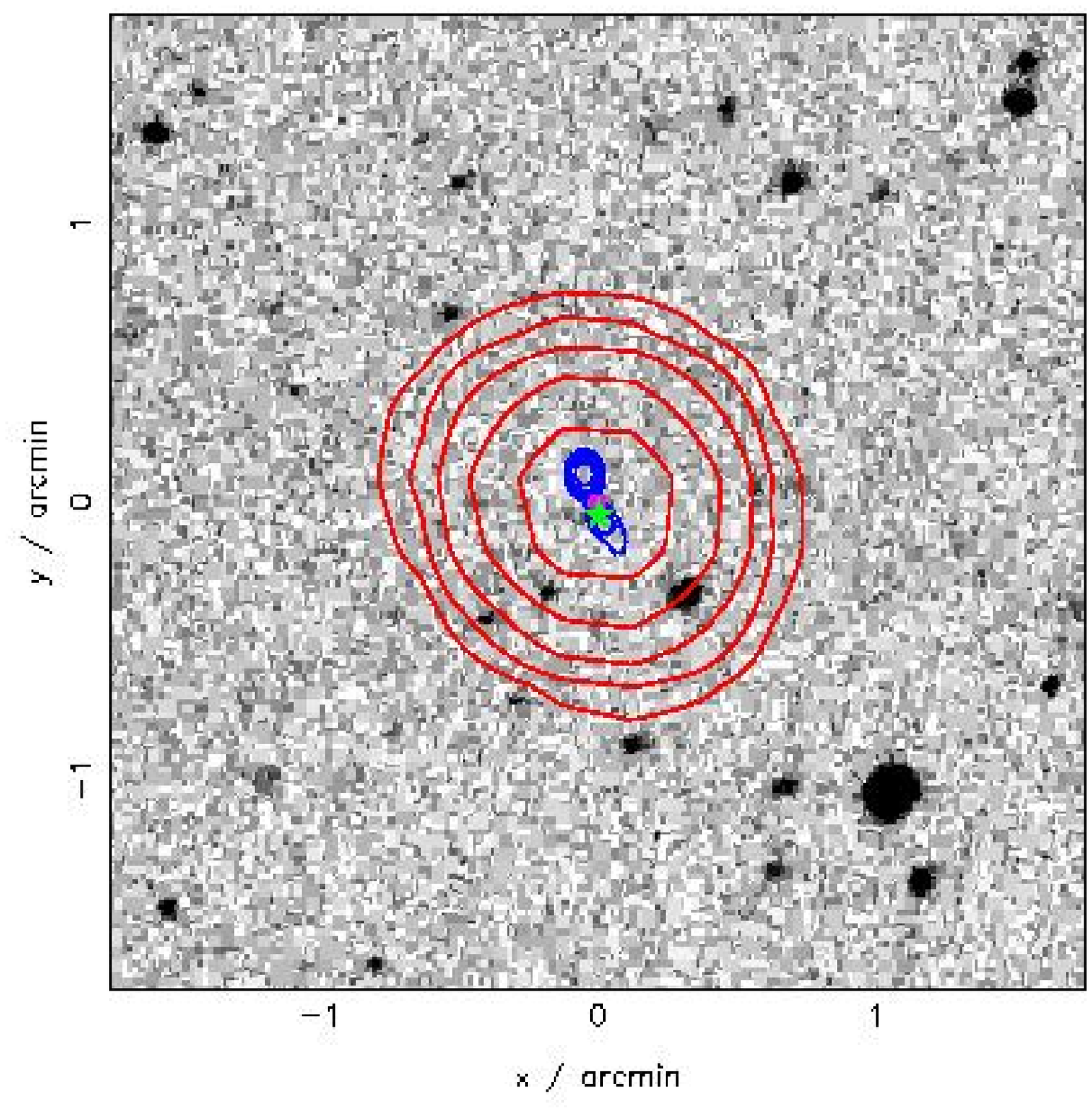}}
      \centerline{C4-041: 1413-0255}
    \end{minipage}
    \vfill
    \begin{minipage}{3cm}     
      \mbox{}
      \centerline{\includegraphics[scale=0.26,angle=270]{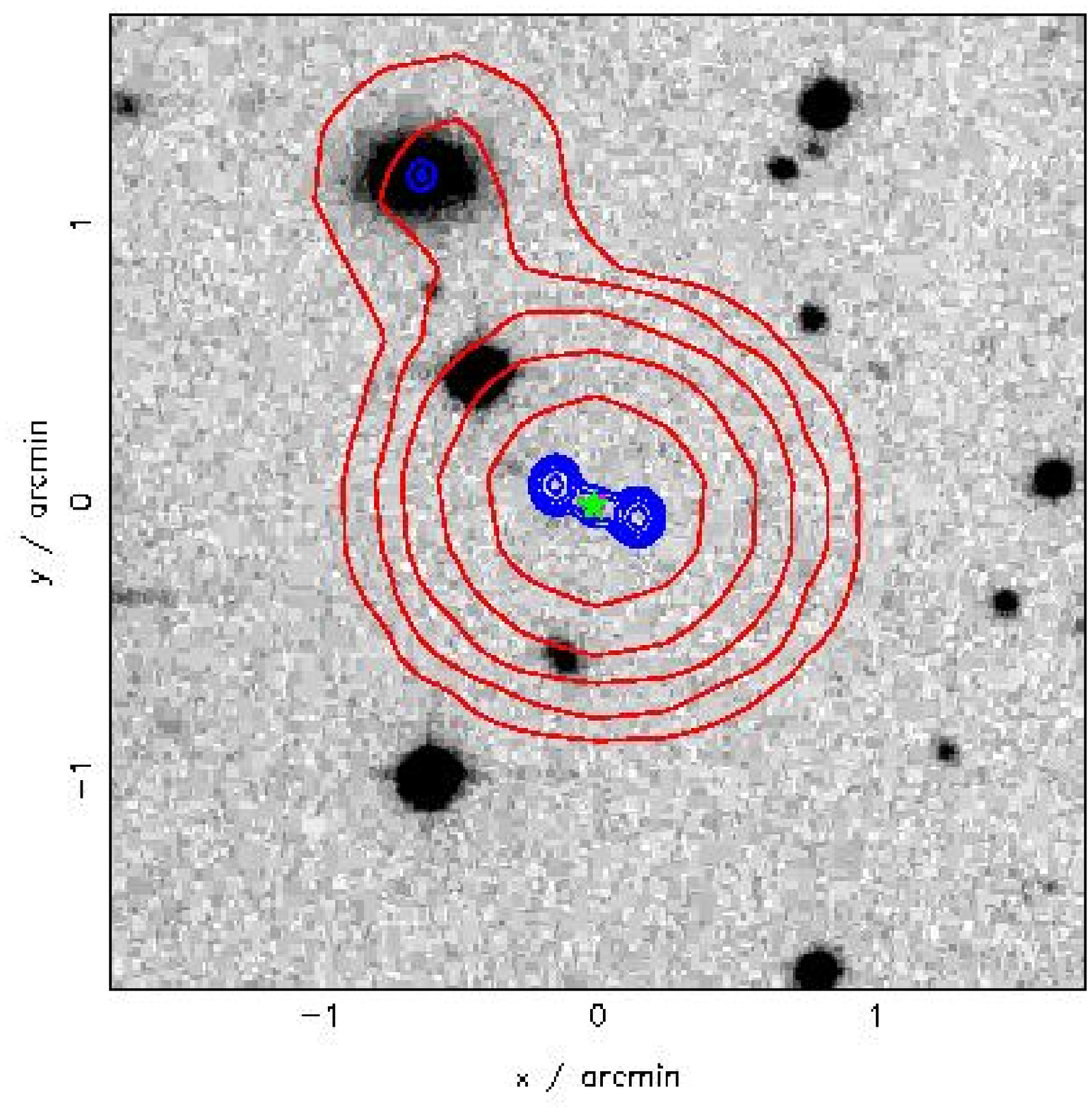}}
      \centerline{C4-043: TXS 1411+019}
    \end{minipage}
    \hspace{3cm}
    \begin{minipage}{3cm}
      \mbox{}
      \centerline{\includegraphics[scale=0.26,angle=270]{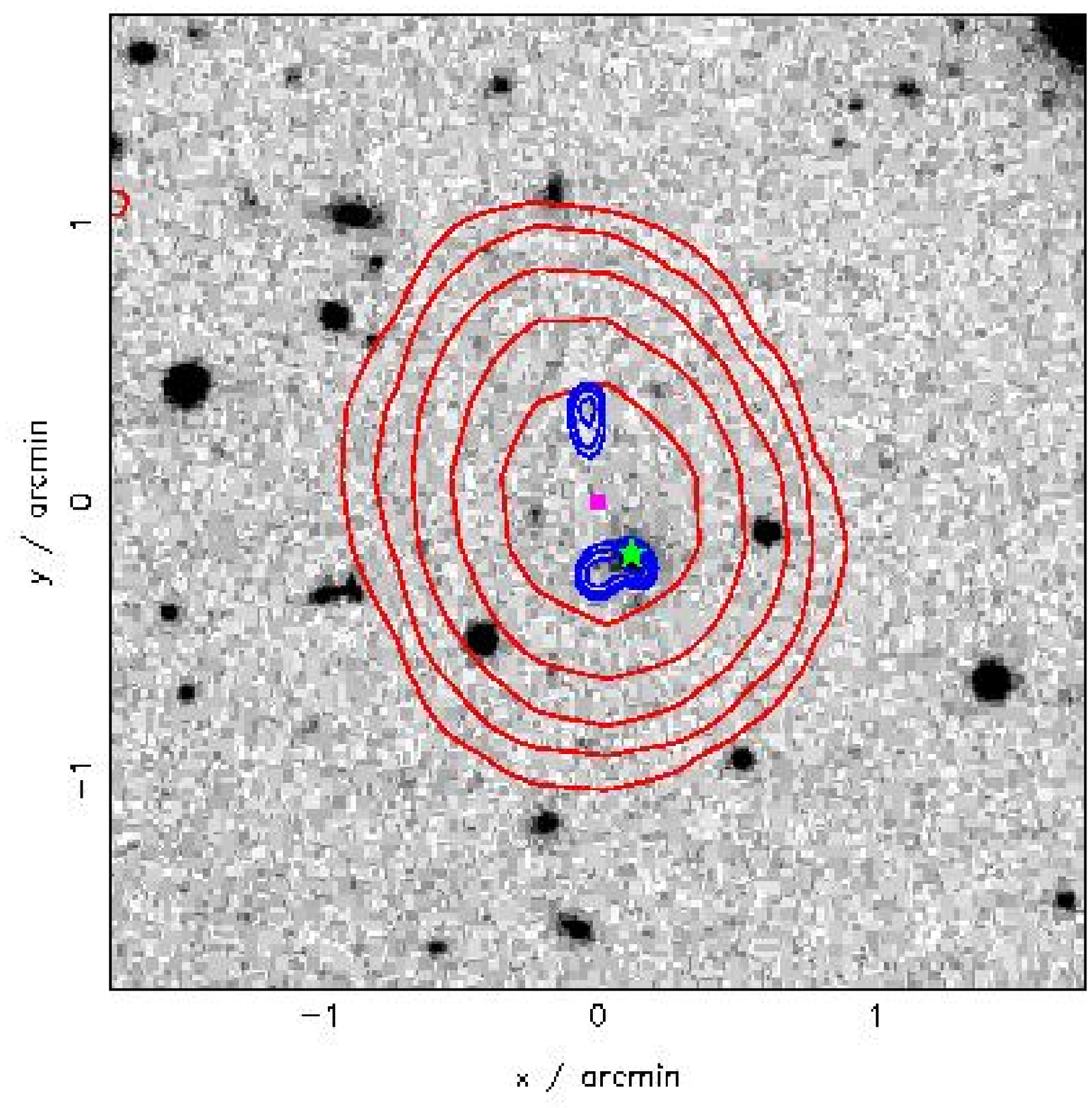}}
      \centerline{C4-044: 1414+0182}
    \end{minipage}
    \hspace{3cm}
    \begin{minipage}{3cm}
      \mbox{}
      \centerline{\includegraphics[scale=0.26,angle=270]{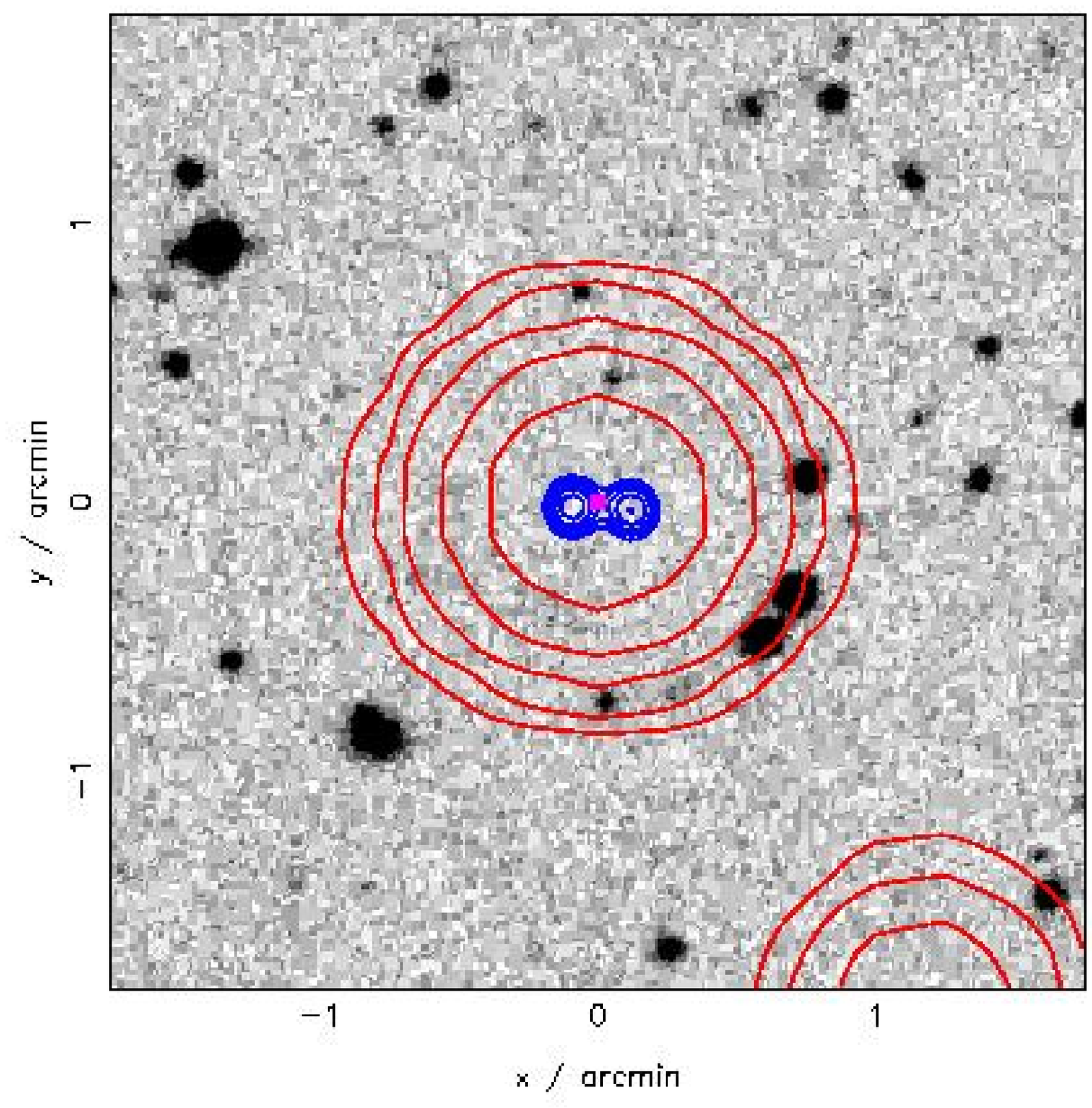}}
      \centerline{C4-045: TXS 1411+002}
    \end{minipage}
    \vfill
    \begin{minipage}{3cm}     
      \mbox{}
      \centerline{\includegraphics[scale=0.26,angle=270]{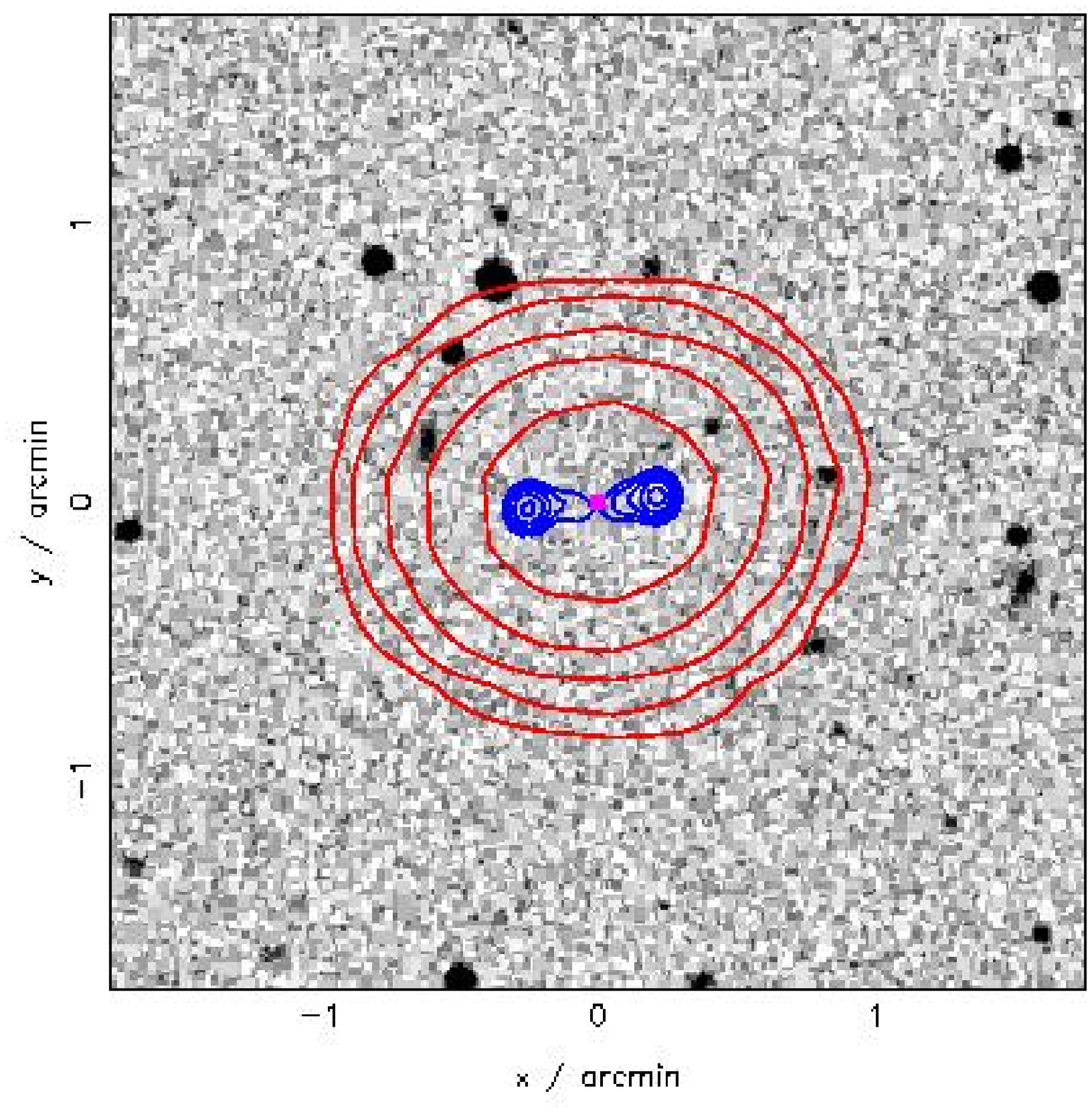}}
      \centerline{C4-046: TXS 1412+031}
    \end{minipage}
    \hspace{3cm}
    \begin{minipage}{3cm}
      \mbox{}
      \centerline{\includegraphics[scale=0.26,angle=270]{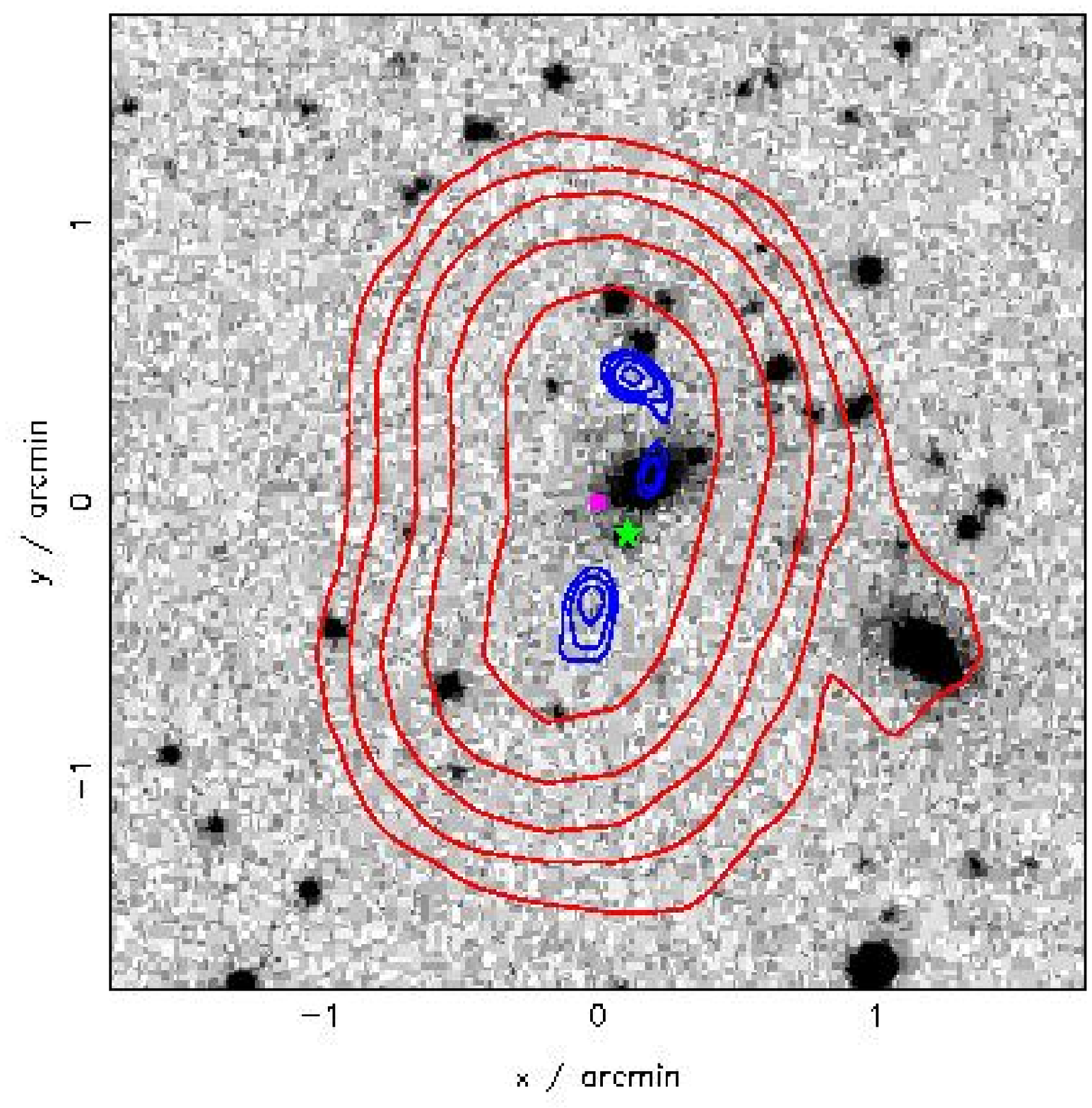}}
      \centerline{C4-047: LEDA 184576}
    \end{minipage}
    \hspace{3cm}
    \begin{minipage}{3cm}
      \mbox{}
      \centerline{\includegraphics[scale=0.26,angle=270]{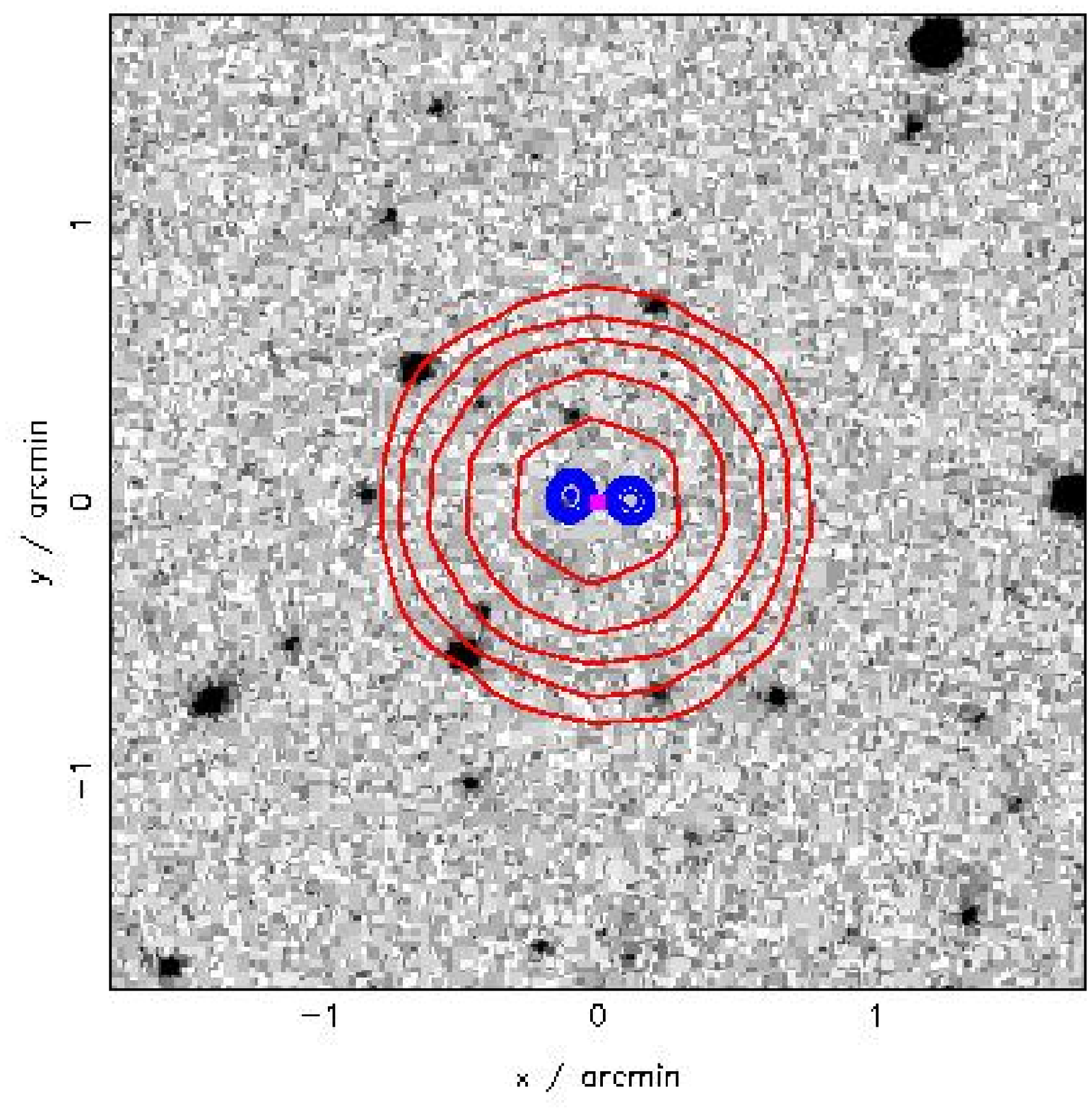}}
      \centerline{C4-048: 1415+0060}
    \end{minipage}
    \vfill
    \begin{minipage}{3cm}      
      \mbox{}
      \centerline{\includegraphics[scale=0.26,angle=270]{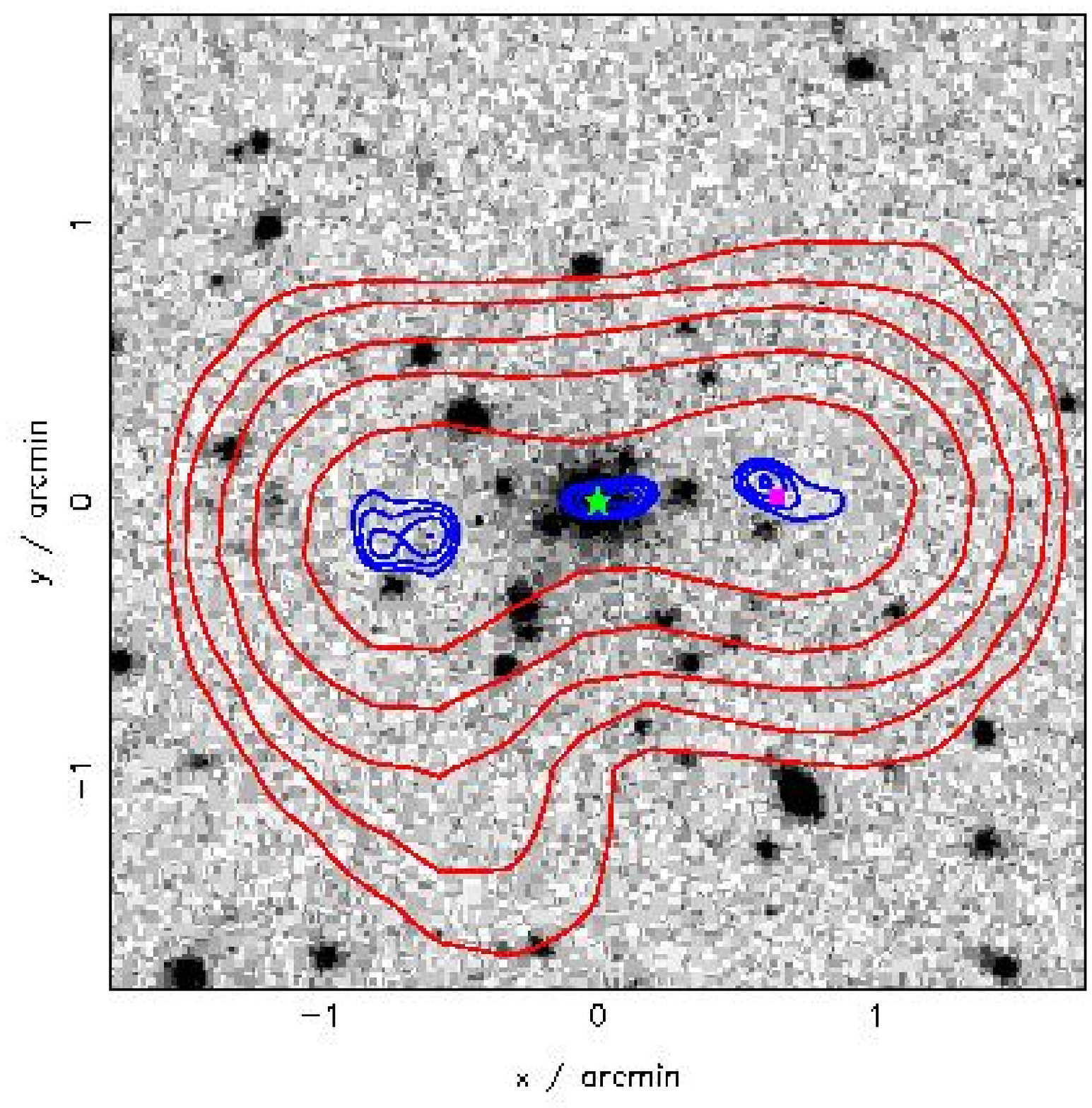}}
      \centerline{C4-049: N274Z243}
    \end{minipage}
    \hspace{3cm}
    \begin{minipage}{3cm}
      \mbox{}
      \centerline{\includegraphics[scale=0.26,angle=270]{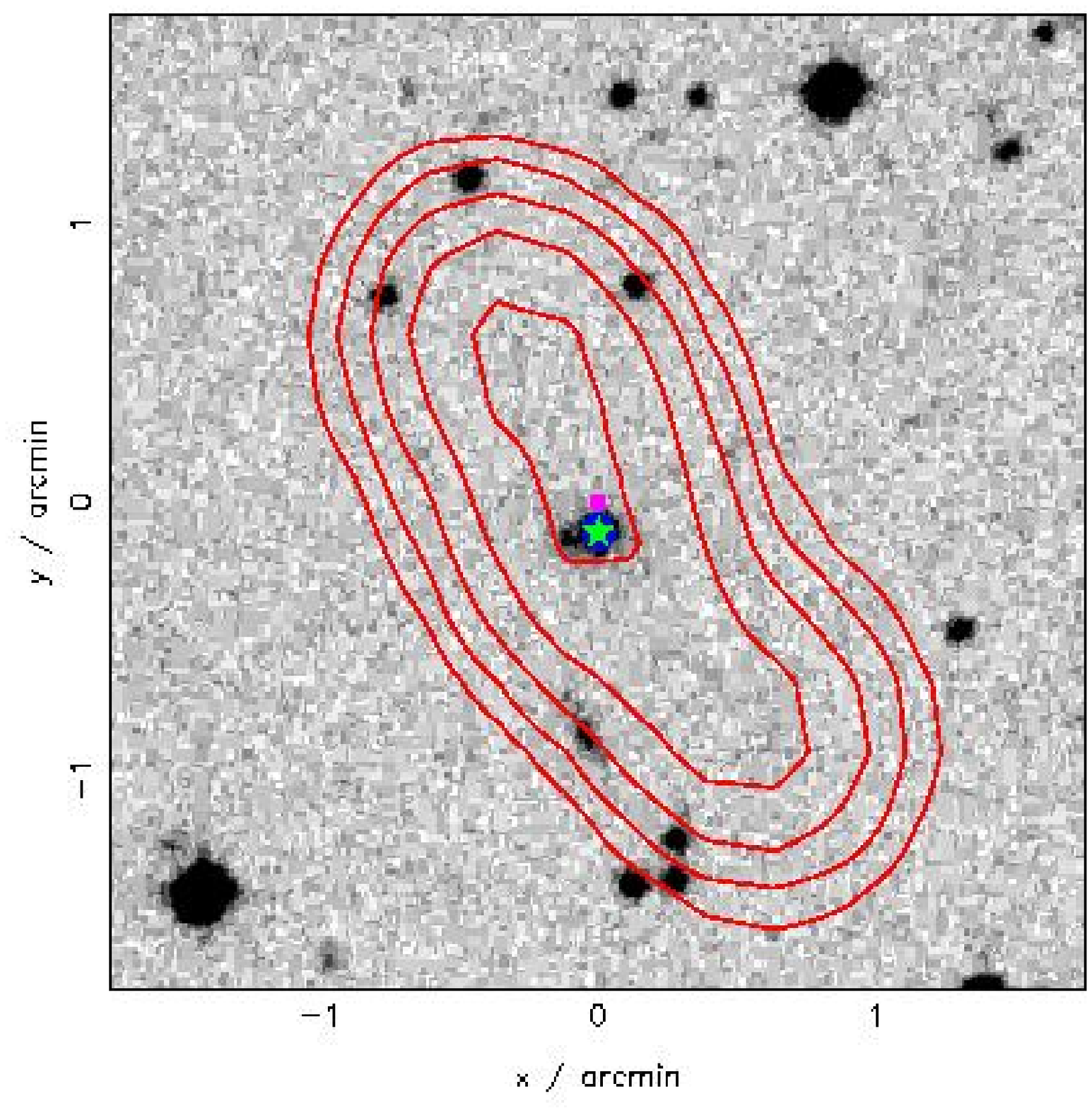}}
      \centerline{C4-050: N342Z086}
    \end{minipage}
    \hspace{3cm}
    \begin{minipage}{3cm}
      \mbox{}
      \centerline{\includegraphics[scale=0.26,angle=270]{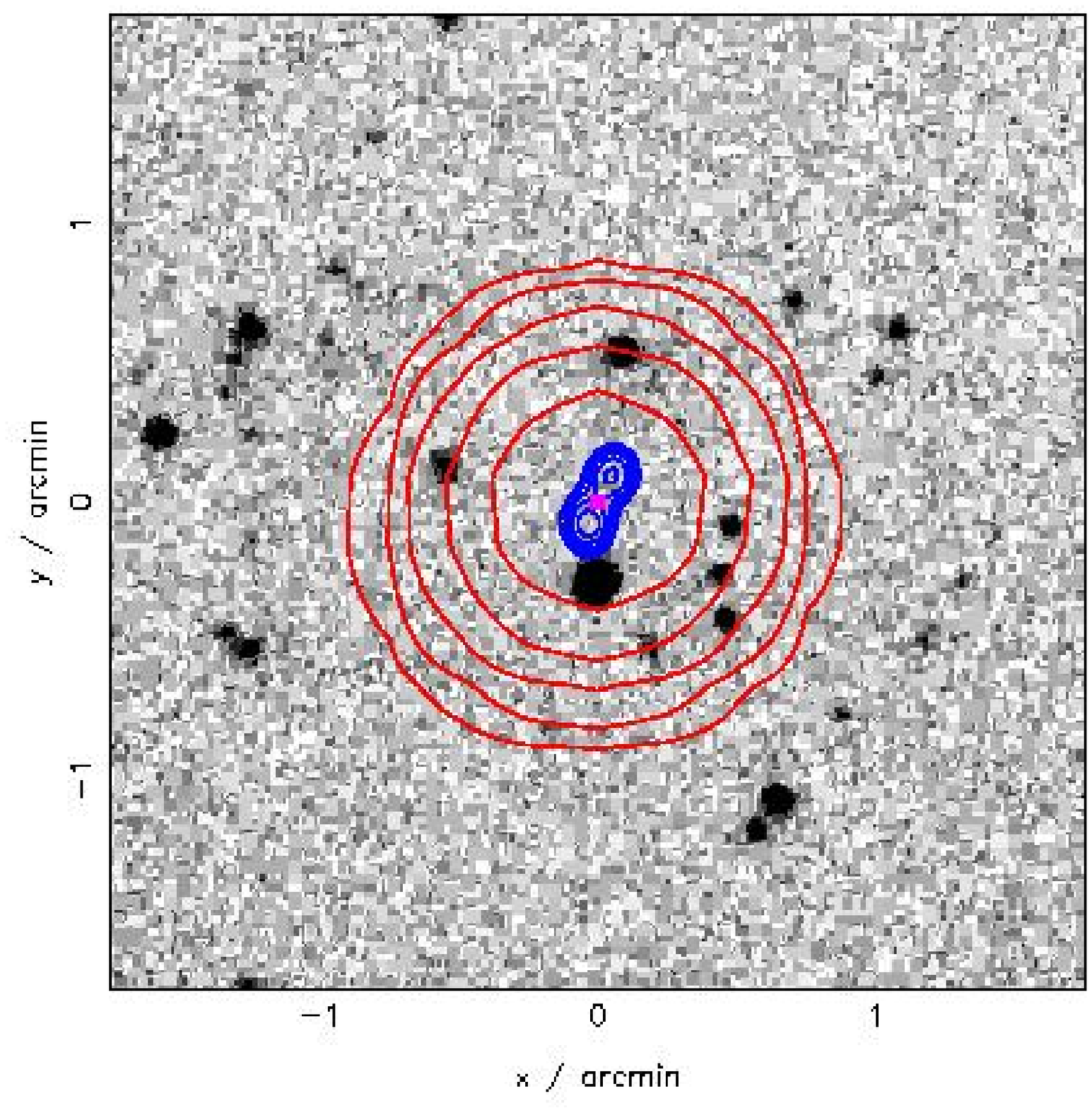}}
      \centerline{C4-051: TXS 1412+026}
    \end{minipage}
  \end{center}
\end{figure}

\begin{figure}
  \begin{center}
    {\bf CoNFIG-4}\\  
  \begin{minipage}{3cm}      
      \mbox{}
      \centerline{\includegraphics[scale=0.26,angle=270]{Contours/C4/053.ps}}
      \centerline{C4-053: TXS 1413+007}
    \end{minipage}
    \hspace{3cm}
    \begin{minipage}{3cm}
      \mbox{}
      \centerline{\includegraphics[scale=0.26,angle=270]{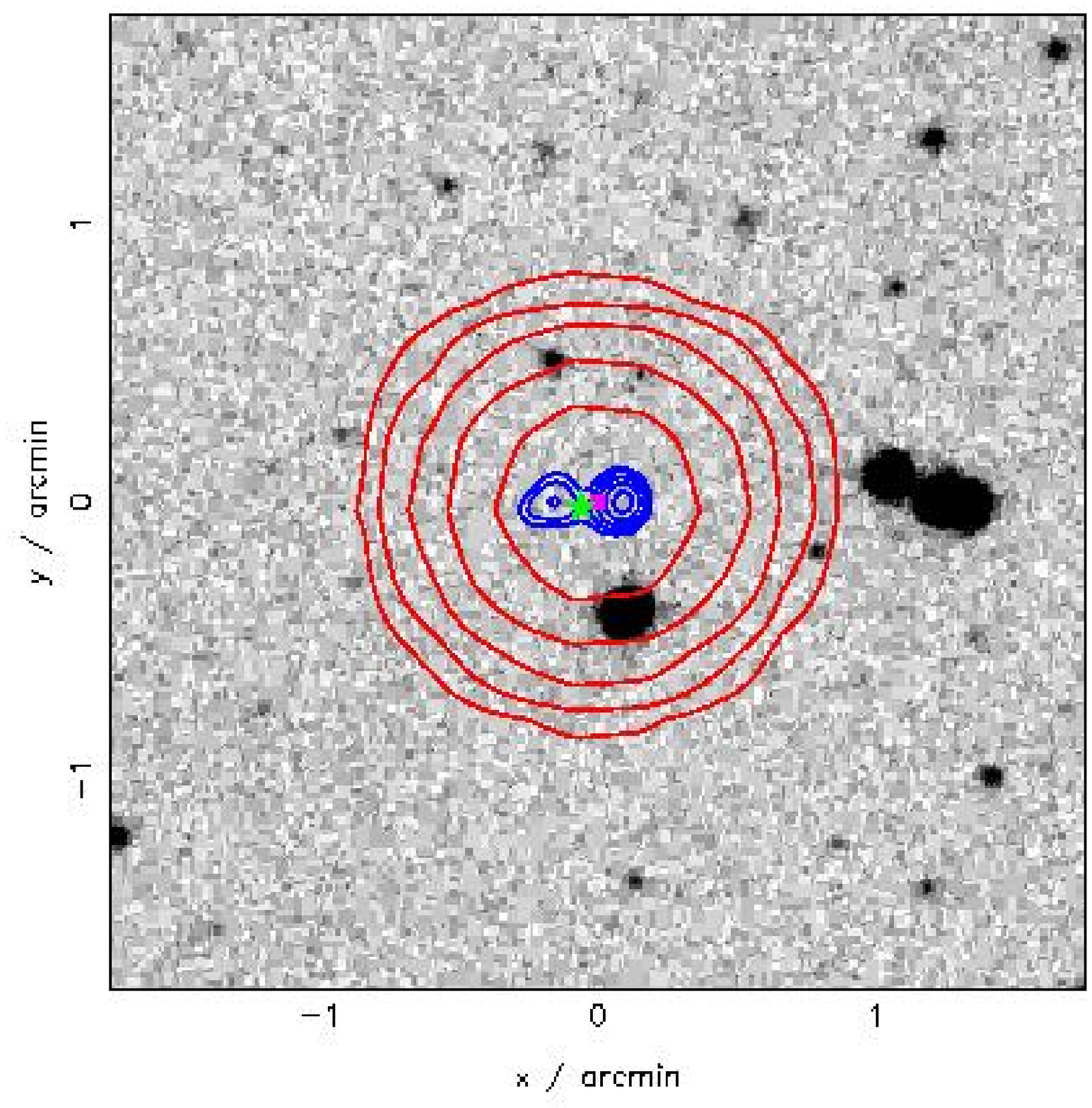}}
      \centerline{C4-054: TXS 1413-011}
    \end{minipage}
    \hspace{3cm}
    \begin{minipage}{3cm}
      \mbox{}
      \centerline{\includegraphics[scale=0.26,angle=270]{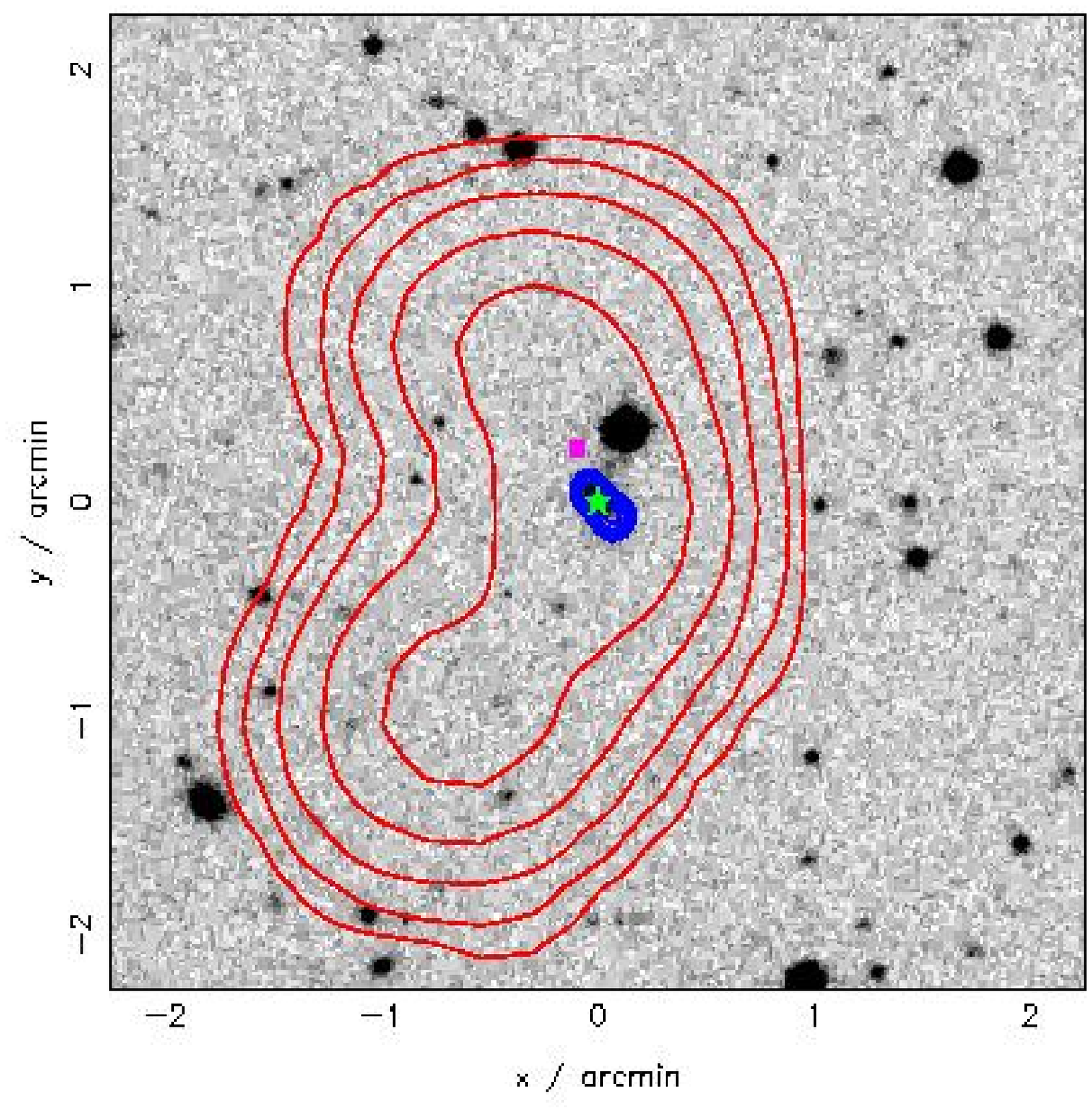}}
      \centerline{C4-055: 1416+0219}
    \end{minipage}
    \vfill
    \begin{minipage}{3cm}     
      \mbox{}
      \centerline{\includegraphics[scale=0.26,angle=270]{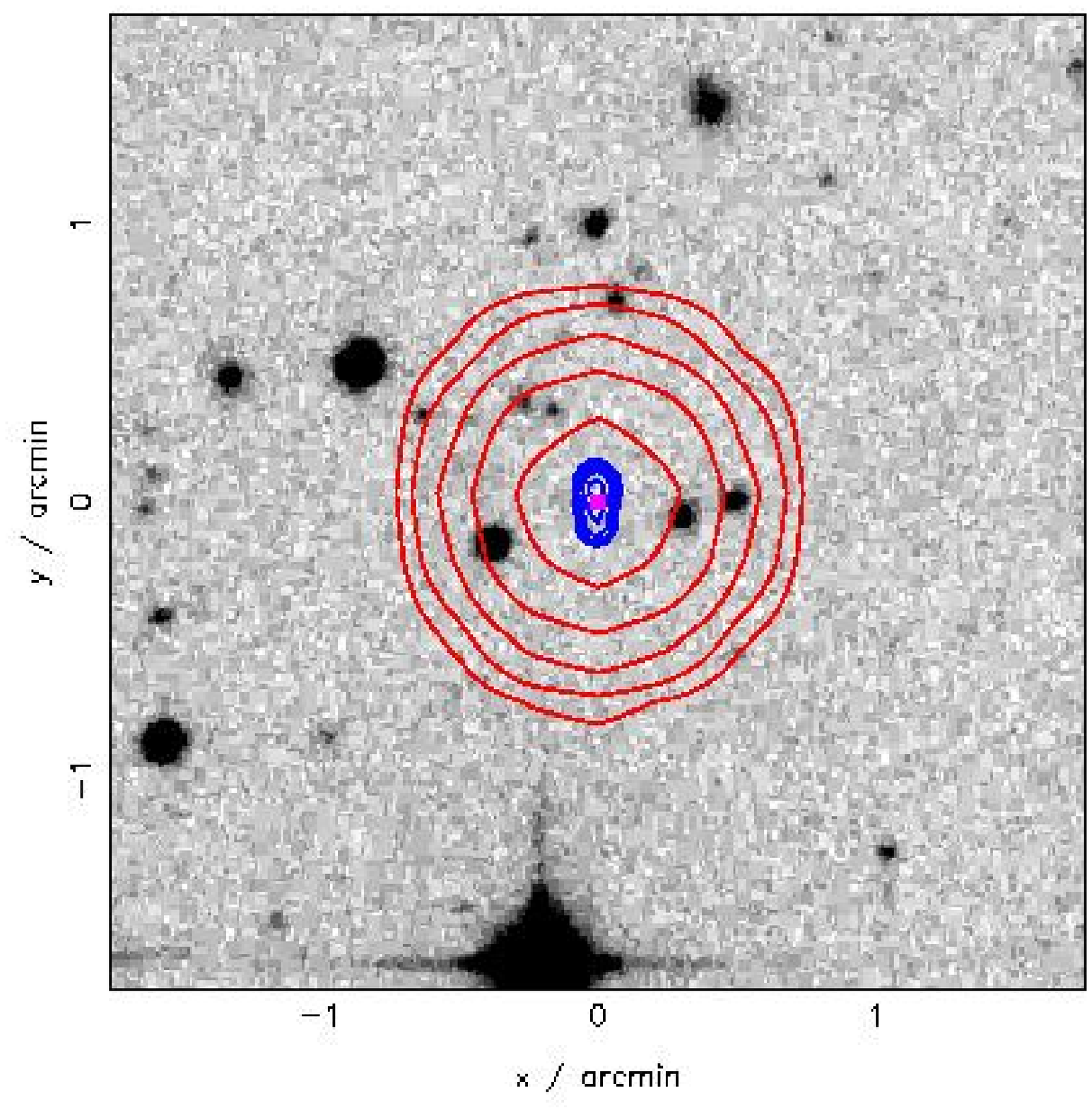}}
      \centerline{C4-058: TXS 1414-007}
    \end{minipage}
    \hspace{3cm}
    \begin{minipage}{3cm}
      \mbox{}
      \centerline{\includegraphics[scale=0.26,angle=270]{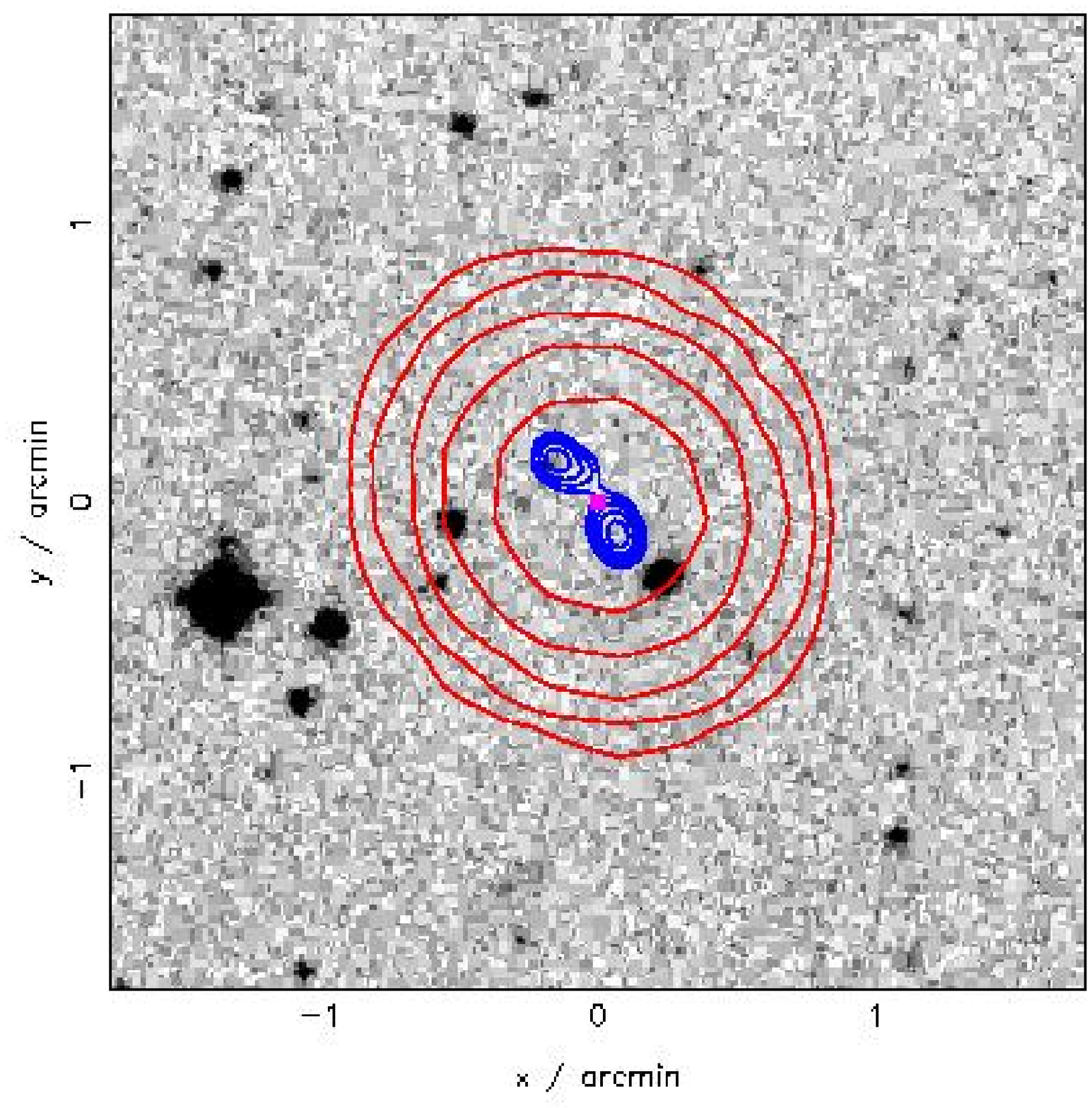}}
      \centerline{C4-059: TXS 1415+008}
    \end{minipage}
    \hspace{3cm}
    \begin{minipage}{3cm}
      \mbox{}
      \centerline{\includegraphics[scale=0.26,angle=270]{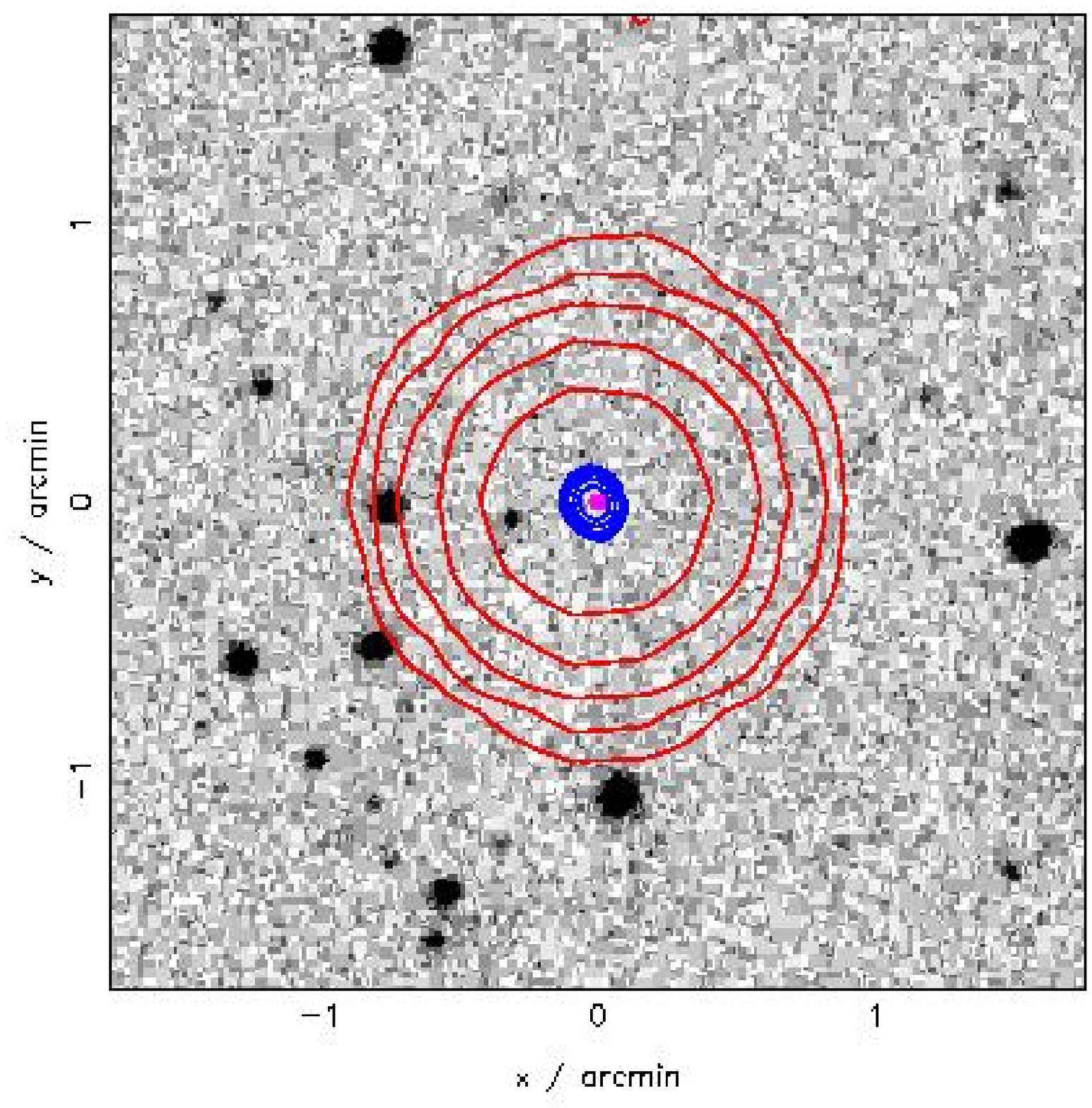}}
      \centerline{C4-060: TXS 1415-011}
    \end{minipage}
    \vfill
    \begin{minipage}{3cm}     
      \mbox{}
      \centerline{\includegraphics[scale=0.26,angle=270]{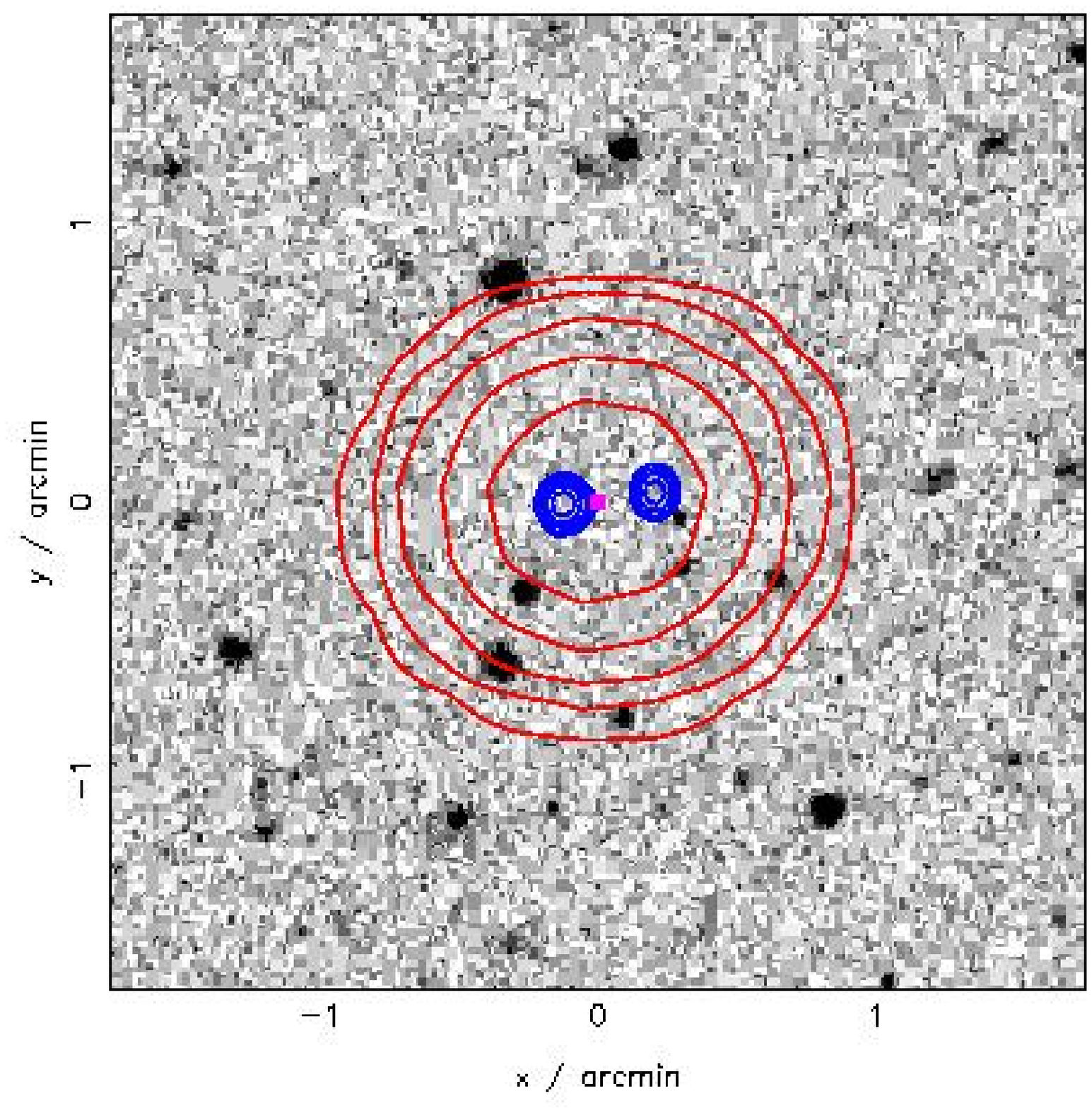}}
      \centerline{C4-061: TXS 1415+013}
    \end{minipage}
    \hspace{3cm}
    \begin{minipage}{3cm}
      \mbox{}
      \centerline{\includegraphics[scale=0.26,angle=270]{Contours/C4/062.ps}}
      \centerline{C4-062: TXS 1415+016}
    \end{minipage}
    \hspace{3cm}
    \begin{minipage}{3cm}
      \mbox{}
      \centerline{\includegraphics[scale=0.26,angle=270]{Contours/C4/063.ps}}
      \centerline{C4-063: TXS 1416-022}
    \end{minipage}
    \vfill
    \begin{minipage}{3cm}      
      \mbox{}
      \centerline{\includegraphics[scale=0.26,angle=270]{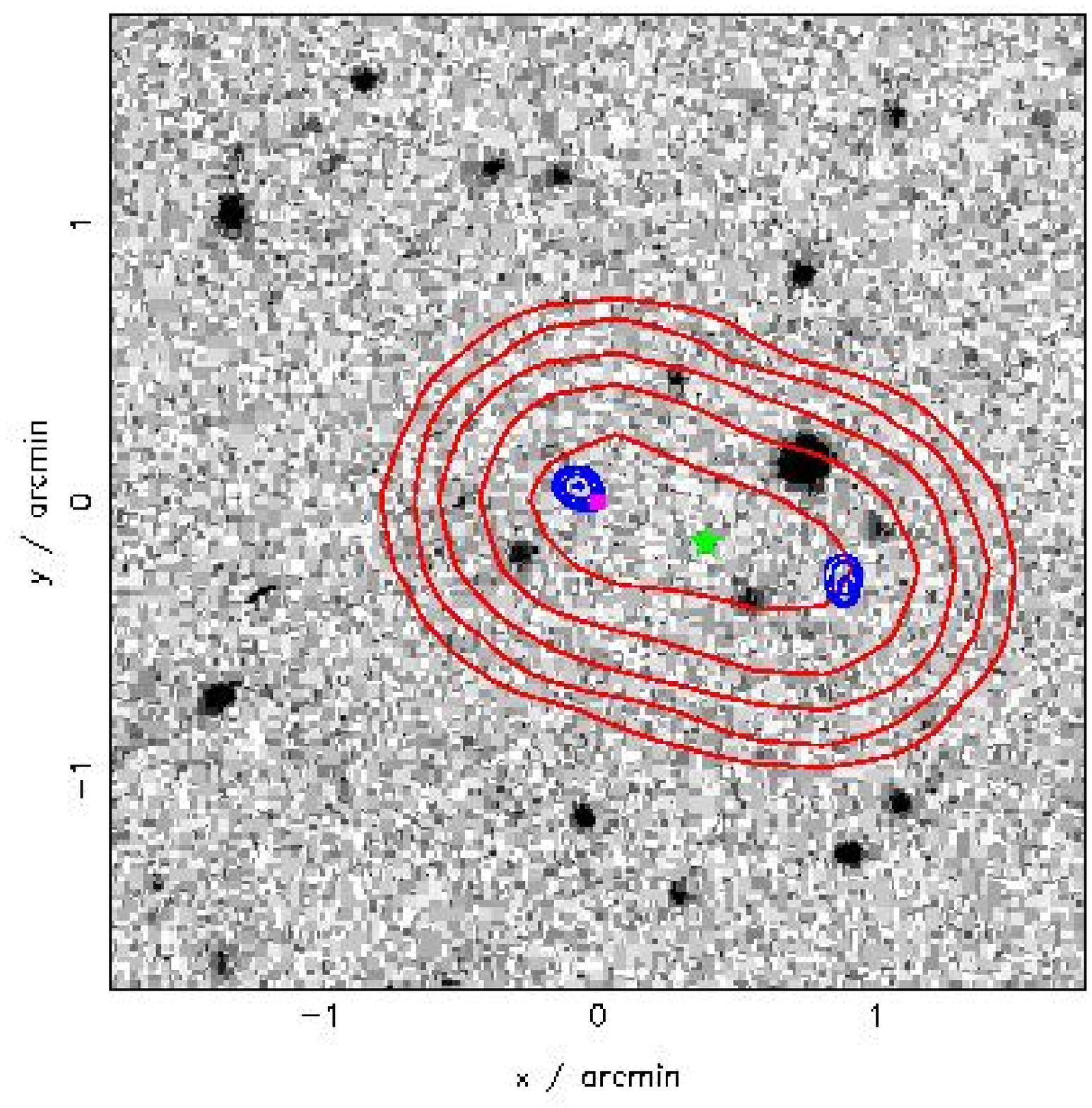}}
      \centerline{C4-064: TXS 1416+006}
    \end{minipage}
    \hspace{3cm}
    \begin{minipage}{3cm}
      \mbox{}
      \centerline{\includegraphics[scale=0.26,angle=270]{Contours/C4/067.ps}}
      \centerline{C4-067: 1419-0324}
    \end{minipage}
    \hspace{3cm}
    \begin{minipage}{3cm}
      \mbox{}
      \centerline{\includegraphics[scale=0.26,angle=270]{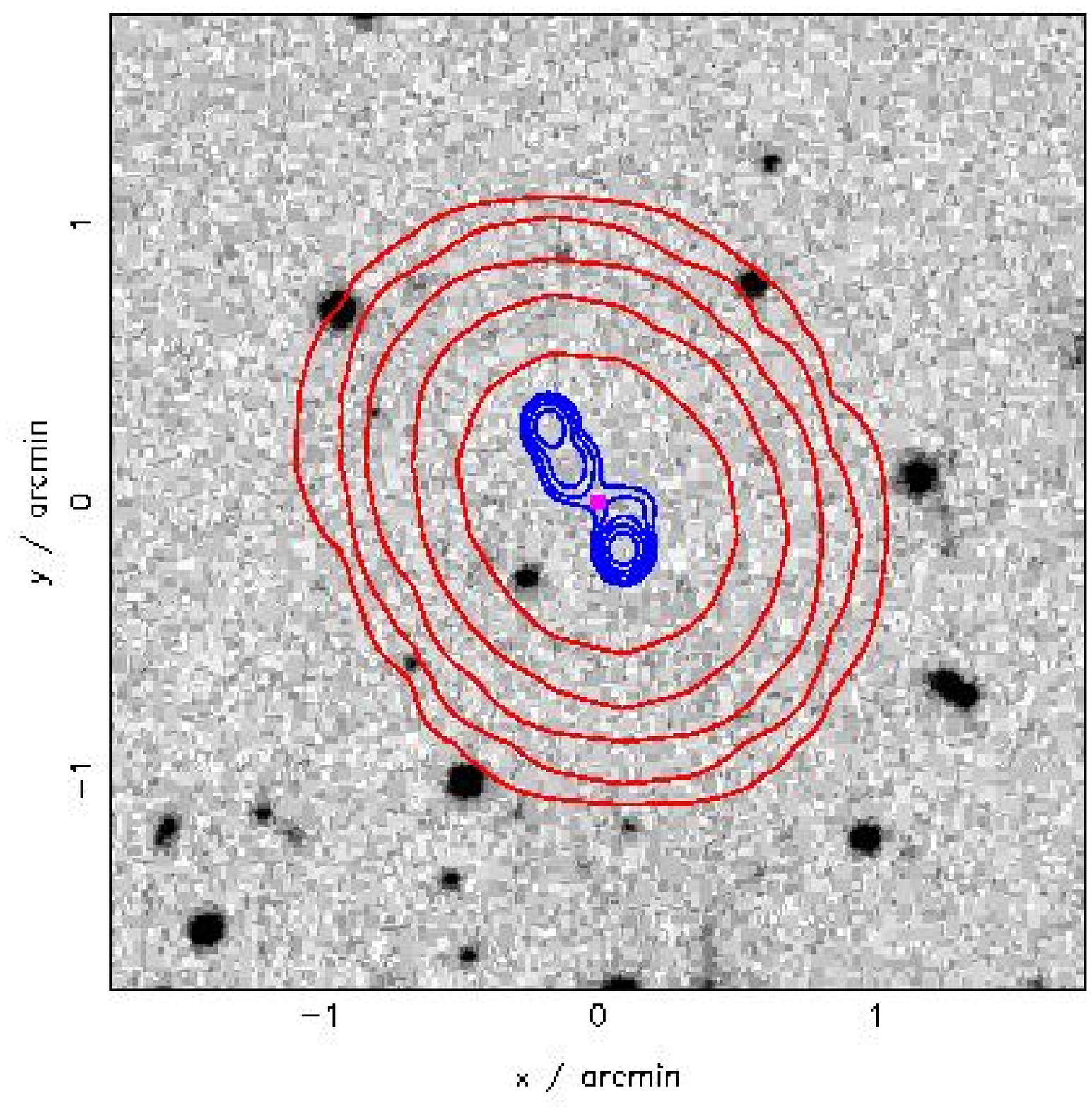}}
      \centerline{C4-068: TXS 1416-000}
    \end{minipage}
  \end{center}
\end{figure}

\begin{figure}
  \begin{center}
    {\bf CoNFIG-4}\\  
  \begin{minipage}{3cm}      
      \mbox{}
      \centerline{\includegraphics[scale=0.26,angle=270]{Contours/C4/070.ps}}
      \centerline{C4-070: 4C -01.33}
    \end{minipage}
    \hspace{3cm}
    \begin{minipage}{3cm}
      \mbox{}
      \centerline{\includegraphics[scale=0.26,angle=270]{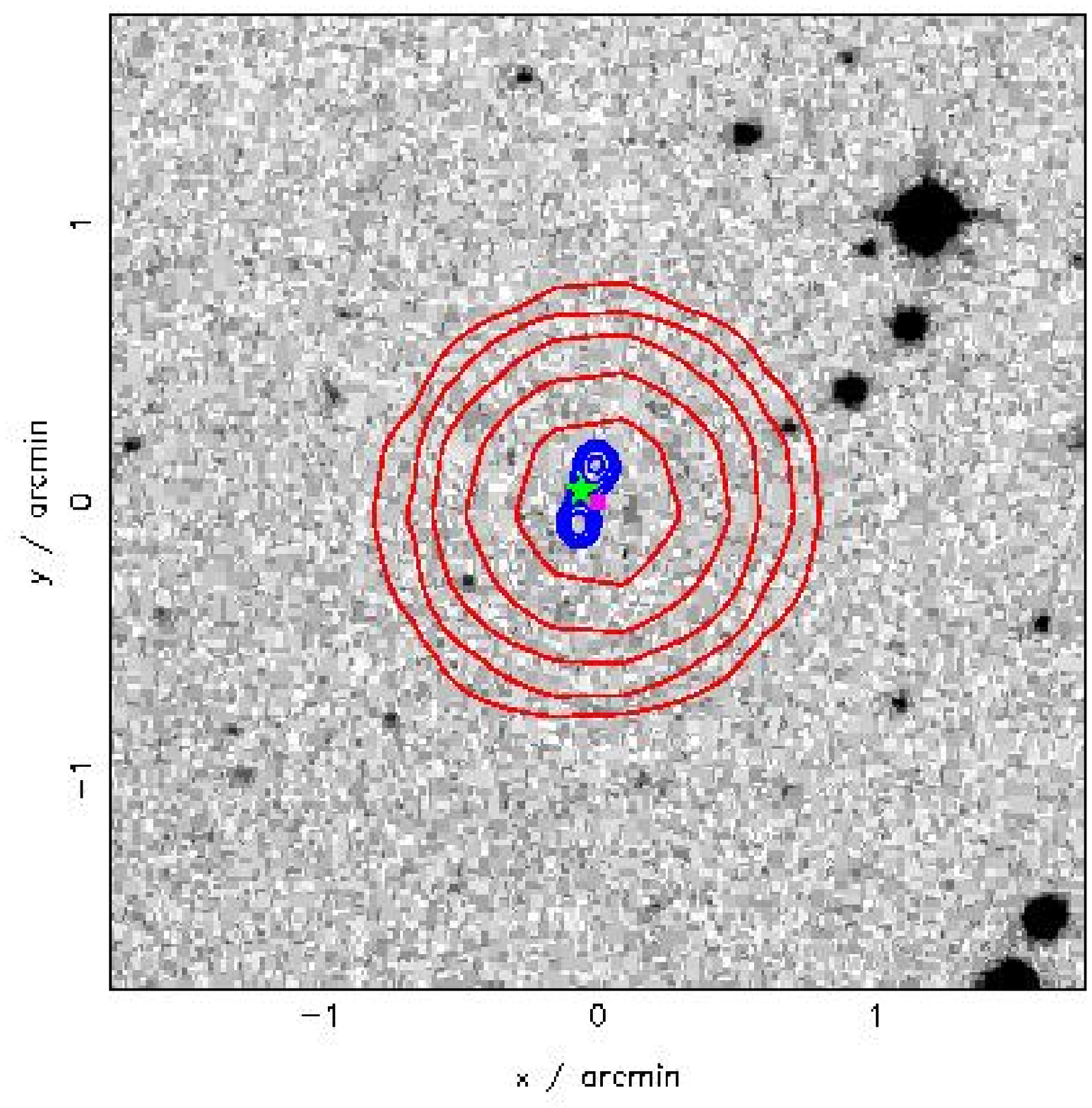}}
      \centerline{C4-071: J141932+00} 
    \end{minipage}
    \hspace{3cm}
    \begin{minipage}{3cm}
      \mbox{}
      \centerline{\includegraphics[scale=0.26,angle=270]{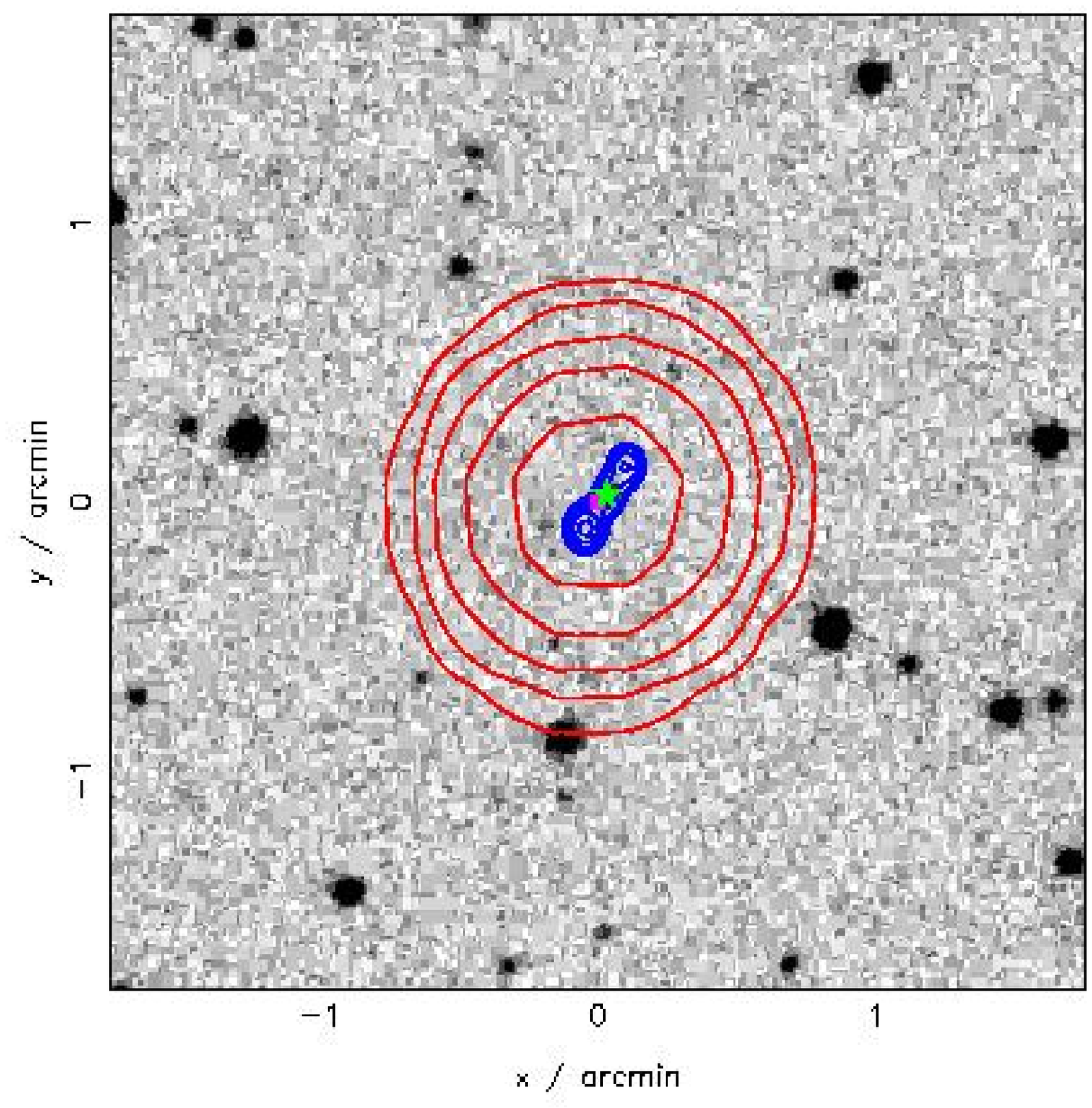}}
      \centerline{C4-072: J142033-00}
    \end{minipage}
    \vfill
    \begin{minipage}{3cm}     
      \mbox{}
      \centerline{\includegraphics[scale=0.26,angle=270]{Contours/C4/075.ps}}
      \centerline{C4-075: TXS 1418+030}
    \end{minipage}
    \hspace{3cm}
    \begin{minipage}{3cm}
      \mbox{}
      \centerline{\includegraphics[scale=0.26,angle=270]{Contours/C4/076.ps}}
      \centerline{C4-076: 4C -02.60}
    \end{minipage}
    \hspace{3cm}
    \begin{minipage}{3cm}
      \mbox{}
      \centerline{\includegraphics[scale=0.26,angle=270]{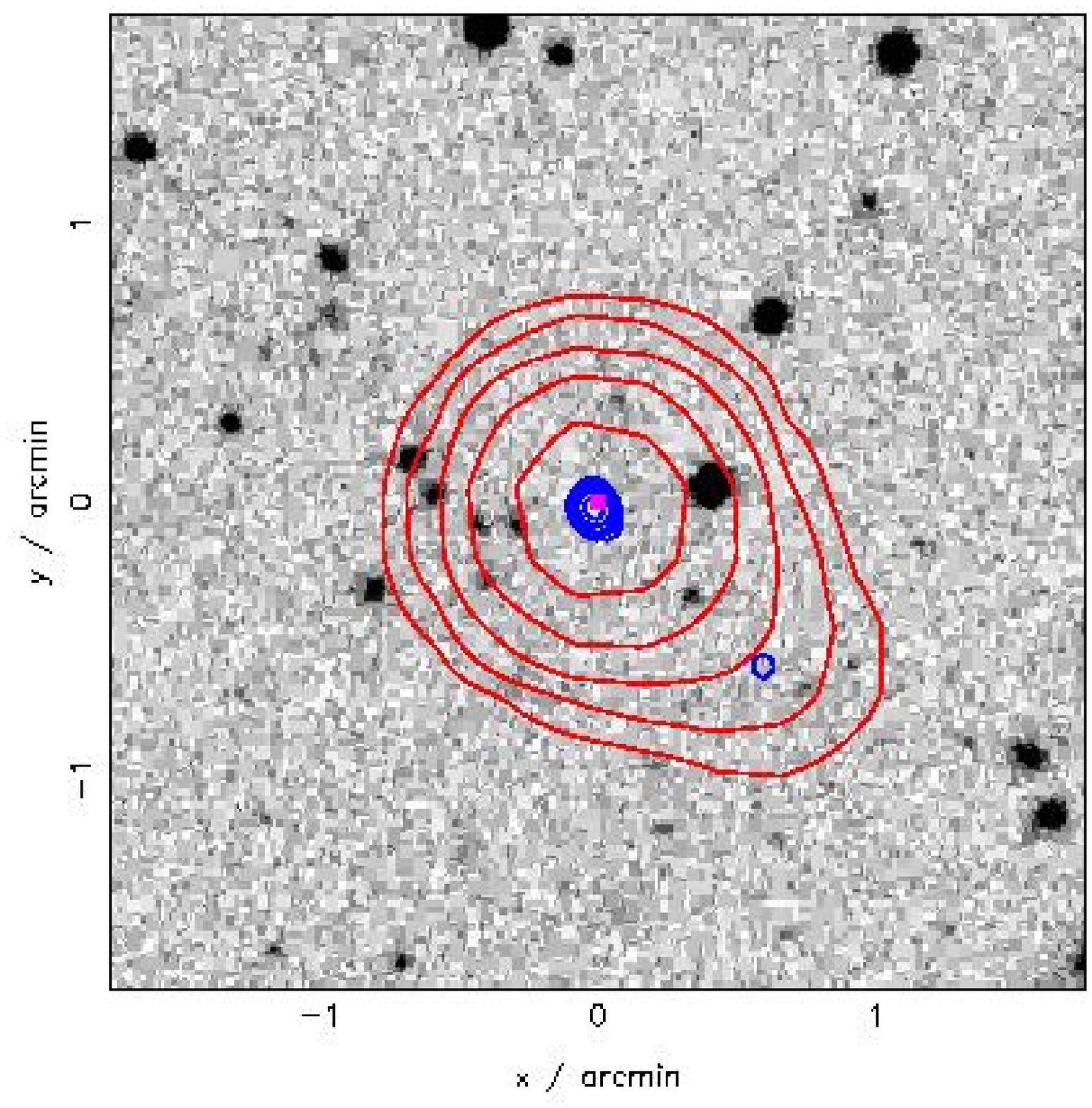}}
      \centerline{C4-077: TXS 1419+016}
    \end{minipage}
    \vfill
    \begin{minipage}{3cm}     
      \mbox{}
      \centerline{\includegraphics[scale=0.26,angle=270]{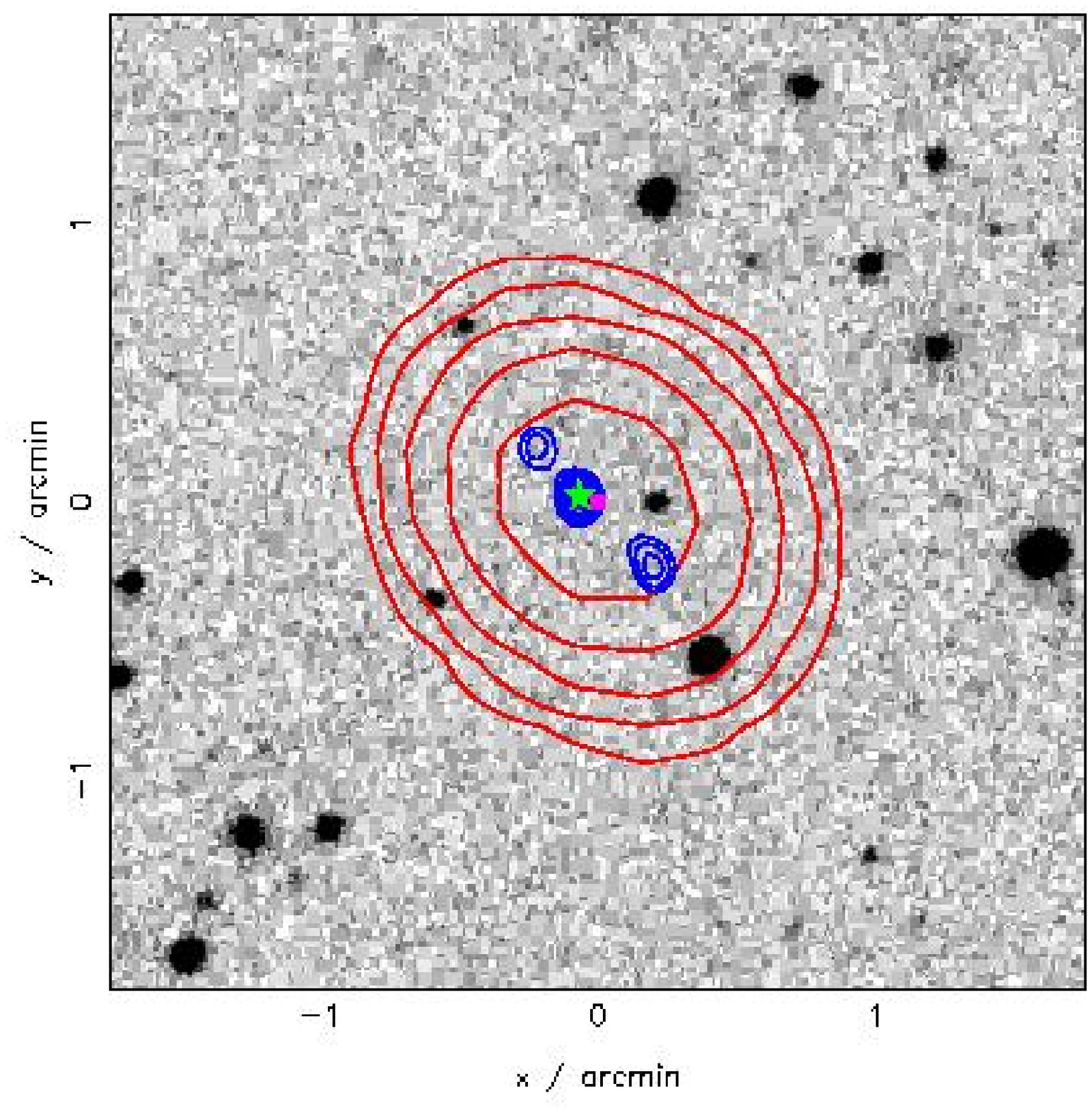}}
      \centerline{C4-078: J142235-01}
    \end{minipage}
    \hspace{3cm}
    \begin{minipage}{3cm}
      \mbox{}
      \centerline{\includegraphics[scale=0.26,angle=270]{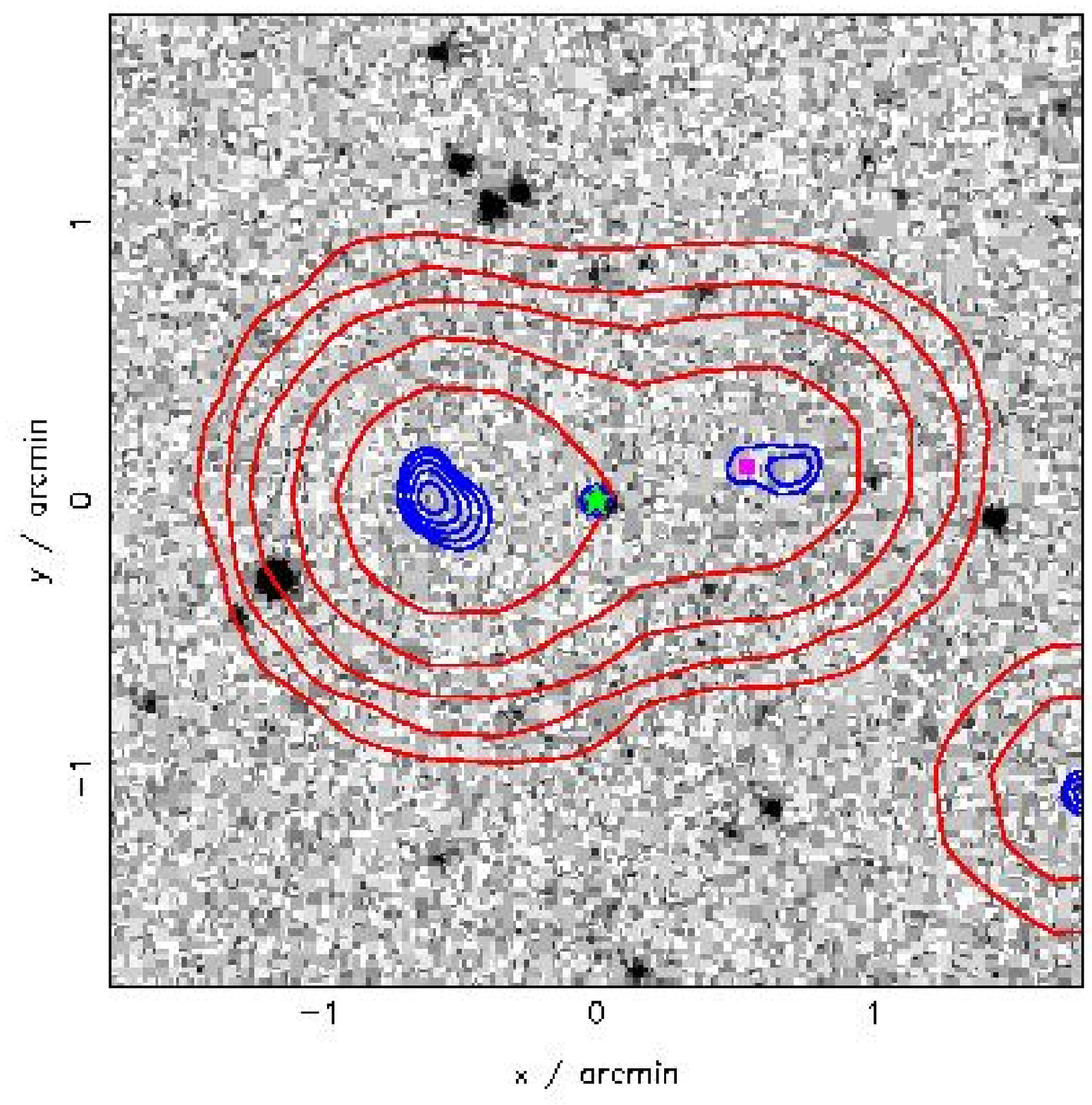}}
      \centerline{C4-080: 1423+0220}
    \end{minipage}
    \hspace{3cm}
    \begin{minipage}{3cm}
      \mbox{}
      \centerline{\includegraphics[scale=0.26,angle=270]{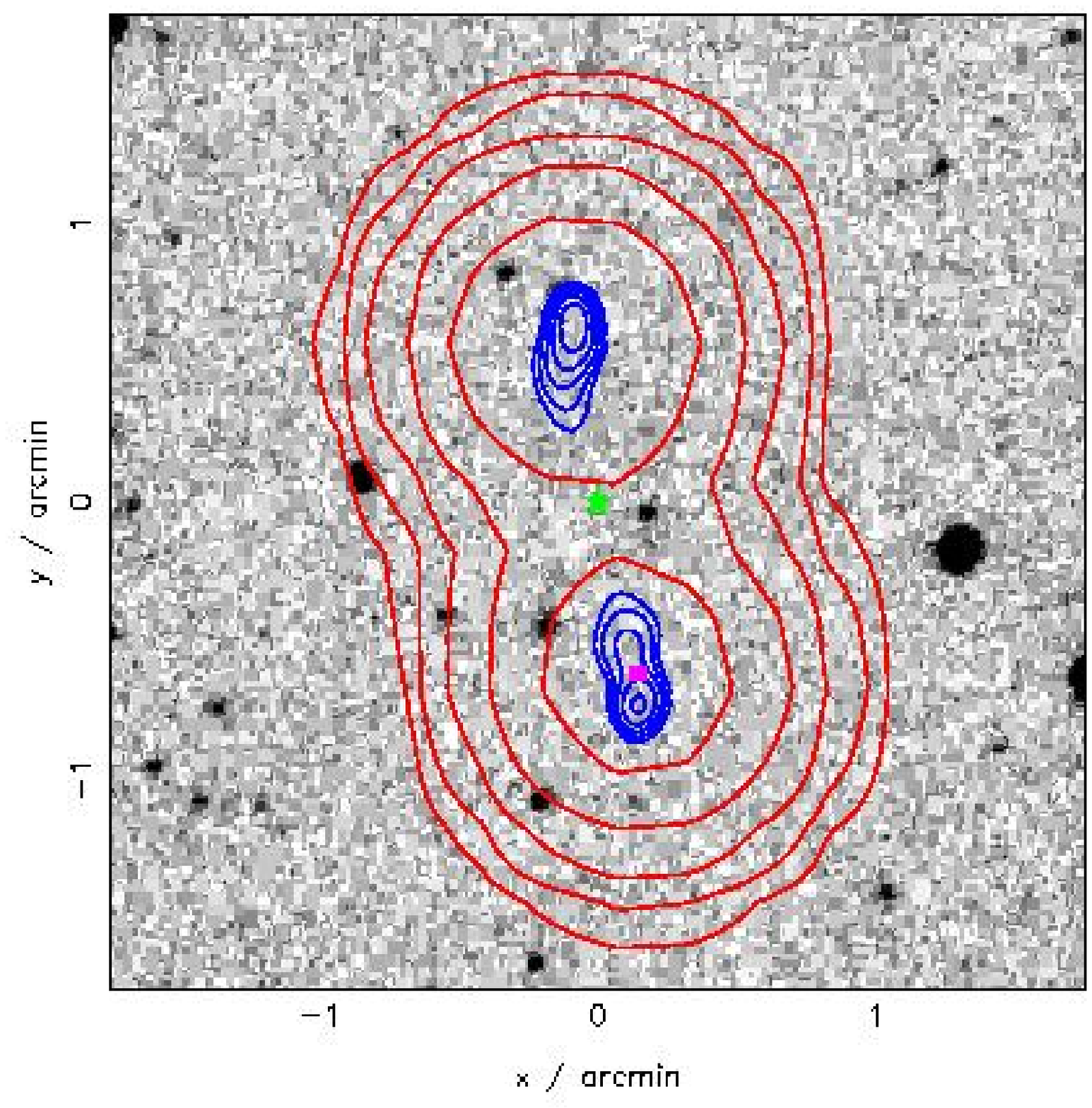}}
      \centerline{C4-081: 4C -00.55}
    \end{minipage}
    \vfill
    \begin{minipage}{3cm}      
      \mbox{}
      \centerline{\includegraphics[scale=0.26,angle=270]{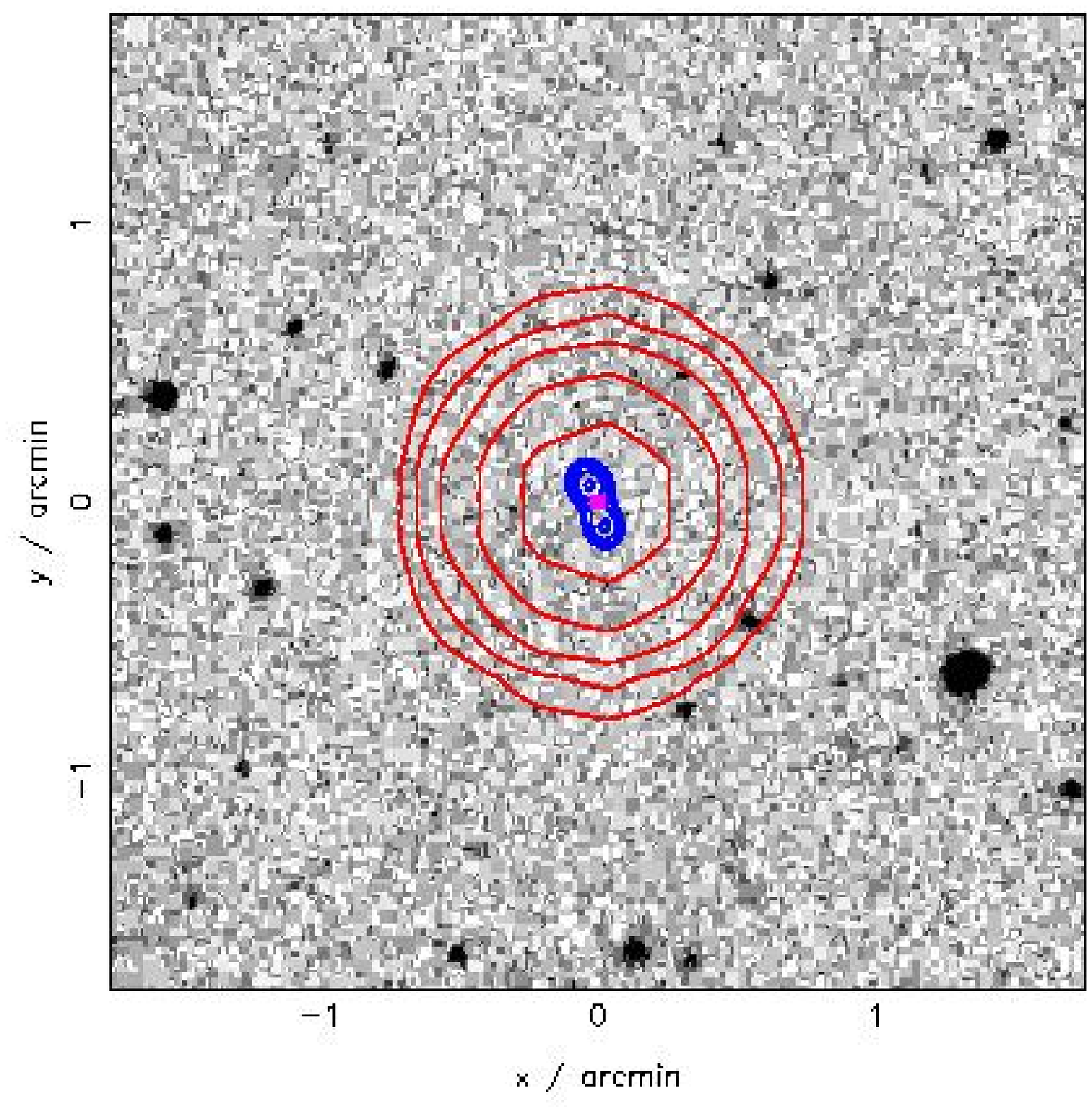}}
      \centerline{C4-082: 1423-0276}
    \end{minipage}
    \hspace{3cm}
    \begin{minipage}{3cm}
      \mbox{}
      \centerline{\includegraphics[scale=0.26,angle=270]{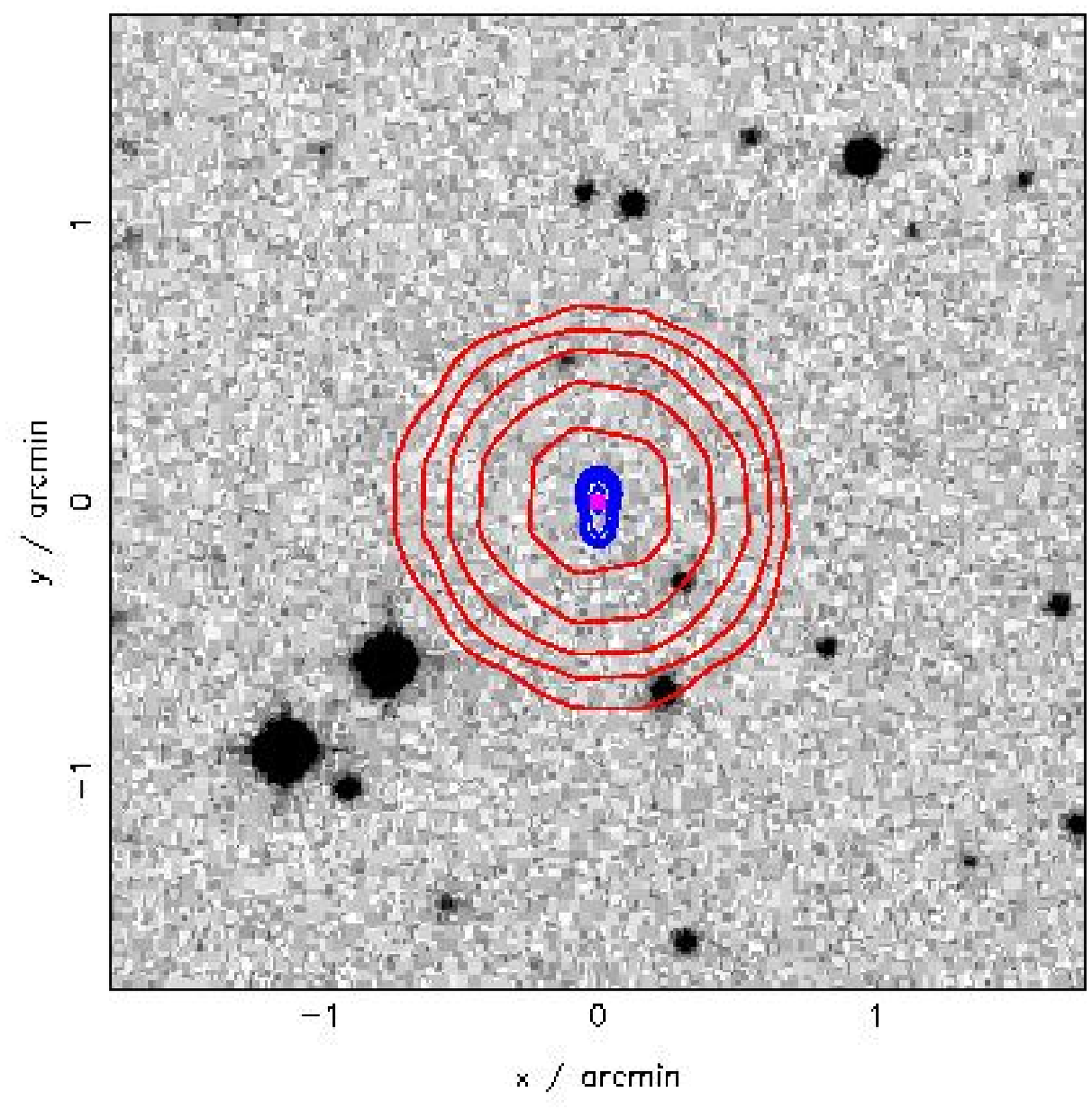}}
      \centerline{C4-083: TXS 1421+015}
    \end{minipage}
    \hspace{3cm}
    \begin{minipage}{3cm}
      \mbox{}
      \centerline{\includegraphics[scale=0.26,angle=270]{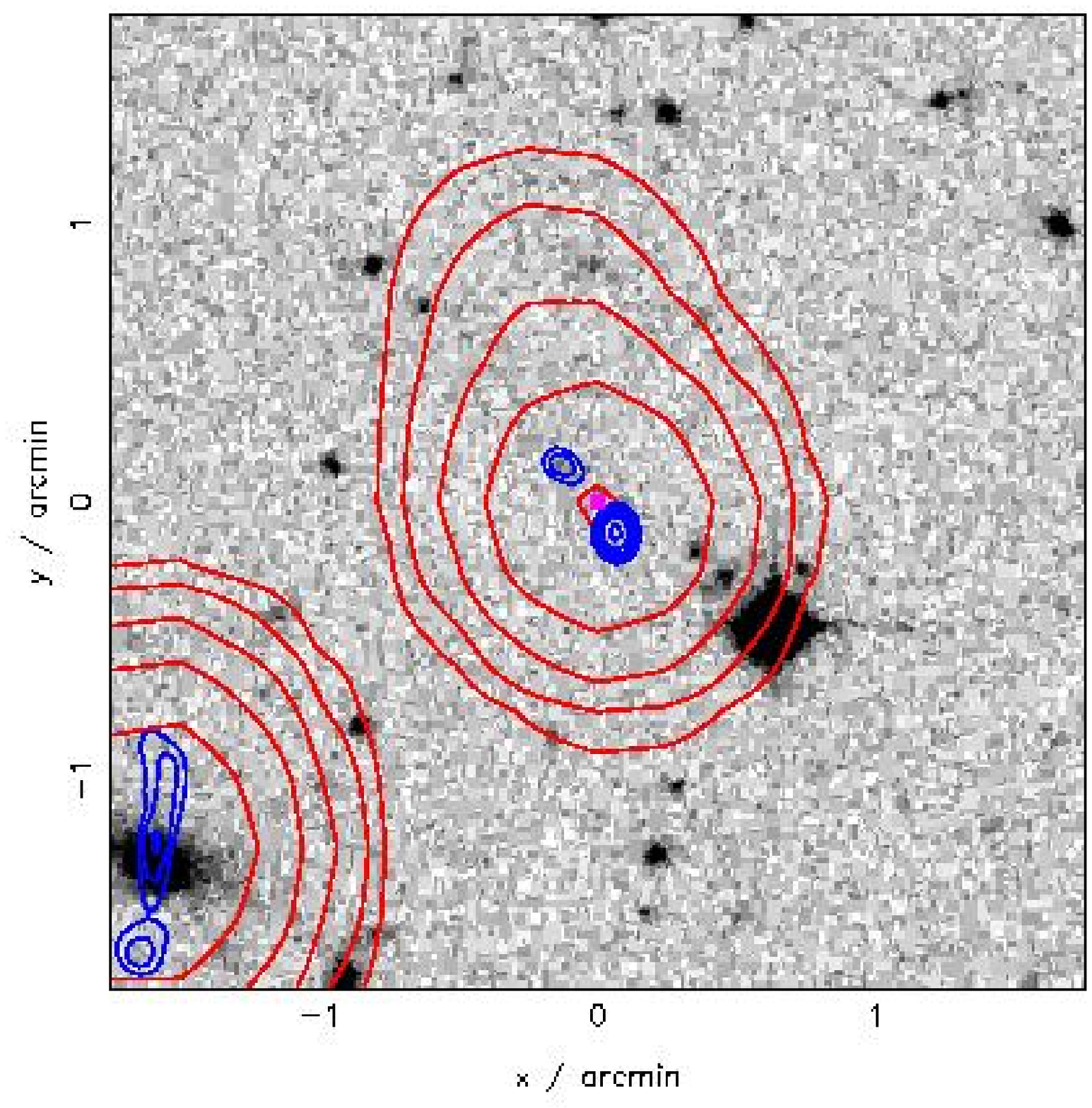}}
      \centerline{C4-084: 1423+0052}
    \end{minipage}
  \end{center}
\end{figure}

\begin{figure}
  \begin{center}
    {\bf CoNFIG-4}\\  
  \begin{minipage}{3cm}      
      \mbox{}
      \centerline{\includegraphics[scale=0.26,angle=270]{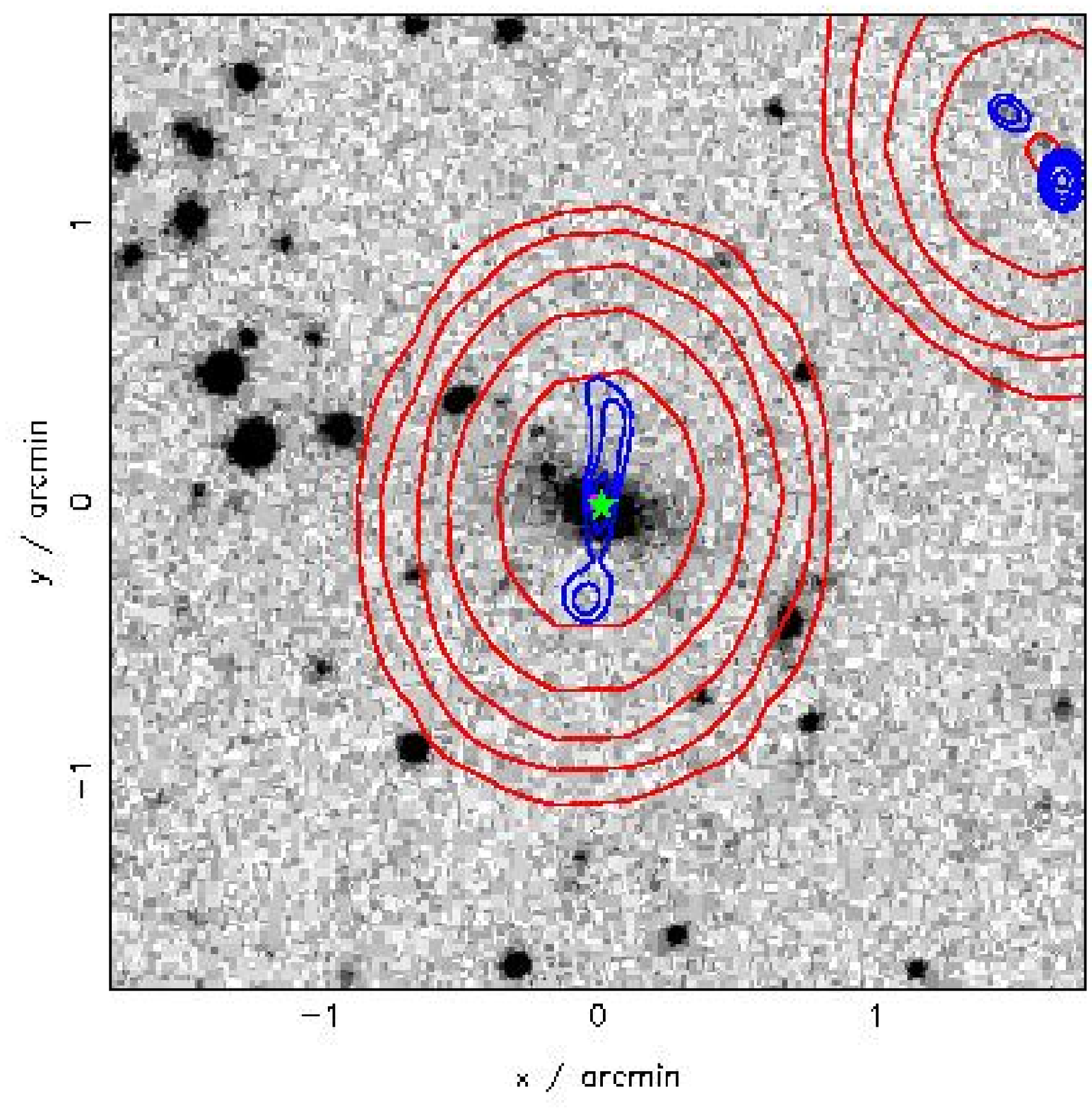}}
      \centerline{C4-085: N344Z154}
    \end{minipage}
    \hspace{3cm}
    \begin{minipage}{3cm}
      \mbox{}
      \centerline{\includegraphics[scale=0.26,angle=270]{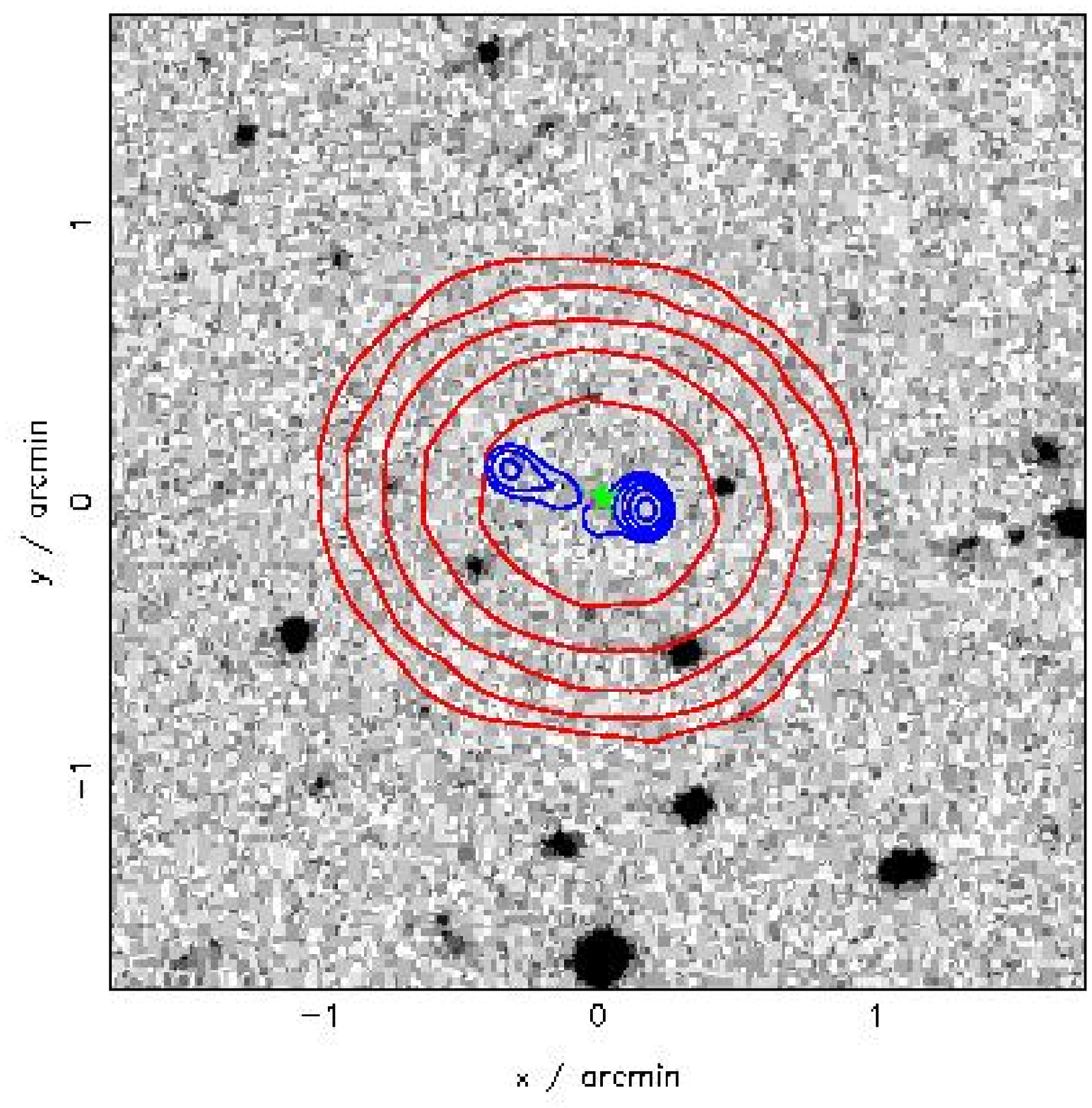}}
      \centerline{C4-087: TXS 1421+006}
    \end{minipage}
    \hspace{3cm}
    \begin{minipage}{3cm}
      \mbox{}
      \centerline{\includegraphics[scale=0.26,angle=270]{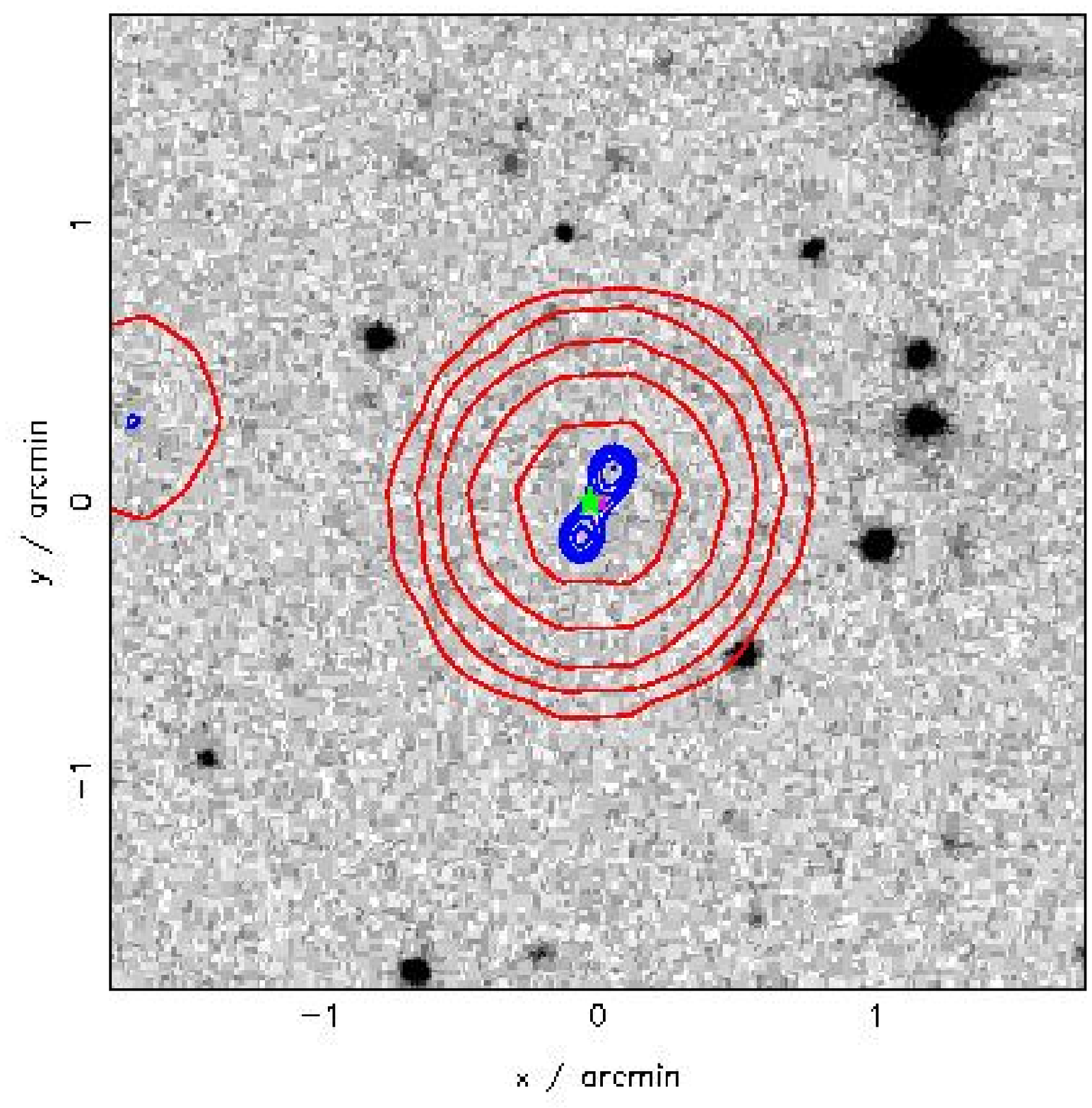}}
      \centerline{C4-088: 1424-0174}
    \end{minipage}
    \vfill
    \begin{minipage}{3cm}     
      \mbox{}
      \centerline{\includegraphics[scale=0.26,angle=270]{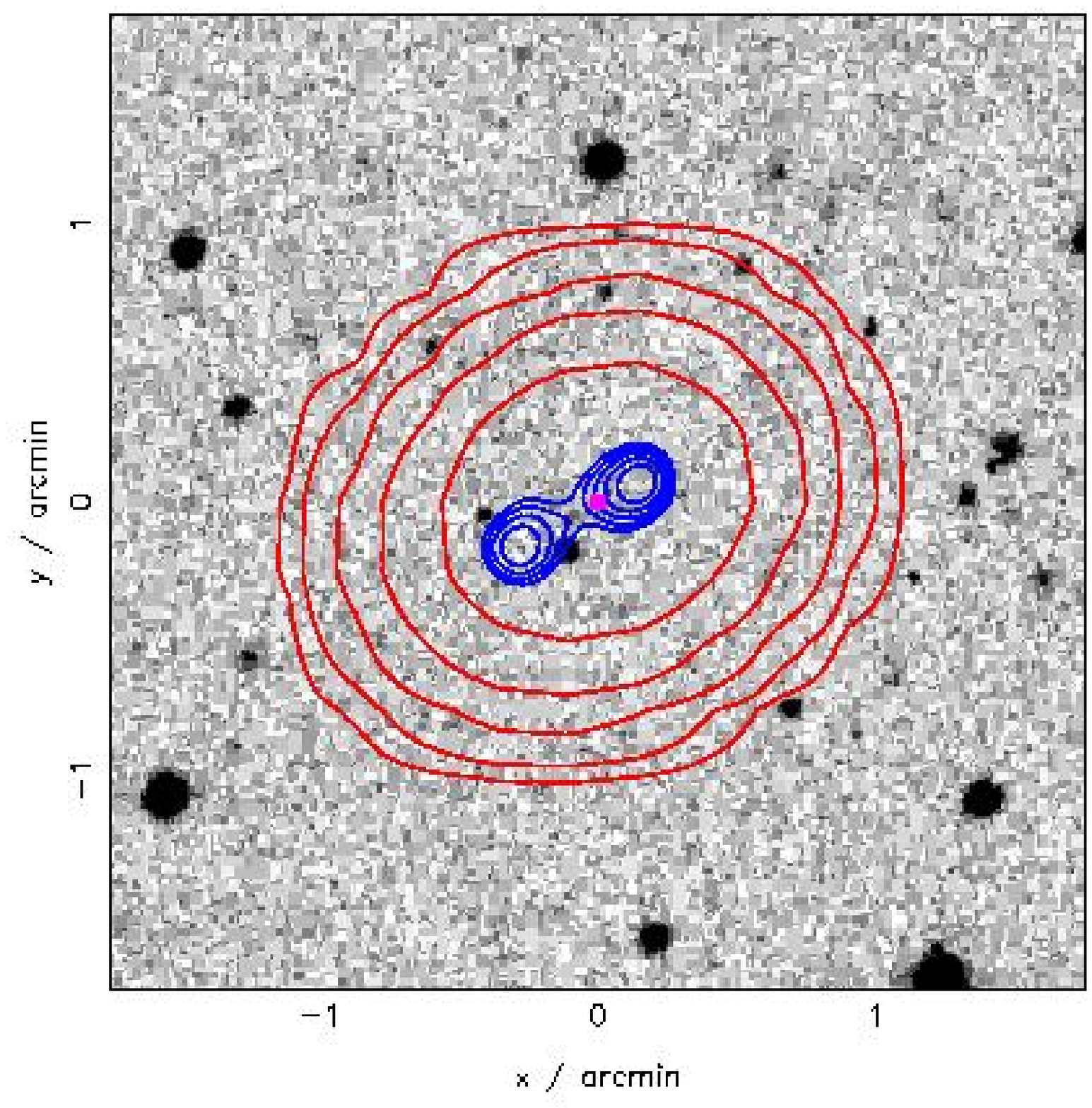}}
      \centerline{C4-089: 4C -03.51}
    \end{minipage}
    \hspace{3cm}
    \begin{minipage}{3cm}
      \mbox{}
      \centerline{\includegraphics[scale=0.26,angle=270]{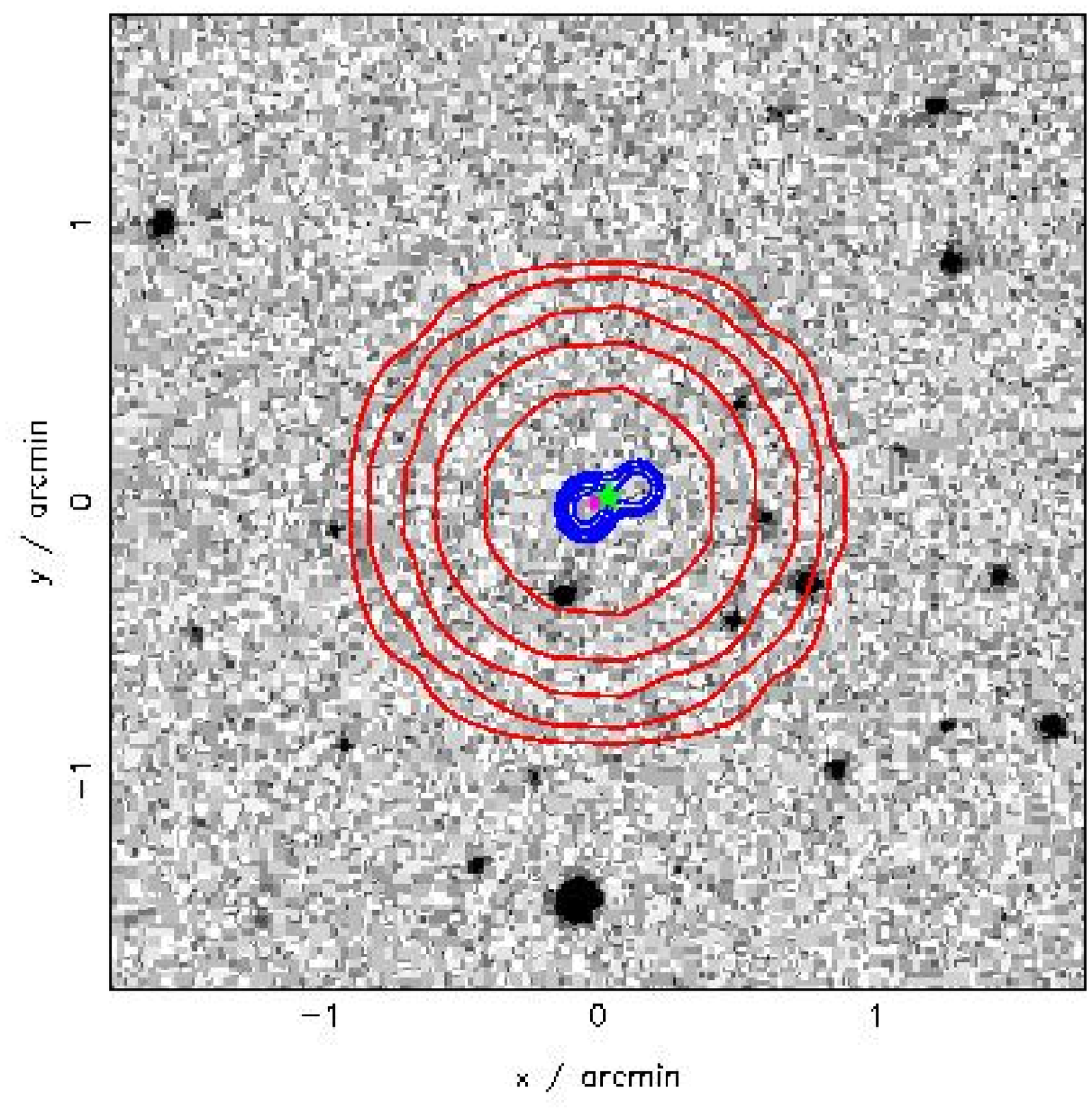}}
      \centerline{C4-090: 1425-0264}
    \end{minipage}
    \hspace{3cm}
    \begin{minipage}{3cm}
      \mbox{}
      \centerline{\includegraphics[scale=0.26,angle=270]{Contours/C4/092.ps}}
      \centerline{C4-092: TXS 1422-010}
    \end{minipage}
    \vfill
    \begin{minipage}{3cm}     
      \mbox{}
      \centerline{\includegraphics[scale=0.26,angle=270]{Contours/C4/094.ps}}
      \centerline{C4-094: 1423-0005}
    \end{minipage}
    \hspace{3cm}
    \begin{minipage}{3cm}
      \mbox{}
      \centerline{\includegraphics[scale=0.26,angle=270]{Contours/C4/095.ps}}
      \centerline{C4-095: TXS 1423+030}
    \end{minipage}
    \hspace{3cm}
    \begin{minipage}{3cm}
      \mbox{}
      \centerline{\includegraphics[scale=0.26,angle=270]{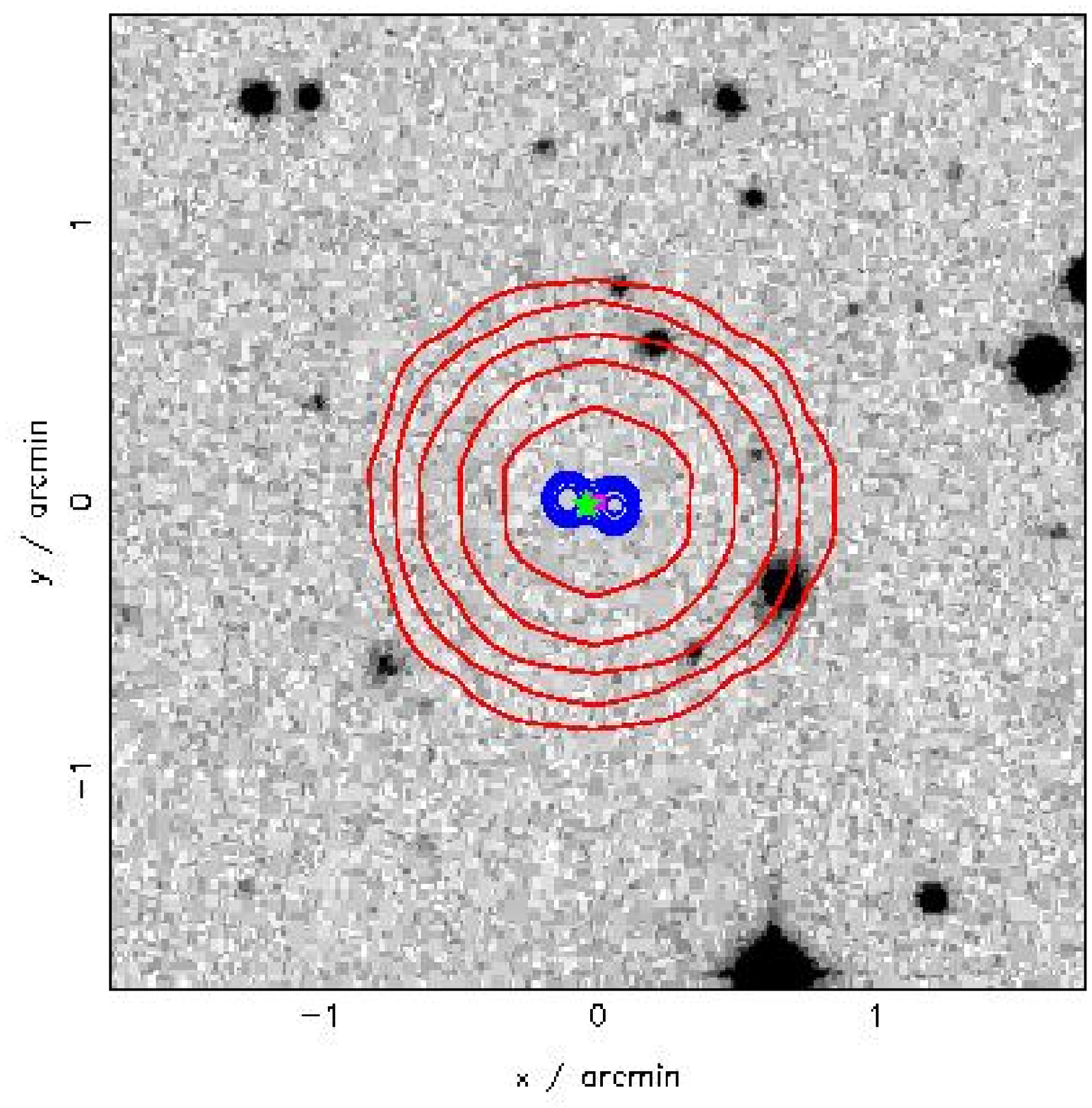}}
      \centerline{C4-096: TXS 1423+018}
    \end{minipage}
    \vfill
    \begin{minipage}{3cm}      
      \mbox{}
      \centerline{\includegraphics[scale=0.26,angle=270]{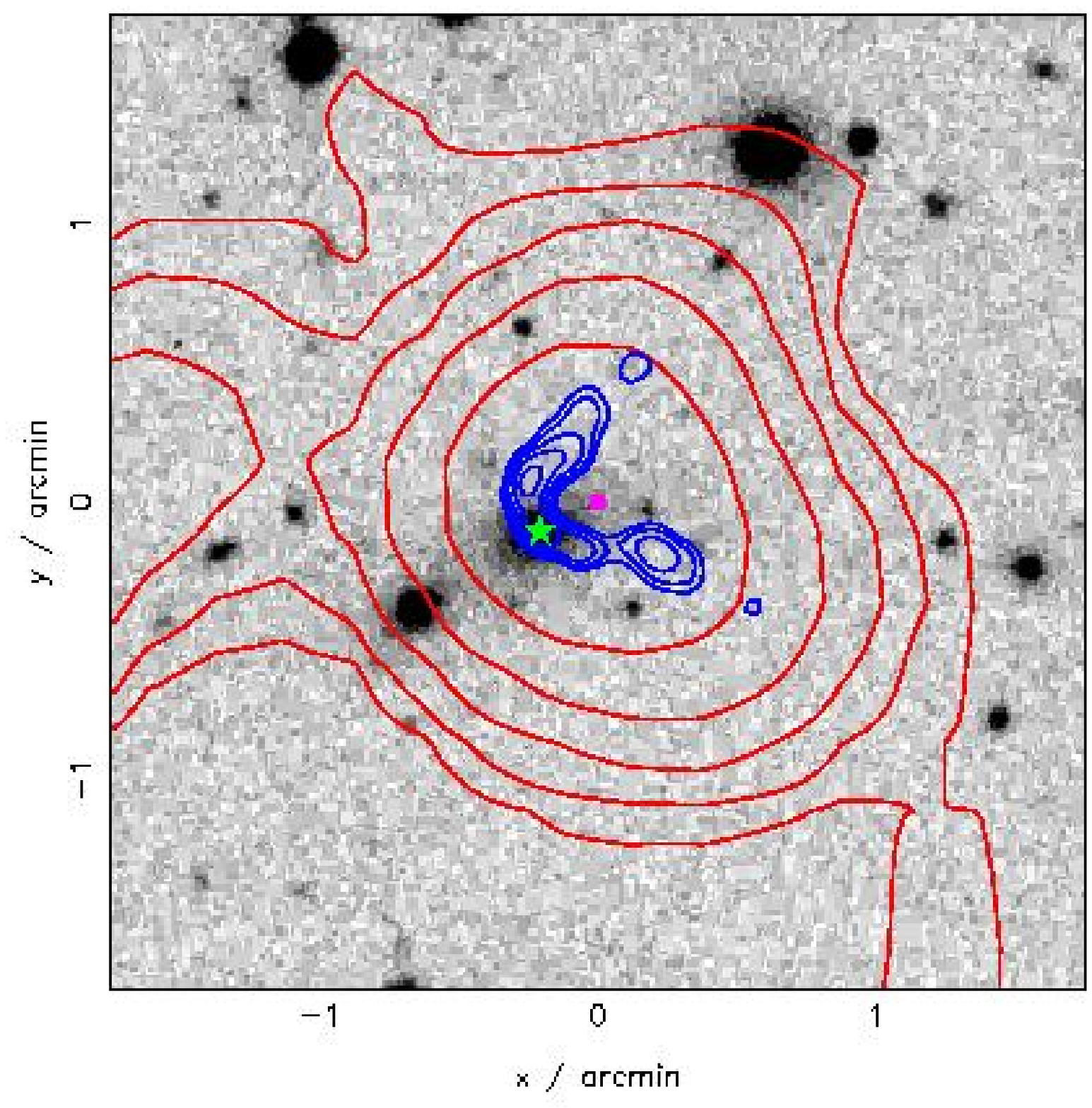}}
      \centerline{C4-098: N344Z014}
    \end{minipage}
    \hspace{3cm}
    \begin{minipage}{3cm}
      \mbox{}
      \centerline{\includegraphics[scale=0.26,angle=270]{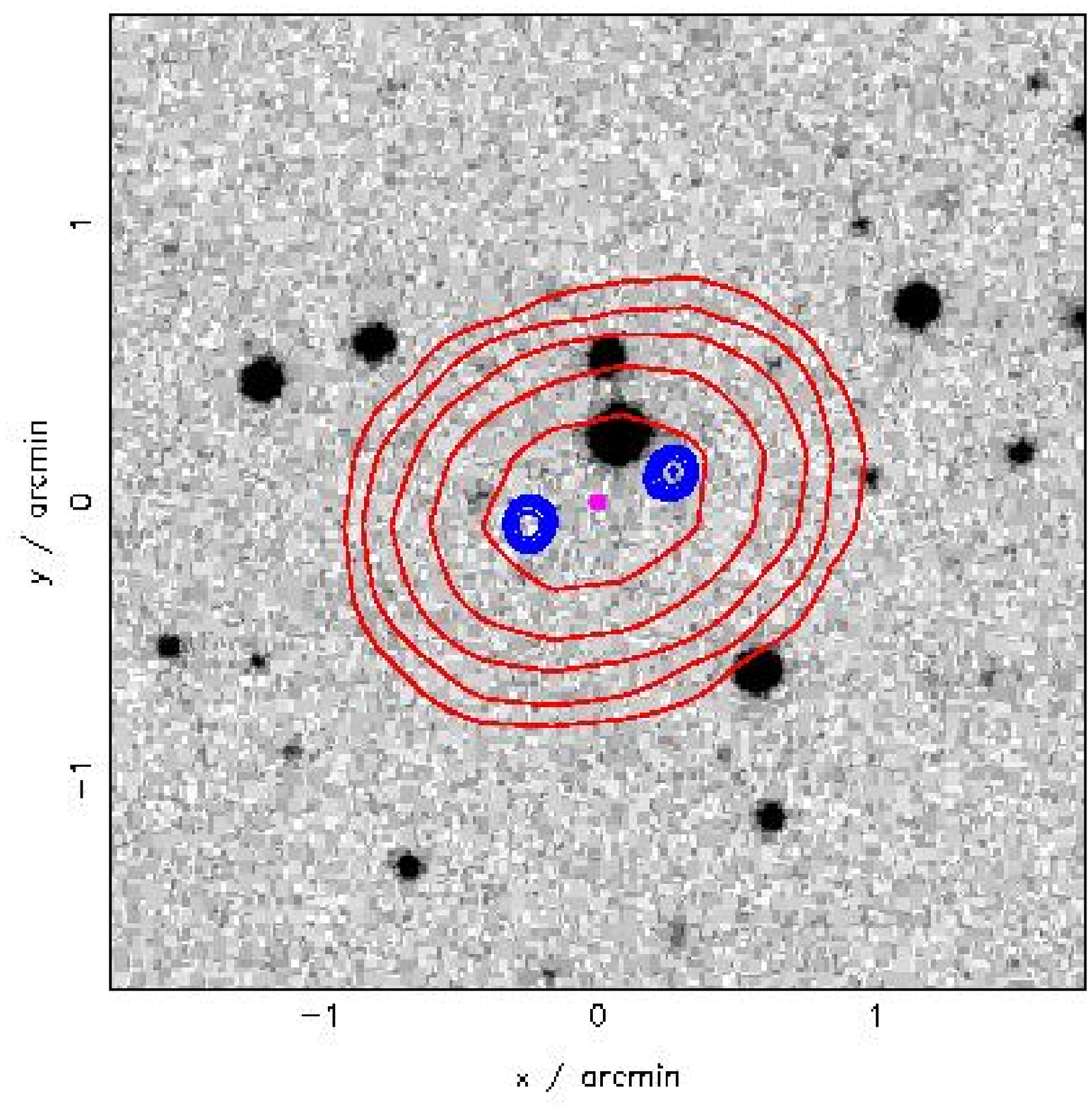}}
      \centerline{C4-100: TXS 1423+019}
    \end{minipage}
    \hspace{3cm}
    \begin{minipage}{3cm}
      \mbox{}
      \centerline{\includegraphics[scale=0.26,angle=270]{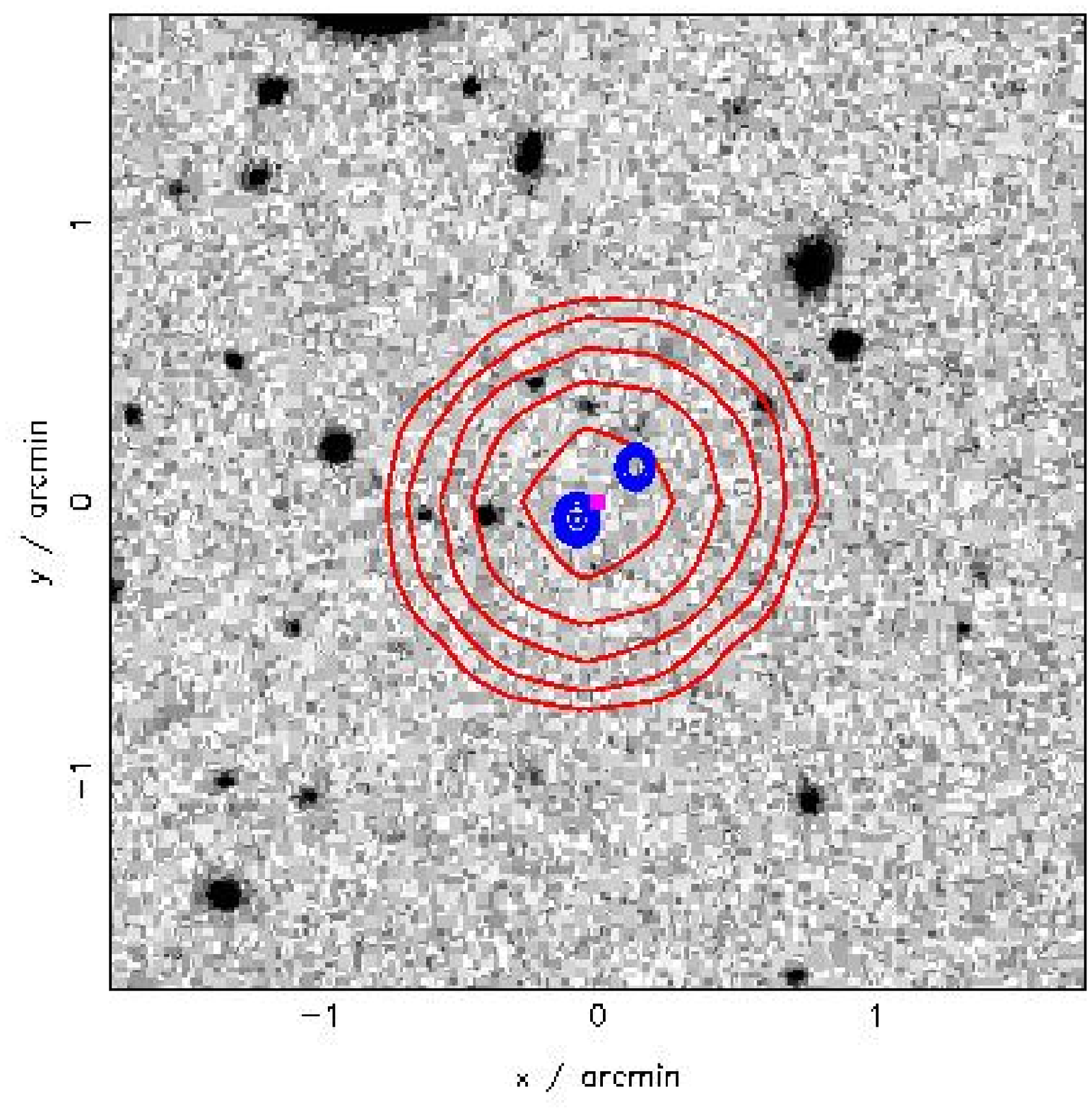}}
      \centerline{C4-102: 1426+0093}
    \end{minipage}
  \end{center}
\end{figure}

\begin{figure}
  \begin{center}
    {\bf CoNFIG-4}\\  
  \begin{minipage}{3cm}      
      \mbox{}
      \centerline{\includegraphics[scale=0.26,angle=270]{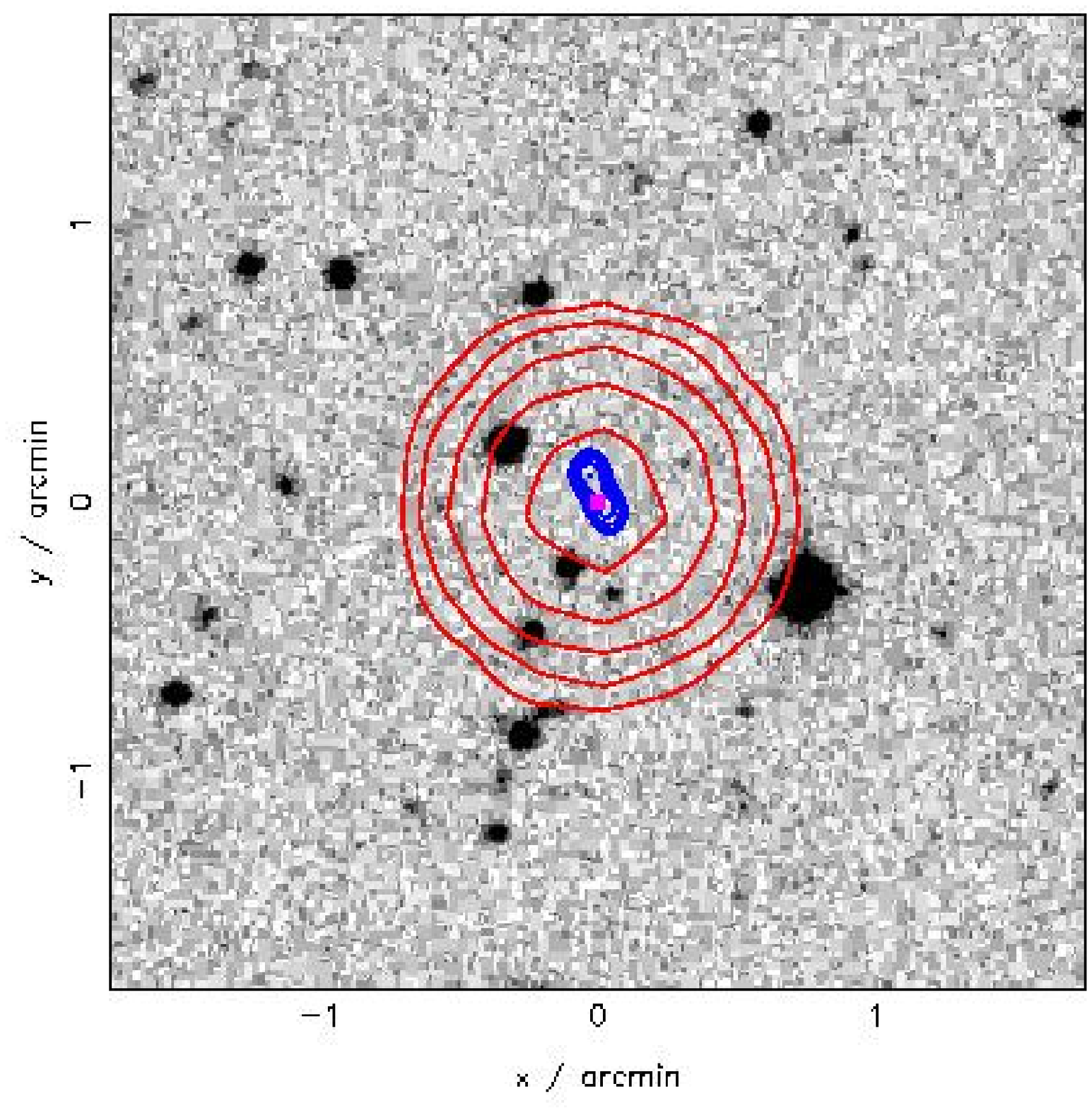}}
      \centerline{C4-105: 1427-0187}
    \end{minipage}
    \hspace{3cm}
    \begin{minipage}{3cm}
      \mbox{}
      \centerline{\includegraphics[scale=0.26,angle=270]{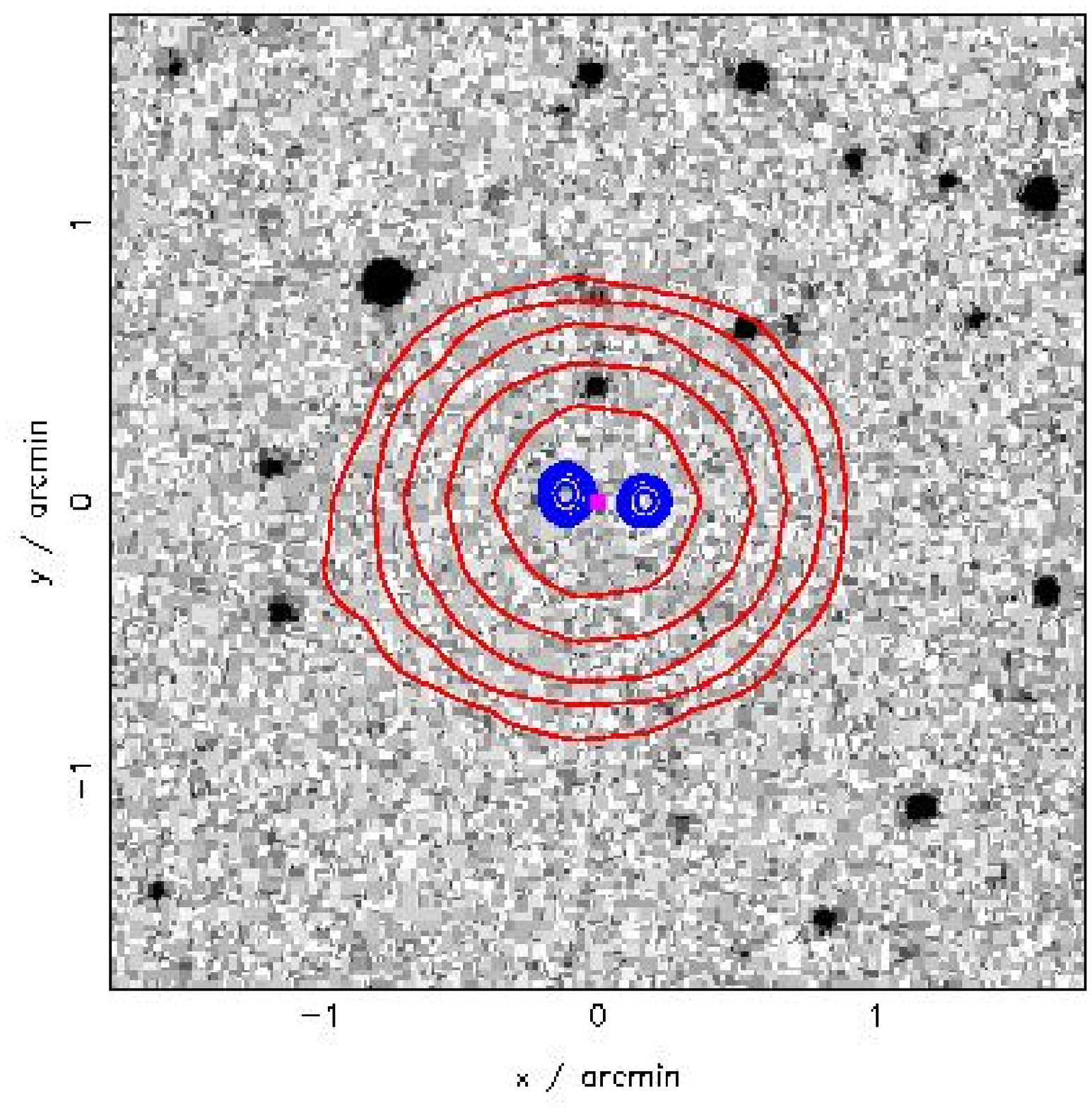}}
      \centerline{C4-106: TXS 1425+005}
    \end{minipage}
    \hspace{3cm}
    \begin{minipage}{3cm}
      \mbox{}
      \centerline{\includegraphics[scale=0.26,angle=270]{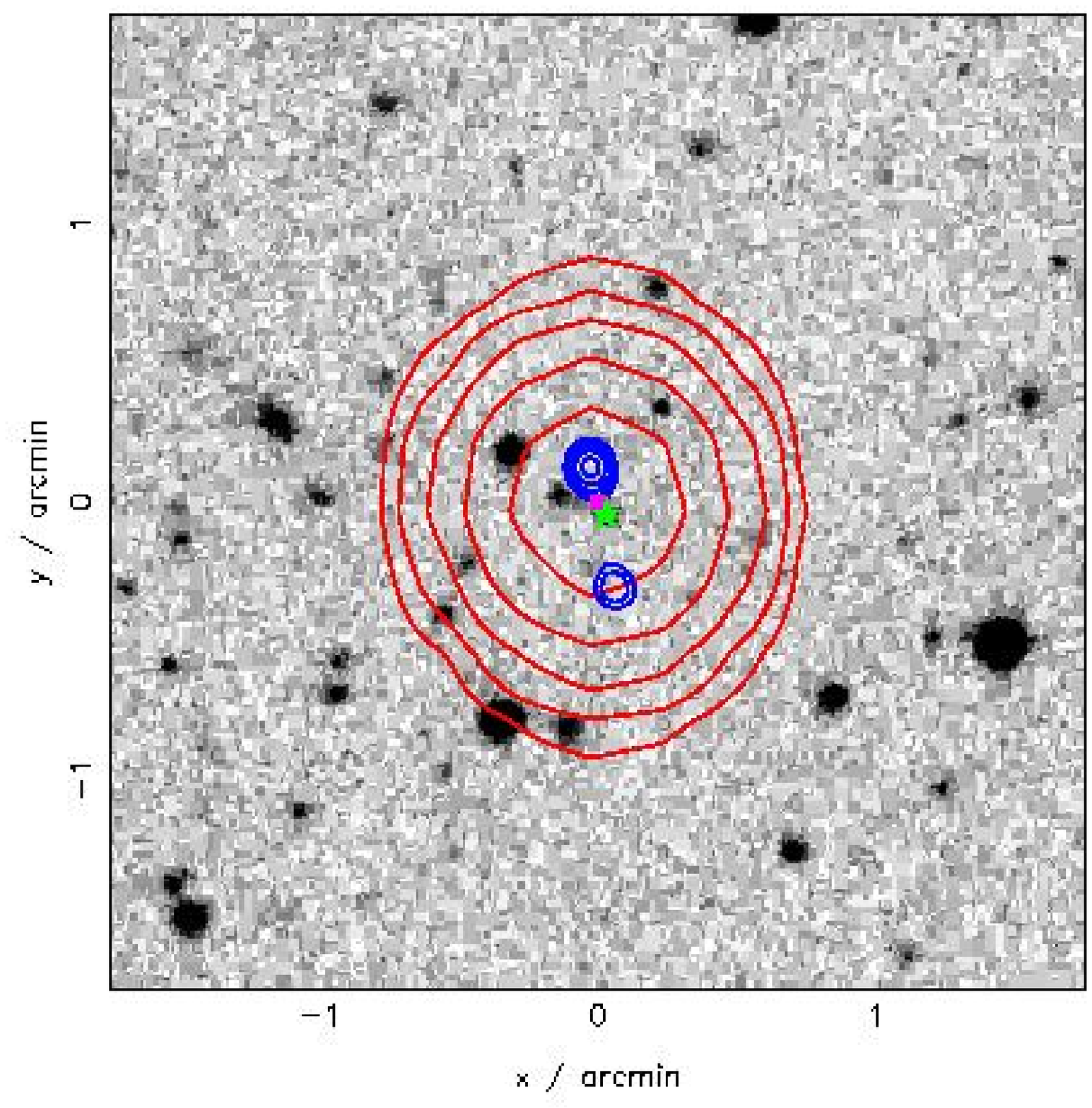}}
      \centerline{C4-107: J142746+00}
    \end{minipage}
    \vfill
    \begin{minipage}{3cm}     
      \mbox{}
      \centerline{\includegraphics[scale=0.26,angle=270]{Contours/C4/115.ps}}
      \centerline{C4-115: TXS 1426+030}
    \end{minipage}
    \hspace{3cm}
    \begin{minipage}{3cm}
      \mbox{}
      \centerline{\includegraphics[scale=0.26,angle=270]{Contours/C4/119.ps}}
      \centerline{C4-119: TXS 1427+009}
    \end{minipage}
    \hspace{3cm}
    \begin{minipage}{3cm}
      \mbox{}
      \centerline{\includegraphics[scale=0.26,angle=270]{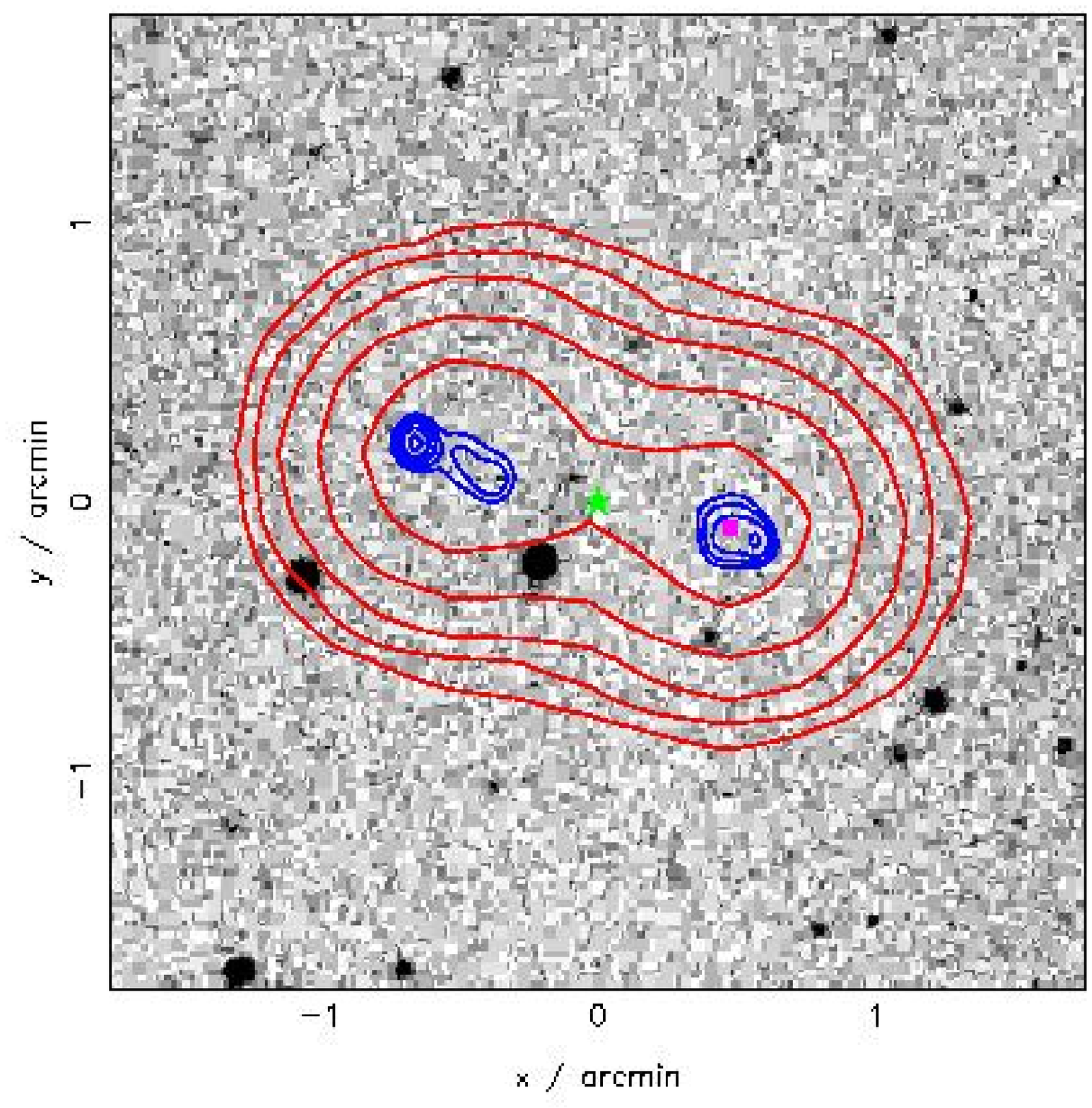}}
      \centerline{C4-120:1430-0192}
    \end{minipage}
    \vfill
    \begin{minipage}{3cm}     
      \mbox{}
      \centerline{\includegraphics[scale=0.26,angle=270]{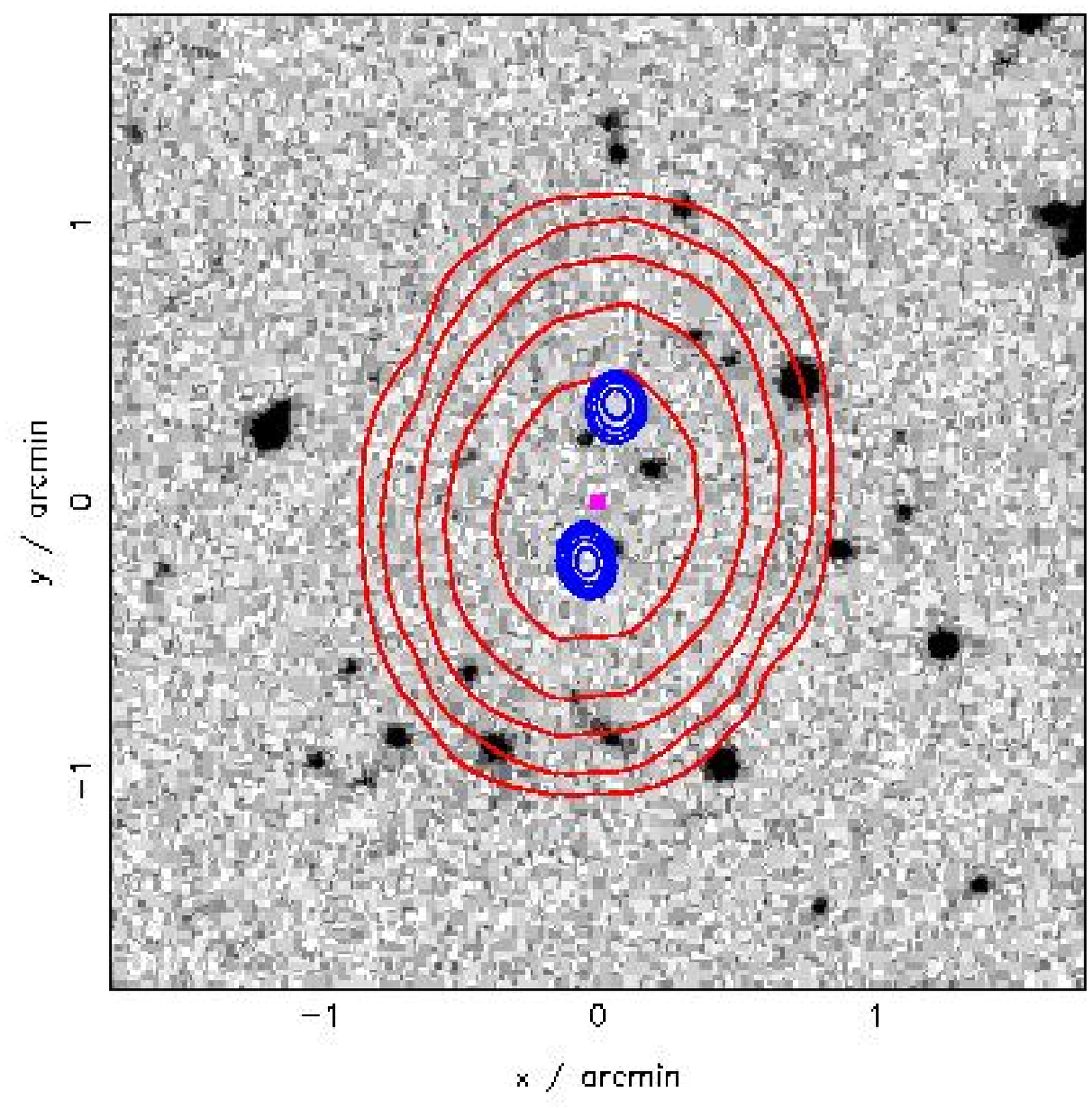}}
      \centerline{C4-122: TXS 1427+012}
    \end{minipage}
    \hspace{3cm}
    \begin{minipage}{3cm}
      \mbox{}
      \centerline{\includegraphics[scale=0.26,angle=270]{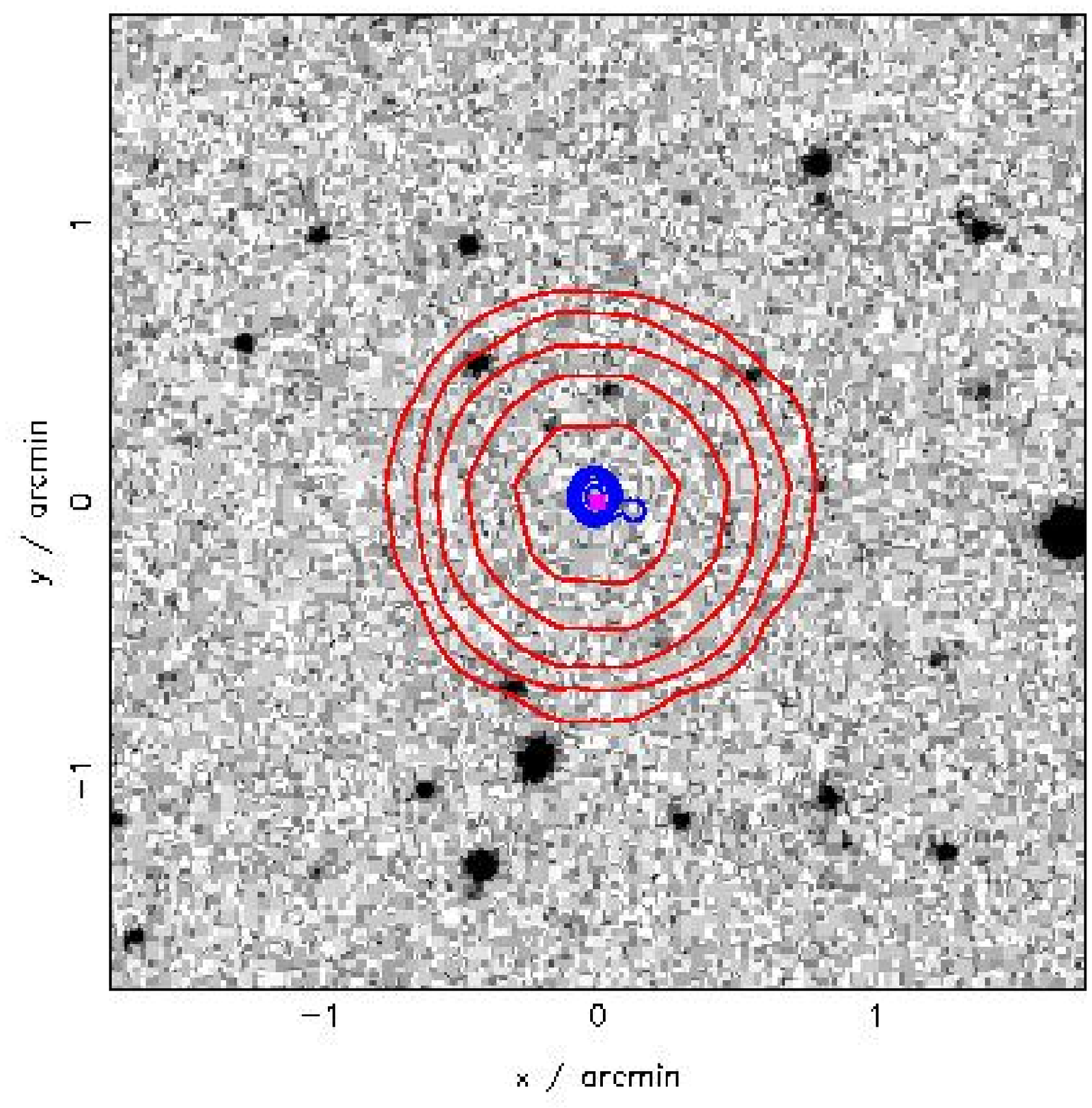}}
      \centerline{C4-124: TXS 1428+007}
    \end{minipage}
    \hspace{3cm}
    \begin{minipage}{3cm}
      \mbox{}
      \centerline{\includegraphics[scale=0.26,angle=270]{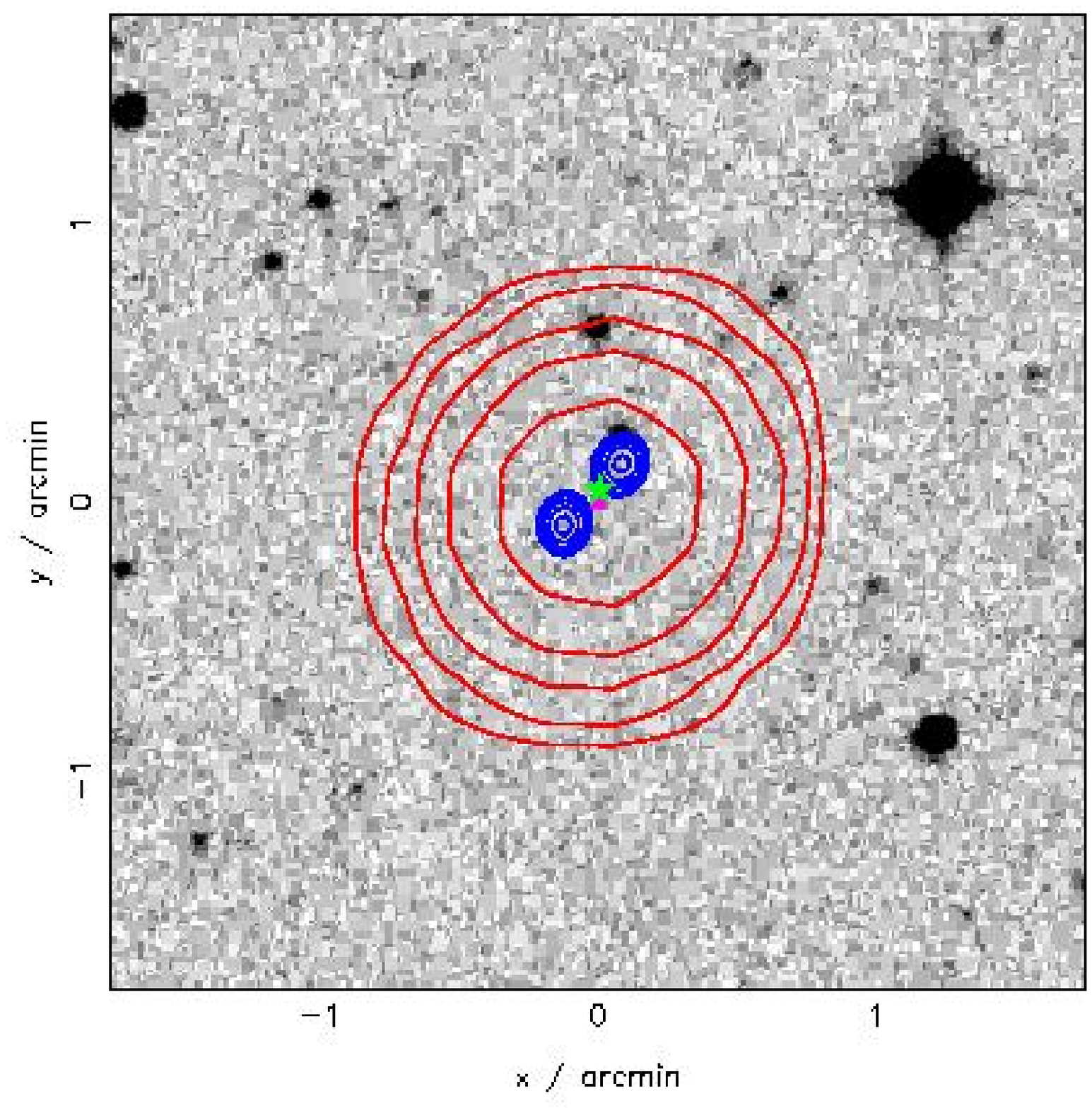}}
      \centerline{C4-125: TXS 1428-013}
    \end{minipage}
    \vfill
    \begin{minipage}{3cm}      
      \mbox{}
      \centerline{\includegraphics[scale=0.26,angle=270]{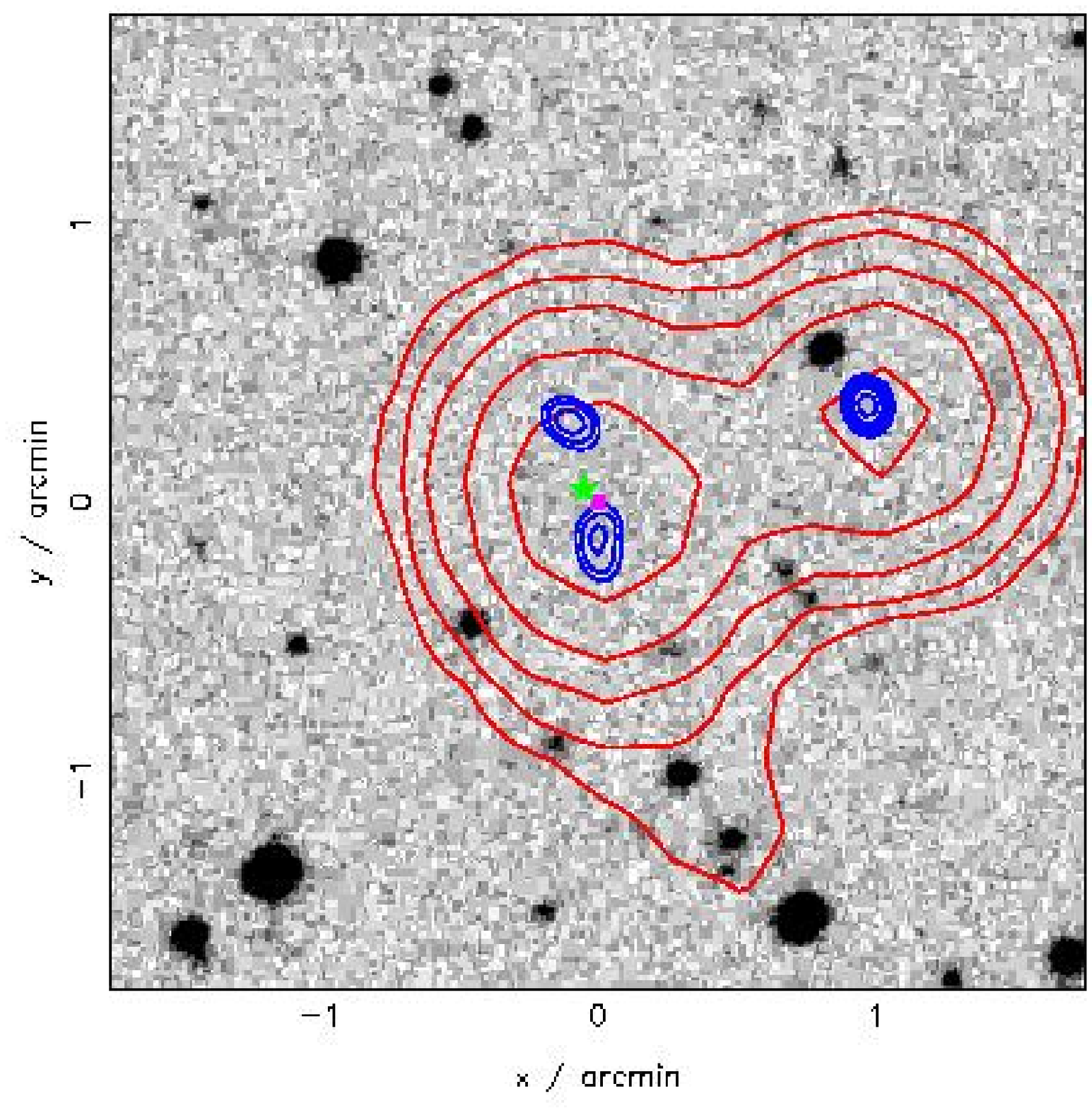}}
      \centerline{C4-128: 1431-0093}
    \end{minipage}
    \hspace{3cm}
    \begin{minipage}{3cm}
      \mbox{}
      \centerline{\includegraphics[scale=0.26,angle=270]{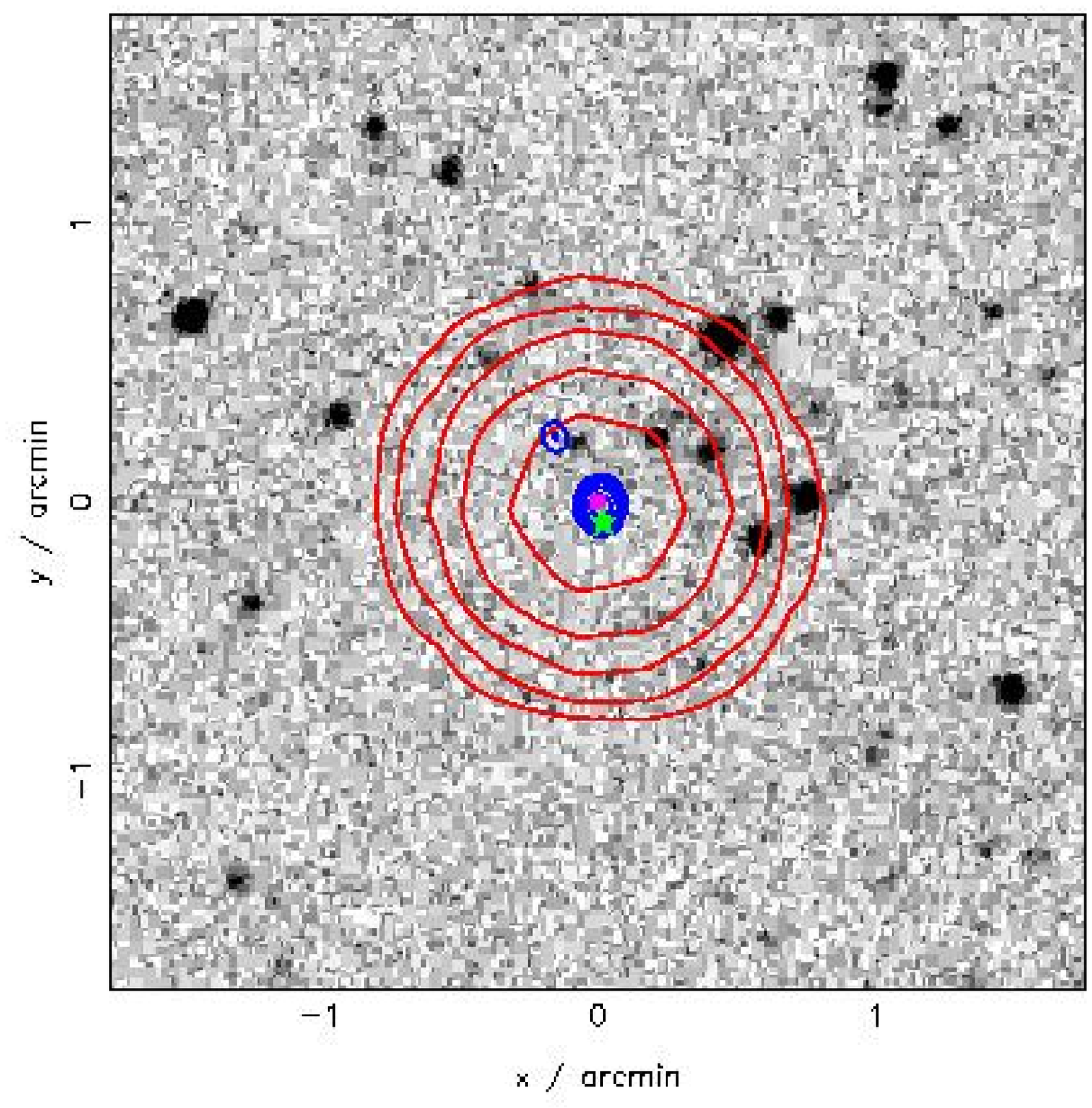}}
      \centerline{C4-129: TXS 1429-006}
    \end{minipage}
    \hspace{3cm}
    \begin{minipage}{3cm}
      \mbox{}
      \centerline{\includegraphics[scale=0.26,angle=270]{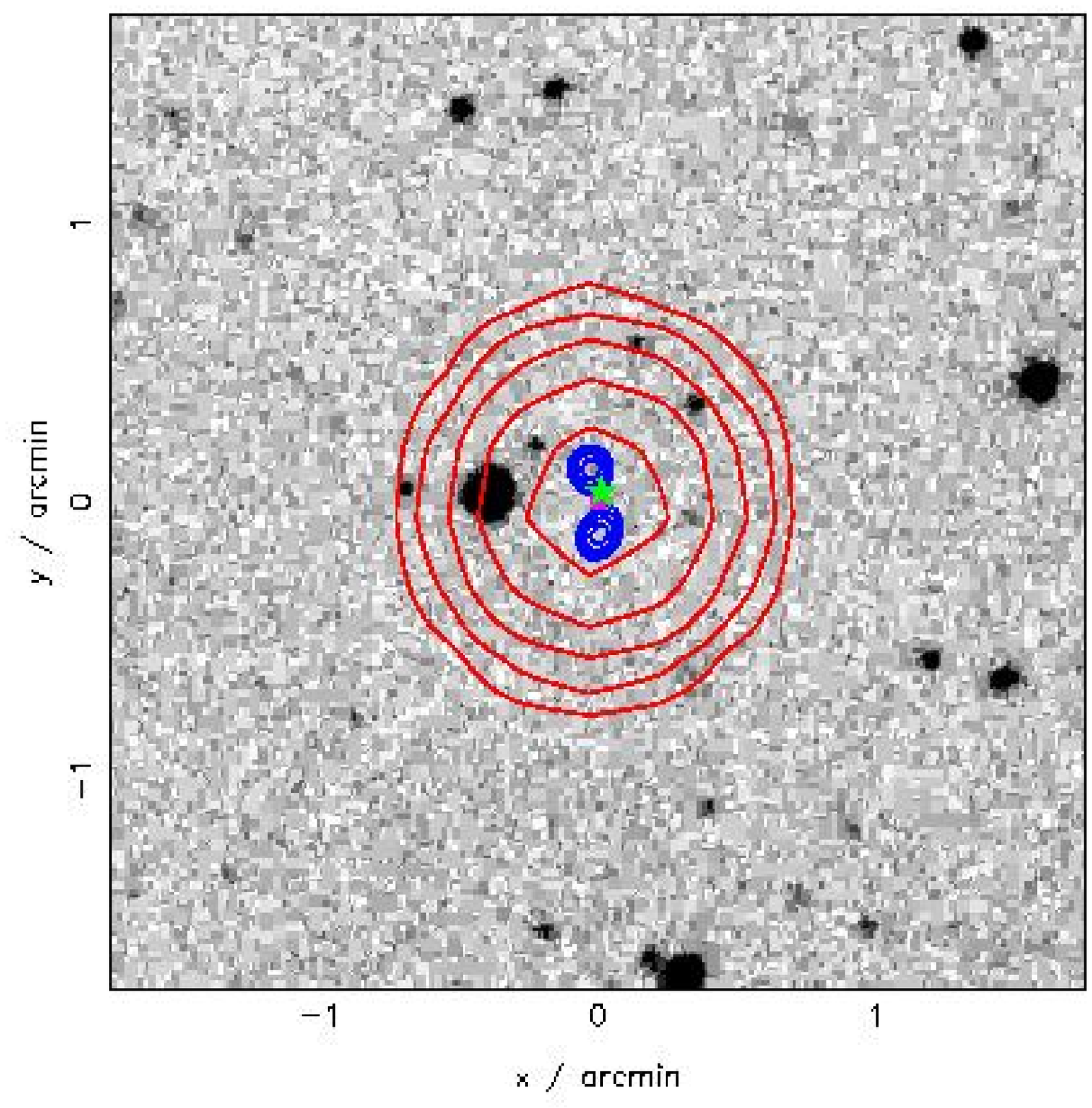}}
      \centerline{C4-131: 1432+0262}
    \end{minipage}
  \end{center}
\end{figure}

\begin{figure}
  \begin{center}
    {\bf CoNFIG-4}\\  
  \begin{minipage}{3cm}      
      \mbox{}
      \centerline{\includegraphics[scale=0.26,angle=270]{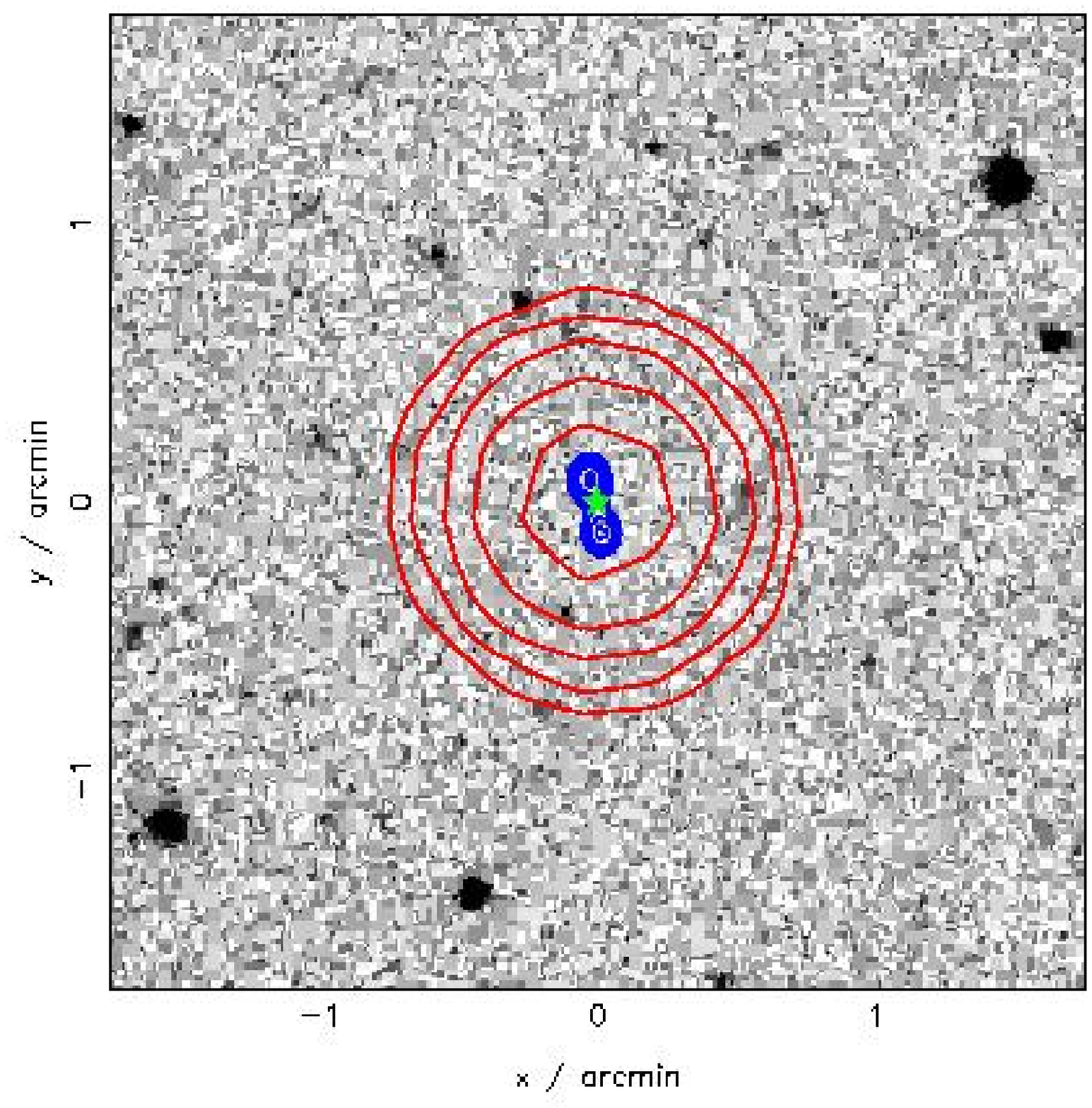}}
      \centerline{C4-133: 1432-0305}
    \end{minipage}
    \hspace{3cm}
    \begin{minipage}{3cm}
      \mbox{}
      \centerline{\includegraphics[scale=0.26,angle=270]{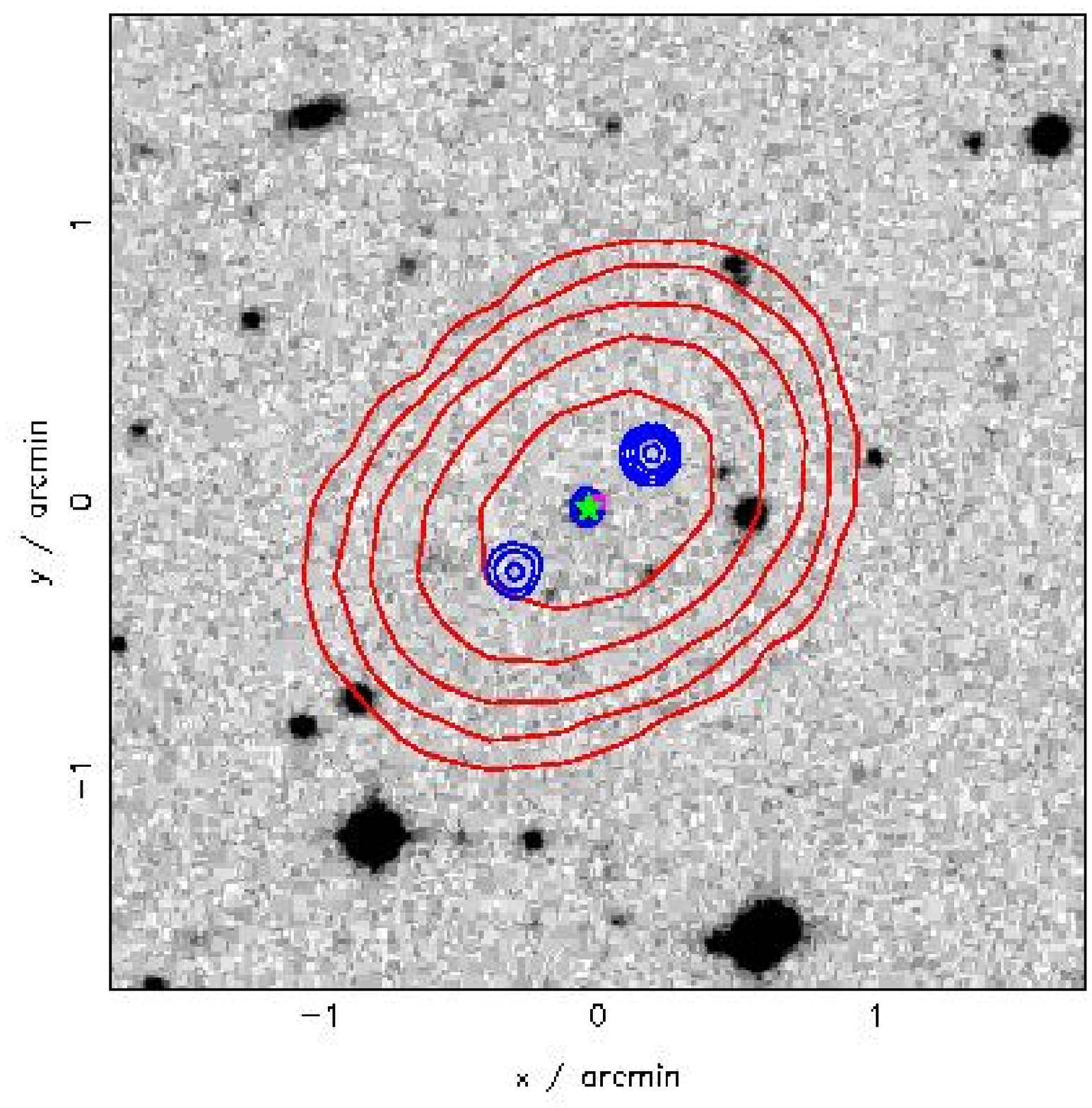}}
      \centerline{C4-134: J143244-00}
    \end{minipage}
    \hspace{3cm}
    \begin{minipage}{3cm}
      \mbox{}
      \centerline{\includegraphics[scale=0.26,angle=270]{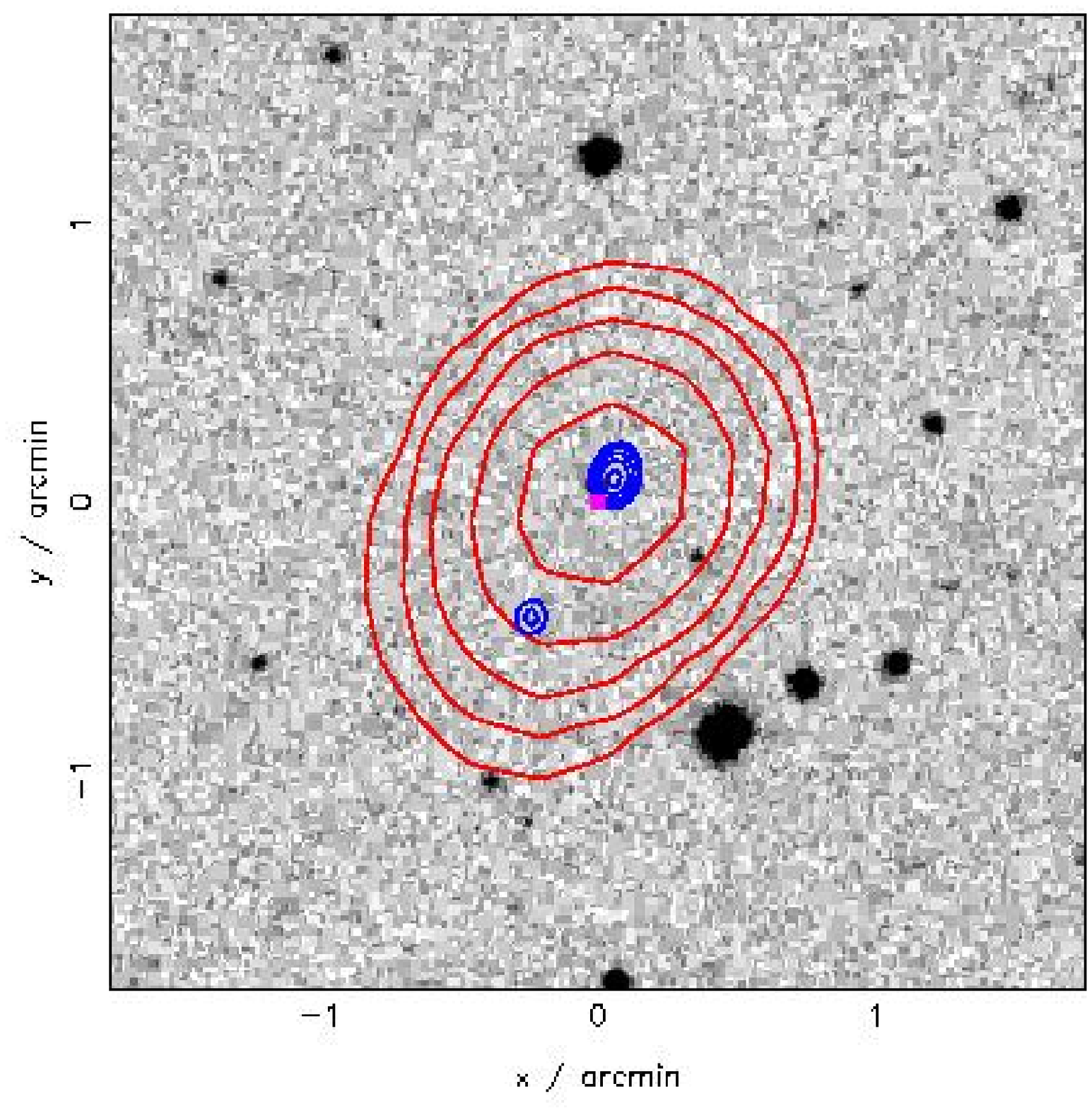}}
      \centerline{C4-135: 1432+0078}
    \end{minipage}
    \vfill
    \begin{minipage}{3cm}     
      \mbox{}
      \centerline{\includegraphics[scale=0.26,angle=270]{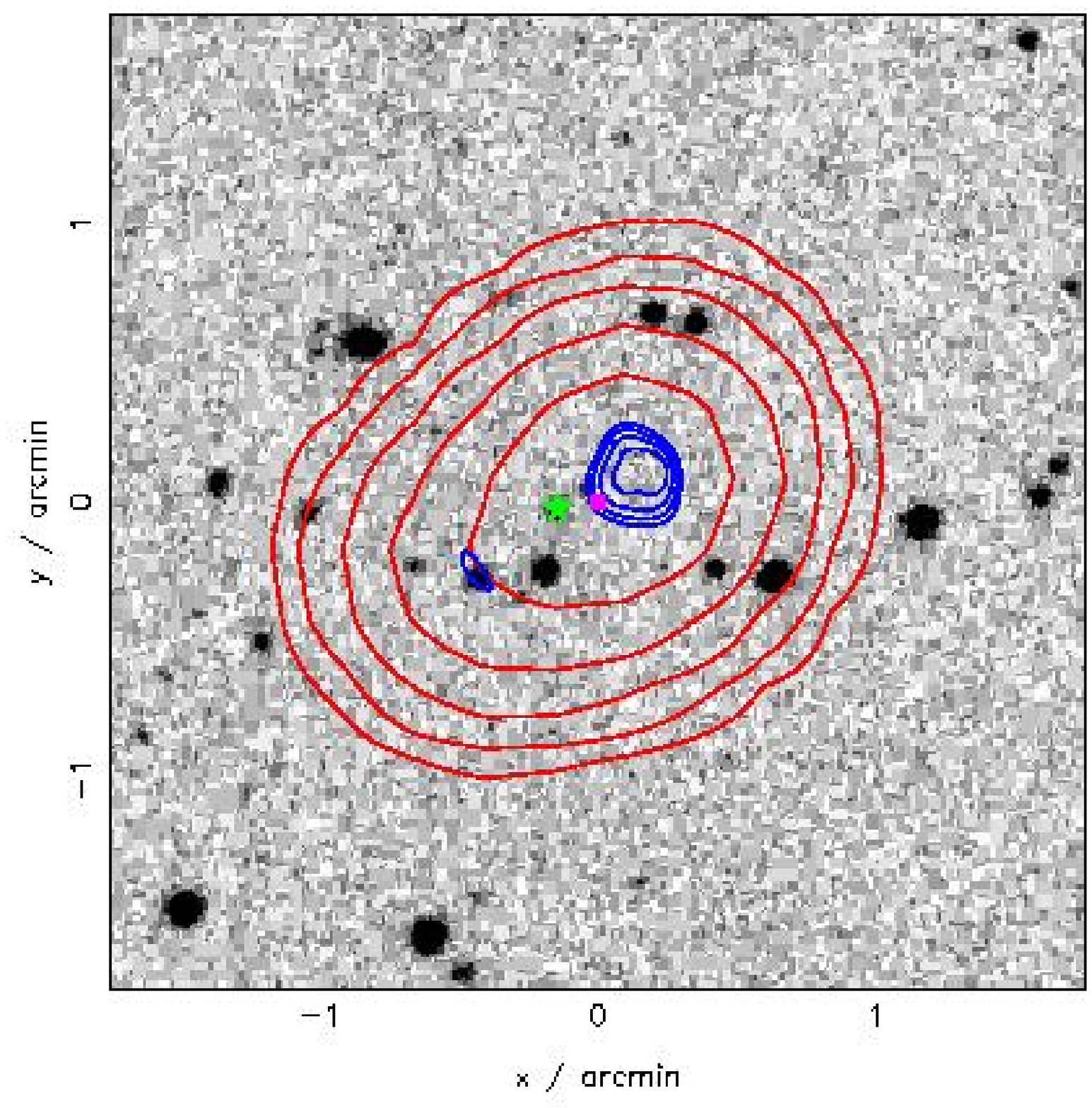}}
      \centerline{C4-137: TXS 1430+011}
    \end{minipage}
    \hspace{3cm}
    \begin{minipage}{3cm}
      \mbox{}
      \centerline{\includegraphics[scale=0.26,angle=270]{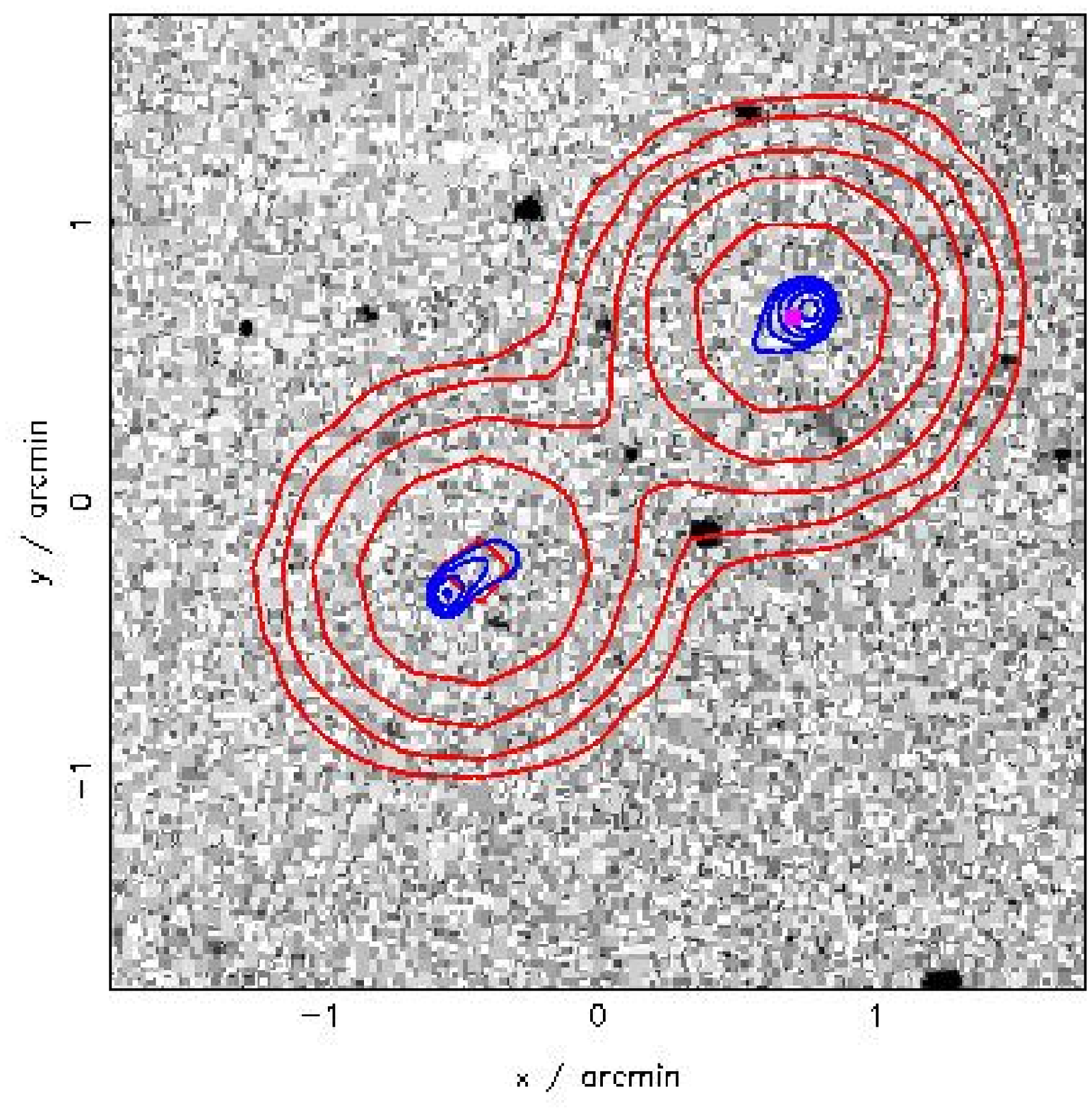}}
      \centerline{C4-138: TXS 1430-002}
    \end{minipage}
    \hspace{3cm}
    \begin{minipage}{3cm}
      \mbox{}
      \centerline{\includegraphics[scale=0.26,angle=270]{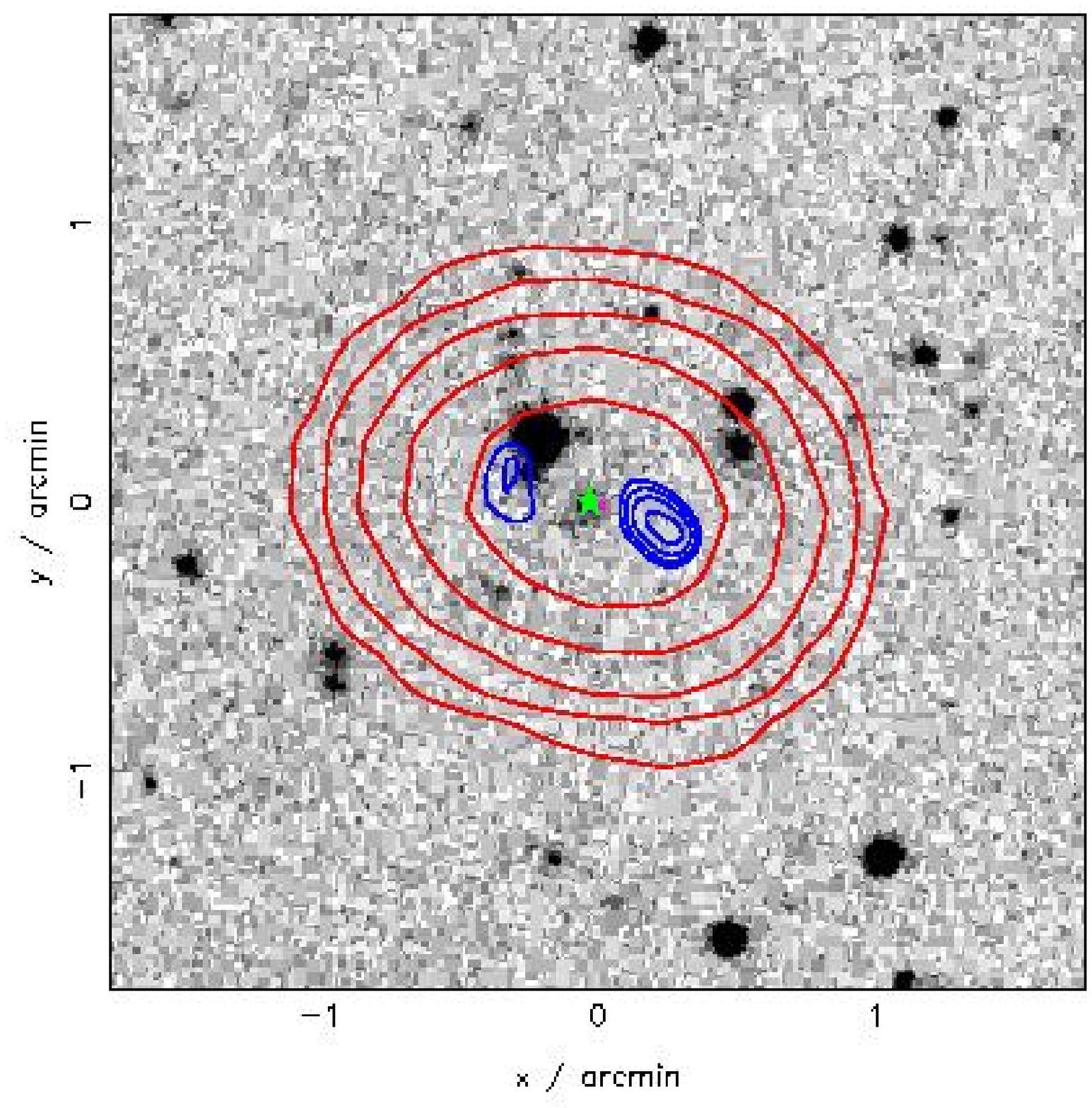}}
      \centerline{C4-142: GB6 B1431+0230}
    \end{minipage}
    \vfill
    \begin{minipage}{3cm}     
      \mbox{}
      \centerline{\includegraphics[scale=0.26,angle=270]{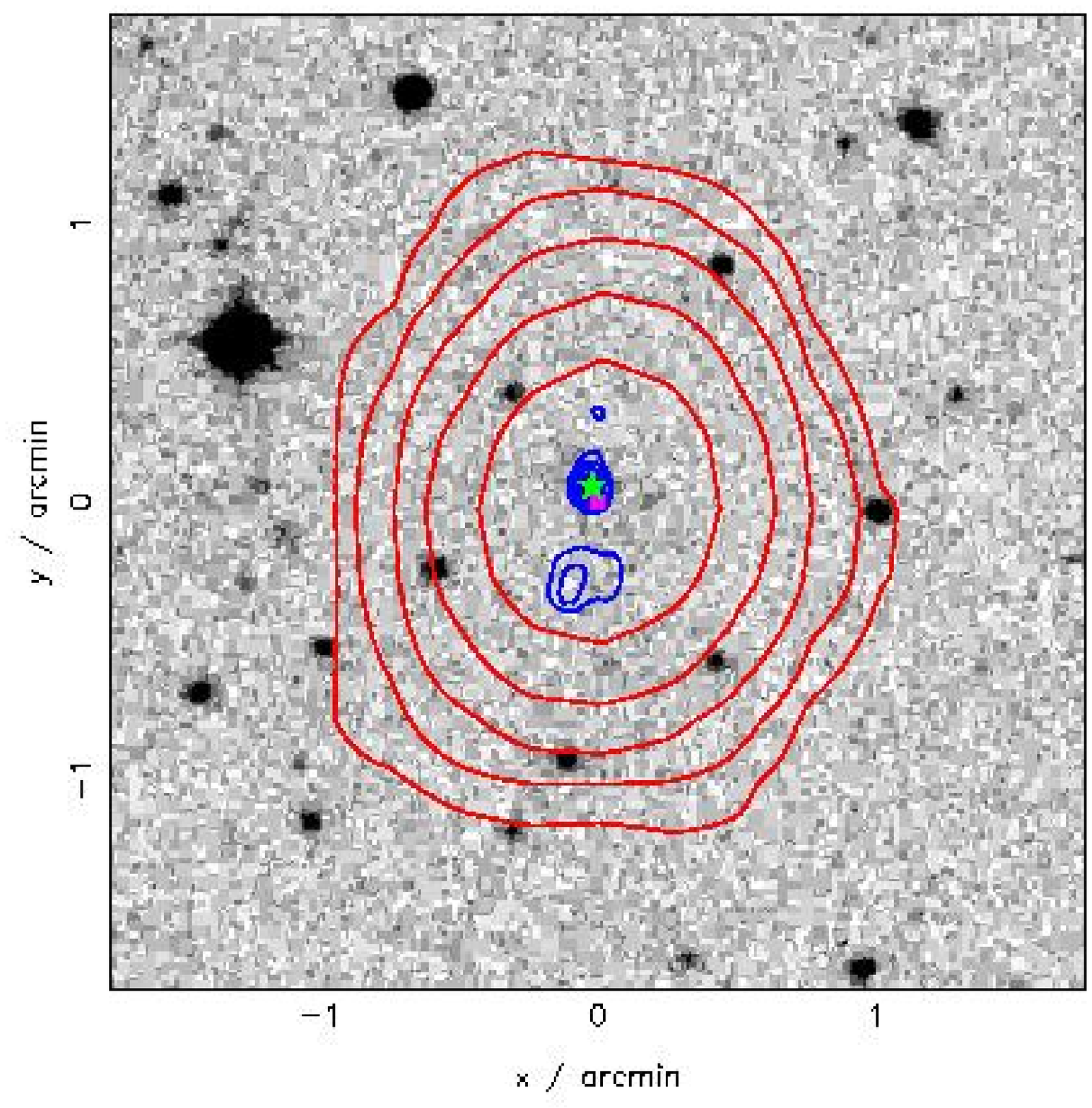}}
      \centerline{C4-143: 1433-0239}
    \end{minipage}
    \hspace{3cm}
    \begin{minipage}{3cm}
      \mbox{}
      \centerline{\includegraphics[scale=0.26,angle=270]{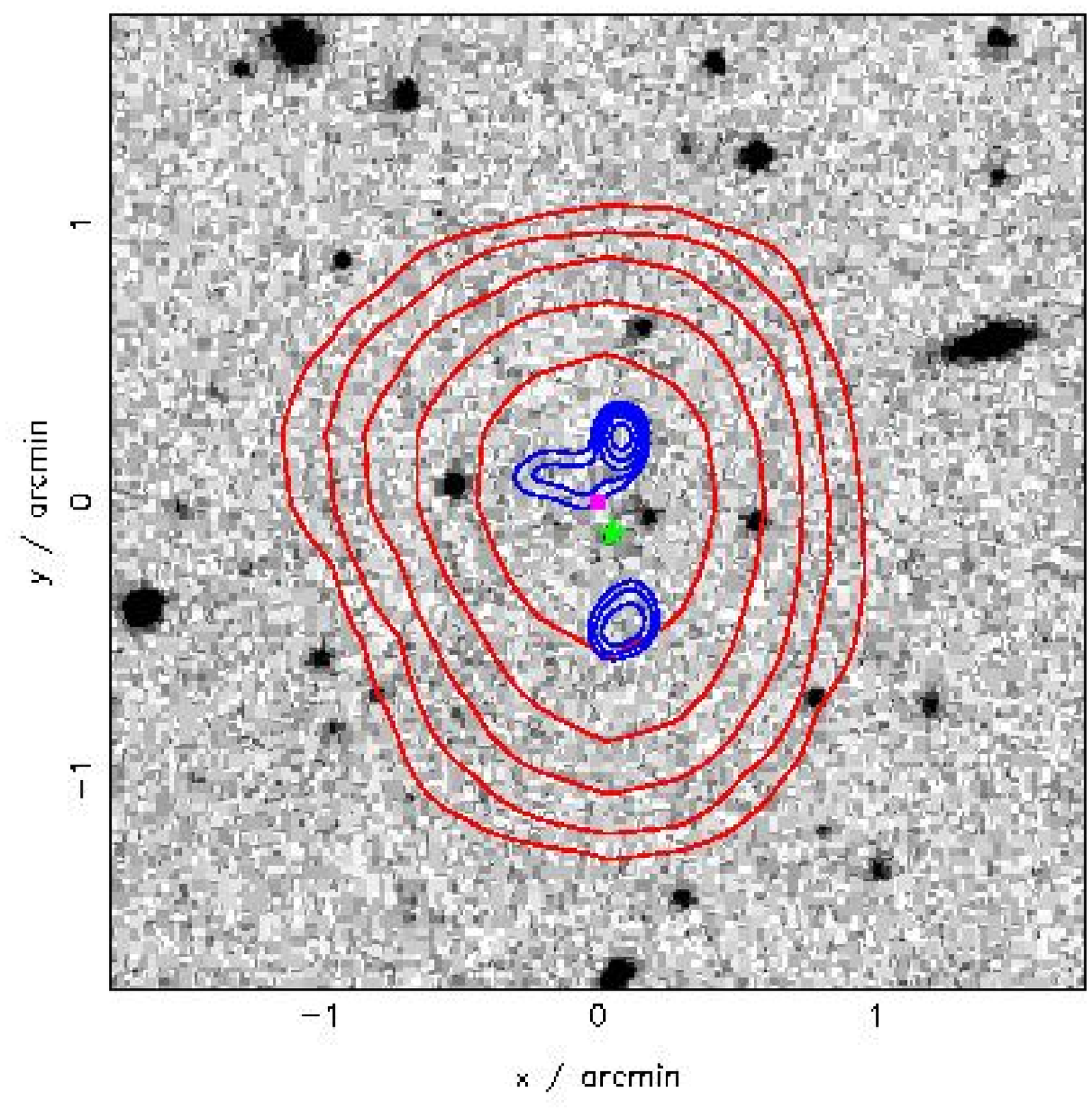}}
      \centerline{C4-145: TXS 1431+008}
    \end{minipage}
    \hspace{3cm}
    \begin{minipage}{3cm}
      \mbox{}
      \centerline{\includegraphics[scale=0.26,angle=270]{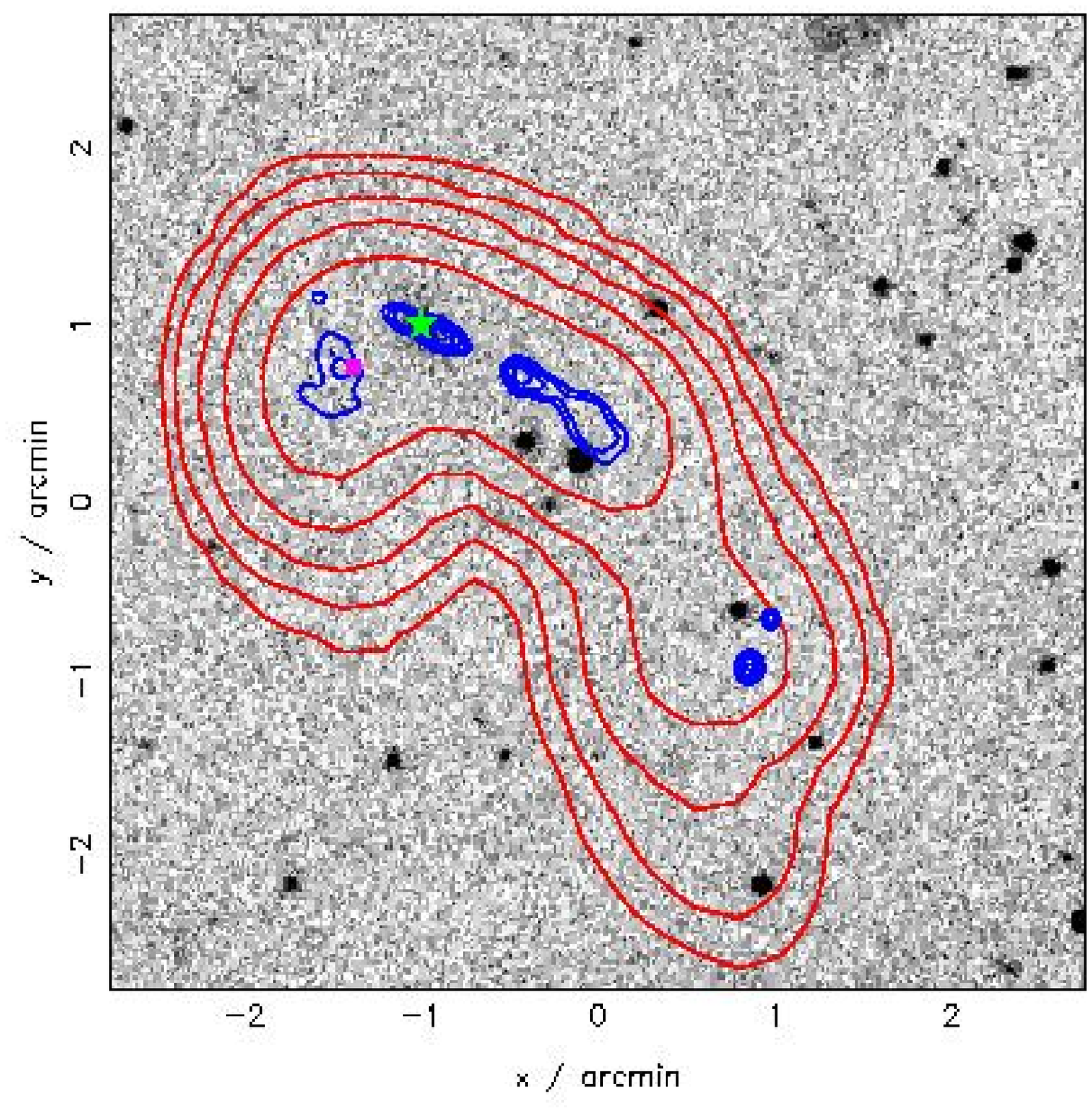}}
      \centerline{C4-146: 1434+0158}
    \end{minipage}
    \vfill
    \begin{minipage}{3cm}      
      \mbox{}
      \centerline{\includegraphics[scale=0.26,angle=270]{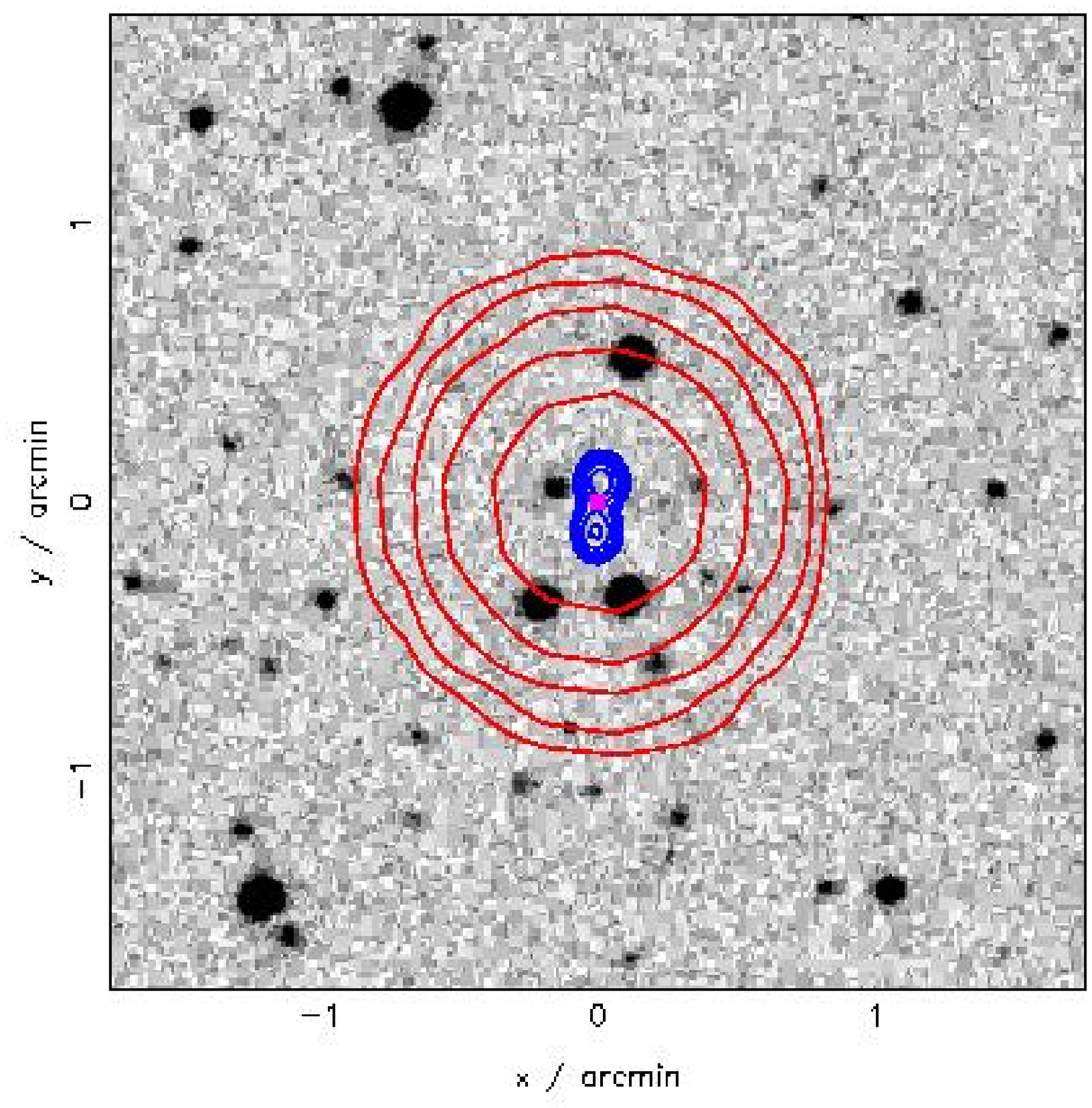}}
      \centerline{C4-148: TXS 1431-001}
    \end{minipage}
    \hspace{3cm}
    \begin{minipage}{3cm}
      \mbox{}
      \centerline{\includegraphics[scale=0.26,angle=270]{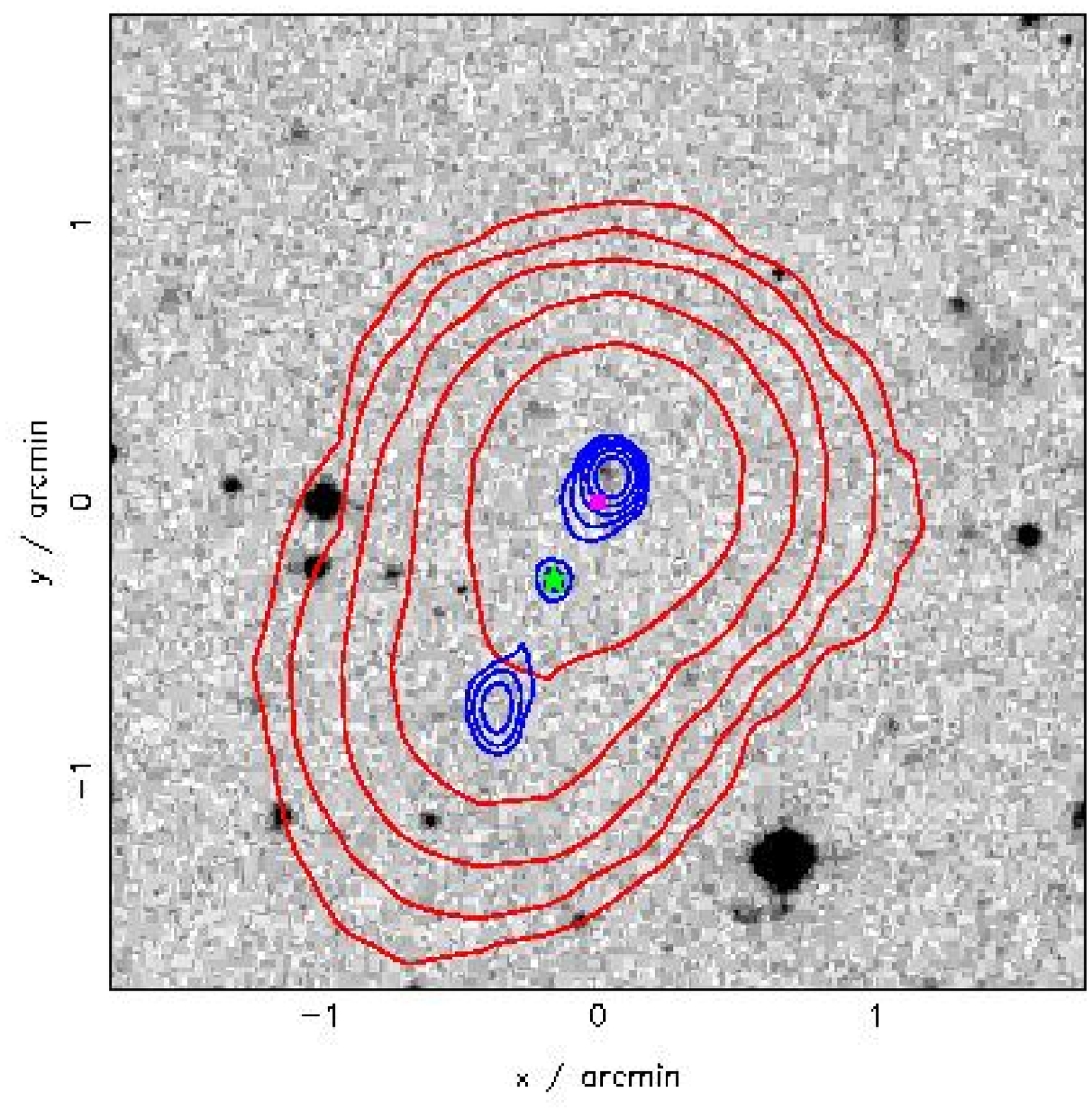}}
      \centerline{C4-149: TXS 1431-011}
    \end{minipage}
    \hspace{3cm}
    \begin{minipage}{3cm}
      \mbox{}
      \centerline{\includegraphics[scale=0.26,angle=270]{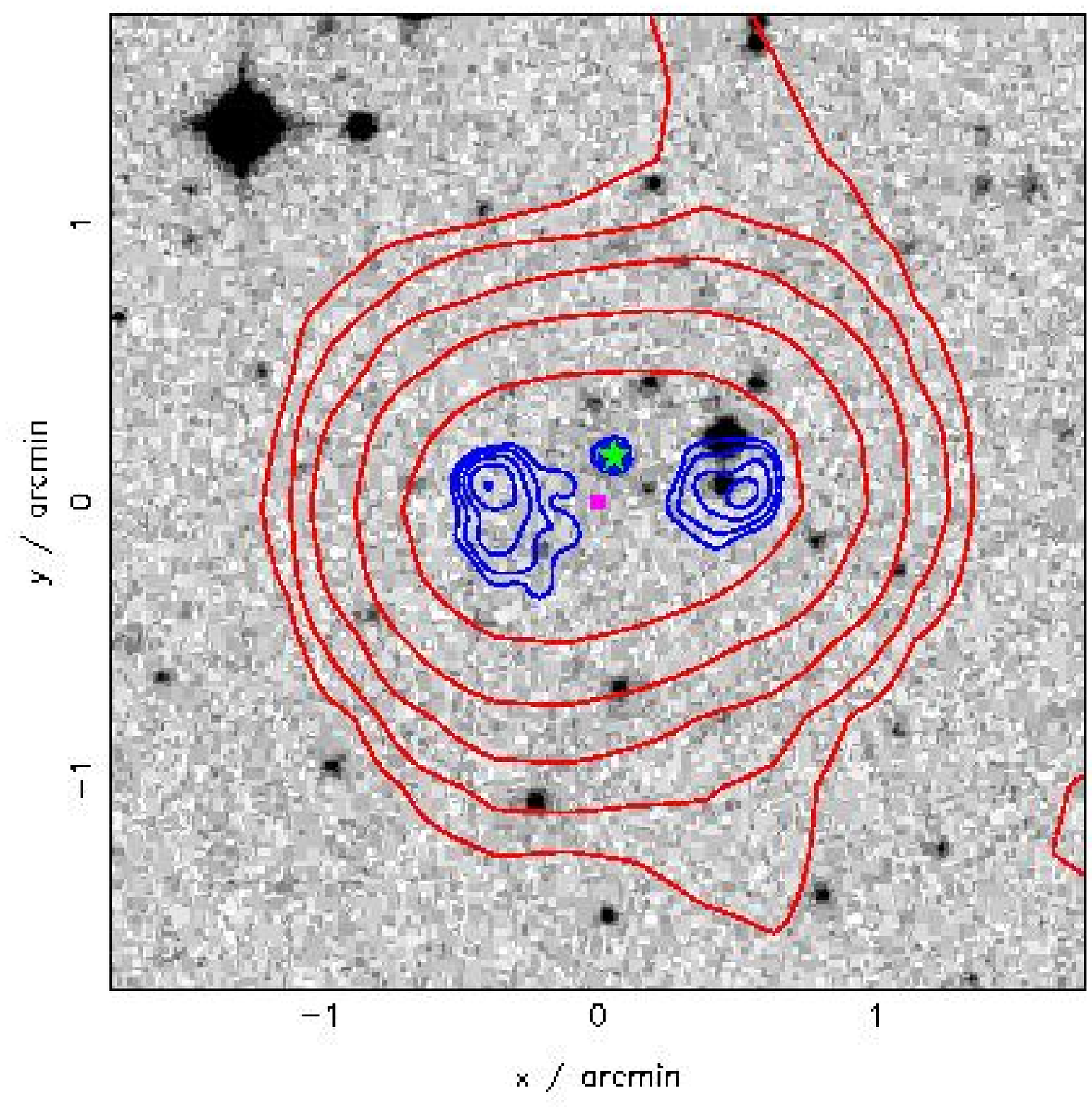}}
      \centerline{C4-150: TXS 1432-020}
    \end{minipage}
  \end{center}
\end{figure}

\begin{figure}
  \begin{center}
    {\bf CoNFIG-4}\\  
  \begin{minipage}{3cm}      
      \mbox{}
      \centerline{\includegraphics[scale=0.26,angle=270]{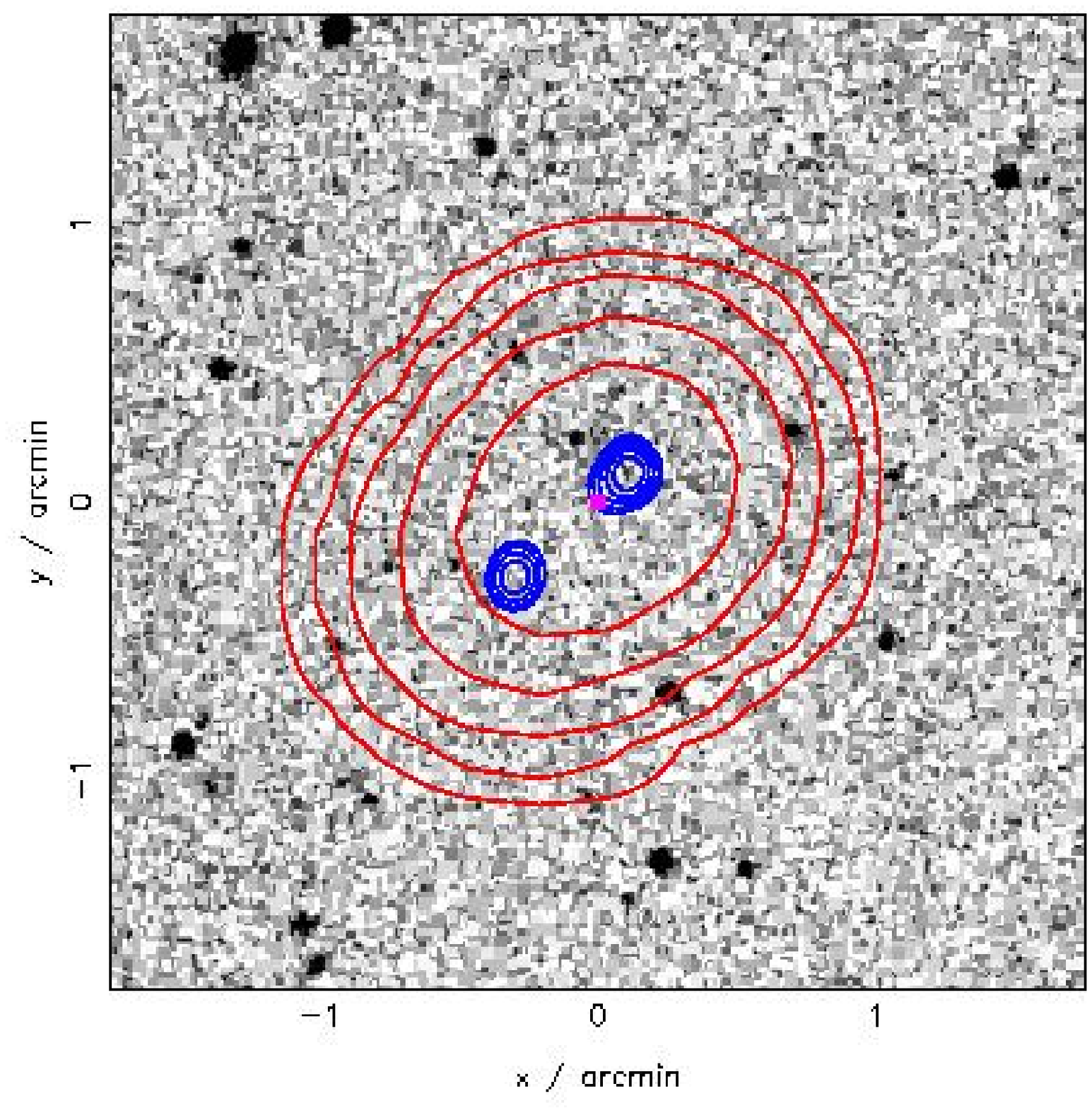}}
      \centerline{C4-151: TXS 1432+028}
    \end{minipage}
    \hspace{3cm}
    \begin{minipage}{3cm}
      \mbox{}
      \centerline{\includegraphics[scale=0.26,angle=270]{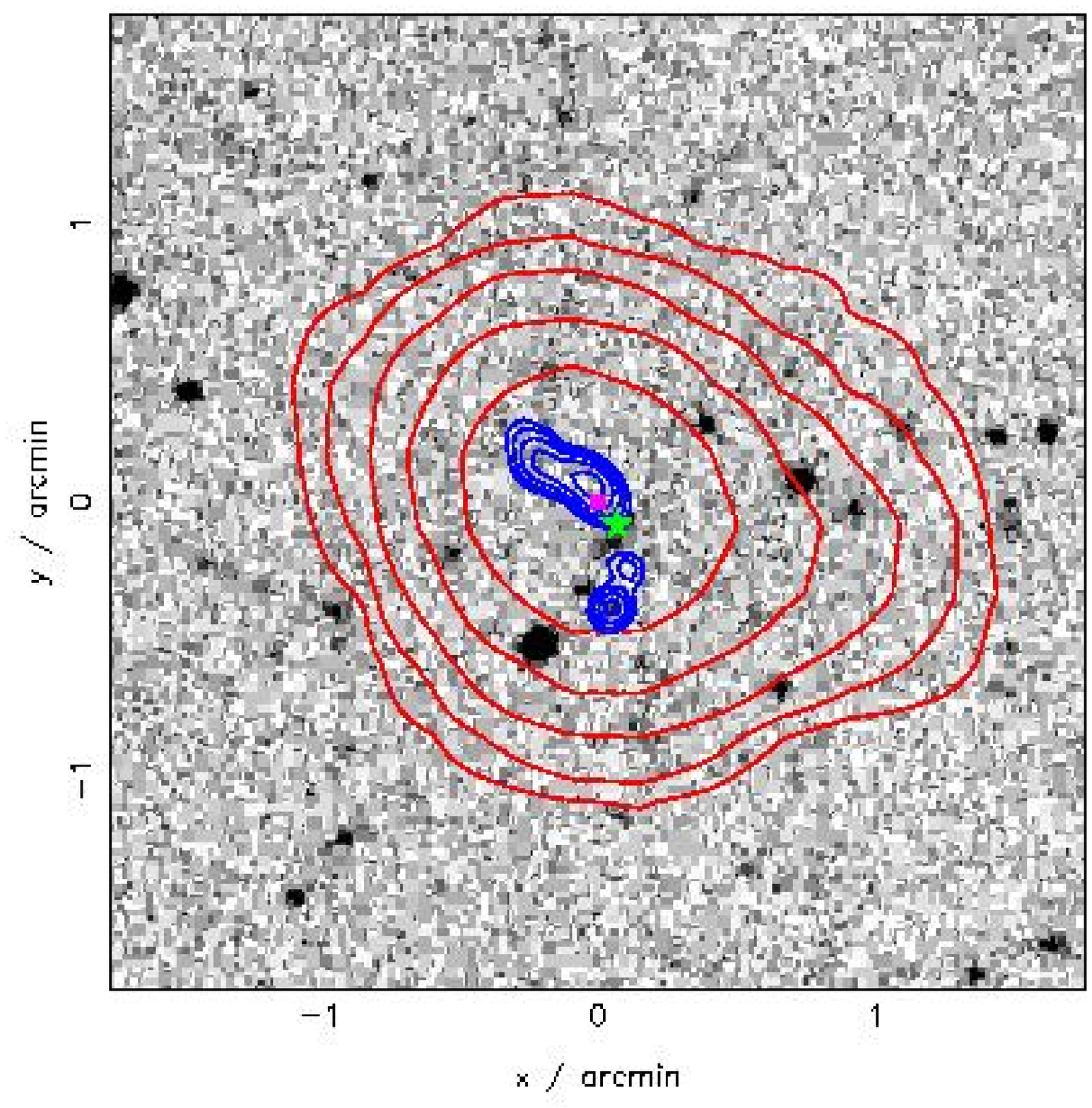}}
      \centerline{C4-152: 1435-0268}
    \end{minipage}
    \hspace{3cm}
    \begin{minipage}{3cm}
      \mbox{}
      \centerline{\includegraphics[scale=0.26,angle=270]{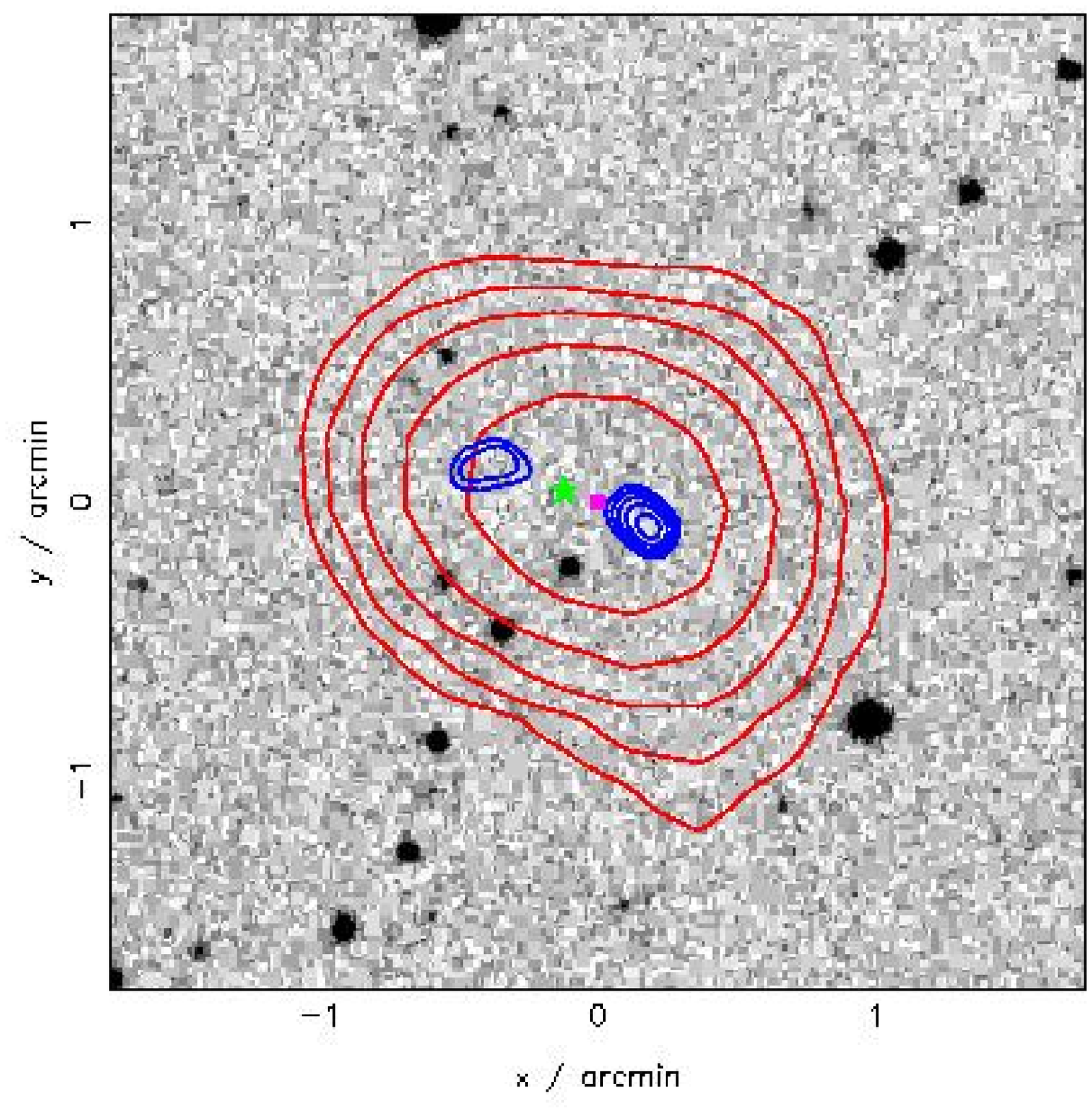}}
      \centerline{C4-153: 1435+0243}
    \end{minipage}
    \vfill
    \begin{minipage}{3cm}     
      \mbox{}
      \centerline{\includegraphics[scale=0.26,angle=270]{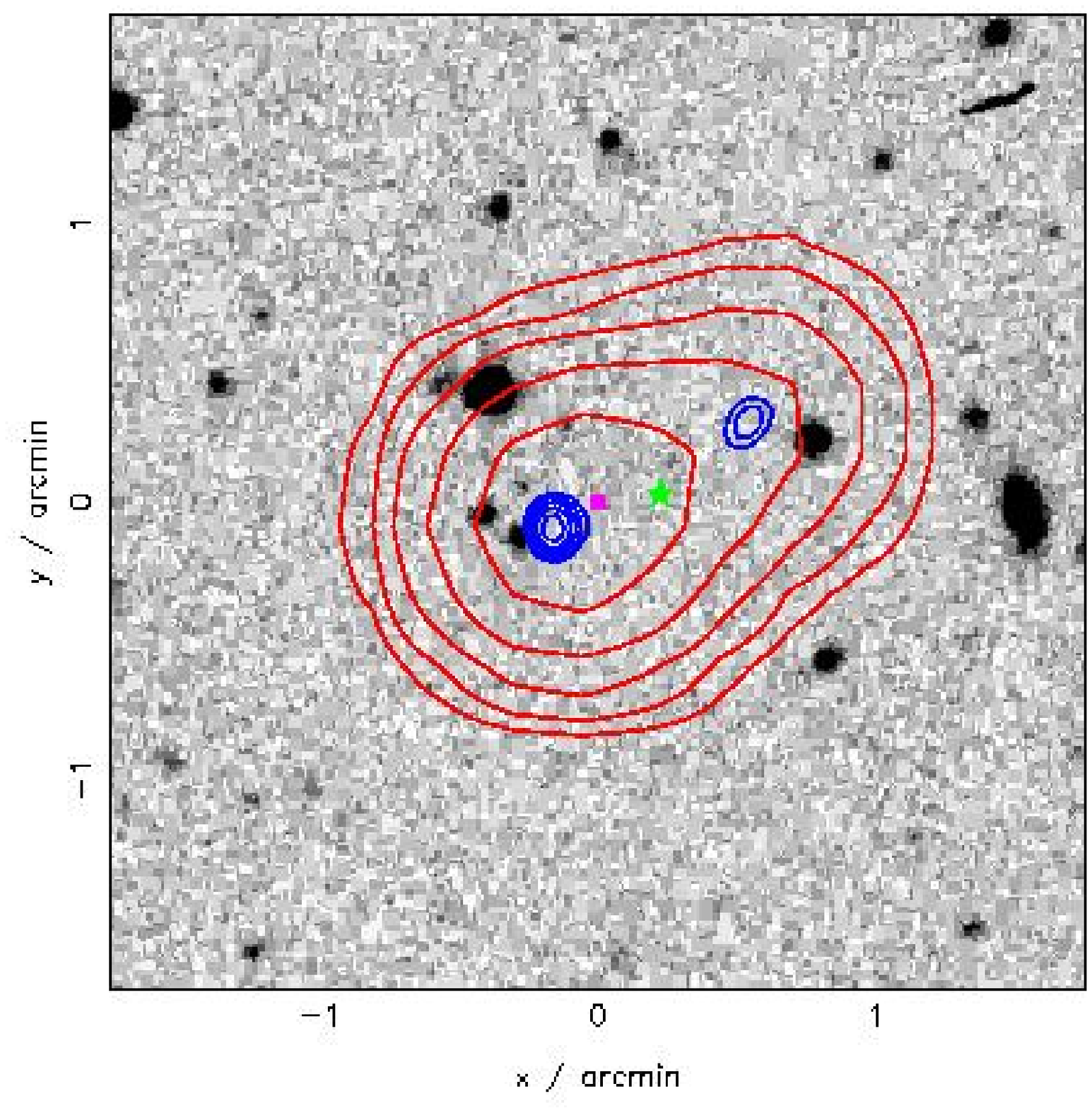}}
      \centerline{C4-154: TXS 1433-015}
    \end{minipage}
    \hspace{3cm}
    \begin{minipage}{3cm}
      \mbox{}
      \centerline{\includegraphics[scale=0.26,angle=270]{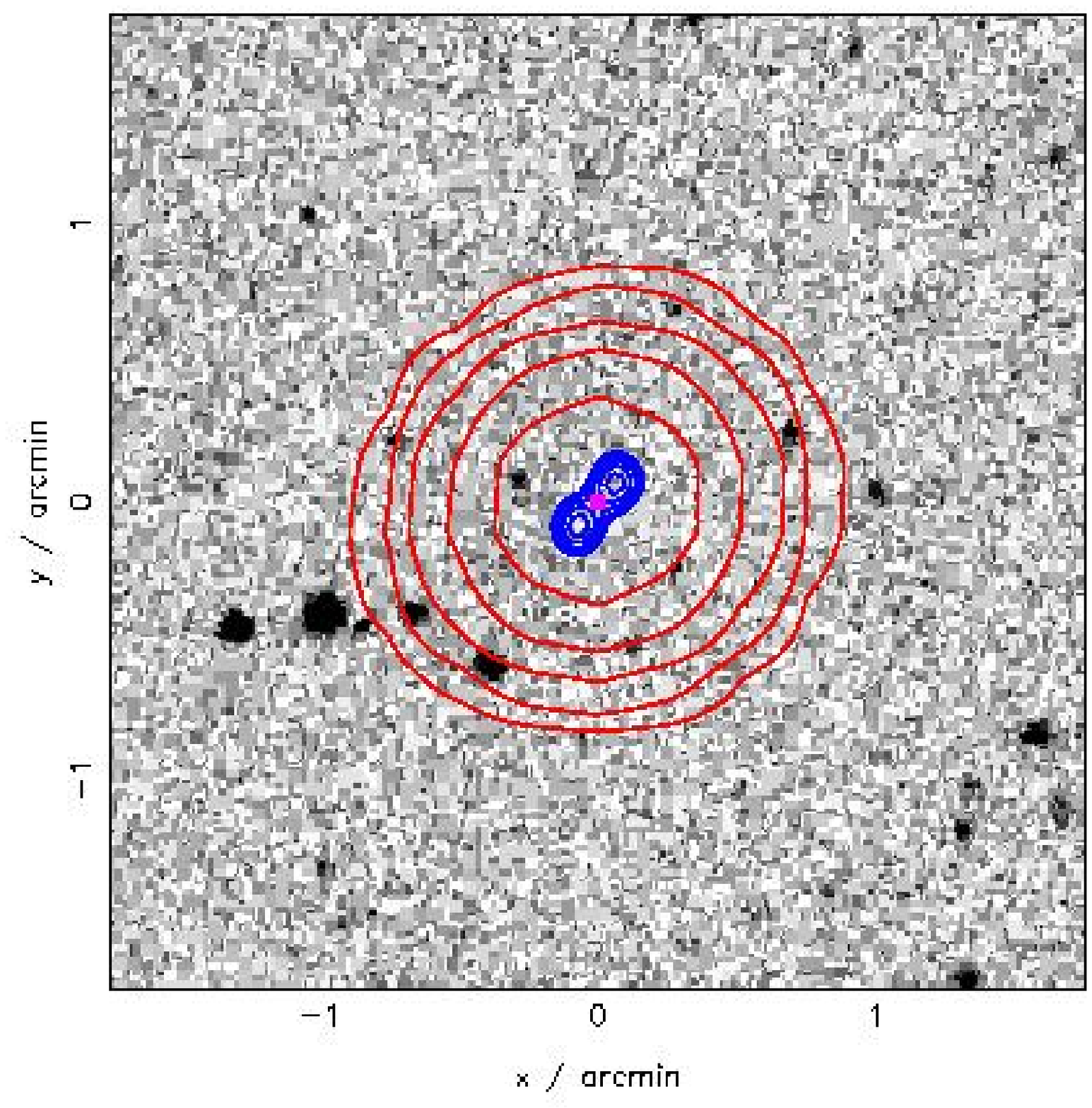}}
      \centerline{C4-160: TXS 1434-028}
    \end{minipage}
    \hspace{3cm}
    \begin{minipage}{3cm}
      \mbox{}
      \centerline{\includegraphics[scale=0.26,angle=270]{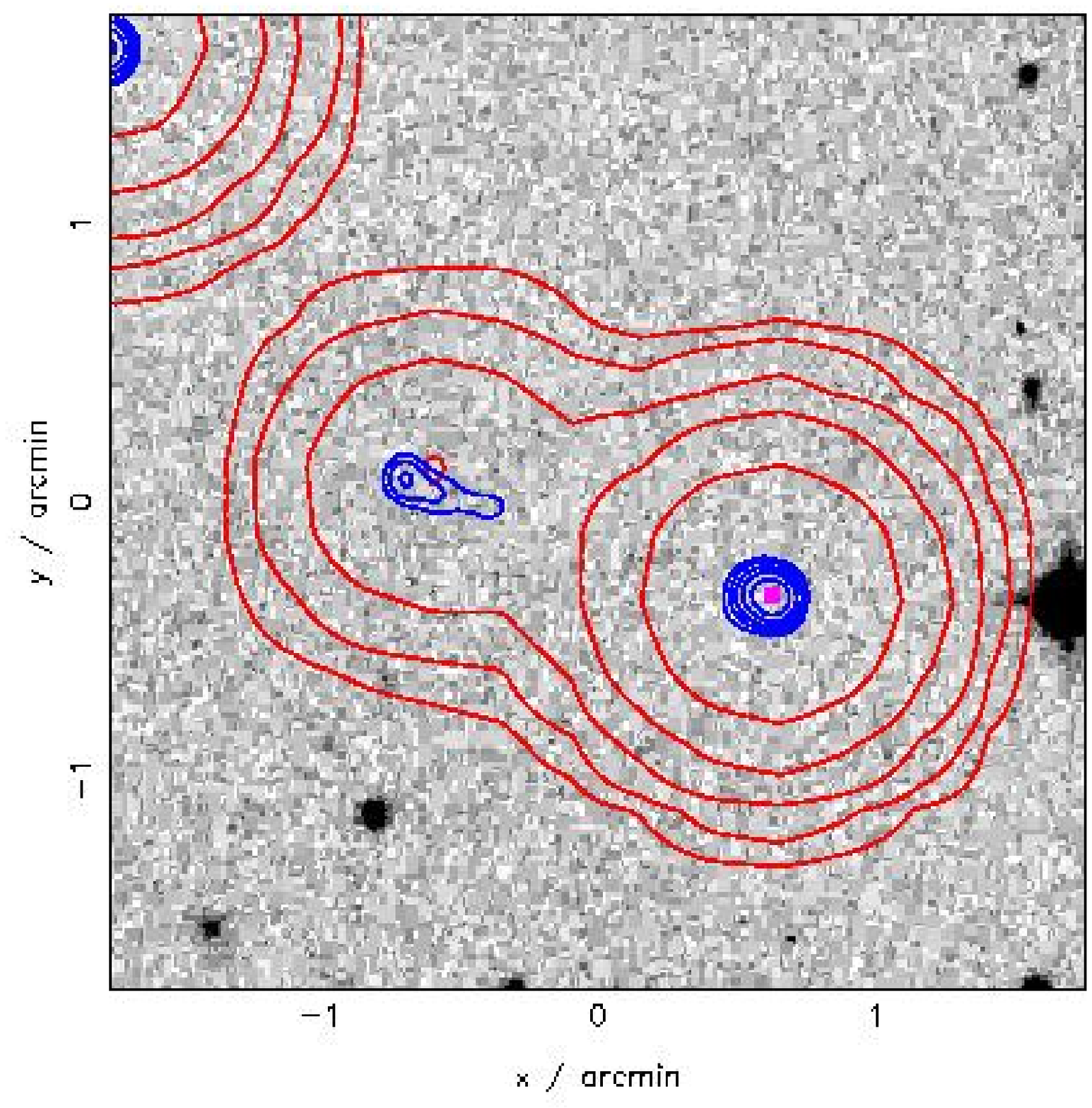}}
      \centerline{C4-162: TXS 1434+019}
    \end{minipage}
    \vfill
    \begin{minipage}{3cm}     
      \mbox{}
      \centerline{\includegraphics[scale=0.26,angle=270]{Contours/C4/163.ps}}
      \centerline{C4-163: 1437+0175}
    \end{minipage}
    \hspace{3cm}
    \begin{minipage}{3cm}
      \mbox{}
      \centerline{\includegraphics[scale=0.26,angle=270]{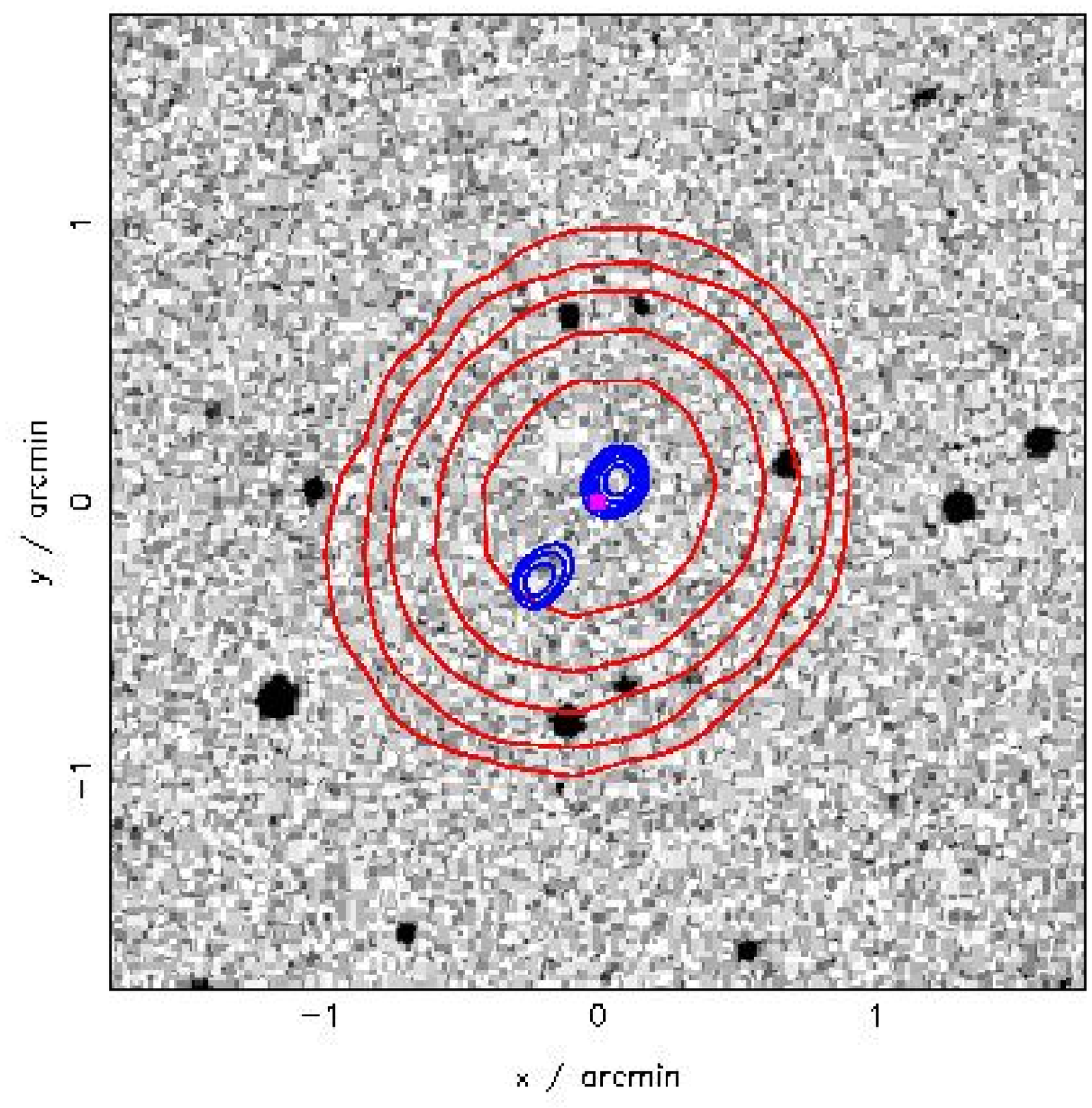}}
      \centerline{C4-164: TXS 1435+031}
    \end{minipage}
    \hspace{3cm}
    \begin{minipage}{3cm}
      \mbox{}
      \centerline{\includegraphics[scale=0.26,angle=270]{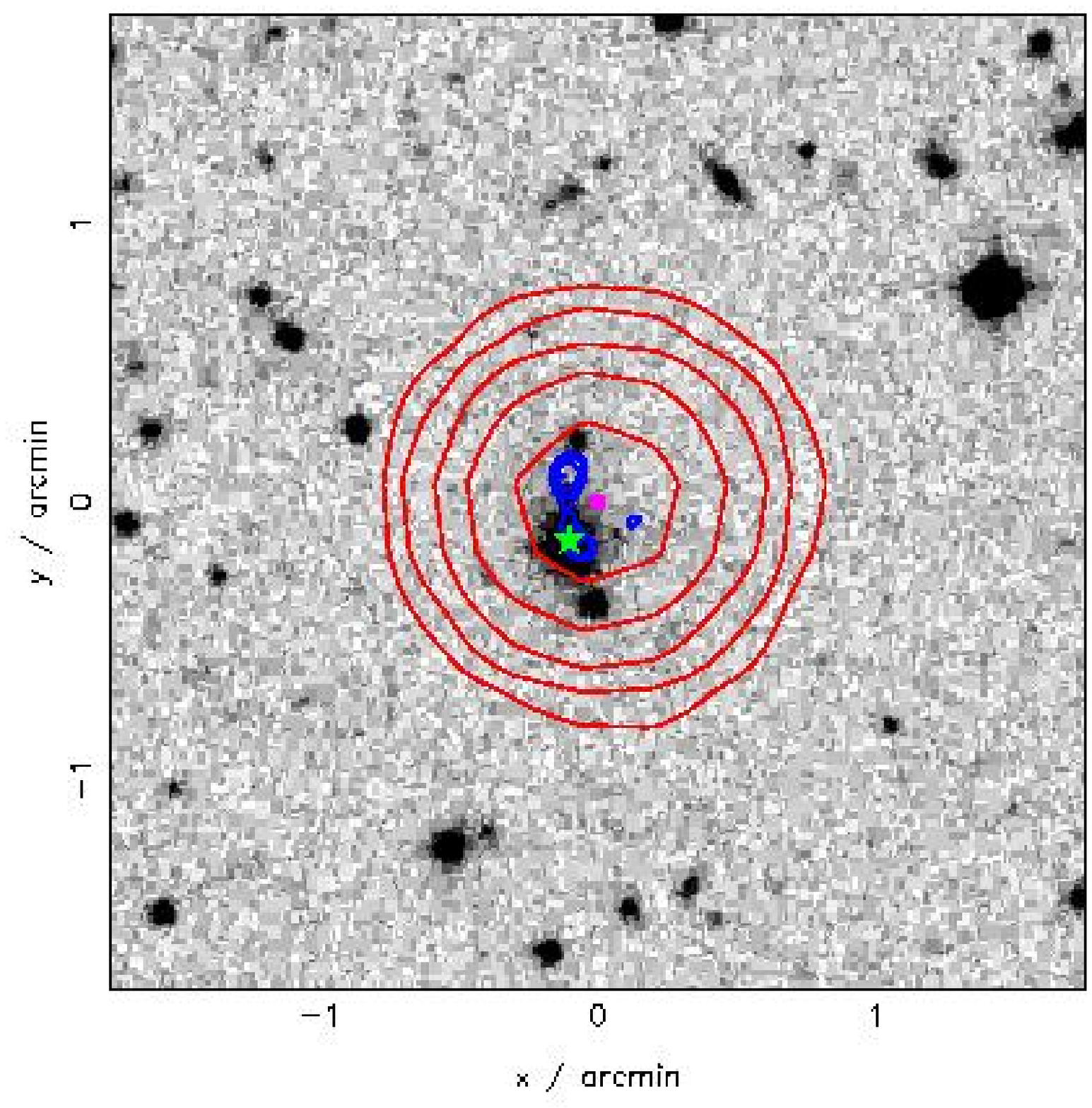}}
      \centerline{C4-166: 1437-0025}
    \end{minipage}
    \vfill
    \begin{minipage}{3cm}      
      \mbox{}
      \centerline{\includegraphics[scale=0.26,angle=270]{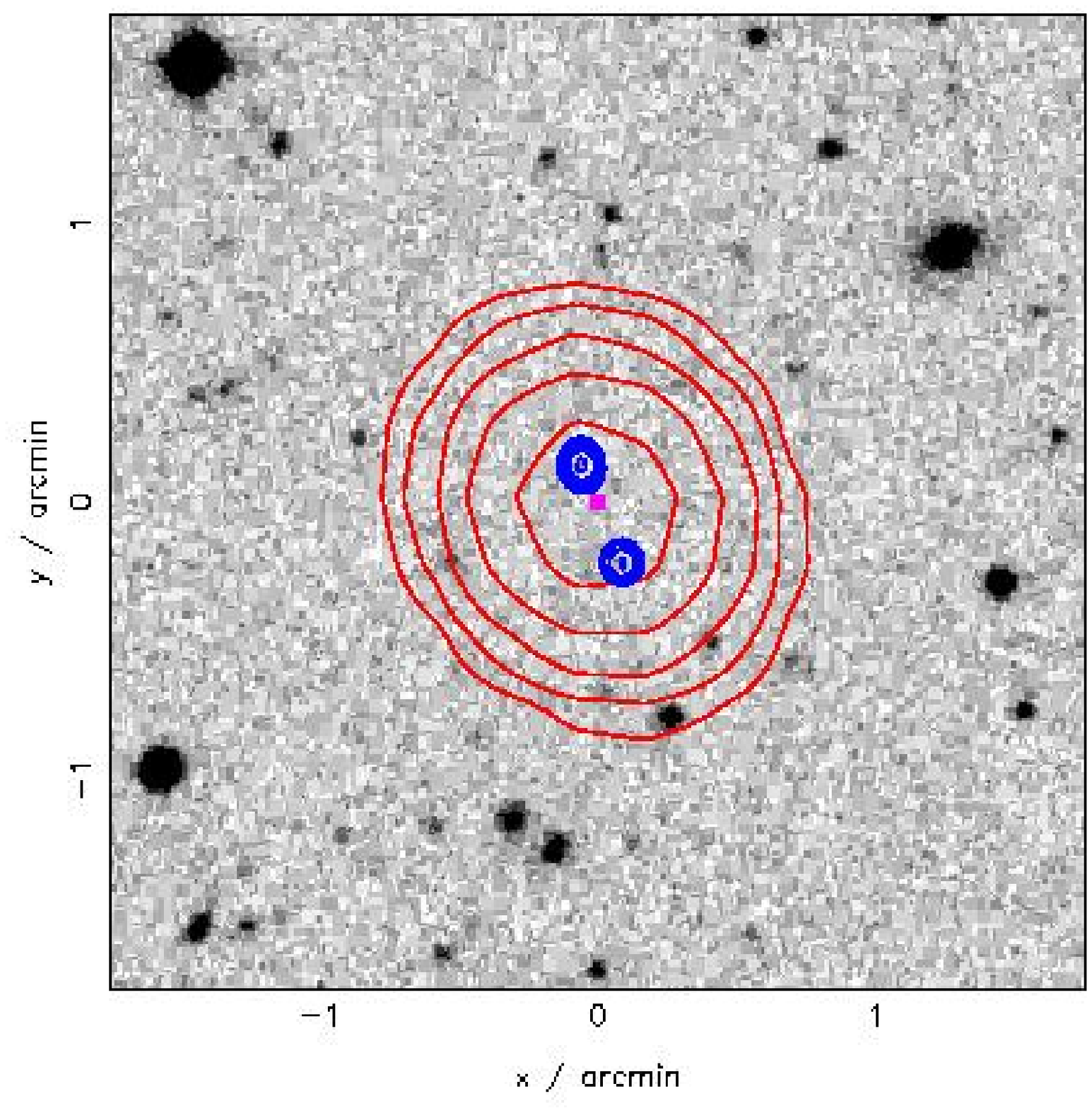}}
      \centerline{C4-168: 1437-0069}
    \end{minipage}
    \hspace{3cm}
    \begin{minipage}{3cm}
      \mbox{}
      \centerline{\includegraphics[scale=0.26,angle=270]{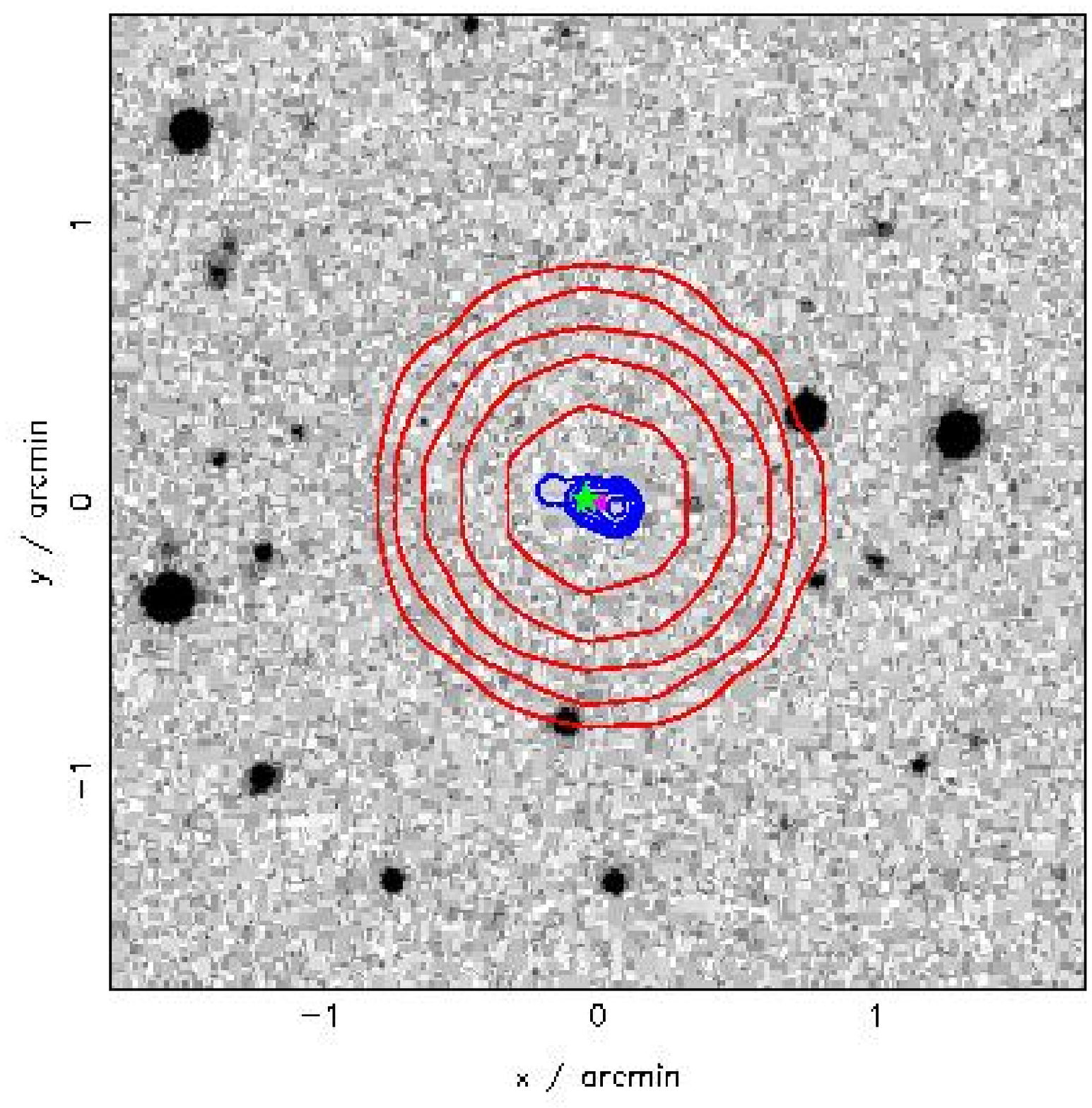}}
      \centerline{C4-169: J143757+01}
    \end{minipage}
    \hspace{3cm}
    \begin{minipage}{3cm}
      \mbox{}
      \centerline{\includegraphics[scale=0.26,angle=270]{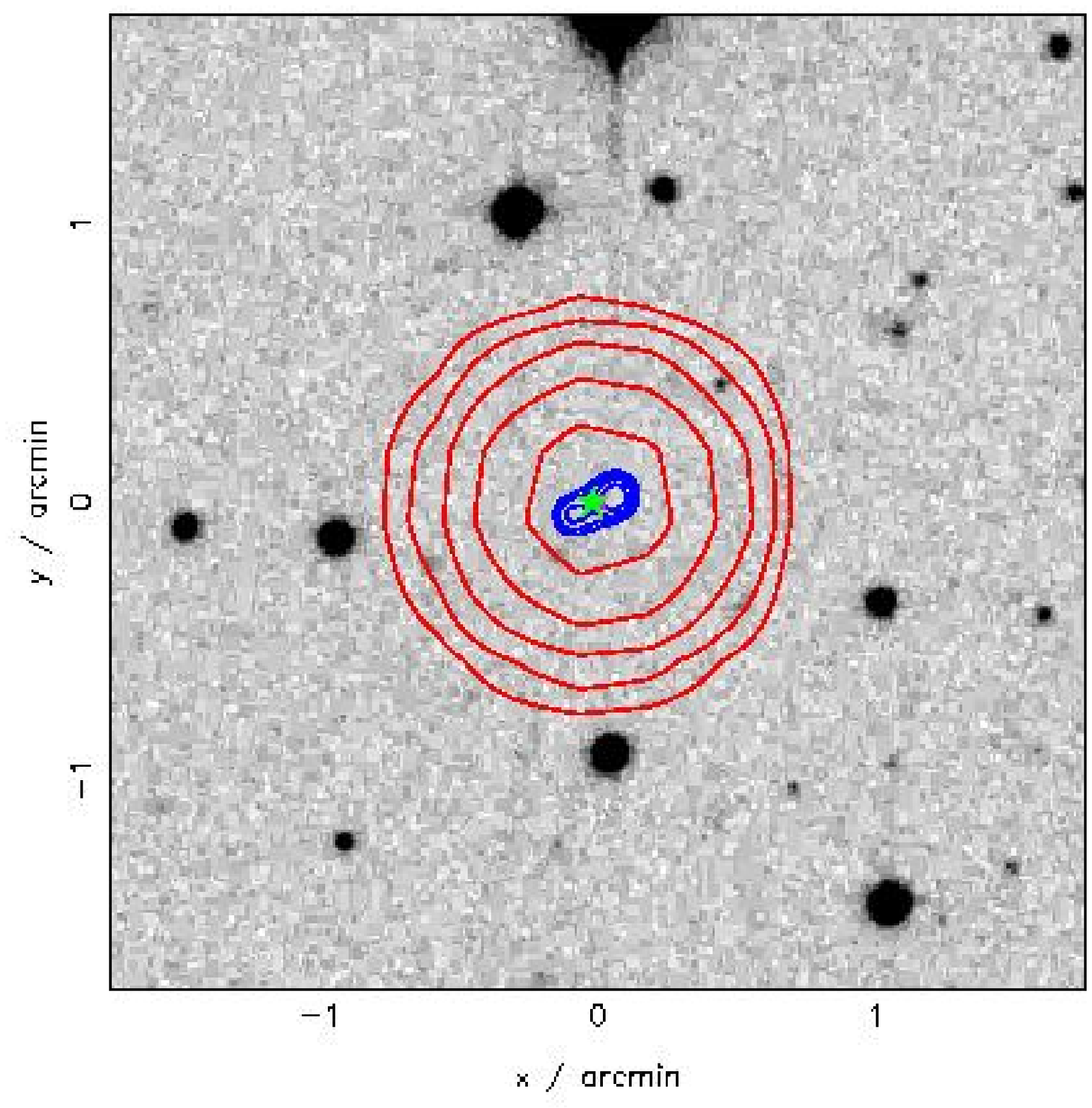}}
      \centerline{C4-170: TXS 1435+028}
    \end{minipage}
  \end{center}
\end{figure}

\begin{figure}
  \begin{center}
    {\bf CoNFIG-4}\\  
  \begin{minipage}{3cm}      
      \mbox{}
      \centerline{\includegraphics[scale=0.26,angle=270]{Contours/C4/175.ps}}
      \centerline{C4-175: TXS 1435+020}
    \end{minipage}
    \hspace{3cm}
    \begin{minipage}{3cm}
      \mbox{}
      \centerline{\includegraphics[scale=0.26,angle=270]{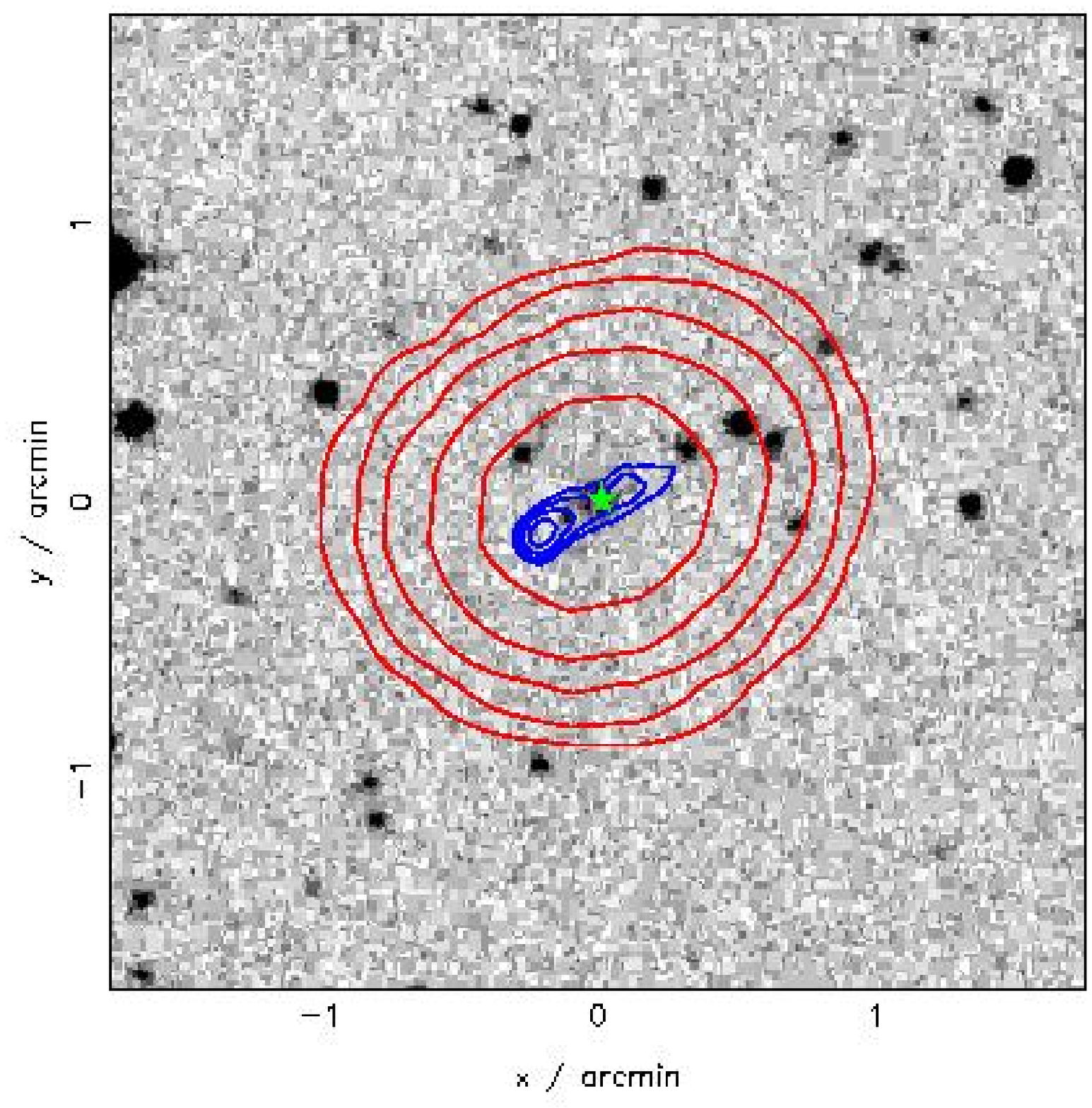}}
      \centerline{C4-176: 1438-0133}
    \end{minipage}
    \hspace{3cm}
    \begin{minipage}{3cm}
      \mbox{}
      \centerline{\includegraphics[scale=0.26,angle=270]{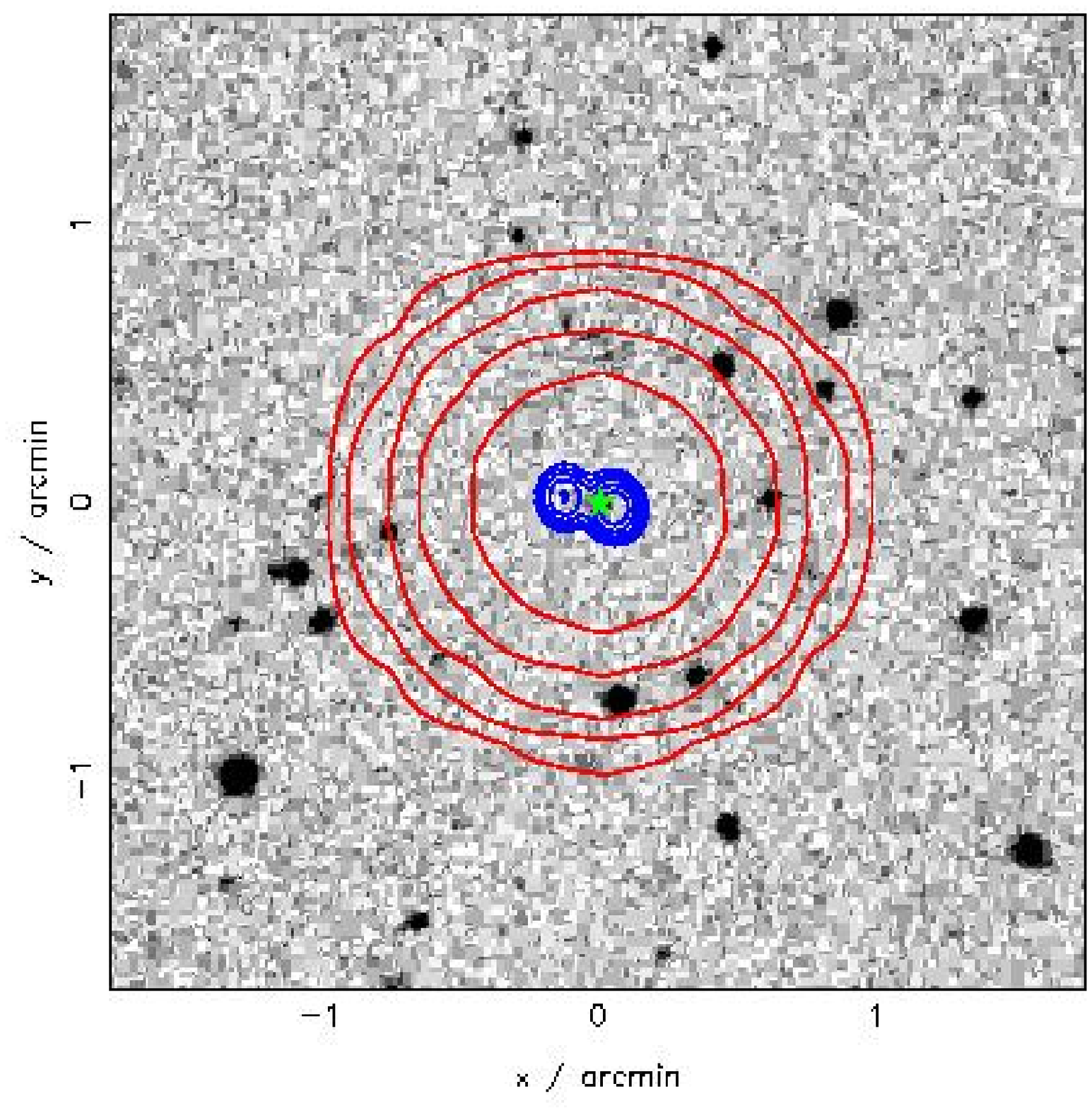}}
      \centerline{C4-177: 4C -02.61}
    \end{minipage}
    \vfill
    \begin{minipage}{3cm}     
      \mbox{}
      \centerline{\includegraphics[scale=0.26,angle=270]{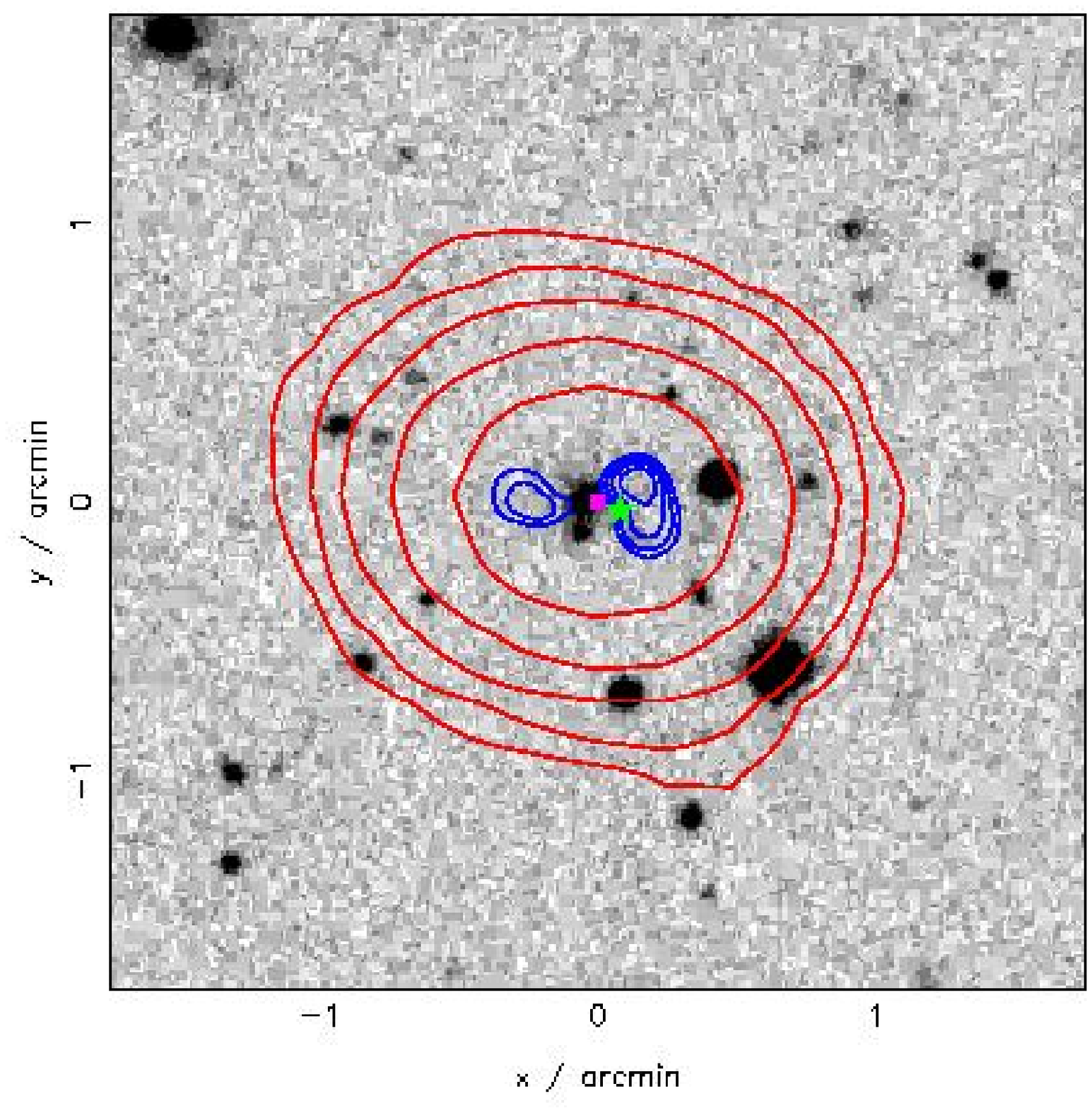}}
      \centerline{C4-178: 1438-0100}
    \end{minipage}
    \hspace{3cm}
    \begin{minipage}{3cm}
      \mbox{}
      \centerline{\includegraphics[scale=0.26,angle=270]{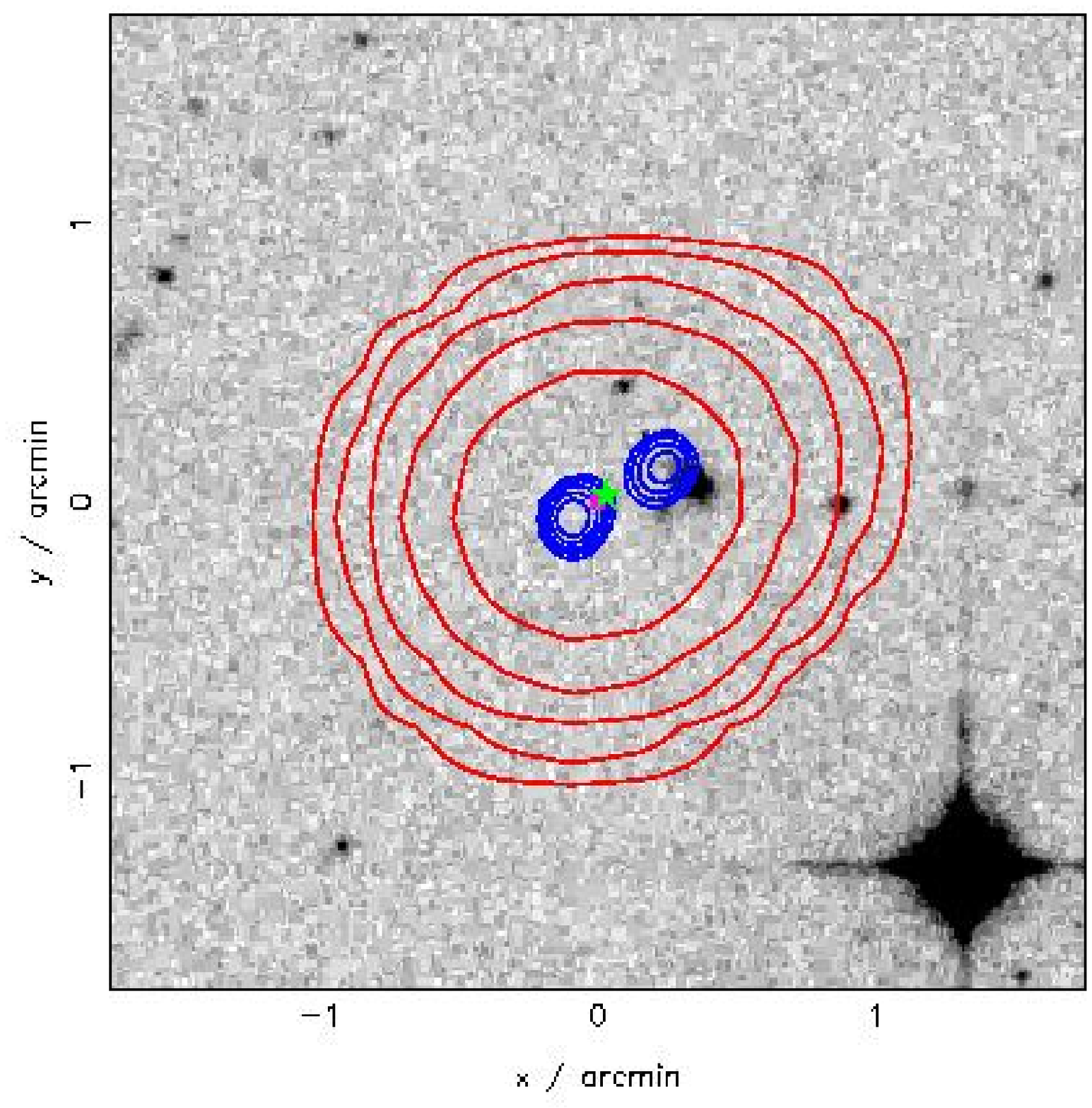}}
      \centerline{C4-179: 4C 00.50}
    \end{minipage}
    \hspace{3cm}
    \begin{minipage}{3cm}
      \mbox{}
      \centerline{\includegraphics[scale=0.26,angle=270]{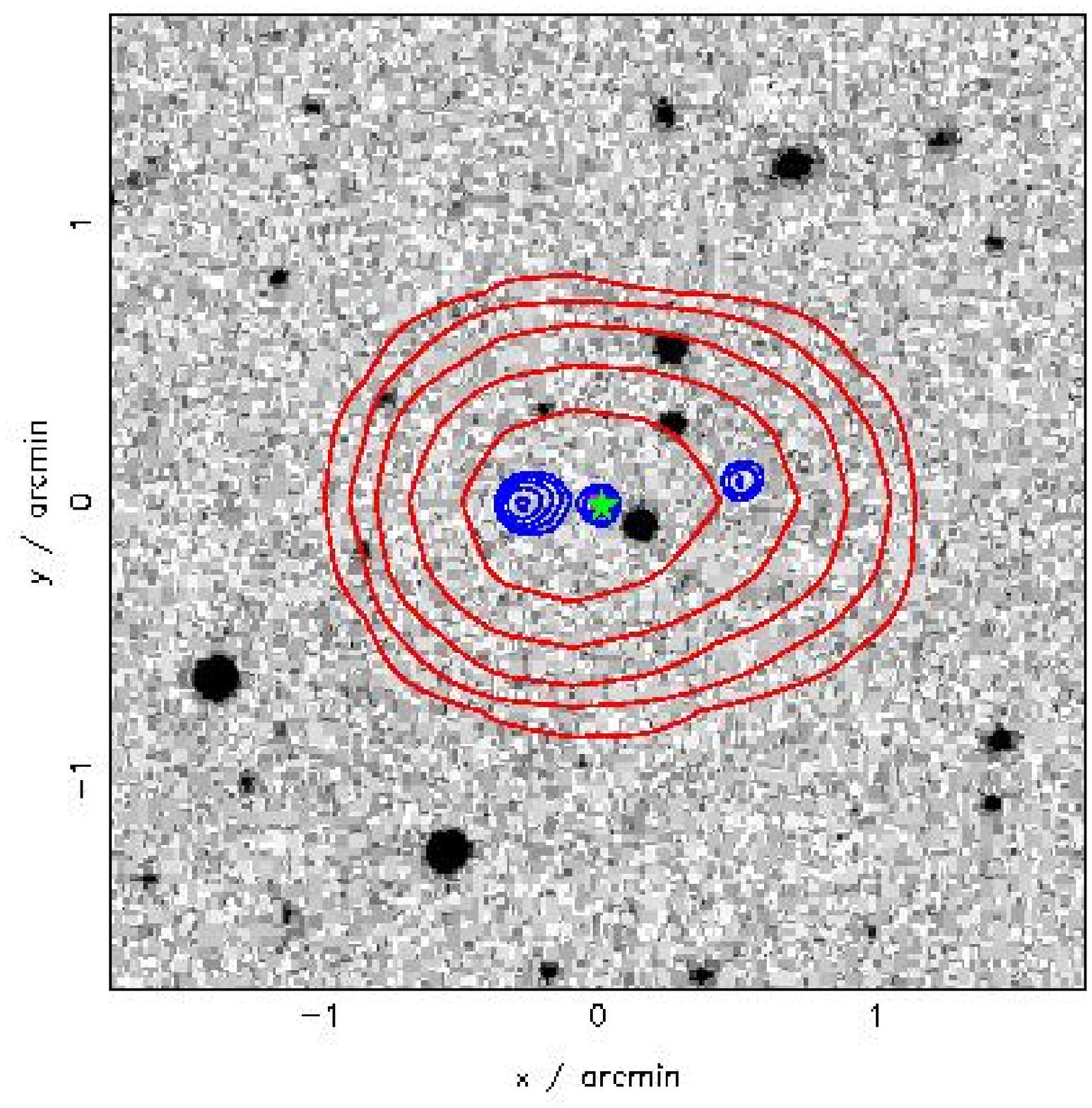}}
      \centerline{C4-180: 1438+0022}
    \end{minipage}
    \vfill
    \begin{minipage}{3cm}     
      \mbox{}
      \centerline{\includegraphics[scale=0.26,angle=270]{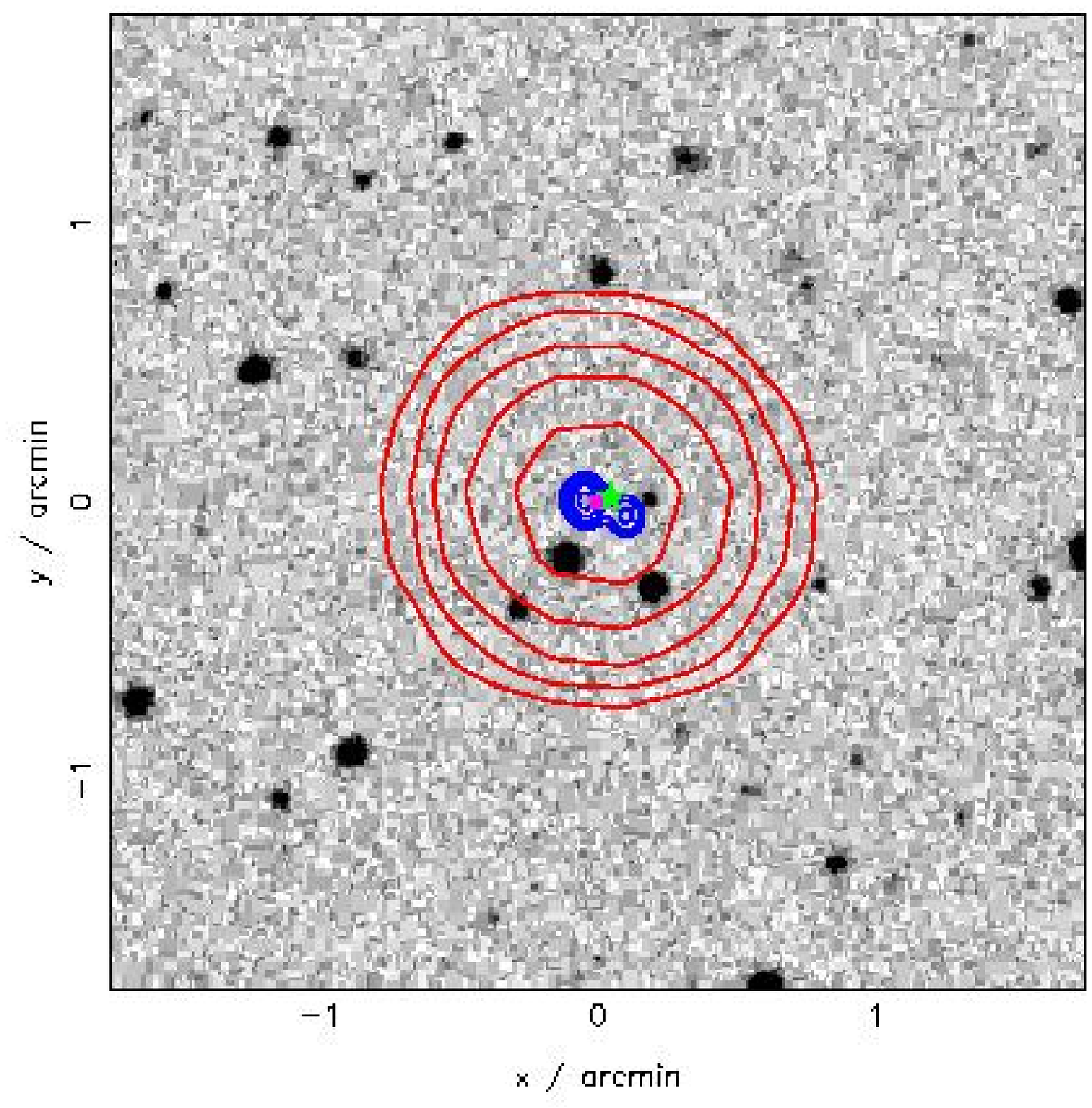}}
      \centerline{C4-182: TXS 1436+011}
    \end{minipage}
    \hspace{3cm}
    \begin{minipage}{3cm}
      \mbox{}
      \centerline{\includegraphics[scale=0.26,angle=270]{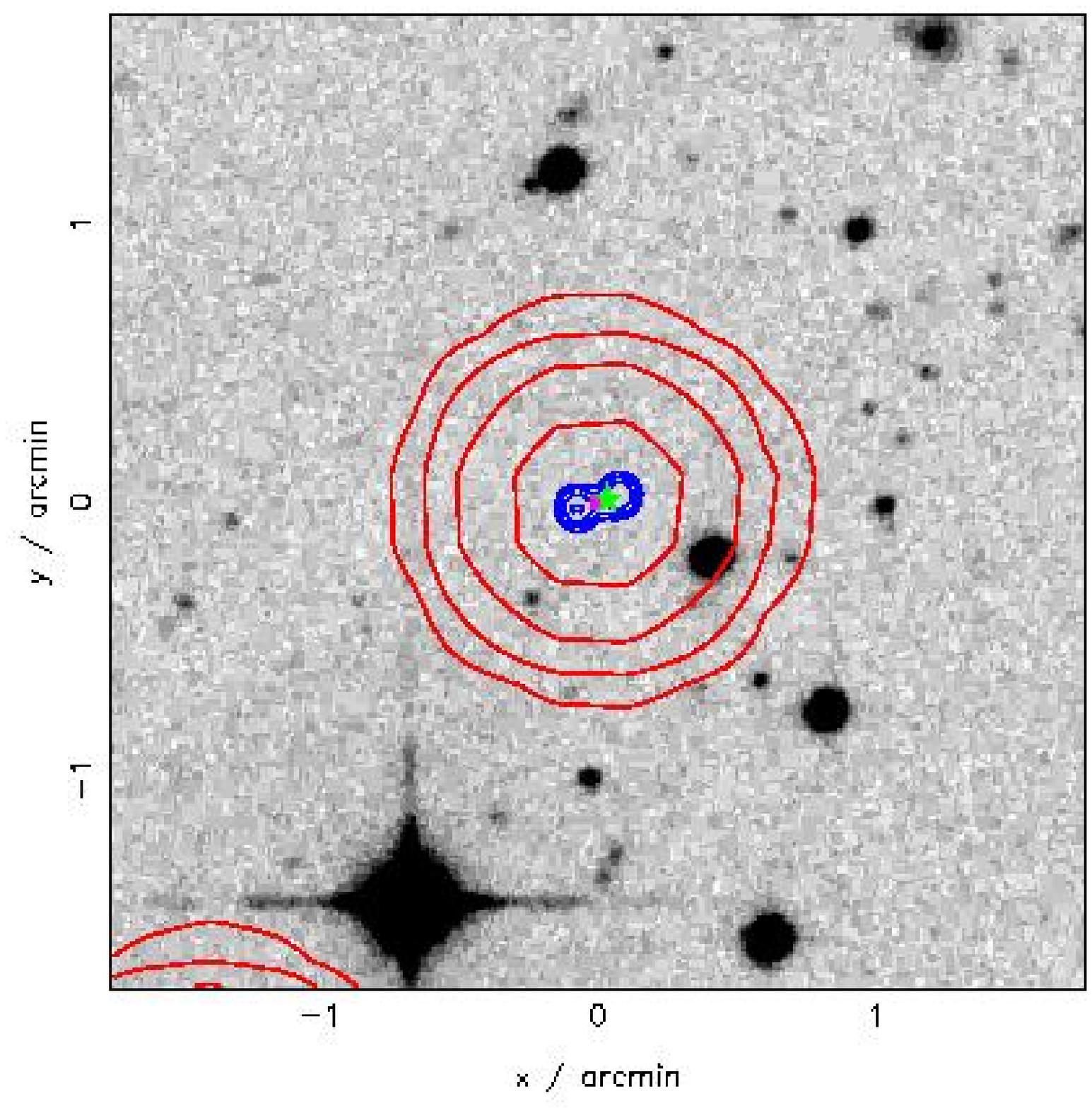}}
      \centerline{C4-183: 1438-0081}
    \end{minipage}
    \hspace{3cm}
    \begin{minipage}{3cm}
      \mbox{}
      \centerline{\includegraphics[scale=0.26,angle=270]{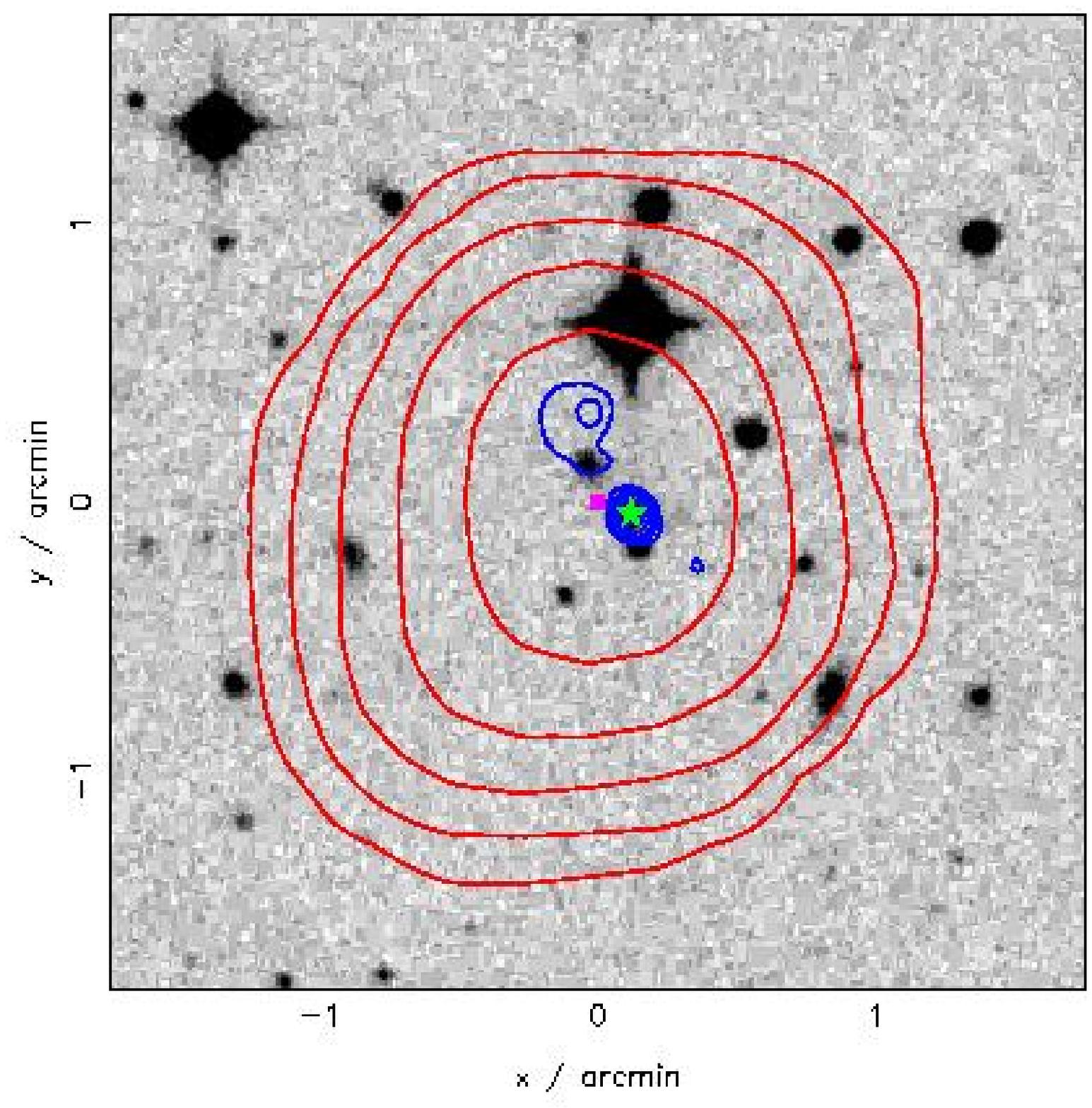}}
      \centerline{C4-184: 1438+0068}
    \end{minipage}
    \vfill
    \begin{minipage}{3cm}      
      \mbox{}
      \centerline{\includegraphics[scale=0.26,angle=270]{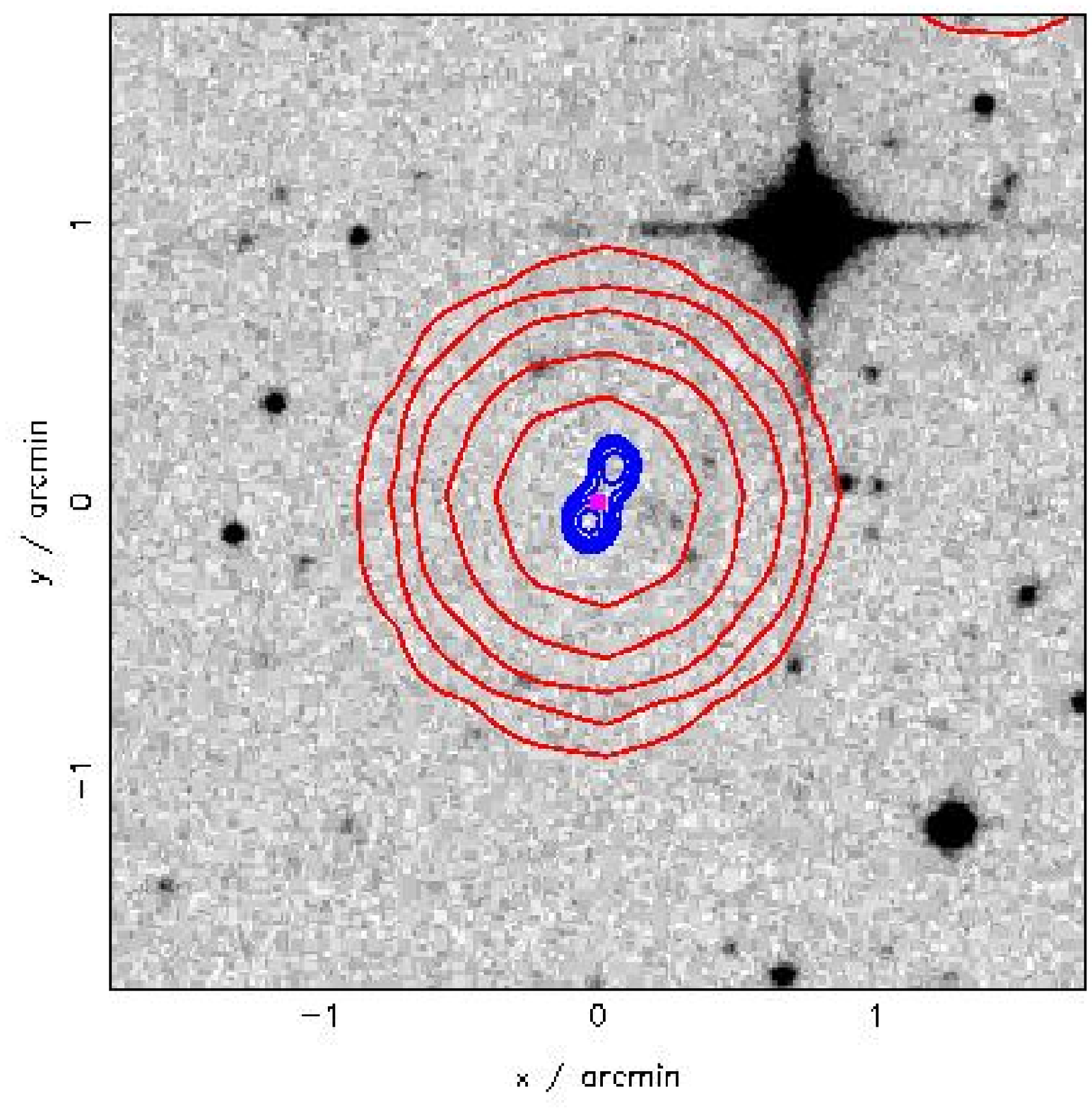}}
      \centerline{C4-185: 1438-0085}
    \end{minipage}
    \hspace{3cm}
    \begin{minipage}{3cm}
      \mbox{}
      \centerline{\includegraphics[scale=0.26,angle=270]{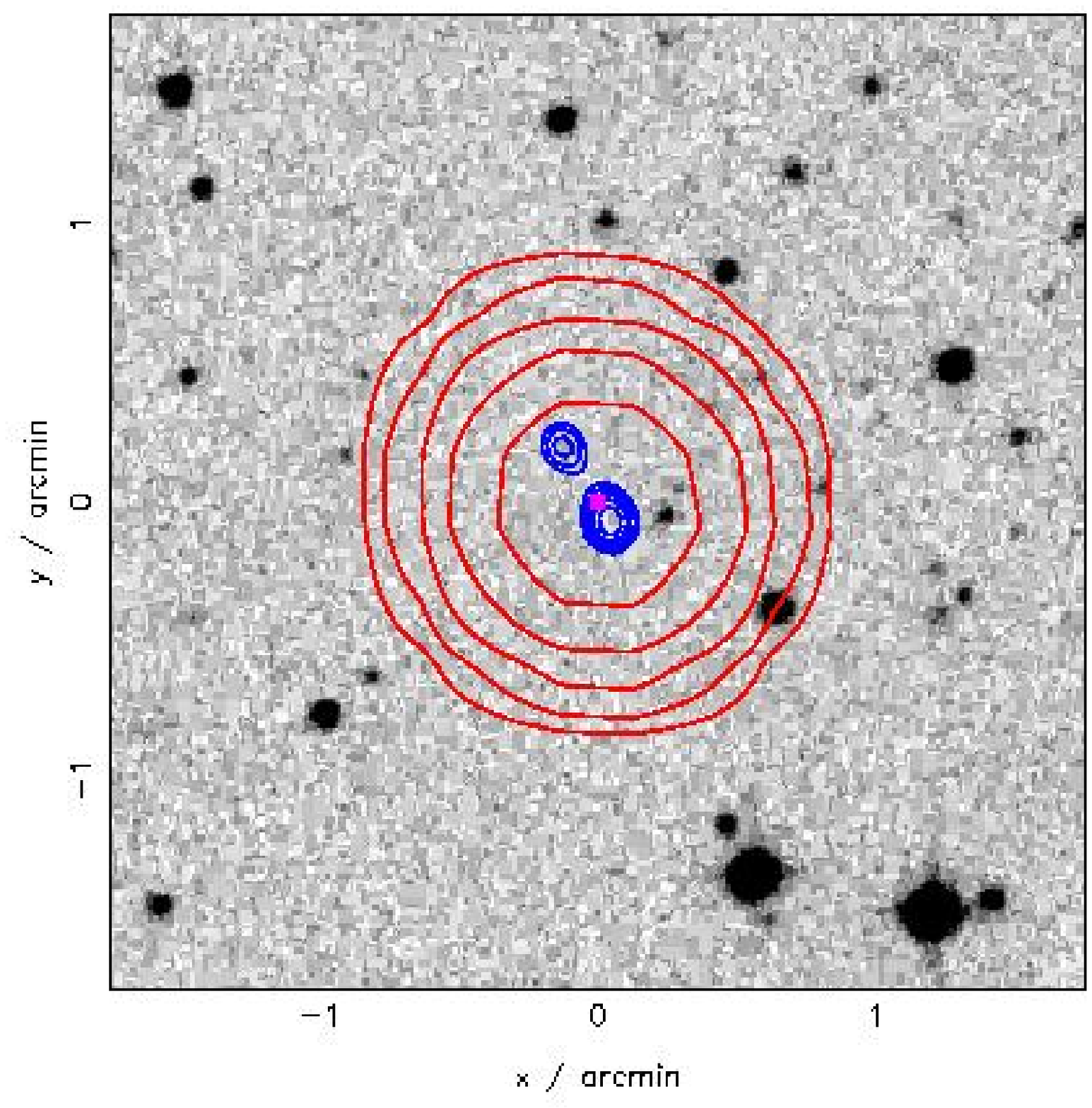}}
      \centerline{C4-187: TXS 1437+009}
    \end{minipage}
    \hspace{3cm}
    \begin{minipage}{3cm}
      \mbox{}
      \centerline{\includegraphics[scale=0.26,angle=270]{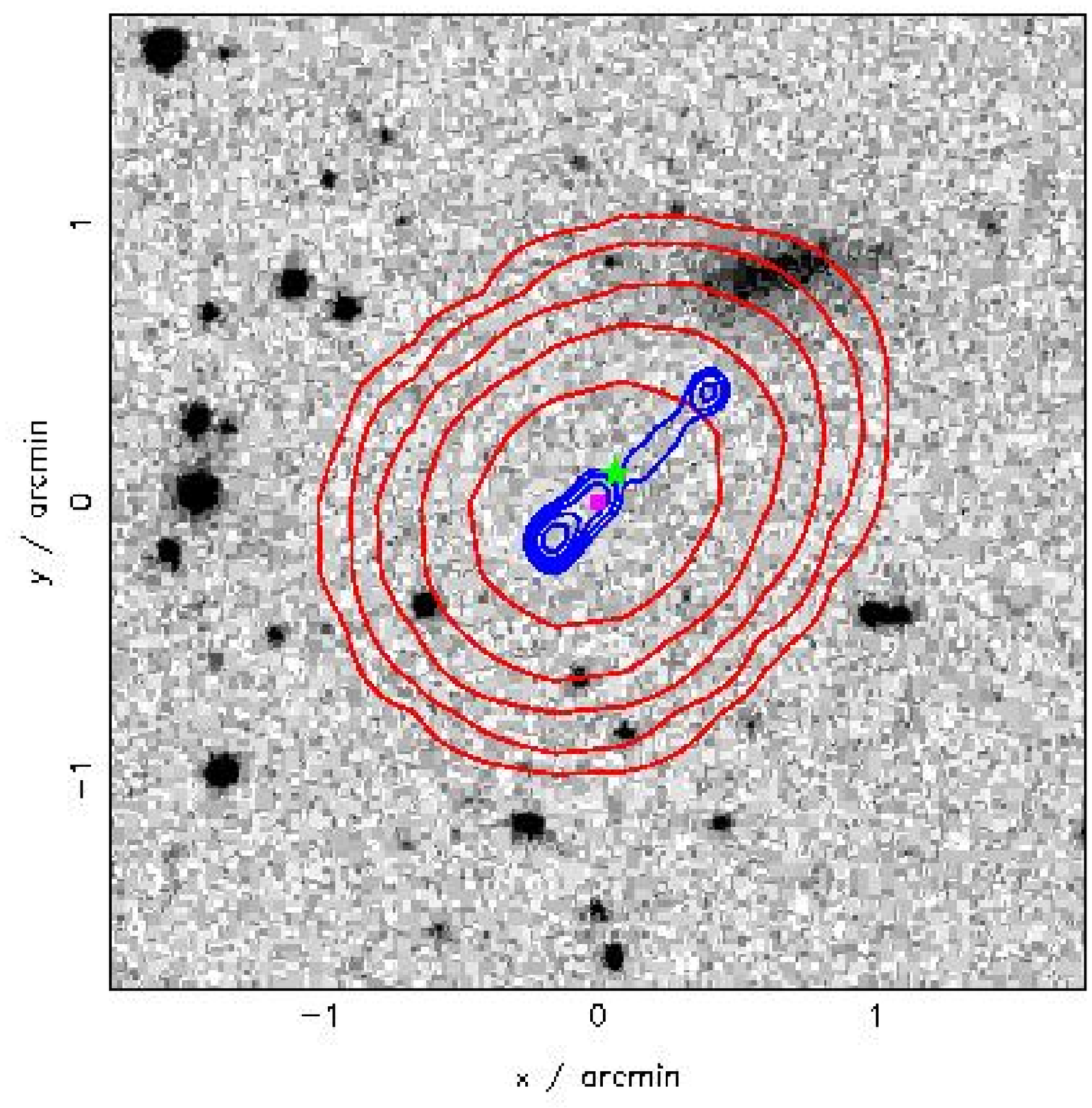}}
      \centerline{C4-188: TXS 1437-001}
    \end{minipage}
  \end{center}
\end{figure}

\begin{figure}
  \begin{center}
  {\bf CENSORS}\\
    \begin{minipage}{3cm}      
      \mbox{}
      \centerline{\includegraphics[scale=0.28,angle=270]{Contours/CENSORS/004.ps}}
      \centerline{CENSORS-004}
    \end{minipage}
    \hspace{2.5cm}
    \begin{minipage}{3cm}
      \mbox{}
      \centerline{\includegraphics[scale=0.28,angle=270]{Contours/CENSORS/005.ps}}
      \centerline{CENSORS-005}
    \end{minipage}
    \hspace{2.5cm}
    \begin{minipage}{3cm}
      \mbox{}
      \centerline{\includegraphics[scale=0.28,angle=270]{Contours/CENSORS/007.ps}}
      \centerline{CENSORS-007}
    \end{minipage}
    \vfill
    \begin{minipage}{3cm}     
      \mbox{}
      \centerline{\includegraphics[scale=0.28,angle=270]{Contours/CENSORS/014.ps}}
      \centerline{CENSORS-014}
    \end{minipage}
    \hspace{2.5cm}
    \begin{minipage}{3cm}
      \mbox{}
      \centerline{\includegraphics[scale=0.28,angle=270]{Contours/CENSORS/015.ps}}
      \centerline{CENSORS-015}
    \end{minipage}
    \hspace{2.5cm}
    \begin{minipage}{3cm}
      \mbox{}
      \centerline{\includegraphics[scale=0.28,angle=270]{Contours/CENSORS/016.ps}}
      \centerline{CENSORS-016}
    \end{minipage}
    \vfill
    \begin{minipage}{3cm}     
      \mbox{}
      \centerline{\includegraphics[scale=0.28,angle=270]{Contours/CENSORS/017.ps}}
      \centerline{CENSORS-017}
    \end{minipage}
    \hspace{2.5cm}
    \begin{minipage}{3cm}
      \mbox{}
      \centerline{\includegraphics[scale=0.28,angle=270]{Contours/CENSORS/020.ps}}
      \centerline{CENSORS-020}
    \end{minipage}
    \hspace{2.5cm}
    \begin{minipage}{3cm}
      \mbox{}
      \centerline{\includegraphics[scale=0.28,angle=270]{Contours/CENSORS/035.ps}}
      \centerline{CENSORS-035}
    \end{minipage}
    \vfill
    \begin{minipage}{3cm}      
      \mbox{}
      \centerline{\includegraphics[scale=0.28,angle=270]{Contours/CENSORS/038.ps}}
      \centerline{CENSORS-038}
    \end{minipage}
    \hspace{2.5cm}
    \begin{minipage}{3cm}
      \mbox{}
      \centerline{\includegraphics[scale=0.28,angle=270]{Contours/CENSORS/039.ps}}
      \centerline{CENSORS-039}
    \end{minipage}
    \hspace{2.5cm}
    \begin{minipage}{3cm}
      \mbox{}
      \centerline{\includegraphics[scale=0.28,angle=270]{Contours/CENSORS/040.ps}}
      \centerline{CENSORS-040}
    \end{minipage}
  \end{center}
\end{figure}

\begin{figure}
  \begin{center}
    {\bf CENSORS}\\  
  \begin{minipage}{3cm}      
      \mbox{}
      \centerline{\includegraphics[scale=0.28,angle=270]{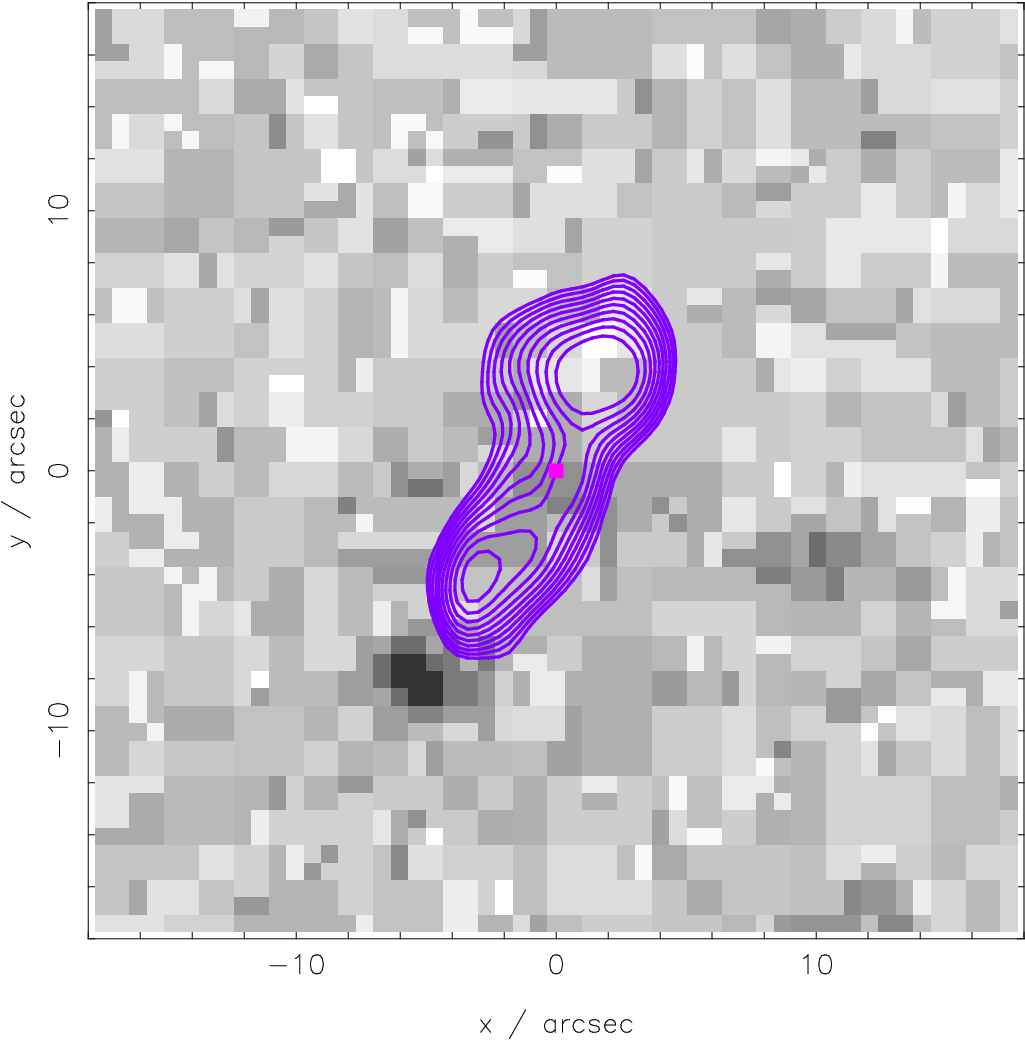}}
      \centerline{CENSORS-043}
    \end{minipage}
    \hspace{2.5cm}
    \begin{minipage}{3cm}
      \mbox{}
      \centerline{\includegraphics[scale=0.28,angle=270]{Contours/CENSORS/045.ps}}
      \centerline{CENSORS-045}
    \end{minipage}
    \hspace{2.5cm}
    \begin{minipage}{3cm}
      \mbox{}
      \centerline{\includegraphics[scale=0.28,angle=270]{Contours/CENSORS/047.ps}}
      \centerline{CENSORS-047}
    \end{minipage}
    \vfill
    \begin{minipage}{3cm}     
      \mbox{}
      \centerline{\includegraphics[scale=0.28,angle=270]{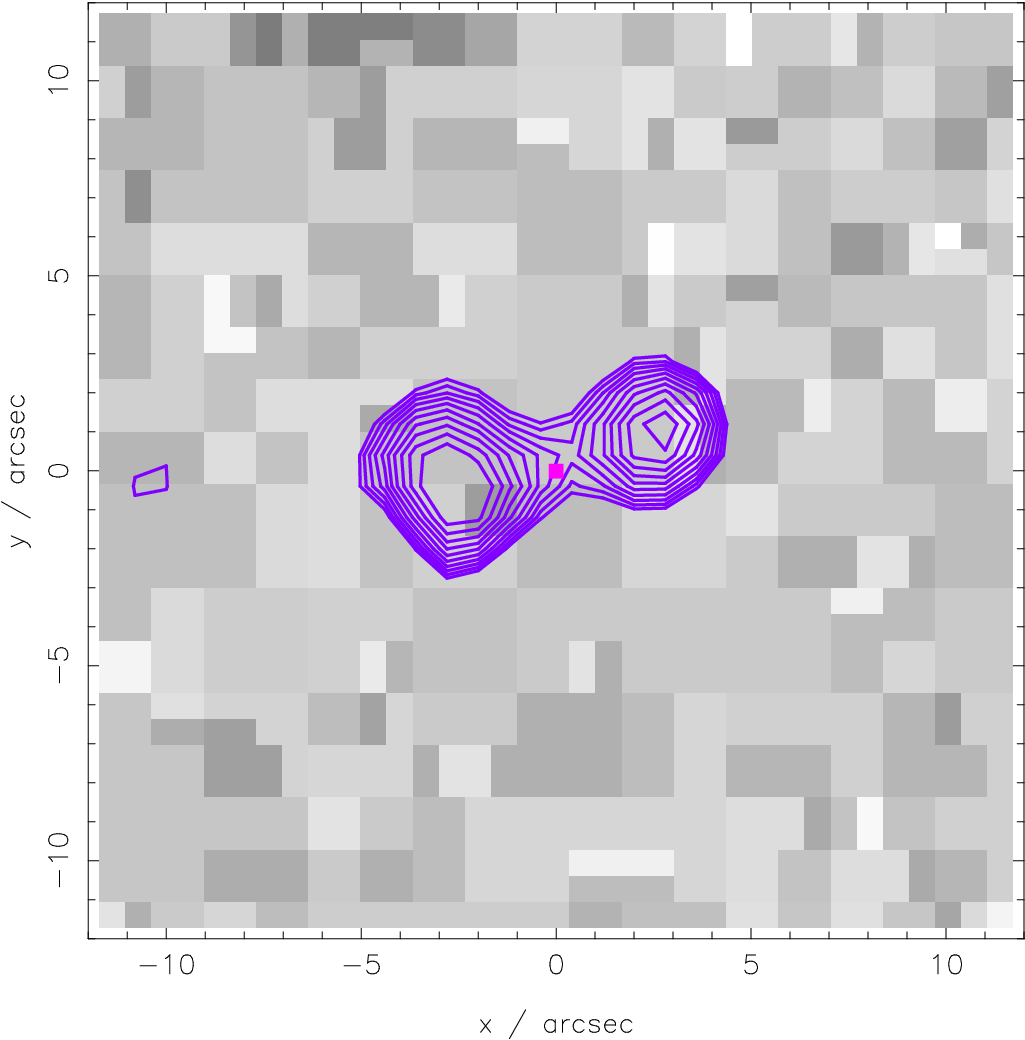}}
      \centerline{CENSORS-050}
    \end{minipage}
    \hspace{2.5cm}
    \begin{minipage}{3cm}
      \mbox{}
      \centerline{\includegraphics[scale=0.28,angle=270]{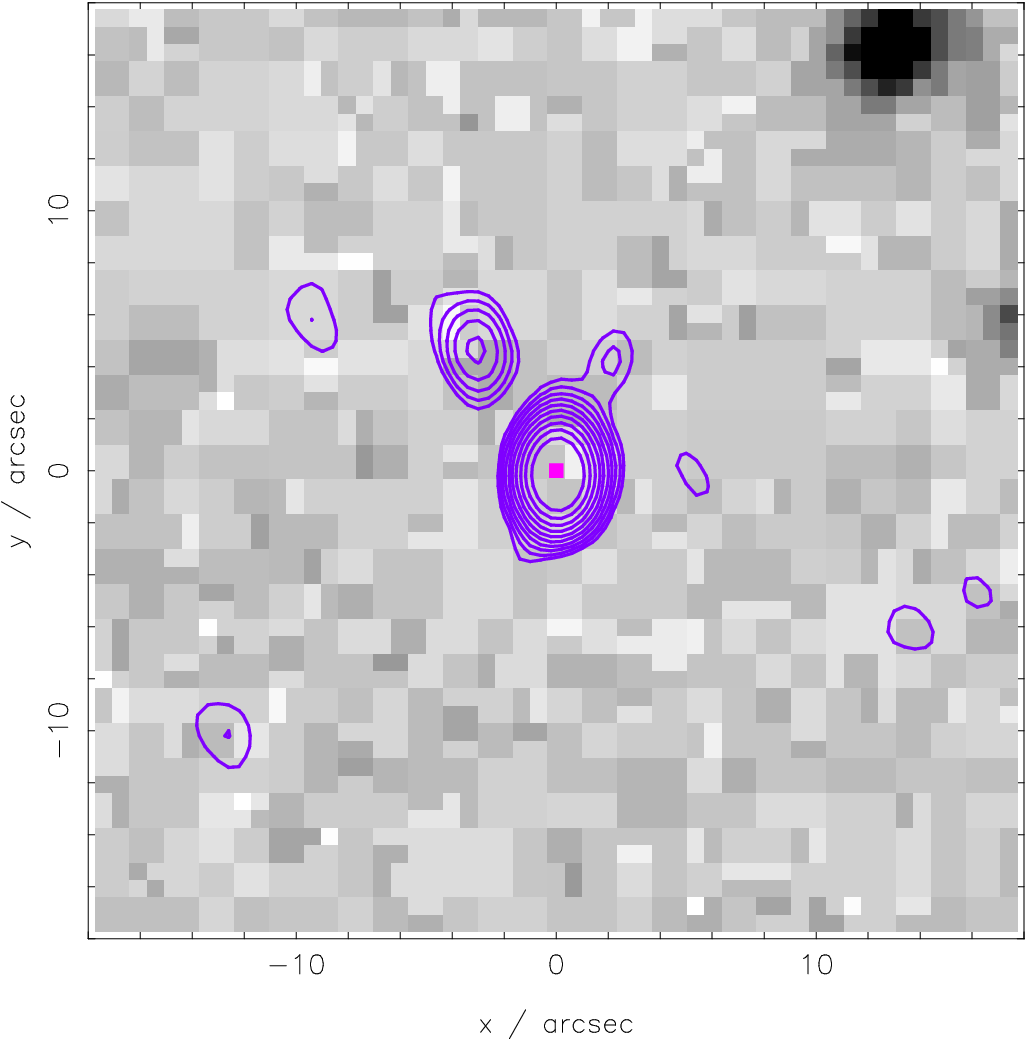}}
      \centerline{CENSORS-051}
    \end{minipage}
    \hspace{2.5cm}
    \begin{minipage}{3cm}
      \mbox{}
      \centerline{\includegraphics[scale=0.28,angle=270]{Contours/CENSORS/053.ps}}
      \centerline{CENSORS-053}
    \end{minipage}
    \vfill
    \begin{minipage}{3cm}     
      \mbox{}
      \centerline{\includegraphics[scale=0.28,angle=270]{Contours/CENSORS/063.ps}}
      \centerline{CENSORS-063}
    \end{minipage}
    \hspace{2.5cm}
    \begin{minipage}{3cm}
      \mbox{}
      \centerline{\includegraphics[scale=0.28,angle=270]{Contours/CENSORS/064.ps}}
      \centerline{CENSORS-064}
    \end{minipage}
    \hspace{2.5cm}
    \begin{minipage}{3cm}
      \mbox{}
      \centerline{\includegraphics[scale=0.28,angle=270]{Contours/CENSORS/065.ps}}
      \centerline{CENSORS-065}
    \end{minipage}
    \vfill
    \begin{minipage}{6cm}      
      \mbox{}
      \centerline{\includegraphics[scale=0.28,angle=270]{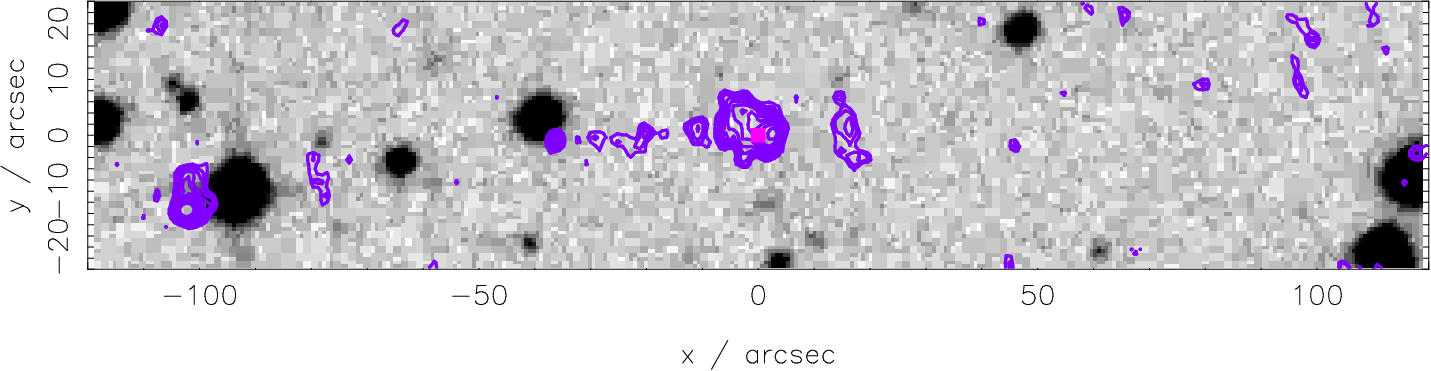}}
      \centerline{CENSORS-066}
    \end{minipage}
    \hspace{2.5cm}
    \begin{minipage}{3cm}
      \mbox{}
      \centerline{\includegraphics[scale=0.28,angle=270]{Contours/CENSORS/071.ps}}
      \centerline{CENSORS-071}
    \end{minipage}
  \end{center}
\end{figure}

\begin{figure}
  \begin{center}
    {\bf CENSORS}\\
    \begin{minipage}{3cm}
      \mbox{}
      \centerline{\includegraphics[scale=0.28,angle=270]{Contours/CENSORS/075.ps}}
      \centerline{CENSORS-075}
    \end{minipage}
    \hspace{2.5cm}
    \begin{minipage}{3cm}      
      \mbox{}
      \centerline{\includegraphics[scale=0.28,angle=270]{Contours/CENSORS/076.ps}}
      \centerline{CENSORS-076}
    \end{minipage}
    \hspace{2.5cm}
    \begin{minipage}{3cm}
      \mbox{}
      \centerline{\includegraphics[scale=0.28,angle=270]{Contours/CENSORS/078.ps}}
      \centerline{CENSORS-078}
    \end{minipage}
    \vfill
    \begin{minipage}{3cm}
      \mbox{}
      \centerline{\includegraphics[scale=0.28,angle=270]{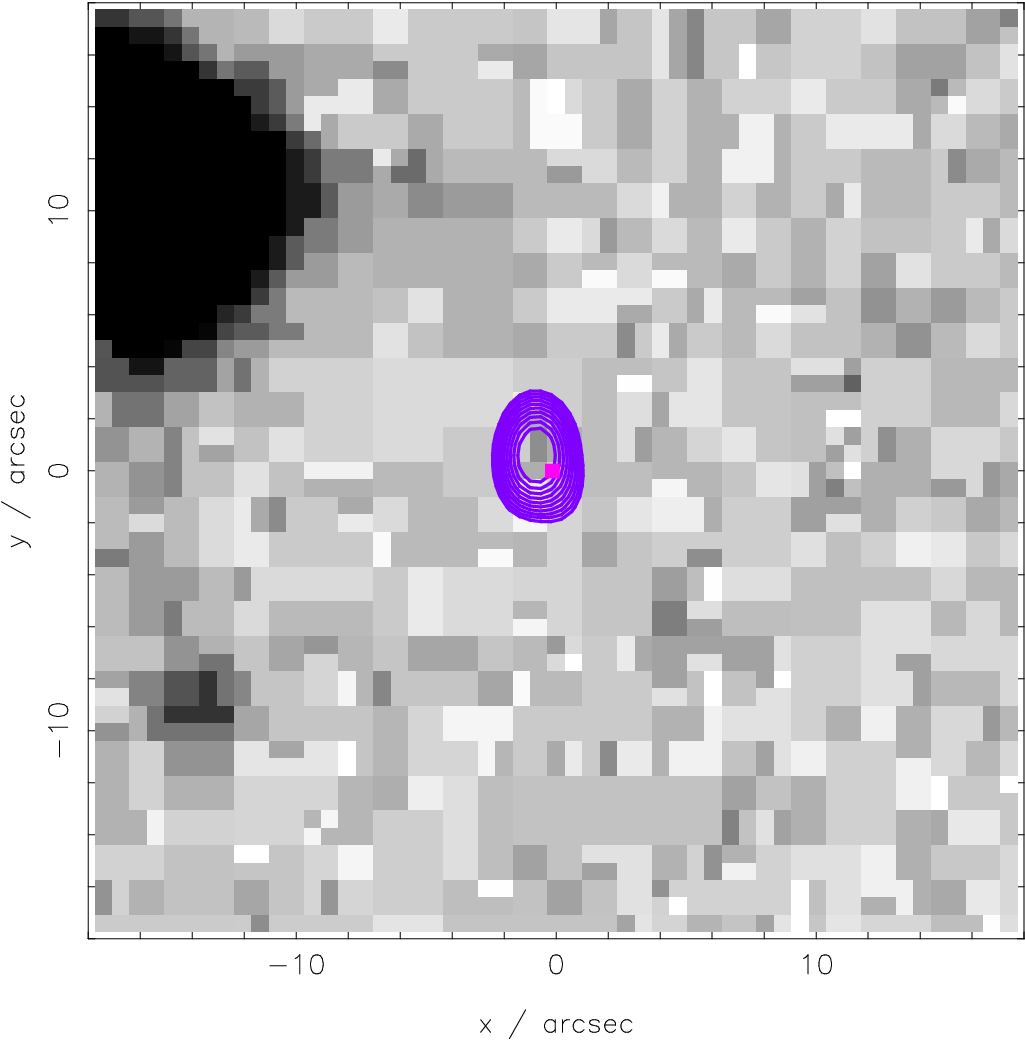}}
      \centerline{CENSORS-079}
    \end{minipage}
    \hspace{2.5cm}
    \begin{minipage}{3cm}     
      \mbox{}
      \centerline{\includegraphics[scale=0.28,angle=270]{Contours/CENSORS/087.ps}}
      \centerline{CENSORS-087}
    \end{minipage}
    \hspace{2.5cm}
    \begin{minipage}{3cm}
      \mbox{}
      \centerline{\includegraphics[scale=0.28,angle=270]{Contours/CENSORS/094.ps}}
      \centerline{CENSORS-094}
    \end{minipage}
    \vfill
    \begin{minipage}{3cm}
      \mbox{}
      \centerline{\includegraphics[scale=0.28,angle=270]{Contours/CENSORS/099.ps}}
      \centerline{CENSORS-099}
    \end{minipage}
    \hspace{2.5cm}
    \begin{minipage}{3cm}     
      \mbox{}
      \centerline{\includegraphics[scale=0.28,angle=270]{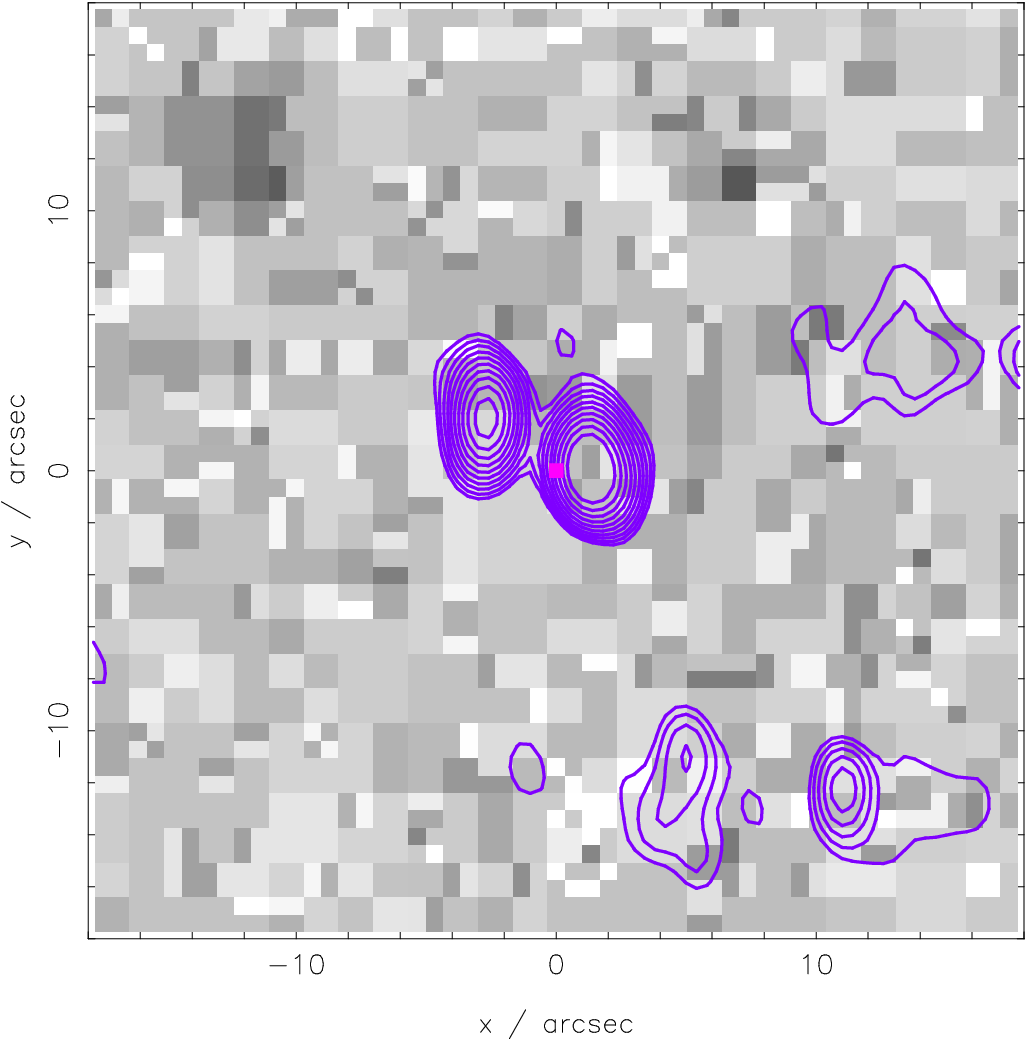}}
      \centerline{CENSORS-100}
    \end{minipage}
    \hspace{2.5cm}
    \begin{minipage}{3cm}
      \mbox{}
      \centerline{\includegraphics[scale=0.28,angle=270]{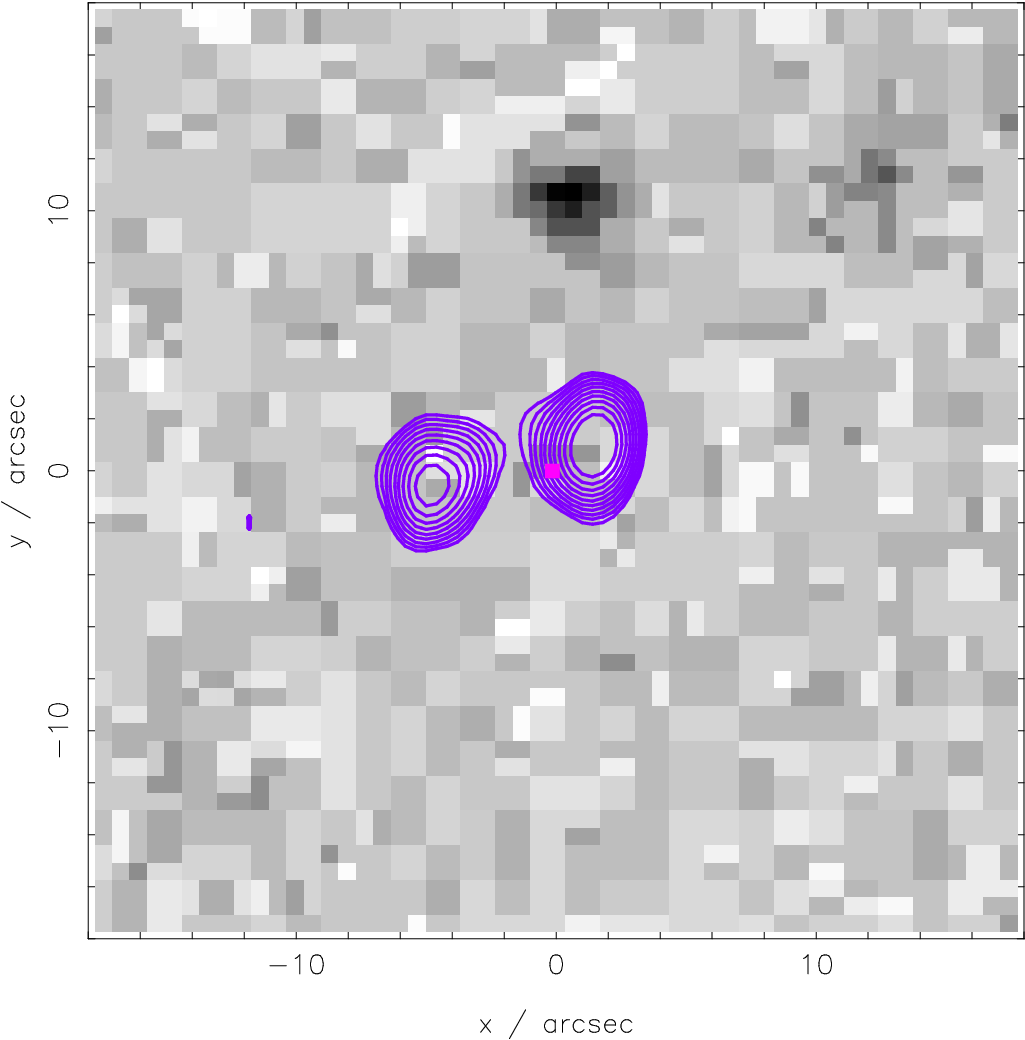}}
      \centerline{CENSORS-105}
    \end{minipage}
    \vfill
    \begin{minipage}{3cm}
      \mbox{}
      \centerline{\includegraphics[scale=0.28,angle=270]{Contours/CENSORS/106.ps}}
      \centerline{CENSORS-106}
    \end{minipage}
    \hspace{2.5cm}
    \begin{minipage}{3cm}      
      \mbox{}
      \centerline{\includegraphics[scale=0.28,angle=270]{Contours/CENSORS/107.ps}}
      \centerline{CENSORS-107}
    \end{minipage}
    \hspace{2.5cm}
    \begin{minipage}{3cm}
      \mbox{}
      \centerline{\includegraphics[scale=0.28,angle=270]{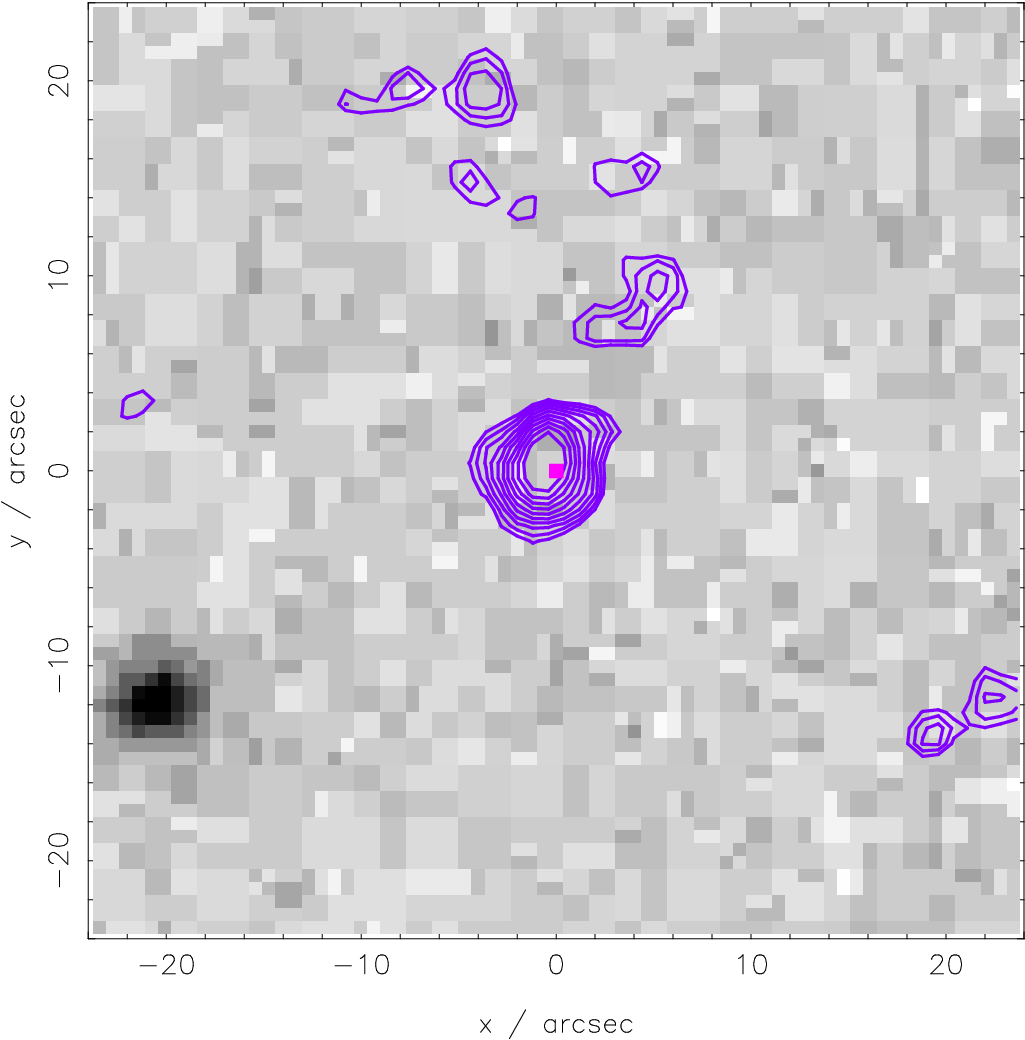}}
      \centerline{CENSORS-109}
    \end{minipage}
  \end{center}
\end{figure}

\begin{figure}
  \begin{center}
    {\bf CENSORS}\\
    \begin{minipage}{3cm}
      \mbox{}
      \centerline{\includegraphics[scale=0.28,angle=270]{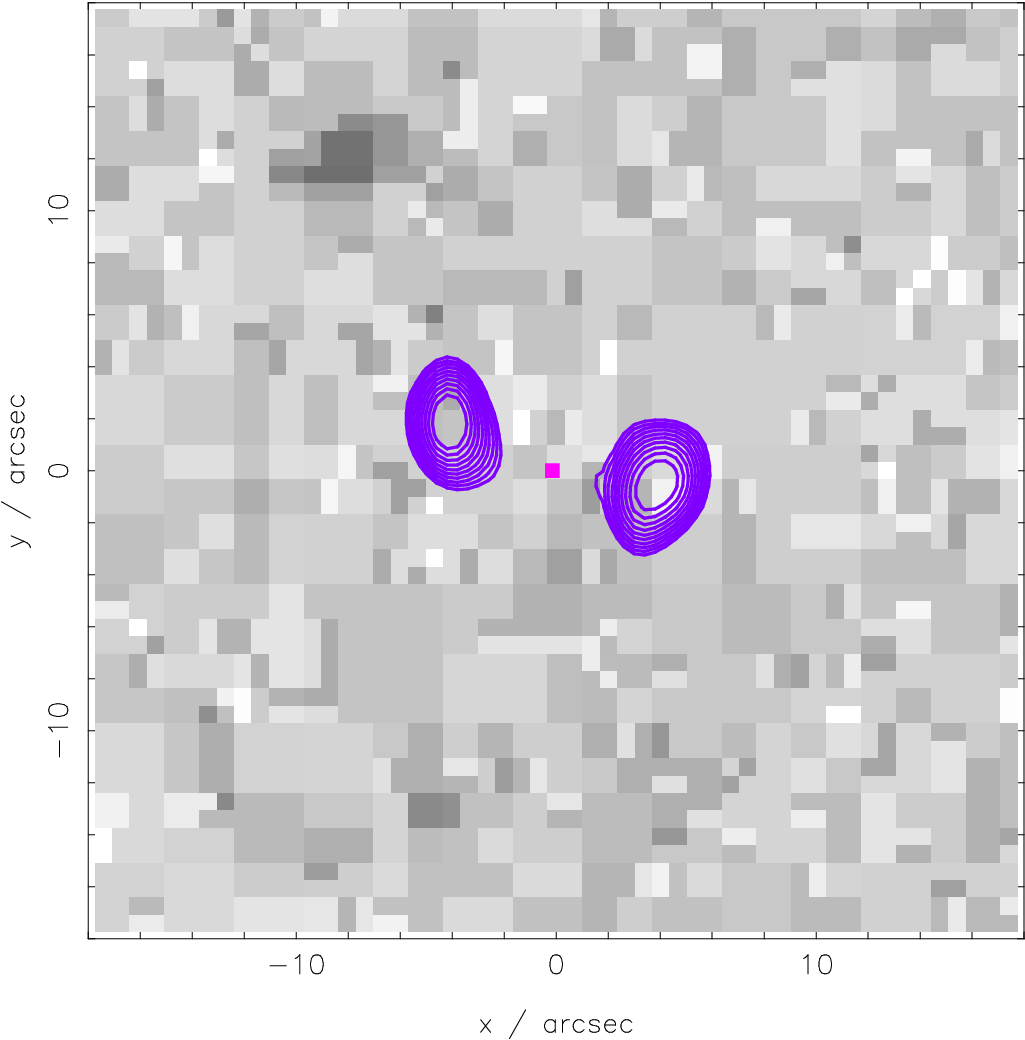}}
      \centerline{CENSORS-117}
    \end{minipage}
    \hspace{2.5cm} 
    \begin{minipage}{3cm}
      \mbox{}
      \centerline{\includegraphics[scale=0.28,angle=270]{Contours/CENSORS/118.ps}}
      \centerline{CENSORS-118}
    \end{minipage}
    \hspace{2.5cm}
    \begin{minipage}{3cm}
      \mbox{}
      \centerline{\includegraphics[scale=0.28,angle=270]{Contours/CENSORS/119.ps}}
      \centerline{CENSORS-119}
    \end{minipage}
    \vfill
    \begin{minipage}{3cm}
      \mbox{}
      \centerline{\includegraphics[scale=0.28,angle=270]{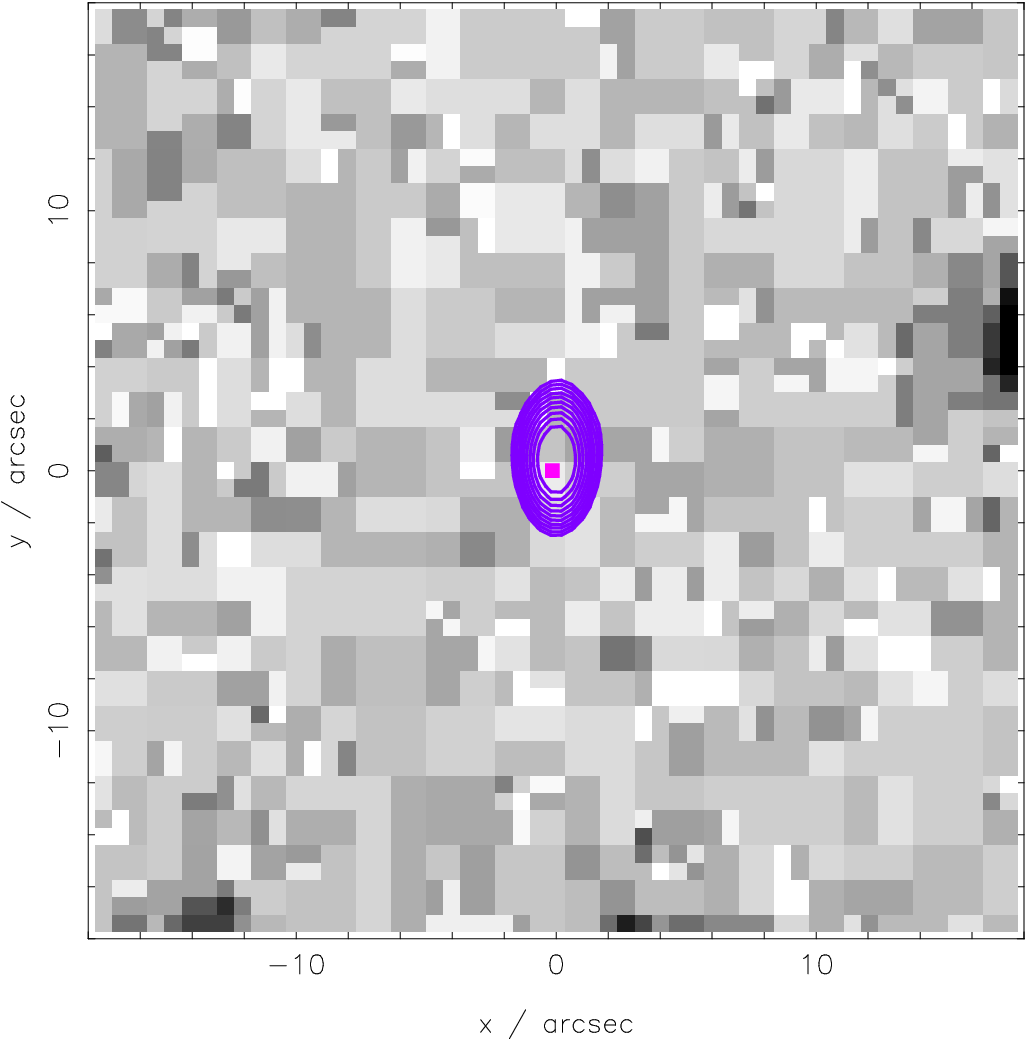}}
      \centerline{CENSORS-132}
    \end{minipage}
    \hspace{2.5cm}
    \begin{minipage}{3cm}     
      \mbox{}
      \centerline{\includegraphics[scale=0.28,angle=270]{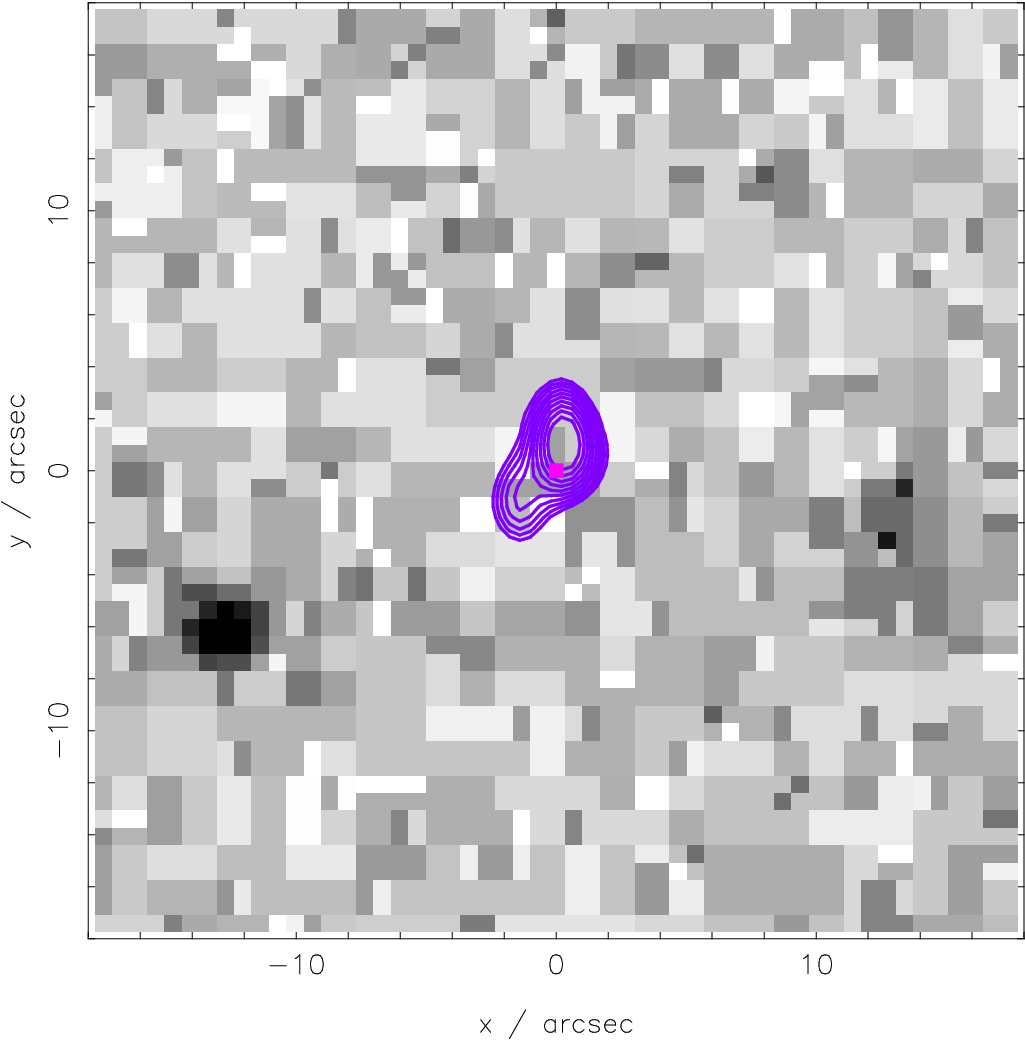}}
      \centerline{CENSORS-136}
    \end{minipage}
  \end{center}
\end{figure}

\label{lastpage}

\end{document}